%% file: main_elsevier.tex
\documentclass[a4paper,fleqn]{cas-sc}

\usepackage[authoryear,longnamesfirst]{natbib}

\usepackage{newunicodechar}
\newunicodechar{ʹ}{'}

\usepackage{eso-pic}
\usepackage{amsmath}
\usepackage{amsfonts}
\usepackage{amssymb}
\usepackage{amsthm}
\usepackage{mathtools}
\usepackage{bbold}

\numberwithin{equation}{section}

\usepackage{savesym}
\savesymbol{div}
\usepackage{physics}
\restoresymbol{TxF}{div}
\usepackage{shuffle}
\usepackage{cancel}

\graphicspath{{./}}
\usepackage{float}
\usepackage[maxfloats=256]{morefloats}
\usepackage{subcaption}

\newcommand{\figscale}{0.70}

\usepackage{tikz}
\usepackage{tikz-3dplot}
\usetikzlibrary{backgrounds,shapes,positioning,calc,intersections,arrows.meta,decorations.markings}

\usepackage[dvipsnames]{xcolor}
\definecolor{mybrown}{RGB}{171,6,44}
\definecolor{mygreen}{RGB}{0,128,0}
\definecolor{pinegreen}{RGB}{62,134,103}
\usepackage{enumitem}
\usepackage{comment}
\usepackage{blkarray}
\usepackage[most]{tcolorbox}
\usepackage{imakeidx}

\DeclareMathOperator{\Li}{Li}

\DeclareMathOperator{\Gr}{Gr}

\newcommand{\longhookrightarrow}{\lhook\joinrel\longrightarrow}
\newcommand{\smallbinom}[2]{\genfrac{(}{)}{0pt}{0}{#1}{#2}}
\newcommand{\conv}{{\rm conv}}
\newcommand{\la}{\langle}
\newcommand{\ra}{\rangle}

\newcommand*{\LABELREVERSE}{{non-simple intersecting}}
\newcommand*{\LABELNONSIMPLE}{{non-simple non-intersecting}}
\newcommand{\Qbar}{\mkern 1.5mu\overline{\mkern-1.5mu Q\mkern-1.5mu}\mkern 1.5mu}

\theoremstyle{remark}
\newtheorem*{remark}{Remark}
\theoremstyle{plain}
\newtheorem{theorem}{Theorem}[section]

\makeatletter
\newtheorem*{rep@theorem}{\rep@title}
\newcommand{\newreptheorem}[2]{%
  \newenvironment{rep#1}[1]{%
    \def\rep@title{#2~\ref{##1}}%
    \begin{rep@theorem}%
  }{%
    \end{rep@theorem}%
  }%
}
\makeatother
\newreptheorem{theorem}{Theorem}
\newreptheorem{cor}{Corollary}
\newreptheorem{proposition}{Proposition}
\theoremstyle{definition}

\newtheorem{eg}[theorem]{Example}

\definecolor{boxgray}{RGB}{235,235,235}

\tcbset{
  importantbox/.style={
    colback=boxgray,
    colframe=black!30
  },
  definitionbox/.style={importantbox},
  resultbox/.style={importantbox}
}

\begin{document}

\AddToShipoutPictureFG*{%
  \AtPageUpperLeft{%
    \put(\LenToUnit{\dimexpr\paperwidth-1.0cm\relax},
         \LenToUnit{-0.8cm}){%
      \makebox[0pt][r]{\footnotesize MPP-2026-137}%
    }%
  }%
}

\makeatletter
\expandafter\gdef\csname r@ch:Introduction\endcsname{{1}{3}}
\expandafter\gdef\csname r@sec:why-N4-SYM\endcsname{{1.2}{3}}
\expandafter\gdef\csname r@sec:The Positive Geometry Program\endcsname{{1.3}{5}}
\expandafter\gdef\csname r@sec:Implications of Positive Geometries\endcsname{{1.4}{6}}
\expandafter\gdef\csname r@sec:Main Contributions\endcsname{{1.5}{8}}
\expandafter\gdef\csname r@fig:curvy_4gon\endcsname{{1}{9}}
\expandafter\gdef\csname r@fig:sing_geom\endcsname{{2}{10}}
\expandafter\gdef\csname r@sec:Outline\endcsname{{1.6}{11}}
\expandafter\gdef\csname r@ch:Scattering Amplitudes\endcsname{{2}{12}}
\expandafter\gdef\csname r@sec:What is an Amplitude?\endcsname{{2.1}{12}}
\expandafter\gdef\csname r@eq:Ainout\endcsname{{2}{13}}
\expandafter\gdef\csname r@eq:feynm_int\endcsname{{5}{13}}
\expandafter\gdef\csname r@eq:mom_cons\endcsname{{12}{15}}
\expandafter\gdef\csname r@eq:helicity_scaling\endcsname{{14}{15}}
\expandafter\gdef\csname r@eq:parity\endcsname{{15}{15}}
\expandafter\gdef\csname r@eg:3pt_ampl\endcsname{{2.1}{16}}
\expandafter\gdef\csname r@eq:A3_general\endcsname{{18}{16}}
\expandafter\gdef\csname r@eq:MHV_3pt\endcsname{{19}{16}}
\expandafter\gdef\csname r@eq:MHVbar_3pt\endcsname{{20}{16}}
\expandafter\gdef\csname r@sec:Recursion Relations\endcsname{{2.2}{16}}
\expandafter\gdef\csname r@eq:An_residue\endcsname{{22}{17}}
\expandafter\gdef\csname r@eq:An_fact\endcsname{{23}{17}}
\expandafter\gdef\csname r@eq:onshell_rec\endcsname{{26}{18}}
\expandafter\gdef\csname r@eq:ij_shift\endcsname{{27}{18}}
\expandafter\gdef\csname r@eq:4pt_ampl\endcsname{{2.2}{19}}
\expandafter\gdef\csname r@eq:hel_4pt_conf\endcsname{{34}{19}}
\expandafter\gdef\csname r@eq:4pt_rec\endcsname{{36}{19}}
\expandafter\gdef\csname r@eq:4pt_rec_1\endcsname{{38}{19}}
\expandafter\gdef\csname r@eq:4pt_square\endcsname{{40}{20}}
\expandafter\gdef\csname r@eq:4pt_angle\endcsname{{41}{20}}
\expandafter\gdef\csname r@eq:eq:PT\endcsname{{42}{20}}
\expandafter\gdef\csname r@eq:An_recursion\endcsname{{43}{20}}
\expandafter\gdef\csname r@fig:single-cut-forward-limit\endcsname{{3}{21}}
\expandafter\gdef\csname r@eq:forward\endcsname{{45}{21}}
\expandafter\gdef\csname r@eq:box_int_1\endcsname{{46}{21}}
\expandafter\gdef\csname r@sec:Maximally Supersymmetric Yang-Mills Theory\endcsname{{2.3}{23}}
\expandafter\gdef\csname r@sec:Maximally supersymmetric Yang-Mills theory\endcsname{{2.3}{23}}
\expandafter\gdef\csname r@fig:helicity\endcsname{{4}{24}}
\expandafter\gdef\csname r@eq:superampl_dec\endcsname{{52}{24}}
\expandafter\gdef\csname r@sec:Momentum Twistors\endcsname{{2.4}{25}}
\expandafter\gdef\csname r@sec:Momentum twistors\endcsname{{2.4}{25}}
\expandafter\gdef\csname r@eq:x_i\endcsname{{54}{25}}
\expandafter\gdef\csname r@fig:mom_tw\endcsname{{5}{26}}
\expandafter\gdef\csname r@eq:iijj\endcsname{{59}{26}}
\expandafter\gdef\csname r@eq:box_diagr\endcsname{{6}{27}}
\expandafter\gdef\csname r@eq:Ibox\endcsname{{60}{27}}
\expandafter\gdef\csname r@eq:int_mom_tw\endcsname{{61}{27}}
\expandafter\gdef\csname r@eq:inf_tw\endcsname{{63}{27}}
\expandafter\gdef\csname r@eq:Ward_SUSY\endcsname{{71}{28}}
\expandafter\gdef\csname r@eq:Atilde\endcsname{{72}{29}}
\expandafter\gdef\csname r@eq:ij_supershift\endcsname{{73}{29}}
\expandafter\gdef\csname r@eq:super_shift\endcsname{{74}{29}}
\expandafter\gdef\csname r@eq:3pt_super_ampl\endcsname{{75}{29}}
\expandafter\gdef\csname r@eq:MHV\endcsname{{77}{29}}
\expandafter\gdef\csname r@eq:ampl_R_dec\endcsname{{78}{29}}
\expandafter\gdef\csname r@fig:graph_0\endcsname{{7}{30}}
\expandafter\gdef\csname r@eq:extract_R\endcsname{{80}{30}}
\expandafter\gdef\csname r@sec:On-Shell Diagrams and Grassmannian Contours\endcsname{{2.5}{30}}
\expandafter\gdef\csname r@eq:mom_cons_lambda\endcsname{{81}{31}}
\expandafter\gdef\csname r@eq:tilde_lambda_perp\endcsname{{84}{31}}
\expandafter\gdef\csname r@eq:MHVbar_Gr\endcsname{{86}{31}}
\expandafter\gdef\csname r@eq:MHV_Gr\endcsname{{87}{31}}
\expandafter\gdef\csname r@eq:edge_integration\endcsname{{88}{31}}
\expandafter\gdef\csname r@fig:kin_planes\endcsname{{8}{32}}
\expandafter\gdef\csname r@eq:Gr_int\endcsname{{91}{32}}
\expandafter\gdef\csname r@eq:Gr_contour\endcsname{{93}{32}}
\expandafter\gdef\csname r@fig:onshell-equivalence-moves\endcsname{{9}{33}}
\expandafter\gdef\csname r@eq:Gr_contour_momentum_twistor\endcsname{{94}{33}}
\expandafter\gdef\csname r@fig:graph_1\endcsname{{10}{34}}
\expandafter\gdef\csname r@eq:C_alpha\endcsname{{95}{34}}
\expandafter\gdef\csname r@eq:Gr_int_alpha\endcsname{{96}{34}}
\expandafter\gdef\csname r@eq:onshell_k2_n4\endcsname{{2.3}{34}}
\expandafter\gdef\csname r@eq:C_par\endcsname{{97}{34}}
\expandafter\gdef\csname r@eq:Gr24_form\endcsname{{98}{35}}
\expandafter\gdef\csname r@eq:onshell_form\endcsname{{101}{35}}
\expandafter\gdef\csname r@eq:nonneg_24\endcsname{{103}{35}}
\expandafter\gdef\csname r@ch:Positive Geometries\endcsname{{3}{36}}
\expandafter\gdef\csname r@sec:Canonical Forms\endcsname{{3.1}{36}}
\expandafter\gdef\csname r@eq:S\endcsname{{110}{37}}
\expandafter\gdef\csname r@eq:bP_dec\endcsname{{115}{37}}
\expandafter\gdef\csname r@eq:PG_triple\endcsname{{116}{38}}
\expandafter\gdef\csname r@eq:res_def\endcsname{{117}{38}}
\expandafter\gdef\csname r@eg:interval\endcsname{{3.1}{38}}
\expandafter\gdef\csname r@eq:omega_ab\endcsname{{119}{38}}
\expandafter\gdef\csname r@fig:half-pizza\endcsname{{11}{39}}
\expandafter\gdef\csname r@eg:half_pizza\endcsname{{3.2}{39}}
\expandafter\gdef\csname r@eq:pizza_form\endcsname{{124}{39}}
\expandafter\gdef\csname r@eq:iota2\endcsname{{131}{40}}
\expandafter\gdef\csname r@eg:simplexes\endcsname{{3.3}{40}}
\expandafter\gdef\csname r@eq:form_simplex_chart\endcsname{{135}{40}}
\expandafter\gdef\csname r@eq:proj_meas\endcsname{{137}{40}}
\expandafter\gdef\csname r@eq:form_simplex\endcsname{{138}{41}}
\expandafter\gdef\csname r@eq:can_form_simplex\endcsname{{139}{41}}
\expandafter\gdef\csname r@sec:Triangulations\endcsname{{3.2}{41}}
\expandafter\gdef\csname r@eq:disk\endcsname{{140}{41}}
\expandafter\gdef\csname r@eq:cancel\endcsname{{141}{41}}
\expandafter\gdef\csname r@eq:tr_pizzas\endcsname{{142}{41}}
\expandafter\gdef\csname r@fig:pent_tr\endcsname{{12}{42}}
\expandafter\gdef\csname r@prop:can_form_tr\endcsname{{3.2.2}{42}}
\expandafter\gdef\csname r@eq:can_form_tr\endcsname{{145}{42}}
\expandafter\gdef\csname r@eq:pent_tr\endcsname{{3.4}{42}}
\expandafter\gdef\csname r@eq:can_form_pent\endcsname{{149}{43}}
\expandafter\gdef\csname r@prop:pol_tr\endcsname{{3.2.3}{43}}
\expandafter\gdef\csname r@eg:pizza_slice\endcsname{{3.5}{44}}
\expandafter\gdef\csname r@eq:pizza_slice\endcsname{{152}{44}}
\expandafter\gdef\csname r@eq:six_cones\endcsname{{153}{44}}
\expandafter\gdef\csname r@eq:ext_tr_pizza\endcsname{{155}{44}}
\expandafter\gdef\csname r@eq:pizza_slice_form\endcsname{{156}{44}}
\expandafter\gdef\csname r@sec:Adjoint Hypersurface\endcsname{{3.3}{44}}
\expandafter\gdef\csname r@eq:form_pq\endcsname{{157}{44}}
\expandafter\gdef\csname r@fig:pizza_tr\endcsname{{13}{45}}
\expandafter\gdef\csname r@eq:deg_q_p\endcsname{{158}{45}}
\expandafter\gdef\csname r@eq:adj_deg\endcsname{{160}{45}}
\expandafter\gdef\csname r@eq:ineq_dim\endcsname{{165}{46}}
\expandafter\gdef\csname r@eq:inter_cond\endcsname{{169}{47}}
\expandafter\gdef\csname r@qu:adj_by_int\endcsname{{3.3.3}{47}}
\expandafter\gdef\csname r@fig:adjoints\endcsname{{14}{48}}
\expandafter\gdef\csname r@sec:Push-Forward and Scattering Equations\endcsname{{3.4}{48}}
\expandafter\gdef\csname r@eq:fiber\endcsname{{177}{48}}
\expandafter\gdef\csname r@eq:push_forward_def\endcsname{{178}{48}}
\expandafter\gdef\csname r@eq:pushforward_coord\endcsname{{180}{48}}
\expandafter\gdef\csname r@eq:Phi\endcsname{{181}{48}}
\expandafter\gdef\csname r@eq:pushforward_heuristic\endcsname{{183}{49}}
\expandafter\gdef\csname r@eq:or_matr\endcsname{{184}{49}}
\expandafter\gdef\csname r@eq:newton_map\endcsname{{185}{49}}
\expandafter\gdef\csname r@thm:newt_map\endcsname{{3.4.2}{49}}
\expandafter\gdef\csname r@eq:newt_push\endcsname{{186}{49}}
\expandafter\gdef\csname r@eq:F_ideal\endcsname{{188}{49}}
\expandafter\gdef\csname r@eq:pushforward_comp\endcsname{{192}{50}}
\expandafter\gdef\csname r@eg:pent_newt_map\endcsname{{3.9}{50}}
\expandafter\gdef\csname r@eq:pent_NPM\endcsname{{194}{50}}
\expandafter\gdef\csname r@eq:stringy:can_forms\endcsname{{197}{51}}
\expandafter\gdef\csname r@eq:In\endcsname{{198}{51}}
\expandafter\gdef\csname r@eq:SCF_int\endcsname{{204}{52}}
\expandafter\gdef\csname r@thm:stringy_field_theory_limit\endcsname{{3.4.4}{52}}
\expandafter\gdef\csname r@eq:stringy_limit\endcsname{{205}{52}}
\expandafter\gdef\csname r@eq:scatt_eqs\endcsname{{208}{53}}
\expandafter\gdef\csname r@eq:Phi_map_2\endcsname{{210}{53}}
\expandafter\gdef\csname r@eq:n5_assoc\endcsname{{214}{53}}
\expandafter\gdef\csname r@sec:Integral Representations\endcsname{{3.5}{54}}
\expandafter\gdef\csname r@eq:biduality\endcsname{{218}{54}}
\expandafter\gdef\csname r@eq:C_cone\endcsname{{221}{55}}
\expandafter\gdef\csname r@eq:dual_pol_cone\endcsname{{222}{55}}
\expandafter\gdef\csname r@eq:Lapl_tr_pol\endcsname{{226}{55}}
\expandafter\gdef\csname r@eq:volume_gauge_S\endcsname{{228}{55}}
\expandafter\gdef\csname r@eq:volume_00\endcsname{{239}{57}}
\expandafter\gdef\csname r@eq:volume_0\endcsname{{241}{57}}
\expandafter\gdef\csname r@eq:cavvv\endcsname{{242}{57}}
\expandafter\gdef\csname r@eq:caval\endcsname{{244}{57}}
\expandafter\gdef\csname r@eq:vol_form\endcsname{{246}{57}}
\expandafter\gdef\csname r@eq:contour_can_form\endcsname{{252}{58}}
\expandafter\gdef\csname r@eq:delta_4\endcsname{{255}{59}}
\expandafter\gdef\csname r@eq:square_contour_integral\endcsname{{256}{59}}
\expandafter\gdef\csname r@eq:4gon_res\endcsname{{259}{59}}
\expandafter\gdef\csname r@ch:Amplituhedra\endcsname{{4}{60}}
\expandafter\gdef\csname r@sec:Hodges and Cyclic Polytopes\endcsname{{4.1}{60}}
\expandafter\gdef\csname r@fig:n6_hodg\endcsname{{15}{61}}
\expandafter\gdef\csname r@eq:A6_YM\endcsname{{260}{61}}
\expandafter\gdef\csname r@fig:Hodg_pol\endcsname{{16}{62}}
\expandafter\gdef\csname r@eq:forms_sum\endcsname{{263}{62}}
\expandafter\gdef\csname r@eq:tetra_function\endcsname{{264}{62}}
\expandafter\gdef\csname r@eq:Q6_form\endcsname{{266}{62}}
\expandafter\gdef\csname r@eq:forms_sum_2\endcsname{{267}{62}}
\expandafter\gdef\csname r@eq:R1\endcsname{{269}{63}}
\expandafter\gdef\csname r@eq:Rinv\endcsname{{270}{63}}
\expandafter\gdef\csname r@eq:R_inv_2\endcsname{{272}{63}}
\expandafter\gdef\csname r@eq:Y0\endcsname{{273}{63}}
\expandafter\gdef\csname r@eq:iijj_2\endcsname{{275}{63}}
\expandafter\gdef\csname r@fig:graph_2\endcsname{{17}{64}}
\expandafter\gdef\csname r@eq:cyclic_positive\endcsname{{276}{64}}
\expandafter\gdef\csname r@eq:nmhv_cyclic_polytope\endcsname{{277}{64}}
\expandafter\gdef\csname r@fig:BCFW_cells_k1\endcsname{{18}{65}}
\expandafter\gdef\csname r@eq:R6\endcsname{{281}{65}}
\expandafter\gdef\csname r@sec:Tree Amplituhedra\endcsname{{4.2}{65}}
\expandafter\gdef\csname r@eq:Pl_emb\endcsname{{287}{66}}
\expandafter\gdef\csname r@eq:Pl_rel\endcsname{{288}{66}}
\expandafter\gdef\csname r@eq:pos_dec\endcsname{{290}{66}}
\expandafter\gdef\csname r@eq:bound_aff_perm\endcsname{{291}{67}}
\expandafter\gdef\csname r@eq:pos_def\endcsname{{292}{67}}
\expandafter\gdef\csname r@eq:top_perm\endcsname{{294}{67}}
\expandafter\gdef\csname r@eq:strand_rule\endcsname{{296}{67}}
\expandafter\gdef\csname r@eq:form_pos_gr\endcsname{{298}{67}}
\expandafter\gdef\csname r@fig:top_cells\endcsname{{19}{68}}
\expandafter\gdef\csname r@eq:meas_Gr\endcsname{{299}{68}}
\expandafter\gdef\csname r@eg:Gr24_strat\endcsname{{4.2.3}{68}}
\expandafter\gdef\csname r@eq:Pl_rel_2\endcsname{{300}{68}}
\expandafter\gdef\csname r@eq:Ca_par\endcsname{{301}{68}}
\expandafter\gdef\csname r@eq:non_cl_pos\endcsname{{302}{68}}
\expandafter\gdef\csname r@eq:alpha_ineq\endcsname{{303}{68}}
\expandafter\gdef\csname r@fig:Gr24_poset\endcsname{{20}{69}}
\expandafter\gdef\csname r@fig:proj_pol\endcsname{{21}{70}}
\expandafter\gdef\csname r@eq:porj_sympl\endcsname{{312}{70}}
\expandafter\gdef\csname r@eq:pos_Z\endcsname{{313}{70}}
\expandafter\gdef\csname r@eq:Ampl_map\endcsname{{314}{71}}
\expandafter\gdef\csname r@eq:def_ampl\endcsname{{315}{71}}
\expandafter\gdef\csname r@eq:bd_24\endcsname{{317}{71}}
\expandafter\gdef\csname r@eq:tw_emb\endcsname{{319}{72}}
\expandafter\gdef\csname r@eq:seg_m1\endcsname{{323}{72}}
\expandafter\gdef\csname r@eg:m=1\endcsname{{4.1}{72}}
\expandafter\gdef\csname r@fig:m1\endcsname{{22}{73}}
\expandafter\gdef\csname r@eq:Yi\endcsname{{327}{73}}
\expandafter\gdef\csname r@eq:tw_cycl\endcsname{{335}{74}}
\expandafter\gdef\csname r@eq:m2_cond\endcsname{{336}{74}}
\expandafter\gdef\csname r@eq:m4_cond\endcsname{{337}{74}}
\expandafter\gdef\csname r@sec:BCFW Recursion and Tiles\endcsname{{4.3}{76}}
\expandafter\gdef\csname r@fig:m1_BCFW_recursion\endcsname{{23}{78}}
\expandafter\gdef\csname r@eq:sigma_m2_k2_n4\endcsname{{352}{78}}
\expandafter\gdef\csname r@eq:pos_m1\endcsname{{355}{78}}
\expandafter\gdef\csname r@eq:Y_m1\endcsname{{356}{78}}
\expandafter\gdef\csname r@fig:m1_BCFW_examples\endcsname{{24}{79}}
\expandafter\gdef\csname r@fig:m1_k2_n4\endcsname{{25}{80}}
\expandafter\gdef\csname r@fig:m2_tile\endcsname{{26}{80}}
\expandafter\gdef\csname r@eq:m2_AT\endcsname{{359}{80}}
\expandafter\gdef\csname r@fig:m2_BCFW\endcsname{{27}{81}}
\expandafter\gdef\csname r@eq:m2_tiling\endcsname{{361}{81}}
\expandafter\gdef\csname r@eg:m2_k2_n5\endcsname{{4.4}{81}}
\expandafter\gdef\csname r@fig:m2_k2_n5\endcsname{{28}{82}}
\expandafter\gdef\csname r@eq:123145\endcsname{{365}{82}}
\expandafter\gdef\csname r@subsec:The A22n tiling and its forms\endcsname{{4.3.4}{82}}
\expandafter\gdef\csname r@eq:m2_alln_tiling\endcsname{{367}{82}}
\expandafter\gdef\csname r@six_invariant\endcsname{{368}{82}}
\expandafter\gdef\csname r@plane_intersection\endcsname{{369}{82}}
\expandafter\gdef\csname r@fig:recursive_A_nk4\endcsname{{29}{83}}
\expandafter\gdef\csname r@four_invariant\endcsname{{370}{83}}
\expandafter\gdef\csname r@eq:BCFW_prod\endcsname{{373}{84}}
\expandafter\gdef\csname r@eq:z_int\endcsname{{374}{84}}
\expandafter\gdef\csname r@eq:terms_n7_k2\endcsname{{378}{84}}
\expandafter\gdef\csname r@fig:bcfw-product-example\endcsname{{30}{85}}
\expandafter\gdef\csname r@sec:Loop Amplituhedra\endcsname{{4.4}{85}}
\expandafter\gdef\csname r@fig:BCFW_loop\endcsname{{31}{86}}
\expandafter\gdef\csname r@eq:loop-grassmannian-matrix\endcsname{{389}{86}}
\expandafter\gdef\csname r@eq:loop_Gr_pos\endcsname{{390}{87}}
\expandafter\gdef\csname r@eq:loop_ampl_map\endcsname{{391}{87}}
\expandafter\gdef\csname r@eq:loop_amplituhedron_def\endcsname{{392}{87}}
\expandafter\gdef\csname r@eq:iso_l1_m2\endcsname{{397}{87}}
\expandafter\gdef\csname r@eq:loop_pos\endcsname{{398}{88}}
\expandafter\gdef\csname r@eq:mult_pos\endcsname{{399}{88}}
\expandafter\gdef\csname r@eq:l1_m3\endcsname{{402}{88}}
\expandafter\gdef\csname r@eq:om_fib\endcsname{{410}{89}}
\expandafter\gdef\csname r@sec:The Four-Point Two-Loop Amplituhedron\endcsname{{4.5}{89}}
\expandafter\gdef\csname r@two_loop_Ampl\endcsname{{415}{89}}
\expandafter\gdef\csname r@two_triangles_with_loops\endcsname{{32}{90}}
\expandafter\gdef\csname r@eq:l1_cons\endcsname{{416}{90}}
\expandafter\gdef\csname r@A_on_triangle\endcsname{{417}{90}}
\expandafter\gdef\csname r@B_on_triangle\endcsname{{419}{90}}
\expandafter\gdef\csname r@line_points\endcsname{{421}{90}}
\expandafter\gdef\csname r@ABCD\endcsname{{422}{90}}
\expandafter\gdef\csname r@eq:Plucker_rel\endcsname{{423}{91}}
\expandafter\gdef\csname r@factorised_ABCD\endcsname{{424}{91}}
\expandafter\gdef\csname r@Deltas\endcsname{{425}{91}}
\expandafter\gdef\csname r@Delta24_2\endcsname{{426}{91}}
\expandafter\gdef\csname r@other_Deltas\endcsname{{427}{91}}
\expandafter\gdef\csname r@sign_regions\endcsname{{429}{91}}
\expandafter\gdef\csname r@Schubert_divisors\endcsname{{430}{91}}
\expandafter\gdef\csname r@eq:l1_fact\endcsname{{432}{91}}
\expandafter\gdef\csname r@AB_CD_divisor\endcsname{{435}{92}}
\expandafter\gdef\csname r@Li_Li_factorisation\endcsname{{436}{92}}
\expandafter\gdef\csname r@L12_factorization\endcsname{{437}{92}}
\expandafter\gdef\csname r@factorization_V12\endcsname{{438}{92}}
\expandafter\gdef\csname r@tab:algebraic_strat_two_loop\endcsname{{1}{93}}
\expandafter\gdef\csname r@fig:residual-arrangement-components\endcsname{{33}{94}}
\expandafter\gdef\csname r@residual_by_drop\endcsname{{442}{94}}
\expandafter\gdef\csname r@canonical_form\endcsname{{443}{94}}
\expandafter\gdef\csname r@integrand_ansatz\endcsname{{444}{94}}
\expandafter\gdef\csname r@table:adjoint_dof\endcsname{{2}{95}}
\expandafter\gdef\csname r@adjoint\endcsname{{445}{95}}
\expandafter\gdef\csname r@first_residual_condition\endcsname{{447}{95}}
\expandafter\gdef\csname r@second_residual_condition\endcsname{{450}{95}}
\expandafter\gdef\csname r@fig:l2_diagr\endcsname{{34}{96}}
\expandafter\gdef\csname r@L2n4_form\endcsname{{454}{96}}
\expandafter\gdef\csname r@ch:Canonical Forms as Dual Volumes\endcsname{{5}{98}}
\expandafter\gdef\csname r@sec:Completely Monotone Positive Geometries\endcsname{{5.1}{98}}
\expandafter\gdef\csname r@eq:canonical_q_over_p_CM\endcsname{{458}{98}}
\expandafter\gdef\csname r@fig:non_convex_polypols\endcsname{{35}{99}}
\expandafter\gdef\csname r@eq:polytope_laplace_recall_CM\endcsname{{460}{99}}
\expandafter\gdef\csname r@eq:polytope_CM_derivative\endcsname{{461}{99}}
\expandafter\gdef\csname r@eq:CMC_cons\endcsname{{462}{99}}
\expandafter\gdef\csname r@eq:CM_def_cone\endcsname{{463}{99}}
\expandafter\gdef\csname r@fig:two-side-by-side\endcsname{{36}{100}}
\expandafter\gdef\csname r@eq:eg_CM\endcsname{{464}{100}}
\expandafter\gdef\csname r@eq:BHWC_laplace\endcsname{{466}{100}}
\expandafter\gdef\csname r@eq:CM_PG_laplace\endcsname{{467}{100}}
\expandafter\gdef\csname r@eq:vol_form_2\endcsname{{468}{101}}
\expandafter\gdef\csname r@eq:fund_sol\endcsname{{473}{101}}
\expandafter\gdef\csname r@fig:half_disk\endcsname{{37}{102}}
\expandafter\gdef\csname r@eq:fund_sol_int_repr\endcsname{{474}{102}}
\expandafter\gdef\csname r@eq:riesz_quadr\endcsname{{478}{102}}
\expandafter\gdef\csname r@eq:spectrahedral_cone_CM\endcsname{{481}{103}}
\expandafter\gdef\csname r@eq:detA\endcsname{{482}{103}}
\expandafter\gdef\csname r@eq:p_pol\endcsname{{483}{103}}
\expandafter\gdef\csname r@sec:Transcendental Measures\endcsname{{5.2}{104}}
\expandafter\gdef\csname r@eq:square_linear_map\endcsname{{491}{105}}
\expandafter\gdef\csname r@eq:square_riesz_kernel\endcsname{{493}{105}}
\expandafter\gdef\csname r@eq:pizza_cone_short\endcsname{{496}{105}}
\expandafter\gdef\csname r@eq:line_conic_denominator\endcsname{{497}{105}}
\expandafter\gdef\csname r@fig:half_pizzas\endcsname{{38}{106}}
\expandafter\gdef\csname r@eq:ice_cream\endcsname{{498}{106}}
\expandafter\gdef\csname r@eq:half_pizza_a_canonical_short\endcsname{{499}{106}}
\expandafter\gdef\csname r@fig:linearquadricmeas1\endcsname{{39(i)}{107}}
\expandafter\gdef\csname r@fig:linearquadricmeas2\endcsname{{39(ii)}{107}}
\expandafter\gdef\csname r@fig:linearquadricmeas3\endcsname{{39(iii)}{107}}
\expandafter\gdef\csname r@fig:linequadric\endcsname{{39}{107}}
\expandafter\gdef\csname r@eq:naive_half_pizza_integral\endcsname{{500}{107}}
\expandafter\gdef\csname r@eq:line_conic_integral_trans\endcsname{{503}{107}}
\expandafter\gdef\csname r@eq:mu_line_conic_arctan_trans\endcsname{{505}{108}}
\expandafter\gdef\csname r@eq:mu_line_conic_tangent_trans\endcsname{{506}{108}}
\expandafter\gdef\csname r@eq:mu_line_conic_log_trans\endcsname{{507}{108}}
\expandafter\gdef\csname r@eq:mu_half_pizza_general_trans\endcsname{{511}{108}}
\expandafter\gdef\csname r@fig:pizza_slice2_geometry\endcsname{{40(i)}{109}}
\expandafter\gdef\csname r@fig:pizza_slice2_dual\endcsname{{40(ii)}{109}}
\expandafter\gdef\csname r@fig:pizza_slice2_measure\endcsname{{40(iii)}{109}}
\expandafter\gdef\csname r@fig:pizza_slice2\endcsname{{40}{109}}
\expandafter\gdef\csname r@eq:measure_triangulation_trans\endcsname{{512}{109}}
\expandafter\gdef\csname r@eq:mu_two_lines_one_conic_trans\endcsname{{514}{109}}
\expandafter\gdef\csname r@eq:mu_one_curvy_s_lines_trans\endcsname{{515}{109}}
\expandafter\gdef\csname r@fig:curvy_pentagon_measure\endcsname{{41}{110}}
\expandafter\gdef\csname r@eq:dual_letters_trans\endcsname{{516}{110}}
\expandafter\gdef\csname r@fig:2con\endcsname{{42}{111}}
\expandafter\gdef\csname r@app:nodal_cubic_measure\endcsname{{5.2.4}{111}}
\expandafter\gdef\csname r@eq:nodal_cubic_appendix\endcsname{{518}{111}}
\expandafter\gdef\csname r@eq:nodal_cubic_canonical_appendix\endcsname{{519}{111}}
\expandafter\gdef\csname r@eq:dual_cubic_quartic_appendix\endcsname{{520}{111}}
\expandafter\gdef\csname r@eq:cubic_measure_period_appendix\endcsname{{522}{111}}
\expandafter\gdef\csname r@fig:nodal_cubic_measure\endcsname{{43}{112}}
\expandafter\gdef\csname r@eq:cubic_measure_elliptic_appendix\endcsname{{525}{112}}
\expandafter\gdef\csname r@sec:Convexity in the Grassmannian\endcsname{{5.3}{112}}
\expandafter\gdef\csname r@eq:plucker_embedding_convexity\endcsname{{527}{113}}
\expandafter\gdef\csname r@eq:convX_def_long\endcsname{{529}{113}}
\expandafter\gdef\csname r@eq:extendably_convex_def_long\endcsname{{530}{113}}
\expandafter\gdef\csname r@eq:positive_grass_extendably_convex\endcsname{{533}{113}}
\expandafter\gdef\csname r@fig:extendable_convexity_curve\endcsname{{44}{114}}
\expandafter\gdef\csname r@eq:exterior_cyclic_polytope_def_long\endcsname{{535}{114}}
\expandafter\gdef\csname r@eq:conv_ampl_exterior_cyclic_long\endcsname{{537}{114}}
\expandafter\gdef\csname r@eq:gr24_plucker_quadric\endcsname{{542}{115}}
\expandafter\gdef\csname r@eq:k_m_2_extendable_convexity_long\endcsname{{544}{115}}
\expandafter\gdef\csname r@eq:schub_hyp\endcsname{{547}{116}}
\expandafter\gdef\csname r@eq:schubert_hyperplanes_bar\endcsname{{549}{116}}
\expandafter\gdef\csname r@eq:schubert_exterior_cyclic_polytope_def\endcsname{{550}{116}}
\expandafter\gdef\csname r@magic_det\endcsname{{552}{116}}
\expandafter\gdef\csname r@eq:schubert_polytope_same_intersection\endcsname{{553}{116}}
\expandafter\gdef\csname r@fig:circuits4\endcsname{{45}{117}}
\expandafter\gdef\csname r@eq:Zmatrix_appendix\endcsname{{555}{117}}
\expandafter\gdef\csname r@eq:wall\endcsname{{557}{117}}
\expandafter\gdef\csname r@fig:c246bases\endcsname{{46}{118}}
\expandafter\gdef\csname r@eq:wedge_wall_hexagon\endcsname{{562}{118}}
\expandafter\gdef\csname r@sec:Duality for Amplituhedra\endcsname{{5.4}{119}}
\expandafter\gdef\csname r@eq:extendable_dual_def\endcsname{{563}{119}}
\expandafter\gdef\csname r@eq:dual_amplituhedron_def\endcsname{{568}{120}}
\expandafter\gdef\csname r@eq:ext_cycl_dual\endcsname{{570}{120}}
\expandafter\gdef\csname r@eq:dual_amplituhedron_ineqs\endcsname{{571}{120}}
\expandafter\gdef\csname r@eq:dual_amplituhedron_ineqs_222\endcsname{{572}{120}}
\expandafter\gdef\csname r@eq:twist_map_def\endcsname{{573}{120}}
\expandafter\gdef\csname r@eq:twist_map\endcsname{{574}{120}}
\expandafter\gdef\csname r@eq:schubert_polytope_dual_twist\endcsname{{575}{120}}
\expandafter\gdef\csname r@eq:dual_mk2\endcsname{{577}{121}}
\expandafter\gdef\csname r@eq:twisted_amplituhedron_sign_conditions\endcsname{{578}{121}}
\expandafter\gdef\csname r@eq:amplituhedron_affine_laplace\endcsname{{579}{121}}
\expandafter\gdef\csname r@eq:amplituhedron_grassmannian_laplace\endcsname{{580}{121}}
\expandafter\gdef\csname r@eq:integral_dual_positive_grassmannian\endcsname{{581}{122}}
\expandafter\gdef\csname r@sec:Aomoto Forms\endcsname{{5.5}{122}}
\expandafter\gdef\csname r@eq:can_pair\endcsname{{583}{122}}
\expandafter\gdef\csname r@eq:aom_equiv\endcsname{{584}{122}}
\expandafter\gdef\csname r@eq:form_sympl_dlog\endcsname{{587}{123}}
\expandafter\gdef\csname r@eq:IntervalCanonicalForm\endcsname{{588}{123}}
\expandafter\gdef\csname r@eq:aomoto_log\endcsname{{589}{123}}
\expandafter\gdef\csname r@eq:AomotoDEschematic\endcsname{{590}{123}}
\expandafter\gdef\csname r@eq:aom_symbol\endcsname{{591}{124}}
\expandafter\gdef\csname r@eq:aomoto_duality\endcsname{{594}{124}}
\expandafter\gdef\csname r@Goncharov_polylogs\endcsname{{595}{124}}
\expandafter\gdef\csname r@G_a1_z\endcsname{{596}{124}}
\expandafter\gdef\csname r@eq:pol_to_gonch\endcsname{{597}{124}}
\expandafter\gdef\csname r@G_in_deltaz\endcsname{{599}{125}}
\expandafter\gdef\csname r@eq:gonch\endcsname{{601}{125}}
\expandafter\gdef\csname r@Goncharov_pairing\endcsname{{602}{125}}
\expandafter\gdef\csname r@fig:Polylogs_plot\endcsname{{47}{126}}
\expandafter\gdef\csname r@G_measure\endcsname{{608}{126}}
\expandafter\gdef\csname r@mu\endcsname{{609}{126}}
\expandafter\gdef\csname r@eq:invariant-X\endcsname{{611}{126}}
\expandafter\gdef\csname r@eq:invariant-G\endcsname{{612}{126}}
\expandafter\gdef\csname r@eq:invariant-symbol\endcsname{{613}{127}}
\expandafter\gdef\csname r@eq:Aomoto_mod\endcsname{{615}{127}}
\expandafter\gdef\csname r@eq:aom_mod_sp\endcsname{{616}{127}}
\expandafter\gdef\csname r@eq:equiv\endcsname{{617}{127}}
\expandafter\gdef\csname r@eq:Au\endcsname{{619}{127}}
\expandafter\gdef\csname r@eq:AomotoBracketCrossRatios\endcsname{{620}{127}}
\expandafter\gdef\csname r@eq:flag-x\endcsname{{621}{127}}
\expandafter\gdef\csname r@eq:general-flag-G\endcsname{{622}{128}}
\expandafter\gdef\csname r@eq:m2-G\endcsname{{624}{128}}
\expandafter\gdef\csname r@eq:ngh_shift\endcsname{{625}{128}}
\expandafter\gdef\csname r@eq:It\endcsname{{626}{128}}
\expandafter\gdef\csname r@eq:Auv\endcsname{{629}{128}}
\expandafter\gdef\csname r@eq:u_gauge\endcsname{{630}{129}}
\expandafter\gdef\csname r@eq:pars_Aom\endcsname{{632}{129}}
\expandafter\gdef\csname r@eq:dual_vol_aomoto\endcsname{{634}{129}}
\expandafter\gdef\csname r@eq:can_vol_par\endcsname{{635}{129}}
\expandafter\gdef\csname r@eq:volume_pairing_duality\endcsname{{636}{129}}
\expandafter\gdef\csname r@fig:half_pizza_pairing\endcsname{{48}{130}}
\expandafter\gdef\csname r@eq:J_ex\endcsname{{637}{130}}
\expandafter\gdef\csname r@eq:J_1\endcsname{{638}{130}}
\expandafter\gdef\csname r@eq:Ja_integral\endcsname{{639}{130}}
\expandafter\gdef\csname r@eq:Ja_result\endcsname{{641}{130}}
\expandafter\gdef\csname r@eq:I_app_naive_pairing\endcsname{{642}{131}}
\expandafter\gdef\csname r@eq:Jtilde_app_result\endcsname{{645}{131}}
\expandafter\gdef\csname r@sec:Chapter 5 open problems\endcsname{{5.6}{131}}
\expandafter\gdef\csname r@ch:From Integrands to Integrals\endcsname{{6}{133}}
\expandafter\gdef\csname r@sec:The Geometry of Integration\endcsname{{6.1}{133}}
\expandafter\gdef\csname r@subsec:Integrals as Period Pairings\endcsname{{6.1.1}{134}}
\expandafter\gdef\csname r@eq:Pi\endcsname{{646}{134}}
\expandafter\gdef\csname r@eq:fibre-Xs\endcsname{{647}{134}}
\expandafter\gdef\csname r@eq:Ds-definition\endcsname{{648}{134}}
\expandafter\gdef\csname r@eq:Bs-definition\endcsname{{649}{134}}
\expandafter\gdef\csname r@eq:I_int_s\endcsname{{650}{134}}
\expandafter\gdef\csname r@eq:dR_cohom\endcsname{{651}{135}}
\expandafter\gdef\csname r@eq:rel_homol\endcsname{{652}{135}}
\expandafter\gdef\csname r@eq:period-pairing\endcsname{{653}{135}}
\expandafter\gdef\csname r@subsec:Landau Varieties as Failure of Local Triviality\endcsname{{6.1.2}{135}}
\expandafter\gdef\csname r@eq:rel_fb\endcsname{{654}{135}}
\expandafter\gdef\csname r@eq:L\endcsname{{655}{135}}
\expandafter\gdef\csname r@eq:Land_crit\endcsname{{659}{136}}
\expandafter\gdef\csname r@subsec:Monodromy Discontinuities and Vanishing Cycles\endcsname{{6.1.3}{136}}
\expandafter\gdef\csname r@eq:monodromy-representation\endcsname{{660}{136}}
\expandafter\gdef\csname r@eq:monodromy-action-integral\endcsname{{661}{136}}
\expandafter\gdef\csname r@eq:variation-cycle\endcsname{{662}{136}}
\expandafter\gdef\csname r@eq:disc-monodromy\endcsname{{663}{136}}
\expandafter\gdef\csname r@eq:PL_var\endcsname{{664}{137}}
\expandafter\gdef\csname r@subsec:Discontinuities as Residues\endcsname{{6.1.4}{137}}
\expandafter\gdef\csname r@eq:S_in_int\endcsname{{665}{137}}
\expandafter\gdef\csname r@eq:pham-disc-residue\endcsname{{666}{137}}
\expandafter\gdef\csname r@eq:local-log-form-residue\endcsname{{668}{137}}
\expandafter\gdef\csname r@eq:variation-torus\endcsname{{669}{137}}
\expandafter\gdef\csname r@eq:local-residue\endcsname{{671}{137}}
\expandafter\gdef\csname r@subsec:Landau Loci versus Actual Singularities\endcsname{{6.1.5}{138}}
\expandafter\gdef\csname r@subsec:Example Aomoto Interval\endcsname{{6.1.6}{138}}
\expandafter\gdef\csname r@eq:I_log\endcsname{{672}{138}}
\expandafter\gdef\csname r@eq:DiBi\endcsname{{674}{138}}
\expandafter\gdef\csname r@eq:DicapBi\endcsname{{675}{139}}
\expandafter\gdef\csname r@eq:aomoto-interval-landau\endcsname{{676}{139}}
\expandafter\gdef\csname r@eq:log_disc_1\endcsname{{678}{139}}
\expandafter\gdef\csname r@eq:Var_S0\endcsname{{679}{139}}
\expandafter\gdef\csname r@eq:disc_log\endcsname{{680}{139}}
\expandafter\gdef\csname r@eq:aomoto-log-residue\endcsname{{681}{139}}
\expandafter\gdef\csname r@eq:aomoto-disc-residue\endcsname{{682}{139}}
\expandafter\gdef\csname r@subsec:Landau Equations for Feynman Integrals\endcsname{{6.1.7}{139}}
\expandafter\gdef\csname r@eq:schematic-loop-integral\endcsname{{683}{139}}
\expandafter\gdef\csname r@eq:prop_D\endcsname{{684}{140}}
\expandafter\gdef\csname r@eq:on-shell-variety-H\endcsname{{686}{140}}
\expandafter\gdef\csname r@eq:PiH\endcsname{{687}{140}}
\expandafter\gdef\csname r@eq:landau-cut-H\endcsname{{688}{140}}
\expandafter\gdef\csname r@eq:pinch-momentum-space\endcsname{{689}{140}}
\expandafter\gdef\csname r@eq:landau-variety-H\endcsname{{692}{140}}
\expandafter\gdef\csname r@eq:F_LE\endcsname{{693}{141}}
\expandafter\gdef\csname r@eq:cutkosky-rule-comment\endcsname{{694}{141}}
\expandafter\gdef\csname r@subsec:Pure Integrals and Canonical Differential Equations\endcsname{{6.1.8}{142}}
\expandafter\gdef\csname r@eq:general-differential-equation-master-integrals\endcsname{{697}{142}}
\expandafter\gdef\csname r@eq:singular-loci-landau-DE\endcsname{{698}{142}}
\expandafter\gdef\csname r@eq:pure-function-differential\endcsname{{699}{142}}
\expandafter\gdef\csname r@eq:canonical-differential-equation-pure-basis\endcsname{{701}{143}}
\expandafter\gdef\csname r@eq:canonical-DE-local-monodromy\endcsname{{702}{143}}
\expandafter\gdef\csname r@subsec:Boxes in Momentum Twistors\endcsname{{6.1.9}{143}}
\expandafter\gdef\csname r@eq:four-mass-box-momentum-twistor-integrand\endcsname{{703}{143}}
\expandafter\gdef\csname r@eq:int_box\endcsname{{704}{144}}
\expandafter\gdef\csname r@eq:Xi_4m\endcsname{{705}{144}}
\expandafter\gdef\csname r@eq:X_i_to_p\endcsname{{706}{144}}
\expandafter\gdef\csname r@eq:hodges-gaussian-momentum-twistor\endcsname{{707}{144}}
\expandafter\gdef\csname r@eq:four-mass-parametric\endcsname{{709}{144}}
\expandafter\gdef\csname r@eq:four-mass-F\endcsname{{710}{144}}
\expandafter\gdef\csname r@eq:four-mass-box-cross-ratios\endcsname{{711}{144}}
\expandafter\gdef\csname r@eq:four-mass-box-discriminant\endcsname{{712}{144}}
\expandafter\gdef\csname r@eq:Delta\endcsname{{713}{144}}
\expandafter\gdef\csname r@eq:box_cut\endcsname{{714}{144}}
\expandafter\gdef\csname r@eq:four-mass-leading-singularity\endcsname{{716}{145}}
\expandafter\gdef\csname r@eq:z-zbar-def\endcsname{{717}{145}}
\expandafter\gdef\csname r@eq:z-zbar-sol\endcsname{{718}{145}}
\expandafter\gdef\csname r@eq:four-mass-box-result\endcsname{{719}{145}}
\expandafter\gdef\csname r@eq:Xi_0m\endcsname{{720}{145}}
\expandafter\gdef\csname r@sec:Leading Singularities and Maximal Cuts\endcsname{{6.2}{146}}
\expandafter\gdef\csname r@subsec:Leading Singularities as Maximal Residues\endcsname{{6.2.1}{146}}
\expandafter\gdef\csname r@eq:LS-four-point-box-onshell\endcsname{{722}{146}}
\expandafter\gdef\csname r@eq:n-box-from-corners\endcsname{{723}{146}}
\expandafter\gdef\csname r@eq:k-box-from-corners\endcsname{{724}{146}}
\expandafter\gdef\csname r@eq:n-k-leading-landau-diagram\endcsname{{726}{146}}
\expandafter\gdef\csname r@subsec:One-Loop Prescriptive Unitarity\endcsname{{6.2.2}{147}}
\expandafter\gdef\csname r@eq:one-loop-integral-reduction-general\endcsname{{728}{147}}
\expandafter\gdef\csname r@eq:one-loop-common-denominator\endcsname{{729}{147}}
\expandafter\gdef\csname r@eq:two-mass-easy-cut-equations\endcsname{{730}{147}}
\expandafter\gdef\csname r@eq:two-mass-easy-cut-solutions\endcsname{{731}{148}}
\expandafter\gdef\csname r@eq:chiral-pentagon-integrand\endcsname{{732}{148}}
\expandafter\gdef\csname r@eq:chiral-pentagon-unit-LS\endcsname{{733}{148}}
\expandafter\gdef\csname r@eq:one-loop-MHV-ratio\endcsname{{734}{148}}
\expandafter\gdef\csname r@eq:chiral-pentagon-integrated\endcsname{{735}{148}}
\expandafter\gdef\csname r@eq:NkMHV-one-loop-chiral-pentagon-expansion\endcsname{{736}{148}}
\expandafter\gdef\csname r@eq:loop-expansion-P-functions\endcsname{{737}{149}}
\expandafter\gdef\csname r@eq:one-loop-ratio-function-general-k\endcsname{{739}{149}}
\expandafter\gdef\csname r@eq:finite-one-loop-ratio-function-general-k\endcsname{{740}{149}}
\expandafter\gdef\csname r@subsec:On-Shell Diagrams and Grassmannian Localization\endcsname{{6.2.3}{150}}
\expandafter\gdef\csname r@eq:cut-onshell-diagrams\endcsname{{741}{150}}
\expandafter\gdef\csname r@eq:onshell-form-before-mcut\endcsname{{742}{150}}
\expandafter\gdef\csname r@eq:bosonic-localization-equations\endcsname{{743}{150}}
\expandafter\gdef\csname r@eq:alpha-solutions\endcsname{{744}{150}}
\expandafter\gdef\csname r@eq:MCut-pushforward\endcsname{{745}{150}}
\expandafter\gdef\csname r@eq:jacobian-onshell\endcsname{{746}{150}}
\expandafter\gdef\csname r@eq:LS_def\endcsname{{747}{150}}
\expandafter\gdef\csname r@eq:hel_Gamma\endcsname{{748}{151}}
\expandafter\gdef\csname r@eq:LS_max_k\endcsname{{749}{151}}
\expandafter\gdef\csname r@subsec:Amplituhedron Boundaries and LS Configurations\endcsname{{6.2.4}{151}}
\expandafter\gdef\csname r@eq:ampl_bd\endcsname{{753}{151}}
\expandafter\gdef\csname r@eq:LS_Res\endcsname{{754}{152}}
\expandafter\gdef\csname r@eq:LS-dictionary\endcsname{{755}{152}}
\expandafter\gdef\csname r@subsec:One-Loop Schubert Problems\endcsname{{6.2.5}{152}}
\expandafter\gdef\csname r@eq:general-one-loop-schubert-cut\endcsname{{756}{152}}
\expandafter\gdef\csname r@eq:zero-mass-box-schubert-solutions\endcsname{{759}{152}}
\expandafter\gdef\csname r@eq:four-point-one-loop-MHV-integrand\endcsname{{761}{153}}
\expandafter\gdef\csname r@eq:one-mass-schubert-solutions\endcsname{{764}{153}}
\expandafter\gdef\csname r@eq:two-mass-easy-schubert-solutions\endcsname{{766}{153}}
\expandafter\gdef\csname r@eq:four-mass-cut-discriminant\endcsname{{769}{154}}
\expandafter\gdef\csname r@eq:eight-point-four-mass-box-LS\endcsname{{774}{154}}
\expandafter\gdef\csname r@eq:4m_AHBCCT_expression\endcsname{{776}{154}}
\expandafter\gdef\csname r@eq:four-mass-psi-factor\endcsname{{777}{154}}
\expandafter\gdef\csname r@subsec:Composite Leading Singularities\endcsname{{6.2.6}{154}}
\expandafter\gdef\csname r@eq:toy_form\endcsname{{778}{154}}
\expandafter\gdef\csname r@eq:double-box-seven-cuts\endcsname{{780}{155}}
\expandafter\gdef\csname r@eq:double-box-mutual-cut\endcsname{{781}{155}}
\expandafter\gdef\csname r@sec:WLwithLI\endcsname{{6.3}{155}}
\expandafter\gdef\csname r@subsec:The Observable\endcsname{{6.3.1}{156}}
\expandafter\gdef\csname r@eq:WL-LI-definition\endcsname{{785}{156}}
\expandafter\gdef\csname r@eq:WL-LI-log-derivative\endcsname{{786}{156}}
\expandafter\gdef\csname r@ampl_def_integrand\endcsname{{788}{156}}
\expandafter\gdef\csname r@ampl_form\endcsname{{789}{156}}
\expandafter\gdef\csname r@eq:log-amplitude-integrand-thesis\endcsname{{790}{157}}
\expandafter\gdef\csname r@eq:WL-LI-from-log-amplitude\endcsname{{791}{157}}
\expandafter\gdef\csname r@eq:WL-LI-perturbative-expansion\endcsname{{792}{157}}
\expandafter\gdef\csname r@eq:Born-WL-LI\endcsname{{793}{157}}
\expandafter\gdef\csname r@fig:negative-geometry-two-loop-relation\endcsname{{49}{158}}
\expandafter\gdef\csname r@subsec:Expansion in Negative Geometries\endcsname{{6.3.2}{158}}
\expandafter\gdef\csname r@loop_positivty\endcsname{{794}{158}}
\expandafter\gdef\csname r@eq:mutual-positivity-negative-geometry-section\endcsname{{795}{158}}
\expandafter\gdef\csname r@eq:two-loop-negative-geometry-relation\endcsname{{796}{158}}
\expandafter\gdef\csname r@eq:negative-geometry-definition\endcsname{{797}{158}}
\expandafter\gdef\csname r@eq:Omega-negative-geometry-expansion\endcsname{{798}{159}}
\expandafter\gdef\csname r@eq:log-negative-geometry-expansion\endcsname{{799}{159}}
\expandafter\gdef\csname r@eq:integrated-negative-geometry-thesis\endcsname{{800}{159}}
\expandafter\gdef\csname r@eq:WL-negative-geometry-expansion-thesis\endcsname{{803}{159}}
\expandafter\gdef\csname r@eq:F4-Born-negative-geometry\endcsname{{804}{160}}
\expandafter\gdef\csname r@eq:n4-two-loop-log-integrand-negative-geometry\endcsname{{805}{160}}
\expandafter\gdef\csname r@subsec:The Structure of the Wilson Loop in Perturbation Theory\endcsname{{6.3.3}{160}}
\expandafter\gdef\csname r@eq:F-decomposition\endcsname{{806}{160}}
\expandafter\gdef\csname r@Figure_Yangian invariants\endcsname{{50}{161}}
\expandafter\gdef\csname r@eq:F-Born-Kermit-expansion\endcsname{{807}{161}}
\expandafter\gdef\csname r@eq:F-Born-standard-Kermit-expansion\endcsname{{808}{161}}
\expandafter\gdef\csname r@eq:Kermit-six-invariant-F\endcsname{{809}{161}}
\expandafter\gdef\csname r@eq:Kermit-four-invariant-F\endcsname{{810}{161}}
\expandafter\gdef\csname r@eq:F4-one-loop-decomposition\endcsname{{814}{162}}
\expandafter\gdef\csname r@eq:F4-one-loop-integrated-function\endcsname{{815}{162}}
\expandafter\gdef\csname r@eq:F5-one-loop-LS-example\endcsname{{816}{162}}
\expandafter\gdef\csname r@eq:F5-Born-LS\endcsname{{817}{162}}
\expandafter\gdef\csname r@eq:Fn-one-loop-chiral-pentagon-expansion\endcsname{{818}{162}}
\expandafter\gdef\csname r@eq:Omega-n-ij-Kermit-sum\endcsname{{819}{162}}
\expandafter\gdef\csname r@tab:LS-count-thesis\endcsname{{3}{163}}
\expandafter\gdef\csname r@subsec:Classification of Leading Singularities\endcsname{{6.3.4}{163}}
\expandafter\gdef\csname r@eq:one-loop-LS-count-thesis\endcsname{{820}{163}}
\expandafter\gdef\csname r@eq:PT\endcsname{{822}{163}}
\expandafter\gdef\csname r@fig:LS-summary-thesis\endcsname{{51}{164}}
\expandafter\gdef\csname r@eq:LS-conformal-invariance-thesis\endcsname{{823}{164}}
\expandafter\gdef\csname r@eq:ecr_cons\endcsname{{824}{164}}
\expandafter\gdef\csname r@LS_simple_formulae\endcsname{{831}{165}}
\expandafter\gdef\csname r@fig:simple-LS-thesis\endcsname{{52}{166}}
\expandafter\gdef\csname r@eg:red\endcsname{{6.7}{166}}
\expandafter\gdef\csname r@eq:2-loop\endcsname{{833}{166}}
\expandafter\gdef\csname r@fig:L2-reverse-incompatible\endcsname{{53}{167}}
\expandafter\gdef\csname r@eq:reverse-incompatible-bracket-thesis\endcsname{{837}{167}}
\expandafter\gdef\csname r@subsec:Taxonomy Compatibility and Examples\endcsname{{6.3.5}{167}}
\expandafter\gdef\csname r@fig:5pt-simple-13-thesis\endcsname{{54}{168}}
\expandafter\gdef\csname r@eg:simple-compatible-n5-l1\endcsname{{6.8}{168}}
\expandafter\gdef\csname r@omega_n5_13\endcsname{{841}{168}}
\expandafter\gdef\csname r@eg:simple-incompatible-n5-l2\endcsname{{6.9}{168}}
\expandafter\gdef\csname r@LSvaluesimpleincompatibleexample\endcsname{{844}{168}}
\expandafter\gdef\csname r@eg:nonsimple-three-loop-star\endcsname{{6.10}{169}}
\expandafter\gdef\csname r@non-simple-LS-cuts-thesis\endcsname{{847}{169}}
\expandafter\gdef\csname r@loop-nonsimple-nonintersecting-thesis\endcsname{{848}{169}}
\expandafter\gdef\csname r@LS-star-thesis\endcsname{{850}{169}}
\expandafter\gdef\csname r@fig:non-simple-LS-thesis\endcsname{{55}{170}}
\expandafter\gdef\csname r@eq:allowed-intersection-thesis\endcsname{{854}{170}}
\expandafter\gdef\csname r@eq:allowed-intersection-on-AB-thesis\endcsname{{855}{170}}
\expandafter\gdef\csname r@eg:reverse-compatible-l1\endcsname{{6.11}{170}}
\expandafter\gdef\csname r@alpha-thesis\endcsname{{857}{170}}
\expandafter\gdef\csname r@fig:reverse-LS-L1-thesis\endcsname{{56}{171}}
\expandafter\gdef\csname r@fig:reverse-incompatible-L2-thesis\endcsname{{57}{171}}
\expandafter\gdef\csname r@eg:reverse-incompatible-l2\endcsname{{6.12}{171}}
\expandafter\gdef\csname r@eg:six-loop-single-kermit\endcsname{{6.13}{171}}
\expandafter\gdef\csname r@AB_L=6.2-thesis\endcsname{{58}{172}}
\expandafter\gdef\csname r@sec:Geometric Landau Analysis\endcsname{{6.4}{172}}
\expandafter\gdef\csname r@subsec:refined-landau-analysis\endcsname{{6.4.1}{172}}
\expandafter\gdef\csname r@tab:all-L2-config-thesis\endcsname{{4}{173}}
\expandafter\gdef\csname r@eq:form_B\endcsname{{861}{173}}
\expandafter\gdef\csname r@fig:aom_tr_num\endcsname{{59}{174}}
\expandafter\gdef\csname r@eq:Izw\endcsname{{862}{174}}
\expandafter\gdef\csname r@eq:disc_res_aom\endcsname{{863}{174}}
\expandafter\gdef\csname r@eq:aom_disc_eg\endcsname{{864}{174}}
\expandafter\gdef\csname r@eq:one-loop-pentagon-cuts\endcsname{{869}{175}}
\expandafter\gdef\csname r@eq:2me_sols\endcsname{{870}{175}}
\expandafter\gdef\csname r@loop_positivty_second\endcsname{{872}{175}}
\expandafter\gdef\csname r@eq:SLS_2me\endcsname{{873}{176}}
\expandafter\gdef\csname r@subsec:geometric-landau-negative-geometries\endcsname{{6.4.2}{176}}
\expandafter\gdef\csname r@fig:laddersgeneral\endcsname{{60}{177}}
\expandafter\gdef\csname r@fig:ladder-LS-L1\endcsname{{61}{177}}
\expandafter\gdef\csname r@eq:ladder-cuts-thesis\endcsname{{875}{177}}
\expandafter\gdef\csname r@eq:ladder-function-thesis\endcsname{{877}{177}}
\expandafter\gdef\csname r@eq:Omega-513-thesis\endcsname{{878}{177}}
\expandafter\gdef\csname r@eq:Omega-613-614-thesis\endcsname{{879}{177}}
\expandafter\gdef\csname r@eq:hel_deg\endcsname{{880}{178}}
\expandafter\gdef\csname r@eq:example-boundary-thesis\endcsname{{881}{178}}
\expandafter\gdef\csname r@fig:L2.II_LD\endcsname{{62}{179}}
\expandafter\gdef\csname r@eq:CD-EF-cuts-thesis\endcsname{{882}{179}}
\expandafter\gdef\csname r@eq:pre-thesis\endcsname{{884}{179}}
\expandafter\gdef\csname r@eq:pinchmat-thesis\endcsname{{885}{179}}
\expandafter\gdef\csname r@fig:subleading-two-loop-diagrams\endcsname{{63}{180}}
\expandafter\gdef\csname r@eq:sing-a-thesis\endcsname{{886}{180}}
\expandafter\gdef\csname r@eq:partphy-thesis\endcsname{{888}{180}}
\expandafter\gdef\csname r@eq:CD-bracket-thesis\endcsname{{889}{180}}
\expandafter\gdef\csname r@eq:sing-b-thesis\endcsname{{891}{180}}
\expandafter\gdef\csname r@eq:ineq-thesis\endcsname{{892}{180}}
\expandafter\gdef\csname r@fig:selections\endcsname{{64}{181}}
\expandafter\gdef\csname r@eq:bad-shrink-thesis\endcsname{{895}{181}}
\expandafter\gdef\csname r@sec:Symbols and Bootstrap\endcsname{{6.5}{182}}
\expandafter\gdef\csname r@subsec:what-is-the-symbol\endcsname{{6.5.1}{182}}
\expandafter\gdef\csname r@eq:euler-dilog-identity\endcsname{{896}{182}}
\expandafter\gdef\csname r@eq:euler-dilog-symbol\endcsname{{897}{182}}
\expandafter\gdef\csname r@subsubsec:symbol-definition-basic-properties\endcsname{{6.5.1.1}{182}}
\expandafter\gdef\csname r@eq:symbol-recursive-definition\endcsname{{898}{182}}
\expandafter\gdef\csname r@eq:symbol-recursive\endcsname{{899}{182}}
\expandafter\gdef\csname r@eq:symbol-general-tensor\endcsname{{900}{183}}
\expandafter\gdef\csname r@eq:symbol-multiplicativity\endcsname{{901}{183}}
\expandafter\gdef\csname r@eq:symbol-log-dilog\endcsname{{903}{183}}
\expandafter\gdef\csname r@eq:symbol-classical-polylog\endcsname{{905}{183}}
\expandafter\gdef\csname r@subsubsec:symbol-integrability-reconstruction\endcsname{{6.5.1.2}{183}}
\expandafter\gdef\csname r@eq:symbol-integrability\endcsname{{908}{183}}
\expandafter\gdef\csname r@subsubsec:symbol-discontinuities-differentiation\endcsname{{6.5.1.3}{184}}
\expandafter\gdef\csname r@eq:symbol-derivative-action\endcsname{{909}{184}}
\expandafter\gdef\csname r@eq:symbol-discontinuity-action\endcsname{{910}{184}}
\expandafter\gdef\csname r@eq:symbol-diff-disc-summary\endcsname{{911}{184}}
\expandafter\gdef\csname r@eg:aomoto-symbol-discontinuity\endcsname{{6.14}{184}}
\expandafter\gdef\csname r@eq:S_I\endcsname{{913}{184}}
\expandafter\gdef\csname r@eq:aomoto-symbol-disc-result\endcsname{{915}{185}}
\expandafter\gdef\csname r@subsubsec:symbol-algebraic-letters\endcsname{{6.5.1.4}{185}}
\expandafter\gdef\csname r@eq:symbol-section-four-mass-delta\endcsname{{917}{185}}
\expandafter\gdef\csname r@eq:symbol-section-z-zbar\endcsname{{918}{185}}
\expandafter\gdef\csname r@eq:four-mass-basic-letters\endcsname{{919}{185}}
\expandafter\gdef\csname r@eq:four-mass-symbol\endcsname{{920}{185}}
\expandafter\gdef\csname r@eq:four-mass-odd-letters\endcsname{{921}{185}}
\expandafter\gdef\csname r@eq:four-mass-double-cover\endcsname{{922}{185}}
\expandafter\gdef\csname r@eq:sqrt-local-expansion\endcsname{{923}{186}}
\expandafter\gdef\csname r@eq:odd-square-root-letter\endcsname{{924}{186}}
\expandafter\gdef\csname r@eq:norm-square-root-letter\endcsname{{925}{186}}
\expandafter\gdef\csname r@eq:fact_sr\endcsname{{926}{186}}
\expandafter\gdef\csname r@eq:norm-definition-cover\endcsname{{930}{186}}
\expandafter\gdef\csname r@subsubsec:symbol-bootstrap-general\endcsname{{6.5.1.5}{187}}
\expandafter\gdef\csname r@eq:symbol-bootstrap-ansatz\endcsname{{932}{187}}
\expandafter\gdef\csname r@subsec:symbol-bootstrap-negative-geometries\endcsname{{6.5.2}{187}}
\expandafter\gdef\csname r@subsubsec:bootstrap-general-strategy\endcsname{{6.5.2.1}{187}}
\expandafter\gdef\csname r@eq:neg-geo-decomposition-bootstrap\endcsname{{933}{187}}
\expandafter\gdef\csname r@eq:neg-geo-symbol-ansatz\endcsname{{934}{188}}
\expandafter\gdef\csname r@subsubsec:bootstrap-differential-equation-constraints\endcsname{{6.5.2.2}{188}}
\expandafter\gdef\csname r@eq:Dij-bootstrap\endcsname{{936}{188}}
\expandafter\gdef\csname r@eq:DE-bootstrap\endcsname{{937}{188}}
\expandafter\gdef\csname r@eq:L2-DE-bootstrap\endcsname{{938}{188}}
\expandafter\gdef\csname r@eq:last-entry-DE-bootstrap\endcsname{{939}{188}}
\expandafter\gdef\csname r@fig:d1diagrams\endcsname{{65}{189}}
\expandafter\gdef\csname r@subsubsec:bootstrap-ladder-alphabets\endcsname{{6.5.2.3}{189}}
\expandafter\gdef\csname r@eq:five-point-two-loop-alphabet-bootstrap\endcsname{{940}{189}}
\expandafter\gdef\csname r@eq:d1-bootstrap\endcsname{{941}{189}}
\expandafter\gdef\csname r@subsubsec:bootstrap-six-point-two-loop\endcsname{{6.5.2.4}{189}}
\expandafter\gdef\csname r@eq:six-point-two-loop-ladder-decomp-bootstrap\endcsname{{942}{189}}
\expandafter\gdef\csname r@fig:d1planar\endcsname{{66}{190}}
\expandafter\gdef\csname r@tab:six-point-two-loop-bootstrap\endcsname{{5}{190}}
\expandafter\gdef\csname r@eq:6ptMand\endcsname{{944}{190}}
\expandafter\gdef\csname r@eq:six-point-spurious-bootstrap\endcsname{{945}{190}}
\expandafter\gdef\csname r@eq:six-point-collinear-bootstrap\endcsname{{946}{190}}
\expandafter\gdef\csname r@tab:six-point-two-loop-free-parameters\endcsname{{6}{191}}
\expandafter\gdef\csname r@fig:FD6pt\endcsname{{67}{191}}
\expandafter\gdef\csname r@subsubsec:bootstrap-five-point-three-loop\endcsname{{6.5.2.5}{191}}
\expandafter\gdef\csname r@eq:five-point-three-loop-decomp-bootstrap\endcsname{{948}{191}}
\expandafter\gdef\csname r@tab:five-point-three-loop-bootstrap\endcsname{{7}{192}}
\expandafter\gdef\csname r@tab:five-point-three-loop-free-parameters\endcsname{{8}{192}}
\expandafter\gdef\csname r@subsec:outlook-geometric-landau-analysis\endcsname{{6.5.3}{192}}
\expandafter\gdef\csname r@triangle_LS\endcsname{{68}{194}}
\expandafter\gdef\csname r@fig:T3ILandau\endcsname{{69(i)}{194}}
\expandafter\gdef\csname r@sub@fig:T3ILandau\endcsname{{(i)}{194}}
\expandafter\gdef\csname r@fig:elliptcishrink\endcsname{{69(ii)}{194}}
\expandafter\gdef\csname r@sub@fig:elliptcishrink\endcsname{{(ii)}{194}}
\expandafter\gdef\csname r@fig:elliptic\endcsname{{69}{194}}
\expandafter\gdef\csname r@ch:positivity-and-cluster-structures\endcsname{{7}{195}}
\expandafter\gdef\csname r@sec:landau-analysis-on-the-grassmannian\endcsname{{7.1}{196}}
\expandafter\gdef\csname r@subsec:From Momenta to Lines in Three-space\endcsname{{7.1.1}{196}}
\expandafter\gdef\csname r@eq:embedding\endcsname{{959}{196}}
\expandafter\gdef\csname r@eq:embedding-inner-product\endcsname{{960}{196}}
\expandafter\gdef\csname r@eq:embedding-distance\endcsname{{961}{196}}
\expandafter\gdef\csname r@eq:external-internal-propagator-embedding\endcsname{{964}{197}}
\expandafter\gdef\csname r@eq:internal-internal-propagator-embedding\endcsname{{965}{197}}
\expandafter\gdef\csname r@eq:face-mass-propagators\endcsname{{966}{197}}
\expandafter\gdef\csname r@eq:projective-embedding-integral\endcsname{{970}{197}}
\expandafter\gdef\csname r@eq:getLM\endcsname{{974}{198}}
\expandafter\gdef\csname r@eq:LMintegral\endcsname{{978}{198}}
\expandafter\gdef\csname r@fig:TP2\endcsname{{70}{199}}
\expandafter\gdef\csname r@eq:momentum-twistor-denominator-general\endcsname{{979}{199}}
\expandafter\gdef\csname r@eg:triple-pentagon-momentum-twistors\endcsname{{7.1}{199}}
\expandafter\gdef\csname r@eq:nine_prop\endcsname{{980}{199}}
\expandafter\gdef\csname r@eq:incid\endcsname{{981}{200}}
\expandafter\gdef\csname r@subsec:incidence-varieties-line-configurations\endcsname{{7.1.2}{200}}
\expandafter\gdef\csname r@eq:chapter7-plucker-quadric\endcsname{{983}{200}}
\expandafter\gdef\csname r@eq:chapter7-line-incidence-pairing\endcsname{{984}{200}}
\expandafter\gdef\csname r@eq:chapter7-incidence-variety\endcsname{{986}{201}}
\expandafter\gdef\csname r@eg:chapter7-three-lines\endcsname{{7.2}{201}}
\expandafter\gdef\csname r@eq:chapter7-K3-decomposition\endcsname{{987}{201}}
\expandafter\gdef\csname r@eq:hodgestar\endcsname{{988}{201}}
\expandafter\gdef\csname r@figure:fourlines\endcsname{{71}{202}}
\expandafter\gdef\csname r@eg:chapter7-triangulated-quadrilateral\endcsname{{7.3}{202}}
\expandafter\gdef\csname r@eq:chapter7-four-components\endcsname{{989}{202}}
\expandafter\gdef\csname r@eq:IK4\endcsname{{990}{202}}
\expandafter\gdef\csname r@eq:dim_CG\endcsname{{991}{203}}
\expandafter\gdef\csname r@eq:chapter7-affine-line\endcsname{{993}{203}}
\expandafter\gdef\csname r@eq:matricesST1\endcsname{{994}{203}}
\expandafter\gdef\csname r@eq:chapter7-distance-equality\endcsname{{995}{203}}
\expandafter\gdef\csname r@eq:chapter7-distance-map\endcsname{{996}{203}}
\expandafter\gdef\csname r@eq:chapter7-VG-distance-fiber-product\endcsname{{997}{203}}
\expandafter\gdef\csname r@eq:chapter7-laman-condition\endcsname{{998}{203}}
\expandafter\gdef\csname r@figure:rigidity\endcsname{{72}{204}}
\expandafter\gdef\csname r@eq:CS\endcsname{{999}{204}}
\expandafter\gdef\csname r@eg:K24\endcsname{{7.5}{205}}
\expandafter\gdef\csname r@eg:K33\endcsname{{7.6}{205}}
\expandafter\gdef\csname r@table:all-connected-graphs\endcsname{{9}{208}}
\expandafter\gdef\csname r@table:triangle-free-graphs\endcsname{{10}{208}}
\expandafter\gdef\csname r@eq:GTideal\endcsname{{1013}{208}}
\expandafter\gdef\csname r@eq:zenodo\endcsname{{1015}{208}}
\expandafter\gdef\csname r@figure:6-node-comps\endcsname{{73}{209}}
\expandafter\gdef\csname r@subsec:projections-leading-singularity-degree\endcsname{{7.1.3}{209}}
\expandafter\gdef\csname r@eq:chapter7-landau-map\endcsname{{1019}{209}}
\expandafter\gdef\csname r@eq:H_cons\endcsname{{1022}{210}}
\expandafter\gdef\csname r@eq:chapter7-LS-degree\endcsname{{1024}{210}}
\expandafter\gdef\csname r@eq:chapter7-multidegree-LS\endcsname{{1026}{211}}
\expandafter\gdef\csname r@eq:chapter7-CI-multidegree\endcsname{{1027}{211}}
\expandafter\gdef\csname r@eg:chapter7-four-mass-box-LS\endcsname{{7.8}{211}}
\expandafter\gdef\csname r@eg:chapter7-triangle-LS-degree\endcsname{{7.9}{212}}
\expandafter\gdef\csname r@eq:MD_K3\endcsname{{1030}{212}}
\expandafter\gdef\csname r@eg:chapter7-triangulated-pentagon-LS\endcsname{{7.10}{212}}
\expandafter\gdef\csname r@subsec:leading-discriminants-superleading-resultants\endcsname{{7.1.4}{212}}
\expandafter\gdef\csname r@eg:chapter7-pfd-two-quadratics\endcsname{{7.11}{213}}
\expandafter\gdef\csname r@eq:Q1Q2\endcsname{{1036}{213}}
\expandafter\gdef\csname r@eq:chapter7-LS-discriminant\endcsname{{1039}{214}}
\expandafter\gdef\csname r@eg:chapter7-four-mass-box-discriminant\endcsname{{7.12}{214}}
\expandafter\gdef\csname r@eq:chapter7-four-mass-box-discriminant\endcsname{{1040}{214}}
\expandafter\gdef\csname r@eq:chapter7-SLS-resultant\endcsname{{1041}{214}}
\expandafter\gdef\csname r@eg:chapter7-pentagon-resultant\endcsname{{7.13}{214}}
\expandafter\gdef\csname r@eq:chapter7-pentagon-resultant\endcsname{{1042}{214}}
\expandafter\gdef\csname r@eq:SLS_deg\endcsname{{1043}{215}}
\expandafter\gdef\csname r@eq:chapter7-LS-discriminant-degree-bound\endcsname{{1044}{215}}
\expandafter\gdef\csname r@eg:chapter7-pentabox-direct-resultant\endcsname{{7.14}{215}}
\expandafter\gdef\csname r@eq:chapter7-four-linear-pentabox\endcsname{{1045}{215}}
\expandafter\gdef\csname r@eq:chapter7-pentabox-quadric-two\endcsname{{1046}{215}}
\expandafter\gdef\csname r@eq:chapter7-ten-by-ten-resultant\endcsname{{1048}{216}}
\expandafter\gdef\csname r@eg:chapter7-triple-pentagon-discriminant\endcsname{{7.15}{216}}
\expandafter\gdef\csname r@eq:SLS_fact\endcsname{{1053}{216}}
\expandafter\gdef\csname r@eg:chapter7-double-pentagon-resultant\endcsname{{7.16}{216}}
\expandafter\gdef\csname r@eg:chapter7-triangle-SLS-resultant\endcsname{{7.17}{217}}
\expandafter\gdef\csname r@subsec:next-to-leading-singularities\endcsname{{7.1.5}{217}}
\expandafter\gdef\csname r@eq:chapter7-multisectional-genus\endcsname{{1058}{218}}
\expandafter\gdef\csname r@eg:chapter7-one-loop-NLS\endcsname{{7.18}{218}}
\expandafter\gdef\csname r@eg:chapter7-NLS-double-box\endcsname{{7.19}{218}}
\expandafter\gdef\csname r@eq:chapter7-gK2\endcsname{{1060}{218}}
\expandafter\gdef\csname r@eg:chapter7-NLS-three-lines\endcsname{{7.20}{219}}
\expandafter\gdef\csname r@sec:recursive-landau-analysis\endcsname{{7.2}{219}}
\expandafter\gdef\csname r@eq:on-shell-space-with-H\endcsname{{1068}{219}}
\expandafter\gdef\csname r@eq:landau-map-with-H\endcsname{{1069}{220}}
\expandafter\gdef\csname r@eq:recursive-L1-L2\endcsname{{1073}{220}}
\expandafter\gdef\csname r@eq:L1-finite-fiber\endcsname{{1075}{220}}
\expandafter\gdef\csname r@eq:general-substitution-map\endcsname{{1076}{220}}
\expandafter\gdef\csname r@eq:targ\endcsname{{1077}{220}}
\expandafter\gdef\csname r@fig:recursive-landau-general\endcsname{{74}{221}}
\expandafter\gdef\csname r@eq:recursive-fiber-decomposition-v2\endcsname{{1078}{221}}
\expandafter\gdef\csname r@eq:branch-locus-recursion\endcsname{{1080}{221}}
\expandafter\gdef\csname r@subsec:box-substitution-maps\endcsname{{7.2.2}{221}}
\expandafter\gdef\csname r@eq:X-of-x-box-v2\endcsname{{1081}{221}}
\expandafter\gdef\csname r@eq:box-quadratic-v2\endcsname{{1082}{221}}
\expandafter\gdef\csname r@eq:box-chain-coefficients-v2\endcsname{{1083}{222}}
\expandafter\gdef\csname r@eq:box-roots-v2\endcsname{{1084}{222}}
\expandafter\gdef\csname r@eq:Lpm-normalization-v2\endcsname{{1085}{222}}
\expandafter\gdef\csname r@eq:four-mass-box-substitution-maps-v2\endcsname{{1086}{222}}
\expandafter\gdef\csname r@eq:fg-polynomial-system-v2\endcsname{{1087}{222}}
\expandafter\gdef\csname r@eq:polynomial-ls-recursion-v2\endcsname{{1088}{222}}
\expandafter\gdef\csname r@eg:pentabox-recursion\endcsname{{7.21}{222}}
\expandafter\gdef\csname r@fig:pentabox-recursion\endcsname{{75}{223}}
\expandafter\gdef\csname r@eq:pentabox-discriminant-prefactor-v2\endcsname{{1093}{223}}
\expandafter\gdef\csname r@eq:pentabox-recursive-formula-v2\endcsname{{1094}{223}}
\expandafter\gdef\csname r@subsec:recursive-factorization-discriminants-resultants\endcsname{{7.2.3}{223}}
\expandafter\gdef\csname r@eq:ls-discriminant-recursion-v2\endcsname{{1095}{223}}
\expandafter\gdef\csname r@eq:box-recursion-ls-v2\endcsname{{1096}{224}}
\expandafter\gdef\csname r@eq:polynomial-sls-recursion-v2\endcsname{{1098}{224}}
\expandafter\gdef\csname r@eg:double-pentagon-recursion\endcsname{{7.22}{224}}
\expandafter\gdef\csname r@eq:pentagon-resultant-normalization-v2\endcsname{{1100}{224}}
\expandafter\gdef\csname r@eq:double-pentagon-resultant-normalization-v2\endcsname{{1101}{224}}
\expandafter\gdef\csname r@eq:double-pentagon-recursive-formula-v2\endcsname{{1102}{224}}
\expandafter\gdef\csname r@eq:sls-resultant-recursion-v2\endcsname{{1103}{224}}
\expandafter\gdef\csname r@eq:box-recursion-sls-v2\endcsname{{1104}{224}}
\expandafter\gdef\csname r@subsec:trees-of-loops-recursion\endcsname{{7.2.4}{225}}
\expandafter\gdef\csname r@eq:tree-multidegree-v2\endcsname{{1106}{225}}
\expandafter\gdef\csname r@eq:tree-iterated-ls-recursion-v2\endcsname{{1108}{225}}
\expandafter\gdef\csname r@eg:path-trees-recursion\endcsname{{7.23}{225}}
\expandafter\gdef\csname r@eg:recursive-triangle\endcsname{{7.24}{226}}
\expandafter\gdef\csname r@eq:recursive-triangle-v2\endcsname{{1113}{226}}
\expandafter\gdef\csname r@subsec:recursive-limitations-extensions\endcsname{{7.2.5}{226}}
\expandafter\gdef\csname r@eq:cycle-fell-definition-v2\endcsname{{1118}{226}}
\expandafter\gdef\csname r@eq:cycle-discriminant-as-univariate-discriminant\endcsname{{1119}{227}}
\expandafter\gdef\csname r@eg:triple-pentagon-cycle\endcsname{{7.25}{227}}
\expandafter\gdef\csname r@eq:f3-recursive-product-v2\endcsname{{1121}{227}}
\expandafter\gdef\csname r@sec:reality-positivity-positroids\endcsname{{7.3}{227}}
\expandafter\gdef\csname r@eq:dual-variables-positive-section\endcsname{{1123}{227}}
\expandafter\gdef\csname r@eq:planar-mandelstam-dual-distance\endcsname{{1124}{227}}
\expandafter\gdef\csname r@eq:positive-momentum-twistors-lines\endcsname{{1125}{228}}
\expandafter\gdef\csname r@eq:dual-distance-four-bracket\endcsname{{1126}{228}}
\expandafter\gdef\csname r@eq:dual-conformal-cross-ratio-positive\endcsname{{1127}{228}}
\expandafter\gdef\csname r@eq:positive-kinematic-space\endcsname{{1128}{228}}
\expandafter\gdef\csname r@eq:positive-four-brackets\endcsname{{1129}{228}}
\expandafter\gdef\csname r@eq:positive-line-configuration\endcsname{{1130}{228}}
\expandafter\gdef\csname r@eq:positive-section-four-mass-gram\endcsname{{1132}{229}}
\expandafter\gdef\csname r@eq:four-mass-cross-ratio-discriminant\endcsname{{1133}{229}}
\expandafter\gdef\csname r@eq:four-mass-cross-ratios\endcsname{{1134}{229}}
\expandafter\gdef\csname r@eq:positive-section-pentagon-gram\endcsname{{1135}{229}}
\expandafter\gdef\csname r@sec:line-configurations-positroids-amplituhedron-map\endcsname{{7.3.3}{230}}
\expandafter\gdef\csname r@eq:component-bicoloring-positive-section\endcsname{{1136}{230}}
\expandafter\gdef\csname r@fig:grassmann-lines-vrc-positive-section\endcsname{{76}{231}}
\expandafter\gdef\csname r@fig:grassmann-graph-k3-positive-section\endcsname{{77}{232}}
\expandafter\gdef\csname r@eq:M_Z\endcsname{{1144}{232}}
\expandafter\gdef\csname r@eq:vrc-landau-fiber-identification\endcsname{{1147}{232}}
\expandafter\gdef\csname r@eq:positroid-kernel-condition\endcsname{{1152}{233}}
\expandafter\gdef\csname r@sec:rationality-cluster-structures\endcsname{{7.4}{235}}
\expandafter\gdef\csname r@fig:cluster-quiver-mutation\endcsname{{78}{236}}
\expandafter\gdef\csname r@subsec:cluster-algebras-amplitude-singularities\endcsname{{7.4.1}{236}}
\expandafter\gdef\csname r@eq:cluster-exchange-relation\endcsname{{1160}{236}}
\expandafter\gdef\csname r@eq:cl_chart\endcsname{{1161}{236}}
\expandafter\gdef\csname r@eq:cluster-torus-dlog-form\endcsname{{1163}{237}}
\expandafter\gdef\csname r@eq:A2_alph\endcsname{{1179}{239}}
\expandafter\gdef\csname r@eq:hexagon-alphabet\endcsname{{1182}{240}}
\expandafter\gdef\csname r@eq:four-mass-discriminant-cluster-discussion\endcsname{{1183}{241}}
\expandafter\gdef\csname r@subsec:rational-landau-singularities-degenerations\endcsname{{7.4.2}{242}}
\expandafter\gdef\csname r@eq:three-mass-perfect-square\endcsname{{1185}{242}}
\expandafter\gdef\csname r@eq:three-mass-solutions\endcsname{{1186}{242}}
\expandafter\gdef\csname r@eq:three-mass-solutions-twistors\endcsname{{1187}{242}}
\expandafter\gdef\csname r@eq:external-incidence-variety-Hu\endcsname{{1191}{243}}
\expandafter\gdef\csname r@eq:L-triangle-degeneration\endcsname{{1197}{244}}
\expandafter\gdef\csname r@eq:rational-discriminant-external-factorization\endcsname{{1206}{245}}
\expandafter\gdef\csname r@eq:rational-discriminant-mixed-factorization\endcsname{{1207}{246}}
\expandafter\gdef\csname r@eq:rational-triangle-eight-solutions\endcsname{{1209}{246}}
\expandafter\gdef\csname r@subsec:cluster-box-promotion-map\endcsname{{7.4.3}{247}}
\expandafter\gdef\csname r@eq:cluster-substitution-map-general\endcsname{{1216}{247}}
\expandafter\gdef\csname r@fig:promotion_4mb_K_3\endcsname{{79}{248}}
\expandafter\gdef\csname r@eq:promo_map\endcsname{{1221}{248}}
\expandafter\gdef\csname r@fig:recursion_3mb\endcsname{{80}{249}}
\expandafter\gdef\csname r@eq:3mb-solutions-cluster-section\endcsname{{1224}{249}}
\expandafter\gdef\csname r@eq:3mb-substitution-maps\endcsname{{1225}{249}}
\expandafter\gdef\csname r@fig:rat-pentabox-cluster\endcsname{{81}{250}}
\expandafter\gdef\csname r@eq:pentabox-cluster-recursion\endcsname{{1226}{250}}
\expandafter\gdef\csname r@eq:pentabox-rational-cluster-factors\endcsname{{1227}{250}}
\expandafter\gdef\csname r@eq:rational-triangle-disc-split-cluster\endcsname{{1228}{250}}
\expandafter\gdef\csname r@eq:discr-tr-fact-explicit\endcsname{{1230}{251}}
\expandafter\gdef\csname r@eq:discr-tr-fact-explicit-mixed\endcsname{{1231}{251}}
\expandafter\gdef\csname r@eq:kissing-triangle-promotion-map\endcsname{{1233}{251}}
\expandafter\gdef\csname r@eq:kissing-triangle-lines\endcsname{{1234}{251}}
\expandafter\gdef\csname r@fig:chain-triangles-cluster-promotion\endcsname{{82}{252}}
\expandafter\gdef\csname r@sec:positivity-cluster-outlook\endcsname{{7.5}{252}}
\expandafter\gdef\csname r@sec:Conclusion\endcsname{{8}{254}}
\bibcite{Abreu:2022mfk}{{1}{2017}{{Abreu et~al.}}{{Abreu, Britto, Duhr and Gardi}}}
\bibcite{AdamsWeinzierl2018}{{2}{2018}{{Adams and Weinzierl}}{{}}}
\bibcite{ait2023linear}{{3}{2023a}{{Ait El~Manssour et~al.}}{{Ait El~Manssour, H{\"a}rk{\"o}nen and Sturmfels}}}
\bibcite{Liner_PDE}{{4}{2023b}{{Ait El~Manssour et~al.}}{{Ait El~Manssour, H{\"a}rk{\"o}nen and Sturmfels}}}
\bibcite{Alday:2011ga}{{5}{2011}{{Alday et~al.}}{{Alday, Buchbinder and Tseytlin}}}
\bibcite{Alday:2009zm}{{6}{2010}{{Alday et~al.}}{{Alday, Henn, Plefka and Schuster}}}
\bibcite{Alday:2013ip}{{7}{2013a}{{Alday et~al.}}{{Alday, Henn and Sikorowski}}}
\bibcite{Alday:2012hy}{{8}{2013b}{{Alday et~al.}}{{Alday, Heslop and Sikorowski}}}
\bibcite{Alday:2007hr}{{9}{2007}{{Alday and Maldacena}}{{}}}
\bibcite{AnastasiouBernDixonKosower2003}{{10}{2003}{{Anastasiou et~al.}}{{Anastasiou, Bern, Dixon and Kosower}}}
\bibcite{AnastasiouBrittoFengKunsztMastrolia2006}{{11}{2007}{{Anastasiou et~al.}}{{Anastasiou, Britto, Feng, Kunszt and Mastrolia}}}
\bibcite{Anastasiou:2003kj}{{12}{2002}{{Anastasiou and Melnikov}}{{}}}
\bibcite{Aneesh:2019cvt}{{13}{2020}{{Aneesh et~al.}}{{Aneesh, Banerjee, Jagadale, Rajan, Laddha and Mahato}}}
\bibcite{Aomoto1982}{{14}{1982}{{Aomoto}}{{}}}
\bibcite{AomotoKita2011}{{15}{2011}{{Aomoto and Kita}}{{}}}
\bibcite{ABHY_original}{{16}{2018a}{{Arkani-Hamed et~al.}}{{Arkani-Hamed, Bai, He and Yan}}}
\bibcite{Positive_geometries}{{17}{2017a}{{Arkani-Hamed et~al.}}{{Arkani-Hamed, Bai and Lam}}}
\bibcite{ArkaniHamedBaumannHillmanJoyceLeePimentel2025}{{18}{2025a}{{Arkani-Hamed et~al.}}{{Arkani-Hamed, Baumann, Hillman, Joyce, Lee and Pimentel}}}
\bibcite{Cosmological_polytopes}{{19}{2017b}{{Arkani-Hamed et~al.}}{{Arkani-Hamed, Benincasa and Postnikov}}}
\bibcite{ArkaniHamed:2010kv}{{20}{2011}{{Arkani-Hamed et~al.}}{{Arkani-Hamed, Bourjaily, Cachazo, Caron-Huot and Trnka}}}
\bibcite{Grassmannian}{{21}{2016}{{Arkani-Hamed et~al.}}{{Arkani-Hamed, Bourjaily, Cachazo, Goncharov, Postnikov and Trnka}}}
\bibcite{ArkaniHamed:2009dn}{{22}{2010a}{{Arkani-Hamed et~al.}}{{Arkani-Hamed, Cachazo, Cheung and Kaplan}}}
\bibcite{ArkaniHamed:2008gz}{{23}{2010b}{{Arkani-Hamed et~al.}}{{Arkani-Hamed, Cachazo and Kaplan}}}
\bibcite{ArkaniHamed:2024surfaceYM}{{24}{2025b}{{Arkani-Hamed et~al.}}{{Arkani-Hamed, Cao, Dong, Figueiredo and He}}}
\bibcite{Cosmoehdra}{{25}{2024}{{Arkani-Hamed et~al.}}{{Arkani-Hamed, Figueiredo and Vaz{\~a}o}}}
\bibcite{Stringy_Can_Forms}{{26}{2021a}{{Arkani-Hamed et~al.}}{{Arkani-Hamed, He and Lam}}}
\bibcite{AHLTBinary}{{27}{2023}{{Arkani-Hamed et~al.}}{{Arkani-Hamed, He, Lam and Thomas}}}
\bibcite{ArkaniHamedHennTrnka2021}{{28}{2022}{{Arkani-Hamed et~al.}}{{Arkani-Hamed, Henn and Trnka}}}
\bibcite{positive_amplitudes}{{29}{2015}{{Arkani-Hamed et~al.}}{{Arkani-Hamed, Hodges and Trnka}}}
\bibcite{ArkaniHamed:2008yf}{{30}{2008}{{Arkani-Hamed and Kaplan}}{{}}}
\bibcite{ALS}{{31}{2021b}{{Arkani-Hamed et~al.}}{{Arkani-Hamed, Lam and Spradlin}}}
\bibcite{Arkani-Hamed:2018rsk}{{32}{2019}{{Arkani-Hamed et~al.}}{{Arkani-Hamed, Langer, Yelleshpur~Srikant and Trnka}}}
\bibcite{Arkani_Hamed_2018}{{33}{2018b}{{Arkani-Hamed et~al.}}{{Arkani-Hamed, Thomas and Trnka}}}
\bibcite{the_amplituhedron}{{34}{2014a}{{Arkani-Hamed and Trnka}}{{}}}
\bibcite{Arkani-Hamed:2013kca}{{35}{2014b}{{Arkani-Hamed and Trnka}}{{}}}
\bibcite{Arkani-Hamed:2017ahv}{{36}{2017}{{Arkani-Hamed and Yuan}}{{}}}
\bibcite{1Garding_1970}{{37}{1970}{{Atiyah et~al.}}{{Atiyah, Bott and Gårding}}}
\bibcite{Baadsgaard:2015twa}{{38}{2015}{{Baadsgaard et~al.}}{{Baadsgaard, Bjerrum-Bohr, Bourjaily, Damgaard and Feng}}}
\bibcite{Banerjee:2018tun}{{39}{2019}{{Banerjee et~al.}}{{Banerjee, Laddha and Raman}}}
\bibcite{baoHe2019m2}{{40}{2019}{{Bao and He}}{{}}}
\bibcite{Beck2004}{{41}{2004}{{Beck}}{{}}}
\bibcite{BeisertEdenStaudacher2006}{{42}{2007}{{Beisert et~al.}}{{Beisert, Eden and Staudacher}}}
\bibcite{Beisert:2010jr}{{43}{2012}{{Beisert et~al.}}{{}}}
\bibcite{BendleBoehmDeckerGeorgoudisPfreundtRahnWasserZhang2020}{{44}{2020}{{Bendle et~al.}}{{Bendle, B{\"o}hm, Decker, Georgoudis, Pfreundt, Rahn, Wasser and Zhang}}}
\bibcite{BenincasaDian2025}{{45}{2025}{{Benincasa and Dian}}{{}}}
\bibcite{Berends:1987me}{{46}{1987}{{Berends and Giele}}{{}}}
\bibcite{BFZ}{{47}{1996}{{Berenstein et~al.}}{{Berenstein, Fomin and Zelevinsky}}}
\bibcite{BerghoffPanzer2025}{{48}{2025}{{Berghoff and Panzer}}{{}}}
\bibcite{Bern:2008qj}{{49}{2008}{{Bern et~al.}}{{Bern, Carrasco and Johansson}}}
\bibcite{Bern:1994zx}{{50}{1994}{{Bern et~al.}}{{Bern, Dixon, Dunbar and Kosower}}}
\bibcite{BernDixonDunbarKosower1994}{{51}{1995}{{Bern et~al.}}{{Bern, Dixon, Dunbar and Kosower}}}
\bibcite{BernDixonKosower1996}{{52}{1996}{{Bern et~al.}}{{Bern, Dixon and Kosower}}}
\bibcite{Bern:1994cg}{{53}{1998}{{Bern et~al.}}{{Bern, Dixon and Kosower}}}
\bibcite{Bern:2007dw}{{54}{2007}{{Bern et~al.}}{{Bern, Dixon and Kosower}}}
\bibcite{BernDixonSmirnov2005}{{55}{2005}{{Bern et~al.}}{{Bern, Dixon and Smirnov}}}
\bibcite{BesierFesti2020}{{56}{2021}{{Besier and Festi}}{{}}}
\bibcite{Besier:2018jen}{{57}{2019}{{Besier et~al.}}{{Besier, van Straten and Weinzierl}}}
\bibcite{BesierWasserWeinzierl2019}{{58}{2020}{{Besier et~al.}}{{Besier, Wasser and Weinzierl}}}
\bibcite{Bjoerk1979}{{59}{1979}{{Bj{\"o}rk}}{{}}}
\bibcite{BjorkenLandau1959}{{60}{1959}{{Bjorken}}{{}}}
\bibcite{Bloch:2005bh}{{61}{2006}{{Bloch et~al.}}{{Bloch, Esnault and Kreimer}}}
\bibcite{bochnak2010real}{{62}{2010}{{Bochnak et~al.}}{{Bochnak, Coste and Roy}}}
\bibcite{BochnakCosteRoy1998}{{63}{1998}{{Bochnak et~al.}}{{Bochnak, Coste and Roy}}}
\bibcite{Boels:2010nw}{{64}{2007}{{Boels et~al.}}{{Boels, Mason and Skinner}}}
\bibcite{BognerWeinzierl}{{65}{2010}{{Bogner and Weinzierl}}{{}}}
\bibcite{BoehmWittmannWuXuZhang2020}{{66}{2020}{{B{\"o}hm et~al.}}{{B{\"o}hm, Wittmann, Wu, Xu and Zhang}}}
\bibcite{BolliniGiambiagi1972}{{67}{1972}{{Bollini and Giambiagi}}{{}}}
\bibcite{Bosch2013AlgebraicGeometry}{{68}{2013}{{Bosch}}{{}}}
\bibcite{Bourjaily:2010kw}{{69}{2010}{{Bourjaily}}{{}}}
\bibcite{Bourjaily:2012gy}{{70}{2012}{{Bourjaily}}{{}}}
\bibcite{BourjailyHerrmannLangerMcLeodTrnka2019}{{71}{2019}{{Bourjaily et~al.}}{{Bourjaily, Herrmann, Langer, McLeod and Trnka}}}
\bibcite{BourjailyHerrmannLangerMcLeodTrnka2020}{{72}{2020}{{Bourjaily et~al.}}{{Bourjaily, Herrmann, Langer, McLeod and Trnka}}}
\bibcite{BourjailyHerrmannTrnka2017}{{73}{2017}{{Bourjaily et~al.}}{{Bourjaily, Herrmann and Trnka}}}
\bibcite{BourjailyKalyanapuramLangerPatatoukos2021}{{74}{2021}{{Bourjaily et~al.}}{{Bourjaily, Kalyanapuram, Langer and Patatoukos}}}
\bibcite{BourjailyTrnka2015}{{75}{2015}{{Bourjaily and Trnka}}{{}}}
\bibcite{BourjailyVerguVonHippel2023}{{76}{2023}{{Bourjaily et~al.}}{{Bourjaily, Vergu and von Hippel}}}
\bibcite{brakensiek2024rigidity}{{77}{2024}{{Brakensiek et~al.}}{{Brakensiek, Dhar, Gao, Gopi and Larson}}}
\bibcite{Brandhuber:2007yx}{{78}{2008}{{Brandhuber et~al.}}{{Brandhuber, Heslop and Travaglini}}}
\bibcite{BrandhuberHeslopTravaglini2009}{{79}{2009}{{Brandhuber et~al.}}{{Brandhuber, Heslop and Travaglini}}}
\bibcite{BT}{{80}{2018}{{Breiding and Timme}}{{}}}
\bibcite{Brink:1976bc}{{81}{1977}{{Brink et~al.}}{{Brink, Schwarz and Scherk}}}
\bibcite{Britto:2004nc}{{82}{2005a}{{Britto et~al.}}{{Britto, Cachazo and Feng}}}
\bibcite{Britto:2004ap}{{83}{2005b}{{Britto et~al.}}{{Britto, Cachazo and Feng}}}
\bibcite{Britto:2005fq}{{84}{2005c}{{Britto et~al.}}{{Britto, Cachazo, Feng and Witten}}}
\bibcite{Broedel:2018iwv}{{85}{2018a}{{Broedel et~al.}}{{Broedel, Duhr, Dulat, Penante and Tancredi}}}
\bibcite{BroedelDuhrDulatTancredi2018}{{86}{2018b}{{Broedel et~al.}}{{Broedel, Duhr, Dulat and Tancredi}}}
\bibcite{Brown:2009qja}{{87}{2009a}{{Brown}}{{}}}
\bibcite{Brown2012}{{88}{2012}{{Brown}}{{}}}
\bibcite{Brown2015}{{89}{2017}{{Brown}}{{}}}
\bibcite{BrownDuhr2020DlogNotPolylog}{{90}{2022}{{Brown and Duhr}}{{}}}
\bibcite{Brown:PG_Hodge}{{91}{2025}{{Brown and Dupont}}{{}}}
\bibcite{Brown2009}{{92}{2009b}{{Brown}}{{}}}
\bibcite{Brown2009_alt1}{{93}{2010}{{Brown}}{{}}}
\bibcite{BrownHennMazzucchelliTrnka2025}{{94}{2025}{{Brown et~al.}}{{Brown, Henn, Mazzucchelli and Trnka}}}
\bibcite{BrownOktemParanjapeTrnka2024}{{95}{2024}{{Brown et~al.}}{{Brown, Oktem, Paranjape and Trnka}}}
\bibcite{bruser2025geometry}{{96}{2025}{{Br{\"u}ser and Weigert}}{{}}}
\bibcite{BFS}{{97}{2021}{{Brysiewicz et~al.}}{{Brysiewicz, Fevola and Sturmfels}}}
\bibcite{branden2014hyperbolicity}{{98}{2014}{{Brändén}}{{}}}
\bibcite{busemann1961convexity}{{99}{1961}{{Busemann}}{{}}}
\bibcite{Cachazo:2012kg}{{100}{2013}{{Cachazo and Geyer}}{{}}}
\bibcite{CachazoHeYuan2014}{{101}{2014}{{Cachazo et~al.}}{{Cachazo, He and Yuan}}}
\bibcite{Cachazo:2008vp}{{102}{2008}{{Cachazo and Skinner}}{{}}}
\bibcite{Cao:2025ymrecursion}{{103}{2025}{{Cao and Zhu}}{{}}}
\bibcite{CapuanoFerroLukowskiPalazio2025}{{104}{2025}{{Capuano et~al.}}{{Capuano, Ferro, {\L }ukowski and Palazio}}}
\bibcite{CapuanoFerroLukowskiPalazioZhang2026GeneralisedClusterAdjacency}{{105}{2026}{{Capuano et~al.}}{{Capuano, Ferro, {\L }ukowski, Palazio and Zhang}}}
\bibcite{CaronHuot:2011kk}{{106}{2011}{{Caron-Huot}}{{}}}
\bibcite{CaronHuotCorreiaGiroux2025}{{107}{2025}{{Caron-Huot et~al.}}{{Caron-Huot, Correia and Giroux}}}
\bibcite{CaronHuotDixonDulatEtAl2020}{{108}{2020}{{Caron-Huot et~al.}}{{Caron-Huot, Dixon, Drummond, Dulat, Foster, G{\"u}rdo{\u g}an, von Hippel, McLeod and Papathanasiou}}}
\bibcite{CaronHuotDixonMcLeodVonHippel2016}{{109}{2016}{{Caron-Huot et~al.}}{{Caron-Huot, Dixon, McLeod and von Hippel}}}
\bibcite{Carrolo:2025pue}{{110}{2025}{{Carr{\^o}lo et~al.}}{{Carr{\^o}lo, Chicherin, Henn, Yang and Zhang}}}
\bibcite{Carrolo:2026qpu}{{111}{2026}{{Carr{\^o}lo et~al.}}{{Carr{\^o}lo, Chicherin, Henn, Yang and Zhang}}}
\bibcite{CataneseHostenKhetanSturmfels2006}{{112}{2006}{{Catanese et~al.}}{{Catanese, Ho\c {s}ten, Khetan and Sturmfels}}}
\bibcite{Catani:2008xa}{{113}{2008}{{Catani et~al.}}{{Catani, Gleisberg, Krauss, Rodrigo and Winter}}}
\bibcite{CGR}{{114}{2019}{{Charlton et~al.}}{{Charlton, Gangl and Radchenko}}}
\bibcite{Chen2013}{{115}{2013}{{Chen et~al.}}{{Chen, Davenport, May, Moreno~Maza, Xia and Xiao}}}
\bibcite{Chen1977}{{116}{1977}{{Chen}}{{}}}
\bibcite{Cheung:2009dc}{{117}{2009}{{Cheung and O'Connell}}{{}}}
\bibcite{Chew:1961ev}{{118}{1962}{{Chew}}{{}}}
\bibcite{neg_geom_pos}{{119}{2025}{{Chicherin et~al.}}{{Chicherin, Henn, Trnka and Zhang}}}
\bibcite{Chicherin:2022zxo}{{120}{2022a}{{Chicherin and Henn}}{{}}}
\bibcite{ChicherinHenn2022}{{121}{2022b}{{Chicherin and Henn}}{{}}}
\bibcite{ChicherinHennMazzucchelliTrnkaYangZhang2026}{{122}{2026}{{Chicherin et~al.}}{{Chicherin, Henn, Mazzucchelli, Trnka, Yang and Zhang}}}
\bibcite{Chicherin:2020umh}{{123}{2021}{{Chicherin et~al.}}{{Chicherin, Henn and Papathanasiou}}}
\bibcite{Choquet}{{124}{1969}{{Choquet}}{{}}}
\bibcite{ColemanNorton1965}{{125}{1965}{{Coleman and Norton}}{{}}}
\bibcite{Collins1984}{{126}{1984}{{Collins}}{{}}}
\bibcite{Conde:2012wb}{{127}{2012}{{Conde and Rajabi}}{{}}}
\bibcite{CorreiaGirouxMizera2026}{{128}{2026}{{Correia et~al.}}{{Correia, Giroux and Mizera}}}
\bibcite{Costa:2011mg}{{129}{2011}{{Costa et~al.}}{{Costa, Penedones, Poland and Rychkov}}}
\bibcite{CoxLittleOShea2005}{{130}{2005}{{Cox et~al.}}{{Cox, Little and O'Shea}}}
\bibcite{Crespo_Ruiz2023}{{131}{2023}{{Crespo~Ruiz and Santos}}{{}}}
\bibcite{Cutkosky1960}{{132}{1960}{{Cutkosky}}{{}}}
\bibcite{damgaardFerroLukowskiParisi2021associahedron}{{133}{2021}{{Damgaard et~al.}}{{Damgaard, Ferro, {\L }ukowski and Parisi}}}
\bibcite{Damgaard:2019ztj}{{134}{2019}{{Damgaard et~al.}}{{Damgaard, Ferro, Lukowski and Parisi}}}
\bibcite{DAndreaDickenstein2001}{{135}{2001}{{D'Andrea and Dickenstein}}{{}}}
\bibcite{De:2024bpk}{{136}{2025}{{De et~al.}}{{De, Pavlov, Spradlin and Volovich}}}
\bibcite{DePokraka2024}{{137}{2024}{{De and Pokraka}}{{}}}
\bibcite{deLoeraRambauSantos2010triangulations}{{138}{2010}{{De~Loera et~al.}}{{De~Loera, Rambau and Santos}}}
\bibcite{DelDuca:2009au}{{139}{2010a}{{Del~Duca et~al.}}{{Del~Duca, Duhr and Smirnov}}}
\bibcite{DelDuca:2010zg}{{140}{2010b}{{Del~Duca et~al.}}{{Del~Duca, Duhr and Smirnov}}}
\bibcite{DennenPrlinaSpradlinStanojevicVolovich2017}{{141}{2017}{{Dennen et~al.}}{{Dennen, Prlina, Spradlin, Stanojevic and Volovich}}}
\bibcite{DennenSpradlinVolovich2016}{{142}{2016}{{Dennen et~al.}}{{Dennen, Spradlin and Volovich}}}
\bibcite{GabriBlog}{{143}{2019}{{Dian}}{{}}}
\bibcite{Dian_2023}{{144}{2023}{{Dian et~al.}}{{Dian, Heslop and Stewart}}}
\bibcite{Dian:2024hil}{{145}{2025}{{Dian et~al.}}{{Dian, Mazzucchelli and Tellander}}}
\bibcite{Dimca2004}{{146}{2004}{{Dimca}}{{}}}
\bibcite{Dirac1936}{{147}{1936}{{Dirac}}{{}}}
\bibcite{Dixon:1996wi}{{148}{1996}{{Dixon}}{{}}}
\bibcite{Dixon:2011xs}{{149}{2011}{{Dixon}}{{}}}
\bibcite{DixonEtAl2017}{{150}{2017a}{{Dixon et~al.}}{{Dixon, Drummond, Harrington, McLeod, Papathanasiou and Spradlin}}}
\bibcite{hexagon}{{151}{2011}{{Dixon et~al.}}{{Dixon, Drummond and Henn}}}
\bibcite{DixonDrummondHenn2011}{{152}{2012}{{Dixon et~al.}}{{Dixon, Drummond and Henn}}}
\bibcite{DixonGurdoganLiuMcLeodWilhelm2023}{{153}{2023}{{Dixon et~al.}}{{Dixon, G{\"u}rdo{\u g}an, Liu, McLeod and Wilhelm}}}
\bibcite{DixonGurdoganMcLeodWilhelm2022}{{154}{2022}{{Dixon et~al.}}{{Dixon, G{\"u}rdo{\u g}an, McLeod and Wilhelm}}}
\bibcite{Dixon:2016apl}{{155}{2017b}{{Dixon et~al.}}{{Dixon, von Hippel, McLeod and Trnka}}}
\bibcite{Dixon:2026ipt}{{156}{2026}{{Dixon et~al.}}{{Dixon, Oktem, Paranjape, Trnka, Xu and Zhang}}}
\bibcite{DlapaHelmerPapathanasiouTellander2023}{{157}{2023}{{Dlapa et~al.}}{{Dlapa, Helmer, Papathanasiou and Tellander}}}
\bibcite{DrtonSturmfelsSullivant2009}{{158}{2009}{{Drton et~al.}}{{Drton, Sturmfels and Sullivant}}}
\bibcite{DrummondFosterGurdoganClusterAdjacency2018}{{159}{2018}{{Drummond et~al.}}{{Drummond, Foster and G{\"u}rdo{\u g}an}}}
\bibcite{DrummondFosterGurdoganHarrington2019}{{160}{2019}{{Drummond et~al.}}{{Drummond, Foster, G{\"u}rdo{\u g}an and Kalousios}}}
\bibcite{DrummondFosterGurdoganKalousios2020}{{161}{2020}{{Drummond et~al.}}{{Drummond, Foster, G{\"u}rdo{\u g}an and Kalousios}}}
\bibcite{DrummondFosterGurrieri2018}{{162}{2021a}{{Drummond et~al.}}{{Drummond, Foster, G{\"u}rdo{\u g}an and Kalousios}}}
\bibcite{DrummondFosterGurdoganKalousios2021}{{163}{2021b}{{Drummond et~al.}}{{Drummond, Foster, G{\"u}rdo{\u g}an and Kalousios}}}
\bibcite{DrummondGurdoganLi2026TropicalSymmetries}{{164}{2026}{{Drummond et~al.}}{{Drummond, G{\"u}rdo{\u g}an and Li}}}
\bibcite{Drummond:2010qh}{{165}{2010}{{Drummond and Ferro}}{{}}}
\bibcite{DrummondHennKorchemskySokatchev2008}{{166}{2008}{{Drummond et~al.}}{{Drummond, Henn, Korchemsky and Sokatchev}}}
\bibcite{DrummondHennKorchemskySokatchev2007}{{167}{2010a}{{Drummond et~al.}}{{Drummond, Henn, Korchemsky and Sokatchev}}}
\bibcite{Drummond:2008vq}{{168}{2010b}{{Drummond et~al.}}{{Drummond, Henn, Korchemsky and Sokatchev}}}
\bibcite{Drummond:2007aua}{{169}{2007}{{Drummond et~al.}}{{Drummond, Henn, Smirnov and Sokatchev}}}
\bibcite{Drummond:2009fd}{{170}{2009}{{Drummond et~al.}}{{Drummond, Henn and Plefka}}}
\bibcite{Duhr:2012fh}{{171}{2012}{{Duhr}}{{}}}
\bibcite{DuhrDulat2019}{{172}{2019}{{Duhr and Dulat}}{{}}}
\bibcite{Duhr:2011zq}{{173}{2012}{{Duhr et~al.}}{{Duhr, Gangl and Rhodes}}}
\bibcite{DuhrMaggioNegaSauerTancrediWagner2025}{{174}{2025}{{Duhr et~al.}}{{Duhr, Maggio, Nega, Sauer, Tancredi and Wagner}}}
\bibcite{Correlahedron}{{175}{2017}{{Eden et~al.}}{{Eden, Heslop and Mason}}}
\bibcite{Eden:2010ce}{{176}{2011}{{Eden et~al.}}{{Eden, Korchemsky and Sokatchev}}}
\bibcite{Eden:1966dnq}{{177}{1966}{{Eden et~al.}}{{Eden, Landshoff, Olive and Polkinghorne}}}
\bibcite{Ehrenpreis1970}{{178}{1970}{{Ehrenpreis}}{{}}}
\bibcite{ES}{{179}{2011}{{Elekes and Sharir}}{{}}}
\bibcite{EllisGieleKunsztMelnikov2009}{{180}{2009}{{Ellis et~al.}}{{Ellis, Giele, Kunszt and Melnikov}}}
\bibcite{Ellis:2011cr}{{181}{2012}{{Ellis et~al.}}{{Ellis, Kunszt, Melnikov and Zanderighi}}}
\bibcite{Ellis:1991qj}{{182}{1996}{{Ellis et~al.}}{{Ellis, Stirling and Webber}}}
\bibcite{ElvangFreedmanKiermaier2009}{{183}{2010}{{Elvang et~al.}}{{Elvang, Freedman and Kiermaier}}}
\bibcite{Elvang:2013cua}{{184}{2015}{{Elvang and Huang}}{{}}}
\bibcite{Engelund:2011fg}{{185}{2012}{{Engelund and Roiban}}{{}}}
\bibcite{EK}{{186}{2017}{{Escobar and Knutson}}{{}}}
\bibcite{evenZoharLakrecParisiTesslerShermanBennettWilliams2023cluster}{{187}{2024a}{{Even-Zohar et~al.}}{{Even-Zohar, Lakrec, Parisi, Sherman-Bennett, Tessler and Williams}}}
\bibcite{higher_m_ampl}{{188}{2025a}{{Even-Zohar et~al.}}{{Even-Zohar, Lakrec, Parisi, Tessler, Sherman-Bennett and Williams}}}
\bibcite{evenZoharLakrecParisiTesslerShermanBennettWilliams2024clusterResults}{{189}{2024b}{{Even-Zohar et~al.}}{{Even-Zohar, Lakrec, Parisi, Tessler, Sherman-Bennett and Williams}}}
\bibcite{evenZoharLakrecTessler2025bcfw}{{190}{2025b}{{Even-Zohar et~al.}}{{Even-Zohar, Lakrec and Tessler}}}
\bibcite{Even-Zohar:2025ngd}{{191}{2026}{{Even-Zohar et~al.}}{{Even-Zohar, Parisi, Sherman-Bennett, Tessler and Williams}}}
\bibcite{Fairlie1962}{{192}{1962}{{Fairlie et~al.}}{{Fairlie, Landshoff, Nuttall and Polkinghorne}}}
\bibcite{Feng:2009ei}{{193}{2010}{{Feng et~al.}}{{Feng, Wang, Wang and Zhang}}}
\bibcite{Ferro:2023qdp}{{194}{2024}{{Ferro et~al.}}{{Ferro, Glew, Lukowski and Stalknecht}}}
\bibcite{Ferro:2024fibration}{{195}{2025}{{Ferro et~al.}}{{Ferro, Glew, Lukowski and Stalknecht}}}
\bibcite{Ferro:2022abq}{{196}{2023}{{Ferro and {\L }ukowski}}{{}}}
\bibcite{ferroLukowskiMoerman2020boundaries}{{197}{2020}{{Ferro et~al.}}{{Ferro, {\L }ukowski and Moerman}}}
\bibcite{Ferro}{{198}{2016}{{Ferro et~al.}}{{Ferro, Lukowski, Orta and Parisi}}}
\bibcite{FevolaMizeraTelen2024}{{199}{2024a}{{Fevola et~al.}}{{Fevola, Mizera and Telen}}}
\bibcite{FevolaMizeraTelenPLD2024}{{200}{2024b}{{Fevola et~al.}}{{Fevola, Mizera and Telen}}}
\bibcite{Fevola:Pos_Geom}{{201}{2025}{{Fevola and Sattelberger}}{{}}}
\bibcite{Feynman:1963ax}{{202}{1963}{{Feynman}}{{}}}
\bibcite{FP}{{203}{2023}{{Fomin and Pylyavskyy}}{{}}}
\bibcite{FominWilliamsZelevinsky2021}{{204}{2021}{{Fomin et~al.}}{{Fomin, Williams and Zelevinsky}}}
\bibcite{FominZelevinsky2002}{{205}{2002}{{Fomin and Zelevinsky}}{{}}}
\bibcite{FominZelevinsky2003}{{206}{2003}{{Fomin and Zelevinsky}}{{}}}
\bibcite{Forde:2007mi}{{207}{2007}{{Forde}}{{}}}
\bibcite{Franco:2014csa}{{208}{2015a}{{Franco et~al.}}{{Franco, Galloni, Mariotti and Trnka}}}
\bibcite{Franco:2014csa_alt1}{{209}{2015b}{{Franco et~al.}}{{Franco, Galloni, Penante and Wen}}}
\bibcite{Franco:2015rma}{{210}{2015c}{{Franco et~al.}}{{Franco, Galloni, Penante and Wen}}}
\bibcite{Fraser}{{211}{2016}{{Fraser}}{{}}}
\bibcite{FrellesvigEtAl2021}{{212}{2022}{{Frellesvig et~al.}}{{Frellesvig, Tommasini and Wever}}}
\bibcite{galashinKarpLam2022ball}{{213}{2022}{{Galashin et~al.}}{{Galashin, Karp and Lam}}}
\bibcite{GalashinLam}{{214}{2020}{{Galashin and Lam}}{{}}}
\bibcite{GalashinLamPositroidCluster}{{215}{2023}{{Galashin and Lam}}{{}}}
\bibcite{Sottile2003SecantConjecture}{{216}{2012}{{Garc{\'i}a-Puente et~al.}}{{Garc{\'i}a-Puente, Hein, Hillar, Mart{\'i}n~del Campo, Ruffo, Sottile and Teitler}}}
\bibcite{Gehrmann:2018yef}{{217}{2018}{{Gehrmann et~al.}}{{Gehrmann, Henn and Lo~Presti}}}
\bibcite{GehrmannRemiddi2000}{{218}{2000}{{Gehrmann and Remiddi}}{{}}}
\bibcite{gkz}{{219}{2009}{{Gelfand et~al.}}{{Gelfand, Kapranov and Zelevinsky}}}
\bibcite{GlewLukowski2025}{{220}{2025}{{Glew and {\L }ukowski}}{{}}}
\bibcite{GlewPokraka2025}{{221}{2025}{{Glew and Pokraka}}{{}}}
\bibcite{GoldenGoncharovSpradlinVerguVolovich2014}{{222}{2014}{{Golden et~al.}}{{Golden, Goncharov, Spradlin, Vergu and Volovich}}}
\bibcite{Goncharov2001}{{223}{2001}{{Goncharov}}{{}}}
\bibcite{Goncharov2005}{{224}{2005}{{Goncharov}}{{}}}
\bibcite{Goncharov:2010jf}{{225}{2010}{{Goncharov et~al.}}{{Goncharov, Spradlin, Vergu and Volovich}}}
\bibcite{GoreskyMacPherson1988}{{226}{1988}{{Goresky and MacPherson}}{{}}}
\bibcite{GoergesNegaTancrediWagner2023}{{227}{2023}{{G{\"o}rges et~al.}}{{G{\"o}rges, Nega, Tancredi and Wagner}}}
\bibcite{M2}{{228}{2026}{{Grayson and Stillman}}{{}}}
\bibcite{GriffithsHarris1978}{{229}{1978}{{Griffiths and Harris}}{{}}}
\bibcite{Gubser:1998bc}{{230}{1998}{{Gubser et~al.}}{{Gubser, Klebanov and Polyakov}}}
\bibcite{Guevara:2026qzd}{{231}{2026}{{Guevara et~al.}}{{Guevara, Lupsasca, Skinner, Strominger and Weil}}}
\bibcite{guler1996barrier}{{232}{1996}{{G{\"u}ler}}{{}}}
\bibcite{GurdoganParisi2020}{{233}{2023}{{G{\"u}rdo\u {g}an and Parisi}}{{}}}
\bibcite{Garding_1951}{{234}{1951}{{Gårding}}{{}}}
\bibcite{Garding_1959}{{235}{1959}{{Gårding}}{{}}}
\bibcite{Guler_Hyperbolic_pol}{{236}{1997}{{Güler}}{{}}}
\bibcite{Haag:1958vt}{{237}{1958}{{Haag}}{{}}}
\bibcite{Hannesdottir:2021kpd}{{238}{2022}{{Hannesdottir et~al.}}{{Hannesdottir, McLeod, Schwartz and Vergu}}}
\bibcite{HannesdottirMcLeodSchwartzVergu2023}{{239}{2023}{{Hannesdottir et~al.}}{{Hannesdottir, McLeod, Schwartz and Vergu}}}
\bibcite{ABJM_amplituhedron}{{240}{2023a}{{He et~al.}}{{He, Huang and Kuo}}}
\bibcite{He:2025correlahedron}{{241}{2026}{{He et~al.}}{{He, Huang and Kuo}}}
\bibcite{He:2024fij}{{242}{2025}{{He et~al.}}{{He, Jiang, Liu and Yang}}}
\bibcite{He:2023exb}{{243}{2023b}{{He et~al.}}{{He, Kuo, Li and Zhang}}}
\bibcite{heKuoZhang2021twistorString}{{244}{2021a}{{He et~al.}}{{He, Kuo and Zhang}}}
\bibcite{HeLiYang2021NotesClusterFeynmanIntegrals}{{245}{2021b}{{He et~al.}}{{He, Li and Yang}}}
\bibcite{He:2021non}{{246}{2021c}{{He et~al.}}{{He, Li and Yang}}}
\bibcite{HeLiYang2021KinematicsClusterFeynmanIntegrals}{{247}{2022}{{He et~al.}}{{He, Li and Yang}}}
\bibcite{He:2015yua}{{248}{2015}{{He and Zhang}}{{}}}
\bibcite{HellerVonManteuffel2022}{{249}{2022}{{Heller and von Manteuffel}}{{}}}
\bibcite{Heller:2019gkq}{{250}{2020}{{Heller et~al.}}{{Heller, von Manteuffel and Schabinger}}}
\bibcite{HelmerPapathanasiouTellander2024}{{251}{2024}{{Helmer et~al.}}{{Helmer, Papathanasiou and Tellander}}}
\bibcite{Helton_Linear}{{252}{2007}{{Helton and Vinnikov}}{{}}}
\bibcite{HenkePapathanasiou2020}{{253}{2020}{{Henke and Papathanasiou}}{{}}}
\bibcite{Henn:2025xrc}{{254}{2025}{{Henn et~al.}}{{Henn, Matija{\v {s}}i{\'c}, Miczajka, Peraro, Xu and Zhang}}}
\bibcite{Henn:CM}{{255}{2025}{{Henn and Raman}}{{}}}
\bibcite{Henn2013}{{256}{2013}{{Henn}}{{}}}
\bibcite{Henn:2014qga}{{257}{2015}{{Henn}}{{}}}
\bibcite{Henn:2019swt}{{258}{2020}{{Henn et~al.}}{{Henn, Korchemsky and Mistlberger}}}
\bibcite{Henn:2014yza}{{259}{2014}{{Henn and Plefka}}{{}}}
\bibcite{Herrmann:2020qlt}{{260}{2021}{{Herrmann et~al.}}{{Herrmann, Langer, Trnka and Zheng}}}
\bibcite{Duhr:2019tlz}{{261}{2020}{{Herrmann and Parra-Martinez}}{{}}}
\bibcite{Herrmann:2022nkh}{{262}{2022}{{Herrmann and Trnka}}{{}}}
\bibcite{Hodges:2009hk}{{263}{2013}{{Hodges}}{{}}}
\bibcite{Hodges:2005bf}{{264}{2005}{{Hodges}}{{}}}
\bibcite{HolleringMazzucchelliParisiSturmfels2025Lines}{{265}{2025}{{Hollering et~al.}}{{Hollering, Mazzucchelli, Parisi and Sturmfels}}}
\bibcite{HolleringMazzucchelliParisiSturmfels2026}{{266}{2026a}{{Hollering et~al.}}{{Hollering, Mazzucchelli, Parisi and Sturmfels}}}
\bibcite{HolleringMazzucchelliParisiSturmfels2026Positivity}{{267}{2026b}{{Hollering et~al.}}{{Hollering, Mazzucchelli, Parisi and Sturmfels}}}
\bibcite{tHooft:1973jz}{{268}{1974}{{'t~Hooft}}{{}}}
\bibcite{Hoermander1990}{{269}{1990}{{H{\"o}rmander}}{{}}}
\bibcite{Huh2013}{{270}{2013}{{Huh}}{{}}}
\bibcite{hormander1}{{271}{1983}{{Hörmander}}{{}}}
\bibcite{ItzyksonZuber}{{272}{1980}{{Itzykson and Zuber}}{{}}}
\bibcite{Jagadale:2020qfa}{{273}{2021}{{Jagadale and Laddha}}{{}}}
\bibcite{Jagadale:2022rbl}{{274}{2022}{{Jagadale and Laddha}}{{}}}
\bibcite{Jagadale:2023hjr}{{275}{2023}{{Jagadale and Laddha}}{{}}}
\bibcite{karp2017sign}{{276}{2017}{{Karp}}{{}}}
\bibcite{karp2020defining}{{277}{2020}{{Karp}}{{}}}
\bibcite{Karp2021Wronskians}{{278}{2021}{{Karp}}{{}}}
\bibcite{KarpPurbhoo2023UniversalPlucker}{{279}{2023}{{Karp and Purbhoo}}{{}}}
\bibcite{KarpWilliams}{{280}{2019}{{Karp and Williams}}{{}}}
\bibcite{Kleiss:1988ne}{{281}{1989}{{Kleiss and Kuijf}}{{}}}
\bibcite{knutsonLamSpeyer2013positroid}{{282}{2013}{{Knutson et~al.}}{{Knutson, Lam and Speyer}}}
\bibcite{KobaNielsen1969}{{283}{1969}{{Koba and Nielsen}}{{}}}
\bibcite{koefler2025taking}{{284}{2025}{{Koefler and Sinn}}{{}}}
\bibcite{Polypols}{{285}{2025}{{Kohn et~al.}}{{Kohn, Piene, Ranestad, Rydell, Shapiro, Sinn, Sorea and Telen}}}
\bibcite{kohn2020projective}{{286}{2020}{{Kohn and Ranestad}}{{}}}
\bibcite{Lukowski:2020bya}{{287}{2020}{{Kojima and Langer}}{{}}}
\bibcite{deKorteYu2026}{{288}{2026}{{de~Korte and Yu}}{{}}}
\bibcite{Kotikov1991}{{289}{1991}{{Kotikov}}{{}}}
\bibcite{Kotikov:2002ab}{{290}{2003}{{Kotikov and Lipatov}}{{}}}
\bibcite{Pos_Certif}{{291}{2019}{{Kozhasov et~al.}}{{Kozhasov, Michałek and Sturmfels}}}
\bibcite{Kristensson:2021ani}{{292}{2021}{{Kristensson et~al.}}{{Kristensson, Wilhelm and Zhang}}}
\bibcite{Kummer_2015}{{293}{2015}{{Kummer}}{{}}}
\bibcite{Kummer_2022}{{294}{2022}{{Kummer and Sinn}}{{}}}
\bibcite{TNN_grassmannian}{{295}{2014}{{Lam}}{{}}}
\bibcite{Lam:_PG_notes}{{296}{2022}{{Lam}}{{}}}
\bibcite{lam2024face}{{297}{2024}{{Lam}}{{}}}
\bibcite{Lam:ModuliSpacesPG}{{298}{2025}{{Lam}}{{}}}
\bibcite{Landau1959}{{299}{1959}{{Landau}}{{}}}
\bibcite{LeePomeransky2013}{{300}{2013}{{Lee and Pomeransky}}{{}}}
\bibcite{LSZ:1955}{{301}{1955}{{Lehmann et~al.}}{{Lehmann, Symanzik and Zimmermann}}}
\bibcite{Leinartas1978}{{302}{1978}{{Leinartas}}{{}}}
\bibcite{lemaire2005regularchains}{{303}{2005}{{Lemaire et~al.}}{{Lemaire, Maza and Xie}}}
\bibcite{Lemmon:2025dhq}{{304}{2025}{{Lemmon and Trnka}}{{}}}
\bibcite{LippstreuSpradlinSrikantVolovich2024ZigguratII}{{305}{2024a}{{Lippstreu et~al.}}{{Lippstreu, Spradlin, Srikant and Volovich}}}
\bibcite{LippstreuSpradlinVolovich2024ZigguratI}{{306}{2024b}{{Lippstreu et~al.}}{{Lippstreu, Spradlin and Volovich}}}
\bibcite{Lisitsyn:2025prd}{{307}{2025}{{Lisitsyn et~al.}}{{Lisitsyn, Oktem, Sherman-Bennett and Trnka}}}
\bibcite{lukowski2019boundaries}{{308}{2022}{{{\L }ukowski}}{{}}}
\bibcite{lukowskiMoerman2021boundaries}{{309}{2021}{{{\L }ukowski and Moerman}}{{}}}
\bibcite{Lukowski:2022fwz}{{310}{2022}{{Lukowski et~al.}}{{Lukowski, Moerman and Stalknecht}}}
\bibcite{lukowski2019cluster}{{311}{2019}{{{\L }ukowski et~al.}}{{{\L }ukowski, Parisi, Spradlin and Volovich}}}
\bibcite{Maazouz:2024qmm}{{312}{2025}{{Maazouz et~al.}}{{Maazouz, Pfister and Sturmfels}}}
\bibcite{MagoSchreiberSpradlinVolovich2019}{{313}{2019}{{Mago et~al.}}{{Mago, Schreiber, Spradlin and Volovich}}}
\bibcite{Maldacena:1997re}{{314}{1998}{{Maldacena}}{{}}}
\bibcite{MS}{{315}{2023}{{Mali{\'c} and Streinu}}{{}}}
\bibcite{mandelshtam2023combinatorics}{{316}{2023}{{Mandelshtam et~al.}}{{Mandelshtam, Pavlov and Pratt}}}
\bibcite{Mandelstam:1982cb}{{317}{1983}{{Mandelstam}}{{}}}
\bibcite{Mangano:1990by}{{318}{1991}{{Mangano and Parke}}{{}}}
\bibcite{Mangano:1987xk}{{319}{1988}{{Mangano et~al.}}{{Mangano, Parke and Xu}}}
\bibcite{marsh2016twists}{{320}{2016}{{Marsh and Scott}}{{}}}
\bibcite{Martin1}{{321}{2003}{{Martin}}{{}}}
\bibcite{Mason:2010yk}{{322}{2010}{{Mason and Skinner}}{{}}}
\bibcite{Mason:2009qx}{{323}{2009}{{Mason and Skinner}}{{}}}
\bibcite{Mastrolia:2009dr}{{324}{2009}{{Mastrolia}}{{}}}
\bibcite{MastroliaMizera2019}{{325}{2019}{{Mastrolia and Mizera}}{{}}}
\bibcite{MatijasicThesis}{{326}{2024}{{Matija{\v {s}}i{\'c}}}{{}}}
\bibcite{repoEffortless}{{327}{2025}{{Matija{\v {s}}i{\'c} and Miczajka}}{{}}}
\bibcite{Euler_Integrals}{{328}{2023}{{Matsubara-Heo et~al.}}{{Matsubara-Heo, Mizera and Telen}}}
\bibcite{MatsubaraHeoTelen2025}{{329}{2025}{{Matsubara-Heo and Telen}}{{}}}
\bibcite{MR}{{330}{2022}{{Matveiakin and Rudenko}}{{}}}
\bibcite{MazloumiXu2025ClusterCosmologicalCorrelators}{{331}{2025}{{Mazloumi and Xu}}{{}}}
\bibcite{MazzucchelliHennAomotoForms}{{332}{2026}{{Mazzucchelli and Henn}}{{}}}
\bibcite{MazzucchelliPavlovWang2025}{{333}{2025}{{Mazzucchelli et~al.}}{{Mazzucchelli, Pavlov and Wang}}}
\bibcite{mazzucchelli2025exterior}{{334}{2025}{{Mazzucchelli and Pratt}}{{}}}
\bibcite{Mazzucchelli:DV}{{335}{2025}{{Mazzucchelli and Raman}}{{}}}
\bibcite{Exponential_varieties}{{336}{2016}{{Michałek et~al.}}{{Michałek, Sturmfels, Uhler and Zwiernik}}}
\bibcite{Minahan:2002ve}{{337}{2003}{{Minahan and Zarembo}}{{}}}
\bibcite{Mizera:2017cqs}{{338}{2018}{{Mizera}}{{}}}
\bibcite{Mizera:2019gea}{{339}{2019}{{Mizera}}{{}}}
\bibcite{Mizera:2021icv}{{340}{2022}{{Mizera and Telen}}{{}}}
\bibcite{moerman2023positive}{{341}{2023}{{Moerman}}{{}}}
\bibcite{mohammadiMoninParisi2020triangulations}{{342}{2020}{{Mohammadi et~al.}}{{Mohammadi, Monin and Parisi}}}
\bibcite{MukhinTarasovVarchenko2009}{{343}{2009}{{Mukhin et~al.}}{{Mukhin, Tarasov and Varchenko}}}
\bibcite{Muller_2017}{{344}{2017}{{Muller and Speyer}}{{}}}
\bibcite{Nair:1988bq}{{345}{1988}{{Nair}}{{}}}
\bibcite{Nakanishi1959}{{346}{1959}{{Nakanishi}}{{}}}
\bibcite{Nakanishi1971}{{347}{1971}{{Nakanishi}}{{}}}
\bibcite{OssermanTrager2019}{{348}{2019}{{Osserman and Trager}}{{}}}
\bibcite{Ossola:2006us}{{349}{2007}{{Ossola et~al.}}{{Ossola, Papadopoulos and Pittau}}}
\bibcite{Palamodov1970}{{350}{1970}{{Palamodov}}{{}}}
\bibcite{Panzer:2014caa}{{351}{2015a}{{Panzer}}{{}}}
\bibcite{Panzer2015}{{352}{2015b}{{Panzer}}{{}}}
\bibcite{ParanjapeSkowronekSpradlinVolovichWeng2026ClusterBootstrapCosmologicalCorrelators}{{353}{2026a}{{Paranjape et~al.}}{{Paranjape, Skowronek, Spradlin, Volovich and Weng}}}
\bibcite{Paranjape:2026kix}{{354}{2026b}{{Paranjape et~al.}}{{Paranjape, Skowronek, Spradlin, Volovich and Weng}}}
\bibcite{parisi2024magic}{{355}{2024}{{Parisi et~al.}}{{Parisi, Sherman-Bennett, Tessler and Williams}}}
\bibcite{parisiShermanBennettWilliams2023m2}{{356}{2023}{{Parisi et~al.}}{{Parisi, Sherman-Bennett and Williams}}}
\bibcite{Parke:1986gb}{{357}{1986}{{Parke and Taylor}}{{}}}
\bibcite{Penrose:1967wn}{{358}{1967}{{Penrose}}{{}}}
\bibcite{Peskin:1995ev}{{359}{1995}{{Peskin and Schroeder}}{{}}}
\bibcite{Pham1967}{{360}{1967}{{Pham}}{{}}}
\bibcite{Pham2011}{{361}{2011}{{Pham}}{{}}}
\bibcite{PoegelWangWeinzierl2023}{{362}{2023}{{P{\"o}gel et~al.}}{{P{\"o}gel, Wang and Weinzierl}}}
\bibcite{Pokraka:2024fao}{{363}{2025}{{Pokraka et~al.}}{{Pokraka, Rajan, Ren, Volovich and Zhao}}}
\bibcite{PST}{{364}{2017}{{Ponce et~al.}}{{Ponce, Sturmfels and Trager}}}
\bibcite{Postnikov:2006kva}{{365}{2006}{{Postnikov}}{{}}}
\bibcite{PrattSodomacoSturmfels2026}{{366}{2026}{{Pratt et~al.}}{{Pratt, Sodomaco and Sturmfels}}}
\bibcite{chowlam}{{367}{2025}{{Pratt and Sturmfels}}{{}}}
\bibcite{PrlinaSpradlinStankowiczStanojevic2018}{{368}{2018a}{{Prlina et~al.}}{{Prlina, Spradlin, Stankowicz and Stanojevic}}}
\bibcite{PrlinaSpradlinStanojevic2018}{{369}{2018b}{{Prlina et~al.}}{{Prlina, Spradlin and Stanojevic}}}
\bibcite{Rajan:2024ome}{{370}{2025}{{Rajan et~al.}}{{Rajan, Sverrisd{\'o}ttir and Sturmfels}}}
\bibcite{Raman:2019utu}{{371}{2019}{{Raman}}{{}}}
\bibcite{Ranestad:adjoint}{{372}{2024}{{Ranestad et~al.}}{{Ranestad, Sinn and Telen}}}
\bibcite{Ranestad:what_is_PG}{{373}{2025}{{Ranestad et~al.}}{{Ranestad, Sturmfels and Telen}}}
\bibcite{Raz}{{374}{2017}{{Raz}}{{}}}
\bibcite{ReedSimon:1979}{{375}{1979}{{Reed and Simon}}{{}}}
\bibcite{Remiddi1997}{{376}{1997}{{Remiddi}}{{}}}
\bibcite{Remiddi:1999ew}{{377}{2000}{{Remiddi and Vermaseren}}{{}}}
\bibcite{Risager:2005vk}{{378}{2005}{{Risager}}{{}}}
\bibcite{Roiban:2004yf}{{379}{2005}{{Roiban et~al.}}{{Roiban, Spradlin and Volovich}}}
\bibcite{Schwartz:2014sze}{{380}{2014}{{Schwartz}}{{}}}
\bibcite{Scott_2014}{{381}{2014}{{Scott and Sokal}}{{}}}
\bibcite{Scott2006}{{382}{2006}{{Scott}}{{}}}
\bibcite{SS}{{383}{}{{Sidman and Lee-St.John}}{{}}}
\bibcite{sinn2015algebraic}{{384}{2015}{{Sinn}}{{}}}
\bibcite{Smirnov2004}{{385}{2004}{{Smirnov}}{{}}}
\bibcite{Smirnov2012}{{386}{2012}{{Smirnov}}{{}}}
\bibcite{Srednicki:2007qs}{{387}{2007}{{Srednicki}}{{}}}
\bibcite{stalknecht2024positive}{{388}{2024}{{Stalknecht}}{{}}}
\bibcite{Stalknecht:2026ylm}{{389}{2026}{{Stalknecht}}{{}}}
\bibcite{Steinmann1960}{{390}{1960}{{Steinmann}}{{}}}
\bibcite{StreaterWightman}{{391}{2000}{{Streater and Wightman}}{{}}}
\bibcite{Sturmfels2002}{{392}{2002}{{Sturmfels}}{{}}}
\bibcite{tHooftVeltman1972}{{393}{1972}{{{'t Hooft} and Veltman}}{{}}}
\bibcite{Telen_PG}{{394}{2025}{{Telen}}{{}}}
\bibcite{telen2026toric}{{395}{2026}{{Telen}}{{}}}
\bibcite{tessler2025notes}{{396}{2025}{{Tessler}}{{}}}
\bibcite{Tsikh1992}{{397}{1992}{{Tsikh}}{{}}}
\bibcite{Vanhove2014}{{398}{2014}{{Vanhove}}{{}}}
\bibcite{Veltman1963}{{399}{1963}{{Veltman}}{{}}}
\bibcite{Veneziano1968}{{400}{1968}{{Veneziano}}{{}}}
\bibcite{Vergu:2025mag}{{401}{2025}{{Vergu}}{{}}}
\bibcite{Wachspress:1975}{{402}{1975}{{Wachspress}}{{}}}
\bibcite{Weinberg:1964ew}{{403}{1964}{{Weinberg}}{{}}}
\bibcite{Weinberg:1965nx}{{404}{1965}{{Weinberg}}{{}}}
\bibcite{Weinberg:1995mt}{{405}{1995}{{Weinberg}}{{}}}
\bibcite{Weinberg:2000cr}{{406}{2000}{{Weinberg}}{{}}}
\bibcite{Weinzierl:2022eaz}{{407}{2022}{{Weinzierl}}{{}}}
\bibcite{Wess:1992cp}{{408}{1992}{{Wess and Bagger}}{{}}}
\bibcite{Whitney1965}{{409}{1965}{{Whitney}}{{}}}
\bibcite{widder2015laplace}{{410}{2015}{{Widder}}{{}}}
\bibcite{Wigner:1939cj}{{411}{1939}{{Wigner}}{{}}}
\bibcite{WilliamsThesis2005}{{412}{2005}{{Williams}}{{}}}
\bibcite{Williams2014ClusterAlgebras}{{413}{2014}{{Williams}}{{}}}
\bibcite{Witten:1995ex}{{414}{1996}{{Witten}}{{}}}
\bibcite{Witten:1998qj}{{415}{1998}{{Witten}}{{}}}
\bibcite{Witten:2003nn}{{416}{2004}{{Witten}}{{}}}
\bibcite{Yang:2022gko}{{417}{2022}{{Yang}}{{}}}
\bibcite{ziegler2012lectures}{{418}{2012}{{Ziegler}}{{}}}
\makeatother
\let\WriteBookmarks\relax
\def\floatpagepagefraction{1}
\def\textpagefraction{.001}

\shorttitle{Positive singularities and volumes}
\shortauthors{E. Mazzucchelli}
\title[mode=title]{Positive Singularities and Volumes in Scattering Amplitudes}

\author[1]{Elia Mazzucchelli}[
  orcid=0009-0007-7116-2743]
\ead{eliam@mpp.mpg.de}
\ead[url]{https://eliamazzucchelli.github.io/}
\affiliation[1]{organization={Max Planck Institute for Physics},
                addressline={Boltzmannstra{\ss}e 8},
                postcode={85748},
                city={Garching bei M\"unchen},
                country={Germany}}

\begin{abstract}
Recent advances have revealed that scattering amplitudes in certain quantum
field theories admit a geometric formulation in terms of positive geometries.
In this framework, amplitudes are encoded by differential forms whose boundary
structure reflects physical principles such as locality, unitarity, and
factorization.

This thesis gives a self-contained introduction to positive geometries, with
particular emphasis on the Amplituhedron, which describes amplitudes in planar
maximally supersymmetric Yang--Mills theory through its canonical form. We
develop three related directions connecting positivity, volumes, and
singularities.

First, we study positivity properties of canonical forms through dual volume
representations. For polytopes, canonical functions compute volumes of dual
polytopes; we extend this picture to nonlinear positive geometries, uncovering
non-negative transcendental measures and links with complete monotonicity.
Second, we investigate loop-level amplitudes via their singularity structure.
Combining Amplituhedron geometry with Landau analysis, we classify leading
singularities of the Wilson loop with Lagrangian insertion and constrain the
possible singular loci. Third, we examine conjectural relations between
Landau singularities, positivity, and cluster algebras. Using momentum-twistor
and Grassmannian methods, we identify recursive structures across loop orders,
prove several infinite families of cases, and propose a strategy toward the
general conjectures.

Overall, the thesis develops the perspective that positive geometry provides a
unifying language for the analytic and geometric organization of scattering
amplitudes, from canonical forms and volumes to the singularities that emerge
after loop integration.
\end{abstract}

\begin{keywords}
scattering amplitudes \sep positive geometry \sep Amplituhedron \sep Positivity \sep Landau singularities  \sep cluster algebras
\end{keywords}

\maketitle
\tableofcontents

\input{Ch1}
\clearpage
\input{Ch2}
\clearpage
\input{Ch3}
\clearpage
\input{Ch4}

\clearpage
\input{Ch5}
\clearpage
\input{Ch6}
\clearpage
\input{Ch7}

\clearpage
\input{Conclusion}
\clearpage
\input{Acknowledgment}


\end{document}

%% file: Ch1.tex

\newpage

\section{Introduction}\label{ch:Introduction}

\subsection{Prelude}

Modern physics is organized around principles. Classical mechanics is built from
the action principle, quantum mechanics from amplitudes and probabilities, and
special relativity from the geometry of spacetime. Quantum field theory brings
these ideas together with locality and symmetry. It is the basic framework for
describing relativistic quantum systems, and in particular for understanding
particle interactions~\cite{Peskin:1995ev,Weinberg:1995mt,Schwartz:2014sze}.

This framework has been extraordinarily successful. Perturbative quantum field
theory underlies the Standard Model description of elementary particles and has
led to precision predictions for collider processes. In this setting, scattering
amplitudes are central objects: they encode the probability amplitudes for
processes in which incoming particles interact and produce outgoing particles.
Together with phase-space integration and appropriate infrared-safe
observables, they provide the bridge between the formalism of quantum field
theory and phenomenological predictions~\cite{Ellis:1991qj,Dixon:1996wi,
Elvang:2013cua,Henn:2014yza}.

At the same time, our understanding of quantum field theory is incomplete in
several important respects. Perturbation theory often produces expressions that
are vastly more complicated than the final answers. Gauge redundancy
introduces many auxiliary degrees of freedom that cancel only after large sums
of terms. Loop amplitudes lead to complicated transcendental functions whose
singularities are difficult to control. More conceptually, the usual formulation
of quantum field theory is tied to locality in spacetime, while a satisfactory
unification with gravity remains beyond the standard perturbative framework.
These issues suggest that the Lagrangian and Feynman-diagram formulation,
although extremely powerful, may not be the only natural way to organize
physical observables.

A more modern amplitudes program takes this possibility seriously. Rather than
starting from off-shell fields and summing diagrams, it studies amplitudes
directly as functions of on-shell kinematic data. General principles such as
Lorentz invariance, locality, unitarity, analyticity, factorization, and
symmetry impose strong constraints on these functions. In favorable theories,
these constraints lead to recursive constructions and to expressions whose
simplicity is hidden in the diagrammatic expansion~\cite{Britto:2004ap,
Britto:2005fq,Bern:2007dw,Elvang:2013cua}.

A particularly revealing example is planar maximally supersymmetric
Yang--Mills theory, $\mathcal{N}=4$ SYM theory in short. Its amplitudes exhibit structures that are invisible in
ordinary Feynman diagrams but become natural in variables such as momentum
twistors and in Grassmannian formulations, which provide geometric languages
for planar kinematics and are reviewed in Chapter~\ref{ch:Scattering Amplitudes}
~\cite{Hodges:2009hk,Mason:2009qx,Grassmannian,the_amplituhedron}. The positive-geometry program can be viewed
as a further step in this direction. It proposes that certain amplitudes and
integrands are most naturally described not as sums of diagrams, but as
canonical differential forms associated with geometric regions. This thesis
studies this reformulation and asks what it teaches us about positivity,
volumes, and singularities, both before and after loop integration.
The intended reader lies at the interface of mathematics and physics. The
exposition is therefore pedagogical in both directions: it introduces the
amplitude-theoretic language needed by mathematicians and the
positive-geometric framework needed by physicists. 

\subsection{Why supersymmetric Yang--Mills theory?}
\label{sec:why-N4-SYM}

The theory studied throughout much of this thesis is planar maximally
supersymmetric Yang--Mills theory, or planar \(\mathcal N=4\) SYM. This is not
because it is a direct model of particle physics: quantum chromodynamics (QCD) is the gauge theory
relevant for strong-interaction phenomenology. Rather, \(\mathcal N=4\) SYM provides a particularly clean setting in which
to uncover structures that are otherwise hidden by the simultaneous presence
of running couplings, masses, complicated colour factors, ultraviolet
renormalization, and complicated transcendental functions at loop level.

Despite these simplifications, \(\mathcal N=4\) SYM is a genuine interacting
four-dimensional non-abelian gauge theory. It contains gauge fields, fermions,
and scalars in a single supersymmetric multiplet, and its coupling does not
run. The theory is ultraviolet finite and exactly conformally invariant,
while its scattering amplitudes retain non-trivial infrared singularities,
factorization channels, and intricate loop corrections
\cite{Brink:1976bc,Mandelstam:1982cb,Beisert:2010jr}. It therefore separates
many universal questions about gauge theory from complications that are
specific to more general models. In the planar limit, colour ordering further
reduces amplitudes to functions with a fixed cyclic ordering of their external
particles. The resulting theory is simple enough for hidden structures to
become visible, but rich enough that the structures discovered are often of
broader relevance.

The importance of the theory extends well beyond perturbation theory. The
AdS/CFT correspondence identifies planar \(\mathcal N=4\) SYM with type-IIB
string theory on \({\rm AdS}_5\times {\rm S}^5\), providing a concrete realization of
gauge/string duality and relating weakly coupled gauge theory to a geometric
description at strong coupling \cite{Maldacena:1997re,Gubser:1998bc,
Witten:1998qj}. The planar theory also exhibits integrability, first uncovered
in its spectral problem and subsequently developed into exact methods for
quantities such as anomalous dimensions
\cite{Minahan:2002ve,BeisertEdenStaudacher2006,Beisert:2010jr}.
These developments established \(\mathcal N=4\) SYM as a setting in which
perturbation theory, conformal field theory, integrability, and string theory
could be studied within one framework.

Scattering amplitudes revealed another, largely independent, source of hidden
simplicity. Early multiloop calculations showed that planar amplitudes were
far more structured than their Feynman-diagram expansions suggested. The two-loop four-particle amplitude was found to be related iteratively to
the one-loop result, leading to the Anastasiou--Bern--Dixon--Kosower (ABDK) relation
and subsequently to the Bern--Dixon--Smirnov (BDS) ansatz for planar
maximally helicity-violating amplitudes
\cite{AnastasiouBernDixonKosower2003,BernDixonSmirnov2005}. Although the BDS
ansatz is not complete starting from six particles, it correctly isolates the
universal exponentiating infrared contribution. Its failure leaves a finite
remainder function depending only on conformal cross-ratios, thereby exposing
a new class of functions constrained by symmetry, singularities, and physical
limits \cite{DrummondHennKorchemskySokatchev2008,Goncharov:2010jf}.

At the same time, on-shell methods changed how amplitudes were constructed.
The Britto--Cachazo--Feng--Witten (BCFW) recursion relations showed that tree amplitudes can be reconstructed
from lower-point amplitudes using complex analysis and their factorization
poles, without referring to individual Feynman diagrams
\cite{Britto:2004ap,Britto:2005fq}. Generalized unitarity similarly turned
loop-level cuts into constructive data. These results suggested that
amplitudes should be regarded as objects determined directly by analyticity,
unitarity, locality, and symmetry, rather than merely as the outcome of
a diagrammatic expansion.

Planarity then revealed a symmetry absent from the conventional Lagrangian
description. Introducing dual coordinates through
\(p_i=x_i-x_{i+1}\), planar amplitudes were found to possess dual conformal and
dual superconformal symmetry
\cite{Drummond:2007aua,Drummond:2008vq}. Hodges' momentum twistors make this
symmetry manifest and turn many kinematic conditions into elementary
incidence relations in projective space \cite{Hodges:2009hk}. The same dual
coordinates define a null polygon, leading to the duality between planar maximally helicity-violating (MHV)
amplitudes and light-like polygonal Wilson loops
\cite{Brandhuber:2007yx,DrummondHennKorchemskySokatchev2007,
DrummondHennKorchemskySokatchev2008}. At strong coupling this polygon becomes
the boundary of the minimal surface appearing in the Alday--Maldacena
description of scattering amplitudes \cite{Alday:2007hr}.

Ordinary and dual superconformal symmetries combine into a Yangian symmetry of
tree amplitudes \cite{Drummond:2009fd}. The corresponding invariants admit a
Grassmannian formulation, whose residues organize tree amplitudes and
leading singularities \cite{ArkaniHamed:2009dn,Grassmannian,Drummond:2010qh}.
This brought together several ingredients recurring throughout this thesis:
on-shell recursion, residues, Grassmannians, positivity, and projective
geometry. Crucially, these were not auxiliary mathematical descriptions
added after the amplitudes had been computed. They provided natural variables
and spaces in which physical constraints became simpler.

These last developments culminated in the positive-geometry formulation of
scattering amplitudes. The positive Grassmannian organizes on-shell diagrams
and their boundary relations, while the Amplituhedron associates planar
\(\mathcal N=4\) SYM tree amplitudes and loop integrands with canonical
differential forms of geometric regions
\cite{Grassmannian,the_amplituhedron,Positive_geometries}.
Factorization and unitarity are then encoded by boundaries and residues,
whereas BCFW representations become triangulations of a single underlying
geometry. This replaces large sums of terms containing spurious singularities
by an object whose physical singularity structure is built into its
definition.

The study of \(\mathcal N=4\) SYM is therefore valuable for two related
reasons. First, it is a solvable laboratory in which general features of
interacting gauge theories---infrared factorization, unitarity, analytic
structure, and transcendental functions---can be examined in unusually pure
form. Many techniques developed there, including generalized unitarity, pure integrals, symbols, and bootstrap methods, have
subsequently influenced precision calculations in more general quantum field
theories. Second, the theory reveals structures that may be fundamental in
their own right: dual conformal symmetry, Yangian invariance, Grassmannian
geometry, cluster algebras, and positive geometries.

The point is not that planar \(\mathcal N=4\) SYM approximates QCD.
Rather, it serves as a highly constrained theoretical laboratory: by
stripping away inessential complications, it lets us ask what scattering
amplitudes fundamentally are. The positive-geometry programme pursued in this thesis
continues this historical development. It asks whether amplitudes and loop
integrands are best understood as canonical forms, and whether their
positivity and singularities can be read directly from the defining geometry.

\subsection{The positive geometry program}
\label{sec:The Positive Geometry Program}

The basic proposal of the positive-geometry program is that certain
quantities in quantum field theory can be interpreted as canonical
differential forms associated with distinguished real regions in complex
algebraic varieties. In this formulation, locality, unitarity, and
factorization are encoded in the boundary structure of the region rather
than imposed term by term in a diagrammatic expansion.

The central mathematical notion is that of a \emph{positive geometry}: a
triple
\[
(\mathcal X,P,\mathbf{\Omega}_P),
\]
where \(\mathcal X\) is a complex algebraic variety, \(P\) is an oriented
real semialgebraic subset of its real points, and \(\mathbf{\Omega}_P\) is
its canonical differential form. The form is rational and characterized
recursively by having only logarithmic singularities on the boundary of
\(P\), with the residue on each boundary component equal to the canonical
form of that boundary
\cite{Positive_geometries,Lam:_PG_notes,Herrmann:2022nkh,
Brown:PG_Hodge,Telen_PG}.

The simplest examples are polytopes. If
\(P\subseteq\mathbb R^m\) is a full-dimensional polytope, its canonical form
has simple poles on the facets, and the residue on each facet is the
canonical form of that facet. Its denominator is therefore controlled by
the boundary hyperplanes, while its numerator removes spurious
singularities. A triangulation computes the same canonical form by
decomposing \(P\) into simpler regions. Already in this elementary setting,
the basic dictionary is visible: boundaries correspond to singularities,
residues to factorization, and triangulations to different representations
of one underlying object.

The relevance of this framework to scattering amplitudes is realized most
fully by the \emph{Amplituhedron}. Its canonical form describes tree
amplitudes and loop integrands in planar \(\mathcal N=4\) supersymmetric
Yang--Mills theory
\cite{Grassmannian,the_amplituhedron,Positive_geometries}. Physical
properties are translated into geometric statements: factorization is
encoded by residues on boundaries, locality by the allowed poles, and
Britto--Cachazo--Feng--Witten recursion by triangulations of the geometry
\cite{Britto:2005fq,Hodges:2009hk,the_amplituhedron}. In this way, the
Amplituhedron replaces a sum of terms containing spurious singularities by a
single geometric object whose canonical form has the required singularity
structure. This gives a geometric explanation for cancellations that are
otherwise visible only after the complete amplitude has been assembled.

Planarity is essential to this particular construction. A fixed cyclic
ordering permits the introduction of dual coordinates and momentum twistors,
and underlies the positive-Grassmannian and positroid structures entering the
Amplituhedron. No comparably complete non-planar analogue is presently used
in this thesis. Nevertheless, several ingredients---including generalized
unitarity, leading singularities, on-shell diagrams, and Grassmannian
residues---remain meaningful beyond the planar limit
\cite{Franco:2014csa_alt1,Franco:2015rma,Lisitsyn:2025prd}. Accordingly,
the results developed below have different ranges of applicability. The
dual-volume questions of
Chapter~\ref{ch:Canonical Forms as Dual Volumes} concern positive geometries
more generally, whereas the specific Amplituhedron and Wilson-loop
constructions of Chapters~\ref{ch:Amplituhedra} and
\ref{ch:From Integrands to Integrals} are planar. The incidence varieties,
discriminants, and resultants studied in
Chapter~\ref{ch:positivity-and-cluster-structures} are algebraic-geometric
constructions of broader scope, while their positivity and
cluster-factorization properties rely strongly on planar
momentum-twistor kinematics.

The relation to quantum chromodynamics is similarly indirect but concrete.
The Amplituhedron is not a positive geometry for generic QCD amplitudes.
However, many methods developed in the planar amplitudes program are now
standard in multiloop quantum field theory. Generalized unitarity organizes
integrands and their cuts, integration-by-parts identities reduce loop
integrals to master integrals, and differential equations determine their
kinematic dependence
\cite{Bern:2007dw,Ellis:2011cr,Kotikov1991,Remiddi1997,
GehrmannRemiddi2000,Henn2013}. Landau equations constrain the possible
singular loci of general Feynman integrals and, in polylogarithmic cases,
help determine their symbol alphabets
\cite{Landau1959,Eden:1966dnq,DennenSpradlinVolovich2016}. Thus the
specific positive geometry is special to planar \(\mathcal N=4\) SYM, while
several analytic and algebraic tools motivated by it apply much more broadly.

The Amplituhedron is also part of a wider family of positive geometries in
physics. The \emph{associahedron} describes tree amplitudes in bi-adjoint
scalar theory, with facets corresponding to factorization channels
\cite{ABHY_original,Stringy_Can_Forms,AHLTBinary}. Related constructions
include Stokes polytopes, accordiohedra, and positive geometries for
higher-valent scalar interactions
\cite{Banerjee:2018tun,Raman:2019utu,Aneesh:2019cvt,
Jagadale:2020qfa,Jagadale:2022rbl,Jagadale:2023hjr}. Other variants include
the momentum Amplituhedron in spinor-helicity space, the correlahedron for
stress-tensor correlators in planar \(\mathcal N=4\) SYM, and the
three-dimensional Amplituhedron associated with planar
\(\mathcal N=6\) Chern--Simons matter theory
\cite{Damgaard:2019ztj,ferroLukowskiMoerman2020boundaries,
Correlahedron,ABJM_amplituhedron,stalknecht2024positive}.

Positive geometries also arise in cosmology. Cosmological polytopes,
cosmohedra, and related constructions encode wavefunction coefficients and
cosmological correlators, making their singularities, recursion relations,
and differential equations accessible through geometry and combinatorics
\cite{Cosmological_polytopes,Cosmoehdra,BenincasaDian2025,
ArkaniHamedBaumannHillmanJoyceLeePimentel2025}. These examples indicate that
positive geometry is not tied to one particular theory, but expresses a
broader mechanism in which physical singularities arise from boundaries of
geometric spaces.

In parallel, a mathematical theory is emerging at the intersection of
algebraic geometry, combinatorics, topology, and analysis
\cite{Positive_geometries,Lam:_PG_notes,Ranestad:what_is_PG,
Brown:PG_Hodge}. For the Amplituhedron, this includes the positive
Grassmannian, positroid stratifications, sign variation, Grassmann polytopes,
and the combinatorics of faces and triangulations
\cite{Postnikov:2006kva,TNN_grassmannian,karp2017sign,
galashinKarpLam2022ball,KarpWilliams,lukowski2019boundaries}. Recent work
has established broad classes of BCFW triangulations and uncovered
cluster-algebraic structures in their tiles
\cite{evenZoharLakrecTessler2025bcfw,
evenZoharLakrecParisiTesslerShermanBennettWilliams2023cluster,
Even-Zohar:2025ngd}. Algebraic approaches to canonical forms, adjoint
hypersurfaces, and interpolation further connect positive geometry with
projective algebraic geometry
\cite{Ranestad:adjoint,Polypols,Telen_PG,koefler2025taking,
Fevola:Pos_Geom}. We develop the definitions and constructions needed here in Chapters~\ref{ch:Positive Geometries}
and~\ref{ch:Amplituhedra}.

\subsection{Implications of positive geometries}
\label{sec:Implications of Positive Geometries}

The power of positive geometry is both conceptual and computational. We now
summarize its main implications for scattering amplitudes in planar
\(\mathcal N=4\) SYM. This overview necessarily uses several terms, most of which will only be defined and explained in later chapters. It is meant as a roadmap: the reader may use it now as a guide to the main themes of the thesis, and return to it later once the technical framework has been developed.

\subsubsection{Triangulations and BCFW expansion}
The first concrete role of positive geometries is to provide efficient ways of constructing tree
amplitudes and loop integrands. In planar \(\mathcal N=4\) SYM, the
Amplituhedron turns the problem of computing an amplitude into the problem of
computing a canonical form. Different triangulations of the geometry give
different formulae for the same object, and BCFW recursion appears as one
such triangulation~\cite{Britto:2004ap,Britto:2005fq,the_amplituhedron,
Grassmannian,evenZoharLakrecTessler2025bcfw}. This gives a geometric
explanation for why intermediate representations may contain spurious poles,
while the full canonical form does not~\cite{Hodges:2009hk}. At loop level, the
same philosophy applies to the all-loop integrand at any loop-order: instead of summing
Feynman diagrams, one computes the rational form associated with a positive
region in the space of loop lines~\cite{ArkaniHamed:2010kv,the_amplituhedron,
Arkani-Hamed:2013kca,Franco:2014csa,Dian:2024hil}.

\subsubsection{Residues and leading singularities}
A second computational output is the organization of residues through
on-shell diagrams and the positive Grassmannian. On-shell diagrams provide
local coordinates on cells of the positive Grassmannian, and their boundary
operations encode physical operations such as factorization, soft limits and
unitarity cuts. The corresponding Grassmannian forms produce Yangian-invariant leading singularities amplitudes~\cite{Postnikov:2006kva,ArkaniHamed:2009dn,
Grassmannian,Bourjaily:2012gy}. In this way, the
positive Grassmannian gives a combinatorial and geometric classification of
many residues that would otherwise arise from complicated contour
computations. This viewpoint is essential throughout the thesis: leading
singularities arise as residues on boundaries and are naturally encoded by
on-shell diagrams. This geometric understanding provides a powerful route to
all-loop classifications of leading singularities. The classification in
this thesis builds on the earlier Grassmannian and on-shell-diagram
description of leading singularities, while specializing it to the Wilson loop
with Lagrangian insertion~\cite{Grassmannian,ArkaniHamed:2009dn,Bourjaily:2012gy,BrownHennMazzucchelliTrnka2025}.

\subsubsection{Local integrands and pure integrals}
Positive geometry also clarifies the choice of useful loop-integrand bases.
In modern amplitude computations one often wants integrands with prescribed
singularity properties: local integrals, \(\mathrm{d}\log\) integrands, pure
integrals, or bases adapted to generalized unitarity cuts. Pure integrals are
normalizations whose \(\epsilon\)-expansion has a uniform transcendental-weight
structure and which often satisfy especially simple differential equations.
The Amplituhedron and related
Grassmannian constructions naturally produce logarithmic forms, often
expressible in favorable coordinates as products of \(\mathrm{d}\log\)'s
~\cite{ArkaniHamed:2010kv,Grassmannian,the_amplituhedron}. This connects
positive geometry to the construction of local integrands and to prescriptive
unitarity, where one chooses integrals whose cuts isolate particular physical
information~\cite{BourjailyTrnka2015,BourjailyHerrmannTrnka2017,
BourjailyHerrmannLangerMcLeodTrnka2019,
BourjailyHerrmannLangerMcLeodTrnka2020}. More broadly, it suggests that
positive geometry may help explain which integrand bases are natural before
integration, rather than merely providing one representation after such a basis
has been chosen~\cite{ArkaniHamedBaumannHillmanJoyceLeePimentel2025,
DePokraka2024,CapuanoFerroLukowskiPalazio2025,GlewPokraka2025}.

\subsubsection{Logarithmic forms and Hodge theory}
The appearance of logarithmic forms also connects positive geometry to the
cohomological structures underlying Feynman integrals. In the
differential-equation approach, one seeks bases of master integrals satisfying
canonical systems of linear differential equations
~\cite{Kotikov1991,Remiddi1997,GehrmannRemiddi2000,Henn2013,Henn:2014yza}.
The role of \(\mathrm{d}\log\)-forms in this construction is closely tied to pure
integrals, uniform transcendental weight, and \(\epsilon\)-factorized
systems in the polylogarithmic setting
~\cite{Duhr:2011zq,Goncharov:2010jf,Henn:2014yza,Dixon:2016apl}; see
Subsection~\ref{subsec:Pure Integrals and Canonical Differential Equations}
for the integrand and differential-equation perspective used later.

From the positive-geometry side, logarithmic singularities are built into the
definition of canonical forms through the recursive residue property. The
formulation of~\cite{Brown:PG_Hodge} recasts positive geometries and canonical
forms in the language of mixed Hodge theory, placing them in the same broad
mathematical landscape as Feynman integrals viewed as periods of algebraic
varieties~\cite{Bloch:2005bh,Vanhove2014,Brown2015}. This perspective becomes
especially relevant beyond the polylogarithmic case, where Hodge-theoretic
information can guide the construction of \(\epsilon\)-factorized
differential equations for elliptic and Calabi--Yau integrals
~\cite{AdamsWeinzierl2018,FrellesvigEtAl2021,
PoegelWangWeinzierl2023,GoergesNegaTancrediWagner2023,
DuhrMaggioNegaSauerTancrediWagner2025}.

\subsubsection{Geometric Landau analysis}
After integration, the central objects are no longer rational forms but
multivalued transcendental functions. Positive geometry still provides useful
information, because the singularities of the integrated function are
constrained by the singularities and boundary structure of the integrand.
Potential singularities of Feynman integrals are detected by Landau analysis,
which describes when the integration contour is pinched by propagator
singularities~\cite{Landau1959,ColemanNorton1965,Eden:1966dnq}. However,
ordinary Landau analysis gives a candidate locus rather than the actual
singularity set: some candidates can be removed by numerator zeros or by
cancellations among different terms in the full integrand. This was already emphasized in the study of Landau
singularities and symbol alphabets of one- and two-loop MHV amplitudes in
planar \(\mathcal N=4\) SYM~\cite{DennenSpradlinVolovich2016}. A geometric
refinement was then proposed using the Amplituhedron: instead of starting
from a particular representation in terms of local Feynman integrals, one
extracts candidate branch points directly from the boundary structure of the
geometry~\cite{DennenPrlinaSpradlinStanojevicVolovich2017}. This provides a
way to distinguish physical singularities from spurious ones, and it was
further developed in all-loop studies of massless planar theories
~\cite{PrlinaSpradlinStanojevic2018}. These ideas underlie later work
relating Amplituhedron boundaries, momentum-Amplituhedron boundaries,
negative geometries, and geometric Landau analysis to the symbol alphabets of
integrated quantities
~\cite{Arkani-Hamed:2018rsk,ferroLukowskiMoerman2020boundaries,
neg_geom_pos,ChicherinHennMazzucchelliTrnkaYangZhang2026}.

\subsubsection{Symbol bootstrap and cluster structures}
Landau analysis is closely related to the symbol bootstrap. When an
integral is expected to be polylogarithmic, part of the integration problem
is reduced to determining its symbol alphabet and the allowed tensors built
from that alphabet. Physical constraints such as first-entry conditions,
Steinmann relations, final-entry conditions, collinear limits, multi-Regge
limits and symmetry can then determine highly non-trivial amplitudes without
performing all integrations directly
~\cite{Goncharov:2010jf,DixonDrummondHenn2011,
CaronHuotDixonMcLeodVonHippel2016,DixonEtAl2017}. Positive geometry enters
this bootstrap program by supplying geometric constraints on which letters
can appear. In planar \(\mathcal N=4\) SYM, many symbol letters are closely
related to cluster coordinates on Grassmannians, and cluster adjacency gives
additional restrictions on which letters can appear next to one another
~\cite{GoldenGoncharovSpradlinVerguVolovich2014,
DrummondFosterGurdoganHarrington2019,
DrummondFosterGurdoganKalousios2020,HenkePapathanasiou2020,
lukowski2019boundaries,Chicherin:2020umh}. Positive geometry therefore supplies part of the bootstrap
data: it constrains the alphabet through boundary and Landau analysis, and
organizes the allowed letters and adjacencies through Grassmannian and
cluster structures.

\subsubsection{Positivity from dual volumes}
Positive geometry also suggests new analytic forms of positivity. For
polytopes, the canonical function admits a dual volume interpretation, an
idea already visible in the description of certain tree-level gluon
amplitudes as volumes of polytopes~\cite{Hodges:2009hk}. More generally, dual
volume representations lead naturally to complete monotonicity and to
Laplace transform formulae involving non-negative measures. Such positivity
properties have recently appeared in quantum field theory, for example for
scalar Feynman integrals in suitable kinematic variables~\cite{Henn:CM}. This
fits into a broader pattern of positivity phenomena in planar
\(\mathcal N=4\) SYM, from positivity of Amplituhedron and Wilson-loop
integrands to observed positivity properties of integrated six-point
amplitudes and finite Wilson-loop quantities
~\cite{positive_amplitudes,Dixon:2016apl,neg_geom_pos,Henn:CM}. For
Amplituhedra, this perspective points toward the search for a dual
Amplituhedron: a dual geometric object whose volume or measure-theoretic
description would make positivity of the canonical function manifest
~\cite{Positive_geometries,Herrmann:2020qlt,Ferro,positive_amplitudes,
mazzucchelli2025exterior}.

\medskip

In summary, positive geometry has allowed several concrete advances on the
physics side by providing the following geometric dictionary:
\[
\begin{array}{ccl}
\text{canonical differential forms} & \leadsto &
\text{planar } \mathcal{N}=4 \text{ tree-amplitudes and loop-integrands},\\[2mm]
\text{triangulations} & \leadsto &
\text{BCFW expansions and cancellation of spurious poles},\\[2mm]
\text{cut structure} & \leadsto &
\text{local integrals and prescriptive unitarity},\\[2mm]
\text{logarithmic forms} & \leadsto &
\text{pure integrals and canonical differential equations},\\[2mm]
\text{dual volumes} & \leadsto &
\text{positivity and complete monotonicity},\\[2mm]
\text{boundaries and residues} & \leadsto &
\text{factorization and leading singularities},\\[2mm]
\text{boundary stratifications} & \leadsto &
\text{geometric Landau analysis},\\[2mm]
\text{Grassmannian geometry} & \leadsto &
\text{positivity and cluster structures}.
\end{array}
\]
The chapters that follow focus on the final four lines of this dictionary: positivity properties from dual volume representations of canonical forms and leading singularities and Landau singularities together with their emergent positivity and cluster structures.

\subsection{Main contributions}
\label{sec:Main Contributions}

This thesis has two complementary goals. The first is pedagogical: to
introduce positive geometries and their emergence in the formulation of
scattering amplitudes in planar maximally supersymmetric Yang--Mills theory
via the Amplituhedron. Along the way, we also review leading singularities
and their relation to on-shell diagrams, as well as the basic language of
symbols and cluster algebras used later in the thesis. The second is to
develop applications of this framework to positivity properties, volume
interpretations, and singularities of loop-level observables. The main
contributions fall into three directions.

\begin{figure}[pos=t]
\centering
\begin{minipage}[c]{0.32\linewidth}
\includegraphics[width=\linewidth]{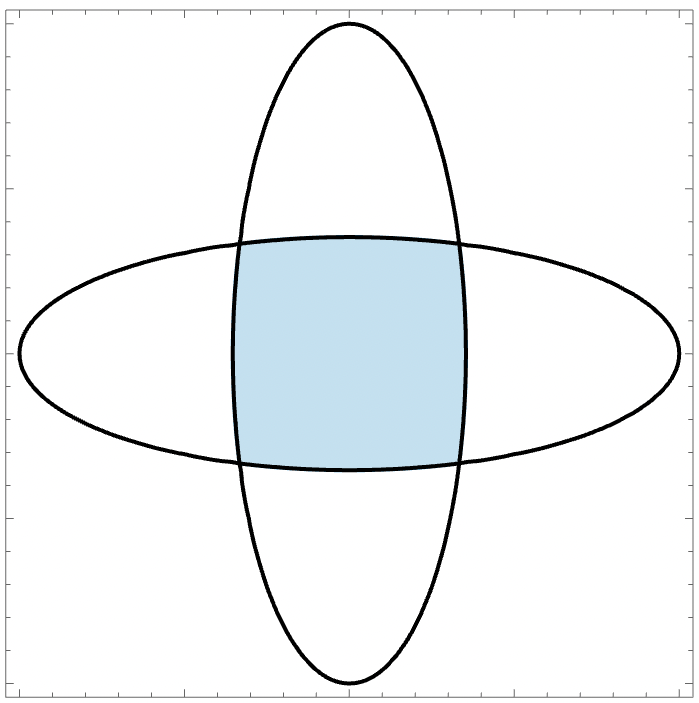}
\end{minipage}
\hfill
\begin{minipage}[c]{0.32\linewidth}
\includegraphics[width=\linewidth]{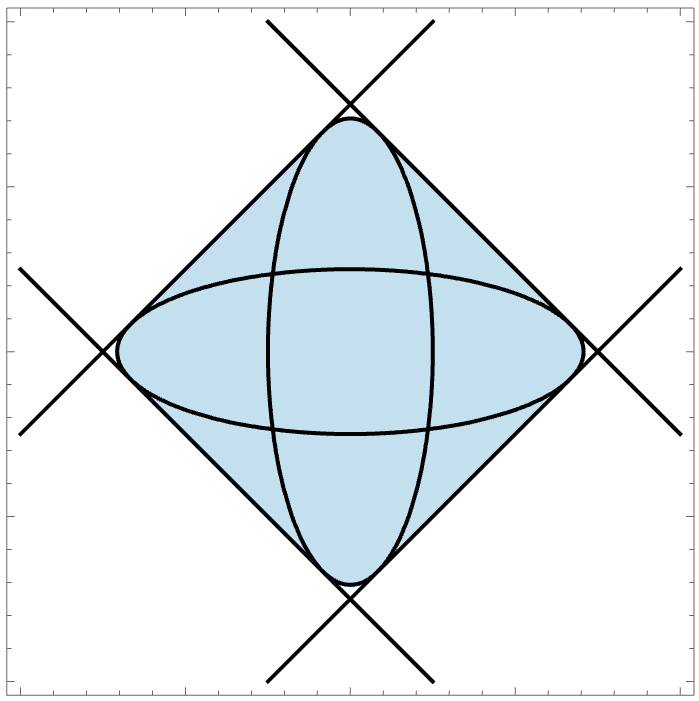}
\end{minipage}%
\hfill
\begin{minipage}[c]{0.35\linewidth}
\includegraphics[width=\linewidth]{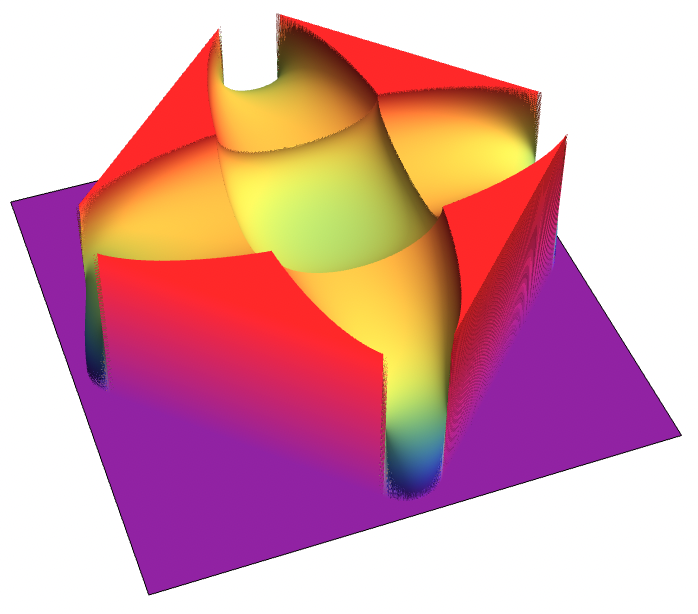}
\end{minipage}
\caption{(Dual volume of a curvy quadrilateral~\cite{Mazzucchelli:DV}) On the left, a curvy
quadrilateral (in blue), viewed as a positive geometry in the plane. In the
middle, its convex dual set. On the right, the graph of the non-negative
measure on the dual set, whose Laplace transform equals the
canonical function of the original geometry.}
\label{fig:curvy_4gon}
\end{figure}

\subsubsection{Duality for positive geometries}

The first contribution concerns the interpretation of canonical forms as
volume functions. For polytopes, this dual-volume interpretation is clear: the canonical form computes the volume of the dual polytope. This realizes, in a precise setting, the idea that amplitudes are volumes in
kinematic space. We extend this picture beyond polytopes.

A central theme is the relation between positive geometries and complete
monotonicity~\cite{Henn:CM}. We show that, for broad classes of examples,
canonical functions can be realized as Laplace transforms of non-negative
measures supported on dual regions~\cite{Mazzucchelli:DV}, see
Figure~\ref{fig:curvy_4gon}. Unlike in the polytope case, these measures are
generally not algebraic: their densities are period functions and may evaluate
to transcendental functions, reflecting the nonlinearity of the original
positive geometry. This provides a mechanism by which canonical forms inherit
strong positivity. We develop explicit examples and an algorithmic approach
for computing measures representing planar geometries bounded by lines and
conics. This leads to a refined notion of duality for positive geometries, in
which the original semialgebraic set and its canonical form are replaced by a
dual geometry equipped with a non-negative, generally transcendental measure.

We then analyze notions of convexity and duality in Grassmannian settings
~\cite{mazzucchelli2025exterior}. These are crucial steps toward constructing
a full dual picture for the Amplituhedron. This leads us to consider the
\textit{exterior cyclic polytope}, a polytope in the Pl\"ucker embedding that
can be regarded as a linear cousin of the Amplituhedron, and to discuss first
steps toward concrete dual-volume constructions for Amplituhedra.

Finally, we discuss positivity of Aomoto forms
~\cite{MazzucchelliHennAomotoForms}. These are polylogarithmic functions
arising from integral pairings of two simplices in projective space. We show
that, on a natural Euclidean region in the space of deformations, Aomoto forms
satisfy complete monotonicity. This gives a broad source of strongly positive
transcendental functions, including examples built from Goncharov
polylogarithms.

\begin{figure}[pos=t]
\centering
\resizebox{\textwidth}{!}{%
\renewcommand{\arraystretch}{1.15}
\begin{tabular}{@{}c@{\hspace{1.5em}}c@{\hspace{1.5em}}c@{\hspace{1.5em}}c@{\hspace{1.5em}}c@{}}
\raisebox{-0.5\height}{%
\begin{minipage}{0.23\textwidth}
\centering
\begin{tikzpicture}[scale=0.4]
    
    \coordinate (i) at (1,-2);
    \coordinate (P) at (0.3,0.05);
    \coordinate (Q) at (-0.45,2.25);
    \coordinate (j) at (-2,2);
    \coordinate (k) at (1,2.5);
    \coordinate (C) at (-0.7,3);
    \coordinate (D) at (1.35,-3);
    \coordinate (E) at (-3.5,1.75);
    \coordinate (F) at (3,2.85);
    \coordinate (A) at (-2.5,-0.5);
    \coordinate (B) at (2.5,0.5);



    \draw[red, thick] (C) -- (D);
    \node[above, red] at (C) {$CD$};

    \draw[red, thick] (E) -- (F);
    \node[above, red] at (E) {$EF$};

    \draw[teal, thick] (A) -- (B);
    \node[above, teal] at (B) {$AB$};

    \fill (i) circle (2pt);
    \fill (j) circle (2pt);
    \fill (k) circle (2pt);
    \fill (P) circle (2pt);
    \fill (Q) circle (2pt);

    \node[right] at (i) {$k$};
    \node[above] at (j) {$i$};
    \node[above] at (k) {$j$};

     \end{tikzpicture}
\end{minipage}%
}
&
\raisebox{-1.5em}{\Large\(\Longrightarrow\)}
&
\raisebox{-0.2\height}{%
\begin{minipage}{0.32\textwidth}
\centering
\includegraphics[width=\linewidth]{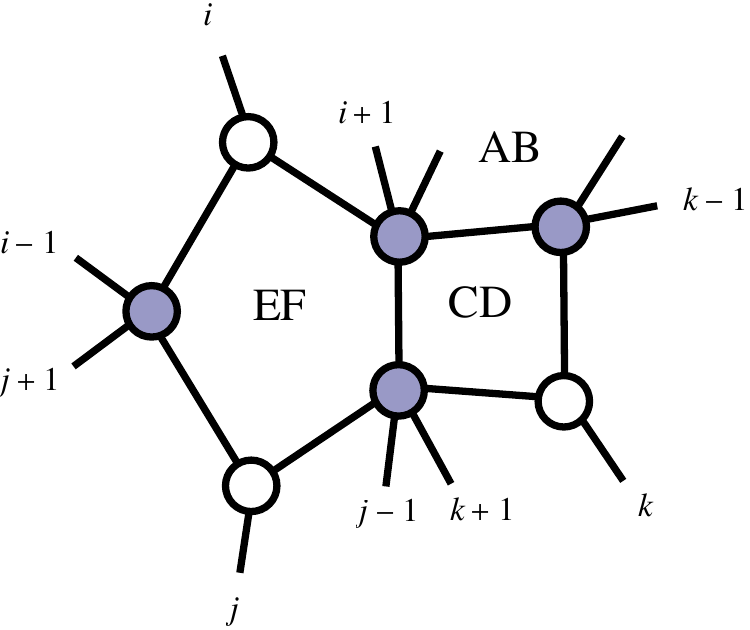}
\end{minipage}%
}
&
\raisebox{-1.5em}{\Large\(\Longrightarrow\)}
&
\raisebox{-1.6\height}{%
\begin{minipage}{0.35\textwidth}
\centering
\resizebox{\linewidth}{!}{%
\(\displaystyle \big\langle k(k-1j)(ii+1)(AB)\big\rangle\)
}
\end{minipage}%
}
\\[1.5ex]
\textbf{Geometric boundary}
&
&
\textbf{Landau diagram}
&
&
\textbf{Landau singularity}
\end{tabular}%
}
\caption{(Geometric Landau analysis~\cite{ChicherinHennMazzucchelliTrnkaYangZhang2026}) The Amplituhedron constrains the
singularities of integrals in \(\mathcal N=4\) SYM through a correspondence
between geometric boundaries and Landau diagrams
~\cite{DennenPrlinaSpradlinStanojevicVolovich2017,
ChicherinHennMazzucchelliTrnkaYangZhang2026}. Boundaries of loop
Amplituhedra are configurations of lines in three-space, while
Landau diagrams are represented by planar bicolored graphs. Solutions of the
Landau equations yield candidate singularities of the integral.}
\label{fig:sing_geom}
\end{figure}

\subsubsection{From integrands to integrals}

The second contribution concerns positive geometries for integrated
quantities. At loop level, positive geometries yield integrands as canonical
forms, while actual amplitudes are obtained by integrating these. The result are multivalued transcendental
functions of the kinematics. Their branch loci, symbols and alphabets are
therefore not visible from the poles of the integrands alone, but are
nevertheless constrained by the same geometric data.

We study this question in planar \(\mathcal N=4\) supersymmetric Yang--Mills
theory and in related finite field-theoretic quantities. Two notions of singularity play
complementary roles. Leading singularities are maximal residues of loop
integrands, defined before integration and naturally described by on-shell
diagrams and positive geometry. Landau singularities, by contrast, are
singularities of the integrated function in external kinematic space. They are
detected by Landau equations, which describe when the integration contour is
pinched by propagator singularities, see Figure~\ref{fig:sing_geom}.

One main result is the all-loop classification of leading singularities of the
Wilson loop with Lagrangian insertion, a finite field-theoretic quantity closely related to
the logarithm of the planar amplitude. We classify these leading singularities
and show how they are organized by the underlying positive geometry
~\cite{BrownHennMazzucchelliTrnka2025}. We also develop geometric Landau
analysis for the same class of observables
~\cite{ChicherinHennMazzucchelliTrnkaYangZhang2026}. The key idea is to
refine the naive Landau locus using the boundary structure of the relevant
positive or negative geometry, thereby constraining the possible symbol
alphabet of the integrated answer.

A detailed understanding of the boundary structure of loop Amplituhedra, and
of the numerators of their canonical forms, is therefore crucial for refining
Landau analysis and may also inform the construction of the relevant space of
logarithmic forms. As a case study, we analyze the two-loop four-point MHV
Amplituhedron, studying its complex and real boundary stratifications with
tools from computer algebra~\cite{Dian:2024hil}.

This computational approach to Amplituhedron boundaries is complemented by a
broader study of hyperplane arrangements in Grassmannians. In
~\cite{MazzucchelliPavlovWang2025}, we study these arrangements with emphasis
on their topology, encoded by Euler characteristics. The topology of these
complements controls the dimensions of natural cohomology groups, and in
related settings such dimensions count master integrals
~\cite{AomotoKita2011,Mizera:2017cqs,Mizera:2019gea,ArkaniHamedHennTrnka2021}.

\subsubsection{Positivity and cluster structures}

The third contribution develops a Grassmannian framework for Landau
singularities in planar kinematics. Momentum twistors turn loop momenta into
lines in projective three-space, so propagator equations become incidence
conditions among lines. This reformulates the usual
Landau analysis in terms of incidence varieties in
products of Grassmannians, whose projections produce discriminants and
resultants describing candidate singular loci.

The foundations for this viewpoint are developed in two steps. First, we study
the algebraic geometry of configurations of lines in three-space with
prescribed incidences. These line-incidence varieties provide the basic
geometric objects underlying the Landau diagrams relevant for planar
amplitudes~\cite{HolleringMazzucchelliParisiSturmfels2025Lines}. Second, we
formulate Landau analysis directly on the Grassmannian: leading and
super-leading Landau singularities are expressed as discriminants and
resultants associated with these incidence varieties
~\cite{HolleringMazzucchelliParisiSturmfels2026}. This is a geometric
language for studying singularities at any loop order.

This formulation is also related to positivity and cluster algebras. In planar \(\mathcal N=4\) SYM, amplitudes are expected to be regular on the positive region of kinematic space, and their rational singularities are conjectured to factorize into cluster variables; related cluster-algebraic structures also occur in individual Feynman-integral families~\cite{Chicherin:2020umh}. We unravel recursive structures in Grassmannian Landau analysis, governed by substitution maps closely related to promotion maps, and explain the emergence of positivity and cluster factorization. For several infinite families of Landau diagrams at arbitrary loop order, we prove the expected positivity and cluster-factorization properties~\cite{HolleringMazzucchelliParisiSturmfels2026Positivity}.

This final contribution connects the two main themes of the thesis. On the
one hand, it uses positive and Grassmannian geometry to control physical
singularities after integration. On the other hand, it shows that these
singularities retain a memory of the same combinatorial structures---positroid
strata, promotion maps and cluster variables---that organize amplitudes at the
integrand level.

\subsection{Outline and reading guide}
\label{sec:Outline}

This thesis is divided into two parts with distinct purposes.
Chapters~\ref{ch:Scattering Amplitudes}--\ref{ch:Amplituhedra} are primarily
pedagogical: they introduce the physical and mathematical background needed
for the rest of the thesis. Chapters~\ref{ch:Canonical Forms as Dual Volumes}--%
\ref{ch:positivity-and-cluster-structures} contain the main research results
and develop three complementary directions concerning positivity, volumes,
and singularities.

\subsubsection{Pedagogical foundations}
Chapter~\ref{ch:Scattering Amplitudes} reviews the \(S\)-matrix perspective,
on-shell kinematics, spinor-helicity variables, recursion relations, and the
special role of planar \(\mathcal N=4\) supersymmetric Yang--Mills theory. It
then introduces momentum twistors, on-shell diagrams, and Grassmannian contour
formulae. These provide the language used throughout the thesis: recursion
relations anticipate triangulations, momentum twistors give natural
coordinates for planar kinematics, and Grassmannian contours connect
residues, positroid cells, and leading singularities.

Chapter~\ref{ch:Positive Geometries} introduces positive geometries and
canonical forms. Starting from polytopes, we discuss logarithmic
singularities, recursive residues, triangulations, adjoint hypersurfaces,
push-forwards, scattering equations, and integral representations. The aim
is to provide a self-contained mathematical framework for the idea that
physical observables may arise as canonical differential forms of geometric
regions.

Chapter~\ref{ch:Amplituhedra} applies this framework to planar
\(\mathcal N=4\) SYM. Beginning with Hodges' polytope picture and cyclic
polytopes, it introduces tree Amplituhedra, BCFW tiles, and loop
Amplituhedra. BCFW recursion becomes a triangulation statement, while loop
Amplituhedra extend the geometry to rational loop integrands. The chapter
ends with the four-point two-loop Amplituhedron as a case study of
boundaries, numerators, and real and complex stratifications. Although this
example anticipates later research questions, the principal purpose of the
chapter is to complete the pedagogical bridge between scattering amplitudes
and positive geometry.

\subsubsection{Research chapters}
Chapter~\ref{ch:Canonical Forms as Dual Volumes} develops the first main
research direction: dual volumes and positivity of canonical forms. We study
complete monotonicity, Laplace-transform representations, and non-negative
measures associated with nonlinear positive geometries. We then discuss
convexity and duality in Grassmannians, first steps toward dual
Amplituhedra, and positivity properties of Aomoto forms.

Chapter~\ref{ch:From Integrands to Integrals} develops the second research
direction, concerning the passage from rational integrands to integrated
transcendental functions. After reviewing symbols, alphabets, leading
singularities, and Landau equations, we study all-loop leading singularities
of Wilson loops with Lagrangian insertion and develop geometric Landau
analysis as a tool for constraining the branch loci and symbol alphabets of
integrated quantities.

Chapter~\ref{ch:positivity-and-cluster-structures} develops the third
research direction. We formulate Landau analysis in momentum-twistor and
Grassmannian language, uncover recursive structures, study the reality and
regularity of candidate singularities in positive kinematics, and analyze
the factorization of rational Landau singularities into cluster variables.
This connects the analytic structure of loop-integrals with the
positive and combinatorial structures of the integrand.

The conclusion summarizes the principal results and discusses future
directions.

The separation between the two parts also provides a reading guide.
Readers seeking a systematic introduction should follow
Chapters~\ref{ch:Scattering Amplitudes}--\ref{ch:Amplituhedra} in order before
turning to the research chapters. Readers already familiar with scattering
amplitudes may consult these chapters selectively and begin directly with
Chapter~\ref{ch:Canonical Forms as Dual Volumes},
\ref{ch:From Integrands to Integrals}, or
\ref{ch:positivity-and-cluster-structures}, according to their interests.

Readers mainly interested in positive geometry and mathematics may begin
with Chapter~\ref{ch:Positive Geometries} and proceed to
Chapter~\ref{ch:Canonical Forms as Dual Volumes}, using
Chapter~\ref{ch:Amplituhedra} for the physical application. Readers interested
in loop amplitudes and their analytic structure should focus on
Chapters~\ref{ch:Amplituhedra}, \ref{ch:From Integrands to Integrals}, and
\ref{ch:positivity-and-cluster-structures}. The more technical examples may
be consulted selectively; their role is to make the general geometric ideas
and the new results concrete.

%% file: Ch2.tex

\section{Scattering amplitudes}
\label{ch:Scattering Amplitudes}

Scattering amplitudes are among the most basic observables in quantum field
theory. They encode transition amplitudes between asymptotic particle states
and, after squaring and integrating over phase space, give physical quantities
such as cross sections and decay rates
~\cite{Peskin:1995ev,Weinberg:1995mt,Schwartz:2014sze}. In perturbation
theory, amplitudes are usually computed from Feynman diagrams. At tree
level this yields rational functions of the external kinematic data, while
at loop level it yields rational differential forms that must be integrated
over loop momenta.

Modern on-shell methods reorganize this perturbative expansion. Rather
than arising from a Lagrangian and summing diagrams, amplitudes
are functions constrained by general physical principles: Lorentz covariance,
locality, unitarity, analyticity, factorization, and symmetry
~\cite{Dixon:1996wi,Bern:2007dw,Elvang:2013cua,Henn:2014yza}. These constraints
are especially powerful in gauge theory, where on-shell methods reveal
structures that are invisible term by term in the Feynman-diagram expansion.
The simplest examples are recursion relations for tree amplitudes, which
construct higher-point amplitudes from lower-point ones using only their pole
structure and factorization residues~\cite{Britto:2004ap,Britto:2005fq}.

The main theory considered in this thesis is planar maximally supersymmetric
Yang--Mills theory, or planar \(\mathcal N=4\) SYM. This theory is a central
testing ground for modern ideas in amplitudes because of its high degree of
symmetry, including superconformal symmetry, dual conformal symmetry, and
Yangian symmetry~\cite{Drummond:2008vq,Drummond:2009fd,Beisert:2010jr}. In
the planar limit, the natural kinematic variables are momentum twistors. They
make dual conformal symmetry manifest and turn propagator conditions into
simple incidence relations in projective space
~\cite{Hodges:2009hk,Mason:2009qx}. These variables are the language in which
the Amplituhedron and the Grassmannian formulation of amplitudes are most
naturally expressed~\cite{Grassmannian,the_amplituhedron}. Consequently, the Amplituhedron construction itself relies essentially on the
special properties of planar \(\mathcal N=4\) SYM, especially planarity,
supersymmetry, and dual conformal symmetry. The broader perspective of the
thesis is that the structures it exposes---canonical forms, boundaries,
residues, triangulations, and positivity---also have analogues in other
positive-geometric formulations of physical observables, such as
associahedra for bi-adjoint scalar amplitudes, momentum-space and ABJM
Amplituhedra, correlator geometries, and cosmological polytopes, as reviewed
in Section~\ref{sec:The Positive Geometry Program}. These are distinct from
the general amplitude methods, such as unitarity cuts, master integrals,
differential equations, and Landau analysis, that connect the discussion to
QCD.

The goal of this chapter is to introduce the amplitude-theoretic language
needed in the rest of the thesis. We begin with scattering amplitudes from
the on-shell point of view, introduce spinor-helicity variables and BCFW
recursion, explain the special role of planar \(\mathcal N=4\) SYM, and then
pass to momentum twistors, on-shell diagrams, and Grassmannian contour
formulae. This provides the bridge from standard scattering amplitudes to the
positive-geometric framework developed in the following chapters.

\medskip

The chapter is structured as follows. 
Section~\ref{sec:What is an Amplitude?} recalls the \(S\)-matrix definition of
scattering amplitudes and introduces the on-shell viewpoint, spinor-helicity
variables, and examples of three- and four-point gluon amplitudes.
Section~\ref{sec:Recursion Relations} explains on-shell recursion, deriving
BCFW recursion from Cauchy's theorem and factorization, and briefly discusses
its loop-integrand analogue. Section~\ref{sec:Maximally Supersymmetric Yang-Mills Theory}
introduces planar \(\mathcal N=4\) SYM, its on-shell supermultiplet,
superamplitudes, helicity sectors, and enhanced symmetries.
Section~\ref{sec:Momentum Twistors} introduces momentum twistors, which solve
the planar kinematic constraints and turn propagator singularities into
incidence conditions in projective space. Finally,
Section~\ref{sec:On-Shell Diagrams and Grassmannian Contours} reviews
on-shell diagrams, positroid cells, and Grassmannian contour formulae, giving
the first bridge from amplitudes to the positive Grassmannian and the
Amplituhedron.

\subsection{What is an amplitude?}\label{sec:What is an Amplitude?}

Quantum field theory (QFT) is the modern framework combining the principles
of special relativity and quantum mechanics~\cite{Peskin:1995ev,Weinberg:1995mt}. It provides the basic language for describing particle
interactions across a wide range of energy scales. In the context of this
work, the main objects of interest are \textit{scattering amplitudes}, which
encode the dynamical information underlying observable quantities in
high-energy particle processes~\cite{Elvang:2013cua,Henn:2014yza}. For the
purposes of this thesis, the important point is that amplitudes can be
studied as functions of external on-shell kinematic data, constrained by
symmetry, locality, unitarity, and analyticity.

In conventional QFT, scattering amplitudes are defined in terms of the
\textit{\(S\)-matrix}~\cite{Weinberg:1995mt,Peskin:1995ev}. This construction
assumes that, at sufficiently early and late times compared to the duration
of the interaction, the dynamics becomes effectively free. More precisely,
one assumes the existence of asymptotic Hilbert spaces
\(\mathcal H_{\rm in}\) and \(\mathcal H_{\rm out}\), spanned by
multi-particle states of free fields. Under the additional assumption of
asymptotic completeness, these asymptotic states describe the full space of
physical scattering states~\cite{Haag:1958vt,ReedSimon:1979}. The
\(S\)-matrix is then a unitary linear map
\begin{equation}
    S:\mathcal H_{\rm in}\longrightarrow \mathcal H_{\rm out} \, ,
\end{equation}
and a scattering amplitude is the corresponding matrix element
\begin{equation}
    A_{{\rm in}\rightarrow {\rm out}}
    =
    \langle {\rm out}|\, S \, |{\rm in}\rangle \, .
    \label{eq:Ainout}
\end{equation}
Its squared modulus, together with the appropriate phase-space measure, gives
the probability density for the process. After integration over phase space,
one obtains physical observables such as decay rates and cross sections.

For a process involving a total of \(n\) external particles, different
\textit{channels} correspond to different partitions of these particles into
incoming and outgoing states. These channels are related by analytic
continuation of a single \(n\)-point amplitude \(A_n\). Equivalently, by
crossing symmetry, an outgoing particle may be regarded as an incoming
antiparticle with reversed momentum, and vice versa. Thus the various
physical scattering processes associated with a fixed set of external
particles correspond to different real kinematic regions of one analytic
function of the external momenta~\cite{Eden:1966dnq,StreaterWightman,Weinberg:1995mt,Peskin:1995ev}.

Amplitudes depend on the external momenta
\(p_i\in\mathbb R^{1,D-1}\), \(i=1,\dots,n\), where
\(\mathbb R^{1,D-1}\) denotes \(D\)-dimensional Minkowski space with
signature \((+,-,\ldots,-)\). They also depend on the quantum numbers of the
external particles, such as spin, helicity, flavour, and colour. It is useful
to separate the identity contribution to the \(S\)-matrix by writing
\(S=\mathbf 1+iT\). The connected part of the scattering process is then
defined by factoring out the overall momentum-conserving delta function:
\begin{equation}
    \langle p_1,\ldots,p_m;{\rm out}|\, T \, |
    p_{m+1},\ldots,p_n;{\rm in}\rangle
    =
    (2\pi)^D \delta^{(D)}
    \left(\sum_{i=1}^n p_i\right)
    A_n(p_1,\ldots,p_n) \, .
\end{equation}
Here \(D\) denotes the dimension of spacetime and \(\delta^{(D)}\) is the
\(D\)-dimensional delta function. By convention, all momenta may be taken to
be outgoing. The amplitude \(A_n\) can be extracted from time-ordered
correlation functions by the Lehmann--Symanzik--Zimmermann (LSZ) reduction
formula, which puts the external legs on shell and removes the corresponding
external propagators~\cite{LSZ:1955}.

A powerful framework for computing amplitudes is \textit{perturbation
theory}. When the relevant interaction is controlled by a small coupling
constant \(g\), observables can be expanded formally around the free theory.
For amplitudes, this gives a loop expansion of the form
\begin{equation}
    A_n
    =
    \sum_{\ell\geq 0} g^{n-2+2\ell}
    A_n^{(\ell)} \, ,
\end{equation}
where \(A_n^{(0)}\) is the tree-level amplitude, and \(A_n^{(\ell)}\) is the
\(\ell\)-loop contribution. Each term is computed by summing all Feynman
diagrams with the prescribed external states and loop order~\cite{Peskin:1995ev,Srednicki:2007qs,Schwartz:2014sze}. The tree-level contribution \(A_n^{(0)}\) gives the
leading semiclassical approximation, while the loop corrections \(A_n^{(\ell)}\),
\(\ell\geq 1\), encode genuinely quantum effects. The efficient computation of
these higher-order contributions is a central problem in perturbative quantum
field theory, both for its intrinsic mathematical structure and for its role
in precision predictions for collider physics~\cite{Ellis:1991qj,Bern:1994zx,Dixon:1996wi,Anastasiou:2003kj,Britto:2004nc,Ellis:2011cr,Elvang:2013cua,Henn:2014yza}.

Feynman diagrams provide a combinatorial representation of the perturbative
expansion. Their external legs correspond to the scattered particles, their
internal edges to propagators, and their vertices to interaction terms in the
Lagrangian. The Feynman rules assign to each diagram an integrand depending
on the external momenta and, at loop level, on internal loop momenta. A
typical \(\ell\)-loop contribution takes the schematic form
\begin{equation}\label{eq:feynm_int}
    \int
    \prod_{a=1}^{\ell}
	    \frac{\mathrm{d}^D k_a}{(2\pi)^D}
    \;
    \frac{N(k_a,p_j)}
    {\prod_e \big(q_e^2-m_e^2+i 0 \big)} \, ,
\end{equation}
where \(k_a\) are loop momenta, \(p_j\) are external momenta, \(N\) is a
theory-dependent numerator, and the denominator factors arise from propagators corresponding to edge $e$ in the Feynman diagram; here the \(q_e\) are integer linear combinations of the \(k_a\)
and \(p_j\). At tree level no loop integrations are present, and the
amplitude is obtained directly as a rational function of the external
kinematic data, up to the spinor, polarization, and group-theoretic structures
appropriate to the theory. At loop level, \(\ell\geq 1\), integrals of the
form~\eqref{eq:feynm_int} define periods of algebraic varieties and exhibit a
rich analytic structure, whose systematic study remains an active area of
research~\cite{Bloch:2005bh,Brown:2009qja,Panzer2015,Abreu:2022mfk}.

Loop integrations may be divergent. Ultraviolet (UV) divergences arise from
large loop momenta and probe the short-distance behavior of the theory,
whereas infrared (IR) divergences arise from soft or collinear regions of
momentum space, especially in theories with massless particles. These IR
divergences are regulator-dependent singularities, for example poles in
dimensional regularization. They should be distinguished from the physical
poles and branch cuts of amplitudes as functions of kinematic invariants,
although the two are related by factorization and unitarity. Regularization
and renormalization provide the systematic framework for defining loop
amplitudes and extracting finite physical observables~\cite{Collins1984,Peskin:1995ev,Weinberg:1995mt,Srednicki:2007qs,ItzyksonZuber}.
IR divergences are also relevant for the theory studied in this work,
\(\mathcal N=4\) super Yang--Mills theory. In principle, these divergences
obscure the assumptions underlying the conventional \(S\)-matrix.
Nevertheless, the high degree of symmetry and special analytic structure of
this theory make it possible to define and study suitably regulated
amplitudes, amplitude integrands, or finite quantities in which the universal
soft and collinear behavior has been subtracted or factorized~\cite{Henn:2014yza,Elvang:2013cua,Bern:1994zx}.

We adopt the perturbative \textit{on-shell} amplitude framework, which treats
scattering amplitudes and loop integrands as functions of external on-shell
data and studies the constraints imposed by locality, unitarity, gauge
invariance, and symmetry~\cite{Britto:2004ap,Britto:2005fq,Elvang:2013cua}.
Its starting point is the classification of the external states themselves.
According to Wigner's classification, elementary particles are associated with
irreducible representations of the Poincar\'e group
\({\rm SO}(1,D-1)\)~\cite{Wigner:1939cj,Weinberg:1995mt}. These fall into two
broad classes:

\begin{itemize}
	\item in the \textit{massive} case, \(m \neq 0\), irreducible
    representations are labelled by the mass \(m\) and the \textit{spin},
    which corresponds to a representation of the little group
    \({\rm SO}(D-1)\). In four dimensions, this reduces to a single quantum
    number \(s \in \tfrac{1}{2}\mathbb{N}\), the \textit{spin};
	\item in the \textit{massless} case, \(m=0\), particles are labelled by
    representations of the little group \({\rm SO}(D-2)\); in four dimensions
    this is characterized by the \textit{helicity}
    \(h \in \tfrac{1}{2}\mathbb{Z}\).
\end{itemize}

In the on-shell framework, scattering amplitudes are defined as functions of
kinematic data subject to the following principles:
\begin{enumerate}
    \item \textbf{On-shell kinematics:} the amplitude depends on momenta \(p_i\)
    satisfying the on-shell conditions
    \begin{equation}
p_i^2 = (p_{i}^0)^2 - (p_{i}^1)^2 - \dots - (p_{i}^{D-1})^2 = m_i^2 \, ,
    \qquad \sum_i p_i = 0 \, ;
\end{equation}
    \item \textbf{Lorentz covariance:} the amplitude transforms covariantly
    under Lorentz transformations, with transformation properties determined
    by the little-group representations, namely spin or helicity, of the
    external states;
    \item \textbf{Analyticity:} the amplitude is an analytic function of the complexified
    kinematic invariants, with singularities only at specific \textit{physical
    poles}~\cite{Eden:1966dnq};
    \item \textbf{Factorization:} on physical poles corresponding to
    intermediate on-shell states, the amplitude factorizes into lower-point
    amplitudes~\cite{Weinberg:1964ew};
    \item \textbf{Symmetry:} the amplitude is further constrained by the
    symmetries of the theory; for instance, in \(\mathcal{N}=4\) SYM, by
    superconformal and dual conformal symmetry~\cite{Beisert:2010jr}.
\end{enumerate}

These principles determine the amplitude uniquely in many cases and provide a
constructive definition independent of an explicit formulation in terms of
asymptotic states. This perspective is closely related to the \(S\)-matrix
bootstrap program, initiated in the 1960s~\cite{Eden:1966dnq,Chew:1961ev},
which aimed to determine scattering amplitudes directly from general
consistency conditions such as analyticity, unitarity, and crossing symmetry.
Modern on-shell methods can be viewed as a realization of this
philosophy, enhanced by additional structures such as factorization, locality,
and, in certain theories, extended symmetries.

\smallskip

We now present an explicit example of amplitudes defined within the
\textit{on-shell} framework. A crucial ingredient in this approach is the
choice of kinematic variables, which can make the underlying principles
manifest and lead to expressions that are both compact and physically
transparent. For massless particles in four spacetime dimensions, a
particularly convenient parametrization is provided by \textit{spinor-helicity}
variables~\cite{Elvang:2013cua,Henn:2014yza}. This formalism exploits the
fact that the Lorentz algebra \(\mathfrak{so}(1,3)\) is isomorphic to
\(\mathfrak{sl}(2,\mathbb{C})\), or equivalently that \({\rm SL}(2,\mathbb{C})\)
is the double cover of \({\rm SO}^+(1,3)\). Analogous constructions in other
dimensions have also been developed~\cite{Cheung:2009dc,Rajan:2024ome,Maazouz:2024qmm,Pokraka:2024fao}.
Given a momentum \(p \in \mathbb{R}^{1,3}\), one associates a \(2 \times 2\)
Hermitian matrix
\begin{equation}
    p^{\alpha \dot{\alpha}} = p^\mu \sigma_\mu^{\alpha\dot{\alpha}} =
    \begin{pmatrix}
        p_0+p_3 & p_1 - i p_2 \\
        p_1 + i p_2 & p_0 - p_3
    \end{pmatrix} \, ,
\end{equation}
where \(\sigma_\mu^{\alpha\dot{\alpha}} = (\mathbb{1}_2, \sigma_1,\sigma_2,\sigma_3)^{\alpha\dot{\alpha}}\)
are the Pauli matrices, with \(\alpha,\dot{\alpha} \in \{1,2\}\). A key
property is 
\begin{equation}
p^2 = \det(p^{\alpha \dot{\alpha}}) \, .
\end{equation}
For massless momenta, \(p^2 = 0\), and hence
\(\det(p^{\alpha \dot{\alpha}}) = 0\). It follows that
\(p^{\alpha \dot{\alpha}}\) has rank one and can be factorized as
\begin{equation}
p^{\alpha \dot{\alpha}} = \lambda^\alpha \widetilde{\lambda}^{\dot{\alpha}} \, ,
\end{equation}
where \(\lambda\) and \(\widetilde{\lambda}\) are two-component \textit{spinors}
or \textit{spinor-helicity variables}. For a real momentum \(p\), the spinors
are related by complex conjugation, \(\widetilde{\lambda}= \pm \lambda^*\), but
in general one allows the momenta to be complex and independent. Applying this construction to each external momentum \(p_i\)
yields a set of spinors \(\lambda_i\) and \(\widetilde{\lambda}_i\) for
\(i=1,\dots,n\).

Lorentz-invariant quantities are then expressed in terms of \textit{spinor
brackets}
\begin{equation}
\la i j \ra := \varepsilon_{\alpha \beta} \lambda_{i}^{\alpha} \lambda_{j}^{\beta} \, ,
\qquad
[ij] := \varepsilon_{\dot{\alpha} \dot{\beta}} \widetilde{\lambda}_{i}^{\dot{\alpha}} \widetilde{\lambda}_{j}^{\dot{\beta}} \, ,
\end{equation}
where \(\varepsilon = \left(\begin{smallmatrix}0 & 1 \\ -1 & 0\end{smallmatrix}\right)\) is the
two-dimensional Levi--Civita tensor. For example, Mandelstam invariants
\(s_{ij}\) become
\begin{equation}
s_{ij} =  (p_i + p_j)^2 = 2\, p_i \cdot p_j = \la ij \ra [ij] \, .
\end{equation}
In particular, both angle and square brackets carry mass dimension one.

Scattering amplitudes are then naturally expressed as functions of spinor
brackets. These variables are not independent, but satisfy quadratic relations
following from momentum conservation,
\begin{equation}\label{eq:mom_cons}
    \sum_{j=1}^n \la i j \ra [j k] = 0 \, ,
\qquad \forall\, i,k \in \{1,\dots,n\} \, .
\end{equation}

A key simplifying feature of spinor-helicity variables is their natural
encoding of helicity. For massless particles in four spacetime dimensions,
the little group is \({\rm SO}(2) \cong {\rm U}(1)\), and acts on the spinors
by rescaling
\begin{equation}
\lambda_i \;\mapsto\; t_i \lambda_i \, ,
\qquad
\widetilde{\lambda}_i \;\mapsto\; t_i^{-1} \widetilde{\lambda}_i \, ,
\end{equation}
for each particle \(i=1,\dots,n\). This transformation leaves the momentum
\(p_i = \lambda_i \widetilde{\lambda}_i\) invariant, but acts non-trivially on
the helicity \(h_i \in \tfrac{1}{2}\mathbb{Z}\) of the particle.
As a consequence, the \(n\)-point amplitude \(A_n\) is covariant under the
action of the little group on each particle \(i\):
\begin{equation}\label{eq:helicity_scaling}
    A_n(t_i \lambda_i,t_i^{-1}\widetilde{\lambda}_i;h_i) =t_i^{-2h_i} A_n( \lambda_i,\widetilde{\lambda}_i;h_i) \, .
\end{equation}
The \textit{parity} transformation, which acts on all momenta as
\(p=(p_0,\vec{p}) \mapsto (p_0,-\vec{p})\), and on spinors as
\(\lambda \leftrightarrow \widetilde{\lambda}\), simultaneously flips the
helicities of all external states and yields 
\begin{equation}\label{eq:parity}
    A_n(1^{h_1} , 2^{h_2}, \dots,n^{h_n}) = A_n(1^{-h_1} , 2^{-h_2}, \dots,n^{-h_n}) \Big|_{\la ij \ra \leftrightarrow [i j] } \, .
\end{equation}

\begin{eg}[Massless three-point tree-level amplitudes]\label{eg:3pt_ampl}
Consider a tree-level amplitude \(A_3(1^{h_1},2^{h_2},3^{h_3})\) for three
massless particles of helicity \(h_i\). Momentum conservation for three
particles imposes that
\begin{equation}
\la 12 \ra [12] = \la 13 \ra [13] = \la 23 \ra [23] = 0  \, .
\end{equation}
As a consequence, the kinematic space factors into two components: either
\begin{equation}
[12]=[23]=[13]=0 \, , \quad \text{or} \quad \la 12 \ra = \la 23 \ra = \la 13 \ra = 0 \, .
\end{equation}
Equivalently, all spinors \(\widetilde{\lambda}_i\) or \(\lambda_i\) are
proportional, respectively. In the former case, \(A_3\) is a rational function of \(\la 12 \ra,\la 23 \ra,\la 13 \ra\) and by~\eqref{eq:helicity_scaling}
\begin{equation}\label{eq:A3_general}
A_3(1^{h_1},2^{h_2},3^{h_3})
= a \,
\la 12 \ra^{h_3 - h_1 - h_2} \,
\la 13 \ra^{h_2 - h_3 - h_1} \,
\la 23 \ra^{h_1 - h_2 - h_3} \, ,
\end{equation}
where \(a\) is a constant. Hence, the three-point amplitude is completely
determined by little-group scaling, in close analogy with three-point
correlation functions in conformal field theory, which are fixed by scaling
dimensions up to an overall constant.

As a concrete example, consider the three-gluon amplitude with helicities
\(h_1=h_2=-1\) and \(h_3=+1\). By dimensional analysis,~\eqref{eq:A3_general}
has mass dimension one, as expected from the cubic interaction term in the
Yang--Mills Lagrangian. We can therefore write
\begin{tcolorbox}[definitionbox]
\textbf{Three-point MHV gluon amplitude.}
\begin{equation}\label{eq:MHV_3pt}
A_3^{\mathrm{YM}}[1^{-},2^{-},3^{+}]
=  \frac{\la 12 \ra^3}{\la 13 \ra \la 23 \ra} \, ,
\end{equation}
\end{tcolorbox}\noindent
where we stripped-off the Yang--Mills coupling constant and color-dependence, as customary for \emph{colored ordered amplitudes}, intruded below. Dimensional
analysis also shows that this amplitude cannot be expressed in terms of square
brackets: replacing angle brackets by square brackets in~\eqref{eq:A3_general}
would yield a function of mass dimension \(-1\), which cannot arise from a
local interaction, i.e. a polynomial in the fields and their derivatives.
Hence,~\eqref{eq:MHV_3pt} is the unique three-gluon amplitude with helicity
configuration \((--+)\). From~\eqref{eq:parity}, the amplitude for the
opposite helicity configuration \((++-)\) is
\begin{tcolorbox}[definitionbox]
\textbf{Three-point anti-MHV gluon amplitude.}
\begin{equation}\label{eq:MHVbar_3pt}
A_3^{\mathrm{YM}}[1^{+},2^{+},3^{-}]
= \frac{[ 12 ]^3}{[13] [23]} \, .
\end{equation}
\end{tcolorbox}\noindent

By a similar argument,
\begin{equation}
    A_3^{\mathrm{YM}}[1^{+},2^{+},3^{+}] = A_3^{\mathrm{YM}}[1^{-},2^{-},3^{-}] = 0 \, .
\end{equation}
For the all-plus amplitude this follows since~\eqref{eq:A3_general} for
\(h_1=h_2=h_3 = 1\) would have mass dimension \(-3\), and similarly for its
square-bracket analogue. Such a term would correspond to a non-local
interaction, which is forbidden by locality.
\end{eg}

We now turn to higher-point amplitudes, where additional principles such as
analyticity and factorization play a crucial role and enforce recursion
relations between amplitudes at different multiplicities.

\subsection{Recursion relations}\label{sec:Recursion Relations}

Scattering amplitudes in perturbation theory are analytic functions of the
external on-shell kinematic data~\cite{Eden:1966dnq}. Prior to performing loop
integrations, each Feynman diagram defines an integrand that is a rational
function of the kinematic invariants built from both external and loop
momenta~\cite{Peskin:1995ev,Weinberg:1995mt}. These rational functions are
constructed according to the Feynman rules by gluing interaction vertices with
propagators. As a consequence of locality, singularities can arise only when
propagators go on shell, namely when internal momenta satisfy \(P^2 = m^2\)
for some mass \(m\) in the spectrum of the theory. The singularity structure
of scattering amplitudes therefore admits a direct physical interpretation:
simple poles correspond to the propagation of intermediate on-shell states,
while branch cuts appearing after loop
integration encode multi-particle production thresholds and the opening of
new physical channels~\cite{Eden:1966dnq,Weinberg:1965nx}.

On-shell recursion relations rely on a simple but powerful interplay between
the analytic structure of scattering amplitudes and general physical
principles such as locality and unitarity~\cite{Britto:2005fq,Britto:2004ap,Eden:1966dnq,Weinberg:1965nx}. At tree level, the analytic structure of
scattering amplitudes is particularly simple: they are meromorphic functions
whose singularities consist only of simple poles associated with physical
factorization channels~\cite{Eden:1966dnq}. In this section we mainly restrict
our attention to tree-level amplitudes for massless particles, and discuss
extensions of recursion relations to loop integrands towards the end.

\subsubsection{Tree-level on-shell recursion}
We now introduce on-shell recursion. Consider a meromorphic deformation
\(\widehat{A}_n(z)\) of an \(n\)-point tree-level amplitude, such that
\(A_n = \widehat{A}_n(0)\), defined for \(z\) on the Riemann sphere
\(\mathbb{P}^1\). This deformation is implemented by shifting the external
momenta \(\widehat{p}_i(z)\) in a way that preserves both the on-shell
conditions \(\widehat{p}_i(z)^2 = 0\) and momentum conservation
\(\sum_{i=1}^n \widehat{p}_i(z) = 0\). Applying Cauchy's residue theorem to
the meromorphic function \(\widehat{A}_n(z)/z\), we obtain
\begin{equation}\label{eq:An_residue}
A_n = {\rm Res}_{z=0} \frac{\widehat{A}_n(z)}{z}
= -\sum_{z_I} {\rm Res}_{z=z_I} \frac{\widehat{A}_n(z)}{z}
- {\rm Res}_{z=\infty} \frac{\widehat{A}_n(z)}{z} \, ,
\end{equation}
where the sum runs over the finite poles \(z_I \in \mathbb{P}^1 \setminus \{0\}\).

By locality, tree-level amplitudes can develop singularities only in the form
of simple poles at \(P_I^2 = 0\), where \(P_I = \sum_{i \in I} p_i\) for some
subset \(I \subseteq \{1,\dots,n\}\). The momentum \(P_I\) corresponds to the
momentum flowing through an internal propagator in a Feynman diagram
contributing to \(A_n\). By unitarity, the amplitude factorizes on such poles
into lower-point amplitudes,
\begin{equation}\label{eq:An_fact}
    {\rm Res}_{P_I^2=0} \, A_n = \sum_{h} A_L(-P_I^{h}) \, A_R(P_I^{-h}) \, ,
\end{equation}
where the sum runs over the helicities \(h\) of the intermediate on-shell
states. This factorization property follows directly from the structure of
perturbation theory: as \(P_I^2 \to 0\), the propagator \(1/P_I^2\) becomes
singular, and the diagram effectively splits into two subdiagrams,
corresponding to the lower-point amplitudes \(A_L\) and \(A_R\).

In order to relate~\eqref{eq:An_residue} and~\eqref{eq:An_fact}, we make two
assumptions:
\begin{enumerate}
    \item First, we assume that the deformation of the kinematics is such that
    \(\widehat{P}_I(z)\) depends linearly on \(z\). This implies that
    \(\widehat{A}_n(z)\) has only simple poles located at points \(z_I\) where
    \(\widehat{P}_I^2(z_I)=0\). For \textit{generic} external kinematics,
    these solutions satisfy \(z_I \neq 0\), and poles associated with
    different propagators occur at distinct points on \(\mathbb{P}^1\).
    \item Second, we assume that \(\widehat{A}_n(z)/z\) has no pole at
    \(z=\infty\). Although \(\widehat{P}_I(z)\) grows linearly with \(z\) and
    therefore does not vanish at infinity, the residue
    \({\rm Res}_{z=\infty} \frac{\widehat{A}_n(z)}{z}\) cannot be interpreted
    in terms of a physical factorization channel as in~\eqref{eq:An_fact}.
    We therefore assume that this contribution vanishes, which is the case,
    for instance, if
\begin{equation}
    \lim_{z \rightarrow \infty }\widehat{A}_n(z) = 0 \quad \implies \quad {\rm Res}_{z=\infty} \frac{\widehat{A}_n(z)}{z} = 0 \, .
\end{equation}
The role of possible boundary contributions at infinity in recursion
relations has been investigated in~\cite{Feng:2009ei,Conde:2012wb,Lemmon:2025dhq},
although a general understanding remains incomplete.
\end{enumerate}

Let us spell out how the propagator factor in the original kinematics arises.
Since \(\widehat{P}_I^2(z)\) is linear in \(z\) and vanishes at \(z=z_I\), we
may write
\begin{equation}
    \widehat{P}_I^2(z) = P_I^2 \left(1-\frac{z}{z_I}\right) \, .
\end{equation}
Thus the residue of \(\widehat{A}_n(z)/z\) at \(z=z_I\) produces the unshifted
propagator factor \(1/P_I^2\), while the remaining factors are the lower-point
amplitudes evaluated on the shifted on-shell kinematics.

Equipped with these assumptions, combining~\eqref{eq:An_residue} with~\eqref{eq:An_fact}, we obtain the \textit{on-shell recursion relations}:
\begin{tcolorbox}[resultbox]
\textbf{Tree-level on-shell recursion.}
\begin{equation}\label{eq:onshell_rec}
    A_n = \sum_I \sum_{h} A_L(-\widehat{P}_I^{\,h};z_I) \, \frac{1}{P_I^2} \, A_R(\widehat{P}_I^{-h};z_I) \, ,
\end{equation}
\end{tcolorbox}\noindent
where the sum runs over all subsets \(I \subseteq \{1,\dots,n\}\) such that
\(\widehat{P}_I^2(z)\) develops a pole at \(z=z_I\), and \(h\) runs over the
helicities of the intermediate on-shell states. This formula holds in any
spacetime dimension and for a broad class of quantum field theories. In gauge
theories,~\eqref{eq:onshell_rec} yields a manifestly gauge-invariant
recursion.

There exist many choices of deformations of the external kinematics in the
literature: the Britto--Cachazo--Feng--Witten (BCFW) shift~\cite{Britto:2005fq,Britto:2004ap}, Risager's all-line shift~\cite{Risager:2005vk}, and more
general multi-line shifts used in the construction of recursion relations and
on-shell representations of amplitudes~\cite{ArkaniHamed:2008gz}. In this
work, we focus on four spacetime dimensions and consider the
Britto--Cachazo--Feng--Witten (BCFW) shift, also known as
\begin{equation}\label{eq:ij_shift}
[i,j\rangle\text{-shift} \quad : \quad
\lambda_i(z) := \lambda_i + z \, \lambda_j \, ,
\qquad
\widetilde{\lambda}_j(z) := \widetilde{\lambda}_j - z \, \widetilde{\lambda}_i \, ,
\end{equation}
for \(i \neq j\) among \(\{1,\dots,n\}\), while keeping all other spinors
unchanged. This deformation preserves both the on-shell conditions and
momentum conservation. Under this shift, an internal momentum
\(\widehat{P}_I(z)\) depends on \(z\) only if exactly one of the legs \(i\) or
\(j\) belongs to the subset \(I\). If both \(i,j \in I\) or both
\(i,j \notin I\), then \(\widehat{P}_I(z) = P_I\) is independent of \(z\) and
therefore does not give rise to a pole of \(\widehat{A}_n(z)\). For a given
choice of shifted legs, the number of poles depends on how many factorization
channels are affected by the deformation. For color-ordered amplitudes, only
factorization channels compatible with the cyclic ordering can appear; this is
why adjacent BCFW shifts lead to particularly simple recursions. In this case
the minimal number of poles, equal to \(n-3\), is obtained when \(i\) and
\(j\) are chosen to be adjacent. In gauge theories, one can show that the
contribution from infinity vanishes for suitable choices of helicities of the
shifted legs~\cite{ArkaniHamed:2008yf}:
\begin{equation}
    [-,-\rangle , [+,+\rangle, [-,+\rangle\text{ -shift} \implies \lim_{z \rightarrow \infty }\widehat{A}_n(z) = 0 \, .
\end{equation}

\subsubsection{Gluon amplitudes and color ordering}
We now apply the recursion relations to compute certain \(n\)-point tree-level
gluon amplitudes in Yang--Mills theory. To this end, we first introduce the
notion of \textit{color ordering}. The full tree-level \(n\)-point amplitude
can be decomposed as
\begin{equation}
    A_n(1,\dots,n) = g_{{\rm YM}}^{\,n-2}\, \sum_{\sigma \in S_{n-1}}  {\rm Tr}(T^{a_1}T^{a_{\sigma(2)}}\cdots T^{a_{\sigma(n)}}) \, A_n[1 , \sigma(2), \cdots ,\sigma(n)] \, ,
\end{equation}
where the sum runs over all permutations of \(\{2,\dots,n\}\), and we have
suppressed the helicity labels of the external particles. The color structure
is fully encoded in the trace factors involving the generators \(T^a\) of the
Lie algebra \(\mathfrak{su}(N)\), while the coefficients \(A_n[\cdots]\) are
individually gauge-invariant \textit{color-ordered amplitudes}. Color-ordered
amplitudes satisfy several important relations:
\begin{enumerate}
    \item \textit{Cyclic invariance and reflection symmetry}~\cite{Mangano:1990by,Dixon:1996wi}:
    \begin{align}
        A_n[1,2,\dots,n] &= A_n[2,3,\dots,n,1] \, , \\
        A_n[1,2,\dots,n] &= (-1)^n A_n[n,\dots,2,1] \, .
    \end{align}

    \item \textit{The Kleiss--Kuijf relations}~\cite{Kleiss:1988ne}:
    \begin{equation}
        A_n[1,\alpha,n,\beta]
        = (-1)^{|\beta|}
        \sum_{\sigma \in \alpha \shuffle \beta^T}
        A_n[1,\sigma,n] \, ,
    \end{equation}
    where \(\alpha\) and \(\beta\) are ordered sets, \(\beta^T\) denotes the reversed ordering of \(\beta\), and \(\shuffle\) denotes all shuffles preserving the relative order within \(\alpha\) and \(\beta^T\). They reduce the number of independent amplitudes to $(n-2)!$.

    \item \textit{The Bern--Carrasco--Johansson (BCJ) relations}~\cite{Bern:2008qj}:
    \begin{equation}
        \sum_{i=2}^{n-1}
        \left( \sum_{j=2}^{i} s_{1j} \right)
        A_n[2,3,\dots,i,1,i+1,\dots,n]
        = 0 \, .
    \end{equation}
   They further reduce the number of independent amplitudes to \((n-3)!\).
\end{enumerate}
In addition, color-ordered amplitudes obey universal factorization, soft, and
collinear limits~\cite{Weinberg:1965nx,Dixon:2011xs}. At loop level, in
addition to the single-trace structure, one must also consider multi-trace
contributions, such as \({\rm Tr}(T^{a_1}\cdots T^{a_k}) \, {\rm Tr}(T^{a_{k+1}}\cdots T^{a_n})\)~\cite{Dixon:1996wi,Dixon:2011xs}. A key feature for
our purposes is that only planar diagrams consistent with the cyclic ordering
contribute to a given color-ordered amplitude. As a consequence, recursion
relations for \(A_n[\cdots]\) take a simpler form and, in certain cases, can
be solved explicitly to yield compact expressions for \(n\)-point amplitudes.
We now illustrate this with a simple example.

\begin{eg}[Four-point tree-level gluon amplitudes]\label{eq:4pt_ampl}
    Consider the four-point tree-level color-ordered gluon amplitude in
    Yang--Mills theory \(A_4[1^{h_1},2^{h_2},3^{h_3},4^{h_4}]\) with
    \(h_i \in \{\pm 1\}\). Up to dihedral symmetry and parity, there are three
    helicity configurations:
\begin{equation}\label{eq:hel_4pt_conf}
(++++) \ , \quad (-+++) \ , \quad (--++) \, .
\end{equation}
For these configurations, the \([1,2\rangle\)-shift does not produce a pole at
infinity in~\eqref{eq:An_residue}. Since only the \(s\)- and \(t\)-channels
are compatible with the planar ordering, and \(s=(p_1+p_2)^2\) is invariant
under the \([1,2\rangle\)-shift, the only pole contributing to~\eqref{eq:An_residue}
arises from the \(t\)-channel,
\begin{equation}
t = P^2 = \langle 23\rangle [23] \, , \qquad P = p_2 + p_3 \, .
\end{equation}
The BCFW recursion relations~\eqref{eq:onshell_rec} then reduce \(A_4\) to
three-point gluon amplitudes:
\begin{equation}\label{eq:4pt_rec}
A_4[1^{h_1},2^{h_2},3^{h_3},4^{h_4}]
= \sum_{h = \pm 1} A_3[\widehat{1}^{h_1},\widehat{P}^{h},4^{h_4}] \, \frac{1}{P^2} \, A_3[-\widehat{P}^{-h},\widehat{2}^{h_2},3^{h_3}] \, ,
\end{equation}
where the hats indicate \([1,2\rangle\)-shifted kinematics as in~\eqref{eq:ij_shift}.
We now show that the color-ordered amplitudes for the first two helicity
configurations in~\eqref{eq:hel_4pt_conf} vanish for generic kinematics. For
the all-plus configuration, this follows immediately from the fact that the
three-point all-plus amplitude vanishes. For the \((-+++)\) configuration, we
proceed as follows. In~\eqref{eq:4pt_rec}, the term with \(h=-1\) yields a
right subamplitude with helicities \((+++)\), which vanishes. Consider
therefore the contribution with \(h=+1\). The pole in the deformation
parameter \(z\) is determined by
\begin{equation}
\widehat{P}^2(z) = \langle \widehat{2}3 \rangle [\widehat{2}3] = \langle 23 \rangle [\widehat{2}3] = 0 \, .
\end{equation}
For generic kinematics, \(\langle 23 \rangle \neq 0\), and hence
\([\widehat{2}3] = 0\). On the other hand, for \(h=+1\) the right three-point
subamplitude has helicity \((-++)\), and by Example~\ref{eg:3pt_ampl} it is
proportional to \([\widehat{2}3]^3\). Therefore, this contribution vanishes at
the pole, and we conclude that the four-point color-ordered amplitude with
helicity configuration \((-+++)\) vanishes.

Consider now the helicity configuration \((--++)\). The contribution with
\(h=-1\) vanishes for the following reason. As discussed above, the residue
enforces \([\widehat{2}3] = 0\). On the other hand, the right subamplitude
\(A_3[-\widehat{P}^{+},\widehat{2}^{-},3^{+}]\) has helicity \((+-+)\) and is
therefore supported on the locus where the corresponding angle brackets
vanish. On this locus, the three associated \(\lambda\)-spinors are
proportional. Combining this with \([\widehat{2}3] = 0\) and momentum
conservation implies that the three corresponding \(\widetilde{\lambda}\)-spinors
are also proportional. Hence, all square brackets among these three
particles vanish and the right subamplitude vanishes. We are therefore
left with the contribution for \(h=+1\):
\begin{equation}\label{eq:4pt_rec_1}
A_3[\widehat{1}^{-},\widehat{P}^{+},4^{+}] \, \frac{1}{P^2} \, A_3[-\widehat{P}^{-},\widehat{2}^{-},3^{+}]
= \frac{[\widehat{P}4]^3}{[\widehat{1} 4] [\widehat{1} \widehat{P}]} \cdot \frac{1}{\langle 23 \rangle [23]} \cdot \frac{\langle \widehat{P} \widehat{2} \rangle^3}{ \langle \widehat{P} 3 \rangle \langle \widehat{2} 3 \rangle} \, .
\end{equation}
Since \(\widetilde{\lambda}_{\widehat{1}}=\widetilde{\lambda}_1\),
\(\widehat{\lambda_{2}}=\lambda_2\), and
\(\langle \widehat{2}3\rangle=\langle 23\rangle\), we can combine pairs of
spinor brackets as
\begin{equation}
[\widehat{1} \widehat{P}] \langle \widehat{P}3 \rangle
= \widetilde{\lambda}_1 \widehat{P}^{\alpha\dot{\alpha}} \lambda_3
= \widetilde{\lambda}_1 (\widehat{p}_2^{\alpha\dot{\alpha}}+p_3^{\alpha\dot{\alpha}}) \lambda_3
= \widetilde{\lambda}_1 \widehat{p}_2^{\alpha\dot{\alpha}} \lambda_3
= [12]\langle 23\rangle \, ,
\end{equation}
and similarly $\langle 2 \widehat{P} \rangle [\widehat{P} 4]
= \langle 23 \rangle [34]$. Substituting into~\eqref{eq:4pt_rec_1}, we obtain
\begin{equation}\label{eq:4pt_square}
A_4[1^{-}2^{-}3^{+}4^{+}] =  \frac{[34]^4}{[12][23][34][41]} \, .
\end{equation}
We can equivalently rewrite this amplitude in terms of angle brackets using
momentum conservation at four points. For instance,~\eqref{eq:mom_cons} for
\(n=4\), \(i=1\) and \(j=3\), implies $\langle 12 \rangle [23] = \langle 14 \rangle [34]$. Using similar relations, one can rewrite~\eqref{eq:4pt_square} as
\begin{equation}\label{eq:4pt_angle}
A_4[1^{-}2^{-}3^{+}4^{+}] =  \frac{\langle 12 \rangle^4}{\langle 12 \rangle \langle 23 \rangle \langle 34 \rangle \langle 41 \rangle} \, .
\end{equation}
The two expressions~\eqref{eq:4pt_angle} and~\eqref{eq:4pt_square} are
related by a parity transformation~\eqref{eq:parity}, combined with the
permutation \((3,4,1,2)\).
\end{eg}

Our example illustrates the four-point instances of general statements about
the \(n\)-point tree-level color-ordered gluon amplitude in Yang--Mills
theory:
\begin{enumerate}
\renewcommand{\labelenumi}{(\roman{enumi})}
    \item the all-plus and all-minus amplitudes vanish for generic kinematics~\cite{Mangano:1990by,Dixon:1996wi};
    \item the single-minus and single-plus amplitudes vanish for generic kinematics~\cite{Mangano:1990by,Dixon:1996wi};
    \item the \textit{maximally helicity violating} (MHV) amplitude is given by the \textit{Parke--Taylor formula}~\cite{Parke:1986gb,Mangano:1990by,Dixon:2011xs}:
    \begin{tcolorbox}[definitionbox]
\textbf{Parke--Taylor formula.}
    \begin{equation}\label{eq:eq:PT}
        A_n[1^{-},2^{-},3^{+},\dots,n^{+}]
        = 
        \frac{\langle 12 \rangle^4}{ \langle 12 \rangle \langle 23 \rangle \cdots \langle n-1\,n \rangle \langle n1 \rangle} \, .
    \end{equation}
\end{tcolorbox}\noindent
\end{enumerate}

The statements (i)--(iii) above can be proved by induction on \(n\geq 3\),
using arguments analogous to those presented in Example~\ref{eq:4pt_ampl}
combined with the following recursion formula. Performing a \([1,2\rangle\)-shift
with \((h_1,h_2)\in\{(+,+),(+,-),(-,-)\}\), one finds:
\begin{align}\label{eq:An_recursion}
A_n&[1^{h_1},2^{h_2},3^{h_3},\dots,n^{h_n}]
=
\sum_{k=4}^{n}
\begin{tikzpicture}[baseline=-0.6ex,scale=0.95]
\node[circle,draw,minimum size=9mm] (L) at (0,0) {$\mathrm{L}$};
\node[circle,draw,minimum size=9mm] (R) at (3.2,0) {$\mathrm{R}$};
\draw (L) -- node[above] {$\widehat{P}_{2k}$} (R);
\draw (L.140) -- ++(-0.95,0.65) node[left] {$\hat{1}^{h_1}$};
\draw (L.170) -- ++(-0.95,0.18) node[left] {$n^{h_n}$};
\draw (L.190) -- ++(-0.95,-0.18) node[left] {$\vdots$};
\draw (L.220) -- ++(-0.95,-0.65) node[left] {$k^{h_k}$};
\draw (R.40) -- ++(0.95,0.65) node[right] {$\hat{2}^{h_2}$};
\draw (R.10) -- ++(0.95,0.18) node[right] {$3^{h_3}$};
\draw (R.-10) -- ++(0.95,-0.18) node[right] {$\vdots$};
\draw (R.-40) -- ++(0.95,-0.65) node[right] {$(k-1)^{h_{k-1}}$};
\end{tikzpicture}
\\[1ex]
&=
\sum_{k=4}^{n}\sum_{h_I=\pm}
\widehat{A}_{n-k+3}[\hat{1}^{h_1},\widehat{P}_{2k}^{\,h_I},k^{h_k},\ldots,n^{h_n}]
\frac{1}{P_{2k}^2}
\widehat{A}_{k-1}[-\widehat{P}_{2k}^{-h_I},\hat{2}^{h_2},3^{h_3},\ldots,(k-1)^{h_{k-1}}] \, ,
\end{align}
where $P_{2k}=\sum_{a=2}^{k-1} p_a \, .$
Indeed, for a color-ordered amplitude only cyclically consecutive
factorization channels contribute. Under the \([1,2\rangle\)-shift, the only
such channels depending on \(z\) are precisely those separating the ordered set
\(\{2,3,\ldots,k-1\}\) from its complement, giving~\eqref{eq:An_recursion}.
The vanishing of the all-plus and single-minus amplitudes follows by
induction, since every term in the recursion contains a lower-point all-plus
or single-minus amplitude, or a three-point amplitude with a forbidden
helicity configuration. For the MHV sector, the same induction leaves only the
helicity assignments for which both lower-point amplitudes are MHV or
three-point anti-MHV. Substituting the induction hypothesis into~\eqref{eq:An_recursion}
then gives the Parke--Taylor formula~\eqref{eq:eq:PT}.

\begin{remark}[Vanishing amplitudes and support]
We emphasize that the vanishing of tree-level color-ordered all-plus and
single-minus gluon amplitudes holds for generic kinematics. In fact, the
all-plus amplitude vanishes identically on the full kinematic space in
four-dimensional Yang--Mills theory, as follows from helicity selection rules
and supersymmetric Ward identities~\cite{Mangano:1990by,Elvang:2013cua}. By
contrast, in~\cite{Witten:2003nn} it was conjectured that the single-minus
amplitude has support on a lower-dimensional locus in kinematic space, and
this was recently confirmed in~\cite{Guevara:2026qzd}. It is an open problem
to incorporate such degenerate kinematic contributions systematically into
recursion relations.
\end{remark}

\medskip

\subsubsection{Loop integrands and forward limits}
It is natural to ask whether recursion relations also hold at loop level. We
briefly discuss the salient aspects of this question and set the stage for the
BCFW recursion of loop integrands, which will be discussed in
Section~\ref{sec:Loop Amplituhedra}. For further details on loop-level
recursion, generalized unitarity, and loop-integrand constructions, see~\cite{Bern:1994zx,Britto:2004ap,Britto:2005fq,ArkaniHamed:2010kv,Grassmannian,Bourjaily:2010kw,Mason:2010yk,Boels:2010nw,Elvang:2013cua,Henn:2014yza}.

\begin{figure}[pos=t]
\centering
\resizebox{0.70\textwidth}{!}{%
\begin{tikzpicture}[
    x=1cm,y=1cm,
    line cap=round,
    line join=round,
    >={Latex[length=3mm,width=2mm]},
    ext/.style={-Latex, line width=0.95pt},
    prop/.style={line width=0.95pt},
    blob/.style={draw=black, fill=gray!20, line width=0.95pt},
    hole/.style={draw=black, fill=white, line width=0.95pt},
    cutr/.style={red, dashed, line width=1.0pt},
    cutb/.style={blue, dashed, line width=1.0pt}
]

\draw[blob] (2.35,3.75) ellipse (1.42 and 1.02);
\draw[hole] (1.55,3.75) ellipse (0.28 and 0.52);
\draw[hole] (2.78,4.16) ellipse (0.56 and 0.27);
\draw[hole] (2.82,3.46) ellipse (0.56 and 0.24);
\draw[ext] (1.437,4.531) -- (0.98,5.88);
\draw[ext] (1.437,2.969) -- (0.78,1.72);
\draw[ext] (3.263,4.531) -- (4.05,5.78);
\draw[ext] (3.263,2.969) -- (4.28,1.80);
\node at (0.96,6.18) {$\hat{1}$};
\node at (0.72,1.42) {$2$};
\node at (4.08,6.15) {$\hat{4}$};
\node at (4.22,1.52) {$3$};
\draw[cutr] (2.02,4.72).. controls (1.93,4.28) and (1.98,3.88)..(2.08,3.52).. controls (2.16,3.16) and (2.14,2.90)..(2.03,2.76);
\draw[->,red,line width=0.95pt] (2.10,3.10) -- (2.10,3.56);
\draw[cutb] (2.49,4.55).. controls (2.74,4.88) and (2.97,4.82)..(3.10,4.48);
\draw[->,blue,line width=0.95pt] (2.80,4.70) -- (2.92,4.58);
\draw[-Latex, line width=1.05pt] (4.82,3.60) -- (6.10,3.60);
\node at (5.55,5.30) {$(i)$};
\draw[blob] (7.55,5.00) circle (0.63);
\draw[hole] (7.55,5.00) circle (0.34);
\draw[blob] (10.03,5.02) ellipse (1.38 and 0.67);
\draw[hole] (10.03,5.28) ellipse (0.54 and 0.23);
\draw[hole] (10.03,4.76) ellipse (0.54 and 0.20);
\draw[prop] (8.18,5.00) -- (8.65,5.00);
\draw[cutr] (8.50,5.48) .. controls (8.34,5.20) and (8.34,4.80) .. (8.50,4.52);
\draw[cutr] (8.70,5.48) .. controls (8.86,5.20) and (8.86,4.80) .. (8.70,4.52);
\draw[->,red,line width=0.95pt] (8.60,5.18) -- (8.60,5.38);
\draw[ext] (7.235,5.546) -- (7.02,6.40);
\draw[ext] (7.235,4.454) -- (6.90,3.68);
\draw[ext] (10.917,5.533) -- (11.08,6.43);
\draw[ext] (10.917,4.507) -- (11.28,3.70);
\node at (7.02,6.68) {$\hat{1}$};
\node at (6.82,3.5) {$2$};
\node at (11.12,6.63) {$\hat{4}$};
\node at (11.15,3.40) {$3$};
\node at (5.62,1.95) {$(ii)$};
\coordinate (L) at (7.58,2.38);
\coordinate (R) at (10.24,2.38);
\path[fill=gray!20, draw=black, line width=0.95pt] (L).. controls (8.30,2.38) and (9.35,2.36) .. (R).. controls (10.70,1.80) and (10.45,0.95) .. (9.30,0.85).. controls (8.30,0.75) and (7.35,1.05) .. (7.25,1.80).. controls (7.18,2.05) and (7.35,2.25) .. (L)-- cycle;
\draw[prop] (L).. controls (8.05,3.18) and (9.45,3.20)..(R);
\draw[prop] (L).. controls (8.30,2.38) and (9.35,2.36)..(R);
\draw[hole] (8.25,1.72) ellipse (0.27 and 0.50);
\draw[hole] (9.45,1.36) ellipse (0.55 and 0.24);
\draw[ext] (L) -- (6.95,2.98);
\draw[ext] (7.55,1.12) -- (7.18,0.28);
\draw[ext] (R) -- (11.42,2.42);
\draw[ext] (9.95,0.98) -- (10.66,0.28);
\node at (6.84,3.10) {$\hat{1}$};
\node at (7.10,0.05) {$2$};
\node at (11.70,2.58) {$\hat{4}$};
\node at (10.78,0.05) {$3$};
\node at (8.72,3.10) {$r$};
\node at (9.67,3.10) {$-r$};
\draw[cutb] (9.12,3.10).. controls (8.95,2.83) and (8.95,2.55)..(9.12,2.32);
\draw[cutb] (9.36,3.10).. controls (9.53,2.83) and (9.53,2.55)..(9.36,2.32);
\draw[->,blue,line width=0.95pt] (9.24,2.82) -- (9.24,3.02);
\end{tikzpicture}%
}
\caption{Two types of contributions arising from a single cut of a loop-dependent propagator. The first in red gives an ordinary factorization channel $(i)$, while the second in blue gives a forward-limit $(ii)$, with legs of momenta \(r\) and \(-r\).}
\label{fig:single-cut-forward-limit}
\end{figure}
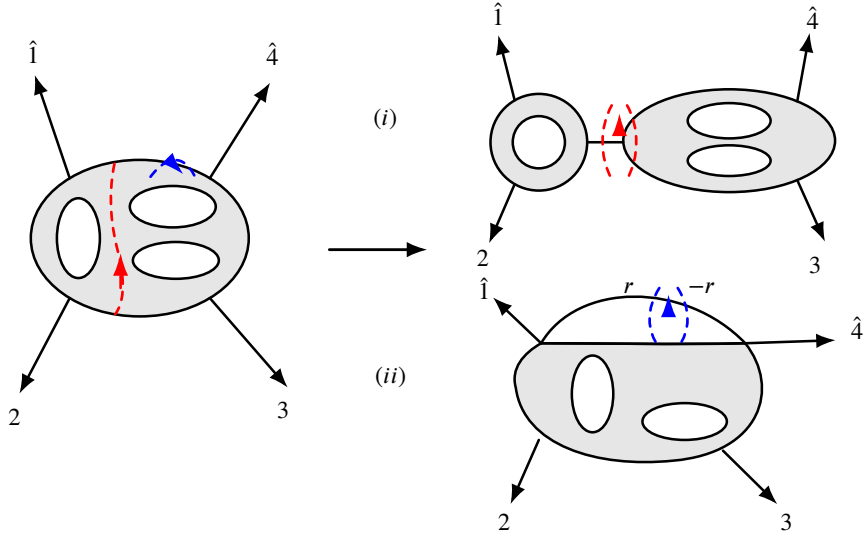

Since integrated loop amplitudes have branch cuts and more complicated
analytic structure, the natural objects to which BCFW reasoning applies are
loop integrands. At the level of individual Feynman diagrams, these are
rational functions with poles on propagator loci of the form appearing
in~\eqref{eq:feynm_int}. Under a BCFW shift of external legs, the shifted loop
integrand develops two qualitatively different types of poles, both schematically represented in Figure~\ref{fig:single-cut-forward-limit}. The first type $(i)$ comes from propagators that are independent of the loop momenta. Their
residues have the familiar factorization form into products of lower-point,
lower-loop integrands. The second type $(ii)$ comes from propagators that depend on
loop momentum. The residue of such a pole is a single-line cut in the diagram,
and is naturally described as a lower-loop integrand with two additional
adjacent legs, evaluated in the \textit{forward limit}
\begin{equation}\label{eq:forward}
  p_i = r, \qquad p_{i+1} = -r, \qquad r^2=0 \, .
\end{equation}
Thus an \(\ell\)-loop \(n\)-point integrand can produce an \((\ell-1)\)-loop
\((n+2)\)-point integrand in a forward limit. This suggests a loop-level
analogue of tree-level recursion, but two subtleties must first be addressed.

\begin{itemize}
  \item First, the loop integrand is not canonically defined before integration.
  Loop momenta are integration variables, so different routings of the same
  loop momentum give equivalent integrated amplitudes but different rational
  functions before integration. For example, the one-loop four-point box in
  Figure~\ref{eq:box_diagr} is written up to normalization as
  \begin{equation}\label{eq:box_int_1}
    I_{{\rm box}}    =
    \frac{1}{
      k^2
      (k-p_1)^2
      (k-p_1-p_2)^2
      (k+p_4)^2
    } \, .
  \end{equation}
  This Feynman diagram, dressed by color and other group-theoretic factors,
  contributes to four-point one-loop amplitudes in several theories. Starting
  from~\eqref{eq:box_int_1}, one can equivalently shift the integration
  variable \(k \mapsto k+p_1\). The resulting expression defines the same
  integral, but under a BCFW shift it can have a different pole structure in
  the shift parameter. This ambiguity is solved in \textit{planar theories},
  where only planar Feynman diagrams contribute, by introducing dual momentum
  coordinates. We discuss these coordinates in Section~\ref{sec:Momentum Twistors}.

  \item Second, the forward limit can be singular. The single-cut residue
  localizes~\eqref{eq:forward}, and certain lower-loop diagrams develop
  singularities in this limit. A typical example occurs when the two forward
  legs attach to the same external line: momentum conservation then produces
  an external-bubble singularity. In the same way, tadpole-like configurations
  can also appear. Although such contributions often integrate to zero in
  massless theories using dimensional regularization, they are present at the
  level of the rational integrand and can contaminate the single cuts. A
  consistent recursion therefore either removes these terms or works in a
  theory where they cancel. More generally, after a regulator and an
  integration prescription have been specified, some integrand-level
  equalities may be understood modulo terms in the kernel of the corresponding
  integration map. Depending on the regulator, contour, observable and
  representation, this kernel can include total derivatives,
  integration-by-parts-exact terms, parity-odd terms, or scaleless integrals;
  the relevant ambiguity must be fixed before comparing two integrands term by
  term.
\end{itemize}

The main theory under consideration in this thesis is planar \(\mathcal{N}=4\)
super Yang--Mills theory, which we introduce in the next section. In this
theory both problems have a resolution~\cite{ArkaniHamed:2010kv,Elvang:2013cua}.
Planarity fixes a preferred routing and hence a canonical four-dimensional
rational loop integrand in dual momentum variables, while
forward-limit contributions from external bubbles and tadpoles cancel by
supersymmetry.

For general gauge theories, loop integrands are less canonical. Generalized
unitarity and integrand reduction construct them by matching cuts and
expanding numerators over a basis of propagator structures~\cite{Bern:1994zx,Ossola:2006us}.
These methods are computationally powerful, especially in QCD, but the
resulting integrand is typically defined only up to the chosen kernel of the
integration map. Related approaches, including the Feynman-tree theorem, loop-tree
duality, forward limits, scattering-equation formulae, and \(Q\)-cut
representations, express loop information in terms of tree-level data and
on-shell cuts~\cite{Feynman:1963ax,Catani:2008xa,Baadsgaard:2015twa,He:2015yua}.
They make factorization more manifest, but forward singularities, tadpoles,
massless bubbles, and scaleless terms remain subtle in non-supersymmetric
Yang--Mills theory.

Surface kinematics offers a recent way to resolve this ambiguity by assigning variables to curves on a punctured surface before imposing ordinary momentum-space identifications. Degenerations that would give (1/0) ambiguities then remain distinct, producing a well-defined planar Yang--Mills integrand with generalized gauge invariance and consistent single-cut factorization~\cite{ArkaniHamed:2024surfaceYM}. Since this integrand vanishes at infinity in suitable surface-kinematic directions, it can be reconstructed from its single-cut residues. The construction is explicit at one loop for all multiplicities and in simple two-loop examples; a related refinement removes selected scaleless terms and gives an algorithmic all-loop recursion, with results through the two-loop five-point pure Yang--Mills integrand and simplified higher-loop cases~\cite{ArkaniHamed:2024surfaceYM,Cao:2025ymrecursion}.

At tree level, BCFW recursion gives different rational representations of the
same amplitude, often involving spurious poles that cancel in the sum. In the
positive-geometric formulation developed later, these representations will be
interpreted as triangulations of a canonical form. At loop level, this is a triangulation of the loop Amplituhedron.

There is an important point we want to raise regarding the concept of a four-dimensional integrand. The loop integrands used here are four-dimensional rational
differential forms. Concretely, in planar kinematics the loop variables are
four-dimensional momenta and the integrand is studied through its poles,
residues and rational dependence on these variables. This is the natural
language for generalized unitarity, forward limits, leading singularities and
the Amplituhedron: all of these constructions ask about the rational form
before the loop integration has been
performed~\cite{Bern:1994zx,ArkaniHamed:2010kv,Grassmannian,the_amplituhedron}.

Loop amplitudes are different objects. Massless loop amplitudes
usually have infrared, and sometimes ultraviolet, divergences and therefore
require a regulator. In dimensional regularization one analytically continues
the loop momenta to $D=4-2\epsilon$
dimensions~\cite{tHooftVeltman1972,BolliniGiambiagi1972}. The strictly
four-dimensional integrand is then the restriction of a possible
\(D\)-dimensional integrand to the four-dimensional subspace. This restriction
can lose information: numerator terms proportional to the
\((-2\epsilon)\)-dimensional components of loop momenta vanish on all strictly
four-dimensional cuts, but may nevertheless contribute after integration. In
particular, evanescent numerator terms can multiply divergent integrals and
leave finite contributions in the \(\epsilon\to0\) expansion. This is one reason
why \(D\)-dimensional unitarity is needed in less supersymmetric theories, where
rational terms are not always determined by four-dimensional cuts
alone~\cite{EllisGieleKunsztMelnikov2009,AnastasiouBrittoFengKunsztMastrolia2006}.

Planar \(\mathcal N=4\) SYM has a special status in this thesis. Its
four-dimensional on-shell data are so constraining that generalized cuts,
leading singularities and logarithmic poles determine a preferred
four-dimensional rational integrand in the planar variables, and the
Amplituhedron gives a geometric construction of precisely this integrand-level
object. The qualification is nevertheless important: the Amplituhedron is a
four-dimensional construction before integration, and the strict
four-dimensional integrand may not contain all regulator-dependent information
that may be relevant for regulated loop amplitudes.

\subsection{Maximally supersymmetric Yang--Mills theory}\label{sec:Maximally Supersymmetric Yang-Mills Theory}\label{sec:Maximally supersymmetric Yang-Mills theory}

Supersymmetry relates bosonic and fermionic degrees of freedom and provides
powerful constraints on quantum field theories. In particular, it improves
ultraviolet behavior, restricts the form of interactions, and often leads to
enhanced symmetries such as conformal invariance. Among supersymmetric gauge
theories in four dimensions, maximally supersymmetric Yang--Mills theory
(\(\mathcal{N}=4\) SYM) occupies a distinguished role: it is a finite,
conformal quantum field theory and serves as a central testing ground for
modern developments in scattering amplitudes and gauge/string duality~\cite{Brink:1976bc,Mandelstam:1982cb,Beisert:2010jr}.
It can be formulated as a four-dimensional Lagrangian field theory or as the
dimensional reduction of ten-dimensional \(\mathcal{N}=1\) super Yang--Mills
theory~\cite{Brink:1976bc,Witten:1995ex}. Here we instead define it through
its particle content and on-shell kinematics.

\subsubsection{Supersymmetry algebra}
Supersymmetry can be understood as an extension of the Poincar\'e symmetry
algebra in \(D\) dimensions to a Lie \textit{superalgebra}, whose generators
are divided into \textit{bosonic} and \textit{fermionic} ones. The bosonic
generators consist of translations \(P_\mu\) and Lorentz transformations
\(M_{\mu\nu}\), forming the Poincar\'e algebra. In four dimensions, the
fermionic generators are spinor-valued supercharges
\begin{equation}
Q^\alpha_A \, , \qquad \Qbar^{\dot\alpha A} \, ,
\end{equation}
with \(A=1,\dots,\mathcal{N}\), and \(\alpha,\dot\alpha \in \{1,2\}\) spinor
indices. The supersymmetry algebra is characterized by the anticommutation
relations
\begin{align}
\{ Q^\alpha_A , \Qbar^{\dot\alpha B} \}
&= 2 \, \delta_A^{\,B} \, \sigma_\mu^{\alpha\dot\alpha} P^\mu \, ,
\qquad
\{ Q_A^\alpha , Q_B^\beta \} = 0 \, ,
\qquad
\{ \Qbar^{\dot\alpha A} , \Qbar^{\dot\beta B} \} = 0 \, ,
\end{align}
where \(\{a,b\}:=ab + ba\). Together with the commutation relations between
the supercharges and the Lorentz generators, this defines the \textit{Poincar\'e
superalgebra}. The index \(A\) labels \(\mathcal{N}\) independent copies of
supersymmetry, so that the algebra contains \(4\mathcal{N}\) real supercharges
in four spacetime dimensions. There is also an \({\rm SU}(\mathcal{N})\)
symmetry acting on the label \(A\), called \textit{R-symmetry}~\cite{Wess:1992cp}.

The action of the supercharges forms an exterior algebra and organizes
particle states into supermultiplets. Starting from a highest-helicity state
\(h_{\text{max}}\), repeated action of the supercharges lowers the helicity in
steps of \(1/2\). For a massless representation with \(\mathcal{N}\)
supersymmetries, one obtains \(2^{\mathcal{N}}\) states whose helicities fill
out an interval
\begin{equation}
h_{\text{max}},\; h_{\text{max}} - \frac{1}{2},\; \dots,\; h_{\text{max}} - \frac{\mathcal{N}}{2} \, .
\end{equation}
In a consistent interacting four-dimensional gauge theory, one requires that
no particle of helicity greater than one appears in the spectrum, since
massless particles with spin greater than one cannot be coupled consistently
without introducing gravity~\cite{Wess:1992cp,Weinberg:2000cr}. Taking
\(h_{\text{max}} = 1\) for a gauge boson gives the bound \(\mathcal{N} \le 4\).
Thus \(\mathcal{N}=4\) is the maximal amount of supersymmetry compatible with
a four-dimensional gauge theory. For \(\mathcal{N}=4\), the helicity range is
precisely from \(+1\) to \(-1\), so the multiplet is CPT self-conjugate and
contains only particles of spin at most one.

\subsubsection{The on-shell supermultiplet}
The corresponding \(\mathcal{N}=4\) supermultiplet is uniquely fixed by
repeated action of the four supercharges. A convenient way to organize the
fields is to assemble all particles into a single CPT self-conjugate
supermultiplet~\cite{Brink:1976bc,Elvang:2013cua}. One introduces Grassmann
variables \(\widetilde{\eta}^A\) with \(\{\widetilde{\eta}^A,\widetilde{\eta}^B\}=0\),
transforming under the R-symmetry group \({\rm SU}(4)\). The multiplet can
then be conveniently encoded in an on-shell superfield
\begin{equation}
\Phi(\widetilde{\eta}) =
g^{+}
+ \widetilde{\eta}^A \psi_A^{+}
+ \frac{1}{2} \widetilde{\eta}^A \widetilde{\eta}^B \phi_{AB}
+ \frac{1}{3!} \widetilde{\eta}^A \widetilde{\eta}^B \widetilde{\eta}^C \epsilon_{ABCD} \bar{\psi}^{-D}
+ \frac{1}{4!} \widetilde{\eta}^A \widetilde{\eta}^B \widetilde{\eta}^C \widetilde{\eta}^D \epsilon_{ABCD} g^{-} \, ,
\end{equation}
where the fields are: a gluon \(g^+\) of helicity \(+1\), four fermions
\(\psi_A^+\) of helicity \(+1/2\), six scalars \(\phi_{AB}\) of helicity
\(0\), four fermions \(\bar{\psi}^{-A}\) of helicity \(-1/2\), and a gluon
\(g^-\) of helicity \(-1\), with \(A,B=1,\dots,4\). All fields transform in
the adjoint representation of \(\mathrm{SU}(N)\)~\cite{Brink:1976bc}.

\subsubsection{Superamplitudes and helicity sectors}
External states are described using on-shell superfields
\(\Phi_i = \Phi(p_i,\widetilde{\eta}_i)\), which can be viewed as
superwavefunctions for the \(i\)-th particle in a scattering process. These
depend on the on-shell momentum \(p_i\) and Grassmann variables
\(\widetilde{\eta}_i^A\). Scattering amplitudes of superfields
\(A_n(\Phi_1,\dots,\Phi_n)\) are known as \textit{superamplitudes} and encode
all component amplitudes within a single object~\cite{Nair:1988bq,Elvang:2013cua}.
Expanding \(A_n\) in the variables \(\widetilde{\eta}_i^A\),
component amplitudes are obtained by selecting appropriate monomials. For
example, the MHV gluon amplitude is
\begin{equation}
A_n(1^+,\dots,i^-,\dots,j^-,\dots,n^+)
=
\left(
\prod_{A=1}^4 \frac{\partial}{\partial \widetilde{\eta}_i^A}
\right) \left(
\prod_{B=1}^4 \frac{\partial}{\partial \widetilde{\eta}_j^B}
\right) A_n(\Phi_1,\dots,\Phi_n)
\Big{|}_{\widetilde{\eta}=0} \, .
\end{equation}

The \({\rm SU}(4)\) R-symmetry imposes strong constraints on the structure of
superamplitudes: \(A_n\) must be a sum of monomials whose total degree in the
\(\widetilde{\eta}\) variables is a multiple of four. More precisely, before
extracting the universal supermomentum-conserving factor, the
\(\mathrm{N}^k\mathrm{MHV}\) sector has total Grassmann degree
\(4(k+2)\)~\cite{Nair:1988bq}. After factoring out this universal degree-eight
factor, the reduced amplitude has Grassmann degree \(4k\). The lowest
non-trivial case, of degree eight, corresponds to the maximally helicity
violating (MHV) amplitudes, while higher-degree sectors (NMHV,
N\(^{2}\)MHV, etc.) encode configurations of increasing helicity.

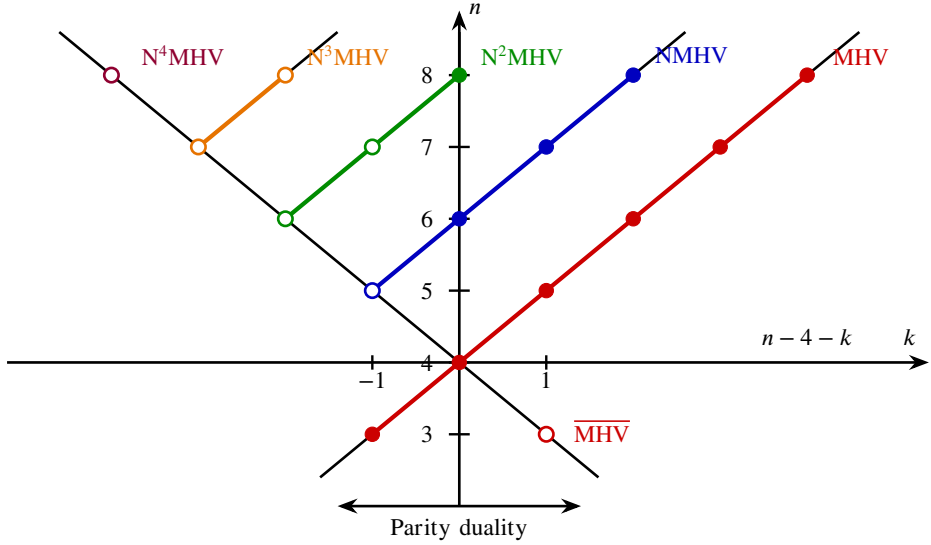
\begin{figure}[pos=t]
\centering
\begin{tikzpicture}[
    x=1.15cm,y=0.95cm,
    >=Stealth,
    axis/.style={-Stealth, line width=1pt},
    edge/.style={line width=1pt, black},
    tick/.style={line width=0.9pt}
]
\colorlet{cMHV}{red!80!black}
\colorlet{cNMHV}{blue!75!black}
\colorlet{cNNMHV}{green!55!black}
\colorlet{cNthree}{orange!90!black}
\colorlet{cNfour}{purple!75!black}
\def\dotr{2.7pt}
\newcommand{\fulldot}[3]{\filldraw[fill=#3,draw=#3] (#1,#2) circle (\dotr);}
\newcommand{\hollowdot}[3]{\draw[draw=#3,line width=1pt,fill=white] (#1,#2) circle (\dotr);}
\newcommand{\halfdot}[3]{\begin{scope}\clip (#1,#2) circle (\dotr);\fill[#3] (#1,#2-\dotr) rectangle (#1+\dotr,#2+\dotr);\end{scope}\draw[draw=#3,line width=1pt,fill=white] (#1,#2) circle (\dotr);}
\draw[axis] (-5.2,4) -- (5.4,4);
\draw[axis] (0,2.0) -- (0,8.9);
\draw[tick] (-1,3.9) -- (-1,4.1);
\draw[tick] ( 1,3.9) -- ( 1,4.1);
\node[below] at (-1,4) {$-1$};
\node[below] at ( 1,4) {$1$};
\foreach \yy in {3,4,5,6,7,8}{\draw[tick] (-0.12,\yy) -- (0.12,\yy);\node[left] at (-0.20,\yy) {$\yy$};}
\node[above] at (0.18,8.72) {$n$};
\node[above] at (4.0,4.12) {$n-4-k$};
\node[above] at (5.18,4.12) {$k$};
\draw[edge] (-4,8) -- (-3,7) -- (-2,6) -- (-1,5) -- (0,4) -- (1,3);
\draw[edge] (-3,7) -- (-2,8);
\draw[edge] (-2,6) -- (-1,7) -- (0,8);
\draw[edge] (-1,5) -- (0,6) -- (1,7) -- (2,8);
\draw[edge] (-1,3) -- (0,4) -- (1,5) -- (2,6) -- (3,7) -- (4,8);
\draw[edge] (-4.6,8.6) -- (-4,8);
\draw[edge] (-2,8) -- (-1.4,8.6);
\draw[edge] ( 0,8) -- ( 0,8.6);
\draw[edge] ( 2,8) -- ( 2.6,8.6);
\draw[edge] ( 4,8) -- ( 4.6,8.6);
\draw[edge] (-1.6,2.4) -- (-1,3);
\draw[edge] ( 1,3) -- ( 1.6,2.4);
\draw[line width=1.6pt,cNthree] (-3,7) -- (-2,8);
\draw[line width=1.6pt,cNNMHV] (-2,6) -- (-1,7) -- (0,8);
\draw[line width=1.6pt,cNMHV]  (-1,5) -- (0,6) -- (1,7) -- (2,8);
\draw[line width=1.6pt,cMHV]   (-1,3) -- (0,4) -- (1,5) -- (2,6) -- (3,7) -- (4,8);
\node[text=cNfour,   anchor=west] at (-3.75,8.28) {$\mathrm{N}^4\mathrm{MHV}$};
\node[text=cNthree,  anchor=west] at (-1.85,8.28) {$\mathrm{N}^3\mathrm{MHV}$};
\node[text=cNNMHV,   anchor=west] at ( 0.15,8.28) {$\mathrm{N}^2\mathrm{MHV}$};
\node[text=cNMHV,    anchor=west] at ( 2.15,8.28) {$\mathrm{NMHV}$};
\node[text=cMHV,     anchor=west] at ( 4.20,8.25) {$\mathrm{MHV}$};
\node[text=cMHV,     anchor=west] at ( 1.22,3.05) {$\overline{\mathrm{MHV}}$};
\hollowdot{-4}{8}{cNfour}
\hollowdot{-3}{7}{cNthree}
\hollowdot{-2}{8}{cNthree}
\hollowdot{-2}{6}{cNNMHV}
\hollowdot{-1}{7}{cNNMHV}
\fulldot{0}{8}{cNNMHV}
\hollowdot{-1}{5}{cNMHV}
\fulldot{0}{6}{cNMHV}
\fulldot{1}{7}{cNMHV}
\fulldot{2}{8}{cNMHV}
\fulldot{-1}{3}{cMHV}
\fulldot{0}{4}{cMHV}
\fulldot{1}{5}{cMHV}
\fulldot{2}{6}{cMHV}
\fulldot{3}{7}{cMHV}
\fulldot{4}{8}{cMHV}
\hollowdot{1}{3}{cMHV}
\draw[<->, line width=1pt] (-1.4,2.00) -- (1.4,2.00);
\node[below] at (0,1.95) {$\mathrm{Parity\ duality}$};
\end{tikzpicture}
\caption{Non-vanishing helicity sectors of \(n\)-point superamplitudes in \(\mathcal{N}=4\) SYM. The horizontal coordinate labels the \(\mathrm{N}^k\mathrm{MHV}\) degree. Parity maps the \(\mathrm{N}^k\mathrm{MHV}\) to the \(\mathrm{N}^{n-k-4}\mathrm{MHV}\) sector, represented by reflection across the vertical axis.}
\label{fig:helicity}
\end{figure}

The integer \(k\) controlling the Grassmann degree labels the helicity sector.
This allows one to decompose the full tree-level superamplitude as
\begin{equation}\label{eq:superampl_dec}
A_n = A_n^{\text{MHV}} + A_n^{\text{NMHV}} + A_n^{\text{N}^2\text{MHV}} + \cdots + A_n^{\overline{\text{MHV}}} \, ,
\end{equation}
where \(A_n^{\text{MHV}}\) has Grassmann degree \(8\), \(A_n^{\text{NMHV}}\)
has degree \(12\), and in general \(A_n^{\text{N}^k\text{MHV}}\) has degree
\(4(k+2)\). The anti-MHV sector \(\overline{\text{MHV}}\) is obtained from the
MHV one by parity. For given $n$, the maximal sector is \(k=n-4\), and parity maps the
\(\text{N}^k\text{MHV}\) sector to the \(\text{N}^{\,n-k-4}\text{MHV}\) sector.
Note that a superamplitude at fixed $k$ packages several component amplitudes. For example, an MHV superamplitude contains the pure-gluon amplitude with two negative-helicity gluons, amplitudes with one negative-helicity gluon and a fermion--antifermion pair, and amplitudes involving pairs of scalars.

\subsubsection{Symmetries and the planar limit}

The symmetries of \(\mathcal{N}=4\) super Yang--Mills theory can be
understood starting from its vacuum structure. In the \(SU(4)_R\)-covariant
notation used above, the six real scalars are packaged into antisymmetric fields
\(\phi_{AB}\). The scalar potential is schematically of the form
\([\phi_{AB},\phi_{CD}]^2\), so supersymmetric vacua are characterized by
commuting scalar expectation values,
\([\langle\phi_{AB}\rangle,\langle\phi_{CD}\rangle]=0\)~\cite{Wess:1992cp}.
At the origin of moduli space, where all scalar vacuum expectation values
vanish, \(\langle\phi_{AB}\rangle=0\), all particles are massless and the
theory is scaleless.

As a consequence, the theory is scale invariant at the classical level.
Remarkably, this symmetry persists quantum mechanically: the beta function
vanishes to all orders, so the coupling does not run and no scale is
generated~\cite{Brink:1976bc,Mandelstam:1982cb}. The theory is therefore
conformally invariant. Namely, it is invariant under the four-dimensional
conformal group \({\rm SO}(2,4)\), containing the Poincar\'e group, dilations,
and special conformal transformations. Supersymmetry further this to
the superconformal symmetry group \({\rm PSU}(2,2|4)\)~\cite{Beisert:2010jr}.

These symmetries have powerful consequences. For instance, superconformal Ward
identities imply that helicity
configurations with fewer than two negative-helicity gluons vanish at all
orders; hence the MHV sector is the first non-trivial
contribution~\cite{Nair:1988bq,Elvang:2013cua}. This in absence of supersymmetry.

A further simplification arises in the planar limit, defined by taking the
rank \(N\) of the gauge group \(\mathrm{SU}(N)\) to infinity while keeping the
`t Hooft coupling \(\lambda = g_{\mathrm{YM}}^2 N\) fixed~\cite{tHooft:1973jz}.
In this limit, only planar Feynman diagrams contribute to the leading
single-trace part of the amplitude. Color-ordered amplitudes therefore become
the natural objects, and the cyclic ordering of external particles becomes
part of the kinematic data. This ordering is essential for introducing dual
coordinates and, subsequently, momentum twistors.
In planar kinematics one introduces dual coordinates \(x_i\) defined below.
Momentum conservation then implies that the \(x_i\) form a closed null polygon.
This null polygon also underlies the amplitude/Wilson-loop duality present in planar \(\mathcal{N}=4\) SYM. The bosonic lightlike Wilson loop captures the planar MHV amplitude, and its supersymmetric extensions relate the full planar superamplitude to a super Wilson loop, or equivalently to a Wilson loop in twistor space~\cite{Alday:2007hr,Brandhuber:2007yx,DrummondHennKorchemskySokatchev2007,
DrummondHennKorchemskySokatchev2008,Mason:2010yk,CaronHuot:2011kk}.
Planar amplitudes exhibit an additional hidden symmetry, known as dual
conformal symmetry, which acts conformally on these dual coordinates~\cite{Drummond:2008vq}.
This symmetry becomes manifest in momentum twistor variables, to be discussed
in the next section, and plays a central role in the modern geometric
formulation of scattering amplitudes. More generally, the combined action of
superconformal and dual conformal symmetries severely restricts the analytic
structure of amplitudes. Their closure gives rise to Yangian symmetry, one of
the main algebraic structures underlying the remarkable simplicity of planar
\(\mathcal{N}=4\) SYM amplitudes~\cite{Beisert:2010jr}.

\subsection{Momentum twistors}\label{sec:Momentum Twistors}\label{sec:Momentum twistors}

We now introduce a natural set of kinematic variables: momentum twistors. They solve momentum conservation automatically, make dual conformal symmetry manifest, and provide the coordinates in which the Amplituhedron geometry will be naturally formulated.

On-shell momenta \(p_i\) satisfy two fundamental constraints: the massless
condition \(p_i^2=0\) and momentum conservation \(\sum_{i=1}^n p_i=0\). In
spinor-helicity variables \((\lambda_i,\widetilde{\lambda}_i)\), the null
condition is solved trivially by writing
\begin{equation}
p_i^{\alpha\dot\alpha}=\lambda_i^\alpha \widetilde{\lambda}_i^{\dot\alpha} \, ,
\end{equation}
but momentum conservation remains non-trivial. Conversely, one can introduce
\textit{dual variables} \(x_i\) defined as
\begin{equation}\label{eq:x_i}
p_i = x_i - x_{i+1}, \qquad x_{n+1} \equiv x_1 \, ,
\end{equation}
in which case momentum conservation is automatic, at the price that the null
condition is no longer manifest.

\subsubsection{Momentum twistors at tree level}
It is therefore natural to seek a parametrization in which both constraints
are solved simultaneously. Such variables were introduced by~\cite{Hodges:2009hk}
and are known as \textit{momentum twistors}. They are based on~\cite{Penrose:1967wn}
twistor theory, which relates points in spacetime to
lines in projective twistor space \(\mathbb{CP}^3\). A twistor is defined as
\begin{equation}
z=(\lambda,\mu)\in\mathbb{CP}^3 \, ,
\end{equation}
with the \textit{incidence relation}
\begin{equation}
\mu_{\dot\alpha} = x_{\dot\alpha\alpha} \lambda^\alpha \, ,
\end{equation}
for fixed \(x \in \mathbb{C}^4\). For fixed \(x\), the spinor \(\lambda\) is a
homogeneous coordinate on \(\mathbb{CP}^1\), and the incidence relation
therefore parametrizes a projective line in \(\mathbb{CP}^3\). Conversely, a
line in twistor space determines a point in spacetime.

\begin{figure}[pos=t]
\centering
\begin{tikzpicture}[
    scale=0.85,
    >=Stealth,
    line cap=round,
    line join=round,
    orangeedge/.style={draw=orange!90!black, line width=1.1pt},
    blackedge/.style={draw=black, line width=1.1pt},
    dashedblack/.style={draw=black, line width=1.1pt, dash pattern=on 4pt off 3pt},
    dashedorange/.style={draw=orange!90!black, line width=1.1pt, dash pattern=on 4pt off 3pt},
    maparrow/.style={->, draw=gray!55, line width=1.0pt}
]
\coordinate (zi)    at (0.0, 0.0);
\coordinate (zip)   at (0.95, 2.25);
\coordinate (zipp)  at (3.75, 3.15);
\coordinate (zim)   at (0.90,-2.15);
\coordinate (zimm)  at (3.75,-3.15);
\coordinate (zippd) at (4.75, 3.12);
\coordinate (zimmd) at (4.75,-3.12);
\draw[blackedge] (zi) -- (zip) -- (zipp);
\draw[blackedge] (zi) -- (zim) -- (zimm);
\draw[dashedblack] (zipp) -- (zippd);
\draw[dashedblack] (zimm) -- (zimmd);
\fill[orange!90!black] (zi)   circle (2.7pt);
\fill[orange!90!black] (zip)  circle (2.7pt);
\fill[orange!90!black] (zipp) circle (2.7pt);
\fill[orange!90!black] (zim)  circle (2.7pt);
\fill[orange!90!black] (zimm) circle (2.7pt);
\node[orange!90!black, left=10pt]  at (zi)   {$z_i$};
\node[orange!90!black, left=14pt]  at (zip)  {$z_{i+1}$};
\node[orange!90!black, above=8pt]  at (zipp) {$z_{i+2}$};
\node[orange!90!black, left=14pt]  at (zim)  {$z_{i-1}$};
\node[orange!90!black, below=8pt]  at (zimm) {$z_{i-2}$};
\coordinate (xi)    at (8.1, 1.15);
\coordinate (xip)   at (9.45, 2.55);
\coordinate (xim)   at (8.1,-1.10);
\coordinate (ximm)  at (9.35,-2.65);
\coordinate (xipd)  at (10.55, 2.48);
\coordinate (ximmd) at (10.45,-2.60);
\draw[orangeedge] (xi) -- (xip);
\draw[orangeedge] (xi) -- (xim);
\draw[orangeedge] (xim) -- (ximm);
\draw[dashedorange] (xip) -- (xipd);
\draw[dashedorange] (ximm) -- (ximmd);
\fill[black] (xi)   circle (2.9pt);
\fill[black] (xip)  circle (2.9pt);
\fill[black] (xim)  circle (2.9pt);
\fill[black] (ximm) circle (2.9pt);
\node[left=10pt]  at (xi)   {$x_i$};
\node[above=8pt]  at (xip)  {$x_{i+1}$};
\node[left=12pt]  at (xim)  {$x_{i-1}$};
\node[below=8pt]  at (ximm) {$x_{i-2}$};
\node[right=10pt] at ($(xi)!0.5!(xip)$) {$p_i$};
\node[left=10pt] at ($(xi)!0.5!(xim)$) {$p_{i-1}$};
\node[right=10pt] at ($(xim)!0.5!(ximm)$) {$p_{i-2}$};
\coordinate (a1) at ($(zi)!0.55!(zip)$);
\coordinate (a2) at ($(zi)!0.58!(zim)$);
\draw[maparrow] (a1) .. controls (3.7,0.75) and (6.2,0.65) .. (xi);
\draw[maparrow] (a2) .. controls (3.5,-0.85) and (6.2,-0.75) .. (xim);
\node[align=center, font=\large] at (1.9,-4.55) {Momentum twistor\\space};
\node[align=center, font=\large] at (9.0,-4.55) {Dual momentum\\space};
\end{tikzpicture}
\caption{Relation between dual momentum space and momentum twistor space. A dual point \(x_i\) corresponds to the line \((z_i z_{i+1})\) in twistor space. Adjacent dual points are null separated because the corresponding lines intersect, and the edge \(p_i=x_i-x_{i+1}\) is represented by the shared twistor \(z_{i+1}\).}
\label{fig:mom_tw}
\end{figure}
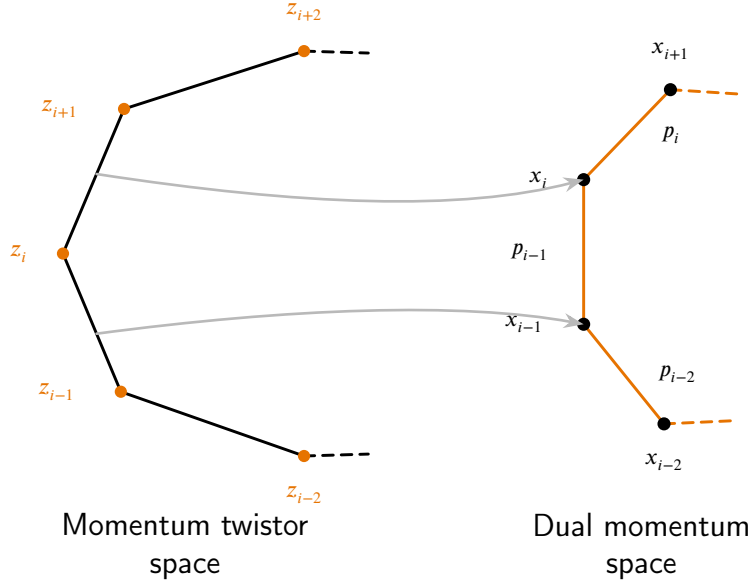

A crucial property is that two spacetime points \(x\) and \(y\) are
null-separated if and only if the corresponding lines in twistor space
intersect. More precisely, for two lines \(L_x=(z_i z_j)\) and
\(L_y=(z_k z_l)\) in \(\mathbb{CP}^3\), dual to points in spacetime $x$ and $y$, respectively, the condition that they intersect is
\begin{equation}
\langle z_i z_j z_k z_l \rangle = 0 \, ,
\end{equation}
and under the twistor correspondence this is equivalent to \((x-y)^2=0\).
Here the four-bracket is the \({\rm SL}(4)\)-invariant 
\begin{equation}
\langle z_i z_j z_k z_l \rangle := \epsilon_{abcd} z_i^a z_j^b z_k^c z_l^d \, ,
\end{equation}
which we often denote simply by \(\langle ijkl \rangle\).

Momentum twistors arise by associating to each dual point \(x_i\) the line
\((z_i,z_{i+1})\) in twistor space. The intersection property ensures that
consecutive points are null separated and hence the null condition is
automatically satisfied, while momentum conservation is manifest by
construction. In this way, a configuration of \(n\) momentum twistors
\(z_i \in \mathbb{CP}^3\) encodes a set of \(n\) on-shell momenta satisfying
all kinematic constraints. This is depicted in Figure~\ref{fig:mom_tw}.

Momentum twistor brackets are related to spinor-helicity invariants by
\begin{equation*}
\begin{aligned}
\langle i j k l\rangle
={}&
\langle ij\rangle[\mu_k\mu_l]
-\langle ik\rangle[\mu_j\mu_l]
+\langle il\rangle[\mu_j\mu_k]
+\langle jk\rangle[\mu_i\mu_l]
-\langle jl\rangle[\mu_i\mu_k]
+\langle kl\rangle[\mu_i\mu_j] \, ,
\end{aligned}
\end{equation*}
where \([\mu_i\mu_j]= \epsilon_{\dot\alpha\dot\beta}
\mu_i^{\dot\alpha}\mu_j^{\dot\beta}\). For special indices, this reduces to
\begin{equation}\label{eq:iijj}
\langle i\, i{+}1\, j\, j{+}1\rangle
=
\langle i\, i{+}1\rangle
\langle j\, j{+}1\rangle \, x_{ij}^2 \, ,
\end{equation}
where \(x_{ij}^2:=(x_i-x_j)^2\). Thus physical propagator singularities
\(x_{ij}^2=0\) become linear incidence conditions in momentum-twistor space:
the two lines \((z_i z_{i+1})\) and \((z_j z_{j+1})\) intersect. For generic
momenta at more than three points, \(\langle i\,i{+}1\rangle \neq 0\), and the
only singularities of the tree-level color-ordered planar amplitude are at
\(x_{ij}^2=0\), or equivalently when some
\(\langle i\, i{+}1\, j\, j{+}1\rangle\) vanishes.

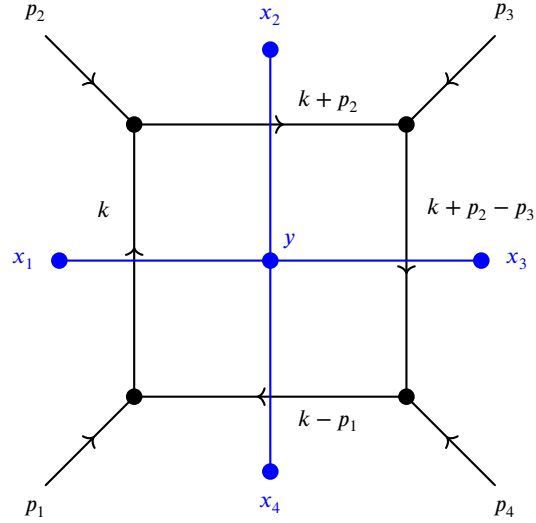
\begin{figure}[pos=t]
\centering
\begin{tikzpicture}[scale=0.9, line cap=round, line join=round]
\tikzset{mom/.style={thick,postaction={decorate},decoration={markings, mark=at position 0.55 with {\arrow{>}}}},dual/.style={blue, thick},dualpt/.style={circle, fill=blue, inner sep=2.2pt},vert/.style={circle, fill=black, inner sep=2.2pt}}
\coordinate (A) at (-2,  2);
\coordinate (B) at ( 2,  2);
\coordinate (C) at ( 2, -2);
\coordinate (D) at (-2, -2);
\coordinate (O) at ( 0,  0);
\coordinate (P2) at (-3.3,  3.3);
\coordinate (P3) at ( 3.3,  3.3);
\coordinate (P4) at ( 3.3, -3.3);
\coordinate (P1) at (-3.3, -3.3);
\coordinate (X2) at ( 0,  3.1);
\coordinate (X3) at ( 3.1,  0);
\coordinate (X4) at ( 0, -3.1);
\coordinate (X1) at (-3.1,  0);
\draw[mom] (A) -- (B);
\draw[mom] (B) -- (C);
\draw[mom] (C) -- (D);
\draw[mom] (D) -- (A);
\node at (0.85, 2.35) {$k+p_2$};
\node at (3.1, 0.75) {$k+p_2-p_3$};
\node at (0.85,-2.35) {$k-p_1$};
\node at (-2.45, 0.75) {$k$};
\node[vert] at (A) {};
\node[vert] at (B) {};
\node[vert] at (C) {};
\node[vert] at (D) {};
\draw[mom] (P2) -- (A);
\draw[mom] (P3) -- (B);
\draw[mom] (P4) -- (C);
\draw[mom] (P1) -- (D);
\node at (-3.45,  3.65) {$p_2$};
\node at ( 3.45,  3.65) {$p_3$};
\node at ( 3.45, -3.65) {$p_4$};
\node at (-3.45, -3.65) {$p_1$};
\draw[dual] (X1) -- (O) -- (X3);
\draw[dual] (X2) -- (O) -- (X4);
\node[dualpt] at (X1) {};
\node[dualpt] at (X2) {};
\node[dualpt] at (X3) {};
\node[dualpt] at (X4) {};
\node[dualpt] at (O)  {};
\node[blue, left=6pt]  at (X1) {$x_1$};
\node[blue, above=6pt] at (X2) {$x_2$};
\node[blue, right=6pt] at (X3) {$x_3$};
\node[blue, below=6pt] at (X4) {$x_4$};
\node[blue, above right=2pt] at (O) {$y$};
\end{tikzpicture}
\caption{A four-point one-loop scalar box Feynman diagram in black, labelled by external momenta \(p_i\) and one loop momentum \(k\). In blue is the planar dual graph, labelled by dual momentum coordinates. The internal region variable is denoted by \(y\).}
\label{eq:box_diagr}
\end{figure}

\subsubsection{Loop variables as lines}
For loop integrals, in addition to external momenta \(p_i\), we also have loop
momenta \(k_a\). The following construction is best suited for \textit{planar}
Feynman diagrams, namely diagrams that can be drawn in a disk with the \(n\)
labelled external edges attached to the boundary in cyclic order and without
self-intersections. For non-planar diagrams, this construction is not available in this simple form, because there is then no globally consistent assignment
of region variables to faces.

We introduce dual loop momenta \(y_a\) for each loop variable \(a=1,\dots,\ell\),
in addition to the external dual variables \(x_i\) in~\eqref{eq:x_i}. For a
given \(n\)-point \(\ell\)-loop planar Feynman diagram, see for example
Figure~\ref{eq:box_diagr} for \(n=4\) and \(\ell=1\), the dual variables
correspond to regions, or faces, of the graph. Equivalently, they correspond
to vertices of the planar dual graph of the Feynman diagram, drawn in blue in
Figure~\ref{eq:box_diagr}. Propagators correspond to edges of the dual graph
and are given by expressions of the form \((x_i-y_a)^2\) or \((y_a-y_b)^2\).
For example, choosing the loop-momentum routing \(k=y-x_1\), the momentum and
dual-momentum representations of the scalar box integrand read
\begin{equation}\label{eq:Ibox}
    \mathcal{I}_{{\rm box}} = \frac{\mathrm{d}^4k}{
      k^2
      (k+p_1)^2
      (k+p_1+p_2)^2
      (k-p_4)^2
    } = \frac{\mathrm{d}^4y}{
      (y-x_1)^2 (y-x_2)^2 (y-x_3)^2 (y-x_4)^2
    } \, .
\end{equation}
Different choices of loop-momentum routing give equivalent integrals, but
different rational functions before integration; the dual variables remove
this ambiguity in the planar case.

Each loop momentum \(y_a\) is associated in twistor space with a line
\(A_aB_a\) in \(\mathbb{P}^3\), where the notation \(A_aB_a\) indicates a
line through two points \(A_a\) and \(B_a\). By~\eqref{eq:iijj}, propagators
take the form \(\langle A_a B_a\, i\, i{+}1\rangle\). Integrating \(y\) over
Minkowski space is translated into integrating over lines \(AB\):
\begin{equation}\label{eq:int_mom_tw}
	\int \mathrm{d}^4 y \quad \longleftrightarrow \quad  \int \frac{\langle AB \, \mathrm{d}^2A  \rangle \langle AB \, \mathrm{d}^2B  \rangle}{\langle A B \, I_\infty \rangle^4 } \, ,
\end{equation}
where \(A,B\in\mathbb{P}^3\) span the line \(AB\), and
\begin{equation}
\langle AB\,\mathrm{d}^2A\rangle
:=
\epsilon_{IJKL}A^I B^J \mathrm{d}A^K\wedge \mathrm{d}A^L,
\qquad
\langle AB\,\mathrm{d}^2B\rangle
:=
\epsilon_{IJKL}A^I B^J \mathrm{d}B^K\wedge \mathrm{d}B^L \, .
\end{equation}
The product in the measure of~\eqref{eq:int_mom_tw} really means
\(\langle AB\,\mathrm{d}^2A\rangle \wedge \langle  AB\,\mathrm{d}^2B\rangle\) is the standard
\(\mathrm{SL}(4)\)-invariant four-form on the space of lines in \(\mathbb{P}^3\). 
The \textit{infinity twistor} is
\begin{equation}\label{eq:inf_tw}
	I_\infty := \begin{pmatrix}
		0 & 0 & 1 & 0 \\  0 & 0 & 0 & 1
	\end{pmatrix} \, ,
\end{equation}
and reflects the choice of compactification of dual momentum space $y$ to the space of lines $AB$.
The denominator involving the infinity twistor fixes the affine Minkowski
measure inside projective twistor space; dual conformally invariant integrands
are precisely those for which this dependence cancels. The infinity twistor is
also used to pass from twistor brackets to spinor-helicity brackets:
\begin{equation}
\langle z_i z_{i+1} \, I_{\infty} \rangle = \langle i\,i{+}1 \rangle \, .
\end{equation}
For example, the box integral~\eqref{eq:Ibox} becomes
\begin{equation}
	\mathcal{I}_{{\rm box}} =
    \langle 12 \rangle \langle 23 \rangle \langle 34 \rangle \langle 41 \rangle \,
    \frac{\langle AB \, \mathrm{d}^2A  \rangle \langle AB \, \mathrm{d}^2B  \rangle}{
     \langle  AB 12 \rangle  \langle  AB 23 \rangle  \langle  AB 34 \rangle  \langle  AB 41 \rangle
    } \, .
\end{equation}
The dependence on the infinity twistor~\eqref{eq:inf_tw} has cancelled out.
This is a characteristic feature of integrands with weight \(-4\) in each
loop variable, as is the case in \(y\) for~\eqref{eq:Ibox}.

\subsubsection{Momentum supertwistors}
Momentum twistors admit a natural supersymmetric extension, needed to encode
the kinematics of supersymmetric theories. We continue to use the
\(\widetilde{\eta}\)-representation introduced in Section~\ref{sec:Maximally Supersymmetric Yang-Mills Theory}.
Supermomentum conservation reads \(\sum_{i=1}^n q_i = 0\), with
\begin{equation}
q_i^{\alpha A} = \lambda_i^\alpha \widetilde{\eta}_i^A \, .
\end{equation}
Introducing dual Grassmann variables \(\theta_i^{\alpha A}\) via
\begin{equation}
q_i^{\alpha A} = \theta_i^{\alpha A} - \theta_{i+1}^{\alpha A} \, ,
\end{equation}
one defines fermionic twistor coordinates
\begin{equation}
\chi_i^A = \theta_i^{\alpha A} \lambda_{i\alpha} \, .
\end{equation}
The resulting \textit{momentum supertwistors} are then
\begin{equation}
\mathcal{Z}_i = (z_i,\chi_i) \in \mathbb{CP}^{3|4} \, ,
\end{equation}
encoding both momentum and supermomentum conservation in a manifestly
invariant way.

A key distinction must then be made between three related symmetry structures.
First, the ordinary superconformal symmetry of \(\mathcal{N}=4\) SYM is the
spacetime supersymmetry algebra \({\rm PSU}(2,2|4)\) acting on the original
on-shell data \((\lambda_i,\widetilde{\lambda}_i,\widetilde{\eta}_i)\); in
these variables its generators are realized as differential operators of
degree up to two~\cite{Witten:2003nn,Beisert:2010jr}. Second, planar
amplitudes exhibit an additional \textit{dual} superconformal symmetry, which
acts naturally on dual variables or, equivalently, linearly on momentum
supertwistors. Its generators form the \(\mathfrak{gl}(4|4)\) superalgebra
and take the simple form
\begin{equation}
J^I{}_J = \sum_{i=1}^n \mathcal{Z}_i^I \frac{\partial}{\partial \mathcal{Z}_i^J} \, ,
\end{equation}
so that dual superconformal invariance becomes manifest~\cite{Drummond:2008vq,Mason:2009qx}.
Finally, the ordinary and dual superconformal symmetries combine in the planar
theory and generate an infinite-dimensional Yangian symmetry. This enlarged
algebra is one of the strongest organizing principles of planar
\(\mathcal{N}=4\) SYM, and underlies the remarkable simplicity of its
amplitudes, their Grassmannian representations, and ultimately the geometry of
the Amplituhedron~\cite{Drummond:2009fd,Grassmannian}.

We now have a natural set of variables for planar \(\mathcal{N}=4\) SYM
amplitudes. Before turning to on-shell diagrams, let us record how
superamplitudes are expressed in momentum-twistor language. From the
supersymmetric Ward identities,
\begin{equation}\label{eq:Ward_SUSY}
Q^\alpha_A \, A_n = 0 \, , \qquad  \Qbar^{\dot{\alpha}A} \, A_n = 0 \, ,
\end{equation}
one can solve the first set of constraints by factoring out a Grassmann delta
function
\begin{equation}\label{eq:Atilde}
\delta^{(8)}(q) = \prod_{A=1}^4 \delta^{(2)}(q^A)
=  \prod_{A=1}^{4}\sum_{1\leq i<j\leq n} \langle ij \rangle \, \widetilde{\eta}^A_i\widetilde{\eta}^A_j\, ,
\qquad
q = \sum_{i=1}^n \lambda_i \widetilde{\eta}_i \, ,
\end{equation}
enforcing supermomentum conservation. One can therefore write the
\(\text{N}^k\text{MHV}\) superamplitude as~\eqref{eq:Atilde} times a function
of momentum supertwistors that is a polynomial of total Grassmann degree
\(4k\) in the fermionic variables \(\chi_i\). The second set of equations
in~\eqref{eq:Ward_SUSY} implies non-trivial differential equations
constraining the form of amplitudes beyond tree level and is
closely related to anomalous dual superconformal symmetry~\cite{Drummond:2008vq,CaronHuot:2011kk}.

The BCFW recursion relation presented in Section~\ref{sec:Recursion Relations}
can be extended to the supersymmetric setting by shifting also the fermionic
variables:
\begin{equation}\label{eq:ij_supershift}
[i,j\rangle\text{-supershift} \ : \quad
\lambda_i(z) := \lambda_i + z \, \lambda_j \, ,
\quad
\widetilde{\lambda}_j(z) := \widetilde{\lambda}_j - z \, \widetilde{\lambda}_i \, ,
\quad
\widetilde{\eta}_j(z) := \widetilde{\eta}_j - z \, \widetilde{\eta}_i \, .
\end{equation}
In momentum-supertwistor variables, the corresponding deformation can be
written as a linear shift
\begin{equation}\label{eq:super_shift}
	\mathcal{Z}_i(z)=\mathcal{Z}_i+ z \, \mathcal{Z}_{j} \, .
\end{equation}
BCFW recursion relations allow one to construct all tree-level amplitudes and
loop-level integrands starting from a small set of building blocks~\cite{Grassmannian,ArkaniHamed:2010kv}.
Using the superspace formalism and spinor-helicity variables, the three-point
superamplitudes are
\begin{tcolorbox}[definitionbox]
\textbf{Three-point superamplitudes.}
\begin{equation}\label{eq:3pt_super_ampl}
A_3^{\text{MHV}} = \frac{\delta^{(4)}(p) \, \delta^{(8)}(q)}{\langle 12\rangle \langle 23\rangle \langle 31\rangle} \, ,
\qquad \qquad
A_3^{\overline{\text{MHV}}}
= \frac{\delta^{(4)}(p) \, \delta^{(4)}\big(\tilde{q}\big)}{[12][23][31]} \, ,
\end{equation}
\end{tcolorbox}\noindent
where
\begin{equation}
	p = \lambda_1 \widetilde{\lambda}_1 + \lambda_2 \widetilde{\lambda}_2 + \lambda_3 \widetilde{\lambda}_3 \ , \quad
    q = \lambda_1 \widetilde{\eta}_1 + \lambda_2 \widetilde{\eta}_2 + \lambda_3 \widetilde{\eta}_3  \ , \quad
    \tilde{q} = [12]\widetilde{\eta}_3 + [23]\widetilde{\eta}_1 + [31]\widetilde{\eta}_2 \, .
\end{equation}
An explicit formula for all tree-level \(A_n^{\text{N}^k\text{MHV}}\)
amplitudes can be obtained via BCFW recursion~\cite{Dixon:2011xs}; see
Section~\ref{sec:BCFW Recursion and Tiles}. For example, the \(n\)-point
tree-level MHV superamplitude~is 
\begin{tcolorbox}[definitionbox]
\textbf{Tree-level MHV superamplitude.}
\begin{equation}\label{eq:MHV}
A_n^{\text{MHV}} = \frac{\delta^{(8)}(q)}{\langle 12\rangle \langle 23\rangle \cdots \langle n1\rangle} \, .
\end{equation}
\end{tcolorbox}\noindent

\subsubsection{Ratio functions and bosonization}
It is convenient for the discussion in the following sections to repackage
superamplitudes in the following way. The \(n\)-point tree-level
\(\text{N}^k\)MHV superamplitude in planar \(\mathcal{N}=4\) SYM can be
written in momentum supertwistors \(\mathcal{Z}_i=(z_i,\chi_i)\), for
\(i=1,\dots,n\), as
\begin{tcolorbox}[definitionbox]
\textbf{Ratio-function decomposition.}
\begin{equation}\label{eq:ampl_R_dec}
	A_n^{\text{N}^k\text{MHV}}(\mathcal{Z}_i)
    = A_n^{{\rm MHV}}(\mathcal{Z}_i) \, \mathcal{R}_{k,n}(\mathcal{Z}_i) \, ,
\end{equation}
\end{tcolorbox}\noindent
where the MHV superamplitude \(A_n^{{\rm MHV}}(\mathcal{Z}_i)\) is given
in~\eqref{eq:MHV}, and \(\mathcal{R}_{k,n}(\mathcal{Z}_i)\) is the \(n\)-point
\(\text{N}^k\text{MHV}\) \textit{ratio function}. The latter is a rational
function in the bosonic momentum twistors \(z_i\) and a polynomial of total
degree \(4k\) in the fermionic momentum twistors \(\chi_i\). The same
decomposition holds for loop-level integrands, which we discuss in
Section~\ref{sec:Loop Amplituhedra}.

For the geometric construction that will follow, it is convenient to
\textit{bosonize} the fermionic degrees of freedom and work in a purely
bosonic kinematic space. For this, we introduce auxiliary fermionic variables
\(\phi^A_a\), with \(a=1,\dots,k\) and \(A=1,\dots,4\), and define
\textit{bosonized momentum twistors}
\begin{equation}
	Z_i = (z_i ,  \phi_1 \cdot  \chi_i , \dots, \phi_k \cdot  \chi_i)  \in \mathbb{P}^{k+3} \, ,
\end{equation}
where \(\phi_a \cdot  \chi_i := \sum_{A=1}^4 \phi_a^A \chi_i^A\). The
non-trivial part of the superamplitude is then recovered by extracting the
appropriate coefficient in the auxiliary variables \(\phi_a^A\):
\begin{equation}\label{eq:extract_R}
		\mathcal{R}_{k,n}(\mathcal{Z}_i)
    = \prod_{a=1}^k \prod_{A=1}^4
    \frac{\partial}{\partial \phi_a^A} \, \mathcal{R}_{k,n}(Z_i) \, .
\end{equation}
In summary, for describing \(\text{N}^k\text{MHV}\) superamplitudes, one can
trade the fermionic degrees of freedom \(\chi_i\), which encode the helicities
of the particles, for bosonic degrees of freedom by extending each bosonic
momentum twistor \(z_i\) by \(k\) additional bosonic coordinates
\(\phi_a \cdot \chi_i\). This is possible because \(\mathcal{R}_{k,n}(\mathcal{Z}_i)\) is a
polynomial of total Grassmann degree \(4k\) in the \(\chi_i\). The kinematic
space for \(n\)-particle \(\text{N}^k\text{MHV}\) superamplitudes in planar
\(\mathcal{N}=4\) SYM is then the configuration space of \(n\) points \(Z_i\)
in \(\mathbb{P}^{k+3}\), equivalently non-zero vectors in \(\mathbb{C}^{k+4}\)
modulo independent rescaling. 

\subsection{On-shell diagrams and Grassmannian contours}\label{sec:On-Shell Diagrams and Grassmannian Contours}

Perturbative scattering amplitudes in QFT are usually computed using Feynman
diagrams. This approach has notable drawbacks: the diagrammatic expansion
obscures gauge invariance and, in special theories, also hides additional
symmetries of amplitudes. For example, as we discussed previously, planar
\(\mathcal{N}=4\) SYM enjoys a hidden dual conformal supersymmetry, which
combines with ordinary superconformal symmetry into an extended Yangian
symmetry. Remarkably, individual terms appearing in BCFW recursion are both
gauge invariant and Yangian invariant. Each term in the recursion can be
associated with an \textit{on-shell diagram}, a combinatorial object closely
related to the \textit{positive Grassmannian}. This sequence of connections is
at the heart of the geometric and combinatorial formulation of scattering
amplitudes in planar \(\mathcal{N}=4\) SYM.

\providecommand{\kp}{k'}

\subsubsection{On-shell diagrams}

\begin{figure}[pos=t]
\centering
\includegraphics[width=0.35\textwidth]{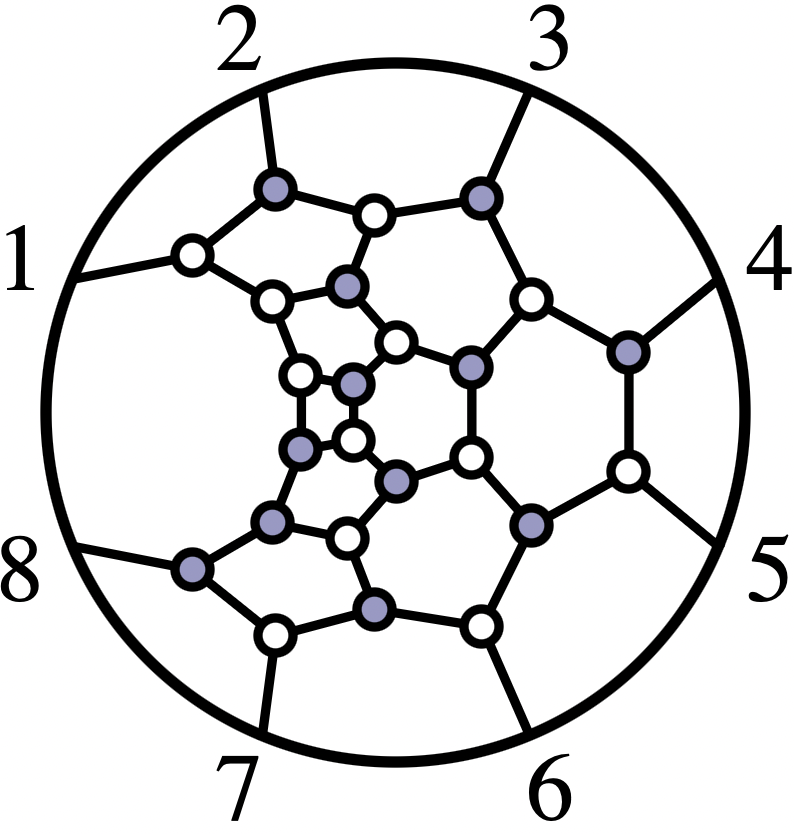}
\caption{On-shell diagram with \(n=8\) and Grassmannian helicity degree
\(\kp=4\)~\cite{Bourjaily:2012gy}.}
\label{fig:graph_0}
\end{figure}

In our setting, on-shell diagrams are planar bipartite graphs, also called plabic graphs, with trivalent vertices colored black and white; see Figure~\ref{fig:graph_0}. They are built by gluing three-valent black and
white vertices, corresponding to three-point \(\text{MHV}\) and
\(\overline{\text{MHV}}\) superamplitudes in~\eqref{eq:3pt_super_ampl}.

To understand this construction, we repackage the kinematic data. We stack
together the spinors \(\lambda_i\), with \(i=1,\dots,n\), in a \(2\times n\)
matrix \(\lambda\), and similarly the \(\widetilde{\lambda}_i\) in a
\(2\times n\) matrix \(\widetilde{\lambda}\). The complexified Lorentz group
\({\rm SL}(2)\) acts by rotating the rows of these matrices. The amplitude is
invariant under this action and covariant under the little group
\({\rm U}(1)^n\). Thus, up to homogeneous degree, the amplitude depends only
on the two-dimensional linear spans of \(\lambda\) and
\(\widetilde{\lambda}\) inside \(\mathbb{C}^n\). Momentum conservation is the
matrix equation
\begin{equation}\label{eq:mom_cons_lambda}
	\lambda \cdot \widetilde{\lambda}^{\top} = 0 \, ,
\end{equation}
which geometrically means that the two-plane spanned by \(\lambda\) is orthogonal to the two-plane spanned by
\(\widetilde{\lambda}\):
\begin{equation}
	\lambda \subseteq \widetilde{\lambda}^{\perp} \, .
\end{equation}
This naturally leads to the Grassmannian \(\Gr(k,n)\), the space of
\(k\)-dimensional linear subspaces of \(\mathbb{C}^n\). A point in
\(\Gr(k,n)\) is an equivalence class of \(\kp\times n\) complex matrices
modulo left multiplication by \({\rm GL}(k)\). In this language, the
\(n\)-point amplitude can be regarded as a function on the product
\(\Gr(2,n)\times \Gr(2,n)\) subject to~\eqref{eq:mom_cons_lambda}.
This space is known as the \textit{spinor-helicity variety}
~\cite{Maazouz:2024qmm}.

Let us revisit the three-point kinematics of
Example~\ref{eg:3pt_ampl} in this language. For \(n=3\), using the Schouten
identity
\begin{equation}
	\widetilde{\lambda}_1[23]
	+\widetilde{\lambda}_2[31]
	+\widetilde{\lambda}_3[12]=0 \, ,
\end{equation}
we can write
\begin{equation}\label{eq:tilde_lambda_perp}
	\widetilde{\lambda}^{\perp}
	=
	\begin{pmatrix}
		[23] & [31] & [12]
	\end{pmatrix} \, .
\end{equation}
Equation~\eqref{eq:mom_cons_lambda} forces \(\lambda\) to degenerate to a
one-dimensional subspace and take the form
\begin{equation}
	\lambda
	=
	\begin{pmatrix}
		[23] & [31] & [12] \\
		0 & 0 & 0
	\end{pmatrix} \, .
\end{equation}

The crucial point in gluing three-point amplitudes is to combine all delta
functions arising from momentum and supermomentum conservation into one global
system. A natural way to do this is to introduce auxiliary variables so that
the local delta-function constraints become linear. Solving all local
conditions is then turned into a single linear-algebra problem. Beautifully,
these auxiliary variables themselves live on Grassmannians. For example, the
three-point anti-MHV amplitude can be rewritten as
\begin{equation}\label{eq:MHVbar_Gr}
	A^{\overline{\text{MHV}}}_3
    =
    \int \mathbf{\Omega}_W \,
    \delta^{1 \times 4}(W \cdot \widetilde{\eta}) \,
    \delta^{1 \times 2}(W \cdot \widetilde{\lambda}) \,
    \delta^{2\times 2}(\lambda \cdot W^\perp) \, ,
    \qquad
    \mathbf{\Omega}_W
    =
	    \frac{\mathrm{d}^{1\times 3} W}{{\rm GL}(1)}\,
    \frac{1}{(1)(2)(3)} \, ,
\end{equation}
where \(W \in \Gr(1,3)\) is a three-vector modulo complex rescaling,
\(\widetilde{\eta}=(\widetilde{\eta}_1,\widetilde{\eta}_2,\widetilde{\eta}_3)\),
and \((i):=W_i\). To see that~\eqref{eq:MHVbar_Gr} is equivalent
to~\eqref{eq:3pt_super_ampl}, one solves the two-dimensional integral over
\(W\) using \(\delta^{1 \times 2}(W \cdot \widetilde{\lambda})\), which fixes
\(W\) up to scaling to be \(W=\widetilde{\lambda}^{\perp}\) as
in~\eqref{eq:tilde_lambda_perp}. Then \((1)=[23]\), \((2)=[31]\), and
\((3)=[12]\), while the delta functions
\(\delta^{1 \times 4}(W \cdot \widetilde{\eta})\) and
\(\delta^{2\times 2}(\lambda \cdot W^\perp)\) become supermomentum and
momentum conservation, respectively. Analogously, the MHV case becomes
\begin{equation}\label{eq:MHV_Gr}
	A^{\text{MHV}}_3
    =
    \int \mathbf{\Omega}_B\,
    \delta^{2 \times 4}(B \cdot \widetilde{\eta})\,
    \delta^{2 \times 2}(B \cdot \widetilde{\lambda}) \,
    \delta^{2\times 1}(\lambda \cdot B^\perp) \, ,
    \qquad
    \mathbf{\Omega}_B
    =
	    \frac{\mathrm{d}^{2\times 3} B}{{\rm GL}(2)}\,
    \frac{1}{(12)(23)(31)} \, ,
\end{equation}
where \(B=(B_1B_2B_3)\) is a \(2\times 3\) matrix representative of a point in
\(\Gr(2,3)\), and \((ij):=\det(B_iB_j)\) is the
\(2\times2\) minor.

We can now glue black and white vertices by introducing an \textit{on-shell
phase space integral}
\begin{equation}\label{eq:edge_integration}
	\int
	\frac{\mathrm{d}^2\lambda_I \mathrm{d}^2 \widetilde{\lambda}_I}{{\rm GL}(1)}
	\mathrm{d}^4\widetilde{\eta}_I
\end{equation}
for every internal edge \(I\) of the diagram. In this way we form an on-shell
diagram \(\Gamma\). Each trivalent vertex contributes a small Grassmannian,
either \(\Gr(1,3)\) or \(\Gr(2,3)\), together with delta functions
imposing the local on-shell constraints. Gluing two legs identifies their
on-shell data and integrates over the shared state. After eliminating the
internal variables, all local constraints combine into a single matrix
\(C\in\Gr(\kp,n)\), whose rows encode the linear relations among the
external data.

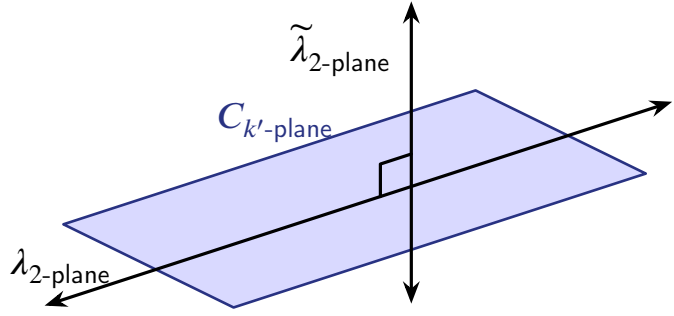
\begin{figure}[pos=t]
\centering
\begin{tikzpicture}[
    scale=1.0,
    line cap=round,
    line join=round,
    >=Stealth
]
\definecolor{planeblue}{RGB}{40,50,130}
\def\ang{18}
\coordinate (O) at (0.55,-0.05);
\coordinate (A) at (-4.05,-0.55);
\coordinate (B) at ($(A)+(5.45,{5.45*tan(\ang)})$);
\coordinate (D) at ($(A)+(2.25,-1.10)$);
\coordinate (C) at ($(B)+(2.25,-1.10)$);
\filldraw[fill=blue!15,draw=planeblue,line width=1pt](A) -- (B) -- (C) -- (D) -- cycle;
\draw[very thick,<->]($(O)+(-4.85,{-4.85*tan(\ang)})$)--($(O)+(3.45,{3.45*tan(\ang)})$);
\draw[very thick,<->]($(O)+(0,-1.55)$)--($(O)+(0,2.45)$);
\def\r{0.43}
\coordinate (P) at ($(O)+(0,\r)$);
\coordinate (Q) at ($(O)+({-\r*cos(\ang)},{-\r*sin(\ang)})$);
\coordinate (R) at ($(O)+(0,\r)+({-\r*cos(\ang)},{-\r*sin(\ang)})$);
\draw[very thick] (P) -- (R) -- (Q);
\node[font=\Large] at (-0.40,1.75) {$\widetilde{\lambda}_{2\text{-plane}}$};
\node[font=\Large] at (-4.10,-1.13) {$\lambda_{2\text{-plane}}$};
\node[text=planeblue, font=\Large] at (-1.25,0.8) {$C_{\kp\text{-plane}}$};
\end{tikzpicture}
\caption{Kinematic configuration of on-shell diagrams from the Grassmannian
perspective. The \(\kp\)-plane \(C\) is orthogonal to the
\(\widetilde{\lambda}\)-plane and contains the \(\lambda\)-plane.}
\label{fig:kin_planes}
\end{figure}

More explicitly, each white vertex imposes one relation among the incident
\(\widetilde{\lambda}\) variables, while a black vertex imposes two relations.
For a graph with \(n_b\) black vertices, \(n_w\) white vertices, and \(n_I\)
internal edges, this gives \(2n_b+n_w\) local constraints. The internal-edge
integrations~\eqref{eq:edge_integration} and the delta functions at each
vertex eliminate one \(\widetilde{\lambda}_I\) and
\(\widetilde{\eta}_I\) per internal edge, leaving
\begin{equation}
	\kp = 2n_b+n_w-n_I
\end{equation}
independent linear relations among the \(n\) external edge variables. These
can be assembled into
\begin{equation}
	C\cdot\widetilde{\lambda}=0,
	\qquad
	C\cdot\widetilde{\eta}=0 \, ,
\end{equation}
where \(C\) is a \(\kp\times n\) complex matrix, with \(n=3n_V-2n_I\) and
\(n_V=n_b+n_w\). Since the constraints are defined up to
\({\rm GL}(\kp)\) transformations, \(C\) defines a point in the Grassmannian
\(\Gr(\kp,n)\). The delta-function constraints on \(\lambda\) are encoded by \(\lambda\cdot C^\perp=0\). Geometrically, the \(\kp\)-plane \(C\) is
orthogonal to \(\widetilde{\lambda}\) and contains
\(\lambda\), as shown in Figure~\ref{fig:kin_planes}.

Putting everything together, to each on-shell diagram we associate an
\textit{on-shell form}
\begin{tcolorbox}[definitionbox]
\textbf{On-shell form.}
\begin{equation}\label{eq:Gr_int}
 	\mathbf{\Omega}_\Gamma
 	=
 	\prod_{\text{internal edges } e}\frac{1}{\mathrm{GL}(1)_e}
	\prod_{w} \mathbf{\Omega}_w
	\prod_{b} \mathbf{\Omega}_b\;
	\delta^{\kp \times 4}(C\cdot \widetilde\eta)\,
	\delta^{\kp \times 2}(C\cdot \widetilde{\lambda})\,
	\delta^{2 \times (n-\kp)}(\lambda\cdot C^\perp) \, .
\end{equation}
\end{tcolorbox}\noindent
Since \(\delta^{\kp \times 4}(C\cdot \widetilde\eta)\) is a polynomial in the
Grassmann variables, similarly to~\eqref{eq:Atilde}, the label \(\kp\)
determines the helicity degree of the on-shell form \(f_\Gamma\). In the convention of Section~\ref{sec:Maximally Supersymmetric Yang-Mills Theory}, an
\(\mathrm{N}^k\mathrm{MHV}\) tree amplitude corresponds to \(\kp=k+2\). The
dimension of the integration in~\eqref{eq:Gr_int} is
\begin{equation}
	 \dim(\Gamma) := 2n_V-n_I = n_I-n_V+n = n_f-1 \, ,
\end{equation}
where we used trivalence of the graph in the second equality and Euler's
formula for planar \(\Gamma\) in the last equality, with \(n_f\) the number of
faces of \(\Gamma\). The total number of bosonic delta functions is \(2n-4\).
If \(\dim(\Gamma)=2n-4\), then the integral is fully localized by the bosonic
delta functions. In this case the on-shell form is a leading
singularity: the integration variables are completely fixed by the bosonic
constraints, leaving an algerbaic function of the external kinematics.

\subsubsection{Equivalence moves and Grassmannian contours}

\begin{figure}[pos=t]
\centering
\resizebox{\textwidth}{!}{%
\begin{tikzpicture}[line cap=round,line join=round,scale=1]
\definecolor{vertexpurple}{RGB}{170,170,220}
\tikzset{panel/.style={draw=black!65, rounded corners=10pt, line width=0.8pt},
edge/.style={black, line width=1.5pt},
bdot/.style={circle, draw=black, fill=vertexpurple, line width=1.3pt, inner sep=3.0pt},
wdot/.style={circle, draw=black, fill=white, line width=1.3pt, inner sep=3.0pt},
title/.style={font=\normalfont\small},
lab/.style={font=\normalsize},
arrowlab/.style={font=\fontsize{20}{20}\selectfont}}
\node[title] at (2.7,4.25) {Merge / unmerge};
\node[title] at (9.75,4.25) {Square move};
\node[title] at (16.6,4.25) {Bubble reduction};
\draw[panel] (0.2,-0.15) rectangle (5.2,3.9);
\draw[panel] (5.4,-0.15) rectangle (14.1,3.9);
\draw[panel] (14.3,-0.15) rectangle (18.9,3.9);
\begin{scope}[shift={(0.95,2.85)}]
  \coordinate (L1) at (0,0);
  \coordinate (L2) at (1.05,0);
  \draw[edge] (-0.55,0) -- (L1);
  \draw[edge] (L1) -- (L2);
  \draw[edge] (L2) -- (1.60,0);
  \draw[edge] (L1) -- +(0,0.72);
  \draw[edge] (L1) -- +(0,-0.72);
  \draw[edge] (L2) -- +(0,0.72);
  \draw[edge] (L2) -- +(0,-0.72);
  \node[bdot] at (L1) {};
  \node[bdot] at (L2) {};
\end{scope}
\node[arrowlab] at (3.2,2.80) {$\leftrightarrow$};
\begin{scope}[shift={(4.45,2.85)}]
  \coordinate (M) at (0,0);
  \draw[edge] (-0.55,0) -- (M) -- (0.55,0);
  \draw[edge] (M) -- +(0,0.72);
  \draw[edge] (M) -- +(0,-0.72);
  \node[bdot] at (M) {};
\end{scope}
\begin{scope}[shift={(0.95,1.05)}]
  \coordinate (L1) at (0,0);
  \coordinate (L2) at (1.05,0);
  \draw[edge] (-0.55,0) -- (L1);
  \draw[edge] (L1) -- (L2);
  \draw[edge] (L2) -- (1.60,0);
  \draw[edge] (L1) -- +(0,0.72);
  \draw[edge] (L1) -- +(0,-0.72);
  \draw[edge] (L2) -- +(0,0.72);
  \draw[edge] (L2) -- +(0,-0.72);
  \node[wdot] at (L1) {};
  \node[wdot] at (L2) {};
\end{scope}
\node[arrowlab] at (3.2,1.00) {$\leftrightarrow$};
\begin{scope}[shift={(4.45,1.05)}]
  \coordinate (M) at (0,0);
  \draw[edge] (-0.55,0) -- (M) -- (0.55,0);
  \draw[edge] (M) -- +(0,0.72);
  \draw[edge] (M) -- +(0,-0.72);
  \node[wdot] at (M) {};
\end{scope}
\def\s{1.35}
\def\l{0.48}
\begin{scope}[shift={(7.00,1.22)}]
  \coordinate (BL) at (0,0);
  \coordinate (BR) at (\s,0);
  \coordinate (TL) at (0,\s);
  \coordinate (TR) at (\s,\s);
  \draw[edge] (TL)--(TR)--(BR)--(BL)--cycle;
  \draw[edge] (TL) -- +(-\l,\l);
  \draw[edge] (TR) -- +(\l,\l);
  \draw[edge] (BR) -- +(\l,-\l);
  \draw[edge] (BL) -- +(-\l,-\l);
  \node[bdot] at (TL) {};
  \node[wdot] at (TR) {};
  \node[wdot] at (BL) {};
  \node[bdot] at (BR) {};
  \node[lab] at (-0.60,\s+0.58) {$1$};
  \node[lab] at (\s+0.60,\s+0.58) {$2$};
  \node[lab] at (\s+0.60,-0.58) {$3$};
  \node[lab] at (-0.60,-0.58) {$4$};
\end{scope}
\node[arrowlab] at (9.75,1.90) {$\leftrightarrow$};
\begin{scope}[shift={(11.00,1.22)}]
  \coordinate (BL) at (0,0);
  \coordinate (BR) at (\s,0);
  \coordinate (TL) at (0,\s);
  \coordinate (TR) at (\s,\s);
  \draw[edge] (TL)--(TR)--(BR)--(BL)--cycle;
  \draw[edge] (TL) -- +(-\l,\l);
  \draw[edge] (TR) -- +(\l,\l);
  \draw[edge] (BR) -- +(\l,-\l);
  \draw[edge] (BL) -- +(-\l,-\l);
  \node[wdot] at (TL) {};
  \node[bdot] at (TR) {};
  \node[bdot] at (BL) {};
  \node[wdot] at (BR) {};
  \node[lab] at (-0.60,\s+0.58) {$1$};
  \node[lab] at (\s+0.60,\s+0.58) {$2$};
  \node[lab] at (\s+0.60,-0.58) {$3$};
  \node[lab] at (-0.60,-0.58) {$4$};
\end{scope}
\begin{scope}[shift={(15.30,1.85)}]
  \coordinate (T) at (0,0.70);
  \coordinate (B) at (0,-0.70);
  \draw[edge] (0,1.55) -- (T);
  \draw[edge] (B) -- (0,-1.55);
  \draw[edge] (T) .. controls (-0.35,0.20) and (-0.35,-0.20) .. (B);
  \draw[edge] (T) .. controls ( 0.35,0.20) and ( 0.35,-0.20) .. (B);
  \node[bdot] at (T) {};
  \node[wdot] at (B) {};
\end{scope}
\node[arrowlab] at (16.6,1.85) {$\to$};
\begin{scope}[shift={(17.95,1.85)}]
  \draw[edge] (0,1.55) -- (0,-1.55);
\end{scope}
\end{tikzpicture}%
}
\caption{Equivalence moves satisfied by on-shell diagrams.}
\label{fig:onshell-equivalence-moves}
\end{figure}
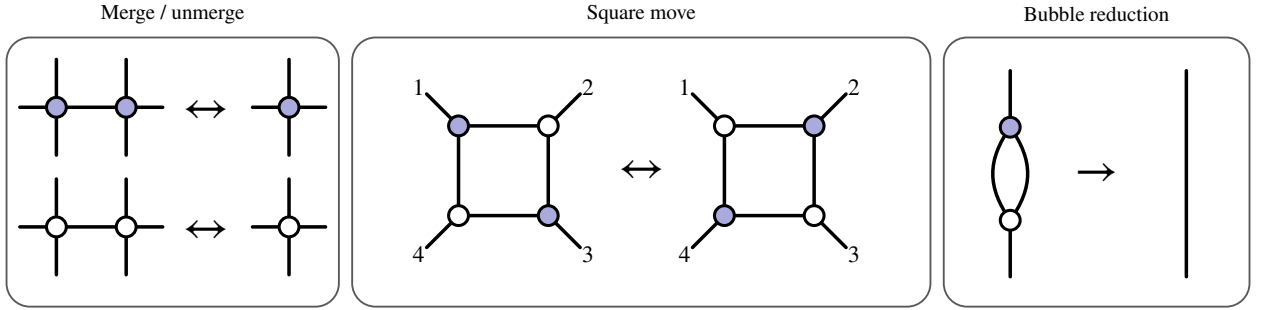

It is relatively straightforward to verify that the on-shell
form~\eqref{eq:Gr_int} of an on-shell diagram \(\Gamma\) is invariant under
the \textit{equivalence moves} shown in
Figure~\ref{fig:onshell-equivalence-moves}. Therefore, the relevant
combinatorial objects encoding on-shell forms are equivalence classes of
on-shell diagrams under these moves.

The integral~\eqref{eq:Gr_int} can be formulated more elegantly as a contour integral in the Grassmannian \(\Gr(\kp,n)\):
\begin{tcolorbox}[definitionbox]
\textbf{Grassmannian contour integral.}
\begin{equation}\label{eq:Gr_contour}
\mathbf{\Omega}_\Gamma
 	= \int_{\Gamma}
\frac{1}{\mathrm{GL}(\kp)}
\frac{
{\mathrm d}^{\kp\times n} C
}{
(12\cdots \kp)(23\cdots \kp{+}1)\cdots(n1\cdots \kp{-}1)
}
\,
\delta^{\kp \times 4}(C\cdot \widetilde\eta)\,
\delta^{\kp \times 2}(C\cdot \widetilde{\lambda})\,
\delta^{2 \times (n-\kp)}(\lambda\cdot C^\perp) \, .
\end{equation}
\end{tcolorbox}\noindent
Here the brackets \((12 \cdots \kp)=\det(C_1C_2\cdots C_{\kp})\) are
consecutive minors of \(C\). The choice of contour specifies which residues of
the Grassmannian form are taken. Different contours correspond to different
on-shell diagrams, and hence to different Yangian invariants. The on-shell
form~\eqref{eq:Gr_int} is obtained from~\eqref{eq:Gr_contour} by choosing the
specific contour associated to the on-shell diagram \(\Gamma\) and fixing the \({\rm GL}(\kp)\) redundancy.

For amplitudes it is often preferable to pass from spinor-helicity variables
to momentum twistors. This change of variables solves momentum conservation
and factors out the universal MHV, or Parke--Taylor, prefactor. More
precisely, an \(\mathrm{N}^k\mathrm{MHV}\) superamplitude is written as
in~\eqref{eq:ampl_R_dec}. In the Grassmannian contour integral
representation~\eqref{eq:Gr_contour}, this removes the two
\(\lambda\)-directions contained in \(C\). After factoring out
\(A_n^{{\rm MHV}}(\mathcal Z_i)\),~\eqref{eq:ampl_R_dec} becomes
\begin{tcolorbox}[definitionbox]
\textbf{Momentum-twistor Grassmannian contour integral.}
\begin{equation}\label{eq:Gr_contour_momentum_twistor}
\int_{\Gamma}
\frac{1}{\mathrm{GL}(k)}
\frac{
{\mathrm d}^{k\times n} C
}{
(12\cdots k)(23\cdots k{+}1)\cdots(n1\cdots k{-}1)
}
\,
\delta^{4k|4k}(C\cdot \mathcal Z) \, ,
\end{equation}
\end{tcolorbox}\noindent
where \(\mathcal Z\) is the matrix whose columns are the \(n\) momentum supertwistors \(\mathcal Z_i=(z_i,\chi_i)\). Thus the momentum-space Grassmannian \(\Gr(\kp,n)\), with \(\kp=k+2\), is replaced by the momentum-twistor Grassmannian \(\Gr(k,n)\). This shifted convention is the one naturally used in the Amplituhedron setting in Chapters~\ref{ch:Amplituhedra} and~\ref{ch:From Integrands to Integrals}.

\subsubsection{Relation to amplitudes and positive geometry}

On-shell diagrams and their forms are relevant for the following reasons:
\begin{enumerate}
    \item They represent individual terms in BCFW recursion for tree
    amplitudes and loop integrands in planar \(\mathcal{N}=4\) SYM. The
    recursion can itself be formulated as combinatorial operations on on-shell
    diagrams~\cite{Britto:2004ap,Britto:2005fq,Grassmannian}; see
    Section~\ref{sec:BCFW Recursion and Tiles}.
    \item Modulo the equivalence moves in
    Figure~\ref{fig:onshell-equivalence-moves}, on-shell diagrams are in
    one-to-one correspondence with positroids~\cite{Postnikov:2006kva}. These
    encode the combinatorics of the positive Grassmannian and, more broadly,
    the Amplituhedron~\cite{the_amplituhedron}; see Chapter~\ref{ch:Amplituhedra}.
    \item On-shell forms are in bijective correspondence with \textit{leading
    singularities} of loop amplitudes in planar \(\mathcal{N}=4\) SYM~\cite{ArkaniHamed:2009dn}. We discuss this correspondence in
    Section~\ref{sec:Leading Singularities and Maximal Cuts}.
    \item On-shell forms are Yangian invariant~\cite{Drummond:2009fd,Drummond:2010qh}, and it is believed that every
    Yangian invariant arises from a Grassmannian integral representation of
    the form~\eqref{eq:Gr_contour} for some choice of contour~\cite{ArkaniHamed:2009dn}.
\end{enumerate}

In this thesis we focus on planar \(\mathcal{N}=4\) SYM, where
on-shell diagrams are especially powerful because they are classified by
positroids and cells of the positive Grassmannian. Extensions including non-planar configurations, reduced supersymmetry, and
gravity, also exist~\cite{Grassmannian,Elvang:2013cua,Cachazo:2012kg,Franco:2014csa,
Franco:2015rma,ABHY_original,Carrolo:2026qpu,Lisitsyn:2025prd}. However, in
these broader settings the full combinatorial and geometric picture is less
complete.

Coming back to the list above, many of these aspects will accompany our
journey through positive geometries, the Amplituhedron, and loop amplitudes.
We conclude this section with some concrete examples, which also allow us to
introduce the \texttt{Mathematica} package \texttt{Positroids.m} of
Bourjaily~\cite{Bourjaily:2012gy}, a useful tool for computations involving
on-shell diagrams and their forms.

\subsubsection{Boundary measurements and examples}
First, we discuss an explicit parametrization of an on-shell diagram \(\Gamma\).
We start by assigning a perfect orientation to \(\Gamma\), that is, we add
arrows to all edges such that each white vertex has one incoming arrow, while
each black vertex has two incoming arrows. The arrows specify the elimination
procedure for internal edge variables described above. Given a perfect
orientation of \(\Gamma\), among the boundary vertices, namely those of degree
one, there are precisely \(\kp\) sources. We assign edge variables \(\alpha_e\)
to each edge \(e\) of \(\Gamma\). Given \(i,j\in\{1,\dots,n\}\) with \(i\) a
source, we can parametrize \(C\) by the boundary measurement
\begin{equation}\label{eq:C_alpha}
	C_{ij}(\alpha) = -\sum_{\gamma \, : \, i \rightarrow j} \prod_{e \in \gamma} \alpha_e  \, ,
\end{equation}
where we sum over all directed paths \(\gamma\) from \(i\) to \(j\) following
the arrows in the graph. For each path we take the product of all edge
variables along the way. By convention \(C_{ii}=1\) and \(C_{ij}=0\) if \(j\)
is a source. The overall signs in this formula depend on the convention for
sources and edge orientations; they do not affect the positroid cell encoded
by the diagram. The matrix \(C(\alpha)\) is called the \textit{boundary
measurement} of \(\Gamma\). In this parametrization,~\eqref{eq:Gr_int}
becomes
\begin{equation}\label{eq:Gr_int_alpha}
 	\mathbf{\Omega}_\Gamma  = \int \prod_{\text{vertices } v} \frac{1}{{\rm GL}(1)_v}
\prod_{\text{edges } e} \frac{\mathrm{d}\alpha_e}{\alpha_e} \,
\delta^{\kp \times 4}(C(\alpha)\cdot \widetilde\eta)\,
\delta^{\kp \times 2}(C(\alpha)\cdot \widetilde{\lambda})\,
\delta^{2 \times (n-\kp)}(\lambda\cdot C(\alpha)^\perp) \, .
\end{equation}

On-shell diagrams are in bijection with several combinatorial objects. A
practical encoding of a \((\kp,n)\) on-shell diagram \(\Gamma\) is its associated
affine permutation \(\sigma\). This is a list of \(n\) non-negative integers
\((\sigma(1),\dots,\sigma(n))\), where \(\sigma(i)\) is obtained by starting
from the boundary vertex \(i\) in \(\Gamma\) and following a left-right path in
\(\Gamma\), turning left at every white vertex and right at every black vertex.
The path ends at some external vertex \(j\); if \(j\geq i\) with respect to
the cyclic ordering, then \(\sigma(i)=j\), while otherwise \(\sigma(i)=j+n\).
Then \(\Gamma\) is equivalently encoded by \(\sigma\)~\cite{Postnikov:2006kva}.

\begin{eg}\label{eq:onshell_k2_n4}
\begin{figure}[pos=t]
\centering
\includegraphics[width=0.30\textwidth]{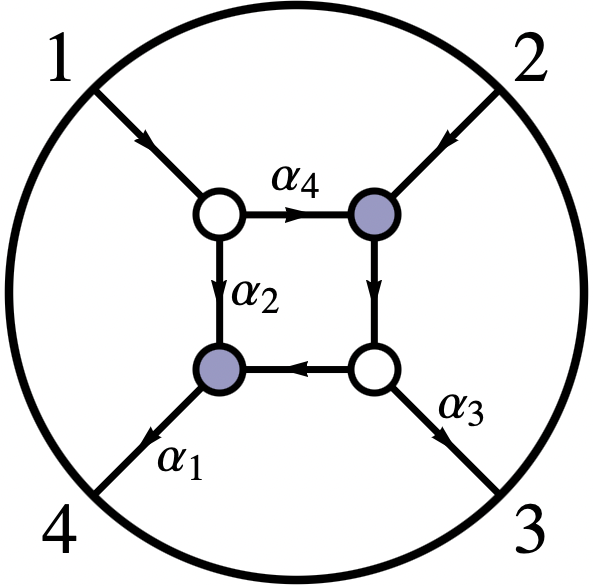}
\caption{On-shell diagram for \(n=4\) and Grassmannian helicity degree \(\kp=2\).}
\label{fig:graph_1}
\end{figure}

Consider the on-shell diagram \(\Gamma\) for \(n=4\) in Figure~\ref{fig:graph_1}.
It is given a perfect orientation, and the non-gauge-fixed edge variables
\(\alpha_1,\dots,\alpha_4\) are labelled. Its associated permutation is
\(\sigma=(3,4,5,6)\). Using the package
\texttt{Positroids.m}~\cite{Bourjaily:2012gy}, one finds helicity degree
\(k'=2\) and boundary measurement matrix
\begin{equation}\label{eq:C_par}
C(\alpha) =  \begin{pmatrix}
	1 & \alpha_2+\alpha_4 & \alpha_2 \alpha_3 & 0 \\
	0 & 1 & \alpha_3 & \alpha_1
\end{pmatrix} \, .
\end{equation}
This \(2\times4\) matrix is obtained from the boundary measurement~\eqref{eq:C_alpha}
by a \({\rm GL}(2)\) transformation and by gauge fixing the variables
corresponding to the unlabelled edges in Figure~\ref{fig:graph_1}.

Let us consider the on-shell function for this diagram. The integral~\eqref{eq:Gr_int_alpha}
is four-dimensional, \(\dim(\Gamma)=2n-4=4\), which coincides with the number
of constraints imposed by the bosonic delta functions. In fact, the delta
function \(\delta^{2\times2}(\lambda \cdot C^\perp)\) fully localizes
\(C=\lambda\). The remaining delta functions are then the supermomentum and
momentum conservation constraints. The differential form in~\eqref{eq:Gr_int_alpha}
can be rewritten in an invariant way as
\begin{equation}\label{eq:Gr24_form}
	\frac{\mathrm{d}\alpha_1 \mathrm{d}\alpha_2 \mathrm{d}\alpha_3 \mathrm{d}\alpha_4}{\alpha_1 \alpha_2 \alpha_3 \alpha_4}
    \quad  \longrightarrow \quad
    \mathbf{\Omega}_\Gamma(C) = \frac{\mathrm{d}^{2 \times 4}C}{{\rm GL}(2)} \frac{1}{(12)(23)(34)(41)}  \, ,
\end{equation}
where the arrow indicates that the first expression equals the pullback of
\(\Omega\) along the parametrization
\begin{equation}
\mathbb{C}^4 \longrightarrow \Gr(2,4),
\qquad
\alpha \longmapsto C(\alpha)
\end{equation}
in~\eqref{eq:C_par}. Here \((ij)\) denotes the determinant of the corresponding
\(2\times2\) minor of \(C\). Thus \(\mathbf{\Omega}_\Gamma\) is a rational differential top-form
on the Grassmannian \(\Gr(2,4)\), defined on the open subset where
\((ij)\neq0\).

The advantage of the invariant perspective is that we can localize \(C=\lambda\)
by dropping the measure from \(\Omega\) and evaluating the remaining factors at
\(\lambda\). The minors become the familiar spinor-helicity brackets:
\begin{equation}
(ij)\big|_{C=\lambda}=\langle ij\rangle \, .
\end{equation}
The reader may check that this result agrees with solving the delta function
\(\delta^{2\times2}(\lambda \cdot C^\perp)\) in the parametrization variables
\(\alpha_1,\dots,\alpha_4\), while carefully including the correct Jacobian
factor. Putting everything together, we find that the on-shell function of the
diagram \(\Gamma\) in Figure~\ref{fig:graph_1}, obtained from the corresponding on-shell form by stripping off the measure, is equal to the tree-level four-point MHV superamplitude:
\begin{equation}\label{eq:onshell_form}
	\Omega_\Gamma
    = \frac{\delta^{(4)}(\lambda \cdot \widetilde{\lambda}) \, \delta^{(8)}(\lambda \cdot \widetilde{\eta})}{\langle 12\rangle \langle 23\rangle \langle 34\rangle \langle 41 \rangle}
    =  A^{{\rm MHV}}_4 \, .
\end{equation}
\end{eg}

In this example the distinction between the spinor-helicity and momentum-twistor conventions is especially simple. The Grassmannian helicity
degree is \(k'=2\), corresponding to \(k=0\) in the usual \(\mathrm{N}^k\mathrm{MHV}\) notation. Passing to momentum twistors solves momentum conservation and factors out the universal MHV, or Parke--Taylor, superamplitude. Thus
\begin{equation}
	A^{\mathrm{MHV}}_4
	=
	\Omega_\Gamma \,,
	\qquad\text{with}\qquad
	\mathcal R_{0,4}=1 \, ,
\end{equation}
and we therefore just computed the only on-shell form contributing to the tree-level four-point MHV amplitude. The momentum-space Grassmannian \(\Gr(k',4)=\Gr(2,4)\) is replaced, after removing the two \(\lambda\)-directions, by the
momentum-twistor Grassmannian \(\Gr(k,4)=\Gr(0,4)\), where the latter is a single point.

This example illustrates on-shell diagrams and their forms and also gives our first encounter with a positive geometry. Ignoring the usual momentum- and
supermomentum-conserving delta functions in~\eqref{eq:onshell_form}, the key
ingredient is the form \(\mathbf{\Omega}_\Gamma\) on \(\Gr(2,4)\) in~\eqref{eq:Gr24_form}.
This is a complex meromorphic form whose poles match the physical poles
\(\langle i\,i{+}1\rangle\) in the Parke--Taylor factor. The remarkable
feature is that \(\mathbf{\Omega}_\Gamma\) is canonically associated with the subset of the
real Grassmannian \(\Gr_{\geq 0}(2,4)\subset \Gr_{\mathbb{R}}(2,4)\)
defined by \(C\in\Gr_{\mathbb{R}}(2,4)\) such that
\begin{equation}\label{eq:nonneg_24}
	(12) \geq 0 \, , \quad
    (13) \geq 0 \, , \quad
    (14) \geq 0\, , \quad
    (23) \geq 0 \, , \quad
    (24) \geq 0 \, , \quad
    (34) \geq 0 \, .
\end{equation}
In the parametrization~\eqref{eq:C_par}, the inequalities~\eqref{eq:nonneg_24}
become \(\alpha_1,\alpha_2,\alpha_3,\alpha_4\geq 0\). Thus the image of the
positive parametrization looks like a four-dimensional positive orthant
\(\mathbb{R}_{\geq 0}^4\), but embedded inside the curved manifold
\(\Gr_{\mathbb{R}}(2,4)\).

The Amplituhedron program is the great extension of this idea: on-shell forms
and Grassmannian contour integrals such as~\eqref{eq:Gr_contour} compute forms
canonically associated with real positive regions in Grassmannians. For the
full tree amplitude and loop integrand, these regions combine into the
\textit{Amplituhedron}. Thus on-shell diagrams form the first bridge from
amplitudes to positive geometry: their combinatorics labels positroid cells,
their forms are logarithmic on these cells, and BCFW recursion assembles them
into amplitudes. The interplay between complex rational forms and real
semialgebraic geometry is the framework of \textit{positive geometries}, which
we introduce next. In Chapter~\ref{ch:Amplituhedra}, this picture is lifted
from the positive Grassmannian to the Amplituhedron.

%% file: Ch3.tex

\section{Positive geometries}\label{ch:Positive Geometries}

The previous chapter introduced the on-shell and Grassmannian language of
scattering amplitudes. We now isolate the mathematical structure that will be used
throughout the rest of the thesis: positive geometries and their canonical forms.
Rather than repeating the general motivation for the positive-geometry program,
already discussed in the introduction, the purpose of this chapter is
to give a self-contained account of the constructions needed later: canonical
forms, triangulations, adjoint hypersurfaces, push-forwards, scattering equations,
and integral representations.

We follow the original semialgebraic definition of positive geometry introduced
in~\cite{Positive_geometries}. For accessible introductions and reviews,
see~\cite{Lam:_PG_notes,Ranestad:what_is_PG,Fevola:Pos_Geom}. The notes
\cite{Lam:_PG_notes} are especially useful as a concise mathematical introduction
to positive geometries and canonical forms,
while~\cite{Ranestad:what_is_PG,Fevola:Pos_Geom} provide broader expository
perspectives and examples. For concrete treatments of polytopes, adjoints, and
plane examples closely related to those appearing in this chapter,
see~\cite{Telen_PG,Polypols,kohn2020projective,bruser2025geometry}. For the
physics-facing side of the subject, including its relation to scattering
amplitudes and the Amplituhedron,
see~\cite{Herrmann:2022nkh,the_amplituhedron,Grassmannian}.

In the semialgebraic formulation, a positive geometry is a triple
\((\mathcal{X},P,\mathbf{\Omega}_P)\), where \(P\) is a real oriented
semialgebraic set inside the real points of a complex variety \(\mathcal{X}\).
The defining property of the canonical top-form \(\mathbf{\Omega}_P\)
is recursive: it has logarithmic singularities along the algebraic boundary of
\(P\), and its residue on each boundary component is the canonical form of that
boundary. This residue property
is the main reason positive geometries are useful in amplitudes. It geometrizes
the same pattern that appears physically as factorization: when a boundary is
approached, the form reduces to the form associated with a lower-dimensional
geometry.

We use the semialgebraic formulation because the real region \(P\), its positivity
properties, and its boundary stratification are essential for the applications in
this thesis. In Chapter~\ref{ch:Amplituhedra}, \(P\) will become an Amplituhedron
region. In Chapter~\ref{ch:Canonical Forms as Dual Volumes}, the same real-region
data will be used to discuss dual volumes and positivity
properties of canonical functions. In Chapters~\ref{ch:From Integrands to Integrals} and~\ref{ch:positivity-and-cluster-structures}, boundary
stratifications will guide the analysis of singularities after loop integration.

A broader Hodge-theoretic formulation of positive geometries, in terms of
genus-zero pairs and canonical maps, was developed in~\cite{Brown:PG_Hodge}. This
perspective does not require a distinguished real semialgebraic region as part of
the initial data. Although we will not use this formulation directly, it provides
an important conceptual extension of the framework and suggests natural points of
contact with logarithmic forms, periods, and the Hodge-theoretic structures
appearing in the study of Feynman integrals.

\medskip

The chapter is structured as follows. Section~\ref{sec:Canonical Forms} gives the
definition of canonical forms and illustrates it with intervals, simplices, and
curved examples. Section~\ref{sec:Triangulations} explains additivity under
triangulations, which is the mathematical analogue of representing the same
amplitude in different on-shell expansions. Section~\ref{sec:Adjoint Hypersurface}
studies the numerator of a canonical form through adjoint hypersurfaces and
residual arrangements. Section~\ref{sec:Push-Forward and Scattering Equations}
introduces push-forwards of canonical forms and relates them to scattering
equations, stringy canonical forms, and moduli-space integrals; for introductory
material on this circle of ideas, see~\cite{Stringy_Can_Forms,Lam:ModuliSpacesPG,CachazoHeYuan2014,ABHY_original}. Finally,
Section~\ref{sec:Integral Representations} discusses integral and dual-volume
representations of canonical functions, preparing the positivity results of
Chapter~\ref{ch:Canonical Forms as Dual Volumes}.

\subsection{Canonical forms}\label{sec:Canonical Forms}

We follow the semialgebraic definition of positive geometry introduced
in~\cite{Positive_geometries}; see also~\cite{Lam:_PG_notes} for a recent review.
We first recall the basic algebraic and semialgebraic notions needed for the
definition. A reader familiar with these notions may move directly to the
definition in~\eqref{eq:PG_triple}. The guiding picture is that \(\mathcal{X}\) is
the complex algebraic space on which the rational form lives, while
\(P\subseteq \mathcal{X}_{\mathbb R}\) is the real semialgebraic set
whose boundary determines the poles and residues of that form.

\subsubsection{Projective and semialgebraic preliminaries}
We denote by \(\mathbb P^n\) the \(n\)-dimensional complex projective space,
\begin{equation}
	\mathbb P^n=(\mathbb C^{n+1}\setminus\{0\})/\sim \, ,
    \qquad
    Y\sim \lambda Y\quad \forall\,\lambda\in\mathbb C^* \, .
\end{equation}
We also consider real projective space, defined in the same way over \(\mathbb
R\), and denoted by \(\mathbb P^n_{\mathbb R}\). We denote points in projective
space by homogeneous coordinates $Y=[Y_0:Y_1:\cdots:Y_n]$, so that
\begin{equation}
	[Y_0:Y_1:\cdots:Y_n]
    =
    [\lambda Y_0:\lambda Y_1:\cdots:\lambda Y_n] \, ,
    \qquad
    \lambda\in\mathbb C^* \, .
\end{equation}
On the affine chart \(Y_0\neq0\), we may fix the rescaling by setting \(Y_0=1\).
The local affine coordinates on this chart are
\begin{equation}
x_i=\frac{Y_i}{Y_0} \, ,\qquad i=1,\dots,n \, .
\end{equation}
The complement of this affine chart is the hyperplane at infinity
\(H_\infty=\{Y_0=0\}\cong\mathbb P^{n-1}\). Thus projective space is obtained by
compactifying affine space by adding \(H_\infty\).

There is are canonical projections over $\mathbb{C}$ and $\mathbb{R}$, respectively,
\begin{equation}
\pi:\mathbb C^{m+1}\setminus\{0\}\longrightarrow\mathbb P^m \, , \qquad \pi_{\mathbb R}:\mathbb R^{m+1}\setminus\{0\}\longrightarrow\mathbb P^m_{\mathbb R} \, .
\end{equation}
A projective variety \(\mathcal{X}\subseteq\mathbb P^m\) is the
projection under \(\pi\) of a cone \(\widehat{\mathcal{X}}\) in \(\mathbb C^{m+1}\)
cut out by finitely many homogeneous polynomials \(f_i\in\mathbb
C[Y_0,\dots,Y_m]\). That is, \(\mathcal{X}=\pi(\widehat{\mathcal{X}}\setminus\{0\})\), where
\begin{equation}
	\widehat{\mathcal{X}}
    =
    \{Y\in\mathbb C^{m+1}: f_1(Y)=\cdots=f_r(Y)=0\} \, .
\end{equation}
If the defining equations can be chosen with real coefficients, we define the real
points \(\mathcal{X}_{\mathbb R}\subseteq\mathbb P^m_{\mathbb R}\) as the
projection of \(\widehat{\mathcal{X}}_{\mathbb R}\setminus\{0\}\) under
\(\pi_{\mathbb R}\), where
\begin{equation}
\widehat{\mathcal{X}}_{\mathbb R}=\widehat{\mathcal{X}}\cap\mathbb R^{m+1} \, .
\end{equation}

A basic semialgebraic cone \(\widehat P\) in \(\mathbb R^{m+1}\) is a
subset defined by homogeneous polynomial equations and inequalities:
\begin{equation}\label{eq:S}
	\widehat P
    =
    \{Y\in\mathbb R^{m+1}:
    f_1(Y)=\cdots=f_r(Y)=0,\,
    g_1(Y)\geq 0,\dots,g_s(Y)\geq 0\} \, ,
\end{equation}
where \(f_i,g_j\in\mathbb R[Y_0,\dots,Y_m]\) are homogeneous polynomials. A
semialgebraic cone is a finite union of basic semialgebraic cones. A semialgebraic
set \(P\) in real projective space \(\mathbb P^m_{\mathbb R}\) is the
image under \(\pi_{\mathbb R}\) of a semialgebraic cone \(\widehat P\)
with the origin removed. We say that a subset of \(\mathbb P^m_{\mathbb R}\) is
very compact if there exists a real hyperplane in \(\mathbb P^m_{\mathbb R}\) not
intersecting it. Equivalently, the subset is contained in a single affine chart.

Let \(\mathcal{X}\subseteq\mathbb P^m\) be a projective variety. There exists a
unique finest decomposition
\begin{equation}
	\mathcal{X}=\mathcal{X}_1\cup\mathcal{X}_2\cup\cdots\cup\mathcal{X}_r
\end{equation}
such that each \(\mathcal{X}_i\) is a non-empty irreducible projective variety and
no \(\mathcal{X}_i\) is contained in \(\mathcal{X}_j\) for \(i\neq j\). This is
called the irreducible decomposition of \(\mathcal{X}\), and \(\mathcal{X}\) is
irreducible if \(r=1\). An irreducible projective variety \(\mathcal{X}\) can have
singular points, for example cusps or self-intersections, but it contains a unique
maximal open smooth complex manifold \(\mathcal{X}_{\rm reg}\subseteq\mathcal{X}\),
which is dense in \(\mathcal{X}\) with respect to the Zariski topology. We define the dimension of \(\mathcal{X}\) to be
the complex dimension of \(\mathcal{X}_{\rm reg}\). The singular locus
\begin{equation}
\mathcal{X}_{\rm sing}:=\mathcal{X}\setminus\mathcal{X}_{\rm reg}
\end{equation}
is itself a projective variety. For us \(\mathcal{X}\) is \emph{normal} if its singular locus has codimension at least two. 

Let \(\mathcal{X}\subseteq\mathbb P^m\) be an irreducible projective variety of
dimension \(d\), defined over \(\mathbb R\), and let \(P \subseteq \mathcal{X}_{\mathbb R}\) be a closed semialgebraic set. We denote by \({\rm int}(P)\) its
interior with respect to the Euclidean topology on \(\mathcal{X}_{\mathbb R}\). We
say that \(P\) is \(d\)-dimensional if \({\rm int}(P) \cap\mathcal{X}_{\rm reg}\)
contains a non-empty open subset of \(\mathcal{X}_{\rm reg}\). We will always
assume that \(P\) is regular: 
\begin{equation}
P=\overline{{\rm int}(P)} \, .
\end{equation}
The topological boundary of \(P\) in \(\mathcal{X}_{\mathbb R}\) is denoted
by \(\partial P\). The algebraic boundary of \(P\) is the Zariski
closure of \(\partial P\):
\begin{equation}
	\partial_a P
    :=
    \overline{\partial P}^{\,\rm Zar}
    \subseteq\mathcal{X} \, .
\end{equation}
This means that \(\partial_a P\) is the smallest projective variety containing \(\partial P\).
If \(\partial P \neq\emptyset\), then its irreducible decomposition
\begin{equation}\label{eq:bP_dec}
	\partial_a P=\mathcal D_1\cup\mathcal D_2\cup\cdots\cup\mathcal D_r
\end{equation}
consists, under the regularity assumptions above, of irreducible
subvarieties \(\mathcal D_i\subseteq\mathcal{X}\) of codimension
one~\cite{sinn2015algebraic}.

\subsubsection{Definition of positive geometry}
We are now ready to define a positive geometry. A positive geometry is a triple
\begin{equation}\label{eq:PG_triple}
	(\mathcal{X},P,\mathbf{\Omega}_P) \, ,
\end{equation}
where:
\begin{itemize}
	\item \(\mathcal{X}\) is a \(d\)-dimensional irreducible normal complex projective variety, defined by homogeneous polynomials with real coefficients;
	\item \(P \subseteq\mathcal{X}_{\mathbb R}\) is a \(d\)-dimensional closed regular semialgebraic subset, equipped with an orientation on \({\rm int}(P)\), and such that \(\partial P \neq\emptyset\);
	\item \(\mathbf{\Omega}_P\) is a rational complex differential \(d\)-form on \(\mathcal{X}\), called the canonical form.
\end{itemize}
The defining property is that \(\mathbf{\Omega}_P\) exists and is unique
with respect to the following recursive conditions:
\begin{enumerate}
	\item If \(d>0\), then \(\mathbf{\Omega}_P\) has only simple poles along the irreducible components \(\mathcal D_1,\dots,\mathcal D_r\) of the algebraic boundary \(\partial_a P\), and is holomorphic elsewhere. Let \(D_i\) be the closure in \(\mathcal D_{i,\mathbb R}\) of the relative interior of \(\mathcal D_i\cap\partial P\). We denote its interior by \(D_{i,>0}\). The orientation on \({\rm int}(P)\) induces an orientation on \(D_{i,>0}\).

    Locally near a smooth point of \(\mathcal D_i\), choose an equation \(f_i=0\) for \(\mathcal D_i\). Since \(\mathbf{\Omega}_P\) has a simple pole on \(\mathcal D_i\), we can write
    \begin{equation}\label{eq:res_def}
    	\mathbf{\Omega}_P
        =
        \frac{\mathrm{d}f_i}{f_i}\wedge\omega_i
        + \text{regular terms},
        \qquad
        {\rm Res}_{\mathcal D_i}\mathbf{\Omega}_P
        :=
        \iota_i^*\omega_i \, ,
    \end{equation}
    where \(\iota_i:\mathcal D_i\hookrightarrow\mathcal{X}\) is the inclusion. The recursive condition requires that
    \begin{equation}
    	(\mathcal D_i,D_i,{\rm Res}_{\mathcal D_i}\mathbf{\Omega}_P)
    \end{equation}
    is a positive geometry for every \(i=1,\dots,r\).
	\item If \(d=0\), then \(\mathcal{X}=P\) is a single point and \(\mathbf{\Omega}_P=\pm1\) specifying the orientation.
\end{enumerate}

\subsubsection{Basic examples}
The simplest positive geometry is therefore a single point. One-dimensional
positive geometries are also quite restricted. If \(d=1\), then normality implies
that \(\mathcal{X}\) is smooth. If \(\mathcal{X}\) had positive genus, it would
admit a nonzero global holomorphic one-form, which could be added to any candidate
canonical form without changing its residues, contradicting uniqueness. Hence
\(\mathcal{X}\simeq\mathbb P^1\). Thus one-dimensional positive geometries are
given by finite unions of closed intervals in \(\mathbb P^1_{\mathbb R}\).

\begin{eg}[\(\mathcal{X}=\mathbb P^1\)]\label{eg:interval}
Let \(P=[a,b]\subseteq\mathbb R=\{[1:x]:x\in\mathbb R\}\subseteq\mathbb
P^1_{\mathbb R}\), oriented from \(a\) to \(b\). Its canonical form, written in
the local affine coordinate \(x\), is
\begin{equation}\label{eq:omega_ab}
	\mathbf\Omega_{[a,b]}(x)=\frac{b-a}{(x-a)(b-x)}\,\mathrm{d}x \, .
\end{equation}
This form has simple poles at the two boundary points \(x=a\) and \(x=b\). At
\(\mathcal D_1=\{x=a\}\), using \(f_1=x-a\), we write
\begin{equation}
\mathbf\Omega_{[a,b]}=\frac{\mathrm{d}f_1}{f_1}\left(\frac{b-a}{b-x}\right) \, .
\end{equation}
Therefore
\begin{equation}
	{\rm Res}_{x=a}\mathbf\Omega_{[a,b]}=+1 \, .
\end{equation}
Similarly, at \(\mathcal D_2=\{x=b\}\), using \(f_2=b-x\), one finds
\begin{equation}
	{\rm Res}_{x=b}\mathbf\Omega_{[a,b]}=-1 \, .
\end{equation}
These signs agree with the induced boundary orientations.
\end{eg}


\begin{figure}[pos=t]
\centering
\begin{tikzpicture}[scale=2.2,line cap=round,line join=round]
\definecolor{mylightgreen}{RGB}{220,245,220}
\definecolor{myred}{RGB}{210,70,70}
\definecolor{myblue}{RGB}{70,155,215}
\fill[mylightgreen] (-1,0) arc[start angle=180,end angle=0,radius=1] -- cycle;
\draw[black, line width=0.9pt] (-1.45,0) -- (1.45,0);
\draw[black, line width=0.9pt] (0,0) circle (1);
\draw[myred, line width=1.6pt] (-1,0) -- (1,0); 
\draw[myblue, line width=1.6pt] (-1,0) arc[start angle=180,end angle=0,radius=1];
\node at (0,0.42) {$P$};
\node at (1.3,0.0) [below] {$\mathcal{D}_1$};
\node at (0.0,-0.83) {$\mathcal{D}_2$};
\node[myred] at (0.5,0.2) {$D_1$};
\node[myblue] at (0,1.20) {$D_2$};
\end{tikzpicture}
\caption{The half-pizza region
\(P\) of Example~\ref{eg:half_pizza}, with ambient
boundary divisors \(\mathcal{D}_1=\{y=0\}\) and
\(\mathcal{D}_2=\{1-x^2-y^2=0\}\), and boundaries 
\(D_1\subseteq\mathcal{D}_1\), \(D_2\subseteq\mathcal{D}_2\).}
\label{fig:half-pizza}
\end{figure}
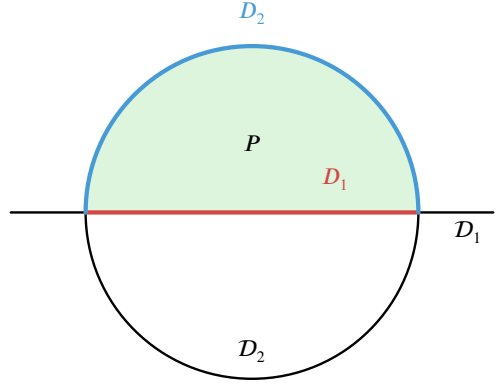

Let us now consider the simplest example of a curved positive geometry.

\begin{eg}[The half-pizza]\label{eg:half_pizza}
Consider the closed semialgebraic set \(P\subseteq\mathbb R^2\) in the
affine chart \([1:x_1:x_2]\) of \(\mathbb P^2\), defined as
\begin{equation}
P=\{ (x_1,x_2) \in \mathbb{R}^2 \,: \, x_2\geq 0,\ f(x_1,x_2):=1-x_1^2-x_2^2\geq 0 \} \, .
\end{equation}
Its interior \({\rm int}(P)\) is defined by the strict inequalities \(x_2>0\) and
\(f(x_1,x_2)>0\). We claim that \(P\) is a positive geometry with canonical form
\begin{equation}\label{eq:pizza_form}
	\mathbf{\Omega}_P(x_1,x_2)=\frac{2\,\mathrm{d}x_1\wedge \mathrm{d}x_2}{x_2\,f(x_1,x_2)} \, .
\end{equation}
There are two boundary divisors,
\begin{equation}
\mathcal D_1=\{x_2=0\},\qquad \mathcal D_2=\{f=0\} \, ,
\end{equation}
containing the boundary geometries \(D_1\) and \(D_2\), which are respectively the
interval \([-1,1]\) and the upper semicircle; see Figure~\ref{fig:half-pizza}.

The residue along \(\mathcal D_1\) is immediate. Since
\begin{equation}
\mathbf{\Omega}_P=\frac{\mathrm{d}x_2}{x_2}\wedge\left(-\frac{2\,\mathrm{d}x_1}{f(x_1,x_2)}\right) \, ,
\end{equation}
we obtain
\begin{equation}
	{\rm Res}_{\mathcal D_1}\mathbf{\Omega}_P
    =
    -\frac{2\,\mathrm{d}x_1}{f(x_1,0)}
    =
    -\frac{2\,\mathrm{d}x_1}{(1+x_1)(1-x_1)} \, .
\end{equation}
This is the canonical form of the interval \(D_1=[-1,1]\) with the boundary
orientation induced by the residue convention~\eqref{eq:res_def}.

The residue along \(\mathcal D_2\) is slightly more interesting. Since
\begin{equation}
\mathrm{d}f=-2x_1\,\mathrm{d}x_1-2x_2\,\mathrm{d}x_2,\qquad \mathrm{d}f\wedge \mathrm{d}x_1=2x_2\,\mathrm{d}x_1\wedge \mathrm{d}x_2 \, ,
\end{equation}
we can rewrite
\begin{equation}
	\mathbf{\Omega}_P=\frac{\mathrm{d}f}{f}\wedge\frac{\mathrm{d}x_1}{x_2^2} \, .
\end{equation}
Therefore
\begin{equation}
	{\rm Res}_{\mathcal D_2}\mathbf{\Omega}_P=\iota_2^*\frac{\mathrm{d}x_1}{x_2^2} \, ,
\end{equation}
where \(\iota_2:\mathcal D_2\hookrightarrow\mathbb P^2\) is the inclusion. Using
\(x_1\) as a coordinate on the conic is misleading globally, because \(x_1\) is not a
global coordinate on \(\mathcal D_2\). Instead, parametrize the conic by
\begin{equation}\label{eq:iota2}
	\iota_2:\mathbb P^1\longrightarrow\mathcal D_2,
    \qquad
    t\longmapsto
    \left(\frac{1-t^2}{1+t^2},\frac{2t}{1+t^2}\right) \, .
\end{equation}
On the upper semicircle, \(t\in[0,\infty]\). A direct computation gives
\begin{equation}
	\iota_2^*\frac{\mathrm{d}x_1}{x_2^2}=-\frac{\mathrm{d}t}{t} \, .
\end{equation}
This is the canonical form of the interval \([0,\infty]\subseteq\mathbb P^1\) with
the orientation induced by the residue convention~\eqref{eq:res_def}. Hence the
residue on \(\mathcal D_2\) is the canonical form of the boundary geometry
\(D_2\), and \(P\) is a positive geometry with canonical
form~\eqref{eq:pizza_form}.
\end{eg}

\subsubsection{Projective simplices}
The simplest class of semialgebraic sets in higher dimensions is given by
projective polytopes. A projective polytope \(P \subseteq\mathbb
P^m_{\mathbb R}\) is the image under the canonical projection of a polyhedral cone
\(\widehat P \subseteq\mathbb R^{m+1}\), namely a finite union of basic
semialgebraic cones as in~\eqref{eq:S}, with no equations and with all
inequalities given by linear forms. By triangulating, every projective polytope is a positive geometry. We first verify this directly for simplices.

\begin{eg}[Projective simplices]\label{eg:simplexes}
Let \(\Delta^m_{\geq 0}\subseteq\mathbb P^m_{\mathbb R}\) be the projective simplex
\begin{equation}
	\Delta^m_{\geq 0}
    =
    \left\{[Y_0:\cdots:Y_m]\in\mathbb P^m_{\mathbb R} \, : \,  Y_i\geq 0\text{ for all }i\right\} \, .
\end{equation}
Its interior \(\Delta^m_{>0}\) is the region where all \(Y_i\) can be chosen
strictly positive. In the affine chart \(Y_0=1\), with local coordinates
\(x_i=Y_i/Y_0\), this becomes
\begin{equation}
x_i\geq 0,
\qquad i=1,\dots,m \, ,
\end{equation}
and its interior is given by \(x_i>0\).

We claim that \(\Delta^m_{\geq 0}\) is a positive geometry in \(\mathbb P^m\) with
canonical form
\begin{equation}\label{eq:form_simplex_chart}
	\mathbf\Omega_{\Delta^m_{\geq 0}}(x)
    =
    \frac{1}{x_1x_2\cdots x_m}\,\mathrm{d}x_1\wedge \mathrm{d}x_2\wedge\cdots\wedge \mathrm{d}x_m
\end{equation}
in the chart \(Y_0=1\). The simplex has \(m+1\) facets. In this chart, the facet
\(Y_0=0\) lies at infinity, while the remaining facets are \(x_i=0\). The
projective semialgebraic set \(\Delta^m_{\geq 0}\) is invariant under permutations
of the \(m+1\) homogeneous coordinates, so it is enough to check the recursive
property on one affine facet and then use symmetry.

For \(m=0\), the simplex is a point and its canonical form is \(+1\). For \(m>0\),
we compute the residue along \(x_m=0\). With the convention~\eqref{eq:res_def},
this gives the canonical form of the corresponding boundary simplex:\begin{equation}
	{\rm Res}_{x_m=0}\mathbf\Omega_{\Delta^m_{\geq 0}}
    =
    (-1)^{m-1}
    \frac{1}{x_1\cdots x_{m-1}}\,
    \mathrm{d}x_1\wedge\cdots\wedge \mathrm{d}x_{m-1} \, .
\end{equation}
By induction, this is the canonical form of the boundary simplex
\(\Delta^{m-1}_{\geq 0}\) with the induced orientation. By symmetry among the
homogeneous coordinates, the same statement holds for every facet. Hence
\(\Delta^m_{\geq 0}\) is a positive geometry.

It is now convenient to introduce a projectively invariant notation. We write
\(Y=[Y_0:Y_1:\cdots:Y_m]\), and denote by \(E_i\) the \(i\)-th standard basis
vector in \(\mathbb R^{m+1}\). For vectors in \(\mathbb C^{m+1}\), let
\(\langle\cdots\rangle\) denote the determinant of the \((m+1)\times(m+1)\) matrix
obtained by stacking the vectors in the brackets. The standard projective top-form
is
\begin{equation}\label{eq:proj_meas}
	\langle Y\,\mathrm{d}^mY\rangle
    =
    \sum_{i=0}^m
    (-1)^i
    Y_i\,
    \mathrm{d}Y_0\wedge\cdots\wedge\widehat{\mathrm{d}Y_i}\wedge\cdots\wedge \mathrm{d}Y_m \, .
\end{equation}
Then the canonical form of the standard simplex can be written as
\begin{equation}\label{eq:form_simplex}
	\mathbf\Omega_{\Delta^m_{\geq 0}}(Y)
    =
    \frac{\langle E_0E_1\cdots E_m\rangle^m\langle Y\,\mathrm{d}^mY\rangle}
    {\langle Y E_1\cdots E_m\rangle
    \langle Y E_0E_2\cdots E_m\rangle
    \cdots
    \langle Y E_0\cdots E_{m-1}\rangle} \, .
\end{equation}
The power \(m\) in the numerator ensures invariance under independent rescalings
of the vertices \(E_i\). The expression is also invariant under rescaling of
\(Y\), and hence descends to a well-defined rational top-form on projective space.
Since it agrees with~\eqref{eq:form_simplex_chart} in the chart \(Y_0=1\), it is
the canonical form of \(\Delta^m_{\geq 0}\).

We can now deduce the canonical form of any projective simplex. Let \(Z\in{\rm
GL}(m+1)\), and let \(Z_i\in\mathbb R^{m+1}\) denote its columns. The image
\(Z(\Delta^m_{\geq 0})\) is the simplex with vertices \(\pi_\mathbb{R}(Z_0),\dots,\pi_\mathbb{R}(Z_m)\). Its
canonical form is obtained by pushing forward \(\mathbf\Omega_{\Delta^m_{\geq 0}}\)
along the projective linear transformation defined by \(Z\):
\begin{tcolorbox}[definitionbox]
\textbf{Canonical form of a projective simplex.}
\begin{equation}\label{eq:can_form_simplex}
	\mathbf\Omega_{Z(\Delta^m_{\geq 0})}(Y)
    =
    Z_*\mathbf\Omega_{\Delta^m_{\geq 0}}(Y)
    =
    \frac{\langle Z_0Z_1\cdots Z_m\rangle^m\langle Y\,\mathrm{d}^mY\rangle}
    {\langle YZ_1\cdots Z_m\rangle
    \langle YZ_0Z_2\cdots Z_m\rangle
    \cdots
    \langle YZ_0\cdots Z_{m-1}\rangle} \, .
\end{equation}
\end{tcolorbox}\noindent
\end{eg}

These examples illustrate the defining features of canonical forms: logarithmic
poles on the algebraic boundary, residues given by canonical forms of
lower-dimensional boundary geometries, and projective covariance. In the next
section we explain how more complicated positive geometries can be built by adding
such forms through triangulations.

\subsection{Triangulations}\label{sec:Triangulations}

A common strategy is to understand a complicated object from simpler constituents.
For positive geometries, this is the idea of triangulations or tilings:
decomposing a semialgebraic region into smaller pieces, each of which is itself a
positive geometry. Canonical forms satisfy an additivity property which often
allows one to compute the canonical form of the whole geometry by adding the
canonical forms of the tiles. We now make this idea precise.

\subsubsection{Pseudo-positive geometries and additivity}

The definition of positive geometry requires nontrivial boundary strata all the
way down to dimension zero. For example, the closed disk
\begin{equation}\label{eq:disk}
P=\{(x_1,x_2)\in\mathbb R^2:f(x_1,x_2)=1-x_1^2-x_2^2\geq 0\}\subseteq\mathbb P^2_{\mathbb R}
\end{equation}
is not a positive geometry in this sense: its algebraic boundary is the circle,
but the circle itself has empty boundary. It is therefore convenient, when
discussing triangulations, to enlarge the class of objects to pseudo-positive
geometries (PPG), defined as positive geometries except that we allow empty
algebraic boundary, in which case the canonical form is declared to be zero.
Similarly, we allow the empty set in any projective variety to be a
pseudo-positive geometry with \(\mathbf\Omega_\emptyset=0\).

We can now say that the two half-pizzas
\begin{equation}\label{eq:cancel}
P^{+}=\{f(x_1,x_2)\geq 0,\ x_2\geq 0\},\qquad
P^{-}=\{f(x_1,x_2)\geq 0,\ -x_2\geq 0\} \, ,
\end{equation}
see Example~\ref{eg:half_pizza}, triangulate the pseudo-positive geometry given by
the disk~\eqref{eq:disk}. This is true at the level of closed semialgebraic sets.
At the level of canonical forms, one expects the additivity relation
\begin{equation}\label{eq:tr_pizzas}
	\mathbf\Omega_{P^{+}}+\mathbf\Omega_{P^{-}}=\mathbf{\Omega}_P=0 \, .
\end{equation}
With the orientations chosen so that the induced orientations on the internal
segment
\begin{equation}
D_1=[-1,1]\subseteq\{x_2=0\}
\end{equation}
are opposite, the residues along the internal boundary cancel and~\eqref{eq:cancel} holds true.

This example also illustrates that positive geometries are not closed under
arbitrary unions. If the orientations of the two half-pizzas are chosen so that
the internal residues do not cancel, the union carries an internal boundary and
the resulting form is weighted. The framework of weighted positive geometries,
introduced in~\cite{Dian_2023}, enlarges the category so that such unions are
allowed and canonical forms satisfy a natural additivity property. We will not use
this formalism here, but it is relevant in loop-level examples, where internal
boundaries can appear. We refer the reader to~\cite{Dian_2023} for this topic, and
point out the topological flavor of the definition of weighted positive
geometries, similar to that of canonical maps in~\cite{Brown:PG_Hodge}.

\subsubsection{Signed triangulations}
We now return to the goal of this section and define a special class of
triangulations called signed triangulations. Let \(\mathcal{X}\) be a
\(d\)-dimensional irreducible normal complex projective variety, and let
\(P\) and \(P_\sigma\), for \(\sigma\) in a finite index set, be closed
oriented semialgebraic subsets of \(\mathcal{X}_{\mathbb R}\). We say that the
\(P_\sigma\) form a \textit{signed triangulation of} \(P\) if
\begin{equation}
\bigcup_\sigma {\rm int}(P_\sigma)
\end{equation}
covers a dense subset of \(P\), and at every point in this union the signed sum of the
local orientations of the pieces containing the point is equal to the orientation
of \({\rm int}(P)\) if the point lies in ${\rm int}(P)$ and to zero otherwise.

\begin{tcolorbox}[resultbox]
\textbf{Canonical-form triangulation.}\label{prop:can_form_tr}
	If the collection \(\{P_\sigma\}\) forms a signed triangulation of \(P\), and \(P\) and all \(P_\sigma\) are pseudo-positive geometries, then
	\begin{equation}\label{eq:can_form_tr}
		\mathbf{\Omega}_P=\sum_a\mathbf\Omega_{P_\sigma} \, .
	\end{equation}
	In this case we also say that \(\{P_\sigma\}\) canonical-form triangulates \(P\).
\end{tcolorbox}\noindent

We call a signed triangulation \(P_\sigma\) of $P$ \emph{unternal}, if \(P_\sigma \subseteq P\) for every $\sigma$, and external otherwise. In~\cite{Positive_geometries}, the authors also define \textit{boundary triangulations} in a recursive way, adapted to the recursive definition of positive geometry, and show that boundary triangulations imply the same additivity of canonical forms without the assumption on $P$ being a pseudo-positive geometry. Note that the absence of internal boundaries and weighted forms in Proposition~\ref{prop:can_form_tr} follows from the assumption that \(P\) is a pseudo-positive geometry.

\subsubsection{Projective polytopes}
The main application of the signed-triangulation property~\eqref{eq:can_form_tr} for us is heuristic:
given a semialgebraic set \(P\) which we do not yet know to be a
pseudo-positive geometry, but which we can triangulate by pseudo-positive
geometries, we can consider the candidate form obtained from the right-hand side
of~\eqref{eq:can_form_tr}. If \(P\) is indeed a pseudo-positive geometry,
then this form must be its canonical form.


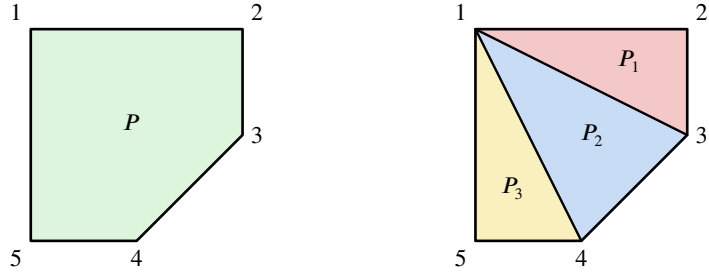
\begin{figure}[pos=t]
\centering
\begin{tikzpicture}[scale=1.4,line cap=round,line join=round]
\definecolor{mylightgreen}{RGB}{220,245,220}
\definecolor{myred}{RGB}{245,200,200}
\definecolor{myblue}{RGB}{200,220,245}
\definecolor{myyellow}{RGB}{250,240,190}
\tikzset{boundary/.style={draw=black, line width=1pt},diag/.style={draw=black, line width=0.9pt},vlabel/.style={font=\small},plabel/.style={font=\small}}
\begin{scope}[shift={(0,0)}]
  \coordinate (A1) at (-1,1);\coordinate (A2) at (1,1);\coordinate (A3) at (1,0);\coordinate (A4) at (0,-1);\coordinate (A5) at (-1,-1);
  \fill[mylightgreen] (A1)--(A2)--(A3)--(A4)--(A5)--cycle;
  \draw[boundary] (A1)--(A2)--(A3)--(A4)--(A5)--cycle;
  \node[vlabel, above left] at (A1) {$1$};\node[vlabel, above right] at (A2) {$2$};\node[vlabel, right] at (A3) {$3$};\node[vlabel, below] at (A4) {$4$};\node[vlabel, below left] at (A5) {$5$};
  \node at (-0.05,0.12) {$P$};
\end{scope}
\begin{scope}[shift={(4.2,0)}]
  \coordinate (B1) at (-1,1);\coordinate (B2) at (1,1);\coordinate (B3) at (1,0);\coordinate (B4) at (0,-1);\coordinate (B5) at (-1,-1);
  \fill[myred] (B1)--(B2)--(B3)--cycle;\fill[myblue] (B1)--(B3)--(B4)--cycle;\fill[myyellow] (B1)--(B4)--(B5)--cycle;
  \draw[boundary] (B1)--(B2)--(B3)--(B4)--(B5)--cycle;
  \draw[diag] (B1)--(B3);\draw[diag] (B1)--(B4);
  \node[vlabel, above left] at (B1) {$1$};\node[vlabel, above right] at (B2) {$2$};\node[vlabel, right] at (B3) {$3$};\node[vlabel, below] at (B4) {$4$};\node[vlabel, below left] at (B5) {$5$};
  \node[plabel] at (0.45,0.70) {$P_1$};\node[plabel] at (0.10,0.00) {$P_2$};\node[plabel] at (-0.65,-0.5) {$P_3$};
\end{scope}
\end{tikzpicture}
\caption{The pentagon \(P\) of Example~\ref{eq:pent_tr} with its fan triangulation on the right.}
\label{fig:pent_tr}
\end{figure}

\begin{eg}[Triangulation of a pentagon]\label{eq:pent_tr}
Consider the pentagon \(P\subseteq\mathbb R^2\) given by the convex hull
of the five clockwise-ordered points
\begin{equation}
(-1,1),\quad (1,1),\quad (1,0),\quad (0,-1),\quad (-1,-1) \, ,
\end{equation}
labelled \(1,\dots,5\), respectively; see Figure~\ref{fig:pent_tr}. Equivalently,
\(P\) is cut out by
\begin{equation}
x_1+1\geq 0,\quad x_2+1\geq 0,\quad 1-x_1+x_2\geq 0,\quad 1-x_1\geq 0,\quad 1-x_2\geq 0 \, .
\end{equation}
We triangulate \(P\) into the three triangles
\begin{equation}
[1,2,3],\qquad [1,3,4],\qquad [1,4,5] \, ,
\end{equation}
as in Figure~\ref{fig:pent_tr}. The canonical form of each triangle \([a,b,c]\) is
computed using~\eqref{eq:can_form_simplex}, in the affine chart \(Y=[1:x_1:x_2]\).
Stripping off the common measure \(\mathrm{d}x_1\wedge \mathrm{d}x_2\), one obtains
\begin{equation*}
\scalebox{0.8}{$
\begin{aligned}
&
\begin{array}[t]{@{}c@{\;\;+\;\;}c@{\;\;+\;\;}c@{}}
\ [1,2,3] & [1,3,4] & [1,4,5] \\[2mm]
\displaystyle
\frac{2}{(1-x_1)(1-x_2)(x_1+2x_2-1)}
&
\displaystyle
\frac{9}{(-x_1+x_2+1)(-x_1-2x_2+1)(2x_1+x_2+1)}
&
\displaystyle
\frac{2}{(x_1+1)(x_2+1)(-2x_1-x_2-1)}
\end{array}
\end{aligned}
$}
\end{equation*}
The poles on the internal diagonals cancel in the sum, leaving poles only on the
five boundary lines of the pentagon. This yields
\begin{equation}\label{eq:can_form_pent}
    \mathbf{\Omega}_P
    =
    \frac{x_1x_2+3x_1-3x_2-5}{(x_1+1)(x_2+1)(1-x_1+x_2)(1-x_1)(1-x_2)}\,\mathrm{d}x_1\wedge \mathrm{d}x_2 \, .
\end{equation}
To check that this is the correct canonical form, one verifies that each of the
five residues gives the canonical form of the corresponding edge. For example, on
the edge \(x_2=-1\), corresponding to the segment from \((-1,-1)\) to \((0,-1)\),
one obtains
\begin{equation}
	\operatorname*{Res}_{x_2=-1}\mathbf{\Omega}_P=\frac{-\mathrm{d}x_1}{x_1(x_1+1)} \, ,
\end{equation}
which correctly reproduces the form of the segment, with the induced orientation.
\end{eg}

In a similar spirit, we use an interior-triangulation argument to prove that projective
polytopes are positive geometries. A projective polytope
\(P \subseteq\mathbb P^m_{\mathbb R}\) is a semialgebraic subset for which
there exists an affine chart \(\mathbb R^m\subseteq\mathbb P^m_{\mathbb R}\) such
that the image of \(P\) in this chart is the convex hull of finitely many
points \(Z_i\in\mathbb R^m\). Equivalently, a projective polytope is the
projection in \(\mathbb P^m_{\mathbb R}\) of a strictly convex polyhedral cone in
\(\mathbb R^{m+1}\) under the canonical projection \(\pi_{\mathbb{R}} \colon \mathbb
R^{m+1}\setminus\{0\}\to\mathbb P^m_{\mathbb R}\).

The following standard result is a useful consequence of triangulation.
\begin{tcolorbox}[resultbox]
\textbf{Projective polytopes.}\label{prop:pol_tr}
Let \(P \subseteq\mathbb P^m_{\mathbb R}\) be an oriented projective polytope and let \(\{P_\sigma\}\) be a triangulation of \(P\) into simplices, all oriented by the induced orientation from \(P\). Then \(P\) is a positive geometry with canonical form given by~\eqref{eq:can_form_tr}. In particular, every projective polytope is a positive geometry.
\end{tcolorbox}

\noindent\textit{Sketch of proof.} Recall that by Example~\ref{eg:simplexes} every projective simplex is a positive geometry. The oriented triangulation \(\{P_\sigma\}\) gives a signed triangulation of \(P\). We consider the sum of the simplex canonical forms as in~\eqref{eq:can_form_tr} and argue that it satisfies the recursive residue property, by induction on the dimension \(m\). For \(m=0\), there is nothing to prove.

Let \(m>0\). Pick any facet \(D\) of \(P\), supported on a hyperplane
\(\mathcal D\subseteq\mathbb P^m\). The only forms among
\(\mathbf\Omega_{P_\sigma}\) contributing to \({\rm Res}_{\mathcal
D}\mathbf{\Omega}_P\) are those for which \(P_\sigma\cap D\)
contains an \((m-1)\)-dimensional simplex. Denote the set of such indices by
\(\Sigma)\). By standard properties of triangulations of polytopes, the collection 
\(\{P_\sigma\cap D\}_{\sigma \in \Sigma}\) is a triangulation of the facet
\(D\subseteq\mathcal D\cong\mathbb P^{m-1}\) into projective simplices. Therefore
\begin{equation}
	{\rm Res}_{\mathcal D}\mathbf{\Omega}_P
    =
    \sum_{\sigma \, \in \,  \Sigma}{\rm Res}_{\mathcal D}\mathbf\Omega_{P_\sigma}
    =
    \sum_{\sigma \, \in \,  \Sigma}\mathbf\Omega_{P_\sigma\cap D}
    =
    \mathbf\Omega_D \, .
\end{equation}
In the second equality we used that projective simplices are positive geometries,
and in the last equality the induction hypothesis, as \(D\) is an
\((m-1)\)-dimensional projective polytope.

\subsubsection{A nonlinear example}

\begin{figure}[pos=t]
\begin{minipage}[c]{0.32\linewidth}
\includegraphics[width=\linewidth]{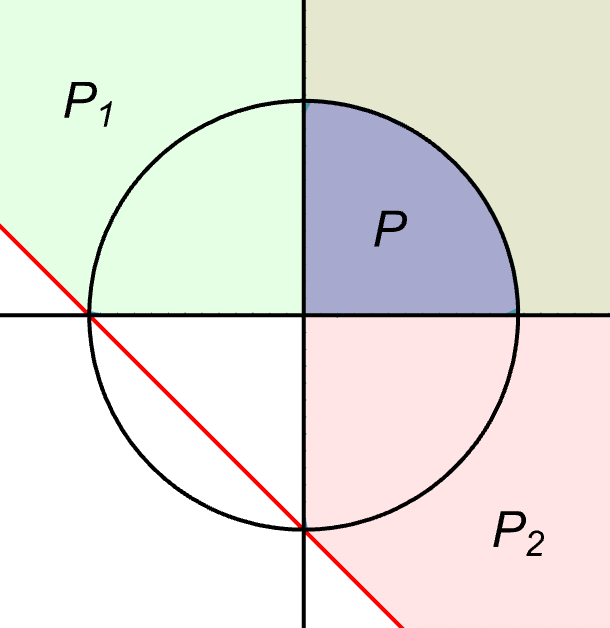}
\end{minipage}\hfill
\begin{minipage}[c]{0.327\linewidth}
\includegraphics[width=\linewidth]{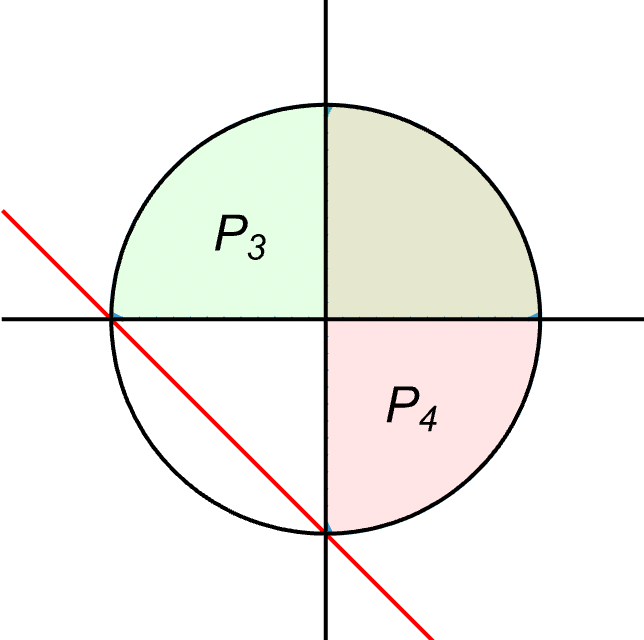}
\end{minipage}\hfill
\begin{minipage}[c]{0.325\linewidth}
\includegraphics[width=\linewidth]{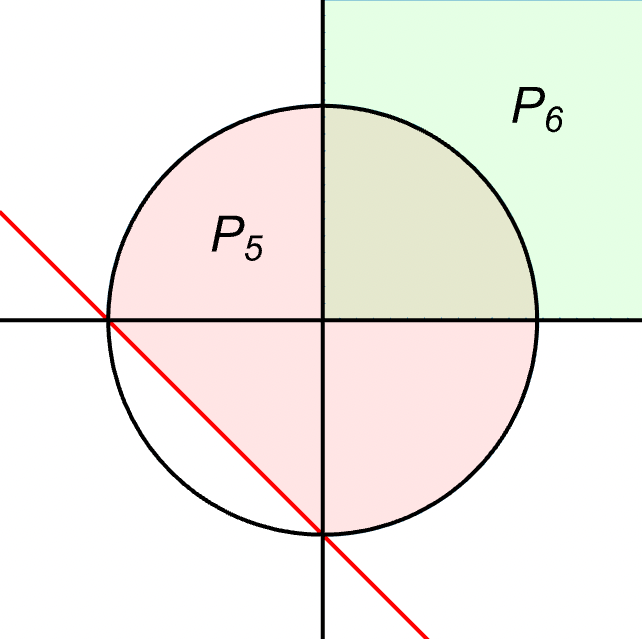}
\end{minipage}
\caption{The signed triangulation $P_\sigma$ in~\eqref{eq:six_cones} for the pizza slice \(P\) in~\eqref{eq:pizza_slice}. The red line corresponds to the numerator of the canonical form of \(P\) in~\eqref{eq:pizza_slice_form}.}
\label{fig:pizza_tr}
\end{figure}

\begin{eg}[External triangulation of a pizza slice]\label{eg:pizza_slice}
In Example~\ref{eg:half_pizza} we considered the half-pizza positive geometry. We
now consider a slice of a pizza. Let \(P\subseteq\mathbb R^2\), in affine
coordinates \([1:x_1:x_2]\) on \(\mathbb P^2\), be
\begin{equation}\label{eq:pizza_slice}
	P=\{x_1\geq 0,\ x_2\geq 0,\ 1-x_1^2-x_2^2\geq 0\} \, .
\end{equation}
Its interior \({\rm int}(P)\) is defined by \(x_1>0\), \(x_2>0\), and \(1-x_1^2-x_2^2>0\). We
claim that \(P\) is a positive geometry and determine its form by
triangulation.

Consider the following seven closed semialgebraic sets in \(\mathbb R^2\):
\begin{equation}\label{eq:six_cones}
\begin{aligned}
P_1 &= \{x_1+x_2+1\geq 0,\ x_1\geq 0\}, &
P_2 &= \{x_1+x_2+1\geq 0,\ x_2\geq 0\} \, , \\
P_3 &= \{1-x_1^2-x_2^2\geq 0,\ x_1\geq 0\}, &
P_4 &= \{1-x_1^2-x_2^2\geq 0,\ x_2\geq 0\} \, , \\
P_5 &= \{x_1+x_2+1\geq 0,\ 1-x_1^2-x_2^2\geq 0\}, &
P_6 &= \{x_1\geq 0,\ x_2\geq 0\} \, , \\
P_7 &= \{x_1+x_2+1\geq 0\}.
\end{aligned}
\end{equation}
The first six regions are either triangles in projective space, with the line at
infinity included as a boundary component, or half-pizzas, and hence are positive
geometries; see Figure~\ref{fig:pizza_tr}. For example, \(P_6\) is the
standard two-dimensional simplex, with boundaries \(x_1=0\), \(x_2=0\), and the line
at infinity. The last region \(P_7\) is a pseudo-positive geometry
with vanishing canonical form.
We orient \((P_1,\dots,P_7)\) with signs
\begin{equation}
(-,-,+,+,-,+,+) \, ,
\end{equation}
where \(+\) indicates the orientation induced from the standard form \(\mathrm{d}x_1\wedge
\mathrm{d}x_2\) on \(\mathbb R^2\). Then \(\{P_\sigma\}_{\sigma=1,\dots,7}\) forms an
external triangulation of \(P\). Hence, if
\(P\) is a positive geometry, then its canonical form is equal to the corresponding signed sum. Stripping off the common measure \(\mathrm{d}x_1\wedge \mathrm{d}x_2\), we get
\begin{equation}\label{eq:ext_tr_pizza}
\resizebox{0.65\textwidth}{!}{$
\begin{aligned}
\Omega_{P}
&=
-\frac{1}{x_1(x_1+x_2+1)}
-\frac{1}{x_2(x_1+x_2+1)}
+\frac{2}{x_1(1-x_1^2-x_2^2)}
+\frac{2}{x_2(1-x_1^2-x_2^2)}
\\
&\hspace{1.2cm}
-\frac{2}{(x_1+x_2+1)(1-x_1^2-x_2^2)}
+\frac{1}{x_1x_2}
+0
=
\frac{x_1+x_2+1}{x_1x_2(1-x_1^2-x_2^2)}.
\end{aligned}
$}
\end{equation}
Thus the candidate canonical form is
\begin{equation}\label{eq:pizza_slice_form}
\mathbf{\Omega}_P=\frac{x_1+x_2+1}{x_1x_2(1-x_1^2-x_2^2)}\,\mathrm{d}x_1\wedge \mathrm{d}x_2 \, .
\end{equation}
To show that \(P\) is actually a positive geometry, one checks that the
residues are the canonical forms of the boundary geometries, using the same
techniques as in Example~\ref{eg:half_pizza}.
\end{eg}

Let us look at~\eqref{eq:ext_tr_pizza}. We computed the form
\(\mathbf{\Omega}_P\) by breaking up the geometry into seven pieces, but
the resulting expression is quite simple. In the denominator there are three
polynomial factors, corresponding to the boundary components of \(P\),
while the numerator is linear. The vanishing of the numerator defines a line,
depicted in red in Figure~\ref{fig:pizza_tr}. This line passes through two special
points, \((-1,0)\) and \((0,-1)\), arising as intersections of boundary components
but lying outside the positive region \(P\). These two points are residual
intersections of the boundary divisors: they arise from intersections of boundary
components but do not lie on the positive region. These ``bad'' points uniquely
fix the red line, and the aim of the next section is to explain this observation.

\subsection{Adjoint hypersurface}\label{sec:Adjoint Hypersurface}

\subsubsection{Canonical numerators}
Let \(\mathcal{X}=\mathbb P^m\) and let \((\mathcal{X},P,\mathbf{\Omega}_P)\) be a positive geometry. In projective
coordinates \(Y=[Y_0:\cdots:Y_m]\), the canonical form has the structure
\begin{tcolorbox}[definitionbox]
\textbf{Canonical numerator.}
\begin{equation}\label{eq:form_pq}
	\mathbf{\Omega}_P(Y)=\frac{q(Y)}{p(Y)}\,\langle Y\,\mathrm{d}^mY\rangle \, ,
\end{equation}
\end{tcolorbox}\noindent
where \(q,p\in\mathbb C[Y_0,\dots,Y_m]\) are homogeneous polynomials, and
\(\langle Y\,\mathrm{d}^mY\rangle\) is the standard projective measure on \(\mathbb P^m\),
see~\eqref{eq:proj_meas}. We call \(q(Y)/p(Y)\) the canonical function of
\(P\).

Since \(\mathbf{\Omega}_P\) is well-defined on projective space, it is
invariant under rescaling \(Y\mapsto\lambda Y\), with \(\lambda\in\mathbb C^*\).
The measure \(\langle Y\,\mathrm{d}^mY\rangle\) has projective weight \(m+1\), and
therefore the degrees of \(q\) and \(p\) must satisfy
\begin{tcolorbox}[resultbox]
\textbf{Degree constraint.}
\begin{equation}\label{eq:deg_q_p}
	\deg(q)+m+1=\deg(p) \, .
\end{equation}
\end{tcolorbox}\noindent
Moreover, \(\mathbf{\Omega}_P\) has simple poles only along the algebraic
boundary \(\partial_a P =\bigcup_i\mathcal D_i\). Taking \(p\) to be the
reduced defining equation of the algebraic boundary, we may write
\begin{equation}
p(Y)=\prod_i p_i(Y) \, ,
\end{equation}
where \(p_i\) are homogeneous irreducible polynomials cutting out the irreducible
components \(\mathcal D_i\). We therefore obtain
\begin{equation}\label{eq:adj_deg}
	\deg(q)=\sum_i\deg(p_i)-m-1 \, .
\end{equation}

Since \(p\) encodes the algebraic boundary of \(P\), it is natural to ask
whether \(q\) is also determined by geometric data. Let us explore this question
with an example.

\begin{eg}[Numerator of the pizza slice]
Consider the pizza-slice positive geometry $P$ of Example~\ref{eg:pizza_slice}. According to~\eqref{eq:form_pq}, in the
affine chart \([1:x_1:x_2]\) we have
\begin{equation}
p(x_1,x_2)=x_1x_2(1-x_1^2-x_2^2) \, .
\end{equation}
The degree count in~\eqref{eq:adj_deg} gives $\deg(q)=1$,
so the numerator is linear, as found by triangulation in~\eqref{eq:ext_tr_pizza}.
Suppose we did not know this linear polynomial, and let us determine it
geometrically.

The three boundary divisors are
\begin{equation}
\mathcal D_1=\{1-x_1^2-x_2^2=0\},\qquad \mathcal D_2=\{x_2=0\},\qquad \mathcal D_3=\{x_1=0\} \, .
\end{equation}
Consider the boundary divisor \(\mathcal D_2=\{x_2=0\}\). The residue of
\(\mathbf{\Omega}_P\) along \(\mathcal D_2\) should be the canonical form
of the segment \(D_2= P \cap \mathcal D_{2,\mathbb R}=[0,1]\times\{0\}\).
However, after restricting the denominator \(p(x_1,x_2)=x_1x_2(1-x_1^2-x_2^2)\) to \(x_2=0\),
one finds possible poles at \(x_1=0,\pm1\). The points \(x_1=0\) and \(x_1=1\) are
genuine boundary points of the segment \(D_2\), while \(x_1=-1\) is residual: it
belongs to the intersection \(\mathcal D_1\cap\mathcal D_2=\{(1,0),(-1,0)\}\), but
does not lie on the boundary of \(P \). Therefore the numerator must vanish
at \((-1,0)\) in order to cancel this unwanted pole in the residue.

Similarly, considering the boundary divisor \(\mathcal D_3=\{x_1=0\}\), the residue
should be the canonical form of the segment \(D_3=\{0\}\times[0,1]\). Restricting
the denominator to \(x_1=0\) produces possible poles at \(x_2=0,\pm1\). The point
\(x_2=-1\) is residual, so the numerator must also vanish at \((0,-1)\). Since \(q\)
is linear, its vanishing locus is a line. There is a unique line through the two
residual points \((-1,0)\) and \((0,-1)\), namely
\begin{equation}
q(x_1,x_2)=x_1+x_2+1 \, .
\end{equation}
This is the red line in Figure~\ref{fig:pizza_tr}; the result matches~\eqref{eq:ext_tr_pizza}.
\end{eg}

In the argument above, we explored lower-dimensional intersections of the
algebraic boundary and determined which of them lie on the actual semialgebraic
boundary and which do not. The latter give unwanted poles in residues of the
canonical form. These unwanted poles must be compensated by zeros of the numerator
\(q\). If there are enough such vanishing, or interpolation, conditions, they can
uniquely determine \(q\), and hence the canonical form~\eqref{eq:form_pq}. Solving
the interpolation conditions is a linear algebra problem: the unknowns are the
coefficients of a homogeneous polynomial \(q\) of fixed degree~\eqref{eq:adj_deg}.

\subsubsection{Residual arrangements}
To formalize this procedure, we first stratify the algebraic boundary
\(\partial_a P \). We take the \emph{strata} to be the irreducible components of
all non-empty intersections of boundary divisors, grouped by codimension. Thus the
codimension-one strata are the irreducible components \(\mathcal D_i=\mathcal
D_i^{(1)}\) of the algebraic boundary. For \(r>1\), the codimension-\(r\) strata
are the irreducible components of codimension \(r\) obtained by intersecting
collections of boundary divisors and lower-dimensional strata.\footnote{This
definition assumes that the strata arising from repeated intersections behave
normally. More generally, singular loci of codimension one can contribute to the
stratification. For examples and related discussions,
see~\cite{koefler2025taking,Telen_PG,Polypols}.} We denote the codimension-\(r\)
strata by $\mathcal D^{(r)}_\alpha$.
In this way we obtain collections of complex projective varieties
\begin{equation}
\{\mathcal D^{(1)}_\alpha\}\supset\{\mathcal D^{(2)}_\alpha\}\supset\dots\supset\{\mathcal D^{(m)}_\alpha\} \, ,
\end{equation}
where the last collection consists of points.

We now partition these strata into boundary and residual strata. The intersection of a
codimension-\(r\) stratum \(\mathcal D^{(r)}_\alpha\) with
\(P\) has dimension bounded by
\begin{equation}\label{eq:ineq_dim}
\dim_{\mathbb R}\left(\partial P \cap\mathcal D^{(r)}_{\alpha,\mathbb R}\right) \leq m-r \, .
\end{equation}
The stratum \(\mathcal D^{(r)}_\alpha\) is called a \textit{boundary} if the inequality in~\eqref{eq:ineq_dim} is an equality, and it is called \textit{residual} otherwise. In particular, if the intersection in~\eqref{eq:ineq_dim} is empty, then \(\mathcal D^{(r)}_\alpha\) is residual. The \emph{residual arrangement of}
\(P\) is the union of all residual strata:
\begin{equation}
\mathcal R(P)=\bigcup_{\mathcal D^{(r)}_\alpha\ \mathrm{residual}}\mathcal D^{(r)}_\alpha \, .
\end{equation}

\begin{eg}[Residual arrangement of the pizza slice]
Let \(P\) be the pizza slice of Example~\ref{eg:pizza_slice}; see also
Figure~\ref{fig:pizza_tr}. Its residual arrangement is
\begin{equation}
	\mathcal R(P)=\{(-1,0),(0,-1)\} \, .
\end{equation}
\end{eg}

\subsubsection{Adjoint hypersurfaces}
Let now \((\mathcal{X},P,\mathbf{\Omega}_P)\) be a positive
geometry. We define the adjoint hypersurface \(A_{P}\) to be the Zariski
closure of the vanishing locus of the canonical form. For a nonzero canonical form,
\(A_{P}\) is either empty or a hypersurface. If \(\mathcal{X}=\mathbb
P^m\), then after writing
\begin{equation}
\mathbf{\Omega}_P(Y)=\frac{q(Y)}{p(Y)}\langle Y\,\mathrm{d}^mY\rangle \, ,
\end{equation}
the adjoint hypersurface is simply the projective hypersurface defined by the vanishing locus of $q$. Its degree is fixed by~\eqref{eq:adj_deg}.

More generally, the canonical form is determined by its poles,
\(\partial_a P \), and its zeros, \(A_{P}\). The degree of
\(A_{P}\) is determined by the degree of \(\partial_a P \) and the
embedding data of \(\mathcal{X}\subseteq\mathbb P^N\). A general expectation is
that
\begin{equation}\label{eq:inter_cond}
	\mathcal R(P)\subseteq A_{P} \, .
\end{equation}
Under smoothness assumptions on the stratification, this containment is proved
in~\cite[Prop.~1]{Lam:_PG_notes}, and it has been observed to hold in many more
general settings.
A natural question is:
\begin{tcolorbox}[resultbox]
\textbf{Interpolation question.}\label{qu:adj_by_int}
Is \(A_{P}\) fully determined by its degree and by interpolating the residual arrangement, as in~\eqref{eq:inter_cond}?
\end{tcolorbox}\noindent
The answer is affirmative when \(P\) is a projective
polytope~\cite{kohn2020projective,bruser2025geometry}. In this case, if the
hyperplane arrangement
\begin{equation}
\partial_a P =H_1\cup\cdots\cup H_r
\end{equation}
in \(\mathbb P^m\) is \emph{simple}, meaning that at most \(m\) hyperplanes among the
\(H_i\) pass through any point in \(\mathbb P^m\), then the residual strata impose
ordinary vanishing conditions. For non-simple arrangements, several residual
strata collide, and the adjoint is characterized by higher-order vanishing
conditions; see~\cite{bruser2025geometry} for details.

\begin{figure}[pos=t]
\begin{minipage}[c]{0.32\linewidth}
\includegraphics[width=\linewidth]{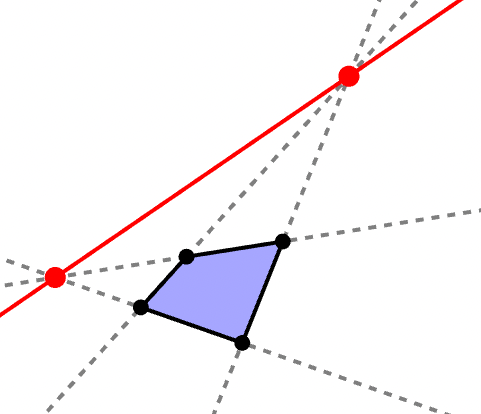}
\end{minipage}\hfill
\begin{minipage}[c]{0.327\linewidth}
\includegraphics[width=\linewidth]{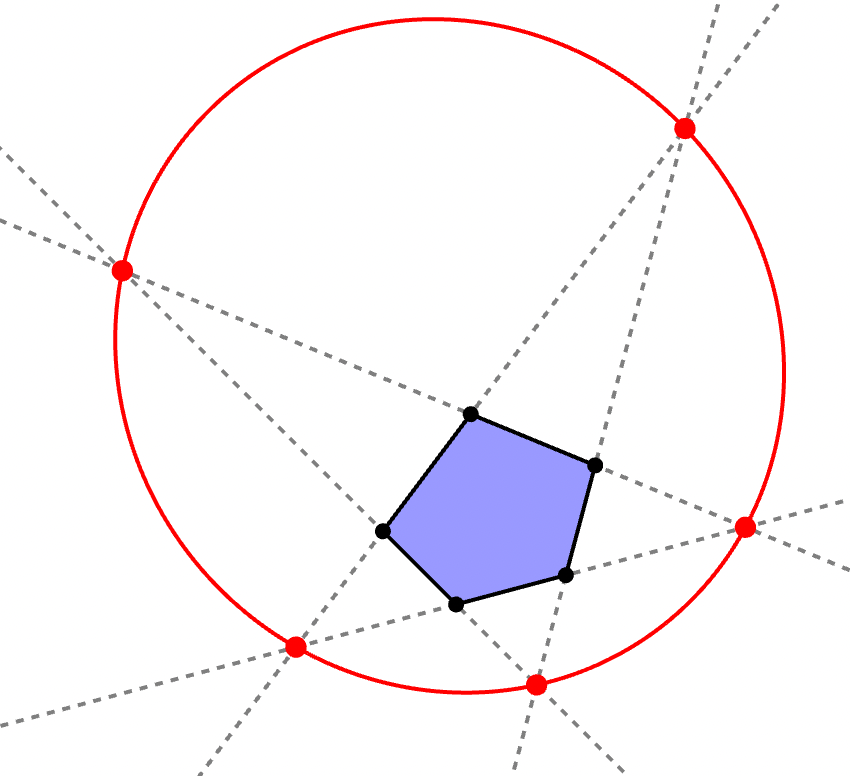}
\end{minipage}\hfill
\begin{minipage}[c]{0.325\linewidth}
\includegraphics[width=\linewidth]{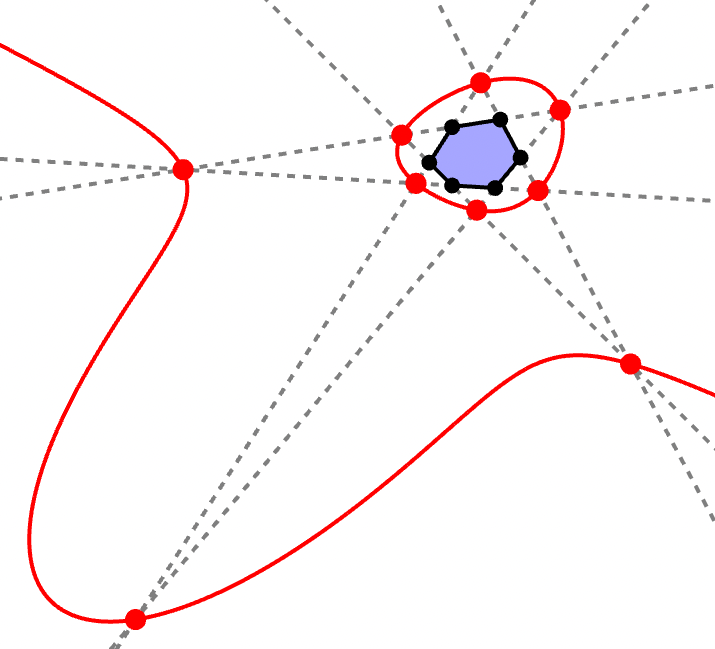}
\end{minipage}
\caption{Adjoint curves, shown in red, of \(n\)-gons, shown in blue, for \(n=4,5,6\).}
\label{fig:adjoints}
\end{figure}

\begin{eg}[Adjoints of polygons]
Let \(P\) be a convex \(n\)-gon in the projective plane \(\mathbb P^2\),
with vertices \(Z_i\in\mathbb R^3\) for \(i=1,\dots,n\). Assuming the vertices are
cyclically labelled, the facets of \(P\) are the lines \(L_i\) defined by
\begin{equation}
\ell_i(Y)=\langle Y\,Z_i\,Z_{i+1}\rangle=0,
\qquad i=1,\dots,n \, ,
\end{equation}
with indices understood cyclically. By~\eqref{eq:form_pq}, we can write
\begin{equation}
	\mathbf{\Omega}_P(Y)=\frac{q(Y)}{\ell_1(Y)\ell_2(Y)\cdots\ell_n(Y)}\langle Y\,\mathrm{d}^2Y\rangle \, .
\end{equation}
Since all \(\ell_i(Y)\) are linear, the degree formula~\eqref{eq:adj_deg} implies
that \(q(Y)\) is a homogeneous polynomial in \(Y=[Y_0:Y_1:Y_2]\) of degree
\(n-3\).

The vector space of homogeneous polynomials of degree \(n-3\) in three variables
has dimension $\binom{n-1}{2}$.
Up to overall scale, this gives
\begin{equation}
\binom{n-1}{2}-1=\frac{n(n-3)}{2}
\end{equation}
degrees of freedom. The residual arrangement consists of all points
\begin{equation}
\mathcal
R(P) = \left\{ L_i\cap L_j \, : \quad
\qquad |i-j|>1 \right \}\, ,
\end{equation}
where cyclically adjacent lines are excluded. There are precisely \(n(n-3)/2\)
such points. Hence the adjoint polynomial \(q\) cuts out the unique curve of
degree \(n-3\) passing through the residual points.
Computing
\begin{equation}
q(Y)=\sum_{|I|=n-3}\alpha_IY^I
\end{equation}
is therefore a linear algebra problem: each interpolation condition \(q(R)=0\),
for \(R\in\mathcal R(P)\), gives a linear equation in the coefficients
\(\alpha_I\). Up to an overall scale, \(q\) is fixed by solving this linear
system. In Figure~\ref{fig:adjoints} we illustrate the adjoint curves for
\(n=4,5,6\) computed by this method.
\end{eg}

The adjoint of a polytope was first introduced by
Wachspress~\cite{Wachspress:1975} in his construction of Wachspress coordinates, a
rational generalization of classical barycentric coordinates. Recently, this
construction has been reinterpreted in a broader algebro-geometric framework, in
connection with positive geometries and related
constructions~\cite{telen2026toric}. For example, affirmative answers to
the interpolation quation hold for a class of positive geometries in the plane
called polypoles~\cite{Polypols} and for certain
Amplituhedra~\cite{Ranestad:adjoint,Dian:2024hil,positive_amplitudes}.

\subsection{Push-forward and scattering equations}
\label{sec:Push-Forward and Scattering Equations}

\subsubsection{Push-forwards of meromorphic forms}
We introduced the class of objects called positive geometries. We now introduce
maps between them. Let
\begin{equation}
\Phi:\mathcal{X}_1\longrightarrow\mathcal{X}_2
\end{equation}
be a surjective meromorphic map between complex manifolds of the same dimension
\(m\), and let \(\mathbf\Omega\) be a meromorphic top-form on \(\mathcal{X}_1\).
A generic point
\(y \in\mathcal{X}_2\) has a finite fiber
\begin{equation}\label{eq:fiber}
	\Phi^{-1}(y)=\{x^{(1)},\dots,x^{(d)}\} \, ,
\end{equation}
where the positive integer \(d\) is called the degree of \(\Phi\). There exist open
neighborhoods \(V\subseteq\mathcal{X}_2\) of \(y\) and \(U_i\subseteq\mathcal{X}_1\)
of \(x^{(i)}\) such that \(\Phi|_{U_i}:U_i\to V\) is a biholomorphism for every
\(i=1,\dots,d\). The \emph{push-forward of} \(\mathbf\Omega\) is the meromorphic top-form
on \(\mathcal{X}_2\) defined locally by
\begin{tcolorbox}[definitionbox]
\textbf{Push-forward of a meromorphic form.}
\begin{equation}\label{eq:push_forward_def}
	(\Phi_*\mathbf\Omega)|_V
	:=
	\sum_{i=1}^d (\Phi|_{U_i}^{-1})^*(\mathbf\Omega|_{U_i}) \, .
\end{equation}
\end{tcolorbox}\noindent
Equivalently, in local coordinates \(x=(x_1,\dots,x_m)\) on \(\mathcal{X}_1\) and
\(y=(y_1,\dots,y_m)\) on \(\mathcal{X}_2\), if
\begin{equation}
\mathbf\Omega=\Omega(x)\,\mathrm{d}x_1\wedge\cdots\wedge \mathrm{d}x_m \, ,
\end{equation}
then
\begin{equation}\label{eq:pushforward_coord}
	\Phi_*\mathbf\Omega
	=
	\left(
	\sum_{x \, \in \, \Phi^{-1}(y)}
	\frac{\Omega(x)}{\det\left(\frac{\partial\Phi}{\partial x}\right)(x)}
	\right) \mathrm{d}y_1\wedge\cdots\wedge \mathrm{d}y_m \, .
\end{equation}
The expression is initially defined away from critical values, and then extends
meromorphically.

Let $(\mathcal{X}_1,P_1)$ and $(\mathcal{X}_2,P_2)$ be two pseudo-positive geometries of the same dimension \(m\). A \emph{morphism}
\begin{equation}\label{eq:Phi}
	\Phi:(\mathcal{X}_1,P_1)\longrightarrow(\mathcal{X}_2,P_2)
\end{equation}
is a rational map \(\Phi:\mathcal{X}_1\to\mathcal{X}_2\) such that its restriction
to the interiors
\begin{equation}
\Phi|_{{\rm int}(P_1)}:{\rm int}(P_1) \longrightarrow {\rm int}(P_2)
\end{equation}
is an orientation-preserving diffeomorphism and extends continuously to a map from
${\rm int}(P_1)$ to ${\rm int}(P_2)$. If \(\Phi\) is an isomorphism of varieties, we
call it an \emph{isomorphism of pseudo-positive geometries}. Two pseudo-positive
geometries are \emph{isomorphic} if such a map exists.

A heuristic principle~\cite[Heuristic 4.1]{Positive_geometries} is that morphisms
preserve canonical forms. Given a morphism \(\Phi\) as in~\eqref{eq:Phi}, one
expects
\begin{tcolorbox}[resultbox]
\textbf{Push-forward heuristic.}
\begin{equation}\label{eq:pushforward_heuristic}
	\Phi_*\mathbf\Omega_{P_1}=\mathbf\Omega_{P_2} \, .
\end{equation}
\end{tcolorbox}\noindent
This property is not proven in complete generality, but it holds in many
examples~\cite{Positive_geometries}. If \(\Phi\) is an isomorphism of
pseudo-positive geometries, then~\eqref{eq:pushforward_heuristic} is simply a
change of variables. We now discuss a special class of maps from simplices to
projective polytopes, and their push-forwards.

\subsubsection{Newton polytope maps}
Let \(P \subseteq\mathbb P^m_{\mathbb R}\) be a projective polytope, given
by the convex hull of points \(Z_i\in\mathbb P^m_{\mathbb R}\) for
\(i=1,\dots,n\). We assume that there exists an \(m\times n\) integer matrix
\(z\), whose columns we denote by \(z_i\in\mathbb Z^m\), such that, writing $Z_i=[1:z_i]\in\mathbb P^m_{\mathbb R}$,
\begin{equation}\label{eq:or_matr}
	{\rm sign}\big(\langle Z_{i_0}\cdots Z_{i_m}\rangle\big)
	=
	{\rm sign}\big(\langle \widetilde Z_{i_0}\cdots\widetilde Z_{i_m}\rangle\big),
	\qquad 1\leq i_0<\cdots<i_m\leq n \, .
\end{equation}
The configurations \(Z\) and \(\widetilde Z\) are then said to have the same
\emph{oriented matroid}. Technically, this assumption means that the oriented matroid of \(Z\) admits
an integral realization by points \(\widetilde Z_i=[1:z_i]\) in an affine chart.

The corresponding \emph{Newton polytope map} is naturally defined on the positive torus
\(\mathbb R^m_{>0}\), whose projective closure is the standard simplex
\(\Delta^m_{\geq 0}\):
\begin{equation}\label{eq:newton_map}
	\Phi:\mathbb R^m_{>0}\longrightarrow P,
	\qquad
	x\longmapsto\left[\sum_{i=1}^n x^{z_i}Z_i\right] \, ,
\end{equation}
where $x^{z_i}:=x_1^{z_{1i}}\cdots x_m^{z_{mi}}$.
After choosing the appropriate projective compactification, this map extends to a
morphism of positive geometries. The main result is the following.

\begin{tcolorbox}[resultbox]
\textbf{Newton polytope map.}\label{thm:newt_map}
If~\eqref{eq:or_matr} holds, then the Newton polytope map \(\Phi\) in~\eqref{eq:newton_map} defines a morphism of positive geometries and preserves canonical forms:
\begin{equation}\label{eq:newt_push}
	\Phi_*\left(\prod_{a=1}^m\frac{\mathrm{d}x_a}{x_a}\right)=\mathbf{\Omega}_P \, .
\end{equation}
\end{tcolorbox}\noindent

\subsubsection{Companion matrices}
Before discussing concrete examples, we explain how to compute push-forwards in
practice. In the definition~\eqref{eq:push_forward_def}, finding the local inverse
branches of \(\Phi\), or equivalently the points in the fiber~\eqref{eq:fiber},
requires solving a system of \(m\) polynomial equations in \(m\) variables. These
equations arise from the rational relations
\begin{equation}
y=\Phi(x)
\end{equation}
in local parametrizations of source and target, after clearing denominators.
Algebraically, the fiber over \(y\) is encoded by a zero-dimensional ideal
\begin{equation}\label{eq:F_ideal}
	I=\langle F_1(x,y),\dots,F_m(x,y)\rangle\subseteq\mathbb C(Y)[x] \, ,
\end{equation}
where \(\mathbb C(y)[x]\) is the ring of polynomials in \(x\) with coefficients
rational in \(y\).

In general one cannot solve this system symbolically in radicals. For fixed
numerical values of \(y\), one can solve the system numerically, but for symbolic
push-forwards it is more efficient to use companion matrices and Stickelberger's
Theorem; see~\cite{stalknecht2024positive,Sturmfels2002} for details. A companion
matrix \(T_i\) is the matrix representing multiplication by \(x_i\) in the
quotient ring
\begin{equation}
\mathbb C(y)[x]/I \, .
\end{equation}
This quotient is a finite-dimensional vector space over \(\mathbb C(y)\). After
choosing a basis, one obtains \(m\) commuting \(d\times d\) matrices
\(T_1,\dots,T_m\), where \(d=\deg(\Phi)\). Stickelberger's Theorem says that the points of \(I\) are the simultaneous eigenvalues of the companion matrices.

We can then evaluate the push-forward using traces. Suppose
\begin{equation}
\mathbf\Omega=\Omega(x)\,\mathrm{d}x_1\wedge\cdots\wedge \mathrm{d}x_m \, ,
\end{equation}
and suppose the fiber is defined by \(F(x,y)=0\). Differentiating \(F(x,y)=0\)
gives
\begin{equation}
F_x\,\mathrm{d}x+F_y\,\mathrm{d}y=0 \, ,
\end{equation}
where \(F_x=\partial F/\partial x\) and \(F_y=\partial F/\partial y\). Hence, up
to an ordering-dependent sign,
\begin{equation}\label{eq:pushforward_comp}
	\Phi_*(\Omega\,\mathrm{d}x_1\wedge\cdots\wedge \mathrm{d}x_m)
	=
	{\rm Tr}\big(\omega(T)\big)\,\mathrm{d}y_1\wedge\cdots\wedge \mathrm{d}y_m,
	\qquad
	\omega=\Omega\,(-1)^m \, \frac{\det(F_y)}{\det(F_x)} \, .
\end{equation}
Here \(\omega(T)\) means that the rational function \(\omega(x,y)\) is evaluated
at \(x_i\mapsto T_i\), while the target coordinates \(y_j\) remain scalars. This
computation can be performed in a computer algebra system, such as
\texttt{Macaulay2}, as we now illustrate.

\begin{eg}[Pentagonal Newton Map]\label{eg:pent_newt_map}
Let \(P\) be the pentagon in \(\mathbb P^2\) given by the convex hull of
the columns of
\begin{equation}
	Z=
	\left(
	\begin{array}{rrrrr}
		 1 &  1 & 1 & 1 &  1 \\
		-1 &  0 & 1 & 1 & -1 \\
		-1 & -1 & 0 & 1 &  1
	\end{array}
	\right) \, .
\end{equation}
This is the same matrix appearing in~\cite[Example 2.1]{Telen_PG}. Consider the
Newton map~\eqref{eq:newton_map} with \(Z_i=\widetilde Z_i=(1,z_i)\):
\begin{equation}\label{eq:pent_NPM}
	\Phi(x_1,x_2)=\frac{1}{x_1x_2}Z_1+\frac{1}{x_2}Z_2+x_1Z_3+x_1x_2Z_4+\frac{x_2}{x_1}Z_5 \, .
\end{equation}
We normalize this three-vector so that its first component is \(1\), and set it
equal to \((1,y_1,y_2)\). We obtain equations of the form \(y_i=G_i/G_0\), where
\(G_i\) and \(G_0\) are polynomials in \(x_1,x_2\). Clearing the denominator
\(G_0\), we get
\begin{equation*}
\begin{aligned}
F_1(x,y)&=G_1(x)-y_1G_0(x)
&&=1-x_1^2x_2+x_2^2-x_1^2x_2^2+y_1(1+x_1+x_1^2x_2+x_2^2+x_1^2x_2^2) \, ,\\[1mm]
F_2(x,y)&=G_2(x)-y_2G_0(x)
&&=1+x_1-x_2^2-x_1^2x_2^2+y_2(1+x_1+x_1^2x_2+x_2^2+x_1^2x_2^2).
\end{aligned}
\end{equation*}
Combining Theorem~\ref{thm:newt_map} with~\eqref{eq:pushforward_comp}, we obtain
\begin{equation}
	\mathbf{\Omega}_P(Y)=\Phi_*\left(\frac{\mathrm{d}x_1\wedge \mathrm{d}x_2}{x_1x_2}\right)={\rm Tr}\big(\omega(T)\big)\,\mathrm{d}y_1\wedge \mathrm{d}y_2 \, .
\end{equation}
In this example, \(\omega\) is obtained from~\eqref{eq:pushforward_comp} with
\(F=(F_1,F_2)\). The companion matrices \(T_1,T_2\) represent multiplication by
\(x_1,x_2\) in the quotient ring
\begin{equation}
\mathbb C(Y_1,Y_2)[x_1,x_2]/\langle F_1,F_2\rangle \, .
\end{equation}
We compute them in \texttt{Macaulay2}~\cite{M2}, using the package
\texttt{RealRoots}:
\begin{verbatim}
needsPackage "RealRoots";

K = frac(QQ[Y1,Y2]);
R = K[x1,x2];

J = ideal(F1,F2);
I = saturate(J,ideal(G0));

T1 = (regularRepresentation(x1, I))_1;
T2 = (regularRepresentation(x2, I))_1;
\end{verbatim}
The command \texttt{saturate} removes the points lying on the vanishing locus of
\(G_0\). This is important because clearing the denominator \(G_0\) may introduce
spurious solutions. The companion matrices have size \(7\times7\), so the degree
of the Newton polytope map~\eqref{eq:pent_NPM} is \(7\). Finally, one evaluates
\(\omega(T_1,T_2,y_1,y_2)\), treating \(y_1,y_2\) as scalars, and takes the trace.
The result is the canonical form of the pentagon computed
in~\eqref{eq:can_form_pent}.
\end{eg}

This example also shows that push-forwards are usually not the most efficient way
to compute canonical forms: they require either solving a zero-dimensional
polynomial system or manipulating large symbolic matrices. For the pentagon, we
could have obtained~\eqref{eq:can_form_pent} much more efficiently by
triangulating it into three triangles, as in Section~\ref{sec:Triangulations}, or
by determining its adjoint numerator from residual points, as in
Section~\ref{sec:Adjoint Hypersurface}. However, push-forwards are conceptually
important because they lead naturally to scattering equations, relating saddle
points of string-like integrals, field-theory scattering amplitudes, and canonical
forms of polytopes~\cite{Stringy_Can_Forms,Lukowski:2022fwz}.

\subsubsection{Stringy canonical forms}
In~\cite{Stringy_Can_Forms}, the authors considered the following class of
integrals, called \emph{stringy canonical forms} or \emph{stringy integrals}:
\begin{tcolorbox}[definitionbox]
\textbf{Stringy canonical form.}
\begin{equation}\label{eq:stringy:can_forms}
	\mathcal I(y,c)=\int_{\mathbb R^m_{>0}}\prod_{i=1}^m\frac{\mathrm{d}x_i}{x_i}\,R(x;y,c),
	\qquad
	R(x;y,c)=\prod_{i=1}^m x_i^{y_i}\prod_I p_I(x)^{-c_I} \, .
\end{equation}
\end{tcolorbox}\noindent
Here \(p_I(x)\in\mathbb C[x_1^{\pm1},\dots,x_m^{\pm1}]\) are finitely many Laurent
polynomials. For suitable real parts of the parameters this integral converges,
and elsewhere it is defined by analytic continuation as a meromorphic function of
\(y=(y_1,\dots,y_m)\) and \(c=(c_I)\). The integral
in~\eqref{eq:stringy:can_forms} is an instance of an \emph{Euler integral};
see~\cite{Euler_Integrals} for an introduction.

The name stringy arises because \(n\)-point tree-level open-string amplitudes
involve worldsheet integrals~\cite{KobaNielsen1969}:
\begin{equation}\label{eq:In}
	\mathcal I_n(s_{ab})=(\alpha')^{n-3}\int_{\mathcal M_{0,n}^+}\frac{\mathrm{d}^{n-3}z}{z_{12}z_{23}\cdots z_{1n}}\prod_{a<b}(z_{ab})^{\alpha's_{ab}} \, ,
\end{equation}
over the positive part \(\mathcal M_{0,n}^+\) of the moduli space \(\mathcal
M_{0,n}\) of a Riemann sphere with \(n\) marked points. Here \(z_i\) are
coordinates of the marked points on \(\mathbb P^1\), \(z_{ab}:=z_b-z_a\),
\(s_{ab}\) are Mandelstam invariants, and the string scale \(\alpha'\) encodes the
size of the strings.

The positive part of the moduli space is defined by \(n\) ordered points on the
real projective line, modulo \({\rm SL}(2,\mathbb R)\):
\begin{equation}
	\mathcal M_{0,n}^+=\{z_1<z_2<\cdots<z_n\}/{\rm SL}(2,\mathbb R) \, .
\end{equation}
Fixing the \({\rm SL}(2,\mathbb R)\)-gauge by
\begin{equation}
z_1=0,\qquad z_2=1,\qquad z_n=\infty \, ,
\end{equation}
we may write the remaining ordered region as $1<z_3<z_4<\cdots<z_{n-1}<\infty$.
It can be parametrized by
\begin{equation}
	z_3=1+x_1,\qquad z_4=1+x_1+x_2,\qquad \dots,\qquad z_{n-1}=1+x_1+\cdots+x_{n-3} \, ,
\end{equation}
with \(x_i>0\). Under this parametrization, \eqref{eq:In} can be written in the
form~\eqref{eq:stringy:can_forms}, with \(m=n-3\), suitable choices of parameters
\(y_i\) and \(c_I\), and polynomials
\begin{equation}
p_{ab}=\sum_{k=a}^{b-1}x_k
\end{equation}
after the appropriate gauge convention. Thus stringy canonical forms generalize
actual string-amplitude integrals.

\begin{eg}[Beta Function]
In 1968 Veneziano~\cite{Veneziano1968} found the formula for the tree-level
four-point open-string amplitude, given, up to conventional shifts, by
\begin{equation}
	B(-\alpha's,-\alpha't),
	\qquad
	B(a,b)=\int_0^1 \mathrm{d}z\,z^{a-1}(1-z)^{b-1}=\frac{\Gamma(a)\Gamma(b)}{\Gamma(a+b)} \, .
\end{equation}
Here \(s=s_{12}\) and \(t=s_{23}\) are Mandelstam invariants. The beta function is
an instance of a stringy canonical form, with \(m=1\) and \(p_1(x)=1+x\):
\begin{equation}\label{eq:SCF_int}
	\int_0^\infty\frac{\mathrm{d}x}{x}\,x^y(1+x)^{-c}
	=
	\int_0^1\frac{\mathrm{d}z}{z(1-z)}\,z^y(1-z)^{c-y}
	=
	B(y,c-y) \, ,
\end{equation}
where in the first equality we used the change of variables \(z=x/(1+x)\).
\end{eg}

An interesting limit of string theory is the regime in which the string length
scale goes to zero, \(\alpha'\to0\), so that strings become effectively
point-like. This is therefore referred to as the \emph{field-theory} or QFT limit. In
this limit, string amplitudes reduce to ordinary quantum field theory amplitudes.
For instance, the \(\alpha'\to0\) limit of tree-level bosonic open-string
amplitudes yields tree-level amplitudes of the planar limit of colored scalars
with cubic interaction, also called \(\mathrm{Tr}(\Phi^3)\)
theory~\cite{CachazoHeYuan2014,Mizera:2017cqs}.

The QFT limit of the stringy
integrals~\eqref{eq:stringy:can_forms} is remarkably geometric~\cite[Claim
1]{Stringy_Can_Forms}:

\begin{tcolorbox}[resultbox]
\textbf{Field-theory limit of stringy integrals.}\label{thm:stringy_field_theory_limit}
\begin{equation}\label{eq:stringy_limit}
	\lim_{\alpha'\to0}(\alpha')^m \, \mathcal I(\alpha'y,\alpha'c)=\Omega_P(y) \, ,
\end{equation}
\end{tcolorbox}\noindent
where \(\Omega_P(y)\) denotes the canonical function of the polytope \(P\), namely
the coefficient of the canonical form after stripping off the standard measure.
The polytope \(P\) is the Minkowski sum of the Newton polytopes of the Laurent
polynomials \(p_I\), each rescaled by \(c_I\). If
\begin{equation}
p_I(x)=\sum_J a_Jx^J,
\qquad
J=(j_1,\dots,j_m)\in\mathbb Z^m,
\qquad
x^J=x_1^{j_1}\cdots x_m^{j_m} \, ,
\end{equation}
then the Newton polytope of \(p_I\) is the convex hull in \(\mathbb R^m\) of the
exponent vectors \(J\) for which \(a_J\neq0\). The Minkowski sum of two sets in
\(\mathbb R^m\) is the set of all pairwise sums of points from the two sets.

\begin{eg}[Field-theory limit of the beta function]
The QFT limit of~\eqref{eq:SCF_int} is
\begin{equation}
	\lim_{\alpha'\to0}\alpha'B(\alpha'y,\alpha'(c-y))=\frac{c}{y(c-y)} \, .
\end{equation}
This is the canonical function of the interval \([0,c]\), whose interior is
\((0,c)\). The interval \([0,1]\) is the Newton polytope of the polynomial
\(p_1(x)=1+x\) in~\eqref{eq:SCF_int}.
\end{eg}

\subsubsection{Scattering equations}
There is another surprising feature of stringy integrals. In the \emph{high-energy
limit} \(\alpha'\to\infty\), the integral \((\alpha')^m\mathcal
I(\alpha'y,\alpha'c)\) is dominated by saddle points, which are solutions to
\begin{tcolorbox}[definitionbox]
\textbf{Scattering equations.}
\begin{equation}\label{eq:scatt_eqs}
	0=x_i\frac{\partial}{\partial x_i}\log R(x;y,c)=y_i-\sum_I c_Ix_i\frac{\partial}{\partial x_i}\log p_I(x) \, ,
\end{equation}
\end{tcolorbox}\noindent
or equivalently
\begin{equation}
	y_i=\sum_I c_I\frac{x_i}{p_I(x)}\frac{\partial p_I}{\partial x_i} \, ,
	\qquad i=1,\dots,m \, .
\end{equation}
Equations~\eqref{eq:scatt_eqs} are called \emph{scattering equations}. They owe their
name to their form for open-string integrals. Cachazo, He, and Yuan
(CHY)~\cite{CachazoHeYuan2014} proposed a universal description of tree-level
scattering amplitudes in arbitrary dimensions as localized integrals over the
moduli space of punctured Riemann spheres, linking string theory, field theory,
and geometry. One incarnation of this formula is for \(\mathrm{Tr}(\Phi^3)\)
theory, where scattering equations express amplitudes as sums over \((n-3)!\)
solutions and produce the canonical form of a polytope: the ABHY
associahedron~\cite{ABHY_original}, named after
Arkani-Hamed, Bai, He, and Yan.

We have now reached the connection to push-forwards. The scattering equations
define 
\begin{equation}\label{eq:Phi_map_2}
	\Phi:\mathbb R^m_{>0}\longrightarrow P \, ,
	\qquad
	x\longmapsto y(x) \, ,
\end{equation}
where \(y(x)\) is determined by~\eqref{eq:scatt_eqs}, and \(P\) is the rescaled
Newton polytope appearing in~\eqref{eq:stringy_limit}. After compactification,
this map extends to a morphism of positive geometries. It is equivalent to the
Newton polytope map~\eqref{eq:newton_map}, and the canonical form of \(P\) is
obtained as the push-forward of the canonical form of the positive orthant:
\begin{equation}
\mathbf{\Omega}_P=\Phi_*\left(\prod_{i=1}^m\frac{\mathrm{d}x_i}{x_i}\right) \, .
\end{equation}

\begin{eg}[Five-point ABHY]
From~\eqref{eq:In}, the five-point (\(n=5\)) open-string integral is an example
of~\eqref{eq:stringy:can_forms} with \(m=2\) and
\begin{equation}
p_{13}=1+x_1 \, ,
\qquad
p_{24}=1+x_2 \, ,
\qquad
p_{14}=1+x_1+x_2 \, ,
\end{equation}
see~\cite[Example 7.2]{Stringy_Can_Forms}. The polytope \(P\)
in~\eqref{eq:Phi_map_2} is the Minkowski sum of the Newton polytopes of the
\(p_I\), rescaled by the corresponding parameters \(c_I\). These are two segments
for \(p_{13}\) and \(p_{24}\), and one triangle for \(p_{14}\). Their Minkowski
sum is a pentagon in \(\mathbb R^2\).

In coordinates \(y=(y_1,y_2)\), define
\begin{equation}
\begin{aligned}
	Y_1&=y_1 \, ,\\
	Y_2&=y_2 \, ,\\
	Y_3&=c_{13}+c_{14}-y_1 \, ,\\
	Y_4&=c_{13}+c_{24}+c_{14}-y_1-y_2 \, ,\\
	Y_5&=c_{24}+c_{14}-y_2.
\end{aligned}
\end{equation}
The pentagon \(P\) is the region where all \(Y_i\geq 0\). We invite the reader to
use the method presented in Example~\ref{eg:pent_newt_map} to verify the following
push-forward formula, where \(\Phi\) is defined by the scattering
equations~\eqref{eq:scatt_eqs}:
\begin{equation}\label{eq:n5_assoc}
	\Phi_*\left(\frac{\mathrm{d}x_1\wedge \mathrm{d}x_2}{x_1x_2}\right)
	=
	\left(\frac{1}{Y_1Y_2}+\frac{1}{Y_2Y_3}+\frac{1}{Y_3Y_4}+\frac{1}{Y_4Y_5}+\frac{1}{Y_1Y_5}\right)\mathrm{d}Y_1\wedge \mathrm{d}Y_2
	=
	\mathbf{\Omega}_P \, .
\end{equation}
In this case, the scattering equations have \((n-3)!=2\) solutions, and they are
an instance of the CHY formula. The pentagon \(P\) is the two-dimensional ABHY
associahedron, and its canonical form~\eqref{eq:n5_assoc} gives the five-point
amplitude in planar \(\mathrm{Tr}(\Phi^3)\) theory. In this theory, tree-level
Feynman diagrams are trivalent trees. At \(n=5\), there are five trivalent trees
with five cyclically labelled external edges. Each of the five terms in
\(\mathbf{\Omega}_P\) in~\eqref{eq:n5_assoc} corresponds to one such diagram.
\end{eg}

\subsubsection{Maximum-likelihood degree and twisted cohomology}
The push-forward~\eqref{eq:push_forward_def} along the scattering-equation map
\(\Phi\) in~\eqref{eq:Phi_map_2} can be expressed as a sum over \(d\) terms, where
\(d\) is the degree of \(\Phi\), or equivalently the number of solutions to the
scattering equations~\eqref{eq:scatt_eqs}. This number is called the
maximum-likelihood degree, or ML degree, of the space
\begin{equation}
\mathcal{U}=(\mathbb C^*)^m\setminus\left\{\prod_Ip_I=0\right\} \, .
\end{equation}
For the \(n\)-point ABHY associahedron, the ML degree is
\((n-3)!\)~\cite{ABHY_original}. This terminology comes from algebraic statistics,
where the ML degree measures the intrinsic algebraic complexity of
maximum-likelihood estimation on
\(\mathcal{U}\)~\cite{CataneseHostenKhetanSturmfels2006,DrtonSturmfelsSullivant2009}.
Furthermore,~\cite{Huh2013} proved that, if \(\mathcal{U}\) is a smooth \emph{very affine
variety}, its ML degree equals its signed topological Euler characteristic, thereby
giving a topological meaning to this quantity.

More recently, this picture has been related to twisted de Rham cohomology of
\(\mathcal{U}\). In this setting, one studies differential forms twisted by the multivalued
integrand \(R(x;y,c)\), with singularities along \(\{\prod_Ip_I=0\}\). The
dimension of the relevant twisted cohomology group is generically equal to the ML
degree~\cite{MatsubaraHeoTelen2025,AomotoKita2011}. This connection is important
in physics, where twisted cohomology organizes the vector spaces of Euler, string,
and Feynman integrals. A choice of basis is often called a basis of master
integrals~\cite{Mizera:2017cqs,MastroliaMizera2019}.

\subsection{Integral representations}\label{sec:Integral Representations}

We now present further representations of canonical forms, related to integral
transforms. The main example is the following: the canonical function of a
projective polytope can be written as a Laplace transform over the dual cone, or
equivalently as the volume of a dual convex body. This point of view will be
important later, when we return to dual-volume representations for more general
positive geometries.

\subsubsection{Convex duality}
We begin by recalling convex duality for cones. As in the definition of
semialgebraic sets in projective space, it is convenient to work one dimension
higher. Let \(C\subseteq\mathbb R^{m+1}\) be a cone. We define its closed dual
cone by
\begin{equation}
	C^* := \{W\in\mathbb R^{m+1}:W\cdot Y\geq 0\text{ for all }Y\in C\} \, ,
\end{equation}
where $W\cdot Y=\sum_{i=0}^m W_iY_i$ is the standard inner product on \(\mathbb R^{m+1}\). 
We say that \(C\) is pointed if it contains no line, equivalently
\begin{equation}
C\cap(-C)=\{0\} \, .
\end{equation}
If \(C\) is full-dimensional, pointed, closed, and convex, then so is \(C^*\).
Biduality yields
\begin{equation}\label{eq:biduality}
	(C^*)^*=C \, .
\end{equation}
For an arbitrary cone \(C\), the bidual gives the closed convex cone generated by \(C\).

Let \(P\subseteq\mathbb P^m_{\mathbb R}\) be very compact, meaning that there exists a hyperplane not intersecting $P$. Then one can choose a
pointed cone $P\subseteq\mathbb R^{m+1}$ such that
\begin{equation}
\pi^{-1}(P)=\widehat P \cup(-\widehat P) \, ,
\end{equation}
where \(\pi:\mathbb R^{m+1}\setminus\{0\}\to\mathbb P^m_{\mathbb R}\) is the
canonical projection. The choice of \(\widehat P\) is determined up to an overall
sign. We define the projective dual of \(P\) by
\begin{equation}
P^*:=\pi(\widehat P^*) \, ,
\end{equation}
which is independent of this sign.

Convex duality acts particularly nicely on projective polytopes. Let
\(P\subseteq\mathbb P^m_{\mathbb R}\) be a projective polytope, and choose a
closed pointed cone over it of the form
\begin{equation}\label{eq:C_cone}
	\widehat P={\rm cone}(Z_1,\dots,Z_n):=\left\{\sum_{i=1}^n\lambda_iZ_i:\lambda_i\geq 0\right\}\subseteq\mathbb R^{m+1} \, ,
\end{equation}
where \(Z_i\in\mathbb R^{m+1}\). We assume that \(\widehat P\) is
full-dimensional. Its dual cone is then
\begin{equation}\label{eq:dual_pol_cone}
	\widehat P^*=\{W\in\mathbb R^{m+1}:W\cdot Z_i \geq 0,\ i=1,\dots,n\} \, .
\end{equation}
Equation~\eqref{eq:C_cone} is the vertex representation of the cone, while
\eqref{eq:dual_pol_cone} is the facet representation of its dual. By biduality,
vertices of \( P\) correspond to facets of \( P^*\), and facets of
\( P\) correspond to rays of \( P^*\).

If \(F\) is a face of \(\widehat P\) generated by the vectors
\(Z_i\) for \(i\in I\), then the dual face is
\begin{equation}
	F^\vee:=\{W\in\widehat P^\vee:W\cdot Z_i=0\text{ for all }i\in I\} \, .
\end{equation}
The assignment \(F\mapsto F^\vee\) gives an inclusion-reversing correspondence
between faces of \(\widehat P\) and faces of \(\widehat P^\vee\):
\begin{equation}
	F_1\subseteq F_2\qquad\Longleftrightarrow\qquad F_2^\vee\subseteq F_1^\vee \, .
\end{equation}

\subsubsection{Laplace transform representation}
Let \(P\subseteq\mathbb P^m_{\mathbb R}\) be a projective polytope. We write its
canonical form as
\begin{equation}
	\mathbf{\Omega}_P(Y)=\Omega_P(Y)\,\langle Y\,\mathrm{d}^mY\rangle \, ,
\end{equation}
where \(\Omega_P(Y)\) is the canonical function. The canonical function is
homogeneous of degree \(-(m+1)\) in \(Y\). The first main
integral representation is that \(\Omega_P\) is the Laplace transform of the
characteristic function of the dual cone \(\widehat P^*\):
\begin{tcolorbox}[resultbox]
\textbf{Laplace representation for polytopes.}
\begin{equation}\label{eq:Lapl_tr_pol}
	\Omega_P(Y)= \int_{\widehat P^*}e^{-Y\cdot W}\,\mathrm{d}^{m+1}W,
	\qquad
	Y\in {\rm int}(\widehat P) \, ,
\end{equation}
\end{tcolorbox}\noindent
where the interior ${\rm int}(\widehat P)$ of $\widehat{P}$ is an open cone.
For \(Y\notin  {\rm int}(\widehat P)\), the Laplace integral need not converge, so the integral representation is naturally defined on the open cone \({\rm int}(\widehat P)\). Its right-hand side then analytically continues to the rational canonical function.

An interesting way of rewriting~\eqref{eq:Lapl_tr_pol} is to use polar coordinates
\(W=t\overline W\), with \(t>0\) and \(\overline W\in S^m\), with $S^m\subset \mathbb{R}^{m+1}$ the $m$-dimensional unit sphere. The measure
transforms as
\begin{equation}
\mathrm{d}^{m+1}W= \frac{1}{m!} \, t^m\,\mathrm{d}t\,\langle\overline W\,\mathrm{d}^m\overline W\rangle \, .
\end{equation}
Performing the \(t\)-integral gives
\begin{tcolorbox}[resultbox]
\textbf{Spherical gauge.}
\begin{equation}\label{eq:volume_gauge_S}
	\Omega_P(Y)=\int_{\widehat P^*\cap S^m}\frac{\langle\overline W\,\mathrm{d}^m\overline W\rangle}{(Y\cdot\overline W)^{m+1}} \, .
\end{equation}
\end{tcolorbox}\noindent
Depending on the normalization convention for \(\langle\overline W\,\mathrm{d}^m\overline
W\rangle\), the factor \(m!\) may be absorbed into the measure. This formula can
be interpreted as fixing the projective scaling freedom of \(W\) by the nonlinear
condition \(\overline W\in S^m\).

\begin{eg}[Simplices]
Let \(P\) be a projective simplex, and choose a cone over it
\begin{equation}
\widehat P={\rm cone}(Z_0,\dots,Z_m) \, .
\end{equation}
The dual cone is again simplicial:
\begin{equation}
\widehat P^*={\rm cone}(W_0,\dots,W_m) \, ,
\end{equation}
where \(W_i\) is the inward-pointing normal vector to the facet opposite \(Z_i\).
Equivalently, \(W_i\) is characterized, up to positive rescaling, by
\begin{equation}
	W_i\cdot Z_j=0\quad\text{for }j\neq i,
	\qquad
	W_i\cdot Z_i>0 \, .
\end{equation}
The linear map
\begin{equation}
\Phi:\mathbb R^{m+1}_{\geq 0}\longrightarrow\widehat P^*,
\qquad
(t_0,\dots,t_m)\longmapsto\sum_{i=0}^m t_iW_i
\end{equation}
gives a change of variables in~\eqref{eq:Lapl_tr_pol}. Therefore
\begin{align}
	\Omega_P(Y)
	&=
	\int_{\widehat P^*}e^{-Y\cdot W}\,\mathrm{d}^{m+1}W\nonumber\\
	&=
	\langle W_0W_1\cdots W_m\rangle
	\int_{\mathbb R^{m+1}_{\geq 0}}
	\exp\left(-\sum_{i=0}^m t_i(Y\cdot W_i)\right)dt_0\cdots \mathrm{d}t_m\nonumber\\
	&=
	\frac{\langle W_0W_1\cdots W_m\rangle}{(Y\cdot W_0)(Y\cdot W_1)\cdots(Y\cdot W_m)} \, .
\end{align}
The orientation fixes the sign of the determinant. This is precisely the canonical
function of a projective simplex, equivalently the coefficient of the canonical
form in~\eqref{eq:can_form_simplex}.
\end{eg}

Let us now explain why~\eqref{eq:Lapl_tr_pol} satisfies the recursive residue
property. The integral converges for every \(Y\in {\rm int}(\widehat P)\), by definition of
the dual cone. Let \(F\) be a facet of \(\widehat P\), and let \(W_F\) be the
corresponding ray of the dual cone, i.e. the inward-pointing normal vector to
\(F\). Near the facet \(Y\cdot W_F=0\), the dual cone splits locally as a product
of the ray generated by \(W_F\) and the dual cone of the facet. Integrating along
this ray gives
\begin{equation}
\int_0^\infty e^{-t(Y\cdot W_F)}\,\mathrm{d}t=\frac{1}{Y\cdot W_F} \, .
\end{equation}
Thus the Laplace transform has a simple pole along the facet \(F\), and its
residue is the Laplace transform associated with the boundary polytope \(F\). By
induction on the dimension, the Laplace transform satisfies the same recursive
residue property as the canonical form. This proves~\eqref{eq:Lapl_tr_pol} for
projective polytopes. For more details on the proof see~\cite{Positive_geometries}.

\subsubsection{Dual-volume representation}
We now reinterpret~\eqref{eq:Lapl_tr_pol} as a volume formula. Choose an affine
chart and write \(Y=(1,x)\in\mathbb R^{m+1}\). We identify the affine polytope
\(P\subseteq\mathbb R^m\) with the slice \(\widehat P\cap\{Y_0=1\}\). For a convex
set \(K\subseteq\mathbb R^m\), define its support function by
\begin{equation}
h_K(u):=\sup_{z\in K} u\cdot z \, .
\end{equation}
The dual cone of \(\widehat P={\rm cone}\{(1,z):z\in P\}\) can be written as
\begin{equation}
\widehat P^*=\{(w_0,w)\in\mathbb R^{m+1}: w_0\geq h_P(-w)\} \, .
\end{equation}
Therefore, for \(x\in{\rm int}(P)\),
\begin{align}
	\Omega_P(1,x)
	&=
	\int_{\widehat P^*}e^{-(1,x)\cdot(w_0,w)}\,\mathrm{d}w_0\,\mathrm{d}^m w \nonumber\\
	&=
	\int_{\mathbb R^m}
	\left(\int_{h_P(-w)}^\infty e^{-w_0-x\cdot w}\,\mathrm{d}w_0\right)
	\mathrm{d}^m w \nonumber\\
	&=
	\int_{\mathbb R^m}e^{-h_P(-w)-x\cdot w}\,\mathrm{d}^m w \, .
\end{align}
Since
\begin{equation}
h_P(-w)+x\cdot w=h_{P-x}(-w) \, ,
\end{equation}
we obtain
\begin{equation}\label{eq:volume_00}
	\Omega_P(1,x)=\int_{\mathbb R^m}e^{-h_{P-x}(-w)}\,\mathrm{d}^m w \, .
\end{equation}

We now use a standard volume identity, written in the sign convention adapted to
the previous formula. If \(K\subseteq\mathbb R^m\) is a convex body with
\(0\in{\rm int}(K)\), define
\begin{equation}
K^\circ:=\{w\in\mathbb R^m: w\cdot z\geq -1 \text{ for all } z\in K\} \, .
\end{equation}
Then
\begin{equation}\label{eq:volume_0}
	\int_{\mathbb R^m}e^{-h_K(-w)}\,\mathrm{d}^m w=m!\,{\rm Vol}(K^\circ) \, .
\end{equation}
Indeed, this follows from Cavalieri's principle:
\begin{align}\label{eq:cavvv}
	\int_{\mathbb R^m}e^{-h_K(-w)}\,\mathrm{d}^m w
	&=
	\int_0^\infty e^{-t}\,\mathrm{d}\,{\rm Vol}\{w:h_K(-w)\leq t\} \, .
\end{align}
With the above convention for \(K^\circ\),
\begin{equation}
\{w:h_K(-w)\leq t\}=tK^\circ \, ,
\end{equation}
and hence
\begin{equation}\label{eq:caval}
{\rm Vol}\{w:h_K(-w)\leq t\}=t^m{\rm Vol}(K^\circ) \, .
\end{equation}
Therefore
\begin{equation}
\int_0^\infty e^{-t}\,\mathrm{d}\left(t^m{\rm Vol}(K^\circ)\right)
=m!\,{\rm Vol}(K^\circ) \, .
\end{equation}
Substituting this in~\eqref{eq:caval} yields~\eqref{eq:cavvv}.
Combining~\eqref{eq:volume_00} and~\eqref{eq:volume_0}, we obtain
\begin{tcolorbox}[resultbox]
\textbf{Dual-volume formula.}
\begin{equation}\label{eq:vol_form}
	\Omega_P(1,x)=m! \, {\rm Vol}\big((P-x)^\circ\big) \, .
\end{equation}
\end{tcolorbox}\noindent
Thus the canonical function of a polytope computes the volume of the polar dual of
the translated polytope \(P-x\).

Equivalently, one may fix the projective scaling of the dual cone by an affine slice by writing $W=w_0(1,w)$. In the slice where $Y=(1,x)$, $P$ corresponds to $\widehat{P} \cap \{Y_0=1\}$. Then, on the dual cone $w_0>0$ and $z \cdot w \geq -1$ for every $z \in P$, so that $\widehat{P}^* \cap \{W_0=1\}$ id identified with $P^\circ$. Hence, starting from the Laplace integral formula~\eqref{eq:Lapl_tr_pol}, the measure becomes \(\mathrm d^{m+1}W=w_0^m\,\mathrm dw_0\,\mathrm d^m w\), and integrating out $w_0$ we obtain
\begin{equation}
	\Omega_P(1,x)= \int_{P^\circ}\frac{\mathrm{d}^mw}{(1+x\cdot w)^{m+1}} \, ,
\end{equation}
This should be compared with~\eqref{eq:volume_gauge_S}: there the projective freedom was fixed by the nonlinear slice \(\widehat P^*\cap S^m\), whereas here it is fixed by an affine slice of the dual cone.

\subsubsection{Contour representation}
We now move to an equivalent contour-integral representation for the canonical
function of a polytope. Let $\widehat{P}$ be a cone generated by $n$ vectors $Z_i \in \mathbb{R}^{m+1}$ as in~\eqref{eq:C_cone}.
The Laplace integral~\eqref{eq:Lapl_tr_pol} is an integral over
\(\mathbb R^{m+1}\) against the characteristic function of the dual cone:
\begin{equation}
\Omega_P(Y)=\int_{\mathbb R^{m+1}}e^{-Y\cdot W}\,\chi_{\widehat P^*}(W)\,\mathrm{d}^{m+1}W \, .
\end{equation}
Using~\eqref{eq:dual_pol_cone}, this characteristic function is
\begin{equation}
\chi_{\widehat P^*}(W)=\prod_{j=1}^n\theta(W\cdot Z_j) \, ,
\end{equation}
where \(\theta(t)\) is the Heaviside function. We use the distributional
representation
\begin{equation}
\theta(t)=\frac{1}{2\pi i}\int_{\mathbb R}\frac{\mathrm{d}c}{c-i\varepsilon}\,e^{ict},
\qquad \varepsilon>0 \, ,
\end{equation}
together with the Fourier representation
\begin{equation}
\delta^{m+1}(U)=\frac{1}{(2\pi)^{m+1}}\int_{\mathbb R^{m+1}}e^{iU\cdot W}\,\mathrm{d}^{m+1}W \, .
\end{equation}
Up to the conventional normalization determined by these Fourier choices, one
obtains
\begin{tcolorbox}[resultbox]
\textbf{Contour representation.}
\begin{equation}\label{eq:contour_can_form}
\begin{aligned}
	\Omega_P(Y)
	&=
	\frac{1}{(2\pi i)^{n-m-1}}
	\int_{\mathbb R^n}
	\frac{\mathrm{d}^nc}{(c_1-i\varepsilon_1)\cdots(c_n-i\varepsilon_n)}
	\,
	\delta^{m+1}\left(Y-\sum_{j=1}^n c_jZ_j\right)
	\\
	&=:
	\int_\Gamma
	\frac{\mathrm{d}^{\,n-m-1}c}{c_1\cdots c_n}\,
	\delta^{m+1}\left(Y-\sum_{j=1}^n c_jZ_j\right).
\end{aligned}
\end{equation}
\end{tcolorbox}\noindent
The last expression is symbolic and defined by the previous line: the delta
functions localize \(m+1\) of the \(c\)-variables, leaving an
\((n-m-1)\)-dimensional contour integral of the rational function \(1/(c_1\cdots
c_n)\).

\begin{eg}[Simplices]
The simplest case of~\eqref{eq:contour_can_form} occurs for projective simplices,
when \(n=m+1\). The delta functions fully localize the variables \(c_0,\dots,c_m\)
in $Y=\sum_{j=0}^m c_jZ_j$. The solution is
\begin{equation}
	c_j=
	\frac{\langle Y\,Z_0\cdots\widehat{Z_j}\cdots Z_m\rangle}{\langle Z_j\,Z_0\cdots\widehat{Z_j}\cdots Z_m\rangle} \, ,
\end{equation}
with signs determined by the ordering in the brackets. The Jacobian from the delta
functions is \(\langle Z_0\cdots Z_m\rangle\), and~\eqref{eq:contour_can_form}
reproduces the canonical function of the simplex~\eqref{eq:can_form_simplex}.
\end{eg}

More generally, for a suitable choice of contour, the
formula~\eqref{eq:contour_can_form} expresses the canonical function as a sum of
canonical functions of simplices, precisely as in a triangulation of the polytope.
The choice of contour specifies a triangulation, and triangulation independence is
reflected by the global residue theorem, a multidimensional generalization of
Cauchy's theorem relating residues of a meromorphic form. We now illustrate this.

\begin{eg}[A Square]
Consider a square $P={\rm conv}(Z_1,Z_2,Z_3,Z_4)$
in $\mathbb{P}^2$, so that \(n=4\) and \(m=2\). The delta-function
constraint in~\eqref{eq:contour_can_form} is
\begin{equation}
	Y=c_1Z_1+c_2Z_2+c_3Z_3+c_4Z_4 \, .
\end{equation}
Equivalently, contracting with pairs of vertices gives
\begin{equation}\label{eq:delta_4}
\begin{aligned}
	\langle Y12\rangle&=c_3\langle123\rangle+c_4\langle124\rangle \, ,\\
	\langle Y23\rangle&=c_1\langle123\rangle+c_4\langle234\rangle \, ,\\
	\langle Y13\rangle&=-c_2\langle123\rangle+c_4\langle134\rangle.
\end{aligned}
\end{equation}
Here we wrote \(i\) for \(Z_i\) inside brackets. We solve the constraints by using
them to eliminate \(c_1,c_2,c_3\). The remaining contour integral is a
one-dimensional integral in \(c_4\):
\begin{equation}\label{eq:square_contour_integral}
\Omega_P(Y)=\frac{1}{2\pi i}\int_\Gamma\frac{\mathrm{d}c_4}{\ell_1(c_4)\ell_2(c_4)\ell_3(c_4)\ell_4(c_4)} \, ,
\end{equation}
where
\begin{align}
\ell_1(c_4)&=\langle234\rangle c_4-\langle Y23\rangle+i\langle123\rangle\varepsilon_1,\nonumber\\
\ell_2(c_4)&=\langle134\rangle c_4-\langle Y13\rangle-i\langle123\rangle\varepsilon_2,\nonumber\\
\ell_3(c_4)&=\langle124\rangle c_4-\langle Y12\rangle+i\langle123\rangle\varepsilon_3,\nonumber\\
\ell_4(c_4)&=c_4-i\varepsilon_4 \, .
\end{align}
Since the integrand scales as \(|c_4|^{-4}\) for \(|c_4|\to\infty\), there is no
pole at infinity. We label the four finite poles by \(1,2,3,4\), according to
which \(c_j\) vanishes. Assuming all ordered brackets
\begin{equation}
\langle Z_{i_1}Z_{i_2}Z_{i_3}\rangle>0,
\qquad 1\leq i_1<i_2<i_3\leq4 \, ,
\end{equation}
one finds that the poles labelled \(1\) and \(3\) lie in the lower half-plane,
while the poles labelled \(2\) and \(4\) lie in the upper half-plane. Computing
residues, the pole labelled \(1\), for example, gives the canonical function of
the triangle \({\rm conv}(Z_2,Z_3,Z_4)\), which we denote by \([2,3,4]\). Closing
the contour in the upper half-plane gives the residues at \(2\) and \(4\), while
closing it in the lower half-plane gives the residues at \(1\) and \(3\). Thus
\begin{equation}\label{eq:4gon_res}
	\Omega_P=[2,3,4]+[1,2,4]=[1,3,4]+[1,2,3] \, .
\end{equation}
This is exactly the equality between the two triangulations of a square, obtained
by drawing either the diagonal \(2\text{--}4\) or the diagonal \(1\text{--}3\).
The equality between the two residue sums follows from Cauchy's theorem.
\end{eg}

There is another integral representation for polytope canonical functions,
which we do not discuss here; see~\cite[Section 7.4.7]{Positive_geometries}. We
have restricted this section to projective polytopes. Later, the Grassmannian
contour formula will play the role of~\eqref{eq:contour_can_form} for the
Amplituhedron, and Chapter~\ref{ch:Canonical Forms as Dual Volumes} returns to
volume representations for nonlinear positive geometries.

%% file: Ch4.tex

\section{Amplituhedra}\label{ch:Amplituhedra}

In the previous chapter we introduced positive geometries and their canonical
forms in an abstract setting. We now specialize this framework to the main
example relevant for this thesis: the Amplituhedron. This is a semialgebraic set in the Grassmannian conjectured to be a positive geometry, first introduced by Arkani-Hamed and Trnka in 2013 to describe tree amplitudes
and loop integrands in planar \(\mathcal N=4\) super Yang--Mills theory
directly in terms of geometry~\cite{the_amplituhedron}. Its canonical form
reproduces the amplitude or integrand, while its boundaries encode physical
singularities and factorization channels.

The Amplituhedron is built from the same ingredients that of
Chapter~\ref{ch:Scattering Amplitudes}: momentum twistors, the positive Grassmannian, and on-shell
diagrams. In this sense, it is the geometric completion of the on-shell story.
The Grassmannian formula and on-shell diagrams provide local coordinates,
residues, and Yangian invariants
~\cite{Grassmannian}, while the Amplituhedron packages
these data into a single positive region whose canonical form has the correct
singularity structure. For introductory accounts of the Amplituhedron and its
role in scattering amplitudes, see
~\cite{Grassmannian,stalknecht2024positive,Herrmann:2022nkh}.
For mathematical perspectives on positive Grassmannians, total positivity,
sign variation, and amplituhedral geometry, see
~\cite{Postnikov:2006kva,TNN_grassmannian,Lam:_PG_notes,KarpWilliams,GalashinLam}.

A useful way to enter the subject is through polytopes. Hodges observed that
certain tree-level gluon amplitudes admit a volume interpretation in terms of
polytopes in momentum-twistor space~\cite{Hodges:2009hk}. This idea already
contains several features of the later Amplituhedron: positivity, projective
geometry, canonical forms, and the appearance of spurious singularities that
cancel in the final answer. Cyclic polytopes provide the simplest
amplituhedral examples and serve as a bridge between the polytopal positive
geometries of Chapter~\ref{ch:Positive Geometries} and the Grassmannian images studied in this chapter.

The tree Amplituhedron generalizes this picture by replacing a polytope with
the image of the positive Grassmannian under a linear map determined by
external momentum twistors~\cite{the_amplituhedron}. Its canonical form
computes tree-level amplitudes in planar \(\mathcal N=4\) SYM. Triangulations
of the Amplituhedron give different representations of the same form, and the
BCFW expansion becomes a distinguished class of triangulations
~\cite{Britto:2005fq,Britto:2004ap,the_amplituhedron,ABHY_original}.
This is the concrete realization of the triangulation principle
discussed in Section~\ref{sec:Triangulations}. Recent progress on
triangulations, sign-flip descriptions, cluster structures, and boundary
stratifications has made this geometry increasingly explicit
~\cite{ABHY_original,GalashinLam,Damgaard:2019ztj,Lukowski:2020bya,Ranestad:adjoint,Dian:2024hil,positive_amplitudes}.

At loop level, the Amplituhedron is enlarged by additional variables
representing loop momenta as lines in momentum-twistor space. Its canonical
form gives the all-loop integrand of planar \(\mathcal N=4\) SYM
~\cite{the_amplituhedron}. The loop Amplituhedron is less studied than the tree
geometry: its boundary structure is richer and its triangulations are not known in general. These
features make loop Amplituhedra a natural testing ground for the algebraic tools of Chapter~\ref{ch:Positive Geometries}, especially adjoint hypersurfaces, residual arrangements, and boundary stratifications. 
The goal of this chapter is therefore twofold. First, we introduce the
Amplituhedron as the central example of a positive geometry in scattering
amplitudes. Second, the later chapters by emphasize the aspects of the Amplituhedron geometry that control singularities and the combinatorics of recursion relations: canonical forms, triangulations and boundaries.

\medskip

The chapter is organized as follows. Section~\ref{sec:Hodges and Cyclic Polytopes}
begins with Hodges' polytope picture and cyclic polytopes, which provide the
simplest examples connecting amplitudes to volumes. Section~\ref{sec:Tree Amplituhedra}
introduces tree Amplituhedra as positive Grassmannian images and explains how
their canonical forms reproduce tree amplitudes. Section~\ref{sec:BCFW Recursion and Tiles}
discusses BCFW recursion as a triangulation of the tree Amplituhedron and
introduces the corresponding tiles. Section~\ref{sec:Loop Amplituhedra}
turns to loop level, starting from BCFW recursion for the loop-integrand.
Finally, Section~\ref{sec:The Four-Point Two-Loop Amplituhedron} studies the
four-point two-loop Amplituhedron as a concrete example, focusing on its
boundary structure and the numerator of the canonical form. This case study will be important later, when we use boundaries of Amplituhedra to constrain leading singularities and Landau singularities of loop-amplitudes.

\subsection{Cyclic polytopes}\label{sec:Hodges and Cyclic Polytopes}

This section gives a historical entry point to the Amplituhedron through the
work of Hodges~\cite{Hodges:2009hk}. Hodges observed that certain tree-level
gluon amplitudes admit a description in terms of volumes of projective
polytopes in momentum-twistor space. This idea was one of the precursors to
the Amplituhedron of Arkani-Hamed and Trnka~\cite{the_amplituhedron}.
Hodges' original motivation was to understand the cancellation of
\textit{spurious poles} in compact, gauge-invariant expressions for
tree-level gluon amplitudes. In the supersymmetric setting, the relevant
polytopes become special cases of the tree Amplituhedron.

\subsubsection{Hodges' six-point example}

Let us first consider six-particle NMHV amplitudes in Yang--Mills theory, without supersymmetry.
There are three independent helicity configurations modulo parity and
dihedral symmetry. We focus on the six-point color-ordered gluon amplitude
with helicity configuration \((---+++)\), which take the form
\begin{equation}\label{eq:A6_YM}
\begin{aligned}
	A_6&[1^-,2^-,3^-,4^+,5^+,6^+] = \\[6pt]
		& \frac{\langle 12\rangle^4\langle 23\rangle^4}
{\langle 12\rangle\langle 23\rangle\langle 34\rangle
 \langle 45\rangle\langle 56\rangle\langle 61\rangle} 
\frac{1}{\langle 1235\rangle}
\left(
\frac{\langle 1356\rangle^3}
{\langle 2356\rangle\langle 1236\rangle\langle 1256\rangle}
+\frac{\langle 1345\rangle^3}
{\langle 1234\rangle\langle 1245\rangle\langle 2345\rangle}
\right) \, .
\end{aligned}
\end{equation}
Here \(\langle ij\rangle\) denotes the spinor-helicity angle bracket, while
\(\langle ijkl\rangle=\langle z_i z_j z_k z_l\rangle\) denotes a
momentum-twistor four-bracket. We suppress this distinction in the notation
when no confusion should arise.

\begin{figure}[pos=t]
\centering
\begin{tikzpicture}[
    thick,
    line cap=round,
    line join=round,
    every node/.style={font=\small},
    blob/.style={circle,draw,minimum size=8mm,inner sep=0pt},
    vertex/.style={circle,fill,inner sep=1.1pt}
]


\begin{scope}[shift={(0,3.8)}, scale=0.82]

\node[vertex] (v1) at (0,0) {};
\node[vertex] (v2) at (1.2,0) {};
\node[vertex] (v3) at (2.4,0) {};
\node[vertex] (v4) at (3.6,0) {};

\draw (v1) -- (v2);
\draw (v2) -- (v3);
\draw (v3) -- (v4);

\draw (v1) -- ++(-0.85,0.65) node[left] {$1^-$};
\draw (v1) -- ++(-0.85,-0.65) node[left] {$2^-$};
\draw (v2) -- ++(0,0.95) node[above] {$3^-$};
\draw (v3) -- ++(0,-0.95) node[below] {$4^+$};
\draw (v4) -- ++(0.85,0.65) node[right] {$5^+$};
\draw (v4) -- ++(0.85,-0.65) node[right] {$6^+$};

\node at (2,-2.2) {(i)};

\end{scope}

\begin{scope}[shift={(1.8,-1.4)}, scale=0.82]

\node[vertex] (c) at (0,0) {};
\node[vertex] (u) at (0,1.15) {};
\node[vertex] (dl) at (-1.0,-0.75) {};
\node[vertex] (dr) at (1.0,-0.75) {};

\draw (c) -- (u);
\draw (c) -- (dl);
\draw (c) -- (dr);

\draw (u) -- ++(-0.75,0.65) node[left] {$1^-$};
\draw (u) -- ++(0.75,0.65) node[right] {$2^-$};

\draw (dl) -- ++(-0.95,-0.20) node[left] {$3^-$};
\draw (dl) -- ++(-0.45,-0.85) node[left] {$6^+$};

\draw (dr) -- ++(0.95,-0.20) node[right] {$4^+$};
\draw (dr) -- ++(0.45,-0.85) node[right] {$5^+$};

\node at (0,-2.0) {(ii)};

\end{scope}


\begin{scope}[shift={(8.7,3.8)}, scale=0.72]

\node[blob] (L) at (0,0) {};
\node[blob] (R) at (4.6,0) {};

\draw (L) -- (R);
\node at (1.55,0.36) {$+$};
\node at (3.05,0.36) {$-$};

\draw (L) -- ++(-1.9, 1.1) node[left] {$2^-$};
\draw (L) -- ++(-1.9,-1.1) node[left] {$\hat{3}^{\, -}$};

\draw (R) -- ++( 1.9, 1.1) node[right] {$\hat{4}^{\, +}$};
\draw (R) -- ++( 1.9,-1.1) node[right] {$5^+$};
\draw (R) -- ++( 0.0, 2.2) node[above] {$1^-$};
\draw (R) -- ++( 0.0,-2.2) node[below] {$6^+$};

\node at (2.3,-2.6) {(a)};

\end{scope}

\begin{scope}[shift={(8.7,-1.3)}, scale=0.72]

\node[blob] (L2) at (0,0) {};
\node[blob] (R2) at (4.6,0) {};

\draw (L2) -- (R2);
\node at (1.55,0.36) {$+$};
\node at (3.05,0.36) {$-$};

\draw (L2) -- ++(-1.9, 1.1) node[left] {$1^-$};
\draw (L2) -- ++(-1.9,-1.1) node[left] {$6^+$};
\draw (L2) -- ++( 0.0, 2.2) node[above] {$2^-$};
\draw (L2) -- ++( 0.0,-2.2) node[below] {$\hat{3}^{\, -}$};

\draw (R2) -- ++(1.9, 1.1) node[right] {$\hat{4}^{\, +}$};
\draw (R2) -- ++(1.9,-1.1) node[right] {$5^+$};

\node at (2.3,-2.5) {(b)};

\end{scope}

\end{tikzpicture}
\caption{On the left, (i) and (ii) are two of the many Feynman diagrams contributing to
\(A_6[1^-,2^-,3^-,4^+,5^+,6^+]\). On the right, (a) and (b) are the only two BCFW
diagrams contributing to the \([3,4\rangle\)-shift recursion.}
\label{fig:n6_hodg}
\end{figure}
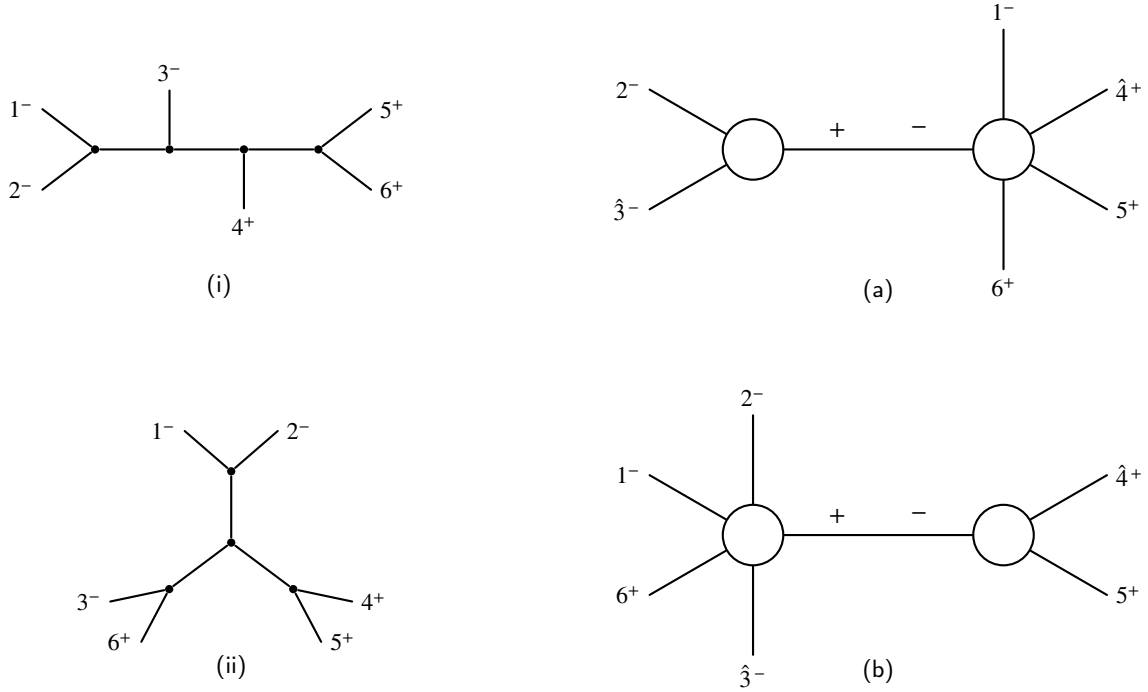

The configuration \((---+++)\) is a \textit{split-helicity} configuration,
meaning that the negative-helicity gluons and positive-helicity gluons appear
in consecutive blocks in the color ordering. Equivalent representations of
~\eqref{eq:A6_YM} were first obtained in~\cite{Mangano:1987xk,Berends:1987me},
while the particular form displayed above appeared in
~\cite{Roiban:2004yf,Britto:2004ap}. A direct evaluation from Feynman diagrams
is cumbersome: many planar tree diagrams with cubic and quartic interactions
contribute to \(A_6[1^-,2^-,3^-,4^+,5^+,6^+]\). In~\cite{De:2024bpk}, the
authors provide \texttt{Mathematica} code for such computations. By contrast,
a BCFW \([3,4\rangle\)-shift yields an expression involving only two terms
~\cite{Britto:2004ap,Britto:2005fq,Hodges:2005bf}; see
Figure~\ref{fig:n6_hodg}.

\subsubsection{Spurious poles and polytope volumes}

The crucial point is that the pole \(\langle 1235\rangle\) in
~\eqref{eq:A6_YM} is \textit{spurious}: on its vanishing locus, the term in
parentheses in~\eqref{eq:A6_YM} also vanishes, so the amplitude is finite
away from its physical poles. This follows from the Schouten identity
\begin{equation}
	\langle abcd\rangle \langle abef\rangle
+\langle abce\rangle \langle abfd\rangle
+\langle abcf\rangle \langle abde\rangle
=0 \, ,
\end{equation}
which is a consequence of the linear dependence of any five vectors in a
four-dimensional vector space. With
\begin{equation}
a=3,\qquad b=5,\qquad c=1,\qquad d=2,\qquad e=4,\qquad f=6 \, ,
\end{equation}
the Schouten identity relates the two terms in parentheses in
~\eqref{eq:A6_YM} and shows that their numerator vanishes when
\(\langle 1235\rangle=0\).

\begin{figure}[pos=t]
\centering
\begin{tikzpicture}[line cap=round,line join=round,scale=1.0]

\begin{scope}[shift={(7,0.1)}, every node/.style={font=\large}]
  \coordinate (A) at (0.0,0.0);    
  \coordinate (B) at (1.4,2.6);    
  \coordinate (C) at (3.3,2.9);    
  \coordinate (D) at (5.1,1.5);    
  \coordinate (E) at (4.5,-0.4);   
  \coordinate (F) at (2.5,2.0);    

  \draw[very thick] (A)--(B)--(C)--(D)--(E)--cycle;

  \draw[very thick] (A)--(D);
  \draw[very thick] (B)--(F)--(C);
  \draw[very thick] (F)--(E);

  \node[below left]  at (A) {145};
  \node[above left]  at (B) {165};
  \node[above]       at (C) {365};
  \node[right]       at (D) {345};
  \node[below right] at (E) {134};
  \node[below left]  at (F) {136};
\end{scope}

\begin{scope}[shift={(0,0)}, every node/.style={font=\Large}]
  \coordinate (v1) at (0.0,1.1);
  \coordinate (v4) at (1.6,3.0);
  \coordinate (v3) at (3.2,2.1);
  \coordinate (v5) at (3.4,0.8);
  \coordinate (v6) at (1.8,-0.25);

  \draw[very thick] (v1)--(v4)--(v3)--(v5)--(v6)--cycle;

  \draw[very thick] (v4)--(v5);
  \draw[very thick] (v1)--(v5);

  \draw[very thick,dashed] (v6)--(v3);
  \draw[very thick] (v1)--(v3);

  \node[left]  at (v1) {1};
  \node[above] at (v4) {4};
  \node[right] at (v3) {3};
  \node[right] at (v5) {5};
  \node[below] at (v6) {6};
\end{scope}

\end{tikzpicture}
\caption{On the left, the polytope whose canonical function computes the
non-Parke--Taylor part of \(A_6[1^-,2^-,3^-,4^+,5^+,6^+]\); on the right, its convex dual polytope \(P_6\)~\cite{Hodges:2009hk}.}
\label{fig:Hodg_pol}
\end{figure}
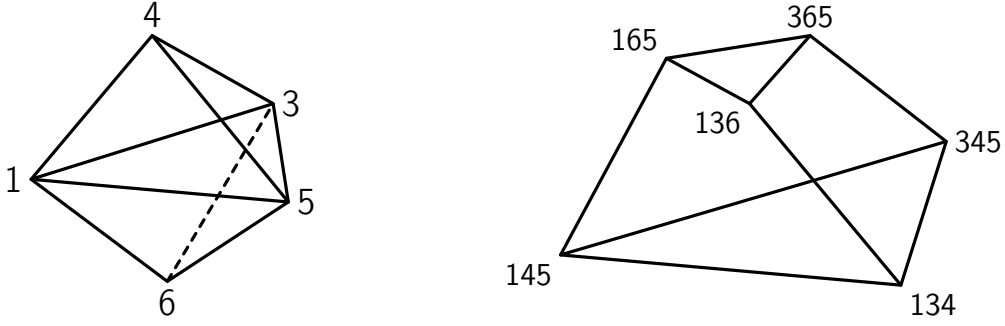

Hodges~\cite{Hodges:2009hk} explained this cancellation geometrically.
Stripping off the first rational factor in~\eqref{eq:A6_YM}, we can write the
remaining part as
\begin{equation}\label{eq:forms_sum}
	[1356](2) + [1345](2) = [13456](2) \, ,
\end{equation}
where
\begin{equation}\label{eq:tetra_function}
	[abcd](x)
	=
	\frac{\langle abcd\rangle^3}
	{\langle xabc\rangle
	 \langle xbcd\rangle
	 \langle xcda\rangle
	 \langle xdab\rangle}
\end{equation}
is the canonical function of the three-dimensional projective simplex given
by the convex hull of \(z_a,z_b,z_c,z_d\), see
~\eqref{eq:can_form_simplex}. In the second equality in~\eqref{eq:forms_sum},
we used additivity of canonical forms with respect to triangulations, as
discussed in Section~\ref{sec:Triangulations}. Equation~\eqref{eq:forms_sum}
is therefore the canonical function of a projective polytope in
\(\mathbb P^3\), given by the convex hull of the five points
\begin{equation}
z_1,\ z_3,\ z_4,\ z_5,\ z_6 \, .
\end{equation}
This polytope is a bipyramid with triangular base on the vertices
\(z_1,z_3,z_5\); see Figure~\ref{fig:Hodg_pol}. In Hodges' notation, this
polytope is the convex dual of \(P_6\). As discussed in
Section~\ref{sec:Integral Representations}, the dual-volume integral in
~\eqref{eq:volume_gauge_S}, performed over \(P_6\), yields the canonical
function \([13456]\):
\begin{equation}\label{eq:Q6_form}
	[13456](2)
	=
	\int_{P_6}
	\frac{\langle y\,\mathrm{d}^3y\rangle}{(z_2\cdot y)^4} \, .
\end{equation}

The cancellation of the spurious pole \(\langle 1235\rangle\) in
~\eqref{eq:A6_YM} is now geometric. The plane
\(\langle x135\rangle=0\) in \(\mathbb P^3\) is not a facet of the polytope
\([13456]\). Hence the canonical function \([13456](x)\) has no pole along
this locus, and neither does its evaluation at \(x=z_2\).

This geometric understanding also explains why the same amplitude can be
represented in different ways. For example, the polytope \([13456]\) can be
triangulated into three simplices instead of two, yielding
\begin{equation}\label{eq:forms_sum_2}
	[13456](2)
	=
	[1346](2)+[3546](2)+[5146](2) \, .
\end{equation}
This gives an equivalent representation of~\eqref{eq:A6_YM},
corresponding to a different BCFW shift than the \([3,4\rangle\)-shift.

Analogously to the \(n=6\) case, Hodges~\cite{Hodges:2009hk} constructed a
polytope \(P_n\) in \(\mathbb P^3\) whose dual volume computes the NMHV gluon
amplitude with split-helicity configuration
\begin{equation}
(---+\cdots+) \, .
\end{equation}
Moreover, he explained how to extend this construction to the NMHV
superamplitude of \(\mathcal N=4\) SYM. The core idea is to add one extra
bosonic dimension corresponding to the fermionic degrees of freedom. In this
way, one moves from ordinary momentum-twistor space to bosonized
super-momentum-twistor space. The resulting geometry is a four-dimensional
projective polytope, while the split-helicity gluon amplitudes are recovered
from suitable three-dimensional projections.

\subsubsection{Cyclic polytopes}

As explained in Chapter~\ref{ch:Scattering Amplitudes}, see
~\eqref{eq:ampl_R_dec}, the \(n\)-point tree-level NMHV sector corresponds to
\(k=1\). It is encoded by the ratio function $\mathcal R^{(1)}_n(\mathcal Z_1,\dots,\mathcal Z_n)$, which is rational in the bosonic momentum twistors \(z_i\) and polynomial of
degree \(4\) in the fermionic variables \(\chi_i\). Super-BCFW recursion gives
\begin{tcolorbox}[definitionbox]
\textbf{BCFW representation of the tree-level NMHV ratio function.}

\begin{equation}\label{eq:R1}
	\mathcal{R}_{1,n}
	=
	\sum_{i=2}^{n-2}
	\sum_{j=i+2}^{n-1}
	[1,i,i+1,j,j+1] \, .
\end{equation}
\end{tcolorbox}\noindent
Each \([a,b,c,d,e]\) is an \textit{\(R\)-invariant}, given in
super-momentum twistors by
\begin{equation}\label{eq:Rinv}
[a,b,c,d,e]
=
\frac{
\delta^{(4)}\big(
\langle abcd\rangle \chi_e
+ \langle bcde\rangle \chi_a
+ \langle cdea\rangle \chi_b
+ \langle deab\rangle \chi_c
+ \langle eabc\rangle \chi_d
\big)
}{
\langle abcd\rangle
\langle bcde\rangle
\langle cdea\rangle
\langle deab\rangle
\langle eabc\rangle
} \, .
\end{equation}
The terminology reflects the fact that~\eqref{eq:Rinv} is invariant under the
R-symmetry \({\rm SU}(4)\) acting on the \(\chi_i\). More strongly, each
\(R\)-invariant is Yangian invariant.

We now rewrite~\eqref{eq:Rinv} in terms of bosonized momentum twistors
\begin{equation}
Z_i=(z_i,\phi\cdot\chi_i)\in\mathbb P^4,
\qquad
\phi\cdot\chi_i:=\sum_{A=1}^4\phi^A\chi_i^A \, ,
\end{equation}
as introduced in Section~\ref{sec:Momentum Twistors}. The determinant $\langle Z_aZ_bZ_cZ_dZ_e\rangle$
is a \(5\times5\) determinant and is linear in the auxiliary Grassmann
variables \(\phi^A\). Equivalently, the coefficient of
\(\phi^1\phi^2\phi^3\phi^4\) in $\langle Z_aZ_bZ_cZ_dZ_e\rangle^4$
reproduces the Grassmann numerator of the \(R\)-invariant, up to the
conventional normalization of the fermionic delta function. Thus the
bosonized form of the \(R\)-invariant is
\begin{equation}\label{eq:R_inv_2}
	[a,b,c,d,e]
	=
	\left.
	\frac{\langle abcde\rangle^4}
	{\langle Yabcd\rangle
	 \langle Ybcde\rangle
	 \langle Ycdea\rangle
	 \langle Ydeab\rangle
	 \langle Yeabc\rangle}
	\right|_{Y=Y_0} \, ,
\end{equation}
where all brackets in~\eqref{eq:R_inv_2} are five-brackets of bosonized
twistors \(Z_i\), and
\begin{equation}\label{eq:Y0}
Y_0=[0:0:0:0:1]
\end{equation}
is the reference point selecting the original fermionic component. Indeed, for
example,
\begin{equation}
\langle Y_0 \,d e a b\rangle
=
\langle d e a b\rangle
\end{equation}
is the ordinary momentum-twistor four-bracket. Therefore the
\(R\)-invariant~\eqref{eq:R_inv_2} is the canonical function of the
four-dimensional simplex with vertices $Z_a,Z_b,Z_c,Z_d,Z_e\in\mathbb P^4$
evaluated at \(Y=Y_0\).

By relaxing \(Y=Y_0\) in~\eqref{eq:R_inv_2}, we extend
\(\mathcal R_{1,n}\) to a function $\mathcal R_{1,n}(Y;Z)$ of a generic point \(Y\in\mathbb P^4\). The resulting function is a sum of
canonical functions of four-dimensional projective simplices as in~\eqref{eq:R1}. A natural
question is whether these simplices triangulate a single polytope in
\(\mathbb P^4\).

A first hint comes from the possible singularities of tree-level amplitudes
in planar \(\mathcal N=4\) SYM. They can occur only when planar propagators
go on shell. In momentum-twistor variables, these loci are
\begin{equation}\label{eq:iijj_2}
	\langle z_i z_{i+1} z_j z_{j+1}\rangle
	=
	\langle Y_0 \,i\,i{+}1\,j\,j{+}1\rangle
	=0 \, .
\end{equation}
There are \(n(n-3)/2\) such planar channels. It turns out that there is a
unique combinatorial type of projective polytope in \(\mathbb P^4\) whose
facets are of the form~\eqref{eq:iijj_2}: the \emph{cyclic polytope} $C_{4,n}(Z)$
with vertices \(Z_1,\dots,Z_n\). This follows from Gale's evenness
criterion~\cite{ziegler2012lectures}. In order for all \(Z_i\) to be vertices
of the polytope, and for the desired brackets to define its facets, one
imposes the positivity conditions
\begin{equation}\label{eq:cyclic_positive}
	\langle Z_a Z_b Z_c Z_d Z_e\rangle>0,
	\qquad
	1\leq a<b<c<d<e\leq n \, .
\end{equation}
The cyclic polytope \(C_{4,n}(Z)\) is simplicial: each facet is a simplex.
Every simplicial polytope admits simple triangulations: one may
fix a vertex \(v\) and take the collection of simplices spanned by \(v\) and
the facets not adjacent to \(v\).

The \(R\)-invariants in~\eqref{eq:R1} give precisely such a triangulation of
\(C_{4,n}(Z)\), with fixed vertex \(Z_1\). By additivity of canonical forms
under triangulations, see Section~\ref{sec:Triangulations}, we find
\begin{tcolorbox}[resultbox]
\textbf{Cyclic polytope and NMHV ratio function.}

\begin{equation}\label{eq:nmhv_cyclic_polytope}
	\mathcal R_{1,n}(Y;Z)
	=
	\Omega_{C_{4,n}(Z)}(Y) \, .
\end{equation}
\end{tcolorbox}\noindent
Evaluating at \(Y=Y_0\) as in~\eqref{eq:Y0} recovers the NMHV ratio function. We can
also fix any other vertex instead of \(Z_1\), obtaining a different
triangulation of the same cyclic polytope. These different triangulations
correspond to different BCFW representations.

\subsubsection{Examples: \(n=5\) and \(n=6\)}

\begin{figure}[pos=t]
\centering
\includegraphics[width=0.3\textwidth]{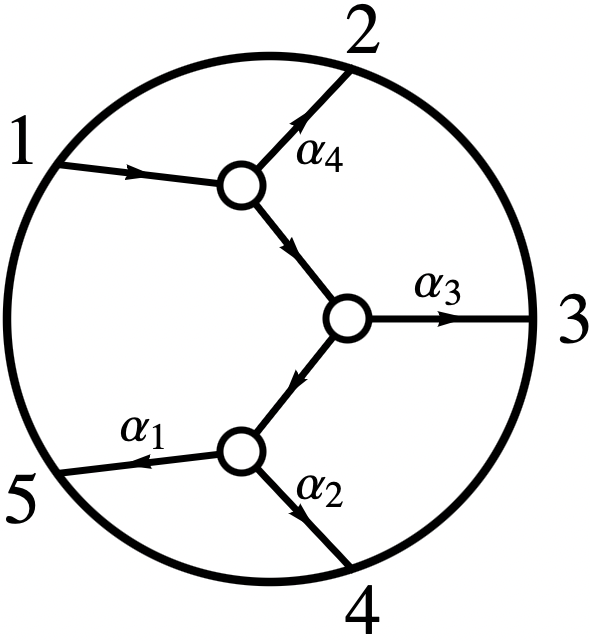}
\caption{On-shell diagram for the \(R\)-invariant \([1,2,3,4,5]\).}
\label{fig:graph_2}
\end{figure}

For \(n=5\), the cyclic polytope \(C_{4,5}(Z)\) is a simplex, and
\begin{equation}
\mathcal R^{(1)}_5(Y;Z)=[1,2,3,4,5] \, .
\end{equation}
The BCFW recursion yields a single on-shell diagram, shown in
Figure~\ref{fig:graph_2}. In a suitable perfect orientation, the boundary
measurement gives a \(1\times5\) matrix
\begin{equation}
C(\alpha)=(1,\alpha_4,\alpha_3,\alpha_2,\alpha_1) \, ,
\end{equation}
with all \(\alpha_i \geq 0\). The map $Y=\sum_{i=1}^5 C_i(\alpha)Z_i$ parametrizes the simplex \([1,2,3,4,5]\). The on-shell form is
\begin{equation}
\frac{\mathrm{d}\alpha_1}{\alpha_1}
\frac{\mathrm{d}\alpha_2}{\alpha_2}
\frac{\mathrm{d}\alpha_3}{\alpha_3}
\frac{\mathrm{d}\alpha_4}{\alpha_4} \, ,
\end{equation}
which is the canonical form of the simplex in this parametrization.

\begin{figure}[pos=t]
\centering
\includegraphics[width=0.8\textwidth]{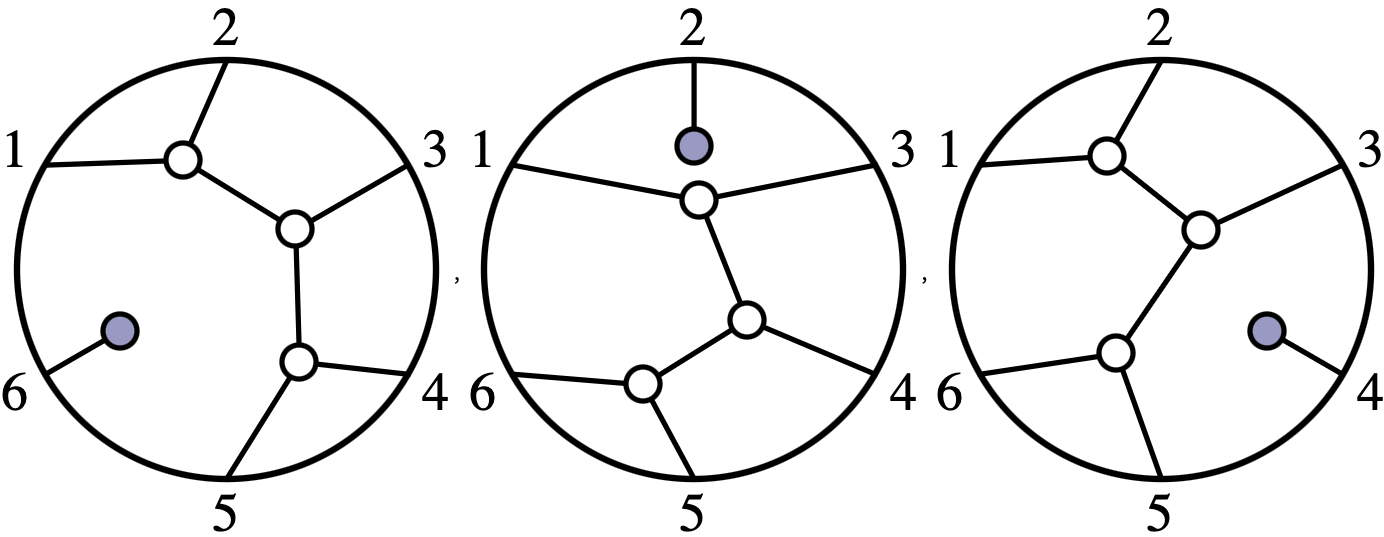}
\caption{Plabic graphs for the BCFW terms in the six-point NMHV ratio
function \(\mathcal R^{(1)}_6\).}
\label{fig:BCFW_cells_k1}
\end{figure}

For \(n=6\), the BCFW \([1,2\rangle\)- and \([2,3\rangle\)-shifts yield two
decompositions into three terms:
\begin{equation}\label{eq:R6}
\begin{aligned}
	\mathcal R^{(1)}_6(Y;Z)
	=
	\Omega_{C_{4,6}(Z)}(Y)
	&=
	[1,2,3,4,5]
	+
	[1,2,3,5,6]
	+
	[1,3,4,5,6]
	\\
	&=
	[2,3,4,6,1]
	+
	[2,3,4,5,6]
	+
	[2,4,5,6,1] .
\end{aligned}
\end{equation}
This representation is vastly simpler than a direct Feynman-diagram expansion
of the same superamplitude. Geometrically,~\eqref{eq:R6} arises from two
different triangulations of the cyclic polytope \(C_{4,6}(Z)\) into
simplices. The on-shell diagrams corresponding to the three terms in the
first row of~\eqref{eq:R6} are displayed in Figure~\ref{fig:BCFW_cells_k1}.
The on-shell form of each diagram is the canonical form of the associated
simplex, namely the corresponding \(R\)-invariant.

\subsubsection{Recovering Hodges' polytope}

Recall that the split-helicity amplitude \((---+++)\) is extracted from the
NMHV superamplitude by taking the coefficient of
\begin{equation}
\prod_{A=1}^4 \eta_1^A\eta_2^A\eta_3^A \, .
\end{equation}
Geometrically, for this split-helicity component, the extraction can be
understood in terms of the \emph{vertex figure of \(C_{4,6}(Z)\) at \(Z_2\)}. This
is the three-dimensional projective polytope obtained by taking a local slice
of \(C_{4,6}(Z)\) near \(Z_2\).

The facets of the vertex figure come from the facets of \(C_{4,6}\) adjacent
to \(Z_2\):
\begin{equation}
	[1234],
	\quad
	[1245],
	\quad
	[1256],
	\quad
	[2345],
	\quad
	[2356],
	\quad
	[2361] \, .
\end{equation}
After deleting the common label \(2\), they become the facets of the
three-dimensional polytope dual to Hodges' \(P_6\). Its canonical function is
the one appearing in~\eqref{eq:Q6_form}, and hence it gives the nontrivial
part of the six-point split-helicity amplitude \((---+++)\) in
~\eqref{eq:A6_YM}.

This form can be obtained directly from the triangulations in~\eqref{eq:R6}.
One keeps only the simplices containing \(Z_2\), deletes the label \(2\), and
then evaluates the resulting canonical function at \(Y=Z_2\). Applying this
procedure to the first row of~\eqref{eq:R6} gives~\eqref{eq:forms_sum}, while
applying it to the second row gives~\eqref{eq:forms_sum_2}.

Similarly, the vertex figure at \(Z_i\) of \(C_{4,n}(Z)\) yields the dual
polytope of Hodges' \(P_n\), whose canonical function yields the
\(n\)-point NMHV split-helicity amplitude with negative-helicity gluons at
\(i-1,i,i+1\). Other split-helicity configurations, in which the
negative-helicity gluons are not consecutive, do not appear to have such a
simple interpretation in terms of vertex figures of \(C_{4,n}(Z)\). In the
following sections we therefore focus on the geometry associated with the full
superamplitude.

\subsection{Tree Amplituhedra}\label{sec:Tree Amplituhedra}

\subsubsection{The positive Grassmannian}

We begin by recalling the positive Grassmannian and some elements of its
combinatorics; see~\cite{Postnikov:2006kva,TNN_grassmannian,Lam:_PG_notes}
for introductory accounts. The Grassmannian
\(\Gr_{\mathbb K}(k,n)\) is the space of \(k\)-dimensional linear
subspaces of \(\mathbb K^n\), where for us \(\mathbb K=\mathbb R\) or
\(\mathbb C\). Equivalently, it is the space of \((k-1)\)-dimensional
projective linear subspaces of \(\mathbb P^{n-1}_{\mathbb K}\). We denote the
complex Grassmannian by $\Gr(k,n):=\Gr_{\mathbb C}(k,n)$.
In particular, $\Gr(1,n)=\mathbb P^{n-1}$.

One may identify \(\Gr_{\mathbb K}(k,n)\) with the space of
\(k\times n\) matrices of rank \(k\), modulo left multiplication by
\({\rm GL}(k,\mathbb K)\). From this description one sees that
\(\Gr_{\mathbb K}(k,n)\) is a smooth manifold of dimension \(k(n-k)\).
On the open chart where the first \(k\) columns form an invertible matrix, the
\({\rm GL}(k)\)-action can be used to fix this block to the identity. This
gives the local parametrization
\begin{equation}
	\mathbb{K}^{k(n-k)} \hookrightarrow \Gr(k,n),
	\qquad
	X=(x_{ij}) \mapsto
	\begin{pmatrix}
1 & 0 & \cdots & 0 & x_{11} & x_{12} & \cdots & x_{1,\,n-k} \\
0 & 1 & \cdots & 0 & x_{21} & x_{22} & \cdots & x_{2,\,n-k} \\
\vdots & \vdots & \ddots & \vdots & \vdots & \vdots & \ddots & \vdots \\
0 & 0 & \cdots & 1 & x_{k1} & x_{k2} & \cdots & x_{k,\,n-k}
\end{pmatrix} \, .
\end{equation}

We now recall how \(\Gr(k,n)\) is realized as a complex projective
variety. Denote by \([n]=\{1,\dots,n\}\), and let
\(\binom{[n]}{k}\) be the set of \(k\)-element subsets of \([n]\), always
written in increasing order. Let \(C\) be a \(k\times n\) matrix and let
\(I=\{i_1<\cdots<i_k\}\in\binom{[n]}{k}\). We write
\begin{equation}
p_I(C)=\langle C_{i_1}\cdots C_{i_k}\rangle
       =\langle i_1,\dots,i_k\rangle
\end{equation}
for the determinant of the \(k\times k\) matrix formed by the columns
\(C_{i_1},\dots,C_{i_k}\). These determinants are the
\textit{Plücker coordinates} of \(C\). Under left multiplication
\(C\mapsto U\cdot C\), with \(U\in{\rm GL}(k)\), every Plücker coordinate
rescales by
\begin{equation}
p_I(U\cdot C)=\det(U)\,p_I(C) \, .
\end{equation}
Thus the Plücker coordinates define the \textit{Plücker embedding}
\begin{equation}\label{eq:Pl_emb}
	\Gr(k,n)
	\hookrightarrow
	\mathbb{P}^{\binom{n}{k}-1},
	\qquad
	C \mapsto \left(p_I(C)\right)_{I\in\binom{[n]}{k}} \, .
\end{equation}
The image is cut out by the quadratic \textit{Plücker relations}
\begin{equation}\label{eq:Pl_rel}
\sum_{r=1}^{k+1} (-1)^{r-1}\,
p_{I\cup\{j_r\}}\,
p_{J\setminus\{j_r\}} = 0 \, ,
\end{equation}
where \(I\in\binom{[n]}{k-1}\), \(J=\{j_1<\cdots<j_{k+1}\}\), and any
Plücker coordinate with repeated indices is understood to vanish.

There is a distinguished semialgebraic subset of
\(\Gr_{\mathbb R}(k,n)\), called the \textit{positive Grassmannian},
defined by
\begin{tcolorbox}[definitionbox]
\textbf{Positive Grassmannian.}

\begin{equation}
	\Gr_{>0}(k,n)
	:=
	\left\{
	C\in\Gr_{\mathbb R}(k,n):
	p_I(C)>0
	\quad
	\forall\, I\in \smallbinom{[n]}{k}
	\right\} \, .
\end{equation}
\end{tcolorbox}\noindent
Similarly, the \textit{non-negative Grassmannian}
\(\Gr_{\geq 0}(k,n)\) is defined by requiring all Plücker coordinates to
be non-negative. For \(k=1\), these are the open and closed standard
projective simplices in \(\mathbb P^{n-1}\). For higher \(k\), they are
curved analogues of simplices inside the real Grassmannian
\(\Gr_{\mathbb R}(k,n)\).

\subsubsection{Positroid varieties and cells}

The non-negative Grassmannian is the Euclidean closure of
\(\Gr_{>0}(k,n)\) inside \(\Gr_{\mathbb R}(k,n)\)
~\cite{Postnikov:2006kva}, and it is homeomorphic to a closed ball
~\cite{galashinKarpLam2022ball}. Moreover, it admits a decomposition into
positroid cells:
\begin{equation}\label{eq:pos_dec}
	\Gr_{\geq 0}(k,n)
	=
	\bigcup_{\sigma}\Pi_{\sigma,>0} \, .
\end{equation}
We now recall one convenient way to label these cells.

A function \(\sigma:\mathbb Z\to\mathbb Z\) is called a
\textit{bounded affine permutation of type \((k,n)\)} if it is a bijection
satisfying
\begin{equation}\label{eq:bound_aff_perm}
	\sum_{i=1}^n\bigl(\sigma(i)-i\bigr)=kn,
	\qquad
	\sigma(i+n)=\sigma(i)+n,
	\qquad
	i\leq \sigma(i)\leq i+n \, ,
\end{equation}
for every \(i\in\mathbb Z\). The finite data
\((\sigma(1),\dots,\sigma(n))\) determines the whole affine permutation by
periodicity.
To such a \(\sigma\), one can associate an open stratum
\begin{equation}\label{eq:pos_def}
	\left\{
	C\in\Gr(k,n):
	C_i\in{\rm span}(C_{i+1},C_{i+2},\dots,C_{\sigma(i)})
	\right\} \, ,
\end{equation}
where column indices are understood cyclically. This is a compact way of
encoding the corresponding cyclic rank conditions. The \textit{positroid
variety} \(\Pi_{\sigma}\) is the Zariski closure of~\eqref{eq:pos_def}.
The associated \emph{positroid cell} is
\begin{equation}
\Pi_{\sigma,\geq 0}:=\Pi_{\sigma}\cap\Gr_{\geq 0}(k,n) \, ,
\end{equation}
and we also write $\Pi_{\sigma,>0}$ for its relative interior.

For example, the bounded affine permutation
\begin{equation}\label{eq:top_perm}
	\sigma_{k,n}
	=
	(k+1,k+2,\dots,n,n+1,\dots,n+k)
\end{equation}
is of type \((k,n)\). For this permutation, no Plücker coordinate is forced
to vanish, and its positive part is precisely the top cell:
\begin{equation}
	\Pi_{\sigma_{k,n},> 0}
	=
	\Gr_{> 0}(k,n) \, .
\end{equation}

\begin{figure}[pos=t]
\centering

\newcommand{\topcellpanel}[4]{%
\begin{minipage}[t]{0.31\textwidth}
\centering
\parbox[c][2.2em][c]{\linewidth}{\centering
\(\Gr_{>0}(#1,#2)\)
}%

\vspace{0.5em}

\includegraphics[width=\linewidth,height=0.23\textheight,keepaspectratio]{#3}

\vspace{0.5em}

\parbox[c][2.2em][c]{\linewidth}{\centering
\(\displaystyle \sigma=#4\)
}%
\end{minipage}%
}

\topcellpanel{2}{5}{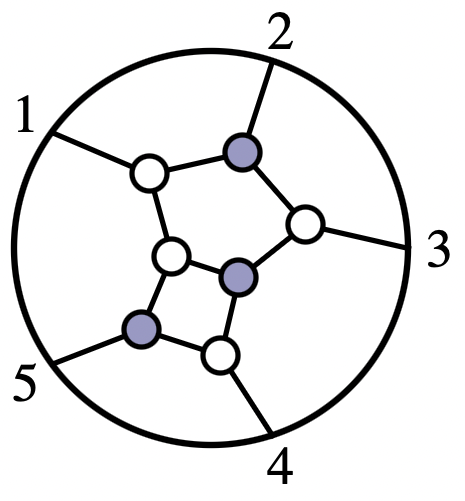}{(3,4,5,6,7)}
\hfill
\topcellpanel{3}{6}{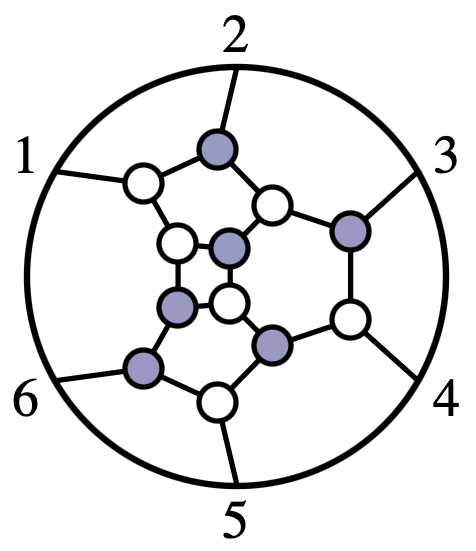}{(4,5,6,7,8,9)}
\hfill
\topcellpanel{4}{7}{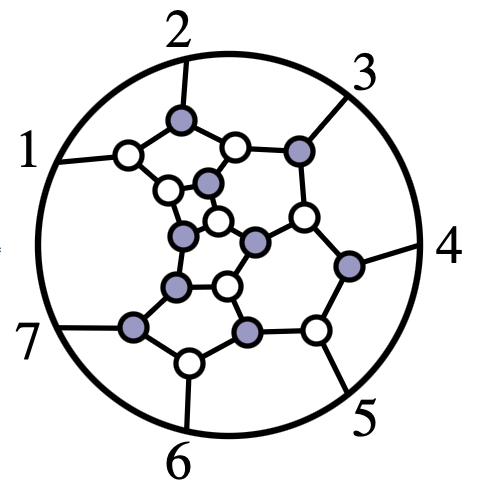}{(5,6,7,8,9,10,11)}

\caption{Plabic graphs for some top cells and their bounded affine permutations.}
\label{fig:top_cells}
\end{figure}

Positroids are equivalently labelled by bounded affine permutations, or by
equivalence classes of reduced plabic graphs, namely planar bicolored graphs
embedded in a disk~\cite{Postnikov:2006kva}. We already encountered  a refined version of these graphs: the on-shell
diagrams in Section~\ref{sec:On-Shell Diagrams and Grassmannian Contours}, with the additional property that the allter are three-valent. Because of the connection to on-shell forms and the discussion in Section~\ref{sec:On-Shell Diagrams and Grassmannian Contours}, we will be mainly working with on-shell diagrams.
The permutation associated with an on-shell diagram is obtained by following
strands: starting from an external leg \(i\), move through the graph by
turning left at every white vertex and right at every black vertex. If the
strand exits at external leg \(j\), then the decorated permutation sends
\begin{equation}\label{eq:strand_rule}
\sigma(i)=j
\quad\text{if } j>i,
\qquad
\sigma(i)=j+n
\quad\text{if } j\leq i \, .
\end{equation}
Examples of on-shell diagrams associated with top-cell permutations of the form
~\eqref{eq:top_perm} are shown in Figure~\ref{fig:top_cells}.

Closed positroid cells define positive geometries
~\cite{knutsonLamSpeyer2013positroid,Grassmannian}: for every \(\sigma\)
as in~\eqref{eq:bound_aff_perm},
\begin{tcolorbox}[resultbox]
\textbf{Positroid cell as a positive geometry.}

\begin{equation}
	\left(
	\Pi_{\sigma},
	\Pi_{\sigma,\geq 0},
	\mathbf{\Omega}_{\sigma}
	\right)
	\quad
	\text{is a positive geometry} \, .
\end{equation}
\end{tcolorbox}\noindent
Their canonical forms can be obtained by taking iterated residues of the
canonical form of the non-negative Grassmannian. With our normalization, this
top form is
\begin{tcolorbox}[definitionbox]
\textbf{Canonical form of the positive Grassmannian.}

\begin{equation}\label{eq:form_pos_gr}
	\mathbf{\Omega}_{k,n}(C)
	=
	\frac{
	\bigwedge_{a=1}^k
	\langle C\,\mathrm{d}^{\,n-k}C_a\rangle
	}{
	\prod_{i=1}^n
	\langle i,i+1,i+2,\dots,i+k-1\rangle
	} \, ,
\end{equation}
\end{tcolorbox}\noindent
where we take indices modulo $n$, and the factors in the measure are
\begin{equation}\label{eq:meas_Gr}
	\langle C\,\mathrm{d}^{\,n-k}C_a\rangle
	:=\frac{1}{(n-k)!} \, 
	\varepsilon_{i_1\cdots i_n}
	C_1^{i_1}\cdots C_k^{i_k}\,
	\mathrm{d}C_a^{i_{k+1}}\wedge\cdots\wedge
	\mathrm{d}C_a^{i_n} \, ,
\end{equation}
where we use Einstein summation convention. Notice that the rational form~\eqref{eq:form_pos_gr} has constant numerator. In this sense, the
non-negative Grassmannian, and more generally positroid cells, are
\emph{simplex-like} positive geometries.

The Grassmannian \(\Gr(k,n)\), and more generally positroid varieties, carry a cluster structure~\cite{Scott2006,Muller_2017,GalashinLamPositroidCluster}. We will discuss cluster algebras in Section~\ref{sec:rationality-cluster-structures}; for now we only record one relevant feature. There is a preferred collection of parametrizations of the non-negative Grassmannian and its positroid cells by \textit{cluster charts}. These charts may be obtained from on-shell diagrams, as explained in Section~\ref{sec:On-Shell Diagrams and Grassmannian Contours}. Each such chart yields a collection of \emph{cluster coordinates} $\alpha_i$, as many as the dimension of the positroid variety $\Pi_\sigma$ under consideration. The positive part of the chart is where all cluster coordinates are positive, giving a parametrization of the positive part of the variety. The canonical form $\mathbf{\Omega}_\sigma$ becomes just the wedge product of all $\mathrm{d}\log(\alpha_i)$. Different choices of plabic graph give different cluster coordinates, and hence different positive parametrizations of the positroid variety.

\subsubsection{Example: \(\Gr_{\geq 0}(2,4)\).}
\label{eg:Gr24_strat}

	The Grassmannian \(\Gr(2,4)\) parametrizes lines in
	\(\mathbb P^3\). Under the Plücker embedding it is a hypersurface in
	\(\mathbb P^5\), cut out by the single Plücker relation
	\begin{equation}\label{eq:Pl_rel_2}
		\langle 12 \rangle \, \langle 34 \rangle
		+
		\langle 23 \rangle \, \langle 14 \rangle
		-
		\langle 13 \rangle \, \langle 24 \rangle
		=0 \, .
	\end{equation}
	Recall that we already encountered \(\Gr_{\geq 0}(2,4)\) in
	Example~\ref{eq:onshell_k2_n4}. The plabic graph in
	Figure~\ref{fig:graph_1} is associated with the top-dimensional cell
	\(\Gr_{>0}(2,4)\).

	Consider the parametrization
	\begin{equation}\label{eq:Ca_par}
	C(\alpha)
	=
	\begin{pmatrix}
		1 & 0 & -\alpha_1 & -\alpha_2  \\
		0 & 1 & \alpha_3 & \alpha_4
	\end{pmatrix} \, ,
	\end{equation}
	where the signs are chosen so that the Plücker coordinates take a simple
	form:
	\begin{equation}\label{eq:non_cl_pos}
	\begin{gathered}
		\langle 12\rangle=1,
		\qquad
		\langle 13\rangle=\alpha_3,
		\qquad
		\langle 14\rangle=\alpha_4,
		\\
		\langle 23\rangle=\alpha_1,
		\qquad
		\langle 24\rangle=\alpha_2,
		\qquad
		\langle 34\rangle=\alpha_2\alpha_3-\alpha_1\alpha_4.
	\end{gathered}
	\end{equation}
	This parametrizationis not a cluster chart, but it is useful to explore the boundary structure of $\Gr_{\geq 0}(2,4)$.  
	The non-negative Grassmannian is then described by
	\begin{equation}\label{eq:alpha_ineq}
		\alpha_1,\alpha_2,\alpha_3,\alpha_4\geq 0,
		\qquad
		\alpha_2\alpha_3-\alpha_1\alpha_4\geq 0 \, .
	\end{equation}
	The canonical form~\eqref{eq:form_pos_gr} becomes
	\begin{equation}
		\mathbf{\Omega}_{2,4}
		=
		\frac{
		\mathrm{d}\alpha_1\,\mathrm{d}\alpha_2\,\mathrm{d}\alpha_3\,\mathrm{d}\alpha_4
		}{
		\alpha_1\,\alpha_4\,
		(\alpha_2\alpha_3-\alpha_1\alpha_4)
		} \, ,
	\end{equation}
	up to the orientation convention. Notice that the denominator contains only
	the boundary divisors of the top cell; the inequalities
	\(\alpha_2\geq 0\) and \(\alpha_3\geq 0\) cut out lower-dimensional
	boundary strata rather than facets of the top cell.

	This parametrization does not reach the boundary \(\langle12\rangle=0\),
	nor lower-dimensional faces contained in it. Let us instead examine some
	boundaries that are visible in this chart. Consider the facet $\langle23\rangle=\alpha_1=0 $. On this boundary,
	\begin{equation}
\langle34\rangle=\alpha_2\alpha_3
\end{equation}
	factors into two components, \(\alpha_2=0\) and \(\alpha_3=0\). This is a
	distinctive feature of curved geometries: restrictions of boundary divisors
	can become reducible because of Plücker relations. By contrast, for
	projective polytopes, intersections of supporting hyperplanes are linear
	and irreducible.

\begin{figure}[pos=t]
\centering
\includegraphics[width=1.0\textwidth]{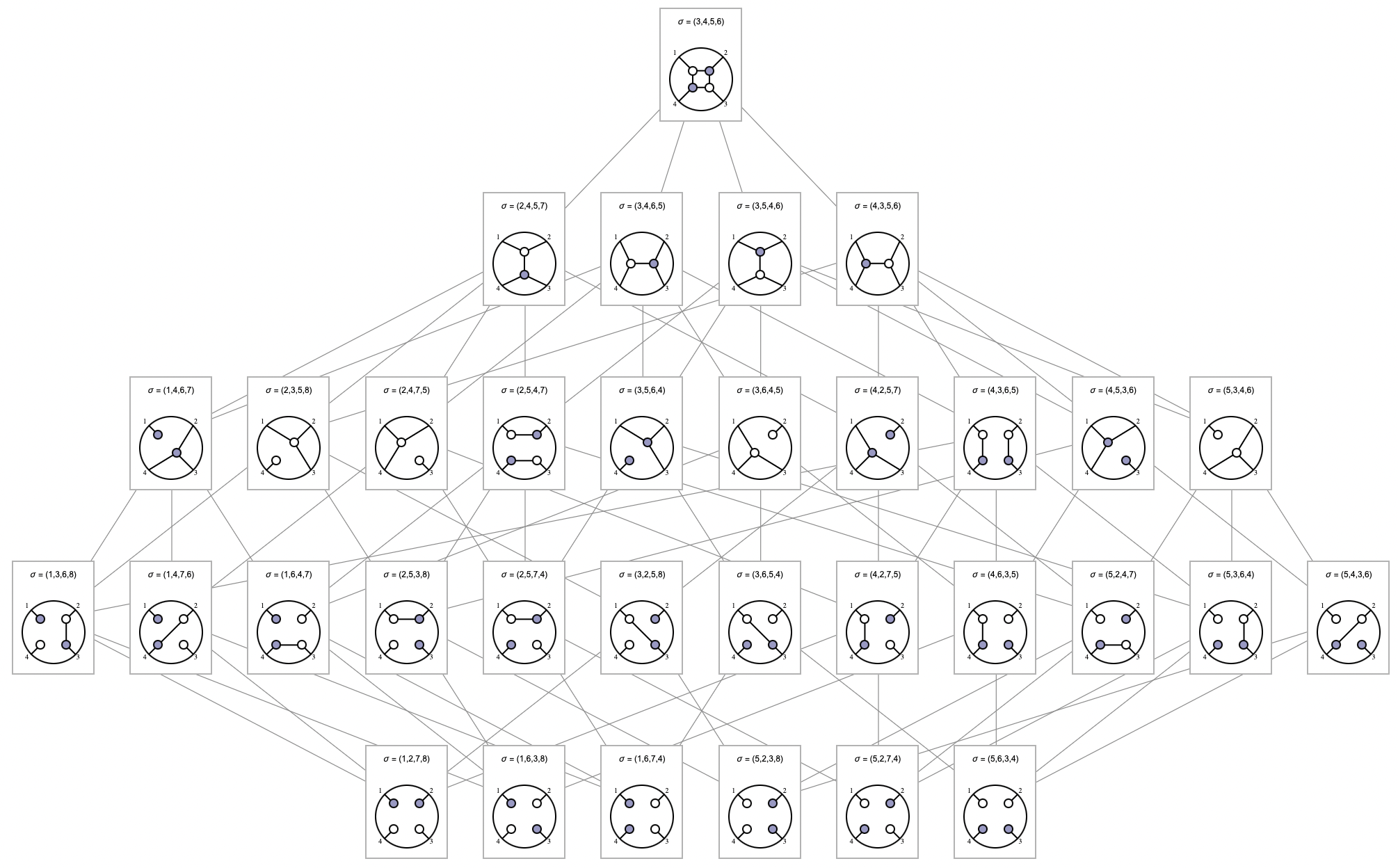}
\caption{Positroid stratification and boundary poset of
\(\Gr_{\geq 0}(2,4)\).}
\label{fig:Gr24_poset}
\end{figure}

	By dihedral symmetry, there are \(8\) codimension-two faces coming
	from components of
	\begin{equation}
\langle i{-}1,i\rangle=\langle i,i{+}1\rangle=0 \, ,
\end{equation}
	and two more coming from the loci
	\begin{equation}
\langle23\rangle=\langle14\rangle=0,
	\qquad
	\langle12\rangle=\langle34\rangle=0 \, .
\end{equation}
	Thus there are \(10\) codimension-two faces; see the third row of
	Figure~\ref{fig:Gr24_poset}.

	To explore the full boundary stratification computationally, one starts
	from the top-cell 
	\begin{equation}
\sigma=(3,4,5,6)
\end{equation}
	for \(k=2\) and \(n=4\). The command
	\texttt{boundary[$\sigma$]} in \texttt{Positroids.m}~\cite{Bourjaily:2012gy} outputs 	\begin{equation}
		(3,4,6,5),
		\qquad
		(2,4,5,7),
		\qquad
		(4,3,5,6),
		\qquad
		(3,5,4,6) \, .
	\end{equation}
	These correspond to the boundaries
	\(\langle i,i+1\rangle=0\) for \(i=1,\dots,4\), respectively. For
	example, for the first permutation, \(\sigma(1)=2\) implies that \(C_1\)
	is proportional to \(C_2\), and hence \(\langle12\rangle=0\).

	By repeated application of \texttt{boundary[]} one obtains the full
	positroid stratification of \(\Gr_{\geq 0}(2,4)\). It consists of
	\(33\) cells, whose count is recorded in the \(f\)-vector
	\begin{equation}
		f=(f_0,f_1,f_2,f_3,f_4)=(6,12,10,4,1) \, ,
	\end{equation}
	where \(f_r\) denotes the number of cells of dimension \(r\). The full
	list of cells and the boundary poset is displayed in
	Figure~\ref{fig:Gr24_poset}. The poset arranges cells vertically by
	codimension, starting with the top-dimensional cell at the top and ending
	with the zero-dimensional cells at the bottom. An edge between cells of
	codimension \(r\) and \(r+1\) indicates that the latter is a boundary of
	the former. This result appears already in the thesis of~\cite{WilliamsThesis2005}. 
We will later return to the positroid stratification of
	\(\Gr_{\geq 0}(2,4)\) from the geometric perspective of incidences of
	lines in \(\mathbb P^3\).

\subsubsection{The Amplituhedron map}

We now introduce the Amplituhedron. The guiding idea is that Amplituhedra are
images of the non-negative Grassmannian, in the same way that projective
polytopes are images of higher-dimensional simplices.

\begin{figure}[pos=t]
\centering
\begin{tikzpicture}[
    line join=round,
    line cap=round,
    >=Latex,
    vertex/.style={circle,fill=black,inner sep=1.6pt},
    lab/.style={font=\Large},
    edge/.style={black,thick},
    hidden/.style={black,thick,dashed},
    proj/.style={->,thick,gray!70}
]

\coordinate (A) at (0.0,3.8);
\coordinate (B) at (-2.4,0.2);
\coordinate (C) at (1.8,-1.0);
\coordinate (D) at (3.4,1.4);

\fill[blue!25]   (A) -- (B) -- (C) -- cycle;
\fill[orange!30] (A) -- (C) -- (D) -- cycle;

\draw[edge]   (A) -- (B) -- (C) -- cycle;
\draw[edge]   (A) -- (C);
\draw[edge]   (A) -- (D) -- (C);
\draw[hidden] (B) -- (D);

\node[vertex,label=above:$1$] at (A) {};
\node[vertex,label=left:$2$]  at (B) {};
\node[vertex,label=below:$3$] at (C) {};
\node[vertex,label=right:$4$] at (D) {};

\draw[proj] (4.5,1.2) -- node[above,black,font=\LARGE] {$\widetilde{Z}$} (6.2,1.2);

\coordinate (pA) at (7.3,2.7);
\coordinate (pB) at (6.9,-0.6);
\coordinate (pC) at (10.8,-0.6);
\coordinate (pD) at (11.3,1.8);

\fill[blue!25]   (pA) -- (pB) -- (pC) -- cycle;
\fill[orange!30] (pA) -- (pC) -- (pD) -- cycle;

\draw[edge] (pA) -- (pB) -- (pC) -- (pD) -- cycle;
\draw[edge] (pA) -- (pC);

\node[vertex,label=above:$Z_1$] at (pA) {};
\node[vertex,label=below:$Z_2$] at (pB) {};
\node[vertex,label=below:$Z_3$] at (pC) {};
\node[vertex,label=right:$Z_4$] at (pD) {};

\end{tikzpicture}
\caption{A three-dimensional simplex \(\Delta^3\) projects to a quadrilateral
in the plane under the linear map \(\widetilde{Z}\) in~\eqref{eq:porj_sympl}
with \(n=4\). The two colored facets of the simplex project to the two
triangles in the quadrilateral, colored accordingly.}
\label{fig:proj_pol}
\end{figure}
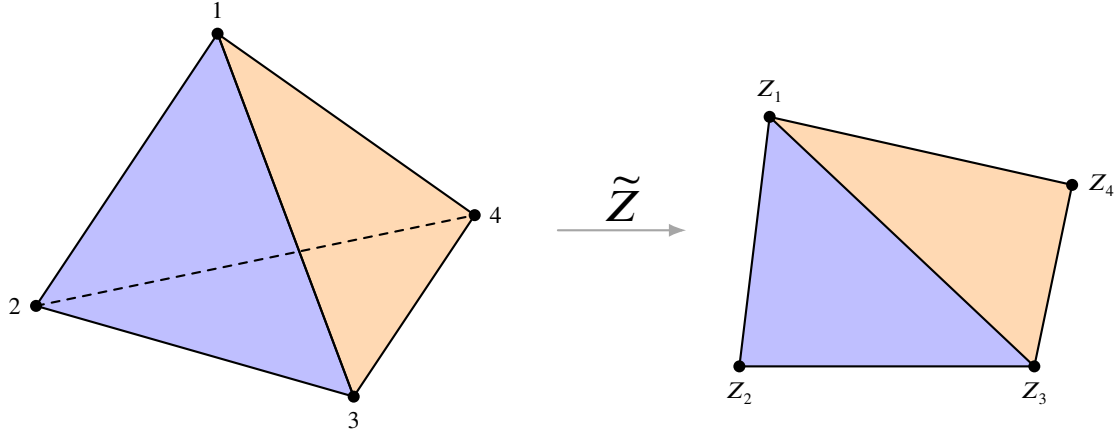

Indeed, any projective polytope
\begin{equation}
P={\rm conv}(Z_1,\dots,Z_n)\subseteq\mathbb P^m
\end{equation}
can be realized as the image of the standard projective simplex
\(\Delta^{n-1}=\mathbb P^{n-1}_{\geq 0}\) under the linear projection
\begin{equation}\label{eq:porj_sympl}
	\widetilde{Z}:\Delta^{n-1}\longrightarrow P,
	\qquad
	C\longmapsto Y:=C\cdot Z^\top=\sum_{i=1}^n C_iZ_i \, ,
\end{equation}
where \(Z=(Z_1,\dots,Z_n)\) is the \((m+1)\times n\) matrix formed by the
vertices. Surjectivity of \(\widetilde Z\) is precisely the statement that
every point of \(P\) is a convex combination of its vertices. If \(n>m+1\),
then \(\widetilde Z\) is not injective. It extends to a rational map $P^{n-1}\dashrightarrow\mathbb P^m$
defined away from the projectivized kernel $P(\ker Z)$ which has dimension \(n-m-2\).

In Section~\ref{sec:Hodges and Cyclic Polytopes}, we saw that tree-level
NMHV superamplitudes in planar \(\mathcal N=4\) SYM are encoded by canonical
functions of projective polytopes in \(\mathbb P^4\). Moreover, requiring
these polytopes to have physical boundaries $\langle Y\,i\,i{+}1\,j\,j{+}1\rangle=0$
led us to cyclic polytopes, defined by positivity conditions of the form
\begin{equation}\label{eq:pos_Z}
	\langle Z_{i_1}Z_{i_2}\cdots Z_{i_{k+m}}\rangle>0,
	\qquad
	1\leq i_1<i_2<\cdots<i_{k+m}\leq n \, .
\end{equation}
For NMHV amplitudes, this corresponds to \(k=1\) and \(m=4\).

This is a special case of a more general construction. Fix non-negative
integers \(k,m,n\) with \(m\geq1\) and \(n\geq k+m\). Let \(Z\) be a real
\((k+m)\times n\) matrix with positive maximal minors as in~\eqref{eq:pos_Z}. Consider the \textit{Amplituhedron map}
\begin{tcolorbox}[definitionbox]
\textbf{Tree Amplituhedron map.}

\begin{equation}\label{eq:Ampl_map}
	\widetilde{Z}:\Gr(k,n)\dashrightarrow\Gr(k,k+m),
	\qquad
	C\longmapsto Y:=C\cdot Z^\top \, .
\end{equation}
\end{tcolorbox}\noindent
This is a rational map on the complex Grassmannian, but positivity of \(Z\)
ensures that it is well-defined on \(\Gr_{\geq 0}(k,n)\). The \textit{tree Amplituhedron} is defined as

\begin{tcolorbox}[definitionbox]
\textbf{Tree Amplituhedron.}

\begin{equation}\label{eq:def_ampl}
	\mathcal A_{k,m,n}(Z)
	=
	\widetilde{Z}\big(\Gr_{\geq 0}(k,n)\big)
	\subseteq
	\Gr_{\mathbb R}(k,k+m) \, .
\end{equation}
\end{tcolorbox}\noindent
The case of main physical interest is \(m=4\), the value relevant to
four-dimensional planar \(\mathcal N=4\) SYM in momentum-twistor variables.
The parameter \(k\) labels the helicity sector \({\rm N}^k{\rm MHV}\), and
\(n\) is the number of external particles. The central conjecture is that
\(\mathcal A_{k,4,n}(Z)\) is a positive geometry whose canonical form computes
tree-level amplitudes in planar \(\mathcal N=4\) SYM. We return to this
conjecture at the end of the section.

One can also consider analogues of~\eqref{eq:def_ampl} after relaxing the
positivity assumption~\eqref{eq:pos_Z} on \(Z\). Given a \((k+m)\times n\)
matrix \(Z\), a sufficient condition for the map~\eqref{eq:Ampl_map} to be
well-defined on \(\Gr_{\geq 0}(k,n)\) is that there exists a
\(k\times(k+m)\) real matrix \(M\) such that \(M\cdot Z\) has positive
\(k\times k\) minors~\cite{TNN_grassmannian}. In that case, the image of a
positroid cell under \(\widetilde Z\) is called a \textit{Grassmann
polytope}, or \textit{grasstope}~\cite{Positive_geometries}. These objects
remain more mysterious than Amplituhedra. They were first investigated
in~\cite{TNN_grassmannian}, later in~\cite{karp2017sign,karp2020defining}, and in
the case \(m=1\) in~\cite{mandelshtam2023combinatorics}.

\subsubsection{Semialgebraic descriptions and sign flips}

By the Tarski--Seidenberg theorem~\cite{BochnakCosteRoy1998}, the
Amplituhedron is a semialgebraic set in
\(\Gr_{\mathbb R}(k,k+m)\). Nevertheless, finding an explicit
description in terms of polynomial equalities and inequalities is a difficult
and important problem.

To illustrate the issue, consider the case \(m=k=2\), \(n=4\). The
Amplituhedron \(\mathcal A_{2,2,4}(Z)\) is a copy of
\(\Gr_{\geq 0}(2,4)\), since \(Z\) is an isomorphism; one may take \(Z\)
to be the identity matrix. Recall that \(\Gr_{\geq 0}(2,4)\) is defined
by the inequalities
\begin{equation}
\langle ij\rangle\geq 0,
\qquad
1\leq i<j\leq4 \, .
\end{equation}
Although this gives six inequalities, its boundary has only four
codimension-one components, as we saw in Example~\ref{eg:Gr24_strat}. This is
a consequence of the Plücker relation~\eqref{eq:Pl_rel_2}. Since all
Plücker coordinates are non-negative, the condition
\(\langle13\rangle=0\) forces at least two terms in~\eqref{eq:Pl_rel_2} to
vanish. Thus \(\langle13\rangle=0\) cuts out a codimension-two boundary, not
a facet. The same holds for \(\langle24\rangle=0\). The four
codimension-one boundaries are instead given by
\begin{equation}\label{eq:bd_24}
	\langle12\rangle=0,
	\qquad
	\langle23\rangle=0,
	\qquad
	\langle34\rangle=0,
	\qquad
	\langle14\rangle=0 \, .
\end{equation}

Suppose we did not know the inequalities defining
\(\Gr_{\geq 0}(2,4)\), but only knew the boundary equations
~\eqref{eq:bd_24}. A natural guess would be to impose fixed signs on these
four cyclic minors. However, the set where these four minors are positive is
too large. Indeed, by the Plücker relation, its interior has two disconnected
components, cut out by
\begin{equation}
\langle13\rangle,\langle24\rangle>0 \quad \text{and} \quad \langle13\rangle,\langle24\rangle<0 \, .
\end{equation}
Since in the interior \(\Gr_{>0}(2,4)\), all cyclic minors in
~\eqref{eq:bd_24} are strictly positive, these components are disconnected.
They meet only in codimension two, along \(\langle13\rangle=0\) or
\(\langle24\rangle=0\). Thus one needs an additional inequality, for example
\(\langle13\rangle\geq 0\), or equivalently \(\langle24\rangle\geq 0\), since
one implies the other by the Plücker relation. Similar phenomena occur for
more general Amplituhedra.

\subsubsection{Twistor coordinates and the B-Amplituhedron}

It is useful to introduce equations adapted to the Amplituhedron. Consider
the map
\begin{equation}\label{eq:tw_emb}
	\Gr(k,k+m)
	\hookrightarrow
	\Gr(m,n),
	\qquad
	Y\longmapsto z:=Y^\perp Z \, ,
\end{equation}
where \(Y^\perp\in\Gr(m,k+m)\) denotes the orthogonal complement of
\(Y\) with respect to the standard inner product. Here \(Z\) is the
\((k+m)\times n\) external data matrix, so \(Y^\perp Z\) is an \(m\times n\)
matrix and hence determines a point of \(\Gr(m,n)\).

There are distinguished functions on \(\Gr(k,k+m)\), called
\textit{twistor coordinates}, given by $\langle Y\,Z_{i_1}\cdots Z_{i_m}\rangle$. This denotes the determinant of the \((k+m)\times(k+m)\) matrix formed by a
basis of the \(k\)-plane \(Y\), together with the \(m\) columns
\(Z_{i_1},\dots,Z_{i_m}\). The importance of~\eqref{eq:tw_emb} is that
\begin{equation}
\langle Y\,Z_{i_1}\cdots Z_{i_m}\rangle
=
p_I(z),
\qquad
I=\{i_1,\dots,i_m\}\in \smallbinom{[n]}{m} \, ,
\end{equation}
where \(p_I(z)\) is the corresponding Plücker coordinate of
\(z=Y^\perp Z\)~\cite{KarpWilliams,lam2024face}.

For \(m=4\), when \(Z\) is the matrix of bosonized momentum supertwistors,
the image of
\begin{equation}
Y_0=[\mathbf 0_{k\times4}\,|\,\mathbf 1_k]
\end{equation}
under the twistor embedding is the \(4\times n\) matrix of ordinary bosonic
momentum twistors \(z_i\). Thus the twistor coordinates become the usual
momentum-twistor brackets:
\begin{equation}
\langle Y_0\,Z_{i_1}Z_{i_2}Z_{i_3}Z_{i_4}\rangle
=
\langle z_{i_1}z_{i_2}z_{i_3}z_{i_4}\rangle \, .
\end{equation}
The image of the Amplituhedron under~\eqref{eq:tw_emb} is called the
\textit{B-Amplituhedron} \(\mathcal B_{k,m,n}(Z)\). The Amplituhedron and the
B-Amplituhedron are isomorphic~\cite[Proposition 3.12]{KarpWilliams}.
The advantage of the B-Amplituhedron is that it is often better suited for
describing the polynomial inequalities defining the geometry.

\subsubsection{The case \(m=1\)}

Let us first consider \(m=1\), which was studied in detail by Karp and
Williams~\cite{KarpWilliams}. In this case, the B-Amplituhedron
\(\mathcal B_{k,1,n}(Z)\) is determined by the condition that the sequence
\begin{equation}\label{eq:seg_m1}
	(\langle Y1\rangle,\langle Y2\rangle,\dots,\langle Yn\rangle)
	\quad
	\text{has \(k\) sign flips} \, .
\end{equation}
Since all equations are homogeneous, there is an overall sign ambiguity,
which may be fixed by requiring \(\langle Y1\rangle>0\). If none of the
entries in~\eqref{eq:seg_m1} vanish, this condition cuts out the interior of
\(\mathcal B_{k,1,n}(Z)\). If some entries vanish, the correct condition is
that the maximal number of sign flips, after replacing each zero by either
sign, is equal to \(k\). For example, the sign pattern $(++0+)$ has maximal number of flips equal to \(2\), obtained by replacing the zero
with a minus sign.

The Amplituhedron \(\mathcal A_{k,1,n}(Z)\) lies in
\begin{equation}
\Gr(k,k+1)= (\mathbb P^k)^\vee \, ,
\end{equation}
whose points are hyperplanes in \(\mathbb P^k\). Identifying
\((\mathbb P^k)^*\cong\mathbb P^k\) using the standard inner product, the
twistor coordinate \(\langle Yi\rangle\) becomes the hyperplane equation
\begin{equation}
\langle Yi\rangle=Y^\perp\cdot Z_i \, .
\end{equation}
Since \(Z\) is positive, these hyperplanes define a cyclic hyperplane
arrangement in \(\mathbb P^k\). The sign-flip condition~\eqref{eq:seg_m1}
is equivalent to the statement that \(\mathcal A_{k,1,n}(Z)\) is the union
of the bounded chambers of this arrangement.

\begin{eg}[\(m=1,k=2\)]\label{eg:m=1}
	Consider the Amplituhedron \(\mathcal A_{2,1,n}(Z)\). Its combinatorics is
	independent of the chosen positive matrix \(Z\)~\cite{KarpWilliams}.
	We choose
	\begin{equation}
Z_i=(1,\cos\theta_i,\sin\theta_i),
	\qquad
	\theta_i=\frac{\pi(i-1)}{2(n-1)},
	\qquad
	i=1,\dots,n \, ,
\end{equation}
 producing particularly simple figures. We fix the affine chart $Y^\perp=[1:x:y]$
	of \(\mathbb P^2\). Each vector \(Z_i\) defines a line
	\begin{equation}\label{eq:Yi}
		\langle Yi\rangle
		=
		Y^\perp\cdot Z_i
		=
		x\cos\theta_i+y\sin\theta_i+1=0 \, .
	\end{equation}
	Together, these lines form a \emph{cyclic line arrangement} in \(\mathbb R^2\).

	Each connected component \(R\) in the complement of these lines is called
	a \textit{chamber}. It has a sign vector
	\begin{equation}
v(R)\in\{\pm1\}^n
\end{equation}
	recording the signs of the hyperplane equations~\eqref{eq:Yi} at any point
	of \(R\). By the sign-flip condition~\eqref{eq:seg_m1}, the Amplituhedron
	\(\mathcal A_{2,1,n}(Z)\) is the union of all closed chambers whose sign
	vector has \(k=2\) sign flips, after fixing the projective ambiguity by
	\(\langle Y1\rangle>0\).

	For example, \(\mathcal A_{2,1,4}(Z)\) is the union of the following three
	chambers in the cyclic arrangement of four lines:
	\begin{equation}
(+-++),
	\qquad
	(+--+),
	\qquad
	(++-+) \, ,
\end{equation}
and is shown on the left panel of Figure~\ref{fig:m1}.
	Similarly, for \(n=5\), the Amplituhedron \(\mathcal A_{2,1,5}(Z)\) is
	the union of the six bounded chambers shown in the right panel of
	Figure~\ref{fig:m1}.

	\begin{figure}[pos=t]
    \centering
    \includegraphics[width=0.5\textwidth]{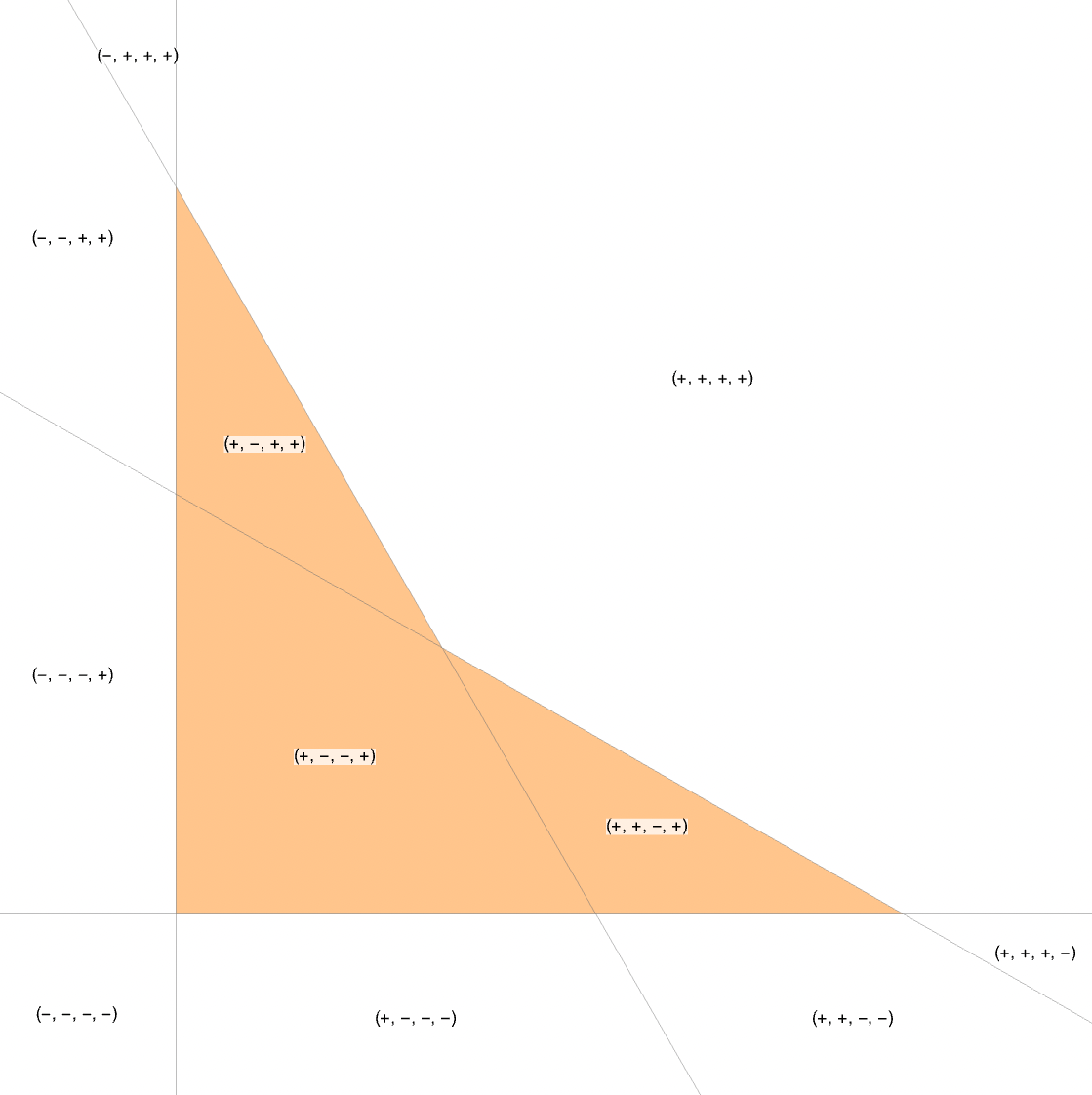}\hfill
    \includegraphics[width=0.5\textwidth]{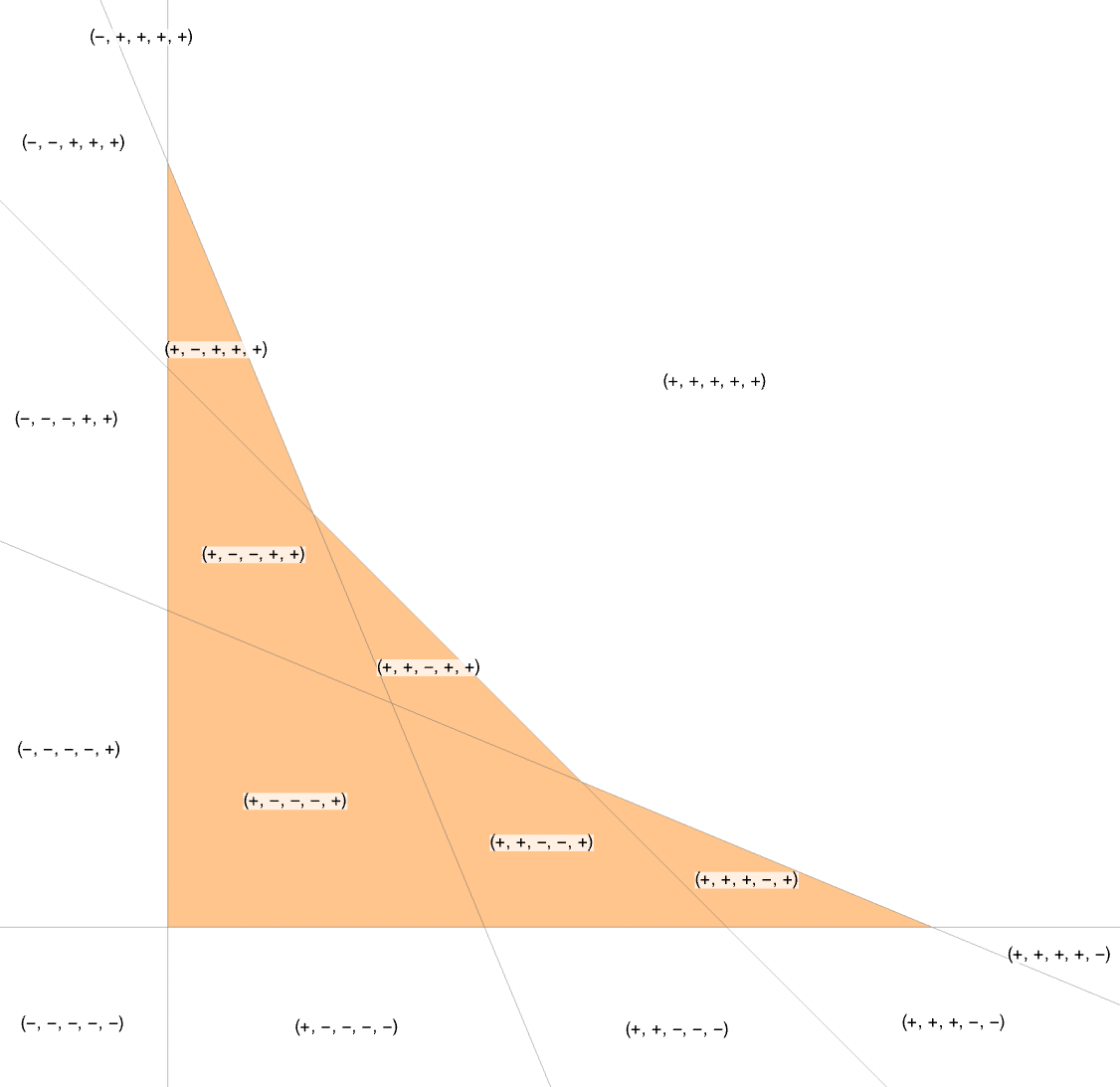}
    \caption{The Amplituhedra \(\mathcal A_{2,1,4}(Z)\) and
    \(\mathcal A_{2,1,5}(Z)\), respectively, given by the colored bounded
    chambers of cyclic hyperplane arrangements. The colored chambers are
    precisely those satisfying the sign-flip condition~\eqref{eq:seg_m1}.}
    \label{fig:m1}
	\end{figure}
\end{eg}

Amplituhedra for \(m=1\) are relatively simple objects. As
Figure~\ref{fig:m1} shows, they are not convex, and hence not projective
polytopes. Nevertheless, being unions of bounded chambers of hyperplane
arrangements, they can be triangulated into simplices. In
~\cite{KarpWilliams}, the authors studied these triangulations and the
topology of \(m=1\) Amplituhedra, proving in particular that they are
homeomorphic to closed balls and equipping them with CW structures. These
results support, and in this setting establish, the positive-geometry
structure of \(m=1\) Amplituhedra.

\subsubsection{Positive projections and higher \(m\)}

To study inequalities for \(m>1\), we follow the idea of \emph{positive projections} from~\cite{Arkani_Hamed_2018}. Let
\begin{equation}
\Pi:\mathbb R^{k+m}\longrightarrow \mathbb R^{k+m'}
\end{equation}
be a linear projection, with \(m'<m\), represented by a
\((k+m')\times(k+m)\) matrix. Given a \((k+m)\times n\) positive
matrix $Z$, define its projection as
\begin{equation}
Z'=\Pi Z \, .
\end{equation}
We assume that \(Z'\) has positive maximal minors. Then
\begin{equation}
Y=CZ^\top\in\mathcal A_{k,m,n}(Z)
\quad\Longrightarrow\quad
Y'=C(Z')^\top=Y\Pi^\top
\in
\mathcal A_{k,m',n}(Z') \, .
\end{equation}

A natural question is whether the converse holds: if all suitable positive
projections of a point \(Y\in\Gr(k,k+m)\) lie in the corresponding lower
\(m\) Amplituhedra, must \(Y\) lie in \(\mathcal A_{k,m,n}(Z)\)? This is a
subtle question, discussed for instance in~\cite[Problem 3.14]{KarpWilliams}.
The principle is known in the cases \(m=1\) and \(m=2\), and is expected to
capture the physical \(m=4\) Amplituhedron. It provides a practical way of
constructing natural inequalities.

The relevant projections are special. If \(m>2\), one projects through the
two-plane \(Z_iZ_{i+1}\) for each \(i=1,\dots,n\). If \(m=2\), one projects
through the point \(Z_i\). By projecting through a linear subspace \(U\), we
mean projecting onto the quotient by \(U\), or equivalently onto a chosen
complement to \(U\). These projections preserve positivity of the external
data and reduce the problem recursively to \(m=1\).

Consider \(m=2\), and let \(Z\) be a positive \((k+2)\times n\) matrix.
Projecting through \(Z_1\) gives a positive \((k+1)\times n\) matrix \(Z'\),
since
\begin{equation}
	\langle Z'_{i_1}\cdots Z'_{i_{k+1}}\rangle
	=
	\langle Z_1 Z_{i_1}\cdots Z_{i_{k+1}}\rangle
	>0 \, .
\end{equation}
The projection \(Y'\) of a point \(Y\in\Gr(k,k+2)\) through \(Z_1\) lies
in the \(m=1\) Amplituhedron \(\mathcal A_{k,1,n}(Z')\) if and only if its
twistor brackets satisfy the sign-flip condition~\eqref{eq:seg_m1}. These
brackets satisfy
\begin{equation}
\langle Y'Z_i'\rangle=\langle Y\,1\,i\rangle \, .
\end{equation}
Thus the \(m=1\) sign-flip condition becomes the statement that $(\langle Y12\rangle,\langle Y13\rangle,\dots,\langle Y1n\rangle)
$ has \(k\) sign flips.
Repeating the same argument for the projection through each \(Z_i\), and
using the \emph{twisted cyclic symmetry}
\begin{equation}\label{eq:tw_cycl}
	Z_{i+n}=(-1)^{k-1}Z_i \, ,
\end{equation}
one obtains \(n\) cyclically shifted sign-flip
conditions. These conditions are equivalent to the following description of
the interior of the \(m=2\) Amplituhedron:
\begin{tcolorbox}[definitionbox]
\textbf{Sign-flip characterization for \(m=2\).}

\begin{equation}\label{eq:m2_cond}
\begin{aligned}
	&\langle Y\,i\,i{+}1\rangle>0,
	\qquad i=1,\dots,n \, ,
	\\
	&(\langle Y12\rangle,\langle Y13\rangle,\dots,\langle Y1n\rangle)
	\quad
	\text{has \(k\) sign flips}.
\end{aligned}
\end{equation}
\end{tcolorbox}\noindent
Here the indices are understood cyclically using~\eqref{eq:tw_cycl}. These
conditions have been proven to specify the interior of the \(m=2\)
Amplituhedron~\cite[Theorem 5.1]{parisiShermanBennettWilliams2023m2}, and
the \(n\) adjacent twistor coordinates $\langle Y\,i\,i{+}1\rangle$ cut out the boundary components of
\(\mathcal A_{k,2,n}\)~\cite[Lemma 2.12]{koefler2025taking}.

Now let \(m=4\). One projects through \(Z_iZ_{i+1}\), reducing to \(m'=2\).
The analysis is analogous and leads to the following conditions
~\cite{Arkani_Hamed_2018}:
\begin{tcolorbox}[definitionbox]
\textbf{Sign-flip characterization for \(m=4\).}

\begin{equation}\label{eq:m4_cond}
\begin{aligned}
	&\langle Y\,i\,i{+}1\,j\,j{+}1\rangle>0,
	\qquad
	\text{for non-adjacent } i,j \in [n] \, ,
	\\
	&(\langle Y1234\rangle,\langle Y1235\rangle,\dots,\langle Y123n\rangle)
	\quad
	\text{has \(k\) sign flips}.
\end{aligned}
\end{equation}
\end{tcolorbox}\noindent
The indices are again understood cyclically, with the appropriate twisted
cyclic signs. In~\cite[Corollary 8.8]{evenZoharLakrecTessler2025bcfw}, the
authors proved that the vanishing of the $n(n-3)/2$
twistor coordinates $\langle Y\,i\,i{+}1\,j\,j{+}1\rangle$ cuts out the boundary components of \(\mathcal A_{k,4,n}(Z)\). However, the
statement that the semialgebraic set defined by~\eqref{eq:m4_cond} equals
the interior of the Amplituhedron has not been proven explicitly in the
literature, although it is expected to follow from recently proven results
about tilings of \(m=4\) Amplituhedra.

More generally, a semialgebraic description for arbitrary \(m\) was proposed
in~\cite[Equation 5.14]{Arkani_Hamed_2018}. The resulting
semialgebraic set is proven to contain the Amplituhedron
\cite[Corollary 3.21]{KarpWilliams} for every \(k,m,n\). On the other
hand, for \(m=6\) there seem to be additional boundary components beyond the
vanishing loci appearing in that description. Thus a complete semialgebraic
description of Amplituhedra remains open in full generality.

\subsubsection{Main conjectures}

We now state the conjectures at the core of the Amplituhedron program
~\cite{the_amplituhedron}. The first is a purely mathematical statement.

\begin{tcolorbox}[resultbox]
\textbf{Tree Amplituhedra are positive geometries.}
For every positive matrix \(Z\), the tree Amplituhedron defines a positive
geometry
\begin{equation}
	\left(
	\Gr(k,k+m),
	\mathcal A_{k,m,n}(Z),
	\mathbf{\Omega}_{k,m,n}
	\right) \, .
\end{equation}

\end{tcolorbox}\noindent

The second conjecture concerns triangulations of Amplituhedra and will be
made more precise in the next section.

\begin{tcolorbox}[resultbox]
\textbf{BCFW tiling.}
Consider a BCFW recursion for the tree-level \(n\)-point
\({\rm N}^k{\rm MHV}\) superamplitude in planar \(\mathcal N=4\) SYM. Let
\(\Sigma\) be the corresponding collection of plabic graphs, or equivalently
positroid cells. Then, for \(m=4\), these cells tile the Amplituhedron:
\begin{equation}
	\mathcal A_{k,4,n}(Z)
	=
	\bigcup_{\sigma\in\Sigma}
	\mathcal A_{\sigma}(Z) \, ,
\end{equation}
where \(\mathcal A_{\sigma}(Z)\) is the injective image of the corresponding
positroid cell under the Amplituhedron map~\eqref{eq:Ampl_map}.

\end{tcolorbox}\noindent

Since BCFW recursion is a well-established method for computing tree-level
superamplitudes, these conjectures imply the expected relation between
scattering amplitudes and canonical forms.

\begin{tcolorbox}[resultbox]
\textbf{Ratio function vs canonical function.}
Tree-level scattering amplitudes in planar \(\mathcal N=4\) SYM are encoded
by the \(m=4\) Amplituhedron:
\begin{equation}
	\mathcal R_{k,n}(Z)
	=
	\Omega_{k,4,n}(Y_0;Z) \, ,
\end{equation}
where \(\mathcal R^{k,n}\) is the ratio function in~\eqref{eq:ampl_R_dec},
\(\Omega_{k,4,n}\) is the canonical function of the \(m=4\) Amplituhedron, and
\begin{equation}
Y_0=[\mathbf 0_{k\times4}\,|\,\mathbf 1_k] \, .
\end{equation}

\end{tcolorbox}\noindent

There are also loop Amplituhedra, which encode loop-level integrands of
planar \(\mathcal N=4\) SYM. The same conjectural picture extends
to loop level, as we will discuss in Section~\ref{sec:Loop Amplituhedra}.

We summarize some of what is currently known about tree Amplituhedra.

\begin{itemize}
    \item \(k=1\): In this case
    \begin{equation}
\mathcal A_{1,m,n}(Z)=C_{m,n}(Z)
\end{equation}
    is the \(m\)-dimensional cyclic polytope with \(n\) vertices. Its
    triangulations are well understood~\cite{deLoeraRambauSantos2010triangulations},
    and it defines a positive geometry by Proposition~\ref{prop:pol_tr}. For
    \(m=4\), we saw in Section~\ref{sec:Hodges and Cyclic Polytopes} that the
    canonical function of \(C_{4,n}(Z)\) encodes NMHV superamplitudes, and
    that BCFW expansions correspond to triangulations of this cyclic
    polytope.

    \item \(n=k+m\): The Amplituhedron $A_{k,m,k+m}(Z)$ is a copy of the non-negative Grassmannian
    \(\Gr_{\geq 0}(k,k+m)\), since the Amplituhedron map
~\eqref{eq:Ampl_map} is an isomorphism. As discussed above, this is a
    positive geometry. Moreover, it is simplex-like, meaning that its
    canonical form has constant numerator. From the perspective of
    triangulations and tilings, it should therefore be viewed as an
    irreducible object. Physically, for \(m=4\), the canonical form
    reproduces the \(\overline{\rm MHV}\) superamplitude~\cite{Mason:2009qx}.
    Recall that the helicity degree \(k\) is bounded by \(0\leq k\leq n-4\),
    and \(k=n-4\) corresponds to the \(\overline{\rm MHV}\) sector.

    \item \(m=1\): The Amplituhedron \(\mathcal A_{k,1,n}(Z)\) is the union
    of bounded chambers of a cyclic hyperplane arrangement. In
~\cite{KarpWilliams}, the authors characterized its triangulations
    and topology. These results establish the positive-geometry structure for
    \(m=1\).

    \item \(m=2\): The Amplituhedron \(\mathcal A_{k,2,n}(Z)\) has been
    extensively studied~\cite{lukowski2019boundaries,baoHe2019m2}. Its
    boundary stratification is understood in terms of the positroid
    stratification of the non-negative Grassmannian
    \(\Gr_{\geq 0}(2,n)\) in the twistor embedding~\cite{lam2024face}.
    In~\cite{parisiShermanBennettWilliams2023m2}, the authors classified all
    tiles and tilings and uncovered a connection to the hypersimplex; they
    also proved the inequality description~\eqref{eq:m2_cond}. Although each
    tile, being the positive part of a cluster variety, is a positive geometry
~\cite{parisiShermanBennettWilliams2023m2}, the positive-geometry
    structure of the full Amplituhedron \(\mathcal A_{k,2,n}(Z)\) is not yet
    established in general. For \(m=k=2\), it was proven in
~\cite{Ranestad:adjoint} by studying the residual arrangement and showing
    the existence of a unique adjoint hypersurface interpolating it; see
    Section~\ref{sec:Adjoint Hypersurface}.

    \item \(m=4\): The Amplituhedron \(\mathcal A_{k,4,n}(Z)\) is the
    physical tree Amplituhedron for planar \(\mathcal N=4\) SYM. BCFW
    recursion yields tilings by injective images of positroid cells, as shown
    in~\cite{evenZoharLakrecTessler2025bcfw}. We discuss this in more detail
    in the following section. The BCFW tiles also admit a cluster-algebraic
    description in terms of compatible \(\Gr(4,n)\) cluster variables and
    product promotion
~\cite{evenZoharLakrecParisiTesslerShermanBennettWilliams2023cluster,evenZoharLakrecParisiTesslerShermanBennettWilliams2024clusterResults}.
    Tilings give, in principle, a way to compute the canonical form of
    \(\mathcal A_{k,4,n}(Z)\). Other approaches to computing these forms
    include fiber-based methods and pushforwards of positroid-cell forms
~\cite{mohammadiMoninParisi2020triangulations}, as well as sign-flip
    triangulations~\cite{Lukowski:2020bya}. As a computational tool, we mention
    the \texttt{Mathematica} package \texttt{amplituhedronBoundaries}
~\cite{lukowskiMoerman2021boundaries}, built on \texttt{positroids.m}~\cite{Bourjaily:2012gy}, for
    computing boundary stratifications of Amplituhedra.
\end{itemize}

\subsubsection{The momentum Amplituhedron}

We conclude this section by mentioning the \textit{momentum Amplituhedron}.
This is a close cousin of the Amplituhedron discussed above. It is expected
to be a positive geometry defined directly in spinor-helicity space, rather
than in momentum-twistor space. It was introduced
in~\cite{Damgaard:2019ztj}, and its canonical form is
expected to compute tree-level amplitudes in planar \(\mathcal N=4\) SYM.

The momentum Amplituhedron is complementary to the usual Amplituhedron: the
latter naturally computes ratio functions in momentum-twistor variables,
while the momentum Amplituhedron keeps the physical spinor-helicity variables
manifest. Its boundary stratification was studied
in~\cite{ferroLukowskiMoerman2020boundaries}. Further developments relate the
momentum Amplituhedron to the kinematic associahedron
~\cite{damgaardFerroLukowskiParisi2021associahedron}, twistor-string maps
~\cite{heKuoZhang2021twistorString}, and loop-level generalizations
~\cite{Ferro:2022abq}. For an introduction to the momentum
Amplituhedron, see~\cite{stalknecht2024positive,moerman2023positive}.

\subsection{BCFW recursion and tiles}\label{sec:BCFW Recursion and Tiles}

\subsubsection{Positroid tiles and tilings}

In the previous section we introduced the Amplituhedron
\(\mathcal A_{k,m,n}(Z)\) as the image of the non-negative Grassmannian
\(\Gr_{\geq 0}(k,n)\) under the map induced by a positive matrix \(Z\),
see~\eqref{eq:Ampl_map}. We also reviewed sign-flip descriptions for
\(m=1,2,4\). In this section we discuss tilings of Amplituhedra in these
cases. By a tiling we mean a collection of positroid cells in
\(\Gr_{\geq 0}(k,n)\) whose images under the Amplituhedron map are
injective, pairwise non-overlapping on their interiors, and cover the
Amplituhedron after taking closures. For \(m=4\), these tilings are closely
related to on-shell diagrams and BCFW recursion.

Let us make this notion precise. Recall the positroid stratification
\begin{equation}
\Gr_{\geq 0}(k,n)=\bigcup_{\sigma}\Pi_{\sigma,>0}
\end{equation}
from~\eqref{eq:pos_dec}, where each open positroid cell is labelled by a
bounded affine permutation \(\sigma\) of type \((k,n)\), see
~\eqref{eq:bound_aff_perm}. Given a positive \((k+m)\times n\) matrix \(Z\),
consider the Amplituhedron map \(\widetilde Z\) in~\eqref{eq:Ampl_map}. For a
bounded affine permutation \(\sigma\), define
\begin{equation}
\mathcal A_{\sigma,>0}
:=
\widetilde Z(\Pi_{\sigma,>0})
\end{equation}
to be the open positroid image, or open grasstope, associated with \(\sigma\).
We write \(\mathcal A_{\sigma}\) for the Euclidean closure of
\(\mathcal A_{\sigma,>0}\) in \(\Gr_{\mathbb R}(k,k+m)\).

We call \(\mathcal A_{\sigma}\) a \textit{positroid tile} if
\begin{equation}
\dim \Pi_\sigma=km
\end{equation}
and \(\widetilde Z\) is injective on \(\Pi_{\sigma,>0}\). A collection
\(\{\mathcal A_\sigma\}_{\sigma\in\Sigma}\) of positroid tiles is called a
\textit{tiling} of the Amplituhedron \(\mathcal A_{k,m,n}(Z)\) if the open
tiles \(\mathcal A_{\sigma,>0}\) are pairwise disjoint and the closed tiles
cover the Amplituhedron:
\begin{equation}
\mathcal A_{k,m,n}(Z)
=
\bigcup_{\sigma\in\Sigma}\mathcal A_\sigma \, .
\end{equation}

\begin{eg}[$k=1$]
	In the case \(k=1\), one has
\begin{equation}
\Gr_{\geq 0}(1,n)=\Delta^{n-1} \, ,
\end{equation}
the \((n-1)\)-dimensional standard projective simplex. The Amplituhedron
\begin{equation}
\mathcal A_{1,m,n}(Z)=C_{m,n}(Z)
\end{equation}
is the \(m\)-dimensional cyclic polytope on \(n\) vertices. Positroid tiles
are \(m\)-dimensional faces of the simplex whose images are \(m\)-simplices in
\(C_{m,n}(Z)\). Thus, for \(k=1\), positroid tilings are ordinary
triangulations of the cyclic polytope~\cite{deLoeraRambauSantos2010triangulations}.

\end{eg}

\subsubsection{Tiles for \(m=1\)}

To gain intuition for positroid tiles and tilings, it is useful to start with
\(m=1\). The combinatorial description of tiles and tilings in this case,
together with the relevant proofs, can be found in~\cite{KarpWilliams}.
As discussed in Example~\ref{eg:m=1}, each Amplituhedron
\(\mathcal A_{k,1,n}(Z)\) is the union of bounded chambers of a cyclic
hyperplane arrangement; see, for example, Figure~\ref{fig:m1}. Positroid
tiles are especially simple in this case: they correspond to bounded chambers
in the hyperplane arrangement, or equivalently to sign regions satisfying the
sign-flip condition~\eqref{eq:seg_m1}.

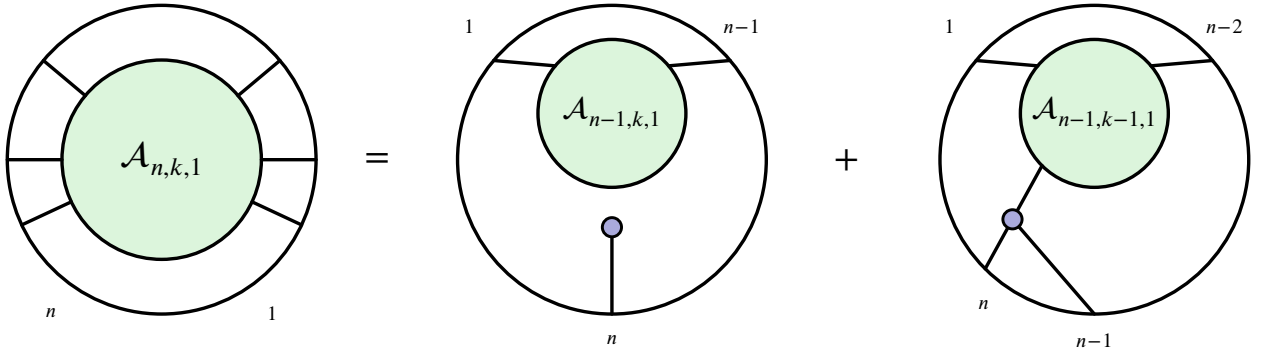
\begin{figure}[pos=t]
\centering
\resizebox{\textwidth}{!}{%
\begin{tikzpicture}[line cap=round,line join=round,scale=1]

\definecolor{vertexpurple}{RGB}{170,170,220}
\definecolor{Alightgreen}{RGB}{220,245,220}

\tikzset{
  circ/.style={black, line width=1.5pt},
  edge/.style={black, line width=1.5pt},
  bdot/.style={circle, draw=black, fill=vertexpurple, line width=1.3pt, inner sep=3.0pt},
  wdot/.style={circle, draw=black, fill=white, line width=1.3pt, inner sep=3.0pt},
  Aregion/.style={draw=black, fill=Alightgreen, line width=1.5pt}
}

\begin{scope}[shift={(0,0)}]

  \def\R{2.4}
  \def\r{1.55}

  \draw[circ] (0,0) circle (\R);
  \draw[Aregion] (0,0) circle (\r);

  \foreach \ang in {140,40,205,335,180,0} {
    \draw[edge] (\ang:\r) -- (\ang:\R);
  }

  \node[font=\fontsize{16}{16}\selectfont] at (0,0) {$\mathcal{A}_{n,k,1}$};
  \node at (-1.72,-2.38) {$n$};
  \node at ( 1.72,-2.38) {$1$};

\end{scope}

\node[font=\fontsize{16}{16}\selectfont] at (3.35,0) {$=$};

\begin{scope}[shift={(7.0,0)}]

  \def\R{2.4}
  \def\r{1.15}
  \coordinate (c) at (0,0.72);

  \draw[circ] (0,0) circle (\R);
  \draw[Aregion] (c) circle (\r);

  \draw[edge] ($(c)+(140:\r)$) -- (140:\R);
  \draw[edge] ($(c)+(40:\r)$)  -- (40:\R);

  \draw[edge] (270:\R) -- (0,-1.05);
  \node[bdot] at (0,-1.05) {};

  \node[font=\fontsize{14}{14}\selectfont] at (0,0.72) {$\mathcal{A}_{n-1,k,1}$};
  \node at (-2.22,2.08) {$1$};
  \node at ( 2.02,2.08) {$n\!-\!1$};
  \node at (0,-2.80) {$n$};

\end{scope}

\node[font=\fontsize{16}{16}\selectfont] at (10.65,0) {$+$};

\begin{scope}[shift={(14.5,0)}]

  \def\R{2.4}
  \def\r{1.15}
  \coordinate (c) at (0,0.72);

  \draw[circ] (0,0) circle (\R);
  \draw[Aregion] (c) circle (\r);

  \draw[edge] ($(c)+(140:\r)$) -- (140:\R);
  \draw[edge] ($(c)+(40:\r)$)  -- (40:\R);

  \coordinate (iL) at ($(c)+(225:\r)$);
  \coordinate (oL) at (225:\R);
  \coordinate (b)  at ($(iL)!0.52!(oL)$);

  \draw[edge] (iL) -- (b) -- (oL);
  \draw[edge] (b) -- (270:\R);

  \node[bdot] at (b) {};

  \node[font=\fontsize{14}{14}\selectfont] at (0,0.72) {$\mathcal{A}_{n-1,k-1,1}$};
  \node at (-2.25,2.08) {$1$};
  \node at ( 2.02,2.08) {$n\!-\!2$};
  \node at (-1.72,-2.25) {$n$};
  \node at (0,-2.80) {$n\!-\!1$};

\end{scope}

\end{tikzpicture}%
}
\caption{BCFW-like recursion for the \(m=1\) Amplituhedron.}
\label{fig:m1_BCFW_recursion}
\end{figure}

There is a BCFW-like recursion for the \(m=1\) tiles, building plabic graphs
with increasing \(n\) and \(k\) from simpler ones:
\begin{enumerate}
 \item Start from the plabic graph with one boundary vertex incident to a
 black lollipop. This is the graph for \(n=1\) and \(k=0\).
 \item Given a plabic graph with \(n-1\) boundary vertices produced by the
 recursion, add a new boundary vertex \(n\), incident either to a black
 lollipop or to the edge adjacent to boundary vertex \(n-1\), as in
 Figure~\ref{fig:m1_BCFW_recursion}.
\end{enumerate}
The first operation preserves \(k\), while the second increases \(k\) by one.

Let \(\Gamma\) be a plabic graph with \(n\) boundary vertices produced by this
recursion. Then \(k\) equals \(n\) minus the number of connected components of
\(\Gamma\) after removing the boundary of the disk. We say that \(\Gamma\) is of type
\((k,n)\).
To such a graph \(\Gamma\), we associate a sign vector
\begin{equation}
s=s(\Gamma)\in\{\pm1\}^n
\end{equation}
as follows. Set \(s_1=+1\). For \(i>1\), define \(s_i=s_{i-1}\) if the
boundary vertices \(i\) and \(i-1\) do not share an internal vertex, and
define \(s_i=-s_{i-1}\) otherwise. The resulting sign vector has \(k\) sign
flips. Conversely, every sign vector with \(k\) sign flips arises from a
plabic graph of type \((k,n)\).

Each such graph is equivalently encoded by its bounded affine permutation $\sigma$, obtained by the strand rule explained around~\eqref{eq:strand_rule}.
Each BCFW graph \(\Gamma\) of type \((k,n)\) yields a positroid tile of
\(\mathcal A_{k,1,n}(Z)\), associated with the permutation \(\sigma(\Gamma)\). The
collection of all such tiles gives a tiling
\begin{equation}
	\mathcal A_{k,1,n}(Z)
	=
	\bigcup_{\sigma\in\Sigma_{k,1,n}}
	\mathcal A_\sigma \, ,
\end{equation}
where \(\Sigma_{k,1,n}\) is a collection of permutations produced by the
\(m=1\) recursion. This collection has cardinality
\begin{equation}
|\Sigma_{k,1,n}|=\binom{n-1}{k} \, .
\end{equation}

\begin{figure}[pos=t]
\centering
\includegraphics[width=1\textwidth]{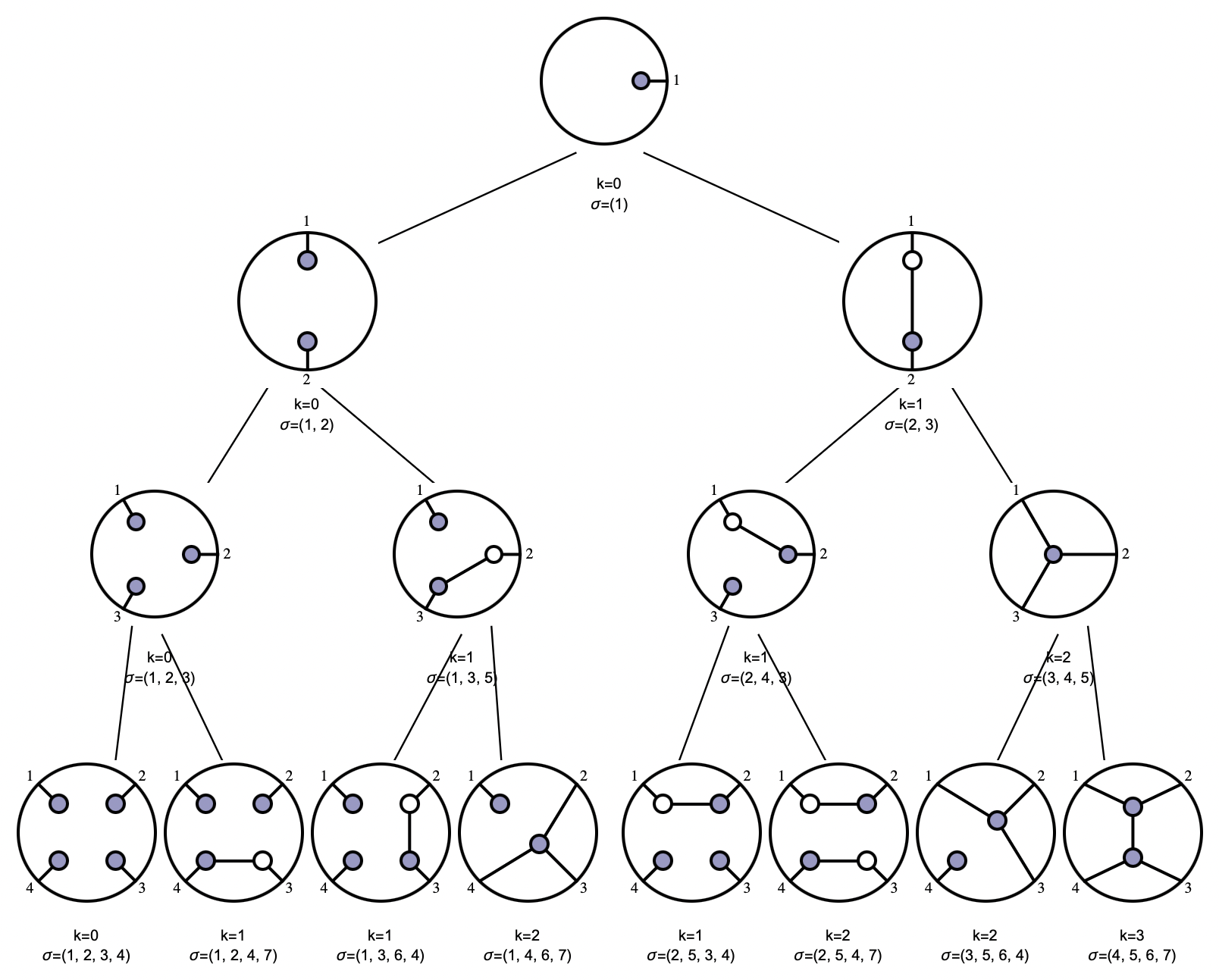}
\caption{BCFW recursion on tiles for \(m=1\), up to \(n=4\);
see~\cite[Figure 6]{KarpWilliams}.}
\label{fig:m1_BCFW_examples}
\end{figure}

\begin{figure}[pos=t]
\centering
\includegraphics[width=0.8\textwidth]{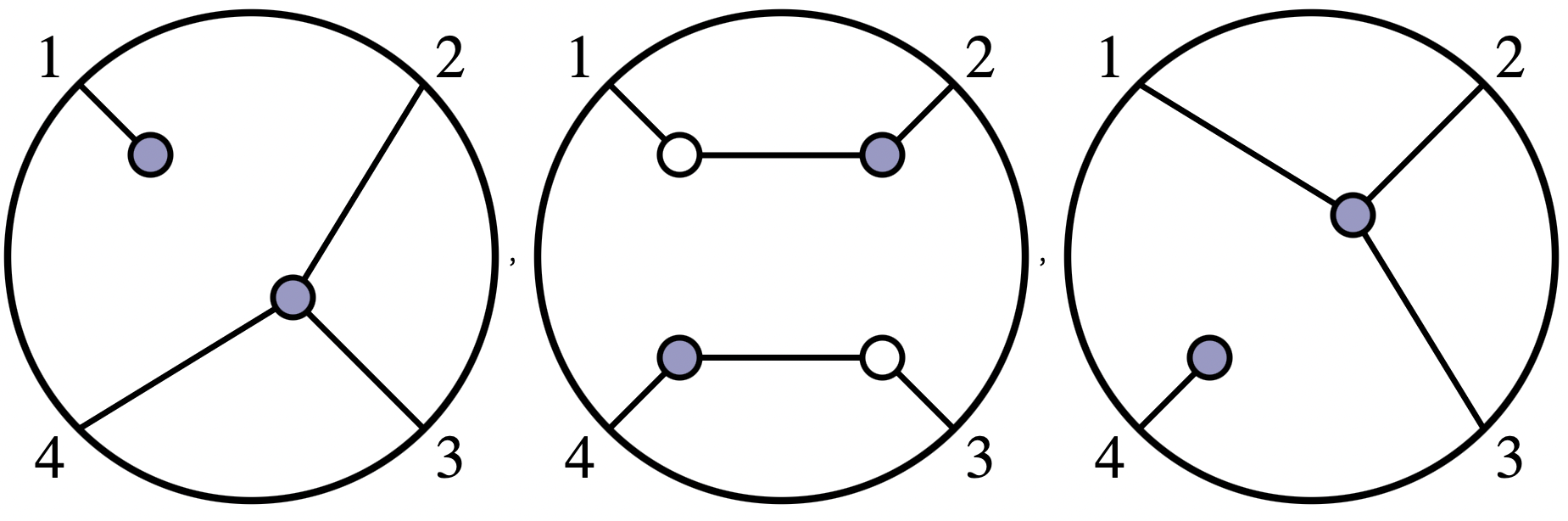}
\caption{Plabic graphs for the BCFW tiles of \(\mathcal A_{2,1,4}(Z)\),
associated with the permutations~\eqref{eq:sigma_m2_k2_n4}.}
\label{fig:m1_k2_n4}
\end{figure}

\begin{eg}[$m=1$, \(k=2\), \(n=4\)]
	In this case there are three BCFW tiles, displayed in
	Figure~\ref{fig:m1_k2_n4}. They are associated with the bounded affine
	permutations
	\begin{equation}\label{eq:sigma_m2_k2_n4}
		 (1,4,6,7),
		 \qquad
		 (2,5,4,7),
		 \qquad
		 (3,5,6,4) \, .
    \end{equation}
    The corresponding sign vectors are
    \begin{equation}
		 (++-+),
		 \qquad
		 (+--+),
		 \qquad
		 (+-++) \, ,
    \end{equation}
    which are precisely the three bounded chambers shown on the left of
    Figure~\ref{fig:m1}.

    Let us look at the tile associated with
    \begin{equation}
\sigma=(2,5,4,7) \, .
\end{equation}
    This is an affine permutation of type \((2,4)\). A parametrization
    of the corresponding positroid cell is obtained from \texttt{Positroids.m}~\cite{Bourjaily:2012gy}
    by \texttt{permToMatrix[\{2, 5, 4, 7\}]}:
    \begin{equation}\label{eq:pos_m1}
    	C(\alpha)
    	=
    	\begin{pmatrix}
    		1 & \alpha_2 & 0 & 0 \\
    		0 & 0 & 1 & \alpha_1
    	\end{pmatrix} \, .
    \end{equation}
    The cell \(\Pi_{\sigma,>0}\) is parametrized by
    \(\alpha_1,\alpha_2>0\), and therefore has dimension \(2=km\). Its image under the Amplituhedron map is parametrized by
    \begin{equation}\label{eq:Y_m1}
    	Y=C(\alpha)\cdot Z^\top
    	=
    	\begin{pmatrix}
    		Z_1+\alpha_2 Z_2 \\
    		Z_3+\alpha_1 Z_4
    	\end{pmatrix} \, ,
    \end{equation}
    which is a point of \(\Gr(2,3)\). One verifies that \(Y\) lies in the
    chamber with sign vector \((+--+)\). For example, with the cyclic bracket
    convention used here,
    \begin{equation}
    \begin{aligned}
    	\langle Y1\rangle
    	&=
    	\alpha_2\bigl(\langle123\rangle+\alpha_1\langle234\rangle\bigr)>0,
    	\\
    	\langle Y2\rangle
    	&=
    	-\langle123\rangle-\alpha_1\langle124\rangle<0,
    \end{aligned}
    \end{equation}
    and similarly \(\langle Y3\rangle<0\) and \(\langle Y4\rangle>0\), using
    positivity of \(Z\) and \(\alpha_i>0\).

    Thus \(\mathcal A_\sigma\) is the chamber \((+--+)\). As one can see from
    the left panel of Figure~\ref{fig:m1}, this chamber has four facets, so it
    is not a triangle. More generally, positroid tiles for \(m=1\) are not
    necessarily simplices, even though the whole Amplituhedron can be viewed
    as a union of chambers in a hyperplane arrangement. The reason is that the
    tiles are injective images of nonlinear positroid cells in
    \(\Gr_{\geq 0}(k,n)\).

    We invite the reader to check that the pushforward of the form $\mathrm{d}\log(\alpha_1) \mathrm{d}\log(\alpha_2) $ of \(\Pi_{\sigma,>0}\) under~\eqref{eq:Y_m1} yields the canonical
    form of the quadrilateral chamber \((+--+)\).
\end{eg}

\subsubsection{Tiles for \(m=2\)}

Let us now discuss positroid tiles and tilings for the \(m=2\)
Amplituhedron. The main reference for tiles and tilings is
~\cite{parisiShermanBennettWilliams2023m2}, while the BCFW-like recursion was
introduced in~\cite{baoHe2019m2}. The story has beautiful combinatorics and
connections to the hypersimplex and the tropical Grassmannian. Here we focus
on the combinatorial description of the tiles and tilings, and on their
characterization in terms of signs of twistor coordinates.

Recall that the \(m=2\) Amplituhedron \(\mathcal A_{k,2,n}(Z)\) is
characterized by the conditions in~\eqref{eq:m2_cond}. The key idea is to
interpret the twistor coordinates $\langle Yij\rangle$
as arcs or diagonals of a convex \(n\)-gon \(P_n\), whose vertices are labelled
cyclically from \(1\) to \(n\).

\begin{figure}[pos=t]
\centering
\begin{tikzpicture}[scale=0.7,line cap=round,line join=round]
\usetikzlibrary{calc}

\definecolor{vertexpurple}{RGB}{170,170,220}
\colorlet{trianglefill}{black!14}
\colorlet{diaggray}{black!35}

\tikzset{
  polyedge/.style={diaggray, line width=1.2pt},
  diagedge/.style={diaggray, line width=1.2pt},
  plabicedge/.style={black, line width=1.5pt},
  boundarydot/.style={
    circle, draw=black, fill=vertexpurple,
    line width=1.3pt, minimum size=8pt, inner sep=0pt
  },
  whitevertex/.style={
    circle, draw=black, fill=white,
    line width=1.3pt, minimum size=8pt, inner sep=0pt
  }
}

\coordinate (v1) at ( 90:4.2);
\coordinate (v2) at ( 50:4.15);
\coordinate (v3) at ( 10:4.10);
\coordinate (v4) at (-30:4.15);
\coordinate (v5) at (-70:4.10);
\coordinate (v6) at (-110:4.10);
\coordinate (v7) at (-150:4.15);
\coordinate (v8) at (170:4.10);
\coordinate (v9) at (130:4.10);

\coordinate (m189) at ($(v8)!0.5!(v9)$);
\coordinate (w189) at ($(v1)!0.6667!(m189)$);

\coordinate (m178) at ($(v7)!0.5!(v8)$);
\coordinate (w178) at ($(v1)!0.6667!(m178)$);

\coordinate (m234) at ($(v3)!0.5!(v4)$);
\coordinate (w234) at ($(v2)!0.6667!(m234)$);

\coordinate (m247) at ($(v4)!0.5!(v7)$);
\coordinate (w247) at ($(v2)!0.6667!(m247)$);

\coordinate (m467) at ($(v6)!0.5!(v7)$);
\coordinate (w467) at ($(v4)!0.6667!(m467)$);

\fill[trianglefill] (v1)--(v8)--(v9)--cycle;
\fill[trianglefill] (v1)--(v7)--(v8)--cycle;
\fill[trianglefill] (v2)--(v3)--(v4)--cycle;
\fill[trianglefill] (v2)--(v4)--(v7)--cycle;
\fill[trianglefill] (v4)--(v6)--(v7)--cycle;

\draw[polyedge] (v1)--(v2)--(v3)--(v4)--(v5)--(v6)--(v7)--(v8)--(v9)--cycle;

\draw[diagedge] (v8)--(v1);
\draw[diagedge] (v7)--(v1);
\draw[diagedge] (v7)--(v2);
\draw[diagedge] (v2)--(v4);
\draw[diagedge] (v7)--(v4);
\draw[diagedge] (v6)--(v4);

\draw[plabicedge] (w189)--(v1);
\draw[plabicedge] (w189)--(v8);
\draw[plabicedge] (w189)--(v9);

\draw[plabicedge] (w178)--(v1);
\draw[plabicedge] (w178)--(v7);
\draw[plabicedge] (w178)--(v8);

\draw[plabicedge] (w234)--(v2);
\draw[plabicedge] (w234)--(v3);
\draw[plabicedge] (w234)--(v4);

\draw[plabicedge] (w247)--(v2);
\draw[plabicedge] (w247)--(v4);
\draw[plabicedge] (w247)--(v7);

\draw[plabicedge] (w467)--(v4);
\draw[plabicedge] (w467)--(v6);
\draw[plabicedge] (w467)--(v7);

\foreach \p in {v1,v2,v3,v4,v5,v6,v7,v8,v9}{
  \node[boundarydot] at (\p) {};
}

\foreach \p in {w189,w178,w234,w247,w467}{
  \node[whitevertex] at (\p) {};
}

\node[above=4pt]         at (v1) {$1$};
\node[above right=4pt]   at (v2) {$2$};
\node[right=6pt]         at (v3) {$3$};
\node[right=6pt]         at (v4) {$4$};
\node[below right=4pt]   at (v5) {$5$};
\node[below left=4pt]    at (v6) {$6$};
\node[left=6pt]          at (v7) {$7$};
\node[left=6pt]          at (v8) {$8$};
\node[above left=4pt]    at (v9) {$9$};

\end{tikzpicture}
\caption{A bicolored triangulation \(\mathcal T\) of type \((5,9)\), given by
five black triangles in a triangulation of a nonagon. In black is the
associated plabic graph, with polygon diagonals shown in light gray.}
\label{fig:m2_tile}
\end{figure}
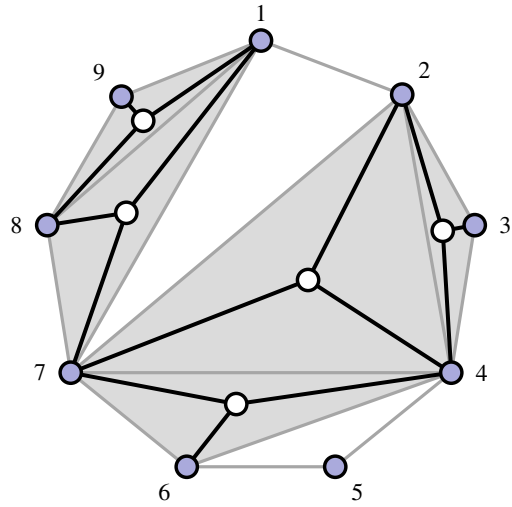

A \textit{bicolored triangulation of type \((k,n)\)} is a triangulation
\(\mathcal T\) of \(P_n\) in which \(k\) triangles are colored black and the
remaining triangles are colored white. Forgetting the internal diagonals that
separate regions of the same color gives a \textit{bicolored subdivision} of
type \((k,n)\). Given a bicolored triangulation \(\mathcal T\), we build a
labelled bipartite graph \(\mathcal{G}(\mathcal T)\) as follows: place black boundary
vertices labelled \(1,\dots,n\) at the vertices of the polygon, and place a
trivalent white vertex in the middle of each black triangle, connected to the
three vertices of that triangle; see Figure~\ref{fig:m2_tile}. To determine
the corresponding bounded affine permutation, one completes the graph by
attaching a pending edge from each boundary vertex to the outer boundary
circle. Isolated black boundary vertices are then interpreted as black
lollipops.

Let \(\overline{\mathcal T}\) be a bicolored subdivision of type \((k,n)\). A
directed arc \(i\to j\) of \(P_n\) is called \textit{compatible} with
\(\overline{\mathcal T}\) if it does not cross regions of different colors.
It is called \textit{black} if it bounds a black region of
\(\overline{\mathcal T}\). For example, in Figure~\ref{fig:m2_tile}, the arc
\(3\to7\) is compatible, while \(1\to3\) is not; the arcs \(2\to7\) and
\(2\to4\) are black, while \(1\to2\) is not.

For every compatible arc \(i\to j\), let
\(a_{\overline{\mathcal T}}(i,j)\) be the number of black triangles of
\(\overline{\mathcal T}\) lying to the left of the directed arc \(i\to j\).
For instance, in Figure~\ref{fig:m2_tile},
\begin{equation}
a_{\overline{\mathcal T}}(3,7)=2,
\qquad
a_{\overline{\mathcal T}}(4,5)
=
a_{\overline{\mathcal T}}(4,6)
=
0 \, .
\end{equation}
To every bicolored subdivision \(\overline{\mathcal T}\) of type \((k,n)\),
we associate the semialgebraic set
\begin{equation}\label{eq:m2_AT}
	\mathcal A_{\overline{\mathcal T}}
	:=
	\left\{
	Y\in\Gr(k,k+2):
	(-1)^{a_{\overline{\mathcal T}}(i,j)}
	\langle Yij\rangle
	\geq 0
	\quad
	\text{for every black arc } i\to j
	\right\} \, .
\end{equation}

Given a bicolored triangulation \(\mathcal T\), the corresponding plabic graph
\(\mathcal{G}(\mathcal T)\) determines a bounded affine permutation
\(\sigma(\mathcal T)\) of type \((k,n)\), by the usual strand rule, and hence a positroid cell
\(\Pi_{\sigma(\mathcal T),\geq 0}\). The main result for \(m=2\) is that the
image
\begin{equation}
\mathcal A_{\overline{\mathcal T}}
=
\widetilde Z(\Pi_{\sigma(\mathcal T),\geq 0})
\end{equation}
is a positroid tile, and every positroid tile arises in this way. Moreover,
\(\mathcal A_{\sigma(\mathcal T)}\) depends only on the underlying bicolored
subdivision \(\overline{\mathcal T}\), and is equal to
\(\mathcal A_{\overline{\mathcal T}}\) as defined in~\eqref{eq:m2_AT}.

Tilings are also described naturally in this language. Fix a triangulation
\(T\) of \(P_n\). Let \(\Sigma_{k,2,n}(T)\) be the collection of bicolored
triangulations of type \((k,n)\) obtained by coloring \(k\) triangles of
\(T\) black. Then
\begin{equation}\label{eq:m2_tiling}
	\mathcal A_{k,2,n}(Z)
	=
	\bigcup_{\mathcal T\in \Sigma_{k,2,n}(T)}
	\mathcal A_{\overline{\mathcal T}}
\end{equation}
is a tiling of the \(m=2\) Amplituhedron. Since an \(n\)-gon triangulation has
\(n-2\) triangles, every tiling of this type contains
\begin{equation}
| \Sigma_{k,2,n}(T)|=\binom{n-2}{k}
\end{equation}
tiles. Moreover, it was shown in~\cite{parisi2024magic} that every tiling of
the \(m=2\) Amplituhedron arises from BCFW recursion.

\begin{figure}[pos=t]
\centering
\resizebox{\textwidth}{!}{%
\begin{tikzpicture}[line cap=round,line join=round,scale=1]

\definecolor{vertexpurple}{RGB}{170,170,220}
\definecolor{Alightgreen}{RGB}{220,245,220}

\tikzset{
  circ/.style={black, line width=1.5pt},
  edge/.style={black, line width=1.5pt},
  bdot/.style={circle, draw=black, fill=vertexpurple, line width=1.3pt, inner sep=3.0pt},
  wdot/.style={circle, draw=black, fill=white, line width=1.3pt, inner sep=3.0pt},
  Aregion/.style={draw=black, fill=Alightgreen, line width=1.5pt}
}

\begin{scope}[shift={(0,0)}]

  \def\R{2.4}
  \def\r{1.55}

  \draw[circ] (0,0) circle (\R);
  \draw[Aregion] (0,0) circle (\r);

  \foreach \ang in {140,40,205,335,180,0} {
    \draw[edge] (\ang:\r) -- (\ang:\R);
  }

  \node[font=\fontsize{16}{16}\selectfont] at (0,0) {$\mathcal{A}_{n,k,2}$};

  \node at (-2.15,2.08) {$2$};
  \node at ( 1.90,2.08) {$n\!-\!2$};
  \node at (-1.72,-2.08) {$1$};
  \node at ( 1.72,-2.08) {$n\!-\!1$};
  \node at (0,-2.78) {$n$};

\end{scope}

\node[font=\fontsize{16}{16}\selectfont] at (3.25,0) {$=$};

\begin{scope}[shift={(7.0,0)}]

  \def\R{2.4}
  \def\r{1.1}
  \coordinate (c) at (0,0.75);

  \draw[circ] (0,0) circle (\R);
  \draw[Aregion] (c) circle (\r);

  \draw[edge] ($(c)+(140:\r)$) -- (140:\R);
  \draw[edge] ($(c)+(40:\r)$)  -- (40:\R);

  \draw[edge] (270:\R) -- (0,-1.2);
  \node[bdot] at (0,-1.2) {};

  \node at (-2.15,2.05) {$1$};
  \node at ( 1.90,2.10) {$n\!-\!1$};
  \node at (0,-2.75) {$n$};

  \node[font=\fontsize{14}{14}\selectfont] at (0,0.75) {$\mathcal{A}_{n-1,k,2}$};

\end{scope}

\node[font=\fontsize{16}{16}\selectfont] at (10.55,0) {$+$};

\begin{scope}[shift={(14.4,0)}]

  \def\R{2.4}
  \def\r{1.1}
  \coordinate (c) at (0,0.75);

  \draw[circ] (0,0) circle (\R);
  \draw[Aregion] (c) circle (\r);

  \draw[edge] ($(c)+(140:\r)$) -- (140:\R);
  \draw[edge] ($(c)+(40:\r)$)  -- (40:\R);

  \coordinate (iL) at ($(c)+(230:\r)$);
  \coordinate (iR) at ($(c)+(-50:\r)$);
  \coordinate (oL) at (230:\R);
  \coordinate (oR) at (-50:\R);

  \coordinate (bL) at ($(iL)!0.48!(oL)$);
  \coordinate (bR) at ($(iR)!0.48!(oR)$);
  \coordinate (w)  at (0,-1.18);

  \draw[edge] (iL) -- (bL) -- (oL);
  \draw[edge] (iR) -- (bR) -- (oR);
  \draw[edge] (bL) -- (w) -- (bR);
  \draw[edge] (w) -- (270:\R);

  \node[bdot] at (bL) {};
  \node[bdot] at (bR) {};
  \node[wdot] at (w) {};

  \node at (-2.20,2.05) {$2$};
  \node at ( 1.85,2.10) {$n\!-\!2$};
  \node at (-1.75,-2.02) {$1$};
  \node at ( 1.72,-2.10) {$n\!-\!1$};
  \node at (0,-2.78) {$n$};

  \node[font=\fontsize{14}{14}\selectfont] at (0,0.75) {$\mathcal{A}_{n-1,k-1,2}$};

\end{scope}

\end{tikzpicture}%
}
\caption{BCFW-like recursion for the \(m=2\) Amplituhedron;
see~\cite{baoHe2019m2}.}
\label{fig:m2_BCFW}
\end{figure}
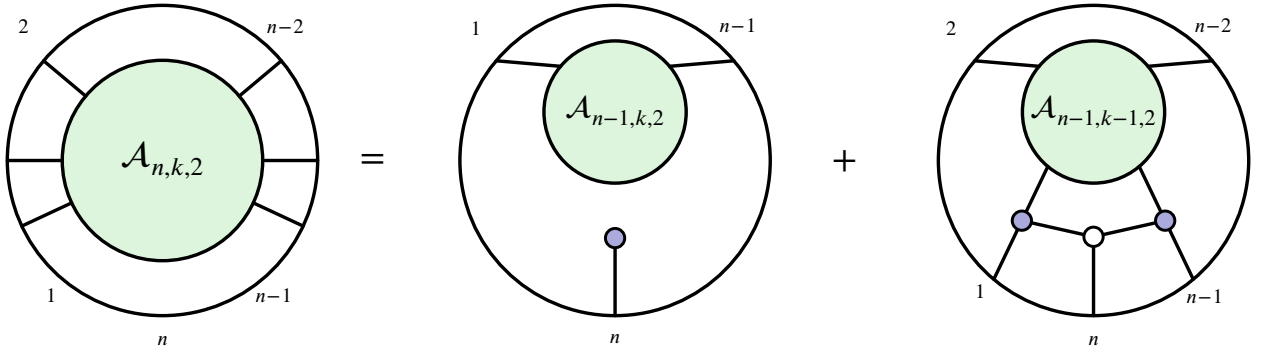

The BCFW-like recursion for the \(m=2\) Amplituhedron was introduced
in~\cite{baoHe2019m2} and is shown in Figure~\ref{fig:m2_BCFW}. From the
viewpoint of bicolored triangulations, the recursion is immediate: a
bicolored triangulation of type \((k,n)\) of \(P_n\) is obtained either from a
bicolored triangulation of type \((k,n-1)\) of \(P_{n-1}\) by adding the white
triangle \(\{1,n-1,n\}\), or from one of type \((k-1,n-1)\) by adding the
black triangle \(\{1,n-1,n\}\). These two possibilities correspond to the two
terms on the right-hand side of Figure~\ref{fig:m2_BCFW}.

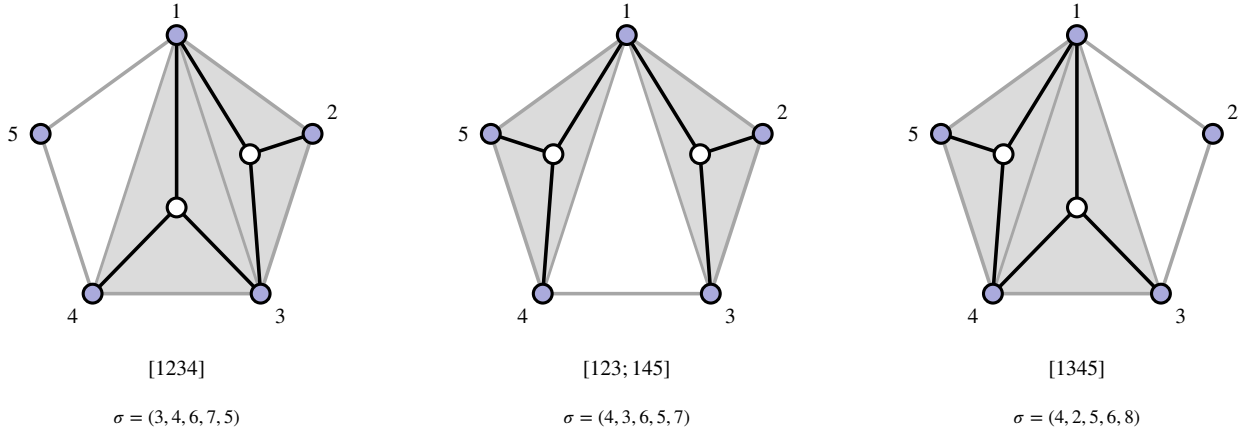
\begin{figure}[pos=t]
\centering
\resizebox{\textwidth}{!}{%
\begin{tikzpicture}[scale=0.9,line cap=round,line join=round]
\usetikzlibrary{calc}

\definecolor{vertexpurple}{RGB}{170,170,220}
\colorlet{trianglefill}{black!14}
\colorlet{diaggray}{black!35}

\tikzset{
  polyedge/.style={diaggray, line width=1.5pt},
  diagedge/.style={diaggray, line width=1.2pt},
  plabicedge/.style={black, line width=1.5pt}
}

\def\vrad{4.4pt}

\newcommand{\PentagonSetup}{
  \coordinate (v1) at ( 90:2.35);
  \coordinate (v2) at ( 18:2.35);
  \coordinate (v3) at (-54:2.35);
  \coordinate (v4) at (-126:2.35);
  \coordinate (v5) at (162:2.35);

  \coordinate (m23) at ($(v2)!0.5!(v3)$);
  \coordinate (m34) at ($(v3)!0.5!(v4)$);
  \coordinate (m45) at ($(v4)!0.5!(v5)$);

  \coordinate (w123) at ($(v1)!0.6667!(m23)$);
  \coordinate (w134) at ($(v1)!0.6667!(m34)$);
  \coordinate (w145) at ($(v1)!0.6667!(m45)$);
}

\newcommand{\PentagonFrame}{
  \draw[polyedge] (v1)--(v2)--(v3)--(v4)--(v5)--cycle;

  \draw[diagedge] (v1)--(v3);
  \draw[diagedge] (v1)--(v4);
}

\newcommand{\BoundaryLabels}{
  \node[above=4pt]       at (v1) {$1$};
  \node[above right=4pt] at (v2) {$2$};
  \node[below right=4pt] at (v3) {$3$};
  \node[below left=4pt]  at (v4) {$4$};
  \node[left=6pt]        at (v5) {$5$};
}

\begin{scope}[shift={(0,0)}]
  \PentagonSetup

  \fill[trianglefill] (v1)--(v2)--(v3)--(v4)--cycle;

  \PentagonFrame

  \draw[plabicedge] (w123)--(v1);
  \draw[plabicedge] (w123)--(v2);
  \draw[plabicedge] (w123)--(v3);

  \draw[plabicedge] (w134)--(v1);
  \draw[plabicedge] (w134)--(v3);
  \draw[plabicedge] (w134)--(v4);

  \foreach \p in {w123,w134}{
    \filldraw[fill=white,draw=black,line width=1.3pt] (\p) circle (\vrad);
  }

  \foreach \p in {v1,v2,v3,v4,v5}{
    \filldraw[fill=vertexpurple,draw=black,line width=1.3pt] (\p) circle (\vrad);
  }

  \BoundaryLabels
  \node at (0,-3.15) {$[1234]$};
  \node[align=center] at (0,-3.95)
    {\footnotesize $\sigma=(3, 4, 6, 7, 5)$};
\end{scope}

\begin{scope}[shift={(7.4,0)}]
  \PentagonSetup

  \fill[trianglefill] (v1)--(v2)--(v3)--cycle;
  \fill[trianglefill] (v1)--(v4)--(v5)--cycle;

  \PentagonFrame

  \draw[plabicedge] (w123)--(v1);
  \draw[plabicedge] (w123)--(v2);
  \draw[plabicedge] (w123)--(v3);

  \draw[plabicedge] (w145)--(v1);
  \draw[plabicedge] (w145)--(v4);
  \draw[plabicedge] (w145)--(v5);

  \foreach \p in {w123,w145}{
    \filldraw[fill=white,draw=black,line width=1.3pt] (\p) circle (\vrad);
  }

  \foreach \p in {v1,v2,v3,v4,v5}{
    \filldraw[fill=vertexpurple,draw=black,line width=1.3pt] (\p) circle (\vrad);
  }

  \BoundaryLabels
  \node at (0,-3.15) {$[123;145]$};
  \node[align=center] at (0,-3.95)
    {\footnotesize $\sigma=(4, 3, 6, 5, 7)$};
\end{scope}

\begin{scope}[shift={(14.8,0)}]
  \PentagonSetup

  \fill[trianglefill] (v1)--(v3)--(v4)--(v5)--cycle;

  \PentagonFrame

  \draw[plabicedge] (w134)--(v1);
  \draw[plabicedge] (w134)--(v3);
  \draw[plabicedge] (w134)--(v4);

  \draw[plabicedge] (w145)--(v1);
  \draw[plabicedge] (w145)--(v4);
  \draw[plabicedge] (w145)--(v5);

  \foreach \p in {w134,w145}{
    \filldraw[fill=white,draw=black,line width=1.3pt] (\p) circle (\vrad);
  }

  \foreach \p in {v1,v2,v3,v4,v5}{
    \filldraw[fill=vertexpurple,draw=black,line width=1.3pt] (\p) circle (\vrad);
  }

  \BoundaryLabels
  \node at (0,-3.15) {$[1345]$};
  \node[align=center] at (0,-3.95)
    {\footnotesize $\sigma=(4, 2, 5, 6, 8)$};
\end{scope}

\end{tikzpicture}%
}
\caption{All bicolored subdivisions of type \((2,5)\) for the underlying
triangulation \(T=\{13,14\}\) of the pentagon, together with their plabic
graphs and associated permutations.}
\label{fig:m2_k2_n5}
\end{figure}

\begin{eg}[\(m=2\), \(k=2\), \(n=5\)]\label{eg:m2_k2_n5}
	The Amplituhedron \(\mathcal A_{2,2,5}(Z) \subseteq \Gr_{\mathbb R}(2,4)\) is four-dimensional. Fix the fan
	triangulation $T=\{13,14\}$	of a pentagon. There are three bicolored triangulations of type \((2,5)\)
	over \(T\), shown in Figure~\ref{fig:m2_k2_n5}. They yield the tiling
	\begin{equation}
		\mathcal A_{2,2,5}(Z)
		=
		[1234]\cup[123;145]\cup[1345] \, ,
	\end{equation}
	where we identify each tile with its label.

	The notation \([a_1a_2a_3;b_1b_2b_3]\) denotes the tile associated with the
	bicoloring in which the triangles $\{a_1,a_2,a_3\}$ and $\{b_1,b_2,b_3\}$
	are black. In the special case where two adjacent black triangles combine
	into a quadrilateral, we use the shorter notation \([abcd]\).

	A parametrization of the tile \([1234]\), obtained with
	\texttt{Positroids.m}~\cite{Bourjaily:2012gy}, is
	\begin{equation}
		C(\alpha)
		=
		\begin{pmatrix}
   		1 & \alpha_2+\alpha_4 & \alpha_2 \alpha_3 & 0 & 0 \\
   		0 & 1 & \alpha_3 & \alpha_1 & 0
   		\end{pmatrix} \, ,
	\end{equation}
	which is essentially the same as~\eqref{eq:C_par}. Therefore \([1234]\)
	is isomorphic to \(\Gr_{\geq 0}(2,4)\).

	The tile \([123;145]\) is more interesting. According to~\eqref{eq:m2_AT},
	it is cut out by
	\begin{equation}\label{eq:123145}
		\langle Y12\rangle\geq 0,
		\quad
		\langle Y23\rangle\geq 0,
		\quad
		\langle Y13\rangle\leq0,
		\quad
		\langle Y14\rangle\leq0,
		\quad
		\langle Y45\rangle\geq 0,
		\quad
		\langle Y15\rangle\geq 0 \, .
	\end{equation}
	The six inequalities correspond to the black arcs of the bicolored
	subdivision associated with \([123;145]\). One checks that this tile has
	six boundary components, one for the vanishing of each inequality
	in~\eqref{eq:123145}.
\end{eg}

\subsubsection{The \(\mathcal A_{2,2,n}\) tiling and its forms}\label{subsec:The A22n tiling and its forms}

The tiles for $m=k=2$ and their canonical forms will play a crucial rule in the classification of leading singularities of the Wilson loop with Lagrangian insertion in Section~\ref{sec:WLwithLI}. 

The discussion of Example~\ref{eg:m2_k2_n5} extends to all
\(n\geq4\). Fix the triangulation
\begin{equation}
T_n=\{13,14,\dots,1(n-1)\}
\end{equation}
of the \(n\)-gon. Then
\begin{equation}\label{eq:m2_alln_tiling}
	\mathcal A_{2,2,n}(Z)
	=
	\bigcup_{2\leq i<j\leq n-1}
	[1,i,i+1;1,j,j+1] \, .
\end{equation}

It was shown in~\cite[Section 6]{parisiShermanBennettWilliams2023m2} that
tiles for the \(m=2\) Amplituhedron can be identified with positive parts of
cluster varieties. From~\cite[Section 5.7]{Positive_geometries}, it follows
that these tiles are positive geometries. Their canonical forms have been
determined in~\cite{Positive_geometries}; see also~\cite{lukowski2019cluster}.

We now specialize to \(m=k=2\), where a point
\(Y\in\Gr(2,4)\) may be represented by a line
\(AB\subseteq\mathbb P^3\). By abuse of notation, we denote both a tile and its
canonical function by the same bracket. For two black triangles \(\{a,b,c\}\)
and \(\{d,e,f\}\), one has the so called
\begin{tcolorbox}[definitionbox]
\textbf{Kermit forms.}

\begin{equation}\label{six_invariant}
    [abc;def]
    =
    \frac{
    \langle AB\,(abc)\cap(def)\rangle^2
    }{
    \langle ABab\rangle
    \langle ABbc\rangle
    \langle ABca\rangle
    \langle ABde\rangle
    \langle ABef\rangle
    \langle ABfd\rangle
    } \, .
\end{equation}
\end{tcolorbox}\noindent
Here
\begin{equation}\label{plane_intersection}
\begin{aligned}
\langle AB\,(abc)\cap(def)\rangle
&:=
\langle Aabc\rangle\langle Bdef\rangle
-
\langle Babc\rangle\langle Adef\rangle
\\
&=
\langle ABab\rangle\langle cdef\rangle
+
\langle ABbc\rangle\langle adef\rangle
+
\langle ABca\rangle\langle bdef\rangle.
\end{aligned}
\end{equation}
The special case of adjacent triangles gives, with our orientation convention,
\begin{equation}\label{four_invariant}
    [abcd]
    :=
    [abc;cda]
    =
    -
    \frac{
    \langle abcd\rangle^2
    }{
    \langle ABab\rangle
    \langle ABbc\rangle
    \langle ABcd\rangle
    \langle ABda\rangle
    } \, .
\end{equation}

By additivity of canonical forms under tilings, see
Section~\ref{sec:Triangulations}, one expects the canonical form of
\(\mathcal A_{2,2,n}(Z)\) to be the sum of the forms appearing in
~\eqref{eq:m2_alln_tiling}. For \(m=k=2\), this expectation is proven to be
true because the Amplituhedron is known to be a positive geometry
~\cite{Ranestad:adjoint}, and then Proposition~\ref{prop:can_form_tr} applies.

\subsubsection{BCFW recursion for \(m=4\)}

We now move to \(m=4\). In
Section~\ref{sec:On-Shell Diagrams and Grassmannian Contours}, we discussed
on-shell diagrams obtained by gluing three-point amplitudes. In planar
\(\mathcal N=4\) SYM, the super-BCFW recursion can be formulated entirely in
terms of on-shell diagrams. The reason is that the supershift
in~\eqref{eq:super_shift} corresponds diagrammatically to attaching a
\textit{BCFW bridge}~\cite{Grassmannian}. Iterating the recursion down
to three-point amplitudes produces a collection of on-shell diagrams whose
forms sum to the tree-level superamplitude at fixed \(n\) and \(k\).

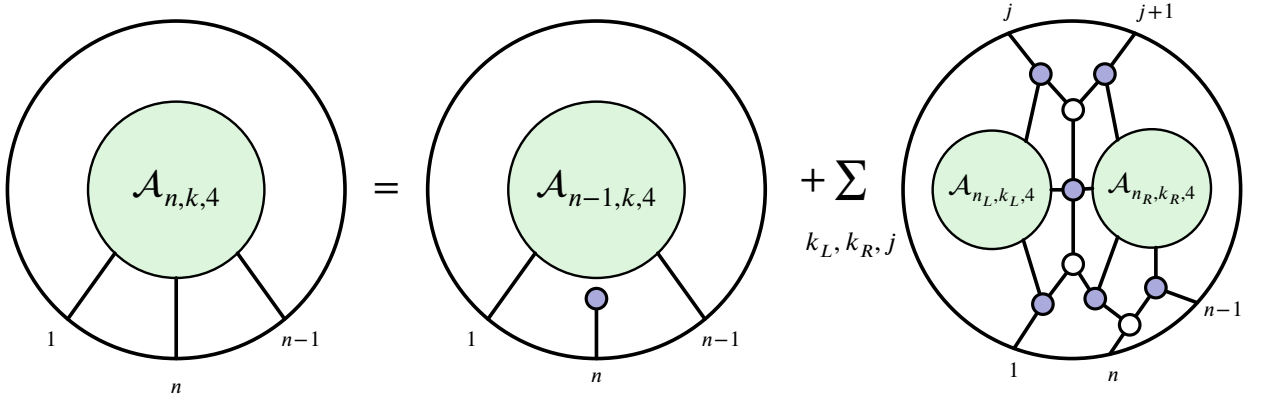
\begin{figure}[pos=t]
\centering
\resizebox{\textwidth}{!}{%
\begin{tikzpicture}[line cap=round,line join=round,scale=1]

\definecolor{vertexpurple}{RGB}{170,170,220}
\definecolor{Alightgreen}{RGB}{220,245,220}

\tikzset{
  circ/.style={black, line width=1.5pt},
  edge/.style={black, line width=1.5pt},
  bdot/.style={circle, draw=black, fill=vertexpurple, line width=1.3pt, inner sep=3.0pt},
  wdot/.style={circle, draw=black, fill=white, line width=1.3pt, inner sep=3.0pt},
  Aregion/.style={draw=black, fill=Alightgreen, line width=1.0pt}
}

\begin{scope}[shift={(0,0)}]

  \def\R{2.45}
  \def\r{1.28}

  \draw[circ] (0,0) circle (\R);
  \draw[Aregion] (0,0) circle (\r);

  \node[font=\fontsize{16}{16}\selectfont] at (0,0) {$\mathcal{A}_{n,k,4}$};

  \draw[edge] ($(0,0)+(226:\r)$) -- ($(0,0)+(230:\R)$);
  \draw[edge] ($(0,0)+(270:\r)$) -- ($(0,0)+(270:\R)$);
  \draw[edge] ($(0,0)+(314:\r)$) -- ($(0,0)+(310:\R)$);



  \node at (230:2.82) {$1$};
  \node at (270:2.88) {$n$};
  \node at (310:2.82) {$n\!-\!1$};

\end{scope}

\node[font=\fontsize{16}{16}\selectfont] at (3.05,0) {$=$};

\begin{scope}[shift={(6.10,0)}]

  \def\R{2.45}
  \def\r{1.28}

  \draw[circ] (0,0) circle (\R);
  \draw[Aregion] (0,0) circle (\r);

  \node[font=\fontsize{16}{16}\selectfont] at (0,0) {$\mathcal{A}_{n-1,k,4}$};

  \draw[edge] ($(0,0)+(226:\r)$) -- ($(0,0)+(230:\R)$);
  \draw[edge] ($(0,0)+(314:\r)$) -- ($(0,0)+(310:\R)$);

  \coordinate (dOut) at (270:\R);
  \coordinate (dV)   at (0,-1.58);
  \draw[edge] (dOut) -- (dV);
  \node[bdot] at (dV) {};



  \node at (230:2.82) {$1$};
  \node at (270:2.72) {$n$};
  \node at (310:2.82) {$n\!-\!1$};

\end{scope}

\node[font=\fontsize{16}{16}\selectfont] at (9.55,0.14) {$+ \sum$};
\node[font=\fontsize{11}{11}\selectfont] at (9.8,-0.8) {$k_L,k_R,j$};

\begin{scope}[shift={(13.00,0)}]

  \def\R{2.45}
  \def\rd{0.86}

  \draw[circ] (0,0) circle (\R);

  \coordinate (CL) at (-1.16,0.00);
  \coordinate (CR) at ( 1.16,0.02);

  \draw[Aregion] (CL) circle (\rd);
  \draw[Aregion] (CR) circle (\rd);

  \node[font=\fontsize{12}{12}\selectfont] at (CL) {$\mathcal{A}_{n_L,k_L,4}$};
  \node[font=\fontsize{12}{12}\selectfont] at (CR) {$\mathcal{A}_{n_R,k_R,4}$};



  \coordinate (aB)   at (112:\R);
  \coordinate (bB)   at ( 68:\R);
  \coordinate (oneB) at (210:\R);
  \coordinate (nB)   at (250:\R);
  \coordinate (dB)   at (283:\R);
  \coordinate (cB)   at (318:\R);

  \coordinate (Ltop)   at ($(CL)+(55:\rd)$);
  \coordinate (Lmid)   at ($(CL)+(0:\rd)$);
  \coordinate (LdownR) at ($(CL)+(-58:\rd)$);

  \coordinate (Rtop)   at ($(CR)+(125:\rd)$);
  \coordinate (Rmid)   at ($(CR)+(180:\rd)$);
  \coordinate (RdownL) at ($(CR)+(-125:\rd)$);
  \coordinate (Rdown)  at ($(CR)+(-86:\rd)$);

  \coordinate (tL) at (-0.45, 1.68);
  \coordinate (tR) at ( 0.48, 1.68);
  \coordinate (tW) at ( 0.02, 1.16);
  \coordinate (cM) at ( 0.02, 0.00);
  \coordinate (bW) at ( 0.02,-1.08);
  \coordinate (bL) at (-0.42,-1.66);

  \coordinate (bR) at ( 0.34,-1.58);
  \coordinate (dW) at ( 0.84,-1.96);
  \coordinate (cV) at ( 1.22,-1.42);

  \draw[edge] (aB) -- (tL);
  \draw[edge] (bB) -- (tR);
  \draw[edge] (tL) -- (tW);
  \draw[edge] (tR) -- (tW);
  \draw[edge] (tW) -- (cM);

  \draw[edge] (cM) -- (Lmid);
  \draw[edge] (cM) -- (Rmid);

  \draw[edge] (tL) -- (Ltop);
  \draw[edge] (tR) -- (Rtop);

  \draw[edge] (cM) -- (bW);
  \draw[edge] (bW) -- (bL);
  \draw[edge] (bW) -- (bR);

  \draw[edge] (nB) -- (bL);
  \draw[edge] (bL) -- (LdownR);

  \draw[edge] (bR) -- (RdownL);
  \draw[edge] (bR) -- (dW);
  \draw[edge] (dB) -- (dW);
  \draw[edge] (dW) -- (cV);
  \draw[edge] (cV) -- (Rdown);
  \draw[edge] (cV) -- (cB);

  \node[bdot] at (tL) {};
  \node[bdot] at (tR) {};
  \node[wdot] at (tW) {};
  \node[bdot] at (cM) {};
  \node[wdot] at (bW) {};
  \node[bdot] at (bL) {};
  \node[bdot] at (bR) {};
  \node[wdot] at (dW) {};
  \node[bdot] at (cV) {};

  \node at ($(aB)+(0,0.30)$) {$j$};
  \node at ($(bB)+(0.3,0.30)$) {$j\!+\!1$};
  \node at ($(nB)+(0,-0.30)$) {$1$};
  \node at ($(dB)+(0.05,-0.32)$) {$n$};
  \node at ($(cB)+(0.38,-0.08)$) {$n\!-\!1$};

\end{scope}

\end{tikzpicture}%
}
\caption{BCFW recursion for the tree-level \(m=4\) Amplituhedron;
see~\cite{evenZoharLakrecParisiTesslerShermanBennettWilliams2023cluster}.}
\label{fig:recursive_A_nk4}
\end{figure}

Geometrically, this collection of on-shell diagrams gives a tiling of the
\(m=4\) Amplituhedron. This was conjectured in~\cite{the_amplituhedron}
and proven in~\cite{evenZoharLakrecTessler2025bcfw}. The diagrammatic
recursion is shown in Figure~\ref{fig:recursive_A_nk4}. In the summation term,
the helicity degrees satisfy
\begin{equation}
k_L+k_R=k-1 \, ,
\end{equation}
and the bridge contributes the remaining unit of helicity degree. The index
\(j\) ranges over the allowed factorization channels, with
\begin{equation}
2\leq j\leq n-2 \, ,
\end{equation}
subject to the condition that the left and right subamplitudes lie in allowed
helicity sectors.

Tiles for the \(m=4\) Amplituhedron arising from BCFW recursion have a
combinatorial characterization in terms of chord diagrams. To each chord
diagram one associates a bounded affine permutation, and hence an on-shell
diagram. A convenient parametrization of the corresponding positroid cell is
given by \textit{domino variables}. We refer the reader to~\cite{evenZoharLakrecTessler2025bcfw,evenZoharLakrecParisiTesslerShermanBennettWilliams2023cluster} for the
details. There are also tiles and tilings for \(m=4\) that do not arise from
BCFW recursion, such as the \emph{spurion tile} studied in~\cite{evenZoharLakrecParisiTesslerShermanBennettWilliams2024clusterResults}.
Here we focus on the structure of the recursion and on simple examples.

\subsubsection{The BCFW product}

Let us now explain the operation in the last diagram of
Figure~\ref{fig:recursive_A_nk4}. At the level of on-shell diagrams, this
operation fuses two diagrams \(\Gamma_L\) and \(\Gamma_R\) by inserting a bridge between
them. It is called the \textit{BCFW product}~\cite{ArkaniHamed:2010kv}. At
the level of on-shell forms \(\mathcal Y_L\) and \(\mathcal Y_R\), 
\begin{equation}\label{eq:BCFW_prod}
\begin{aligned}
\left(
\mathcal Y_L \underset{\mathrm{BCFW}}{\otimes} \mathcal Y_R
\right)&(1,\ldots,n)
:=
[n{-}1,n,1,j,j{+}1]\,
\mathcal Y_R(1,\ldots,j,I)\,
\mathcal Y_L(I,j{+}1,\ldots,n{-}1,\widehat n),
\\
\widehat n
&=
(n{-}1,n)\cap(j,j{+}1,1),
\qquad
I
=
(j,j{+}1)\cap(n{-}1,n,1).
\end{aligned}
\end{equation}
Here the labels denote bosonic momentum twistors \(z_i\), and
\([a,b,c,d,e]\) is the R-invariant in~\eqref{eq:Rinv}. The notation
\((ab)\cap(cde)\) denotes the intersection of the line spanned by \(a,b\)
with the plane spanned by \(c,d,e\). Using the \emph{Grassmann--Cayley algebra}, which is the exterior algebra encoding linear spans and intersections of linear subspaces, this intersection can be written as
\begin{equation}\label{eq:z_int}
	(ab)\cap(cde)
	=
	z_a\langle bcde\rangle
	-
	z_b\langle acde\rangle \, .
\end{equation}

The BCFW product can be lifted to bosonized momentum supertwistors by allowing
a dependence on \(Y\in\Gr(k,k+4)\), as in
Section~\ref{sec:Hodges and Cyclic Polytopes}, where R-invariants were
identified with canonical functions of simplices. In the bosonized setting,
the same formula is interpreted after replacing the ordinary four-brackets
\(\langle abcd\rangle\) by twistor coordinates
\begin{equation}
\langle Yabcd\rangle \, .
\end{equation}
Thus the labels \(i\) in~\eqref{eq:BCFW_prod} are understood as bosonized
twistors \(Z_i\), and the intersection formula~\eqref{eq:z_int} is interpreted
with this replacement.

Tiles and tilings for the \(m=4\) Amplituhedron are closely connected with the
cluster algebra of \(\Gr(4,n)\)
~\cite{Grassmannian,evenZoharLakrecParisiTesslerShermanBennettWilliams2023cluster}. Remarkably,
the BCFW product~\eqref{eq:BCFW_prod} defines an operation on cluster
algebras, belonging to a class of maps called \textit{cluster promotion maps}
~\cite{evenZoharLakrecParisiTesslerShermanBennettWilliams2023cluster,Even-Zohar:2025ngd}.

The fundamental building block of the BCFW recursion is the R-invariant
~\eqref{eq:R_inv_2}. It is the simplest non-trivial ratio function, appearing
at \(n=5\) and \(k=1\). As we saw in
Section~\ref{sec:Hodges and Cyclic Polytopes}, it is the canonical function
of a four-dimensional projective simplex. Its on-shell diagram is shown in
Figure~\ref{fig:graph_2}, and appears as the core bridge separating the two
green blobs in the last diagram of Figure~\ref{fig:recursive_A_nk4}.

\subsubsection{Examples from BCFW recursion}

The simplest non-trivial example of the \(m=4\) recursion is the case
\(n=6\), \(k=1\), which was discussed in~\eqref{eq:R6}. The BCFW recursion
produces the three familiar R-invariants, whose on-shell diagrams are shown in
Figure~\ref{fig:BCFW_cells_k1}. Diagrammatically, they arise from the
boundary term and the factorization terms in
Figure~\ref{fig:recursive_A_nk4}, after contracting the trivial three- and
four-point subdiagrams. Geometrically, these three terms are precisely the
three simplices in a BCFW triangulation of the cyclic polytope
\(C_{4,6}(Z)\).

The \(m=4\) BCFW recursion of Figure~\ref{fig:recursive_A_nk4} is implemented
in \texttt{Positroids.m}~\cite{Bourjaily:2012gy}. For example, for \(n=7\) and physical helicity
degree \(k=2\), the package convention is to input
\begin{equation}
\kp =k+2=4 \, .
\end{equation}
Calling
\begin{equation}
\texttt{bcfwTermNames[7,4]}
\end{equation}
outputs the terms in the BCFW recursion, expressed using the BCFW product
~\eqref{eq:BCFW_prod}:
\begin{equation}\label{eq:terms_n7_k2}
\resizebox{0.65\textwidth}{!}{$
\begin{aligned}
&A_{6}^{(4)} \otimes A_{3}^{(1)},&
&A_{5}^{(3)} \otimes A_{4}^{(2)},&
&A_{4}^{(2)} \otimes A_{5}^{(3)} \, ,\\
&A_{3}^{(2)} \otimes \left(A_{5}^{(3)} \otimes A_{3}^{(1)}\right),&
&A_{3}^{(2)} \otimes \left(A_{4}^{(2)} \otimes A_{4}^{(2)}\right),&
&A_{3}^{(2)} \otimes \left(A_{3}^{(2)} \otimes A_{5}^{(2)}\right).
\end{aligned}
$}
\end{equation}
Here the superscript is the package's helicity label \(\kp=k+2\). The BCFW
product is not associative, so the parentheses in~\eqref{eq:terms_n7_k2} are
part of the data.

\begin{figure}[pos=t]
\centering

\begin{tabular}{c@{\quad}c@{\quad}c@{\quad}c@{\quad}c}
\begin{minipage}{0.26\textwidth}
\centering
\includegraphics[width=\linewidth]{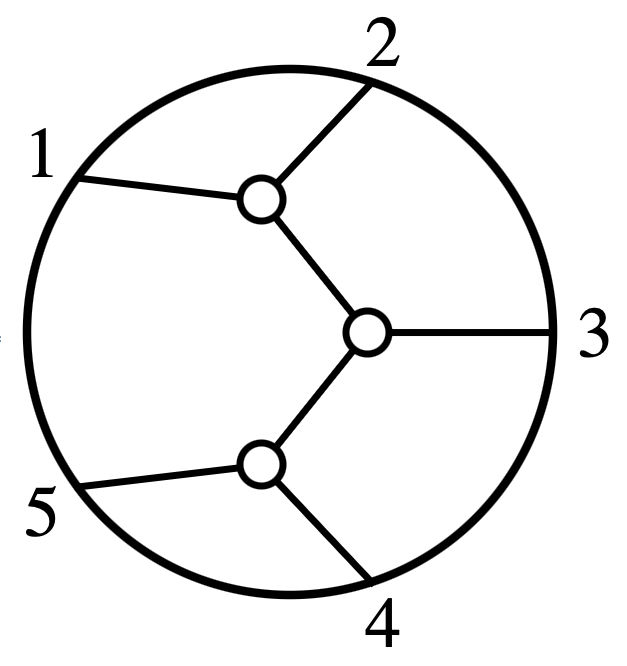}\\[-1mm]
{\small $(4,5,6,7,8)$}
\end{minipage}
&
{\Large $\underset{\mathrm{BCFW}}{\otimes}$}
&
\begin{minipage}{0.26\textwidth}
\centering
\includegraphics[width=\linewidth]{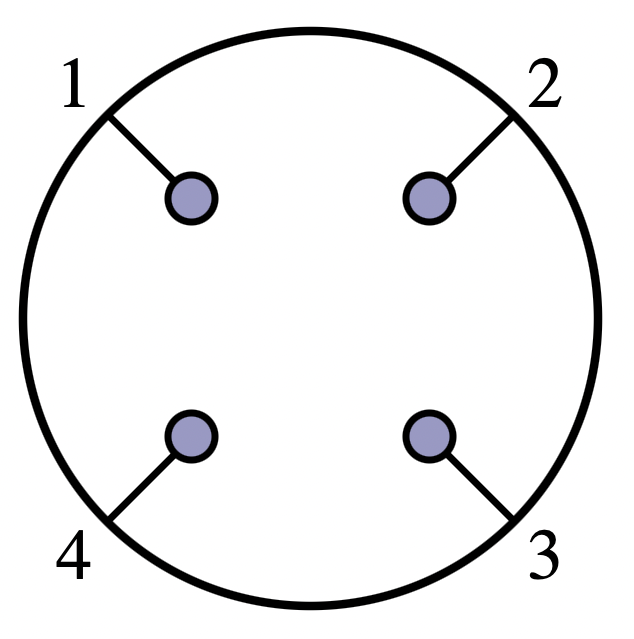}\\[-1mm]
\vspace{2mm}
{\small $(3,4,5,6)$}
\end{minipage}
&
{\Large $=$}
&
\begin{minipage}{0.26\textwidth}
\centering
\includegraphics[width=\linewidth]{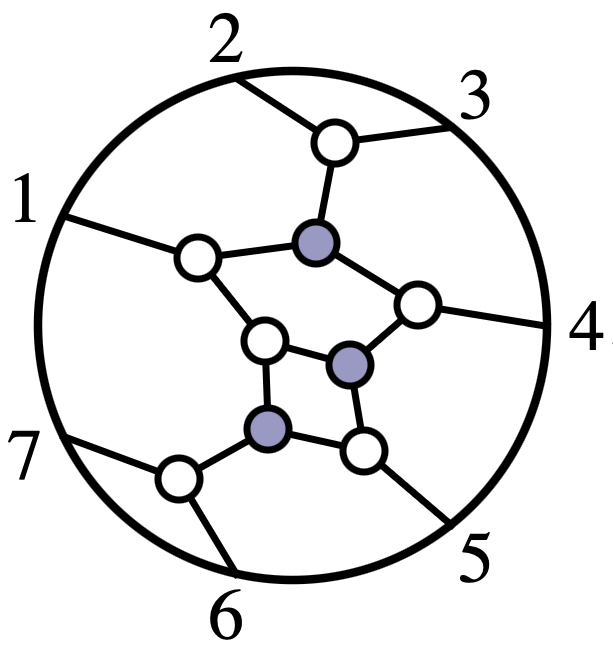}\\[-1mm]
{\small $(4,6,7,9,8,10,12)$}
\end{minipage}
\end{tabular}

\caption{BCFW product corresponding to the second term in~\eqref{eq:terms_n7_k2},
contributing to the recursion for \(n=7\) and \(k=2\).}
\label{fig:bcfw-product-example}

\end{figure}

If one is interested only in the number of terms, the command
\begin{equation}
\texttt{termsInBCFW[7,4]}
\end{equation}
returns \(6\). More importantly, the command
\begin{verbatim}
treeContour[n, k+2] // dualGrassmannian
\end{verbatim}
outputs a list of bounded affine permutations corresponding to the on-shell
diagrams in the BCFW recursion. For \(n=7\) and \(k=2\), one obtains the six
permutations
\begin{equation}
\begin{aligned}
& (5,6,7,9,10,8,11) \, ,\quad
  (4,6,7,9,8,10,12) \, ,\quad
  (3,6,7,8,9,11,12) \, ,
\\
& (5,7,6,8,10,9,11) \, ,\quad
  (4,7,6,8,9,10,12) \, ,\quad
  (5,7,8,6,9,10,11) \, .
\end{aligned}
\end{equation}
Let us examine the second permutation:
\begin{equation}
(4,6,7,9,8,10,12) \, .
\end{equation}
As suggested by the second term in~\eqref{eq:terms_n7_k2}, this term arises
from the BCFW product of a five-point invariant and the unique four-point
\(k=0\) invariant. Its on-shell diagram is shown in
Figure~\ref{fig:bcfw-product-example}. The corresponding on-shell function is
\begin{equation}
[1,2,3,4,5]\underset{\mathrm{BCFW}}{\otimes} 1
=
[6,7,1,4,5]\,[1,2,3,4,I],
\qquad
I=(67)\cap(451) \, ,
\end{equation}
where the unique Yangian invariant at \(n=4\) and \(k=0\) is the constant function~\(1\).

In the bosonized \(k=2\) setting, this intersection can be expanded using the
same Grassmann--Cayley rule as in~\eqref{eq:z_int}. For instance, for any
appropriate five labels \(a,b,c,d,e\),
\begin{equation}
	\langle Iabcde\rangle
	=
	\langle6abcde\rangle\langle Y7451\rangle
	-
	\langle7abcde\rangle\langle Y6451\rangle \, .
\end{equation}
This gives an explicit rational expression for the corresponding BCFW tile
form in bosonized momentum-twistors.

\subsection{Loop Amplituhedra}\label{sec:Loop Amplituhedra}

\subsubsection{Loop-level BCFW recursion }

In the previous sections we discussed tree Amplituhedra. The terminology
comes from the fact that, for \(m=4\), the Amplituhedron encodes tree-level
scattering amplitudes in planar \(\mathcal N=4\) SYM. A central role was
played by BCFW recursion, which is implemented geometrically by a tiling of
the \(m=4\) tree Amplituhedron by positroid images.

We now move to loop level. As discussed in
Section~\ref{sec:Recursion Relations}, tree-level BCFW recursion reconstructs
amplitudes from their factorization singularities, which arise when an
internal propagator goes on shell~\cite{Britto:2004ap,Britto:2005fq}. As we discussed in Section~\ref{sec:Recursion Relations}. At
loop level, analogous recursion relations exist for the integrand, but two
new issues appear. First, the loop integrand is not canonically defined in a
generic quantum field theory, because different Feynman diagrams involve
different choices of loop-momentum routing. Second, in addition to ordinary
factorization channels, one must include forward-limit contributions, where
two internal on-shell legs are identified with opposite momenta.

Both issues are especially well behaved in planar \(\mathcal N=4\) SYM. The
first is resolved by passing to dual momentum coordinates and momentum
twistors, which remove the ambiguity in the parametrization of planar Feynman integrands.
The second is controlled by supersymmetry: after summing over the full
on-shell supermultiplet, problematic forward-limit singularities such as
external bubbles and tadpoles cancel~\cite{ArkaniHamed:2010kv}. Thus, in
planar \(\mathcal N=4\) SYM, one can define a loop integrand unambiguously at
fixed \(n\), helicity degree \(k\), and loop order \(\ell\).

\begin{figure}[pos=t]
\centering
\resizebox{\textwidth}{!}{%
\begin{tikzpicture}[line cap=round,line join=round,scale=1]

\definecolor{vertexpurple}{RGB}{170,170,220}
\definecolor{Alightgreen}{RGB}{220,245,220}
\definecolor{loopblue}{RGB}{70,155,215}
\definecolor{bridgered}{RGB}{210,70,70}

\tikzset{
  circ/.style={black, line width=1.5pt},
  edge/.style={black, line width=1.5pt},
  rededge/.style={draw=bridgered, line width=1.5pt},
  blueedge/.style={draw=loopblue, line width=1.5pt},
  bdot/.style={circle, draw=black, fill=vertexpurple, line width=1.3pt, inner sep=3.0pt},
  wdot/.style={circle, draw=black, fill=white, line width=1.3pt, inner sep=3.0pt},
  Aregion/.style={draw=black, fill=Alightgreen, line width=1.0pt}
}

\begin{scope}[shift={(0,0)}]

  \def\R{2.42}
  \def\r{1.05}

  \draw[circ] (0,0) circle (\R);
  \draw[Aregion] (0,0) circle (\r);

  \node[font=\fontsize{16}{16}\selectfont] at (0,0) {$\mathcal{A}_{n}^{(k),\ell}$};

  \draw[edge] ($(0,0)+(142:\r)$) -- ($(0,0)+(142:\R)$);
  \draw[edge] ($(0,0)+( 38:\r)$) -- ($(0,0)+( 38:\R)$);
  \draw[edge] ($(0,0)+(198:\r)$) -- ($(0,0)+(198:\R)$);
  \draw[edge] ($(0,0)+(246:\r)$) -- ($(0,0)+(246:\R)$);
  \draw[edge] ($(0,0)+(314:\r)$) -- ($(0,0)+(314:\R)$);


  \node at (-1.25,-2.5) {$1$};
  \node at ( 1.7,-2.05) {$n$};

\end{scope}

\node[font=\fontsize{16}{16}\selectfont] at (3.10,0) {$=$};

\node[font=\fontsize{16}{16}\selectfont] at (4.55,0.18) {$\sum$};
\node[font=\fontsize{13}{13}\selectfont,align=left] at (4.80,-1.46) {%
$\ell_L+\ell_R=\ell$\\
$k_L+k_R=k+1$\\
$n_L+n_R=n+2$
};

\begin{scope}[shift={(8.55,0)}]

  \def\R{2.52}
  \def\rxA{0.86}
  \def\ryA{0.66}

  \draw[circ] (0,0) circle (\R);

  \coordinate (CL) at (-1.14,0.18);
  \coordinate (CR) at ( 1.14,0.18);

  \coordinate (Ltop)  at ($(CL)+({\rxA*cos(105)},{\ryA*sin(105)})$);
  \coordinate (Lside) at ($(CL)+(-\rxA,0)$);
  \coordinate (Lbot)  at ($(CL)+(0,-\ryA)$);
  \coordinate (Lred)  at ($(CL)+(\rxA,0)$);

  \coordinate (Rtop)  at ($(CR)+({\rxA*cos(75)},{\ryA*sin(75)})$);
  \coordinate (Rside) at ($(CR)+(\rxA,0)$);
  \coordinate (Rbot)  at ($(CR)+(0,-\ryA)$);
  \coordinate (Rred)  at ($(CR)+(-\rxA,0)$);

  \coordinate (LW) at (-1.14,-1.56);
  \coordinate (RB) at ( 1.14,-1.56);

  \draw[edge] (-1.00, 2.30) -- (Ltop);
  \draw[edge] ( 1.00, 2.30) -- (Rtop);
  \draw[edge] (-2.35,-0.76) -- (Lside);
  \draw[edge] ( 2.35,-0.76) -- (Rside);

  \draw[edge] (-1.74,-2.00) -- (LW)
    node[pos=0.20, below left=1pt] {$1$};
  \draw[edge] ( 1.80,-1.90) -- (RB)
    node[pos=0.20, below right=1pt] {$n$};
  \draw[edge] (LW) -- (RB);
  \draw[edge] (LW) -- (Lbot);
  \draw[edge] (RB) -- (Rbot);

  \draw[Aregion] (CL) ellipse [x radius=\rxA, y radius=\ryA];
  \draw[Aregion] (CR) ellipse [x radius=\rxA, y radius=\ryA];

  \node[font=\fontsize{12}{12}\selectfont] at (CL) {$\mathcal{A}_{n_L}^{(k_L),\ell_L}$};
  \node[font=\fontsize{12}{12}\selectfont] at (CR) {$\mathcal{A}_{n_R}^{(k_R),\ell_R}$};

  \draw[rededge] (Lred) -- (Rred);

  \node[wdot] at (LW) {};
  \node[bdot] at (RB) {};



\end{scope}

\node[font=\fontsize{16}{16}\selectfont] at (11.95,0) {$+$};

\begin{scope}[shift={(15.85,0)}]

  \def\R{2.52}
  \def\r{1.00}

  \draw[circ] (0,0) circle (\R);
  \draw[Aregion] (0,0) circle (\r);

  \node[font=\fontsize{14}{14}\selectfont] at (0,0) {$\mathcal{A}_{n+2}^{(k+1),\ell-1}$};

  \draw[edge] ($(0,0)+(140:\r)$) -- ($(0,0)+(140:\R)$);
  \draw[edge] ($(0,0)+( 40:\r)$) -- ($(0,0)+( 40:\R)$);
  \draw[edge] ($(0,0)+(192:\r)$) -- ($(0,0)+(192:\R)$);
  \draw[edge] ($(0,0)+(348:\r)$) -- ($(0,0)+(348:\R)$);

  \coordinate (oneB) at ($(0,0)+(228:\R)$);
  \coordinate (nB)   at ($(0,0)+(312:\R)$);

  \coordinate (Wc) at (-1.00,-1.22);
  \coordinate (Bc) at ( 1.00,-1.22);

  \coordinate (GW) at ($(0,0)+(236:\r)$);
  \coordinate (GB) at ($(0,0)+(304:\r)$);

  \coordinate (BL) at ($(0,0)+(245:\r)$);
  \coordinate (BR) at ($(0,0)+(295:\r)$);

  \draw[edge] (oneB) -- (Wc);
  \draw[edge] (nB) -- (Bc);

  \node at ($(oneB)+(-0.16,-0.12)$) {$1$};
  \node at ($(nB)+( 0.16,-0.12)$) {$n$};

  \draw[edge] (Wc) -- (GW);
  \draw[edge] (Bc) -- (GB);

  \draw[edge] (Wc) .. controls (-0.45,-1.98) and (0.45,-1.98) .. (Bc);

  \draw[blueedge]
    (BL) .. controls (-0.96,-1.56) and (0.96,-1.56) .. (BR);


  \node[wdot] at (Wc) {};
  \node[bdot] at (Bc) {};

\end{scope}

\end{tikzpicture}%
}
\caption{Diagrammatic BCFW recursion for loop-level integrands in $\mathcal{N}=4$ SYM. }
\label{fig:BCFW_loop}
\end{figure}
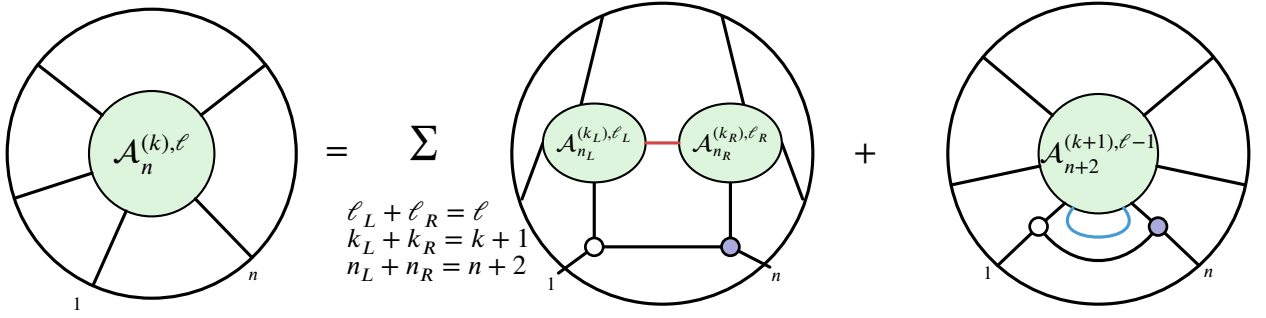

This loop integrand is a rational function of the external bosonized momentum
twistors \(Z_i\in\mathbb P^{k+3}\), for \(i=1,\dots,n\), together with
\(\ell\) loop lines \(AB_a\). It satisfies a BCFW recursion whose diagrammatic
form is shown in Figure~\ref{fig:BCFW_loop}; see
~\cite{ArkaniHamed:2010kv,Grassmannian}. The first term on the
right-hand side is the loop-level factorization. The second
term is the forward-limit contribution, which lowers the loop order by one
and raises the helicity degree by one.

Similarly to the BCFW product~\eqref{eq:BCFW_prod} at tree level, the forward
limit has both a diagrammatic and an algebraic interpretation: two momentum
twistors are fused into a single loop line. We refer to
~\cite{Elvang:2013cua,ArkaniHamed:2010kv} for the physical construction, and
to~\cite{tessler2025notes} for a more mathematical formulation.

The loop-Amplituhedron conjecture is the loop-level extension of the tree
story. It predicts that there exists a positive geometry whose canonical form
is the full loop integrand of planar \(\mathcal N=4\) SYM, and whose tilings
are produced by loop-level BCFW recursion.

\subsubsection{The loop Grassmannian and the loop Amplituhedron}

We now introduce the loop Amplituhedron. Let \(k,m,n\) be non-negative
integers as in the tree case, and let \(\ell\geq 0\) be the loop order. We
first define the \(\ell\)-loop Grassmannian
\begin{equation}
\Gr(k,n;\ell) \, ,
\end{equation}
which is naturally a subvariety of
\begin{equation}
\Gr(k,n)\times\Gr(2,n)^\ell \, .
\end{equation}
A point of \(\Gr(k,n;\ell)\) is a tuple
\begin{equation}
(C,\{D^{(l)}\}_{a=1}^{\ell}) \, ,
\end{equation}
where \(C\in\Gr(k,n)\) is the tree part and
\(D^{(a)}\in\Gr(2,n)\) are the loop parts, subject to the condition that
each \(D^{(a)}\) lies in the orthogonal complement of \(C\). Equivalently,
after choosing matrix representatives, one requires
\begin{equation}
D^{(a)}\cdot C^\top=0,
\qquad
a=1,\dots,\ell \, .
\end{equation}
Thus
\begin{equation}
\dim \Gr(k,n;\ell)
=
k(n-k)+2\ell(n-k-2) \, .
\end{equation}
We represent a point of \(\Gr(k,n;\ell)\) by stacking the loop matrices
above the tree matrix:
\begin{equation}
\left(C,\{D^{(a)}\}\right)
=
\left(
\begin{array}{c}
D^{(1)}
\\
D^{(2)}
\\
\vdots
\\
D^{(\ell)}
\\
C
\end{array}
\right).
\label{eq:loop-grassmannian-matrix}
\end{equation}
The positive part \(\Gr_{>0}(k,n;\ell)\) is defined by requiring the
maximal minors of every matrix obtained by stacking any subset of the
\(D^{(a)}\)'s on top of \(C\) to be positive:
\begin{equation}
C,
\qquad
\left(
\begin{array}{c}
D^{(a_1)}
\\
C
\end{array}
\right),
\qquad
\left(
\begin{array}{c}
D^{(a_1)}
\\
D^{(a_2)}
\\
C
\end{array}
\right),
\qquad
\cdots,
\qquad
\left(
\begin{array}{c}
D^{(a_1)}
\\
\vdots
\\
D^{(a_r)}
\\
C
\end{array}
\right)
\label{eq:loop_Gr_pos}
\end{equation}
for every choice of distinct loop indices \(a_1,\dots,a_r\). Here \(r\) ranges
over the values for which the displayed matrix has at most \(n\) rows. The
non-negative part \(\Gr_{\geq 0}(k,n;\ell)\) is the Euclidean closure of
\(\Gr_{>0}(k,n;\ell)\) in the corresponding real product of
Grassmannians.

We extend the Amplituhedron map~\eqref{eq:Ampl_map} to the loop Grassmannian
by
\begin{tcolorbox}[definitionbox]
\textbf{Loop Amplituhedron map.}

\begin{equation}\label{eq:loop_ampl_map}
\begin{aligned}
	\widetilde Z:
	\Gr(k,n;\ell)
	&\longrightarrow
	\Gr(k,k+m;\ell),
	\\
	\left(C,\{D^{(a)}\}\right)
	&\longmapsto
	\left(
	Y,\{AB_a\}
	\right)
	:=
	\left(
	C\cdot Z^\top,
	\{D^{(a)}\cdot Z^\top\}_{a=1}^{\ell}
	\right).
\end{aligned}
\end{equation}
\end{tcolorbox}\noindent
The \(\ell\)-loop Amplituhedron is then defined by
\begin{tcolorbox}[definitionbox]
\textbf{Loop Amplituhedron.}

\begin{equation}\label{eq:loop_amplituhedron_def}
	\mathcal A_{k,m,n}^{(\ell)}(Z)
	:=
	\widetilde Z\bigl(\Gr_{\geq 0}(k,n;\ell)\bigr) \, .
\end{equation}
\end{tcolorbox}\noindent
It is a full-dimensional semialgebraic subset of
\(\Gr(k,k+m;\ell)\), of dimension
\begin{equation}
\dim \mathcal A_{k,m,n}^{(\ell)}(Z) = mk+2\ell(m-2) \, .
\end{equation}
The loop Grassmannian and the loop Amplituhedron carry a natural
\(S_\ell\)-symmetry, which permutes the \(\ell\) loop variables. This symmetry
is expected from the \(\ell\)-loop integrand of planar \(\mathcal N=4\) SYM.

Loop Grassmannians and loop Amplituhedra can be defined more generally, with
loop spaces of dimensions other than two; see
\cite[Section 6.4]{Positive_geometries}. As for the tree case, for scattering amplitudes in four
dimensions, the relevant case is \(m=4\), and where each loop is represented by a line \(AB_a\) in momentum-twistor space.

Let us record a few useful limiting cases.

\begin{itemize}
	\item If \(\ell=0\), then
	\begin{equation}
\mathcal A_{k,m,n}^{(0)}(Z)
	=
	\mathcal A_{k,m,n}(Z) \, ,
\end{equation}
	so the loop Amplituhedron reduces to the tree Amplituhedron.

	\item If \(n=k+m\), then \(\widetilde Z\) is an isomorphism and
	\begin{equation}
\mathcal A_{k,m,k+m}^{(\ell)}(Z)
	\cong
	\Gr_{\geq 0}(k,k+m;\ell) \, .
\end{equation}

	\item If \(k=0\) and \(\ell=1\), then
	\begin{equation}
\Gr(0,n;1)\cong\Gr(2,n) \, ,
\end{equation}
	and hence
	\begin{equation}\label{eq:iso_l1_m2}
		\mathcal A_{0,m,n}^{(1)}(Z)
		\cong
		\mathcal A_{2,m-2,n}(Z) \, .
	\end{equation}
	In particular, for \(m=4\), the one-loop MHV Amplituhedron is identified
	with the tree-level \(m=2\), \(k=2\) Amplituhedron. We will use this identification later.
\end{itemize}

\subsubsection{Inequalities and the fibration viewpoint}

The sign-flip description of the tree Amplituhedron in~\eqref{eq:m4_cond}
extends naturally to loop level. For the physical case \(m=4\), the loop
Amplituhedron \(\mathcal A_{k,4,n}^{(\ell)}(Z)\) is described by the
tree-level conditions~\eqref{eq:m4_cond}, together with the loop-positivity
conditions
\begin{equation}\label{eq:loop_pos}
\begin{aligned}
&\langle Y\,AB_a\,i\,i{+}1\rangle\geq 0,
\\
&
\big(
\langle Y\,AB_a\,12\rangle,
\langle Y\,AB_a\,13\rangle,
\dots,
\langle Y\,AB_a\,1n\rangle
\big)
\quad
\text{has \(k+2\) sign flips,}
\end{aligned}
\end{equation}
for every \(a\in[\ell]\), and the multi-loop positivity conditions
\begin{equation}\label{eq:mult_pos}
	\langle Y\,AB_{a}\,AB_{b}\rangle\geq 0 \, ,
\end{equation}
for every pair \(a,b\in[\ell]\).

The conditions~\eqref{eq:loop_pos} are precisely the \(m=2\) sign-flip
conditions~\eqref{eq:m2_cond}, with \(Y\) replaced by the quotient geometry
seen by the loop line \(AB_l\) modulo the tree plane \(Y\). This leads to the
following useful viewpoint. There is a natural projection forgetting the loop
variables:
\begin{equation}
	\pi_{\rm tree}:
	\mathcal A_{k,m,n}^{(\ell)}(Z)
	\longrightarrow
	\mathcal A_{k,m,n}^{(0)}(Z),
	\qquad
	\left(Y,\{AB_a\}\right)
	\longmapsto
	Y \, .
\end{equation}
For fixed \(Y\), the fiber \(\pi_{\rm tree}^{-1}(Y)\) has dimension
\(2\ell(m-2)\), and is cut out by the loop-positivity and multi-loop
positivity conditions~\eqref{eq:loop_pos} and~\eqref{eq:mult_pos} for fixed $Y$.

The fibration method aims to decompose the tree-level Amplituhedron into
chambers over which the fibers have constant combinatorial type. The canonical
form of the full loop Amplituhedron is then obtained by summing, over all
chambers, the wedge product of the chamber form with the canonical form of the
corresponding fiber. This strategy has been developed mainly for the momentum
Amplituhedron
~\cite{Damgaard:2019ztj,Ferro:2022abq,Ferro:2023qdp,Ferro:2024fibration},
and also appears in related work on the ABJM Amplituhedron and the
Correlahedron~\cite{parisiShermanBennettWilliams2023m2,ABJM_amplituhedron,He:2023exb,He:2025correlahedron}.
We now illustrate the idea with a toy example.

\begin{eg}[\(k=1\), \(m=3\), \(n=4\), \(\ell=1\)]
The tree Amplituhedron \(\mathcal A_{1,3,4}^{(0)}(Z)\) is 
\begin{equation}
\Delta={\rm conv}(Z_1,Z_2,Z_3,Z_4)\subseteq\mathbb P^3 \, .
\end{equation}
The one-loop Amplituhedron \(\mathcal A_{1,3,4}^{(1)}(Z)\) consists of pairs
\((Y,AB)\), where \(Y\in\Delta\) and \(AB\subseteq\mathbb P^3\) is a line,
subject to the sign-flip condition
\begin{equation}\label{eq:l1_m3}
	\big(
	\langle Y\,AB\,1\rangle,
	\langle Y\,AB\,2\rangle,
	\langle Y\,AB\,3\rangle,
	\langle Y\,AB\,4\rangle
	\big)
	\quad
	\text{has \(3\) sign flips.}
\end{equation}
Let \(\Pi=(YAB)\) be the plane spanned by \(Y\) and the line \(AB\), and set
\begin{equation}
\lambda_i:=\langle Y\,AB\,i\rangle \, .
\end{equation}
Up to an overall projective sign,~\eqref{eq:l1_m3} fixes the sign pattern of
\((\lambda_1,\lambda_2,\lambda_3,\lambda_4)\) to be
\begin{equation}
(+,-,+,-) \, .
\end{equation}
Geometrically, this means that the plane \(\Pi\) separates
\(\{Z_1,Z_3\}\) from \(\{Z_2,Z_4\}\). Hence \(\Pi\) intersects the four edges
\((12)\), \((23)\), \((34)\), and \((41)\) of the tetrahedron, while it
misses \((13)\) and \((24)\). The four boundary components of the fiber over
fixed \(Y\) occur when some \(\lambda_i=0\), equivalently when \(\Pi\)
contains one of the vertices \(Z_i\).

We now compute the canonical form using the fibration. Write
\begin{equation}
Y=\sum_{i=1}^{4} c_i Z_i,
\qquad
c_i>0 \, .
\end{equation}
Since \(\langle YABY\rangle=0\), the four \(\lambda_i\)'s obey
\begin{equation}
	c_1\lambda_1+c_2\lambda_2+c_3\lambda_3+c_4\lambda_4=0 \, .
\end{equation}
Because \(\lambda_1,\lambda_3>0\) and \(\lambda_2,\lambda_4<0\), define
\begin{equation}
	\alpha
	=
	\frac{c_1\lambda_1}{c_1\lambda_1+c_3\lambda_3},
	\qquad
	\beta
	=
	\frac{-c_2\lambda_2}{-c_2\lambda_2-c_4\lambda_4} \, .
\end{equation}
Then \(0<\alpha<1\) and \(0<\beta<1\). Thus the fiber is identified with the
square \((0,1)^2\). Its canonical form is
\begin{equation}
\mathrm{d}\log\frac{\alpha}{1-\alpha}
\wedge
\mathrm{d}\log\frac{\beta}{1-\beta} \, .
\end{equation}
Substituting back
\begin{equation}
c_1=\frac{\langle Y234\rangle}{\langle1234\rangle},
\qquad
c_2=\frac{\langle Y341\rangle}{\langle1234\rangle},
\qquad
c_3=\frac{\langle Y412\rangle}{\langle1234\rangle},
\qquad
c_4=\frac{\langle Y123\rangle}{\langle1234\rangle} \, ,
\end{equation}
one obtains
\begin{equation}\label{eq:om_fib}
	\Omega_{\rm fib}
	=
	\mathrm{d}\log
	\frac{
	\langle Y234\rangle\,\langle YAB1\rangle
	}{
	\langle Y412\rangle\,\langle YAB3\rangle
	}
	\wedge
	\mathrm{d}\log
	\frac{
	\langle Y341\rangle\,\langle YAB2\rangle
	}{
	\langle Y123\rangle\,\langle YAB4\rangle
	} \, .
\end{equation}
The canonical form of the tetrahedron \(\Delta\) is
\begin{equation}
	\Omega_{\rm tree}
	=
	\frac{
	\langle1234\rangle^{3}\,\langle Y\,\mathrm{d}^3Y\rangle
	}{
	\langle Y123\rangle
	\langle Y234\rangle
	\langle Y341\rangle
	\langle Y412\rangle
	} \, .
\end{equation}
Therefore the full one-loop canonical form is
\begin{equation}
\Omega^{(1)}_{1,3,4}(Y,AB)
=
\Omega_{\rm tree}(Y)\wedge\Omega_{\rm fib}(Y,AB) \, .
\end{equation}

In this example, the whole tree space \(\Delta\) is a single chamber: the
fiber has the same combinatorial type over every point \(Y\in\Delta\). In
general, one must subdivide the tree Amplituhedron into several
chambers so that the fibers have constant canonical form.
\end{eg}

The previous example is close to the physical case. For \(m=4\), the one-loop
Amplituhedron with \(k=1\) consists of pairs \((Y,AB)\), where \(Y\) lies in
the four-dimensional cyclic polytope \(C_{4,n}(Z)\) and \(AB\) is a line in
\(\mathbb P^4\), subject to loop positivity~\eqref{eq:loop_pos}. Its
canonical form computes the one-loop NMHV integrand in planar
\(\mathcal N=4\) SYM. For \(n=5\), the tree polytope is a simplex and the
analysis is similar to the example above. At six points, the fibration
already requires a non-trivial chamber decomposition; see
\cite[Equation (7.12)]{Ferro:2023qdp} for an explicit result.

\subsection{The four-point two-loop Amplituhedron}
\label{sec:The Four-Point Two-Loop Amplituhedron}

In this section we study one loop Amplituhedron in detail. This section serves as a
concrete model for the boundary and residual arrangements that will reappear
in the later analysis of singularities of loop amplitudes. We set
\begin{equation}
n=4,\qquad k=0,\qquad m=4,\qquad \ell=2 \, .
\end{equation}
This is the Amplituhedron for the four-point two-loop MHV integrand in planar
\(\mathcal N=4\) SYM. The results in this section are based on
~\cite{Dian:2024hil}. Our goal is to illustrate the intricate boundary
structure of this loop Amplituhedron, which reflects the singularities of the
integrand from the point of view of iterated residues. We also introduce
computational tools for studying the relevant complex and real algebraic
geometry. Figures in this section are adapted from~\cite{Dian:2024hil}.
Finally, we use the adjoint method of
Section~\ref{sec:Adjoint Hypersurface} to determine the canonical form
uniquely, and hence recover the previously known two-loop integrand. This extends the
one-loop MHV analysis of~\cite{Ranestad:adjoint}, equivalently the case
\(m=k=2\) and \(\ell=0\).

Throughout this section we abbreviate
\begin{equation}
\mathcal A_4^{(\ell)}:=\mathcal A_{0,4,4}^{(\ell)}(Z) \, .
\end{equation}
Since \(n=k+m\), the two-loop Amplituhedron \(\mathcal A_4^{(2)}\) is an
isomorphic copy of the two-loop non-negative Grassmannian
\(\Gr_{\geq 0}(0,4;2)\). After projecting out the trivial tree part, this
is the eight-dimensional semialgebraic set
\begin{equation}\label{two_loop_Ampl}
	\mathcal A_4^{(2)}(Z)
	\cong
	\left\{
	(AB,CD)\in \mathcal A_4^{(1)}\times \mathcal A_4^{(1)}
	:
	\langle ABCD\rangle\geq 0
	\right\} \, ,
\end{equation}
where \(\mathcal A_4^{(1)}\) is an isomorphic copy of
\(\Gr_{\geq 0}(2,4)\).

The inequalities cutting out the interior of \(\mathcal A_4^{(1)}\) are
obtained from the isomorphism~\eqref{eq:iso_l1_m2} and the sign-flip
conditions~\eqref{eq:m2_cond}:
\begin{equation}\label{eq:l1_cons}
\begin{aligned}
	&\langle AB12\rangle>0 \, ,\
	\langle AB23\rangle>0 \, ,\
	\langle AB34\rangle>0 \, ,\
	\langle AB14\rangle>0 \, ,\
    \langle AB13\rangle<0 \, ,\
	\langle AB24\rangle<0 \, .
\end{aligned}
\end{equation}
The (closed) Amplituhedron is obtained by replacing strict inequalities by
non-strict ones.

\subsubsection{A geometric picture of the one-loop space}

To visualize the two-loop space, we first recall a useful geometric picture
of \(\mathcal A_4^{(1)}\). Fix an affine chart
\(\mathbb R^3\subseteq\mathbb P^3\) containing the four external points
\(Z_i\). Since a projective line is determined by any two distinct points on
it, we choose one point on the line \(AB\) to be
\begin{equation}\label{A_on_triangle}
    A
    :=
    AB\cap(-412)
    =
    -Z_4\langle AB12\rangle
    +Z_1\langle AB42\rangle
    +Z_2\langle AB14\rangle \, .
\end{equation}
Here \((ij)\) denotes the line through \(Z_i,Z_j\), and \((ijk)\) denotes the
plane through \(Z_i,Z_j,Z_k\). By~\eqref{eq:l1_cons}, the coefficients in
~\eqref{A_on_triangle} are positive. Hence \(A\) lies in the triangle
\begin{equation}
T_1={\rm conv}(-Z_4,Z_1,Z_2) \, .
\end{equation}
Similarly, we choose
\begin{equation}\label{B_on_triangle}
    B
    :=
    AB\cap(234)
    =
    Z_2\langle AB34\rangle
    +Z_3\langle AB42\rangle
    +Z_4\langle AB23\rangle \, .
\end{equation}
Again, the Amplituhedron inequalities imply that \(B\) lies in the triangle
\begin{equation}
T_2={\rm conv}(Z_2,Z_3,Z_4) \, .
\end{equation}
The minus sign in front of \(Z_4\) in~\eqref{A_on_triangle} is essential; see
~\cite{GabriBlog} for further details and a visualization. Thus the one-loop
four-point Amplituhedron admits the following line-geometry description:
\begin{equation}\label{line_points}
    AB\in\mathcal A_4^{(1)}
    \iff
    A\in T_1
    \quad\text{and}\quad
    B\in T_2 \, .
\end{equation}

\begin{figure}[pos=t]
\centering
\resizebox{\textwidth}{!}{%
\begin{tikzpicture}[line cap=round,line join=round,>=Latex]

    \colorlet{lineblue}{blue!70}
    \colorlet{linered}{red!80}

    \begin{scope}[xshift=-6.15cm,scale=1.5,transform shape]

        \coordinate (L1)    at (-3,2);
        \coordinate (L2)    at (0,0);
        \coordinate (Lm4)   at (-3,-2);
        \coordinate (L2A)   at (-3,0);
        \coordinate (Lm4A)  at (-1.5,1);
        \coordinate (LA)    at (-2,0);

        \draw[thick] (L1) -- (L2) -- (Lm4) -- cycle;
        \draw[blue,thick] (L2) -- (L2A);
        \draw[blue,thick] (Lm4) -- (Lm4A);

        \node[above] at (L1) {$1$};
        \node[above] at (L2) {$2$};
        \node[below] at (Lm4) {$-4$};
        \node[below right] at (-1.3,-1) {$T_1$};

        \node[above left] at (LA) {\textcolor{blue}{$A$}};
        \fill[blue] (LA) circle (2pt);

        \coordinate (L3)    at (3,2);
        \coordinate (L4)    at (3,-2);
        \coordinate (L2B)   at (3,0);
        \coordinate (L4B)   at (1.5,1);
        \coordinate (LB)    at (2,0);
        \coordinate (LP1)   at (1.2,-0.8);
        \coordinate (LP2)   at (3,1);

        \draw[blue,thick] (L4) -- (L4B);
        \draw[blue,thick] (L2) -- (L2B);
        \draw[red,thick]  (LP1) -- (LP2);
        \draw[thick] (L2) -- (L3) -- (L4) -- cycle;

        \node[above] at (L3) {$3$};
        \node[below] at (L4) {$4$};
        \node[below left] at (1.3,-1) {$T_2$};

        \node[above] at (LB) {\textcolor{blue}{$B$}};
        \fill[blue] (LB) circle (2pt);

        \coordinate (LC) at (-1.8,-0.8);
        \coordinate (LD) at ( 1.8,-0.8);

        \node[above right] at (LC) {\textcolor{red}{$C$}};
        \fill[red] (LC) circle (2pt);

        \node[below right] at (LD) {\textcolor{red}{$D$}};
        \fill[red] (LD) circle (2pt);

        \node[right,font=\small] at (1.65,1.2) {\textcolor{red}{{\scriptsize $(ABC)\cap(234)$}}};

    \end{scope}

    \begin{scope}[xshift=6.15cm,scale=1.5,transform shape]

        \coordinate (R1)   at (-3,2);
        \coordinate (R2)   at (0,0);
        \coordinate (Rm4)  at (-3,-2);
        \coordinate (R3)   at (3,2);
        \coordinate (R4)   at (3,-2);

        \draw[thick] (R1) -- (R2) -- (Rm4) -- cycle;

        \draw[thick] (R2) -- (R3) -- (R4) -- cycle;

        \node[above] at (R1) {$1$};
        \node[above] at (R2) {$2$};
        \node[below] at (Rm4) {$-4$};
        \node[above] at (R3) {$3$};
        \node[below] at (R4) {$4$};

        \node at (-1.1,-1.3) {$T_1$};
        \node at (1.1,-1.3) {$T_2$};

        \coordinate (RA) at (-2.05,0.08);
        \coordinate (RB) at ( 2.05,0.08);

        \fill[blue] (RA) circle (2.5pt);
        \fill[blue] (RB) circle (2.5pt);

        \node[blue] at (-2.2,0.43) {$A$};
        \node[blue] at ( 2.2,0.43) {$B$};

        \def\rloop{0.62}

        \draw[linered,thick] (RA) circle (\rloop);
        \draw[linered,thick,-{Latex[length=2.2mm,width=1.6mm]}]
            ($(RA)+(20:\rloop)$)
            arc[start angle=20,end angle=78,radius=\rloop];
        \node[linered] at (-2.48,1.0) {$C(t)$};

        \draw[linered,thick] (RB) circle (\rloop);
        \draw[linered,thick,-{Latex[length=2.2mm,width=1.6mm]}]
            ($(RB)+(160:\rloop)$)
            arc[start angle=160,end angle=218,radius=\rloop];
        \node[linered] at (2.48,1.0) {$D(t)$};

    \end{scope}

\end{tikzpicture}%
}
\caption{Left: illustration of one point \((AB,CD)\) in \(\mathcal{A}^{(2)}_4\) according to \eqref{line_points}. For fixed \(A,B,C\), condition \eqref{ABCD} determines on which side of the line \((ABC)\cap(234)\) the point \(D\) must lie in \(T_2\). The blue lines represent the vanishing of \(\Delta^{24}_i\) for \(i\in\{1,\dots,4\}\), see \eqref{Delta24_2} and \eqref{Deltas}. Right: a non-trivial loop in the interior of \(\mathcal{A}^{(2)}_4\), such that \(C(t),D(t)\) revolves around $AB$.}
\label{two_triangles_with_loops}
\end{figure}
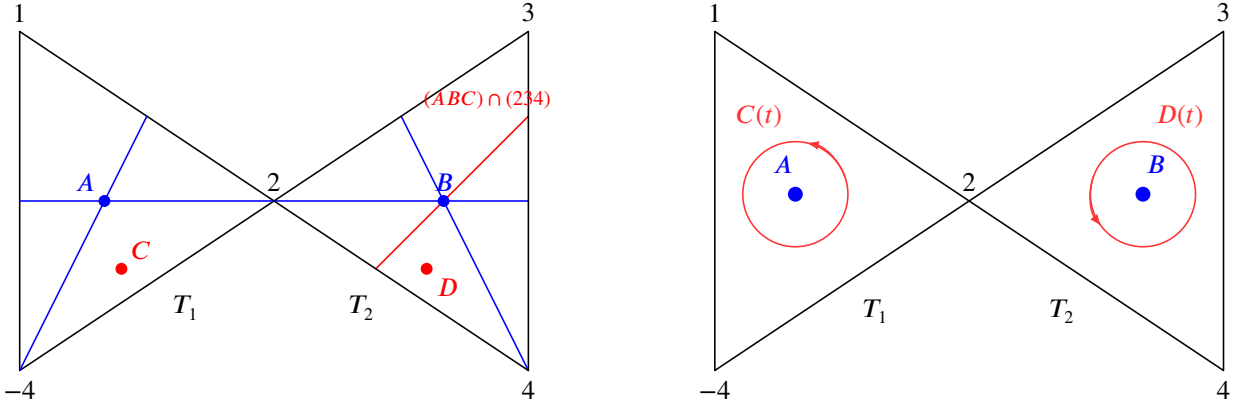

This picture generalizes immediately to \(\mathcal A_4^{(\ell)}\): one
introduces \(\ell\) lines, or equivalently \(\ell\) points in \(T_1\) and
\(\ell\) points in \(T_2\). The new ingredient is the multi-loop positivity
condition~\eqref{eq:mult_pos}, which imposes mutual positivity between each
pair of loop lines. For \(\ell=2\), we have two lines, represented by
\(A,C\in T_1\) and \(B,D\in T_2\), subject to the single inequality
\begin{equation}\label{ABCD}
\begin{aligned}
        \langle ABCD\rangle
        =&\ 
        \langle AB12\rangle\langle CD34\rangle
        + \langle AB34\rangle\langle CD12\rangle
        + \langle AB23\rangle\langle CD14\rangle
        \\
        &+
        \langle AB14\rangle\langle CD23\rangle
        - \langle AB13\rangle\langle CD24\rangle
        - \langle AB24\rangle\langle CD13\rangle
        \geq 0.
\end{aligned}
\end{equation}
This condition is invariant under \(\mathbb Z_2\times D_4\), where
\(\mathbb Z_2\) exchanges \(AB\) and \(CD\), and \(D_4\) is the dihedral
symmetry acting on the four external labels.
Using the Plücker relation
\begin{equation}\label{eq:Plucker_rel}
	\langle AB12\rangle\langle AB34\rangle
	+
	\langle AB14\rangle\langle AB23\rangle
	=
	\langle AB13\rangle\langle AB24\rangle \, ,
\end{equation}
and similarly for \(CD\), one can rewrite~\eqref{ABCD} as
\begin{equation}\label{factorised_ABCD}
       \langle ABCD\rangle
       =
       \frac{
       \Delta^{24}_2\Delta^{24}_4-\Delta^{24}_1\Delta^{24}_3
       }{
       \langle AB24\rangle\langle CD24\rangle
       }
       \geq 0 \, ,
\end{equation}
where
\begin{equation}\label{Deltas}
    \Delta^{24}_i
    :=
    \langle AB\,i\,i{+}1\rangle\langle CD24\rangle
    -
    \langle AB24\rangle\langle CD\,i\,i{+}1\rangle \, .
\end{equation}
Geometrically,
\begin{equation}\label{Delta24_2}
    \Delta^{24}_2
    =
    -\langle AB,2,(CD)\cap(234)\rangle
    =
    -\langle AB2D\rangle \, ,
\end{equation}
where in the last equality we used~\eqref{B_on_triangle} for \(CD\). Similarly,
after imposing~\eqref{A_on_triangle} and~\eqref{B_on_triangle}, one obtains
\begin{equation}\label{other_Deltas}
    \Delta^{24}_1=-\langle AB2C\rangle,
    \qquad
    \Delta^{24}_3=\langle AB4D\rangle,
    \qquad
    \Delta^{24}_4=\langle AB4C\rangle \, .
\end{equation}
Thus, for fixed \(A\) and \(B\), the sign of \(\Delta^{24}_2\) is determined
by which side of the line \((2B)\) the point \(D\) lies on inside \(T_2\). This is illustrated on the left of Figure~\ref{two_triangles_with_loops}.
The sign regions of
\begin{equation}
(\Delta^{24}_1,\Delta^{24}_2,\Delta^{24}_3,\Delta^{24}_4)
\end{equation}
subdivide each triangle into four smaller triangles. The positivity condition
~\eqref{factorised_ABCD} allows \(12\) of the \(16\) possible sign regions.
The four excluded regions are
\begin{equation}\label{sign_regions}
    (+---),
    \qquad
    (+-++),
    \qquad
    (-+--),
    \qquad
    (-+++) \, .
\end{equation}

\subsubsection{Complex boundary strata}

Let us now describe the algebraic boundary strata of
\(\mathcal A_4^{(2)}\), using the terminology of
Section~\ref{sec:Adjoint Hypersurface}. First recall the strata of
\(\mathcal A_4^{(1)}\cong\Gr_{\geq 0}(2,4)\). These were described in
\cite[Table 1]{Ranestad:adjoint} and can be written in terms of the \emph{Schubert
varieties}
{\small
\begin{equation}\label{Schubert_divisors}
\begin{aligned}
L_i &:= \{AB\cap(i\,i{+}1)\neq\emptyset\} \, , \qquad V_i &:= \{Z_i\in AB\} \, ,\qquad P_i &:= \{AB\subseteq(i{-}1\,i\,i{+}1)\} \, ,
\end{aligned}
\end{equation}}
where \(i\in\{1,\ldots,4\}\) is taken modulo \(4\).

The strata and boundary poset of
\(\mathcal A_4^{(1)}\cong\Gr_{\geq 0}(2,4)\) appeared in
Figure~\ref{fig:Gr24_poset}. The four codimension-one strata are the
\(L_i\)'s. The ten codimension-two strata are
\begin{equation}
V_i,\qquad P_i,\qquad L_1L_3,\qquad L_2L_4 \, .
\end{equation}
From now on we omit intersection symbols between strata for readability. The
nonlinear boundary structure of \(\Gr_{\geq 0}(2,4)\) is reflected in the
fact that repeated intersections may factor into several components. For
example,
\begin{equation}\label{eq:l1_fact}
	L_1L_2=V_2\cup P_2 \, .
\end{equation}
Indeed, a line intersecting both \((12)\) and \((23)\) either passes through
\(Z_2\) or lies in the plane \((123)\). These two components are both
two-dimensional and distinct; they intersect along the codimension-one locus
\(V_2P_2\), consisting of lines through \(Z_2\) inside the plane \((123)\).

The factorization~\eqref{eq:l1_fact} can be verified computationally in
\texttt{Macaulay2}. Fix \(Z\) to be the identity matrix and use Plücker
coordinates \(p_{ij}\). Then:
\begin{verbatim}
R = QQ[p12,p13,p14,p23,p24,p34];

IGr = ideal(p12*p34 + p14*p23 - p13*p24);
I = ideal(p34,p14);

decompose(IGr + I)
\end{verbatim}
Here \(I_{\Gr}\) is the Plücker ideal of \(\Gr(2,4)\). The ideal
\(I_{\Gr}+I\) describes \(L_1L_2\), since
\begin{equation}
L_1=\{\langle AB12\rangle=p_{34}=0\},
\qquad
L_2=\{\langle AB23\rangle=p_{14}=0\} \, .
\end{equation}
The command \texttt{decompose} returns
\begin{verbatim}
{ideal (p34, p24, p14), ideal (p34, p14, p13)}
\end{verbatim}
The first component corresponds to \(P_2\), while the second corresponds to
\(V_2\).

By~\eqref{two_loop_Ampl}, the algebraic boundary of
\(\mathcal A_4^{(2)}\) contains the product stratification coming from two
copies of \(\mathcal A_4^{(1)}\). We use superscripts \((1)\) and \((2)\) to
indicate the \(AB\) and \(CD\) factors, respectively. Product strata are
those obtained only from the varieties~\eqref{Schubert_divisors} in each loop
factor. The product stratification contributes eight boundary components:
\begin{equation}
L_i^{(\ell)},
\qquad
i=1,\ldots,4,
\qquad
\ell=1,2 \, .
\end{equation}
There is one additional component, given by the vanishing of the mutual positivity bracket:
\begin{equation}\label{AB_CD_divisor}
    L(1,2)
    :=
    \{AB,CD\subseteq\mathbb P^3:AB\cap CD\neq\emptyset\}
    \subseteq \Gr(2,4)^2 \, .
\end{equation}
Thus \(\partial_a\mathcal A_4^{(2)}\) has nine boundary components in total.

The non-product strata, namely those involving \(L(1,2)\), require further
factorizations. For example, on \(L_1^{(1)}L_1^{(2)}\),
\(\langle ABCD\rangle\) becomes
\begin{equation}\label{Li_Li_factorisation}
    \langle ABCD\rangle
    =
    \frac{
    \Delta^{24}_2\Delta^{24}_4
    }{
    \langle AB24\rangle\langle CD24\rangle
    } \, ,
\end{equation}
which reveals the geometric factorization
\begin{equation}\label{L12_factorization}
    L(1,2)L_1^{(1)}L_1^{(2)}
    =
    V_1(1,2)\cup P_1(1,2) \, .
\end{equation}
Here \(V_1(1,2)\) is the component where the three lines \(AB\), \(CD\), and
\((12)\) meet in one point, while \(P_1(1,2)\) is the component where they are
coplanar. More generally,
\begin{equation}\label{factorization_V12}
\begin{aligned}
    V_{i-1}(1,2)V_i(1,2)
        &= V_i^{(1)}V_i^{(2)} \cup P_i^{(12)} \, ,\\
    P_{i-1}(1,2)P_i(1,2)
        &= P_i^{(1)}P_i^{(2)} \cup V_i^{(12)}.
\end{aligned}
\end{equation}
Here \(V_i^{(12)}\) is the two-dimensional component where
\(AB=CD\) passes through \(Z_i\), and \(P_i^{(12)}\) is the two-dimensional
component where
\begin{equation}
AB=CD\subseteq(i{-}1\,i\,i{+}1) \, .
\end{equation}
The components on the right-hand side of~\eqref{factorization_V12} have
different dimensions, reflecting the non-generic nature of these
intersections.

Using the symmetries and these incidence factorizations, one computes all
strata of \(\partial_a\mathcal A_4^{(2)}\). The computation can also be
performed fully in \texttt{Macaulay2}~\cite{M2}. The row labelled
``Components'' in Table~\ref{tab:algebraic_strat_two_loop} gives the number
of strata in each codimension. A complete list appears in
\cite[Appendix A]{Dian:2024hil}. Since all strata are normal, the
stratification can be constructed by taking irreducible components of repeated intersections only.

\begin{table}[pos=h]
    \centering
    \begin{tabular}{l|c|c|c|c|c|c|c|c}
        Codimension: & 1 & 2 & 3 & 4 & 5 & 6 & 7 & 8 \\
        \hline
        Components: & 9 & 44 & 144 & 324 & 450 & 370 & 168 & 36 \\
        \hline
        Boundaries: & 9 & 44 & 144 & 286 & 356 & 306 & 156 & 34 \\
        \hline
        Residual: & 0 & 0 & 0 & 38 & 94 & 64 & 12 & 2 \\
        \hline
        Regions: & 9 & 52 & 176 & 326 & 416 & 342 & 156 & 34
    \end{tabular}
    \caption{Number of strata in each codimension of the algebraic boundary
    of \(\mathcal A_4^{(2)}\), decomposed into boundary and residual
    components, together with the number of real regions.}
    \label{tab:algebraic_strat_two_loop}
\end{table}

\subsubsection{Real strata and residual components}

So far we have discussed the complex boundary stratification. However,
\(\mathcal A_4^{(2)}\) is a real semialgebraic set. From the point of view of
positive geometries, it is essential to determine which complex strata meet
the real Amplituhedron in the expected dimension. This produces the
decomposition into boundary and residual strata.

Already the top cell of \(\mathcal A_4^{(2)}\) is interesting. Although the
ordinary non-negative Grassmannian is homeomorphic to a closed ball, loop
positive Grassmannians need not have this property. In fact, the interior of
\(\mathcal A_4^{(2)}\) is not simply connected. A non-trivial loop is easy to
describe: keep the line \(AB\) fixed and let \(CD\) revolve around it, as on the right in
Figure~\ref{two_triangles_with_loops}. This loop cannot be shrunk without
forcing \(AB\) and \(CD\) to intersect, which is precisely the boundary
condition \(L(1,2)\). See~\cite[Theorem 4.1]{Dian:2024hil} for details.

For higher codimension, a real stratum is obtained by intersecting a complex
stratum with \(\mathcal A_4^{(2)}\). This can be studied computationally with
the library \texttt{RegularChains}~\cite{lemaire2005regularchains} in
\texttt{Maple}. Using \texttt{LazyRealTriangularize}~\cite{Chen2013}, one can
for example call:
\begin{verbatim}
with(RegularChains):
R := PolynomialRing([p12, p34, p23, p14, p13, p24]):

F := [p12*p34 - p13*p24 + p14*p23, p12, p23]:
N := [p34, p14, -p13, -p24]:
P := []:
H := []:

dec := RealTriangularize(F, N, P, H, R):
Display(dec, R):
\end{verbatim}
Here \(F\) contains polynomial equalities, \(N\) contains non-negative
inequalities, \(P\) contains strict inequalities, and \(H\) contains
non-vanishing conditions. The command produces a decomposition into regular
semialgebraic sets, equivalently into cells. In the example above, the
intersection of the relevant complex stratum with
\(\Gr_{\geq 0}(2,4)\) decomposes into \(12\) cells. Among these, two are
two-dimensional, corresponding exactly to \(V_2\) and \(P_2\). Hence
\(V_2\) and \(P_2\) are genuine \emph{boundary strata} of
\(\Gr_{\geq 0}(2,4)\), see Section~\ref{sec:Adjoint Hypersurface} for the relevant definition.

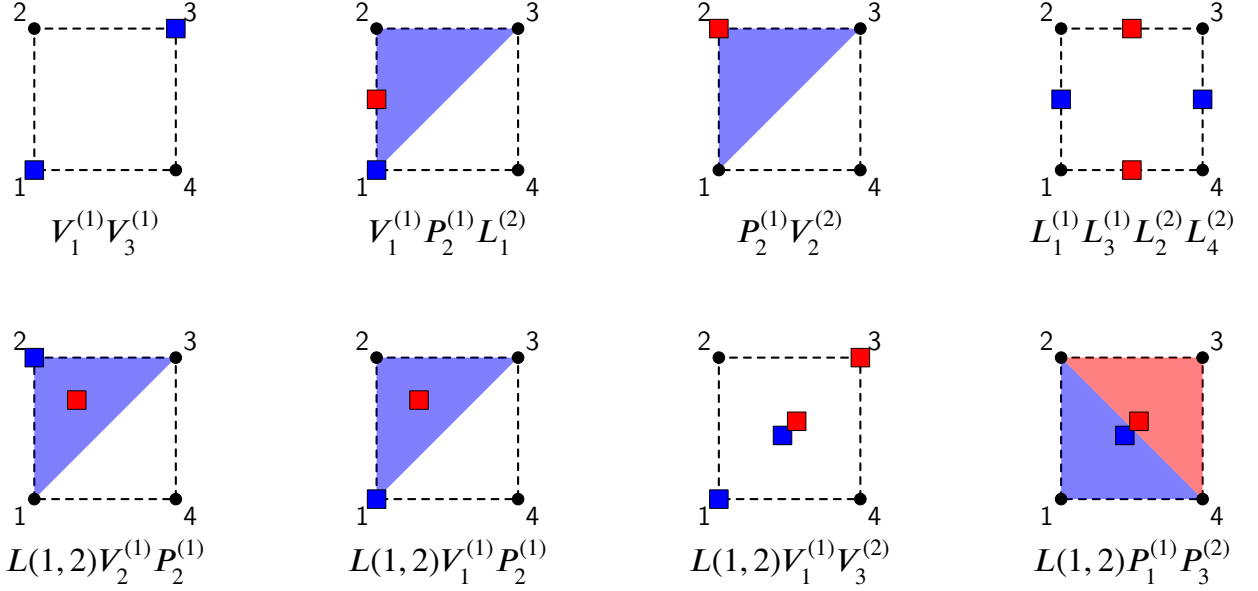
\begin{figure}[pos=t]
\centering
\resizebox{\textwidth}{!}{%
\begin{tikzpicture}[line cap=round,line join=round]

\tikzset{
  vtx/.style={draw,circle,fill=black,inner sep=0pt,minimum size=5pt},
  bmark/.style={draw,rectangle,fill=blue,inner sep=0pt,minimum size=8pt},
  rmark/.style={draw,rectangle,fill=red,inner sep=0pt,minimum size=8pt}
}

\def\s{2.15}

\def\xsep{5.2}
\def\ysep{-5.0}

\newcommand{\resbox}[5]{%
\begin{scope}[shift={(#1,#2)}]
  \draw[dashed,line width=0.9pt]
    (0,0) -- (0,\s) -- (\s,\s) -- (\s,0) -- cycle;

  #4

  \foreach \x/\y in {0/0,\s/0,\s/\s,0/\s}{
    \node[vtx] at (\x,\y) {};
  }

  #5

  \node[below left,font=\large]  at (0,0) {1};
  \node[above left,font=\large]  at (0,\s) {2};
  \node[above right,font=\large] at (\s,\s) {3};
  \node[below right,font=\large] at (\s,0) {4};

  \node[font=\Large] at ({0.5*\s},-0.95) {$#3$};
\end{scope}
}


\resbox{0}{0}
  {V_1^{(1)}V_3^{(1)}}
  {}
  {
    \node[bmark] at (0,0) {};
    \node[bmark] at (\s,\s) {};
  }

\resbox{\xsep}{0}
  {V_1^{(1)}P_2^{(1)}L_1^{(2)}}
  {
    \fill[blue,opacity=0.5] (0,0) -- (0,\s) -- (\s,\s) -- cycle;
  }
  {
    \node[bmark] at (0,0) {};
    \node[rmark] at (0,{0.5*\s}) {};
  }

\resbox{{2*\xsep}}{0}
  {P_2^{(1)}V_2^{(2)}}
  {
    \fill[blue,opacity=0.5] (0,0) -- (0,\s) -- (\s,\s) -- cycle;
  }
  {
    \node[rmark] at (0,\s) {};
  }

\resbox{{3*\xsep}}{0}
  {L_1^{(1)}L_3^{(1)}L_2^{(2)}L_4^{(2)}}
  {}
  {
    \node[rmark] at ({0.5*\s},0) {};
    \node[rmark] at ({0.5*\s},\s) {};
    \node[bmark] at (0,{0.5*\s}) {};
    \node[bmark] at (\s,{0.5*\s}) {};
  }


\resbox{0}{\ysep}
  {L(1,2)V_2^{(1)}P_2^{(1)}}
  {
    \fill[blue,opacity=0.5] (0,0) -- (0,\s) -- (\s,\s) -- cycle;
  }
  {
    \node[bmark] at (0,\s) {};
    \node[rmark] at ({0.3*\s},{0.7*\s}) {};
  }

\resbox{\xsep}{\ysep}
  {L(1,2)V_1^{(1)}P_2^{(1)}}
  {
    \fill[blue,opacity=0.5] (0,0) -- (0,\s) -- (\s,\s) -- cycle;
  }
  {
    \node[bmark] at (0,0) {};
    \node[rmark] at ({0.3*\s},{0.7*\s}) {};
  }

\resbox{{2*\xsep}}{\ysep}
  {L(1,2)V_1^{(1)}V_3^{(2)}}
  {}
  {
    \node[bmark] at (0,0) {};
    \node[rmark] at (\s,\s) {};
    \node[bmark] at ({0.45*\s},{0.45*\s}) {};
    \node[rmark] at ({0.55*\s},{0.55*\s}) {};
  }

\resbox{{3*\xsep}}{\ysep}
  {L(1,2)P_1^{(1)}P_3^{(2)}}
  {
    \fill[blue,opacity=0.5] (\s,0) -- (0,0) -- (0,\s) -- cycle;
    \fill[red,opacity=0.5]  (\s,0) -- (\s,\s) -- (0,\s) -- cycle;
  }
  {
    \node[bmark] at ({0.45*\s},{0.45*\s}) {};
    \node[rmark] at ({0.55*\s},{0.55*\s}) {};
  }

\end{tikzpicture}%
}
\caption{Irreducible components of the residual arrangement. The diagrams illustrate the corresponding Schubert conditions.}
\label{fig:residual-arrangement-components}
\end{figure}

In general, the real dimension of a stratum can be smaller than its complex
dimension. Such strata are residual. For \(\mathcal A_4^{(1)}\), the residual
arrangement is empty, as is true for any ordinary non-negative Grassmannian.
In contrast, the residual arrangement of \(\mathcal A_4^{(2)}\) is
four-dimensional. Its irreducible components are shown in
Figure~\ref{fig:residual-arrangement-components}, together with all elements
in their orbits under the symmetry group of \(\mathcal A_4^{(2)}\). We use
the graphical notation for Schubert conditions introduced in
~\cite{Ranestad:adjoint}. Blue denotes conditions on \(AB\), red denotes
conditions on \(CD\). A small colored box denotes an incidence condition,
either with a line \((i\,i{+}1)\) or with a point \(i\), while a shaded
triangle denotes a coplanarity condition.

For example, consider the four configurations in the first row of
Figure~\ref{fig:residual-arrangement-components}. The bracket
\(\langle ABCD\rangle\) reduces respectively to
\begin{equation}
-\langle AB24\rangle\langle CD13\rangle
\end{equation}
for the first three cases, and to
\begin{equation}
-\langle AB24\rangle\langle CD13\rangle
-
\langle AB13\rangle\langle CD24\rangle
\end{equation}
in the fourth case. These expressions are incompatible with the positivity
conditions defining \(\mathcal A_4^{(2)}\), and hence the corresponding
complex strata do not meet the Amplituhedron in the expected dimension.

The strata in the second row behave differently. For instance, on
\(L(1,2)V_2^{(1)}P_2^{(1)}\), 
\begin{equation}\label{residual_by_drop}
    \langle ABCD\rangle
    =
    \langle AB34\rangle\langle CD12\rangle
    +
    \langle AB14\rangle\langle CD23\rangle
    =
    0 \, .
\end{equation}
Since all brackets appearing here are non-negative on
\(\mathcal A_4^{(2)}\), this forces a drop in real dimension. The real
intersection is three-dimensional rather than four-dimensional, and the
stratum is residual. The same happens for the other strata in the
second row of Figure~\ref{fig:residual-arrangement-components}.

\subsubsection{The canonical form and the adjoint polynomial}

We now turn to the canonical form. The \(\ell\)-loop
\(n\)-point MHV Amplituhedron \(\mathcal A_n^{(\ell)}\), with \(k=0\) and
\(m=4\), is expected to have a canonical form on
\(\Gr(2,4)^\ell\) as
\begin{equation}\label{canonical_form}
    \mathbf{\Omega}_{n}^{(\ell)}
    =
    \prod_{a=1}^{\ell}
    \langle A_aB_a\,\mathrm{d}^2A_a\rangle
    \langle A_aB_a\,\mathrm{d}^2B_a\rangle
    \cdot
    \Omega_n^{(\ell)} \, .
\end{equation}
Here \(\Omega_n^{(\ell)}\) is a rational function, homogeneous of degree zero
in the external twistors \(Z_i\), and of degree \(-4\) in each loop line
\(A_aB_a\). It should have logarithmic poles only along the algebraic boundary
of \(\mathcal A_n^{(\ell)}\). Thus we write
\begin{equation}\label{integrand_ansatz}
    \Omega_n^{(\ell)}
    =
    \frac{
    \mathcal N_n^{(\ell)}
    }{
    \mathcal D_n^{(\ell)}
    },
    \qquad
    \mathcal D_n^{(\ell)}
    =
    \prod_{a=1}^{\ell}
    \prod_{i=1}^{n}
    \langle A_aB_a\,i\,i{+}1\rangle
    \prod_{a<b}
    \langle A_aB_aA_{b}B_{b}\rangle \, .
\end{equation}
The numerator \(\mathcal N_n^{(\ell)}\) is homogeneous of degree
\(n+\ell-5\) in each loop line \(A_aB_a\), and is called the
\textit{adjoint polynomial} of \(\mathcal A_n^{(\ell)}\).

The number of degrees of freedom of \(\mathcal N_n^{(\ell)}\) grows rapidly:
for fixed \(\ell\) it grows polynomially with \(n\), while increasing
\(\ell\) produces very large spaces of possible numerators; see
Table~\ref{table:adjoint_dof}. Therefore direct interpolation becomes
impractical at high multiplicity or high loop order. Conceptually, however,
one may hope for an all-point, all-loop argument analogous to the tree-level
proof of~\cite{Ranestad:adjoint}.

\begin{table}[pos=t]
\centering
\begin{tabular}{|c|c|c|c|}
  \hline
   \(n \setminus \ell\) & 1 & 2 & 3 \\
  \hline
   4 & 0 & \textbf{20} & 1539 \\
  \hline
   5 & 5 & 209 & 22099 \\
  \hline
   6 & 19 & 1274 & 198484 \\
  \hline
\end{tabular}
\caption{Number of degrees of freedom of the adjoint polynomial
\(\mathcal N_n^{(\ell)}\), modulo an overall multiplicative constant;
see~\cite[Section 5.1]{Dian:2024hil}. For \((n,\ell)=(4,2)\), this number is
\(20\).}
\label{table:adjoint_dof}
\end{table}

For \(\mathcal A_4^{(2)}\), the numerator is bihomogeneous of bidegree
\((1,1)\) on \(\Gr(2,4)^2\). We now show~\cite[Thm 5.1]{Dian:2024hil}:
\begin{tcolorbox}[resultbox]
\textbf{Two-loop four-point adjoint interpolation.}
There is, up to scale, a unique bihomogeneous polynomial of bidegree
\((1,1)\) on \(\Gr(2,4)^2\) that vanishes on the residual arrangement of
\(\mathcal{A}^{(2)}_4\) displayed in
Figure~\ref{fig:residual-arrangement-components}.
\end{tcolorbox}\noindent
A general Ansatz is
\begin{equation}\label{adjoint}
\mathcal N_4^{(2)}(AB,CD)
=
\sum_{i,j=1}^{6}
\gamma_{ij}\,
\langle AB\,Y_i\rangle
\langle CD\,Y_j\rangle \, ,
\end{equation}
where
\begin{equation}
    Y_1=(12),\quad
    Y_2=(23),\quad
    Y_3=(34),\quad
    Y_4=(41),\quad
    Y_5=(13),\quad
    Y_6=(24) \, ,
\end{equation}
and \(\gamma_{ij}=\gamma_{ji}\in\mathbb C\). Thus the polynomial has \(21\)
symmetric coefficients, or \(20\) degrees of freedom modulo scale.

We impose vanishing on the four-dimensional irreducible components of the
residual arrangement. On \(V_1^{(1)}V_3^{(1)}\), we have \(AB=13\), hence
\begin{equation}\label{first_residual_condition}
    \mathcal N_4^{(2)}(13,CD)
    =
    \sum_{j=1}^{6}
    \gamma_{6j}\,
    \langle1234\rangle
    \langle CD\,Y_j\rangle \, .
\end{equation}
Vanishing for all \(CD\in\Gr(2,4)\) forces
\begin{equation}
\gamma_{6j}=0
\qquad
\text{for all }j \, .
\end{equation}
By dihedral symmetry and exchange symmetry of the two loop lines, this also
implies
\begin{equation}
\gamma_{j6}=\gamma_{j5}=\gamma_{5j}=0 \, .
\end{equation}

Next consider the residual component \(V_1^{(1)}P_2^{(1)}L_1^{(2)}\). There
\begin{equation}\label{second_residual_condition}
    \mathcal N_4^{(2)}(AB,CD)
    =
    \sum_{j=2}^{6}
    \gamma_{3j}\,
    \langle AB34\rangle
    \langle CD\,Y_j\rangle
    +
    \gamma_{6j}\,
    \langle AB24\rangle
    \langle CD\,Y_j\rangle \, .
\end{equation}
Together with the previous constraints, vanishing gives
\begin{equation}
\gamma_{3j}=0,
\qquad
j=2,\ldots,6 \, .
\end{equation}
Applying the symmetric configurations leaves only
\begin{equation}
\gamma_{13}=\gamma_{31},
\qquad
\gamma_{24}=\gamma_{42}
\end{equation}
possibly nonzero. These remaining terms already vanish on the residual
components \(P_2^{(1)}V_2^{(2)}\) and
\(L_1^{(1)}L_3^{(1)}L_2^{(2)}L_4^{(2)}\).

It remains to impose vanishing on \(L(1,2)V_2^{(1)}P_2^{(1)}\). Using
~\eqref{residual_by_drop}, the restriction is
\begin{equation}
     (\gamma_{13}-\gamma_{24})\,
     \langle AB34\rangle
     \langle CD12\rangle = 0 \, .
\end{equation}
Hence \(\gamma_{24}=\gamma_{13}\). One checks that the resulting polynomial
also vanishes on \(L(1,2)V_1^{(1)}P_2^{(1)}\) and on the remaining residual
strata in Figure~\ref{fig:residual-arrangement-components}. Therefore
\begin{tcolorbox}[definitionbox]
\textbf{Two-loop four-point numerator.}
\begin{equation}\label{L2n4_form}
\begin{aligned}
    \mathcal N_4^{(2)}(AB,CD)
    =
    \langle1234\rangle^3
    \big(
    &\langle AB12\rangle\langle CD34\rangle
    +\langle AB34\rangle\langle CD12\rangle
    \\
    &+\langle AB14\rangle\langle CD23\rangle
    +\langle AB23\rangle\langle CD14\rangle
    \big).
\end{aligned}
\end{equation}
\end{tcolorbox}\noindent
The normalization \(\langle1234\rangle^3\) is fixed by imposing projective
scaling invariance in the external twistors \(Z_i\). The resulting form
~\eqref{integrand_ansatz} agrees with the known four-point two-loop MHV
integrand; see, for instance, \cite[Eq.~(2)]{Dian_2023}.

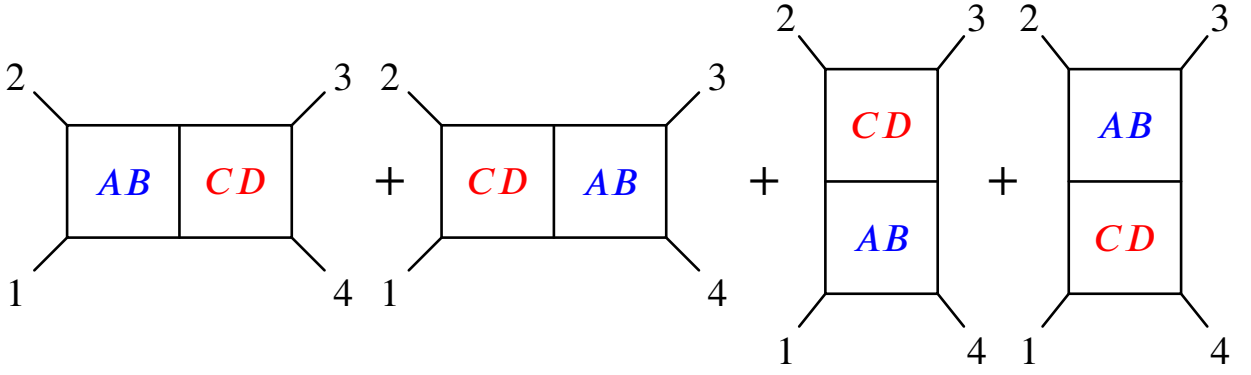
\begin{figure}
\centering
\resizebox{\textwidth}{!}{%
\begin{tikzpicture}[line cap=round,line join=round]

\tikzset{
  boxdiag/.style={thick},
  ext/.style={thick},
  plus/.style={font=\Large},
  lab/.style={font=\large}
}

\def\plusy{1.2}

\begin{scope}[yshift=0.6cm]
    \draw[boxdiag] (0,0) rectangle (2.4,1.2);
    \draw[boxdiag] (1.2,0) -- (1.2,1.2);

    \draw[ext] (0,1.2) -- (-0.35,1.55);
    \draw[ext] (0,0) -- (-0.35,-0.35);
    \draw[ext] (2.4,1.2) -- (2.75,1.55);
    \draw[ext] (2.4,0) -- (2.75,-0.35);

    \node[lab] at (-0.55,1.72) {$2$};
    \node[lab] at (-0.55,-0.58) {$1$};
    \node[lab] at (2.95,1.72) {$3$};
    \node[lab] at (2.95,-0.58) {$4$};

    \node[blue, lab] at (0.6,0.6) {$AB$};
    \node[red,  lab] at (1.8,0.6) {$CD$};
\end{scope}

\node[plus] at (3.45,\plusy) {$+$};

\begin{scope}[xshift=4.0cm,yshift=0.6cm]
    \draw[boxdiag] (0,0) rectangle (2.4,1.2);
    \draw[boxdiag] (1.2,0) -- (1.2,1.2);

    \draw[ext] (0,1.2) -- (-0.35,1.55);
    \draw[ext] (0,0) -- (-0.35,-0.35);
    \draw[ext] (2.4,1.2) -- (2.75,1.55);
    \draw[ext] (2.4,0) -- (2.75,-0.35);

    \node[lab] at (-0.55,1.72) {$2$};
    \node[lab] at (-0.55,-0.58) {$1$};
    \node[lab] at (2.95,1.72) {$3$};
    \node[lab] at (2.95,-0.58) {$4$};

    \node[red,  lab] at (0.6,0.6) {$CD$};
    \node[blue, lab] at (1.8,0.6) {$AB$};
\end{scope}

\node[plus] at (7.45,\plusy) {$+$};

\begin{scope}[xshift=8.1cm]
    \draw[boxdiag] (0,0) rectangle (1.2,2.4);
    \draw[boxdiag] (0,1.2) -- (1.2,1.2);

    \draw[ext] (0,2.4) -- (-0.28,2.75);
    \draw[ext] (1.2,2.4) -- (1.48,2.75);
    \draw[ext] (0,0) -- (-0.28,-0.35);
    \draw[ext] (1.2,0) -- (1.48,-0.35);

    \node[lab] at (-0.42,2.95) {$2$};
    \node[lab] at (1.62,2.95) {$3$};
    \node[lab] at (-0.42,-0.60) {$1$};
    \node[lab] at (1.62,-0.60) {$4$};

    \node[red,  lab] at (0.6,1.8) {$CD$};
    \node[blue, lab] at (0.6,0.6) {$AB$};
\end{scope}

\node[plus] at (10.0,\plusy) {$+$};

\begin{scope}[xshift=10.7cm]
    \draw[boxdiag] (0,0) rectangle (1.2,2.4);
    \draw[boxdiag] (0,1.2) -- (1.2,1.2);

    \draw[ext] (0,2.4) -- (-0.28,2.75);
    \draw[ext] (1.2,2.4) -- (1.48,2.75);
    \draw[ext] (0,0) -- (-0.28,-0.35);
    \draw[ext] (1.2,0) -- (1.48,-0.35);

    \node[lab] at (-0.42,2.95) {$2$};
    \node[lab] at (1.62,2.95) {$3$};
    \node[lab] at (-0.42,-0.60) {$1$};
    \node[lab] at (1.62,-0.60) {$4$};

    \node[blue, lab] at (0.6,1.8) {$AB$};
    \node[red,  lab] at (0.6,0.6) {$CD$};
\end{scope}

\end{tikzpicture}%
}
\caption{The four two-loop four-point box diagrams contributing to the MHV integrand.
Each is associated to a monomial in the adjoint polynomial~\eqref{L2n4_form}.}
\label{fig:l2_diagr}
\end{figure}

Expanding the numerator~\eqref{L2n4_form} gives four terms, corresponding to
the four double-box diagrams in Figure~\ref{fig:l2_diagr}. This relation
between numerator monomials and Feynman diagrams is special to this example.
In general, the connection between the adjoint polynomial and a diagrammatic
expansion is much less direct. Geometrically, each double box gives the
canonical form of a semialgebraic region in \(\Gr(2,4)^2\), and their
union gives an external triangulation of \(\mathcal A_4^{(2)}\).

\subsubsection{Weighted residues and topology}

Finally, let us emphasize several features of \(\mathcal A_4^{(2)}\) that
distinguish it from tree-level Amplituhedra. As mentioned above, the interior
of \(\mathcal A_4^{(2)}\) is not simply connected. Moreover, some
higher-codimension real strata are disconnected, or have interiors that are
not simply connected; see~\cite[Section 4]{Dian_2023}. This makes it more
subtle to construct a CW structure on \(\mathcal A_4^{(2)}\), and first steps
in this direction were taken in the same reference.

This topological complexity is reflected in the residues of the canonical
form. In particular, the presence of internal boundaries and non-unit maximal
residues prevents \(\mathcal A_4^{(2)}\) from being a positive geometry in
the strict sense~\cite{Dian_2023}.
To see this explicitly, fix \(Z\) to be the \(4\times4\) identity matrix and
parametrize
\begin{equation}
\begin{pmatrix}
A \\ B
\end{pmatrix}
=
\begin{pmatrix}
1 & \alpha_1 & 0 & -\alpha_2 \\
0 & \alpha_3 & 1 & \alpha_4
\end{pmatrix},
\qquad
\begin{pmatrix}
C \\ D
\end{pmatrix}
=
\begin{pmatrix}
1 & \beta_1 & 0 & -\beta_2 \\
0 & \beta_3 & 1 & \beta_4
\end{pmatrix} \, .
\end{equation}
In these coordinates, the four-point two-loop MHV canonical function becomes
\begin{equation}
\Omega^{(2)}_4
=
-\frac{
\beta_1\alpha_4+\alpha_1\beta_4+\beta_2\alpha_3+\alpha_2\beta_3
}{
\alpha_1\beta_1\alpha_2\beta_2\alpha_3\beta_3\alpha_4\beta_4
\big((\alpha_1-\beta_1)(\alpha_4-\beta_4)+(\alpha_2-\beta_2)(\alpha_3-\beta_3)\big)
} \, .
\end{equation}
Consider the following two sequences of residues:
\begin{equation}
\begin{aligned}
&
\operatorname*{Res}_{\alpha_4=0}
\operatorname*{Res}_{\beta_1=0}
\operatorname*{Res}_{\beta_4=0}
\operatorname*{Res}_{\alpha_1=0}
\operatorname*{Res}_{\alpha_2=0}
\operatorname*{Res}_{\alpha_3=0}
\operatorname*{Res}_{\beta_2=0}
\operatorname*{Res}_{\beta_3=0}
\Omega^{(2)}_4
=1,
\\[1.2em]
&
\operatorname*{Res}_{\beta_4=0}
\operatorname*{Res}_{\beta_1=0}
\operatorname*{Res}_{\alpha_4=\beta_4}
\operatorname*{Res}_{\alpha_1=\beta_1}
\operatorname*{Res}_{\alpha_2=0}
\operatorname*{Res}_{\alpha_3=0}
\operatorname*{Res}_{\beta_2=0}
\operatorname*{Res}_{\beta_3=0}
\Omega^{(2)}_4
=-2.
\end{aligned}
\end{equation}
The second maximal residue has weight \(-2\), rather than \(\pm1\). On the
other hand, the form is already fixed by interpolation on the residual
arrangement, or equivalently by the requirement that it have logarithmic
singularities only along the algebraic boundary of \(\mathcal A_4^{(2)}\).
Thus this weighted residue cannot be removed by modifying the numerator while
preserving the required singularity structure.

It is expected that this issue is resolved by enlarging the category from
positive geometries to \textit{weighted positive geometries}~\cite{Dian_2023},
a class also appearing in~\cite{Brown:PG_Hodge}. In this broader sense, loop
Amplituhedra are expected to be weighted positive geometries. Proving this
requires checking the recursive residue property on all strata. This is known
for the MHV one-loop Amplituhedron at arbitrary \(n\)~\cite{Ranestad:adjoint};
partial two-loop results appear in~\cite[Section~9.1]{Franco:2014csa}, but a
complete proof for \(\mathcal A_4^{(2)}\) and beyond remains open.

Natural next steps are higher loop order and higher multiplicity. For arbitrary
\(\ell\), one would like to determine the residual arrangement and prove that
interpolation on it fixes the adjoint hypersurface. Some \(\ell=3\) results
appear in~\cite[Section~9.1]{Franco:2014csa}, and all-loop boundary
information was obtained in~\cite{the_amplituhedron,Arkani-Hamed:2018rsk}. It
would be particularly useful to extract further all-loop information about the
four-point MHV integrand directly from Amplituhedron geometry~\cite{Stalknecht:2026ylm}.

%% file: Ch5.tex
\section{Canonical forms as dual volumes}
\label{ch:Canonical Forms as Dual Volumes}

Positive geometries attach rational differential forms to real geometric
regions. In the previous chapters, the canonical form was characterized mainly
by its poles and residues: it has logarithmic singularities on the boundary,
and its residues are canonical forms of boundary geometries. This recursive
description is the one most directly related to locality, factorization and
triangulations. However, in the simplest examples there is another
interpretation of the same object. For convex polytopes, the canonical
function is also a volume function: its value at a point computes the volume
of a suitable polar dual region.

A central mystery in the positive-geometry program is how far this dual
interpretation extends. Hodges' polytope for tree-level gluon amplitudes fits
naturally into the polytope story, giving a literal meaning to the idea that
amplitudes are volumes. For the Amplituhedron, however, the analogous
``dual Amplituhedron'' is still not understood. The difficulty is structural:
the Amplituhedron is not usually a polytope, and in general it lives in a
Grassmannian rather than in projective space. Thus ordinary polar duality
does not immediately apply.

The question is not merely formal. A dual-volume representation would imply
strong positivity properties of the canonical function. For cones over
projective geometries, this is closely related to \emph{complete monotonicity}, or
equivalently to writing the canonical function as a Laplace transform of a
non-negative measure supported on a dual cone. For polytopes this measure is
constant on the dual cone. For positive geometries with curved boundaries,
the same idea persists, but the measure becomes non-trivial and often
transcendental. This leads naturally to hyperbolic polynomials, fundamental
solutions of constant-coefficient PDEs, spectrahedral cones, and positive
measures.

This chapter develops this dual viewpoint in three directions. First, we
study completely monotone positive geometries in projective space and compute
explicit dual measures for planar examples, including regions bounded by
lines and conics and a nodal cubic. Second, we discuss how to formulate
convexity and duality in the Grassmannian using the Plücker embedding. This
leads to the \emph{exterior cyclic polytope} and to a candidate dual Amplituhedron;
in the case \(k=m=2\), the dual is again an Amplituhedron with twisted
external data. Third, we study Aomoto forms, which arise from pairing
positive geometries by integration. These provide a bridge from rational
canonical forms to polylogarithmic functions and give a simple setting where
duality, positivity and transcendental functions can be studied together.

\medskip

The chapter is structured as follows.
Section~\ref{sec:Completely Monotone Positive Geometries} introduces complete
monotonicity and its relation to dual-volume representations, hyperbolicity
and spectrahedra. Section~\ref{sec:Transcendental Measures} computes
representing measures for several planar positive geometries.
Section~\ref{sec:Convexity in the Grassmannian} introduces extendable
convexity and the exterior cyclic polytope. Section~\ref{sec:Duality for Amplituhedra}
defines the extendable dual Amplituhedron and studies the twist map.
Section~\ref{sec:Aomoto Forms} discusses Aomoto forms, Goncharov
polylogarithms and positivity of canonical pairings. Finally,
Section~\ref{sec:Chapter 5 open problems} collects the open problems raised here.

\subsection{Completely monotone positive geometries}
\label{sec:Completely Monotone Positive Geometries}

In this section we relate the duality problem for positive geometries to a
strong analytic positivity property: \textit{complete monotonicity}. This
property has recently appeared in perturbative quantum field theory; for
example, under some assumptions, scalar Feynman integrals are completely monotone functions of
Euclidean kinematic variables~\cite{Henn:CM}. In the case of polytopes, the
connection with positive geometries is already visible from the standard
dual-volume formula for canonical functions. The extension to nonlinear
positive geometries was developed in~\cite{Mazzucchelli:DV}, on which this
section is based.
Figures in this section and the following section are adapted from~\cite{Mazzucchelli:DV}
unless explicitly indicated otherwise.

Throughout the section we consider full-dimensional positive geometries
\((\mathbb P^m,P,\mathbf{\Omega}_P)\) in projective space. We denote by
\(\widehat P\subseteq \mathbb R^{m+1}\) the \emph{open} affine pointed cone over \(P\), previously denoted by ${\rm int}(\widehat{P})$ in Section~\ref{sec:Integral Representations}. Similarly, we identify $P$ with the interior of the semialgebraic set. After choosing a projective
volume form and stripping it off, the canonical form gives a homogeneous
rational function
\begin{equation}\label{eq:canonical_q_over_p_CM}
    \Omega_P(x)=\frac{q(x)}{p(x)} \, .
\end{equation}
Here \(p\) cuts out the algebraic boundary \(\partial_a P\), while \(q\) cuts
out the adjoint hypersurface \(A_P\), see Section~\ref{sec:Adjoint Hypersurface}.
By projective invariance, the degrees of \(p\) and \(q\) are related as in
~\eqref{eq:deg_q_p}.

\subsubsection{Positive convexity and complete monotonicity}
\begin{figure}[pos=t]
\begin{minipage}[c]{0.33\linewidth}
\includegraphics[width=\linewidth]{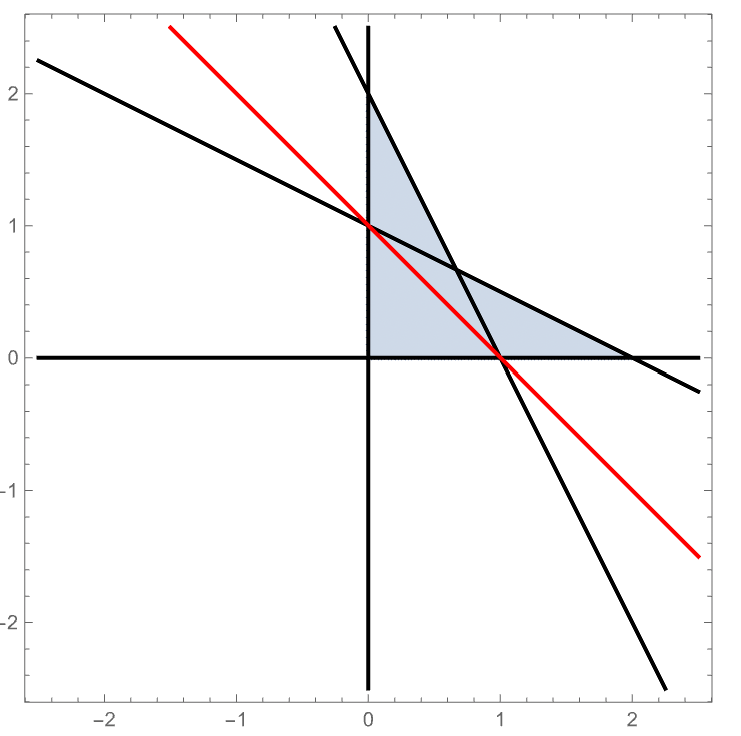}
\end{minipage}
\hfill
\begin{minipage}[c]{0.33\linewidth}
\includegraphics[width=\linewidth]{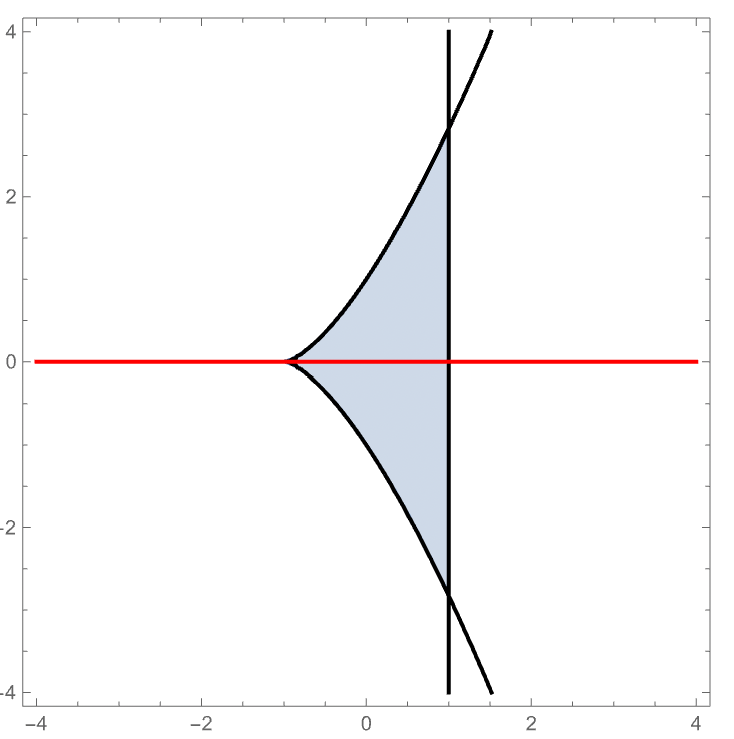}
\end{minipage}%
\hfill
\begin{minipage}[c]{0.33\linewidth}
\includegraphics[width=\linewidth]{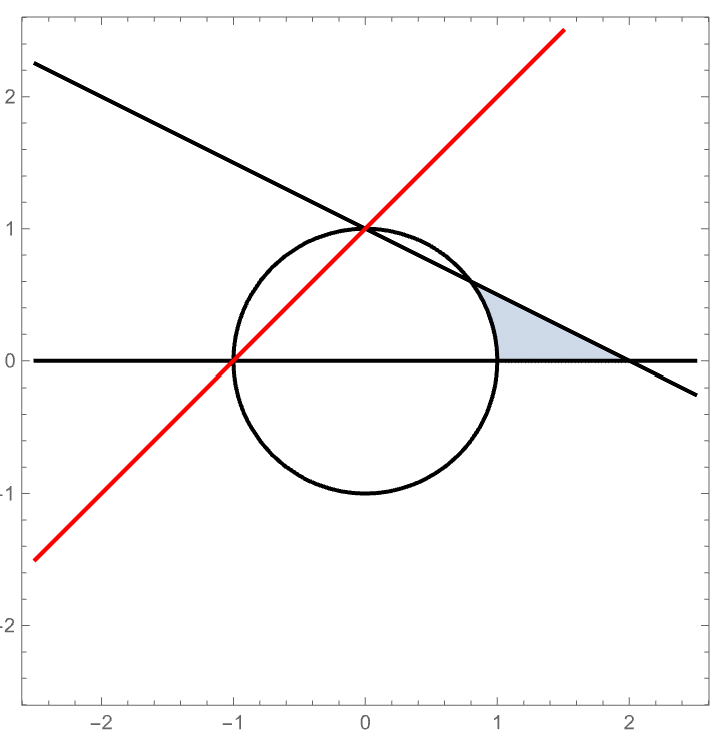}
\end{minipage}%
\caption{Three positive geometries in \(\mathbb R^2\subseteq\mathbb P^2\),
given in an affine chart by the shaded regions. The components of the
algebraic boundary are shown in black, and the adjoint hypersurface in red.
The first two examples are not positively convex, because either the
algebraic boundary or the adjoint hypersurface intersects the interior of the
region. The last example is positively convex but not convex. None of these
examples is completely monotone, since completely monotone positive
geometries are necessarily convex.}
\label{fig:non_convex_polypols}
\end{figure}

There is a first, weaker positivity notion for positive geometries. Following
~\cite{Positive_geometries}, we say that \(P\) is \textit{positively convex} if
the sign of the canonical form can be chosen so that \(\Omega_P\) is
positive and regular on \( P\). In terms of
~\eqref{eq:canonical_q_over_p_CM}, this means that neither the algebraic
boundary nor the adjoint hypersurface meets the interior of \(P\):
\begin{equation}
    \partial_a P\cap P=\emptyset,
    \qquad
    A_P\cap P=\emptyset \, .
\end{equation}
For polyhedra in an affine chart, positive convexity is equivalent to ordinary
convexity. For geometries with higher-degree boundaries this equivalence
fails: a positive geometry may be positively convex without being convex, as
illustrated in Figure~\ref{fig:non_convex_polypols}.

For projective polytopes, positive convexity follows from the stronger
dual-volume representation. If \(P\) is a projective polytope, then
\begin{equation}\label{eq:polytope_laplace_recall_CM}
    \Omega_P(x)
    =
    \int_{\widehat P^*}
    e^{-x \cdot y }\,\mathrm{d}^{m+1}y \, ,
    \qquad
    x\in\widehat P \, .
\end{equation}
As recalled in Section~\ref{sec:Integral Representations}, this integral
computes the volume of the polar dual polytope \((P-x)^\circ\) in an affine
chart. More importantly for us, it implies an infinite family of positivity
inequalities. For \(x,v_1,\ldots,v_r\in\widehat P\), differentiating under the
integral sign gives
\begin{equation}\label{eq:polytope_CM_derivative}
\begin{aligned}
    (-1)^rD_{v_1}\cdots D_{v_r} \, \Omega_P(x)
    &=
    \int_{\widehat P^*}
    ( v_1 \cdot y)\cdots( v_r \cdot y )
    \, e^{-x \cdot y}\,\mathrm{d}^{m+1}y \,.
\end{aligned}
\end{equation}
Since \(v_i\in\widehat P\) and \(y\in\widehat P^*\), every factor
\( v_i \cdot y\) is non-negative. Hence
\begin{equation}\label{eq:CMC_cons}
    (-1)^rD_{v_1}\cdots D_{v_r} \, \Omega_P(x)\geq 0 \, .
\end{equation}
Thus the canonical function of a polytope is not only positive: it is
completely monotone.

\begin{figure}[pos=t]
\centering
\begin{minipage}[t]{0.48\linewidth}
\centering
\includegraphics[width=\linewidth,height=0.32\textheight,keepaspectratio]{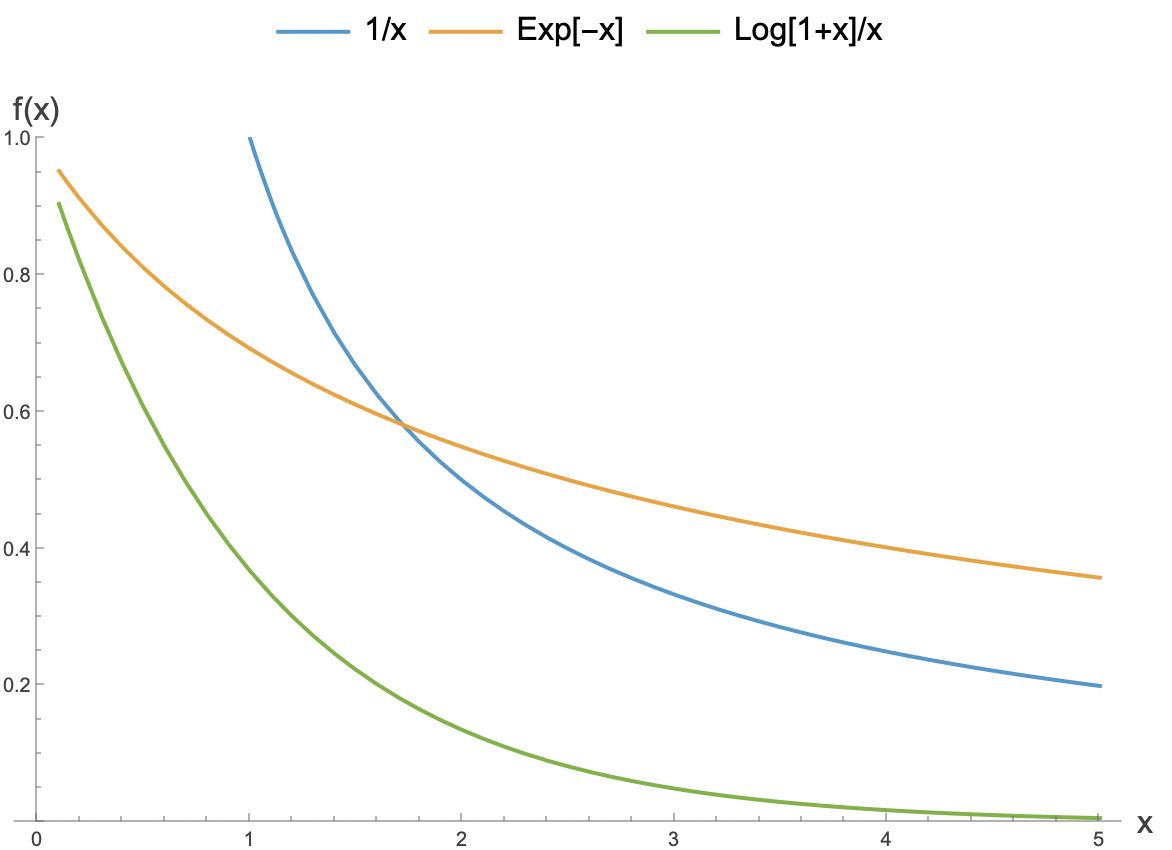}

\smallskip
\makebox[\linewidth][c]{(i)}
\end{minipage}
\hfill
\begin{minipage}[t]{0.42\linewidth}
\centering
\includegraphics[width=\linewidth,height=0.32\textheight,keepaspectratio]{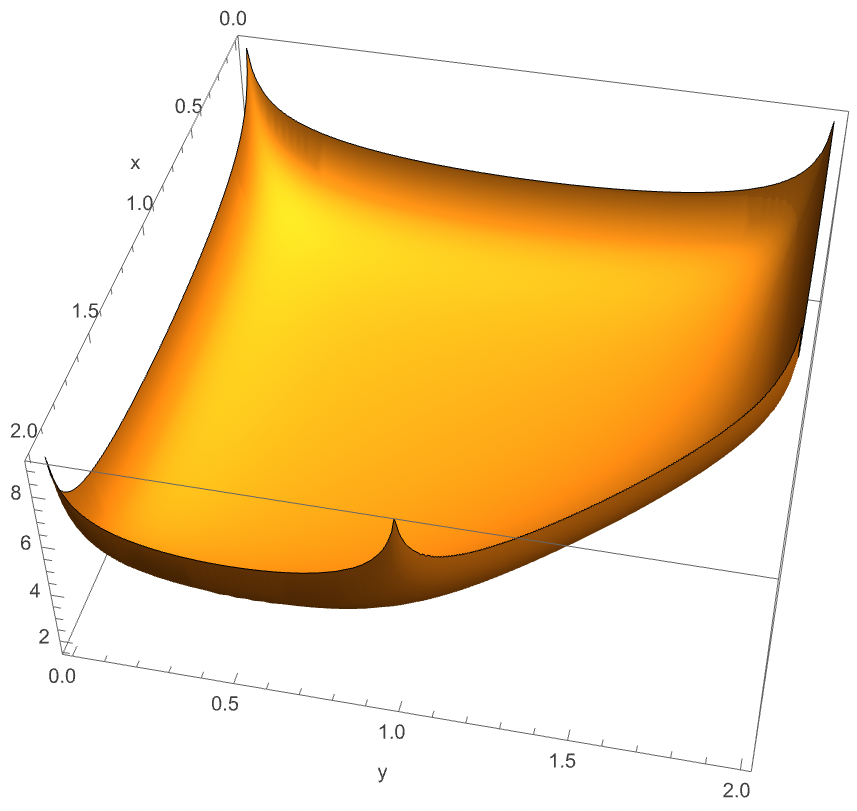}

\smallskip
\makebox[\linewidth][c]{(ii)}
\end{minipage}
\caption{(i): Examples of completely monotone functions on
\((0,\infty)\). (ii): The canonical function of a pentagon, as in~\eqref{eq:can_form_pent}, gives a multivariate completely monotone function
on the cone over the pentagon.}
\label{fig:two-side-by-side}
\end{figure}

Let \(C\subseteq\mathbb R^{m+1}\) be an open convex cone. A smooth function
\(f:C\to\mathbb R\) is called \textit{completely monotone} on \(C\) if
\begin{tcolorbox}[definitionbox]
\textbf{Complete monotonicity on a cone.}
\begin{equation}\label{eq:CM_def_cone}
    (-1)^rD_{v_1}\cdots D_{v_r} \, f(x)\geq 0
\end{equation}
\end{tcolorbox}\noindent
for all \(r\geq 0\) and all \(x,v_1,\ldots,v_r\in C\). We say that \(f\) is
\textit{absolutely monotone} if the same inequalities hold without the factor
\((-1)^r\). The condition with \(r=0\) says that \(f\geq 0\). The condition with
\(r=1\) says that \(f\) decreases along every direction in \(C\). The higher
conditions impose an infinite hierarchy of convexity-type inequalities.

In one variable, \(C=(0,\infty)\), the condition becomes
\begin{equation}\label{eq:eg_CM}
    (-1)^r f^{(r)}(x)\geq 0 \, ,
    \qquad
    x>0 \, ,\quad r\geq 0 \, \, .
\end{equation}
Basic examples are
\begin{equation}
    \frac{1}{x+a} \, ,
    \qquad
    e^{-ax} \, ,
    \qquad
    \frac{\log(x+1+a)}{x} \, ,
    \qquad a\geq 0 \, ,
\end{equation}
see Figure~\ref{fig:two-side-by-side}.
The space of completely monotone functions is closed under positive linear
combinations, pointwise products, uniform limits, positive linear
reparametrizations, and composition with absolutely monotone functions; see
~\cite{Scott_2014} for further details.

The following theorem gives a characterization of complete monotonicity~\cite{widder2015laplace,Choquet}.
\begin{tcolorbox}[resultbox]
\textbf{Bernstein--Hausdorff--Widder--Choquet.} A smooth function \(f\) on an open convex cone \(C\subseteq\mathbb R^{m+1}\) is
completely monotone if and only if there exists a unique non-negative Borel
measure \(\mathrm{d}\mu\), supported on the dual cone \(C^*\), such that
\begin{equation}\label{eq:BHWC_laplace}
    f(x)
    =
    \int_{C^*} e^{-x \cdot y}\,\mathrm{d}\mu(y) \, ,
    \qquad
    x\in C \, .
\end{equation}
\end{tcolorbox}\noindent
Thus complete monotonicity is precisely the analytic condition for the
existence of a positive Laplace-transform representation on the dual cone. In
the examples below the representing measure is often absolutely continuous
with respect to Lebesgue measure on \(C^*\); if this is the case, we write
\(\mathrm{d}\mu(y)=\mu(y)\,\mathrm{d}y\).

Motivated by~\eqref{eq:polytope_laplace_recall_CM}, we call a full-dimensional
projective positive geometry \(P\subseteq\mathbb P^m\) \textit{completely
monotone} if, after choosing the sign of the canonical form, the canonical
function \(\Omega_P\) is completely monotone on the cone \(\widehat P\). By the
previous theorem, this is equivalent to the existence of a non-negative measure
\(\mu_P\) supported on \(\widehat P^*\) such that
\begin{tcolorbox}[definitionbox]
\textbf{Completely monotone positive geometry.}
\begin{equation}\label{eq:CM_PG_laplace}
    \Omega_P(x)
    =
    \int_{\widehat P^*}
    e^{- x \cdot y}\,
    \mu_P(y)\,\mathrm{d}^{m+1}y \, ,
    \qquad
    x\in\widehat P \, .
\end{equation}
\end{tcolorbox}\noindent
In particular, a completely monotone positive geometry is convex and
positively convex. The converse is false: strict positivity of the canonical
function is much weaker, as in the last example of
Figure~\ref{fig:non_convex_polypols}.

Formula~\eqref{eq:CM_PG_laplace} gives the desired dual-volume interpretation.
In an affine chart $x$, 
\begin{equation}\label{eq:vol_form_2}
    \Omega_P(1,x)
    = m! \,
   \operatorname{Vol}_{\mu_P}\bigl((P-x)^\circ\bigr) \, ,
\end{equation}
where the volume is computed with respect to the density \(\mu_P\). Thus a
completely monotone positive geometry has a dual description
\begin{equation}
    (\mathbb P^m,P,\mathbf{\Omega}_P)
    \quad\longleftrightarrow\quad
    (\mathbb P^m,P^*,\mu_P) \, .
\end{equation}
This also has a probabilistic interpretation: \(\Omega_P\) is the \emph{moment
generating function} of the measure \(\mu_P\). It also defines a \emph{barrier
function} for the pointed cone \(\widehat P\), relevant for interior-point
methods in convex optimization~\cite{guler1996barrier}. Moreover, the triple
\((P^*,\mu_P,\iota_P)\), where
\(\iota_P:P^*\hookrightarrow\mathbb P^m_{\mathbb R}\), is a \emph{rational
exponential family} in the sense of~\cite{Exponential_varieties}, with rational
partition function \(\Omega_P\).

\subsubsection{Support and positivity of the inverse transform}
Let \(P\) be a positive geometry with canonical function \(q/p\). Deciding whether \(P\) is completely monotone requires understanding the inverse Fourier--Laplace transform of \(q/p\).

\begin{tcolorbox}[resultbox]
\textbf{Inverse-transform problem.}
The problem has two parts:
\begin{enumerate}[label=(\roman*)]
    \item the inverse transform of \(q/p\) must be supported on the dual cone
    \(\widehat P^*\);
    \item the inverse transform must be a non-negative measure.
\end{enumerate}
\end{tcolorbox}\noindent
The first condition is controlled by hyperbolicity. The second is a harder
measure-theoretic positivity question. We will solve it for an important class
of examples, namely simplex-like minimal spectrahedra.

A homogeneous polynomial \(p\in\mathbb R[x_0,\ldots,x_m]\) is
\textit{hyperbolic} with respect to \(e\in\mathbb R^{m+1}\) if \(p(e)>0\) and,
for every \(x\in\mathbb R^{m+1}\), the polynomial
\begin{equation}
    t\longmapsto p(x+te)
\end{equation}
has only real roots. If \(p\) is hyperbolic with respect to \(e\), then it is
hyperbolic with respect to every point in the connected component \(C\) of
\(\{p\neq0\}\) containing \(e\). This component is an open convex cone, called
the \textit{hyperbolicity cone} of \(p\). Equivalently, \(p\) is hyperbolic on
\(C\) if and only if
\begin{equation}
    p(z)\neq0
    \qquad
    \text{for all } z\in C+i\mathbb R^{m+1} \, .
\end{equation}
In this case we call the real vanishing locus of \(p\) a hyperbolic
hypersurface, and \(C\) its hyperbolicity region.

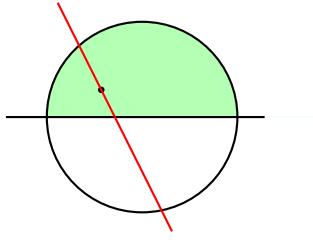
\begin{figure}[pos=t]
\centering
\begin{tikzpicture}[scale=1.8]

  \fill[green!30] (-1,0) arc[start angle=180, end angle=0, radius=0.7] -- (1,0) -- (-1,0) -- cycle;

  \draw[thick] (-1,0) arc[start angle=180, end angle=540, radius=0.7];
  \draw[thick] (-1.3,0) -- (0.6,0);

  \coordinate (P) at (-0.6,0.2);
  \filldraw[black] (P) circle (0.02);

  \coordinate (Q1) at ($(P)!-1.3!(-1,1)$);
  \coordinate (Q2) at ($(P)!0.8!(-1,1)$);

  \draw[red, thick,name path=redline] (Q1) -- (Q2);

\end{tikzpicture}
\caption{The polynomial
\(p_0(x)=x_2(x_0^2-x_1^2-x_2^2)\) is hyperbolic with hyperbolicity region
given by the half-pizza \(P_0\), shown in green in the affine chart. Any line
through a point of \(P_0\) meets the algebraic boundary of \(P_0\), shown in black, in three real points,
counted with multiplicity.}
\label{fig:half_disk}
\end{figure}

For example, for \(-1<a<1\), the polynomial \(p_a\) in
~\eqref{eq:line_conic_denominator} is hyperbolic with hyperbolicity cone
\(\widehat P_a\). At \(a=0\), this can be visualized as in
Figure~\ref{fig:half_disk}: every real line through a point of the
hyperbolicity region meets the algebraic boundary in real points, counted with
multiplicity.

\subsubsection{Hyperbolicity and fundamental solutions}
The theory of hyperbolic polynomials has its origin in the theory of
constant-coefficient partial differential equations and the well-posedness of
the Cauchy problem~\cite{Garding_1959,1Garding_1970,hormander1,Guler_Hyperbolic_pol}.
To a homogeneous polynomial \(p\), one associates the differential operator
\begin{equation}
    p(\partial)
    =
    p\!\left(
    \frac{\partial}{\partial x_0},
    \ldots,
    \frac{\partial}{\partial x_m}
    \right) \, .
\end{equation}
A fundamental solution is a distribution \(E\) satisfying
\begin{equation}\label{eq:fund_sol}
    p(\partial)E=\delta \, ,
\end{equation}
where \(\delta\) is the Dirac distribution at the origin of $\mathbb{R}^{m+1}$.

When \(p\) is hyperbolic, the fundamental solution relevant for us is
supported on a proper cone~\cite[Theorem 2.2]{Guler_Hyperbolic_pol}.

\begin{tcolorbox}[resultbox]
\textbf{Hyperbolic fundamental solution.}
Let \(p\in\mathbb R[x_0,\ldots,x_m]\) be hyperbolic with hyperbolicity cone
\(C\). Then there is a unique fundamental solution \(E\) of
\(p(\partial)E=\delta\) supported on the dual cone \(C^*\). It is given by
\begin{equation}\label{eq:fund_sol_int_repr}
    E(y)
    =
    (2\pi)^{-m-1}
    \int_{\mathbb R^{m+1}}
    e^{i\,  y \cdot \xi} \, 
    p_-(\xi)^{-1}\,\mathrm{d}\xi \, ,
\end{equation}
where
\begin{equation}
    p_-(\xi)^{-1}
    :=
    \lim_{t\to0^+}p(\xi-it e)^{-1},
    \qquad e\in C \, .
\end{equation}
\end{tcolorbox}\noindent
There is an analogous formula for \(p^{-\alpha}\),
whose inverse transform is called the \textit{Riesz measure} or
\textit{Riesz kernel} of \(p^{-\alpha}\); see
G\(\mathring{\rm a}\)rding~\cite[Theorem 3.1]{Garding_1951} and
\cite[Theorem 4.8]{Pos_Certif}. For \(\alpha=1\), the Riesz kernel agrees with
the fundamental solution above.

A basic example is the Lorentz form
\begin{equation}
    p(x)=x_0^2-x_1^2-\cdots-x_m^2 \, ,
\end{equation}
with hyperbolicity cone
\begin{equation}
    C=
    \{x\in\mathbb R^{m+1}:p(x)>0,\ x_0>0\} \, .
\end{equation}
By~\cite[Proposition 5.6]{Scott_2014}, \(p^{-\alpha}\) is completely monotone
on \(C\) for every \(\alpha>(m-1)/2\). Its Riesz kernel on \(C^*=C\) is
\begin{equation}\label{eq:riesz_quadr}
  \mu_\alpha(y)
  =
  \left(
    \pi^{\frac{m-1}{2}}
    2^{2\alpha-1}
    \Gamma(\alpha)
    \Gamma\!\left(\alpha-\frac{m-1}{2}\right)
  \right)^{-1}
  \left(
    y_0^2-y_1^2-\cdots-y_m^2
  \right)^{\alpha-\tfrac{m+1}{2}} \, .
\end{equation}
At exceptional values of \(\alpha\), this expression has to be interpreted
distributionally. For example, in \(3+1\) dimensions the fundamental solution
of the wave operator localizes on the light cone:
\begin{equation}
  \left(
    \partial_0^2-\partial_1^2-\partial_2^2-\partial_3^2
  \right)
  \left(
    \frac{1}{2\pi}
    \theta(y_0)
    \delta(y_0^2-y_1^2-y_2^2-y_3^2)
  \right)
  =
  \delta(y) \, .
\end{equation}
This support property is special to the hyperbolic case. For comparison,
elliptic fundamental solutions are not supported on proper cones. For instance,
the fundamental solution of the Laplacian in dimension \(m+1\), for \(m\geq2\),
is proportional to \(|x|^{1-m}\) and is supported on all of
\(\mathbb R^{m+1}\).

\subsubsection{The hyperbolicity criterion}
We now return to complete monotonicity. The following theorem gives a necessary
condition for a rational function to be completely monotone~\cite[Thm 3.16]{Mazzucchelli:DV}.

\begin{tcolorbox}[resultbox]
\textbf{Hyperbolicity criterion for CM.} Let \(p,q\in\mathbb R[x_0,\ldots,x_m]\) be homogeneous, coprime polynomials
which are positive on an open convex cone \(C\). If
\begin{equation}
    f=\left(\frac{q}{p}\right)^\alpha
\end{equation}
is completely monotone on \(C\) for some \(\alpha>0\), then \(p\) is
hyperbolic, and its hyperbolicity cone contains \(C\).
\end{tcolorbox}\noindent
The idea is simple. By the Laplace-transform representation
~\eqref{eq:BHWC_laplace}, a completely monotone function extends holomorphically
to the tube domain \(C+i\mathbb R^{m+1}\). If \(p\) vanished somewhere in this
tube domain, the pole of \(q/p\) would have to be cancelled by a zero of \(q\).
Since \(p\) and \(q\) are coprime, this cannot happen identically. Hence
\(p\) has no zeros on \(C+i\mathbb R^{m+1}\), which is equivalent to
hyperbolicity.

This has the following consequence~\cite[Cor 3.28]{Mazzucchelli:DV}.
\begin{tcolorbox}[resultbox]
\textbf{CM positive geometries are hyperbolic.}  If \((\mathbb P^m,P)\) is a completely monotone positive geometry, then its algebraic boundary is hyperbolic and \(P\) is one of its hyperbolicity regions. 
\end{tcolorbox}\noindent
Indeed, if \(p\) is the denominator of
\(\Omega_P=q/p\), then \(p\) cuts out \(\partial_aP\), and the theorem shows
that the hyperbolicity cone of \(p\) contains \(\widehat P\). Since \(P\) is a
connected component of the complement of the algebraic boundary~\cite{sinn2015algebraic}, the
hyperbolicity region is precisely \(\widehat P\).

This motivates the following terminology. We call a projective positive
geometry \((\mathbb P^m,P)\) \textit{hyperbolic} if its algebraic boundary is
cut out by a hyperbolic polynomial and \(\widehat P\) is one of the
corresponding hyperbolicity cones. Every completely monotone positive geometry
is hyperbolic. Thus hyperbolicity is the support condition in the inverse
Fourier--Laplace transform problem above.

\subsubsection{Spectrahedra and positivity of the measure}
Hyperbolicity alone does not imply complete monotonicity: it controls the
support of the inverse transform, but not its sign. To obtain positivity of
the representing measure, we encounter spectrahedra.

A \textit{spectrahedral cone} is a cone of the form
\begin{equation}\label{eq:spectrahedral_cone_CM}
    C
    =
    \left\{
    x\in\mathbb R^{m+1}:
    A(x):=x_0A_0+\cdots+x_mA_m\succ0
    \right\} \, ,
\end{equation}
where the \(A_i\) are real symmetric \(d\times d\) matrices, and
\(A(x)\succ0\) means that \(A(x)\) is positive definite. Finite intersections
of spectrahedral cones are again spectrahedral, by taking block-diagonal
matrices.

The determinant
\begin{equation}\label{eq:detA}
    p(x)=\det A(x)
\end{equation}
is naturally associated with the spectrahedral cone. The algebraic boundary of
\(C\) is contained in the vanishing locus of \(p\). If the algebraic boundary
is exactly the vanishing locus of \(p\), we call the spectrahedral cone
\textit{minimal}. In this case \(C\) is a hyperbolicity cone of \(p\).

Polytopal cones are elementary examples. If
\begin{equation}\label{eq:p_pol}
    p(x)=\prod_{i=1}^r \ell_i(x)
\end{equation}
is a product of real linear forms, then its hyperbolicity cones are
polyhedral. Such a cone is spectrahedral, since \(p\) is the determinant of the
diagonal matrix with entries \(\ell_1,\ldots,\ell_r\).

The ice-cream cone \(C\) in~\eqref{eq:ice_cream} is also spectrahedral. Indeed,
\begin{equation}
    A(x)
    =
    \begin{pmatrix}
        x_0-x_2 & x_1\\
        x_1 & x_0+x_2
    \end{pmatrix}
\end{equation}
is positive definite if and only if
\begin{equation}
    x_0+x_2>0,
    \qquad
    x_0^2-x_1^2-x_2^2>0 \, .
\end{equation}
These conditions are equivalent to~\eqref{eq:ice_cream}. It follows that
non-empty intersections of cones bounded by linear forms and Lorentz quadrics
in three variables are spectrahedral; these will be the main examples studied in the next section. 

Not every hyperbolic polynomial admits a symmetric determinantal
representation. For example, the specialized V\'amos polynomial
\begin{equation}
    q(x)
    =
    x_1^2x_2^2+
    4(x_1+x_2+x_3+x_4)
    (x_1x_2x_3+x_1x_2x_4+x_1x_3x_4+x_2x_3x_4)
\end{equation}
is hyperbolic with respect to \(e=(1,1,0,0)\), but no power of \(q\)
admits a symmetric determinantal representation~\cite{Kummer_2015}. Its
hyperbolicity cone is nevertheless spectrahedral. This phenomenon is related
to the \emph{generalized Lax conjecture}, which predicts that every hyperbolicity
cone is spectrahedral, or equivalently that after multiplication by a suitable
auxiliary hyperbolic polynomial, a hyperbolic polynomial admits a definite
determinantal representation~\cite{Helton_Linear}. The conjecture is known for
polynomials in three variables~\cite{Helton_Linear} and for elementary
symmetric polynomials~\cite{branden2014hyperbolicity}.

The relevance of spectrahedra is that determinantal powers have positive
Riesz measures~\cite[Corollary 4.2]{Pos_Certif}.

\begin{tcolorbox}[resultbox]
\textbf{Determinantal complete monotonicity.} Let \(C=\{A(x)\succ0\}\) be a spectrahedral cone, with \(A(x)\) a real
symmetric \(d\times d\) matrix pencil. Then
\begin{equation}
    x \longmapsto \det(A(x))^{-\alpha}
\end{equation}
is completely monotone on \(C\) for
\begin{equation}
    \alpha\in\left\{0,\tfrac12,1,\ldots,\tfrac{d-2}{2}\right\}
    \quad\text{or}\quad
    \alpha>\tfrac{d-2}{2} \, .
\end{equation}
\end{tcolorbox}\noindent
The proof uses positivity properties of the Wishart distribution on the space
of real symmetric matrices; see~\cite{Pos_Certif,Scott_2014} and the review in
~\cite{Mazzucchelli:DV}. Geometrically, the associated Riesz measure can be
computes volumes of \emph{spectrahedral shadows}; see~\cite[Eq.~(34)]{Mazzucchelli:DV}.

This gives a useful class of completely monotone positive geometries. Suppose
that \(P\) is a simplex-like minimal spectrahedral positive geometry, so that
its algebraic boundary is cut out by a determinant \(p=\det A(x)\), and the
adjoint numerator is constant. Then the canonical function is, up to
normalization, \(1/p\). The theorem above guarantees that its Riesz measure is
non-negative. Therefore \(P\) is completely monotone~\cite{Mazzucchelli:DV}.

\begin{tcolorbox}[resultbox]
\textbf{Simplex-like minimal spectrahedra.}
Simplex-like minimal spectrahedral positive geometries are completely
monotone. 
\end{tcolorbox}\noindent

We have therefore solved the two parts of the inverse-transform problem for
this class of examples. Hyperbolicity gives the correct support on the dual
cone, and the determinantal spectrahedral structure gives positivity of the
Riesz measure. In the next section we compute the representing measure
\(\mu_P\) explicitly.

\subsection{Transcendental measures}
\label{sec:Transcendental Measures}

In the previous section we saw that complete monotonicity is equivalent to the
existence of a non-negative measure on the dual cone whose Laplace transform is
the canonical function. We now turn to the problem of computing this measure.
The main point is that, even when the canonical function is rational, the
representing measure need not be algebraic or piecewise constant.

For polytopes the answer is simple: the representing measure is the
characteristic function of the dual cone. For positive geometries with curved
boundaries, this simplicity is lost. Already for regions in the projective
plane bounded by lines and conics, the measure involves inverse trigonometric
functions and logarithms. For more complicated hyperbolic boundaries, elliptic
periods can appear, as for the nodal cubic example~\cite{Mazzucchelli:DV}.

\subsubsection{The polytope case and the role of the numerator}

Let \(P\subseteq\mathbb P^m\) be a projective polytope, and write its canonical
function as $\Omega_P(x)=p(x)/q(x)$, where \(p\) is the product of the facet-defining linear forms and \(q\) is the adjoint numerator. The Laplace representation recalled in~\eqref{eq:polytope_laplace_recall_CM} says that the inverse transform of the canonical function is the characteristic function on the dual polytope, identically equal to 1 on $\widehat{P}^*$.

It is useful, however, to separate the denominator $p$ from the numerator $q$. The
Riesz kernel of \(p^{-1}\) is generally not the characteristic function of the
dual cone. Instead, it is a piecewise polynomial function, or, for general
powers of the linear factors in $p$, an Aomoto--Gelfand hypergeometric function~\cite[Theorems~3.3 and~7.4]{Pos_Certif}. The adjoint numerator \(q\) acts on this Riesz kernel by the differential operator \(q(\partial)\), and in the
polytope case this operation precisely produces the characteristic function of
\(\widehat P^\ast\).

For example, consider the cone over the square
\begin{equation}
    \widehat P=
    \{x_0+x_1>0,\ x_0-x_1>0,\ x_0+x_2>0,\ x_0-x_2>0\}
    \subseteq\mathbb R^3 \, .
\end{equation}
Its algebraic boundary is cut out by
\begin{equation}
    p(x)
    =
    (x_0+x_1)(x_0-x_1)(x_0+x_2)(x_0-x_2) \, .
\end{equation}
Starting from the elementary Laplace representation of \(p^{-1}\), and setting
\begin{equation}
\label{eq:square_linear_map}
    y_0=t_1+t_2+t_3+t_4,\qquad
    y_1=t_1-t_2,\qquad
    y_2=t_3-t_4 \, ,
\end{equation}
one finds that the support is the dual cone
\begin{equation}
    \widehat P^\ast=\{y_0\geq |y_1|+|y_2|\} \, .
\end{equation}
The Riesz kernel of \(p^{-1}\) is, up to the Jacobian, the length of the
one-dimensional fiber of the map~\eqref{eq:square_linear_map}. Explicitly,
\begin{equation}
    \nu(y)
    =
    \frac14\bigl(y_0-|y_1|-|y_2|\bigr)_+ ,
    \label{eq:square_riesz_kernel}
\end{equation}
where \(r_+:=\max(r,0)\). The canonical function of the square is instead
\begin{equation}
    \Omega_P(x)=\frac{4x_0}{p(x)} \, .
\end{equation}
Therefore the representing measure of \(\Omega_P\) is obtained by applying
\(4\partial_{y_0}\) to~\eqref{eq:square_riesz_kernel}, giving
\begin{equation}
    4\partial_{y_0}\nu(y)=1 \, ,
\end{equation}
for every $y \in \widetilde{P}^*$.
This example illustrates a general phenomenon: the denominator controls the
support and analytic type of the Riesz kernel, while the adjoint numerator
selects the canonical measure.

\subsubsection{Half-pizzas}

\begin{figure}[pos=t]
\centering
\resizebox{\figscale\textwidth}{!}{%
\begin{minipage}{\textwidth}
\centering

\subcaptionbox{\(a=1/2\)}[0.48\linewidth]{%
\includegraphics[width=\linewidth]{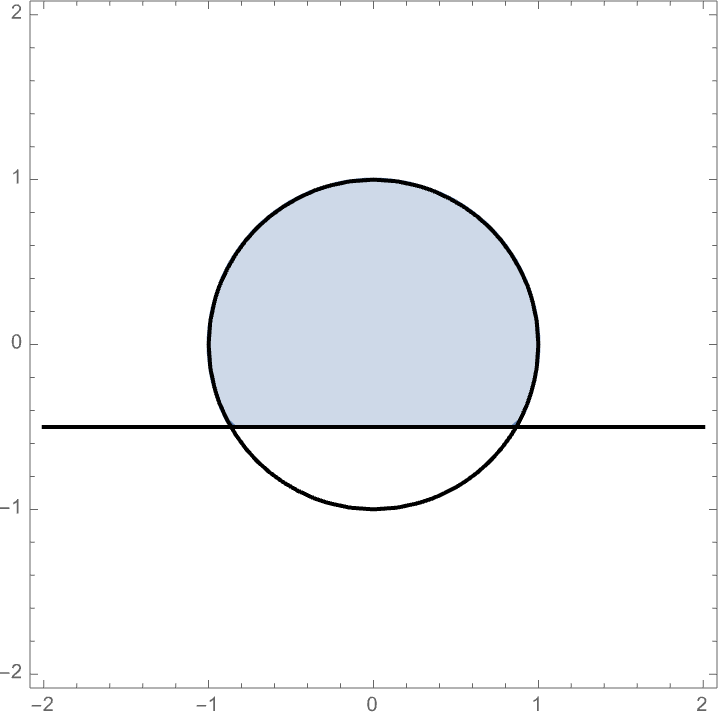}}
\hfill
\subcaptionbox{dual for \(a=1/2\)}[0.48\linewidth]{%
\includegraphics[width=\linewidth]{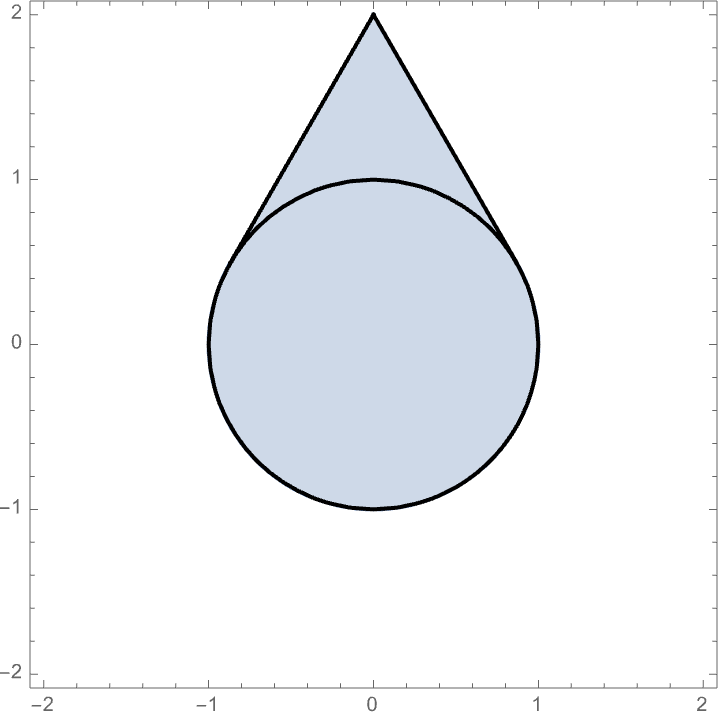}}

\vspace{0.5cm}

\subcaptionbox{\(a=1\)}[0.48\linewidth]{%
\includegraphics[width=\linewidth]{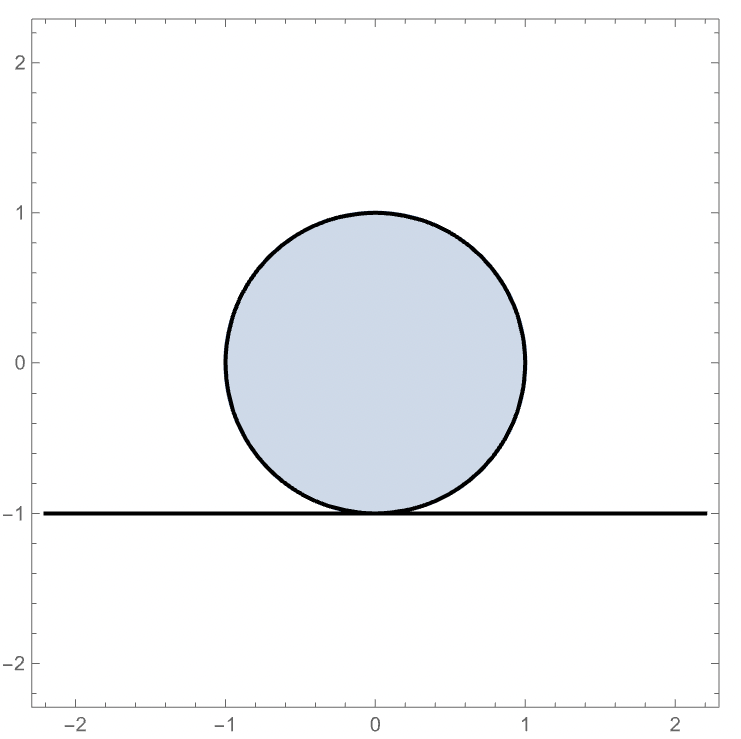}}
\hfill
\subcaptionbox{\(a=3/2\)}[0.48\linewidth]{%
\includegraphics[width=\linewidth]{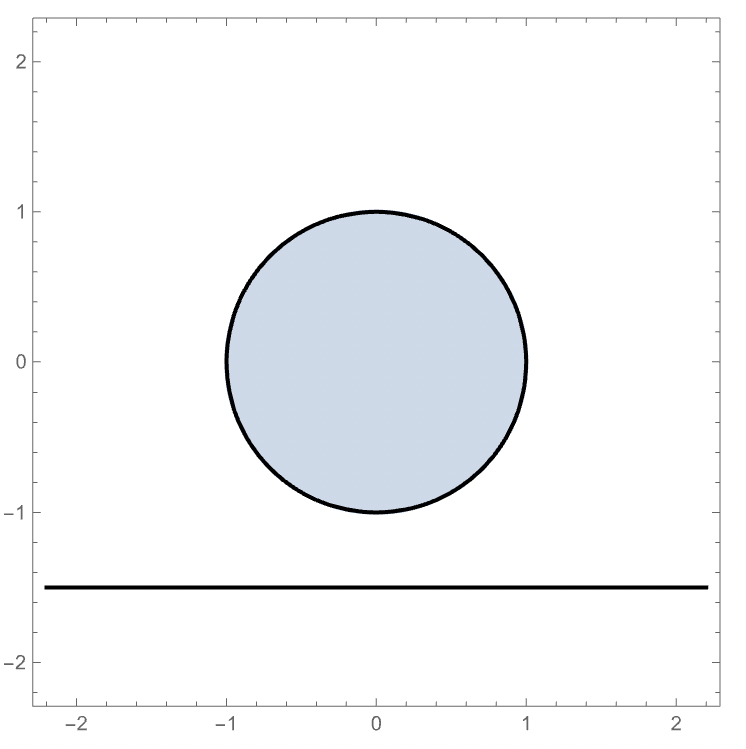}}

\end{minipage}%
}
\caption{The cones \(\widehat P_a\) from~\eqref{eq:pizza_cone_short}, shown on
the slice \(x_0=1\), for different values of \(a\). For \(a=1/2\) we also show
the dual cone \(\widehat P_a^*\).}
\label{fig:half_pizzas}
\end{figure}

Projective polytopes are completely monotone positive geometries. The first
question is whether there are genuinely nonlinear examples. Consider a one-parameter family of ``half-pizzas'', similar to
Example~\ref{eg:half_pizza}. For \(-1<a<1\), define
\begin{equation}\label{eq:pizza_cone_short}
    \widehat P_a
    =
    \left\{
    x\in\mathbb R^3:
    x_0^2-x_1^2-x_2^2>0,\quad
    x_2+a \, x_0>0,\quad
    x_0>0
    \right\} \, .
\end{equation}
Its algebraic boundary is cut out by
\begin{equation}\label{eq:line_conic_denominator}
    p_a(x)
    =
    (x_0^2-x_1^2-x_2^2)(x_2+a \, x_0) \, .
\end{equation}
For \(a=0\) one recovers the standard half-pizza. For \(-1<a<1\), the line
meets the conic in two real points. At \(a=1\) it is tangent to the conic, and
for \(a>1\) it does not meet the conic in real points. We also write
\begin{equation}\label{eq:ice_cream}
    C=
    \left\{
    x_0^2-x_1^2-x_2^2>0,\quad x_0>0
    \right\}
\end{equation}
for the ice-cream cone. For \(-1<a<1\), the dual cone \(\widehat P_a^*\) is the
convex hull of \(C^*=C\) and the ray spanned by \((a,0,1)\); see
Figure~\ref{fig:half_pizzas}.

For \(-1<a<1\), this is a positive geometry in \(\mathbb P^2\), with canonical
function
\begin{equation}\label{eq:half_pizza_a_canonical_short}
    \Omega_a(x)
    =
    \frac{2\sqrt{1-a^2}}
    {(x_0^2-x_1^2-x_2^2)(x_2+a \, x_0)} \, .
\end{equation}
The normalization is so that the residues at the vertices are
\(\pm1\). Using the results related to hyperbolicity and spectrahedra discussed in the previous subsections:
\begin{tcolorbox}[resultbox]
\textbf{Half-pizzas are CM.}
 The posisitve geometry \(P_a\) is completely monotone for~\({-1<a<1}\). 
\end{tcolorbox}\noindent
The constant measure on \(\widehat P_a^\ast\) does not
reproduce the canonical function. Instead, \begin{equation}
    \int_{\widehat P_a^\ast}
    e^{-\langle x,y\rangle}\,\mathrm{d}^3y
    =
    \Omega_a(x)
    +
    \frac{2}{(x_0^2-x_1^2-x_2^2)^{3/2}}
    \arctan
    \left(
    \frac{x_2+a \, x_0}
    {\sqrt{1-a^2}\sqrt{x_0^2-x_1^2-x_2^2}}
    \right).
    \label{eq:naive_half_pizza_integral}
\end{equation}
Thus the naive volume has the desired canonical function as its rational part,
but also contains a transcendental correction. This computation was obtained
in~\cite[Section~10]{Positive_geometries} as the large-\(n\) limit of
canonical forms of polygons inscribed in the half-pizza. Therefore, even when a
curved positive geometry is a limit of polytopes, the limiting constant-measure dual volume need not be the canonical function of the
limiting curved geometry.

The correct measure for the canonical function~\eqref{eq:half_pizza_a_canonical_short} is obtained by convolving the Riesz
measure of the Lorentz quadratic form with that of the linear factor
\(x_2+a \, x_0\). More precisely,
\begin{equation}
    \mathrm{d}\mu_1(y)
    =
    \frac{1}{2\pi}
    (y_0^2-y_1^2-y_2^2)^{-1/2}
    \chi_{C^\ast}(y)\,\mathrm{d}^3y \, ,
\end{equation}
where $\chi_{C^\ast}$ is the characteristic function of the cone $C^*$, while
\begin{equation}
    \mathrm{d}\mu_2(y)
    =
    \int_0^\infty
    \delta^{(3)}\!\left(y-t(a,0,1)\right)\,\mathrm{d}t \, .
\end{equation}
Their convolution gives
\begin{equation}
    \mu_a(y)
    =
    2\sqrt{1-a^2}
    \int_{\mathcal R_a(y)}
    \frac{\mathrm{d}t}
    {2\pi\sqrt{(y_0-a \, t)^2-(y_2-t)^2-y_1^2}},
    \label{eq:line_conic_integral_trans}
\end{equation}
where
\begin{equation}
    \mathcal R_a(y)
    =
    \left\{
    t\geq 0:
    y_0-a \, t>0,\quad
    (y_0-a \, t)^2-(y_2-t)^2-y_1^2>0
    \right\} \, .
\end{equation}
Evaluating this elementary period gives different formulas according to the
relative position of the line and the conic.

\begin{figure}[pos=t]
\centering
\subcaptionbox{$a=1/2$\label{fig:linearquadricmeas1}}[0.31\linewidth]{%
  \includegraphics[width=\linewidth]{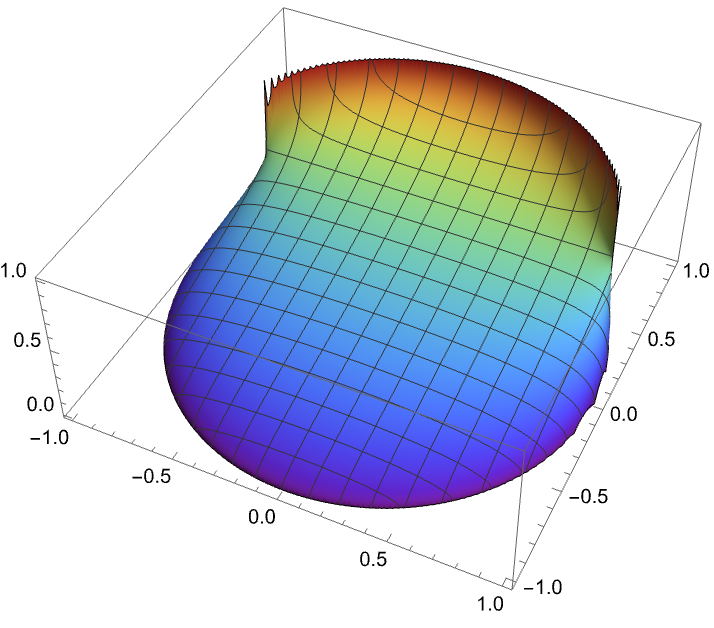}%
}
\hfill
\subcaptionbox{$a=1$\label{fig:linearquadricmeas2}}[0.31\linewidth]{%
  \includegraphics[width=\linewidth]{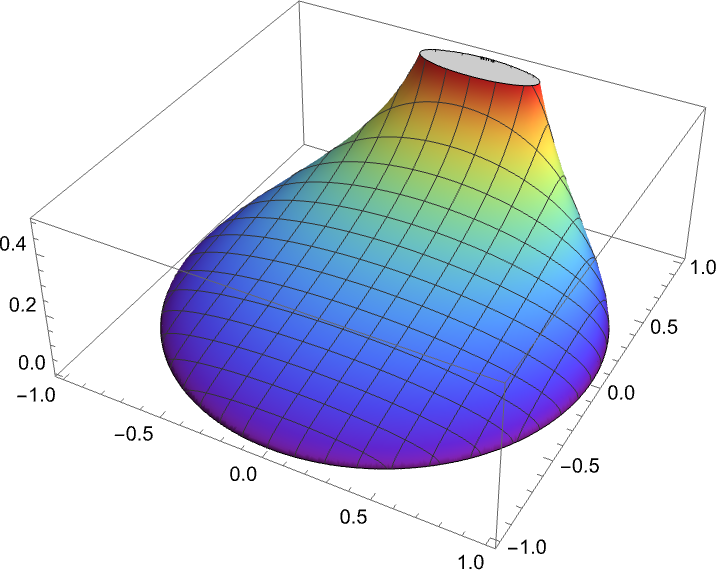}%
}
\hfill
\subcaptionbox{$a=10$\label{fig:linearquadricmeas3}}[0.31\linewidth]{%
  \includegraphics[width=\linewidth]{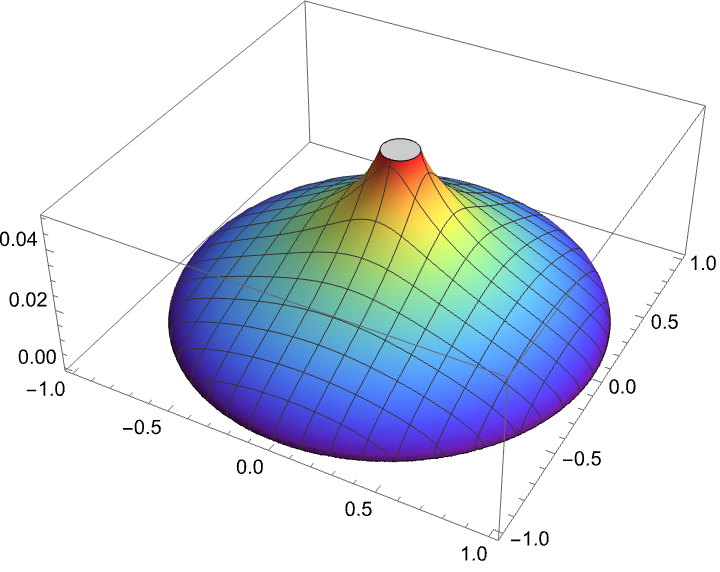}%
}
\caption{The graph of the measure \(\mu_a\) for different relative positions
of the line and the conic, plotted on the slice \(C^\ast\cap\{y_0=1\}\).
For \(-1<a<1\), the density is an arctangent function. At \(a= 1\), it becomes
algebraic. For \(a>1\), it is logarithmic.}
\label{fig:linequadric}
\end{figure}

For \(-1<a<1\), one obtains
\begin{tcolorbox}[resultbox]
\textbf{Half-pizza measure.}
\begin{equation}
\label{eq:mu_line_conic_arctan_trans}
\mu_a(y)=
\begin{cases}
\displaystyle
\frac12+
\frac1\pi
\arctan\!\left(
\frac{y_2-a \, y_0}
{\sqrt{(1-a^2)(y_0^2-y_1^2-y_2^2)}}
\right),
& y\in C^\ast \, ,\\[4mm]
1,
& y\in \widehat P_a^\ast\setminus C^\ast .
\end{cases}
\end{equation}
\end{tcolorbox}\noindent
Thus the density is constant equal to one on the horn of the dual region
outside the dual conic, and is an arctangent function inside \(C^\ast\). In
particular, it is non-negative on \(\widehat P_a^\ast\).

At the tangent value \(a=1\), the quadratic polynomial in the integration
variable degenerates, and one finds
\begin{equation}
\label{eq:mu_line_conic_tangent_trans}
    \mu_{1}(y)
    =
    \frac{\sqrt{y_0^2-y_1^2-y_2^2}}
    {2\pi (y_0-y_2)},
    \qquad y\in C^\ast \, .
\end{equation}
For \(a>1\), the line does not intersect the real conic and 
\begin{equation}
\label{eq:mu_line_conic_log_trans}
    \mu_a(y)
    =
    \frac{1}{4\pi\sqrt{a^2-1}}
    \log
    \left(
    \frac{
    a \, y_0-y_2+
    \sqrt{(a^2-1)(y_0^2-y_1^2-y_2^2)}
    }{
    a \, y_0-y_2-
    \sqrt{(a^2-1)(y_0^2-y_1^2-y_2^2)}
    }
    \right) \, .
\end{equation}
The cases \(a\geq1\) should be viewed as useful limiting or analytic
continuations of the product of a Lorentz quadratic and a line. The
half-pizza positive geometry itself corresponds to \(0\leq a<1\). This example
already displays the main phenomenon of the section: a rational canonical
function can have a transcendental density in its dual-volume representation.

The same measure can also be interpreted as the fundamental solution of the
hyperbolic operator \(p_a(\partial)\):
\begin{equation}
    p_a(\partial)\,\mu_a=\delta \, .
\end{equation}
Away from the origin, and away from the singular support of the measure, it
satisfies the homogeneous equation. Thus the dual-volume measure also has a
PDE interpretation, controlled by the characteristic variety \(p_a=0\)~\cite{Ehrenpreis1970,Palamodov1970,Hoermander1990,Bjoerk1979,ait2023linear}.

\subsubsection{Polycons and triangulations}

The half-pizza has a coordinate-free formulation. Let
\(Q\subseteq\mathbb P^2\) be a non-degenerate real conic with equation
\(q(x)= x \cdot (A x)\), and let \(L\) be a line with equation
\(\ell(x)= \ell \cdot x\). Assume that \(q(x)>0\) defines the
hyperbolicity cone \(C\), and that \(L\) intersects \(Q\) in two real points.
The cone over this more general half-pizza is
\begin{equation}
    \widehat P
    =
    \{q(x)>0,\ \ell(x)>0,\ x_0>0\} \, , 
    \subseteq C
\end{equation}
and defines a positive geometry with canonical function
\begin{equation}
    \Omega_P(x)=\frac{c(\ell,q)}{\ell(x)q(x)} \, ,
\end{equation}
where the constant \(c(\ell,q)\) fixes the residue normalization. The dual cone
\(\widehat P^\ast\) is the convex hull of \(C^\ast\) and the ray spanned by
\(\ell\). If \(q^\ast(y)= y \cdot (A^{-1}y)\), and if \(\ell^\ast\) is
the line dual to the pair of tangents from \(\ell\) to the dual conic, then
the representing measure is
\begin{tcolorbox}[resultbox]
\textbf{General half-pizza measure.}
\begin{equation}
\label{eq:mu_half_pizza_general_trans}
\mu_P(y)=
\begin{cases}
\displaystyle
\frac12+
\frac1\pi
\arctan\!\left(\frac{\ell^\ast(y)}{\sqrt{q^\ast(y)}}\right),
& y\in C^\ast \, ,\\[3mm]
1,
& y\in \widehat P^\ast\setminus C^\ast .
\end{cases}
\end{equation}
\end{tcolorbox}\noindent
This is the building block for positive geometries bounded by one conic
and several lines.
Following~\cite{Polypols}, we call such a region a \textit{polycon}. More
precisely, a polycon is a full-dimensional semialgebraic set in the
hyperbolicity region of a conic, whose boundary consists of linear edges and
arcs of the conic. If it has \(r\) conic arcs and \(s\) linear edges, we say
that it has \emph{type} \((r,s)\).

The measure of a polycon can be computed by signed triangulations of canonical
forms as in Section~\ref{sec:Triangulations}. One finds an external triangulation of \(P\) into convex pieces \(P_i\),
each either a triangle or a half-pizza of type \((1,1)\). Here external means
that each \(P_i\) contains \(P\) and is itself convex. Consequently,
\(P_i^\ast\subseteq P^\ast\), so the corresponding triangulation is internal on
the dual side. On the level of canonical functions, and hence of measures, one
has
\begin{tcolorbox}[resultbox]
\textbf{Signed measure triangulation.}
\begin{equation}
\label{eq:measure_triangulation_trans}
    \Omega_P=\sum_i\varepsilon_i\,\Omega_{P_i}
    \quad \Longrightarrow \quad
    \mu_P=\sum_i\varepsilon_i\,\mu_{P_i} \, ,
\end{equation}
\end{tcolorbox}\noindent
where \(\varepsilon_i\in\{\pm1\}\) are the signs required for the
canonical-form triangulation. The implication follows from uniqueness of the
Laplace representation. Thus canonical-form triangulations translate directly
into formulas for the dual measure.

\begin{figure}[pos=t]
\centering
\subcaptionbox{Geometry \(P\)\label{fig:pizza_slice2_geometry}}[0.31\linewidth]{%
  \includegraphics[width=\linewidth]{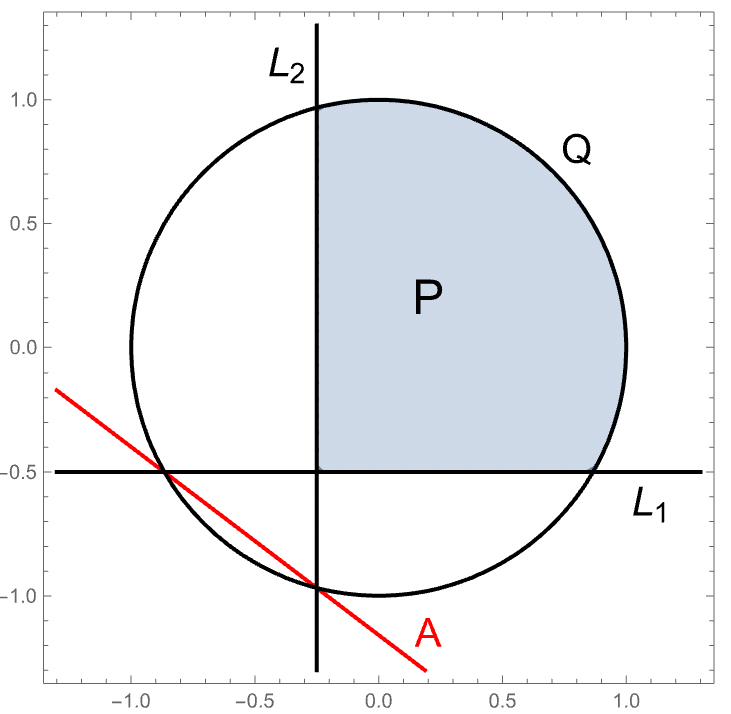}%
}
\hfill
\subcaptionbox{Dual \(P^\ast\)\label{fig:pizza_slice2_dual}}[0.31\linewidth]{%
  \includegraphics[width=\linewidth]{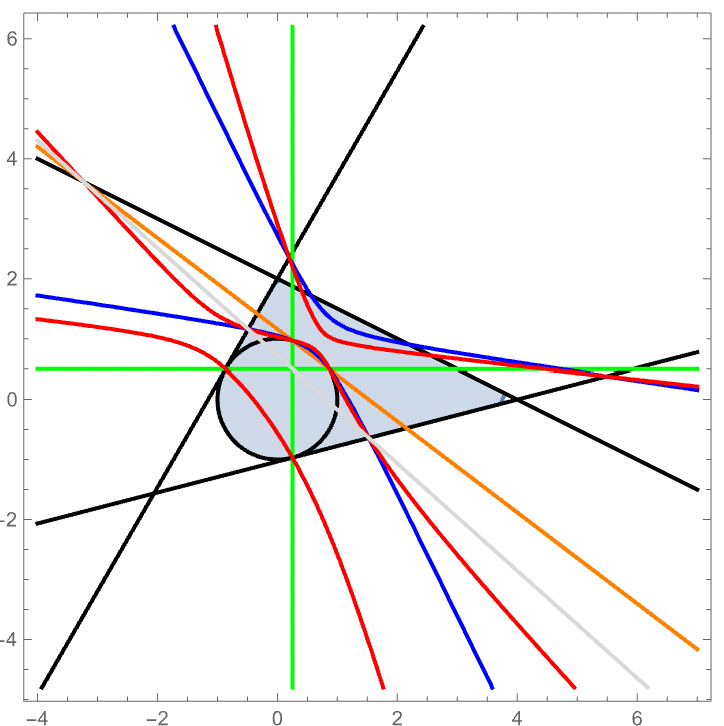}%
}
\hfill
\subcaptionbox{Measure \(\mu_P\)\label{fig:pizza_slice2_measure}}[0.34\linewidth]{%
  \includegraphics[width=\linewidth]{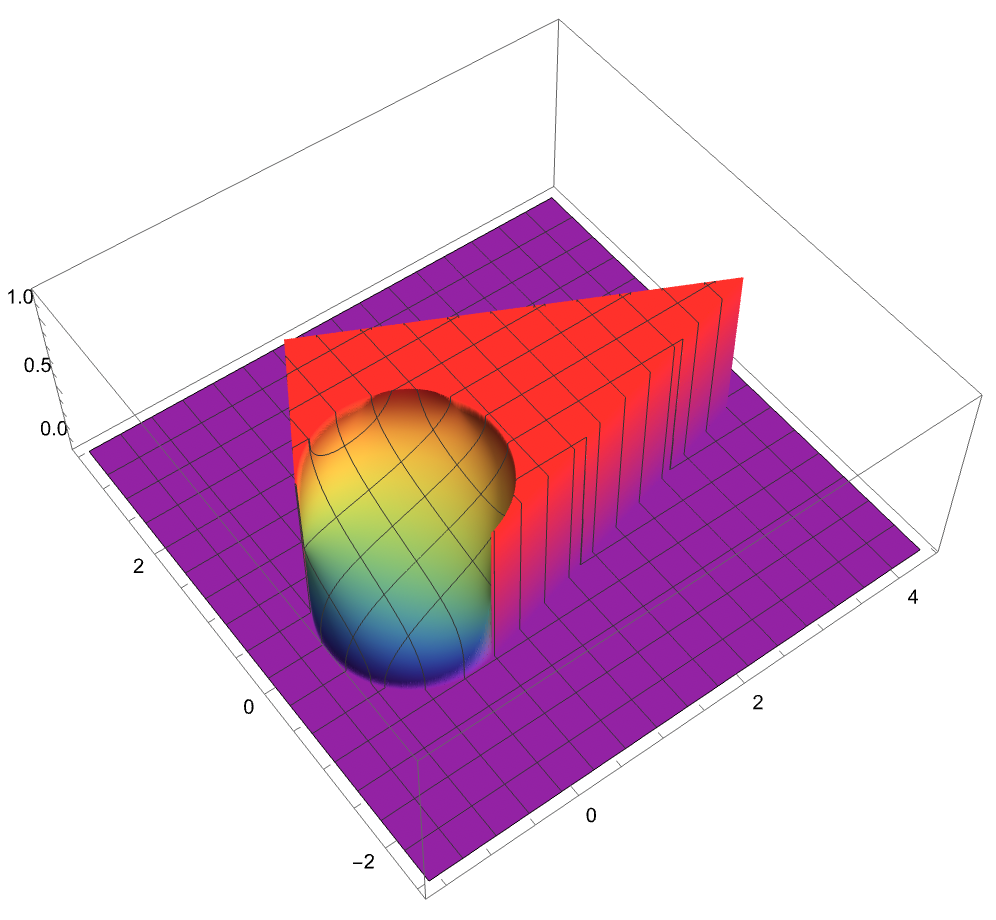}%
}
\caption{A positive geometry \(P\) bounded by a conic and two lines. The
adjoint line is shown in red. The dual picture contains additional curves
governing the analytic structure of the measure. The graph on the right shows
the representing measure \(\mu_P\).}
\label{fig:pizza_slice2}
\end{figure}

For example, suppose \(P\) is bounded by one conic \(q=0\) and two lines
\(\ell_1=0\), \(\ell_2=0\). Its canonical function has the form
\begin{equation}
    \Omega_P(x)
    =
    \frac{\alpha(x)}
    {\ell_1(x)\ell_2(x)q(x)} \, ,
\end{equation}
where \(\alpha(x)\) defines the adjoint line. Using a signed triangulation into
half-pizzas and triangles as in Example~\ref{eg:pizza_slice}, one obtains, for
\(y\in C^\ast\),
\begin{equation}
\label{eq:mu_two_lines_one_conic_trans}
\begin{aligned}
    \mu_P(y)
    =
    \frac12
    &+\frac1\pi
    \arctan\!\left(\frac{\ell_1^\ast(y)}{\sqrt{q^\ast(y)}}\right)
    +\frac1\pi
    \arctan\!\left(\frac{\ell_2^\ast(y)}{\sqrt{q^\ast(y)}}\right)  -\frac1\pi
    \arctan\!\left(\frac{\alpha^\ast(y)}{\sqrt{q^\ast(y)}}\right),
\end{aligned}
\end{equation}
while \(\mu_P(y)=1\) for \(y \in \widehat P^\ast\setminus C^\ast\). Combining
the arctangents produces a single arctangent, together with an integer-valued locally constant term that fixes the branch. The
additional curves appearing in this argument are not part of the original
canonical form, but they control the analytic structure of the representing measure.

\begin{figure}[pos=t]
\centering
\begin{minipage}[c]{0.31\textwidth}
\centering
\includegraphics[width=\linewidth]{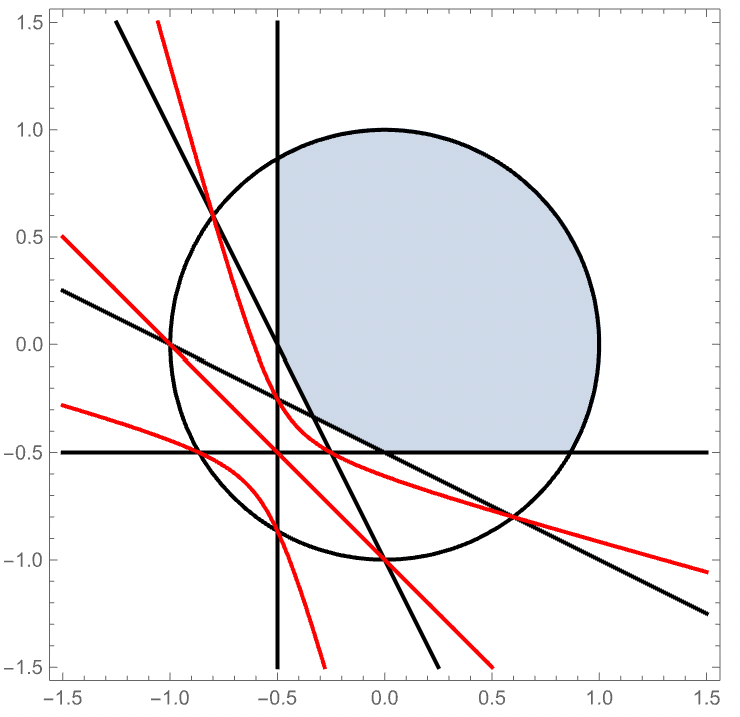}
\end{minipage}
\hfill
\begin{minipage}[c]{0.31\textwidth}
\centering
\includegraphics[width=\linewidth]{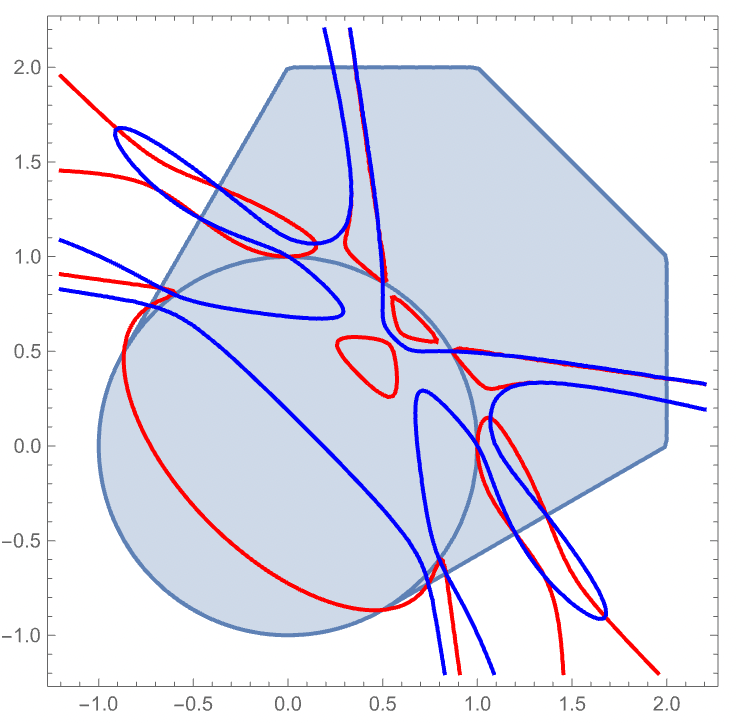}
\end{minipage}
\hfill
\begin{minipage}[c]{0.35\textwidth}
\centering
\includegraphics[width=\linewidth]{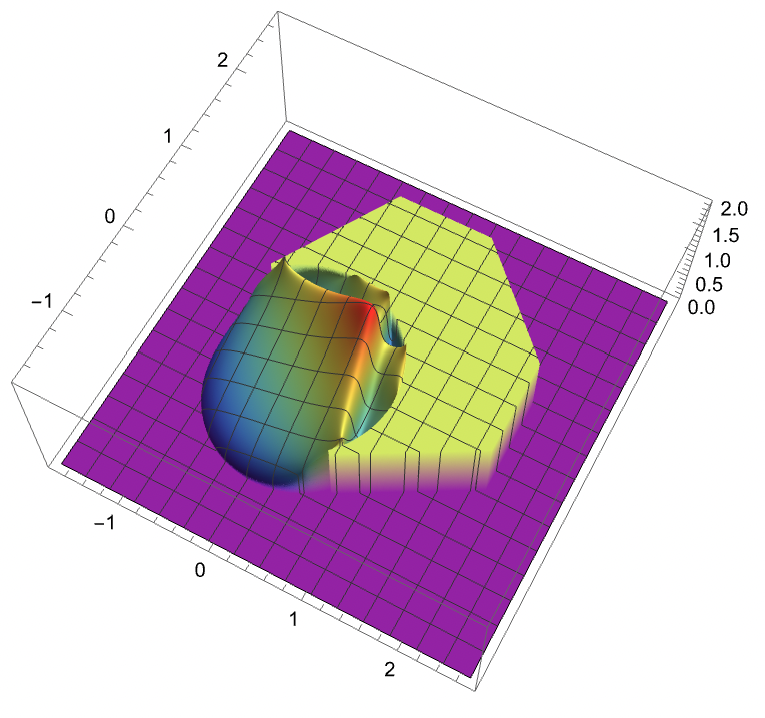}
\end{minipage}%
\caption{A polycon \(P=P(1,4)\), its dual \(P^\ast\), and the graph of the
representing measure. The additional curves in the dual picture are the
vanishing loci of the dual letters \(f_P\) and \(g_P\) appearing in~\eqref{eq:dual_letters_trans}.}
\label{fig:curvy_pentagon_measure}
\end{figure}

For a minimally curved polycon \(P(1,s)\), with one conic arc and \(s\) linear
edges, the formula is explicit. If \(\ell_i\) are the
boundary lines, ordered cyclically, then for \(y \in C^\ast\)
\begin{equation}
\label{eq:mu_one_curvy_s_lines_trans}
    \mu_{P(1,s)}(y)
    =
    \frac12
    +\frac1\pi\sum_{i=1}^s
    \arctan\!\left(\frac{\ell_i^\ast(y)}{\sqrt{q^\ast(y)}}\right)
    -
    \frac1\pi\sum_{i=1}^{s-1}
    \arctan\!\left(\frac{\ell_{i,i+1}^\ast(y)}{\sqrt{q^\ast(y)}}\right) \, .
\end{equation}
Here \(\ell_{i,i+1}\) is the line through the two intersection points with the
conic that do not lie on the chosen curved edge. On
\(\widehat P^\ast\setminus C^\ast\), the measure is again equal to one. An example for $s=4$ is appears in Figure~\ref{fig:curvy_pentagon_measure}.

Although non-negativity is not obvious from
~\eqref{eq:mu_one_curvy_s_lines_trans}, it follows by induction from the
arctangent addition formula and the geometry of tangent lines to the dual
conic. Decomposing a general polycon into such pieces gives the following
result~\cite[Theorem~4.5]{Mazzucchelli:DV}.

\begin{tcolorbox}[resultbox]
\textbf{Single-conic polycons are CM.} Every polycon bounded by one conic and finitely many lines is a completely
monotone positive geometry. Its representing measure can be obtained from
signed canonical-form triangulations.
\end{tcolorbox}\noindent

After repeated use of the arctangent addition formula, the measure of such a
polycon can be written on \(C^\ast\) as
\begin{tcolorbox}[resultbox]
\textbf{Arctangent measure for single-conic polycons.}
\begin{equation}
\label{eq:dual_letters_trans}
    \mu_P(y)
    =
    \frac12+
    \frac1\pi
    \arctan\!\left(
    \frac{f_P(y)}
    {g_P(y)\sqrt{q^\ast(y)}}
    \right)
    +k(y) \, ,
\end{equation}
\end{tcolorbox}\noindent
where \(f_P\) and \(g_P\) are homogeneous polynomials and
\(k(y)\in\mathbb Z\) is locally constant away from their vanishing loci. The role of $k(y)$ is to specify the correct branch of the inverse tangent. We
call \(f_P\) and \(g_P\) the \textit{dual letters} of the polycon. They play
for the representing measure a role analogous to symbol letters for
polylogarithmic functions: they describe the hypersurfaces where the analytic
behavior changes.

The previous construction extends algorithmically to positive geometries
bounded by several conics. Let \(Q_1,\ldots,Q_t\) be conics with
hyperbolicity cones \(C_1,\ldots,C_t\), and assume
\begin{equation}
    C=\bigcap_{j=1}^t C_j
\end{equation}
is non-empty. One can consider positive geometries inside \(\mathbb P(C)\)
whose boundary consists of arcs of the \(Q_j\)'s and finitely many line
segments. Under mild transversality assumptions, these are positive geometries
by the theory of polypols~\cite{Polypols}. A necessary condition for complete
monotonicity is that the region lie in a hyperbolicity region, hence inside~\(C\).

\begin{figure}[pos=t]
\centering
\begin{minipage}[c]{0.32\linewidth}
\includegraphics[width=\linewidth]{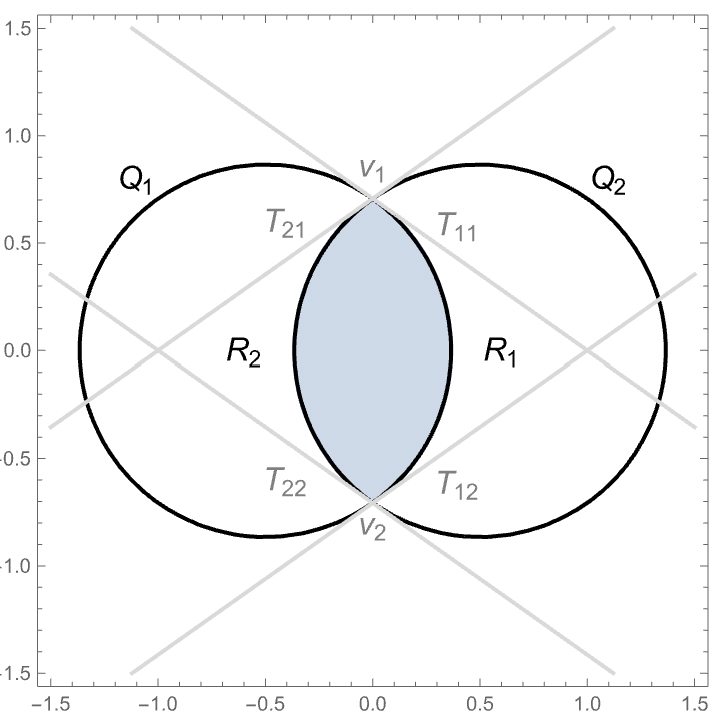}
\end{minipage}
\hfill
\begin{minipage}[c]{0.32\linewidth}
\includegraphics[width=\linewidth]{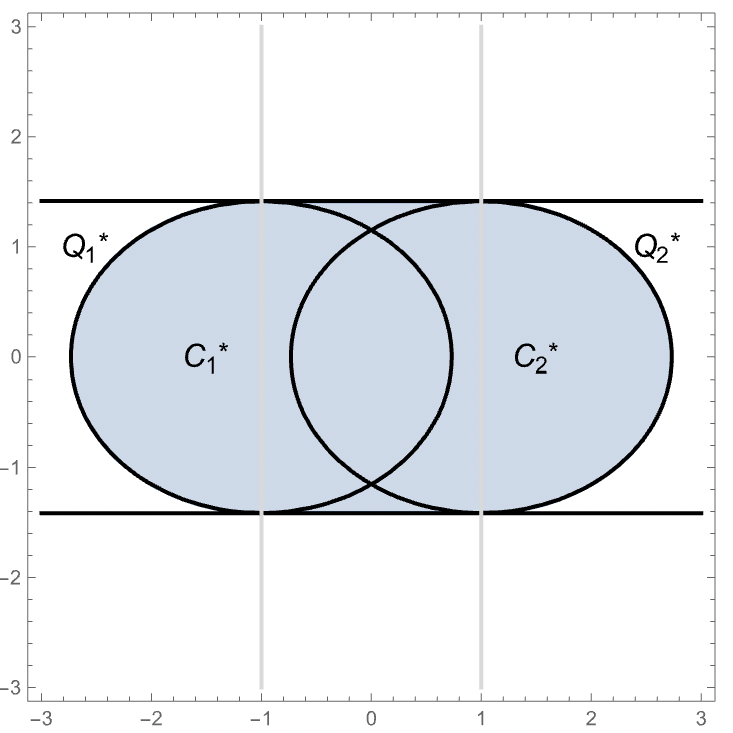}
\end{minipage}%
\hfill
\begin{minipage}[c]{0.35\linewidth}
\includegraphics[width=\linewidth]{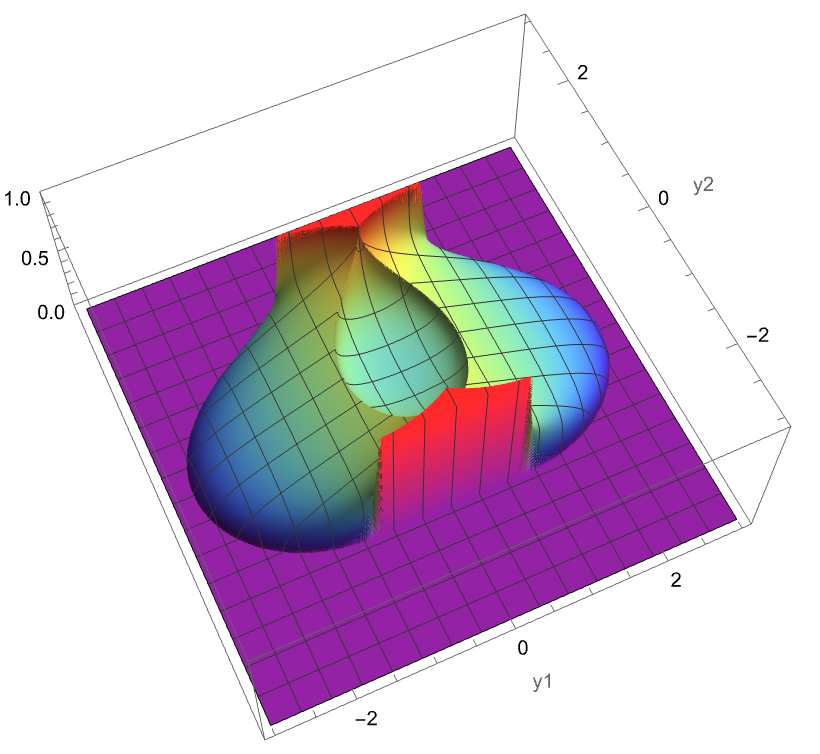}
\end{minipage}
\caption{On the left, a curvy two-gon $P$ bounded by two conics. This is a positive geometry and has an adjoint curve given by the line at infinity ${x_3=0}$. In the middle, its dual $P^*$ and on the right the plot of the measure $\mu_P$ for the canonical function of $P$. Note that $\mu_P$ is non-negative, and hence $(\mathbb{P}^2,P)$ is a completely monotone positive geometry. }
\label{fig:2con}
\end{figure}

The same signed-triangulation strategy gives candidate measures: one replaces
arcs of one conic by tangent line segments, producing larger convex pieces,
and repeats until the problem is reduced to one-conic polycons and ordinary
polygons. In all examples computed, the resulting signed sums are
non-negative. An example with two conics and no lines appears in Figure~\ref{fig:2con}. This motivates the following expectation:
\begin{tcolorbox}[resultbox]
\textbf{Multiple-conic polycons are CM.} Hyperbolic polycons bounded by lines and conics are completely monotone.
\end{tcolorbox}\noindent
This expectation will reappear among the open problems collected at the end of the chapter.

\subsubsection{The nodal cubic}
\label{app:nodal_cubic_measure}

So far we discussed positive geometries
bounded by lines and conics, for which the representing measures are built
from arctangent and logarithmic functions. Here we describe a simple example
where an elliptic period appears. This example is from~\cite[Section 4.5]{Mazzucchelli:DV}.

Consider the cubic
\begin{equation}
\label{eq:nodal_cubic_appendix}
    p(x)
    =
    -2x_2^2x_0+(x_0-x_1)(x_0+x_1)^2
    =
    \det
    \begin{pmatrix}
        2x_0 & -x_1-x_0 & 0\\
        -x_1-x_0 & x_1+x_0 & x_2\\
        0 & x_2 & x_1+x_0
    \end{pmatrix} \, .
\end{equation}
Its real vanishing locus has a node, and \(p\) is hyperbolic with respect to a
cone \(\widehat P\) containing \((1,0,0)\). The corresponding positive
geometry \(P\subseteq\mathbb P^2\) has canonical function
\begin{equation}
\label{eq:nodal_cubic_canonical_appendix}
    \Omega_P(x)=\frac4{p(x)} \, .
\end{equation}
Since \(\widehat P\) is a minimal spectrahedral cone, our previous results imply that \(P\) is completely monotone.

The dual cone \(\widehat P^\ast\) has algebraic boundary with two components.
One is the line dual to the node of the cubic. The other is the dual variety
of the cubic, a quartic:
\begin{equation}
\label{eq:dual_cubic_quartic_appendix}
   q(y)
   =
   -27 y_2^2y_0^2+16y_2^4
    -18y_2^2y_1y_0+13y_2^2y_1^2
    +8y_1^3y_0+8y_1^4 \, .
\end{equation}
Inside \(\widehat P^\ast\) there is a distinguished region \(L\), called a
\textit{lacuna}, on which the fundamental solution, and hence the representing
measure, is constant. In this example,
\begin{equation}
	L=
    \{q(y)<0,\ 0<y_1<y_0,\ -y_0<y_2<y_0,\ y_0>0\} \, .
\end{equation}

\begin{figure}[pos=t]
\centering
\begin{minipage}[c]{0.31\textwidth}
\centering
\includegraphics[width=\linewidth]{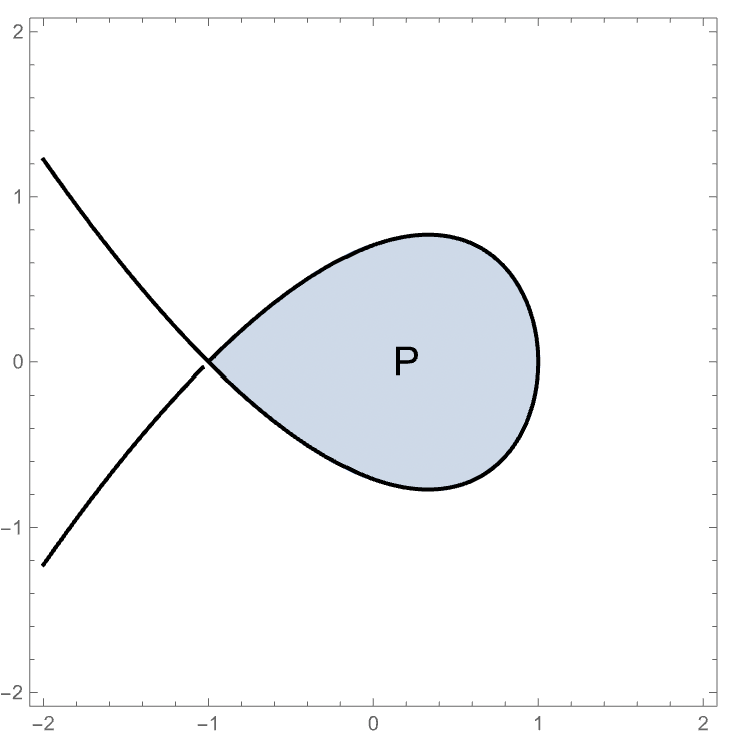}
\end{minipage}
\hfill
\begin{minipage}[c]{0.31\textwidth}
\centering
\includegraphics[width=\linewidth]{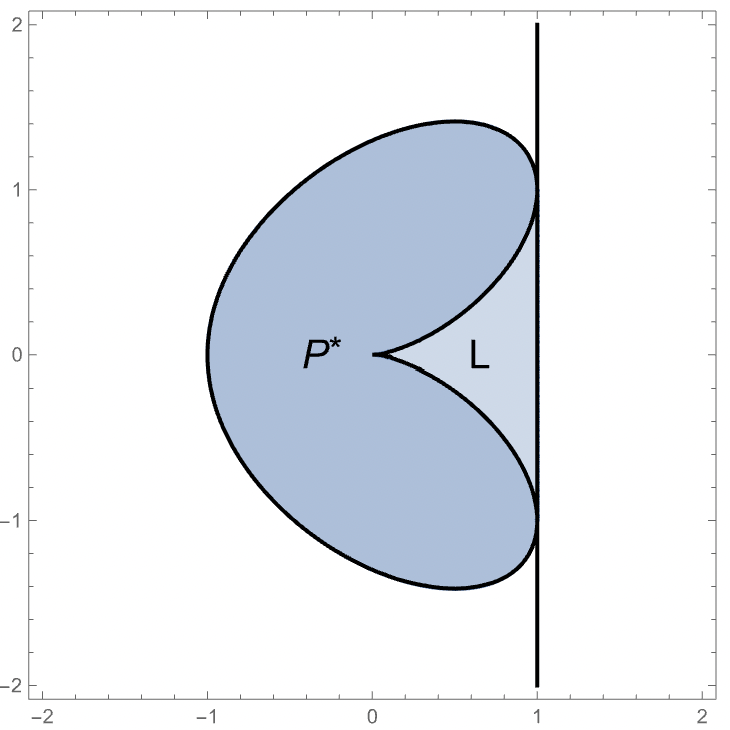}
\end{minipage}
\hfill
\begin{minipage}[c]{0.35\textwidth}
\centering
\includegraphics[width=\linewidth]{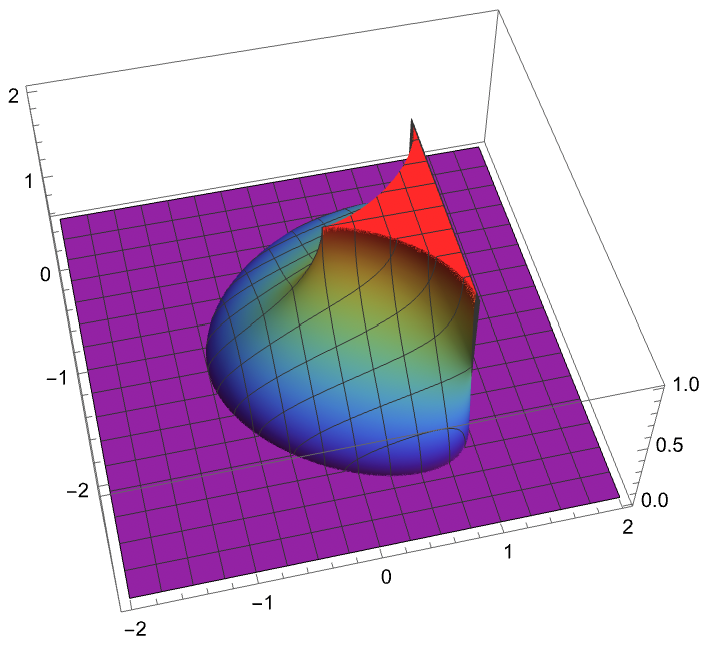}
\end{minipage}%
\caption{A nodal cubic positive geometry, its dual region, and the graph of
the representing measure. The lightly shaded region in the dual picture is
the lacuna \(L\), where the measure is constant. Away from the lacuna, the
measure is governed by elliptic periods.}
\label{fig:nodal_cubic_measure}
\end{figure}

Using the determinantal representation in~\eqref{eq:nodal_cubic_appendix} as the push-forward of the \emph{Wishart} distribution on the space of symmetric $3 \times 3$ matrices, see~\cite{Mazzucchelli:DV} for ore details. This gives
\begin{equation}
\label{eq:cubic_measure_period_appendix}
    \mu_P(y)
    =
    \int_{\mathcal R(y)}
    \frac{4r\,\mathrm{d}r\,\mathrm{d}t}
    {\pi^2\sqrt{(1-t^2)\,h(t,r,y)}} \, ,
\end{equation}
where
\begin{equation}
	h(t,r,y)
    =
    -y_2^2
    -4r^2\bigl(r^2-y_1-r t\sqrt{2(y_0-y_1)}\bigr) \, ,
\end{equation}
and \(\mathcal R(y)\) is the region where \(r>0\), \(0<t<1\), and
\(h(t,r,y)>0\). The quartic
~\eqref{eq:dual_cubic_quartic_appendix} appears as the discriminant of
\(h(1,r,y)\) with respect to \(r\).

Numerically, the integral vanishes outside \(\widehat P^\ast\), as predicted
by hyperbolicity. On the closure of the lacuna $L$,
\begin{equation}
	\mu_P(y)=1 \, .
\end{equation}
Outside the lacuna, the integration region is bounded by the only two real positive roots
\(r_-(y)<r_+(y)\) of $h(1,r,y)$ in $r$, and the integral over \(t\) can be performed. One obtains
\begin{equation}
\label{eq:cubic_measure_elliptic_appendix}
    \mu_P(y)
    =
    \int_{r_-(y)}^{r_+(y)}
    \frac{2^{5/4}\,\mathrm{d}r}
    {\pi^2\sqrt r\,(y_0-y_1)^{1/4}}
     \,
     K\!\left(
    \frac{h(1,r,y)}
    {8r^3\sqrt{2(y_0-y_1)}}
    \right) \, ,
\end{equation}
where \(K\) is the complete elliptic integral of the first kind. The semialgebraic set $P$, its dual, and a plot of the graph of $\mu_P$ are given in Figure~\ref{fig:nodal_cubic_measure}.

The examples in this section show that the dual-volume representation of a
canonical function may reveal analytic structure invisible from the rational form itself. For polycons, it is built from arctangent and logarithmic functions. For the nodal cubic, the dual measure is governed by elliptic periods, with singular support controlled by the dual algebraic boundary and a lacuna on which the measure is constant. In the next section we move to the study of duality for positive geometries that are not full-dimensional in projective space, but rather semialgebraic sets inside projective varieties.

\subsection{Convexity in the Grassmannian}
\label{sec:Convexity in the Grassmannian}

In the previous sections we studied duality for positive geometries in
projective space. There, the essential input was ordinary convexity: a
convex projective region has a polar dual, and the canonical function may be
interpreted as a volume, or more generally as a Laplace transform with respect
to a measure on the dual cone. To approach an analogous question for
Amplituhedra, one first faces a basic obstruction: Amplituhedra live in
Grassmannians, and there is no intrinsic notion of convexity on a
Grassmannian comparable to convexity in projective space.

The purpose of this section is to explain a useful extrinsic substitute. The
Grassmannian has a canonical projective embedding, the Pl\"ucker embedding.
We may therefore take convex hulls in the ambient Pl\"ucker space and then
intersect back with the Grassmannian. This leads to the notion of
\textit{extendable convexity}. The main result reviewed here is that the
tree Amplituhedron \(\mathcal A_{2,2,n}(Z)\subseteq \Gr(2,4)\) is extendably
convex~\cite{mazzucchelli2025exterior}. This provides a bridge between
ordinary convex geometry and Amplituhedron geometry, and it prepares the
definition of dual Amplituhedra in Section~\ref{sec:Duality for Amplituhedra}.

Throughout this section we only consider tree Amplituhedra. Recall that
\(\mathcal A_{k,m,n}(Z)\) is a \(km\)-dimensional semialgebraic subset of
\(\Gr(k,k+m)\), defined as the image of \(\Gr_{\geq 0}(k,n)\) under the
Amplituhedron map associated with a positive \((k+m)\times n\) matrix \(Z\).

\subsubsection{Extendable convexity}

In the following we denote the real Grassmannian with its Pl\"ucker embedding as
\begin{equation}
    \Gr(k,k+m)
    \longhookrightarrow
    \mathbb P\bigl(\bigwedge^k \mathbb R^{k+m}\bigr) \, .
\end{equation}
where the map is given by
\begin{equation}
\label{eq:plucker_embedding_convexity}
    \operatorname{span}\{C_1,\ldots,C_k\}
    \longmapsto
    C_1\wedge\cdots\wedge C_k \, .
\end{equation}
Here \(C_1\wedge\cdots\wedge C_k\) is understood projectively. In coordinates,
if \(C_i=\sum_j C_{ij}e_j\), then the coefficient of
\(e_{i_1}\wedge\cdots\wedge e_{i_k}\) is the Pl\"ucker coordinate $p_I(C)$ with $I=\{i_1,\dots,i_k\}$.
This is the same embedding as in~\eqref{eq:Pl_emb}.

We first recall projective convexity. Recall that a subset
\(S\subseteq\mathbb P^N_{\mathbb R}\) is called very compact if there
exists a real hyperplane \(H\subseteq\mathbb P^N_{\mathbb R}\) with
\(H\cap S=\emptyset\). Choosing such a hyperplane identifies
\(\mathbb P^N_{\mathbb R}\setminus H\) with an affine space. If \(S\) is
connected and very compact, one can take its ordinary convex hull in this
affine chart. The result is independent of the choice of \(H\)
\cite[Lemma~2.2]{Kummer_2022}; we denote it by
\begin{equation}
    \conv(S)\subseteq\mathbb P^N_{\mathbb R} \, .
\end{equation}
Equivalently, this construction can be described in terms of cones
one dimension higher.

Now let \(X \subseteq\mathbb P^N_{\mathbb R}\) be an embedded real projective
variety, and let \(S\subseteq X\) be connected, semialgebraic, and
very compact subset of the real points of $X$. In the notation of Chapter~\ref{ch:Positive Geometries}, $X=\mathcal{X}_\mathbb{R}$. We define
\begin{tcolorbox}[definitionbox]
\textbf{Convex hull inside a variety.}
\begin{equation}
\label{eq:convX_def_long}
    \conv_X(S):=
    X \cap\conv(S) \, .
\end{equation}
\end{tcolorbox}\noindent
We say that \(S\) is \textit{extendably convex} in \(X\) if
\begin{equation}
\label{eq:extendably_convex_def_long}
    S=\conv_X(S) \, .
\end{equation}
Thus convexity is not intrinsic to \(X\), but it depends on the choice of embedding $X \subseteq P^N_{\mathbb R}$. This
notion was first considered by Busemann~\cite{busemann1961convexity}.

When \(X=\mathbb P^N_{\mathbb R}\), this reduces to ordinary
projective convexity. In general, it has many of the formal properties one
expects from a convex hull operation. For connected very compact
semialgebraic sets \(S,T\subseteq X\), one has monotonicity, idempotence,
antiexchange, and equivariance under linear automorphisms preserving \(X\)
\cite[Propositions~3.4--3.5]{mazzucchelli2025exterior}. In particular,
\begin{equation}
    S\subseteq T
    \quad\Longrightarrow\quad
    \conv_X(S)\subseteq\conv_X(T) \, ,
\end{equation}
and
\begin{equation}
    \conv_X(\conv_X(S))=\conv_X(S) \, .
\end{equation}

\begin{figure}[pos=t]
\centering
\begin{minipage}[c]{0.31\textwidth}
\centering
\includegraphics[width=\linewidth]{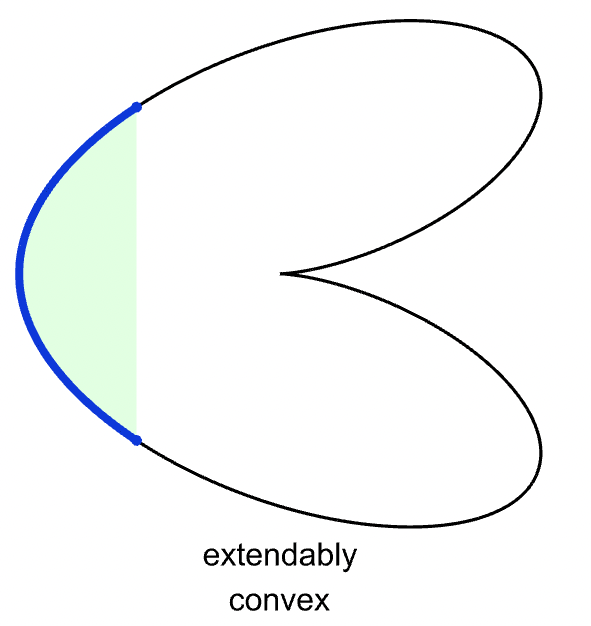}
\end{minipage}
\hspace{0.6in}
\begin{minipage}[c]{0.31\textwidth}
\centering
\includegraphics[width=\linewidth]{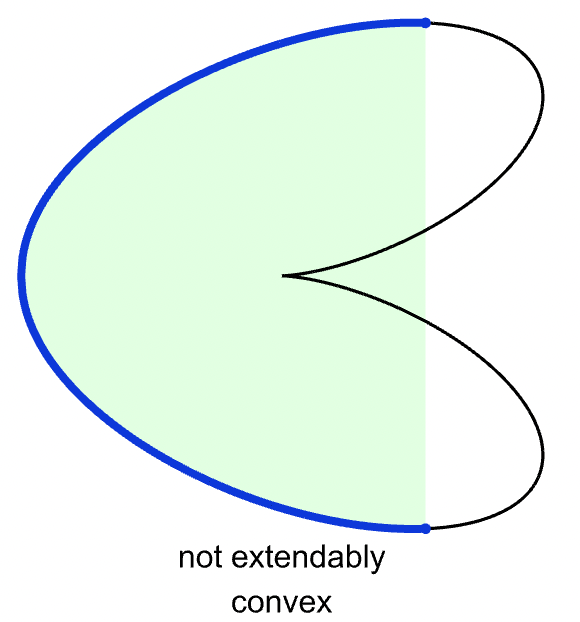}
\end{minipage}
\caption{Illustration of extendable convexity for a curvy segment, shown in
blue, on the real locus of a plane quartic curve, shown in black. The green
region is the ordinary convex hull in the affine chart.}
\label{fig:extendable_convexity_curve}
\end{figure}

A basic example is the non-negative Grassmannian. In the Pl\"ucker embedding,
\(\Gr_{\geq 0}(k,n)\) is the intersection of \(\Gr(k,n)\) with the standard
simplex:
\begin{equation}
\label{eq:positive_grass_extendably_convex}
    \Gr_{\geq 0}(k,n)
    =
    \Gr(k,n)\cap\Delta^N,
    \qquad
    N=\smallbinom{n}{k}-1 \, ,
\end{equation}
where \(\Delta^N\) is cut out by the non-negativity of all Pl\"ucker
coordinates. Hence \(\Gr_{\geq 0}(k,n)\) is extendably convex. This is a
genuinely extrinsic statement: it depends on the Pl\"ucker embedding. It
is also different from other possible notions such as geodesic convexity.

\subsubsection{The exterior cyclic polytope}

We now apply this idea to Amplituhedra. The Amplituhedron map
\(\widetilde Z:\Gr(k,n)\to\Gr(k,k+m)\) extends linearly to the corresponding
Pl\"ucker spaces:
\begin{equation}
\begin{aligned}
    \bigwedge^k Z:
    \mathbb P\!\bigl(\bigwedge^k\mathbb R^n\bigr)
    &\dashrightarrow
    \mathbb P\!\bigl(\bigwedge^k\mathbb R^{k+m}\bigr) \, ,\\
    e_{i_1}\wedge\cdots\wedge e_{i_k}
    &\longmapsto
    Z_{i_1}\wedge\cdots\wedge Z_{i_k}.
\end{aligned}
\end{equation}
For \(I=\{i_1<\cdots<i_k\}\), we write $Z_I:=Z_{i_1}\wedge\cdots\wedge Z_{i_k} $.
The ambient convex hull of the Amplituhedron is the following polytope.

\begin{tcolorbox}[definitionbox]
\textbf{Exterior cyclic polytope.}
\begin{equation}
\label{eq:exterior_cyclic_polytope_def_long}
    C_{k,m,n}(Z)
    :=
    \bigwedge^k Z(\Delta^N)
    =
    \conv\{Z_I:\ I\in\tbinom{[n]}{k}\}
    \subseteq
    \mathbb P\!\bigl(\bigwedge^k\mathbb R^{k+m}\bigr) \, .
\end{equation}
\end{tcolorbox}\noindent
We call \(C_{k,m,n}(Z)\) the \textit{exterior cyclic polytope}. For \(k=1\),
this is the ordinary cyclic polytope. For \(k>1\), it is a linear extension of
the Amplituhedron in Pl\"ucker space.

Since the Amplituhedron is the image of the non-negative Grassmannian, and
\(\Delta^N\) is the ambient convex hull of \(\Gr_{\geq 0}(k,n)\), we have
\begin{equation}
    \mathcal A_{k,m,n}(Z)\subseteq C_{k,m,n}(Z) \, .
\end{equation}
By monotonicitiy of the convex hull, we deduce the following.
\begin{tcolorbox}[resultbox]
\textbf{Convex hull of the Amplituhedron.}
\begin{equation}
\label{eq:conv_ampl_exterior_cyclic_long}
    \conv\bigl(\mathcal A_{k,m,n}(Z)\bigr)
    =
    C_{k,m,n}(Z) \, .
\end{equation}
\end{tcolorbox}\noindent
Thus the question of extendable convexity becomes the question of whether the
reverse inclusion holds after intersecting with the Grassmannian:
\begin{equation}
    \Gr(k,k+m)\cap C_{k,m,n}(Z)
    \overset{?}{=}
    \mathcal A_{k,m,n}(Z) \, .
\end{equation}

There are a few elementary cases. For \(k=1\), the Amplituhedron is the
ordinary cyclic polytope, hence is convex. For \(n=k+m\), the Amplituhedron is
isomorphic to \(\Gr_{\geq 0}(k,k+m)\), hence is extendably convex by
~\eqref{eq:positive_grass_extendably_convex}. For \(m=1\), after the
identification
\begin{equation}
    \Gr(k,k+1)\cong\Gr(1,k+1)\cong\mathbb P^k \, ,
\end{equation}
extendable convexity is ordinary projective convexity. In this case the
Amplituhedron is a union of bounded chambers of a cyclic hyperplane
arrangement and is not convex.

Extendable convexity is therefore restrictive. If an Amplituhedron is extendably convex, then its boundary must arise from
hyperplane sections of the Grassmannian in the Pl\"ucker embedding. This is
consistent with the known boundary equations for \(m=2\) and \(m=4\), namely
\(\langle Y\,i\,i{+}1\rangle=0\) and
\(\langle Y\,i\,i{+}1\,j\,j{+}1\rangle=0\), respectively. For larger \(m\),
higher-degree boundary components are expected~\cite{higher_m_ampl}, so
extendable convexity is too restrictive for the Amplituhedron itself. In such
cases, the more linear object
\begin{equation}
    \Gr(k,k+m)\cap C_{k,m,n}(Z)
\end{equation}
may still be interesting in its own right.

\subsubsection{The case \(k=m=2\)}

The first genuinely nontrivial case is
\begin{equation}
    \mathcal A_{2,2,n}(Z)\subseteq\Gr(2,4) \, .
\end{equation}
Here \(\Gr(2,4)\) is the Grassmannian of lines in $\mathbb{P}^3$. In
the Pl\"ucker embedding $\mathbb{P}^5$
it is the smooth quadric hypersurface cut out by the vanishing of
\begin{equation}
\label{eq:gr24_plucker_quadric}
    p_{12}p_{34}+p_{14}p_{23}-p_{13}p_{24} \, .
\end{equation}
The exterior cyclic polytope \(C_{2,2,n}(Z)\subseteq\mathbb P^5\) is the convex
hull of the \(\binom n2\) points
\begin{equation}
    Z_i\wedge Z_j,
    \qquad 1\leq i<j\leq n \, .
\end{equation}

The main result is the following~\cite[Theorem~6.12]{mazzucchelli2025exterior}.

\begin{tcolorbox}[resultbox]
\textbf{Extendable convexity of \(\mathcal A_{2,2,n}\).} For every positive \(4\times n\) matrix \(Z\),
\begin{equation}
\label{eq:k_m_2_extendable_convexity_long}
    \mathcal A_{2,2,n}(Z)
    =
    \Gr(2,4)\cap C_{2,2,n}(Z) \, .
\end{equation}
Equivalently, the \(k=m=2\) Amplituhedron is extendably convex in
\(\Gr(2,4)\).
\end{tcolorbox}\noindent
This theorem implies that \(\mathcal A_{2,2,n}(Z)\) is a basic semialgebraic
set cut out by linear inequalities in Pl\"ucker coordinates, together with
the Pl\"ucker quadric~\eqref{eq:gr24_plucker_quadric}. We now explain which
linear inequalities are relevant.

In general, proving an identity of the form
\begin{equation}
    S=X\cap P
\end{equation}
for a semialgebraic subset \(S\subseteq X\) and an ambient polytope
\(P\subseteq\mathbb P^N\) involves several separate issues. Even when
\(S\subseteq X\cap P\), the set \(X\cap P\) may have additional boundary
components, extra connected components, or lower-dimensional pieces. These
pathologies are controlled by the general criterion of
\cite[Lemma~3.11]{mazzucchelli2025exterior}. In the present case, the needed
boundary, connectedness, and regularity checks are carried out explicitly in~\cite{mazzucchelli2025exterior}. Here we focus on the boundary and inequality description.

For \(m=1,2,4\), the boundary of \(\mathcal A_{k,m,n}(Z)\) is contained in
the boundary of the exterior cyclic polytope \(C_{k,m,n}(Z)\)
\cite[Proposition~5.6]{mazzucchelli2025exterior}. For example, when \(m=2\),
the boundary is cut out by the equations
\begin{equation}
    \langle Y\,i\,i{+}1\rangle=0,
    \qquad i=1,\ldots,n \, .
\end{equation}
Each such hyperplane contains enough vertices \(Z_I\) of
\(C_{k,2,n}(Z)\), and positivity of \(Z\) ensures that all remaining vertices
lie on the same side. Hence these are facet hyperplanes of the exterior
cyclic polytope $C_{k,2,n}(Z)$.
However, \(C_{2,2,n}(Z)\) usually has many more facets than the Amplituhedron
has boundaries. The relevant ones are the \textit{Schubert facets}. A Schubert
hyperplane in
\(\mathbb P(\bigwedge^k\mathbb R^{k+m})\) is a hyperplane of the form
\begin{equation}\label{eq:schub_hyp}
    \{y:\  y \cdot W^\perp =0\},
    \qquad
    W\in\Gr(m,k+m) \, ,
\end{equation}
where in the scalar product $y \cdot W^\perp$ we identify the orthogonal space $W^\perp \in \Gr(k,k+m)$ with its image in the Plücker embedding \(\mathbb P(\bigwedge^k\mathbb R^{k+m})\). We say that the hyperplane~\eqref{eq:schub_hyp} is dual to $W$. Note that if $y=Y \in \Gr(k,k+m)$, then we can write
\begin{equation}
	y \cdot W^\perp = \langle Y \, W \rangle \, ,
\end{equation}
where the brackets indicate the determinant of the $(k+m)\times(k+m)$ matrix obtained by stacking together matrix representatives of $Y$ and $W$. This follows from the Laplace expansion of the determinant.

For \(k=m=2\), all Schubert facets of \(C_{2,2,n}(Z)\) are dual to the $\binom{n}{2}$ lines~\cite[Thm 5.5]{mazzucchelli2025exterior}
\begin{equation}
\label{eq:schubert_hyperplanes_bar}
    \overline{ij} := \left(\bar i\cap\bar j \right)^\perp \, ,
    \qquad 1\leq i<j\leq n \, ,
\end{equation}
where $\bar i:=(i{-}1\,i\,i{+}1)$ denotes the plane in $\mathbb{P}^3$ spanned by \(Z_{i-1},Z_i,Z_{i+1}\). Here we use the usual twisted cyclic convention \(Z_{i \pm n}=(-1)^{k-1}Z_i\).

Keeping only these Schubert inequalities gives the
\textit{Schubert exterior cyclic polytope}
\begin{tcolorbox}[definitionbox]
\textbf{Schubert exterior cyclic polytope.}
\begin{equation}
\label{eq:schubert_exterior_cyclic_polytope_def}
    \widetilde C_{2,2,n}(Z)
    :=
    \left\{
    y\in\mathbb P^5:
     y \cdot \overline{ij}^\perp  \geq 0,
    \quad
    1\leq i<j\leq n
    \right\} \, .
\end{equation}
\end{tcolorbox}\noindent
The sign is chosen so that
\begin{equation}
    C_{2,2,n}(Z)\subseteq\widetilde C_{2,2,n}(Z) \, .
\end{equation}
This inclusion follows from positivity properties of determinants of minors
of \(Z\). A representative inequality is
\begin{equation}
\label{magic_det}
        \det
        \begin{pmatrix}
            \langle a\,i_1 i_2 i_3 \rangle
            &
            \langle a\,j_1 j_2 j_3 \rangle \\
            \langle b\,i_1 i_2 i_3 \rangle
            &
            \langle b\,j_1 j_2 j_3 \rangle
        \end{pmatrix}
        =
        \langle ab\, (i_1i_2i_3)\cap(j_1j_2j_3)\rangle
        \geq 0 \, .
\end{equation}
This appears in~\cite[Section~14]{Arkani_Hamed_2018} and a conceptual
proof is that the right-hand side is a cluster variable for~\(\Gr(4,n)\)~\cite[Corollary~4.10]{evenZoharLakrecParisiTesslerShermanBennettWilliams2023cluster}, thus positive for positive $Z$.

The remarkable fact is that the additional non-Schubert inequalities of
\(C_{2,2,n}(Z)\) disappear after restricting to the Pl\"ucker quadric~\cite[Theorem~6.12]{mazzucchelli2025exterior}.

\begin{tcolorbox}[resultbox]
\textbf{Schubert inequalities for \(\mathcal A_{2,2,n}\).} For every positive \(4\times n\) matrix \(Z\),
\begin{equation}
\label{eq:schubert_polytope_same_intersection}
  \mathcal A_{2,2,n}(Z) =   \Gr(2,4)\cap\widetilde C_{2,2,n}(Z) \, .
\end{equation}
\end{tcolorbox}\noindent
Thus the \(k=m=2\) Amplituhedron only sees the Schubert facets of the exterior
cyclic polytope. Equivalently,~\eqref{eq:schubert_exterior_cyclic_polytope_def}
gives a linear inequality description of \(\mathcal A_{2,2,n}(Z)\), agreeing
with the description in~\cite{Herrmann:2020qlt}:
\begin{equation}
	\mathcal{A}_{2,2,n}(Z) = \{Y \in \Gr(2,4) \, : \, \langle Y \, \overline{ij} \rangle \geq 0 \, , \ 1 \leq i < j \leq n \} \, .
\end{equation}

\subsubsection{Combinatorics of exterior cyclic polytopes}

Exterior cyclic polytopes have interesting combinatorics. For ordinary cyclic polytopes, the combinatorial type is determined by the number of vertices and the dimension. For
\(C_{k,m,n}(Z)\) with \(k>1\), the situation is subtler. In fact, denoting by \(W_{k,m,n}\) the linear \emph{matroid} of \(\bigwedge^k Z\) for generic \(Z\), starting already at \(m=2\), it is not
determined solely by the matroid of \(Z\). Nevertheless, the \emph{oriented matroid}
of \(\bigwedge^k Z\), and hence the combinatorial type of \(C_{k,m,n}(Z)\), is constant on regions of the positive Grassmannian
\(\Gr_{>0}(k+m,n)\). The walls between these regions are cut out by
determinants of collections of exterior products.

\begin{figure}[pos=t]
\centering
\includegraphics[width=0.6\linewidth]{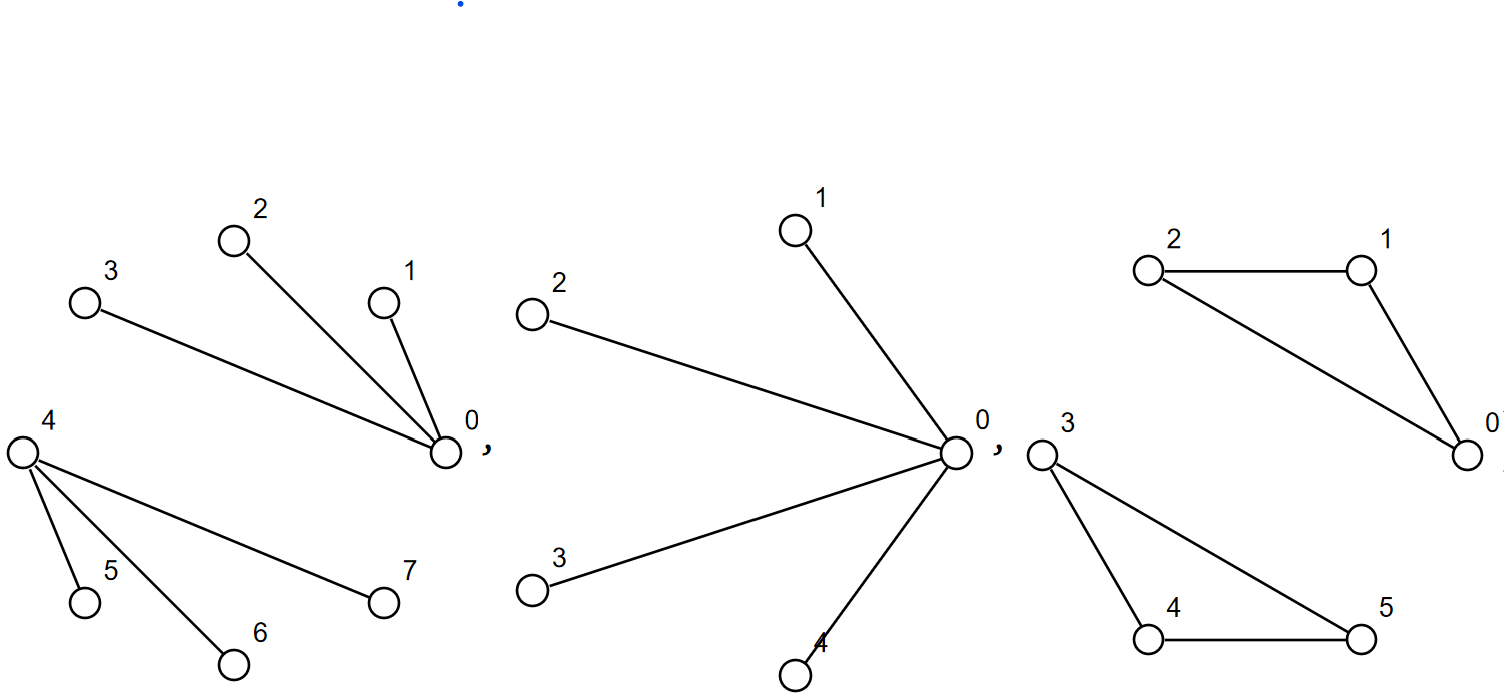}
\caption{Circuits in \(W_{2,2,n}\), up to gluing.}
\label{fig:circuits4}
\end{figure}

Let us comment on the structure of $W_{k,m,n}$ for \(k=m=2\). In this case the columns of \(\bigwedge^2 Z\) are indexed by the edges of the complete graph \(K_n\). This relates the wedge power matroid to graph connectivity, rigidity, and algebraic statistics~\cite{brakensiek2024rigidity,Crespo_Ruiz2023}. The graph interpretation gives a compact way to organize \emph{bases} and \emph{circuits}. One uses two operations on a simple graph \(G\):
\begin{enumerate}[label=(\roman*)]
    \item \textit{Cutting:} if \(e=uv\) is an edge of \(G\), then
    \(\operatorname{Cut}(G,e,v)\) is obtained by adding a vertex \(v'\)
    and replacing \(uv\) by \(uv'\).
    \item \textit{Gluing:} if \(u\) and \(v\) have distance at least three,
    then \(\operatorname{Glue}(G,u,v)\) is obtained by identifying \(u\) and
    \(v\).
\end{enumerate}
Cutting preserves independence, while gluing preserves dependence. Thus
circuits and bases for all \(n\) can be encoded by a small list of graph
types together with these operations. All circuits in \(W_{2,2,n}\) arise by gluing from the three basic circuit types shown in Figure~\ref{fig:circuits4}.

\begin{eg}[$k=m=2$, $n=66$]
In this case we parametrize \(Z\) by the positive Vandermonde matrix
\begin{equation}
\label{eq:Zmatrix_appendix}
Z=
\begin{pmatrix}
1 & 1 & 1 & 1 & 1 & 1  \\
a & b & c & d & e & f  \\
a^2 & b^2 & c^2 & d^2 & e^2 & f^2  \\
a^3 & b^3 & c^3 & d^3 & e^3 & f^3
\end{pmatrix},
\qquad
a<b<c<d<e<f \, .
\end{equation}
Then \(\bigwedge^2 Z\) is a \(6\times15\) matrix whose columns are indexed by
edge \(ij\in\binom{[6]}2\) of the complete graph
\(K_6\). After dividing each column by its common Vandermonde factor, its
maximal minors define the wedge power matroid \(W_{2,2,6}\).
There are \(\binom{15}{6}=5005\) maximal minors. Among these, \(1660\) vanish
and \(3345\) are nonzero. The bases fall into \(12\) symmetry classes,
represented by graphs on six labeled vertices; see Figure~\ref{fig:c246bases}.

\begin{figure}[pos=t]
\centering
\includegraphics[scale=0.2]{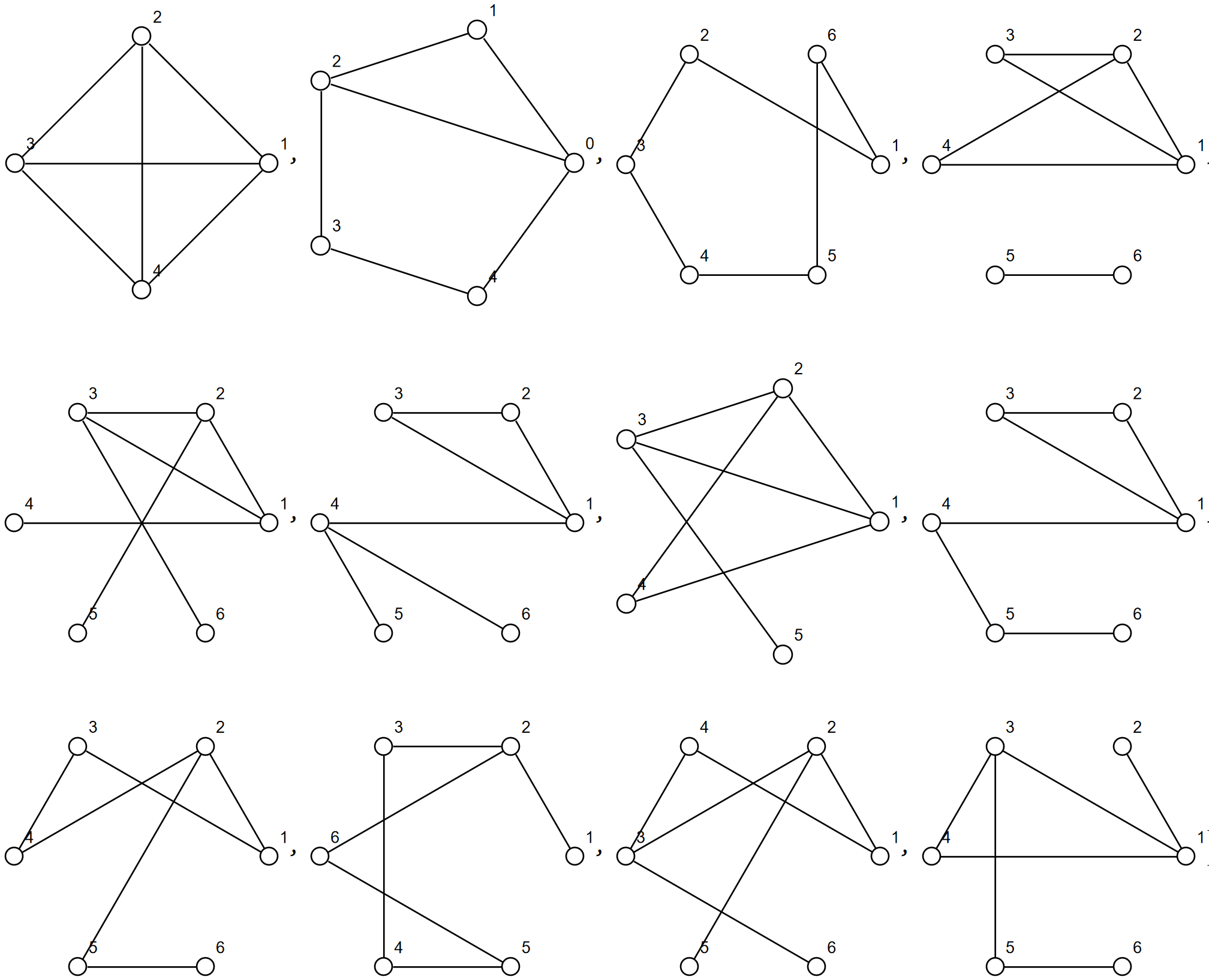}
\caption{Bases of \(W_{2,2,6}\), grouped into symmetry classes.}
\label{fig:c246bases}
\end{figure}

In all but one symmetry class, the sign of the corresponding maximal minor is
fixed by the ordering \(a<b<c<d<e<f\). The exceptional class is represented by
the \(6\)-cycle
\begin{equation}
	 \{16,12,23,34,45,56\} \, .
\end{equation}
For this basis, the maximal minor factors as
\begin{equation}\label{eq:wall}
\begin{aligned}
[16,12,23&,34,45,56]
=
(a-c)(a-d)(a-e)(b-d)(b-e)(b-f)(d-f)(c-e)(c-f)
\\
&\quad\cdot
\bigl(
abd-abe-acd+acf+ade-adf
+bce-bcf-bde+bef+cdf-cef
\bigr).
\end{aligned}
\end{equation}
The final cubic factor can have either sign. For example, if $(a,b,c,d,e)=(1,3,4,7,8)$,
then the cubic is positive for \(f>47/5\), zero for \(f=47/5\), and negative
for \(f<47/5\). Thus the matroid and oriented matroid of \(\bigwedge^2 Z\)
may change as \(Z\) varies over positive matrices.

This change affects the combinatorics of the exterior cyclic polytope. For
\(f=9\), the \(5\)-dimensional polytope \(C_{2,2,6}(Z)\) has \(f\)-vector
\begin{equation}
	(15,75,143,111,30) \, .
\end{equation}
It has \(30\) facets: \(18\) simplices and \(6\) double pyramids over
pentagons. At the wall \(f=47/5\), the three simplex facets
\begin{equation}
	\{12,23,34,45,56\} \, ,\qquad
    \{12,23,34,56,16\} \, ,\qquad
    \{12,16,34,45,56\}
\end{equation}
coalesce into the single facet
\begin{equation}
	\{12,23,34,45,56,16\} \, ,
\end{equation}
which is combinatorially a cyclic polytope \(C_{4,6}\). Crossing the wall
replaces them by 
\begin{equation}
	\{12,16,23,34,45\} \, ,\qquad
    \{12,16,23,45,56\} \, ,\qquad
    \{16,23,34,45,56\} \, .
\end{equation}

\end{eg}

The wall in~\eqref{eq:wall} is, up to permutation, the only one for any $n$~\cite[Thm 4.11]{mazzucchelli2025exterior}.
\begin{tcolorbox}[resultbox]
\textbf{Combinatorial wall for $m=k=2$.}
For any positive $4 \times n$ matrix $Z$, the matroid of $\bigwedge^2 Z$ equals $W_{2,2,n}$ outside the vanishing locus of
\begin{equation}
\label{eq:wedge_wall_hexagon}
\begin{aligned}
   & \qquad \qquad \qquad \det[Z_{12}\ Z_{23}\ Z_{34}\ Z_{45}\ Z_{56}\ Z_{16}] = \\
&\langle1234\rangle\langle1356\rangle\langle2456\rangle
    -\langle1235\rangle\langle1346\rangle\langle2456\rangle
    +\langle1235\rangle\langle1246\rangle\langle3456\rangle ,
\end{aligned}
\end{equation}
and of its permutations.
\end{tcolorbox}\noindent
Curiously, this same expression appears in physics as the algebraic prefactor,
or leading singularity, of the six-dimensional scalar hexagon integral
\cite[Section~2.5, Eq.~(42)]{hexagon}. Crossing the wall~\eqref{eq:wedge_wall_hexagon} changes the face lattice of the exterior cyclic
polytope.

For generic positive \(Z\), the computed \(f\)-vectors of \(C_{2,2,n}(Z)\) are
\[
\begin{array}{c|rrrrrr}
n & f_0 & f_1 & f_2 & f_3 & f_4 & f_5 \\ \hline
5 & 10 & 35  & 55   & 40   & 12  & 1 \\
6 & 15 & 75  & 143  & 111  & 30  & 1 \\
7 & 21 & 147 & 328  & 282  & 82  & 1 \\
8 & 28 & 266 & 664  & 616  & 192 & 1 \\
9 & 36 & 450 & 1217 & 1191 & 390 & 1 .
\end{array}
\]
These computations suggest that, although the combinatorial type can vary with \(Z\), the \(f\)-vector may remains generically constant for positive matrices $Z$.

The construction of \(C_{k,m,n}(Z)\) separates two aspects of the Amplituhedron. The exterior cyclic polytope is an ordinary projective polytope in Pl\"ucker space; the Amplituhedron is obtained by intersecting it with the Grassmannian only in special cases, such as \(k=m=2\). This makes the exterior cyclic polytope a useful linear model for studying convexity, duality and combinatorics before imposing the nonlinear Grassmannian constraint.

\subsection{Duality for Amplituhedra}
\label{sec:Duality for Amplituhedra}

In the previous section we introduced extendable convexity as a way to import
ordinary projective convexity into the Grassmannian through the Plücker
embedding. We now use the same idea to define a dual object. The guiding
question is whether the Amplituhedron admits a dual-volume interpretation
analogous to the one for polytopes and for the projective positive geometries
studied in Sections~\ref{sec:Completely Monotone Positive Geometries}
and~\ref{sec:Transcendental Measures}.

The main construction is the \textit{extendable dual Amplituhedron}. It gives a
candidate support for a dual-volume representation of the Amplituhedron
canonical form. In the special case \(k=m=2\), this dual object can be
described completely: it is again an Amplituhedron, but with the external data
transformed by the twist map. This gives a geometric incarnation of a structure
already familiar from parity duality.

\subsubsection{Extendable duality}

Let \(X\subseteq\mathbb P^N_{\mathbb R}\) be a real projective variety, and let
\(S\subseteq X\) be connected, semialgebraic and very compact. We
define the \textit{extendable dual} of \(S\) inside \(X\) by
\begin{tcolorbox}[definitionbox]
\textbf{Extendable polar dual.}
\begin{equation}
\label{eq:extendable_dual_def}
    S_X^\ast
    :=
    X\cap \operatorname{conv}(S)^\ast
    =
    \left\{
    y\in X \, : \, 
     x \cdot y\geq 0
    \ \text{for all } x\in S
    \right\} \, .
\end{equation}
\end{tcolorbox}\noindent
Here \(\operatorname{conv}(S)^\ast\) is the projective dual of the ambient
convex hull of \(S\). We have identified \((\mathbb P^N)^\vee\) with
\(\mathbb P^N\) using a fixed inner product. Thus, unlike ordinary projective
duality, the construction depends on this choice. By the
Tarski--Seidenberg principle, \(S_X^\ast\) is again semialgebraic
\cite[Section~1.4]{bochnak2010real}.

This duality has the expected formal properties inherited from projective
convexity. It is inclusion-reversing:
\begin{equation}
    S\subseteq T
    \qquad\Longrightarrow\qquad
    T_X^\ast\subseteq S_X^\ast \, ,
\end{equation}
and \(S_X^\ast\) is extendably convex in \(X\). Moreover,
\begin{equation}
    \operatorname{conv}_X(S)
    \subseteq
    (S_X^\ast)_X^\ast \, .
\end{equation}
In contrast with ordinary convex duality, this inclusion can be strict even if
\(S\) is already extendably convex. This reflects the fact that, after taking
the dual in the ambient projective space, we intersect again with the
subvariety \(X\).

For the non-negative Grassmannian this construction is especially simple.
Indeed,
\begin{equation}
    \Gr_{\geq 0}(k,n)
    =
    \Gr(k,n)\cap \Delta^{\binom nk-1} \, ,
\end{equation}
and the standard simplex is projectively self-dual.
Therefore \(\Gr_{\geq 0}(k,n)\) is extendably self-dual inside \(\Gr(k,n)\).

\subsubsection{The extendable dual Amplituhedron}

We now apply this construction to tree Amplituhedra. Recall from
Section~\ref{sec:Convexity in the Grassmannian} that 
\begin{equation}
    \operatorname{conv}
    \bigl(\mathcal A_{k,m,n}(Z)\bigr)
    =
    C_{k,m,n}(Z) \, .
\end{equation}
The extendable dual of the Amplituhedron is therefore
\begin{tcolorbox}[definitionbox]
\textbf{Dual Amplituhedron.}
\begin{equation}
\label{eq:dual_amplituhedron_def}
    \widetilde{\mathcal A}_{k,m,n}(Z)
    :=
    \Gr(k,k+m)
    \cap
    C_{k,m,n}(Z)^\ast \, .
\end{equation}
\end{tcolorbox}\noindent
This object was already considered in
~\cite{Positive_geometries,Arkani_Hamed_2018}, in a closely related form.

The dual polytope \(C_{k,m,n}(Z)^\ast\) is cut out by the vertices of
\(C_{k,m,n}(Z)\). Since these vertices are the Plücker points
\begin{equation}
    Z_I=Z_{i_1}\wedge\cdots\wedge Z_{i_k},
    \qquad
    I=\{i_1<\cdots<i_k\}\in\smallbinom{[n]}{k} \, ,
\end{equation}
we have
\begin{equation}
\label{eq:ext_cycl_dual}
    C_{k,m,n}(Z)^\ast
    =
    \left\{
    y\in\mathbb P\!\bigl(\bigwedge^k\mathbb R^{k+m}\bigr):
      y \cdot Z_I  \geq 0
    \quad
    \text{for all } I\in \smallbinom{[n]}{k}
    \right\} \, .
\end{equation}
Using the standard identification between the Plücker pairing and determinants,
we may equivalently regard the dual Amplituhedron as a subset of
\(\Gr(m,k+m)\):
\begin{equation}
\label{eq:dual_amplituhedron_ineqs}
    \widetilde{\mathcal A}_{k,m,n}(Z)
    =
    \left\{
    Y\in\Gr(m,k+m):
    \langle Y\,i_1\cdots i_k\rangle\geq 0
    \quad
    \text{for all } 1\leq i_1<\cdots<i_k\leq n
    \right\} \, .
\end{equation}
For \(k=m=2\), this becomes
\begin{equation}
\label{eq:dual_amplituhedron_ineqs_222}
    \widetilde{\mathcal A}_{2,2,n}(Z)
    =
    \left\{
    Y\in\Gr(2,4):
    \langle Y\,ij\rangle\geq 0
    \quad
    \text{for all } 1\leq i<j\leq n
    \right\} \, .
\end{equation}

\subsubsection{Twist duality for \(k=m=2\)}

The description above is closely related to the twist map. Let
\(Z\) be a positive $(k+m) \times n$ matrix. For each \(i\in[n]\), define
\begin{equation}
\label{eq:twist_map_def}
    W_i
    :=
    Z_{i-m+1}\wedge\cdots\wedge Z_{i-1}
    \wedge
    Z_{i+1}\wedge\cdots\wedge Z_{i+k-1} \, ,
\end{equation}
where we use the usual twisted cyclic convention $Z_{i \pm n}=(-1)^{k-1}Z_i$.
After identifying \(\bigwedge^{k+m-1}\mathbb R^{k+m}\) with
\(\mathbb R^{k+m}\) using the determinant, the vectors \(W_i\) form a
\((k+m)\times n\) matrix $W=(W_1,\ldots,W_n)$.
The resulting map
\begin{equation}
\label{eq:twist_map}
    \tau \, : \, \operatorname{Mat}_{>0}(k+m,n)\longrightarrow
    \operatorname{Mat}_{>0}(k+m,n),
    \qquad
    Z\longmapsto W \, ,
\end{equation}
where $\operatorname{Mat}_{>0}(k+m,n)$ denotes the set of positive $(k+m)\times n$ matrices. The map $\tau$ is the \textit{twist map}, and in fact preserves total positivity~\cite{marsh2016twists}. This and related maps have been studied in total positivity and cluster algebra theory in~\cite{BFZ,GalashinLam,Muller_2017}, and in the amplitudes literature in connection with parity duality~\cite{Arkani_Hamed_2018}.

For \(k=m=2\), the twist is particularly concrete: $W_i$ corresponds to $\bar i = (i-1\,i\,i+1)$.
On the other hand, by facet description in~\eqref{eq:schubert_exterior_cyclic_polytope_def} we deduce~\cite[Proposition~5.11]{mazzucchelli2025exterior}:
\begin{tcolorbox}[resultbox]
\textbf{Twist duality for exterior cyclic polytopes}.
Let \(Z\in\operatorname{Mat}_{>0}(4,n)\), and let \(W=\tau(Z)\). Then
\begin{equation}
\label{eq:schubert_polytope_dual_twist}
    \widetilde C_{2,2,n}(Z)
    =
    C_{2,2,n}(W)^\ast \, .
\end{equation}
\end{tcolorbox}\noindent
The twist is an involution, up to the usual cyclic convention
~\cite{marsh2016twists}. Thus~\eqref{eq:schubert_polytope_dual_twist} says that
the twist intertwines duality between $C_{2,2,n}(Z)$ and $\widetilde{C}_{2,2,n}(Z)$.

Combining this with the intersection theorem
\begin{equation}
    \Gr(2,4)\cap C_{2,2,n}(Z)
    =
    \Gr(2,4)\cap \widetilde C_{2,2,n}(Z)
    =
    \mathcal A_{2,2,n}(Z) \, ,
\end{equation}
we obtain the corresponding statement for Amplituhedra~\cite[Theorem~6.12]{mazzucchelli2025exterior}.

\begin{tcolorbox}[resultbox]
\textbf{Extendable dual Amplituhedron for \(k=m=2\).}
Let \(Z\in\operatorname{Mat}_{>0}(4,n)\), and let \(W=\tau(Z)\). Then
\begin{equation}
\label{eq:dual_mk2}
    \widetilde{\mathcal A}_{2,2,n}(Z)
    =
    \mathcal A_{2,2,n}(W) \, .
\end{equation}
\end{tcolorbox}\noindent
Thus the extendable dual of the \(k=m=2\) Amplituhedron is again an
Amplituhedron, with external data transformed by the twist map. Equivalently, the interior of
\(\widetilde{\mathcal A}_{2,2,n}(Z)\) is the set of
\(Y\in\Gr(2,4)\) satisfying
\begin{equation}
\label{eq:twisted_amplituhedron_sign_conditions}
\begin{aligned}
    &\langle Y\,12\rangle>0,\quad
    \langle Y\,23\rangle>0,\quad
    \ldots,\quad
    \langle Y\,n{-}1\,n\rangle>0,\quad
    \langle Y\,1n\rangle>0 \, ,\\
    &
    \bigl(
    \langle Y\,12\rangle,\,
    \langle Y\,13\rangle,\,
    \ldots,\,
    \langle Y\,1n\rangle
    \bigr)
    \text{ has zero sign flips.}
\end{aligned}
\end{equation}
From the physics point of view, the appearance of the twist is natural: the
same transformation appears in parity duality, which exchanges complementary
helicity sectors of planar \(\mathcal N=4\) SYM amplitudes
\cite[Section~11]{Arkani_Hamed_2018,Herrmann:2020qlt}.

Computations suggest that the same structure is not restricted to \(k=m=2\).
For example, in the case \((k,m,n)=(3,2,6)\), the exterior cyclic polytope
has \(20\) vertices and \(38\) facets, \(20\) of which are Schubert facets.
Keeping only the Schubert inequalities gives a polytope whose f-vactor is
the reverse of that of \(C_{3,2,6}(Z)\), as expected for a polar dual. This
provides evidence that the Schubert exterior cyclic polytope is dual to the
exterior cyclic polytope of the twisted data beyond \(k=m=2\).

\subsubsection{Toward a dual-volume formula}

The previous construction gives a natural candidate for the support of a
dual-volume representation of the Amplituhedron canonical form. In analogy
with the projective-space formula~\eqref{eq:CM_PG_laplace}, one would like an
identity of the form
\begin{equation}
\label{eq:amplituhedron_affine_laplace}
    \Omega_{k,m,n}(y)
    =
    \int_{\mathbb R^N}
    e^{-y \cdot w}
    \,\mathrm{d}\mu_{k,m,n}(w) \, ,
\end{equation}
where \(y\in\widehat{\mathcal A}_{k,m,n}(Z)\), \(N=\binom{k+m}{m}\), and the integral
is taken inthe cone over the Plücker embedding of \(\Gr(k,k+m)\).

If the representing measure is supported on the Grassmannian, then one may
integrate out the overall scale, as in the projective-space case. This leads
to the projective ansatz
\begin{tcolorbox}[definitionbox]
\textbf{Dual-volume ansatz for Amplituhedra.}
\begin{equation}
\label{eq:amplituhedron_grassmannian_laplace}
    \Omega_{k,m,n}(Y)
    =
    \int_{\widetilde{\mathcal A}_{k,m,n}(Z)}
    \frac{
    \prod_{a=1}^{m}\langle W\,\mathrm{d}^k W_a\rangle
    }
    {\langle Y\,W\rangle^{k+m}}
    \,
    \mu_{k,m,n}(W) \, .
\end{equation}
\end{tcolorbox}\noindent
The support has been restricted to
\(\widetilde{\mathcal A}_{k,m,n}(Z)\), since this is the natural region on
which the denominator \(\langle Y\,W\rangle\) has the correct sign for every
\(Y\in\mathcal A_{k,m,n}(Z)\). A formula of this type was proposed in
\cite[Section~7.4.4]{Positive_geometries}. In the spirit of complete
monotonicity, one would like \(\mu_{k,m,n}\) to be non-negative.

At present, this remains an expectation rather than a theorem. Nevertheless,
there is suggestive evidence. In the half-pizza example
~\eqref{eq:naive_half_pizza_integral}, the naive integral with constant measure
does not give the canonical function itself, but its rational part is exactly
the canonical function. A similar phenomenon appears for the non-negative
Grassmannian.

Let \(n=k+m\), and take \(Z\) to be the identity matrix. Then the Amplituhedron
is the non-negative Grassmannian. Computing
~\eqref{eq:amplituhedron_grassmannian_laplace} with constant yields
\begin{tcolorbox}[definitionbox]
\textbf{Naive dual integral for the positive Grassmannian.}
\begin{equation}
\label{eq:integral_dual_positive_grassmannian}
     \int_{\Gr_{\geq 0}(m,k+m)}
     \frac{
     \prod_{a=1}^{m}\langle W\,\mathrm{d}^k W_a\rangle
     }
     {\langle Y,W\rangle^{k+m}}
     =
     b_{m,k}\,
     \Omega_{k,m,k+m}(Y)\,
     \bigl(1-T_{m,k}(Y)\bigr) \, ,
\end{equation}
\end{tcolorbox}\noindent
where \(b_{m,k}\in\mathbb Q\), \(\Omega_{k,m,k+m}\) is the canonical function
of \(\Gr_{\geq 0}(m,k+m)\), and \(T_{m,k}(Y)\) is purely transcendental. Equation~\eqref{eq:integral_dual_positive_grassmannian} is true for \(m=2\) and \(k=2,3,4,5\), when the integral can be evaluated using the
\texttt{Maple} package \texttt{HyperInt}~\cite{Panzer:2014caa}. In these examples,
\begin{equation}
    b_{2,k}=\frac{(-1)^{k+1}}{(k+1)!\,k!} \, .
\end{equation}
For \(k=2,3\), the transcendental terms can be written in dihedral coordinates
on \(\mathcal M_{0,k+2}\). The relevant point for us is that these terms
behave like moduli-space periods, while the rational part is the canonical
function.

This is the same pattern seen for curved positive geometries: a naive
dual-volume integral produces a rational canonical part together with a
transcendental correction. The corresponding measure-theoretic problem is one
of the main open directions collected at the end of the chapter.

The next section studies a different but related source of transcendental
functions: canonical pairings of positive geometries. These pairings, known in
the simplex case as Aomoto forms, provide a controlled setting in which
canonical forms, duality, positivity and polylogarithmic functions can be
studied simultaneously.

\subsection{Aomoto forms}
\label{sec:Aomoto Forms}

In the previous sections we studied canonical rational
differential forms, and emphasized their interpretation as volumes of dual
objects. This is the geometric picture appropriate to integrands, or to
tree-level amplitudes. Physical observables, however, are obtained after
integration, and the resulting functions are no longer rational in general.
Already at one loop one encounters logarithms and dilogarithms; at higher
orders one finds iterated integrals such as Goncharov polylogarithms and
harmonic polylogarithms~\cite{Goncharov2001,Remiddi:1999ew,Brown:2009qja},
and more generally periods of algebraic varieties
~\cite{Brown:2009qja,Bloch:2005bh,Panzer2015}.

Aomoto forms provide one of the simplest geometric mechanisms by which
rational canonical forms give rise to transcendental functions. They arise by
pairing two positive geometries: one supplies the domain of integration, the
other supplies the canonical form. In the linear case, namely for pairs of
projective polytopes, these pairings reduce to iterated logarithmic integrals.
They include classical polylogarithms and Goncharov polylogarithms, and they retain many structural features of canonical forms: boundary
recursion, triangulation additivity, duality, and positivity.

\subsubsection{Canonical pairings}

Let \(A\) and \(B\) be full-dimensional positive geometries in
\(\mathbb P^m\). We say that the pair \((A,B)\) is \textit{admissible} if the
domain \(A\) does not meet the algebraic boundary \(\partial_a B\) in a way
that produces a non-integrable singularity of \(\mathbf\Omega_B\). For
projective polytopes this means, in particular, that no face of \(A\) lies
entirely on a boundary hyperplane of \(B\).

For an admissible pair, we define the canonical pairing by
\begin{tcolorbox}[definitionbox]
\textbf{Canonical pairing.}
\begin{equation}
    \mathcal I(A,B)
    :=
    \int_A \mathbf\Omega_B \, .
    \label{eq:can_pair}
\end{equation}
\end{tcolorbox}\noindent
In the context of positive geometries this pairing was introduced in
~\cite{Positive_geometries}. When \(A\) and \(B\) are projective simplices, the
functions \(\mathcal I(A,B)\) are the \textit{Aomoto polylogarithms}, or
\textit{Aomoto forms}~\cite{Aomoto1982,Arkani-Hamed:2017ahv}. Aomoto studied
these functions in the early 1980s in connection with hyperlogarithms,
hypergeometric integrals, and addition theorems.

The canonical pairing has three elementary properties. First, reversing the
orientation of either \(A\) or \(B\) changes the sign of \(\mathcal I(A,B)\).
Second, it is projectively covariant:
\begin{equation}
    \mathcal I(g\cdot A,B)
    =
    \mathcal I(A,g^{-1}\cdot B) \, ,
    \qquad
    g\in {\rm PGL}(m+1) \, .
    \label{eq:aom_equiv}
\end{equation}
In particular, it is invariant under simultaneous projective transformations:
\begin{equation}
    \mathcal I(g\cdot A,g\cdot B)=\mathcal I(A,B) \, .
\end{equation}
Third, it is compatible with triangulations. If \(\{A^{(i)}\}\) triangulates
\(A\) and \(\{B^{(j)}\}\) is a canonical-form triangulation of \(B\), then
\begin{equation}
    \mathcal I(A,B)
    =
    \sum_{i,j}
    \mathcal I(A^{(i)},B^{(j)}) \, .
\end{equation}
Thus, for polytopes, one may reduce the study of canonical pairings to the
case where \(A\) and \(B\) are simplices.

\subsubsection{The simplex case}

Let \(A\) and \(B\) be two \(m\)- dimensional simplices in \(\mathbb P^m\). We denote by
\(A_i\) and \(B_i\), for \(i=0,\ldots,m\), the vertices of the two simplices,
and also write \(A\) and \(B\) for the corresponding \((m+1)\times(m+1)\)
matrices. In homogeneous coordinates \(x\), the canonical form of the simplex
\(B\) can be written as a logarithmic form,
\begin{equation}
    \mathbf\Omega_B(x)
    =
    \mathrm{d}\log
    \left(
    \frac{\langle x\,B_2\cdots B_m B_0\rangle}
         {\langle x\,B_1\cdots B_m\rangle}
    \right)
    \wedge
    \cdots
    \wedge
    \mathrm{d}\log
    \left(
    \frac{\langle x\,B_0\cdots B_{m-1}\rangle}
         {\langle x\,B_1\cdots B_m\rangle}
    \right),
    \label{eq:form_sympl_dlog}
\end{equation}
or equivalently by choosing a different distinguished vertex. Hence
\(\mathcal I(A,B)\) is an iterated logarithmic integral of weight \(m\). In
dimension one it gives logarithms, in dimension two dilogarithms, and in
higher dimensions higher polylogarithms.

\begin{eg}[$m=1$]
	The simplest example is the pairing of two intervals in \(\mathbb P^1\). Let
\(B=[b_0,b_1]\). In an affine chart \([1:x]\), its canonical form is
\begin{equation}
    \mathbf\Omega_B(x)
    =
    \frac{b_1-b_0}{(x-b_0)(b_1-x)}\,\mathrm{d}x
    =
    \mathrm{d}\log\left(\frac{x-b_0}{b_1-x}\right).
    \label{eq:IntervalCanonicalForm}
\end{equation}
If \(A=[a_0,a_1]\) is admissible with respect to \(B\), then
\begin{equation}
    \mathcal I_1(A,B)
    =
    \int_{a_0}^{a_1}
    \frac{b_1-b_0}{(x-b_0)(b_1-x)}\,\mathrm{d}x
    =
    \log
    \frac{(a_1-b_0)(b_1-a_0)}
         {(b_1-a_1)(a_0-b_0)} .
    \label{eq:aomoto_log}
\end{equation}
Thus the first Aomoto form is the logarithm of the cross-ratio of the four
endpoints. Its branch points occur when an endpoint of \(A\) collides with an
endpoint of \(B\).

\end{eg}

This behavior persists in higher dimension. The singularities of Aomoto forms
are governed by linear dependencies among the columns of the
\((m+1)\times 2(m+1)\) matrix \([A\mid B]\). For two index sets
\(I,J \subseteq \{0,\ldots,m\}\) with
\(|I|+|J|=m+1\), we write $\langle I \mid J \rangle$ for the determinant of the minor of \([A \mid B]\) formed by the columns
\(I\) of \(A\) and \(J\) of \(B\).

\subsubsection{Differential equations and symbols}

An important feature of Aomoto forms is that they satisfy recursive
differential equations. Differentiating with respect to a vertex of one
simplex, meaning with respect to the coefficients of the vector defining the vertex, lowers the transcendental weight by one and produces lower-dimensional Aomoto forms~\cite{Aomoto1982,Arkani-Hamed:2017ahv}. 

\begin{tcolorbox}[resultbox]
\textbf{Aomoto differential equation.}
Let \(A\) and \(B\) be admissible \(m\)-simplices in \(\mathbb P^m\). Then the
differential of the Aomoto form with respect to a vertex \(B_i\) of \(B\) is
\begin{equation}
    \mathrm{d}_{B_i} \, \mathcal I_m(A,B)
    =
    m
    \sum_{j=0}^{m}
    \mathcal I_{m-1}\bigl(A^{(i,j)},B^{(i)}\bigr)\,
    \mathrm{d}_{B_i}\log \langle \bar j\mid i\rangle \, .
    \label{eq:AomotoDEschematic}
\end{equation}
Here \(\bar j=\{0,\ldots,m\}\setminus\{j\}\). The simplex \(B^{(i)}\) is the
projection of \(B\) away from \(B_i\), while \(A^{(i,j)}\) is obtained by
projecting the lines \(B_iB_k\), for \(k\neq i\), onto the hyperplane spanned
by \(A_0,\ldots,\widehat A_j,\ldots,A_m\).
\end{tcolorbox}\noindent
Equation~\eqref{eq:AomotoDEschematic} is characteristic of a pure-weight
function: differentiation lowers the transcendental weight by one. Repeated
application gives the \emph{symbol} of the Aomoto form. We will review the symbol in Section~\ref{sec:Symbols and Bootstrap}; we now just recall the relevant formula for Aomoto forms. Schematically, replacing in~\eqref{eq:AomotoDEschematic} iterated logarithmic differentials by tensor products, one obtains
\begin{tcolorbox}[resultbox]
\textbf{Aomoto symbol.}
\begin{equation}
\begin{aligned}
    \mathcal S\bigl(\mathcal I_m(A,B)\bigr)
    =
    (-1)^m
    \sum_{\rho,\sigma\in S_{m+1}}
    {\rm sgn}(\rho)\,{\rm sgn}(\sigma)
    \bigotimes_{i=0}^{m-1}
    \langle \rho[0,i]\mid \sigma[i,m]\rangle \, .
\end{aligned}
\label{eq:aom_symbol}
\end{equation}
\end{tcolorbox}\noindent
Here \(S_{m+1}\) is the permutation group on $m+1$ elements, and we abbreviated
\(\rho[0,i]=(\rho(0),\ldots,\rho(i))\),
\(\sigma[i,m]=(\sigma(i),\ldots,\sigma(m))\). The symbol records the
successive logarithmic singularities of the iterated integral. It does not
determine the function uniquely, but it packages its branch structure in a
combinatorial way.

For \(m=1\), this reduces to the symbol of the logarithm in
~\eqref{eq:aomoto_log}. If
\begin{equation}
    [A\mid B]
    =
    \begin{pmatrix}
        1 & 1 & 1 & 1\\
        a_0 & a_1 & b_0 & b_1
    \end{pmatrix} \, ,
\end{equation}
then
\begin{equation}
    \mathcal S\bigl(\mathcal I_1(A,B)\bigr)
    =
    -\langle 0\mid0\rangle
    -\langle 1\mid1\rangle
    +\langle 0\mid1\rangle
    +\langle 1\mid0\rangle \, ,
\end{equation}
which is the symbol of the logarithm of the cross-ratio in~\eqref{eq:aomoto_log}.

\subsubsection{Duality and the antipode}

Aomoto forms also possess a duality inherited from the dual-volume
representation of canonical forms. Suppose \(A\subseteq B\) are projective
polytopes. Using the dual-volume formula for \(\mathbf\Omega_B\), we obtain the following relation:
\begin{tcolorbox}[resultbox]

\textbf{Aomoto duality.}
\begin{equation}
    \mathcal I(A,B)
    =
    \int_A\int_{B^\ast}
    \frac{\langle x\,\mathrm{d}^m x\rangle\,\langle y\,\mathrm{d}^m y\rangle}
         {\langle x,y\rangle^{m+1}}
    =
    \mathcal I(B^\ast,A^\ast).
    \label{eq:aomoto_duality}
\end{equation}
\end{tcolorbox}\noindent
In the second equality we exchanged the order of integration and used
biduality for convex polytopes. At the level of symbols, this duality is
related to the \textit{antipode}: the operation that reverses the order of the
letters in each tensor word. This type of symmetry also appears in
planar \(\mathcal N=4\) SYM, for instance in relations between form-factor and
amplitude symbols~\cite{DixonGurdoganMcLeodWilhelm2022,DixonGurdoganLiuMcLeodWilhelm2023}.

\subsubsection{Goncharov polylogarithms as Aomoto forms}

An important class of iterated integrals appearing in perturbative physics is
given by Goncharov polylogarithms. For \(a_1,\ldots,a_m\in\mathbb C\) and
\(z\in\mathbb C\setminus\{a_i\}\), they are defined recursively:
\begin{equation}
    G(a_1,\ldots,a_m;z)
    :=
    \int_0^z
    \frac{\mathrm{d}t}{t-a_m}\,
    G(a_1,\ldots,a_{m-1};t),
    \label{Goncharov_polylogs}
\end{equation}
with
\begin{equation}
    G(a_1;z)
    =
    \int_0^z \frac{\mathrm{d}t}{t-a_1}
    =
    \log\left(\frac{a_1-z}{a_1}\right).
    \label{G_a1_z}
\end{equation}
The classical polylogarithms are special cases:
\begin{equation}
    {\rm Li}_m(z)
    =
    -G(1,0,\ldots,0;z).
    \label{eq:pol_to_gonch}
\end{equation}

For real parameters satisfying \(0<z<\min(a_i)\), one can rewrite \eqref{Goncharov_polylogs} using the simplex
\begin{equation}
    \delta^{(m)}(z)
    =
    \{0<t_1<\cdots<t_m<z\}\subseteq\mathbb R^m \, ,
\end{equation}
as the following integral:
\begin{equation}
    G(a_1,\ldots,a_m;z)
    =
    \int_{\delta^{(m)}(z)}
    \prod_{i=1}^m
    \frac{\mathrm{d}t_i}{t_i-a_i} \, .
    \label{G_in_deltaz}
\end{equation}
After the change of variables
\begin{equation}
    t_i
    =
    z\,\frac{x_1+\cdots+x_i}{x_1+\cdots+x_m} \, ,
    \qquad x_i>0 \, ,
\end{equation}
this becomes
\begin{equation}
    G(a_1,\ldots,a_m;z)
    =
    \int_{\mathbb R^m_{>0}}
    \prod_{i=1}^m
    \frac{(-1)^m z}
    {-(x_1+\cdots+x_i)z+(x_1+\cdots+x_m)a_i}
    \,\mathrm{d}^m x \, .
    \label{eq:gonch}
\end{equation}
Thus Goncharov polylogarithms are Aomoto forms.

\begin{tcolorbox}[resultbox]
\textbf{Goncharov polylogarithms as Aomoto forms.}
\begin{equation}
    G(a_1,\ldots,a_m;z)
    =
    \mathcal I\bigl(\Delta^m,\Delta^m(a_1,\ldots,a_m;z)^\ast\bigr)
    =
    (-1)^m
    \mathcal I\bigl(\Delta^m(a_1,\ldots,a_m;z),\Delta^m\bigr) \, .
    \label{Goncharov_pairing}
\end{equation}
\end{tcolorbox}\noindent
Here \(\Delta^m(a_1,\ldots,a_m;z)\) is the \textit{Goncharov simplex}. It can
be constructed recursively from 
\begin{equation}
    \Delta^1(a_1;z)=[a_1-z,a_1]
\end{equation}
as
\begin{equation}
    \Delta^m(a_1,\ldots,a_m;z)
    =
    \bigcup_{0\leq t\leq z}
    \{a_m-t\}\times
    \Delta^{m-1}(a_1,\ldots,a_{m-1};t) \, .
\end{equation}
For the special choice \((a_1,\ldots,a_m)=(1,0,\ldots,0)\), this gives the
usual polylogarithmic simplex representing \({\rm Li}_m(z)\).

\subsubsection{Complete and absolute monotonicity}

The representation above also gives positivity properties of Goncharov
polylogarithms. We use the following extension of complete monotonicity. Let
\(U\subseteq\mathbb R^n\) be an open domain and let \(C\subseteq\mathbb R^n\) be an
open convex cone such that \(U+C\subseteq U\). A smooth function \(f:U\to
\mathbb R\) is completely monotone with respect to \(C\) if
\begin{equation}
    (-1)^r D_{v_1}\cdots D_{v_r}f(x)\geq 0
\end{equation}
for all \(x\in U\), \(v_i\in C\), and \(r \in \mathbb{N}\). It is absolutely monotone if
the same inequalities hold without the factor \((-1)^r\). This is a slightly more general definition than the one in Section~\ref{sec:Completely Monotone Positive Geometries}.

\begin{tcolorbox}[resultbox]
\textbf{Complete monotonicity of real Goncharov polylogarithms.}
For \(0<z<\min(a_1,\ldots,a_m)\), the function
\begin{equation}
    (-1)^m G(a_1,\ldots,a_m;z)
\end{equation}
is completely monotone in the variables \(a_1,\ldots,a_m\), and absolutely
monotone in \(z\).
\end{tcolorbox}\noindent
Indeed, using
\begin{equation}
    \frac{1}{a_i-t_i}
    =
    \int_0^\infty e^{-y_i(a_i-t_i)}\,\mathrm{d}y_i \, ,
    \qquad a_i>t_i \, ,
\end{equation}
one obtains
\begin{equation}
    (-1)^mG(a;z)
    =
    \int_{\mathbb R^m_{>0}}
    e^{- y \cdot a}\,
    \mu(y;z)\,\mathrm{d}^m y \, ,
    \label{G_measure}
\end{equation}
where
\begin{equation}
    \mu(y;z)
    =
    \int_{\delta^{(m)}(z)}
    e^{y \cdot t}\,\mathrm{d}^m t
    =
    \int_{\delta^{(m)}(1)}
    z^m e^{z \, y \cdot t}\,\mathrm{d}^m t \, .
    \label{mu}
\end{equation}
The density \(\mu(y;z)\) is positive, so~\eqref{G_measure} is a Laplace
representation in the \(a_i\)'s. Moreover, all derivatives of
\(\mu(y;z)\) with respect to \(z\) are positive for \(z>0\), giving absolute
monotonicity in \(z\). In particular, the classical polylogarithms
\({\rm Li}_m(z)\) are absolutely monotone on \(0<z<1\), as is also evident
from their series expansion
\begin{equation}
    {\rm Li}_m(z)=\sum_{k=1}^{\infty}\frac{z^k}{k^m} \, .
\end{equation}

\begin{figure}[pos=t]
\centering
\includegraphics[width=0.5\linewidth]{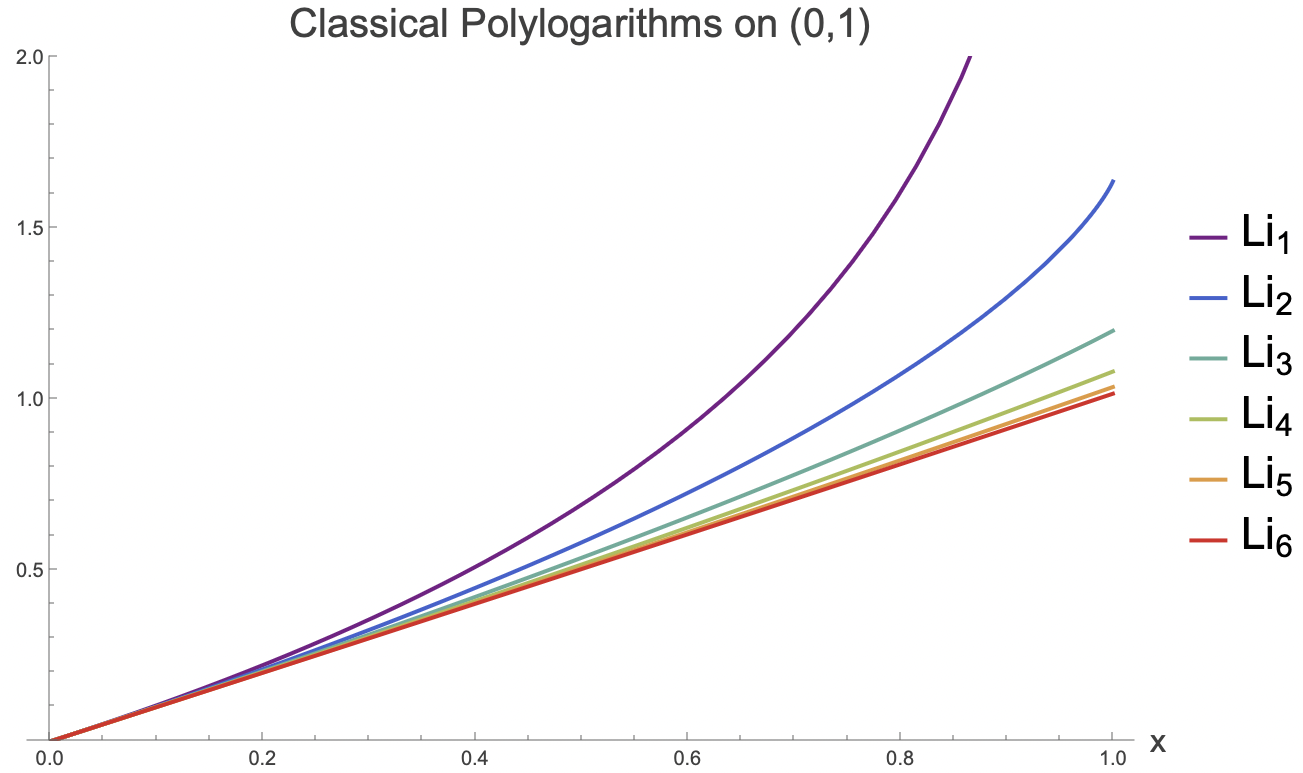}
\caption{The classical polylogarithms \({\rm Li}_m(z)\) for
\(z\in(0,1)\) and \(m=1,\ldots,6\). They are absolutely monotone functions of~\(z\).}
\label{fig:Polylogs_plot}
\end{figure}

\subsubsection{Explicit formulas from Charlton--Gangl--Radchenko}

The symbol~\eqref{eq:aom_symbol} does not by itself give a function.
Remarkably, Charlton--Gangl--Radchenko gave an explicit expression for any
Aomoto form in terms of Goncharov polylogarithms~{\cite[Theorem~7 and Remark~8]{CGR}}. We recall their
formula in our notation.

For subsets \(I,J\subseteq\{1,\ldots,m\}\) with
\(|I|=|J|\), define
\begin{equation}
    X_{I,J}
    :=
    \frac{
    \langle I \mid \{0\}\cup\overline{J}\rangle
    }{
    \langle \{0\}\cup I \mid \overline{J}\rangle
    } \, ,
    \label{eq:invariant-X}
\end{equation}
where \(\overline{J}:=\{1,\ldots,m\}\setminus J\). We have
chosen distinguished vertices \(A_0\) and \(B_0\).

\begin{tcolorbox}[resultbox]
\textbf{Aomoto form in Goncharov polylogs.} The Aomoto form of two admissible \(m\)-simplices can be written as
\begin{equation}
    \mathcal I_m(A,B)
    =
    (-1)^m
    \sum_{\rho,\sigma\in S_m}
    {\rm sgn}(\rho)\,{\rm sgn}(\sigma)\,
    G\!\left(
    X_{\rho[m],\sigma[m]},
    X_{\rho[m-1],\sigma[m-1]},
    \ldots,
    X_{\rho[1],\sigma[1]};
    1
    \right),
    \label{eq:invariant-G}
\end{equation}
where \(\rho[r]=(\rho(1),\ldots,\rho(r))\), and similarly for \(\sigma[r]\).
\end{tcolorbox}\noindent
This is an equality of functions, not merely of symbols. Its symbol is
\begin{tcolorbox}[resultbox]
\textbf{Symbol formula.}
\begin{equation}
    \mathcal S\bigl(\mathcal I_m(A,B)\bigr)
    =
    (-1)^m
    \sum_{\rho,\sigma \, \in \,  S_m}
    {\rm sgn}(\rho)\,{\rm sgn}(\sigma)\,
    X_{\rho[1],\sigma[1]}
    \otimes
    X_{\rho[2],\sigma[2]}
    \otimes
    \cdots
    \otimes
    X_{\rho[m],\sigma[m]} \, .
    \label{eq:invariant-symbol}
\end{equation}
\end{tcolorbox}\noindent

\subsubsection{The Aomoto moduli space and the Euclidean region}

The projective covariance~\eqref{eq:aom_equiv} implies that
\(\mathcal I(A,B)\) depends only on the equivalence class of the matrix
\([A\mid B]\) in \(\Gr(m+1,2m+2)\), together with the ordering of the two sets
of \(m+1\) vertices. Since each vertex is projective, the pairing is also
invariant under rescaling the columns. Thus the natural moduli space is a
decorated version of
\begin{equation}
    X(m+1,2m+2)
    :=
    \Gr(m+1,2m+2)/(\mathbb C^\ast)^{2m+1} \, .
\end{equation}
Quotienting further by permutations of the vertices of the two simplices gives the
\begin{tcolorbox}[definitionbox]
\textbf{Aomoto moduli space.}
\begin{equation}
    X(m+1,2m+2)/(S_{m+1}\times S_{m+1}) \, .
    \label{eq:Aomoto_mod}
\end{equation}
\end{tcolorbox}\noindent
Its dimension is
\begin{equation}\label{eq:aom_mod_sp}
    \dim X(m+1,2m+2)
    =
    (m+1)^2-(2m+1)
    =
    m^2 \, .
\end{equation}

Using projective covariance, we may set \(B=\Delta^m\), or equivalently
\([A\mid B]=[A\mid \mathbf 1_{m+1}]\). The condition \(A\subseteq B\) becomes
componentwise non-negativity of \(A\). Inside this region there is a
distinguished open part where \(A\) is totally positive, namely where all
minors of \(A\) are positive. Equivalently,
\begin{equation}
    A \text{ is totally positive}
    \quad\Longleftrightarrow\quad
    [A\mid D]\in \Gr_{>0}(m+1,2m+2),
    \label{eq:equiv}
\end{equation}
where \(D=\operatorname{diag}(1,-1,1,\ldots,(-1)^m)\). We call this the
\textit{Euclidean region} of~\eqref{eq:aom_mod_sp}.

\begin{tcolorbox}[resultbox]
\textbf{Regularity in the Euclidean region.}
The map
\begin{equation}
    A\longmapsto \mathcal I(A,\mathbf 1_{m+1})
\end{equation}
is regular on the region of totally positive matrices \(A\).
\end{tcolorbox}\noindent

A convenient set of local coordinates is obtained by rescaling rows and
columns so that
\begin{equation}
    A
    =
    \begin{pmatrix}
    1 & 1 & 1 & \cdots & 1\\
    1 & u_{11} & u_{12} & \cdots & u_{1m}\\
    \vdots & \vdots & \vdots & \ddots & \vdots\\
    1 & u_{m1} & u_{m2} & \cdots & u_{mm}
    \end{pmatrix} \, .
    \label{eq:Au}
\end{equation}
The \(m^2\) parameters \(u_{ij}\) are invariant cross-ratios:
\begin{equation}
    u_{ij}
    =
    \frac{
    \det(B_{i\leftarrow A_j})\,
    \det(B_{0\leftarrow A_0})
    }{
    \det(B_{i\leftarrow A_0})\,
    \det(B_{0\leftarrow A_j})
    } \, ,
    \qquad
    1\leq i,j\leq m \, ,
    \label{eq:AomotoBracketCrossRatios}
\end{equation}
where \(B_{i\leftarrow A_j}\) denotes the matrix obtained from \(B\) by
replacing its \(i\)-th column by \(A_j\).

For \(B=\mathbf 1_{m+1}\), the cross-ratios \(X_{I,J}\) in
~\eqref{eq:invariant-X} reduce to \emph{flag-minor} ratios
\begin{equation}
    x_{I,J}
    :=
    \frac{\Delta_{I,J}(A)}
         {\Delta_{\{0\}\cup I,\{0\}\cup J}(A)} \,  ,
    \label{eq:flag-x}
\end{equation}
where \(\Delta_{I,J}(A)\) denotes the minor of \(A\) with row set \(I\) and
column set \(J\). Hence
\begin{equation}
    \mathcal I_m(A,\mathbf 1_{m+1})
    =
    (-1)^m
    \sum_{\rho,\sigma\in S_m}
    {\rm sgn}(\rho)\,{\rm sgn}(\sigma)\,
    G\!\left(
    x_{\rho[m],\sigma[m]},
    x_{\rho[m-1],\sigma[m-1]},
    \ldots,
    x_{\rho[1],\sigma[1]};
    1
    \right).
    \label{eq:general-flag-G}
\end{equation}
Thus Aomoto forms are sums of Goncharov polylogarithms whose letters are
ratios of nested minors. These are cluster-type \(\mathcal{X}\)-coordinates in
positive charts, in agreement with the cluster-polylogarithmic perspective of
Matveiakin and Rudenko~\cite{MR}.

For \(m=2\), there are four rank-one letters and one rank-two letter:
\begin{equation}
    x_{i,j}=\frac{u_{ij}}{u_{ij}-1},
    \qquad
    x_{12,12}
    =
    \frac{u_{11}u_{22}-u_{12}u_{21}}
    {u_{11}u_{22}-u_{12}u_{21}-u_{11}+u_{12}+u_{21}-u_{22}} \, .
\end{equation}
Formula~\eqref{eq:general-flag-G} becomes
\begin{equation}
\begin{aligned}
    \mathcal I_2
    =
    &\,G(x_{12,12},x_{1,1};1)
    -G(x_{12,12},x_{1,2};1)  -G(x_{12,12},x_{2,1};1)
    +G(x_{12,12},x_{2,2};1) \, .
\end{aligned}
\label{eq:m2-G}
\end{equation}
This is a compact dilogarithmic expression for the two-dimensional Aomoto
form.

\subsubsection{Complete monotonicity in positive coordinates}

The coordinates \(u_{ij}\) are natural for the moduli space, but they are not
adapted to complete monotonicity. Already for \(m=1\), the logarithm
\(\log(u_{11})\) is not completely monotone in \(u_{11}\), while it becomes
completely monotone after an inverse change of variables. We now introduce
coordinates in which complete monotonicity is manifest.

The building blocks are neighboring shifts. For \(i=1,\ldots,m\), set
\begin{equation}
\begin{aligned}
    S_i^{(t)}
    &=
    \mathbf 1_{m+1}+tE_{i,i+1},
    \\
    \overline S_i^{(t)}
    &=
    \mathbf 1_{m+1}+tE_{i+1,i},
\end{aligned}
\label{eq:ngh_shift}
\end{equation}
where \(E_{a,b}\) is the elementary matrix. If \(M\) has non-negative minors,
then so do \(M \cdot S_i^{(t)}\) and \(M \cdot \overline S_i^{(t)}\) for \(t>0\). These are
the elementary positive unipotent transformations used in parametrizations of
the totally non-negative Grassmannian~\cite{Postnikov:2006kva}.

Suppose \(A\subseteq\Delta^m\), and let \(A(t)=A \cdot S_i^{(t)}\) remain inside
\(\Delta^m\) for \(t>0\). Using \(\det S_i^{(t)}=1\) and the duality
~\eqref{eq:aomoto_duality}:\begin{equation}
\begin{aligned}
    \mathcal I(A(t),\Delta^m)
    &=
    \int_A\int_{\Delta^m}
    \frac{\langle x\,\mathrm{d}^m x\rangle\,\langle y\,\mathrm{d}^m y\rangle}
    {\langle S_i^{(t)}x,y\rangle^{m+1}}        =
    \int_A\int_{\Delta^m}
    \frac{\langle x\,\mathrm{d}^m x\rangle\,\langle y\,\mathrm{d}^m y\rangle}
    {(\langle x,y\rangle+t\langle E_{i,i+1}x,y\rangle)^{m+1}} .
\end{aligned}
\label{eq:It}
\end{equation}
Since \(A\subseteq\Delta^m\), the factor
\(\langle E_{i,i+1}x,y\rangle=x_{i+1}y_i\) is non-negative. Differentiating
under the integral sign gives the desired sign pattern.

\begin{tcolorbox}[resultbox]
\textbf{Complete monotonicity under neighboring shifts.}
If \(A(t)=A  \cdot S_i^{(t)}\subseteq\Delta^m\) for all \(t>0\), then
\begin{equation}
    t\longmapsto \mathcal I(A(t),\Delta^m)
\end{equation}
is completely monotone on \((0,\infty)\). The same statement holds for the shifts \(\overline S_i^{(t)}\).
\end{tcolorbox}\noindent

This suggests parametrizing the Euclidean region by shifting only \(A\). Let
\begin{equation}
    I_m=(1,2,\ldots,m,1,2,\ldots,m-1,\ldots,1,2,1)
\end{equation}
be the reduced word of length \(N=m(m+1)/2\). Write
\(I_m=(i_1,\ldots,i_N)\), and define
\begin{equation}
    A_m(u,v)
    :=
    \overline S_{i_1}^{(v_1)}
    \cdots
    \overline S_{i_N}^{(v_N)}
    \mathbf 1_{m+1}
    S_{i_N}^{(u_N)}
    \cdots
    S_{i_1}^{(u_1)} .
    \label{eq:Auv}
\end{equation}
All minors of \(A_m(u,v)\) are subtraction-free polynomials in the positive
variables \(u_i,v_i\). There is an \(m\)-dimensional redundancy coming from
diagonal conjugation. We fix 
\begin{equation}
    u_1=\cdots=u_m=1 \, .
    \label{eq:u_gauge}
\end{equation}
This leaves \(m^2\) positive parameters, matching the dimension of the Aomoto
moduli space.

\begin{tcolorbox}[resultbox]
\textbf{Completely monotone coordinates for Aomoto forms.}
Let \(A_m(u,v)\) be defined by~\eqref{eq:Auv}, with the gauge
~\eqref{eq:u_gauge}. Then
\begin{equation}
    (\bar u,v)
    \longmapsto
    \mathcal I_m\bigl(A_m(u,v),\mathbf 1_{m+1}\bigr)
\end{equation}
is completely monotone on \(\mathbb R_{>0}^{m^2}\), where
\(\bar u=(u_{m+1},\ldots,u_N)\) and $u=(1,\dots,1,\bar u)$.
\end{tcolorbox}\noindent

For instance, for \(m=1\) and \(m=2\) one finds
\begin{equation}
\begin{aligned}
A_1
&=
\begin{pmatrix}
1+v_1 & 1\\
v_1 & 1
\end{pmatrix},
\\[2mm]
A_2
&=
\begin{pmatrix}
1 & 1+u_3 & u_3\\
v_1+v_3 &
1+(v_1+v_3)(1+u_3) &
1+u_3(v_1+v_3)\\
v_2v_3 &
v_2\bigl(1+v_3(1+u_3)\bigr) &
1+v_2\bigl(1+v_3u_3\bigr)
\end{pmatrix}.
\end{aligned}
\label{eq:pars_Aom}
\end{equation}
For \(m=1\), the Aomoto logarithm becomes
\begin{equation}
    \mathcal I_1(v_1)=\log\left(1+\frac1{v_1}\right) \, ,
\end{equation}
which is completely monotone for \(v_1>0\).

\subsubsection{Canonical volume pairings}

We end with a speculative extension. The pairing \(\mathcal I(A,B)\) is most
natural for polytopes, where canonical forms have ordinary dual-volume
representations with constant measure. For nonlinear completely monotone
positive geometries, the dual-volume representation involves a non-trivial
measure. If one wants the pairing to be compatible with duality, then the
measure should be included in the definition.

Suppose that \(A\) is a completely monotone positive geometry in
\(\mathbb P^m\), with dual-volume representation of its canonical function given by
\begin{equation}
    \Omega_A(x)
    =
    \int_{A^\ast}
    \mu_A(y)\,
    \frac{\langle y\,\mathrm{d}^m y\rangle}
         {\langle x,y\rangle^{m+1}} \, .
    \label{eq:dual_vol_aomoto}
\end{equation}
If \(B\) is admissible with respect to \(A^\ast\), define the
\textit{canonical volume pairing} by
\begin{tcolorbox}[definitionbox]
\textbf{Canonical volume pairing.}
\begin{equation}
    \mathcal J(A^\ast,B)
    :=
    \int_{A^\ast}
    \mu_A(x)\,\mathbf\Omega_B(x) \, .
    \label{eq:can_vol_par}
\end{equation}
\end{tcolorbox}\noindent

If \(B\) also admits a dual-volume representation and \(A^\ast\subseteq B\), then
\begin{tcolorbox}[resultbox]
\textbf{Duality for canonical volume pairings.}
\begin{equation}
    \mathcal J(A^\ast,B)
    =
    \int_{A^\ast}\int_{B^\ast}
    \mu_A(x)\mu_B(y)
    \frac{\langle x\,\mathrm{d}^m x\rangle\,\langle y\,\mathrm{d}^m y\rangle}
         {\langle x,y\rangle^{m+1}}
    =
    \mathcal J(B^\ast,A) \, .
    \label{eq:volume_pairing_duality}
\end{equation}
\end{tcolorbox}\noindent
For polytopes, where the measures are characteristic functions, this reduces
to the usual Aomoto duality~\eqref{eq:aomoto_duality}.

\begin{figure}[pos=t]
\centering
\begin{minipage}[c]{0.43\linewidth}
\centering
\includegraphics[width=\linewidth]{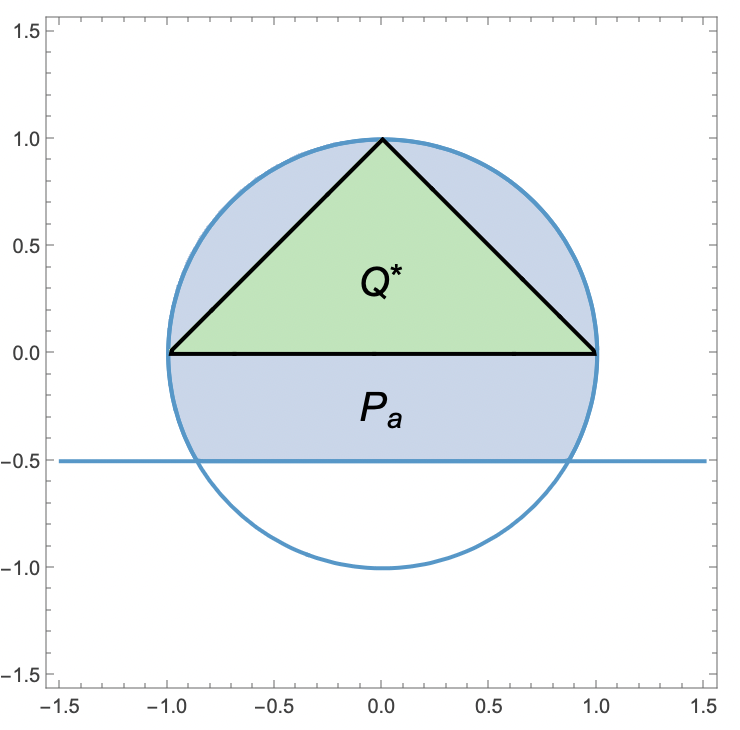}
\end{minipage}
\hspace{0.5in}
\begin{minipage}[c]{0.43\linewidth}
\centering
\includegraphics[width=\linewidth]{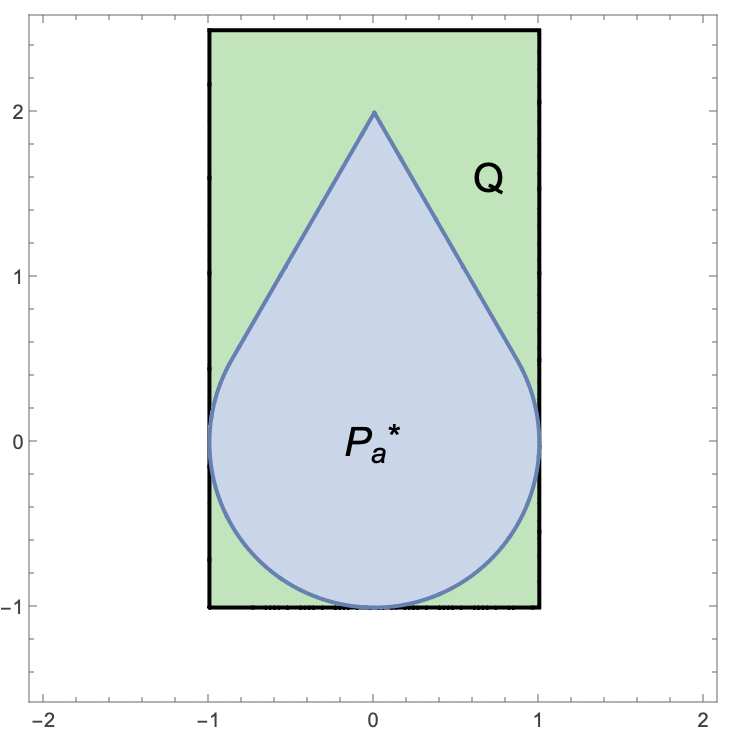}
\end{minipage}%
\caption{On the left: the half-pizza \(P_a\) for \(a=1/2\), shown in blue,
together with the triangle \(Q^\ast\), shown in green. On the right: the dual
picture. The dual \(P_a^\ast\) is bounded by the dual conic together with the
horn coming from the line component of the boundary of \(P_a\), while \(Q\)
contains \(P_a^\ast\). This is the configuration used in~\eqref{eq:J_ex}.}
\label{fig:half_pizza_pairing}
\end{figure}

A simple nonlinear example comes from the half-pizza \(P_a\) of
Sections~\ref{sec:Completely Monotone Positive Geometries} and~\ref{sec:Transcendental Measures}. Let \(Q\) be a triangle containing
\(P_a^\ast\), so that \(Q^\ast\subseteq P_a\). Since \(Q\) is a polytope,
\(\mu_Q=\chi_{Q^\ast}\). Thus
\begin{equation}\label{eq:J_ex}
    \mathcal J(P_a^\ast,Q)
    =
    \int_{P_a^\ast}\mu_a(y)\,\mathbf\Omega_Q(y)
    =
    \int_{Q^\ast}\mathbf\Omega_a(x) \, .
\end{equation}
A convenient choice of \(Q^\ast\) is for example the triangle with vertices
\([1:\pm1:0]\) and \([1:0:1]\), see Figure~\ref{fig:half_pizza_pairing}. Then one obtains
\begin{equation}
    J(a)
    =
    2\int_0^1 \mathrm{d}x_2
    \int_0^{1-x_2} \mathrm{d}x_1\,
    \frac{2\sqrt{1-a^2}}
    {(x_2+a)(1-x_1^2-x_2^2)} \, .
    \label{eq:J_1}
\end{equation}
After the change of variables $ x_2=\frac{1-t^2}{1+t^2}$,
this becomes
\begin{equation}
    J(a)
    =
    \int_0^1
    \frac{4\sqrt{1-a^2}}
    {(a+1)+(a-1)t^2}
    \log\left(\frac{1+t}{1-t}\right)\,\mathrm{d}t .
    \label{eq:Ja_integral}
\end{equation}
For \(0<a<1\), introduce
\begin{equation}
    z
    =
    \frac{1-\sqrt{\frac{1-a}{1+a}}}
         {1+\sqrt{\frac{1-a}{1+a}}} \, .
\end{equation}
Then
\begin{equation}
    J(a)
    =
    \frac{\pi^2}{3}
    +
    4{\rm Li}_2(-z)
    +
    \log^2 z .
    \label{eq:Ja_result}
\end{equation}
Although the first representation of \(J(a)\) involves the transcendental
measure \(\mu_a\), the final answer has weight two rather than the naively
expected weight three. This mirrors the phenomenon already seen in dual-volume
representations: integrating a transcendental measure over a semialgebraic
domain may produce a function of unexpectedly low transcendental weight.

It is also instructive to compare \(J(a)\) with the ``naive'' pairing in which
one integrates over \(P_a^\ast\) with the constant measure rather than the
canonical measure \(\mu_a\). Define
\begin{equation}
\label{eq:I_app_naive_pairing}
    I(a)
    :=
    \int_{P_a^\ast}\int_{Q^\ast}
    \frac{\langle x\,\mathrm{d}^2x\rangle\,\langle y\,\mathrm{d}^2y\rangle}
    {\langle x,y\rangle^3} \, .
\end{equation}
Using the naive dual-volume integral of the half-pizza, as in~\eqref{eq:naive_half_pizza_integral} we find
\begin{equation}
	\int_{P_a^\ast}
    \frac{\langle y\,\mathrm{d}^2y\rangle}
    {\langle x,y\rangle^3}
    =
    \Omega_a(x)+T(x;a) \,.
\end{equation}
By an explicit computation we find 
\begin{equation}
	I(a)=J(a)+\widetilde J(a) \, ,
    \qquad
    \widetilde J(a)
    =
    \int_{Q^\ast}
    \langle x\,\mathrm{d}^2x\rangle\,T(x;a) \, .
\end{equation}
A direct computation gives
\begin{equation}
\label{eq:Jtilde_app_result}
\begin{aligned}
    \widetilde J(a)
    &=
    \frac{\pi^2}{3}
    -
    8\operatorname{Li}_2(-z)
    -
    4\log(1+z)\log z,
    \\[1mm]
    I(a)
    &=
    \frac{2\pi^2}{3}
    -
    4\operatorname{Li}_2(-z)
    +
    \log^2 z
    -
    4\log(1+z)\log z .
\end{aligned}
\end{equation}
Again, the correction term \(\widetilde J(a)\) has weight two rather than the
naively expected weight three. This reflects the same phenomenon seen in
Section~\ref{sec:Transcendental Measures}: the constant-measure volume of
\(P_a^\ast\) has the rational canonical function as its rational part, plus a
transcendental correction. When integrated against \(\mathbf\Omega_Q\), both
pieces evaluate to classical dilogarithms.

This example suggests that canonical volume pairings of nonlinear positive
geometries may provide a natural class of periods interpolating between
Aomoto polylogarithms and the more general periods encountered in perturbative
quantum field theory. They retain a geometric duality, but their symbols and
transcendental weights also remember the non-trivial measures on the dual
positive geometries. A systematic study of such pairings should start with
positive geometries bounded by hyperplanes and quadrics, where the dual
measures are explicit by the methods of
Section~\ref{sec:Transcendental Measures}.

\subsection{Open problems}
\label{sec:Chapter 5 open problems}

The constructions in this chapter leave several concrete problems open. We
collect them here to separate the main results from the directions they
suggest, and we indicate where each problem arises in the preceding sections.

\begin{enumerate}
    \item \textbf{Higher-degree dual measures.}
    Extend the explicit dual-measure construction of
    Section~\ref{sec:Transcendental Measures} beyond regions bounded by lines
    and conics. The examples in
    \eqref{eq:mu_line_conic_arctan_trans}--\eqref{eq:mu_one_curvy_s_lines_trans}
    show that line-conic geometries already produce logarithmic and
    arctangent densities, while the nodal cubic example
    \eqref{eq:nodal_cubic_appendix}--\eqref{eq:cubic_measure_elliptic_appendix}
    shows that elliptic periods can appear in the representing measure; see
    also Figure~\ref{fig:nodal_cubic_measure}. One should understand, for
    higher-degree algebraic boundaries, how the dual boundary, singular
    support, lacunas, and period data of the measure are determined by the
    geometry of \(P\). This problem is motivated by the complete-monotonicity
    criterion in \eqref{eq:CM_PG_laplace} and by the cubic example of
    \cite[Section~4.5]{Mazzucchelli:DV}.

    \item \textbf{Residues and fundamental solutions.}
    Relate the Fourier--Laplace inverse measure of a canonical function to the
    Ehrenpreis--Palamodov representation of the corresponding fundamental
    solution~\cite{Liner_PDE}. The relevant analytic setup was introduced in
    \eqref{eq:fund_sol} and \eqref{eq:fund_sol_int_repr}, where the inverse
    transform of \(1/p\) is interpreted as a fundamental solution of the
    constant-coefficient operator \(p(\partial)\). In examples such as the
    half-pizza, whose cone and canonical function are written in
    \eqref{eq:pizza_cone_short} and \eqref{eq:half_pizza_a_canonical_short},
    this should express the measure on the dual cone in terms of contributions
    supported on the components of the characteristic variety. Such a
    representation would clarify how the dual-volume picture sees the
    recursive residue structure of positive geometries discussed in
    Section~\ref{sec:Adjoint Hypersurface} and used throughout
    Section~\ref{sec:Transcendental Measures}.

    \item \textbf{Multiple-conic polycons.}
    Prove the complete-monotonicity expectation for hyperbolic polycons
    bounded by several conics. The signed-triangulations for the measures was described around \eqref{eq:measure_triangulation_trans}, and the two-conic example is shown
    in Figure~\ref{fig:2con}. In all computed examples the signed sums of
    line-conic and polygonal measures are non-negative, but the general
    positivity mechanism remains to be established. A proof should combine
    the hyperbolicity constraints of
    Section~\ref{sec:Completely Monotone Positive Geometries}, the
    spectrahedral examples in \eqref{eq:spectrahedral_cone_CM}--\eqref{eq:p_pol},
    and the explicit measure formulae of
    Section~\ref{sec:Transcendental Measures}.

    \item \textbf{Grassmannian wall crossing.}
    Describe the stratification of \(\operatorname{Mat}_{>0}(k+m,n)\) by the
    combinatorial type of the exterior cyclic polytope \(C_{k,m,n}(Z)\).
    The relevant construction begins with the Plücker embedding
    \eqref{eq:plucker_embedding_convexity} and the notions of convex hull and
    extendable convexity in \eqref{eq:convX_def_long} and
    \eqref{eq:extendably_convex_def_long}. The exterior cyclic polytope was
    defined in \eqref{eq:exterior_cyclic_polytope_def_long}, and the case
    \(k=m=2\) was studied explicitly through the Vandermonde configuration
    \eqref{eq:Zmatrix_appendix}. In that example, the wall
    given by the expression in
    \eqref{eq:wedge_wall_hexagon}, changes the wedge-power oriented matroid
    and the face lattice of \(C_{2,2,6}(Z)\); see
    Figure~\ref{fig:c246bases}. A systematic description of these walls would give a better understanding of exterior cyclic polytopes and their combinatorics.

    \item \textbf{Tiles, triangulations and adjoints.}
    Compare triangulations of exterior cyclic polytopes with tiles and tilings
    of Amplituhedra, particularly in the tractable case \(k=m=2\)
    \cite{parisiShermanBennettWilliams2023m2}. The exterior cyclic polytope is
    related to the Amplituhedron by
    \eqref{eq:conv_ampl_exterior_cyclic_long}, while the Schubert-polytope
    formulation in \eqref{eq:schub_hyp}--\eqref{eq:schubert_polytope_same_intersection}
    gives a projective model closer to the boundary structure of the
    Amplituhedron. A related problem is to understand how the adjoint
    hypersurfaces of \(C_{2,2,n}(Z)\) and \(\widetilde C_{2,2,n}(Z)\) in
    \(\mathbb P^5\) are related to the adjoint hypersurface of the
    Amplituhedron itself~\cite{Ranestad:adjoint}. This should be compared with
    the general triangulation and adjoint methods of
    Sections~\ref{sec:Triangulations} and~\ref{sec:Adjoint Hypersurface},
    and with the amplituhedral tilings discussed in
    Chapter~\ref{ch:Amplituhedra}.

    \item \textbf{The dual Amplituhedron measure.}
    Construct, or rule out, a non-negative measure \(\mu_{k,m,n}\) on the
    extendable dual Amplituhedron whose Grassmannian Laplace-type integral
    \eqref{eq:amplituhedron_grassmannian_laplace} reproduces the Amplituhedron
    canonical function. The dual geometry was introduced in
    \eqref{eq:extendable_dual_def} and \eqref{eq:dual_amplituhedron_def}; for
    \(m=2\) it is related to the original Amplituhedron by the twist map
    \eqref{eq:twist_map_def}--\eqref{eq:twist_map}. The naive constant-measure
    integral~\eqref{eq:integral_dual_positive_grassmannian} produces an extra
    transcendental correction. The problem is whether this correction can be
    absorbed into a non-trivial measure, or instead obstructs a direct
    dual-volume interpretation.

    \item \textbf{Nonlinear canonical pairings.}
    Determine which canonical volume pairings of nonlinear positive geometries
    evaluate to multiple polylogarithms, satisfy complete or absolute
    monotonicity, and obey antipodal symbol symmetries analogous to Aomoto
    forms. The general pairing was introduced in \eqref{eq:can_pair}; the
    simplex case was related to Aomoto forms in \eqref{eq:aom_equiv} and to
    Goncharov polylogarithms in \eqref{eq:pol_to_gonch}--\eqref{eq:gonch}.
    Hyperplane-quadric geometries provide natural tests, since their dual
    measures are computable by the methods of Section~\ref{sec:Transcendental Measures}.
\end{enumerate}

%% file: Ch6.tex

\section{From Integrands to Integrals}
\label{ch:From Integrands to Integrals}

Loop integration is the point at which the geometric description of amplitudes
passes from rational forms to transcendental functions. Before integration,
loop-level observables in planar \(\mathcal N=4\) SYM are represented by
rational differential forms on loop-momentum space. In favorable cases these
forms are canonical forms of positive geometries, such as the Amplituhedron. Their poles and residues encode physical properties. After integration, the same
observables become multivalued functions of the external kinematics, with branch cuts, monodromies and discontinuities. The main question
of this chapter is how much of the integrand-level geometry survives this
transition.

As discussed in Section~\ref{sec:Recursion Relations}, the
integrand-level constructions used here are primarily four-dimensional,
whereas the integrated amplitude generally requires a regulator.

The classical tool for studying this question is Landau analysis
\cite{Landau1959,Nakanishi1959,Nakanishi1971,Eden:1966dnq}. Landau equations
describe when an integration contour is pinched by the singular locus of the
integrand, and hence give candidate branch loci of the integral. This analytic
description has a topological refinement, going back to Pham and
Picard--Lefschetz theory, in which singularities arise from the failure of
integration cycles to vary locally trivially in families
\cite{Pham1967,Pham2011}. In planar four-dimensional kinematics, momentum
twistors turn propagator equations into incidence conditions among lines in
projective three-space
\cite{ArkaniHamed:2010kv,Arkani-Hamed:2013kca}. This makes Landau analysis
especially compatible with the Amplituhedron and with positive geometry:
instead of studying only the denominators of a chosen integrand representation,
one uses the boundary and residue structure of the geometry
\cite{the_amplituhedron,Positive_geometries,Grassmannian}.

A second theme is the separation between rational and transcendental data.
Leading singularities, namely maximal residues of loop integrands, provide
rational prefactors. The integrated answer is then expressed in terms of
transcendental functions whose branch loci, discontinuities and symbol
alphabets carry the remaining analytic information. Wilson loops with a
Lagrangian insertion provide a particularly useful setting for this separation:
they are finite field-theoretic quantities closely related to the logarithm of the planar
amplitude, and their integrands admit an expansion in negative geometries. In
this expansion, canonical forms determine the leading-singularity prefactors,
while the integrated functions can be constrained by their symbols.

The main point developed in this chapter is that positive
geometries leave a trace after integration. Their boundary
configurations determine leading singularities, produce Landau diagrams,
constrain symbol alphabets and, in favorable cases, reduce the symbol bootstrap
to a finite problem in linear algebra. We illustrate this in detail for ladder
negative geometries. These are building blocks in a geometric expansion of the Wilson loop with Lagrangian insertion. For these examples, geometric Landau analysis predicts the
alphabet, including algebraic letters associated with square-root covers, and
the resulting bootstrap determines the six-point two-loop and five-point
three-loop integrated ladders at symbol level
\cite{ChicherinHennMazzucchelliTrnkaYangZhang2026}.

\medskip

Section~\ref{sec:The Geometry of Integration} develops the general language of
integration as a period pairing. We explain how Landau varieties arise from the
failure of local triviality, how monodromy and vanishing cycles encode
discontinuities, and how residues compute these discontinuities. We then
specialize to Feynman integrals and illustrate the discussion with Aomoto forms
and box integrals in momentum-twistors.
Section~\ref{sec:Leading Singularities and Maximal Cuts} turns to the rational
part of the story: leading singularities as maximal residues of loop
integrands. We review their relation to one-loop prescriptive unitarity,
on-shell diagrams, Amplituhedron boundary
configurations, Schubert problems and composite residues.
Section~\ref{sec:WLwithLI} introduces Wilson loops with a Lagrangian insertion,
their expansion in negative geometries, and the classification of the leading
singularities appearing in this expansion. Section~\ref{sec:Geometric Landau Analysis}
develops geometric Landau analysis for these observables, explaining how
canonical numerators refine naive Landau analysis, how boundary configurations
give Landau diagrams, and how ladder singularities lead to symbol alphabets.
Finally, Section~\ref{sec:Symbols and Bootstrap} reviews the symbol map and
applies the symbol bootstrap to integrated ladder negative geometries. The
chapter ends with an outlook on geometric Landau analysis beyond the ladder
examples.

\subsection{The geometry of integration}
\label{sec:The Geometry of Integration}

We begin with the general mechanism by which rational forms give rise to
multivalued functions. The poles of an integrand are local objects in the
integration variables, whereas the branch points of the integral are global
objects in external kinematic space. Understanding this passage is one of the
central problems in the study of Feynman integrals.

Historically, this problem appears in the analytic S-matrix program and in the
Landau analysis of Feynman integrals
\cite{Landau1959,Nakanishi1959,Nakanishi1971,Eden:1966dnq}. Cutkosky rules
describe discontinuities by putting propagators on shell
\cite{Cutkosky1960}, while the Coleman--Norton theorem interprets physical
Landau singularities as classical on-shell processes in spacetime
\cite{ColemanNorton1965}. The complementary viewpoint, emphasized by Pham, is
topological: singularities occur when integration cycles cannot be transported
trivially in families \cite{Pham1967,Pham2011}. This Picard--Lefschetz
perspective will be the guiding language of this section.

In contemporary on-shell and positive-geometric approaches to amplitudes, the same ideas interact with the geometry of loop
integrands. In planar \(\mathcal N=4\) SYM these integrands are rational forms
with highly constrained singularity structure, often governed by the
Amplituhedron and related positive geometries
\cite{ArkaniHamed:2010kv,the_amplituhedron,Arkani-Hamed:2013kca,Positive_geometries,Grassmannian}.
After integration, however, the answer is a transcendental function whose
singularities and symbol alphabet are constrained not only by the poles of an
integrand representation, but by the topology of the whole integration problem
\cite{DennenSpradlinVolovich2016,DennenPrlinaSpradlinStanojevicVolovich2017,
PrlinaSpradlinStanojevic2018,Arkani-Hamed:2018rsk,
ChicherinHennMazzucchelliTrnkaYangZhang2026}.

The purpose of this section is to set up this bridge. We first describe
integrals as period pairings between differential forms and cycles. We then
explain how Landau varieties arise from the failure of local triviality, how
monodromy and vanishing cycles encode discontinuities, and how residues compute
these discontinuities. We finally specialize the discussion to Feynman
integrals and to box integrals in momentum-twistor space.

\subsubsection{Integrals as period pairings}
\label{subsec:Integrals as Period Pairings}

We start from a general geometric formulation. An integral is a pairing between
a differential form and a cycle, both of which may vary with the external
parameters. The resulting function is analytic as long as this family of
pairings varies locally trivially. Singularities appear precisely when this
local triviality fails. The following overview is brief and is meant to
motivate the geometric language used later in the chapter; for more detailed
accounts, see~\cite{Pham1967,Pham2011,BerghoffPanzer2025}. This is the
Picard--Lefschetz formulation of Landau analysis, and it underlies several
modern approaches to the singularities of Feynman integrals
\cite{Mizera:2021icv,FevolaMizeraTelen2024,HelmerPapathanasiouTellander2024}.

Let \(\mathcal{X}\) be the total space containing both complexified loop variables
and external kinematics, and let \(\mathcal K\) be the space of external
kinematics alone. We consider a holomorphic surjective map forgetting the loop
variables,
\begin{equation}
\label{eq:Pi}
        \Pi:\mathcal{X}\longrightarrow \mathcal K \, .
\end{equation}
A point \(s\in\mathcal K\) specifies the external data, and the fibre
\begin{equation}
\label{eq:fibre-Xs}
        \mathcal{X}_s:=\Pi^{-1}(s)
\end{equation}
is the space of loop-integration variables at fixed kinematics.

Let \(\mathcal D\subseteq\mathcal{X}\) be the singular locus of the integrand. In
Feynman integrals, \(\mathcal D\) is the union of the vanishing loci of the
propagators \cite{Landau1959,Nakanishi1971,Smirnov2004}. In the Amplituhedron
setting, for fixed external data, the analogous role is played by the algebraic
boundary of the loop Amplituhedron, or more generally by the boundary structure
of the positive geometry producing the integrand
\cite{the_amplituhedron,Arkani-Hamed:2013kca,Franco:2014csa,Dian_2023,Dian:2024hil}. We write
\begin{equation}
\label{eq:Ds-definition}
        \mathcal D_s:=\mathcal D\cap\mathcal{X}_s \, .
\end{equation}

Let \(\mathbf{\Omega}\) be a holomorphic top form on
\(\mathcal{X}\setminus\mathcal D\), and denote by \(\mathbf{\Omega}_s\) its
restriction to the fibre \(\mathcal{X}_s\setminus\mathcal D_s\). Finally, let
\(\Gamma_s\) be a family of integration chains in
\(\mathcal{X}_s\setminus\mathcal D_s\). More generally, the chains may have
boundary on a prescribed subspace \(\mathcal B_s\), where
\begin{equation}
\label{eq:Bs-definition}
        \mathcal B_s:=\mathcal B\cap\mathcal{X}_s \, .
\end{equation}

We are interested in integrals of the form
\begin{tcolorbox}[definitionbox]
\textbf{Period pairing.}
\begin{equation}
\label{eq:I_int_s}
        \mathcal I(s)=\int_{\Gamma_s}\mathbf{\Omega}_s \, .
\end{equation}
\end{tcolorbox}\noindent
For a single finite Feynman integral, \(\mathbf{\Omega}_s\) is the usual loop
integrand at fixed external kinematics, and \(\Gamma_s\) is the integration
contour determined by the Feynman \(i\varepsilon\)-prescription. In the
Amplituhedron setting, \(\mathbf{\Omega}_s\) is the canonical form of the loop
Amplituhedron at fixed external data \(Z\), encoding the bosonized momentum
supertwistors. The distinction between the rational integrand and the
transcendental function obtained after integration is one of the central themes
of modern Feynman integral theory
\cite{Panzer2015,Brown2015,Vanhove2014,BroedelDuhrDulatTancredi2018,Euler_Integrals}.

The value of~\eqref{eq:I_int_s} depends only on the de Rham
cohomology class of the form,
\begin{equation}
\label{eq:dR_cohom}
        [\mathbf{\Omega}_s]\in
        H^d(\mathcal{X}_s\setminus\mathcal D_s;\mathbb C) \, ,
\end{equation}
and on the relative homology class of the chain,
\begin{equation}
\label{eq:rel_homol}
        [\Gamma_s]\in
        H_d(\mathcal{X}_s\setminus\mathcal D_s,\,
        \mathcal B_s\setminus\mathcal D_s;\mathbb Z) \, .
\end{equation}
Here \(d\) is the difference between the complex dimensions of \(\mathcal{X}\)
and \(\mathcal K\), equivalently the dimension of the generic fibre
\(\mathcal{X}_s\). Thus the integral is a period pairing,
\begin{equation}
\label{eq:period-pairing}
        H^d(\mathcal{X}_s\setminus\mathcal D_s;\mathbb C)
        \times
        H_d(\mathcal{X}_s\setminus\mathcal D_s,\,
        \mathcal B_s\setminus\mathcal D_s;\mathbb Z)
        \longrightarrow
        \mathbb C \, .
\end{equation}
This period perspective is standard in the modern study of Feynman integrals,
where loop integrals are viewed as periods of algebraic varieties and their
variations are described by differential equations or Gauss--Manin connections
\cite{Bloch:2005bh,Brown2009_alt1,Vanhove2014,Brown2015,MastroliaMizera2019,Euler_Integrals}. We will not focus on solving differential
equations in this chapter, but the language of pure integrals and canonical
differential equations is useful background for the decomposition and bootstrap
constructions below.

\subsubsection{Landau varieties as failure of local triviality}
\label{subsec:Landau Varieties as Failure of Local Triviality}

As \(s\) varies, the fibre \(\mathcal{X}_s\), the divisor \(\mathcal D_s\), and
the chain \(\Gamma_s\) vary with it. If nothing special happens, the topology of
the pair
\begin{equation}
\label{eq:rel_fb}
        (\mathcal{X}_s\setminus\mathcal D_s,\,
        \mathcal B_s\setminus\mathcal D_s)
\end{equation}
is locally constant. In that case the chain \(\Gamma_s\) can be transported
continuously, and the integral defines a holomorphic function of \(s\).
Singular behaviour can occur only when this local triviality fails.
Geometrically, this means that the integration chain becomes trapped by the
singular locus of the integrand, or that the fibre itself degenerates. The locus
in \(\mathcal K\) where this can happen is the \emph{Landau variety},
\begin{equation}
\label{eq:L}
        \mathcal L\subseteq\mathcal K \, .
\end{equation}

There is a precise way to describe this locus. Consider the union
\begin{equation}
        \mathcal D\cup\mathcal B\subseteq\mathcal{X} \, ,
\end{equation}
and assume it admits a \emph{Whitney stratification} \(\mathfrak S\). A Whitney
stratification of a singular space is a decomposition into smooth strata such
that, whenever one stratum lies in the closure of another, their limiting
tangent spaces satisfy Whitney's regularity conditions
\cite{GoreskyMacPherson1988,Dimca2004,HelmerPapathanasiouTellander2024}. This
is the appropriate regularity framework so that contour pinches are detected as
critical values of the projection \(\Pi\) restricted to strata
\(S\in\mathfrak S\),
\begin{equation}
        \Pi|_S:S\longrightarrow\mathcal K \, .
\end{equation}
A critical value of \(\Pi|_S\) is the image of a point at which the differential
of \(\Pi|_S\) is not surjective. We denote the set of critical points by
\begin{equation}
        {\rm Crit}(\Pi|_S)\subseteq S \, .
\end{equation}
Thus, under suitable properness assumptions, the Landau variety may be defined
as
\begin{tcolorbox}[definitionbox]
\textbf{Landau variety.}
\begin{equation}
\label{eq:Land_crit}
        \mathcal L
        =
        \overline{
        \bigcup_{S\in\mathfrak S}
        \Pi\!\left({\rm Crit}(\Pi|_S)\right)
        }
        \subseteq \mathcal K \, .
\end{equation}
\end{tcolorbox}\noindent
This formula assumes that \(\Pi\) is proper, namely that the preimage of every
compact set in \(\mathcal K\) is compact in \(\mathcal{X}\). More generally, if
\(\mathcal{X}\) or \(\mathcal K\) are non-compact, the locus
\eqref{eq:Land_crit} may have to be enlarged by components arising from
singularities at infinity. In physics these are often called second-type Landau
singularities.

For the Feynman-integral and positive-geometric examples used below, the
singular-locus problem is algebraic. The external kinematic space
\(\mathcal K\) is defined by polynomial constraints such as momentum
conservation, on-shell conditions, or their momentum-twistor analogues.
Propagators and boundaries of Amplituhedra are encoded by polynomial
equations. Thus the defining equations of \(\mathcal{X}\), \(\mathcal K\),
\(\mathcal D\), and \(\mathcal B\) are algebraic, and \(\Pi\) is an algebraic map.
Consequently, the Landau variety is itself an algebraic subvariety of
\(\mathcal K\). This is the basis of recent approaches using Landau
discriminants, principal Landau determinants, Whitney stratifications~\cite{Mizera:2021icv,FevolaMizeraTelen2024,FevolaMizeraTelenPLD2024,HelmerPapathanasiouTellander2024,CorreiaGirouxMizera2026}.

Away from \(\mathcal L\), the integration problem is locally trivial. By Thom's
isotopy lemma, the relative homology groups in~\eqref{eq:rel_homol} form a
local system over \(\mathcal K\setminus\mathcal L\)
\cite{GoreskyMacPherson1988,Dimca2004}. Therefore:
\begin{tcolorbox}[resultbox]
\textbf{Holomorphy away from the Landau variety.}
The integral \(\mathcal I(s)\) defines a multivalued holomorphic function on
\(\mathcal K\setminus\mathcal L\).
\end{tcolorbox}\noindent

The Landau variety should be regarded as an upper bound on the possible branch
locus of the integral. In practice, one usually focuses on the codimension-one
components of \(\mathcal L\), since these are the components that can support
branch hypersurfaces of the multivalued function. Higher-codimension components
may affect special loci or intersections of singular hypersurfaces, but they do
not by themselves support generic branch divisors, by the removable
singularities theorem in several complex variables.

\subsubsection{Monodromy, discontinuities, and vanishing cycles}
\label{subsec:Monodromy Discontinuities and Vanishing Cycles}

Fix \(s\in\mathcal K\setminus\mathcal L\) and consider analytic continuation of
\(\mathcal I(s)\) along a path \(\gamma\) in
\(\mathcal K\setminus\mathcal L\). The integration chain \(\Gamma_s\) is
transported to a chain in the final fibre. If \(\gamma\) is closed, the
transported chain need not return to itself. This gives the monodromy action.
The monodromy depends only on the homotopy class of \(\gamma\) with fixed base
point \(s\), and gives a representation
\begin{equation}
\label{eq:monodromy-representation}
        \pi_1(\mathcal K\setminus\mathcal L,s)
        \longrightarrow
        {\rm Aut}\,
        H_d(\mathcal{X}_{s}\setminus\mathcal D_{s},\,
        \mathcal B_{s}\setminus\mathcal D_{s};\mathbb Z),
        \qquad
        [\gamma]\longmapsto M_\gamma \, .
\end{equation}
This induces an action on the integral:
\begin{equation}
\label{eq:monodromy-action-integral}
        \gamma\!\cdot\!\mathcal I(s)
        :=
        \int_{M_\gamma(\Gamma_s)}\mathbf{\Omega}_s \, .
\end{equation}
We define the variation of the chain along \(\gamma\) by
\begin{equation}
\label{eq:variation-cycle}
        {\rm Var}_{\gamma}(\Gamma_s)
        :=
        (M_\gamma-\mathrm{id})\Gamma_s \, .
\end{equation}
The corresponding \emph{discontinuity} is defined as
\begin{tcolorbox}[definitionbox]
\textbf{Discontinuity as monodromy.}
\begin{equation}
\label{eq:disc-monodromy}
        {\rm Disc}_\gamma \, \mathcal I(s)
        :=
        \gamma\!\cdot\!\mathcal I(s)-\mathcal I(s)
        =
        \int_{{\rm Var}_\gamma(\Gamma_s)}\mathbf{\Omega}_s \, .
\end{equation}
\end{tcolorbox}\noindent
A discontinuity therefore measures the failure of the integral to return to the
same value after analytic continuation around the singular locus.

This is the Picard--Lefschetz mechanism \cite{Pham1967,Pham2011}. When a
parameter approaches the Landau variety, certain cycles in the fibre may
collapse to a point, or more generally to a lower-dimensional stratum. These are
the \emph{vanishing cycles} \(\delta_i\). They measure the topology that is
lost at the critical value. If \(\gamma\) is a small loop around a smooth
component of the Landau variety, Picard--Lefschetz theory expresses the
variation of a transported cycle as a sum of vanishing cycles:
\begin{equation}
\label{eq:PL_var}
        {\rm Var}_{\gamma}(\Gamma_s)
        =
        \sum_i n_i\,\delta_i \, ,
\end{equation}
where \(n_i\) is an appropriate intersection number between the transported
cycle and the vanishing cycle. For Feynman integrals, this perspective was
already central in Pham's work and has recently been refined in relative
Picard--Lefschetz theory
\cite{Pham1967,Pham2011,Hannesdottir:2021kpd,BerghoffPanzer2025}.

One important consequence is that iterated discontinuities are constrained by
the possible successive variations of cycles. In relative Picard--Lefschetz
theory, this leads to \emph{hierarchy principles} controlling when iterated variations
must vanish \cite{Hannesdottir:2021kpd,BerghoffPanzer2025}. In perturbative
quantum field theory, a special case is given by the Steinmann relations, which
forbid double discontinuities in overlapping physical channels
\cite{Steinmann1960,Eden:1966dnq}. Their extended versions have become central
constraints in the symbol bootstrap of planar \(\mathcal N=4\) SYM amplitudes
\cite{CaronHuotDixonMcLeodVonHippel2016,DixonEtAl2017}.

\subsubsection{Discontinuities as residues}
\label{subsec:Discontinuities as Residues}

The Picard--Lefschetz viewpoint also explains why discontinuities are dual to
residues of the integrand. Consider a small loop \(\gamma\) around a component
of the Landau variety. Suppose that this component is the image under \(\Pi\) of
critical points of \(\Pi|_S\), for some stratum \(S\) in a Whitney
stratification. Let \(I=\{i_1,\ldots,i_r\}\) be maximal such that
\begin{equation}\label{eq:S_in_int}
        S\subseteq
        \mathcal D_I
        :=
        \mathcal D_{i_1}\cap\cdots\cap\mathcal D_{i_r} \, .
\end{equation}
Here we assume that $\mathcal{D}_I$ is irreducible, otherwise we restrict to an irreducible component.
Then the discontinuity can be expressed as an integral of an iterated residue:
\begin{tcolorbox}[resultbox]
\textbf{Residue formula for discontinuities.}
\begin{equation}
\label{eq:pham-disc-residue}
        {\rm Disc}_{\gamma} \, \mathcal I(s)
        =
        \int_{{\rm Var}_\gamma(\Gamma_s)}
        \mathbf{\Omega}_s
        =
        n(2\pi i)^r
        \int_{\widetilde\Gamma_s}
        {\rm Res}^{r}_{I}\,\mathbf{\Omega}_s \, .
\end{equation}
\end{tcolorbox}\noindent
Here \(n\) is an intersection number, \({\rm Res}^{r}_{I}\mathbf{\Omega}_s\)
is the \(r\)-fold residue of the form along the cut
\(\mathcal D_{i_1,s}\cap\cdots\cap\mathcal D_{i_r,s}\), and
\(\widetilde\Gamma_s\) is the reduced cycle on the cut
\cite{Pham1967,Pham2011,Hannesdottir:2021kpd,BerghoffPanzer2025}.

We can understand this formula locally. Suppose that \(x_1,\ldots,x_d\) are
coordinates on \(\mathcal{X}_s\) and that the cut is given by
\begin{equation}
        x_1=\cdots=x_r=0 \, .
\end{equation}
Assume first that the intersection is transverse. Then the form has the local
shape
\begin{equation}
\label{eq:local-log-form-residue}
        \mathbf{\Omega}_s
        =
        \frac{\mathrm{d}x_1\wedge\cdots\wedge \mathrm{d}x_r}
        {x_1\cdots x_r}
        \wedge
        \omega_s(x)
        +\cdots \, ,
\end{equation}
where the remaining terms are holomorphic along the cut. The variation cycle is
locally 
\begin{equation}
\label{eq:variation-torus}
        {\rm Var}_{\gamma}(\Gamma_s)
        =
        n\,[T^r\times\widetilde\Gamma_s] \, ,
\end{equation}
where \(T^r\) is the small real torus given by
\begin{equation}
        |x_i|=\epsilon_i,
        \qquad
        i=1,\ldots,r \, .
\end{equation}
The integral over \(T^r\) gives \((2\pi i)^r\), and the remaining integral is
over the reduced cycle \(\widetilde\Gamma_s\). The residue evaluates to
\begin{equation}
\label{eq:local-residue}
        {\rm Res}^{r}_{I}\mathbf{\Omega}_s
        =
        \omega_s(0,\ldots,0,x_{r+1},\ldots,x_d) \, .
\end{equation}
Non-transverse intersections require the algebraic definition of the residue,
and global residue theorems extend the classical residue theorem to the
multivariate case, relating maximal residues at different cut solutions
\cite{GriffithsHarris1978,Tsikh1992,CoxLittleOShea2005,Mastrolia:2009dr,MastroliaMizera2019,Euler_Integrals}.

This is the conceptual origin of the relation between cuts, namely discontinuities, and residues. A cut
sets a collection of propagators or boundary equations to zero. A discontinuity
is a period over the corresponding variation cycle. A maximal cut, when it
localizes all integration variables, is a maximal residue. These maximal
residues will be called leading singularities in
Section~\ref{sec:Leading Singularities and Maximal Cuts}.

\subsubsection{Landau loci versus actual singularities}
\label{subsec:Landau Loci versus Actual Singularities}

The Landau variety is an upper bound on the singular locus of an integral. A
component of the Landau variety need not be an actual branch singularity of the
particular period under consideration. There are several reasons for this.

First, the variation cycle may vanish. A component of the Landau variety may
describe a degeneration of the integration problem without producing a
discontinuity of the chosen integral. In the geometric language of
\eqref{eq:disc-monodromy}, the discontinuity is obtained by integrating over
\({\rm Var}_{\gamma}(\Gamma_s)\). Thus the discontinuity vanishes if this
variation cycle is zero in the relevant relative homology group, or if the form
has zero period on it.

Second, the numerator may vanish. The Landau equations only see the denominator
of an integral. A numerator can remove a candidate singularity by vanishing on
the relevant on-shell variety, or by making the corresponding cut invisible to
the integral. This is especially natural in the Amplituhedron and
positive-geometric setting, where canonical-form numerators cancel some boundary
or cut configurations
\cite{DennenPrlinaSpradlinStanojevicVolovich2017,Arkani-Hamed:2018rsk,ChicherinHennMazzucchelliTrnkaYangZhang2026}.

Third, a Landau component may lie on another sheet. A component can be a genuine
singularity of the analytic continuation of the integral, but not of the branch
selected by the physical contour. In physical kinematics this distinction is
controlled by the \(i\varepsilon\)-prescription and by positivity conditions on
the Landau parameters. In the Coleman--Norton interpretation, physical-sheet
singularities correspond to classical on-shell processes with positive proper
times \cite{ColemanNorton1965,Eden:1966dnq,Nakanishi1971}.

Finally, singularities can cancel in sums of integrals. A full amplitude is
often represented as a sum of many individual integrals. Each
may have Landau singularities that cancel in the final amplitude. This is
analogous to spurious poles in a triangulation of a positive geometry: they are
present in a representation but absent in the final object
\cite{the_amplituhedron,DennenSpradlinVolovich2016,DennenPrlinaSpradlinStanojevicVolovich2017,PrlinaSpradlinStanojevic2018,Arkani-Hamed:2018rsk}.

\begin{tcolorbox}[resultbox]
\textbf{Candidate versus actual singularities.}
Landau analysis gives candidate singularities. The actual singularities of a
chosen integral depend on the integration cycle, the numerator, the physical
sheet, and possible cancellations in the final sum.
\end{tcolorbox}\noindent
This distinction is central for the geometric approach. The boundary structure
of a positive geometry, together with the numerator of its canonical form,
refines the naive Landau analysis and discards many spurious candidates. This is
the mechanism behind \emph{geometric Landau analysis}.

\subsubsection{Example: the Aomoto form of an interval}
\label{subsec:Example Aomoto Interval}

Before discussing Feynman integrals, we illustrate the previous formal
discussion with the simplest possible example. Consider the Aomoto form of two
intervals,
\begin{equation}
\label{eq:I_log}
        \mathcal I(s)
        =
        \mathcal I(A(s),B)
        =
        \int_1^s\frac{\mathrm{d}x}{x}
        =
        \log(s) \, ,
\end{equation}
where \(A(s)=\Gamma_s=[1,s]\) and \(B=[0,\infty]\), in the affine chart
\([1:x]\) of \(\mathbb P^1\). It is a special one-dimensional case of the
Aomoto forms discussed in Section~\ref{sec:Aomoto Forms}.

In this example the total space is
\begin{equation}
        \mathcal{X}=\mathbb P^1_x\times\mathcal K \, ,
        \qquad
        \mathcal K=\mathbb P^1_s \, ,
\end{equation}
with projection \(\Pi(x,s)=s\). The components of the singular locus of the
form and of the boundary of the integration chain are
\begin{equation}
\label{eq:DiBi}
        \mathcal D_1=\{x=0\},
        \qquad
        \mathcal D_2=\{x=\infty\},
        \qquad
        \mathcal B_1=\{x=1\},
        \qquad
        \mathcal B_2=\{x=s\} \, .
\end{equation}
The non-empty intersections among these components are
\begin{equation}
\label{eq:DicapBi}
        \mathcal D_1\cap\mathcal B_2=\{(0,0)\},
        \qquad
        \mathcal D_2\cap\mathcal B_2=\{(\infty,\infty)\},
        \qquad
        \mathcal B_1\cap\mathcal B_2=\{(1,1)\} \, .
\end{equation}
On the smooth strata obtained by removing these points, the projection to
\(\mathcal K\) is a submersion. On the zero-dimensional strata in
\eqref{eq:DicapBi}, it cannot be a submersion. Therefore the Landau variety consists of three points:
\begin{equation}
\label{eq:aomoto-interval-landau}
        \mathcal L=\{0,1,\infty\} \, .
\end{equation}

The integral \(\mathcal I(s)=\log(s)\) has logarithmic singularities at
\(s=0\) and \(s=\infty\). The point \(s=1\) is different: there the integration
cycle \([1,s]\) shrinks to a point, but the integral is regular. Thus \(s=1\)
belongs to the Landau variety, in the sense of topological degeneration, but it
is not an actual singularity of the integral. This example already illustrates
the distinction between candidate Landau loci and actual singularities.

Let us compute the discontinuity around \(s=0\). For the principal branch of
the logarithm, the branch cut lies on the negative real axis. If \(s<0\), then
\begin{equation}
        s+i\varepsilon\to |s|e^{i\pi},
        \qquad
        s-i\varepsilon\to |s|e^{-i\pi} \, .
\end{equation}
Hence
\begin{equation}
\label{eq:log_disc_1}
        {\rm Disc}_{\gamma_0}\log s
        =
        \lim_{\varepsilon\to0^+}
        \left(
        \log(s+i\varepsilon)-\log(s-i\varepsilon)
        \right)
        =
        2\pi i \, .
\end{equation}
Topologically, the analytic continuation of the path \([1,s]\) around \(s=0\)
adds a small positively oriented circle \(\delta_0\) around the origin
\(x=0\). Thus
\begin{equation}
\label{eq:Var_S0}
        {\rm Var}_{\gamma_0}([1,s])=[\delta_0] \, ,
\end{equation}
and
\begin{equation}
\label{eq:disc_log}
        {\rm Disc}_{\gamma_0}\log s
        =
        \int_{\delta_0}\frac{\mathrm{d}x}{x}
        =
        2\pi i \, .
\end{equation}
This is the one-dimensional Picard--Lefschetz formula: the vanishing cycle is
the small circle around \(x=0\), and the relevant intersection number with the
interval \([1,s]\) is one, up to orientation.
The residue interpretation is equally simple. With the convention that
\({\rm Res}_{x=0}(\mathrm{d}x/x)=1\), the discontinuity dually becomes a residue,
\begin{equation}
\label{eq:aomoto-log-residue}
        {\rm Disc}_{\gamma_0}\log s
        =
        2\pi i\,{\rm Res}_{x=0}\frac{\mathrm{d}x}{x}
        =
        2\pi i \, .
\end{equation}
More generally, for Aomoto forms, discontinuities are again Aomoto forms on
lower-dimensional faces. If \(A(s)\) becomes incident with a boundary component
\(\mathcal B_i\) of \(B\), then
\begin{tcolorbox}[resultbox]
\textbf{Aomoto discontinuity.}
\begin{equation}
\label{eq:aomoto-disc-residue}
        {\rm Disc}_{\gamma_i}\,\mathcal I(A(s),B)
        =
        2\pi i
        \int_{A(s)\cap\mathcal B_i}
        {\rm Res}_{\mathcal B_i}\,\mathbf{\Omega}_{B} \, .
\end{equation}
\end{tcolorbox}\noindent
This is the simplest incarnation of the residue/discontinuity duality.

\subsubsection{Landau equations for Feynman integrals}
\label{subsec:Landau Equations for Feynman Integrals}

We now specialize the previous discussion to Feynman integrals. A typical
\(\ell\)-loop Feynman integral has the form
\begin{equation}
\label{eq:schematic-loop-integral}
        \mathcal I(s)
        =
        \int_{\Gamma}
        \prod_{a=1}^{\ell} d^D k_a\,
        \frac{N(k,s)}
        {\prod_{e\in G}D_e(k,s)} \, .
\end{equation}
Here \(D\) is the dimension of spacetime, \(s\in\mathcal K\) denotes the
external kinematic data, \(k_a\in\mathbb C^D\) are loop momenta, and \(G\) is
the set of propagator edges of the graph. The inverse propagators have the form
\begin{equation}
\label{eq:prop_D}
        D_e(k,s)=q_e(k,s)^2-m_e^2 \, ,
\end{equation}
where \(q_e(k,s)\) is an affine-linear combination of loop and external
momenta, and \(m_e\) is the corresponding mass. The Feynman
\(i\varepsilon\)-prescription is absorbed into the definition of the integration
contour \(\Gamma\). The numerator \(N(k,s)\) is not relevant for the naive
Landau equations, so in this subsection we temporarily set \(N(k,s)=1\).

The singular hypersurface of the integrand is
\begin{equation}
        \mathcal D
        =
        \bigcup_{e\in G}\{D_e(k,s)=0\} \, .
\end{equation}
A singularity in external kinematics may occur when the loop contour is pinched
by some of these hypersurfaces. There are two distinct conditions. The
first is the \emph{cut condition}: some set of propagators is put on shell. The second
is the \emph{pinch condition}: the on-shell hypersurfaces trap the integration contour,
so that the contour cannot be deformed away.

Let \(H\subseteq G\) be the set of edges put on shell. The corresponding cut
variety is
\begin{equation}
\label{eq:on-shell-variety-H}
        \mathcal D_H
        :=
        \left\{
        (k,s)\in\mathcal{X}:
        D_e(k,s)=0
        \text{ for all } e\in H
        \right\} \, .
\end{equation}
The possible Landau singularities associated with \(H\) are the critical values
of 
\begin{equation}
\label{eq:PiH}
        \Pi_H:=\Pi|_{\mathcal D_H}:
        \mathcal D_H\longrightarrow\mathcal K \, .
\end{equation}
At a smooth point of \(\mathcal D_H\), the cut equations define a transverse
complete intersection in the loop variables precisely when the loop
differentials \(\mathrm{d}_kD_e\), for \(e\in H\), are linearly independent. A Landau
critical point occurs when this transversality fails. Equivalently, there exist
complex coefficients \(\alpha_e\), not all zero, such that
\begin{tcolorbox}[resultbox]
\textbf{Cut and pinch equations.}
\begin{equation}
\label{eq:landau-cut-H}
        D_e(k,s)=0
        \quad \forall e\in H,
        \qquad
        \sum_{e\in H}\alpha_e\,\mathrm{d}_kD_e(k,s)=0 \, .
\end{equation}
\end{tcolorbox}\noindent
The equations \(D_e=0\) impose the cut conditions, while the linear dependence
of the differentials \(\mathrm{d}_kD_e\) is the pinch condition.

In components, for each loop momentum \(k_a\), the pinch condition may be
written as
\begin{equation}
\label{eq:pinch-momentum-space}
        \sum_{e\in H}
        \alpha_e
        \frac{\partial D_e}{\partial k_a^\mu}
        =
        0 \, ,
        \qquad
        a=1,\ldots,\ell \, .
\end{equation}
Equivalently, after choosing an oriented cycle basis for the graph, this becomes
\begin{equation}
        \sum_{e\in C_a}\alpha_e q_e^\mu=0 \, ,
        \qquad
        a=1,\ldots,\ell \, ,
\end{equation}
up to conventional signs.

One often writes the Landau equations with \(H=G\) and allows some
\(\alpha_e\) to vanish. A branch of the solution is then specified by the
subgraph
\begin{equation}
        H=\{e\in G:\alpha_e\neq0\} \, .
\end{equation}
The associated component of the Landau variety is
\begin{equation}
\label{eq:landau-variety-H}
        \mathcal L_H
        :=
        \overline{\Pi_H({\rm Crit}(\Pi_H))}
        \subseteq\mathcal K \, .
\end{equation}
Equations for \(\mathcal L_H\) are obtained by eliminating the loop variables
and the coefficients \(\alpha_e\) from the Landau equations~\eqref{eq:landau-cut-H}.

The same conditions can be expressed in Feynman parameters. Introducing a
parameter \(\alpha_e\) for every propagator, the loop integrations lead to the
Symanzik polynomials \(\mathcal U\) and \(\mathcal F\)
\cite{BognerWeinzierl,Smirnov2004}. Away from the locus \(\mathcal U=0\),
which controls possible singularities at infinity, the finite-distance Landau
equations take the simple form
\begin{tcolorbox}[definitionbox]
\textbf{Parametric Landau equations.}
\begin{equation}
\label{eq:F_LE}
        \mathcal F=0,
        \qquad
        \frac{\partial \mathcal F}{\partial\alpha_e}=0
        \quad \forall e\in G \, .
\end{equation}
\end{tcolorbox}\noindent
The components arising from \(\mathcal U=0\) are the second-type Landau
singularities and require a compactified or stratified analysis
\cite{Fairlie1962,Eden:1966dnq,Nakanishi1971,FevolaMizeraTelen2024,HelmerPapathanasiouTellander2024}.

A related affine formulation is the Lee--Pomeransky representation,
where the two Symanzik polynomials are combined into $\mathcal G=\mathcal U+\mathcal F$ \cite{LeePomeransky2013}. The critical points of \(\mathcal G\) control the
number of master integrals in a sector, which is the rank of the corresponding
local system. This framework underlies modern approaches to Feynman integrals
using twisted cohomology, intersection theory, Landau discriminants, and
principal Landau determinants
\cite{Mizera:2017cqs,MastroliaMizera2019,Euler_Integrals,MatsubaraHeoTelen2025,Mizera:2021icv,FevolaMizeraTelen2024,FevolaMizeraTelenPLD2024,DlapaHelmerPapathanasiouTellander2023,CorreiaGirouxMizera2026}.

A solution branch is called a leading Landau singularity if it comes from a
branch with all \(\alpha_e\neq0\). This terminology should not be confused with
leading singularities of loop integrands, which are maximal residues before
integration. More generally, a subleading Landau singularity is a leading
Landau singularity of a subtopology, obtained by shrinking some propagators, or
equivalently by setting some Feynman parameters to zero.

There is a useful physical interpretation of the coefficients \(\alpha_e\). For
a solution with all \(\alpha_e>0\), the Coleman--Norton theorem states that the
Landau singularity can be interpreted as a classical on-shell process in
spacetime: the internal particles corresponding to the edges in \(H\) propagate
on shell between interaction vertices, with positive proper times
\cite{ColemanNorton1965,Eden:1966dnq}. If the \(\alpha_e\) are not all
positive, the solution may still describe a singularity of the analytic
continuation of the integral, but it need not lie on the physical sheet.

The duality between discontinuities and reisudes becomes, in physical kinematics, the following replacement rule at the integrand-level \cite{Cutkosky1960,Eden:1966dnq,Veltman1963,Elvang:2013cua,Henn:2014yza,Abreu:2022mfk,HannesdottirMcLeodSchwartzVergu2023}:
\begin{tcolorbox}[resultbox]
\textbf{Cutkosky rule.}
\begin{equation}
\label{eq:cutkosky-rule-comment}
        \frac{1}{D_e(k,s)+i0}
        \quad\longrightarrow\quad
        -2\pi i\,\delta^+\!\bigl(D_e(k,s)\bigr) \, .
\end{equation}
\end{tcolorbox}\noindent
Here
\begin{equation}
        \delta^+(q^2-m^2)
        :=
        \theta(q^0)\delta(q^2-m^2)
\end{equation}
imposes the real on-shell condition and selects the positive-energy branch.
Thus the Cutkosky rules are the Lorentzian incarnation of the residue operation
for Feynman integrals. The complex residue specifies how the contour winds
around the propagator divisor, while the \(\delta^+\)-prescription specifies
the reduced cut cycle.

We conclude by emphasizing the limitations of the naive Landau equations. They
may miss critical points at infinity, singular or degenerate strata of the
on-shell variety, and components supported on non-transverse intersections. A
systematic treatment requires compactification and Whitney stratification
\cite{DennenSpradlinVolovich2016,PrlinaSpradlinStanojevic2018,FevolaMizeraTelen2024,HelmerPapathanasiouTellander2024,He:2024fij,Vergu:2025mag}. Moreover, even when a component is present in the Landau
variety, it may be absent from a particular integral for the reasons explained
in Subsection~\ref{subsec:Landau Loci versus Actual Singularities}.

\subsubsection{Differential equations and the Landau variety}
\label{subsec:Pure Integrals and Canonical Differential Equations}

Although the differential-equation method is not a principal subject of this
thesis, we briefly review it because it provides a natural language for several
connections with positive geometry that remain to be explored. Canonical forms,
logarithmic singularities, leading singularities, pure integral bases, and
symbol alphabets all have close counterparts in the differential equations
satisfied by Feynman integrals. In particular, the Landau variety introduced
above constrains the singular divisors of these equations, while canonical
differential equations make the corresponding logarithmic structure manifest.
This suggests the possibility of deriving differential equations, or at least
their alphabets and preferred bases, more directly from positive-geometric
data. For systematic introductions to the differential-equation method, we
refer the reader especially to Henn's lecture notes, as well as to the broader
accounts by Weinzierl and Duhr
\cite{Henn:2014qga,Weinzierl:2022eaz,Duhr:2019tlz}.

The period formulation developed above leads naturally to differential
equations. As the external kinematics \(s\in\mathcal K\) vary, the cohomology
classes of the integrands form a vector bundle over
\(\mathcal K\setminus\mathcal L\), equipped with the Gauss--Manin connection.
Choosing a basis of cohomology classes, or equivalently a basis of master
integrals, turns this connection into a system of first-order differential
equations. Thus Landau analysis and the differential-equation method describe
the same family of periods from complementary perspectives: Landau analysis
identifies where the family can degenerate, while differential equations
describe its variation away from these loci
\cite{Vanhove2014,Brown2015,MastroliaMizera2019,Euler_Integrals}.

Concretely, integration-by-parts identities reduce a family of dimensionally
regulated loop integrals to a finite vector of master integrals
\begin{equation}
        \vec{\mathcal I}(\epsilon,s)
        =
        \bigl(
        \mathcal I_1(\epsilon,s),\ldots,
        \mathcal I_r(\epsilon,s)
        \bigr)^{\mathsf T},
\end{equation}
where \(D=4-2\epsilon\). Differentiation with respect to the external
kinematics, followed by another integration-by-parts reduction, gives
\begin{equation}
\label{eq:general-differential-equation-master-integrals}
        \mathrm d\vec{\mathcal I}(\epsilon,s)
        =
        \mathbf A(\epsilon,s)\,
        \vec{\mathcal I}(\epsilon,s) \, .
\end{equation}
The matrix-valued one-form \(\mathbf A\) represents the Gauss--Manin
connection
\cite{Kotikov1991,Remiddi1997,GehrmannRemiddi2000,Henn2013,
Henn:2014yza}.

The geometric discussion of the Landau variety constrains the singularities
of this system. Away from \(\mathcal L\), the integration problem is locally
trivial, so the Gauss--Manin connection is regular and a basis of periods can
be analytically continued locally. Its genuine singular divisors must
therefore be contained in the codimension-one part of the Landau variety. In a particular basis,
however, the matrix \(\mathbf A\) may also contain \emph{apparent
singularities}: poles introduced by the choice of master integrals that
disappear after a change of basis and do not correspond to non-trivial
monodromy of the periods. Conversely, as emphasized in
Subsection~\ref{subsec:Landau Loci versus Actual Singularities}, a component
of the Landau variety need not be singular for every individual master
integral. The relevant inclusions are therefore schematically
\begin{equation}
\label{eq:singular-loci-landau-DE}
        \left\{\text{branch loci of $\vec{\mathcal I}(\epsilon,s)
$} \right\}
        \subseteq
        \left\{\text{singular divisors of $\mathbf A(\epsilon,s)$} \right\}
        \subseteq
        \mathcal L .
\end{equation}
The first inclusion can be strict because
a variation cycle or residue may vanish for a particular integral; the second
expresses that the Landau variety is the discriminant locus of the underlying
family of integration problems.

In the polylogarithmic setting, one seeks a basis in which this structure is
especially transparent. The relevant analytic objects are \emph{pure
functions}. Informally, a function \(F^{(w)}\) of transcendental weight \(w\)
is pure if its differential lowers the weight by one:
\begin{equation}
\label{eq:pure-function-differential}
        \mathrm dF^{(w)}
        =
        \sum_{\alpha}
        F_{\alpha}^{(w-1)}
        \,\mathrm d\log W_{\alpha}(s) \, ,
\end{equation}
with constant coefficients implicit in the sum. The functions
\(W_{\alpha}(s)\) are called \emph{letters}. Their zeros and poles define the
hypersurfaces around which the corresponding iterated integrals may have
monodromy. Accordingly, the divisors
\begin{equation}
        W_{\alpha}(s)=0
        \qquad\text{or}\qquad
        W_{\alpha}(s)=\infty
\end{equation}
must be compatible with the Landau variety of the integral family. This is
the differential-equation counterpart of using Landau analysis to constrain
a symbol alphabet, a topic we will review in Section~\ref{sec:Symbols and Bootstrap}.

This relation requires some care. Landau analysis predicts candidate
singular hypersurfaces, but it does not by itself determine a unique alphabet.
Several letters can have zeros and poles supported on the same Landau
components, and algebraic letters may live naturally on a finite cover of
kinematic space. For example, a Landau discriminant
\(\Delta(s)=0\) may lead to letters involving
\(\sqrt{\Delta}\), even though the branch hypersurface itself is simply
\(\Delta=0\). Numerators, boundary conditions, and cancellations can also
remove candidate Landau components from the alphabet of a particular
integral. The Landau variety therefore constrains the possible letters, while
the differential equation determines which logarithmic one-forms actually
occur.

A Feynman integral has \emph{uniform transcendental weight} if the
coefficients of its Laurent expansion in \(\epsilon\) have homogeneous weight
order by order. A basis of uniformly transcendental integrals can often be
chosen so that~\eqref{eq:general-differential-equation-master-integrals}
takes the canonical \(\epsilon\)-factorized form
\begin{tcolorbox}[definitionbox]
\textbf{Canonical differential equation.}
\begin{equation}
\label{eq:canonical-differential-equation-pure-basis}
        \mathrm d\vec{\mathcal I}(\epsilon,s)
        =
        \epsilon
        \left(
            \sum_{\alpha}
            A_{\alpha}\,
            \mathrm d\log W_{\alpha}(s)
        \right)
        \vec{\mathcal I}(\epsilon,s) \, .
\end{equation}
\end{tcolorbox}\noindent
Here the matrices \(A_{\alpha}\) are constant, while the letters
\(W_{\alpha}\) encode the singular divisors of the connection. Expanding in
\(\epsilon\), the equation determines each coefficient by iterated
integration of logarithmic one-forms, once boundary data are supplied. The
explicit factor of \(\epsilon\) makes the uniform-weight structure manifest
and separates the constant monodromy data \(A_{\alpha}\) from the geometry of
the singular hypersurfaces \(W_{\alpha}=0\).

The residues of the connection also have a direct monodromy interpretation.
For a small loop around a divisor \(W_{\alpha}=0\), the canonical system has
local monodromy
\begin{equation}
\label{eq:canonical-DE-local-monodromy}
        M_{\alpha}
        =
        \exp\!\left(2\pi i\,\epsilon A_{\alpha}\right).
\end{equation}
This is the differential-equation realization of the monodromy discussed in
Subsection~\ref{subsec:Monodromy Discontinuities and Vanishing Cycles}.
Landau analysis identifies the loci around which monodromy may occur,
Picard--Lefschetz theory describes it through variations of cycles, and the
canonical differential equation represents the same monodromy through the
residue matrices \(A_{\alpha}\).

There are useful integrand-level diagnostics for finding pure bases.
Integrals with constant leading singularities, or with integrands admitting
a \(\mathrm d\log\) representation, are often good candidates for a pure
normalization
\cite{ArkaniHamed:2010kv,Henn2013}. This links the positive-geometric
perspective to differential equations: canonical forms have logarithmic
singularities, while their maximal residues determine leading singularities.
A normalization with constant maximal residues frequently removes algebraic
prefactors and exposes the pure transcendental function. The four-mass box
below provides the simplest example: its scalar normalization has leading
singularity \(1/\sqrt{\Delta}\), whereas multiplication by
\(\sqrt{\Delta}\) produces constant residues and a pure function of weight two.

These implications are useful but not automatic. Constant leading
singularities, \(\mathrm d\log\) integrands, uniform transcendentality, and
canonical differential equations are closely related but distinct
properties. Moreover, not every logarithmic integral is polylogarithmic, and
not every integral family admits a global form of
\eqref{eq:canonical-differential-equation-pure-basis}
\cite{BrownDuhr2020DlogNotPolylog}. Elliptic and more general periods require
generalized notions of purity and more general differential-equation forms
\cite{AdamsWeinzierl2018,FrellesvigEtAl2021,
PoegelWangWeinzierl2023,GoergesNegaTancrediWagner2023,
DuhrMaggioNegaSauerTancrediWagner2025}. Nevertheless, the underlying
geometric relation persists: the differential equations describe the
variation of periods on the complement of the Landau variety, and their
singularities record the degeneration and monodromy of that family.

\subsubsection{Boxes in momentum twistors}
\label{subsec:Boxes in Momentum Twistors}

We now consider one-loop box integrals in momentum-twistor space. They provide
a standard example in the study of Feynman integrals and Landau singularities
\cite{Hodges:2009hk,Goncharov:2010jf,DennenSpradlinVolovich2016,Hannesdottir:2021kpd,Heller:2019gkq}, and they allow us to illustrate how to
pass from an integral representation in momentum twistors to a more standard
parametric representation.

In momentum twistor variables, a loop variable is represented by a line
\(AB\subseteq\mathbb P^3\). We consider a one-loop box integrand of the form
\begin{equation}
\label{eq:four-mass-box-momentum-twistor-integrand}
        \mathbf\Omega_{{\rm box}}(AB;X)
        =
        \frac{
        \langle X_1 X_3\rangle
        \langle X_2 X_4\rangle
        \langle AB\,\mathrm{d}^2A\rangle
        \langle AB\,\mathrm{d}^2B\rangle
        }{
        \langle ABX_1\rangle
        \langle ABX_2\rangle
        \langle ABX_3\rangle
        \langle ABX_4\rangle
        } \, ,
\end{equation}
and the corresponding integral
\begin{equation}
\label{eq:int_box}
        \mathcal I_{{\rm box}}(X)
        =
        \int_{\Gamma_X}\mathbf\Omega_{{\rm box}}(AB;X) \, .
\end{equation}
Here \(X_i\in\mathbb P(\wedge^2\mathbb C^4)\) are the four lines dual to the
dual-momentum points of the box. For the generic four-mass configuration we may
take
\begin{equation}
\label{eq:Xi_4m}
        X_1=(12),
        \qquad
        X_2=(34),
        \qquad
        X_3=(56),
        \qquad
        X_4=(78) \, ,
\end{equation}
where the notation $(ab)$ indicates the projective class of $Z_a \wedge Z_b$, with $Z_a,Z_b$ generic points in $\mathbb{P}^3$.
The four massive corners and the two Mandelstam invariants are encoded by
\begin{equation}
\label{eq:X_i_to_p}
        p_i^2=m_i^2\sim \langle X_iX_{i+1}\rangle,
        \qquad
        s=(p_1+p_2)^2\sim \langle X_1X_3\rangle,
        \qquad
        t=(p_2+p_3)^2\sim \langle X_2X_4\rangle \, .
\end{equation}
The notation \(\sim\) indicates equality up to conventional factors involving
the infinity twistor, as discussed in Section~\ref{sec:Momentum Twistors}.

The Minkowski contour can be described using Hodges' momentum-twistor Gaussian
integral \cite{Hodges:2009hk}. For any bi-twistor \(Q\in\wedge^2\mathbb C^4\),
one has
\begin{equation}
\label{eq:hodges-gaussian-momentum-twistor}
        \int_\Gamma
        \frac{
        \langle AB\,\mathrm{d}^2A\rangle
        \langle AB\,\mathrm{d}^2B\rangle
        }{
        \langle AB\,Q\rangle^4
        }
        =
        \frac{1}{\Gamma(4)}
        \frac{1}{
        \left(\frac12\langle Q\,Q\rangle\right)^2
        } \, .
\end{equation}
Introducing Schwinger parameters \(\alpha_i\), we combine the denominator
factors by writing
\begin{equation}
        Q(\alpha)
        =
        \alpha_1X_1+\alpha_2X_2+\alpha_3X_3+\alpha_4X_4 \, .
\end{equation}
Then, up to normalization, the loop-line integral reduces to the parametric integral
\begin{equation}
\label{eq:four-mass-parametric}
        \mathcal I_{{\rm box}}(X)
        =
        \int_{\mathbb R^4_{\geq 0}}
        \frac{\mathrm{d}^4\alpha}{{\rm GL}(1)}
        \frac{st}{\mathcal F(\alpha)^2} \, ,
\end{equation}
up to a numerical normalization factor. Here
\begin{equation}
\label{eq:four-mass-F}
        \mathcal F(\alpha)
        =
        \frac12\langle Q(\alpha)Q(\alpha)\rangle
        =
        \sum_{1\leq i<j\leq4}
        G_{ij}\alpha_i\alpha_j \, ,
\end{equation}
where $G_{ij}=\langle X_iX_j\rangle$, so $G_{ii}=0$.
Since \(\mathcal F\) is a quadratic form in the $\alpha_i$, the Landau equations imply that the
matrix \(G\), or one of its minors, has a non-zero kernel. 

Introduce the cross-ratios
\begin{equation}
\label{eq:four-mass-box-cross-ratios}
        u=
        \frac{
        \langle X_1X_2\rangle
        \langle X_3X_4\rangle
        }{
        \langle X_1X_3\rangle
        \langle X_2X_4\rangle
        },
        \qquad
        v=
        \frac{
        \langle X_2X_3\rangle
        \langle X_1X_4\rangle
        }{
        \langle X_1X_3\rangle
        \langle X_2X_4\rangle
        } \, .
\end{equation}
We can then write
\begin{equation}
\label{eq:four-mass-box-discriminant}
        \det(G)
        \quad \propto \quad 
        \langle X_1X_3\rangle^2
        \langle X_2X_4\rangle^2
        \Delta,
        \qquad
        \Delta=(1-u-v)^2-4uv \, .
\end{equation}
Thus the leading Landau locus is
\begin{equation}
\label{eq:Delta}
       \det(G) = 0 \quad \iff \quad  \Delta=0 \, .
\end{equation}
This is the four-mass-box branch locus~\cite{Goncharov:2010jf,DennenSpradlinVolovich2016,Hannesdottir:2021kpd}.

The same quantity appears from the maximal residue of the box integrand. The
loop line \(AB\) has four degrees of freedom, so one can impose the four cut
conditions
\begin{equation}
\label{eq:box_cut}
        \langle AB X_i\rangle=0,
        \qquad
        i=1,\ldots,4 \, .
\end{equation}
Geometrically, the solutions are lines in \(\mathbb P^3\) meeting the four
fixed lines \(X_i\). For four generic lines in \(\mathbb P^3\), this is the
classical \emph{Schubert problem} of finding the two transversals to the four lines.
We denote the two solutions by $AB^{(\pm)}(X)$.
The vanishing of \(\Delta\) is precisely the locus where these two transversals collide:
\begin{equation}
        \Delta=0
        \quad\Longleftrightarrow\quad
        AB^{(+)}(X)=AB^{(-)}(X) \, .
\end{equation}
Therefore \(\Delta\) is both the leading Landau singularity and the
discriminant of the Schubert problem of transverse lines to the four external lines $X_i$.

The four-fold residue of the rational form in
\eqref{eq:four-mass-box-momentum-twistor-integrand} at either solution $AB^{(\pm)}(X)$ is
\begin{tcolorbox}[definitionbox]
\textbf{Four-mass leading singularity.}
\begin{equation}
\label{eq:four-mass-leading-singularity}
        {\rm Res}_{AB=AB^{(\pm)}(X)}
        \, \mathbf\Omega_{{\rm box}}
        =
        \pm
        \frac{1}{\sqrt{\Delta}} \, .
\end{equation}
\end{tcolorbox}\noindent
Multiplying the scalar box by \(\sqrt{\Delta}\) gives the \emph{pure} normalization, in which the maximal residues are \(\pm 1\). With this normalization the form in~\eqref{eq:four-mass-box-momentum-twistor-integrand} can be interpreted as the canonical form of a positive geometry in $\Gr(2,4)$.

To write the integrated answer, introduce variables \(z,\bar z\) by
\begin{equation}
\label{eq:z-zbar-def}
        u=z\bar z,
        \qquad
        v=(1-z)(1-\bar z) \, .
\end{equation}
These variables rationalize the square root, in the sense that
\begin{equation}
\label{eq:z-zbar-sol}
        z,\bar z
        =
        \frac{1+u-v\pm\sqrt{\Delta}}{2},
        \qquad
        z-\bar z=\sqrt{\Delta} \, .
\end{equation}
The scalar four-mass box then evaluates to
\begin{tcolorbox}[definitionbox]
\textbf{Four-mass box function.}
\begin{equation}
\label{eq:four-mass-box-result}
        \mathcal I_{4{\rm m}}(z,\bar z)
        =
        \frac{1}{z-\bar z}
        \left[
        2\bigl({\rm Li}_2(z)-{\rm Li}_2(\bar z)\bigr)
        +
        \log(z\bar z)
        \log\!\left(\frac{1-z}{1-\bar z}\right)
        \right] \, ,
\end{equation}
\end{tcolorbox}\noindent
with the branch determined by the \(i\varepsilon\)-prescription
\cite{Bern:1994cg,Goncharov:2010jf,Duhr:2011zq,Elvang:2013cua}. The prefactor
\(1/(z-\bar z)=1/\sqrt{\Delta}\) is the leading singularity of $\mathcal I_{4{\rm m}}$. In the pure normalization, this square-root prefactor is removed, leaving a uniform-weight function.

The massless box is obtained by degenerating the four lines so that adjacent
lines intersect. In the fully massless case we can set
\begin{equation}
\label{eq:Xi_0m}
        X_1=(12),
        \qquad
        X_2=(23),
        \qquad
        X_3=(34),
        \qquad
        X_4=(41) \, ,
\end{equation}
so that all corners are massless, $\langle X_iX_{i+1} \rangle = 0$ for $i=1,\dots,4$.
The two Schubert solutions become rational,
\begin{equation}
        AB=13,
        \qquad
        AB=24 \, .
\end{equation}
The four-mass square root disappears, but infrared regions appear in loop-line
space. The massless scalar box is therefore divergent and must be regulated, for
example in dimensional regularization with \(D=4-2\epsilon\). From the
cohomological viewpoint, the regulated integral belongs to a twisted de Rham
setting, with \(\epsilon\) controlling the local monodromy around boundary
divisors \cite{AomotoKita2011,Euler_Integrals,MastroliaMizera2019}. Different mass regularizations, reached by introducing a small mass for each internal propagator, have also a geometric meaning in momentum twistors, and were considered for example in~\cite{Hodges:2009hk}.

This example anticipates the next section. The same incidence problem that
controls the leading Landau singularity of the four-mass box also controls its
maximal residue. In general, such maximal residues are leading singularities,
and in momentum-twistor space they are governed by Schubert problems and their
higher-loop analogues.

\subsection{Leading singularities and maximal cuts}
\label{sec:Leading Singularities and Maximal Cuts}

The previous section explained how discontinuities of integrals are related to
residues of the integrand. We now focus on the maximal version of this
operation. In four dimensions, a loop integrand with \(\ell\) loops depends on
\(4\ell\) complex loop variables. A maximal cut imposes \(4\ell\) independent
on-shell conditions. If these conditions localize all loop variables to isolated
points, the corresponding multidimensional residues are called leading
singularities.

Leading singularities are important for several reasons. They are the basic
algebraic data probed by generalized unitarity. In prescriptive unitarity, they
can be used as coordinates on the space of integrands. In planar
\(\mathcal N=4\) SYM, they are Yangian invariants and have a natural
Grassmannian description. Finally, in the Amplituhedron, they are boundary
residues of canonical forms and therefore admit a geometric interpretation in
terms of incidence configurations of lines in projective space.

\subsubsection{Leading singularities as maximal residues}
\label{subsec:Leading Singularities as Maximal Residues}

We first fix terminology. In this section, a \emph{leading singularity} means a
maximal-codimension residue of a loop integrand. This should be distinguished
from a \emph{leading Landau singularity}, which is a component of the Landau
variety associated with a Landau branch for which all Feynman parameters are
nonzero. It should also be distinguished from an actual singularity of the
integrated function. These notions are related, but not identical.

In the language of generalized unitarity, cutting an internal propagator puts an
internal particle on shell. In supersymmetric theories, the sum over internal
states is implemented by integrating over the corresponding Grassmann variables.
Thus a maximal cut is obtained by gluing lower-point on-shell amplitudes along
common on-shell legs. At one loop, a maximal cut has the topology of a box. For
a box with corner multiplicities
\(\mathbf n=(n_1,n_2,n_3,n_4)\), one writes schematically
\begin{equation}
\label{eq:LS-four-point-box-onshell}
        {\rm Cut}^{(1)}[{\rm Box}_{\mathbf n}]
        =
        \int
        \prod_{j=1}^{4}
        \mathrm{d}^4\eta_j\,\mathrm{d}q_j
        \prod_{r=1}^{4}
        A^{(0)}_{n_r}
        \bigl(
        q_r,\eta_r;
        -q_{r+1},\eta_{r+1};
        \ldots
        \bigr) \, ,
\end{equation}
where \(r\) is taken modulo \(4\). The measure \(\mathrm{d}q_j\) denotes the
appropriate on-shell measure for the internal momentum \(q_j\). In physical
kinematics it contains a real on-shell delta function, while in the complex
residue interpretation it imposes the complex on-shell condition. The formula
is schematic in that, after imposing the bosonic on-shell constraints, one must
sum over all isolated solutions with the corresponding Jacobian \cite{Nair:1988bq,BernDixonDunbarKosower1994,Britto:2004ap,Britto:2005fq,Grassmannian,Elvang:2013cua,Henn:2014yza}.

The number of external particles of the box cut is
\begin{equation}
\label{eq:n-box-from-corners}
        n=\sum_{r=1}^{4}n_r-8 \, .
\end{equation}
If the tree amplitude at corner \(r\) lies in the
\({\rm N}^{k_r}{\rm MHV}\) sector, the one-loop box cut
contributes to helicity degree
\begin{equation}
\label{eq:k-box-from-corners}
        k=\sum_{r=1}^{4}k_r+2 \, .
\end{equation}

More generally, at \(\ell\) loops a maximal cut is specified by a planar cut
graph \(\mathcal L\). Its edges are the propagators put on shell. We call such
a graph a \emph{leading Landau diagram}. To a leading Landau diagram with
vertices \(v=1,\ldots,V\) and \(I\) internal edges, we attach two integer
vectors
\begin{equation}
        \mathbf n=(n_1,\ldots,n_V),
        \qquad
        \mathbf k=(k_1,\ldots,k_V) \, ,
\end{equation}
encoding the multiplicity and helicity degree of the tree amplitude at each
vertex. The number of external particles and the total helicity degree are
\begin{equation}
\label{eq:n-k-leading-landau-diagram}
        n=\sum_{v=1}^{V}n_v-2I,
        \qquad
        k=\sum_{v=1}^{V}k_v+2V-I-2 \, .
\end{equation}
The second formula follows by comparing Grassmann degrees: the product of the
vertex superamplitudes has degree \(4\sum_v(k_v+2)\), while integration over
the \(I\) internal on-shell superspaces subtracts \(4I\). Equating the result
to \(4(k+2)\) gives \eqref{eq:n-k-leading-landau-diagram}. The corresponding
cut is denoted
\begin{equation}
        {\rm Cut}^{(\ell)}
        [\mathcal L_{\mathbf k,\mathbf n}] \, .
\end{equation}

\subsubsection{One-loop prescriptive unitarity}
\label{subsec:One-Loop Prescriptive Unitarity}

We now explain how leading singularities are used operationally at one loop. By the duality between discontinuities, namely cuts, and residues, discussed in Subsection~\ref{subsec:Discontinuities as Residues}, leading singularities correspond to maximal residues of loop integrands. The practical power of this notion is that, after choosing a
basis of integrands, these residues can determine the coefficients of the
amplitude. This is the idea of \emph{generalized unitarity}, and in a particularly
sharp form, of \emph{prescriptive unitarity}.

Historically, the starting point is ordinary unitarity, which relates
discontinuities of loop amplitudes to products of lower-loop or tree-level
amplitudes by cutting propagators. This follows from \(S\)-matrix unitarity and
is implemented perturbatively by the Cutkosky rules
\cite{Cutkosky1960,Veltman1963,Eden:1966dnq,BernDixonDunbarKosower1994}. Generalized
unitarity refines this idea by imposing several cut conditions simultaneously,
possibly enough to localize the loop momentum completely
\cite{BernDixonDunbarKosower1994,Bern:1994zx,Britto:2004nc,Forde:2007mi,Mastrolia:2009dr,Elvang:2013cua,Henn:2014yza}.
At one loop in four dimensions, the loop momentum has four complex degrees of
freedom. A quadruple cut imposes four on-shell conditions, and, for generic
kinematics, localizes the loop variable to finitely many solutions. The
corresponding residues are the leading singularities.

The practical use of generalized unitarity is tied to integral reduction. Once
a basis of one-loop integrals has been chosen, the amplitude can be written as a
linear combination of basis elements with unknown coefficients, and cuts are
used to determine these coefficients. For a massless four-dimensional theory
with one-loop amplitudes understood in dimensional regularization, integral
reduction takes the schematic form
\begin{equation}
\label{eq:one-loop-integral-reduction-general}
        A_n^{(\ell=1)}
        =
        \sum_i c^{(i)}_4\,\mathcal I^{(i)}_4
        +
        \sum_j c^{(j)}_3\,\mathcal I^{(j)}_3
        +
        \sum_k c^{(k)}_2\,\mathcal I^{(k)}_2
        +
        \cdots \, .
\end{equation}
Here \(\mathcal I_4\), \(\mathcal I_3\), and \(\mathcal I_2\) denote scalar
boxes, triangles, and bubbles, while the dots indicate possible rational terms
and regulator-dependent contributions that are not detected by strictly
four-dimensional cuts.
Ordinary two-particle cuts constrain several coefficients at once. Generalized
cuts impose more on-shell conditions and can isolate individual coefficients.
In particular, quadruple cuts in four dimensions localize the loop momentum and
determine box coefficients directly \cite{Britto:2004nc,Forde:2007mi}.

For planar \(\mathcal N=4\) SYM the one-loop story simplifies dramatically.
The amplitude is cut-constructible in four dimension, rational terms are absent, and the integrated answer can be written in terms of scalar box
integrals only. This is the one-loop manifestation of the no-triangle property,
which follows from improved power counting and supersymmetric cancellations
\cite{BernDixonDunbarKosower1994,Bern:1994zx,BernDixonKosower1996,Bern:2007dw,Elvang:2013cua,Henn:2014yza}. Thus in
\eqref{eq:one-loop-integral-reduction-general} one may set the triangle and
bubble coefficients to zero. The remaining box coefficients are fixed by
quadruple cuts, and are precisely the leading singularities obtained by gluing
four tree amplitudes on the cut.

For the planar integrand, however, it is more natural to use a basis adapted to
momentum twistors and to leading singularities, rather than the scalar-box
basis. The local-integral representation of
\cite{ArkaniHamed:2010kv} replaces scalar boxes by
\emph{chiral} integrands. These integrands are local and dual conformal, and
their numerators are chosen so that they have unit residue on one chosen
quadruple-cut solution and vanish on the other. This is the one-loop prototype
of a prescriptive basis: a basis of integrands chosen so that a specified set
of generalized-unitarity cuts isolates, or diagonalizes, the corresponding
coefficients~\cite{BourjailyTrnka2015,BourjailyHerrmannTrnka2017}.

Let us first recall the global one-loop MHV integrand. The canonical form of
the \(n\)-point one-loop MHV ($k=0$) Amplituhedron gives a pure global integrand of the
form
\begin{equation}
\label{eq:one-loop-common-denominator}
        \mathbf\Omega^{(\ell=1)}_{0,n}(AB;Z)
        =
        \frac{
        \sum_r
        c_r^{(n)}
        \langle AB\,X^r_1\rangle
        \langle AB\,X^r_2\rangle
        \cdots
        \langle AB\,X^r_{n-4}\rangle
        }{
        \langle AB\,12\rangle
        \langle AB\,23\rangle
        \cdots
        \langle AB\,n1\rangle
        }
        \,
        \langle AB\,\mathrm{d}^2A\rangle
        \langle AB\,\mathrm{d}^2B\rangle \, .
\end{equation}
Here the denominator collects all planar one-loop propagators, while the
\(X^r_a\) are fixed bi-twistors specifying the numerator. The numerator has
degree \(n-4\), as required by projective invariance in the loop line \(AB\).
Although \eqref{eq:one-loop-common-denominator} is a pure global integrand, it
is not the most useful representation for integration or for matching cuts. A
basis of simpler local integrands, diagonalizing the relevant maximal residues,
is preferable.

The appropriate basis is given by chiral pentagons. For MHV amplitudes, the only
non-vanishing one-loop leading singularities are the two-mass-easy cuts. These
are the cuts for which the loop line \(AB\) is forced to pass through two
external twistors \(Z_i\) and \(Z_j\). Explicitly, for fixed
\(1\leq i<j\leq n\), the cut equations are
\begin{equation}
\label{eq:two-mass-easy-cut-equations}
        \langle AB\,i{-}1\,i\rangle
        =
        \langle AB\,i\,i{+}1\rangle
        =
        \langle AB\,j{-}1\,j\rangle
        =
        \langle AB\,j\,j{+}1\rangle
        =
        0 \, .
\end{equation}
There are two solutions:
\begin{equation}
\label{eq:two-mass-easy-cut-solutions}
        AB=ij \, ,
        \qquad
        AB=\overline{ij}:=
        (i{-}1\,i\,i{+}1)\cap(j{-}1\,j\,j{+}1) \, ,
\end{equation}
where the latter denoted the line given by the intersection of the planes $(i{-}1\,i\,i{+}1)$ and $(j{-}1\,j\,j{+}1)$ in $\mathbb{P}^3$.
The first solution, \(AB=ij\), is the branch relevant for the MHV leading
singularity. The second solution is the conjugate branch. A scalar box with the
four propagators in \eqref{eq:two-mass-easy-cut-equations} generally has
nonzero residues on both branches. A chiral pentagon is obtained by adding a
fifth propagator and a numerator which kills the unwanted branch.

Let \(X\) be an arbitrary reference line defining the fifth propagator. The
chiral-pentagon form associated with the cut \((i,j)\) is
\begin{equation}
\label{eq:chiral-pentagon-integrand}
        \mathbf\Omega_{ij}^{\rm cp}
        =
        \frac{
        \langle AB\,\overline{ij}\rangle\,
        \langle X\,ij\rangle\,
        \langle AB\,\mathrm{d}^2A\rangle
        \langle AB\,\mathrm{d}^2B\rangle
        }{
        \langle AB\,i{-}1\,i\rangle
        \langle AB\,i\,i{+}1\rangle
        \langle AB\,j{-}1\,j\rangle
        \langle AB\,j\,j{+}1\rangle
        \langle AB\,X\rangle
        } \, .
\end{equation}
The numerator is chosen so that
\begin{equation}
\label{eq:chiral-pentagon-unit-LS}
        {\rm Res}_{AB=ij}\,\mathbf\Omega_{ij}^{\rm cp}=1,
        \qquad
        {\rm Res}_{AB=\overline{ij}}\,
        \mathbf\Omega_{ij}^{\rm cp}=0 \, ,
\end{equation}
up to the orientation convention for the residue. This is why the integral is
called \emph{chiral}: it selects one of the two solutions of the quadruple cut.

Chiral pentagons are therefore better adapted to generalized unitarity than
scalar boxes, because their coefficients are read off from individual leading
singularities rather than from sums of leading singularities
\cite{ArkaniHamed:2010kv,Grassmannian}. They fully
diagonalize the one-loop leading singularities of the amplitude.

For MHV amplitudes, the chiral-pentagon expansion takes an especially simple
form. The non-vanishing leading singularities are all proportional to the
tree-level MHV amplitude. Therefore one obtains the higher-multiplicity
one-loop MHV formula
\begin{tcolorbox}[definitionbox]
\textbf{One-loop MHV ratio.}
\begin{equation}
\label{eq:one-loop-MHV-ratio}
        A^{(\ell=1)}_{0,n}
        =
        A^{(\ell=0)}_{0,n}
        \sum_{1\leq i<j\leq n}
        \mathcal I^{\rm cp}_{ij}\Big|_{X=(1n)} \, .
\end{equation}
\end{tcolorbox}\noindent
Here \(\mathcal I^{\rm cp}_{ij}\) denotes the integral of the chiral-pentagon
form \(\mathbf\Omega^{\rm cp}_{ij}\). For generic \(i,j\), this integral is a
finite dual-conformal function of three cross-ratios:
\begin{equation}
\label{eq:chiral-pentagon-integrated}
\begin{aligned}
        \mathcal I^{\rm cp}_{ij}
        =
        &\log(u)\log(v)
        +{\rm Li}_2(1-u)
        +{\rm Li}_2(1-v)
        +{\rm Li}_2(1-w)
        -{\rm Li}_2(1-uw)
        -{\rm Li}_2(1-vw) \, ,
\end{aligned}
\end{equation}
where
\begin{equation*}
\begin{aligned}
        u
        &=
        \frac{
        \langle X\,i{-}1\,i\rangle
        \langle j\,j{+}1\,i\,i{+}1\rangle
        }{
        \langle X\,i\,i{+}1\rangle
        \langle i{-}1\,i\,j\,j{+}1\rangle
        } \, , \ 
        v
        =
        \frac{
        \langle X\,j\,j{+}1\rangle
        \langle i{-}1\,i\,j{-}1\,j\rangle
        }{
        \langle X\,j{-}1\,j\rangle
        \langle j\,j{+}1\,i{-}1\,i\rangle
        } \, , \ 
        w
        =
        \frac{
        \langle i{-}1\,i\,j\,j{+}1\rangle
        \langle j{-}1\,j\,i\,i{+}1\rangle
        }{
        \langle i{-}1\,i\,j{-}1\,j\rangle
        \langle j\,j{+}1\,i\,i{+}1\rangle
        } .
\end{aligned}
\end{equation*}
The reference line \(X=(1n)\) is a convenient choice in the representation
\eqref{eq:one-loop-MHV-ratio}; the final amplitude is cyclic, although this is
not manifest term by term. The function
\(\mathcal I^{\rm cp}_{ij}\) is pure of transcendental weight two. 

For boundary values of the labels, some chiral pentagons degenerate to box
integrals. For instance, when \(j=i+1\), the pentagon collapses to a box
topology. These boundary integrals contain the soft and collinear infrared
divergences of the massless one-loop amplitude and must be regulated, for
example by dimensional regularization. After integration, the chiral-pentagon
representation is equivalent to the familiar scalar-box expansion, but before
integration it makes locality, dual conformal symmetry, chirality, and the
leading-singularity structure manifest
\cite{ArkaniHamed:2010kv,Elvang:2013cua,Henn:2014yza}.

For higher helicity sectors the same integrand basis can be used, but the
coefficients are no longer constants after factoring out the MHV tree
amplitude. Rather, the one-loop \({\rm N}^k{\rm MHV}\) integrand has the
schematic prescriptive form
\begin{equation}
\label{eq:NkMHV-one-loop-chiral-pentagon-expansion}
        \mathbf{\Omega}^{(\ell=1)}_{k,n}
        =
        \sum_{1\leq i<j\leq n}
        {\rm LS}^{(k)}_{ij}\,
        \mathbf{\Omega}^{\rm cp}_{ij} \, .
\end{equation}
Here \({\rm LS}^{(k)}_{ij}\) is the leading singularity of the
\({\rm N}^k{\rm MHV}\) amplitude on the two-mass-easy cut
\eqref{eq:two-mass-easy-cut-equations}, equivalently the Yangian invariant
computed by the corresponding on-shell diagram. Thus the bosonic chiral
pentagon forms provide the local pure one-loop basis, while all helicity
dependence is carried by the on-shell coefficients. For \(k=0\), these
coefficients reduce to the MHV tree amplitude on the non-vanishing cuts; for
\(k>0\), they are non-trivial Yangian invariants of Grassmann degree \(4k\)
\cite{ArkaniHamed:2010kv,Grassmannian,the_amplituhedron,BourjailyHerrmannTrnka2017}.

For MHV amplitudes the result simplifies further. The relevant leading
singularities are all equal to the tree-level MHV amplitude. After factoring
out \(A^{(\ell=0)}_{0,n}\), the one-loop MHV ratio is therefore a sum of pure
chiral pentagon integrands with unit coefficients on the allowed MHV cuts. For
\(k>0\), the coefficients in
\eqref{eq:NkMHV-one-loop-chiral-pentagon-expansion} are non-trivial Yangian
invariants of Grassmann degree \(4k\)
\cite{ArkaniHamed:2010kv,Grassmannian,the_amplituhedron}.

It is useful to describe the integrated higher-\(k\) structure using
loop-level \textit{ratio functions}. We factor out the tree-level MHV
superamplitude and write the superamplitude as
\begin{equation}
\label{eq:loop-expansion-P-functions}
        A_{k,n}
        =
        A^{(\ell=0)}_{0,n}
        \left(
        \mathcal P^{(\ell=0)}_{k,n}
        +
        \lambda\,
        \mathcal P^{(\ell=1)}_{k,n}(\epsilon)
        +
        \mathcal O(\lambda^2)
        \right) \, ,
\end{equation}
where \(\lambda\) is the planar 't Hooft coupling and
\(\epsilon\) is the infrared regulator. This normalization is standard in
the study of planar \(\mathcal N=4\) SYM superamplitudes and makes dual
superconformal covariance manifest at tree level
\cite{Drummond:2008vq,ArkaniHamed:2009dn,Grassmannian,Elvang:2013cua}.
The tree-level ratio function
\(\mathcal P^{(\ell=0)}_{k,n}\) is the tree-level
\({\rm N}^k{\rm MHV}\) ratio function reviewed in
Section~\ref{sec:On-Shell Diagrams and Grassmannian Contours}.

The one-loop amplitude itself is infrared divergent. In dimensional
regularization, with
\begin{equation}
        D=4-2\epsilon \, ,
\end{equation}
the divergent contribution is universal: it multiplies the tree-level
amplitude and is already present in the MHV sector. More precisely, the
anomalous dual conformal contribution is the same for MHV and non-MHV
superamplitudes, and therefore cancels in a ratio function
\cite{Drummond:2008vq,DrummondHennKorchemskySokatchev2008,BrandhuberHeslopTravaglini2009,ElvangFreedmanKiermaier2009}. One therefore
defines the one-loop finite ratio function by subtracting the universal MHV
infrared-divergent contribution:
\begin{tcolorbox}[definitionbox]
\textbf{One-loop ratio function.}
\begin{equation}
\label{eq:one-loop-ratio-function-general-k}
        \mathcal R^{(\ell=1)}_{k,n}(\epsilon)
        :=
        \mathcal P^{(\ell=1)}_{k,n}(\epsilon)
        -
        \mathcal P^{(\ell=0)}_{k,n}\,
        \mathcal P^{(\ell=1)}_{0,n}(\epsilon) \, .
\end{equation}
\end{tcolorbox}\noindent
This combination is infrared finite as \(\epsilon\to0\), and the finite
quantity
\begin{equation}
\label{eq:finite-one-loop-ratio-function-general-k}
        \mathcal R^{(\ell=1)}_{k,n}(0)
        =
        \lim_{\epsilon\to0}
        \mathcal R^{(\ell=1)}_{k,n}(\epsilon)
\end{equation}
captures the genuinely non-MHV part of the one-loop amplitude after the
universal MHV divergence has been removed.

In the NMHV sector, the ratio functions are sums of \(R\)-invariants dressed
by dual-conformal transcendental functions of cross-ratios. The one-loop
six-point NMHV ratio function is the first nontrivial example and was obtained
from generalized unitarity in early studies of anomalous dual conformal
symmetry. The anomaly here refers to the breaking of dual conformal symmetry
after integration by the infrared regulator; after subtracting the universal
MHV infrared contribution, the ratio function is dual conformal
\cite{Drummond:2008vq,DrummondHennKorchemskySokatchev2008,ElvangFreedmanKiermaier2009}. The all-\(n\) one-loop NMHV ratio function was
then shown to be dual conformal after the infrared subtraction
\cite{ElvangFreedmanKiermaier2009}. Higher-loop six-point NMHV ratio functions
have also been computed, beginning with the two-loop analytic result and later
through bootstrap methods
\cite{DixonDrummondHenn2011,Dixon:2016apl}. We will not need those explicit
formulae here. What matters for the present chapter is the structural lesson:
before integration, generalized unitarity expresses the amplitude in terms of
pure chiral integrands weighted by Yangian-invariant leading singularities;
after integration, the universal infrared-divergent MHV factor is stripped off,
and the finite ratio functions encode the nontrivial helicity dependence.

\begin{tcolorbox}[resultbox]
\textbf{One-loop prescriptive unitarity.}
At one loop, prescriptive unitarity diagonalizes the maximal cuts using a
chiral-pentagon basis. After integration, the universal MHV infrared divergence
is removed by forming ratio functions, leaving finite dual-conformal objects
which carry the non-MHV helicity data.
\end{tcolorbox}\noindent

This one-loop discussion prepares the geometric language of the next
subsections. The quadruple cuts of chiral pentagons are Schubert problems for
the loop line \(AB\) in \(\mathbb P^3\), and the distinction between the two
solutions \(AB=ij\) and \(AB=\overline{ij}\) is the simplest example of how
numerators select some boundaries and remove others. The Grassmannian and
Amplituhedron descriptions developed below make this selection geometric.

\subsubsection{On-shell diagrams and Grassmannian localization}
\label{subsec:On-Shell Diagrams and Grassmannian Localization}

Each tree amplitude appearing at a vertex of a cut graph may be expanded using
BCFW recursion. Iterating the recursion down to three-point amplitudes expresses
each term as an on-shell diagram built from three-point MHV and
\(\overline{\rm MHV}\) vertices
\cite{Britto:2004ap,Britto:2005fq,ArkaniHamed:2009dn,Grassmannian}. Thus a maximal cut can be expanded as a finite sum of on-shell
forms,
\begin{equation}
\label{eq:cut-onshell-diagrams}
        {\rm Cut}^{(\ell)}
        [\mathcal L_{\mathbf k,\mathbf n}]
        =
        \sum_b
        {\rm Tr}\,
        \mathbf\Omega_{\Gamma^{(b)}} \, .
\end{equation}
Here $\Gamma^{(b)}$ are on-shell diagrams obtained by inserting
BCFW expansions into the vertices of the cut graph, and
\(\mathbf\Omega_{\Gamma^{(b)}}\) denotes the corresponding on-shell
form. The symbol \({\rm Tr}\) indicates that the remaining bosonic cut equations
are imposed and that the form is pushed forward to external kinematics, as we explain now.

Recall that we dicussed on-shell diagrams and their forms in Section~\ref{sec:On-Shell Diagrams and Grassmannian Contours}.
Let \(C(\alpha)\) be a positroid parametrization of an on-shell diagram
\(\Gamma\). In the momentum-twistor setting, the associated Grassmannian contour integral takes the form
\begin{equation}
\label{eq:onshell-form-before-mcut}
        \mathbf\Omega_{\Gamma}
        =
        \int
        \prod_{i=1}^{4k}
        \frac{\mathrm{d}\alpha_i}{\alpha_i}\,
        \delta^{(4k)}
        \bigl(C(\alpha)\cdot Z\bigr)\,
        \delta^{(4k)}
        \bigl(C(\alpha)\cdot\chi\bigr) \, ,
\end{equation}
where \(\mathcal Z_i=(Z_i|\chi_i)\) are momentum supertwistors. For a Yangian
invariant of type \((k,n)\), the relevant positroid cell has dimension \(4k\).
For generic external data, the bosonic equations
\begin{equation}
\label{eq:bosonic-localization-equations}
        C(\alpha)\cdot Z=0
\end{equation}
localize the variables \(\alpha_i\) to finitely many solutions
\begin{equation}
\label{eq:alpha-solutions}
        \alpha^{(r)}(Z) \, ,
        \qquad
        r=1,\ldots,\gamma_\Gamma \, .
\end{equation}
The number \(\gamma_\Gamma\) is the degree of the Amplituhedron map restricted
to the positroid cell of \(\Gamma\); it is often called the \emph{intersection number} of the on-shell diagram \cite{Grassmannian,the_amplituhedron,Positive_geometries}.

The fermionic delta functions do not impose additional bosonic constraints, but
are evaluated on the solutions \eqref{eq:alpha-solutions}. Thus
\begin{equation}
\label{eq:MCut-pushforward}
        {\rm Tr}\,\mathbf\Omega_{\Gamma}
        =
        \sum_{r=1}^{\gamma_\Gamma}
        \left.
        \frac{
        \delta^{(4k)}
        \bigl(C(\alpha)\cdot\chi\bigr)
        }{
        \alpha_1\cdots\alpha_{4k}\,
        J(\alpha,Z)
        }
        \right|_{\alpha=\alpha^{(r)}(Z)} \, ,
\end{equation}
where
\begin{equation}
\label{eq:jacobian-onshell}
        J(\alpha,Z)
        =
        \det
        \left(
        \frac{
        \partial\bigl(C(\alpha)\cdot Z\bigr)
        }{
        \partial(\alpha_1,\ldots,\alpha_{4k})
        }
        \right) \, .
\end{equation}
The relevant objects for the following discussion are the individual terms in~\eqref{eq:MCut-pushforward}:
\begin{tcolorbox}[definitionbox]
\textbf{Leading singularity.}
\begin{equation}
\label{eq:LS_def}
        {\rm LS}^{(\ell)}_{k,n}[\Gamma,r]
        :=
        \left.
        \frac{
        \delta^{(4k)}
        \bigl(C(\alpha)\cdot\chi\bigr)
        }{
        \alpha_1\cdots\alpha_{4k}\,
        J(\alpha,Z)
        }
        \right|_{\alpha=\alpha^{(r)}(Z)} \, .
\end{equation}
\end{tcolorbox}\noindent
These are algebraic functions of the external bosonic momentum twistors. They
are often rational, but they need not be; the four-mass box gives the first
standard example of a square-root leading singularity
\cite{Grassmannian,Hodges:2009hk,Goncharov:2010jf}.

The Grassmannian representation makes Yangian invariance manifest. Every
leading singularity of planar \(\mathcal N=4\) SYM is a Yangian invariant, and
a maximal cut is a sum of such invariants, up to the usual overall tree-level
MHV factor \cite{Drummond:2010qh,Grassmannian,ArkaniHamed:2010kv}.

The helicity degree gives an important constraint. If an on-shell diagram
\(\Gamma\) is built from trivalent black and white vertices, let \(n_b\) and
\(n_w\) denote the number of black and white vertices, and let \(n_I\) denote
the number of internal edges. With the conventions of
Section~\ref{sec:On-Shell Diagrams and Grassmannian Contours}, its helicity
degree is
\begin{equation}
\label{eq:hel_Gamma}
        \deg(\Gamma)
        =
        n_b+2n_w-n_I-2 \, .
\end{equation}
Only on-shell diagrams of helicity degree at most \(k\) can contribute to the
\({\rm N}^k{\rm MHV}\) sector:
\begin{tcolorbox}[resultbox]
\textbf{Helicity-degree bound.}
\begin{equation}
\label{eq:LS_max_k}
        {\rm LS}^{(\ell)}_{k,n}[\Gamma,r]
        \text{ contributes to the }
        {\rm N}^k{\rm MHV}
        \text{ superamplitude}
        \ \Longrightarrow \
        \deg(\Gamma)\leq k \, .
\end{equation}
\end{tcolorbox}\noindent
Here by contributing we mean that it appears as a non-vanishing maximal residue of the ${\rm N}^k{\rm MHV}$ superamplitude.

For \(k=0\), the relevant Grassmannian is a point and the corresponding
on-shell function is \(1\), after stripping the tree-level MHV factor. Thus all
leading singularities of MHV superamplitudes in planar \(\mathcal N=4\) SYM
are equal to the tree-level MHV superamplitude:
\begin{equation}
        {\rm LS}^{(\ell)}_{0,n}
        =
        A^{(0)}_{0,n} \, .
\end{equation}
For \(k=1\), the relevant positive Grassmannian is a projective simplex, and
four-dimensional positroid cells have intersection number one. The associated
Yangian invariants are the familiar \(R\)-invariants
\([a,b,c,d,e]\) \cite{Drummond:2008vq,Grassmannian}. Hence all NMHV leading
singularities are integer combinations of \(R\)-invariants, multiplied by the
tree-level MHV superamplitude. The first genuinely algebraic leading
singularities appear in higher helicity sectors, starting with the four-mass
box in the \({\rm N}^2{\rm MHV}\) sector.

\subsubsection{Amplituhedron boundaries and LS configurations}
\label{subsec:Amplituhedron Boundaries and LS Configurations}

We now translate the Grassmannian description into the geometry of the loop
Amplituhedron. Recall from Section~\ref{sec:Loop Amplituhedra} that the
\(\ell\)-loop \({\rm N}^k{\rm MHV}\) Amplituhedron is described by bosonized
momentum supertwistors \(Z_i\in\mathbb C^{k+4}\), a point
\(Y\in\Gr(k,k+4)\), and loop planes
\begin{equation}
        AB_a\in\Gr(2,k+4) \, ,
        \qquad
        a=1,\ldots,\ell \, .
\end{equation}
After fixing \(Y=I_\infty\), or equivalently after projecting through \(Y\), the
external data and loop planes are viewed in the quotient
\begin{equation}
        \mathbb P^3
        \cong
        \mathbb P(\mathbb C^{k+4}/Y) \, .
\end{equation}
By abuse of notation, we continue to denote the projected external points by
\(Z_i\), and the projected loop lines by \(AB_a\).

The propagator divisors of the planar amplitude are the incidence equations
\begin{equation}
\label{eq:ampl_bd}
        \langle AB_a\,ii+1\rangle=0 \, ,
        \qquad
        \langle AB_a\,AB_b\rangle=0 \, .
\end{equation}
Since two lines
\(AB\) and \(CD\) in \(\mathbb P^3\) intersect if and only if $\langle AB\,CD\rangle=0$, maximal-cut equations become incidence problems for configurations of lines in projective three-space.

We call a solution of \(4\ell\) independent incidence equations of the form
\eqref{eq:ampl_bd} a \emph{leading-singularity configuration}, or
\emph{LS configuration}. More precisely, an LS configuration is a
zero-dimensional stratum in the arrangement on \(\Gr(2,4)^\ell\) defined
by the propagator equations \eqref{eq:ampl_bd} for fixed external data $Z_i$. At one loop, this is the
classical Schubert problem of finding lines in \(\mathbb P^3\) meeting four
prescribed lines. 

A leading Landau diagram may impose additional local conditions at trivalent
vertices. If three lines meet pairwise, they may either pass through a common
point or lie in a common plane. These are the two projective possibilities
corresponding to the two three-point amplitudes:
\begin{enumerate}
        \item a white, or \(\overline{\rm MHV}\), vertex corresponds to three
        lines meeting in a common point;
        \item a black, or MHV, vertex corresponds to three lines lying in a
        common plane.
\end{enumerate}
These point/plane alternatives are the momentum-twistor avatars of the
factorizations appearing in on-shell diagrams and in composite residues
\cite{Grassmannian,Dian_2023,Dian:2024hil}.

Up to the standard equivalence moves of on-shell diagrams, choices of residue
orientation, and possible global-residue relations, one obtains the schematic
identification
\begin{tcolorbox}[resultbox]
\textbf{Leading singularities as residues.}
\begin{equation}
\label{eq:LS_Res}
        \left\{
        {\rm LS}^{(\ell)}_{k,n}[\Gamma,r]
        \right\}
        =
        \left\{
        {\rm Res}_{(AB_1^{(r)},\ldots,A B_\ell^{(r)})=
        (AB_1,\ldots,A B_\ell)}
        \mathbf\Omega^{(\ell)}_{k,n}
        (Y,\{AB_a\};Z)
        \bigg|_{Y=I_\infty}
        \right\} \, .
\end{equation}
\end{tcolorbox}\noindent
The set on the left runs over on-shell diagrams and bosonic solutions of their
localization equations. The set on the right runs over LS configurations solving
the corresponding incidence problem. Passing explicitly between the two
descriptions can be done using vector-relation configurations: one assigns
projective points to regions of an on-shell diagram, with linear relations at
the vertices, and recovers the loop lines from suitable joins and intersections
\cite{Grassmannian,HolleringMazzucchelliParisiSturmfels2025Lines}. We come back to this construction in Subsection~\ref{sec:line-configurations-positroids-amplituhedron-map}.

Thus, for fixed \(n\), \(k\), and \(\ell\), the following notions are different
manifestations of the same underlying data:
\begin{tcolorbox}[definitionbox]
\textbf{Leading-singularity dictionary.}
\begin{equation}
\label{eq:LS-dictionary}
\begin{aligned}
        &\text{leading singularities}
        \quad\longleftrightarrow\quad
        \text{maximal cuts}
        \quad\longleftrightarrow\quad
        \text{on-shell diagrams}
        \\[2mm]
        &\quad\longleftrightarrow\quad
        \text{LS configurations}
        \quad\longleftrightarrow\quad
        \text{maximal residues of canonical forms}.
\end{aligned}
\end{equation}
\end{tcolorbox}\noindent
The helicity degree refines this dictionary. LS configurations of too high
helicity degree are residual for the \({\rm N}^k{\rm MHV}\) Amplituhedron:
they solve the incidence equations but do not correspond to boundary strata of
the positive region relevant for helicity \(k\). In this case the numerator of
the canonical form is expected to have vanishing residue on the corresponding
corner, in analogy with residual arrangements and adjoint hypersurfaces
discussed in Section~\ref{sec:Adjoint Hypersurface}.

\subsubsection{One-loop Schubert problems}
\label{subsec:One-Loop Schubert Problems}

We now spell out the one-loop case. A one-loop maximal cut is a box cut:
\begin{equation}
\label{eq:general-one-loop-schubert-cut}
        \langle AB\,X_1\rangle
        =
        \langle AB\,X_2\rangle
        =
        \langle AB\,X_3\rangle
        =
        \langle AB\,X_4\rangle
        =
        0 \, ,
\end{equation}
where \(AB\) is the loop line and \(X_r=(i_r,i_r{+}1)\) with $i_r \in [n]$ are four external lines
dual to planar propagators. Geometrically, this is the Schubert problem of
finding lines \(AB\subseteq\mathbb P^3\) incident to the four fixed lines
\(X_1,\ldots,X_4\). For generic lines, there are two solutions.

The classification of one-loop boxes by mass configuration corresponds
to degenerations of the four lines \(X_r\). A corner of the box is massless
precisely when the adjacent dual lines intersect. The possible one-loop topologies are
\begin{equation}
        0{\rm m},
        \qquad
        1{\rm m},
        \qquad
        2{\rm me},
        \qquad
        2{\rm mh},
        \qquad
        3{\rm m},
        \qquad
        4{\rm m} \, ,
\end{equation}
where \(2{\rm me}\) and \(2{\rm mh}\) denote the \emph{two-mass-easy} and
\emph{two-mass-hard} cases, respectively.

\begin{eg}[Zero-mass cut]
Consider the zero-mass box \({\rm Box}_{(3,3,3,3)}\). This forces \(n=4\), and
hence the cut contributes only to the four-point one-loop amplitude. The four
external lines may be taken to be
\begin{equation}
        X_1=(12) \, ,
        \qquad
        X_2=(23) \, ,
        \qquad
        X_3=(34) \, ,
        \qquad
        X_4=(41) \, .
\end{equation}
The Schubert problem has two solutions,
\begin{equation}
\label{eq:zero-mass-box-schubert-solutions}
        AB^{(1)}=(13) \, ,
        \qquad
        AB^{(2)}=(24) \, .
\end{equation}
They correspond to the two alternating colorings of the box on-shell diagram,
\begin{equation}
        \overline\Gamma^{(1)}
        =
        {\rm Box}_{(w,b,w,b),(3,3,3,3)} \, ,
        \qquad
        \overline\Gamma^{(2)}
        =
        {\rm Box}_{(b,w,b,w),(3,3,3,3)} \, .
\end{equation}
The two diagrams are related by a square move and define the same on-shell form
\cite{Postnikov:2006kva,Grassmannian}. After stripping the tree-level MHV
factor, the corresponding positroid cell is a point, and the stripped leading
singularities are \(1\). This is the same as in Example~\ref{eq:onshell_k2_n4}.

Equivalently, the canonical form of the one-loop four-point MHV Amplituhedron is
\begin{equation}
\label{eq:four-point-one-loop-MHV-integrand}
        \mathbf\Omega^{(1)}_{0,4}(AB;Z)
        =
        \frac{
        \langle1234\rangle^2
        \langle AB\,\mathrm{d}^2A\rangle
        \langle AB\,\mathrm{d}^2B\rangle
        }{
        \langle AB12\rangle
        \langle AB23\rangle
        \langle AB34\rangle
        \langle AB41\rangle
        } \, ,
\end{equation}
and its maximal residues are
\begin{equation}
        {\rm Res}_{AB=13} \, 
        \mathbf\Omega^{(1)}_{0,4}
        =
        {\rm Res}_{AB=24} \, 
        \mathbf\Omega^{(1)}_{0,4}
        =
        1 \, .
\end{equation}
\end{eg}

\begin{eg}[One-mass cut]
Consider a one-mass box. Up to cyclic relabeling, \begin{equation}
        X_1=(12) \, ,
        \qquad
        X_2=(23) \, ,
        \qquad
        X_3=(34) \, ,
        \qquad
        X_4=(45) \, .
\end{equation}
The first three corners are massless, while the fourth is massive. The two
solutions of the Schubert problem are rational:
\begin{equation}
\label{eq:one-mass-schubert-solutions}
        AB^{(w)}=(24) \, ,
        \qquad
        AB^{(b)}
        =
        \overline{2}\cap\overline{4}
        =
        (123)\cap(345) \, .
\end{equation}
where $\overline{i}:=(i{-}1 \, i \, i{+}1)$. If the massive corner has helicity degree \(k_1\), then these two branches have helicity
degrees \(k_1\) and \(k_1+1\), respectively. For the five-point one-loop
amplitude, \(k_1=0\). Thus the first branch contributes to the MHV sector,
while the second branch contributes to the NMHV sector.
\end{eg}

\begin{eg}[Two-mass-easy cut]
For a two-mass-easy box, the two massless corners are opposite. We may take for $1<i<j-2\leq n-2$:
\begin{equation}
        X_1=(i{-}1,i) \, ,
        \qquad
        X_2=(i,i{+}1) \, ,
        \qquad
        X_3=(j{-}1,j) \, ,
        \qquad
        X_4=(j,j{+}1) \, ,
\end{equation}
The two Schubert solutions are
\begin{equation}
\label{eq:two-mass-easy-schubert-solutions}
        AB^{(w)}=(ij) \, ,
        \qquad
        AB^{(b)}
        =
        \overline{i}\cap\overline{j}
        =
        (i{-}1 \, i \, i{+}1)
        \cap
        (j{-}1 \, j \, j{+}1) \, .
\end{equation}
If the two massive corners have helicity degrees \(k_1\) and \(k_3\), then the
two branches have helicity degrees
\begin{equation}
        k_1+k_3 \, ,
        \qquad
        k_1+k_3+2 \, .
\end{equation}
The smallest multiplicity for a two-mass-easy box is \(n=6\), in which case the
massive corners are four-point MHV amplitudes and \(k_1=k_3=0\). The lower
branch therefore gives an MHV leading singularity, while the higher-helicity
branch ${\rm N}^2{\rm MHV}$.
\end{eg}

The two-mass-hard and three-mass cuts are similar. Their Schubert solutions are
still rational and can be constructed by taking intersections of external lines
and planes in \(\mathbb P^3\). The three-mass box first contributes at seven
points and in the NMHV sector. These rational one-loop cuts are the
momentum-twistor form of the standard box coefficients appearing in the
one-loop expansion of planar \(\mathcal N=4\) SYM amplitudes
\cite{Britto:2004ap,Grassmannian,Elvang:2013cua}.

\begin{eg}[Four-mass cut]
The first genuinely algebraic one-loop example is the four-mass box. The four
cut lines \(X_1,X_2,X_3,X_4\) are generic lines in \(\mathbb P^3\). The
conditions 
\begin{equation}
        \langle AB\,X_r\rangle=0 \, ,
        \qquad
        r=1,\ldots,4 \, ,
\end{equation}
have two solutions, the two transversals to the four lines. We denote them by $AB^{(\pm)}(X)$.
Unlike the zero-, one-, two-, and three-mass cases, these solutions are
generically algebraic. In fact, as discussed in Subsection~\ref{subsec:Boxes in Momentum Twistors}, solving this Schubert problem involves a quadratic equation, whose discriminant is
\begin{equation}
\label{eq:four-mass-cut-discriminant}
        \Delta=(1-u-v)^2-4uv \, ,
\end{equation}
where
\begin{equation}
        u=
        \frac{
        \langle X_1X_2\rangle
        \langle X_3X_4\rangle
        }{
        \langle X_1X_3\rangle
        \langle X_2X_4\rangle
        } \, ,
        \qquad
        v=
        \frac{
        \langle X_2X_3\rangle
        \langle X_1X_4\rangle
        }{
        \langle X_1X_3\rangle
        \langle X_2X_4\rangle
        } \, .
\end{equation}
Thus \(\Delta=0\) has three simultaneous interpretations: it is the leading
Landau locus of the four-mass box, the discriminant where the two Schubert
solutions collide, and the square-root locus appearing in the non-rational
Yangian invariant.

The four-mass box first appears at eight points and in the
\({\rm N}^2{\rm MHV}\) sector. For \(n=8\), all four corners are four-point MHV
amplitudes. The corresponding on-shell diagram \(\Gamma_{\rm 4m}\) has affine
permutation
\begin{equation}
        \sigma_{\rm 4m}
        =
        (2,5,4,7,6,9,8,11) \, ,
\end{equation}
and labels an eight-dimensional positroid cell in
\(\Gr_{\geq 0}(2,8)\). A positroid parametrization
\(C_{\rm 4m}(\alpha)\) gives two solutions to the eight bosonic equations
\begin{equation}
        C_{\rm 4m}(\alpha)\cdot Z=0 \, .
\end{equation}
Equivalently, the Amplituhedron map on this positroid cell has degree
\begin{equation}
        \gamma_{\Gamma_{\rm 4m}}=2 \, .
\end{equation}
The corresponding leading singularities have the form
\begin{equation}
\label{eq:eight-point-four-mass-box-LS}
        {\rm LS}^{(1)}_{2,8}[\Gamma_{\rm 4m},\pm]
        =
        \left.
        \frac{
        \delta^{(8)}
        \bigl(C_{\rm 4m}(\alpha)\cdot\chi\bigr)
        }{
        \alpha_1\cdots\alpha_8\,
        J_{\rm 4m}(\alpha,Z)
        }
        \right|_{\alpha=\alpha^{(\pm)}(Z)} \, ,
\end{equation}
where
\begin{equation}
        J_{\rm 4m}(\alpha_\pm,Z)
        =
        \pm\sqrt{\Delta} \, .
\end{equation}
This is the first example of a non-rational Yangian invariant in planar
\(\mathcal N=4\) SYM.

An elegant expression for the same invariant is given in \cite{Grassmannian}:
\begin{equation}
\label{eq:4m_AHBCCT_expression}
        {\rm LS}^{(1)}_{2,8}[\Gamma_{\rm 4m},\pm]
        =
        \left(
        1-
        \frac{
        \langle A\,4\,5\,6\rangle
        \langle B\,8\,1\,2\rangle
        }{
        \langle A\,4\,1\,2\rangle
        \langle B\,8\,5\,6\rangle
        }
        \right)^{-1}
        [A,1,2,3,4]\,
        [B,5,6,7,8]
        \bigg|_{\pm} \, .
\end{equation}
Here \([a,b,c,d,e]\) denotes the usual \(R\)-invariant in~\eqref{eq:Rinv}, and the two branches are
defined by the quadratic incidence equations
\begin{equation}
\label{eq:four-mass-psi-factor}
        A=(78)\cap(56B) \, ,
        \qquad
        B=(34)\cap(12A) \, .
\end{equation}
For example, one may solve these equations by writing $A=Z_7+xZ_8$ and $B=Z_3+yZ_4$. The resulting system in $x,y$ has discriminant \(\Delta\).
\end{eg}

\subsubsection{Composite leading singularities}
\label{subsec:Composite Leading Singularities}

At higher loops, not every leading singularity is obtained by cutting \(4\ell\)
ordinary propagators simultaneously. Some leading singularities are
\emph{composite}: after some propagator residues have been taken, the
remaining lower-dimensional residue form develops additional poles. Taking
the residue at these emergent poles supplies the final localization condition \cite{ArkaniHamed:2009dn,Grassmannian,ArkaniHamed:2010kv,Elvang:2013cua}.

A toy model is the form
\begin{equation}
\label{eq:toy_form}
        \mathbf\Omega
        =
        \frac{\mathrm{d}x\,\mathrm{d}y\,\mathrm{d}z}{x(x+yz)}
\end{equation}
on \(\mathbb C^3\). The denominator has only two irreducible factors. However,
after taking the residue at \(x=0\), the remaining denominator becomes \(yz\).
Thus one may take two further residues and obtain the \emph{composite leading
singularity}
\begin{equation}
        {\rm Res}_{z=0}
        {\rm Res}_{y=0}
        {\rm Res}_{x=0}
        \, \mathbf\Omega
        =
        1 \, .
\end{equation}
The poles at \(y=0\) and \(z=0\) were not manifest before taking the first
residue.

A geometric example occurs in two-loop double-box cuts. Let \(AB\) and \(CD\)
be the two loop lines, and impose seven propagator conditions including
\begin{equation}
\label{eq:double-box-seven-cuts}
        \langle AB\,X_1\rangle
        =
        \langle AB\,X_2\rangle
        =
        \langle AB\,X_3\rangle
        =
        0 \, ,
        \qquad
        \langle CD\,X_4\rangle
        =
        \langle CD\,X_5\rangle
        =
        \langle CD\,X_1\rangle
        =
        0 \, ,
\end{equation}
together with
\begin{equation}
\label{eq:double-box-mutual-cut}
        \langle AB\,CD\rangle=0 \, .
\end{equation}
These seven conditions do not yet localize the two loop lines completely, since
two lines in \(\mathbb P^3\) have eight degrees of freedom. On the
support of suitable cuts involving \(X_1\), the incidence condition
factorizes into two geometric branches: the three lines \(AB\), \(CD\), and
\(X_1\) may be concurrent, or they may be coplanar. Imposing both conditions
produces an additional composite residue. This supplies the eighth localization
condition and gives a leading singularity.

This phenomenon is naturally encoded in the boundary structure of loop
Amplituhedra. The algebraic boundary contains components corresponding not only
to manifest propagator divisors, but also to such factorization loci of
incidence conditions \cite{Dian_2023,Dian:2024hil}. We have seen this explicitely in our four-point two-loop Amplituhedron analysis of Section~\ref{sec:The Four-Point Two-Loop Amplituhedron}. Equivalently, composite
leading singularities are included in the same Grassmannian contour integral
formalism; the relevant contour is an iterated residue rather than a
simultaneous residue on \(4\ell\) manifest propagators
\cite{ArkaniHamed:2009dn,Drummond:2010qh,Grassmannian}.

Let us conclude with some general remarks. Leading singularities are not special to planar \(\mathcal N=4\) SYM. They are
part of generalized unitarity for less-supersymmetric and non-supersymmetric
theories as well. What is special in planar \(\mathcal N=4\) SYM is the
simultaneous availability of momentum twistors, positroid geometry, Yangian
invariance, and the Amplituhedron. Outside the planar maximally supersymmetric
setting, on-shell diagrams and generalized unitarity still exist, but the full
positive-Grassmannian and Amplituhedron structures are not known in the same generality
\cite{Franco:2014csa_alt1,Franco:2015rma,Lisitsyn:2025prd,Carrolo:2026qpu}.

\subsection{Wilson loops with a Lagrangian insertion}
\label{sec:WLwithLI}

We now illustrate the notions introduced so far in this chapter, related to Landau analysis and leading singularities, through a specific finite quantity in planar $\mathcal{N}=4$ SYM theory: the null polygonal Wilson loop with a single Lagrangian insertion. This quantity is closely related to the logarithm of planar MHV amplitudes, admits
a geometric expansion in terms of negative geometries, and has a rich set of
leading singularities classified to all loop orders.

Historically, this observable sits at the intersection of two developments.
The first is the duality between planar MHV amplitudes and null polygonal
Wilson loops, discovered at strong coupling and tested perturbatively at weak
coupling
\cite{Alday:2007hr,Drummond:2007aua,Brandhuber:2007yx,Drummond:2008vq,Eden:2010ce}. The second is the study of correlators of null
polygonal Wilson loops with local operator insertions. These were
introduced as natural AdS/CFT observables and studied at both strong and weak
coupling in \cite{Alday:2011ga,Engelund:2011fg,Alday:2012hy,Alday:2013ip}. The case in which the inserted operator is the Lagrangian is
special, because it computes a coupling derivative of the Wilson loop, and
therefore of the logarithm of the dual amplitude. This idea has also been used
in high-loop studies of the cusp anomalous dimension
\cite{Henn:2019swt}. Recently, Wilson loops with a Lagrangian insertion
were shown to have hidden symmetry properties, a close relation to all-plus
Yang--Mills amplitudes, and a geometric expansion in negative geometries
\cite{ChicherinHenn2022,Chicherin:2022zxo,ArkaniHamedHennTrnka2021,neg_geom_pos,BrownHennMazzucchelliTrnka2025}.
Figures in this section are adapted from these references where they depict the
negative-geometry expansion or the Wilson-loop leading singularities.

This finite quantity provides a bridge
between the geometric integrand-level structures reviewed above and the
transcendental functions obtained after integration. Ordinary massless
amplitudes are infrared divergent and require a regulator, which can obscure the
four-dimensional geometry of the Amplituhedron. By contrast, the Wilson loop
with a Lagrangian insertion is finite in four dimensions while still being
directly related to the loop integrand of the logarithm of the amplitude.
Therefore it is an ideal laboratory for studying how positive geometry controls
integrated amplitudes and related finite quantities.

\subsubsection{Wilson loop with a Lagrangian insertion}
\label{subsec:The Observable}

Let \(x_1,\ldots,x_n\) be the cusps of a null polygon in dual momentum space,
with
\begin{equation}
        (x_i-x_{i+1})^2=0 \, ,
        \qquad
        x_{n+1}\equiv x_1 \, .
\end{equation}
Equivalently, the external momenta are
\begin{equation}
        p_i=x_i-x_{i-1} \, ,
        \qquad
        p_i^2=0 \, ,
        \qquad
        \sum_{i=1}^{n}p_i=0 \, .
\end{equation}
We denote by
\begin{equation}
        W_n(x_1,\ldots,x_n)
\end{equation}
the corresponding \(n\)-sided null polygonal Wilson loop in planar
\(\mathcal N=4\) super Yang--Mills theory. The finite theoretical quantity considered in this
section is the normalized correlator of this Wilson loop with a single
insertion of the Lagrangian density at a point \(x_0\):
\begin{tcolorbox}[definitionbox]
\textbf{Wilson loop with a Lagrangian insertion.}
\begin{equation}
\label{eq:WL-LI-definition}
        F_n(x_1,\ldots,x_n;x_0)
        :=
        \frac{1}{\pi^2}
        \frac{
        \langle W_n(x_1,\ldots,x_n)\,\mathcal L(x_0)\rangle
        }{
        \langle W_n(x_1,\ldots,x_n)\rangle
        } \, .
\end{equation}
\end{tcolorbox}\noindent
The point \(x_0\) is the spacetime point at
which the Lagrangian is inserted.

The normalization by \(\langle W_n\rangle\) is crucial. Null polygonal Wilson
loops have cusp divergences, which are dual to the infrared divergences of
planar massless scattering amplitudes. In the ratio
\eqref{eq:WL-LI-definition}, these universal divergences cancel. Therefore
\(F_n\) is a finite four-dimensional field-theoretic observable
\cite{Alday:2011ga,Alday:2012hy,Alday:2013ip,Henn:2019swt,ChicherinHenn2022,BrownHennMazzucchelliTrnka2025}. It is also dual conformal:
under dual conformal transformations of the points $x_1,\ldots,x_n,x_0$ the quantity transforms covariantly with the appropriate weight at the
insertion point. 
Here ``observable'' is meant in this field-theoretic sense; no claim is made
that planar \(\mathcal N=4\) SYM quantities are experimentally observable in
the ordinary phenomenological sense.
In momentum-twistor variables, the external kinematics is encoded by the $n$ vertices $Z_i \in \mathbb{P}^3$ of the null polygon and a line $ AB=AB_0\subseteq\mathbb P^3$ dual to the Lagrangian insertion point \(x_0\). The remaining points $x_1,\dots,x_n$ become lines $AB_a$, and are treated as loop-integration variables.

The Lagrangian insertion has a simple interpretation in perturbation theory.
Differentiating the logarithm of the Wilson loop with respect to the coupling
brings down an integrated Lagrangian insertion. Schematically,
\begin{equation}
\label{eq:WL-LI-log-derivative}
        g^2\frac{\partial}{\partial g^2}
        \log\langle W_n\rangle
        =
        \int_{AB} \,
        F_n(AB_1,\dots,AB_{\ell};AB) \, ,
\end{equation}
up to the normalization conventions for the coupling and for the Lagrangian.
Thus \(F_n\) can be thought of as the local density whose integration over
the insertion point gives the coupling derivative of the logarithm of $W_n$.

By the amplitude/Wilson-loop duality, the same statement relates
\(F_n\) to the logarithm of the planar MHV amplitude. Let us write the perturbative expansion of the amplitude as
\begin{equation}
        A_n
        =
        1+\sum_{\ell\geq1} g^{2\ell}A_n^{(\ell)}
\end{equation}
be the tree-normalized planar MHV amplitude.
The perturbative contributions are
\begin{equation}\label{ampl_def_integrand}
A_n^{(\ell)} = \int_{AB_1} \dots \int_{AB_\ell} \, \mathbf{\Omega}_n^{(\ell)} \, ,
\end{equation}
where the integrand $\mathbf{\Omega}_n^{(\ell)}$ is the canonical form of the $n$-point one-loop MHV Amplituhedron. It can be written as 
\begin{equation}\label{ampl_form}
     \mathbf{\Omega}_n^{(\ell)} = \prod_{a=1}^{\ell} \, \langle AB_a \, \mathrm{d}^{2}A_a \rangle \, \langle AB_a \, \mathrm{d}^{2}B_a \rangle \, \, \Omega_n^{(\ell)}  \,,
\end{equation}
where $\Omega_n^{(\ell)}$ is a rational function in twistor coordinates in the external momenta $Z_i$ and in the loop variables $AB_a$. 
Similarly, we write the logarithm of the
amplitude at \(\ell\) loops as
\begin{equation}
\label{eq:log-amplitude-integrand-thesis}
        (\log A_n)^{(\ell)}
        =
        \int_{AB_1}\cdots\int_{AB_\ell}
        \widetilde{\mathbf\Omega}_n^{(\ell)} \, ,
\end{equation}
where the integrand \(\widetilde{\mathbf\Omega}_n^{(\ell)}\) will be discussed later. Then, the \(\ell\)-loop
contribution to the Wilson loop with Lagrangian insertion is obtained from the
\((\ell+1)\)-loop logarithmic integrand by leaving one loop variable
unintegrated:
\begin{tcolorbox}[resultbox]
\textbf{Relation to the logarithm of the amplitude.}
\begin{equation}
\label{eq:WL-LI-from-log-amplitude}
        F_n^{(\ell)}(AB;Z)
        =
        \int_{AB_1}\cdots\int_{AB_\ell}
        \widetilde{\mathbf\Omega}_n^{(\ell+1)}
        (AB,AB_1,\ldots,AB_\ell;Z) \, .
\end{equation}
\end{tcolorbox}\noindent
The line \(AB=AB_0\) represents the insertion point \(x_0\) in momentum-twistor
space and is kept fixed. The remaining lines $AB_1,\ldots,AB_\ell$ are integrated over the whole Minkowski spape contours.

We use the perturbative expansion
\begin{equation}
\label{eq:WL-LI-perturbative-expansion}
        F_n
        =
        \sum_{\ell\geq 0}
        g^{2\ell+2}
        F_n^{(\ell)} \, .
\end{equation}
With this convention, \(F_n^{(0)}\) is the Born-level contribution. Since
\eqref{eq:WL-LI-from-log-amplitude} has no integrated loop variables when
\(\ell=0\), the Born-level observable is simply the one-loop MHV integrand:
\begin{equation}
\label{eq:Born-WL-LI}
        F_n^{(0)}
        =
        \widetilde\Omega_n^{(1)}
        =
        \Omega_n^{(1)} \, .
\end{equation}

The quantity \(F_n\) has several remarkable properties:
\begin{enumerate}
        \item \textit{Finiteness.}
        It is finite in four dimensions. This allows us to compare directly the
        four-dimensional geometry of the integrand with the analytic structure
        of the integrated function. No dimensional regulator is needed for the
        integrations in \eqref{eq:WL-LI-from-log-amplitude}. This is a
        significant advantage over ordinary massless scattering amplitudes,
        whose infrared divergences require a regularization scheme and thereby
        obscure some of the four-dimensional geometric structure.

        \item \textit{Relation to the logarithm of the amplitude.}
        The quantity \(F_n\) is directly related to the logarithm of the
        planar MHV amplitude. The Lagrangian insertion is obtained by
        differentiating the Wilson loop, or equivalently the dual amplitude,
        with respect to the coupling. Consequently, \(F_n^{(\ell)}\) is obtained
        from the \((\ell+1)\)-loop integrand of the logarithm of the amplitude
        by keeping one loop variable unintegrated. Thus, as we will see, \(F_n\) inherits the
        Amplituhedron origin of the MHV integrand.

        \item \textit{Dual conformal invariance.}
        The quantity \(F_n\) is dual conformal. One may use dual conformal
        symmetry to send the insertion point \(x_0\), or equivalently the line
        \(AB_0\), to infinity. In that conformal frame, \(F_n\) depends on the
        same number of kinematic variables as an ordinary \(n\)-point scattering
        amplitude in a theory without dual conformal symmetry. This observation is related to the next point.

        \item \textit{Relation to all-plus Yang--Mills amplitudes.}
        The quantity \(F_n\) is conjecturally related to all-plus amplitudes
        in pure Yang--Mills theory. More precisely, after sending the
        Lagrangian insertion point \(x_0\) to infinity, the
        maximal-transcendental part of the \((\ell+1)\)-loop all-plus
        Yang--Mills amplitude is conjectured to coincide with \(F_n^{(\ell)}\)
        \cite{ChicherinHenn2022,Chicherin:2022zxo}. This relation has been
        checked against the available one- and two-loop all-plus data, as well
        as the three-loop four-point all-plus amplitude. It is striking because
        it relates a finite dual-conformal observable in planar
        \(\mathcal N=4\) SYM to amplitudes in a theory without dual conformal
        symmetry.

        \item \textit{Positivity.}
        The quantity \(F_n\) exhibits nontrivial positivity properties. When
        evaluated in the Amplituhedron region, the known integrated
        results have a uniform sign \cite{neg_geom_pos,Henn:CM}. Similar positivity phenomena had previously
        been observed for finite parts of scattering amplitudes in planar
        \(\mathcal N=4\) SYM \cite{positive_amplitudes,Dixon:2016apl}. From
        the point of view of this thesis, this suggests that positivity is not
        merely a property of canonical forms before integration, but may survive
        the passage to integrated functions in a highly constrained way.
\end{enumerate}

\begin{tcolorbox}[resultbox]
\textbf{Properties of $F_n$.}
The Wilson loop with a Lagrangian insertion is finite, dual conformal, directly
related to the logarithm of the planar MHV amplitude, conjecturally related to
the maximal-transcendental part of all-plus Yang--Mills amplitudes, and exhibits
positivity properties in the Amplituhedron region.
\end{tcolorbox}\noindent

\subsubsection{Expansion in negative geometries}
\label{subsec:Expansion in Negative Geometries}

We now explain how the logarithmic integrand appearing in
\eqref{eq:WL-LI-from-log-amplitude} can be organized geometrically. The
ordinary MHV Amplituhedron gives the integrand of the amplitude itself. The
logarithm of the amplitude, however, is naturally organized by a different
decomposition, involving geometries in which some of the mutual positivity
conditions between loop variables are reversed. These are the
\emph{negative geometries} of \cite{ArkaniHamedHennTrnka2021}. They will be the
basic building blocks for the Wilson loop with Lagrangian insertion.

Recall that the \(\ell\)-loop MHV Amplituhedron
\(\mathcal A_n^{(\ell)}\) is described by \(\ell\) loop lines $AB_a\in\Gr(2,4)$ with $a=1,\dots,\ell$ 
together with the external data \(Z_1,\ldots,Z_n \in \mathbb{P}^3\). Each line \(AB_a\) is
required to lie in the one-loop MHV Amplituhedron \(\mathcal A_n^{(1)}\), namely
\begin{equation}\label{loop_positivty}
\begin{aligned}
    &\langle AB_a \, 12 \rangle > 0 \ , \dots, \langle AB_a \, n-1n \rangle > 0 \ , \ \langle AB_a \, 1 n \rangle > 0 \, , \\
    & (\langle AB_a \, 12 \rangle , \dots, \langle AB_a \, 1n \rangle) \text{ has two sign-flips,}
\end{aligned}
\end{equation}
and different loop lines satisfy mutual positivity conditions of the form
\begin{equation}
\label{eq:mutual-positivity-negative-geometry-section}
        \langle AB_a\,AB_b\rangle > 0 \, ,
        \qquad
        1\leq a<b\leq \ell \, .
\end{equation}
The canonical form of this geometry is the \(\ell\)-loop MHV integrand
\(\mathbf\Omega_n^{(\ell)}\), or, after stripping the standard loop measures as in~\eqref{ampl_form},
the canonical function \(\Omega_n^{(\ell)}\).

\begin{figure}[pos=t]
        \centering
         \includegraphics[width=.72\textwidth]{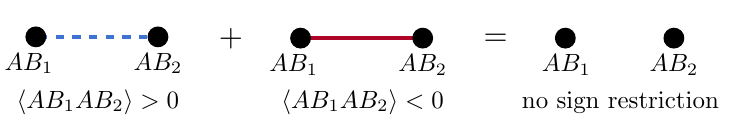}
        \caption{
        The basic two-loop relation behind negative geometries. The product of
        two one-loop Amplituhedra is decomposed according to the sign of the
        mutual bracket \(\langle AB_1 AB_2\rangle\). The positive region is the
        ordinary two-loop Amplituhedron (blue link); the negative region, with
        \(\langle AB_1 AB_2\rangle<0\), is the simplest negative geometry (red link).
        }
        \label{fig:negative-geometry-two-loop-relation}
\end{figure}

The simplest instance of the negative-geometry construction occurs at two
loops. The product of two one-loop Amplituhedra, $ \mathcal A_n^{(1)}\times\mathcal A_n^{(1)}$,
is divided by the sign of the mutual bracket
\(\langle AB_1\,AB_2\rangle\). One region is the ordinary two-loop
Amplituhedron, where $\langle AB_1\,AB_2\rangle$ is positive.
The case $n=4$ and $\ell=2$ was discussed earlier in the thesis as the simplest non-trivial loop Amplituhedron.
The complementary region is defined by
$\langle AB_1\,AB_2\rangle$ being negative.
This complementary region is the first negative geometry. Thus, schematically,
one has a relation of canonical forms
\begin{equation}
\label{eq:two-loop-negative-geometry-relation}
        \mathbf{\Omega}_n^{(1)}(AB_1) \wedge
        \mathbf{\Omega}_n^{(1)}(AB_2)
        =
        \mathbf{\Omega}_n^{(2)}(AB_1,AB_2)
        +
        \mathbf{\Omega}_{n,\{12\}}^{(2)}(AB_1,AB_2) \, ,
\end{equation}
where \(\{12\}\) denotes the graph with two vertices joined by one negative edge. This equation is depicted graphically in Figure~\ref{fig:negative-geometry-two-loop-relation}. This is the elementary identity from which the all-loop expansion of $\widetilde{\mathbf{\Omega}}_n^{(\ell)}$ in~\eqref{eq:log-amplitude-integrand-thesis} is built.

More generally, let \(G\) be a graph on $\ell$ labelled vertices, identified with its edge-set. We associate to the vertex \(a\) the loop line \(AB_a\). The negative geometry associated with \(G\) is obtained from the product of \(\ell\) one-loop Amplituhedra by imposing one additional mutual negativity condition for every edge of the graph:
\begin{equation}
\label{eq:negative-geometry-definition}
        AB_a\in\mathcal A_n^{(1)}
        \quad\text{for all }a,
        \qquad
        \langle AB_a\,AB_b\rangle<0
        \quad\text{for every edge }ab \in G \, ,
\end{equation}
Equivalently, the graph records which mutual brackets have been made negative.
We assume that every space of the form~\eqref{eq:negative-geometry-definition} is a (weighted) positive geometry, and denote its canonical form $ \mathbf{\Omega}_{n,G}^{(\ell)}$.
In this notation, repeatedly applying the two-loop relation
\eqref{eq:two-loop-negative-geometry-relation} gives the signed graph expansion
\begin{equation}
\label{eq:Omega-negative-geometry-expansion}
        \mathbf{\Omega}_n^{(\ell)}
        =
        \sum_{G}
        (-1)^{|G|} \, 
        \mathbf{\Omega}_{n,G}^{(\ell)} \, ,
\end{equation}
where the sum runs over graphs on \(\ell\) vertices, including the graph with no edges. This should be viewed as a geometric inclusion--exclusion
identity as~\eqref{eq:two-loop-negative-geometry-relation}. 

The logarithm of the amplitude selects the connected graphs in this expansion.
This is the same combinatorial mechanism by which the logarithm of a partition
function selects connected diagrams. Thus the canonical function of the
\(\ell\)-loop logarithmic integrand is
\begin{tcolorbox}[definitionbox]
\textbf{Expansion of the logarithmic integrand.}
\begin{equation}
\label{eq:log-negative-geometry-expansion}
        \widetilde{\mathbf{\Omega}}_n^{(\ell)}
        =
        \sum_{G\ {\rm connected}}
        (-1)^{|G|} \, 
        \mathbf{\Omega}_{n,G}^{(\ell)} \, ,
\end{equation}
\end{tcolorbox}\noindent
where the sum runs over connected graphs on $\ell$ vertices.

It is important to point out that the negative-geometry expansion is not a Feynman-diagram expansion, but a purely geometric one. A key advantage is that the corresponding integrated objects are term by term infrared finite in four dimensions~\cite{ArkaniHamedHennTrnka2021,neg_geom_pos,BrownHennMazzucchelliTrnka2025}.

We now connect this expansion to the Wilson loop with a Lagrangian insertion. In \(F_n^{(\ell)}\), one line $AB=AB_0$ of $\widetilde{\mathbf{\Omega}}^{(\ell+1)}_n$ is left unintegrated. Therefore the relevant negative geometries are labelled by connected graphs with one marked vertex $AB$. For a connected marked graph \(G\), we define the corresponding
\emph{integrated negative geometry} by
\begin{tcolorbox}[definitionbox]
\textbf{Integrated negative geometry.}
\begin{equation}
\label{eq:integrated-negative-geometry-thesis}
        \mathcal F_{n,G}^{(\ell)}(AB;Z)
        :=
        \int_{AB_1}\cdots\int_{AB_\ell}
        \mathbf\Omega_{n,G}^{(\ell+1)}
        (AB,AB_1,\ldots,AB_\ell;Z) \, .
\end{equation}
\end{tcolorbox}\noindent
Here \(\mathbf\Omega_{n,G}^{(\ell+1)}\) is the canonical form of the
negative geometry associated with \(G\), and each integration contour is
the Minkowski-space contour in momentum twistors discussed in
Subsection~\ref{subsec:Boxes in Momentum Twistors}. A key property of this
construction is that mutual negativity removes the collinear regions
responsible for infrared divergences. To see this, consider a collinear region
where a loop line \(AB_1\) passes through an external point \(Z_i\) and lies
in the plane \((i{-}1 \,  i \, i{+}1)\). We can parametrize it as
\begin{equation}
        AB_1=(Z_i,-Z_{i-1}+\alpha\, Z_{i+1}) \, .
\end{equation}
The one-loop positivity condition~\eqref{loop_positivty} forces
\(\alpha>0\). For any other loop line \(AB_a\), \begin{equation}
        \langle AB_1 \, AB_a \rangle
        =
        \langle AB_a \, i{-}1 i \rangle
        +
        \alpha \, \langle AB_a \, i i{+}1 \rangle \, .
\end{equation}
The one-loop positivity conditions~\eqref{loop_positivty} for \(AB_a\) make
the brackets on the right positive, whereas the negative-geometry condition
forces \(\langle AB_1 \, AB_a \rangle<0\) whenever the vertex associated with
\(AB_1\) is connected to another vertex of \(G\). Thus, for a connected
graph, integrated negative geometries are free of collinear divergences and are
infrared finite in four dimensions~\cite{ArkaniHamedHennTrnka2021}.

By~\eqref{eq:log-negative-geometry-expansion}, the full Wilson loop with
Lagrangian insertion is obtained by
\begin{equation}
\label{eq:WL-negative-geometry-expansion-thesis}
        F_n^{(\ell)}
        =
        \sum_{\substack{G \ {\rm connected}}}
        \frac{(-1)^{|G|}}{|\operatorname{Aut}(G)|}
        \,
        \mathcal F_{n,G}^{(\ell)} \, ,
\end{equation}
where the sum is over unlabelled graphs on $\ell+1$ vertices with a marked vertex. If the sum is over labelled graphs, then one omits the factor $|\operatorname{Aut}(G)|^{-1}$.

\begin{eg}[Four-point one- and two-loop integrands]
	At four points, the logarithmic integrand and the negative-geometry expansion
are particularly explicit. The Born-level contribution is the one-loop
four-point canonical function,
\begin{equation}
\label{eq:F4-Born-negative-geometry}
        F_4^{(0)}(AB;Z)
        =
        \Omega_4^{(1)}(AB;Z)
        =
        -\frac{\langle1234\rangle^2}
        {\langle AB\,12\rangle
         \langle AB\,23\rangle
         \langle AB\,34\rangle
         \langle AB\,41\rangle} \, ,
\end{equation}
corresponding to the massless box integrand discussed in Subsection~\ref{subsec:Boxes in Momentum Twistors}.

At the next order, \(F_4^{(1)}\) is obtained from the two-loop logarithmic
integrand by keeping \(AB\) fixed and integrating over a second line, which
we denote by \(CD\). The rational function of the logarithmic integrand can be computed using~\eqref{eq:two-loop-negative-geometry-relation} from that of the four-point two-loop Amplituhedron discussed earlier in the thesis. This yields
\begin{equation}
\label{eq:n4-two-loop-log-integrand-negative-geometry}
    \widetilde{\Omega}^{(2)}_4
    =
    \frac{
    -\langle 1234 \rangle^3
    \left(
    \langle AB\,13\rangle \langle CD\,24\rangle
    +
    \langle AB\,24\rangle \langle CD\,13\rangle
    \right)
    }{
    \langle AB\,12\rangle
    \langle AB\,23\rangle
    \langle AB\,34\rangle
    \langle AB\,14\rangle
    \langle AB\,CD\rangle
    \langle CD\,12\rangle
    \langle CD\,23\rangle
    \langle CD\,34\rangle
    \langle CD\,14\rangle
    } \, .
\end{equation}
\end{eg}

Let us close this subsection by summarizing what is currently known at the
level of \emph{integrands}. The original negative-geometry construction gives a simple all-loop formula for four-point negative geometries whose underlying graph is a tree~\cite{ArkaniHamedHennTrnka2021}. This simplicity is special, and the formula obtained there does not immediately generalize to higher \(n\). The integrand of $F^{(\ell=1)}_n$ and the integrated answer are both known, since $F^{(\ell=1)}_n$ admits an expansion into chiral pentagons, obtain by a generalized unitarity method. We will present the formulae in the next subsection.
Beyond two loops, the main explicit all-loop family is provided by ladder geometries, which we discuss in Subsection~\ref{subsec:geometric-landau-negative-geometries}. More general graph topologies, especially graphs with cycles have been recently studied at four points~\cite{BrownOktemParanjapeTrnka2024}. In principle, integrands of negative geometries at low loop order can be constructed by generalized unitarity, once all the leading singularities for a given graph topology are classified. This further motivates studying leading singularities of $\mathcal F_{n,G}^{(\ell)}$.

\subsubsection{The structure of the Wilson loop in perturbation theory}
\label{subsec:The Structure of the Wilson Loop in Perturbation Theory}

One of the main motivations for studying the Wilson loop with a Lagrangian
insertion is that it allows us to use the geometry of the loop integrand to make
predictions about the integrated answer. The negative-geometry expansion gives a
finite four-dimensional integrand for \(F_n^{(\ell)}\), and the leading
singularities of this integrand control the algebraic prefactors of the
transcendental functions appearing after integration.

The expected perturbative structure is 
\begin{tcolorbox}[resultbox]
\textbf{Perturbative decomposition.}
\begin{equation}
\label{eq:F-decomposition}
        F_n^{(\ell)}
        =
        \sum_s
        \Omega_{n,s}\,
        f_{n,s}^{(\ell)} \, ,
\end{equation}
\end{tcolorbox}\noindent
where the functions \(f_{n,s}^{(\ell)}\) are expected to be pure functions of
transcendental weight \(2\ell\), while the coefficients \(\Omega_{n,s}\) are
algebraic functions of the external twistors \(Z_i\) and of the insertion line
\(AB\). The index \(s\) labels a basis of leading singularities. This basis
may depend on the loop order, although the all-loop classification described
below shows that it stabilizes in a precise sense from $\ell \geq 2$.

For integrals evaluating to multiple polylogarithms, a decomposition of the form
\eqref{eq:F-decomposition} is natural from the point of view of the pure bases
and canonical differential equations reviewed in
Subsection~\ref{subsec:Pure Integrals and Canonical Differential Equations}
\cite{Henn2013}. If one can choose a basis of
integrals of uniform transcendental weight, then the nontrivial algebraic
dependence is carried by their leading singularities. From the integrand point
of view, the same structure follows from the possibility of expanding the
integrand in a local basis with unit leading singularities
\cite{ArkaniHamed:2010kv}. Thus the leading singularities of the integrand of
\(F_n^{(\ell)}\) are expected to furnish the prefactors
\(\Omega_{n,s}\) in \eqref{eq:F-decomposition}.
This expectation is also the starting point for the bootstrap methods discussed in Section~\ref{sec:Symbols and Bootstrap}.

Let us first explain the lowest-order case. At Born level there are no
integrated loop variables. By \eqref{eq:WL-LI-from-log-amplitude}, \(F_n^{(0)}\) is equal to the canonical function of the one-loop MHV
Amplituhedron, evaluated on the insertion line \(AB\). This Amplituhedron is equivalent to the $m=k=2$ case discussed in Subsection \ref{subsec:The A22n tiling and its forms}, and a BCFW-like triangulation yields
\begin{equation}
\label{eq:F-Born-Kermit-expansion}
        F_n^{(0)}
        =
        \Omega_n^{(1)}
        =
        \sum_{\Delta_1,\Delta_2\subseteq T}
        [\Delta_1;\Delta_2] \, ,
\end{equation}
where \(T\) is a triangulation of the \(n\)-gon, and the sum runs over
non-overlapping pairs of triangles in \(T\). The whole sum in~\eqref{eq:F-Born-Kermit-expansion} is independent of the choice of $T$, reflecting the triangulation independence of the canonical form. Choosing the triangulation from the vertex \(1\), this becomes
\begin{equation}
\label{eq:F-Born-standard-Kermit-expansion}
        F_n^{(0)}
        =
        \sum_{i=2}^{n-3}
        \sum_{j=i+2}^{n-1}
        [1\,i\,i{+}1;1\,j\,j{+}1] \, .
\end{equation}
In both expressions above we expressed the result in terms of \textit{Kermits}. For two
non-overlapping triangles $\Delta_r=\{a_r,b_r,c_r\}$ with $r=1,2$, the associated Kermit function is
\begin{equation}
\label{eq:Kermit-six-invariant-F}
        [a_1b_1c_1;a_2b_2c_2]
        =
        \frac{
        \langle AB\,(a_1b_1c_1)\cap(a_2b_2c_2)\rangle^2
        }{
        \langle AB\,a_1b_1\rangle
        \langle AB\,b_1c_1\rangle
        \langle AB\,a_1c_1\rangle
        \langle AB\,a_2b_2\rangle
        \langle AB\,b_2c_2\rangle
        \langle AB\,a_2c_2\rangle
        } \, .
\end{equation}
If the two triangles share an edge and form a quadrilateral
\(\{a,b,c,d\}\), this reduces to the four-boundary Kermit
\begin{equation}
\label{eq:Kermit-four-invariant-F}
        [abcd]
        =
        -\frac{\langle abcd\rangle^2}
        {\langle AB\,ab\rangle
         \langle AB\,bc\rangle
         \langle AB\,cd\rangle
         \langle AB\,\mathrm{d}a\rangle} \, .
\end{equation}
These Kermit functions turn out to be the building blocks of the leading-singularity prefactors that appear in \eqref{eq:F-decomposition}. They are the canonical functions of tiles for the $m=k=2$ Amplituehdron, as well as the Yangian invariants for $m=k=2$~\cite{lukowski2019cluster}.

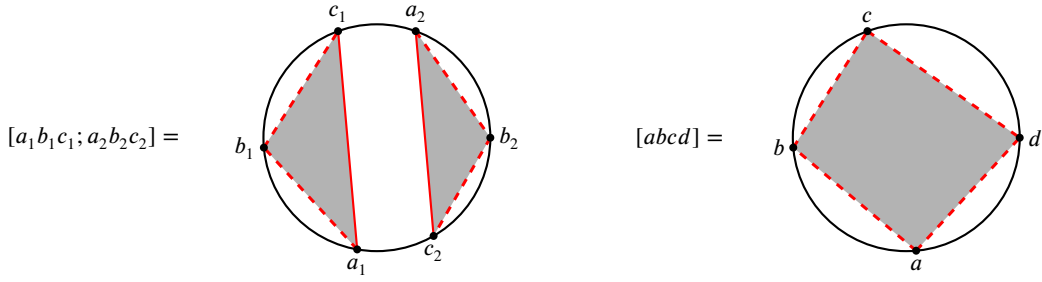
\begin{figure}[pos=t]
\centering
\begin{tikzpicture}[scale = 0.5]
        
    \begin{scope}

    \node at (-7.5,0) {$[a_1 b_1 c_1;a_2 b_2 c_2] = $};
    
    \draw[thick] (0,0) circle(3cm);

    \coordinate (i) at (-100:3cm);
    \coordinate (j) at (185:3cm);
    \coordinate (k) at (110:3cm);
    \coordinate (l) at (70:3cm);
    \coordinate (n) at (-60:3cm);
    \coordinate (m) at (0:3cm);
    
    \node at (i) [below ] {$a_1$};
    \node at (j) [left] {$b_1$};
    \node at (k) [above ] {$c_1$};
    \node at (l) [above ] {$a_2$};
    \node at (m) [right ] {$b_2$};
    \node at (n) [below ] {$c_2$};

    \fill[black, opacity=0.3] (i) -- (j) -- (k) -- cycle;
    
    \fill[black, opacity=0.3]  (l) -- (n) -- (m) -- cycle;

    \draw[line width=0.4mm, red, dashed] (i) -- (j);
    \draw[line width=0.4mm, red, dashed] (l) -- (m);
    \draw[line width=0.4mm, red, dashed] (j) -- (k);
    \draw[line width=0.4mm, red, dashed] (m) -- (n);
    \draw[line width=0.4mm, red, thick] (i) -- (k);
    \draw[line width=0.4mm, red, thick] (l) -- (n);

     \foreach \p in {i,j,k,l,m,n} {
        \fill (\p) circle(3pt);
    }
    \end{scope}


    \begin{scope}[xshift = 14 cm]

    \node at (-6,0) {$[abcd] = $};
    
    \draw[thick] (0,0) circle(3cm);

    \coordinate (i) at (-85:3cm);
    \coordinate (j) at (185:3cm);
    \coordinate (k) at (110:3cm);
    \coordinate (l) at (0:3cm);
    
    \node at (i) [below ] {$a$};
    \node at (j) [left] {$b$};
    \node at (k) [above ] {$c$};
    \node at (l) [right ] {$d$};

    \fill[black, opacity=0.3] (i) -- (j) -- (k) -- (l) -- cycle;

    \draw[line width=0.4mm, red, dashed] (i) -- (j);
    \draw[line width=0.4mm, red, dashed] (j) -- (k);
    \draw[line width=0.4mm, red, dashed] (k) -- (l);
    \draw[line width=0.4mm, red, dashed] (i) -- (l);

     \foreach \p in {i,j,k,l} {
        \fill (\p) circle(3pt);
    }
    \end{scope}

\end{tikzpicture}
\caption{Pictorial representation of bicolored subdivisions of type $(2,n)$ associated to Kermit forms at $n$ points, where the circle represents an $n$-gon. See also Subsection \ref{subsec:The A22n tiling and its forms}. }
\label{Figure_Yangian invariants}
\end{figure}

As we discussed, at higher perturbative orders \(F_n^{(\ell)}\) is obtained by integrating the negative-geometry expansion of the \((\ell+1)\)-loop integrand of the amplitude's logarithm over the lines $AB_1,\ldots,AB_\ell$, while keeping \(AB\) fixed. We define \emph{leading singularities} as follows, by distinguishing between:
\begin{itemize}
    \item a \textit{leading singularity value}, or simply leading singularity, which is the algebraic function in $Z_i$ and $AB$ defined as a (maximal) residue of the integrand, and
    \item a \textit{leading singularity configuration}, which refers to a solution for the loop lines $AB_1$, \dots, $AB_{\ell}$ where the integrand has a pole of (maximal) order $4\ell$.
\end{itemize}
Here by integrand we mean $\widetilde{\mathbf{\Omega}}^{(\ell+1)}_n(AB,AB_1,\dots,AB_\ell)$ in~\eqref{eq:log-negative-geometry-expansion} for the full Wilson loop with Lagrangian insertion \(F_n^{(\ell)}\), or $\mathbf{\Omega}_{n,G}(AB,AB_1,\dots,AB_\ell)$ as in~\eqref{eq:integrated-negative-geometry-thesis} for an individual integrated negative geometry $\mathcal{F}_{n,G}$. Both forms are considered with respect to $AB_1,\dots,AB_\ell$.
Since \(AB\) remains unconstrained, the result is an algebraic function of \(AB\) and \(Z\). In the perturbative decomposition \eqref{eq:F-decomposition}, these functions are precisely the candidate prefactors \(\Omega_{n,s}\). 

\begin{eg}[$n=4$ and $n=5$ leading singularities]
	Let us illustrate this structure at four points. At \(n=4\) and
\(\ell=1\), the integrand is the two-loop logarithmic integrand in~\eqref{eq:n4-two-loop-log-integrand-negative-geometry}, where \(CD\) is the line to be integrated. Localizing \(CD\) on either of the two zero-dimensional one-loop Amplituhedron boundaries,
\begin{equation}
        CD=13,
        \qquad
        CD=24 \, ,
\end{equation}
gives the same leading-singularity value, \begin{equation}
        {\rm Res}_{CD=13}\,\widetilde{\mathbf{\Omega}}_4^{(2)} = 
        {\rm Res}_{CD=24}\,\widetilde{\mathbf{\Omega}}_4^{(2)}
        =
        \mathbf{\Omega}_4^{(1)} \, .
\end{equation}
Thus distinct leading-singularity configurations may give the same
leading-singularity value.

The integrated answer is correspondingly simple. There is only one independent
leading singularity at four points,
\begin{equation}
        \Omega_4^{(1)}=[1234] \, .
\end{equation}
Therefore
\begin{equation}
\label{eq:F4-one-loop-decomposition}
        F_4^{(1)}
        =
        \Omega_4^{(1)}\,f_4^{(1)} \, ,
\end{equation}
where the weight-two function is
\begin{equation}
\label{eq:F4-one-loop-integrated-function}
        f_4^{(1)}
        =
        \log^2 z+\pi^2,
        \qquad
        z=
        \frac{
        \langle AB\,12\rangle
        \langle AB\,34\rangle
        }{
        \langle AB\,14\rangle
        \langle AB\,23\rangle
        } \, .
\end{equation}
This is the first example of the structure in
\eqref{eq:F-decomposition}: an algebraic, actually rational, leading-singularity prefactor
multiplying a pure transcendental function. In fact, at four points the
leading-singularity space remains one-dimensional at all loop orders, generated
by \(\Omega_4^{(1)}=[1234]\). This agrees with the explicit four-point results
known up to three loops \cite{Alday:2012hy,Alday:2013ip,Henn:2019swt}. The
tree-graph sector of the negative-geometry expansion also admits an all-loop
resummation related to the cusp anomalous dimension
\cite{ArkaniHamedHennTrnka2021}.

The first richer example is the pentagonal Wilson loop. The observable
\(F_5^{(\ell)}\) has been computed through \(\ell=2\) in
\cite{Chicherin:2022zxo}. At one loop, there are five leading singularities,
one of which may be written as
\begin{equation}
\label{eq:F5-one-loop-LS-example}
        \Omega_5(13)
        =
        [1234]+[123;145] \, ,
\end{equation}
together with its four cyclic shifts of the indices $i=1,\dots,5$. At two loops, there are six linearly
independent leading singularities. A convenient basis is obtained by adding 
\begin{equation}
\label{eq:F5-Born-LS}
        \Omega_5^{(1)}
        =
        [1234]+[123;145]+[1345] \, .
\end{equation}
Thus the five-point example already shows that the leading-singularity space
can grow with loop order. It also shows that the leading singularities are
naturally expressed as linear combinations of Kermit functions.
\end{eg}

At one loop, there is a representation of $F_n^{(\ell=1)}$ at any number of points in terms of chiral pentagons~\cite{ChicherinHenn2022}:
\begin{tcolorbox}[definitionbox]
\textbf{One-loop chiral-pentagon expansion.}
\begin{equation}
\label{eq:Fn-one-loop-chiral-pentagon-expansion}
        F_n^{(1)}
        =
        \sum_{1 \leq i< j \leq n}
        \Omega_{n}(ij)\,
        \mathcal{I}^{\rm cp}_{ij} \, .
\end{equation}
\end{tcolorbox}\noindent
Here the functions \(\mathcal{I}^{\rm cp}_{ij}(AB)\) are the integrated chiral pentagons, and are pure functions of weight two given in Subsection~\ref{subsec:One-Loop Prescriptive Unitarity}. The one-loop leading singularities can be expressed as sums of Kermits: \begin{equation}
\label{eq:Omega-n-ij-Kermit-sum}
        \Omega_{n}(ij)
        =
        \sum_{\substack{
        \Delta_1\subseteq P_1,\ \Delta_2\subseteq P_2\\
        \Delta_1,\Delta_2\subseteq T}}
        [\Delta_1;\Delta_2] \, ,
\end{equation}
where the diagonal \((ij)\) divides a regular \(n\)-gon $P_n$ into two polygons \(P_1\) and \(P_2\), and \(T\) is any triangulation of $P_n$ containing the arc \((ij)\). This formula is independent of the triangulation $T$, provided it contains $(ij)$. 

Beyond $\ell=1$, individual negative geometries have
also been integrated. At four points, all two-loop negative graphs were studied
in \cite{ArkaniHamedHennTrnka2021}. Still at four points, many higher-loop graphs have been recently evaluated in~\cite{Dixon:2026ipt}. At five points, important two-loop building
blocks include the ladder and triangle negative geometries, whose integrated
forms were analyzed from the point of view of positivity in
\cite{neg_geom_pos}. The next complete observable is the six-point two-loop
Wilson loop with a Lagrangian insertion, which was computed at symbol level in
\cite{Carrolo:2025pue}. Individual ladder negative geometries at six points and two loops, and at five points and three loops, have also been computed at symbol level using geometric Landau analysis and symbol bootstrap, which are discussed in Section~\ref{sec:Symbols and Bootstrap}.

\subsubsection{Classification of leading singularities}
\label{subsec:Classification of Leading Singularities}

We now summarize the all-loop classification of leading singularities of the
Wilson loop with a Lagrangian insertion. The result, proven in
\cite{BrownHennMazzucchelliTrnka2025}, confirms and extends the conjectural
picture that emerged from the explicit tree-level and one-loop analysis of
\cite{ChicherinHenn2022}. The basic surprise is that, although the leading
singularities of \(F_n^{(\ell)}\) are computed from residues of an
\((\ell+1)\)-loop logarithmic integrand, their values are always controlled by
one-loop Amplituhedron geometry, or equivalently by the \(m=k=2\)
Amplituhedron.

The main result is the following.
\begin{tcolorbox}[resultbox]
\textbf{Leading Singularities from Kermits.}
All leading singularities of \(F_n^{(\ell)}\), for \(n\geq4\) and
\(\ell\geq1\), can be expressed as linear combinations of Kermit forms
\eqref{six_invariant}.
\end{tcolorbox}\noindent
Equivalently, the algebraic prefactors in the perturbative decomposition
\eqref{eq:F-decomposition} belong to the vector space generated by canonical
functions of one-loop Amplituhedron tiles. This is a strong statement because
the residues are taken in higher-loop negative geometries. A priori, they could
have produced much more complicated algebraic functions of \(AB\) and the
external data.

The first consequence is that the vector space generated by leading
singularities stabilizes at two loops. The dimensions of the vector spaces of
leading singularities are
\begin{tcolorbox}[definitionbox]
\textbf{Dimension of the leading-singularity space.}
\begin{equation}
\label{eq:one-loop-LS-count-thesis}
       \ell=1:
       \quad
       \frac{n(n-3)}{2},
       \qquad
       \ell\geq2:
       \quad
       \frac{(n-1)(n-2)^2(n-3)}{12} \, .
\end{equation}
\end{tcolorbox}\noindent
These quantities are displayed in Table~\ref{tab:LS-count-thesis} for small $n$.
Thus at higher loops there are new leading-singularity configurations, but they
do not produce new independent leading-singularity values beyond the Kermit
span.

\begin{table}[pos=t]
    \centering
    \begin{tabular}{c|cccccccc}
        \(n\) & 5 & 6 & 7 & 8 & 9 & 10 & 11 & 12 \\
        \hline
        \(\ell=1\) & 5 & 9 & 14 & 20 & 27 & 35 & 44 & 54 \\
        \(\ell\geq2\) & 6 & 20 & 50 & 105 & 196 & 336 & 540 & 825
    \end{tabular}
    \caption{
    Number of linearly independent leading singularities of \(F_n^{(\ell)}\).
    At one loop the space is smaller, while from \(\ell=2\) onward it coincides
    with the full Kermit space.
    }
    \label{tab:LS-count-thesis}
\end{table}

The second consequence concerns the hidden momentum-space conformal symmetry
found in \cite{ChicherinHenn2022}. In the conformal frame in which the
Lagrangian insertion point \(x_0\) is sent to infinity, or equivalently
\begin{equation}
        AB\longrightarrow I_\infty \, ,
\end{equation}
all leading singularities multiplied by the Parke--Taylor factor
\begin{equation}
\label{eq:PT}
        {\rm PT}_n
        =
        \frac{1}
        {
        \langle 12\rangle
        \langle 23\rangle
        \cdots
        \langle n{-}1\,n\rangle
        \langle n1\rangle
        }
\end{equation}
are conformally invariant:
\begin{tcolorbox}[definitionbox]
\textbf{Conformal invariance at infinity.}
\begin{equation}
\label{eq:LS-conformal-invariance-thesis}
  \sum_{i=1}^{n}
  \frac{\partial^2}
  {\partial \lambda^{\alpha}_i
   \partial \tilde{\lambda}^{\dot{\alpha}}_i}
  \left(
        {\rm PT}_n\,
        \Omega_{n,s}\Big|_{AB\to I_\infty}
  \right)
  =
  0 \, .
\end{equation}
\end{tcolorbox}\noindent
The differential operator in~\eqref{eq:LS-conformal-invariance-thesis}
corresponds to the generator of special conformal transformations in
spinor-helicity variables. All other generators of the conformal group act
trivially, since the Parke--Taylor-dressed leading singularity is already
translation invariant, Lorentz invariant, and has the correct little-group and
dilatation weights. Checking the vanishing of
\eqref{eq:LS-conformal-invariance-thesis} is enough for ordinary momentum-space conformal invariance.

\begin{figure}[pos=t]
\centering
\begin{tikzpicture}[scale=0.5]
    \draw[thick] (0,0) circle(3cm);

    \coordinate (i) at (-130:3cm);
    \coordinate (j) at (110:3cm);
    \coordinate (k) at (60:3cm);
    \coordinate (l) at (-50:3cm);

    \coordinate (a) at (100:3cm);
    \coordinate (c) at (75:3cm);
    \coordinate (d) at (-60:3cm);
    \coordinate (f) at (-90:3cm);

    \node at (i) [below left] {$i$};
    \node at (j) [above left] {$j$};
    \node at (k) [above right] {$k$};
    \node at (l) [below right] {$l$};

    \draw[line width=0.4mm, red, thick] (i) -- (j);
    \draw[line width=0.4mm, red, dashed] (k) -- (l);

    \draw[line width=0.4mm, red, dashed] (a) -- (f);
    \draw[line width=0.4mm, red, thick] (c) -- (d);

    \foreach \p in {i,j,k,l,a,c,d,f} {
        \fill (\p) circle(2.5pt);
    }
\end{tikzpicture}

\caption{
Representation of a one-loop Amplituhedron geometry with extra sign
conditions. A solid arc \((ij)\) denotes
\(\langle AB\,ij\rangle<0\), while a dashed arc \((kl)\) denotes
\(\langle AB\,kl\rangle>0\). Every leading singularity of the Wilson loop with
a Lagrangian insertion is a sum of canonical functions of geometries of this
type.
}
\label{fig:LS-summary-thesis}
\end{figure}
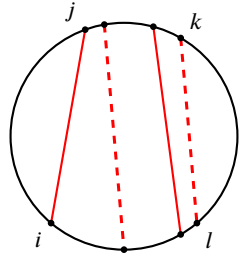

The main result follows from a stronger geometric statement. Consider the
one-loop Amplituhedron \(\mathcal A_n^{(1)}\) in the variable \(AB\). We may
refine this geometry by imposing extra sign conditions
\begin{equation}
\label{eq:ecr_cons}
        \langle AB\,i_rj_r\rangle>0
        \qquad\text{or}\qquad
        \langle AB\,i_rj_r\rangle<0 \, ,
\end{equation}
where the arcs \((i_rj_r)\) form a non-crossing collection in the \(n\)-gon.
We represent such a geometric space as in
Figure~\ref{fig:LS-summary-thesis}, by a collection of arcs that are solid if
they correspond to a negative condition, and dashed otherwise. The main
geometric result is then:
\begin{tcolorbox}[resultbox]
\textbf{Wilson-loop leading singularities.}
Every leading singularity of \(F_n^{(\ell)}\), for \(n\geq4\) and
\(\ell\geq1\), is a sum of canonical functions of one-loop Amplituhedron
geometries in the line \(AB\), with additional sign constraints
\(\langle AB\,i_rj_r\rangle\gtrless0\) associated with non-crossing arcs of the
\(n\)-gon.
\end{tcolorbox}\noindent
This is the formulation that connects directly to the \(m=2\) Amplituhedron and
its tilings by bicolored subdivisions
\cite{parisiShermanBennettWilliams2023m2,lukowski2019boundaries}. Once a
leading singularity has been reduced to the canonical function of such a
refined one-loop geometry, the known tiling results for the \(m=2\)
Amplituhedron express it as a sum of Kermit forms.

The proof of the classification is organized by a taxonomy of
leading-singularity configurations. Recall that these are configurations of
\(\ell\) lines \(AB_a\) in \(\mathbb P^3\), fully localized by solving
incidence conditions involving the \(n\) external points
\(Z_i\in\mathbb P^3\) and the external line \(AB\). These incidences have the
form
\begin{equation}
        \langle AB_a\,i\,i{+}1\rangle=0,
        \qquad
        \langle AB_a\,AB_b\rangle=0,
        \qquad
        \langle AB_a\,AB\rangle=0 \, ,
\end{equation}
possibly together with composite conditions arising from iterated residues, or
equivalently from factorizations of the corresponding polynomials.

The configurations are divided into three families:
\begin{enumerate}
    \item \textit{Simple configurations}: every loop line \(AB_a\), for
    \(a=1,\ldots,\ell\), localizes to
    \begin{equation}
        AB_a=i_a j_a \, ,
    \end{equation}
    with \(i_a<j_a\) and \(i_a,j_a\in\{1,\ldots,n\}\).

    \item \textit{Non-simple configurations}: at least one loop line is
    different from \(ij\). We further subdivide this case according to the
    relative configuration of the loop lines \(AB_a\) and the distinguished
    line \(AB\):
    \begin{enumerate}[label=(\roman*)]
        \item If
        \begin{equation}
                \langle AB\,AB_a\rangle\neq0
                \qquad
                \text{for every }a=1,\ldots,\ell \, ,
        \end{equation}
        we call such configurations \textit{\LABELNONSIMPLE}.

        \item If instead at least one of the loop lines intersects the
        distinguished line, namely 
        \begin{equation}
                \langle AB_a\,AB\rangle=0
                \qquad
                \text{for at least one }a \, ,
        \end{equation}
        we call such configurations \textit{\LABELREVERSE}.
    \end{enumerate}
\end{enumerate}
The first family is analogous to maximal-codimension boundaries of the ordinary
loop Amplituhedron \cite{Arkani-Hamed:2013kca}. The other two families are
specific to the present problem: non-simple configurations arise because mutual
positivity conditions between loop lines have been replaced by mutual
negativity conditions, while the intersecting case also uses the special role
of the unintegrated line \(AB\).

The second layer of the taxonomy is compatibility. Notice that, in a
leading-singularity configuration, the localized lines \(AB_a\) are functions
of the external data \(Z_i\) and of \(AB\). We then define:
\begin{enumerate}[label=(\alph*)]
    \item \textit{Compatible configurations}: every pair of loop lines in the
    configuration satisfies
    \begin{equation}
        \langle AB_a\,AB_b\rangle\geq 0
    \end{equation}
    for every \(AB\in\mathcal A_n^{(1)}\).

    \item \textit{Incompatible configurations}: all other configurations,
    those for which at least one pair of loop lines satisfies
    \begin{equation}
        \langle AB_a\,AB_b\rangle<0
    \end{equation}
    for at least one point \(AB\in\mathcal A_n^{(1)}\).
\end{enumerate}
For simple configurations this has a familiar combinatorial meaning: a simple
configuration is compatible if and only if the arcs \((i_a j_a)\) are
non-crossing. This is closely related to the notion of compatibility in cluster
algebras of type \(A_n\)~\cite{lukowski2019cluster}.

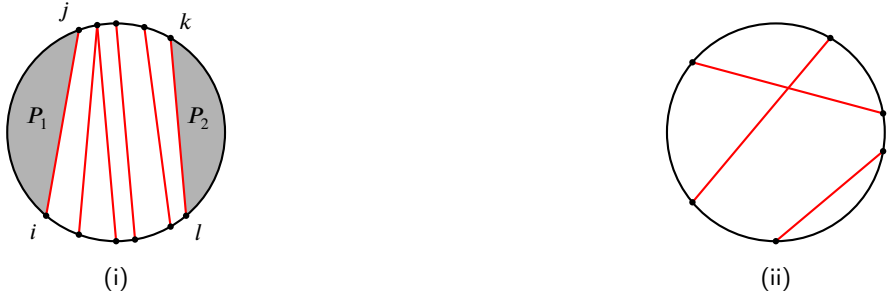
\begin{figure}[pos=t]
\centering

\begin{minipage}{0.47\textwidth}
\centering
\begin{tikzpicture}[scale=0.48]
    \path[use as bounding box] (-3.6,-4.4) rectangle (3.6,3.6);

    \draw[thick] (0,0) circle(3cm);

    \coordinate (i) at (-130:3cm);
    \coordinate (j) at (110:3cm);
    \coordinate (k) at (60:3cm);
    \coordinate (l) at (-50:3cm);

    \coordinate (a) at (100:3cm);
    \coordinate (b) at (90:3cm);
    \coordinate (c) at (75:3cm);
    \coordinate (d) at (-60:3cm);
    \coordinate (e) at (-80:3cm);
    \coordinate (f) at (-90:3cm);
    \coordinate (g) at (-110:3cm);

    \node at (i) [below left] {$i$};
    \node at (j) [above left] {$j$};
    \node at (k) [above right] {$k$};
    \node at (l) [below right] {$l$};

    \fill[black, opacity=0.3] (i) -- (j) arc (110:230:3cm) -- cycle;
    \fill[black, opacity=0.3] (l) -- (k) arc (60:-50:3cm) -- cycle;

    \draw[line width=0.4mm, red, thick] (i) -- (j);
    \draw[line width=0.4mm, red, thick] (k) -- (l);

    \draw[line width=0.4mm, red, thick] (a) -- (f);
    \draw[line width=0.4mm, red, thick] (a) -- (g);
    \draw[line width=0.4mm, red, thick] (b) -- (e);
    \draw[line width=0.4mm, red, thick] (c) -- (d);

    \foreach \p in {i,j,k,l,a,b,c,d,e,f,g} {
        \fill (\p) circle(2.5pt);
    }

    \node at (170:2.2cm) {$P_1$};
    \node at (10:2.3cm) {$P_2$};

    \node at (0,-4.05) {(i)};
\end{tikzpicture}
\end{minipage}
\hfill
\begin{minipage}{0.47\textwidth}
\centering
\begin{tikzpicture}[scale=0.48]
    \path[use as bounding box] (-3.6,-4.4) rectangle (3.6,3.6);

    \draw[thick] (0,0) circle(3cm);

    \coordinate (x) at (-140:3cm);
    \coordinate (y) at (60:3cm);
    \coordinate (w) at (140:3cm);
    \coordinate (z) at (10:3cm);
    \coordinate (r) at (-10:3cm);
    \coordinate (s) at (-90:3cm);

    \draw[line width=0.4mm, red, thick] (x) -- (y);
    \draw[line width=0.4mm, red, thick] (w) -- (z);
    \draw[line width=0.4mm, red, thick] (r) -- (s);

    \foreach \p in {x,y,w,z,r,s} {
        \fill (\p) circle(2.5pt);
    }

    \node at (0,-4.05) {(ii)};
\end{tikzpicture}
\end{minipage}

\caption{
Two simple leading-singularity configurations. In (i), all arcs, shown in red,
are non-crossing, so the configuration is compatible. In (ii), the arcs cross,
so the configuration is incompatible. In (i), the shaded regions are the
polygons \(P_1\) and \(P_2\) bounded by the extremal arcs \((ij)\) and
\((kl)\). The canonical function of the corresponding geometry, given in
\eqref{LS_simple_formulae}, depends only on these two extremal arcs.
}
\label{fig:simple-LS-thesis}
\end{figure}

\begin{tcolorbox}[resultbox]
\textbf{Compatible configurations.}
Simple compatible leading-singularity configurations are represented by
partial triangulations of the \(n\)-gon.
\end{tcolorbox}\noindent
These are an important family of leading singularities: they span the linear
space of leading singularities of \(F_n^{(\ell)}\), and their values have
particularly simple expressions in terms of Kermits. Indeed, any simple
compatible leading-singularity configuration has two extremal arcs \((ij)\)
and \((kl)\), as in Figure~\ref{fig:simple-LS-thesis}. The associated
leading-singularity value depends only on these two arcs, and is given by
\begin{tcolorbox}[definitionbox]
\begin{equation}
\label{LS_simple_formulae}
    \Omega_n(ij,kl)
    =
    \sum_{\substack{
        \Delta_1,\Delta_2\subseteq \overline T\\
        \Delta_1\subseteq P_1,\ \Delta_2\subseteq P_2
    }}
    [\Delta_1;\Delta_2] \, .
\end{equation}
\end{tcolorbox}\noindent
Here the sum is over non-overlapping triangles
\(\Delta_1=\{a_1,b_1,c_1\}\) and
\(\Delta_2=\{a_2,b_2,c_2\}\) with arcs in \(\overline T\), and the polygons
\(P_1\) and \(P_2\) are bounded by \((ij)\) and \((kl)\), respectively, as in
Figure~\ref{fig:simple-LS-thesis}. In the special case \((ij)=(kl)\), one
recovers the expression for \(\Omega_n(ij)\) in
\eqref{eq:Omega-n-ij-Kermit-sum}. From
\eqref{LS_simple_formulae}, it is clear that all simple compatible leading
singularities saturate at \(\ell=2\), since they are specified by only two
arcs.

The proof of the classification has three ingredients. The first concerns the
exclusion of incompatible configurations.
\begin{tcolorbox}[resultbox]
\textbf{Reduction result.}
Incompatible leading-singularity configurations of \(F_n^{(\ell)}\) either
evaluate to zero, or reduce to sums of leading-singularity values of lower-loop
compatible configurations.
\end{tcolorbox}\noindent
Thus incompatible configurations do not give new leading-singularity values of
the full Wilson loop. This is crucial because individual negative geometries can
have complicated non-rational leading-singularity values. These cancel out in
the signed summation over negative geometries in
\eqref{eq:Omega-negative-geometry-expansion} defining the integrand of
\(F_n^{(\ell)}\). Let us illustrate this.

\begin{eg}[Examples of reduction]\label{eg:red}
The simplest example occurs at \(n=5\) and \(\ell=2\). Consider the simple
configuration
\begin{equation}
        AB_1=13 \, ,
        \qquad
        AB_2=24 \, .
\end{equation}
It is incompatible, since $\langle1324\rangle<0$. In the negative-geometry expansion of \(F_5^{(2)}\), graphically given by
\begin{equation}
\label{eq:2-loop}
\begin{tabular}{cc}
 \includegraphics[width=8cm]{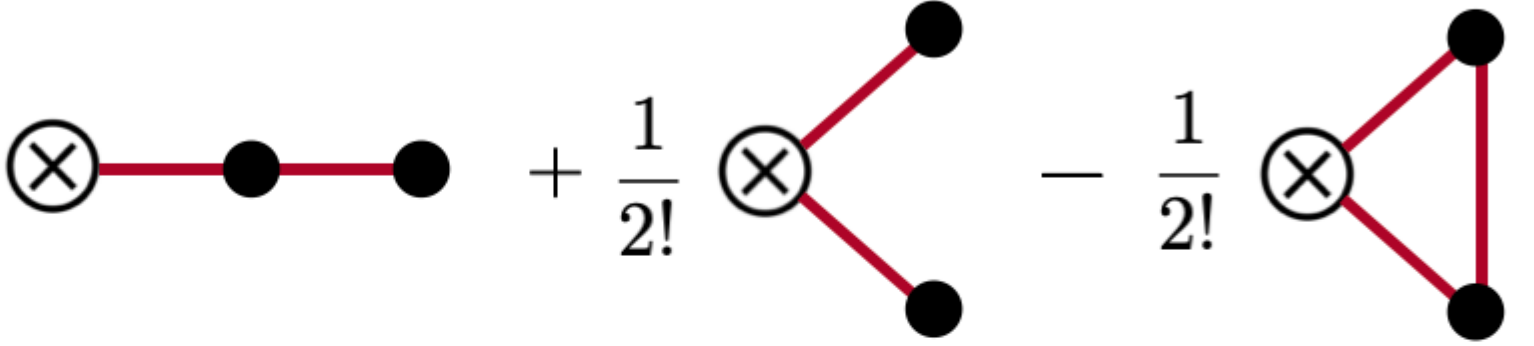}
\end{tabular}
\end{equation}
this leading-singularity value cancels between the ladder graph with the marked
point in the middle and the triangle graph, which enter with opposite signs.
The remaining contribution comes from the two labelled ladder graphs with the
marked point at an end, and the resulting value is
\begin{equation}
        \Omega_5(13)+\Omega_5(24) \, .
\end{equation}
This is a sum of leading singularities associated with simple compatible
configurations at one loop, see \eqref{LS_simple_formulae}, exactly as
predicted by the reduction result.

\begin{figure}[pos=t]
\centering
\begin{tikzpicture}[scale=0.6]

    \coordinate (k) at (1,-2);
    \coordinate (P) at (0.17,0.05);
    \coordinate (Q) at (-1.12,-0.22);
    \coordinate (R) at (-1.25,2.18);
    \coordinate (S) at (-0.72,2.33);
    \coordinate (j) at (-2,2);
    \coordinate (j_plus) at (0,2.5);
    \coordinate (i) at (-1,-3);
    \coordinate (C) at (-1,3);
    \coordinate (D) at (1.4,-3);
    \coordinate (A) at (-2.5,-0.5);
    \coordinate (B) at (2.5,0.5);
    \coordinate (E) at (-1.3,3);
    \coordinate (F) at (-0.95,-3.8);

    \draw[dashed, gray] (j) -- (j_plus);

    \draw[red, thick] (C) -- (D);
    \node[right, red] at (C) {$AB_2$};

    \draw[red, thick] (E) -- (F);
    \node[left, red] at (E) {$AB_1$};

    \draw[teal, thick] (A) -- (B);
    \node[above, teal] at (B) {$AB$};

    \foreach \p in {i,P,Q,R,S,j,j_plus,k} {
        \fill (\p) circle (2pt);
    }

    \node[right] at (i) {$i$};
    \node[left] at (j) {$j$};
    \node[right] at (j_plus) {$j+1$};
    \node[right] at (k) {$k$};

    \begin{scope}[xshift=8cm]
        \node at (4.5,0) {$\bigcup \quad \dots$};

        \draw[thick] (0,0) circle(2cm);

        \coordinate (i) at (-100:2cm);
        \coordinate (j) at (130:2cm);
        \coordinate (k) at (100:2cm);
        \coordinate (l) at (-20:2cm);

        \node at (i) [below] {$i$};
        \node at (j) [above left] {$j$};
        \node at (k) [above] {$j+1$};
        \node at (l) [below right] {$k$};

        \draw[line width=0.4mm, red, dashed] (i) -- (j);
        \draw[line width=0.4mm, red, thick] (i) -- (k);
        \draw[line width=0.4mm, red, thick] (l) -- (j);
        \draw[line width=0.4mm, red, dashed] (l) -- (k);
        \draw[line width=0.4mm, red, thick] (l) -- (i);

        \foreach \p in {i,j,k,l} {
            \fill (\p) circle(2.5pt);
        }
    \end{scope}
\end{tikzpicture}
\caption{
An example of a \textit{\LABELREVERSE} incompatible leading-singularity
configuration at \(\ell=2\), where the bracket
\(\langle AB_1\,AB_2\rangle\), which is a function of \(AB\), does not have a
fixed sign for \(AB\in\mathcal A_n^{(1)}\).
}
\label{fig:L2-reverse-incompatible}
\end{figure}
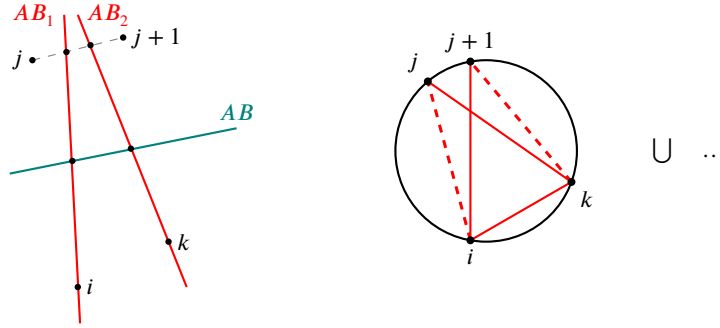

A second example illustrates why compatibility is more subtle for non-simple
configurations. At \(\ell=2\), one may have a configuration in which the two
localized lines intersect and depend on \(AB\). For example,
\begin{equation}
        AB_1=i,\bigl(j+\alpha(j+1)\bigr),
        \qquad
        AB_2=\bigl(j+\beta(j+1)\bigr),k \, ,
\end{equation}
where
\begin{equation}
        \alpha
        =
        \frac{\langle AB\,ij\rangle}
        {\langle AB\,j{+}1\,i\rangle}
\end{equation}
is a function of \(AB\), fixed by imposing
\(\langle AB_1\,AB\rangle=0\), and analogously for \(\beta\). Then
\begin{equation}
\label{eq:reverse-incompatible-bracket-thesis}
        \langle AB_1\,AB_2\rangle
        =
        (\beta-\alpha)\,
        \langle i\,j\,j{+}1\,k\rangle \, .
\end{equation}
Even though the one-loop Amplituhedron conditions for \(AB_1\) and \(AB_2\)
force \(\alpha>0\) and \(\beta>0\), as \(AB\) varies in
\(\mathcal A_n^{(1)}\), the bracket in
\eqref{eq:reverse-incompatible-bracket-thesis} need not have a fixed sign. If
it becomes negative somewhere in \(\mathcal A_n^{(1)}\), the configuration is incompatible, and hence does not contribute a new independent
leading-singularity value. This example explains why compatibility must be
defined as a condition over the whole one-loop Amplituhedron for \(AB\). This example illustrated in Figure~\ref{fig:L2-reverse-incompatible}.
\end{eg}

The next two ingredients concern the classification of the remaining compatible
configurations. They constitute the most technically challenging part of the
proof, and we will illustrate them by examples in the next subsection. 

The last ingredient is the passage from the geometric space of a one-loop Amplituhedron
with extra conditions coming from a signed partial triangulation to its
canonical function, expressed as an expansion in Kermit forms.

\begin{tcolorbox}[resultbox]
\textbf{From triangulations to Kermits.}
For every signed partial triangulation of the \(n\)-gon as in Figure~\ref{fig:LS-summary-thesis}, the corresponding geometric
space can be tiled into Kermits.
\end{tcolorbox}\noindent
This follows quite directly from the recent results on the \(m=2\)
Amplituhedron tiles and tilings, and from their combinatorial encoding by
bicolored subdivisions
\cite{parisiShermanBennettWilliams2023m2,lukowski2019boundaries}.

Putting all ingredients together gives the logical chain
\begin{tcolorbox}[resultbox]
\textbf{Classification chain.}
\begin{equation}
\begin{aligned}
        &\text{leading singularities of the Wilson loop with Lagrangian insertion}
        \longrightarrow \\
        &\text{compatible leading-singularity configurations}
        \longrightarrow
        \text{signed partial triangulations}
        \longrightarrow \\
        &\text{one-loop Amplituhedron geometries in }AB
        \longrightarrow
        \text{linear combinations of Kermit forms}.
\end{aligned}
\end{equation}
\end{tcolorbox}\noindent
This is the mechanism behind the all-loop classification. The next subsection
will illustrate the remaining classification steps through examples.

\subsubsection{Taxonomy of leading singularities and compatibility}
\label{subsec:Taxonomy Compatibility and Examples}

We now illustrate the taxonomy introduced in the previous subsection. The point
of the examples below is twofold. First, they show explicitly why one has to
distinguish between leading-singularity configurations and
leading-singularity values. Individual negative geometries may have residues
which are nonzero, non-Kermit-like, and even non-conformal after
\(AB\to I_\infty\). These residues are not, however, new leading singularities
of the full Wilson loop \(F_n^{(\ell)}\). Second, the examples explain how the
technical classification works in practice: after incompatible configurations
are removed by the reduction result, the remaining configurations give signed
partial triangulations of the \(n\)-gon, and hence canonical functions of
one-loop Amplituhedron geometries in \(AB\).

\begin{eg}[Simple compatible configuration at \(n=5,\ell=1\)]
\label{eg:simple-compatible-n5-l1}

Let us start with the simplest nontrivial compatible example. Take \(n=5\) and
\(\ell=1\). There is only one one-loop negative geometry, namely the ladder.
The nonzero simple compatible configuration is
\begin{equation}
        AB_1=13 \, .
\end{equation}
The associated geometry is the one-loop Amplituhedron
\(\mathcal A_5^{(1)}\) in the line \(AB\), with \begin{equation}
        \langle AB\,13\rangle<0 \, .
\end{equation}
Using the one-loop formula from \eqref{LS_simple_formulae}, the corresponding
leading-singularity value is
\begin{equation}
\label{omega_n5_13}
        \Omega_5(13)
        =
        [1234]+[123;145] \, .
\end{equation}
This is one of the five one-loop leading singularities of \(F_5^{(1)}\), see Figure~\ref{fig:5pt-simple-13-thesis}.
\begin{figure}[pos=t]
\centering

\begin{minipage}{0.31\textwidth}
\centering
\begin{tikzpicture}[scale=1.15]
    \path[use as bounding box] (-1.4,-1.55) rectangle (1.4,1.45);

    \draw[thick] (90:1)
        \foreach \x in {162,234,306,18}
            { -- (\x:1) } -- cycle;

    \coordinate (D) at (18:1);
    \coordinate (C) at (90:1);
    \coordinate (B) at (162:1);
    \coordinate (A) at (234:1);
    \coordinate (E) at (306:1);

    \draw[red, thick] (A) -- (C);

    \node[below] at (A) {1};
    \node[left] at (B) {2};
    \node[above] at (C) {3};
    \node[right] at (D) {4};
    \node[below] at (E) {5};

    \node at (0,-1.38) {(a)};
\end{tikzpicture}
\end{minipage}
\hfill
\begin{minipage}{0.31\textwidth}
\centering
\begin{tikzpicture}[scale=1.15]
    \path[use as bounding box] (-1.4,-1.55) rectangle (1.4,1.45);

    \draw[thick] (90:1)
        \foreach \x in {162,234,306,18}
            { -- (\x:1) } -- cycle;

    \coordinate (D) at (18:1);
    \coordinate (C) at (90:1);
    \coordinate (B) at (162:1);
    \coordinate (A) at (234:1);
    \coordinate (E) at (306:1);

    \fill[black, opacity=0.3] (A) -- (B) -- (C) -- (D) -- cycle;
    \draw[red, dashed] (A) -- (D);

    \node at (0,-1.38) {(b)};
\end{tikzpicture}
\end{minipage}
\hfill
\begin{minipage}{0.31\textwidth}
\centering
\begin{tikzpicture}[scale=1.15]
    \path[use as bounding box] (-1.4,-1.55) rectangle (1.4,1.45);

    \draw[thick] (90:1)
        \foreach \x in {162,234,306,18}
            { -- (\x:1) } -- cycle;

    \coordinate (D) at (18:1);
    \coordinate (C) at (90:1);
    \coordinate (B) at (162:1);
    \coordinate (A) at (234:1);
    \coordinate (E) at (306:1);

    \fill[black, opacity=0.3] (A) -- (B) -- (C) -- cycle;
    \fill[black, opacity=0.3] (A) -- (D) -- (E) -- cycle;

    \draw[red, thick] (A) -- (D);
    \draw[red, thick] (A) -- (C);

    \node at (0,-1.38) {(c)};
\end{tikzpicture}
\end{minipage}

\caption{
A simple compatible leading-singularity configuration at \(n=5\) and
\(\ell=1\). The configuration in (a) gives the geometry
\(\mathcal A_5^{(1)}\cap\{\langle AB\,13\rangle<0\}\), which decomposes into
the two Kermit geometries shown in (b) and (c). Its value is
\(\Omega_5(13)=[1234]+[123;145]\).
}
\label{fig:5pt-simple-13-thesis}
\end{figure}
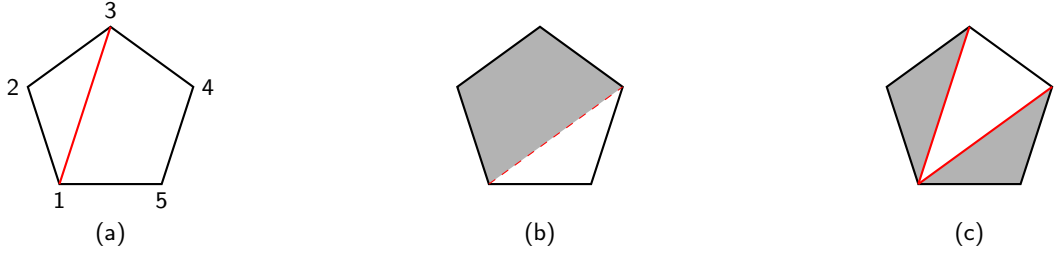
\end{eg}

\begin{eg}[Simple incompatible configuration at \(n=5,\ell=2\)]
\label{eg:simple-incompatible-n5-l2}

For an example of a simple but incompatible configuration we move to two loops.
Consider
\begin{equation}
        AB_1=13,
        \qquad
        AB_2=24 \, .
\end{equation}
We explained in Example~\ref{eg:red} how this incompatible configuration cancels in the sum over negative geometries. On the other hand, this LS configuration and its value are relevant for individual negative geometries, in particular for those associated to both the second and third graphs in~\eqref{eq:2-loop}. The one-loop geometry in \(AB\) in this case has extra conditions
\begin{equation}
        \langle AB\,13\rangle<0 \, ,
        \qquad
        \langle AB\,24\rangle<0 \, .
\end{equation}
Its canonical function can be computed by expanding in Kermits and parametrizing each space. The result is
\begin{equation}
\label{LSvaluesimpleincompatibleexample}
\begin{aligned}
        \Omega_5(13,24)        =
        &\frac{
        \langle AB\,(123)\cap(451)\rangle\,
        \langle AB\,(234)\cap(451)\rangle
        }{
        \langle AB\,14\rangle
        \langle AB\,23\rangle
        \langle AB\,13\rangle
        \langle AB\,15\rangle
        \langle AB\,45\rangle
        \langle AB\,24\rangle
        }
        +[1234] \, .
\end{aligned}
\end{equation}
This value has the typical undesirable features of incompatible
configurations. First, it is not expressible purely in terms of Kermit forms.
Second, in the frame \(AB\to I_\infty\), the first term becomes
\begin{equation}
        \frac{[25][35]}
        {\langle13\rangle\langle14\rangle\langle24\rangle} \, ,
\end{equation}
which is not annihilated by the conformal generator in~\eqref{eq:LS-conformal-invariance-thesis}.
Thus this residue is neither a Kermit nor conformally invariant.

We now turn to the taxonomy of non-simple configurations. The first relevant
statement removes all \LABELNONSIMPLE{} configurations from the list of
possible new leading singularities of the full observable.

\begin{tcolorbox}[resultbox]
\textbf{Classification result.}
All \LABELNONSIMPLE{} leading-singularity configurations are incompatible.
\end{tcolorbox}\noindent
Recall that \LABELNONSIMPLE{} configurations are non-simple configurations in
which no localized loop line intersects \(AB\). The idea of the proof is as
follows. Since \(AB\) does not participate in the localization, a genuinely
\LABELNONSIMPLE{} configuration must contain some line which is not an external
chord. Such a line must be fixed by incidence with other loop lines. If it
intersects only one other loop line, the one-loop Amplituhedron boundary
conditions force it to degenerate to a chord, except in empty cases. If it is
fixed by two or more other loop lines, the relevant auxiliary Schubert problems
force either a degeneration to a simple configuration or a pair of crossing
chords. The latter have negative mutual bracket, hence give an incompatible
configuration. This is the geometric reason why all \LABELNONSIMPLE{}
configurations are excluded.

\begin{eg}[\LABELNONSIMPLE{} configurations in star-graphs]
\label{eg:nonsimple-three-loop-star}

Consider the three-loop negative geometry whose graph is a star on four vertices with a marked pending vertex. Take
\(n=5\), see also Figure~\ref{fig:non-simple-LS-thesis}. We localize
\begin{equation}
        AB_1=24 \, ,
        \qquad
        AB_2=35 \, ,
\end{equation}
and fix \(AB_3\) by imposing
\begin{equation}
\label{non-simple-LS-cuts-thesis}
        \langle AB_3\,AB_1\rangle
        =
        \langle AB_3\,AB_2\rangle
        =
        \langle AB_3\,51\rangle
        =
        \langle AB_3\,12\rangle
        =
        0 \, .
\end{equation}
The Schubert problem \eqref{non-simple-LS-cuts-thesis} has two solutions; the
one lying inside $\mathcal{A}^{(1)}_5$ is
\begin{equation}
\label{loop-nonsimple-nonintersecting-thesis}
        AB_3=(124)\cap(135) \, .
\end{equation}
The resulting geometry in the unintegrated line \(AB\) is the one-loop
Amplituhedron with 
\begin{equation}
        \langle AB\,AB_3\rangle<0 \, .
\end{equation}
Its canonical function can be computed to be
\begin{equation}
\label{LS-star-thesis}
        \frac{
        \langle1235\rangle
        \langle1245\rangle
        \langle AB\,(123)\cap(451)\rangle
        }{
        \langle AB\,12\rangle
        \langle AB\,23\rangle
        \langle AB\,45\rangle
        \langle AB\,51\rangle
        \langle AB\,(124)\cap(135)\rangle
        } \, .
\end{equation}
This is a nonzero residue of the star-negative geometry. Nevertheless, the
configuration is incompatible, because
\begin{equation}
        \langle AB_1\,AB_2\rangle
        =
        \langle2435\rangle<0 \, .
\end{equation}
Thus \eqref{LS-star-thesis} is not a new leading singularity of the full Wilson
loop. It is an example of the kind of non-Kermit residue that the
classification result removes.
\end{eg}

The \LABELNONSIMPLE{} sector can become much more complicated. Consider the
negative geometry associated with the five-loop star, for \(n\geq8\). We may
localize
\begin{equation}
        AB_1=13 \, ,
        \qquad
        AB_2=24 \, ,
        \qquad
        AB_3=57 \, ,
        \qquad
        AB_4=68 \, ,
\end{equation}
and fix \(AB_5\) to be one of the two lines intersecting all four of these
lines. This is a generic Schubert problem in \(\Gr(2,4)\), and solving it
requires a nontrivial quadratic equation. Hence the location of \(AB_5\) is
not rational in the external data.

Numerically one finds that one of the two solutions for \(AB_5\) can lie in
\(\mathcal A_n^{(1)}\), so the corresponding individual negative geometry has
a nonzero residue. The associated value is the canonical function of the
one-loop Amplituhedron in \(AB\) with the additional condition
\begin{equation}
        \langle AB\,AB_5\rangle<0 \, .
\end{equation}
Again, however, according to the classification result implies that this
\LABELNONSIMPLE{} configuration is incompatible, which is in fact the case as $\langle AB_1 \, AB_2 \rangle < 0$. Such examples show that
individual negative geometries can have complicated LS configurations, including
non-rational solutions, and these cancel out in the full Wilson loop.

\begin{figure}[pos=t]
\centering
\begin{tikzpicture}[scale=0.5]

    \begin{scope}
    \node at (0,-5) {(a)};
    \node at (5,-5) {\includegraphics[scale=0.5]{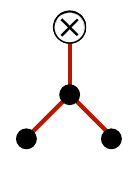}};

    \coordinate (k) at (1,-2);
    \coordinate (i) at (2.4,1.3);
    \coordinate (l) at (-2,-0.77);
    \coordinate (C) at (-1,2);
    \coordinate (D) at (1.5,-3);
    \coordinate (E) at (-2.5,-1);
    \coordinate (F) at (2.8,1.5);
    \coordinate (G) at (-2.5,1);
    \coordinate (H) at (2.8,0.3);
    \coordinate (A) at (3,-2.5);
    \coordinate (B) at (5,-0.5);
    \coordinate (R) at (-0.09,0.14);
    \coordinate (T) at (-0.35,0.7);
    \coordinate (j) at (2.4,0.35);
    \coordinate (m) at (-2,0.92);

    \draw[red, thick] (C) -- (D);
    \node[above, red] at (C) {$AB_3$};

    \draw[red, thick] (E) -- (F);
    \node[left, red] at (E) {$AB_1$};

    \draw[red, thick] (G) -- (H);
    \node[left, red] at (G) {$AB_2$};

    \draw[teal, thick] (A) -- (B);
    \node[right, teal] at (B) {$AB$};

    \foreach \p in {i,j,l,R,T,m,k} {
        \fill (\p) circle (2pt);
    }

    \node[below left] at (k) {1};
    \node[above] at (i) {4};
    \node[below] at (l) {2};
    \node[below] at (j) {5};
    \node[below] at (m) {3};
    \end{scope}

    \begin{scope}[xshift=15cm]
    \node at (0,-5) {(b)};
    \node at (5,-5) {\includegraphics[scale=0.5]{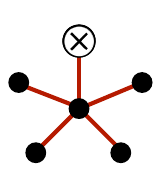}};

    \coordinate (i) at (2.4,1.3);
    \coordinate (l) at (-2,-0.77);
    \coordinate (C) at (-1,2);
    \coordinate (D) at (0.5,-4);
    \coordinate (E) at (-2.5,-1);
    \coordinate (F) at (2.8,1.5);
    \coordinate (G) at (-2.5,1);
    \coordinate (H) at (2.8,0.3);
    \coordinate (A) at (4,-2);
    \coordinate (B) at (6,0);
    \coordinate (R) at (-0.5,-0.05);
    \coordinate (T) at (-0.66,0.75);
    \coordinate (j) at (2.4,0.35);
    \coordinate (m) at (-2,0.92);
    \coordinate (A3) at (-2.5,-4);
    \coordinate (B3) at (2.8,-1.5);
    \coordinate (A4) at (-2.5,-2);
    \coordinate (B4) at (2.8,-2.7);
    \coordinate (i1) at (-2,-3.8);
    \coordinate (i2) at (-2,-2.1);
    \coordinate (i3) at (2.3,-1.7);
    \coordinate (i4) at (2.3,-2.65);
    \coordinate (j1) at (0.1,-2.35);
    \coordinate (j2) at (0.2,-2.8);

    \draw[red, thick] (C) -- (D);
    \node[above, red] at (C) {$AB_5$};

    \draw[red, thick] (E) -- (F);
    \node[above left, red] at (E) {$AB_3$};

    \draw[red, thick] (G) -- (H);
    \node[left, red] at (G) {$AB_4$};

    \draw[teal, thick] (A) -- (B);
    \node[right, teal] at (B) {$AB$};

    \draw[red, thick] (A3) -- (B3);
    \node[left, red] at (A3) {$AB_1$};

    \draw[red, thick] (A4) -- (B4);
    \node[below left, red] at (A4) {$AB_2$};

    \foreach \p in {i,j,l,R,T,m,i1,i2,i3,i4,j1,j2} {
        \fill (\p) circle (2pt);
    }

    \node[above] at (i) {7};
    \node[above] at (l) {5};
    \node[below] at (j) {8};
    \node[above] at (m) {6};
    \node[below] at (i1) {1};
    \node[below] at (i2) {2};
    \node[above] at (i3) {3};
    \node[below] at (i4) {4};
    \end{scope}
\end{tikzpicture}
\caption{
Two \LABELNONSIMPLE{} configurations: (a) An $\ell=3$ star at \(n\geq5\),
and (b) an $\ell=5$ star at \(n\geq8\). They are both incompatible. }
\label{fig:non-simple-LS-thesis}
\end{figure}
\end{eg}

We now move to the genuinely relevant non-simple configurations. Since the
\LABELNONSIMPLE{} class is incompatible, the only non-simple configurations
that can contribute new values are those in which at least one localized loop
line intersects \(AB\). These are the \LABELREVERSE{} configurations.

\begin{tcolorbox}[resultbox]
\textbf{Signed partial triangulations.}
All compatible \LABELREVERSE{} leading singularities correspond to unions of
signed partial triangulations of the \(n\)-gon.
\end{tcolorbox}\noindent
The proof is technical, but its organizing idea is simple. As the
localized loop lines are solved, one produces intersection points in
\(\mathbb P^3\). The first crucial observation is that the Amplituhedron conditions force these points to be always rational in $Z_i$ and $AB$. These points can however be nested: a line may pass through a point which is itself constructed as an intersection of earlier lines and planes. The proof assigns a \textit{length} to such intersection points and controls the configuration inductively. Compatibility rules out the non-rational Schubert solutions and also removes the ``non-MHV'' intersection points of the form \((PQ)\cap(RST)\). What remains are intersection points built recursively from
\begin{equation}
\label{eq:allowed-intersection-thesis}
        (PQ)\cap(ABR)
        =
        P\,\langle AB\,QR\rangle
        -
        Q\,\langle AB\,PR\rangle \, ,
\end{equation}
and
\begin{equation}
\label{eq:allowed-intersection-on-AB-thesis}
\begin{aligned}
        AB\cap(PQR)
        &=
        P\,\langle AB\,QR\rangle
        +
        Q\,\langle AB\,RP\rangle
        +
        R\,\langle AB\,PQ\rangle .
\end{aligned}
\end{equation}
These are used to explain why the final conditions on \(AB\) factor into signs of
ordinary brackets \(\langle AB\,ij\rangle\). Compatibility also forces the
corresponding arcs to be non-crossing. Hence the resulting \(AB\)-geometry is a
union of signed partial triangulations.

\begin{eg}[\LABELREVERSE{} compatible configuration at \(\ell=1\)]
\label{eg:reverse-compatible-l1}

Consider the one-loop ladder and impose that the localized line \(AB_1\)
intersects the distinguished line \(AB\). In order to be fully localized,
\(AB_1\) must also lie on a one-dimensional boundary of
\(\mathcal A_n^{(1)}\). Such boundaries have the form
\begin{equation}
        AB_1
        =
        i,\bigl(j+\alpha(j+1)\bigr),
        \qquad
        \alpha>0 \, ,
\end{equation}
with \(i<j\). The line lies in the plane \((ijj{+}1)\). Requiring
\(\langle AB_1\,AB\rangle=0\) forces \(AB_1\) to pass through the point
\(AB\cap(ijj{+}1)\), and fixes
\begin{equation}
\label{alpha-thesis}
        \alpha
        =
        \frac{\langle AB\,ij\rangle}
        {\langle AB\,j{+}1\,i\rangle} \, .
\end{equation}
The positivity condition \(\alpha>0\) is therefore equivalent to the statement
that the two brackets in \eqref{alpha-thesis} have the same sign. Hence the
associated \(AB\)-geometry decomposes into two pieces as in Figure~\ref{fig:reverse-LS-L1-thesis}. The LS value is a special case of that in the first row of Table~\ref{tab:all-L2-config-thesis}. This is the simplest example of a \LABELREVERSE{} compatible configuration
giving a union of signed partial triangulations.

\begin{figure}[pos=t]
\centering
\begin{tikzpicture}[scale=0.5]

    \node at (0,-6) {(a)};

    \coordinate (i) at (1,-2);
    \coordinate (i_plus) at (0,0);
    \coordinate (j) at (-2,2);
    \coordinate (j_plus) at (0,2.5);
    \coordinate (C) at (-1.5,3);
    \coordinate (D) at (1.5,-3);
    \coordinate (A) at (-2.5,-0.5);
    \coordinate (B) at (2.5,0.5);
    \coordinate (R) at (-1.1,2.23);

    \draw[dashed, gray] (j) -- (j_plus);

    \draw[red, thick] (C) -- (D);
    \node[above, red] at (C) {$AB_1$};

    \draw[teal, thick] (A) -- (B);
    \node[above, teal] at (B) {$AB$};

    \foreach \p in {i,i_plus,j,j_plus,R} {
        \fill (\p) circle (2pt);
    }

    \node[right] at (i) {$i$};
    \node[above left] at (j) {$j$};
    \node[above right] at (j_plus) {$j+1$};

    \begin{scope}[xshift=9cm]
    \node at (0,-3.5) {$\langle AB\,ij\rangle>0$,};
    \node at (0,-4.5) {$\langle AB\,ij{+}1\rangle<0$,};
    \node at (4,-6) {(b)};

    \draw[thick] (0,0) circle(2cm);

    \coordinate (i) at (-100:2cm);
    \coordinate (j) at (130:2cm);
    \coordinate (k) at (100:2cm);

    \node at (i) [below] {$i$};
    \node at (j) [above left] {$j$};
    \node at (k) [above] {$j+1$};

    \draw[line width=0.4mm, red, dashed] (i) -- (j);
    \draw[line width=0.4mm, red, thick] (i) -- (k);

    \foreach \p in {i,j,k} {
        \fill (\p) circle(2.5pt);
    }
    \end{scope}

    \begin{scope}[xshift=16cm]
    \node at (0,-3.5) {$\langle AB\,ij\rangle<0$,};
    \node at (0,-4.5) {$\langle AB\,ij{+}1\rangle>0$,};
    \node at (-3.5,0) {$\bigcup$};

    \draw[thick] (0,0) circle(2cm);

    \coordinate (i) at (-100:2cm);
    \coordinate (j) at (130:2cm);
    \coordinate (k) at (100:2cm);

    \node at (i) [below] {$i$};
    \node at (j) [above left] {$j$};
    \node at (k) [above] {$j+1$};

    \draw[line width=0.4mm, red, thick] (i) -- (j);
    \draw[line width=0.4mm, red, dashed] (i) -- (k);

    \foreach \p in {i,j,k} {
        \fill (\p) circle(2.5pt);
    }
    \end{scope}
\end{tikzpicture}
\caption{
A \LABELREVERSE{} compatible configuration at \(\ell=1\). The line
configuration in \(\mathbb P^3\) is shown in (a). The corresponding
\(AB\)-geometry is a union of one-loop Amplituhedra with extra
signed arc conditions, shown in~(b).
}
\label{fig:reverse-LS-L1-thesis}
\end{figure}
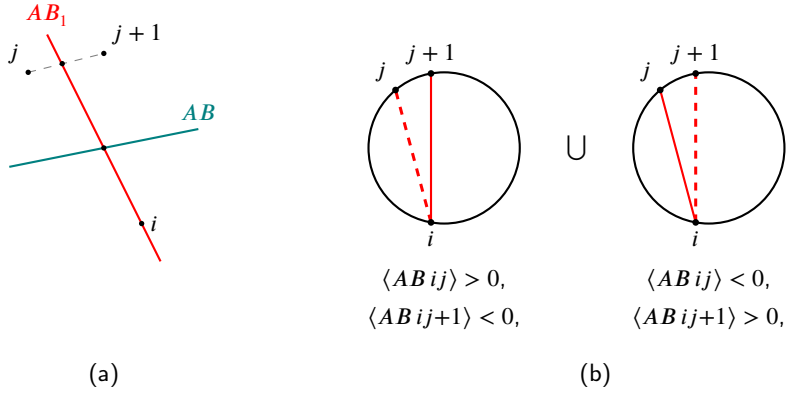
\end{eg}

\begin{eg}[\LABELREVERSE{} incompatible configuration at \(\ell=2\)]
\label{eg:reverse-incompatible-l2}

The first incompatible \LABELREVERSE{} configurations occur at two loops. They
can be obtained by taking two one-loop intersecting configurations of the type
above, with 
\begin{equation}
        i<k<j<l<i \, ,
\end{equation}
see Figure~\ref{fig:reverse-incompatible-L2-thesis}.
Equivalently, the two induced pairs of signed arcs cross. The corresponding
\(AB\)-geometry can be decomposed into pieces associated with crossing signed arcs, but
the resulting canonical functions have the same undesirable features as the
simple incompatible value \eqref{LSvaluesimpleincompatibleexample}: they are
not sums of Kermits and are not conformally invariant after
\(AB\to I_\infty\). These configurations are removed by the reduction
result.

\begin{figure}[pos=t]
\centering
\begin{tikzpicture}[scale=0.5]

    \coordinate (i) at (1,-2);
    \coordinate (P) at (0,0);
    \coordinate (Q) at (-1.25,-0.25);
    \coordinate (R) at (-1.1,2.2);
    \coordinate (S) at (1.87,2.84);
    \coordinate (j) at (-2,2);
    \coordinate (j_plus) at (0,2.5);
    \coordinate (k) at (-3,-2);
    \coordinate (l) at (1,3);
    \coordinate (l_plus) at (2.7,2.7);
    \coordinate (C) at (-1.5,3);
    \coordinate (D) at (1.5,-3);
    \coordinate (A) at (-2.5,-0.5);
    \coordinate (B) at (2.5,0.5);
    \coordinate (E) at (2.5,3.5);
    \coordinate (F) at (-3.5,-2.5);

    \draw[dashed, gray] (j) -- (j_plus);
    \draw[dashed, gray] (l) -- (l_plus);

    \draw[red, thick] (C) -- (D);
    \node[above, red] at (C) {$AB_1$};

    \draw[red, thick] (E) -- (F);
    \node[above, red] at (E) {$AB_2$};

    \draw[teal, thick] (A) -- (B);
    \node[above, teal] at (B) {$AB$};

    \foreach \p in {i,P,Q,R,S,j,j_plus,k,l,l_plus} {
        \fill (\p) circle (2pt);
    }

    \node[right] at (i) {$i$};
    \node[above left] at (j) {$j$};
    \node[above] at (j_plus) {$j+1$};
    \node[above left] at (k) {$k$};
    \node[above] at (l) {$l$};
    \node[above right] at (l_plus) {$l+1$};

    \begin{scope}[xshift=9cm]
    \node at (4.5,0) {$\bigcup \quad \dots$};

    \draw[thick] (0,0) circle(2cm);

    \coordinate (i) at (-100:2cm);
    \coordinate (j) at (130:2cm);
    \coordinate (k) at (100:2cm);
    \coordinate (l) at (50:2cm);
    \coordinate (m) at (20:2cm);
    \coordinate (n) at (200:2cm);

    \node at (i) [below] {$i$};
    \node at (j) [above left] {$j$};
    \node at (k) [above] {$j+1$};
    \node at (l) [above right] {$l$};
    \node at (m) [above right] {$l+1$};
    \node at (n) [left] {$k$};

    \draw[line width=0.4mm, red, dashed] (i) -- (j);
    \draw[line width=0.4mm, red, thick] (i) -- (k);
    \draw[line width=0.4mm, red, thick] (n) -- (l);
    \draw[line width=0.4mm, red, dashed] (n) -- (m);

    \foreach \p in {i,j,k,l,m,n} {
        \fill (\p) circle(2.5pt);
    }
    \end{scope}
\end{tikzpicture}
\caption{
A \LABELREVERSE{} incompatible configuration at \(\ell=2\). 
}
\label{fig:reverse-incompatible-L2-thesis}
\end{figure}
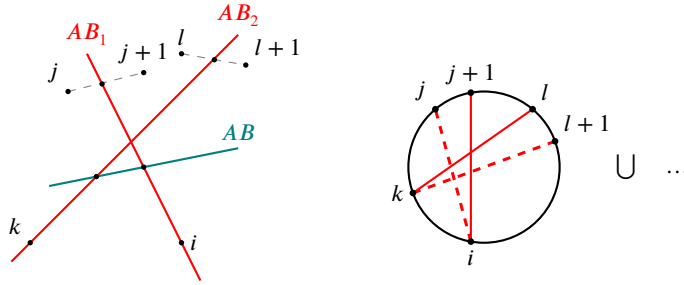
\end{eg}

\begin{eg}[A six-loop configuration giving a single Kermit]
\label{eg:six-loop-single-kermit}

A useful high-loop compatible example occurs at \(\ell=6\). The configuration
in Figure~\ref{AB_L=6.2-thesis} gives a signed partial triangulation for which
all sign patterns except one correspond to empty geometries. The nonempty piece
is a single bicolored triangulation of type \((2,n)\), see the discussion of Kermit triangulations above, hence a single Kermit. The leading-singularity value is
\begin{equation}
        [ijk;abc] \, .
\end{equation}
This example shows that a nontrivial higher-loop localization can reduce
to one elementary Kermit form.

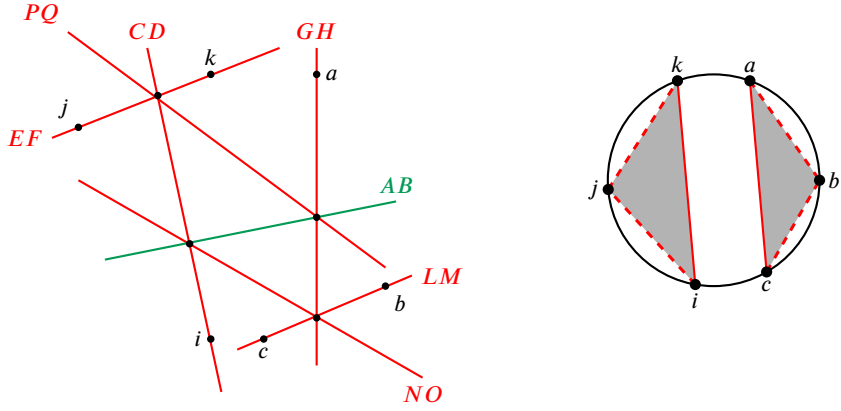
\begin{figure}[pos=t]
\centering
\begin{tikzpicture}[scale=0.7]

    \coordinate (i) at (-0.5,-2);
    \coordinate (W) at (-0.9,-0.2);
    \coordinate (Z) at (1.5,0.3);
    \coordinate (R) at (-1.5,2.6);
    \coordinate (S) at (1.5,-1.6);
    \coordinate (j) at (-3,2);
    \coordinate (k) at (-0.5,3);
    \coordinate (a) at (1.5,3);
    \coordinate (b) at (2.8,-1);
    \coordinate (c) at (0.5,-2);
    \coordinate (C) at (-1.7,3.5);
    \coordinate (D) at (-0.3,-3);
    \coordinate (A) at (-2.5,-0.5);
    \coordinate (B) at (3,0.6);
    \coordinate (G) at (1.5,3.5);
    \coordinate (H) at (1.5,-2.5);
    \coordinate (E) at (-3.5,1.8);
    \coordinate (F) at (0.8,3.5);
    \coordinate (L) at (3.3,-0.8);
    \coordinate (M) at (0,-2.2);
    \coordinate (O) at (-3,1);
    \coordinate (N) at (3.5,-2.73);
    \coordinate (P) at (-3.2,3.8);
    \coordinate (Q) at (2.8,-0.65);




    \draw[red, thick] (C) -- (D);
    \node[above, red] at (C) {$CD$};

    \draw[red, thick] (G) -- (H);
    \node[above, red] at (G) {$GH$};

    \draw[red, thick] (E) -- (F);
    \node[left, red] at (E) {$EF$};

    \draw[red, thick] (L) -- (M);
    \node[right, red] at (L) {$LM$};

    \draw[red, thick] (N) -- (O);
    \node[below, red] at (N) {$NO$};

    \draw[red, thick] (P) -- (Q);
    \node[above left, red] at (P) {$PQ$};

    \draw[ForestGreen, thick] (A) -- (B);
    \node[above, ForestGreen] at (B) {$AB$};

    \foreach \p in {i,W,Z,R,S,j,k,a,b,c} {
        \fill (\p) circle (2pt);
    }

    \node[left] at (i) {$i$};
    \node[above left] at (j) {$j$};
    \node[above] at (k) {$k$};
    \node[right] at (a) {$a$};
    \node[below right] at (b) {$b$};
    \node[below] at (c) {$c$};

    \begin{scope}[scale=1,xshift=9cm,yshift=1cm]
    \draw[thick] (0,0) circle(2cm);

    \coordinate (i) at (-100:2cm);
    \coordinate (j) at (185:2cm);
    \coordinate (k) at (110:2cm);
    \coordinate (l) at (70:2cm);
    \coordinate (n) at (-60:2cm);
    \coordinate (m) at (0:2cm);

    \node at (i) [below] {$i$};
    \node at (j) [left] {$j$};
    \node at (k) [above] {$k$};
    \node at (l) [above] {$a$};
    \node at (m) [right] {$b$};
    \node at (n) [below] {$c$};

    \fill[black, opacity=0.3] (i) -- (j) -- (k) -- cycle;
    \fill[black, opacity=0.3] (l) -- (n) -- (m) -- cycle;

    \draw[line width=0.4mm, red, dashed] (i) -- (j);
    \draw[line width=0.4mm, red, dashed] (l) -- (m);
    \draw[line width=0.4mm, red, dashed] (j) -- (k);
    \draw[line width=0.4mm, red, thick] (i) -- (k);
    \draw[line width=0.4mm, red, thick] (l) -- (n);
    \draw[line width=0.4mm, red, dashed] (n) -- (m);

    \foreach \p in {i,j,k,l,m,n} {
        \fill (\p) circle(3pt);
    }
    \end{scope}
\end{tikzpicture}
\caption{
A leading-singularity configuration at \(\ell=6\) yielding a single Kermit. }
\label{AB_L=6.2-thesis}
\end{figure}
\end{eg}

We close the subsection with the complete two-loop picture. At \(\ell=2\), all
leading-singularity configurations of the negative geometries appearing in
\eqref{eq:2-loop} can be listed explicitly. The analysis starts from the
boundary stratification of the one-loop Amplituhedron and imposes all possible
two-loop cut conditions, as in the examples above. In addition to the
configurations displayed in Table~\ref{tab:all-L2-config-thesis}, one also has
the simple configurations \(AB_1=ij\), \(AB_2=kl\), and the one-loop
\LABELREVERSE{} configuration of Figure~\ref{fig:reverse-LS-L1-thesis} with an
additional simple line \(AB_2=kl\).

By the reduction result, only compatible configurations contribute new leading
singularities of the full \(F_n^{(2)}\). This excludes simple configurations
with crossing arcs, and similarly excludes configurations in which the extra
arc \(kl\) crosses one of the signed arcs induced by a \LABELREVERSE{}
configuration. In Table~\ref{tab:all-L2-config-thesis}, the only incompatible
entries are the configurations in the second row for index orderings such that
the triangles \(\{i,j,j+1\}\) and \(\{k,l,l+1\}\) overlap in the \(n\)-gon.
The compatible entries evaluate to simple compatible leading singularities
\eqref{LS_simple_formulae}, and hence to Kermit forms.

\begin{table}[pos=t]
    \centering
    \begin{tabular}{c|c|c}
    Line configuration & Signed partial triangulations & Value \\
    \hline\hline
        \includegraphics[scale=0.35]{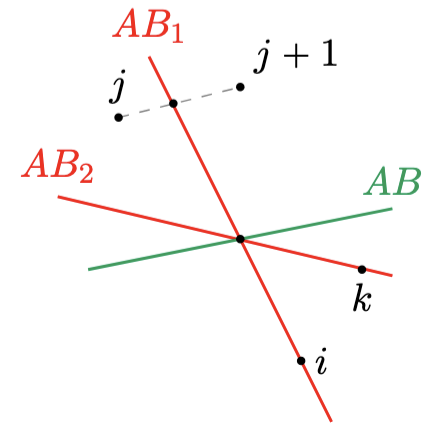}
        &
        \includegraphics[scale=0.38]{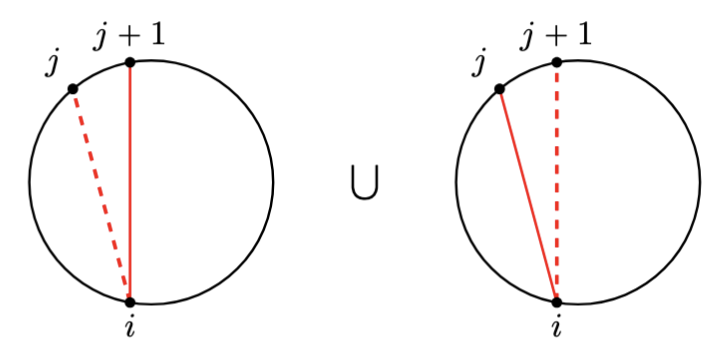}
        &
        \includegraphics[scale=0.65]{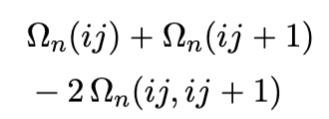}
        \\
        \hline
        \includegraphics[scale=0.65]{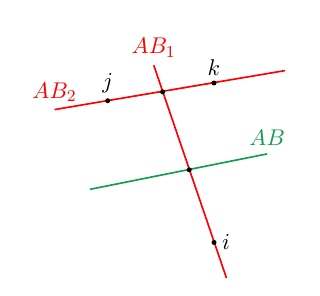}
        &
        \includegraphics[scale=0.65]{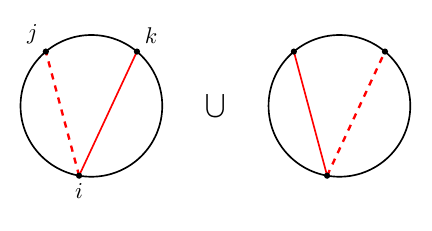}
        &
        \includegraphics[scale=0.65]{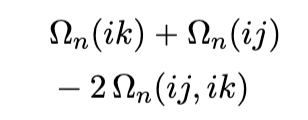}
        \\
        \hline
        \includegraphics[scale=0.65]{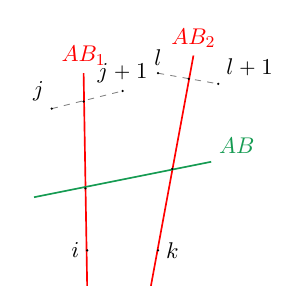}
        &
        \includegraphics[scale=0.65]{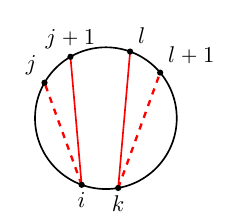}
        &
        \includegraphics[scale=0.65]{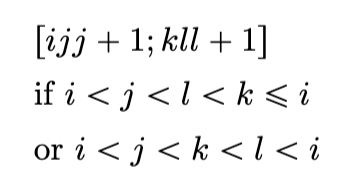}
        \\
        \hline
        \includegraphics[scale=0.35]{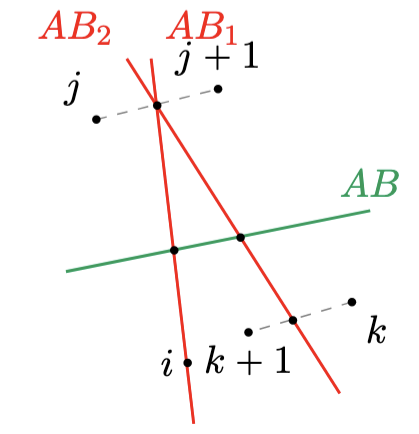}
        &
        \includegraphics[scale=0.65]{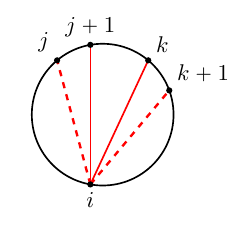}
        &
        \includegraphics[scale=0.65]{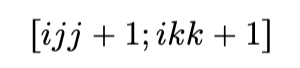}
        \\
    \end{tabular}
    \caption{
    Leading-singularity configurations of \(F_n^{(2)}\). The line
    configurations are associated to signed partial triangulations, which in
    turn can be evaluated in terms of simple compatible leading singularities
    \eqref{LS_simple_formulae} and Kermit forms \eqref{eq:Kermit-six-invariant-F}.
    }
    \label{tab:all-L2-config-thesis}
\end{table}

In the next sections we use this information as input for geometric Landau
analysis and symbol bootstrap. The leading singularities determine the allowed
algebraic prefactors in the ansatz for the integrated observable; the remaining
problem is to determine the pure functions multiplying them.

\subsection{Geometric Landau analysis}
\label{sec:Geometric Landau Analysis}

The singularities of integrated amplitudes are constrained, but not completely determined,
by the singularities of their integrands. As discussed in Section~\ref{sec:The Geometry of Integration}, the usual Landau equations detect possible branch
loci by studying when the integration contour is pinched by propagator singularities
\cite{Landau1959,Eden:1966dnq}. However, this analysis depends only on the denominator
structure of a chosen integrand representation, and therefore often produces more
candidate singularities than are present in the final function. This overprediction is
particularly visible in planar $\mathcal N=4$ SYM, where the Amplituhedron suggests a
more intrinsic way of extracting branch-point information directly from geometry
\cite{DennenPrlinaSpradlinStanojevicVolovich2017,PrlinaSpradlinStankowiczStanojevic2018}.
Figures in this section are adapted from those works and from
\cite{ChicherinHennMazzucchelliTrnkaYangZhang2026} where they illustrate
geometric Landau analysis.

From the positive-geometric point of view, this refinement is natural. A canonical form is
not specified only by its poles. Its numerator is expected to be uniquely determined by the requirement that residues
occur only on genuine boundaries of the geometry, and this numerator can cancel residues
on residual or spurious loci. Since residues and discontinuities are dual to one another,
such cancellations can remove the associated discontinuities after integration. Thus the
boundary structure of the geometry, together with the numerator of the canonical form,
provides information which is invisible to a naive Landau analysis based only on
propagator denominators.

In this section we explain this mechanism and its consequences for the \emph{symbol bootstrap}.
We first discuss simple examples in which numerators remove residual discontinuities,
starting with the Aomoto form of a square and then recalling analogous cancellations for
Amplituhedron integrands
\cite{DennenPrlinaSpradlinStanojevicVolovich2017,PrlinaSpradlinStankowiczStanojevic2018}.
We then turn to the refined Landau analysis of negative geometries, where Landau diagrams
are associated to geometric boundaries and the boundary structure determines which
Landau solutions are physical and which are spurious
\cite{ChicherinHennMazzucchelliTrnkaYangZhang2026}. We finish by explaining how the output of this analysis is converted into a
symbol alphabet. The bootstrap computation is then
carried out in Section~\ref{sec:Symbols and Bootstrap}.

\subsubsection{Refined Landau analysis}
\label{subsec:refined-landau-analysis}

\newcommand{\messagebox}[1]{
\begin{center}
\fbox{
\begin{minipage}{0.92\textwidth}
#1
\end{minipage}
}
\end{center}
}

\begin{figure}[pos=t]
\centering
\begin{minipage}{0.42\linewidth}
\centering
\includegraphics[width=0.9\linewidth]{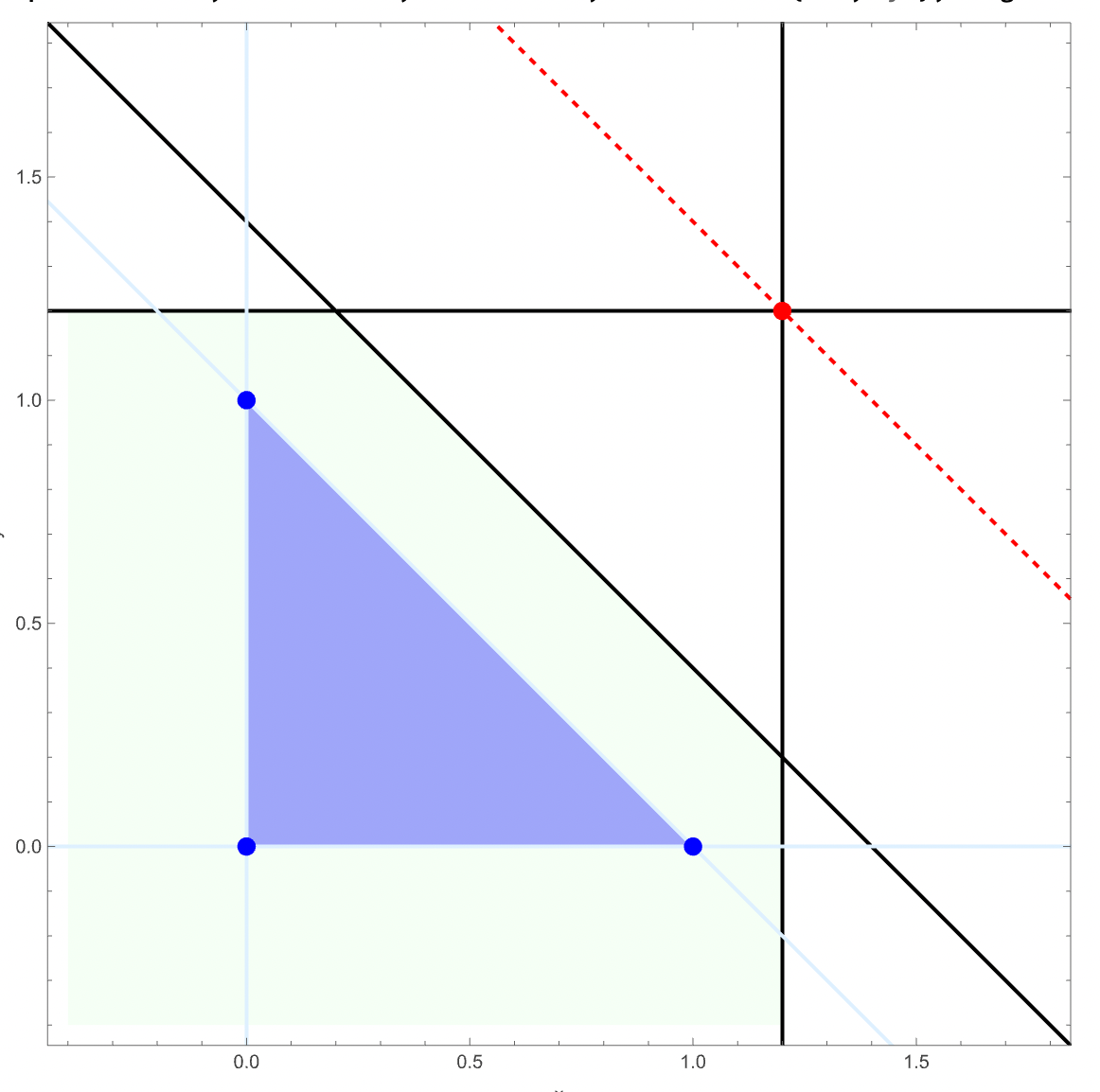}\\[0.2em]
\text{(a)}
\end{minipage}
\qquad
\begin{minipage}{0.42\linewidth}
\centering
\includegraphics[width=0.9\linewidth]{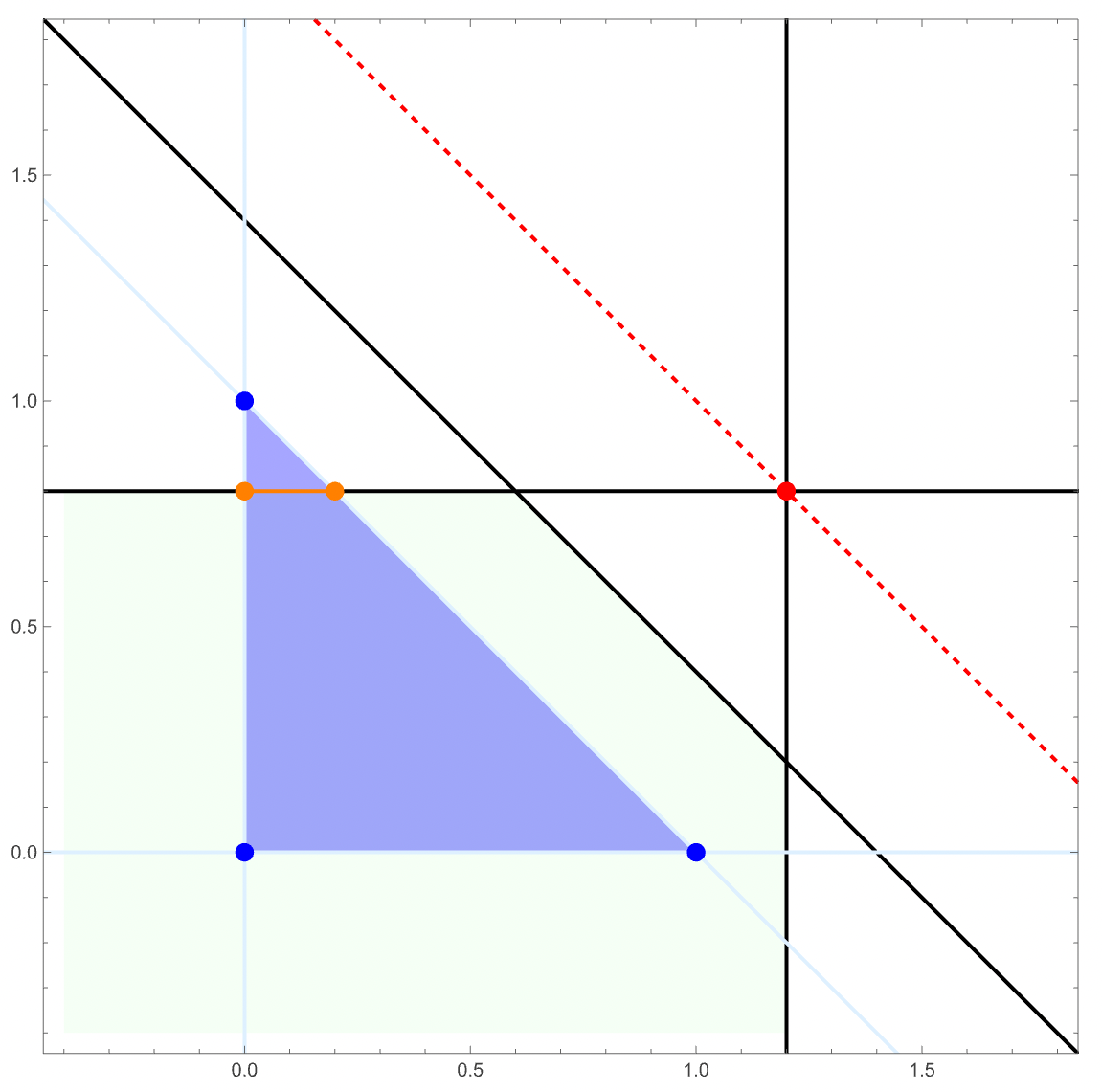}\\[0.2em]
\text{(b)}
\end{minipage}
\caption{Aomoto pairing of the triangle \(A\) (in blue) with the unbounded
quadrilateral \(B=B(z,w)\) (in green). The linear numerator of the canonical
form of \(B\) is the red-dashed line. Panel (a) shows a generic configuration
yielding the integral in~\eqref{eq:Izw}, while panel (b) shows the
configuration used to compute a first discontinuity in~\eqref{eq:aom_disc_eg}.}
\label{fig:aom_tr_num}
\end{figure}

We illustrate how the vanishing of the integrand's numerator can remove potential singularities. Consider the Aomoto form of a quadrilateral paired with a triangle in the plane. We work in affine coordinates $[1:x_1:x_2]$ and choose the
integration region $A$ to be the fixed triangle with vertices $(0,0)$, $(1,0)$
and $(0,1)$. The quadrilateral $B=B(y,w,z)$ defining the integration form is the
unbounded quadrilateral containing $A$ with boundary 
\begin{equation}
  x_1=z,\qquad x_2=w,\qquad x_1+x_2=y \, ,
\end{equation}
where $1<z,w<y$. These are illustrated in Figure~\ref{fig:aom_tr_num}(a) for some fixed values of parameters. Projectively, $B$ is a quadrilateral whose fourth boundary
component is the line at infinity in this chart. Its canonical form is
\begin{equation}\label{eq:form_B}
\boldsymbol{\Omega}_B(y,w,z)
=
\frac{z+w-x_1-x_2}{(z-x_1)(w-x_2)(y-x_1-x_2)}
\, \mathrm{d}x_1 \wedge \mathrm{d}x_2 \, .
\end{equation}
The important feature is the non-trivial linear numerator. It vanishes on the
line passing through the residual point $(z,w)$ and parallel to the edge of
$B$ supported on $x_1+x_2=y$.

The canonical pairing is
\begin{equation}\label{eq:Izw}
\begin{aligned}
I(y,w,z)
&:=\mathcal{I}(A,B(y,w,z))
=
\int_A \boldsymbol{\Omega}_{B(y,w,z)} \\
&=
\int_0^1 \mathrm{d}x_1
\int_0^{1-x_1} \mathrm{d}x_2\,
\frac{z+w-x_1-x_2}{(z-x_1)(w-x_2)(y-x_1-x_2)}
\\
&=2 i\pi \log(w)
+\log(y-w)\log\!\left(\frac{w}{y(w-1)}\right)
+\log(y-1)\log\!\left(\frac{(y-z)(y-w)(w-1)}
{e^{2 i\pi}w^2z}\right)  \\
&\qquad
+\log(y) \, \log\!\left(\frac{wz}{y-z}\right)
-\operatorname{Li}_2\!\left(\frac{w-1}{y-1}\right)
+\operatorname{Li}_2\!\left(\frac{y-1}{w}\right)
-\operatorname{Li}_2\!\left(\frac{y}{w}\right)  \\
&\qquad
+\operatorname{Li}_2\!\left(\frac{w}{y-1}\right)
-\operatorname{Li}_2\!\left(\frac{y-1}{y-z}\right)
+\operatorname{Li}_2\!\left(\frac{y}{y-z}\right).
\end{aligned}
\end{equation}
The function $I(y,w,z)$ is transcendental of weight two, as expected from Section~\ref{sec:The Geometry of Integration}, and is real valued in the region $1<z,w<y$, after
choosing the appropriate branch.

We recall the duality between discontinuities and residues for period
pairings. For Aomoto forms, this takes the form
\begin{equation}\label{eq:disc_res_aom}
{\rm Disc}_{\gamma_i}\,\mathcal I(A,B(s))
=
2\pi i
\int_{A \cap \mathcal B_i(s)}
{\rm Res}_{\mathcal B_i}\,\boldsymbol{\Omega}_{B} \, ,
\end{equation}
where $\gamma_i$ is a small loop around the corresponding branch locus and
$\mathcal B_i$ is the boundary component of $B$ involved in the discontinuity.
Compared to~\eqref{eq:aomoto-disc-residue}, we have shifted the dependence on
parameters $s$ to the polytope $B$.

In the present example, branch points occur for instance at $z=1$ and $w=1$,
where a boundary component of $B(y,w,z)$ touches a vertex of $A$. Consider the
discontinuity across the cut $0<w<1$. The relevant boundary component is the
line $x_2=w$, whose intersection with $A$ is the interval
\begin{equation}\label{eq:aom_disc_eg}
  A\cap \{x_2=w\}=[0,1-w]\times \{w\} \, ,
\end{equation}
which is the orange segment in Figure~\ref{fig:aom_tr_num}.(b).
Using~\eqref{eq:disc_res_aom}, we obtain
\begin{equation}
\frac{1}{2\pi i}{\rm Disc}_{0<w<1}\, I(y,w,z)
=
\int_{0}^{1-w}
{\rm Res}_{x_2=w}\,\boldsymbol{\Omega}_{B(y,w,z)}
=
\int_{0}^{1-w} \frac{\mathrm{d}x_1}{x_1+w-y}
=
\log\!\left(\frac{y-1}{y - w}\right) \, .
\end{equation}
We will see later how the same result can be read directly from
the symbol of~\eqref{eq:Izw}.

The point of this example is that the numerator of
\eqref{eq:form_B} vanishes at the point $(z,w)$. This point is an intersection
of two boundary lines of $B$, namely $x_1=z$ and $x_2=w$, but it is not a vertex of
the quadrilateral. In the terminology of the discussion of residual arrangements and adjoint hypersurfaces of Section~\ref{sec:Adjoint Hypersurface}, it is a \emph{residual} point.
A Landau analysis based only on the denominator of~\eqref{eq:form_B} would
therefore include the simultaneous vanishing of $z-x_1$ and $w-x_2$ as a candidate
codimension-two singularity. However, the canonical numerator removes the
corresponding residue.

Indeed, after the first discontinuity we obtained a function independent of
$z$. Hence the second discontinuity vanishes:
\begin{equation}
{\rm Disc}_{0<z<1-w}{\rm Disc}_{0<w<1} \, I(y,w,z)
=
{\rm Disc}_{0<z<1-w}
\, \log\!\left(\frac{y-1}{y - w}\right)
=
0 \, .
\end{equation}
This agrees with the residue interpretation, since
\begin{equation}
{\rm Res}_{x_1=z}{\rm Res}_{x_2=w}\,\boldsymbol{\Omega}_{B(y,w,z)}
=
0 \, .
\end{equation}
Thus the denominator arrangement predicts a candidate double discontinuity,
but the canonical numerator kills the corresponding maximal residue. This is
the basic mechanism behind refined Landau analysis: not every solution of the
Landau equations gives an actual singularity of the period; the solution must
also be compatible with the residue structure of the canonical form.
\begin{tcolorbox}[resultbox]
\textbf{Numerator vanishing removes discontinuities.}
\begin{equation}
\begin{aligned}
	&\text{Vanishing of the integrand numerator}
\quad\longrightarrow\quad
\text{vanishing residue} \\
&\quad\longrightarrow\quad
\text{vanishing discontinuity}.
\end{aligned}
\end{equation}
\end{tcolorbox}
\noindent

The same phenomenon appears for the amplitude in planar $\mathcal{N}=4$ SYM, where the integrand is encoded by the canonical form of the loop Amplituhedron. The vanishing of the numerator in the canonical form removes potential singularities arising from Landau analysis. Taking the vanishing of the numerator into account, yields a \textit{refined}, or \textit{geometric}, Landau analysis. This method has been first applied to amplitudes in~\cite{DennenSpradlinVolovich2016,DennenPrlinaSpradlinStanojevicVolovich2017,PrlinaSpradlinStankowiczStanojevic2018,PrlinaSpradlinStanojevic2018}, and more recently, to negative geometries in~\cite{ChicherinHennMazzucchelliTrnkaYangZhang2026,Paranjape:2026kix}.

To give a concrete example, discussed in~\cite{DennenPrlinaSpradlinStanojevicVolovich2017}, consider the pentagon cut 
\begin{equation}\label{eq:one-loop-pentagon-cuts}
\begin{aligned}
  \langle AB\, i-1 i\rangle
  &=
  \langle AB\, i i+1\rangle
  =
  \langle AB\, j-1 j\rangle
  =
  \langle AB\, j j+1\rangle=
  \langle AB\, k k+1\rangle
  =
  0 .
\end{aligned}
\end{equation}
According to our previous discussion on the Landau variety in Section~\ref{sec:The Geometry of Integration}, the equations~\eqref{eq:one-loop-pentagon-cuts} cut out a codimension-five component in the space of both external kinematics $Z_i$ and the loop variable $AB$. The image of this locus under the projection to the external kinematics $Z$ has codimension one, since for generic $Z_i$ there exist no line $AB$ solving all five equations~\eqref{eq:one-loop-pentagon-cuts}. However, on a codimension-one locus in the $Z_i$ such overconstrained Schubert problem admits a solution. Since this locus has codimension one in the external kinematic space, it defines a component of the Landau variety. To compute this locus, we solve the first four equations in~\eqref{eq:one-loop-pentagon-cuts}. This is the two-mass-easy Schubert problem discussed in Subsection~\ref{subsec:One-Loop Schubert Problems}. Its two solutions are
\begin{equation}\label{eq:2me_sols}
	AB=ij \, ,\qquad AB=\overline{ij} = (i-1 \, i \, i+1) \cap (j-1 \, j \, j+1) \, .
\end{equation}
The fifth equation $\langle AB \, kk+1 \rangle$ in~\eqref{eq:one-loop-pentagon-cuts} is then satisfied if and only if
\begin{equation}
	\langle i\, j\, k\, k+1\rangle
  \,
  \langle \overline{ij}\, k\, k+1\rangle
  =
  0 \, .
\end{equation}
This equations defines the \textit{super-leading Landau singularity}, where super means that we imposed five cut conditions in four-dimensions.

Here comes the important observation. Recall that the one-loop MHV Amplituhedron $\mathcal{A}^{(1)}_{n}$ is defined by the space of lines in $\mathbb{P}^3$ subject to the conditions
\begin{equation}\label{loop_positivty_second}
\begin{aligned}
    &\langle AB \, 12 \rangle > 0 \ , \dots, \langle AB \, n-1n \rangle > 0 \ , \ \langle AB \, 1 n \rangle > 0 \, , \\
    & (\langle AB \, 12 \rangle , \dots, \langle AB \, 1n \rangle) \text{ has two sign-flips.}
\end{aligned}
\end{equation}
Among the two solutions in~\eqref{eq:2me_sols}, the first one lies on the boundary of the Amplituhedron while the second does not. To see this, the first can be written as $AB=D \cdot Z^\top$ where $D$ is a $2 \times n$ matrix with an identity block at columns $i$ and $j$ and zeroes elsewhere. Since $D$ clearly lies on the boundary of the positive Grassmannian $\Gr_{\geq 0}(2,n)$, $AB$ lies on the boundary of $\mathcal{A}^{(1)}_{n}$. On the other hand, the second solution~\eqref{eq:2me_sols} violates the sign-flip condition in~\eqref{loop_positivty_second}. This can be verified explicitly, or argued from the fact that $\langle a b \, \overline{ij} \rangle \geq 0$ for every distribution of indices, since this bracket defines a cluster variable for $\Gr_{\geq 0}(4,n)$~\cite{GoldenGoncharovSpradlinVerguVolovich2014}. 

It follows that the numerator of the canonical form of $\mathcal{A}^{(1)}_{n}$ vanishes on the second solution in~\eqref{eq:2me_sols}. This fact has two implications. First, the full one-loop MHV amplitude has vanishing leading singularity on the solution $AB=\overline{ij} $. This explains why the expansion in chiral pentagons discussed in Section~\ref{subsec:One-Loop Prescriptive Unitarity} is particularly well-suited for the MHV sector, see~\cite{ArkaniHamed:2010kv}, as it is designed to have vanishing residues on the anti-chiral solutions $\overline{ij} $. Second, according to the discussion above, the candidate Landau singularity
\begin{equation}\label{eq:SLS_2me}
	\langle \overline{ij}\, k\, k+1\rangle
\end{equation}
is absent in the MHV amplitude. The amplitude has vanishing discontinuity on this locus, because of the vanishing of the canonical form's numerator.

It is important to point out that this analysis depends on the helicity sector. For higher helicities $k \geq 1$, the second solution in~\eqref{eq:2me_sols} lies on the boundary of the one-loop ${\rm N}^k{\rm MHV}$ Amplituhedron. This means that the associated leading singularity, and super-leading Landau singularity~\eqref{eq:SLS_2me} potentially contribute to this helicity sector. In general, on-shell diagrams efficiently detect if a boundary stratum, or cut, is relevant for the ${\rm N}^k{\rm MHV}$ sector~\cite{DennenPrlinaSpradlinStanojevicVolovich2017,PrlinaSpradlinStankowiczStanojevic2018}. This is encoded in the helicity degree of the on-shell diagram. For the two-mass-easy cut discussed above, we described this procedure already in Subsection~\ref{subsec:On-Shell Diagrams and Grassmannian Localization}. We come back to this precise connection in the next section, when we discuss this refined Landau analysis for negative geometries. 

This Amplituhedron-refined Landau analysis approach for amplitudes was extended
to two loops and other helicity sectors. The two-loop NMHV case was studied
in~\cite{PrlinaSpradlinStankowiczStanojevic2018}. All-loop aspects of Landau
singularities in massless planar theories were analyzed
in~\cite{PrlinaSpradlinStanojevic2018,LippstreuSpradlinVolovich2024ZigguratI,LippstreuSpradlinSrikantVolovich2024ZigguratII}. Purely geometric studies of boundaries of loop Amplituhedra are relevant in this context. The deepest cuts of MHV at all loop orders and multiplicities were studied in~\cite{Arkani-Hamed:2018rsk}. These developments show that positive (and negative) geometries provide a geometric filter on Landau analysis.

In the remainder of this section we apply the same philosophy to negative
geometries. The geometry is different, but the guiding principle is the same:
the Landau equations give candidate singularities, while the boundary and
residue structure of the relevant geometry decides which of them survive.

\subsubsection{Landau analysis for negative geometries}
\label{subsec:geometric-landau-negative-geometries}

We now apply the refined Landau philosophy to the negative geometries introduced in
Section~\ref{sec:WLwithLI}. The previous discussion of the Wilson loop with a Lagrangian insertion
showed that its integrand admits an expansion into negative geometries, and that its
leading singularities organize the algebraic prefactors of the integrated answer. In the
present section we focus on a particularly tractable family: ladder negative geometries.
They already contain the main new features of the general problem, but are simple enough
that their leading singularities and Landau singularities can be classified explicitly. These results are part of~\cite{ChicherinHennMazzucchelliTrnkaYangZhang2026}.

The guiding principle is the same as for the Amplituhedron. Ordinary Landau analysis
starts from a graph and studies the singularities obtained by cutting propagators and
solving the pinch equations. Geometric Landau analysis adds one more step: the cut
configuration must be compatible with the geometry. Equivalently, it must define a
boundary on which the canonical form has a non-vanishing residue. If a cut configuration
is residual, the numerator of the canonical form vanishes on it, and the corresponding
Landau singularity is spurious.

The negative-geometry setting differs from amplitudes in one important respect. The
insertion line $AB$ is not integrated over, and should be regarded as part of the external
kinematic data. As a consequence, cut diagrams involving $AB$ may carry higher minimal
helicity weight when interpreted as ordinary on-shell diagrams. Nevertheless, on-shell
diagrams still provide the right combinatorial language for deciding which relaxations of a
leading Landau diagram should be kept. We will describe this selection rule after recalling
the leading singularities of ladder geometries.

\begin{figure}[pos=t]
\centering
\begin{tikzpicture}[scale=0.8]
    \draw[mybrown, line width=2pt] (0,0) -- (3,0); 
    \draw[mybrown, line width=2pt] (6,0)--(9,0);
    \draw[mybrown, line width=2pt, dashed] (6,0)--(3,0);

    \draw[fill=white] (0,0) circle (4pt);
    \draw (0,0) -- (45:4pt);
    \draw (0,0) -- (135:4pt);
    \draw (0,0) -- (-45:4pt);
    \draw (0,0) -- (-135:4pt);

    \filldraw (3,0) circle (4pt);
    \filldraw (6,0) circle (4pt);
    \filldraw (9,0) circle (4pt);

    \node[anchor=north] at (0,-0.2) {\small{$AB_0$}};
    \node[anchor=north] at (3,-0.2) {\small{$AB_1$}};
    \node[anchor=north] at (6,-0.2) {\small{$AB_{\ell{-}1}$}};
    \node[anchor=north] at (9,-0.2) {\small{$AB_\ell$}};
\end{tikzpicture}
\caption{Graph associated to the $\ell$-loop ladder negative geometry. The marked
line is the insertion line $AB=AB_0$, while $AB_1,\ldots,AB_\ell$ are integrated.}
\label{fig:laddersgeneral}
\end{figure}
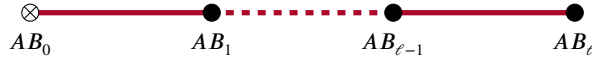

We denote the external insertion line by $AB=AB_0$ and the integrated loop lines of an $\ell$-loop ladder by
\begin{equation}
  AB_1=CD \, ,\qquad
  AB_2=EF \, ,\qquad
  AB_3=GH \, ,\qquad \ldots \, .
\end{equation}
The negative-geometry graph of the ladder is shown in Figure~\ref{fig:laddersgeneral}. We focus on the
\emph{connected} leading singularity configurations, namely those involving all adjacent
multi-loop cuts
\begin{equation}\label{eq:ladder-cuts-thesis}
  \langle AB\,CD\rangle
  =
  \langle CD\,EF\rangle
  =
  \langle EF\,GH\rangle
  =
  \cdots
  =
  0 \, .
\end{equation}
If one of these cuts is absent, the configuration reduces to a lower-loop case.

The all-loop classification of leading-singularity configurations for ladders proceeds by
localizing the loop lines one by one. Each line in $\mathbb P^3$ has four degrees of
freedom. In a maximal cut configuration, the line $(AB)_a$ is constrained by
conditions of the form
\begin{equation}
  \langle AB_a\, i\,i+1\rangle=0 \, ,
  \qquad
  \langle AB_a\, AB_{a+1}\rangle=0 \, .
\end{equation}
Before imposing all multi-loop cuts, each loop line is already forced onto a
lower-dimensional boundary of a one-loop MHV Amplituhedron. These one-loop boundaries
are known explicitly, and positivity gives strong restrictions on which combinations are
allowed. This gives an inductive classification of the ladder leading singularities
\cite{BrownHennMazzucchelliTrnka2025,ChicherinHennMazzucchelliTrnkaYangZhang2026}.

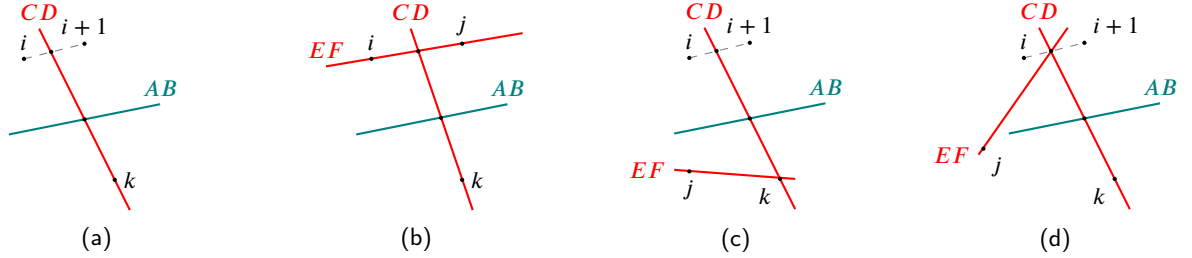
\begin{figure}[pos=t]
\centering

\begin{minipage}{0.23\linewidth}
\centering
\begin{tikzpicture}[scale=0.4]
    \coordinate (i) at (1,-2);
    \coordinate (ip) at (0,0);
    \coordinate (j) at (-2,2);
    \coordinate (jp) at (0,2.5);
    \coordinate (C) at (-1.5,3);
    \coordinate (D) at (1.5,-3);
    \coordinate (A) at (-2.5,-0.5);
    \coordinate (B) at (2.5,0.5);
    \coordinate (R) at (-1.1,2.23);

    \draw[dashed, gray] (j) -- (jp);

    \draw[red, thick] (C) -- (D);
    \node[above, red] at (C) {$CD$};

    \draw[teal, thick] (A) -- (B);
    \node[above, teal] at (B) {$AB$};

    \fill (i) circle (2pt);
    \fill (ip) circle (2pt);
    \fill (j) circle (2pt);
    \fill (jp) circle (2pt);
    \fill (R) circle (2pt);

    \node[right] at (i) {$k$};
    \node[above] at (j) {$i$};
    \node[above] at (jp) {$i+1$};
\end{tikzpicture}\\[0.2em]
\text{(a)}
\end{minipage}
\hfill
\begin{minipage}{0.23\linewidth}
\centering
\begin{tikzpicture}[scale=0.4]
    \coordinate (i) at (1,-2);
    \coordinate (P) at (0.3,0.05);
    \coordinate (Q) at (-0.45,2.25);
    \coordinate (j) at (-2,2);
    \coordinate (k) at (1,2.5);
    \coordinate (C) at (-0.7,3);
    \coordinate (D) at (1.35,-3);
    \coordinate (E) at (-3.5,1.75);
    \coordinate (F) at (3,2.85);
    \coordinate (A) at (-2.5,-0.5);
    \coordinate (B) at (2.5,0.5);

    \draw[red, thick] (C) -- (D);
    \node[above, red] at (C) {$CD$};

    \draw[red, thick] (E) -- (F);
    \node[above, red] at (E) {$EF$};

    \draw[teal, thick] (A) -- (B);
    \node[above, teal] at (B) {$AB$};

    \fill (i) circle (2pt);
    \fill (j) circle (2pt);
    \fill (k) circle (2pt);
    \fill (P) circle (2pt);
    \fill (Q) circle (2pt);

    \node[right] at (i) {$k$};
    \node[above] at (j) {$i$};
    \node[above] at (k) {$j$};
\end{tikzpicture}\\[0.2em]
\text{(b)}
\end{minipage}
\hfill
\begin{minipage}{0.23\linewidth}
\centering
\begin{tikzpicture}[scale=0.4]
    \coordinate (i) at (1,-2);
    \coordinate (ip) at (0,0);
    \coordinate (j) at (-2,2);
    \coordinate (jp) at (0,2.5);
    \coordinate (l) at (-2,-1.75);
    \coordinate (C) at (-1.5,3);
    \coordinate (D) at (1.5,-3);
    \coordinate (E) at (-2.5,-1.7);
    \coordinate (F) at (1.5,-2);
    \coordinate (A) at (-2.5,-0.5);
    \coordinate (B) at (2.5,0.5);
    \coordinate (R) at (-1.1,2.23);

    \draw[dashed, gray] (j) -- (jp);

    \draw[red, thick] (C) -- (D);
    \node[above, red] at (C) {$CD$};

    \draw[red, thick] (E) -- (F);
    \node[left, red] at (E) {$EF$};

    \draw[teal, thick] (A) -- (B);
    \node[above, teal] at (B) {$AB$};

    \fill (i) circle (2pt);
    \fill (ip) circle (2pt);
    \fill (j) circle (2pt);
    \fill (jp) circle (2pt);
    \fill (l) circle (2pt);
    \fill (R) circle (2pt);

    \node[below left] at (i) {$k$};
    \node[above] at (j) {$i$};
    \node[above] at (jp) {$i+1$};
    \node[below] at (l) {$j$};
\end{tikzpicture}\\[0.2em]
\text{(c)}
\end{minipage}
\hfill
\begin{minipage}{0.23\linewidth}
\centering
\begin{tikzpicture}[scale=0.4]
    \coordinate (i) at (1,-2);
    \coordinate (ip) at (0,0);
    \coordinate (j) at (-2,2);
    \coordinate (jp) at (0,2.5);
    \coordinate (k) at (-3.34,-1);
    \coordinate (C) at (-1.5,3);
    \coordinate (D) at (1.5,-3);
    \coordinate (E) at (-3.5,-1.2);
    \coordinate (F) at (-0.55,3);
    \coordinate (A) at (-2.5,-0.5);
    \coordinate (B) at (2.5,0.5);
    \coordinate (R) at (-1.1,2.23);

    \draw[dashed, gray] (j) -- (jp);

    \draw[red, thick] (C) -- (D);
    \node[above, red] at (C) {$CD$};

    \draw[red, thick] (E) -- (F);
    \node[left, red] at (E) {$EF$};

    \draw[teal, thick] (A) -- (B);
    \node[above, teal] at (B) {$AB$};

    \fill (i) circle (2pt);
    \fill (ip) circle (2pt);
    \fill (j) circle (2pt);
    \fill (jp) circle (2pt);
    \fill (k) circle (2pt);
    \fill (R) circle (2pt);

    \node[below left] at (i) {$k$};
    \node[above] at (j) {$i$};
    \node[above right] at (jp) {$i+1$};
    \node[below right] at (k) {$j$};
\end{tikzpicture}\\[0.2em]
\text{(d)}
\end{minipage}

\caption{Leading-singularity configurations for ladder negative geometries at one and
two loops. Panel (a) is the one-loop configuration. Panels (b)--(d) are at two loops.}
\label{fig:ladder-LS-L1}
\end{figure}

At one loop there is a single type of connected ladder leading singularity configuration. At two loops there are three types, shown in Figure~\ref{fig:ladder-LS-L1}. A full classification at all loops was proven in~\cite{ChicherinHennMazzucchelliTrnkaYangZhang2026}. These diagrams represent the leading singularity line configurations in momentum-twistor space. 
The important point for the integrated observable is not only the list of configurations,
but also their values. The ladder integral admits a decomposition
\begin{tcolorbox}[definitionbox]
\textbf{Ladder decomposition.}
\begin{equation}\label{eq:ladder-function-thesis}
  F^{(\ell)}_{n,\mathrm{ladder}}
  =
  \sum_{1\leq i<j\leq n}
  \Omega_{n}(ij)\, f^{(\ell)}_{n,ij} \, ,
\end{equation}
\end{tcolorbox}\noindent
where the sum runs over the diagonals $(ij)$ of the $n$-gon, and
$f^{(\ell)}_{n,ij}$ are expected to be pure functions of transcendental weight $2\ell$.
The rational prefactors $\Omega_{n}(ij)$ are the leading singularities of the ladder, and are the same as in the ladder leading-singularity decomposition~\eqref{eq:ladder-function-thesis}.
For instance, at five points the independent leading singularities are 
\begin{equation}\label{eq:Omega-513-thesis}
  \Omega_{5}(13)
  =
  [1234]+[123;145] \, ,
\end{equation}
together with its cyclic images. At six points there are two dihedral orbits, represented by
\begin{equation}\label{eq:Omega-613-614-thesis}
\begin{aligned}
  \Omega_{6}(13)
  &=
  [1234]+[123;145]+[123;156] \, ,
  \\
  \Omega_{6}(24)
  &=
  [123;145]+[123;156]+[1345]+[134;156] \, .
\end{aligned}
\end{equation}
In general, there are $n(n-3)/2$ arcs on an $n$-gon, but only $\left\lfloor \frac{n}{2} \right\rfloor - 1$ independent orbits under the dihedral symmetry. These are a subset of the LS of $F_n^{(\ell)}$.

Although the LS configurations become more complicated at higher loop order, the
LS values of ladders saturate: they are always represented by the
functions $\Omega_{n}(ij)$ in~\eqref{eq:Omega-n-ij-Kermit-sum}. This is the ladder analogue of
the leading-singularity organization of the full Wilson loop discussed in Section~\ref{sec:WLwithLI}.
We now explain how geometric Landau analysis constrains the singularities, and hence
the symbol alphabets, of the functions $f^{(\ell)}_{n,ij}$ in~\eqref{eq:ladder-function-thesis}.

The procedure follows the Amplituhedron analysis introduced in~\cite{DennenSpradlinVolovich2016,DennenPrlinaSpradlinStanojevicVolovich2017,PrlinaSpradlinStankowiczStanojevic2018,PrlinaSpradlinStanojevic2018}:

\begin{enumerate}
  \item Start from the leading-singularity configurations. These are the
  maximal-codimension boundaries of the geometry; they lie on the boundary of the positive region.

  \item Associate to each leading-singularity configuration its leading Landau diagrams. The edges of such a diagram encode an independent subset of the cut conditions satisfied by the line configuration.

  \item Consider all sub-leading diagrams obtained by relaxing some cut conditions.
  A relaxation is kept only if it remains compatible with the geometry. Equivalently,
  the corresponding cut locus must be a physical boundary, not a residual one. Here by physical we mean that the cut conditions define a boundary of the geometry.

  \item For every physical diagram, solve the Landau equations. The resulting loci in
  external kinematic space are the candidate branch points of the integrated functions.
\end{enumerate}
The selection in the third step can be implemented using the numerator of the canonical form, as in the Aomoto example, or combinatorially through on-shell diagrams. The latter
implementation is purely combinatorial, and hence especially useful in practice. A Landau diagram can be bi-colored. Black three-valent vertices encode coplanarity conditions, while white three-valent vertices concurrency. In the following, we always color black also higher-valent vertices, just for the sake of counting the helicity degree. The helicity degree of a colored leading Landau diagram $\mathcal{L}$ is given by
\begin{equation}\label{eq:hel_deg}
	\deg(\mathcal{L}) := n_w + 2n_b  - n_I - 2 \, ,
\end{equation}
and determines whether the cut can appear in the geometry under consideration. The helicity count in~\eqref{eq:hel_deg} is a special case of the on-shell helicity-degree formula~\eqref{eq:hel_Gamma}. For a given Landau diagram, its minimal helicity weight is obtained by coloring all three-valent vertices white. This however does not necessarily correspond to an actual solution to the cut equations, or equivalently, to a non-vanishing residue; for example, a massless box with three white vertices yields a zero on-shell form.

For amplitudes, the rule is that a Landau diagram can contribute to an $\mathrm{N}^k\mathrm{MHV}$ amplitude only if it admits a bi-coloring of helicity degree at most $k$. For ladder negative geometries, one must modify this statement slightly because the line $AB$ is an external parameter rather than an integrated loop line, an LS configuration can yield landau diagrams of higher helicity weight from the beginning. We find the following rule.

\begin{tcolorbox}[resultbox]
\textbf{Geometric selection rule.}
Start from a bi-colored leading Landau diagram $\mathcal L$ with minimal helicity
weight $k=\deg(\mathcal{L})$. Discard any sub$^r$-leading diagram obtained from $\mathcal L$ by pinching
$r$ propagators if its minimal helicity degree is $k'>k$.
\end{tcolorbox}
\noindent
This rule keeps precisely the relaxations which behave like genuine boundaries of the
negative geometry. If the helicity weight increases, the relaxation is residual: the
corresponding numerator vanishes on the cut locus, and the Landau singularity is
spurious. This is the negative-geometry version of the Amplituhedron selection rule
\cite{PrlinaSpradlinStankowiczStanojevic2018,ChicherinHennMazzucchelliTrnkaYangZhang2026}.

\begin{figure}[pos=t]
\centering
\begin{minipage}{0.35\linewidth}
\centering
\includegraphics[width=\linewidth]{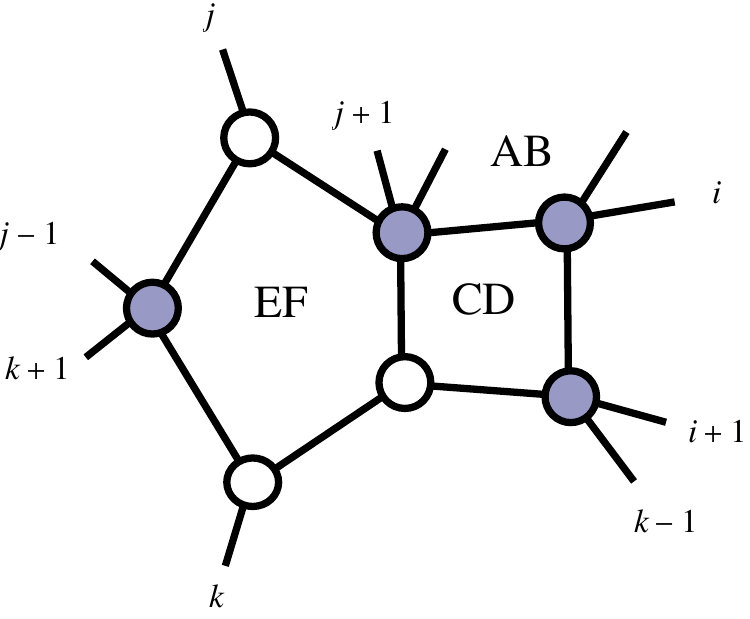}\\[0.2em]
\text{(a)}
\end{minipage}
\qquad
\begin{minipage}{0.35\linewidth}
\centering
\includegraphics[width=\linewidth]{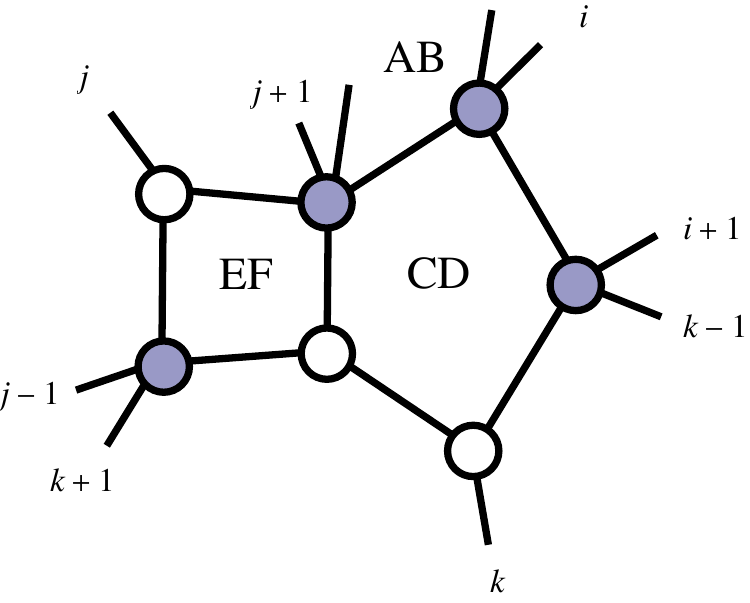}\\[0.2em]
\text{(b)}
\end{minipage}
\caption{Leading Landau diagrams associated with the leading-singularity configuration
in Figure~\ref{fig:ladder-LS-L1}(c). Both have helicity degree one.}
\label{fig:L2.II_LD}
\end{figure}

We now work through the method in an explicit two-loop ladder example. We choose the
leading-singularity configuration shown in Figure~\ref{fig:ladder-LS-L1}(c). The cut
conditions are
\begin{equation}\label{eq:example-boundary-thesis}
\begin{aligned}
  \langle EF\,k{-}1\,k\rangle
  &=
  \langle EF\,k\,k{+}1\rangle
  =
  \langle EF\,j{-}1\,j\rangle
  =
  \langle EF\,j\,j{+}1\rangle
  =
  0 \, ,
  \\
  \langle CD\,k{-}1\,k\rangle
  &=
  \langle CD\,k\,k{+}1\rangle
  =
  \langle CD\,i\,i{+}1\rangle
  =
  \langle CD\,AB\rangle
  =
  \langle CD\,EF\rangle
  =
  0 \, .
\end{aligned}
\end{equation}
Every line in $\mathbb P^3$ has four degrees of freedom, so at first sight the line
$CD$ appears overconstrained. However, on the support of $\langle CD\,EF\rangle=0$,
only three among
\begin{equation}\label{eq:CD-EF-cuts-thesis}
  \langle EF\,k{-}1\,k\rangle \, ,\qquad
  \langle EF\,k\,k{+}1\rangle \, ,\qquad
  \langle CD\,k{-}1\,k\rangle \, ,\qquad
  \langle CD\,k\,k{+}1\rangle
\end{equation}
are independent. Geometrically, any three of these conditions, together with
$\langle CD\,EF\rangle=0$, force both lines $CD$ and $EF$ to pass through the point
$k$. Up to relabelling, the configuration is the maximal cut of either of the two
leading Landau diagrams in Figure~\ref{fig:L2.II_LD}.

Let us solve the Landau equations for the diagram in Figure~\ref{fig:L2.II_LD}. We illustrate here how to solve the classical Landau equations, starting from the dual momentum formulation and then passing to momentum twistors. There is a nice formulation of this problem directly in momentum twistors, or equivalently, on the Grassmannian. We return to the Grassmannian formulation in the next chapter. In dual momentum space, let $y_{CD}$ and $y_{EF}$ be the points associated with the loop lines $CD$ and $EF$. The cut equations and pinch equations are
{\small
\begin{equation*}
\begin{split}
  &
  \alpha_{1}(y_{CD}-x_k)^2
  =
  \alpha_{2}(y_{CD}-x_{i+1})^2
  =
  \alpha_{3}(y_{CD}-x_{AB})^2
  =
  \beta(y_{CD}-y_{EF})^2
  =
  0,
  \\
  &
  \alpha_{4}(y_{EF}-x_k)^2
  =
  \alpha_{5}(y_{EF}-x_{k+1})^2
  =
  \alpha_{6}(y_{EF}-x_j)^2
  =
  \alpha_{7}(y_{EF}-x_{j+1})^2
  =
  0,
  \\
  &
  \alpha_{1}(y_{CD}-x_k)^\mu
  +
  \alpha_{2}(y_{CD}-x_{i+1})^\mu
  +
  \alpha_{3}(y_{CD}-x_{AB})^\mu
  +
  \beta(y_{CD}-y_{EF})^\mu
  =
  0,
  \\
  &
  \alpha_{4}(y_{EF}-x_k)^\mu
  +
  \alpha_{5}(y_{EF}-x_{k+1})^\mu
  +
  \alpha_{6}(y_{EF}-x_j)^\mu
  +
  \alpha_{7}(y_{EF}-x_{j+1})^\mu
  +
  \beta(y_{CD}-y_{EF})^\mu
  =
  0 .
\end{split}
\end{equation*}
}
The leading Landau singularity lies on the branch where all
$\alpha_i$ and $\beta$ are nonzero.

We first solve the equations for $EF$. Since we are in four dimensions, five vectors are
linearly dependent. Therefore the $EF$ pinch equation does not impose an additional
constraint on the external kinematics. The four cut equations for $EF$ give two solutions,
\begin{equation}
  EF=jk \, ,
  \qquad
  EF=\overline{jk} = (j{-} 1 \, j \, j{+}1) \cap (k{-}1 \, k \, k{+}1) \, .
\end{equation}
The second solution is the $\overline{\mathrm{MHV}}$ solution and is not supported by the one-loop MHV geometry of the $EF$ line, namely, it does not lie on the boundary of the positive region defined by the one-loop MHV Amplituedron. Thus the geometric selection keeps $EF=jk$. This is the same two-mass-easy box Schubert problem encountered in the previous subsection.

Substituting this solution into the remaining cut conditions gives
\begin{equation}\label{eq:pre-thesis}
  \beta\langle CD\,jk\rangle
  =
  \alpha_1\langle CD\,k{-}k\rangle
  =
  \alpha_2\langle CD\,ii{+}1\rangle
  =
  \alpha_3\langle CD\,AB\rangle
  =
  0 \, .
\end{equation}
The pinch equation for $CD$ now gives a condition on the external kinematics. Taking
scalar products with the four lightlike vectors appearing in the $CD$ pinch equation
gives 
\begin{equation}\label{eq:pinchmat-thesis}
  \left(
  \begin{matrix}
  0 & 0 & \langle j\,k\,i\,i{+}1\rangle & \langle j\,k\,AB\rangle\\
  0 & 0 & \langle k{-}1\,k\,i\,i{+}1\rangle & \langle k{-}1\,k\,AB\rangle\\
  \langle j\,k\,i\,i{+}1\rangle
  & \langle k{-}1\,k\,i\,i{+}1\rangle
  & 0
  & \langle i\,i{+}1\,AB\rangle\\
  \langle j\,k\,AB\rangle
  & \langle k{-}1\,k\,AB\rangle
  & \langle i\,i{+}1\,AB\rangle
  & 0
  \end{matrix}
  \right)
  \left(
  \begin{matrix}
  \beta\\
  \alpha_1\\
  \alpha_2\\
  \alpha_3
  \end{matrix}
  \right)
  =
  0 \, .
\end{equation}
A nonzero solution exists if and only if the determinant of this matrix vanishes. This
gives the (square of the) leading Landau singularity
\begin{equation}\label{eq:sing-a-thesis}
  \langle k\, (k{-}1\,j)\, (i\,i{+}1)\, (AB)\rangle=0 \, ,
\end{equation}
where we use the short-hand notation
\begin{equation}
	\langle a(bc)(de)(fg)\rangle:=\langle abde\rangle\langle acfg\rangle-\langle acde\rangle\langle abfg\rangle \, .
\end{equation}
Equation~\eqref{eq:sing-a-thesis} is a physical singularity of the two-loop ladder, and it is therefore expected to define an actual singularity of the integral.

\begin{figure}[pos=t]
\centering
\begin{minipage}{0.35\linewidth}
\centering
\includegraphics[width=\linewidth]{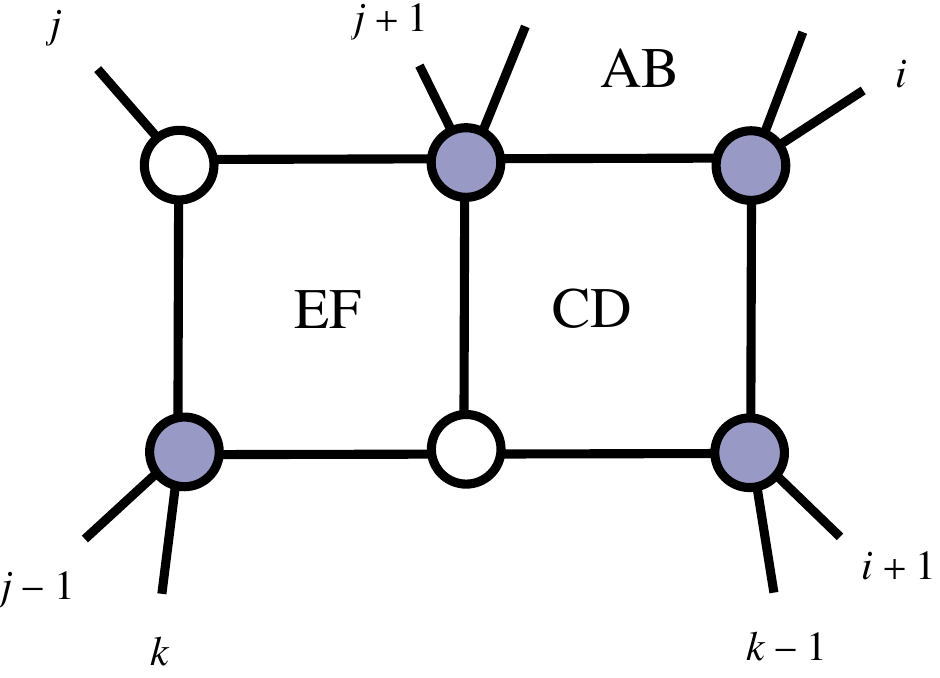}\\[0.2em]
\text{(a)}
\end{minipage}
\qquad
\begin{minipage}{0.35\linewidth}
\centering
\includegraphics[width=\linewidth, trim=0 -0.2cm 0 0]{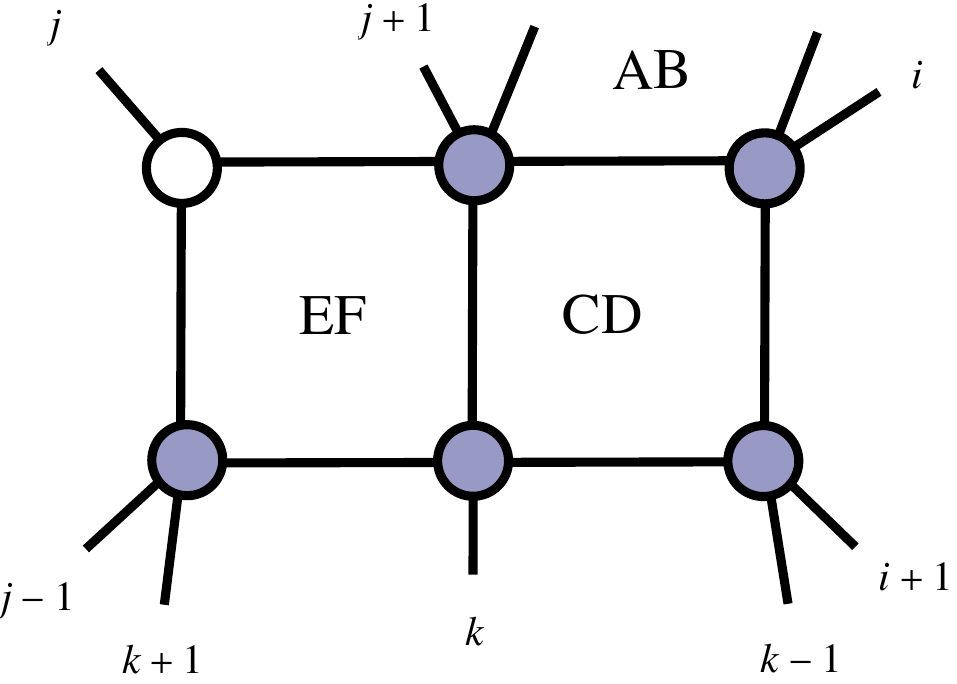}\\[0.2em]
\text{(b)}
\end{minipage}
\caption{Two sub-leading diagrams obtained from the leading Landau diagrams in
Figure~\ref{fig:L2.II_LD}. Diagram (a) gives a physical singularity. Diagram (b)
gives a residual singularity and is removed by the on-shell-diagram selection rule.}
\label{fig:subleading-two-loop-diagrams}
\end{figure}

The next step is to sub-leading solutions to the Landau equations. This means setting some Landau parameters to zero, or equivalently shrinking some propagators in the leading Landau diagram. Consider first the relaxation obtained by shrinking the propagator
$\langle EF\,k\,k{+}1\rangle$. The resulting diagram is shown in
Figure~\ref{fig:subleading-two-loop-diagrams}. It has seven cut conditions, and the loop
lines are localized to a one-dimensional boundary,
\begin{equation}\label{eq:partphy-thesis}
  EF=j\,\star \, ,
  \qquad
  CD=(AB\,\star)\cap(i\,i{+}1\,\star) \, ,
  \qquad
  Z_\star=Z_{k-1}+\alpha Z_k \, .
\end{equation}
The pinch equation is equivalent to
\begin{equation}\label{eq:CD-bracket-thesis}
  \langle CD\, j\, (k{-}1\,k)\cap\bar{j}\rangle
  =
  \langle EF\,i\,i{+}1\rangle
  \langle k{-}1\,k\,AB\rangle
  -
  \langle EF\,AB\rangle
  \langle i\,i{+}1\,k{-}1\,k\rangle
  =
  0 \, ,
\end{equation}
where recall that $\bar j=(j-1 \, j \, j+1)$.
This fixes
\begin{equation}
  Z_\star=(k{-}1\,k)\cap\bar{j} \, ,
\end{equation}
and gives the Landau singularity
\begin{equation}\label{eq:sing-b-thesis}
  \langle \star\, (k{-}1\,j)\, (i\,i{+}1)\, (AB)\rangle=0 \, .
\end{equation}

\begin{figure}[pos=t]
\centering

\begin{minipage}[t]{0.31\linewidth}
\centering
\vphantom{\begin{tikzpicture}[scale=0.75]
    \coordinate (A) at (0,-0.35);
    \coordinate (B) at (4.3,2.4);
\end{tikzpicture}}
\begin{tikzpicture}[scale=0.75]
    \coordinate (A) at (0,0.85);
    \coordinate (B) at (1.85,2.1);
    \coordinate (C) at (1.9,0.05);
    \coordinate (D) at (2,2);

    \fill[gray!30] (A) to[out=45,in=200] (B)
                   to[out=-65,in=85] (C)
                   to[out=180,in=-40] (A);

    \draw[thick,magenta] (-0.3,0.5) to[out=50,in=200] (2.3,2.3);
    \draw[thick] (1.8,2.3) to[out=-70,in=80] (1.8,-0.4);
    \draw[thick,magenta] (2.3,0) to[out=180,in=-50] (-0.3,1.2);

    \fill[teal] (A) circle (3pt) node[left] {$x$};
    \fill[teal] (B) circle (3pt);
    \fill[teal] (C) circle (3pt);

    \node[below] at (-0.2,0.5) {\textbf{$S_1$}};
    \node[left] at (-0.3,1.3) {\textbf{$S_2$}};
    \node at (1.2,1) {\Huge $\mathcal{A}$};
\end{tikzpicture}\\[0.2em]
\(\text{\text{(a)}}\)
\end{minipage}
\hfill
\begin{minipage}[t]{0.31\linewidth}
\centering
\begin{tikzpicture}[scale=0.75]
    \coordinate (A) at (0,0.85);
    \coordinate (B) at (1.85,2.1);
    \coordinate (C) at (1.9,0.05);
    \coordinate (D) at (3.8,1);

    \fill[gray!30] (A) to[out=45,in=200] (B)
                   to[out=-65,in=85] (C)
                   to[out=180,in=-40] (A);

    \draw[thick] (-0.3,0.5) to[out=50,in=200] (2.3,2.3);
    \draw[thick] (1.8,2.3) to[out=-70,in=80] (1.8,-0.4);
    \draw[thick,magenta] (4,1.5) to[out=250,in=0] (2.3,0) to[out=180,in=-50] (-0.3,1.2);
    \draw[thick,magenta] (1.45,2.4) to[out=-40,in=180] (4.3,1);

    \fill[teal] (A) circle (3pt);
    \fill[teal] (B) circle (3pt);
    \fill[teal] (C) circle (3pt);
    \fill[orange] (D) circle (3pt) node[below right] {$x$};

    \node[left] at (-0.3,1.3) {\textbf{$S_1$}};
    \node[left] at (1.45,2.4) {\textbf{$\bar{S}_1$}};
    \node at (1.2,1) {\Huge $\mathcal{A}$};
\end{tikzpicture}\\[0.2em]
\(\text{\text{(b)}}\)
\end{minipage}
\hfill
\begin{minipage}[t]{0.31\linewidth}
\centering
\begin{tikzpicture}[scale=0.75]
    \coordinate (A) at (0,0.85);
    \coordinate (B) at (1.85,2.1);
    \coordinate (C) at (1.9,0.05);
    \coordinate (D) at (3.8,1);

    \fill[gray!30] (A) to[out=45,in=200] (B)
                   to[out=-65,in=85] (C)
                   to[out=180,in=-40] (A);

    \draw[thick] (-0.3,0.5) to[out=50,in=200] (2.3,2.3);
    \draw[thick] (1.8,2.3) to[out=-70,in=80] (1.8,-0.4);
    \draw[thick] (2.3,0) to[out=180,in=-50] (-0.3,1.2);

    \draw[thick,magenta] (1.45,2.4) to[out=-40,in=180] (4.3,1);
    \draw[thick,magenta] (1.3,-0.35) to[out=40,in=250] (4,1.4);

    \fill[teal] (A) circle (3pt);
    \fill[teal] (B) circle (3pt);
    \fill[teal] (C) circle (3pt);
    \fill[magenta] (D) circle (3pt) node[below right] {$x$};

    \node[left] at (1.45,2.4) {\textbf{$\bar{S}_1$}};
    \node[left] at (1.3,-0.35) {\textbf{$S_1$}};
    \node at (1.2,1) {\Huge $\mathcal{A}$};
\end{tikzpicture}\\[0.2em]
\(\text{\text{(c)}}\)
\end{minipage}

\caption{
Three possible geometric pictures for the location of a solution to the Landau equations with respect to the positive geometry \(\mathcal{A}\). The solutions \(x\) shown in panels (a) and (b) are \textit{physical}, since they lie at least on a boundary component of \(\mathcal{A}\). For example, panel (a) may depict a leading singularity configuration defining a vertex of the geometry. Panel (c), instead, is \textit{spurious}, as it lies on the intersection of only residual components.}
\label{fig:selections}
\end{figure}
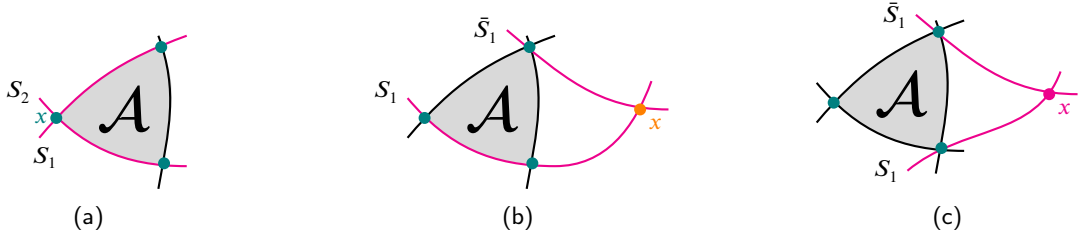

The singularity~\eqref{eq:sing-b-thesis} illustrates an important subtlety. The
solution of the pinch equations may itself lie outside the geometry. Indeed,
on the solution above one finds
\begin{equation}\label{eq:ineq-thesis}
  \langle EF\,k\,k{+}1\rangle
  =
  \langle j\, (k{-}1\,k)\cap\bar{j}\, k\,k{+}1\rangle
  =
  \langle j\,\bar{k}\rangle\langle k\,\bar{j}\rangle \, ,
\end{equation}
which is strictly negative for separated indices $k,j \in [n]$ with $|j-k|>1$, by positivity of the external data $Z_i$.
Nevertheless, the cut locus~\eqref{eq:partphy-thesis} has a positive
one-dimensional component, namely the region with $\alpha>0$. Thus the cut locus is a
boundary of the geometry, and the numerator does not vanish identically on it. The
singularity~\eqref{eq:sing-b-thesis} is therefore physical. This geometric situation is
represented in Figure~\ref{fig:selections}(b).

By contrast, shrinking the propagator $\langle EF\,k{-}1\,k\rangle$ gives the diagram
in Figure~\ref{fig:subleading-two-loop-diagrams}(b). The resulting colored Landau diagram has larger helicity degree than the leading diagram from which it came. According to the selection rule, it is residual. Solving its Landau equations gives a four-mass-box square root. More
explicitly, one finds
\begin{equation}
  EF=j,\big((k\,k{+}1)\cap\bar{j}\big) \, ,
\end{equation}
while $CD$ solves the four cut conditions
\begin{equation}
  \langle CD\, j((k\,k{+}1)\cap\bar{j})\rangle
  =
  \langle CD\,k{-}1\,k\rangle
  =
  \langle CD\,i\,i{+}1\rangle
  =
  \langle CD\,AB\rangle
  =
  0 \, .
\end{equation}
The associated square-root singularity is $\sqrt{\Delta}$, where
\begin{equation}\label{eq:bad-shrink-thesis}
  \Delta=(1-u-v)^2-4uv \, ,
\end{equation}
with
\begin{equation*}
\begin{aligned}
  u
  &=
  \frac{
  \langle k{-}1\,k\,i\,i{+}1\rangle
  \langle ((k\,k{+}1\,j)\cap\bar{j})\,AB\rangle}
  {
  \langle k{-}1\,k\,AB\rangle
  \langle ((k\,k{+}1\,j)\cap\bar{j})\,i\,i{+}1\rangle} \, , \ 
  v
  =
  \frac{
  \langle i\,i{+}1\,AB\rangle
  \langle k{-}1\,k\,((k\,k{+}1\,j)\cap\bar{j})\rangle}
  {
  \langle k{-}1\,k\,AB\rangle
  \langle ((k\,k{+}1\,j)\cap\bar{j})\,i\,i{+}1\rangle} \, .
\end{aligned}
\end{equation*}
For generic indices this is a candidate algebraic singularity. However, it is
spurious for the integrated ladder negative geometry: the corresponding cut locus is residual, it does not intersect the ladder negative geometry in the correct real dimension, and therefore the canonical numerator vanishes on it.

The possible geometric situations are summarized in Figure~\ref{fig:selections}. A
leading singularity configuration such as~\eqref{eq:sing-a-thesis} lies on an actual vertex of the
geometry. A configuration solving a physical sub-leading singularity such as~\eqref{eq:sing-b-thesis} may be
obtained by intersecting a physical boundary $S_1$ with a further pinch condition
$\bar S_1$; even if the final point lies outside the positive region, the cut locus $S_1$
is a boundary and the singularity survives. A residual sub-leading singularity such as
\eqref{eq:bad-shrink-thesis} arises when the cut locus itself is not a boundary, so the numerator
kills the residue.

\begin{tcolorbox}[resultbox]
\textbf{Physical and spurious Landau singularities.}
A Landau singularity is physical if its cut locus is a boundary of the geometry.
It can be spurious even when the Landau equations have a non-trivial solution.
\end{tcolorbox}
\noindent
This is the same numerator mechanism as in the Aomoto and Amplituhedron examples,
now implemented in the ladder negative geometry. We stress that in the schematic picture in Figure~\ref{fig:selections} we represented by $x$ the configurations of loop lines solving the Landau equations, by $\mathcal{A}$ the negative geometry region, and by $S_i$ and $\bar{S}_i$ the components obtained from the cut and pinch conditions, respectively. In general, the pinch conditions do not define boundaries of $\mathcal{A}$. The relevant criterion for a singularity to be physical is that, before imposing the pinch conditions, it lies on a boundary component of $\mathcal{A}$.

Applying the above procedure to all two-loop leading-singularity configurations in
Figure~\ref{fig:ladder-LS-L1} gives a conjecturally complete list of physical singularities
for two-loop ladder negative geometries at arbitrary multiplicity $n$~\cite[Sec 4.2]{ChicherinHennMazzucchelliTrnkaYangZhang2026}. 

\begin{tcolorbox}[resultbox]
\textbf{Two-loop ladder alphabets.}
At $\ell=2$ and arbitrary $n$, geometric Landau analysis gives a finite conjectural list of ladder singularities. This list is the starting point for the symbol alphabet of the pure functions $f^{(2)}_{n,ij}$.
\end{tcolorbox}
\noindent
The next subsection reviews the symbol technology needed to turn this information into
an actual bootstrap ansatz.

\subsection{Symbols and bootstrap}
\label{sec:Symbols and Bootstrap}

The previous subsections explained how geometric Landau analysis produces a
candidate set of branch loci for an integrated observable. For the ladder
negative geometries this output will become the starting point for the
construction of a symbol alphabet. We now explain what this means.

\subsubsection{What is the symbol?}
\label{subsec:what-is-the-symbol}

The symbol is a way of extracting the iterated logarithmic structure of a
polylogarithmic function, and in particular the information relevant for its
discontinuities and differential equations. Its origin lies
in the study of iterated integrals and multiple polylogarithms, going back to
Chen's iterated integrals and Goncharov's theory of multiple polylogarithms and
motivic coproducts
\cite{Chen1977,Goncharov2001,Goncharov2005,Brown2009,Brown2012}. In the
amplitudes literature, the symbol was introduced as a practical tool for
handling the transcendental functions appearing in Wilson loops and scattering
amplitudes in planar \(\mathcal N=4\) SYM
\cite{Goncharov:2010jf,Duhr:2011zq}. It quickly became central to the
\textit{bootstrap program}, where one tries to reconstruct integrated amplitudes
from their expected analytic properties rather than from a direct integration
\cite{DixonDrummondHenn2011,CaronHuotDixonMcLeodVonHippel2016,DixonEtAl2017}.

Mathematically, the symbol is the maximally iterated part of the coproduct on
the Hopf algebra of multiple polylogarithms. Chen iterated integrals form a
shuffle algebra, and their motivic refinements carry a coproduct which
decomposes a weight-\(k\) function into lower-weight pieces
\cite{Chen1977,Goncharov2001,Goncharov2005,Brown2009,Brown2012}. The symbol is
the component of this coproduct in which all factors have weight one; since
weight-one polylogarithms are logarithms, it can be represented by tensor words
in logarithmic letters. Thus the symbol remembers the iterated \(\mathrm{d}\log\)
structure of a function, while forgetting symbol-kernel terms such as powers of
\(\pi\) and zeta values. This makes it powerful for studying discontinuities,
differential equations, and functional identities in amplitudes
\cite{Goncharov:2010jf,Duhr:2011zq,Duhr:2012fh}, but also means that it is not
the full analytic function. The ordinary symbol calculus used below belongs to
the polylogarithmic sector; more general Chen iterated integrals, including
elliptic and modular examples, require enlarged coaction or symbol formalisms
\cite{BrownDuhr2020DlogNotPolylog,Broedel:2018iwv,Kristensson:2021ani}.

A first reason for the usefulness of symbols is that they simplify complicated functional identities between polylogarithms. For example,
Euler's dilogarithm identity
\begin{equation}
\label{eq:euler-dilog-identity}
        \Li_2(x)+\Li_2(1-x)
        +\log x\,\log(1-x)
        -\frac{\pi^2}{6}
        =
        0
\end{equation}
becomes, at symbol level, the elementary cancellation
\begin{equation}
\label{eq:euler-dilog-symbol}
        -(1-x)\otimes x
        -
        x\otimes(1-x)
        +
        x\otimes(1-x)
        +
        (1-x)\otimes x
        =
        0 \, .
\end{equation}
The constant \(\pi^2/6\) is invisible to the symbol. In this way, large
expressions built from polylogarithms often recognize as equivalent, or
reduce to much simpler representatives, by comparing their symbols. This was
one of the key insights in the study of the two-loop six-particle remainder
function in planar \(\mathcal N=4\) SYM: expressions which are extremely
complicated at the level of functions become manageable at symbol level, and
can then be integrated back to compact formulae in terms of
polylogarithms
\cite{DelDuca:2009au,DelDuca:2010zg,Goncharov:2010jf}.

\paragraph{Definition and basic properties}
\label{subsubsec:symbol-definition-basic-properties}

We now recall the definition of the symbol in the form used below. Let \(F\) be
a pure polylogarithmic function of weight \(k\). This means, roughly, that each
differentiation lowers the transcendental weight by one. The \textit{symbol of
\(F\)}, denoted \(\mathcal S(F)\), is defined recursively as follows. Assume that
\begin{equation}
\label{eq:symbol-recursive-definition}
        \mathrm{d} F
        =
        \sum_{\alpha} F^{(\alpha)}\, \mathrm{d}\log W_{\alpha} \, ,
\end{equation}
where each \(F^{(\alpha)}\) has weight \(k-1\), then
\begin{equation}
\label{eq:symbol-recursive}
        \mathcal S(F)
        =
        \sum_{\alpha}
        \mathcal S\!\left(F^{(\alpha)}\right)\otimes W_{\alpha} \, .
\end{equation}
The recursion starts from \(\mathcal S(\log W)=W\). The functions \(W_{\alpha}\)
are called \textit{letters}, and the set of all letters appearing in a class of
functions is called the \textit{symbol alphabet}.

Equivalently, a \textit{weight-\(k\) symbol} is a finite sum
\begin{equation}
\label{eq:symbol-general-tensor}
        \mathcal S(F)
        =
        \sum c_{a_1,\ldots,a_k}\,
        W_{a_1}\otimes W_{a_2}\otimes\cdots\otimes W_{a_k} \, ,
\end{equation}
where usually \(c_{a_1,\ldots,a_k}\in \mathbb Q\). The entries are
multiplicative classes of functions, rather than ordinary vectors of functions.
Thus multiplication inside one entry becomes addition,
\begin{equation}
\label{eq:symbol-multiplicativity}
        \cdots\otimes (W_1W_2)\otimes\cdots
        =
        \cdots\otimes W_1\otimes\cdots
        +
        \cdots\otimes W_2\otimes\cdots \, ,
\end{equation}
and constants vanish,
\begin{equation}
        \cdots\otimes c \otimes\cdots =0,
        \qquad c\in \mathbb C^* \, .
\end{equation}
Thus letters should be understood as elements of the multiplicative group of
rational or algebraic functions of the external kinematics, modulo non-zero
complex constants. The tensor notation is therefore slightly misleading: a
symbol alphabet is a generating set, not necessarily a basis in the usual
linear-algebraic sense. A useful consequence is that multiplicative relations
among letters become linear relations among symbol words.

The simplest examples are
\begin{equation}
\label{eq:symbol-log-dilog}
        \mathcal S(\log x)=x,
        \qquad
        \mathcal S(\operatorname{Li}_2(x))
        =
        -(1-x)\otimes x \, .
\end{equation}
The second identity follows from
\begin{equation}
        \mathrm{d}\,\operatorname{Li}_2(x)
        =
        -\log(1-x)\,\mathrm{d}\log x \, .
\end{equation}
More generally,
\begin{equation}
\label{eq:symbol-classical-polylog}
        \mathcal S(\operatorname{Li}_m(x))
        =
        -(1-x)\otimes
        \underbrace{x\otimes\cdots\otimes x}_{m-1\text{ times}} \, .
\end{equation}

\begin{tcolorbox}[resultbox]
\textbf{Symbols.}
The symbol records the iterated logarithmic singularities of a
polylogarithmic function. Its letters encode the logarithmic arguments,
and the order of the tensor entries records the order in which discontinuities or derivatives probe them.
\end{tcolorbox}
\noindent

On the other hand, the symbol captures only part of the information of a
polylogarithmic function. For example,
\begin{equation}
        \mathcal S(\operatorname{Li}_2(x))
        =
        \mathcal S(\operatorname{Li}_2(x)+c\,\pi^2),
        \qquad c\in\mathbb Q \, ,
\end{equation}
because constants of weight two are invisible to the symbol. At higher weight
there are further symbol-kernel terms, such as zeta values multiplying
lower-weight functions. Thus reconstructing an actual analytic function from a
symbol requires additional information.

\paragraph{Integrability and reconstruction}
\label{subsubsec:symbol-integrability-reconstruction}

Also, not every tensor of the form~\eqref{eq:symbol-general-tensor} is the
symbol of a function. The obstruction is the equality of mixed partial
derivatives. If
\begin{equation}
        S=
        \sum c_{a_1,\ldots,a_k}\,
        W_{a_1}\otimes\cdots\otimes W_{a_k} \, ,
\end{equation}
then \(S\) is \textit{integrable} only if, for every adjacent pair of entries,
\begin{equation}
\label{eq:symbol-integrability}
        \sum c_{a_1,\ldots,a_k}\,
        \mathrm{d}\log W_{a_i}\wedge
        \mathrm{d}\log W_{a_{i+1}}\,
        W_{a_1}\otimes\cdots\widehat{W_{a_i}}
        \cdots\widehat{W_{a_{i+1}}}\cdots\otimes W_{a_k}
        =
        0 \, .
\end{equation}
Here the hats mean that the two entries are omitted. This condition is the
symbol-level version of the statement that there exists a function \(F\) whose
total differential gives the lower-weight functions in
\eqref{eq:symbol-recursive-definition}. Equivalently,
\eqref{eq:symbol-integrability} guarantees that the resulting iterated integral
along a path away from the divisors \(\{W_a=0\}\) is independent of the chosen
path, up to monodromies around these divisors. Hence an integrable symbol is
precisely one which is the symbol of a genuine multivalued polylogarithmic
function. Integrability~\eqref{eq:symbol-integrability} is a central constraint
in the symbol bootstrap.

To reconstruct an analytic function from a symbol is in general not a purely
mechanical task \cite{Duhr:2012fh,DuhrDulat2019}. One first checks the
integrability condition~\eqref{eq:symbol-integrability}. Second, one chooses a
class of functions in which to integrate the symbol. In many amplitude
applications this class is given by Goncharov polylogarithms
\cite{Goncharov2001,Duhr:2011zq}, with arguments constrained by the alphabet
under consideration. Third, one fixes terms beyond the symbol, namely
lower-weight functions multiplied by transcendental constants and purely
constant terms, by imposing boundary conditions, regularity conditions,
collinear limits, or known values in special kinematics
\cite{DixonDrummondHenn2011,CaronHuotDixonMcLeodVonHippel2016,DixonEtAl2017}.
Computer packages such as \texttt{PolyLogTools} and \texttt{HyperInt} provide
practical implementations of parts of this program  \cite{DuhrDulat2019,Panzer:2014caa,Panzer2015}.

\paragraph{Discontinuities and differentiation}
\label{subsubsec:symbol-discontinuities-differentiation}

The recursive definition makes the action of differentiation on
\eqref{eq:symbol-general-tensor} transparent:
\begin{equation}
\label{eq:symbol-derivative-action}
        \mathcal S(\mathrm \mathrm{d}F)
        =
        \sum c_{a_1,\ldots,a_k}\,
        W_{a_1}\otimes\cdots\otimes W_{a_{k-1}}\,
        \mathrm{d}\log W_{a_k} \, .
\end{equation}
Thus derivatives act on the last entry of the symbol. Discontinuities instead
act as
\begin{equation}
\label{eq:symbol-discontinuity-action}
        \mathcal S\!\left(
        \frac{1}{2\pi i}\operatorname{Disc}_{W_b=0}F
        \right)
        =
        \sum
        c_{b,a_2,\ldots,a_k}\,
        W_{a_2}\otimes\cdots\otimes W_{a_k} \, ,
\end{equation}
where the sum runs over \(a_2,\ldots,a_k\) and $b$ is fixed.

\begin{tcolorbox}[definitionbox]
\textbf{Symbol dictionary.}
\begin{equation}
\label{eq:symbol-diff-disc-summary}
\begin{array}{ccl}
        \text{discontinuities}
        &\longleftrightarrow&
        \text{first entries of the symbol},
        \\[2mm]
        \text{derivatives}
        &\longleftrightarrow&
        \text{last entries of the symbol}.
\end{array}
\end{equation}
\end{tcolorbox}
\noindent

\begin{eg}[An Aomoto example]
\label{eg:aomoto-symbol-discontinuity}
Consider again the Aomoto pairing from
Subsection~\ref{subsec:refined-landau-analysis}, with \(A\) being the triangle
with vertices \((0,0)\), \((1,0)\) and \((0,1)\), and
\begin{equation}
        I(y,w,z)
        =
        \int_A
        \frac{z+w-x_1-x_2}
        {(z-x_1)(w-x_2)(y-x_1-x_2)}
        \,\mathrm{d}x_1\wedge \mathrm{d}x_2 \, .
\end{equation}
One can compute its symbol using the general formula for symbols of Aomoto
forms, or by applying the symbol map to the explicit analytic expression given
in~\eqref{eq:Izw}. We can perform the latter with the \texttt{Mathematica}
package \texttt{PolyLogTools}~\cite{DuhrDulat2019} by calling
\begin{verbatim}
        ComputeSymbol[F] // SymbolExpand // SymbolFactor
\end{verbatim}
for \(F=I(y,w,z)\). This outputs
\begin{equation}
\label{eq:S_I}
\begin{aligned}
        \mathcal S(I)
        ={}&
        (1-y)\otimes \frac{(y-w)(y-z)}{wz}
        +y\otimes \frac{wz}{(y-w)(y-z)}
        +(1-w)\otimes \frac{y-1}{w-y}
        \\
        &\qquad
        +(1-z)\otimes \frac{y-1}{z-y}
        +w\otimes \frac{w-y}{y-1}
        +z\otimes \frac{z-y}{y-1}.
\end{aligned}
\end{equation}
We can extract the symbol alphabet for this function by calling
\begin{verbatim}
        GetSymbolAlphabet[SymbolExpand[F]] // DeleteDuplicates
\end{verbatim}
whose output is
\begin{equation}
        \{y, 1-y, w, 1-w, z, 1-z, y-w, y-z\} \, .
\end{equation}
We can also check that this symbol satisfies the integrability
conditions~\eqref{eq:symbol-integrability} by calling
\begin{verbatim}
        IntegrabilityCondition[F, 1] // Simplify
\end{verbatim}
which indeed outputs \(0\) in our example.

In Subsection~\ref{subsec:refined-landau-analysis}, we computed, using the
duality between discontinuities and residues, that the discontinuity of $I(y,w,z)$ across the
singular component \(1-w=0\) is
\begin{equation}
\label{eq:aomoto-symbol-disc-result}
        \frac{1}{2\pi i}
        \operatorname{Disc}_{w=1} I(y,w,z)
        =
        \log\!\left(\frac{y-1}{y-w}\right) \, .
\end{equation}
This is in agreement with applying
\eqref{eq:symbol-discontinuity-action} to~\eqref{eq:S_I}: the operation picks
out the terms whose first entry is \(1-w\) and removes this first entry,
yielding
\begin{equation}
        \mathcal S\!\left(
        \frac{1}{2\pi i}\operatorname{Disc}_{w=1}I
        \right)
        =
        \frac{y-1}{w-y} \, .
\end{equation}
This agrees with the symbol of~\eqref{eq:aomoto-symbol-disc-result}.
\end{eg}

\paragraph{Algebraic letters and square-root covers}
\label{subsubsec:symbol-algebraic-letters}

An important distinction between Aomoto forms in the example above and more
general amplitudes is that the latter may require alphabets containing
non-rational letters. In some cases one can rationalize the relevant square
roots by a change of variables, but this is not always possible globally
\cite{Besier:2018jen,BesierWasserWeinzierl2019,BesierFesti2020}. The basic
example showing the appearance of a non-rational singularity is the four-mass
box of Subsection~\ref{subsec:Boxes in Momentum Twistors}. There the integral had a square-root singularity $\sqrt{\Delta}$ associated to the leading Landau
singularity 
\begin{equation}
\label{eq:symbol-section-four-mass-delta}
        \Delta=(1-u-v)^2-4uv \, .
\end{equation}
The natural variables \(z,\bar z\) provide a rational parametrization:
\begin{equation}
\label{eq:symbol-section-z-zbar}
        u=z\bar z \, ,
        \qquad
        v=(1-z)(1-\bar z) \, ,
        \qquad
        \sqrt{\Delta}=z-\bar z \, .
\end{equation}
The integrated four-mass box contains classical polylogarithms in \(z\) and
\(\bar z\), as in~\eqref{eq:four-mass-box-result}. Its alphabet contains the
letters
\begin{equation}
\label{eq:four-mass-basic-letters}
        \{z,\bar z,1-z,1-\bar z\} \, .
\end{equation}
These are algebraic functions of the rational cross-ratios \(u,v\). After
stripping off the leading-singularity normalization \((z-\bar z)^{-1}\), the
symbol of the pure four-mass box in~\eqref{eq:four-mass-box-result} can be written in the compact form
\begin{equation}
\label{eq:four-mass-symbol}
        \mathcal S((z-\bar z) \, \mathcal{I}_{4m})
        =
        u\otimes \frac{1-z}{1-\bar z}
        -
        v\otimes \frac{z}{\bar z} \, .
\end{equation}
The ratios in this formula are usually called \textit{algebraic letters},
\begin{equation}
\label{eq:four-mass-odd-letters}
        \frac{z}{\bar z}
        =
        \frac{1+u-v+\sqrt{\Delta}}
             {1+u-v-\sqrt{\Delta}} \, ,
        \qquad
        \frac{1-z}{1-\bar z}
        =
        \frac{1-u+v-\sqrt{\Delta}}
             {1-u+v+\sqrt{\Delta}} \, .
\end{equation}
They are also called \textit{odd letters}, because they are inverted under the
map \(\sqrt{\Delta}\mapsto -\sqrt{\Delta}\). At symbol level, inversion is
equivalent to a minus sign, since \(W^{-1}\) represents \(-W\).

Let us explain the geometric meaning of this square root. The Landau
discriminant \(\Delta\) in~\eqref{eq:symbol-section-four-mass-delta} is the
locus where the two solutions of the Schubert problem, or maximal cut, for the
four-mass box collide. Geometrically, it is the branch divisor of the double
cover of kinematic space $\mathcal{K}=\mathbb{C}^2$ given by
\begin{equation}
\label{eq:four-mass-double-cover}
        \widetilde{\mathcal K}_{\Delta}
        =
        \{(u,v,r)\in \mathbb C^3\mid r^2=\Delta(u,v)\}
        \longrightarrow
        \mathbb C^2,
        \qquad
        (u,v,r)\longmapsto (u,v) \, .
\end{equation}
On this cover the two Schubert solutions are separated. Going around the
branch divisor \(\Delta=0\) in the kinematic space \(\mathbb C^2\) exchanges
the two sheets, or equivalently sends \(r\mapsto -r\).

Analytically, a square-root Landau singularity means that the local coordinate
on the natural cover is not \(\Delta\), but \(r=\sqrt{\Delta}\). Near a generic
point of the discriminant, \(r=0\), the integral can be expanded in powers of
\(r\) and logarithms of rational functions of \(r\). Schematically one finds a
local behaviour of the form
\begin{equation}
\label{eq:sqrt-local-expansion}
        \mathcal I_{\rm 4m}(u,v)
        \rightarrow
        \sum_{m\geq 0}
        r^m
        \Big[
        A_m(u,v)
        +
        B_m(u,v)\log R(u,v,r)
        \Big]
        +\cdots \, ,
\end{equation}
where the dots indicate terms analytic in the same local variables, the
functions \(A_m(u,v)\), \(B_m(u,v)\) are local holomorphic \emph{germs} away from
other Landau divisors, and \(R(u,v,r)\) is a rational function on the cover.
The monodromy around \(\Delta=0\) acts by \(r\mapsto -r\), and hence exchanges
the two local branches. In the coordinates \(z,\bar z\), this becomes \(z\leftrightarrow \bar z\).

The logarithmic arguments \(R(u,v,r)\) are not arbitrary rational functions on
the cover. Their zeros and poles project to subvarieties of the original
kinematic space, and these projections must be contained in the full Landau
variety. Otherwise the corresponding symbol letter would introduce a new
logarithmic singularity not predicted by Landau analysis. This gives a simple
compatibility condition for square-root letters
\cite{DlapaHelmerPapathanasiouTellander2023,MatijasicThesis,CorreiaGirouxMizera2026}. Explicitly, if \(w=w(u,v)\) is a rational
function on the base, then one can form the algebraic letter
\begin{equation}
\label{eq:odd-square-root-letter}
        \frac{w+r}{w-r} \, ,
\end{equation}
which is inverted by \(r\mapsto -r\). If this expression is to define a letter
of the integral, then the zeros of its numerator and denominator are candidate
singularities. Hence their projections to the base must be contained in the
Landau variety. Algebraically, the relevant projected divisor is controlled by
\begin{equation}
\label{eq:norm-square-root-letter}
        (w+r)(w-r)=w^2-\Delta \, .
\end{equation}
Thus the compatibility condition is, up to non-vanishing factors,
\begin{equation}
\label{eq:fact_sr}
        w^2-\Delta
        \ \propto\
        \prod_a W_a^{m_a} \, ,
\end{equation}
where the \(W_a\) are rational letters obtained from the Landau analysis. In
other words, the \emph{norm} of the upstairs function \(w+r\) must factor over the allowed rational singularities downstairs.

In the four-mass box, the two algebraic letters in
\eqref{eq:four-mass-odd-letters} are obtained by taking
\begin{equation}
        w_1=1+u-v \, ,
        \qquad
        w_2=1-u+v \, .
\end{equation}
One computes
\begin{equation}
        w_1^2-\Delta=4u \, ,
        \qquad
        w_2^2-\Delta=4v \, ,
\end{equation}
whose vanishing loci \(\{u=0\}\) and \(\{v=0\}\) are Landau divisors of the
four-mass box. Therefore, we can write
\begin{equation}
        \frac{w_1+r}{w_1-r}
        =
        \frac{1+u-v+\sqrt{\Delta}}
             {1+u-v-\sqrt{\Delta}}
        =
        \frac{z}{\bar z} \, ,
        \qquad
        \frac{w_2-r}{w_2+r}
        =
        \frac{1-u+v-\sqrt{\Delta}}
             {1-u+v+\sqrt{\Delta}}
        =
        \frac{1-z}{1-\bar z} \, .
\end{equation}

This mechanism generalizes to higher finite-degree covers. Let
\(\pi:\widetilde{\mathcal K}\to\mathcal K\) be an algebraic cover of kinematic
space of degree \(\gamma \in \mathbb{N}\), and let \(\phi\) be a rational function on
\(\widetilde{\mathcal K}\). The \textit{norm} of \(\phi\) with respect to the
cover \(\pi\) is defined by
\begin{equation}
\label{eq:norm-definition-cover}
        \operatorname{Norm}_{\widetilde{\mathcal K}/\mathcal K}(\phi)(s)
        =
        \prod_{\tilde s \, \in \, \pi^{-1}(s)} \phi(\tilde s) \, ,
\end{equation}
for a generic point \(s\in\mathcal K\). Equivalently, this is the field-theoretic norm of $\phi$ with respect to the field extension $\mathbb{C}(\widetilde{\mathcal{K}})/ \mathbb{C}(\mathcal{K})$ from commutative algebra~\cite{Bosch2013AlgebraicGeometry}.

Being symmetric in the points of a generic fibre, the norm descends to a
rational function on \(\mathcal K\). In the square-root case
\(\phi=w+r\), this gives precisely
\eqref{eq:norm-square-root-letter}. The higher-degree analogue of
\eqref{eq:fact_sr} is that the numerator and denominator of
\(\operatorname{Norm}_{\widetilde{\mathcal K}/\mathcal K}(\phi)\) should be
supported on the Landau variety downstairs.

The construction of algebraic letters from Landau data is implemented
algorithmically in the \texttt{Mathematica} package
\texttt{Effortless.m}~\cite{repoEffortless}. Given a square-root Landau
discriminant \(\Delta\) and a list of rational letters \(W_a\), the package
searches for low-complexity rational functions \(w\), represented by
polynomial numerator and denominator ans\"atze of bounded degree, such that~\eqref{eq:fact_sr} holds true.
Thus the search is not over all rational functions on the cover, but a bounded-degree search guided by the structure of the Landau variety.

\paragraph{The symbol bootstrap}
\label{subsubsec:symbol-bootstrap-general}

We can now state the idea of the \textit{symbol bootstrap}. This is a method
for determining the symbol of an integral from its expected analytic
properties, assuming that the integral evaluates to polylogarithmic functions
of a given weight. The examples relevant here
are finite quantities in planar \(\mathcal N=4\) SYM, where the expected
transcendental weight at loop order \(\ell\) is \(2\ell\)
\cite{Kotikov:2002ab,ArkaniHamed:2010kv,Henn2013}. The bootstrap begins
with a finite alphabet
\begin{equation}
        \mathbb A=\{W_a\}_a \, ,
\end{equation}
whose letters are obtained from Landau analysis, geometric Landau analysis,
cluster structures, or a combination of these inputs. One then writes the most
general weight-\(2\ell\) tensor
\begin{equation}
\label{eq:symbol-bootstrap-ansatz}
        S
        =
        \sum_{a_1,\ldots,a_{2\ell}}
        c_{a_1,\ldots,a_{2\ell}}
        W_{a_1}\otimes\cdots\otimes W_{a_{2\ell}} \, ,
\end{equation}
with unknown rational coefficients \(c_{a_1,\ldots,a_{2\ell}}\in\mathbb Q\).
The goal is to fix these coefficients by imposing a sequence of constraints such as:
\begin{enumerate}
        \item \emph{integrability}~\eqref{eq:symbol-integrability}, which
        ensures that the tensor \(S\) is the symbol of some function;

        \item \emph{first-entry conditions}, which restrict branch cuts to the
        expected singularities;

        \item \emph{symmetries}, such as dihedral symmetry, parity, or
        transformation properties under changes of branch, often referred to
        as \emph{Galois symmetries};

        \item \emph{physical limits} as collinear and multi-Regge
        limits, or known lower-loop behaviour;

        \item \emph{hierarchy principle}, imposing sequential-discontinuity constraints such as Steinmann
        relations, which forbid certain iterated cuts;

        \item \emph{cluster adjacency}, when available, which requires adjacent
        symbol letters to be compatible cluster coordinates.
\end{enumerate}
We return to the last point and the relation to cluster algebras in Section~\ref{sec:rationality-cluster-structures}.

Historically, the most successful applications of this strategy have been in
planar \(\mathcal N=4\) SYM. The six-point amplitude was bootstrapped to high
loop order using the hexagon alphabet, Steinmann constraints, final-entry
conditions, collinear limits and multi-Regge data
\cite{DixonDrummondHenn2011,CaronHuotDixonMcLeodVonHippel2016}. The
seven-point amplitude led to the heptagon bootstrap and to the discovery of
strong cluster adjacency constraints
\cite{DixonEtAl2017,DrummondFosterGurdoganHarrington2019}. More recent work has
extended these ideas to larger alphabets, algebraic letters, Feynman-integral
families, Wilson-loop observables, and negative geometries
\cite{DlapaHelmerPapathanasiouTellander2023,He:2021non,Chicherin:2020umh,ChicherinHennMazzucchelliTrnkaYangZhang2026}.

The lesson for the next subsection is the following. Geometric Landau analysis
does not directly give the integrated function. Rather, it gives the geometric
input for the first stages of the bootstrap: the possible branch loci, the
algebraic covers, and the letters that should appear in the symbol. We now
apply this philosophy to ladder negative geometries.

\subsubsection{Bootstrapping integrated negative geometries}
\label{subsec:symbol-bootstrap-negative-geometries}

\paragraph{General strategy}
\label{subsubsec:bootstrap-general-strategy}

We now apply the symbol technology reviewed above to the integrated negative
geometries associated with ladder graphs. The point of this subsection is not
only to present two explicit bootstrap computations, but also to explain how
the geometric information discussed earlier survives integration. The
canonical form of a negative geometry gives rational leading singularities,
while geometric Landau analysis predicts the possible branch loci and algebraic
covers of the pure functions multiplying them. The symbol bootstrap combines
these two types of data into an ansatz for the integrated observable. In
particular, geometric Landau analysis provides a refined alphabet from the
start, greatly reducing the size of the bootstrap problem. The results
summarized below are those of
\cite{ChicherinHennMazzucchelliTrnkaYangZhang2026}.

Integrated ladder negative geometries have the form
\begin{equation}
\label{eq:neg-geo-decomposition-bootstrap}
        F_{n,{\rm ladder}}^{(\ell)}
        =
        \sum_{1 \leq i<j \leq n}
        \Omega_n(ij)\, f^{(\ell)}_{n,ij} \, ,
\end{equation}
where \(\Omega_n(ij)\) is the corresponding leading singularity, and
\(f^{(\ell)}_{n,ij}\) is a pure function of transcendental weight \(2\ell\).
The role of the bootstrap is to determine the symbols of these functions.
Thus, after determining an alphabet \(\mathbb A=\{W_a\}\), we make the ansatz
\begin{equation}
\label{eq:neg-geo-symbol-ansatz}
        \mathcal S\!\left(f^{(\ell)}_{n,ij}\right)
        =
        \sum_{a_1,\ldots,a_{2\ell}}
        c_{a_1,\ldots,a_{2\ell}}\,
        W_{a_1}\otimes\cdots\otimes W_{a_{2\ell}},
        \qquad
        c_{a_1,\ldots,a_{2\ell}}\in\mathbb Q \, .
\end{equation}
The alphabet is obtained from the geometric Landau analysis of the negative
geometry described in the previous sections. Rational Landau singularities give
rational letters, while square-root Landau singularities are converted into
algebraic letters by the norm construction explained in
Subsection~\ref{subsec:what-is-the-symbol}. In the computations discussed
below, all algebraic singularities are of square-root type; no higher-degree
covers are needed.

Bootstrapping negative geometries is useful for two related reasons. First, it
gives explicit integrated data for the negative-geometry expansion of the
Wilson loop with a Lagrangian insertion, and hence gives insight into the
singularities and alphabet of the latter. Second, it provides a controlled
setting in which to compare geometric singularity data with known
Feynman-integral function spaces. In the examples below, the Landau diagrams
contain non-planar subtopologies, but the final symbols nevertheless localize
in planar function spaces. This suggests that the geometry imposes strong
constraints on the analytic structure after integration.

\paragraph{Differential-equation constraints}
\label{subsubsec:bootstrap-differential-equation-constraints}

A particularly important additional constraint for the bootstrap comes from a
special second-order differential equation for the functions
\(f^{(\ell)}_{n,ij}\). These equations apply whenever the underlying negative
geometry graph has the unintegrated loop \(AB\) as a pending node. They
originate from the fundamental solution of the d'Alembertian,
\begin{equation}
        \Box_{x_0}\frac{1}{(x_0-x_*)^2}
        =
        -4i\pi^2\delta^4(x_0-x_*) \, ,
\end{equation}
where \(x_0\) is dual to \(AB\). Choosing \(x_*\) to be the dual point
associated with the line \((ij)\) in momentum-twistor space, one obtains the
operator
\begin{equation}
\label{eq:Dij-bootstrap}
        \mathcal D_{ij}
        :=
        (x_0-x_*)^2\,
        \Box_{x_0}\,
        \frac{1}{(x_0-x_*)^2} \, .
\end{equation}
For ladder negative geometries, this operator acts recursively in the loop
order as
\begin{equation}
\label{eq:DE-bootstrap}
\begin{aligned}
        \mathcal D_{ij}\, f^{(\ell)}_{n,ij}
        &=
        -4\sum_{kl}
        \overline{\mathcal C}_{kl,ij}\,
        f^{(\ell-1)}_{n,kl},
        \qquad \ell>1,
        \\
        \mathcal D_{ij}\, f^{(1)}_{n,ij}
        &=
        4\,\overline{\mathcal C}_{i+1\,j+1,ij} \,  .
\end{aligned}
\end{equation}
The coefficients \(\overline{\mathcal C}_{kl,ij}\) are rational functions of
the unintegrated line \(AB\) and of the external data \(Z_i\)~\cite{ChicherinHennMazzucchelliTrnkaYangZhang2026}. Equivalently, they
are obtained from the coefficients of the canonical form of the ladder negative
geometry, whose all-\(n\), all-\(\ell\) expression was described in
\cite{GlewLukowski2025,neg_geom_pos}. The implementation used in
\cite{ChicherinHennMazzucchelliTrnkaYangZhang2026} is provided as a
\texttt{Mathematica} code.

At two loops, the recursive equation reduces to
\begin{equation}
\label{eq:L2-DE-bootstrap}
        \mathcal D_{ij}\, f^{(2)}_{n,ij}
        =
        -4\sum_{kl}
        \overline{\mathcal C}_{kl,ij}\,
        \mathcal I^{\rm cp}_{kl} \, ,
\end{equation}
where \(\mathcal I^{\rm cp}_{kl}\) are the integrated chiral pentagons defined
in~\eqref{eq:chiral-pentagon-integrated}.

The differential equation is especially useful at symbol level. Acting with a
second-order operator on a weight-\(2\ell\) symbol can produce terms of
weights \(2\ell-1\) and \(2\ell-2\). The right-hand side of
\eqref{eq:DE-bootstrap} has weight \(2\ell-2\). Therefore the weight
\(2\ell-1\) contribution must vanish. In particular, if \(S\) is an allowed
last entry, viewed as a multiplicative combination of letters, it must satisfy
\begin{equation}
\label{eq:last-entry-DE-bootstrap}
        \mathcal D_{ij}\log S=0 \, .
\end{equation}
This condition substantially reduces the last-entry space before one constructs
the integrable symbols.

\paragraph{Ladder alphabets from geometric Landau analysis}
\label{subsubsec:bootstrap-ladder-alphabets}

The geometric Landau analysis of two-loop ladders gives a conjectural list of
physical singularities at all multiplicities. Specializing this list to five
points gives precisely the subset of the planar pentagon-function alphabet
needed for the two-loop ladder. In the notation of
\cite{Gehrmann:2018yef}, the relevant letters are
\begin{equation}
\label{eq:five-point-two-loop-alphabet-bootstrap}
        \{W_1,\ldots,W_5,
          W_{11},\ldots,W_{20},
          W_{26},\ldots,W_{30}\} \, .
\end{equation}
The letters \(W_6,\ldots,W_{10}\), corresponding to \(s_{12}+s_{23}\) and its
cyclic images, do not appear because the relevant Landau diagrams contain no
double-triangle cuts. This agrees with the direct integration of the five-point
two-loop ladder negative geometry in \cite{neg_geom_pos}.

At six points and two loops, the geometric Landau analysis gives an alphabet of
\(157\) independent letters. This alphabet is a subset of the \(245\)-letter
planar two-loop six-particle alphabet of \cite{Henn:2025xrc}. Although some of
the Landau diagrams contain non-planar subtopologies, no genuinely non-planar
letter is needed. The \(157\) letters split into \(95\) parity-even letters and
\(62\) parity-odd letters. Some of the odd letters are rationalized in momentum
twistor variables, while others arise from genuine square-root singularities
and are constructed by the algebraic-letter procedure described above. We refer to~\cite{ChicherinHennMazzucchelliTrnkaYangZhang2026} for the precise convention on their numbering.

\begin{figure}[pos=t]
    \centering
    \begin{subfigure}[t]{0.30\linewidth}
        \centering
        \includegraphics[width=\linewidth]{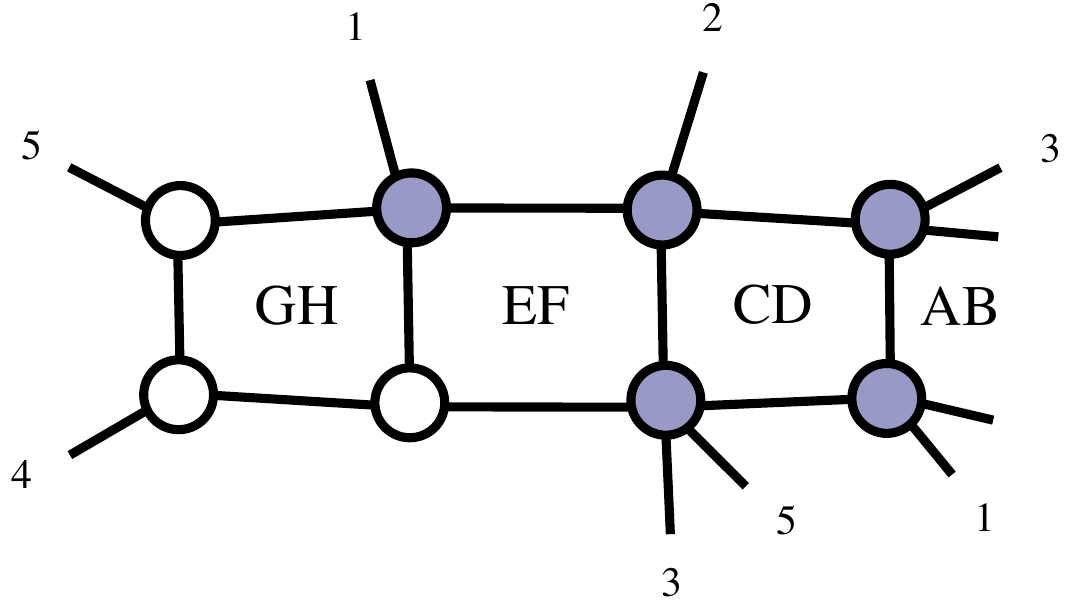}
        \caption{}
    \end{subfigure}
    \qquad
    \begin{subfigure}[t]{0.30\linewidth}
        \centering
        \includegraphics[width=\linewidth, trim=-1.5cm 0.5cm 0 0]{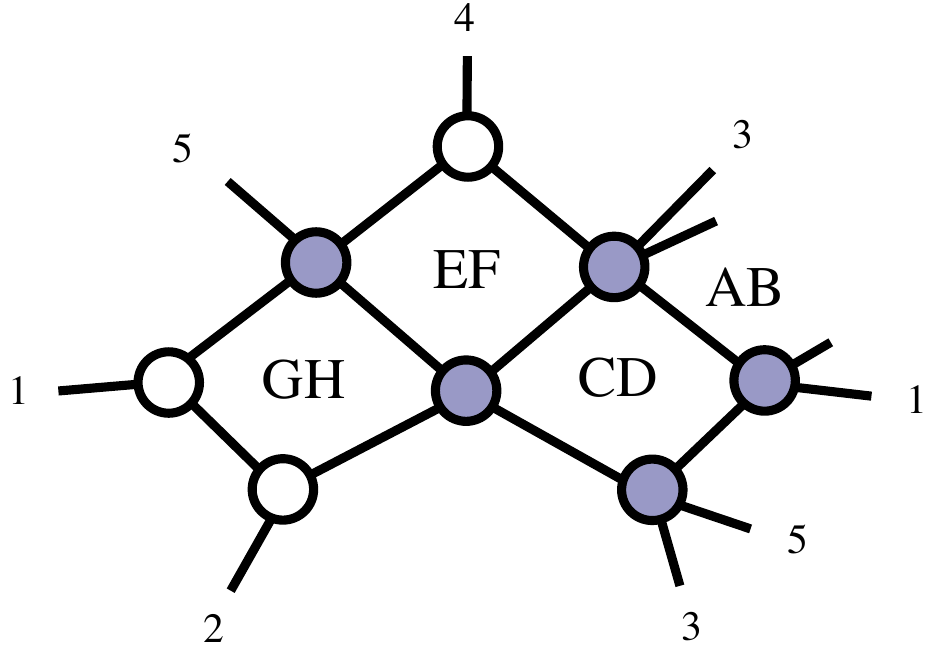}
        \caption{}
    \end{subfigure}
    \qquad
    \begin{subfigure}[t]{0.25\linewidth}
        \centering
        \includegraphics[width=\linewidth]{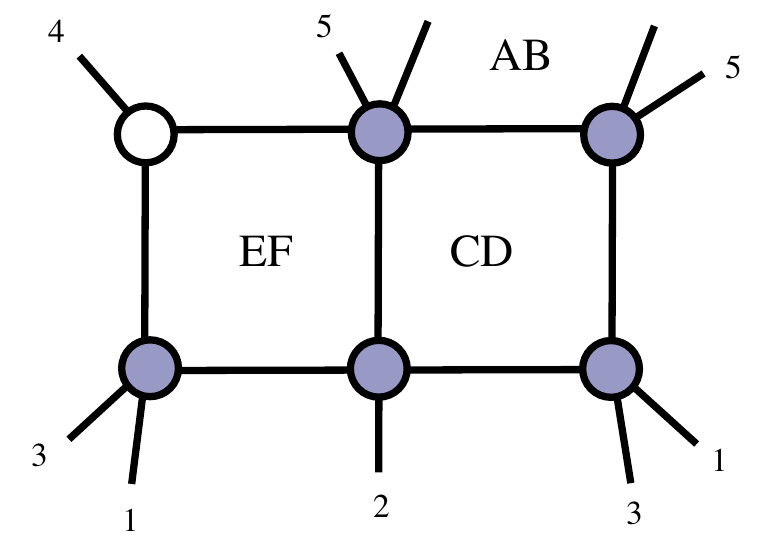}
        \caption{}
    \end{subfigure}

    \caption{Sub\(^{2}\)-leading Landau diagrams (a) and (b), and the
    equivalent two-loop non-planar topology (c) for the five-point three-loop
    ladder negative geometry yielding \eqref{eq:d1-bootstrap}.}
    \label{fig:d1diagrams}
\end{figure}

\begin{figure}[pos=t]
    \centering
    \includegraphics[width=0.4\linewidth]{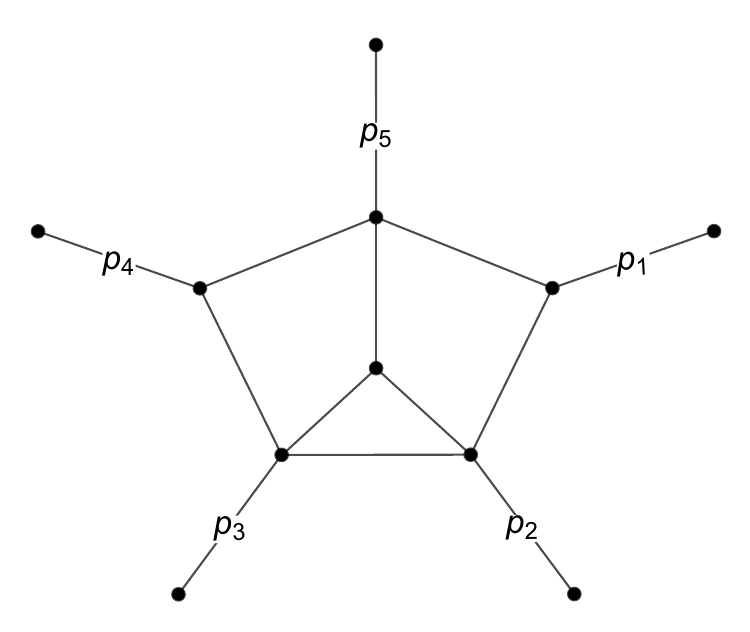}
    \caption{A three-loop planar Feynman integral yielding \(1/d_1\) as its
    leading singularity.}
    \label{fig:d1planar}
\end{figure}

At five points and three loops, the two-loop pentagon alphabet is no longer
sufficient. The geometric Landau analysis of the diagrams in
Figure~\ref{fig:d1diagrams} produces a new square root, together with its cyclic
images,
\begin{equation}
\label{eq:d1-bootstrap}
        d_1=
        \sqrt{
        (s_{12}s_{15}-s_{12}s_{23}+s_{23}s_{34}+s_{15}s_{45})^2
        -4s_{12}s_{15}s_{45}(s_{15}-s_{23})
        } \, .
\end{equation}
Recall that the Mandelstam invariants are $s_{ij}=(p_i+p_j)^2$.
This singularity appears in subtopologies of the three-loop ladder Landau
diagrams. It is absent from the two-loop ladder, as follows from the geometric
Landau analysis, but becomes relevant at three loops. The same singularity
\(d_1\) can also be obtained from planar topologies at five points and three
loops~\cite{neg_geom_pos}, for instance from the leading Landau equation
of the wheel topology in Figure~\ref{fig:d1planar}.

Applying \texttt{Effortless.m}, or equivalently a Schubert analysis as in
\cite{He:2024fij}, gives four odd letters for each cyclic image of \(d_1\).
Taking all cyclic images into account, this gives \(25\) new letters. Together
with the \(20\) two-loop pentagon letters, these form the \(45\)-letter
alphabet used for the five-point three-loop bootstrap.

\paragraph{The six-point two-loop ladder}
\label{subsubsec:bootstrap-six-point-two-loop}

We first consider the six-point two-loop ladder. Its decomposition from
\eqref{eq:neg-geo-decomposition-bootstrap} is
\begin{equation}
\label{eq:six-point-two-loop-ladder-decomp-bootstrap}
\begin{aligned}
F_{6,{\rm ladder}}^{(2)}
={}&
\Omega_6(13) f^{(2)}_{6,13}
+\Omega_6(14) f^{(2)}_{6,14}
+ \dots \, ,
\end{aligned}
\end{equation}
where the dots indicate terms related by dihedral symmetry.
The computation is therefore reduced to two independent pure functions: \(f^{(2)}_{6,13}\) and \(f^{(2)}_{6,14}\).

The starting point is the \(157\)-letter alphabet described above. Since the
integrated functions are dual conformally invariant, they depend only on
dimensionless ratios of kinematic invariants. We therefore remove one overall
scale and work with \(156\) dimensionless letters. The first entries are
restricted to the physical channels,
\begin{equation}
        \left\{
        \left[\frac{s_{i,i+1}}{s_{12}}\right],
        \left[\frac{s_{i,i+1,i+2}}{s_{12}}\right]
        \right\} \, ,
\end{equation}
where the Mandelstam invariants are
\begin{equation}
\label{eq:6ptMand}
s_{12}
=
\frac{\langle6123\rangle}
     {\langle61AB\rangle\langle23AB\rangle} \, ,
\qquad
s_{123}
=
\frac{\langle6134\rangle}
     {\langle61AB\rangle\langle34AB\rangle} \, ,
\end{equation}
together with their cyclic images. There are eight resulting independent
weight-one symbols. The last-entry condition
\eqref{eq:last-entry-DE-bootstrap} gives \(63\) allowed last entries for
\(f^{(2)}_{6,13}\) and \(58\) for \(f^{(2)}_{6,14}\). With these restrictions,
the recursive construction of integrable symbols gives the dimensions shown in
Table~\ref{tab:six-point-two-loop-bootstrap}.

\begin{table}[pos=t]
\centering
\begin{tabular}{c|c|c|c|c}
\hline
weight & 1 & 2 & 3 & 4 \\ \hline
\(\#\) symbols for \(f^{(2)}_{6,13}\) & 8 & 53 & 343 & 675 \\
\(\#\) symbols for \(f^{(2)}_{6,14}\) & 8 & 53 & 343 & 540 \\
\hline
\end{tabular}
\caption{Dimensions of the integrable symbol spaces used in the six-point
two-loop bootstrap after imposing first-entry, dimensionlessness and
last-entry constraints.}
\label{tab:six-point-two-loop-bootstrap}
\end{table}

The next condition is the cancellation of spurious poles. The leading
singularity \(\Omega_6(ij)\) has a pole at \(\langle ABij\rangle=0\), but this
pole is absent from the full ladder. Therefore the corresponding pure function
must vanish in this limit:
\begin{equation}
\label{eq:six-point-spurious-bootstrap}
        \mathcal S\!\left(f^{(2)}_{6,ij}\right)
        \Big|_{s_{ij}\to0}
        =
        0 \, .
\end{equation}
This condition reduces the number of free parameters to \(100\) for
\(f^{(2)}_{6,13}\) and \(96\) for \(f^{(2)}_{6,14}\).

The remaining freedom is fixed by the collinear limit. We use the standard
momentum-twistor parametrization~\cite{CaronHuot:2011kk} by parameters $\epsilon,\tau$:
\begin{equation}
\label{eq:six-point-collinear-bootstrap}
        Z_6\to
        Z_5
        -\frac{\langle1235\rangle}{\langle1234\rangle}\epsilon Z_4
        +\frac{\langle4523\rangle}{\langle4123\rangle}\epsilon\tau Z_1
        +\frac{\langle3451\rangle}{\langle3421\rangle}\epsilon^2 Z_2 \, ,
\end{equation}
and take \(\epsilon\to0\). The result must be finite, independent of the
auxiliary parameter \(\tau\), and equal to the known five-point two-loop
ladder. This gives five non-trivial inhomogeneous constraints. Three of them
already fix the ansatz uniquely. The reduction of the ansatz is summarized in
Table~\ref{tab:six-point-two-loop-free-parameters}.

\begin{table}[pos=t]
\centering
\begin{tabular}{l|c|c}
\hline
constraint & \(f^{(2)}_{6,13}\) & \(f^{(2)}_{6,14}\) \\ \hline
integrable weight-four ansatz & 675 & 540 \\
spurious-pole cancellation & 100 & 96 \\
collinear matching & 0 & 0 \\
\hline
\end{tabular}
\caption{Number of free parameters in the six-point two-loop bootstrap after
successive constraints. The final zero means that the symbol is uniquely
fixed.}
\label{tab:six-point-two-loop-free-parameters}
\end{table}

\begin{figure}[pos=t]
    \centering
    \begin{subfigure}[t]{0.35\linewidth}
        \centering
        \includegraphics[width=\linewidth]{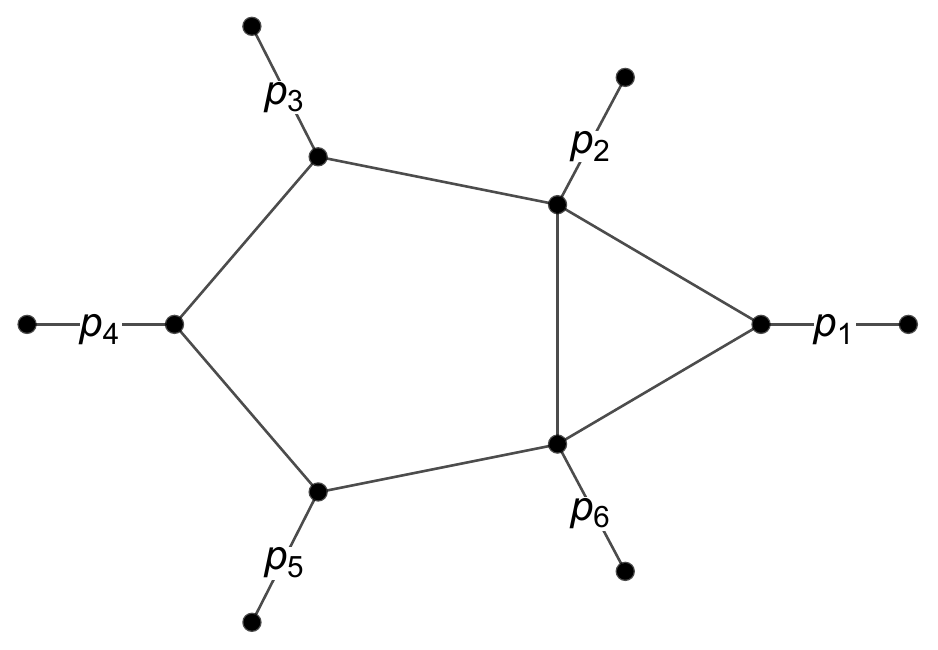}
        \caption{\(W_{102},W_{104}\)}
    \end{subfigure}
    \qquad
    \begin{subfigure}[t]{0.35\linewidth}
        \centering
        \includegraphics[width=\linewidth]{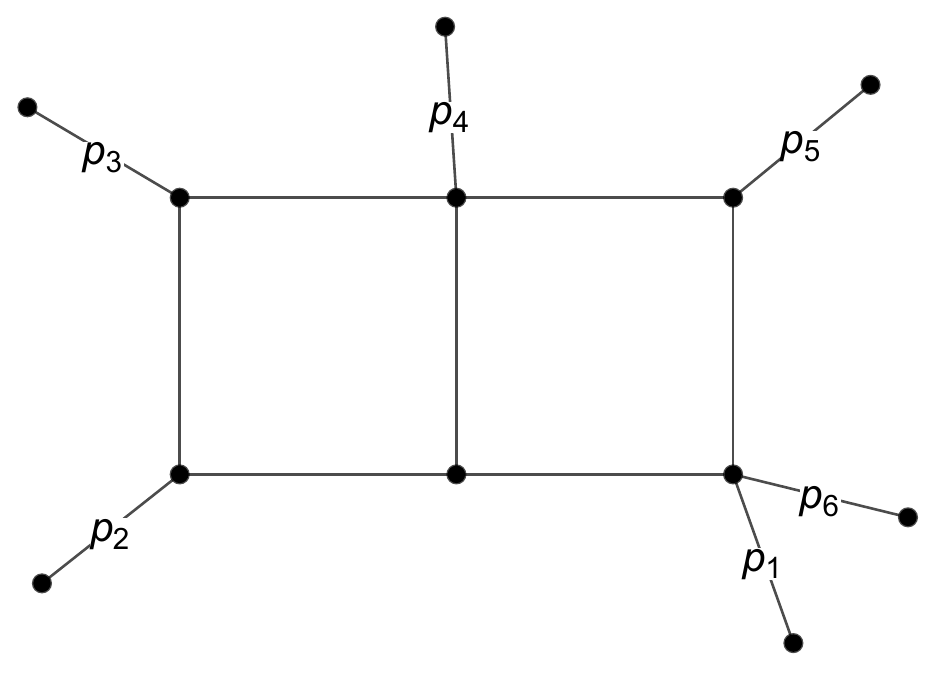}
        \caption{\(W_{81},W_{87}\)}
    \end{subfigure}

    \vspace{2mm}

    \begin{subfigure}[t]{0.35\linewidth}
        \centering
        \includegraphics[width=\linewidth]{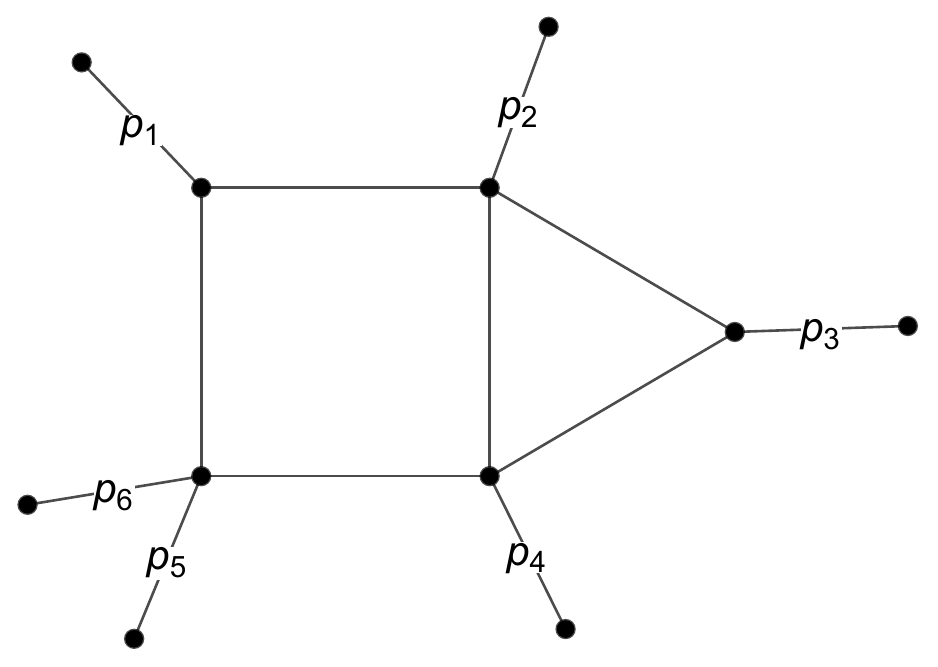}
        \caption{\(W_{58},W_{68}\)}
    \end{subfigure}
    \qquad
    \begin{subfigure}[t]{0.35\linewidth}
        \centering
        \includegraphics[width=\linewidth]{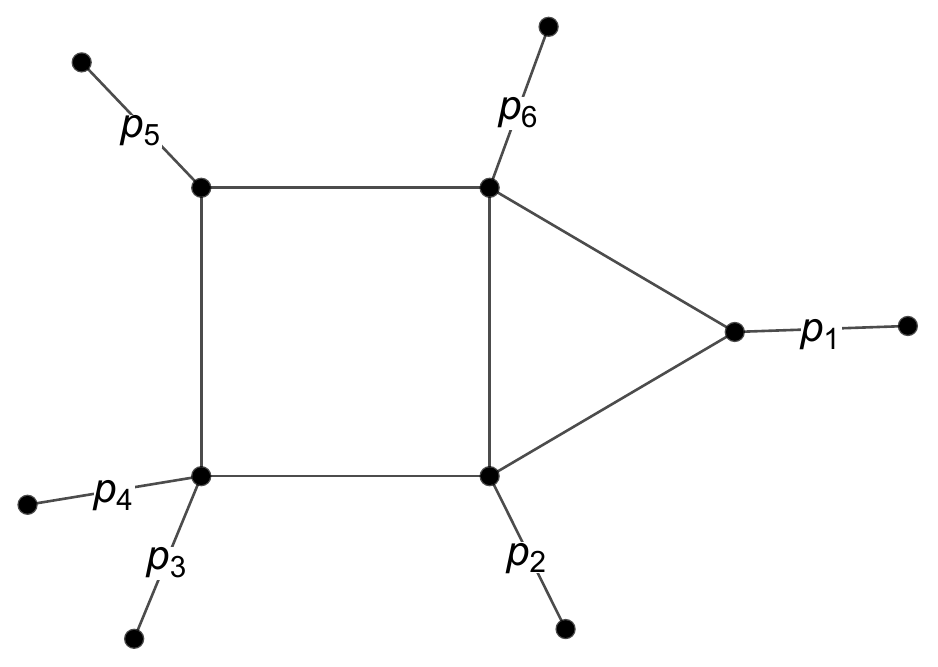}
        \caption{\(W_{60},W_{66}\)}
    \end{subfigure}

    \caption{Four planar Feynman integral sectors and their reflections
    yielding genuine two-loop singularities in \(f_{6,13}^{(2)}\).}
    \label{fig:FD6pt}
\end{figure}

The alphabets actually appearing in the final symbols are smaller than the
starting alphabet. The symbol of \(f^{(2)}_{6,13}\) contains \(84\) letters.
Among these, eight rational letters are genuine two-loop letters,
\begin{equation}
        \{W_{58},W_{60},W_{66},W_{68},
          W_{81},W_{87},W_{102},W_{104}\} \, ,
\end{equation}
in the notation of \cite{Henn:2025xrc}. These are associated with the planar
Feynman integral sectors displayed in Figure~\ref{fig:FD6pt}. By contrast,
\(f^{(2)}_{6,14}\) contains \(58\) letters, all of which are already one-loop
singularities. The final symbols have \(22\) and \(20\) independent last
entries, respectively. Both functions lie in the six-point two-loop planar
function space of \cite{Henn:2025xrc}, in agreement with the
the hexagonal Wilson-loop analysis of \cite{Carrolo:2025pue}.

\paragraph{The five-point three-loop ladder}
\label{subsubsec:bootstrap-five-point-three-loop}

We now consider the five-point three-loop ladder. The previous bootstrap based
only on the two-loop planar pentagon alphabet is not sufficient. The geometric
Landau analysis instead gives a \(45\)-letter alphabet: the \(20\) two-loop
planar pentagon letters, together with \(25\) odd letters associated with the
square roots \(\mathrm{d}_i\) and their cyclic images. Removing one overall scale, we
work with \(44\) dimensionless letters.

The leading-singularity decomposition is
\begin{equation}
\label{eq:five-point-three-loop-decomp-bootstrap}
        F_{5,{\rm ladder}}^{(3)}
        =
        \Omega_5(13)f^{(3)}_{5,13}
        +\dots \, .
\end{equation}
Cyclicity determines all terms from \(f^{(3)}_{5,13}\).
The five-point bootstrap differs from the six-point one in an important way:
lower-point limits are not sufficiently constraining. The d'Alembertian
equation must therefore be imposed more directly. The first entries are the
four physical two-particle channels,
\begin{equation}
        \left\{
        \frac{s_{i,i+1}}{s_{12}}
        \right\} \, .
\end{equation}
The last entries are restricted by \(\mathcal D_{13}\log S=0\), leaving
\(17\) allowed combinations.

The differential equation also constrains the last two entries. For a
weight-six symbol, consider the \((4,2)\) component of the coproduct,
\begin{equation}
        \mathcal S\!\left(f^{(3)}_{5,13}\right)
        =
        \sum_i S^{(i)}_4\otimes S^{(i)}_2 \, .
\end{equation}
The operator \(\mathcal D_{13}\) acts on the weight-two component. The
right-hand side of~\eqref{eq:DE-bootstrap} involves five rational functions
\(\overline{\mathcal C}_{13,kl}\), of which four are linearly independent over
\(\mathbb Q\). Let us denote these four independent functions by \(l_a\),
\(a=1,\ldots,4\). We build an ansatz for \(S^{(i)}_2\) using arbitrary first
entries from the \(44\)-letter alphabet and second entries from the \(17\)
allowed last entries. This gives \(227\) integrable weight-two symbols.
Imposing the differential equation produces four inhomogeneous solutions
\(U_a\) and \(135\) homogeneous solutions \(V_I\),
\begin{equation}
        \mathcal D_{13}U_a=l_a,
        \qquad
        \mathcal D_{13}V_I=0,
        \qquad
        a=1,\ldots,4,\quad I=1,\ldots,135 \, .
\end{equation}
Thus the weight-two part is restricted to the span of the \(U_a\) and \(V_I\).
Using this information, we recursively construct the integrable symbol space
up to weight six. The dimensions are shown in
Table~\ref{tab:five-point-three-loop-bootstrap}.
\begin{table}[pos=t]
\centering
\begin{tabular}{c|c|c|c|c|c|c}
\hline
weight & 1 & 2 & 3 & 4 & 5 & 6 \\ \hline
\(\#\) symbols & 4 & 15 & 61 & 279 & 1380 & 503 \\
\hline
\end{tabular}
\caption{Dimensions of the integrable symbol spaces used in the five-point
three-loop bootstrap after imposing the first-entry condition and the
differential-equation constraints on the last entries and last two entries.}
\label{tab:five-point-three-loop-bootstrap}
\end{table}
The full d'Alembertian equation is then imposed on the weight-six ansatz. One
finds
\begin{equation}
        \mathcal D_{13}\,
        \mathcal S\!\left(f^{(3)}_{5,13}\right)
        =
        \sum_{a=1}^4 l_a\, S^{(a)}_4 \, .
\end{equation}
Matching this expression to the known two-loop ladders on the right-hand side
of~\eqref{eq:DE-bootstrap} fixes \(407\) of the \(503\) coefficients. The
remaining \(96\) coefficients are fixed by spurious-pole cancellation as~\eqref{eq:six-point-spurious-bootstrap}. The counting of the degrees of
freedom is given in Table~\ref{tab:five-point-three-loop-free-parameters}.

\begin{table}[pos=t]
\centering
\begin{tabular}{l|c}
\hline
constraint & \(f^{(3)}_{5,13}\) \\ \hline
integrable weight-six ansatz & 503 \\
d'Alembertian equation & 96 \\
spurious-pole cancellation & 0 \\
\hline
\end{tabular}
\caption{Reduction of free parameters in the five-point three-loop bootstrap.}
\label{tab:five-point-three-loop-free-parameters}
\end{table}

Together with cyclicity, this determines the symbol of
\(F_{5,{\rm ladder}}^{(3)}\). The result has the correct soft and collinear
limits, reducing to the known four-point three-loop ladder. It also exhibits a
notable simplification. The bootstrap requires the full \(45\)-letter alphabet,
but the last entries of the final symbol involve only two-loop planar pentagon
letters. For \(f^{(3)}_{5,13}\), the \(10\) last entries are
\begin{equation}
\begin{gathered}
        \frac{W_2}{W_1},\quad
        \frac{W_4}{W_1},\quad
        \frac{W_5}{W_3},\quad
        \frac{W_{11}}{W_1},\quad
        \frac{W_{14}}{W_1},\quad
        \frac{W_{17}W_{26}}{W_3},
        \\
        \frac{W_3W_{27}}{W_{18}},\quad
        \frac{W_3W_{28}}{W_{19}},\quad
        \frac{W_{20}W_{29}}{W_3},\quad
        \frac{W_1W_{30}}{W_{16}} .
\end{gathered}
\end{equation}
The new three-loop letters do appear, but only in the middle of the symbol:
they occur in the third, fourth and fifth entries, and all \(45\) letters
appear in the fourth entry.

\subsubsection{Outlook}
\label{subsec:outlook-geometric-landau-analysis}

The results of this chapter suggest that geometric Landau analysis provides a
useful bridge between the integrand-level geometry of scattering amplitudes and
the analytic structure of the integrated functions. Starting from the negative
geometry expansion of the Wilson loop with a Lagrangian insertion, we used
boundary configurations of the relevant geometries to compute leading
singularities, construct Landau diagrams, and extract symbol alphabets. For
ladder negative geometries this strategy was strong enough to determine, at
symbol level, the six-point two-loop and five-point three-loop integrated
ladders \(F_{6,\mathrm{ladder}}^{(2)}\) and
\(F_{5,\mathrm{ladder}}^{(3)}\)
\cite{ChicherinHennMazzucchelliTrnkaYangZhang2026}. More generally, this gives
evidence that positive and negative geometries do not merely organize
integrands: they also leave a visible imprint on the singularities, alphabets,
and symbol constraints of the integrated observables.

We close with several directions in which this picture could be developed
further.

\begin{itemize}

\item \textbf{Leading singularities beyond ladders.}
Leading singularities are special contour integrals of loop integrands. After
integration, they appear as rational prefactors multiplying pure
transcendental functions. For ordinary amplitudes, leading singularities are
Yangian invariants and are associated with cells of the positive Grassmannian.
They also exhibit cluster-algebraic structures which are closely correlated
with analogous structures in the symbol alphabet
\cite{lukowski2019cluster,GurdoganParisi2020}. The leading singularities of the
Wilson loop with a Lagrangian insertion are different, but still highly
structured. For the full observable \(F_n^{(\ell)}\), they can be written in
terms of simple kermit building blocks. The hidden conformal symmetry of these
building blocks was conjectured in \cite{ChicherinHenn2022} and proven for all
\(n\) and \(\ell\) in
\cite{ChicherinHennMazzucchelliTrnkaYangZhang2026}.

For individual negative geometries the situation is subtler: the leading
singularities are generally more complicated and need not be conformal. It
would be useful to classify them in families, such as tree graphs or graphs
with bounded valence, and to bound the complexity of the line configurations,
the number of independent values, and the denominators of the resulting
rational functions. Such a classification could make the integrated expansion
of \(F_n^{(\ell)}\) more systematic.

\item \textbf{Geometric Landau analysis for individual pure functions.}
In the ladder examples studied above, the full alphabet predicted by geometric
Landau analysis is often larger than the alphabet actually used by a single
pure function. For instance, the six-point two-loop ladder starts from a
\(157\)-letter alphabet, but the final symbols of \(f^{(2)}_{6,13}\) and
\(f^{(2)}_{6,14}\) involve much smaller subsets. This suggests that there may
exist additional geometric selection rules for the alphabet of an individual
transcendental piece. A natural question is whether each summand
\begin{equation}
        \Omega_{n,s}\, f^{(\ell)}_{n,s}
\end{equation}
admits its own positive or negative geometry whose canonical form produces
precisely the integrand associated with that summand. If so, one could apply
geometric Landau analysis directly to the individual pure function rather than
to the larger observable. This would refine the symbol alphabet, reduce the
bootstrap ansatz, and perhaps explain the observed sparsity of the final
symbols. This problem is already interesting for ladders, where the integrand
is known explicitly, and should be a useful testing ground for a more targeted
geometric bootstrap.

\item \textbf{Beyond MHV.}
Another important direction is to extend negative geometries beyond MHV degree. For MHV
amplitudes, the relation between the logarithm of the amplitude, the Wilson
loop with a Lagrangian insertion, and the negative-geometry expansion is
relatively well developed. For NMHV and higher sectors, the analogous relation
is much less clear. A possible starting point is to define a negative geometry
in which each loop line lives in the NMHV one-loop Amplituhedron, while the
mutual negativity conditions are modified to
\begin{equation}
        \langle YABCD\rangle<0 \, .
\end{equation}
The canonical form of such a geometry should be a supersymmetric function. If
its integral is infrared finite, it could be compared with supersymmetric
Wilson-loop computations. This would test whether the geometric Landau
analysis developed here extends naturally to non-MHV observables.

\item \textbf{General negative geometries and elliptic cuts.}
The ladder cases considered in this chapter remain polylogarithmic at symbol
level. For more complicated negative geometries, however, Landau diagrams may
contain elliptic subtopologies. This raises the question of whether individual
integrated negative geometries can involve elliptic or higher-genus analogues
of multiple polylogarithms. If such cuts survive integration, then the ordinary
symbol alphabet described in this chapter would have to be replaced by an
elliptic or more general coaction/symbol structure
\cite{Broedel:2018iwv,Kristensson:2021ani}.

A first example is the three-loop triangle with one handle in
Figure~\ref{triangle_LS}. At five points its leading Landau diagram contains an
elliptic subtopology after a degeneration, shown in Figure~\ref{fig:elliptic}.
It remains unclear whether this elliptic cut appears in the integrated negative
geometry itself or cancels in the full observable.

\begin{figure}[pos=t]
\centering
\begin{tikzpicture}[baseline=(current bounding box.south),scale=0.8]

    \draw[mybrown, line width=2pt] (0,0) -- (3,0) -- (6,1.5)--(6,-1.5)--(3,0);

    \draw[fill=white] (0,0) circle (4pt);
    \draw (0,0) -- (45:4pt);
    \draw (0,0) -- (135:4pt);
    \draw (0,0) -- (-45:4pt);
    \draw (0,0) -- (-135:4pt);

    \filldraw (3,0) circle (4pt);
    \filldraw (6,1.5) circle (4pt);
    \filldraw (6,-1.5) circle (4pt);
    \node[anchor=north] at (0,-0.1) {\small{$AB$}};
    \node[anchor=north] at (3,-0.1) {\small{$CD$}};
    \node[anchor=south] at (6,1.6) {\small{$EF$}};
    \node[anchor=north] at (6,-1.6) {\small{$GH$}};
\end{tikzpicture}
\quad\quad\quad\quad
\begin{tikzpicture}[scale = 0.45]

    \begin{scope}[xshift=0cm]
    \coordinate (i) at (1,-2);
    \coordinate (P) at (0.3,0.05);
    \coordinate (Q) at (-0.45,2.25);
    \coordinate (j) at (-2,2);
    \coordinate (k) at (1,2.5);
    \coordinate (C) at (-0.7,3);
    \coordinate (D) at (1.35,-3);
    \coordinate (E) at (-3.5,1.75);
    \coordinate (F) at (3,2.85);
    \coordinate (A) at (-0.5,-0.1);
    \coordinate (B) at (2.5,0.5);
    \coordinate (G) at (-2.5,-2.5);
    \coordinate (H) at (-0.15,3);
    \coordinate (l) at (-2.3,-2);

    \draw[red, thick] (C) -- (D);
    \node[above, red] at (C) {$CD$};

    \draw[red, thick] (E) -- (F);
    \node[above, red] at (E) {$EF$};

    \draw[red, thick] (G) -- (H);
    \node[left, red] at (G) {$GH$};

    \draw[teal, thick] (A) -- (B);
    \node[above, teal] at (B) {$AB$};

    \fill (i) circle (2pt);
    \fill (j) circle (2pt);
    \fill (k) circle (2pt);
    \fill (P) circle (2pt);
    \fill (Q) circle (2pt);
    \fill (l) circle (2pt);

    \node[above right] at (i) {$k$};
    \node[above] at (j) {$i$};
    \node[above] at (k) {$j$};
    \node[right] at (l) {$l$};
    \end{scope}

\end{tikzpicture}
\caption{The graph associated to the three-loop triangle with one handle (left) and
one representative leading-singularity line configuration (right).}
\label{triangle_LS}
\end{figure}
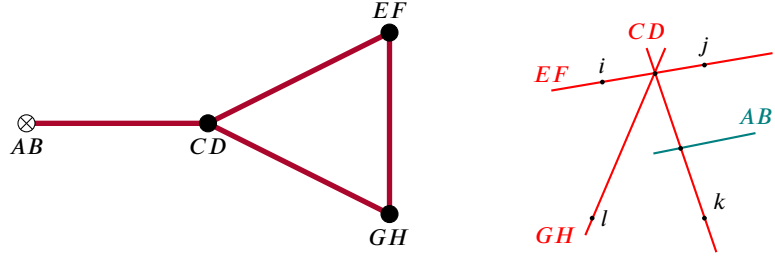

\begin{figure}[pos=t]
    \centering
    \begin{subfigure}[t]{0.35\linewidth}
        \centering
        \includegraphics[width=\linewidth]{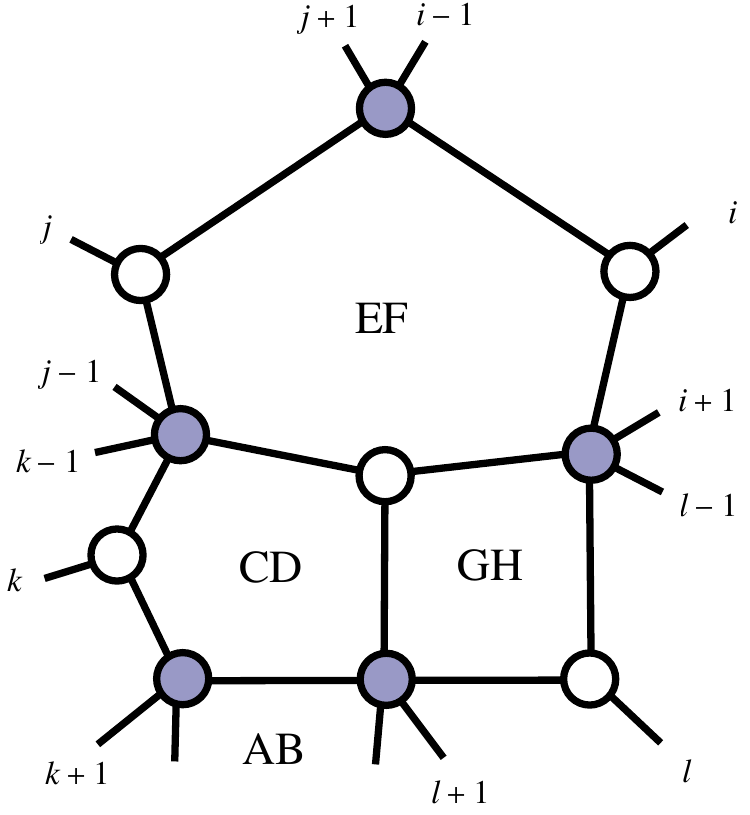}
        \caption{}
        \label{fig:T3ILandau}
    \end{subfigure}
    \qquad
    \begin{subfigure}[t]{0.35\linewidth}
        \centering
        \includegraphics[width=\linewidth]{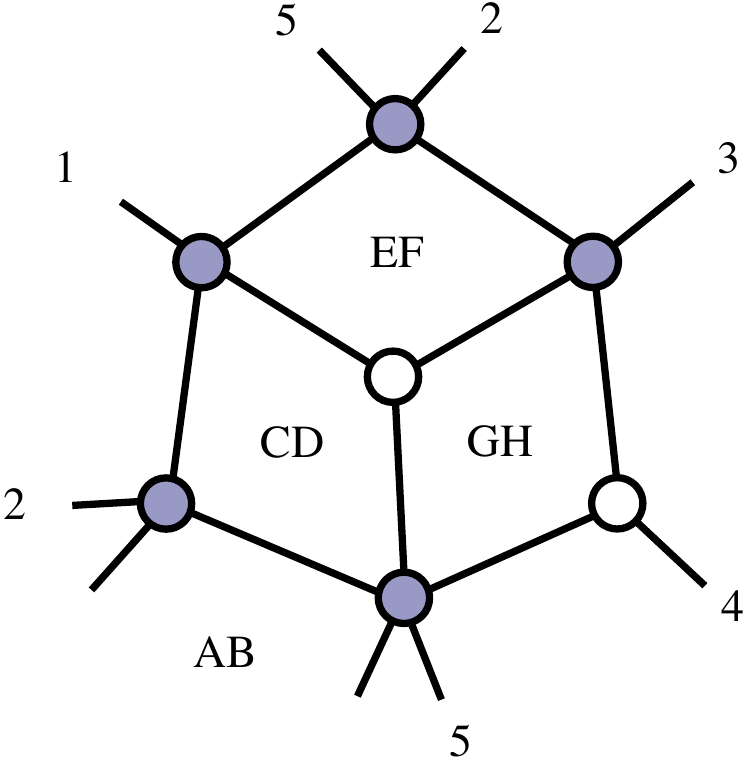}
        \caption{}
        \label{fig:elliptcishrink}
    \end{subfigure}

    \caption{A leading Landau diagram for the three-loop triangle with one
    handle, shown in Figure~\ref{fig:T3ILandau}, and its elliptic
    subtopology at \(n=5\), shown in Figure~\ref{fig:elliptcishrink}.}
    \label{fig:elliptic}
\end{figure}

\item \textbf{Toward a geometric bootstrap.}
The computations in this chapter used geometric Landau analysis as input to the
symbol bootstrap: geometry supplied the alphabet, while integrability,
first-entry conditions, the d'Alembertian equation, spurious-pole cancellation,
and physical limits fixed the symbol. A more ambitious geometric bootstrap
would let the same positive or negative geometry control the alphabet, allowed
words, adjacency relations, and boundary data. This is closely related to the
cluster structures discussed next: in favorable cases, cluster adjacency and
Landau compatibility appear to be two faces of one geometry.

\end{itemize}

%% file: Ch7.tex

\section{Positivity and cluster structures}
\label{ch:positivity-and-cluster-structures}

The previous chapter studied how positive and negative geometries constrain loop-level integrals. Starting from rational canonical forms, we saw that integration produces
multivalued functions whose singularities are encoded by Landau loci, symbol alphabets,
and algebraic covers. Geometric Landau analysis uses the boundary structure of the
underlying geometry to refine the set of possible singularities, and in favorable cases
this information is strong enough to bootstrap integrated functions.

In this chapter we study the singularities themselves. The guiding question is why the
singularities produced by Landau analysis in planar kinematics often have unexpectedly
rigid algebraic and positive structure. This question sits at the intersection of
several developments in planar \(\mathcal N=4\) SYM. Classical Landau analysis describes
possible branch loci of Feynman integrals through pinch singularities of integration
contours
\cite{Landau1959,BjorkenLandau1959,Nakanishi1959,Nakanishi1971,Pham1967,ColemanNorton1965}. In planar kinematics, momentum twistors turn loop
momenta into lines in \(\mathbb P^3\), and propagator equations into incidence
conditions. This made it possible to apply Landau analysis directly in momentum-twistor
variables and to compare the resulting loci with symbol alphabets of amplitudes
\cite{DennenSpradlinVolovich2016}. The Amplituhedron then suggested a
geometric refinement: branch loci should be read from boundary configurations of the
relevant positive geometry
\cite{DennenPrlinaSpradlinStanojevicVolovich2017,PrlinaSpradlinStanojevic2018,PrlinaSpradlinStankowiczStanojevic2018}.

At the same time, a different set of observations pointed to cluster algebras. Symbol
alphabets of planar \(\mathcal N=4\) SYM amplitudes are often built from cluster
coordinates of Grassmannians, and their symbols obey adjacency constraints closely
related to Steinmann-type restrictions on sequential discontinuities
\cite{GoldenGoncharovSpradlinVerguVolovich2014,DrummondFosterGurdoganClusterAdjacency2018,DrummondFosterGurdoganHarrington2019,DrummondFosterGurdoganKalousios2020,CaronHuotDixonDulatEtAl2020}. Related cluster patterns also appear in leading
singularities and Landau singularities
\cite{lukowski2019cluster,GurdoganParisi2020}. These observations suggest that the
cluster structure of amplitudes is not only a property of final integrated functions,
but is already visible in the geometry of their cuts and singular loci.

The aim of this chapter is to explain a mechanism behind this phenomenon. We reformulate
Landau analysis as the algebraic geometry of line configurations in \(\mathbb P^3\). A
Landau diagram determines an incidence variety in a product of Grassmannians; projecting
this variety to the space of external kinematics produces the corresponding Landau
singularity. Leading singularities are controlled by finite maps and their
discriminants, while more degenerate, or superleading, singularities are described by
resultants. This gives a coordinate-free way to study the algebraic complexity of Landau
singularities.

This Grassmannian formulation makes positivity visible. The external kinematics of
planar \(\mathcal N=4\) SYM has a distinguished positive region, and the Amplituhedron
is built from positive external data. It is therefore natural to ask whether Landau
singularities are regular on this region, and whether the recursive operations producing
large families of diagrams preserve positivity. For important families, such as trees of
loops, the answer is governed by explicit recursive substitution maps. These maps are
closely related to positroids and to promotion maps in Grassmannian cluster algebras.

The results reviewed in this chapter are based on the Grassmannian formulation of Landau
analysis and its applications to positivity and cluster structures developed in
\cite{HolleringMazzucchelliParisiSturmfels2025Lines,HolleringMazzucchelliParisiSturmfels2026,HolleringMazzucchelliParisiSturmfels2026Positivity}.
Figures in this chapter are adapted from those references unless explicitly indicated
otherwise.
They provide the final step in the singularity story of this thesis: after
Chapter~\ref{ch:From Integrands to Integrals} explained how singularities enter
integrated functions and symbol alphabets, we study why these singularities
themselves exhibit positivity, rationality, and cluster factorization.

\medskip

\noindent
Section~\ref{sec:landau-analysis-on-the-grassmannian} formulates Landau analysis in
momentum-twistor and Grassmannian language. We explain how loop momenta become lines in
\(\mathbb P^3\), how propagator equations become incidence conditions, and how
line-incidence varieties project to Landau singularities in external kinematic space. We
also introduce leading singularity degrees, leading discriminants, and superleading
resultants.

Section~\ref{sec:recursive-landau-analysis} develops recursive Landau analysis. We
explain how some families of Landau diagrams can be built by gluing box subdiagrams, and how
this operation induces explicit substitutions on the external data. We then apply this
recursion to trees of loops, obtaining families of singularities at arbitrary loop
order.

Section~\ref{sec:reality-positivity-positroids} studies the behavior of these recursive
constructions on the positive region. We introduce positive box substitutions, prove
positivity properties for trees of loops, and relate the resulting line configurations
to positroid cells and to the Amplituhedron map.

Section~\ref{sec:rationality-cluster-structures} turns to rationality and cluster
structures. We first discuss rational Landau singularities and rational degenerations,
then recall the cluster-algebraic structures relevant for planar kinematics. Finally, we
describe cluster promotion maps and state the cluster conjecture for trees of loops.

Section~\ref{sec:positivity-cluster-outlook} closes with open questions, including the
extension of these methods beyond trees of loops, the role of higher-degree covers and
non-rational singularities, and the relation between Grassmannian Landau analysis and
symbol alphabets of full amplitudes.

\subsection{Landau analysis on the Grassmannian}
\label{sec:landau-analysis-on-the-grassmannian}

\subsubsection{From momenta to lines in three-space}
\label{subsec:From Momenta to Lines in Three-space}

The choice of coordinates used to parametrize kinematic space is often crucial for
making formulae simple and structural properties manifest. We first saw this in
Chapter~\ref{ch:Scattering Amplitudes}, where spinor-helicity variables encode massless spin
states and lead to the famously compact Parke--Taylor formula for the \(n\)-point
tree-level color-ordered MHV gluon amplitude in Yang--Mills theory~\cite{Parke:1986gb}.
At the same time, this formula is the canonical function of a positive Grassmannian
geometry, realizing a first connection between scattering amplitudes and positive geometry.

An even more geometric simplification comes from passing from spinor-helicity variables
to twistors and momentum twistors. These variables encode momenta and dual momenta in
four dimensions by projective geometry. In planar kinematics, dual points are
represented by lines in projective three-space, and propagator conditions become
incidence relations. This projective viewpoint is especially well suited to Landau
analysis. Ordinary Minkowski space is non-compact, and singularities at infinite loop
momentum have to be treated separately in the usual affine formulation. These are the
so-called second-type Landau singularities. After compactification, finite and infinite
regions of momentum space are treated uniformly inside projective space. Thus the
projective formulation naturally incorporates the boundary at infinity into the same
algebraic framework as the ordinary propagator singularities.

Before specializing to momentum twistors, let us recall the embedding-space version of
this construction. This gives a useful projective compactification of dual momentum
space. We first work in \(D\)-dimensional Minkowski space
\(\mathbb R^{1,D-1}\), equipped with the metric
\begin{equation}
        \eta=\operatorname{diag}(1,-1,\ldots,-1) \, .
\end{equation}
For planar graphs, we introduce dual momentum coordinates \(x_i \in \mathbb R^{1,D-1}\) by
\begin{equation}
        p_i=x_{i+1}-x_i,
        \qquad i=1,\ldots,n \, ,
\end{equation}
with indices understood cyclically. The external mass is
\begin{equation}
        p_i^2=m_i^2 \, .
\end{equation}
We embed a dual point \(x_i\), together with a face-mass parameter
\(\delta_i\) whose meaning we clarify shortly, into projective embedding space
\(\mathbb P^{D+1}\) by
\begin{equation}
\label{eq:embedding}
        (x_i^\mu,\delta_i)
        \longmapsto
        X_i=[X_i^\mu:X_i^+:X_i^-]
        =
        [x_i^\mu:x_i^2+\delta_i^2:1] \, ,
\end{equation}
where \(\mu=0,1,\dots,D-1\). The point \(X_i\) denotes the equivalence class defined up
to an overall non-zero rescaling. As usual, this construction is first performed over
the real numbers, and later over the complex numbers. The real embedding space is
equipped with the bilinear form
\begin{equation}
\label{eq:embedding-inner-product}
        ( X,Y )
        :=
        -2X^\mu Y^\nu\eta_{\mu\nu}
        +X^+Y^-+X^-Y^+ \, ,
\end{equation}
of signature \((2,D)\). With this convention one finds
\begin{equation}
\label{eq:embedding-distance}
        (X_i,X_j)
        =
        (x_i-x_j)^2+\delta_i^2+\delta_j^2 \, .
\end{equation}
In particular,
\begin{equation}
        (X_i,X_i)=2\delta_i^2 \, ,
        \qquad
        (X_i,X_{i+1})
        =
        p_i^2+\delta_i^2+\delta_{i+1}^2
        =
        m_i^2+\delta_i^2+\delta_{i+1}^2 \, .
\end{equation}
Thus the ordinary massless dual points are recovered by setting
\(\delta_i=0\), in which case \(X_i\) lies on the projective null cone. Note
that, because the \(X_i\) are defined projectively, for every massive \(X_i\) with
\(\delta_i\neq0\) we may choose the normalization \((X_i,X_i)=1\).

The same construction applies to dual momenta of loop variables. Consider an
\(n\)-point \(\ell\)-loop planar Feynman diagram. The faces of the graph are
labelled by dual momenta. External faces carry the points \(x_i\) for
\(i=1,\dots,d\), while internal faces carry dual loop points \(y_a\), with
\(a=1,\ldots,\ell\). We embed the latter as
\begin{equation}
        Y_a=[y_a^\mu:y_a^2+\epsilon_a^2:1],
        \qquad
        (Y_a,Y_a)=2\epsilon_a^2 \, .
\end{equation}
An edge separating an external face \(x_i\) from an internal face \(y_a\) corresponds to
the propagator condition
\begin{equation}
\label{eq:external-internal-propagator-embedding}
        (X_i,Y_a)
        =
        (x_i-y_a)^2+\delta_i^2+\epsilon_a^2 \, .
\end{equation}
Similarly, an edge separating two internal faces \(y_a\) and \(y_b\) gives
\begin{equation}
\label{eq:internal-internal-propagator-embedding}
        (Y_a,Y_b)
        =
        (y_a-y_b)^2+\epsilon_a^2+\epsilon_b^2 \, .
\end{equation}
Thus this embedding formalism naturally describes propagators whose squared masses are
induced by face parameters:
\begin{equation}
\label{eq:face-mass-propagators}
        M_{ia}^2=\delta_i^2+\epsilon_a^2 \, ,
        \qquad
        \widetilde M_{ab}^2=\epsilon_a^2+\epsilon_b^2 \, .
\end{equation}
Here \(M_{ia}\) is the mass of the edge separating the external face \(i\) and the loop
face \(a\), while \(\widetilde M_{ab}\) is the mass of the edge separating two loop
faces. This makes clear both the usefulness and the limitation of the construction: the
masses are not arbitrary, but satisfy relations dictated
by~\eqref{eq:face-mass-propagators} and by the form of the graph.

The embedding~\eqref{eq:embedding} is also not completely canonical. It depends on the
choice of an affine chart in the projective compactification, or equivalently on the
choice of a point, or dually a hyperplane, at infinity. Let
\begin{equation}
        I_\infty=[0:\ldots:0:1:0]\in\mathbb P^{D+1}
\end{equation}
be such a point. With the bilinear form~\eqref{eq:embedding-inner-product}, the affine
chart is the open set
\begin{equation}
        (X,I_\infty)\neq0 \, .
\end{equation}
In the normalization used in~\eqref{eq:embedding}, we have
\((X_i,I_\infty)=1\). Thus ordinary dual momentum space is recovered as the
open chart obtained by fixing the scale with respect to \(I_\infty\). Points satisfying
\((X,I_\infty)=0\) lie at infinity.

A planar Feynman integral can therefore be rewritten in embedding space as an integral
over projective loop variables \(Y_a\), one for each internal face of the dual graph. In
the face-mass embedding these variables satisfy
\((Y_a,Y_a)=2\epsilon_a^2\); in the massless case they lie on the projective
null cone. Schematically, an affine \(\ell\)-loop scalar integral of the form
\begin{equation}
        \int
        \prod_{a=1}^{\ell} d^D y_a\,
        \frac{N(y,x)}
        {\prod_e q_e(y,x)}
\end{equation}
is replaced by a projective integral of the form
\begin{equation}
\label{eq:projective-embedding-integral}
        \int
        \prod_{a=1}^{\ell} \omega(Y_a)\,
        \frac{N(Y,X,I_\infty)}
        {
        \prod_{(i,a) \, \in \,  G_{\rm ext}} (X_i,Y_a)
        \prod_{(a,b) \, \in \,  G_{\rm int}} (Y_a,Y_b)
        \prod_{a=1}^{\ell} (Y_a,I_\infty)^{\nu_a}
        } \, .
\end{equation}
Here \(G_{\rm ext}\) denotes edges separating an external face from an internal face,
\(G_{\rm int}\) denotes edges separating two internal faces, and
\(\omega(Y_a)\) is the projective measure on compactified dual momentum space.
The displayed formula is schematic: the numerator and the powers \(\nu_a\) are so
that the integrand has homogeneous degree zero in each projective variable \(Y_a\).

If this degree balance holds with \(\nu_a=0\) and with numerator independent of
\(I_\infty\), then the integral is intrinsically projective and does not depend on the
choice of point at infinity. This is the case for dual-conformal integrals. If the
degree balance requires factors of
\((Y_a,I_\infty)\), then the integral depends explicitly on the chosen affine
chart. The hyperplane
\begin{equation}
        (Y_a,I_\infty)=0
\end{equation}
is then part of the divisor arrangement relevant for Landau analysis. Pinches involving
this divisor are the projective incarnation of second-type Landau singularities,
associated in affine coordinates with infinite loop momentum. Examples of such Feynman
diagrams at one loop are scalar bubbles, triangles and pentagons with constant
numerators.

As the discussion of the point at infinity suggests, the embedding-space formalism is
especially natural for conformal, and in particular dual-conformal, integrals. The
reason is that the conformal group of
\(D\)-dimensional spacetime acts linearly on the projective embedding space:
ordinary conformal transformations are realized as the action of
\(\operatorname{SO}(2,D)\) on the null cone modulo rescalings
\cite{Dirac1936,Costa:2011mg}. In this language, conformal covariance becomes
ordinary projective homogeneity. This perspective is particularly well suited to planar
\(\mathcal N=4\) SYM, where loop integrands exhibit dual conformal symmetry
\cite{Drummond:2008vq,Drummond:2007aua}. In massive or Higgs-regulated versions of the
theory, the same symmetry can be made manifest by extending dual coordinates by mass
variables, as in the embedding-space description of the Coulomb branch and its relation
to the AdS/CFT picture
\cite{Alday:2009zm}.

The embedding-space discussion explains why projective compactification is the natural
setting. We now specialize to the massless four-dimensional case, where this
compactification has an especially concrete realization in terms of lines in \(\mathbb
P^3\). In the massless planar case relevant for ordinary planar
\(\mathcal N=4\) SYM, all face masses vanish,
\begin{equation}
        \delta_i=0 \, ,
        \qquad
        \epsilon_a=0 \, .
\end{equation}
Thus all embedded points lie on the projective null cone and all propagator conditions
reduce to projective incidence conditions. In four dimensions, the complexified
conformal compactification is the Klein quadric: over
\(\mathbb C\), the projective null cone in \(\mathbb P^5\) is identified with
\(\Gr(2,4)\), the space of lines in \(\mathbb P^3\). Thus each
dual point \(x_i\), or loop dual point \(y_a\), is represented by a line,
\begin{equation}
        M_i\subseteq\mathbb P^3,
        \qquad
        L_a\subseteq\mathbb P^3 \, .
\end{equation}
The external lines \(M_i\) are the usual momentum-twistor lines
\((Z_iZ_{i+1})\), but we use the notation \(M_i\) in this chapter to treat
external and loop lines uniformly.

This embedding is given by mapping dual momenta
\(x_i=(x_i^0,x_i^1,x_i^2,x_i^3)\) and \(y_a\) to the row spans of the
\(2\times4\) matrices
\begin{equation}
\label{eq:getLM}
M_i \,\, = \,\, \begin{pmatrix}
1 & 0 & x_i^0+x_i^3 & x_i^1 - $i$ \, x_i^2 \\
0 & 1 & x_i^1+ $i$ \, x_i^2 & x_i^0 - x_i^3
\end{pmatrix}
\quad {\rm and} \quad L_a \,\, = \,\, \begin{pmatrix}
1 & 0 & y_a^0+y_a^3 & y_a^1 - $i$ \, y_a^2 \\
0 & 1 & y_a^1+ $i$ \, y_a^2 & y_a^0 - y_a^3
\end{pmatrix} \, .
\end{equation}
The key property is as follows: the determinant of any \(4\times4\) matrix obtained by
stacking two of the \(2\times4\) matrices vanishes if and only if the two lines
intersect in \(\mathbb P^3\). This in turn is equivalent to the vanishing of the
corresponding massless propagator. Concretely,
\begin{equation}
        \langle L_a M_i\rangle
        :=
        \det
        \begin{pmatrix}
        L_a\\
        M_i
        \end{pmatrix}
        =
        (x_i-y_a)^2
        =
        (x_i^0-y_a^0)^2
        -(x_i^1-y_a^1)^2
        -(x_i^2-y_a^2)^2
        -(x_i^3-y_a^3)^2 \, ,
\end{equation}
and similarly \(\langle L_a L_b\rangle=(y_a-y_b)^2\).

The embedding~\eqref{eq:getLM} also depends on the choice of an affine chart of
\(\Gr(2,4)\), equivalently on a line at infinity. For the chart used
above, this line can be taken to be
\begin{equation}
        I_\infty=
        \begin{pmatrix}
        0 & 0 & 1 & 0\\
        0 & 0 & 0 & 1
        \end{pmatrix} \, .
\end{equation}
Complexified Minkowski space is identified with the affine chart
\begin{equation}
        \{L\in\Gr(2,4)\mid
        \langle L\,I_\infty\rangle\neq0\} \, .
\end{equation}
The divisor \(\langle L\,I_\infty\rangle=0\) is the boundary at infinity.

A general planar Feynman graph with massless propagators can then be written in
momentum-twistor variables as
\begin{equation}
     \label{eq:LMintegral}
        \mathcal{I}(\mathbf{M})\,\,=\int_\Gamma \frac{N(\mathbf{L};\mathbf{M})}{D(\mathbf{L};\mathbf{M})} \, {\rm d}\mu(\mathbf{L}) \, .
    \end{equation}
This is now a function of the \(d\) external lines
\(\mathbf{M}=(M_1,\ldots,M_d)\in\Gr(2,4)^d\). One integrates
over the \(\ell\) loop lines
\(\mathbf{L}=(L_1,\ldots,L_\ell)\in\Gr(2,4)^\ell\), along a
cycle \(\Gamma\) representing the Minkowski contour with the Feynman prescription. The
denominator has the form
\begin{equation}
\label{eq:momentum-twistor-denominator-general}
        D(\mathbf L;\mathbf M)
        =
        \prod_{(i,a) \, \in \, G_{\rm ext}}
        \langle L_a M_i\rangle
        \prod_{(a,b) \, \in \, G_{\rm int}}
        \langle L_a L_b\rangle \, ,
\end{equation}
possibly multiplied, in non-dual-conformal cases, by factors involving the line at
infinity, such as \(\langle L_a I_\infty\rangle\). In what follows we will be concerned
with integrals that have dual conformal symmetry, so their integrands are independent of
\(I_\infty\). However, most results can be extended to the
non-dual-conformally-invariant setting by considering
\(I_\infty\) as part of the external data, namely as an additional line among
the \(M_i\). The only subtlety is that \(I_\infty\) can appear simultaneously in several
factors \(\langle L_a I_\infty\rangle\), so the resulting external line configurations
are not generic from the point of view of the incidence varieties below.

\begin{figure}[pos=t]
    \centering
\begin{tikzpicture}[scale = .8, every node/.style={font=\small}]

    \begin{scope}[very thick,decoration={markings,mark=at position 0.25 with {\arrow{>}}, mark = at position .75 with {\arrow{>}}}] 
        \draw[postaction={decorate}, black, thick] (-2.5, 2)--(2.5, 2);
        \draw[postaction={decorate}, black, thick] (4, 0)--(1.5, -3);
        \draw[postaction={decorate}, black, thick] (-1.5, -3)--(-4, 0);
    \end{scope}

    \begin{scope}[very thick,decoration={markings,mark=at position 0.5 with {\arrow{>}}}] 
    
        \draw[postaction={decorate}, ->, black, thick] (0,0)--(0,3);
        \draw[postaction={decorate}, ->, black, thick] (0,0)--(3.7,-2.3);
        \draw[postaction={decorate}, ->, black, thick] (0,0)--(-3.7,-2.3);
        \draw[postaction={decorate}, black, thick] (2.5, 2)--(4, 0);
        \draw[postaction={decorate}, black, thick] (1.5, -3)--(-1.5,-3);
        \draw[postaction={decorate}, black, thick] (-4, 0)--(-2.5, 2);
    \end{scope}

    \draw[->, black, thick] (-2.5, 2)--(-3.3, 2.8);
    \draw[->, black, thick] (2.5, 2)--(3.3, 2.8);
    \draw[->, black, thick] (-4, 0)--(-5,0);
    \draw[->, black, thick] (4, 0)--(5,0);
    \draw[->, black, thick] (-1.5, -3)--(-2, -4);
    \draw[->, black, thick] (1.5, -3)--(2, -4);

    \node at (-1.8, .8) {$y_1$};
    \node at (1.8, .8) {$y_2$};
    \node at (0, -1.7) {$y_3$};
    \node at (-4, -1) {$x_1$};
    \node at (4, -1) {$x_6$};
    \node at (-4, 1.1) {$x_2$};
    \node at (4, 1.1) {$x_5$};
    \node at (-1.5, 2.5) {$x_3$};
    \node at (1.5, 2.5) {$x_4$};
    \node at (2.6, -2.8) {$x_7$};
    \node at (-2.6, -2.8) {$x_9$};
    \node at (0, -3.5) {$x_8$};
\end{tikzpicture}
\begin{tikzpicture}
\node at (0,0) {};
\begin{scope}[shift = {(0, .5)}]
    

    \draw[black, thick] (2, 3)--(4, 3);
    \draw[black, thick] (2, 3)--(3, 1.5);
    \draw[black, thick] (4, 3)--(3, 1.5);

    \draw[black, thick] (2, 3)--(1, 3);
    \draw[black, thick] (2, 3)--(1.3, 3.7);
    \draw[black, thick] (2, 3)--(2, 4);

    \draw[black, thick] (4, 3)--(4, 4);
    \draw[black, thick] (4, 3)--(5, 3);
    \draw[black, thick] (4, 3)--(4.7, 3.7);

    \draw[black, thick] (3, 1.5)--(4, .7);
    \draw[black, thick] (3, 1.5)--(2, .7);
    \draw[black, thick] (3, 1.5)--(3, .6);
    
    \node at (1.85, 2.70) {$L_1$};
    \node at (4.20, 2.70) {$L_2$};
    \node at (3.43, 1.55) {$L_3$};

    \node at (2, 4.15) {$M_3$};
    \node at (.75, 3) {$M_1$};
    \node at (1.10, 3.9) {$M_2$};

    \node at (4, 4.15) {$M_4$};
    \node at (5.25, 3) {$M_6$};
    \node at (4.9, 3.9) {$M_5$};

    \node at (4.25, .55) {$M_7$};
    \node at (3, .4) {$M_8$};
    \node at (1.75, .55) {$M_9$};

    \end{scope}

\end{tikzpicture}
\caption{The triple pentagon labeled by dual variables (left).
The dual graph labeled by momentum twistors (right).
These are lines in \(3\)-space with prescribed incidences.}
\label{fig:TP2}
\end{figure}
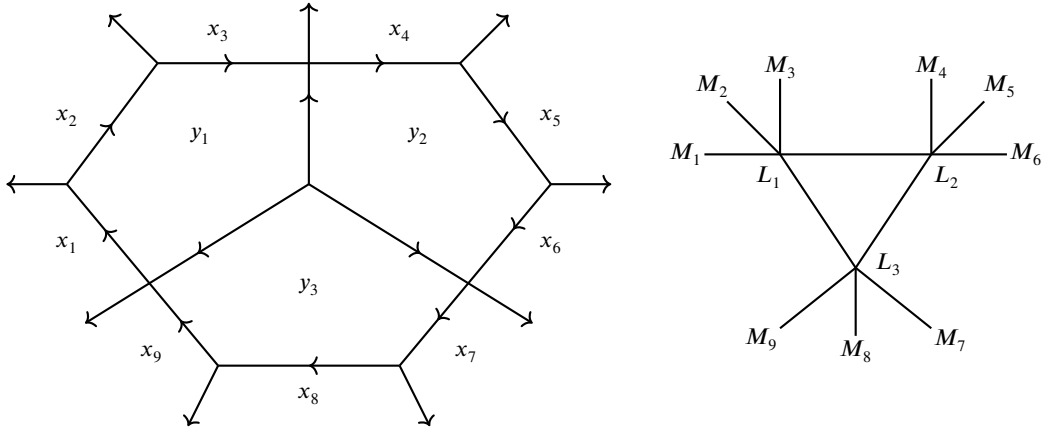

\begin{eg}[Triple pentagon]
\label{eg:triple-pentagon-momentum-twistors}
Consider the scalar Feynman diagram given by the triple pentagon on the left
in Figure~\ref{fig:TP2}. It is a three-loop diagram, whose dual loop momenta
\(y_1,y_2,y_3\) correspond to internal regions of the diagram, and whose
external dual momenta \(x_1,\ldots,x_9\) correspond to the nine external
regions. Passing to momentum twistors gives three loop lines
\(L_1,L_2,L_3\) and nine external lines \(M_1,\ldots,M_9\). The twelve
propagators become the twelve incidence brackets
{\small
\begin{equation}\label{eq:nine_prop}
\begin{aligned}
&
\langle L_1M_1\rangle
\langle L_1M_2\rangle
\langle L_1M_3\rangle
\langle L_2M_4\rangle
\langle L_2M_5\rangle
\langle L_2M_6\rangle
\\[-0.2em]
&\qquad\qquad\times
\langle L_3M_7\rangle
\langle L_3M_8\rangle
\langle L_3M_9\rangle
\langle L_1L_2\rangle
\langle L_1L_3\rangle
\langle L_2L_3\rangle \, .
\end{aligned}
\end{equation}
}
Each factor vanishes exactly when the corresponding pair of lines intersects
in \(\mathbb P^3\). The graph encoding the possible incidences is the planar
dual to the original Feynman diagram, and is depicted on the right of
Figure~\ref{fig:TP2}.
\end{eg}

As discussed above, the integral in
\eqref{eq:LMintegral} may be divergent as written. In planar
\(\mathcal N=4\) SYM, such divergences are infrared in origin and arise from
soft or collinear regions of loop momentum
\cite{Alday:2007hr,DrummondHennKorchemskySokatchev2008}. One therefore
usually defines amplitudes with a regulator, or studies infrared-finite quantities such
as ratios and finite remainders
\cite{BernDixonSmirnov2005,Drummond:2007aua}.

The Landau analysis below should be understood in this regulated or algebraic sense. It
identifies candidate singular loci of the integrand family, and these loci are largely
independent of the particular regulator. Dimensional regularization changes the
dimension and local exponents of the integral, but the basic Landau equations are still
governed by propagator vanishing and pinch conditions
\cite{Landau1959,Nakanishi1971,Eden:1966dnq,Smirnov2012}. A fully geometric formulation
of this statement, especially in the presence of infrared and second-type singularities,
remains subtle and constitutes an important open problem.

Following the momentum-twistor approach to Landau analysis
\cite{DennenSpradlinVolovich2016,DennenPrlinaSpradlinStanojevicVolovich2017,PrlinaSpradlinStanojevic2018}, a Landau diagram specifies a stratum of the
vanishing locus of the denominator~\eqref{eq:momentum-twistor-denominator-general}. In
momentum-twistor variables, this means that a collection of pairs of lines is required
to intersect. Thus the on-shell spaces relevant for Landau analysis are precisely
incidence varieties of line configurations in \(\mathbb P^3\). The graph encoding these
incidences has vertices corresponding to the lines
\(L_a\) and \(M_i\), and edges corresponding to a subset of factors in
\eqref{eq:momentum-twistor-denominator-general}. The graph obtained by
considering all factors is the dual graph to the original Feynman diagram. By slight
abuse of notation, we will often refer to this dual graph itself as the
Landau diagram.

\begin{tcolorbox}[resultbox]
\textbf{On-shell spaces as incidence varieties.}
In momentum-twistor variables, loop and external dual points become lines
\(L_a,M_i\subseteq\mathbb P^3\), respectively, and propagator conditions become
incidence equations:
\begin{equation}\label{eq:incid}
        \langle L_a M_i\rangle=0
        \quad\text{and}\quad
        \langle L_a L_b\rangle=0 \, .
\end{equation}
Thus Landau diagrams specify graphs of incidence conditions among lines, and
the corresponding on-shell spaces are varieties of line configurations in
\(\mathbb P^3\) subject to the corresponding incidences.
\end{tcolorbox}

This picture is also naturally connected to the Amplituhedron. As reviewed in
Chapter~\ref{ch:Amplituhedra}, for fixed multiplicity \(n\), loop order \(\ell\), and
helicity degree \(k\), the planar \(\mathcal N=4\) SYM loop integrand is encoded by the
canonical form of the \(\ell\)-loop Amplituhedron, and the integrated amplitude is
obtained by integrating this form over the loop variables. As emphasized in
Section~\ref{sec:Recursion Relations}, the Landau geometry used in this
chapter is formulated at this four-dimensional integrand level; regulator-dependent
information relevant after integration is not part of the incidence geometry
itself. In the MHV case \(k=0\), the
loop variables are lines \(L_a=AB_a\subseteq\mathbb P^3\), for \(a=1,\ldots,\ell\),
while the external dual points are the momentum-twistor lines \(M_i=(Z_iZ_{i+1})\), for
\(i=1,\ldots,n\). The basic boundary divisors of the loop Amplituhedron include then precisely the line-incidence equations appearing
in~\eqref{eq:momentum-twistor-denominator-general}. For \(k>0\), the same description
holds after projecting through the fixed \(k\)-plane \(Y\), as in the definition of the
\(m=4\) Amplituhedron; see Section~\ref{sec:Loop Amplituhedra}. Thus line-incidence varieties
associated with graphs of such boundary conditions naturally appear in the
stratification of the algebraic boundary of loop Amplituhedra. From the Amplituhedron
side, these varieties organize boundary strata of the canonical form and hence the
possible singular loci of the integrand. From the Landau-analysis side, the same
varieties describe the on-shell spaces obtained by setting propagators to zero.
Understanding their dimension, components, and degenerations is therefore a necessary
step toward understanding both the structure of the integrand numerator
in~\eqref{eq:LMintegral} and the Landau singularities of the corresponding integrated
functions.

The following subsection is devoted to the geometric and algebraic properties of these
incidence varieties of lines in \(\mathbb P^3\). These varieties are the intrinsic
on-shell spaces underlying Landau analysis in momentum-twistor variables and appear
naturally in the algebraic boundary of loop Amplituhedra.

\subsubsection{Incidence varieties of line configurations}
\label{subsec:incidence-varieties-line-configurations}

We now isolate the geometric object that underlies the momentum-twistor formulation of
Landau analysis: configurations of lines in projective three-space satisfying prescribed
incidence relations. The results summarized here are part
of~\cite{HolleringMazzucchelliParisiSturmfels2025Lines}.

A line
\(X\subseteq\mathbb P^3\) is a point of the Grassmannian
\(\Gr(2,4)\). We represent it by a \(2\times4\) matrix
\(\mathbf X\), whose two rows span the corresponding two-dimensional subspace
of \(\mathbb C^4\). Its Plücker coordinates are the \(2\times2\) minors of
\(\mathbf X\):
\begin{equation}
        X=[x_{12}:x_{13}:x_{14}:x_{23}:x_{24}:x_{34}]
        \in\mathbb P^5 \, .
\end{equation}
These coordinates satisfy the Plücker quadric
\begin{equation}
\label{eq:chapter7-plucker-quadric}
        \langle X\, X \rangle
        :=
        x_{12}x_{34}-x_{13}x_{24}+x_{14}x_{23}
        =
        0 \, .
\end{equation}
If \(X\) and \(Y\) are two lines, represented by \(2\times4\) matrices
\(\mathbf X\) and \(\mathbf Y\), then they intersect in \(\mathbb P^3\) if
and only if the \(4\times4\) matrix obtained by stacking \(\mathbf X\) and
\(\mathbf Y\) has determinant zero. In Plücker coordinates this determinant
is the bilinear form
\begin{equation}
\label{eq:chapter7-line-incidence-pairing}
\begin{aligned}
        \langle X\, Y \rangle
        &:=
        \det
        \begin{pmatrix}
        \mathbf X\\
        \mathbf Y
        \end{pmatrix}
        =
        x_{12}y_{34}
        -x_{13}y_{24}
        +x_{14}y_{23}
        +x_{23}y_{14}
        -x_{24}y_{13}
        +x_{34}y_{12} \, .
\end{aligned}
\end{equation}
Therefore,
\begin{equation}
        \langle X\, Y \rangle=0
        \qquad\Longleftrightarrow\qquad
        X\cap Y\neq\varnothing \, .
\end{equation}

\paragraph{Incidence varieties associated with graphs}

We now study configurations of many lines. Fix a graph \(G\) with vertex set
$[\ell]:=\{1,\ldots,\ell\}$. We identify $G$ with its set of edges. The vertex \(i\)
labels a line \(X_i\subseteq\mathbb P^3\), and an edge
\(ij\in G\) imposes the incidence condition \(\langle X_i \, X_j \rangle =0\). This defines a projective variety in the product of Grassmannians $\Gr(2,4)^\ell$.
Let $\mathbb C[X_1,\ldots,X_\ell]$ denote the polynomial ring in the \(6\ell\) Plücker
coordinates of the \(\ell\) lines. We always work modulo the Plücker equations $\langle
X_i  X_i \rangle =0$ for every $i=1,\dots,\ell$. Then the graph \(G\) defines the ideal
$I_G$ in the ring $\mathbb C[X_1,\ldots,X_\ell]$ modulo the Plücker relations, generated
by $\langle X_i \, X_j \rangle $ for every $ij \in G$. The vanishing locus of this ideal
defines the following variety.

\begin{tcolorbox}[definitionbox]
\textbf{Incidence variety.}
The \emph{incidence variety} associated with a graph \(G\) is
\begin{equation}
\label{eq:chapter7-incidence-variety}
        V_G
        =
        \left\{
        (X_1,\ldots,X_\ell)\in \Gr(2,4)^\ell
        \;:\;
        \langle X_i \, X_j \rangle =0 \text{ for every } ij\in G
        \right\} \, .
\end{equation}
It parametrizes configurations of \(\ell\) lines in \(\mathbb P^3\) whose
incidences contain the graph \(G\).
\end{tcolorbox}
\noindent

The phrase ``contain the graph \(G\)'' is important. The equations defining
\(V_G\) impose incidences along the edges of \(G\), but they do not forbid
additional incidences. A point of \(V_G\) may also satisfy \(\langle X_i \, X_j \rangle
=0\) for a non-edge \(ij\notin G\). In fact, for any graph $G'$ containing $G$ we have an inclusion $V_{G'}
\subseteq V_G$. If we want the incidence graph to be exactly \(G\), we must remove these
extra incidence loci. Algebraically, one considers the
\textit{realization} $W_G$ of $G$, defined as the Zariski closure of the configurations whose incidence graph is exactly \(G\). As we will see, for some $G$ this space may be empty.

For Landau analysis, \(V_G\) is the algebraic on-shell space obtained by imposing the
cut equations associated with a graph. The realization \(W_G\) is the component relevant
to configurations with no extra incidences. Extra components of \(V_G\) correspond to
degenerations of the cut equations and can lead to additional singular loci. Thus the
natural objects for Landau analysis are the realization spaces over all subgraphs $G$ of
the graph encoding the propagator factors in the denominator. In principle, a full
Landau analysis should consider a stratification of the full polar locus of the
integrand, as in the Whitney-stratified approach to Landau
singularities~\cite{HelmerPapathanasiouTellander2024}. For instance, if the spaces
\(W_G\) are singular, as they generally are, then their singular loci should be treated
as separate strata.

We now discuss the first examples. They already illustrate two phenomena that make
incidence varieties nontrivial: reducibility and non-reduced scheme structure.

\begin{eg}[Three incident lines]
\label{eg:chapter7-three-lines}
Let \(G=K_3\), the complete graph on three vertices. Then \(V_{K_3}\)
parametrizes triples of pairwise intersecting lines in \(\mathbb P^3\).
There are two basic ways for three lines in projective three-space to be
pairwise incident: they can be concurrent, namely pass through a common point, or coplanar, namely lie in a common plane. 
Thus the incidence variety has two irreducible components,
\begin{equation}
\label{eq:chapter7-K3-decomposition}
        V_{K_3}
        =
        V_{[3]}\cup V^*_{[3]} \, ,
\end{equation}
where \(V_{[3]}\) is the component of concurrent triples, while \(V^*_{[3]}\) is the component of coplanar triples. Note that both components lie in the realization $W_{K_3}$, so $V_{K_3}=W_{K_3}$. The two components are exchanged by projective duality, where the latter exchanges points with planes by taking orthogonal complements, and acts in coordinates~\eqref{eq:chapter7-plucker-quadric} as
\begin{equation}
\label{eq:hodgestar} [x_{12}: x_{13}: x_{14}: x_{23}: x_{24}: x_{34} ]
\quad \mapsto \quad [x_{34}: -x_{24}: x_{23}: x_{14} :- x_{13}: x_{12} ] \, . 
\end{equation}
The decomposition in~\eqref{eq:chapter7-K3-decomposition} can be checked in \texttt{Macaulay2} by calling
\begin{verbatim}
R = QQ[x12,x13,x14,x23,x24,x34,
       y12,y13,y14,y23,y24,y34,
       z12,z13,z14,z23,z24,z34];

XX = x12*x34 - x13*x24 + x14*x23;
YY = y12*y34 - y13*y24 + y14*y23;
ZZ = z12*z34 - z13*z24 + z14*z23;

XY = x12*y34 - x13*y24 + x14*y23
   + x23*y14 - x24*y13 + x34*y12;
XZ = x12*z34 - x13*z24 + x14*z23
   + x23*z14 - x24*z13 + x34*z12;
YZ = y12*z34 - y13*z24 + y14*z23
   + y23*z14 - y24*z13 + y34*z12;

I = ideal(XX,YY,ZZ,XY,XZ,YZ);
decompose I
\end{verbatim}
where the equations $XX$ are the Plücker relations~\eqref{eq:chapter7-plucker-quadric} and the incidence equations $XY$ are those in~\eqref{eq:chapter7-line-incidence-pairing}.
\end{eg}

For three lines, the ideal generated by the three pairwise incidence equations is
\textit{radical} but not \textit{prime}. The two primes correspond to the two components
in~\eqref{eq:chapter7-K3-decomposition}. The defining ideals of these components have
additional generators beyond the pairwise incidence quadrics; geometrically, these extra
equations enforce the stronger condition of concurrence or coplanarity.

The next example shows that even a small graph can have several components.

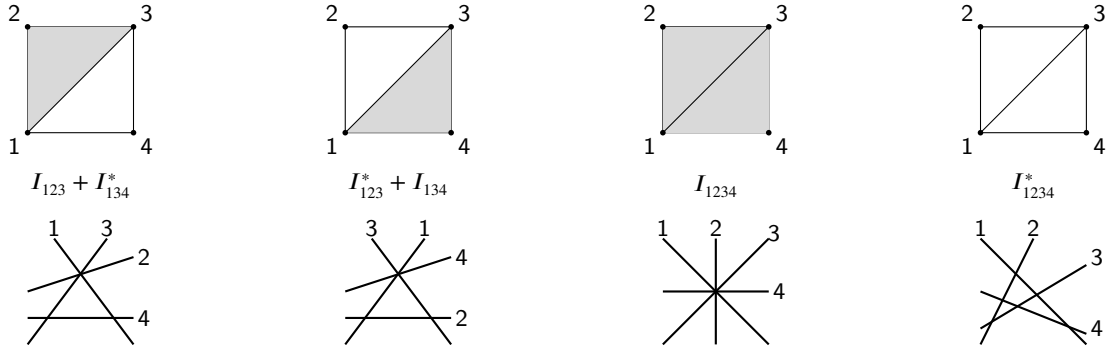
\begin{figure}[pos=t]
	\centering
\begin{tikzpicture}[scale = .7, every node/.style={font=\small}]

    \def\colsep{6} 
    
    \def\dotradius{1.5pt} 
    \def\labelgap{0.25} 
    
    \draw (0,0) rectangle (2,2);
    \fill[gray!30] (0,2) -- (2,2) -- (0,0) -- cycle;
    \foreach \x/\y in {0/0,0/2,2/0,2/2} {
        \fill[black] (\x,\y) circle (\dotradius);
    }
    
    \node at (-\labelgap,2+\labelgap) {2};
    \node at (2+\labelgap,2+\labelgap) {3};
    \node at (2+\labelgap,-\labelgap) {4};
    \node at (-\labelgap,-\labelgap) {1};
    \draw (0,0) -- (2,2);
    
    \node at (1,-1.0) {\(I_{123} + I^*_{134}\)};

    \begin{scope}[shift = {(0, -4)}]
    \draw[black, thick] (0, 0) -- (1.5,2);
    \draw[black, thick] (2, 0) -- (.5,2);
    \draw[black, thick] (0,.5) -- (2, .5);
    \draw[black, thick] (0, 1) -- (2, 1.65);
    
    \node at (1.5, 2.2) {3};
    \node at (.5, 2.2) {1};
    \node at (2.2, .5) {4};
    \node at (2.2, 1.65) {2};
    
    \end{scope}

    \draw (\colsep,0) rectangle (\colsep+2,2);
    \fill[gray!30] (\colsep + 0,0) -- (\colsep+2,0) -- (\colsep+2,2) -- cycle;
    \foreach \x/\y in {0/0,0/2,2/0,2/2} {
        \fill[black] (\colsep+\x,\y) circle (\dotradius);
    }
    
    \node at (\colsep-\labelgap,2+\labelgap) {2};
    \node at (\colsep+2+\labelgap,2+\labelgap) {3};
    \node at (\colsep+2+\labelgap,-\labelgap) {4};
    \node at (\colsep-\labelgap,-\labelgap) {1};
    \draw (\colsep,0) -- (\colsep+2,2);
    
    \node at (\colsep+1,-1.0) {\(I^*_{123} + I_{134}\)};
    
    \begin{scope}[shift = {(\colsep, -4)}]
    \draw[black, thick] (0, 0) -- (1.5,2);
    \draw[black, thick] (2, 0) -- (.5,2);
    \draw[black, thick] (0,.5) -- (2, .5);
    \draw[black, thick] (0, 1) -- (2, 1.65);
    
    \node at (1.5, 2.2) {1};
    \node at (.5, 2.2) {3};
    \node at (2.2, .5) {2};
    \node at (2.2, 1.65) {4};
    
    \end{scope}

    \draw (2*\colsep,0) rectangle (2*\colsep+2,2);
    \fill[gray!30] (2*\colsep,0) rectangle (2*\colsep+2,2);
    \foreach \x/\y in {0/0,0/2,2/0,2/2} {
        \fill[black] (2*\colsep+\x,\y) circle (\dotradius);
    }
    \node at (2*\colsep-\labelgap,2+\labelgap) {2};
    \node at (2*\colsep+2+\labelgap,2+\labelgap) {3};
    \node at (2*\colsep+2+\labelgap,-\labelgap) {4};
    \node at (2*\colsep-\labelgap,-\labelgap) {1};
    \draw (2*\colsep,0) -- (2*\colsep+2,2);
    
    \node at (2*\colsep+1,-1.0) {\(I_{1234}\)};
    
    \begin{scope}[shift = {(2*\colsep, -4)}]
    \draw[black, thick] (0, 0) -- (2,2);
    \draw[black, thick] (2, 0) -- (0,2);
    \draw[black, thick] (0, 1) -- (2, 1);
    \draw[black, thick] (1, 2) -- (1, 0);
    
    \node at (0, 2.2) {1};
    \node at (1, 2.2) {2};
    \node at (2.1, 2.1) {3};
    \node at (2.2, 1.0) {4};
    
    \end{scope}

    \draw (3*\colsep,0) rectangle (3*\colsep+2,2);
    \foreach \x/\y in {0/0,0/2,2/0,2/2} {
        \fill[black] (3*\colsep+\x,\y) circle (\dotradius);
    }
    \node at (3*\colsep-\labelgap,2+\labelgap) {2};
    \node at (3*\colsep+2+\labelgap,2+\labelgap) {3};
    \node at (3*\colsep+2+\labelgap,-\labelgap) {4};
    \node at (3*\colsep-\labelgap,-\labelgap) {1};
    \draw (3*\colsep,0) -- (3*\colsep+2,2);
    
    \node at (3*\colsep+1,-1.0) {\(I^*_{1234}\)};
    
    \begin{scope}[shift = {(3*\colsep, -4)}]
    \draw[black, thick] (0, 2) -- (2,0);
    \draw[black, thick] (1, 2) -- (0,0);
    \draw[black, thick] (0, 0.3) -- (2, 1.5);
    \draw[black, thick] (0,1) -- (2, .2);
    
    \node at (1, 2.2) {2};
    \node at (0, 2.2) {1};
    \node at (2.2, 1.65) {3};
    \node at (2.2, .3) {4};

    \end{scope}

\end{tikzpicture}
\caption{The four irreducible components in the incidence variety
for the graph $G$ in Example \ref{eg:chapter7-triangulated-quadrilateral}.
The last row shows the configurations of four lines given by each component.}
\label{figure:fourlines}
\end{figure}

\begin{eg}[A triangulated quadrilateral]
\label{eg:chapter7-triangulated-quadrilateral}
Let \(G\) be the complete graph \(K_4\) on four vertices but with the edge \(\{2,4\}\) removed.
Equivalently, \(G\) is a quadrilateral with one diagonal. It contains two
triangles, \(\{1,2,3\}\) and \(\{1,3,4\}\), sharing the edge \(\{1,3\}\).
Each triangle of pairwise incident lines can be either concurrent or
coplanar. Thus the two triangles can be colored in two possible ways: black
for concurrent and white for coplanar. We warn the reader that this is the convention followed in~\cite{HolleringMazzucchelliParisiSturmfels2025Lines}, but it is opposite to the one used previously in connection to Landau diagrams and on-shell graphs.

 Since there are two triangles, one
obtains four natural components:
\begin{equation}
\label{eq:chapter7-four-components}
        V_G
        =
        \left( V_{123}\cap V_{134} \right)
        \;\cup\;
        \left(V^*_{123}\cap V^*_{134}\right)
        \;\cup\;
        \left(V_{123}\cap V^*_{134}\right)
        \;\cup\;
        \left(V^*_{123}\cap V_{134}\right) \, .
\end{equation}
This is illustrated in Figure~\ref{figure:fourlines}.
The last two components are the realization \(W_G\): in these configurations
the missing edge \(\{2,4\}\) is generically not an incidence.
\end{eg}

Complete graphs give another important family. For \(G=K_\ell\), the variety
\(V_G\) contains \(\ell\)-tuples of pairwise intersecting lines. For all
\(\ell\geq 3\), it contains the component of concurrent lines $V_{[\ell]}$ and the
component of coplanar $V^*_{[\ell]}$ lines. For \(\ell\geq4\), the scheme structure
becomes subtler: the ideal
\(I_{K_\ell}\) need not be radical, and embedded associated primes can appear. 
Specifically, for $\ell=4$, a computation reveals the minimal primary decomposition:
\begin{equation}
\label{eq:IK4}
 I_{K_4} \,\, = \,\, I_{[4]} \,\,\cap\, \, I^*_{[4]}\, \,\cap\, \, \bigl(\,I_{K_4} + (I_{[4]})^2 + (I^*_{[4]})^2 \,\bigr) \, . 
 \end{equation}
The prime ideals $I_{[\ell]}$ and $I^*_{[\ell]}$ and their varieties serve as basic
building blocks in the following.

It is an interesting question to understand how and if the non-trivial scheme structure
is relevant for the Landau analysis; can multiplicities and embedded components affect
how degenerations are organized, even when the final singular locus is studied
set-theoretically?

\paragraph{Dimension and the rigidity-theory viewpoint}

A naïve first guess would be that every edge of \(G\) imposes one independent condition.
Since \(\Gr(2,4)\) has dimension \(4\), the ambient space
\(\Gr(2,4)^\ell\) has dimension \(4\ell\). Thus one might expect the
codimension of $V_G$ in \(\Gr(2,4)^\ell\) to be equal to the number of
edges of $G$, $|G|$. This is true for many sparse graphs, but it fails when the
incidence conditions are dependent. For example,
\begin{equation}\label{eq:dim_CG}
	\dim(V_{K_\ell}) = 2 \ell + 3 \, ,
\end{equation}
so the codimension is $2\ell-3$, even though $|K_\ell|=\binom{\ell}{2}$. In fact, for
$\ell \geq 3$,
\begin{equation}
	V_{K_\ell} = V_{[\ell]} \cup V_{[\ell]}^* \, ,
\end{equation}
generalizing~\eqref{eq:chapter7-K3-decomposition}. Then, a configuration of $\ell$
concurrent lines has $2\ell+3$ degrees of freedom, since one can pick the concurrency
point $p$ in $\mathbb{P}^3$, which gives 3 degrees of freedom, and any other $\ell$
points in the orthogonal complement of $p$, giving $2\ell$ extra degrees of freedom.

The correct notion of independence has a beautiful connection to the \textit{rigidity
theory} of graphs in the plane. To see this connection we move to an affine chart of
\(\Gr(2,4)\), so that
\begin{equation}
\label{eq:chapter7-affine-line}
        \mathbf X_i
        =
        \begin{pmatrix}
        1 & 0 & \alpha^{(i)}_1 & \alpha^{(i)}_2\\
        0 & 1 & \alpha^{(i)}_3 & \alpha^{(i)}_4
        \end{pmatrix} \implies \langle X_i \, X_j \rangle
        =
        \det
        \begin{pmatrix}
        \alpha^{(i)}_1-\alpha^{(j)}_1
        &
        \alpha^{(i)}_2-\alpha^{(j)}_2
        \\
        \alpha^{(i)}_3-\alpha^{(j)}_3
        &
        \alpha^{(i)}_4-\alpha^{(j)}_4
        \end{pmatrix} \, .
\end{equation}
Now make a linear change of variables which identifies the four affine coordinates
$\alpha^{(i)}_1,\dots,\alpha^{(i)}_4$ of a line $X_i$ with a pair of points
$(s^{(i)},t^{(i)})\in\mathbb C^2 \times \mathbb C^2$:
\begin{equation}
\label{eq:matricesST1}  \alpha_1^{(i)} = s_2^{(i)} - t_2^{(i)}, \quad \alpha_2^{(i)} = t_1^{(i)} - s_1^{(i)},  \quad \alpha_3^{(i)} = s_1^{(i)} + t_1^{(i)} , \quad  \alpha_4^{(i)} = s_2^{(i)} + t_2^{(i)}  \, ,
\end{equation}
In these coordinates, the incidence condition~\eqref{eq:chapter7-affine-line}
becomes
\begin{equation}
\label{eq:chapter7-distance-equality}
        \|s^{(i)}-s^{(j)}\|^2=\|t^{(i)}-t^{(j)}\|^2 \, ,
\end{equation}
where $\|\cdot\|$ denotes the standard norm on $\mathbb{R}^2$. Thus a configuration of
lines satisfying incidences along \(G\) is equivalent, on this affine chart, to a pair
of planar point configurations.

This is the Elekes--Sharir framework~\cite{ES,Raz,Martin1}. Consider the map
\begin{equation}
\label{eq:chapter7-distance-map}
        \delta_G:
        (\mathbb C^2)^\ell
        \longrightarrow
        \mathbb C^{|G|} , \quad s \mapsto \big(
        \|s_i-s_j\|^2
        \big)_{ij\in G} \, .
\end{equation}
Then, on the affine chart,
\begin{equation}
\label{eq:chapter7-VG-distance-fiber-product}
        V_G
        =
        \{(s,t)\in(\mathbb C^2)^\ell\times(\mathbb C^2)^\ell
        \;:\;
        \delta_G(s)=\delta_G(t)\} \, .
\end{equation}
In other words, \(V_G\) is a fiber product of two copies of the distance map
in~\eqref{eq:chapter7-distance-map}. The image and fibers of the map $\delta_G$ are
central objects in rigidity theory \cite{MS,SS}: the dimension of the image of
$\delta_G$ is equal to the \emph{rigidity rank} of $G$. The advantage of the
rigidity-theory viewpoint is that it gives a combinatorial test for when the incidence
equations should be regarded as independent. The columns of the Jacobian matrix $J_\ell$
of $\delta_{K_\ell}$ define the \textit{rigidity matroid} $\mathcal{R}_\ell$.
Informally, a matroid is a combinatorial structure that abstracts the notion of linear
independence: it remembers which subsets of a given set behave like independent sets of
vectors. We can than regard $G$ as an element of this matroid. Independency, in the
matroid sense, of $G$ in $\mathcal{R}_\ell$ is thus related to independence of the
incidence conditions encoded by its edges, and hence on the dimension of $V_G$.

\begin{figure}[pos=t]
	\centering
	\includegraphics[scale = .35]{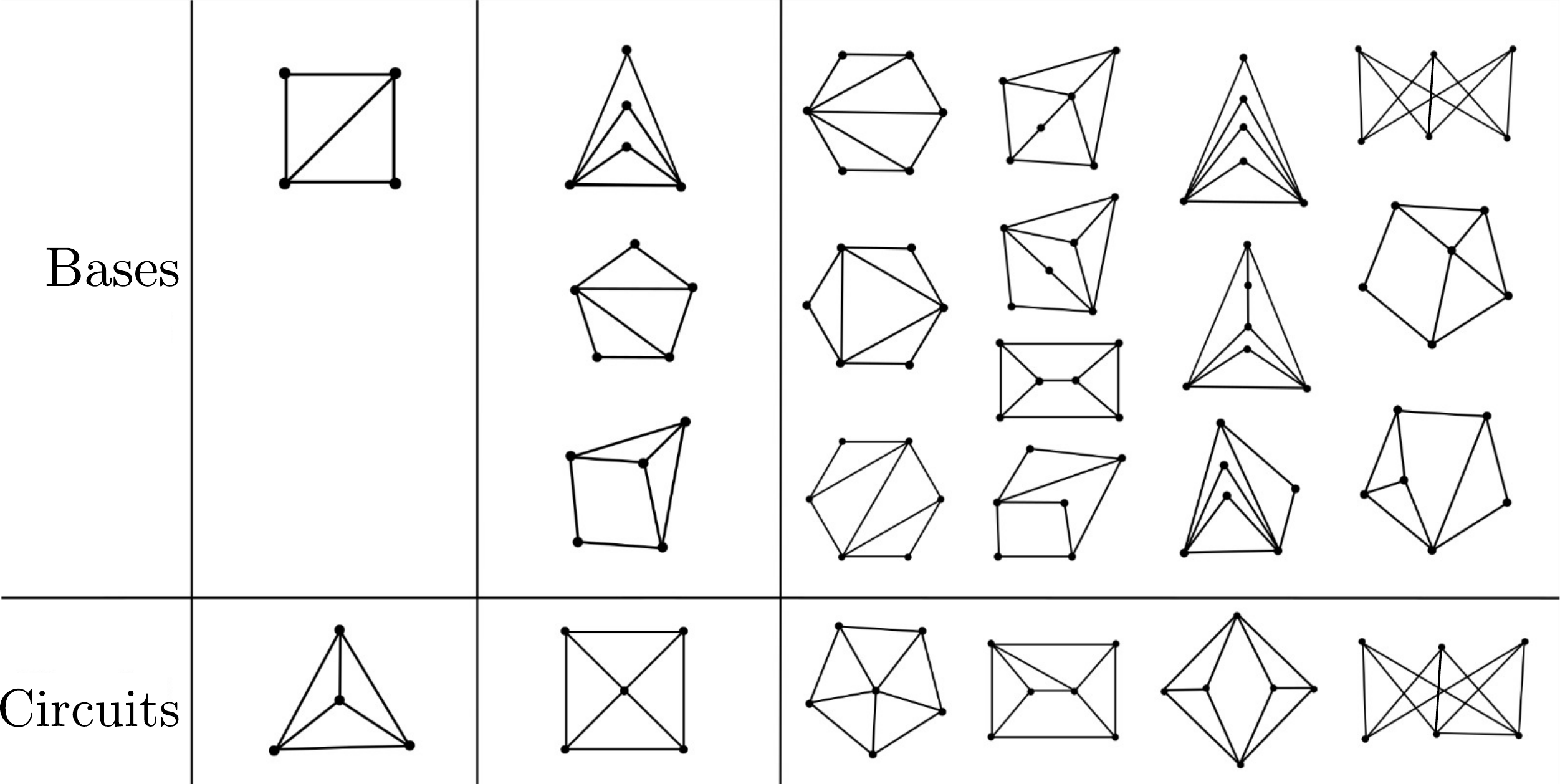}
	\caption{Bases and spanning circuits of the rigidity matroid for
        \(\ell = 4,5,6\).}
	\label{figure:rigidity}
\end{figure}

The Geiringer--Laman theorem gives a purely combinatorial criterion for independence in
the rigidity matroid, namely for when all incidence conditions from $G$ are independent.
A graph \(G\) is independent in $\mathcal{R}_\ell$ if and only if every induced subgraph
satisfies
\begin{equation}
\label{eq:chapter7-laman-condition}
        |G[S]|\leq a|S|-b 
\end{equation}
with $(a,b)=(2,3)$. Here $S$ in any subset of $[\ell]$, and $G[S]$ denotes the
\textit{induced subgraph} from $G$ on $S$, namely all edges in $G$ whose endpoints lie
in $S$. More generally, $G$ is said to be $(a,b)$-sparse
if~\eqref{eq:chapter7-laman-condition} holds true for any induced subgraph. In other
words, \(G\) is independent if and only if it is \((2,3)\)-sparse. In this case the
rigidity rank is \(|G|\). A graph is instead called a \emph{rigidity circuit} if it is
minimally dependent: its edges are dependent in the rigidity matroid, but every proper
subgraph is independent. Figure~\ref{figure:rigidity} shows the unlabeled rigidity basis
and circuits for \(\ell\leq6\).

We define a graph $G$ to be \emph{contraction-stable},
if for every subset of the vertices $S \subseteq [\ell]$, we have that
\begin{tcolorbox}[definitionbox]
\textbf{Contraction-stability.}
\begin{equation}\label{eq:CS}
	|G_S|+4(|S|-1) \geq |G| \, ,
\end{equation}
\end{tcolorbox}\noindent
where $G_S$ is the graph obtained from $G$ by identifying all vertices in $S$ while
merging parallel edges, and deleting resulting loops. If this inequality is strict, we
say that $G$ is \textit{strictly contraction-stable}. The idea is that the
inequality~\eqref{eq:CS} detects when configurations of lines in $V_G$ where the lines
within $S$ coincide may have enough degrees of freedom to constitute an irreducible
component of $V_G$.

\begin{tcolorbox}[resultbox]
\textbf{Dimension and complete intersections.}
The rigidity rank controls the dimension of the generic incidence problem:
\begin{equation}
        \operatorname{codim}(X_G)
        =
        \operatorname{rank}(G) \, ,
\end{equation}
For the full incidence variety \(V_G\), one has
\begin{equation}
        \operatorname{codim}(V_G)
        \leq
        \operatorname{rank}(G)
        \leq
        \min\{|G|,2\ell-3\} \, .
\end{equation}
Moreover, \(V_G\) is a complete intersection if and only if \(G\) is
\((2,3)\)-sparse and \textit{contraction-stable}.
\end{tcolorbox}\noindent
A variety is said to be a complete intersection if it can be cut out by the smallest
possible number of equations, namely by as many independent equations as its
codimension. Here \(X_G\) denotes the Zariski closure of configurations of distinct
lines satisfying the incidences of \(G\). This is the generic, non-degenerate part of
the incidence problem. The full variety \(V_G\) may have extra components coming from
coincidences of lines or from forced additional incidences. The extra graph-theoretic
condition of contraction-stability rules out precisely those degenerations which would
lower the codimension of \(V_G\).

\begin{eg}[Trees and triangles]
If \(G\) is a tree, then \(G\) is \((2,3)\)-sparse and strictly contraction stable. Hence
\begin{equation}
        \operatorname{codim}(V_G)=\ell-1,
        \qquad
        \dim(V_G)=3\ell+1 \, .
\end{equation}
By the edge count, \(G=K_3\) is also independent, and the codimension of $V_G$ is as expected equal to $3$. On the other hand, $G=K_4$ is dependent as $|G|=6>2 \ell -3 = 5$. In fact, from~\eqref{eq:dim_CG} we have that $ \operatorname{codim}(V_{K_4}) = 5 < |G|=6$. All complete graphs $K_\ell$ for $\ell \geq 4$ are dependent.
\end{eg}

\begin{eg}[\(K_{2,4}\)]\label{eg:K24}
Consider the bipartite graph \(K_{2,4}\) with parts $\{1,2\}$ and $\{3,4,5,6\}$. This is the graph containing all edges joining its parts, and no more. This is \((2,3)\)-sparse and contraction-stable, so
\(V_{K_{2,4}}\) is a complete intersection. It is a reducible variety of codimension $8$ in $\Gr(2,4)^6$. It factorizes into two irreducible components, both of codimension~$8$, as follows:
\begin{equation}
	V_{K_{2,4}} = X_{K_{2,4}} \cup Y_{K_{2,4}} \, .
\end{equation} 
The first component $X_{K_{2,4}}=W_{K_{2,4}}$ is the realization of $K_{2,4}$, in which generic lines are distinct, and encodes the classical Schubert problem on $\Gr(2,4)$ of two transversals to other given four lines. The component $Y_{K_{2,4}}$ instead consist of configurations of six lines where the two lines corresponding to the four-valent vertices of $K_{2,4}$ coincide. It is a simple check that such configurations have $6\cdot 4-8=16$ degrees of freedom.

Let now $G$ be the graph obtained from $K_{2,4}$ by adding the edge $12$. Then $W \subseteq V_G$, and therefore $V_G$ has still codimension $8<|G|=9$. Hence, $V_G$ is not a complete intersection. In fact, $G$ is not contraction stable, since the inequality in~\eqref{eq:CS} is violated for $S=\{1,2\}$. On the other hand, by the bound on the edge count, $G$ is independent in the rigidity matroid. This example shows the importance of the condition of contraction-stability in the criterion for complete intersections.
\end{eg}

\paragraph{Reducibility and components}

We now turn to irreducibility. A variety is said to be irreducible if it cannot be
written as the union of two strictly smaller varieties. This property is relevant for
the Landau analysis, as different irreducible components of the polar divisor defined by
the integrand have to be treated as different strata in the analysis of singularities.

The varieties \(V_G\) can be reducible, as shown by $K_3$ in
Example~\ref{eg:chapter7-three-lines} and the triangulation of a square in
Example~\ref{eg:chapter7-triangulated-quadrilateral}. More generally, if $G$ contains a
triangle $K_3$ as a subgraph, it will always be reducible. Indeed it will always contain
one component where the lines corresponding to the subgraph $K_3$ are concurrent, and
one where they are coplanar. On the other hand, there are also triangle-free graphs that
are reducible, such as $K_{2,4}$ discussed in Example~\ref{eg:K24}. In this case, a
further component arose from configurations of lines where two lines were coinciding.
There are also new components arising from other incidence patterns specific to lines in
$\mathbb{P}^3$, as shown by the following example.

\begin{eg}[$K_{3,3}$]\label{eg:K33}
	Let $G=K_{3,3}$ be the bipartite graph with parts $\{1,3,5\}$ and $\{2,4,6\}$. This graph satisfies $|G|=9=2\ell-3$ and is a rigidity basis. It contains no triangle, but the variety $V_{K_{3,3}}$ has three irreducible components, all of codimension 9:
	\begin{equation}
	V_{K_{3,3}} = V_{[6]} \cup V_{[6]}^* \cup Y_{K_{3,3}} \, .
	\end{equation} 
	Generic configurations of six lines in $Y_{K_{3,3}}$ intersect precisely according to the edges of $K_{3,3}$. Such tuples of lines are pairs of three lines $1,3,5$ and $2,4,6$ lying on opposite rulings of a quadratic surface in $\mathbb{P}^3$. 
	
	It is interesting to study this component in terms of the Amplituhedron. The graph $K_{3,3}$ is non-planar, and so we expect the six-loop Amplituhedron's form to have vanishing residue on a boundary associated to $Y_{K_{3,3}}$. This is because non-planar Landau diagrams should have vanishing leading singularity in planar $\mathcal{N}=4$ SYM. We can confirm this geometrically as follows. Up to a projective
transformation, any smooth real quadric with two real rulings may be brought to the standard form, with rulings parametrized by
\begin{equation}
        R(s)=
        \begin{pmatrix}
        1 & s & 0 & 0\\
        0 & 0 & 1 & s
        \end{pmatrix},
        \qquad
        S(t)=
        \begin{pmatrix}
        1 & 0 & t & 0\\
        0 & 1 & 0 & t
        \end{pmatrix} \, .
\end{equation}
A generic point in $V_{K_{3,3}}$ can be parametrized as $R(s_i)$, for distinct $s_i$, and $i=1,3,5$ and $S(t_j)$ for $j=2,4,6$.
Then \(\langle R(s_i)S(t_j)\rangle=0\), giving the \(K_{3,3}\) incidences,
but the remaining brackets have opposite signs,
\begin{equation}
        \langle R(s_i)R(s_j)\rangle=-(s_i-s_j)^2,
        \qquad
        \langle S(t_i)S(t_j)\rangle=(t_i-t_j)^2 \, .
\end{equation}
Thus the nonzero mutual brackets cannot all be positive. On the other hand, the mutual loop positivity conditions of the Amplituhedron require the bracket between any pair of loop lines to be positive~\cite{the_amplituhedron}. Consequently, although \(V_{K_{3,3}}\) is cut out by algebraic boundary equations of the six-loop Amplituhedron, it is residual: its strict real locus has no open positive part. Therefore we expect the numerator of the canonical form to vanish on this component. The same argument works for any bipartite graph $K_{a,b}$ as soon as one part has size at least three and the other has size at least two. In particular, the irreducible component $Y_{K_{2,4}}$ of $V_{K_{2,4}}$ encountered in Example~\ref{eg:K24} is residual for the six-loop Amplituhedron, despite the fact that $K_{2,4}$ is planar. This indicates that irreducible components of on-shell varieties relevant for Landau analysis in the Amplituhedron may be labeled simply by bicolorings, in agreement with the construction of colored Landau diagrams in geometric Landau analysis.
	\end{eg}

For our discussion, it is useful to also consider the \textit{strict realization}
\(Y_G\), the Zariski closure of configurations of distinct lines satisfying exactly the
incidences of \(G\), and no others. Thus, we have the inclusions
\begin{equation}
        Y_G\subseteq X_G\subseteq V_G,
        \qquad
        Y_G\subseteq W_G\subseteq V_G \, .
\end{equation}
For Landau analysis, \(Y_G\) is the component corresponding to the intended cut graph,
while the remaining components of \(V_G\) describe degenerate cuts or more singular
boundary strata, and should be considered separately in the analysis.

\begin{tcolorbox}[resultbox]
\textbf{Irreducibility criterion.}
The incidence variety \(V_G\) is irreducible if and only if \(G\) is
\((2,4)\)-sparse and strictly contraction-stable. In this case
\begin{equation}
        V_G=W_G=X_G=Y_G \, ,
\end{equation}
and this variety is a unirational complete intersection.
\end{tcolorbox}
Unirational means that there exists rational parametrization from an affine space
covering an open dense part of the variety.

The condition \((2,4)\)-sparse is stronger than \((2,3)\)-sparse and rules out the most
basic sources of concurrence/coplanarity decompositions. Strict contraction-stability
rules out additional components coming from collisions of lines. For example, $K_{2,4}$
in Example~\ref{eg:K24} is $(2,4)$-sparse but not strictly contraction-stable. Thus the
theorem says that irreducibility requires both: no dense subgraphs forcing classical
incidence decompositions, and no hidden degenerations from coincident lines.

A large and important class where the components can be described explicitly comes from
\textit{outerplanar} graphs, namely subgraphs of triangulations of polygons. Let \(G\)
be a subset of the edge graph of a triangulation of an \(\ell\)-gon, and let $\tau(G)$
be the number of triangles of $G$. We already encountered such an example when $G$ was
the triangulation of a square in Example~\ref{eg:chapter7-triangulated-quadrilateral},
where $\ell=4$ and $\tau(G)=2$. Any such graph is independent in the rigidity matroid, and if it is a full triangulation then it is a basis. The following result says that they give
complete intersections and describes their irreducible components.

\begin{tcolorbox}[resultbox]
\textbf{Outerplanar graphs.}
If \(G\) is an outerplanar subgraph of a triangulation of an \(\ell\)-gon, then
\begin{equation}
        V_G
        \quad\text{has}\quad
        2^{\tau(G)}
        \quad\text{irreducible components} \, .
\end{equation}
All components have codimension \(2\ell-3\), and they are indexed by bicolorings of the triangles of $G$.
\end{tcolorbox}
The geometric rule is simple: adjacent triangles of the same color merge into larger
concurrent or coplanar regions. The resulting components are intersections of varieties
\(V_\sigma\) and \(V_\tau^*\), where \(V_\sigma\) means that the lines indexed by
\(\sigma\) are concurrent and \(V_\tau^*\) means that the lines indexed by \(\tau\) are
coplanar, and $\sigma$ and $\tau$ are subpolygons of $G$ of the same color.

Wheel graphs give another useful family. The wheel \(W_\ell\) consists of a cycle of
length \(\ell-1\) together with one central vertex connected to every vertex of the
cycle. Its \(\ell-1\) triangular faces can again be colored black or white, and the
components are obtained by merging adjacent triangles of the same color. There is
however one coloring which is forbidden: if all triangles but one have the same color,
then the corresponding expected component is empty. In fact, the colored region without
the lonely triangle already involves all $\ell$ vertices, and leaves no choice for the
lines to arrange according to its coloring. Since there are $2(\ell-1)$ such forbidden
colorings, we obtain:
\begin{tcolorbox}[resultbox]
\textbf{Wheel graphs.}
For the wheel graph \(W_\ell\), the incidence variety \(V_{W_\ell}\) has
\begin{equation}
        2^{\ell-1}-2(\ell-1)
\end{equation}
irreducible components.
\end{tcolorbox}

For example, \(W_4=K_4\). For larger \(\ell\), the wheel graphs are rigidity circuits,
hence they are minimally dependent. They provide a controlled family where the failure
of independence produces many components. The variety $V_{W_\ell}$ is also not a
complete intersection; the more the coloring alternates, the smaller is the dimension of
the corresponding component. In the even case \(\ell\geq 6\), the strict realization of
the wheel is empty: every component forces at least one extra incidence. That is,
for even \(\ell\geq6\) one cannot color the triangles in $W_\ell$ by alternating colors
without repetition.

An interesting problem, both from a purely geometric but also from the physics point of
view, is determining which graph $G$ has a realization ($W_G \not= \emptyset$) or a
strict realization ($Y_G \neq \emptyset$). The smallest graph that is not strictly
realizable is $G=K_{2,3}'$, defined as $K_{2,3}$ with an additional edge between the two
three-valent vertices. For this graph
\begin{equation}
	Y_{K_{2,3}'} = \emptyset , \qquad W_{K_{2,3}'} \neq \emptyset \, .
\end{equation} 
This follows by a geometric argument. In $Y_{K_{2,3}}$, with vertices 1 and 2 being
three-valent, lines 1 and 2 lie on the opposite ruling defined by the lines 3,4,5.
Imposing the incidence between 1 and 2 forces them be the same.

\begin{eg}[Forbidden Incidences]
The smallest non-realizable triangle-free graph $H$ is $K_{4,4}$ with one edge removed, for which $Y_H=W_H=\emptyset$. In general, these problems are related to \emph{incidence theorems}~\cite{FP}.
    Other graphs $G$ with $Y_G = \emptyset$ are the wheel $W_\ell$ with even $\ell \geq 6$. There exist
         many graphs which are not strictly realizable. For example,         
         let $G$ be obtained by gluing $r \geq 3$ complete graphs $K_{\ell_i}$ with $\ell_i \geq 3$ and $i=1,\dots,r$ along a common edge. Then, $Y_G = \emptyset$, but $W_G \neq \emptyset$. The resulting component is $W_{K_{2,3}}$ and has the expected codimension $|K_{2,3}'|=7$.
\end{eg}

We conclude by conjecturing a sufficient condition for strict realizability.

\begin{tcolorbox}[resultbox]
\textbf{Strict realizability conjecture.}
    Let $G$ be a $K_{2,3}^{'}$-free graph which is
        independent in $\mathcal{R}_\ell$. Then $Y_G \neq \emptyset$.
\end{tcolorbox}

The lesson for Landau analysis is that the graph of a cut controls the component
structure of the on-shell space. Sparse and contraction-stable graphs behave as
expected; dense graphs, wheels, and special bipartite graph such as \(K_{2,4}\) produce
additional branches. These branches are not rare pathologies, but features of
higher-loop factorizations of on-shell spaces. Such spaces appear in the algebraic
boundary of loop Amplituhedra, and hence in the Landau analysis of planar amplitudes in
$\mathcal{N}=4$ SYM. However, as shown in Example~\ref{eg:K33}, some components of
on-shell spaces, even if they arise from planar graphs, can be residual. This means that
the Amplituhedron canonical form is expected to have vanishing residue on them. In the
language of geometric Landau analysis~\cite{DennenPrlinaSpradlinStanojevicVolovich2017},
such residual components should not contribute to the Landau singularities of the
integral.

\paragraph{Efficient coordinates and computation}

The Plücker description is intrinsic, but it is not always computationally efficient.
The ideal \(I_G\) lives in \(6\ell\) variables, together with
\(\ell\) Plücker quadrics. Even for moderate \(\ell\), primary decompositions
in these coordinates become difficult.

The first simplification is to pass to an affine chart, as in
\eqref{eq:chapter7-affine-line}. On this chart, each line is described by
four affine coordinates, and the incidence equation becomes the determinant
condition~\eqref{eq:chapter7-affine-line}. This reduces the number of variables from
\(6\ell\) to \(4\ell\). Importantly, this does not lose the primary decomposition: after
homogenization and saturation, one recovers the corresponding decomposition in Plücker
coordinates.

A second simplification uses coordinates adapted to a spanning tree of the graph \(G\).
Let \(G\) be a connected graph on \([\ell]\), and choose a spanning tree \(T\subseteq
G\). For every tree edge \(e\in T\), introduce a
\(2\times 2\) matrix \(U_e\) of new variables, and let \(\mathbb{C}[X]\) be
the polynomial ring in the \(4(\ell-1)\) entries of these matrices. The basic tree
equations are the quadrics \(\det(U_e)=0\). If \(g\in G\setminus T\), then
\(T\cup\{g\}\) contains a unique cycle; writing
\(e_1,\ldots,e_r\) for the tree edges on this cycle, we set
\begin{equation}
        V_g \,=\, X_{e_1}+\cdots+X_{e_r} \, .
\end{equation}
The spanning-tree coordinate ideal associated with \((G,T)\) is therefore
\begin{equation}
\label{eq:GTideal}
I_{G,T}
=
\bigl\langle \det(U_e): e\in T\bigr\rangle
+
\bigl\langle \det(V_g): g\in G\setminus T\bigr\rangle
\subseteq \mathbb{C}[X] \, .
\end{equation}
Thus the incidence equations are expressed entirely as determinant conditions on
\(2\times2\) matrices attached to the tree and to the fundamental cycles determined by
the remaining edges. For \(G=K_3\), this gives the compact system
\begin{equation}
        \det(U_1)=0 \, ,\qquad
        \det(U_2)=0\, ,\qquad
        \det(U_1+U_2)=0 \, ,
\end{equation}
which makes the two components of \(V_{K_3}\) transparent.

\begin{table}[pos=t]
	\centering
	\small
	\begin{tabular}{|c|p{2.2cm}|p{2.0cm}|p{1.8cm}|p{1.8cm}|}
		\hline
		$\ell$ & Connected Graphs & Complete Intersection & Irreducible & Realizable ($W_G \not= \emptyset$) \\
		\hline
		\hline
		4 & 6     & 5      & 3  &   6 \\
		\hline
		5 & 21    & 16    & 6  &  21 \\
		\hline
		6 & 112  & 69   & 17 &  103 \\
		\hline
		7 & 853 & 379 & 52 &  681 \\
		\hline
	\end{tabular}
	\caption{Numerical irreducible decomposition of $V_G$ for connected graphs $G$ with $\ell = 4,5,6,7$.
	The four columns report the number of all graphs $G$
	 up to isomorphism, the number with $V_G$ a complete intersection, irreducible, and with non-empty $W_G$.}
	\label{table:all-connected-graphs}
\end{table}

\begin{table}[pos=t]
	\centering
	\small
	\begin{tabular}{|c|p{2.2cm}|p{2.0cm}|c|c|}
		\hline
		$\ell$ & $\mbox{Triangle-Free}$ Graphs & Complete Intersection & Irreducible & Realizable \\
		\hline
		\hline
		4 & 3    &   3       & 3     &  3 \\
		\hline
		5 & 6    &   6       & 6     &  6 \\
		\hline
		6 & 19  &   19      & 17   &  19 \\
		\hline
		7 & 59  &   57      & 52  &  59 \\
		\hline
		8 &267 &   254   & 219 &  266 \\
		\hline
	\end{tabular}
	\caption{Numerical irreducible decomposition of $V_G$ for triangle-free connected graphs $G$ with $\ell = 5,6,7,8$.
	The numbers in the columns report the same properties as in Table \ref{table:all-connected-graphs}.}
	\label{table:triangle-free-graphs}
\end{table}

\begin{tcolorbox}[resultbox]
\textbf{Computational approach.}
For incidence varieties \(V_G\), primary decompositions can be computed in
affine coordinates and then homogenized back to Plücker coordinates. This
makes symbolic computation feasible without changing the underlying
projective geometry. Spanning-tree coordinates are particularly well suited.
\end{tcolorbox}
\noindent

\begin{figure}[pos=t]
	\centering
	\includegraphics[scale = .35]{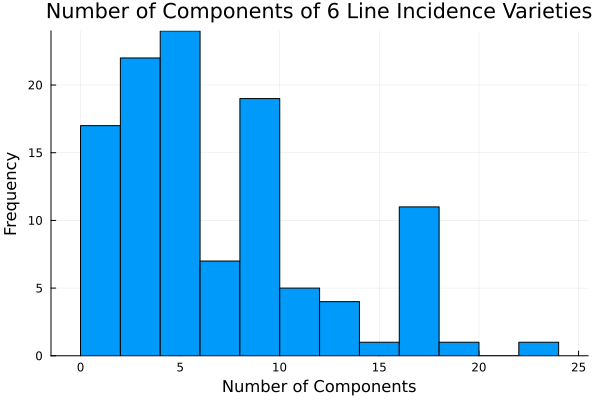} \includegraphics[scale = .35]{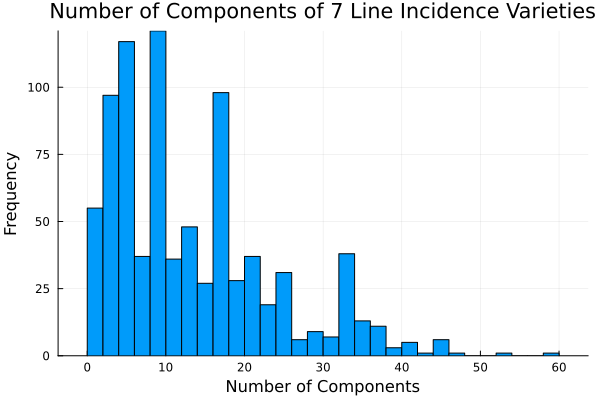}
	\caption{Histogram for the number of components of $V_G$ for connected $G$ with $\ell=6,7$.}
	\label{figure:6-node-comps}
\end{figure}

The computational part of
\cite{HolleringMazzucchelliParisiSturmfels2025Lines} uses these coordinate
systems to study all connected graphs up to eight vertices. Symbolic computations were
performed with \texttt{Macaulay2}~\cite{M2}, and numerical irreducible decompositions
with
\texttt{HomotopyContinuation.jl}~\cite{BT}. The resulting census shows that
the geometry of \(V_G\) becomes intricate very quickly. Even for graphs with seven
vertices, examples appear with many irreducible components. The results are shown in
Tables
\ref{table:all-connected-graphs} and~\ref{table:triangle-free-graphs}. The histograms  in Figure \ref{figure:6-node-comps} 
display the number of components for all incidence varieties for $\ell \in \{6,7\}$. For
example, the graph $G$ with $\ell=7$ and maximal number of irreducible components has
$58$ irreducible components. Our census for small graphs is made available on {\tt
Zenodo} at the supplementary materials website
\begin{equation}
\label{eq:zenodo}
   \hbox{\url{https://zenodo.org/records/17708048}.}
\end{equation}

The computational message is important for the rest of the chapter. Grassmannian Landau
analysis often produces large systems of incidence equations. Without using the geometry
of \(V_G\), these systems can be hard to analyze. The line-variety framework provides
both theoretical criteria and practical coordinates for deciding which components and
dimensions should appear.

\subsubsection{Projections and leading singularity degree}
\label{subsec:projections-leading-singularity-degree}

The previous subsection studied incidence varieties \(V_G\) representing on-shell spaces
relevant for the Landau analysis of integrals in momentum twistors. We now consider
projections to the external data. This is the step where the geometry of line
configurations becomes Landau analysis: the singular loci in kinematic space arise from
critical points of projections from on-shell spaces to the external variables. This is
the same mechanism already discussed in Chapter~\ref{ch:From Integrands to Integrals}:
Landau singularities occur when the fiber of the on-shell space over the external
kinematics becomes non-generic, for instance because the genus of a curve changes,
several solutions collide, or an overdetermined system suddenly acquires a
solution~\cite{Landau1959,Eden:1966dnq,Pham2011,Hannesdottir:2021kpd,HelmerPapathanasiouTellander2024,FevolaMizeraTelen2024}.

Let \(G\) be a graph on the loop vertices \([\ell]=\{1,\ldots,\ell\}\). We write
\(X_1,\ldots,X_\ell\) for the corresponding loop lines, so that
\(V_G\subseteq \Gr(2,4)^\ell\) is cut out by the incidence
conditions \(\langle X_i \, X_j \rangle =0\) for \(ij\in G\). We now choose a vector
\begin{equation}
        u=(u_1,\ldots,u_\ell)\in\mathbb N^\ell \, ,
        \qquad
        |u|:=u_1+\cdots+u_\ell \, .
\end{equation}
The extended graph \(\mathcal{L}=G_u\) is obtained from \(G\) by attaching \(u_i\)
external leaves to the vertex \(i\). We denote the external lines by
\begin{equation}
        \mathbf M=(M_1,\ldots,M_d) \, ,
        \qquad d=|u| \, .
\end{equation}
Each external leaf attached to \(i\) imposes an incidence condition
\(\langle X_i \, M_j \rangle =0\). Thus \(G_u\) defines an incidence variety
\begin{equation}
        V_{G_u}\subseteq
        \Gr(2,4)^\ell\times\Gr(2,4)^d \, .
\end{equation}
The associated \emph{Landau map} is the projection that forgets the loop lines:
\begin{tcolorbox}[resultbox]
\textbf{Landau map.}
\begin{equation}
\label{eq:chapter7-landau-map}
        \psi_u:
        V_{G_u}
        \longrightarrow
        \Gr(2,4)^d \, ,
        \qquad
        (\mathbf X,\mathbf M)\longmapsto \mathbf M \, .
\end{equation}
\end{tcolorbox}\noindent
For fixed external data \(\mathbf M\), the fiber
\(\psi_u^{-1}(\mathbf M)\) is the space of loop-line configurations
satisfying all internal and external incidence conditions. In physical terms, it is the
cut solution space associated with the corresponding Landau diagram. The graph \(G_u\)
should be understood as the planar dual of a Landau diagram: its vertices are dual
regions and its edges are propagators put on shell. The vertices of $G$ correspond to
dual momentum variables.

\begin{tcolorbox}[definitionbox]
\textbf{Landau map.}
The Landau map is the projection from the on-shell incidence variety to the
space of external data. Its fibers are cut solution spaces. The corresponding component of the Landau variety
is the locus where these fibers become non-generic.
\end{tcolorbox}
\noindent

Let $ m=\dim V_G .$ The expected dimension of the generic fiber of \(\psi_u\) is
\(m-d\). This gives a useful hierarchy. If \(d=m\), the generic fiber is finite; this is
the
\emph{leading} case. If \(d=m+1\), the system is overdetermined and the
generic fiber is empty; this is the \emph{superleading} case. If \(d=m-1\), the generic
fiber is a curve; this is the \emph{next-to-leading} case.
In the case of isolated solutions, the latter define maximal residues of the integrand and appear as
algebraic prefactors or maximal iterated discontinuities of the integrated
function~\cite{Cachazo:2008vp,ArkaniHamed:2009dn,Hannesdottir:2021kpd}. We warn the
reader that in physics the word ``leading'' may also mean just the maximal number of
available cuts, give a specific Feyman integral or topology under consideration, even if
the cuts are not enough to localize the loop momenta. In our terminology these would be
subleading cases.

\begin{tcolorbox}[resultbox]
\textbf{Leading, superleading, and next-to-leading.}
Let \(m=\dim V_G\) and \(d=|u|\). The Landau map
\(\psi_u:V_{G_u}\to\Gr(2,4)^d\) has expected fiber dimension
\(m-d\). The case \(d=m\) is leading and its generic fiber consist of finitely many points;
the case \(d=m+1\) is superleading and its fiber is generically empty; the case \(d=m-1\) is next-to-leading and its generic fiber is a curve, etc.
\end{tcolorbox}
\noindent

A complete Landau analysis of an integral of the form
\begin{equation}
        \mathcal{I}(\mathbf M)
        =
        \int_\Gamma
        \frac{N(\mathbf L;\mathbf M)}
        {D(\mathbf L;\mathbf M)}
        \,{\rm d}\mu(\mathbf L) \, ,
\end{equation}
with
\begin{equation}
        D(\mathbf L;\mathbf M)
        =
        \prod_{(i,a) \, \in \, G_{\rm ext}}
        \langle L_a M_i\rangle
        \prod_{(a,b) \, \in \,  G_{\rm int}}
        \langle L_a L_b\rangle \, ,
\end{equation}
would require studying the polar locus defined by the vanishing of the full denominator
\(D(\mathbf L;\mathbf M)\). Let
\(G=G_{\rm int}\cup G_{\rm ext}\). This polar locus is naturally stratified by
the incidence varieties \(V_H\), where \(H\) ranges over subgraphs of \(G\). In the
language of Landau analysis, one studies the projection of these on-shell strata to the
external data; the possible singularities are contained in the corresponding sets of
critical values~\cite{Panzer2015,HelmerPapathanasiouTellander2024}. One may restrict
attention to subgraphs \(H\subseteq G\) that contain both internal and external
incidence conditions~\cite{BerghoffPanzer2025}:
\begin{equation}
\label{eq:H_cons}
        H\cap G_{\rm int}\neq\emptyset,
        \qquad
        H\cap G_{\rm ext}\neq\emptyset \, .
\end{equation}
Indeed, if \(H\) contains only internal edges, the corresponding incidence conditions do
not constrain the external kinematics. If \(H\) contains only external edges, the locus
is already a boundary condition on the external data rather than a singularity produced
by integrating over loop variables. The subgraphs satisfying~\eqref{eq:H_cons} are
therefore the ones relevant for non-trivial loop-dependent Landau singularities.

More precisely, one should refine \(V_H\) to its strict realization \(Y_H\), and then to
the irreducible components \(Y_{H,\sigma}\). These components can themselves be
singular, and their singular loci may contribute further strata. The appropriate global
object is therefore a Whitney stratification of the polar locus~\cite{Whitney1965,GoreskyMacPherson1988,HelmerPapathanasiouTellander2024}.
If \(\mathcal L(\mathcal{I})\) denotes the full Landau variety of the integral, then the
contribution of the smooth strata considered here can be summarized schematically as
\begin{equation}
        \bigcup_{H\subseteq G}
        \bigcup_{\sigma}
        \operatorname{CritVal}\!\left(
        \psi_{Y_{H,\sigma}}
        \right)
        \subseteq
        \mathcal L(\mathcal I) \, ,
\end{equation}
where \(\psi_{Y_{H,\sigma}}\) is the projection from \(Y_{H,\sigma}\) to the external
data, and the union may be restricted to subgraphs satisfying
\eqref{eq:H_cons}. The leading, superleading and next-to-leading diagrams
studied below are distinguished pieces of this more general stratified projection
problem.

In the following we will be mainly concerned with the leading and superleading cases. In
the leading case, the generic fiber consists of finitely many points. These points are
the leading singularity configurations associated to the (dual) Landau diagram \(G_u\).
\begin{tcolorbox}[resultbox]
\textbf{Leading-singularity degree.}
We define the \emph{leading-singularity (LS) degree} by
\begin{equation}
\label{eq:chapter7-LS-degree}
        \gamma_u
        =
        \left|\psi_u^{-1}(\mathbf M)\right|
        \qquad
        \text{for generic }\mathbf M\in\Gr(2,4)^d \, ,
\end{equation}
\end{tcolorbox}\noindent
where the cardinality is counted with multiplicity. Equivalently,
\(\gamma_u\) is the degree of the generically finite Landau map. In the
language of Schubert calculus, it counts the number of configurations of $\ell$ loop
lines satisfying the internal incidences \(G\) and the external Schubert conditions
specified by \(u\).

The LS degrees for fixed \(G\) are packaged by the \textit{multidegree} of \(V_G\). Let
\begin{equation}
        [V_G]\in
        H^*((\mathbb P^5)^\ell,\mathbb Z)
        =
        \mathbb Z[t_1,\ldots,t_\ell]/\langle t_1^6,\ldots,t_\ell^6\rangle
\end{equation}
be the class of $V_G$ in the cohomology or Chow ring of $(\mathbb{P}^5)^\ell$. If
\(m=\dim V_G\), then
\begin{equation}
\label{eq:chapter7-multidegree-LS}
        [V_G]
        =
        \sum_{u\in\mathbb N^\ell \, : \,  |u|=m}
        \gamma_u\,
        t_1^{5-u_1}\cdots t_\ell^{5-u_\ell} \, .
\end{equation}
Thus the coefficient of \(t_1^{5-u_1}\cdots t_\ell^{5-u_\ell}\) is precisely the LS
degree of the extended graph \(G_u\). This gives a compact generating function for all
leading diagrams obtained from the same loop-incidence graph
\(G\).

When \(V_G\) is a complete intersection, the multidegree has a particularly simple form.
In that case the Plücker quadric for the \(i\)-th line contributes
\(2t_i\), while each incidence equation \(\langle X_i \, X_j \rangle =0\) contributes \(t_i+t_j\).
Therefore
\begin{tcolorbox}[definitionbox]
\textbf{Complete-intersection multidegree.}
\begin{equation}
\label{eq:chapter7-CI-multidegree}
        [V_G]
        =
        2^\ell
        \prod_{i=1}^{\ell} t_i
        \prod_{ij\in G}(t_i+t_j) \, .
\end{equation}
\end{tcolorbox}\noindent
Expanding this polynomial gives all LS degrees \(\gamma_u\) for the leading extensions
\(G_u\).

\begin{tcolorbox}[resultbox]
\textbf{LS degrees from multidegrees.}
For a fixed loop-incidence graph \(G\), the multidegree \([V_G]\) is the
generating function of leading-singularity degrees. If \(V_G\) is a complete
intersection, it is given by~\eqref{eq:chapter7-CI-multidegree}.
\end{tcolorbox}
\noindent
If $V_G$ is not a complete intersection, its LS degree or that of individual terms for
fixed $u$ can be computed using tools from numerical algebraic geometry. For all
connected graphs up to eight vertices, the dimensions, irreducible decompositions, and
multidegrees of the incidence varieties \(V_G\) were computed
in~\cite{HolleringMazzucchelliParisiSturmfels2025Lines}. The data are available at the
supplementary Zenodo repository~\eqref{eq:zenodo}. These data are useful beyond the
line-variety problem itself: they give a catalogue of possible leading Landau diagrams,
including planar and non-planar dual loop graphs, together with their LS degrees. In
this sense the Zenodo tables provide an enumerative atlas of planar leading Landau diagrams.

\begin{eg}[The four-mass box]
\label{eg:chapter7-four-mass-box-LS}
Let \(G=K_1\) be the graph with one loop vertex and no internal edges. Then
\(V_G=\Gr(2,4)\), so \(m=4\). The leading extension is
\(u=(4)\): one asks for a line \(X\) incident to four generic external lines
\(M_1,M_2,M_3,M_4\). This is the classical Schubert problem of transversals to
four lines in \(\mathbb P^3\), which has two solutions:
\begin{equation}
        \gamma_{(4)}=2 \, .
\end{equation}
This is the Grassmannian version of the four-mass box maximal cut. The two
solutions are the two branches of the four-mass box. Their collision is
detected by the discriminant
\begin{equation}
        \Delta_{\rm 4m}
        =
        \det\big(\langle M_iM_j\rangle\big)_{i,j=1}^4 \, ,
\end{equation}
which is the square-root discriminant discussed in
Subsection~\ref{subsec:Boxes in Momentum Twistors}.
\end{eg}

This example also explains the relation with the usual notion of degree of an on-shell
diagram. In the earlier on-shell-diagram language, the degree counts the number of
solutions of a maximal cut after imposing the on-shell conditions. Here the same number
is the degree of the Landau map
\(\psi_u\). Thus the LS degree is the Grassmannian-incidence version of the
degree of an on-shell diagram. In the four-mass box, this degree is \(2\), and the
corresponding branch cover is precisely the two-sheeted cover defined by
\(\sqrt{\Delta_{\rm 4m}}\). This information was then used to construct the algebraic
letters necessary for the symbol of the four-mass box integral, as explained in
Section~\ref{sec:Symbols and Bootstrap}.

\begin{eg}[The triangle \(K_3\)]
\label{eg:chapter7-triangle-LS-degree}
Let \(G=K_3\). Then \(V_G=V_{[3]}\cup V^*_{[3]}\), where the two components
parametrize triples of concurrent and coplanar lines. Both components have
dimension \(9\), and \(V_G\) has multidegree
\begin{equation}	[V_{K_3}]
\label{eq:MD_K3}
        =
        8\,t_1t_2t_3(t_1+t_2)(t_1+t_3)(t_2+t_3) = 
        16\,t_1^2t_2^2t_3^2
        +
        8\sum_{\pi\in S_3}
        t_{\pi(1)}^3t_{\pi(2)}^2t_{\pi(3)} \, .
\end{equation}
Thus the only leading extensions, up to permutation, are
\(u=(3,3,3)\) and \(u=(4,3,2)\). For \(u=(3,3,3)\), the LS degree is \(16\);
for \(u=(4,3,2)\), the LS degree is \(8\). Each number is twice the
contribution of one irreducible component, because the concurrent and coplanar components are exchanged by projective duality. The graph $G_u$ for \(u=(3,3,3)\) is shown on the right in Figure~\ref{fig:TP2}. Given $9$ general lines ${\bf M}=(M_1,\ldots,M_9)$ in $\mathbb{P}^3$, there are precisely $8$ concurrent triples ${\bf L}=(L_1,L_2,L_3)$ which satisfy the $9$ Schubert conditions they impose, namely the vanishing of each term in~\eqref{eq:nine_prop}. Similarly, there are $8$ coplanar triples ${\bf L}$ satisfying these equations, so $\gamma_u=8+8=16$. These ${\bf L}$ form a {\em Cayley octad}, namely they are found by intersecting three quadratic surfaces in $\mathbb{P}^3$. See \cite{BourjailyVerguVonHippel2023}.
\end{eg}

Equation~\eqref{eq:MD_K3} follows from a general formula~\cite{PST,EK}:
\begin{equation}
	[V_{[\ell]}] \, = \,  t_1^3 t_2^3 \cdots t_\ell^3 \,\cdot \,
\biggl( 4 \cdot \sum_{(i,j)} t_i^{-2} t_j^{-1} \, + \,
8 \cdot \! \sum_{\{i,j,k\}} t_i^{-1} t_j^{-1} t_k^{-1} \biggr) \, .
\end{equation}  
The multidegree of $[V_{[\ell]}]$ is the same as that of $[V_{[\ell]}]$. This is because
projective duality in $\mathbb{P}^3$, expressed in Plücker coordinates
as~\eqref{eq:hodgestar}, mapping a linear subspace to its orthogonal complement, is an
involution both on the irreducible decomposition of $V_G$. In this case this duality
swappes the components $V_{[\ell]}$ with $V_{[\ell]}^*$.

More generally, the multidegree can also be refined component by component. If
\begin{equation}
        V_G=\bigcup_\sigma V_{G,\sigma}
\end{equation}
is an irreducible decomposition, then one can compute the multidegrees
\([V_{G,\sigma}]\) separately. Their sum gives \([V_G]\), counted with
scheme-theoretic multiplicities. This is important for Landau analysis because different
components of \(V_G\) correspond to different branches of the cut equations. Some
components may be physical, while others may be residual or degenerate. The LS degree of
\(G_u\) may therefore split into separate contributions from different branches.

\begin{eg}[A triangulated pentagon]
\label{eg:chapter7-triangulated-pentagon-LS}
Let \(G\) be the triangulated pentagon
\begin{equation}
        G=\{12,23,34,45,15,13,14\} \, .
\end{equation}
This graph is a rigidity basis, so \(V_G\) is a complete intersection of
dimension \(13\). The formula~\eqref{eq:chapter7-CI-multidegree} gives
\begin{equation}
        [V_G]
        =
        32\, t_1t_2t_3t_4t_5
        (t_1+t_2)(t_1+t_3)(t_1+t_4)(t_1+t_5)
        (t_2+t_3)(t_3+t_4)(t_4+t_5) \, .
\end{equation}
By the triangulated-polygon result above, $V_G$ has eight irreducible components, corresponding to bicolorings of
the three triangles of the triangulation. One may compute the multidegree of
each component separately; their sum recovers the complete-intersection
multidegree above. Thus the same leading diagram has several branches, and
the LS degree of any extension \(G_u\) decomposes into contributions from
these branches.
\end{eg}

In summary, the LS degree is the first invariant of the Landau map. It counts the
generic number of cut solutions, recovers the degree of familiar on-shell problems such
as the four-mass box, and organizes the sheets of the algebraic cover of kinematic
space. The next invariant is the locus where this cover degenerates. This leads to the
corresponding leading and superleading Landau singularities. In our language these are
discriminants and resultants, respectively, which we discuss next.

\subsubsection{Leading discriminants and superleading resultants}
\label{subsec:leading-discriminants-superleading-resultants}

The previous subsection attached to a Landau map its first numerical invariant, the LS
degree. This counts the number of points in a generic leading fiber. The next question
is where this finite cover degenerates. In the leading case, the relevant locus is the
branch locus of the Landau map: it is where two or more leading singularities collide,
or where the fiber otherwise ceases to have its generic cardinality. In the superleading
case, the generic fiber is empty, and the relevant locus is instead the set of external
data for which the overdetermined incidence problem acquires a solution. These two
situations are described respectively by discriminants and resultants.

\begin{tcolorbox}[resultbox]
\textbf{Discriminants and resultants of the Landau map.}
In the leading case, the Landau map is generically finite, and the
\emph{LS discriminant} is the branch locus where the finite fiber becomes
ramified or non-generic. In the superleading case, the generic fiber is empty,
and the \emph{SLS resultant} is the locus where the overdetermined incidence
problem has a solution.
\end{tcolorbox}
\noindent

This terminology is chosen to emphasize the analogy with classical elimination theory. A
univariate polynomial has a discriminant, which vanishes when two roots collide. A
collection of too many equations has a resultant, which vanishes when the equations
admit a common zero. Here the roots are replaced by configurations of loop lines, and
the coefficients of the polynomial system are the external momentum-twistor data. Thus
the Landau map packages the same algebraic phenomena in a geometric way: discriminants
detect ramification of a finite cover, while resultants detect the non-emptiness of an
overdetermined
fiber~\cite{gkz,OssermanTrager2019,PrattSodomacoSturmfels2026,HolleringMazzucchelliParisiSturmfels2026}.

Let us first recall why these loci are physically relevant. Leading singularities are
maximal residues of the loop integrand. In generalized unitarity, maximal cuts localize
the loop variables and determine coefficients of integral bases; in many explicit
integrated answers the same algebraic data appear as prefactors multiplying pure
transcendental functions~\cite{Cachazo:2008vp,ArkaniHamed:2009dn}. In the Wilson-loop
and negative-geometry setting discussed in Chapter~\ref{ch:From Integrands to Integrals}, we distinguished between leading-singularity configurations, namely points
of the localized cut, and leading-singularity values, namely the algebraic functions
obtained as residues of the integrand. The latter are precisely the candidate prefactors
multiplying the pure functions in the decomposition of the integrated observable.

A useful elementary analogy is partial fraction decomposition. In a rational function,
algebraic prefactors often appear when one decomposes the integrand into simpler pieces.
The denominators of the coefficients then contain discriminants and resultants. This is
familiar in one variable, and it also underlies recent algorithmic approaches to
multivariate partial fraction decompositions in scattering-amplitude and
Feynman-integral
computations~\cite{deKorteYu2026,HellerVonManteuffel2022,BoehmWittmannWuXuZhang2020,BendleBoehmDeckerGeorgoudisPfreundtRahnWasserZhang2020,Beck2004,Leinartas1978}.

\begin{eg}[Two quadratic factors]
\label{eg:chapter7-pfd-two-quadratics}
Consider two generic quadratic polynomials in one variable \(x\):
\begin{equation}
        Q_1(x)=a_2x^2+a_1x+a_0 \, ,
        \qquad
        Q_2(x)=b_2x^2+b_1x+b_0 \, .
\end{equation}
Let their roots be \(\alpha_{\pm}\) and
\(\beta_{\pm}\), respectively, so that
\(Q_1(x)=a_2(x-\alpha_{+})(x-\alpha_{-})\) and
\(Q_2(x)=b_2(x-\beta_{+})(x-\beta_{-})\). Over the algebraic closure, one has
the partial fraction decomposition given by
\begin{equation}
\label{eq:Q1Q2}
\begin{aligned}
        \frac{1}{Q_1(x)Q_2(x)}
        &=
        \sum_{\epsilon=\pm}
        \frac{\epsilon\, a_2^2\, Q_2(\alpha_{-\epsilon})}
        {\sqrt{\operatorname{Disc}(Q_1)}\,\operatorname{Res}(Q_1,Q_2)}
        \frac{1}{x-\alpha_\epsilon}
        \\
        &\hspace{0.7cm}
        +
        \sum_{\epsilon=\pm}
        \frac{\epsilon\, b_2^2\, Q_1(\beta_{-\epsilon})}
        {\sqrt{\operatorname{Disc}(Q_2)}\,\operatorname{Res}(Q_1,Q_2)}
        \frac{1}{x-\beta_\epsilon} \, ,
\end{aligned}
\end{equation}
where the familiar discriminant
\begin{equation}
        \operatorname{Disc}(Q_1)
        =
        a_2^2(\alpha_+-\alpha_-)^2
        =
        a_1^2-4a_2a_0
\end{equation}
detects the collision of roots of \(Q_1\), and similarly for \(Q_2\), while
the resultant detects when a root of \(Q_1\) collides with a root of \(Q_2\):
\begin{equation}
\begin{aligned}
        \operatorname{Res}(Q_1,Q_2)
        &=
        a_2^2b_2^2
        \prod_{\epsilon,\eta=\pm}(\alpha_\epsilon-\beta_\eta)
        \\
        &=
        a_2^2 b_0^2
        - a_1 a_2 b_0 b_1
        + a_0 a_2 b_1^2
        + a_1^2 b_0 b_2
        - 2 a_0 a_2 b_0 b_2
        - a_0 a_1 b_1 b_2
        + a_0^2 b_2^2 .
\end{aligned}
\end{equation}
The integral of~\eqref{eq:Q1Q2} over \(x\) can then be expressed as a
combination of logarithms, or equivalently of logarithms and arctangents after
pairing conjugate roots, with algebraic arguments and prefactors. The point
we want to highlight is that discriminants and resultants appear in the
denominators of the algebraic prefactors, so they encode the ``leading''
singular structure of the integrated result. Their emergence appears
algebraically through the partial fraction decomposition. The Landau
discriminants and resultants below are higher-dimensional versions of the
same phenomenon, where partial fraction decompositions are non-unique and are
the subject of ongoing research.
\end{eg}

We now return to line configurations. Let \(G_u\) be a leading Landau diagram, so that
\(|u|=\dim V_G\). The Landau map
\(\psi_u:V_{G_u}\rightarrow\Gr(2,4)^d\) is generically finite.
We define the LS discriminant \(\Delta_{G,u}\) by the condition
\begin{tcolorbox}[definitionbox]
\textbf{Leading singularity (LS) discriminant.}
\begin{equation}
\label{eq:chapter7-LS-discriminant}
        \Delta_{G,u}(\mathbf M)=0
        \qquad\Longleftrightarrow\qquad
        \left|\psi_u^{-1}(\mathbf M)\right|
        <
        \gamma_u \, .
\end{equation}
\end{tcolorbox}\noindent
Hence,
\(\Delta_{G,u}\) detects where the finite set of
leading-singularity configurations degenerates.

\begin{eg}[Four-mass box]
\label{eg:chapter7-four-mass-box-discriminant}
Let \(G=K_1\) and \(u=(4)\). This is the four-mass box diagram discussed in
Example~\ref{eg:chapter7-four-mass-box-LS}. It has LS degree
\(\gamma_u=2\), and its LS discriminant encodes the collision of the two
Schubert solutions \(X^{(\pm)}(\mathbf M)\) for fixed external data
\(\mathbf M=(M_1,M_2,M_3,M_4)\). Therefore
\begin{equation}
\label{eq:chapter7-four-mass-box-discriminant}
        \Delta_{K_1,(4)}(\mathbf M)
        =
        \det
        \big(
        \langle M_iM_j\rangle
        \big)_{i,j=1}^{4} \, .
\end{equation}
\end{eg}

The superleading case is similar. Let
\(|u|=\dim V_G+1\). Then the system of incidence equations has one condition
too many. The SLS resultant \(R_{G,u}\) is the equation of the locus where this
overdetermined system has a solution:
\begin{tcolorbox}[definitionbox]
\textbf{Superleading singulairty (SLS) resultant.}
\begin{equation}
\label{eq:chapter7-SLS-resultant}
        R_{G,u}(\mathbf M)=0
        \qquad\Longleftrightarrow\qquad
        \psi_u^{-1}(\mathbf M)\neq\emptyset \, .
\end{equation}
\end{tcolorbox}\noindent

\begin{eg}[Pentagon resultant]
\label{eg:chapter7-pentagon-resultant}
For \(G=K_1\), the superleading case is \(u=(5)\). A generic line in
\(\mathbb P^3\) cannot meet five generic external lines
\(M_1,\ldots,M_5\). The condition that such a common transversal exists is
the vanishing of the \(5\times5\) incidence Gram determinant:
\begin{equation}
\label{eq:chapter7-pentagon-resultant}
        R_{K_1,(5)}(\mathbf M)
        =
        \det
        \big(
        \langle M_iM_j\rangle
        \big)_{i,j=1}^{5} \, .
\end{equation}
This is the SLS resultant of the one-loop pentagon incidence problem.
Equivalently, it is the condition that the five Schubert divisors in
\(\Gr(2,4)\) have a common point.
\end{eg}

The examples above are the one-loop prototypes. The same definitions apply to any graph
\(G\). Algebraically, SLS resultants are \textit{multigraded Chow forms}, while LS
discriminants are \textit{multigraded Hurwitz forms}
\cite{OssermanTrager2019,PrattSodomacoSturmfels2026,HolleringMazzucchelliParisiSturmfels2026}. This gives both an
intrinsic definition and a method for computing their degrees.

\begin{tcolorbox}[definitionbox]
\textbf{Chow and Hurwitz forms.}
For a superleading diagram \(G_u\), the SLS resultant \(R_{G,u}\) is the
multigraded Chow form of \(V_G\) with respect to the Schubert conditions
specified by \(u\). For a leading diagram \(G_u\), the LS discriminant
\(\Delta_{G,u}\) is the corresponding multigraded Hurwitz form.
\end{tcolorbox}
\noindent

The Chow-form interpretation gives a particularly clean degree formula
\cite{OssermanTrager2019,HolleringMazzucchelliParisiSturmfels2026}. If
\(|u|=\dim V_G+1\), then \(R_{G,u}\) is homogeneous in the Plücker coordinates
of every external line attached to vertex \(i\).

\begin{tcolorbox}[resultbox]
\textbf{Degree of SLS resultants.}
SLS resultant degrees are exactly LS degrees:
\begin{equation}\label{eq:SLS_deg}
        \deg_{M_i} R_{G,u}=\gamma_{u-e_i}
\end{equation}
for an external line \(M_i\) attached to vertex \(i\). LS discriminant degrees
are computed as multigraded Hurwitz degrees. Formula
\eqref{eq:chapter7-LS-discriminant-degree-bound} gives the expected degree;
extraneous factors may have to be removed.
\end{tcolorbox}
\noindent
Here \(e_i\) is the standard basis vector and \(\gamma_{u-e_i}\) is the LS degree of the
leading diagram obtained by removing one external condition from vertex \(i\). Thus the
degrees of the SLS resultant are controlled by the LS degrees computed from the
multidegree of \(V_G\).

The analogous statement for LS discriminants is slightly subtler. Suppose
\(V_G\) is a complete intersection. The expected multidegree
\((b_1,\ldots,b_\ell)\) of the Hurwitz form, hence of the LS discriminant, is
bounded by~\cite{PrattSodomacoSturmfels2026,HolleringMazzucchelliParisiSturmfels2026}
\begin{tcolorbox}[resultbox]
\textbf{Degree bound for LS discriminants.}
\begin{equation}
\label{eq:chapter7-LS-discriminant-degree-bound}
        b_i
        \leq
        2\gamma_u
        +
        \sum_{\substack{j\in[\ell] \, :\\ u_j+\delta_{ij}>0}}
        \gamma_{u+e_i-e_j}
        \bigl(
        1-u_j-\delta_{ij}+\operatorname{degree}_G(i)
        \bigr),
        \qquad i=1,\ldots,\ell \, .
\end{equation}
\end{tcolorbox}
\noindent
Here \(\delta_{ij}\) is Kronecker's delta and
\(\operatorname{degree}_G(i)\) is the valence of vertex \(i\) in \(G\).
The right-hand side is the expected degree of the LS discriminant. In computations, it
can include extraneous factors coming from the embedding or from special features of the
incidence variety. After removing those factors, one obtains the intrinsic LS
discriminant. In~\cite[Section 6]{PrattSodomacoSturmfels2026} the authors developed software in the computer
algebra system
\texttt{Macaulay2} for computing the expected degree
\eqref{eq:chapter7-LS-discriminant-degree-bound} of \(\Delta_{G,u}\) for a
fixed graph \(G\) and \(u\) ranging over all terms in the multidegree
\([V_G]\).

We now discuss explicit computations from this perspective.

\begin{eg}[Pentabox LS discriminant]
\label{eg:chapter7-pentabox-direct-resultant}
The first non-trivial leading example beyond the four-mass box is the
pentabox with \(G=K_2\) the one-edge graph and \(u=(4,3)\). Let
\(A,B,C,D,E,F,G\) be seven external lines. One first introduces an
intermediate line \(X\) meeting \(A,B,C,D\). The four incidence conditions
\(AX=BX=CX=DX=0\) are linear in the Plücker coordinates of \(X\), while
\(XX=0\) is the Plücker quadric. In the following we drop the brackets from
the product notation for readability. The remaining condition is that the two
transversals to \(E,F,G,X\) collide, namely the four-mass-box discriminant for
\((E,F,G,X)\). Thus the pentabox LS discriminant is computed by eliminating
\(X\) from four linear equations and two quadrics.

Concretely, write \(x_1,\ldots,x_6\) for homogeneous Plücker coordinates of
\(X\). The four incidence conditions are linear:
\begin{equation}
\label{eq:chapter7-four-linear-pentabox}
\begin{matrix}
        AX &=& AB x_2 + AC x_3 + AD x_4 + AE x_5 + AG x_6 &=&0 \, ,\\
        BX &=& AB x_1 + BC x_3 + BD x_4 + BE x_5 + BG x_6 &=&0 \, ,\\
        CX &=& AC x_1 + BC x_2 + CD x_4 + CE x_5 + CG x_6 &=&0 \, ,\\
        DX &=& AD x_1 + BD x_2 + CD x_3 + DE x_5 + DG x_6 &=&0 \, .
\end{matrix}
\end{equation}
The first quadratic is \(XX=0\). The second quadratic is
\begin{equation}
\label{eq:chapter7-pentabox-quadric-two}
        \det
        \begin{pmatrix}
        0 & EF & EG & EX\\
        EF & 0 & FG & FX\\
        EG & FG & 0 & GX\\
        EX & FX & GX & 0
        \end{pmatrix}
        =0 \, .
\end{equation}
Therefore the pentabox LS discriminant is obtained from the resultant
\begin{equation}
        \operatorname{Res}_{1,1,1,1,2,2}
\end{equation}
of four linear forms and two quadrics in the six Plücker variables of \(X\).
This resultant has an explicit determinantal formula due to
D'Andrea and Dickenstein~\cite{DAndreaDickenstein2001}. If \(\mathbf T\) is
the coefficient matrix of the four linear forms and \(\mathbf U,\mathbf V\)
are the symmetric matrices defining the two quadrics, then
\begin{equation}
\label{eq:chapter7-ten-by-ten-resultant}
        \operatorname{Res}_{1,1,1,1,2,2}(\mathbf T,\mathbf U,\mathbf V)
        =
        \det
        \begin{pmatrix}
        \mathbf B & \mathbf T^t\\
        \mathbf T & 0
        \end{pmatrix} \, ,
\end{equation}
where \(\mathbf B\) is the \(6\times6\) Bezoutian matrix. This is a
\(10\times10\) determinant, a polynomial of degree
\((4,4,4,4,2,2)\) in the coefficients of the four linear forms and two
quadrics~\cite{DAndreaDickenstein2001,HolleringMazzucchelliParisiSturmfels2026}.

After substituting the incidence data, the numerator has degree \(26\) in the
brackets among \(A,\ldots,G\), and it is multihomogeneous of degree
\begin{equation}
        (8,8,8,8,8,4,8)
\end{equation}
in the seven external Plücker vectors. The computation also produces an
extraneous factor, namely the square of the Gram determinant of
\(A,B,C,D\), of multidegree
\begin{equation}
        (4,4,4,4,4,0,4) \, .
\end{equation}
After removing this factor, the intrinsic pentabox LS discriminant has
multidegree
\begin{equation}
        (4,4,4,4,4,4,4) \, .
\end{equation}
\end{eg}

The previous example illustrates two useful points. First, non-trivial LS discriminants
can be computed by standard elimination after introducing intermediate line variables.
Second, coordinate computations may produce extraneous factors. These factors often have
a clear geometric meaning: here they come from the four-mass box subproblem determined
by the first four external lines.

\begin{eg}[Triple pentagon LS discriminant]
\label{eg:chapter7-triple-pentagon-discriminant}
Let \(G=K_3\) and \(u=(3,3,3)\). This is the triple-pentagon leading diagram.
The incidence variety \(V_{K_3}\) has two irreducible components:
\(V_{[3]}\), where the three loop lines are concurrent, and \(V^*_{[3]}\),
where they are coplanar. For generic external data, the leading fiber has
\(8+8\) points, eight on each component, so the LS degree is \(16\).

The LS discriminant has multidegree
\begin{equation}
        (48,48,48)
\end{equation}
in the three groups of external data. It factors into three irreducible
pieces. The first is the LS discriminant of the concurrent component
\(V_{[3]}\), of degree \((16,16,16)\). The second is the LS discriminant of
the coplanar component \(V^*_{[3]}\), also of degree \((16,16,16)\). The third
is the square of a mixed discriminant of degree \((8,8,8)\). The mixed factor
vanishes when a solution on \(V_{[3]}\) merges with a solution on
\(V^*_{[3]}\).
\end{eg}

This example is important because it shows what happens when the incidence variety is
reducible. The LS discriminant is not simply the product of the discriminants of the
irreducible components. There can also be mixed factors, corresponding to collisions
between solutions lying on different components. For \(K_3\), the mixed factor is
squared. The concurrent component also has a classical interpretation: its Schubert
problem amounts to intersecting three quadratic surfaces in \(\mathbb P^3\), a configuration yielding a Cayley-octad. The degree \(16\) of the component discriminant agrees with the classical discriminant degree of three quadrics in \(\mathbb P^3\).

The SLS resultant is the multigraded Chow form of \(V_G\). Since Chow forms are
multiplicative over irreducible components, if
\(V_G=\bigcup_\sigma V_{G,\sigma}\), then
\begin{equation}
\label{eq:SLS_fact}
        R_{G,u}
        =
        \prod_\sigma R_{G,u,\sigma} \, ,
\end{equation}
up to powers reflecting scheme-theoretic multiplicities. For the outerplanar examples
considered here, the components
\(V_{G,\sigma}\) are indexed by bicolorings of the triangles of \(G\), and the
resultant factors accordingly.

\begin{eg}[Double pentagon resultant]
\label{eg:chapter7-double-pentagon-resultant}
Let \(G=K_2\) and \(u=(4,4)\). This is the two-loop superleading diagram dual to the double pentagon. The generic fiber is
empty: two incident loop lines cannot generically meet four prescribed
external lines each. The SLS resultant \(R_{K_2,(4,4)}\) is an irreducible
polynomial in the eight external lines. It is homogeneous of degree \(4\) in
each of the eight external Plücker vectors:
\begin{equation}
        \deg R_{K_2,(4,4)}
        =
        (4,4,4,4;\,4,4,4,4) \, .
\end{equation}
By the degree formula~\eqref{eq:SLS_deg}, each entry is
the LS degree obtained by removing one of the eight external
conditions.
\end{eg}

\begin{eg}[Triangle SLS resultant]
\label{eg:chapter7-triangle-SLS-resultant}
Let \(G=K_3\) and \(u=(4,3,3)\). This is superleading, since the generic fiber
is empty. Because \(V_{K_3}=V_{[3]}\cup V^*_{[3]}\), the SLS resultant is the
product of two irreducible factors, one for the concurrent component and one
for the coplanar component. Each factor has degree
\begin{equation}
        (8,8,8,8;\,4,4,4;\,4,4,4)
\end{equation}
in the ten external lines, grouped according to the three vertices. Therefore
the full SLS resultant has degree
\begin{equation}
        (16,16,16,16;\,8,8,8;\,8,8,8) \, .
\end{equation}
These numbers are precisely the LS degrees \(\gamma_{u-e_i}\), as predicted
by~\eqref{eq:SLS_deg}.
\end{eg}

The degree formulas above are not only theoretical. They are implemented using the
computational framework for multigraded Hurwitz and Chow forms developed
in~\cite{PrattSodomacoSturmfels2026}, together with symbolic computations in
\texttt{Macaulay2}~\cite{M2}. In practice, the multidegree of \(V_G\) gives
the LS degrees, the Chow-form formula gives the degrees of SLS resultants, and the
Hurwitz-form formula gives the expected degrees of LS discriminants. The examples above
are useful checks: the four-mass box and pentagon give the classical Gram determinants,
the pentabox gives a \(10\times10\) determinant after elimination, the triple pentagon
shows mixed factors for reducible incidence varieties, and the double pentagon gives an
irreducible two-loop SLS resultant of uniform degree \(4\) in each external line.

The leading and superleading cases correspond to zero-dimensional behavior of the Landau
map: finite fibers or generically empty fibers. The next case is qualitatively
different. When the generic fiber is a curve, the relevant singularities are controlled
not by collisions of finitely many points, but by degenerations of the curve itself.
This is the next-to-leading case, to which we now turn.

\subsubsection{Next-to-leading singularities}
\label{subsec:next-to-leading-singularities}

Let \(G\subseteq\binom{[\ell]}{2}\) be a loop-incidence graph, and let
\(m=\dim V_G\). We again attach \(u_i\) external lines to the \(i\)-th loop
line, with \(u=(u_1,\ldots,u_\ell)\) and \(|u|=\sum_i u_i\). The generic fiber of the Landau map has expected dimension \(m-|u|\). The next-to-leading case is the
first positive-dimensional case:
\begin{equation}
        |u|=m-1 \, .
\end{equation}
For generic external data \(\mathbf M\), the fiber
\(\psi_u^{-1}(\mathbf M)\) is then a projective curve. The corresponding
\emph{NLS discriminant} is the hypersurface in external kinematic space where
this curve becomes singular.

\begin{tcolorbox}[resultbox]
\textbf{Next-to-leading singularities.}
In the NLS case the generic fiber of the Landau map is a curve. The NLS
discriminant is the locus in external kinematics where this curve degenerates,
for example by acquiring a node, cusp, or reducible component.
\end{tcolorbox}
\noindent

The basic numerical invariant of an NLS diagram is the genus of the generic fiber. If
the generic fiber is smooth and irreducible, this is the usual genus of the
corresponding Riemann surface. More generally, the generic fiber may be reducible, and
we use its geometric genus, namely the sum of the genera of its irreducible components
corrected by the number of connected components. We call this number \(g_u\) the \emph{NLS
genus}.

\begin{tcolorbox}[definitionbox]
\textbf{NLS genus.}
Let \(G_u\) be next-to-leading, so that \(|u|=\dim V_G-1\). The NLS genus
\(g_u\) is the geometric genus of the generic curve fiber
\(\psi_u^{-1}(\mathbf M)\).
\end{tcolorbox}
\noindent

The genera for all next-to-leading extensions of a fixed graph \(G\) can be packaged
into a generating function, called the \emph{multisectional genus}:
\begin{equation}
\label{eq:chapter7-multisectional-genus}
        g(G)
        =
        \sum_{|u|=\dim V_G-1}
        g_u\,t_1^{5-u_1}\cdots t_\ell^{5-u_\ell} \, .
\end{equation}
This invariant is computed from the same intersection-theoretic framework used for
Hurwitz forms. In particular, the software developed for multigraded Hurwitz forms
in~\cite[Section~6]{PrattSodomacoSturmfels2026} also computes these genera.

\begin{eg}[A single loop line]
\label{eg:chapter7-one-loop-NLS}
Consider the one-loop triangle, $G=k_1$ and $u=(3)$. The fiber of the Landau map consists of all lines
\(X\subseteq\mathbb P^3\) intersecting three fixed external lines
\(M_1,M_2,M_3\). These three Schubert conditions cut out a plane in the
Plücker space \(\mathbb P^5\). Intersecting this plane with $\Gr(2,4)$ gives a plane conic. Hence the NLS genus is
\begin{equation}
        g_{(3)}=0 \, .
\end{equation}
The NLS discriminant is the condition that this conic degenerates into two
lines. It can be computed by solving the three linear equations
\(M_1X=M_2X=M_3X=0\), substituting into the Plücker quadric \(XX=0\), and
taking the determinant of the resulting quadratic form. The result is a polynomial of multidegree \((3,3,3)\) in the three external lines. This is the Landau singularity associated to a one-loop triangle with generic massive corners.
\end{eg}

The next simplest example already exhibits elliptic geometry. 

\begin{eg}[The double box]
\label{eg:chapter7-NLS-double-box}
Consider the double-box
Landau diagram, for which \(G=K_2\) and $u=(3,3)$. Then the multisectional genus of $G$
is
\begin{equation}
\label{eq:chapter7-gK2}
        g(K_2)
        =
        t_1^4-t_1^3t_2+t_1^2t_2^2-t_1t_2^3+t_2^4 \, .
\end{equation}
The coefficient of \(t_1^2t_2^2\) records the genus of the curve associated to
next-to-leading singularity of the double-box. We find \(g_{(3,3)}=1\), which is the
genus of an elliptic curve. This agrees with the elliptic leading singularity found for
the double-box with massive corners in~\cite{BourjailyKalyanapuramLangerPatatoukos2021}.
This singularity appears for the first time in the $n=12$-point $\ell=3$-loop ${\rm
N}^3MHV$ amplitude in planar $\mathcal{N}=4$ SYM, giving the first example of an
amplitude in planar $\mathcal{N}=4$ SYM whose function space is not polilogarithmic.

We explain how to compute the NLS discriminant for this case. Take six external lines \(A,B,C;E,F,G\). The fiber consists
of pairs of incident loop lines \((X,Y)\) such that
\begin{equation}
        AX=BX=CX=0,
        \qquad
        EY=FY=GY=0,
        \qquad
        XY=0 \, .
\end{equation}
Geometrically, the three lines \(A,B,C\) determine a unique quadric surface in
\(\mathbb P^3\), and the three lines \(E,F,G\) determine a second quadric
surface. The NLS fiber is the intersection of these two quadrics. Hence the
generic fiber is a smooth complete intersection of two quadrics in
\(\mathbb P^3\), and therefore an elliptic curve:
\begin{equation}
        g_{(3,3)}=1 \, .
\end{equation}
\end{eg}

The NLS discriminant of the double box is the condition that this elliptic curve becomes
singular. Let \(U\) and \(V\) be the symmetric \(4\times4\) matrices defining the two
quadrics determined by \(A,B,C\) and \(E,F,G\). Then the pencil of quadrics is \(U+tV\),
and the singular members of the pencil are detected by the quartic
\begin{equation}
        f(t)=\det(U+tV) \, .
\end{equation}
The double-box NLS discriminant is the discriminant of this quartic.

\begin{tcolorbox}[resultbox]
\textbf{Double-box NLS discriminant.}
For the double box \(G=K_2\), \(u=(3,3)\), the NLS discriminant is the mixed
discriminant of the two quadrics determined by the triples of external lines
\((A,B,C)\) and \((E,F,G)\). Equivalently, it is the univariate discriminant in $t$ of
\(\det(U+tV)\).
\end{tcolorbox}
\noindent

The resulting polynomial has degree \(12\) in the Plücker coordinates of each of the six
external lines. The entries of \(U\) are explicit trilinear functions of the Plücker
coordinates of \(A,B,C\), and the entries of \(V\) are obtained by replacing \(A,B,C\)
with \(E,F,G\). This gives a concrete formula for the NLS discriminant, although the
fully expanded polynomial is far too large to be useful in print. Indeed, the
discriminant of a general quartic \(\det(U+tV)\) already has tens of millions of
monomials in the entries of the two symmetric matrices~\cite{BFS}.

There is an important contrast between \(u=(3,3)\) and other NLS extensions of the same
graph. For example, take again \(G=K_2\), but now \(u=(4,2)\). The coefficient of
\(t_1t_2^3\) in~\eqref{eq:chapter7-gK2} gives
\begin{equation}
        g_{(4,2)}=-1 \, .
\end{equation}
This negative genus is not a contradiction. It reflects the fact that the generic fiber
is reducible: in this case it is the disjoint union of two conics, each of genus \(0\).
The geometric genus of a disjoint union of two rational curves is \(0+0-1=-1\).

\begin{eg}[Three loop lines]
\label{eg:chapter7-NLS-three-lines}
Let \(G=\{12,23\}\) be the chain on three vertices and let \(u=(3,3,3)\).
The three Schubert conditions at each vertex cut out a \(\mathbb P^2\) inside
the corresponding Plücker \(\mathbb P^5\). The generic fiber is therefore a
curve inside \((\mathbb P^2)^3\), cut out by the three Plücker quadrics and
the two incidence equations \(X_1X_2=X_2X_3=0\). This is a smooth irreducible
complete-intersection curve of genus
\begin{equation}
        g_{(3,3,3)}=5 \, .
\end{equation}
Now let \(G=K_3\) and \(u=(3,3,2)\). The incidence variety \(V_{K_3}\) has two
components, the concurrent component \(V_{[3]}\) and the coplanar component
\(V^*_{[3]}\). The generic NLS fiber is the disjoint union of two genus-five
curves, one on each component. Hence its NLS genus is
\begin{equation}
        5+5-1=9 \, .
\end{equation}
\end{eg}

\subsection{Recursive Landau analysis}
\label{sec:recursive-landau-analysis}

The previous section described Landau singularities as singularities of the Landau map.
In the leading case the generic fiber is finite, and the branch locus is encoded by the
LS discriminant. In the superleading case the generic fiber is empty, and the locus
where a solution appears is encoded by an SLS resultant. We now explain why many of
these objects admit a recursive structure.

The guiding idea is simple. Suppose that a subdiagram is already leading. For fixed
external data its loop lines are then finitely many algebraic functions of that data.
Each of these solutions can be inserted into the rest of the diagram as an effective
external line. Thus a large Landau problem is reduced to a finite collection of smaller
Landau problems. The most important instance for us is recursion by four-mass boxes:
solving a box produces two transversals, the two solutions to the classical Schubert
problem of four lines in $\mathbb{P}^3$. These result in two substitution maps that will
later be responsible for positivity in Section~\ref{sec:reality-positivity-positroids}
and, after rational degeneration, for cluster promotion maps in
Section~\ref{sec:rationality-cluster-structures}. This recursion, physically in the
order of loops, is at the core for computations and proofs about singularities at all
loop orders, and extends beyond the leading and superleading cases discussed here.

The results in this section summarize the recursive structures developed in
\cite[Section~7]{HolleringMazzucchelliParisiSturmfels2026} and in the shorter account
\cite{HolleringMazzucchelliParisiSturmfels2026Positivity}. Related but different recursive
Landau methods in momentum space were developed in
\cite{CaronHuotCorreiaGiroux2025}. We present the construction in a form adapted to
the notation of this chapter. In particular, $G$ denotes the dual loop graph,
$u=(u_1,\ldots,u_\ell)$ records the number of external lines attached to the
loop vertices, and $\mathcal L=G_u$ denotes the corresponding Landau diagram. When we
allow the external lines to satisfy additional incidence relations. Such degenerations
and their physical meaning will become clear below. Let $H_u$ be a graph on the $|u|$
external vertices, encoding the incidences imposed among the external lines. We write
\begin{equation}
V_{H_u}\subseteq \Gr(2,4)^{|u|}
\end{equation}
for the corresponding external incidence variety; if no external incidences are imposed,
then $H_u=\emptyset$ and $V_{H_u}=\Gr(2,4)^{|u|}$. The on-shell space of
the degenerated Landau diagram
$\mathcal L=G_u\cup H_u$ is
\begin{equation}
V_{\mathcal L}
=
\left\{
(\mathbf L,\mathbf M)\in V_G\times V_{H_u}
:
\langle L_iM_a\rangle=0
\text{ for every pendant edge of }G_u
\right\} .
\label{eq:on-shell-space-with-H}
\end{equation}
The Landau map is the projection to the external data,
\begin{tcolorbox}[definitionbox]
\textbf{Landau map with external incidences.}
\begin{equation}
\psi_{\mathcal L}:V_{\mathcal L}\longrightarrow V_{H_u},
\qquad
(\mathbf L,\mathbf M)\longmapsto \mathbf M .
\label{eq:landau-map-with-H}
\end{equation}
\end{tcolorbox}\noindent
Thus the non-generic fiber locus of $\psi_{\mathcal L}$ is a subset of
$V_{H_u}$, and in the generic case $H_u=\emptyset$ it is a subset of
$\Gr(2,4)^{|u|}$, which matches our previous definition.

\subsubsection{Geometric recursion}

Let the loop vertices of $G$ in $\mathcal{L}$ as above be partitioned into two non-empty
subsets
\begin{equation}
[\ell]=V_1\cup V_2 \, .
\end{equation}
Let $G_1$ and $G_2$ be the induced subgraphs on $V_1$ and $V_2$, and let $E$ be the set
of edges of $G$ connecting $V_1$ to $V_2$. Thus
\begin{equation}
G=G_1\cup G_2\cup E \, .
\end{equation}
We split the external valency vector and the external data accordingly:
\begin{equation}
u=(u_1,u_2),
\qquad
\mathbf M=(\mathbf M^{(1)},\mathbf M^{(2)}) \, .
\end{equation}
The corresponding diagrams are
\begin{equation}
\mathcal L_1=G_{1,u_1}\cup H_{u_1} \, ,
\qquad
\mathcal L_2=G_{2,u_2}\cup H_{u_2}\cup E .
\label{eq:recursive-L1-L2}
\end{equation}
The edges $E$ are included in $\mathcal L_2$ because, after solving
$\mathcal L_1$, the lines on the $V_1$ side become fixed lines that are incident
to the remaining loop lines.

We say that $\mathcal L$ is \emph{reducible with respect to} $\mathcal L_1$ if
$\mathcal L_1$ is leading, namely if
\begin{equation}
|u_1|=\dim V_{G_1} \, .
\end{equation}
For generic $\mathbf M^{(1)}$, the fiber of the Landau map
$\psi_1=\psi_{\mathcal L_1}$ is finite. We write
\begin{equation}
\psi_1^{-1}(\mathbf M^{(1)})
=
\left\{
\mathbf L^{(1)}_1(\mathbf M^{(1)}),\ldots,
\mathbf L^{(1)}_{\gamma_1}(\mathbf M^{(1)})
\right\},
\label{eq:L1-finite-fiber}
\end{equation}
where $\gamma_1$ is the LS degree of $\mathcal L_1$. Each point of this fiber defines a
substitution map
\begin{tcolorbox}[definitionbox]
\textbf{Recursive substitution map.}
\begin{equation}
\varphi_r:
\Gr(2,4)^{|u|}
\dashrightarrow
\Gr(2,4)^{|E|+|u_2|} \, ,
\qquad
\mathbf M=(\mathbf M^{(1)},\mathbf M^{(2)})
\longmapsto
\bigl(\mathbf L^{(1)}_r(\mathbf M^{(1)}),\mathbf M^{(2)}\bigr) \, .
\label{eq:general-substitution-map}
\end{equation}
\end{tcolorbox}\noindent

The dashed arrow indicates that the entries of
\(\mathbf L^{(1)}_r(\mathbf M^{(1)})\) are, in general, algebraic functions of
\(\mathbf M^{(1)}\). For a general leading subdiagram \(\mathcal L_1\), these
solutions need not be expressible by radicals; hence \(\varphi_r\) should be understood
as a locally chosen branch of the finite algebraic correspondence defined by the Landau
map \(\psi_{\mathcal L_1}\). Such a branch is well defined only on a simply connected
open subset of
\begin{equation}\label{eq:targ}
	V_{H_{u_1}}\setminus \{\Delta_{\mathcal L_1}=0\} \, ,
\end{equation}
or of \(\Gr(2,4)^{|u_1|}\setminus \{\Delta_{\mathcal L_1}=0\}\) in the
absence of external incidence constraints. The label \(r\) therefore has no global
meaning: it is fixed only after choosing a simply connected domain in~\eqref{eq:targ},
where the points of the finite fiber of \(\psi_{\mathcal L_1}\) can be followed without
monodromy.

\begin{tcolorbox}[resultbox]
\textbf{Geometric recursion.}
If $\mathcal L$ is reducible with respect to the leading subdiagram
$\mathcal L_1$, then the fiber of the full Landau map decomposes as
\begin{equation}
\psi_{\mathcal L}^{-1}(\mathbf M)
=
\bigcup_{\mathbf L^{(1)}\in\psi_1^{-1}(\mathbf M^{(1)})}
\left\{
(\mathbf L^{(1)},\mathbf L^{(2)})
:
\mathbf L^{(2)} \, \in \, 
\psi_2^{-1}\bigl(\varphi_{\mathbf L^{(1)}}(\mathbf M)\bigr)
\right\} .
\label{eq:recursive-fiber-decomposition-v2}
\end{equation}
In words, one first solves the leading subdiagram $\mathcal L_1$; each solution
is then substituted into the remaining Landau problem $\mathcal L_2$ as external data.
\end{tcolorbox}\noindent
Indeed, the incidence equations of $\mathcal L$ split into equations internal to
$G_1$, equations internal to $G_2$, and the connecting equations associated with
$E$. Once a solution of the first group has been chosen, the connecting
equations involve the remaining loop variables only through incidence with fixed lines.
These fixed lines are precisely the extra external data $\mathbf{M}^{(2)}$ of the
remaining problem.

\begin{figure}[pos=t]
    \centering
    \includegraphics[width=1.00\textwidth]{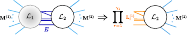}
    \caption{A Landau diagram reducible with respect to
    $\mathcal L_1=G_{1,u_1}\cup H_{u_1}$. Solving the leading subdiagram
    produces finitely many configurations $\mathbf L^{(1)}_r$. These are
    inserted into the residual diagram $\mathcal L_2=G_{2,u_2}\cup H_{u_2}\cup E$
    as effective external lines.}
    \label{fig:recursive-landau-general}
\end{figure}

The same statement has a useful interpretation at the level of branch loci. Let
\begin{equation}
\mathfrak B_{\mathcal L}\subseteq V_{H_u}
\end{equation}
denote the non-generic fiber locus of the Landau map
$\psi_{\mathcal L}:V_{\mathcal L}\to V_{H_u}$. Thus, when
$H_u=\emptyset$, this is a subset of $\Gr(2,4)^{|u|}$. Then a point
$\mathbf M=(\mathbf M^{(1)},\mathbf M^{(2)})\in V_{H_u}$ can be non-generic for the full
map in two ways. Either $\mathbf M^{(1)}$ is already a branch point of
$\psi_1$, or for at least one branch
$\mathbf L^{(1)}_r(\mathbf M^{(1)})$ the substituted data
$\varphi_r(\mathbf M)$ is non-generic for the residual map $\psi_2$. Thus,
set-theoretically,
\begin{equation}
\mathfrak B_{\mathcal L}
=
\pi_1^{-1}\bigl(\mathfrak B_{\mathcal L_1}\bigr)
\cup
\bigcup_{r=1}^{\gamma_1}
\varphi_r^{-1}\bigl(\mathfrak B_{\mathcal L_2}\bigr)
 .
\label{eq:branch-locus-recursion}
\end{equation}
Here $\pi_1:V_{H_u}\to V_{H_{u_1}}$ is the projection forgetting
$\mathbf M^{(2)}$, and $\mathfrak B_{\mathcal L_2}$ is understood inside the
target of the residual map $\psi_2$. This formulation is slightly more general than the
discriminant and resultant formulas below. It only requires the first piece $\mathcal
L_1$ to be leading. The residual piece $\mathcal L_2$ may itself be leading,
superleading, subleading, etc. 

\subsubsection{Recursion by boxes}
\label{subsec:box-substitution-maps}

We now specialize the preceding discussion to a basic seed of the recursion: the
four-mass box. This is the simplest possible seed, for which $G_1=K_1$ is the one-vertex
graph. Let $\mathbf M^{(1)}=(A,B,C,D)$ be four generic external lines in $\mathbb P^3$. There are $\gamma_1=2$ transversal lines
$L^{\pm}(A,B,C,D)$ to $A,B,C,D$, which we now
write explicitly.

We write intersections and joins of linear spaces in $\mathbb P^3$ using the
\emph{Grassmann--Cayley algebra}. Concatenation denotes join and $\cap$ denotes intersection.
Choose two points $d_1,d_2$ spanning the line $D$, so $D=d_1d_2$. Then
\begin{equation}
X(x)=\bigl(p(x)B\bigr)\cap\bigl(p(x)C\bigr)
\label{eq:X-of-x-box-v2}, \qquad p(x)=d_1+x d_2 \, ,
\end{equation}
is a line meeting $B$, $C$, and $D$. The remaining incidence with $A$ gives a quadratic
equation
\begin{equation}
f(x)=\langle A \, X(x)\rangle=a_2x^2+a_1x+a_0 \, .
\label{eq:box-quadratic-v2}
\end{equation}
Its coefficients are \textit{chain polynomials}
\cite{HolleringMazzucchelliParisiSturmfels2026Positivity}:
\begin{equation}
a_2=\langle d_2 A|B|d_2 C\rangle \, ,
\quad
a_1=\langle d_1 A|B|d_2 C\rangle \,+ \,
     \langle d_2 A|B|d_1 C\rangle \, ,
\quad
a_0=\langle d_1 A|B|d_1 C\rangle \, ,
\label{eq:box-chain-coefficients-v2}
\end{equation}
where $\langle abc|de|fgh\rangle
:=
\langle abcd\rangle\langle efgh\rangle
-
\langle abce\rangle\langle dfgh\rangle \, .$

The two roots of $f(x)$ are
\begin{equation}
x^{(\pm)}=
\frac{-a_1\pm\sqrt{a_1^2-4a_0a_2}}{2a_2} \, ,
\label{eq:box-roots-v2}
\end{equation}
and hence the two transversals are $X^{\pm}=X(x^{\pm})$. We use the normalized
representatives
\begin{tcolorbox}[definitionbox]
\textbf{Normalized four-mass box solutions.}
\begin{equation}
\begin{aligned}
	&L^{\pm}(A,B,C,D)
=
\varepsilon(A,B,C,D)\cdot \bigl(p(x^{(\pm)})B\bigr)\cap\bigl(p(x^{(\pm)})C\bigr) \, ,
\\
&\text{with } \ \varepsilon(A,B,C,D)=
\frac{2a_2}{\sqrt{\langle BC\rangle\langle BD\rangle\langle CD\rangle}} \, .
\end{aligned}
\label{eq:Lpm-normalization-v2}
\end{equation}
\end{tcolorbox}\noindent
With this normalization, the Pl\"ucker coordinates of $L^\pm$ have degree
$(1,1,1,1)$ in the external lines $A,B,C,D$. As we will see, with this normalization the box recursion produces no extra prefactor.

\begin{tcolorbox}[definitionbox]
\textbf{Four-mass box substitution maps.}
The two solutions of the four-mass box define algebraic maps
\begin{equation}
\varphi_\pm:
\Gr(2,4)^{4+d_2}
\dashrightarrow
\Gr(2,4)^{1+d_2} \, ,
\quad
(A,B,C,D,\mathbf M^{(2)})
\longmapsto
\bigl(L^\pm(A,B,C,D),\mathbf M^{(2)}\bigr) \, .
\label{eq:four-mass-box-substitution-maps-v2}
\end{equation}
\end{tcolorbox}\noindent

To apply this in practice, it is both useful and instructive to work out a context-free
toy example. Consider the system of two polynomials in two variables
\begin{equation}
\begin{aligned}
f(x)&=a_2x^2+a_1x+a_0 \, ,\\[2mm]
g(x,y)&=b_{22}x^2y^2+b_{21}x^2y+b_{12}xy^2+b_{20}x^2+b_{11}xy
+b_{02}y^2+b_{10}x+b_{01}y+b_{00} \, .
\end{aligned}
\label{eq:fg-polynomial-system-v2}
\end{equation}
Solving $f(x)=0$ gives the roots $x^{(+)}$ and $x^{(-)}$. Evaluating the
$y$-discriminant of $g$ on these two roots and multiplying gives
\begin{equation}
\left.\operatorname{Disc}_y(g)\right|_{x=x^{(+)}}
\cdot
\left.\operatorname{Disc}_y(g)\right|_{x=x^{(-)}}
=
a_2^{-4}\operatorname{Res}_x\bigl(f,\operatorname{Disc}_y(g)\bigr)
=
a_2^{-4}\operatorname{Disc}(f,g) \, .
\label{eq:polynomial-ls-recursion-v2}
\end{equation}
The polynomial $\operatorname{Disc}(f,g)$ is irreducible, has $150$ terms, and has
bidegree $(4,4)$ in the coefficients of $f$ and $g$. This elementary identity is the
algebraic prototype for the pentabox recursion~\cite{gkz}.

\begin{eg}[The pentabox]
\label{eg:pentabox-recursion}
Let $G=K_2$ and $u=(4,3)$. Denote the two loop lines by
$\mathbf L=(X,Y)$ and the external data by
\begin{equation}
\mathbf M=(A,B,C,D,E,F,G) \, ,
\qquad
\mathbf M^{(1)}=(A,B,C,D) \, ,
\qquad
\mathbf M^{(2)}=(E,F,G) \, .
\end{equation}
The line $X$ is the four-mass-box line meeting $A,B,C,D$. To parametrize the
residual line, write $G=g_1g_2$ and set
\begin{equation}
q(y)=g_1+yg_2,
\qquad
Y(y)=\bigl(q(y)E\bigr)\cap\bigl(q(y)F\bigr) \, .
\end{equation}
Then $Y(y)$ meets $E,F,G$, and the remaining incidence between the two loop
lines is
\begin{equation}
g(x,y)=\langle X(x)Y(y)\rangle \, .
\end{equation}
The system is exactly of the form in
\eqref{eq:fg-polynomial-system-v2}. Therefore
\begin{equation}
\operatorname{Disc}_x(f)
=a_1^2-4a_0a_2
=\Delta_{K_1,(4)}(A,B,C,D) \, ,
\end{equation}
and
\begin{equation}
\operatorname{Disc}(f,g)
=
\langle BC\rangle^2\langle BD\rangle^2\langle CD\rangle^2
\cdot
\Delta_{K_2,(4,3)}(\mathbf M) \, .
\label{eq:pentabox-discriminant-prefactor-v2}
\end{equation}
This formula is verified by checking that the degrees in the external lines
\(\mathbf{M}\) match on the two sides of each equation.
After the normalization \eqref{eq:Lpm-normalization-v2}, this prefactor is
removed by the two substitutions. The pentabox discriminant is
\begin{equation}
\Delta_{K_2,(4,3)}(\mathbf M)
=
\prod_{\sigma \, \in \, \{+,-\}}
\Delta_{K_1,(4)}\bigl(\varphi_\sigma(\mathbf M)\bigr).
\label{eq:pentabox-recursive-formula-v2}
\end{equation}
Thus the two-loop problem is reduced to two one-loop box discriminants. 

From~\eqref{eq:four-mass-box-substitution-maps-v2}, even though the four-mass box substitution maps evaluate a function to an algebraic expression in $A,B,C,D$, the product over both roots is a symmetric function in the roots, and hence a rational function in $\mathbf{M}$. In principle, this rational function may have extraneous factors compared to the desired discriminant. The normalization in~\eqref{eq:Lpm-normalization-v2} fixes the projective scaling of the two Schubert solutions so that each Pl\"ucker coordinate of \(L^\pm\) has degree \((1,1,1,1)\) in the four external lines. This choice is conventional but not arbitrary: it is permitted because a line in \(\mathbb P^3\) is defined only projectively, and it is chosen so that the four-mass box substitution produces no extra prefactor. This can be verified, for example, by counting the degrees in the external data $\mathbf{M}$ on both sides of~\eqref{eq:pentabox-recursive-formula-v2}.
\end{eg}

\begin{figure}[pos=t]
    \centering
    \includegraphics[width=0.72\textwidth]{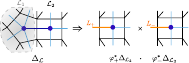}
    \caption{Recursion for the LS discriminant of the pentabox. The left
    four-mass box has two solutions $L^+$ and $L^-$. Substituting them into the
    residual one-loop problem gives the two factors in
    \eqref{eq:pentabox-recursive-formula-v2}.}
    \label{fig:pentabox-recursion}
\end{figure}

\subsubsection{Recursive factorization of discriminants and resultants}
\label{subsec:recursive-factorization-discriminants-resultants}

The pentabox computation is a special case of a general factorization principle. Assume
first that $\mathcal L$ and $\mathcal L_2$ are leading. Let the finite fiber of
$\mathcal L_1$ be written as in \eqref{eq:L1-finite-fiber}, and let
$\varphi_r$ be the corresponding substitution maps.

\begin{tcolorbox}[resultbox]
\textbf{Recursive factorization of LS discriminants.}
If the leading diagram $\mathcal L$ is reducible with respect to the leading
subdiagram $\mathcal L_1$, then
\begin{equation}
\Delta_{\mathcal L}(\mathbf M)
=
\mathcal E(\mathbf M^{(1)})
\prod_{r=1}^{\gamma_1}
\Delta_{\mathcal L_2}\bigl(\varphi_r(\mathbf M)\bigr).
\label{eq:ls-discriminant-recursion-v2}
\end{equation}
The factor $\mathcal E(\mathbf M^{(1)})$ is rational in the Pl\"ucker coordinates
of the external lines $\mathbf M^{(1)}$ and reflects the projective normalization of the algebraic
solutions of $\mathcal L_1$.
\end{tcolorbox}\noindent
The product in \eqref{eq:ls-discriminant-recursion-v2} is symmetric in the points of the
finite fiber of $\psi_1$, and therefore descends from the algebraic cover on which the
individual solutions are defined to a rational function of the original external data.
Note that we already encountered the product expression on the right hand side
of~\eqref{eq:ls-discriminant-recursion-v2} in Chapter~\ref{ch:From Integrands to Integrals}, where we discussed branch covers and algebraic letters. In fact, that
product is the norm of $\Delta_{\mathcal{L}_2}$ with respect to the universal cover of $\Gr(2,4)^{|u|} \setminus \{\Delta_{\mathcal{L}_1} = 0\}$.

For the four-mass box the normalization \eqref{eq:Lpm-normalization-v2} removes the
ambiguity $\mathcal E(\mathbf M^{(1)})$.
\begin{tcolorbox}[resultbox]
\textbf{Box recursion for LS discriminants.}
If $\mathcal L$ is leading and reducible with respect to a four-mass box, then
\begin{equation}
\Delta_{\mathcal L}(\mathbf M)
=
\prod_{\sigma\in\{+,-\}}
\Delta_{\mathcal L_2}\bigl(\varphi_\sigma(\mathbf M)\bigr),
\label{eq:box-recursion-ls-v2}
\end{equation}
with $\varphi_\pm$ given by \eqref{eq:four-mass-box-substitution-maps-v2}.
\end{tcolorbox}\noindent

The same mechanism applies to superleading diagrams, where resultants replace
discriminants. We start again from a context-free polynomial model. Add a quadratic
\begin{equation}
h(y)=c_2y^2+c_1y+c_0
\end{equation}
to the polynomial system in~\eqref{eq:fg-polynomial-system-v2}. The resultant
$\operatorname{Res}(f,g,h)$ is irreducible, has $1340$ terms, and has degree
$(4,4,4)$ in the coefficients of $f,g,h$. Solving $f(x)=0$ first gives
\begin{equation}
\left.\operatorname{Res}_y(g,h)\right|_{x=x^{(+)}}
\cdot
\left.\operatorname{Res}_y(g,h)\right|_{x=x^{(-)}}
=
a_2^{-4}\operatorname{Res}_y\bigl(\operatorname{Res}_x(f,g),h\bigr)
=
a_2^{-4}\operatorname{Res}(f,g,h).
\label{eq:polynomial-sls-recursion-v2}
\end{equation}
This is the resultant analogue of the discriminant identity
\eqref{eq:polynomial-ls-recursion-v2}; both identities are standard examples of elimination by resultants~\cite{gkz}.

\begin{eg}[Pentagon and double-pentagon resultants]
\label{eg:double-pentagon-recursion}
We use the same parametrizations $X(x)$ and $Y(y)$ as in the pentabox example,
but now impose one additional external incidence. Set
\begin{equation}
f(x)=\langle A \, X(x)\rangle,
\qquad
g(x,y)=\langle X(x) \, Y(y)\rangle,
\qquad
h(y)=\langle H \, Y(y)\rangle \, .
\end{equation}
Then the one-loop pentagon and the two-loop double pentagon have SLS resultants
normalized by
\begin{equation}
\operatorname{Res}_x\bigl(f,\langle EX\rangle\bigr)
=
\langle BC\rangle\langle CD\rangle\langle BD\rangle
\cdot R_{K_1,(5)}(A,B,C,D,E)
\label{eq:pentagon-resultant-normalization-v2}
\end{equation}
and
\begin{equation}
\operatorname{Res}(f,g,h)
=
\langle BC\rangle^2\langle BD\rangle^2\langle CD\rangle^2
\langle EF\rangle^2\langle EG\rangle^2\langle FG\rangle^2
\cdot R_{K_2,(4,4)}(\mathbf M).
\label{eq:double-pentagon-resultant-normalization-v2}
\end{equation}
Again, these equations are proven by counting degrees on both sides.
After applying the same box normalization as before, one obtains
\begin{equation}
R_{K_2,(4,4)}(\mathbf M)
=
\prod_{\sigma\in\{+,-\}}
R_{K_1,(5)}\bigl(\varphi_\sigma(\mathbf M)\bigr).
\label{eq:double-pentagon-recursive-formula-v2}
\end{equation}
Thus the double-pentagon SLS resultant is reduced to two one-loop pentagon
resultants.
\end{eg}

\begin{tcolorbox}[resultbox]
\textbf{Recursive factorization of SLS resultants.}
If the superleading diagram $\mathcal L$ is reducible with respect to the
leading subdiagram $\mathcal L_1$, 
\begin{equation}
R_{\mathcal L}(\mathbf M)
=
\mathcal E(\mathbf M^{(1)})
\prod_{r=1}^{\gamma_1}
R_{\mathcal L_2}\bigl(\varphi_r(\mathbf M)\bigr).
\label{eq:sls-resultant-recursion-v2}
\end{equation}
In particular, if $\mathcal L_1$ is a four-mass box and the maps are normalized
as in \eqref{eq:Lpm-normalization-v2}, then
\begin{equation}
R_{\mathcal L}(\mathbf M)
=
\prod_{\sigma\in\{+,-\}}
R_{\mathcal L_2}\bigl(\varphi_\sigma(\mathbf M)\bigr).
\label{eq:box-recursion-sls-v2}
\end{equation}
\end{tcolorbox}\noindent
Equations \eqref{eq:box-recursion-ls-v2} and
\eqref{eq:box-recursion-sls-v2} are the forms of recursion that will be used
below. They turn the algebraic square root of the four-mass box into a controlled
two-branch substitution. The individual factors may be algebraic, but the product over
both branches is rational.

\subsubsection{Trees of loops by recursion}
\label{subsec:trees-of-loops-recursion}

The box recursion is especially effective when the dual loop graph $G$ is a tree. In
that case $V_G$ is a complete intersection and
\begin{equation}
\dim V_G=3\ell+1 \, .
\end{equation}
Equivalently, its multidegree is
\begin{equation}
[V_G]
=
2^\ell\prod_{i=1}^{\ell}t_i
\prod_{ij \, \in \, G}(t_i+t_j).
\label{eq:tree-multidegree-v2}
\end{equation}
A leading vector $u$ is a term in this multidegree, so $|u|=3\ell+1$. The tree structure
implies that at least one vertex has $u_i=4$. To see this, choosing a monomial in
\eqref{eq:tree-multidegree-v2} amounts to choosing one endpoint for each edge of the
tree. If $m_i$ is the number of edges chosen at vertex $i$, then $u_i=4-m_i$. Every
orientation of a tree has a source, so some $m_i=0$, and hence some
$u_i=4$. That vertex is a four-mass box and thus we can apply the recursion.

\begin{tcolorbox}[definitionbox]
\textbf{Trees of loops.}
If $G$ is a tree, every leading Landau diagram $G_u$ can be reduced by iterated
four-mass box substitutions. One repeatedly chooses a vertex with $u_i=4$,
solves the corresponding box, substitutes $L^+$ or $L^-$ into the adjacent
vertex, and continues on the smaller tree. Hence the discriminant is a product over sign choices made during this successive removal of four-mass-boxes.
\end{tcolorbox}\noindent
More explicitly, choose an order
\(v_1,\ldots,v_{\ell-1}\) in which four-mass-box leaves are successively removed from the tree, leaving a final one-loop seed $v_\ell$. For a sign vector
$\sigma=(\sigma_1,\ldots,\sigma_{\ell-1})\in\{+,-\}^{\ell-1}$, let
\begin{equation}
\Phi_\sigma
=
\varphi_{\sigma_{\ell-1}}\circ\cdots\circ\varphi_{\sigma_1}
\end{equation}
be the corresponding iterated substitution map. Then
\begin{equation}
\Delta_{G,u}(\mathbf M)
=
\prod_{\sigma\in\{+,-\}^{\ell-1}}
\Delta_{\mathrm{seed}}\bigl(\Phi_\sigma(\mathbf M)\bigr),
\label{eq:tree-iterated-ls-recursion-v2}
\end{equation}
where $\Delta_{\mathrm{seed}}$ is the one-loop discriminant obtained at $v_\ell$.
Different choices of successive resolution give different presentations of the same
rational function. A statement analogous to~\eqref{eq:tree-iterated-ls-recursion-v2}
holds for SLS resultants of superleading trees.

\begin{eg}[Paths and their roots combinatorics]
\label{eg:path-trees-recursion}
Let $G=P_\ell$ be a path on $\ell$ vertices with $u=(4,3,\dots,3)$. The first step of the recursion is the four-mass box computation. After vertex 1 is replaced by $L^+$ or $L^-$, the adjacent vertex receives this line as an additional external condition. The residual diagram is again a smaller path. Iterating along a path with $\ell$ vertices gives $2^{\ell-1}$ one-loop factors, each evaluated on a different branch of the successive box square roots.

We therefore label the solutions by sign vectors
\begin{equation}
        \widetilde{L}^{(\epsilon)}, \qquad
        \epsilon=(\epsilon_1,\ldots,\epsilon_{\ell-1})
        \in \{\pm\}^{\ell-1} \, .
\end{equation}
The Galois group is the iterated wreath product of \(\ell-1\) copies of
\(S_2\). Equivalently, it is the automorphism group of the binary tree of
successive quadratic choices. This tree has \(2^{\ell-1}\) leaves, and these
leaves can be identified with the vertices of the hypercube
\(Q_{\ell-1}\). Under this identification, two solutions are joined by an
edge precisely when their sign vectors differ in exactly one entry.

The LS discriminant should therefore be built from the codimension-one
collisions corresponding to edges of this hypercube, not from all pairs of
solutions. We identify $Q_{\ell-1}$ with its $(\ell-1)2^{\ell-2}$ edges. Thus, up to possible extraneous factors, one expects
\begin{equation}
\Delta_{P_\ell,u}
\;\propto\;
\prod_{(\epsilon, \epsilon') \, \in \,  Q_{\ell-1}}
\langle
\widetilde{L}^{(\epsilon)}\,\widetilde{L}^{(\epsilon')}
\rangle \, .
\end{equation}
Equivalently, the product is over pairs
\(\epsilon,\epsilon'\in\{\pm\}^{\ell-1}\) differing at only one slot. Pairs differing in more than one slot, differ in more than one independent quadratic choice, so forcing the corresponding solutions to collide generally imposes several
branch degenerations at once; these are higher-codimension loci and should not
give new LS factors.

For \(\ell=2\), there is one quadratic choice and hence two solutions $\widetilde{L}^{(\pm )}$. The hypercube \(Q_1\) is a single edge, so
\begin{equation}
\Delta_{K_2,(4,3)}
\;\propto\;
\langle \widetilde{L}^{(+)}\,\widetilde{L}^{(-)}\rangle \, .
\end{equation}
Note that $\widetilde{L}^{(\pm)}$ are not the four-mass box solutions $L^{\pm}$. The latter are the solutions to the Schubert problem for $M_1,\dots,M_4$, while the former for the Schubert problem for $L^{(\pm)},M_5,M_6,M_7$.

For \(\ell=3\), there are two quadratic choices and four solutions $\widetilde{L}^{(\epsilon)}$ with $\epsilon \in \{\pm \}^2$.
They form the vertices of the square \(Q_2\), and the LS discriminant is
expected to be the product over its four edges:
\begin{equation}
\Delta_{P_3,(4,3,3)}
\;\propto\;
\langle \widetilde{L}^{(++)}\widetilde{L}^{(-+)}\rangle
\langle \widetilde{L}^{(+-)}\widetilde{L}^{(--)}\rangle
\langle \widetilde{L}^{(++)}\widetilde{L}^{(+-)}\rangle
\langle \widetilde{L}^{(-+)}\widetilde{L}^{(--)}\rangle \, .
\end{equation}
The two remaining pairings,
\(\langle \widetilde{L}^{(++)}\widetilde{L}^{(--)}\rangle\) and
\(\langle \widetilde{L}^{(+-)}\widetilde{L}^{(-+)}\rangle\), are diagonals of the square. They
correspond to pairs of solutions differing in two quadratic choices, and hence
should represent simultaneous degenerations rather than new codimension-one LS factors.
\end{eg}

\begin{eg}[A recursive triangle]
\label{eg:recursive-triangle}
The box recursion is not restricted to trees. Let $G=K_3$ and
$u=(4,3,2)$. The first vertex is a four-mass box, so
\eqref{eq:box-recursion-ls-v2} gives
\begin{equation}
\Delta_{K_3,(4,3,2)}(\mathbf M)
=
\prod_{\sigma\in\{+,-\}}
\Delta_{K_2,(4,3)}\bigl(\varphi_\sigma(\mathbf M)\bigr),
\label{eq:recursive-triangle-v2}
\end{equation}
using the normalization \eqref{eq:Lpm-normalization-v2}. The remaining subproblem
is the pentabox. The same argument applies to a cycle with multidegree
\begin{equation}
u=(4,3,\ldots,3,2,3,\ldots,3) \, ,
\end{equation}
because the four-valent vertex again provides a box seed.
\end{eg}

\subsubsection{Limitations and extensions}
\label{subsec:recursive-limitations-extensions}

Not every leading diagram contains a four-mass box. The simplest systematic family of
non-reducible examples is given by cycles with three external lines at each vertex.

\begin{tcolorbox}[definitionbox]
\textbf{Cycles without a box seed.}
Let $G$ be the $\ell$-cycle and let
\begin{equation}
u=(3,\ldots,3) \, .
\end{equation}
Then the diagram is leading but has no four-valent vertex. Its LS discriminant
can nevertheless be computed by cutting the cycle. After cutting, one obtains a
path SLS resultant depending on a single parameter $x$; the cycle discriminant is the univariate discriminant in $x$ of this resultant, see~\eqref{eq:cycle-discriminant-as-univariate-discriminant}.
\end{tcolorbox}\noindent
Here is the construction. Let
$M^{(i)}_1,M^{(i)}_2,M^{(i)}_3$ be the three external lines attached to vertex
$i$. Parametrize the loop line at the last vertex by
\begin{equation}
L_\ell(x)=\bigl(p(x)M^{(\ell)}_2\bigr)\cap
\bigl(p(x)M^{(\ell)}_3\bigr) \, ,
\qquad
p(x)=u+xv \, ,
\qquad
M^{(\ell)}_1=uv \, .
\end{equation}
Cutting the cycle at this vertex turns it into a path $\widetilde G$ on
$\ell-1$ vertices. The line $L_\ell(x)$ is inserted as an extra external line
at both ends of the path. With
\begin{equation}
\widetilde u=(4,3,\ldots,3,4) \, ,
\end{equation}
define
\begin{equation}
f_\ell(x)
=
R_{\widetilde G,\widetilde u}
\bigl(
L_\ell(x),
\mathbf M^{(1)},\ldots,\mathbf M^{(\ell-1)},
L_\ell(x)
\bigr) \, .
\label{eq:cycle-fell-definition-v2}
\end{equation}
This is a univariate polynomial of degree $2^{\ell+1}$ in $x$. The original cycle
becomes non-generic precisely when two roots of $f_\ell(x)$ collide. Thus, up to
an extraneous factor,
\begin{equation}
\Delta_{G,(3,\ldots,3)}(\mathbf M)
=
\operatorname{Disc}_x\bigl(f_\ell(x)\bigr) \, .
\label{eq:cycle-discriminant-as-univariate-discriminant}
\end{equation}
This gives a determinantal representation of the cycle discriminant as the determinant
of the Sylvester matrix of $f_\ell$ and $\partial_x f_\ell$, of size $2^{\ell+2}-1$.
Moreover, one can apply the four-mass box recursion for resultants
to~\eqref{eq:cycle-fell-definition-v2} down to a pentagon. In this way $F_\ell(x)$ becomes the determinant of a $5 \times 5$ matrix with polynomial
entries in $\mathbf{M}$.

\begin{eg}[The triple-pentagon cycle]
\label{eg:triple-pentagon-cycle}
For $\ell=3$, the cycle is $G=K_3$ with $u=(3,3,3)$. The polynomial
$f_3(x)$ factors into two irreducible polynomials of degree eight. These two
factors correspond to the two components of $V_{K_3}$, of either concurrent or coplanar triples of lines. Its discriminant contains the contribution of the concurrent component, the contribution of the coplanar component, and a mixed part.

In the notation above,
\begin{equation}
f_3(x)
=
R_{K_2,(4,4)}
\bigl(L_3(x),\mathbf M^{(1)},\mathbf M^{(2)},L_3(x)\bigr) \, .
\end{equation}
Using the double-pentagon recursion
\eqref{eq:double-pentagon-recursive-formula-v2}, this can be written as
\begin{equation}
f_3(x)
=
\prod_{\sigma\in\{+,-\}}
R_{K_1,(5)}
\bigl(L^{(\sigma)}(x),\mathbf M^{(1)},L_3(x)\bigr),
\label{eq:f3-recursive-product-v2}
\end{equation}
where
\begin{equation}
L^{(\sigma)}(x)=L^{(\sigma)}\bigl(L_3(x),\mathbf M^{(2)}\bigr)
\end{equation}
is the box solution obtained after inserting $L_3(x)$ into the middle vertex.
Each factor in \eqref{eq:f3-recursive-product-v2} is a one-loop pentagon
resultant, equivalently a $5\times5$ Gram determinant.
 Thus even this non-reducible leading problem is controlled by the same box substitutions after
one cuts the cycle and passes through an SLS path problem. 
\end{eg}

\subsection{Reality, positivity, and positroids}
\label{sec:reality-positivity-positroids}

The recursive formulas of Section~\ref{sec:recursive-landau-analysis} produce Landau
singularities from smaller ones by substitution maps. We now ask whether these
constructions preserve the distinguished positive region of planar kinematics. This
question is motivated by a central expectation in planar
\(\mathcal N=4\) SYM: amplitudes, after choosing an infrared-finite normalization such as a remainder
function or ratio function, are expected to be regular on positive kinematic space. In
the language of Landau analysis, this means that the branch loci predicted by Landau
equations should not meet the positive region.

This expectation has several origins. The positive Grassmannian first entered amplitudes
through on-shell diagrams and the Grassmannian contour formula for leading singularities
\cite{Grassmannian,Postnikov:2006kva,ArkaniHamed:2009dn,Cachazo:2008vp}.
It was then geometrized by the Amplituhedron, whose canonical form gives planar
\(\mathcal N=4\) SYM tree amplitudes and loop integrands
\cite{the_amplituhedron,Positive_geometries,positive_amplitudes}. At the level
of integrated functions, explicit bootstrap data and Landau analyses suggest that
physical singularities are absent from the positive domain and that symbol letters and
algebraic branch loci have definite sign there
\cite{DennenPrlinaSpradlinStanojevicVolovich2017,Dixon:2016apl,DixonEtAl2017,CaronHuotDixonMcLeodVonHippel2016,ChicherinHennMazzucchelliTrnkaYangZhang2026}.
This positivity is closely tied to cluster structures: cluster variables give many of
the letters appearing in planar \(\mathcal N=4\) SYM amplitudes, and cluster adjacency
constrains which letters may appear next to one another
\cite{GoldenGoncharovSpradlinVerguVolovich2014,DrummondFosterGurdoganClusterAdjacency2018,DrummondFosterGurdoganHarrington2019,DrummondFosterGurdoganKalousios2020,CaronHuotDixonDulatEtAl2020}.
The cluster aspect will be discussed in
Section~\ref{sec:rationality-cluster-structures}; here we focus on the reality and
positivity mechanism behind the absence of Landau singularities from positive
kinematics.

Let us recall the kinematic region. In planar massless kinematics one introduces dual
variables \(x_i\), with cyclic indices, by
\begin{equation}
 p_i=x_i-x_{i+1} \, ,
 \qquad
 p_i^2=x_{i,i+1}^2=0 \, ,
 \qquad
 \sum_{i=1}^n p_i=0 \, .
\label{eq:dual-variables-positive-section}
\end{equation}
Planar Mandelstam variables are then dual distances
\begin{equation}
 s_{i\cdots j-1}=(p_i+p_{i+1}+\cdots+p_{j-1})^2=x_{ij}^2 \, .
\label{eq:planar-mandelstam-dual-distance}
\end{equation}
Momentum twistors, introduced in Section~\ref{sec:Momentum Twistors}, solve these
constraints by writing
\begin{equation}
 z_i=(\lambda_i,\mu_i) \, ,
 \qquad
 \mu_i=x_i\lambda_i \, ,
 \qquad
 M_i=(z_iz_{i+1}) \, .
\label{eq:positive-momentum-twistors-lines}
\end{equation}
Thus the dual point \(x_i\) is represented by the line \(M_i\subseteq\mathbb P^3\).
After choosing a line at infinity \(I_\infty\), the dual distances are recovered from
four-brackets by
\begin{equation}
 x_{ij}^2=
 \frac{\langle i\,i{+}1\,j\,j{+}1\rangle}
 {\langle i\,i{+}1\,I_\infty\rangle
  \langle j\,j{+}1\,I_\infty\rangle} \, .
\label{eq:dual-distance-four-bracket}
\end{equation}
In dual-conformal quantities the dependence on \(I_\infty\) cancels. For example, for
cyclically ordered indices one obtains cross-ratios of the form
\begin{equation}
 u_{ij;kl}
 =
 \frac{x_{ij}^2x_{kl}^2}{x_{ik}^2x_{jl}^2}
 =
 \frac{\langle i\,i{+}1\,j\,j{+}1\rangle
        \langle k\,k{+}1\,l\,l{+}1\rangle}
       {\langle i\,i{+}1\,k\,k{+}1\rangle
        \langle j\,j{+}1\,l\,l{+}1\rangle} \, .
\label{eq:dual-conformal-cross-ratio-positive}
\end{equation}
The positive kinematic region is the image of the positive Grassmannian:
\begin{equation}
 \mathcal M_n^{>0}
 =
 \operatorname{Conf}_n^{>0}(\mathbb P^3)
 =
 \Gr_{>0}(4,n)/(\mathbb R_{>0})^n \, .
\label{eq:positive-kinematic-space}
\end{equation}
Equivalently, \(\mathbf{z}=(z_1,\ldots,z_n)\) can be represented by a real
\(4\times n\) matrix whose ordered maximal minors are all positive:
\begin{equation}
 \langle i j k l\rangle>0
 \qquad
 \text{for all }1\leq i<j<k<l\leq n \, .
\label{eq:positive-four-brackets}
\end{equation}
In ordinary Mandelstam language, this selects the connected component in which all
planar dual distances \(x_{ij}^2\), and hence all planar channels, have a fixed sign
determined by the affine normalization in~\eqref{eq:dual-distance-four-bracket}, and all
dual-conformal cross-ratios such as~\eqref{eq:dual-conformal-cross-ratio-positive} are
positive and finite. This is a \emph{Euclidean} analytic region rather than a physical
scattering region with a chosen split into incoming and outgoing particles.

The line configurations used in this chapter are slightly more general than the strict
polygonal configuration \(M_i=(z_iz_{i+1})\). We will call an ordered configuration of
lines \(\mathbf M=(M_1,\ldots,M_d)\) positive if there are columns \(z_1,\ldots,z_n\) of
a matrix in \(\Gr_{>0}(4,n)\) and indices \(p_a\) such that
\begin{equation}
 M_a=(z_{p_a}z_{p_a+1}) .
\label{eq:positive-line-configuration}
\end{equation}
When adjacent lines are required to intersect, as in a graph \(H_u\) of external
incidences, this simply means that some of these consecutive pairs share a column. 

\begin{tcolorbox}[resultbox]
\textbf{Expectation: regularity on positive kinematics.}
It is expected, but not proved in this thesis, that planar \(\mathcal N=4\) SYM amplitudes, after an infrared-finite
normalization such as a remainder function or ratio function, are
regular on \(\mathcal M_n^{>0}\). In Landau language, their
physical branch loci should not intersect the positive kinematic region.
Equivalently, all Landau singularities that survive in the integrated observable
should have definite nonzero sign on \(\mathcal M_n^{>0}\).
\end{tcolorbox}

The mathematical form of this expectation is a reality statement for Schubert problems.
The fiber of a leading Landau map is a finite incidence problem for lines
in \(\mathbb P^3\). Asking whether all its points are real for positive external data
places our problem in the tradition of real Schubert calculus~\cite{MukhinTarasovVarchenko2009,Sottile2003SecantConjecture,Karp2021Wronskians,KarpPurbhoo2023UniversalPlucker}.
The conjectures below are momentum-twistor analogues of these phenomena for leading Landau singularities.
\begin{tcolorbox}[definitionbox]
\textbf{Reality conjecture.}
Let \(\mathcal L=G_u\cup H_u\) be a planar leading Landau diagram, and let
\(\mathbf M\in V_{H_u}\) be positive. Then the fiber
\(\psi_{\mathcal L}^{-1}(\mathbf M)\) has cardinality equal to the LS degree of $\mathcal{L}$ and all of its points are real.
\end{tcolorbox}\noindent
If \(V_G\) is reducible, this statement is understood component-wise: for a component
\(V_{G,\sigma}\), the corresponding component of the Landau map should have a fully real
fiber over positive data. This refinement is important for outerplanar graphs, where
components are indexed by bicolorings of triangles.

Reality is related to positivity of the defining equations of the branch locus. A
rational function \(f\in\mathbb R(\Gr(k,n))\) is called
\textit{Grassmann copositive} if it is regular on \(\Gr_{>0}(k,n)\)
and has no zeros there. Since the positive Grassmannian is connected, the sign of such a
function is fixed after an overall normalization.

\begin{tcolorbox}[definitionbox]
\textbf{Positivity conjecture.}
If \(\mathcal L\) is planar leading, then the LS discriminant
\(\Delta_{\mathcal L}\) is Grassmann copositive. If \(\mathcal L\) is planar
superleading, then the SLS resultant \(R_{\mathcal L}\) is Grassmann copositive.
\end{tcolorbox}\noindent
Again, for reducible \(V_G\) this statement is meant component by component.

The reality conjecture implies the positivity conjecture for leading diagrams: if all
points of the fiber remain real and distinct throughout the positive region, then the
branch divisor cannot meet that region. 

We make the following remark: the component-wise formulation is essential. If \(V_G\) is reducible, the discriminant
of the union of components contains additional factors measuring collisions between
fibers lying on different irreducible components. Such mixed discriminants should
not be expected to be Grassmann copositive. A toy model is the reducible polynomial
\(f_1(x)f_2(x)\), for which
\begin{equation}
\operatorname{Disc}(f_1f_2)
=
\operatorname{Disc}(f_1)\operatorname{Disc}(f_2)\operatorname{Res}(f_1,f_2)^2 \, .
\end{equation}
The first two factors detect collisions inside one component, while the resultant
detects collisions between the two components. This mixed factor can vanish even in a
region where all roots remain real; it only records a change in the labelling of roots
between components. In Landau analysis, such factors may arise from strata supported on
intersections of irreducible components. 

\subsubsection{Positive box substitution}

The simplest test case is the four-mass box. Let \(G=K_1\) and \(u=(4)\), so that a
point in the Landau fiber is a line meeting four external lines
\(M_1,M_2,M_3,M_4\). For generic data there are two such transversals. The LS
discriminant is the incidence Gram determinant
\begin{equation}
 \Delta_{K_1,(4)}(M_1,M_2,M_3,M_4)
 =
 \det\bigl(\langle M_iM_j\rangle\bigr)_{i,j=1}^4 .
\label{eq:positive-section-four-mass-gram}
\end{equation}
Equivalently, in the usual one-loop four-mass box variables, one may write the
square-root discriminant as
\begin{equation}
 \Delta_{\mathrm{4m}}
 =
 (1-u-v)^2-4uv,
\label{eq:four-mass-cross-ratio-discriminant}
\end{equation}
where
\begin{equation}
 u=
 \frac{\langle M_1M_2\rangle\langle M_3M_4\rangle}
      {\langle M_1M_3\rangle\langle M_2M_4\rangle},
 \qquad
 v=
 \frac{\langle M_2M_3\rangle\langle M_1M_4\rangle}
      {\langle M_1M_3\rangle\langle M_2M_4\rangle} .
\label{eq:four-mass-cross-ratios}
\end{equation}
Then, $\Delta_{K_1,(4)}$ and $\Delta_{\mathrm{4m}}$ differ only by overall factors, where the latter are non-vanishing for generic data $M_i$.
On the positive four-mass region, we may write
\(M_i=(z_{2i-1}z_{2i})\) with \(\mathbf z=(z_1,\ldots,z_8)\in\Gr_{>0}(4,8)\).
Then the definitions in~\eqref{eq:four-mass-cross-ratios} immediately give
\(u,v>0\), and the Pl\"ucker relations imply \(1-u-v>0\). However, this is
still not enough to prove
\(\Delta_{\mathrm{4m}}=(1-u-v)^2-4uv>0\). The stronger Grassmann
copositivity of \(\Delta_{K_1,(4)}\) was proven in~\cite[Section~4]{ALS} by more refined
algebraic manipulations. Consequently the two Schubert solutions of the four-mass box do
not collide on \(\mathcal M_{>0}\), and the algebraic square-root branch locus of the
four-mass box lies outside the positive kinematic region.

The one-loop superleading analogue is the pentagon. For \(G=K_1\) and
\(u=(5)\), a generic line in \(\mathbb P^3\) cannot meet five generic external
lines. The SLS resultant is
\begin{equation}
 R_{K_1,(5)}(M_1,\ldots,M_5)
 =
 \det\bigl(\langle M_iM_j\rangle\bigr)_{i,j=1}^5 .
\label{eq:positive-section-pentagon-gram}
\end{equation}
This determinant is also Grassmann copositive. In
\cite[Proposition~9.2]{HolleringMazzucchelliParisiSturmfels2026}, this is
verified by substituting a positive parametrization of the relevant positive
Grassmannian; the resulting polynomial has \(5917\) monomials, all with positive
coefficients.

We now turn from positive functions to positive maps. Recall the four-mass box
substitution maps \(\varphi_\pm\) from Section~\ref{sec:recursive-landau-analysis}. They
are obtained by solving the Schubert problem of four lines
\((A,B,C,D)\), giving two transversals \(L^+(A,B,C,D)\) and
\(L^-(A,B,C,D)\), and then inserting one of them into a residual Landau problem.
A substitution map is called copositive if it sends positive line configurations to
positive line configurations.

\begin{tcolorbox}[resultbox]
\textbf{Positive box substitution.}
The four-mass box substitution maps \(\varphi_+\) and \(\varphi_-\) are
copositive. Hence, whenever
\((M_1,\ldots,M_d)\) is positive, the configurations obtained by replacing
\((M_1,M_2,M_3,M_4)\) by either box solution \(L^+\) or \(L^-\) are again
positive.
\end{tcolorbox}

The proof is most transparent after translating the two box solutions into vectors in a
positive matrix. If
\(M_i=(z_{2i-1}z_{2i})\) for \(i=1,2,3,4\), then the two transversals are spanned
by pairs of vectors \((v^+,w^+)\) and \((v^-,w^-)\), which are algebraic functions of
the columns \(z_1,\ldots,z_8\). The statement is that, for positive
\((z_1,\cdots,z_{2d})\), the matrices obtained by replacing the first eight
columns by \((v^\pm,w^\pm)\) and keeping the remaining columns positive represent
positive line configurations. In
Section~\ref{sec:line-configurations-positroids-amplituhedron-map} we explain this again
from the viewpoint of positroid promotion maps, which is the conceptual reason for
positivity.

\subsubsection{Positivity of trees of loops}

We now combine the positive box substitution with the recursive formulas of
Section~\ref{sec:recursive-landau-analysis}. If the dual loop graph \(G\) is a tree,
every leading multidegree \(u\) has a leaf vertex carrying four external conditions.
Removing that leaf is precisely the four-mass box recursion. Since each removal
preserves positivity, the recursion can be iterated preserving positivity.

\begin{tcolorbox}[resultbox]
\textbf{Reality and positivity for trees.}
Let \(\mathcal L=G_u\cup H_u\) be a planar leading Landau diagram with \(G\) being a tree. If \(\mathbf M\in V_{H_u}\) is positive, then every point
of \(\psi_{\mathcal L}^{-1}(\mathbf M)\) is real. Consequently,
\(\Delta_{\mathcal L}\) is Grassmann copositive. The same recursive argument
proves Grassmann copositivity of SLS resultants for tree graphs.
\end{tcolorbox}

Beyond trees, the Positivity Conjecture is also supported by numerical evidence. In
\cite[Section~9]{HolleringMazzucchelliParisiSturmfels2026}, the Reality
Conjecture was tested for all planar graphs \(G\) with \(\ell=3,4\) vertices and all
multidegrees \(u\) with nonzero LS degree. For each such case,
\(10^5\) random positive external configurations were sampled, and the fibers of
the Landau map were solved numerically using \texttt{HomotopyContinuation.jl}
\cite{BT}; all solutions were real. Copositivity of substitution maps was also
checked for non-tree seeds, including \(G=K_2\) with \(u=(4,3)\) and \(G=K_3\) with
\(u=(3,3,3)\). Such seeds can then be used recursively to generate further families of
positive Landau diagrams.

This completes the positivity part of the story for trees, but it does not yet explain
why the box substitution maps preserve positivity. For that we pass to positroids,
vector-relation configurations, and the Amplituhedron map.

\subsubsection{Line configurations, positroids, and the Amplituhedron map}
\label{sec:line-configurations-positroids-amplituhedron-map}

We now reinterpret the Landau map in the language of positroid varieties. This connects
the present chapter with the on-shell-diagram and Amplituhedron constructions reviewed
in Sections~\ref{sec:On-Shell Diagrams and Grassmannian Contours} and~\ref{sec:Tree Amplituhedra}. In the amplitude literature, planar on-shell diagrams are plabic graphs.
As we have seen, they parametrize positroid cells in the positive Grassmannian, and
their canonical forms compute Yangian invariants and leading singularities
\cite{Grassmannian,Postnikov:2006kva,ArkaniHamed:2009dn}. In the present Landau
problem, the finite fibers of \(\psi_{\mathcal L}\) are also leading singularities, or
more precisely, \textit{leading singularity configurations}: they are maximal-cut
solutions of a line-incidence problem. The point of Sections~10 and~11 of
\cite{HolleringMazzucchelliParisiSturmfels2026} is that these two appearances of
positroids are the same. The Landau map associated with an irreducible component of a
line-incidence variety can be identified with the Amplituhedron map restricted to an
associated positroid variety.

Let \(G\) be outerplanar. As in Section~\ref{sec:landau-analysis-on-the-grassmannian},
irreducible components of
\(V_G\) are indexed by bicolorings \(\sigma\) of the triangular faces: black
means that the corresponding lines are concurrent, while white means that they are
coplanar. We write
\begin{equation}
 V_{G,\sigma,u}\subseteq V_{G_u}
\label{eq:component-bicoloring-positive-section}
\end{equation}
for the component of the Landau on-shell space with external multiplicity \(u\). The
degree of the corresponding Landau map \(\psi_{G,\sigma,u}\) is the component-wise LS
degree \(\gamma_{G,\sigma,u}\). This is the same enumerative number that appeared in
Subsection~\ref{subsec:projections-leading-singularity-degree} as a coefficient in the
multidegree of the line-incidence variety.

The relation between line configurations and positroids, and in turn with on-shell
diagrams, can be made very explicit using \textit{vector-relation configurations}
(VRCs). This is the point where the present Landau story meets the language reviewed in
Sections~\ref{sec:On-Shell Diagrams and Grassmannian Contours} and~\ref{sec:Tree Amplituhedra}: on-shell diagrams are represented by plabic graphs, plabic graphs encode
positroid varieties, and positroid varieties carry finite maps whose degrees count
leading singularities. In the present setting, the same combinatorics is reconstructed from the line-incidence problem.

\begin{figure}[pos=t]
    \centering
    \includegraphics[width=0.95\textwidth]{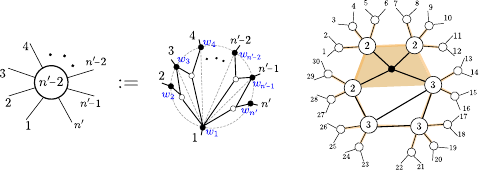}
    \caption{Left: a vertex of a Grassmann graph of degree \(n'\) and helicity
    \(h(v)=n'-2\). Center: plabic graphs for the positroid
    \(U_{n'-2,n'}\). Right: a Grassmann graph
    \(\mathcal G_{G,\sigma,u}\) associated with an outerplanar graph \(G_u\),
    given by a subdivision of a hexagon, with the bicoloring shown in orange.}
    \label{fig:grassmann-lines-vrc-positive-section}
\end{figure}

We first recall the relevant notions. A \textit{Grassmann graph} is a planar
bicolored graph with ordered boundary vertices, in which some vertices are allowed to be
decorated by local Grassmannian pieces. In our construction, the basic local pieces are
plabic graphs for uniform positroids
\(U_{n'-2,n'}\). We draw such as a single vertex of degree \(n'\) and
helicity
\begin{equation}
h(v)=n'-2 \, .
\end{equation}
This notation is illustrated on the left of
Figure~\ref{fig:grassmann-lines-vrc-positive-section}, while the middle of the same
figure shows the corresponding plabic graphs for \(U_{n'-2,n'}\).

Fix an irreducible component \(V_{G,\sigma}\subseteq V_G\), where \(\sigma\) denotes the
bicoloring data specifying the component when \(V_G\) is reducible. Let
\(u=(u_1,\ldots,u_\ell)\), and write \(d=|u|\). We construct a Grassmann graph $\mathcal G_{G,\sigma,u}$ as follows. On each vertex \(i\) of the dual loop graph \(G\), place a vertex
\(v_i\) of \(\mathcal G_{G,\sigma,u}\). If several vertices of \(G\) belong to
the same black region of the bicoloring \(\sigma\), connect the corresponding vertices
\(v_i\) to a common black vertex. For every edge \(ij\) of \(G\) which does not lie in
such a black region, connect \(v_i\) and \(v_j\) by an edge. Finally, for each external
line attached to vertex \(i\), attach a white tripod to \(v_i\). After all these
operations, assign to each vertex \(v_i\) the helicity
\begin{equation}
h(v_i)=\deg(v_i)-2 \, .
\end{equation}
The right panel of Figure~\ref{fig:grassmann-lines-vrc-positive-section} shows this
construction in an outerplanar example.

The positroid \(\Pi_{G,\sigma,u}\) is read from the Grassmann graph
\(\mathcal G_{G,\sigma,u}\) by its bounded affine permutation, exactly as for
plabic graphs and on-shell diagrams in Section~\ref{sec:On-Shell Diagrams and Grassmannian Contours}. Starting from a boundary vertex \(i\), one follows the strand
into the graph. At a vertex of helicity \(h(v)\), the strand turns \(h(v)\) steps
clockwise. The strand eventually exits at another boundary vertex, denoted
\(\pi_{\mathcal G}(i)\). The resulting permutation
\begin{equation}
\pi_{\mathcal G}=(\pi_{\mathcal G}(1),\ldots,\pi_{\mathcal G}(2d))
\end{equation}
encodes a positroid variety of type \((k,n)\), with \(n=2d\),
\begin{equation}
\Pi_{G,\sigma,u}\subseteq \Gr(k,n) \, .
\end{equation}
The boundary vertices come in pairs, one pair for each external line
\begin{equation}
M_a=(z_{2a-1}z_{2a}) \, .
\end{equation}
This gives a practical bridge to computations: once \(\pi_{\mathcal G}\) is known, the
graph, its rank, helicity, and positroid parametrizations can be obtained using the
package \texttt{Positroids.m}, for instance through commands such as \texttt{permK} and
\texttt{permToMatrix}~\cite{Bourjaily:2012gy}.

\begin{figure}[pos=t]
    \centering
    \includegraphics[width=0.9\textwidth]{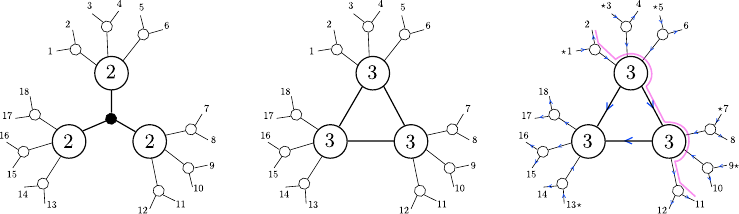}
    \caption{Plabic graphs \(\mathcal G_{G,\sigma,u}\) for the two components
    of the triangle incidence variety. The left graph corresponds to the
    concurrent component \(\sigma={\bf b}\), while the middle graph corresponds
    to the coplanar component \(\sigma={\bf w}\).}
    \label{fig:grassmann-graph-k3-positive-section}
\end{figure}

We now recall the VRC construction. Given a Grassmann graph \(\mathcal G\), a
vector-relation configuration assigns vectors \(v_b\in\mathbb C^4\) and weights
\(r_e\in\mathbb C^*\) to every black vertex \(b\) and edge \(e\) of
\(\mathcal G\), respectively. At every white vertex \(w\), these data satisfy a
linear relation
\begin{equation}
\sum_{e=(wb)} r_e\,v_b=0 \, .
\end{equation}
VRCs are considered modulo the natural \(\operatorname{GL}(4)\)-action and rescalings
preserving the linear relations. If \(z_1,\ldots,z_{2d}\) are the boundary vectors, then
\begin{equation}
\mathbf z=(z_1,\ldots,z_{2d})
\end{equation}
represents a point of \(\Gr(4,2d)\), and the external line configuration
is
\begin{equation}\label{eq:M_Z}
\mathbf M_{\mathbf z}=(M_1,\ldots,M_d),
\qquad
M_a=(z_{2a-1}z_{2a}) \, .
\end{equation}
We denote by \(\mathcal C^{\mathbf z}(\mathcal G)\) the set of VRCs with fixed boundary
data \(\mathbf z\).

The role of the local piece at \(v_i\) is to collect all vectors representing incidence
points on the loop line \(L_i\). These include the vectors assigned to intersections
\(L_i\cap L_j\) with neighboring loop lines, and the vectors assigned to intersections
\(L_i\cap M_a\) with external lines attached to \(i\). The VRC relations in the local
uniform piece force all these points to lie in a common two-dimensional subspace of
\(\mathbb C^4\); after projectivization, this subspace is the line \(L_i\subseteq\mathbb
P^3\).

An internal edge between \(v_i\) and \(v_j\) is represented by the same vector in the
two local pieces, so this vector lies on both \(L_i\) and \(L_j\), imposing
\begin{equation}
\langle L_iL_j\rangle=0 \, .
\end{equation}
Similarly, a white tripod attached to \(v_i\) contains a boundary pair
\((z_{2a-1},z_{2a})\) spanning \(M_a\), and the VRC relation forces the
corresponding incidence vector to lie both on \(L_i\) and on \(M_a\), imposing
\begin{equation}
\langle L_iM_a\rangle=0 \, .
\end{equation}
Thus a VRC with fixed boundary data produces a configuration of lines with the
prescribed incidences, namely a point in the fiber of the Landau map
\(\psi_{G,\sigma,u}\). The Grassmann graph is therefore a combinatorial device
for building the intersection points of the line configuration. This is the role played
by the colored Landau diagrams from
Subsection~\ref{subsec:projections-leading-singularity-degree}; the two pictures are
planar dual to one another.

Conversely, given a line configuration in the fiber of the Landau map, the intersection
points \(L_i\cap L_j\) and \(L_i\cap M_a\), together with the linear dependencies in
each local uniform piece, reconstruct a VRC. This gives the identification
\begin{equation}
\psi_{G,\sigma,u}^{-1}(\mathbf M_{\mathbf z})
\simeq
\mathcal C^{\mathbf z}(\mathcal G_{G,\sigma,u}).
\label{eq:vrc-landau-fiber-identification}
\end{equation}
This is the precise sense in which the Landau map for a component of a line-incidence
variety is the Amplituhedron map restricted to a positroid variety.

\begin{tcolorbox}[resultbox]
\textbf{Landau fibers as VRCs.}
For fixed positive data \(\mathbf z\), the fiber of the Landau map on
the component \(V_{G,\sigma,u}\) is identified with the space of
vector-relation configurations of the associated Grassmann graph
\(\mathcal G_{G,\sigma,u}\) with boundary \(\mathbf z\), as in
\eqref{eq:vrc-landau-fiber-identification}. Thus the line-incidence problem is
equivalent to a positroid intersection problem.
\end{tcolorbox}\noindent

In summary, a Grassmann graph \(\mathcal G_{G,\sigma,u}\) is essentially a cut diagram:
it encodes the incidences of lines prescribed by \(G_u\) together with the choice of
irreducible component, or bicoloring, \(\sigma\). From
\(\mathcal G_{G,\sigma,u}\) we build an on-shell diagram, equivalently the
positroid \(\Pi_{G,\sigma,u}\), by inserting the natural subdiagrams of maximal
helicity, corresponding to \(U_{n'-2,n'}\) in
Figure~\ref{fig:grassmann-lines-vrc-positive-section}, at vertices of degree larger than
three.

This is dual to the operation of inserting subdiagrams of minimal helicity in the
original Landau diagram, which determines the helicity sector to which the cut
contributes. Finally, from the resulting on-shell diagram one reconstructs the fiber of
the Landau map explicitly by constructing the VRC space
\(\mathcal C^{\mathbf z}(\mathcal G_{G,\sigma,u})\). This VRC space is a
collection of points in \(\mathbb P^3\) subject to linear relations encoding the desired
incidences of lines. Remarkably, it is also the fiber of the Amplituhedron map
restricted to \(\Pi_{G,\sigma,u}\).

Let \(\mathbf z=(z_1,\ldots,z_n)\) be the \(4\times n\) matrix defining the external
lines \(\mathbf M\) as in~\eqref{eq:M_Z}. Consider the linear section of
\(\Gr(k,n)\) cut out by
\begin{equation}
C \cdot \mathbf z^\top=0 \, .
\end{equation}
Equivalently, the row span of \(C\) is a \(k\)-plane contained in
\begin{equation}
\mathbf z^\perp=\ker(\mathbf z)\subseteq\mathbb C^n \, ,
\end{equation}
so the relevant intersection is
\begin{equation}
\Gr(k,\mathbf z^\perp)\cap \Pi_{G,\sigma,u} \, .
\end{equation}
This is the fiber of the usual \(m=4\) Amplituhedron map written in coordinates adapted
to the target \(k\)-plane \(Y\). Indeed, if \(Z\) is the usual positive
\((k+4)\times n\) matrix and \(Y\in\Gr(k,k+4)\), then projecting
the columns of \(Z\) to the quotient
\(\mathbb C^{k+4}/Y\simeq\mathbb C^4\) gives precisely the matrix
\(\mathbf z\). The condition that \(C\) maps to \(Y\) is then equivalent to
\(C\mathbf z^\top=0\). The result of~\cite{Even-Zohar:2025ngd} identifies this
linear section with the VRC space:
\begin{equation}
\mathcal C^{\mathbf z}(\mathcal G_{G,\sigma,u})
\simeq
\Gr(k,\mathbf z^\perp)\cap \Pi_{G,\sigma,u} \, .
\end{equation}
When \(\dim \Pi_{G,\sigma,u}=4k\), this intersection is finite, and its cardinality is
the intersection number of the positroid variety, or equivalently the degree of the
restricted Amplituhedron map.

\begin{tcolorbox}[resultbox]
\textbf{Landau map as Amplituhedron map.}
Given a component \(V_{G,\sigma,u}\) and external data \(\mathbf M_{\mathbf z}\)
defined by a positive \(4\times n\) matrix \(\mathbf z\) as in~\eqref{eq:M_Z},
the fiber of the Landau map can be identified with that of the Amplituhedron map
restricted to \(\Pi_{G,\sigma,u}\):
\begin{equation}
 \psi_{G,\sigma,u}^{-1}(\mathbf M_{\mathbf z})
 \simeq
 \Gr(k,\mathbf z^\perp)\cap \Pi_{G,\sigma,u}.
\label{eq:positroid-kernel-condition}
\end{equation}
Hence, the LS degree \(\gamma_{G,\sigma,u}\) equals the intersection number of
\(\Pi_{G,\sigma,u}\), and the LS discriminant \(\Delta_{G,\sigma,u}\) equals the
Hurwitz--Lam form of \(\Pi_{G,\sigma,u}\).
\end{tcolorbox}

\begin{eg}[The one-loop box and pentagon]
Let \(G=K_1\). For the leading one-loop box, \(u=(4)\), the Landau map asks for
lines meeting four prescribed external lines \(M_1,\ldots,M_4\). The associated
positroid is obtained from the uniform rank-two positroid \(U_{2,4}\) by
duplicating each boundary element, so it lives in a Grassmannian with eight
boundary labels. Its Hurwitz--Lam form is the LS discriminant
\(\Delta_{K_1,(4)}\) of the four-line Schubert problem, given in
\eqref{eq:positive-section-four-mass-gram}.

For the superleading one-loop pentagon, \(u=(5)\), the associated positroid is
obtained in the same way from \(U_{2,5}\), now with ten boundary labels. Its
Chow--Lam form is the five-line SLS resultant \(R_{K_1,(5)}\), namely the
determinant in \eqref{eq:positive-section-pentagon-gram}. Thus the two
elementary one-loop Schubert problems already realize the two basic positroid
invariants: the Hurwitz--Lam form for the leading box, and the Chow--Lam form
for the superleading pentagon.
\end{eg}

\begin{eg}[The triangle]
Let \(G=K_3\) and \(u=(3,3,3)\). The incidence variety \(V_G\) has two
irreducible components: the component of three concurrent lines and the component
of three coplanar lines. The two bicolorings \(\sigma={\bf b}\) and
\(\sigma={\bf w}\) therefore give two positroid varieties of type \((k,n)\),
where \(n=18\):
\begin{equation}
\Pi_{G,{\bf b},u},
\qquad
\Pi_{G,{\bf w},u} \, .
\end{equation}
These correspond to the Grassmann graphs shown in
Figure~\ref{fig:grassmann-graph-k3-positive-section}, which in turn are
constructed from \(G_u\) and \(\sigma\) as explained above. The left graph
represents the concurrent component, while the middle graph represents the
coplanar component.

The coplanar component is encoded by the permutation obtained from the strand
rules:
\begin{equation}
\pi_{G,{\bf w},u}
=
(2,11,4,15,6,1,8,17,10,3,12,7,14,5,16,9,18,13) \, .
\end{equation}
For instance, the strand starting at boundary vertex \(2\) exits at boundary
vertex \(11\), so \(\pi_{G,{\bf w},u}(2)=11\), as dictated by the strand rule.
The concurrent component is encoded analogously by the permutation
\(\pi_{G,{\bf b},u}\) obtained from the left graph in
Figure~\ref{fig:grassmann-graph-k3-positive-section} by the same rule.

From the permutations we compute the helicities of the positroids with
\texttt{Positroids.m} as
\begin{equation}
\texttt{permK[}\pi_{G,{\bf w},u}\texttt{]}
\quad\leadsto\quad 6,
\qquad
\texttt{permK[}\pi_{G,{\bf b},u}\texttt{]}
\quad\leadsto\quad 5 \, .
\end{equation}
Hence, for the coplanar component,
\begin{equation}
\Pi_{G,{\bf w},u}\subseteq \Gr(6,18),
\qquad
\dim \Pi_{G,{\bf w},u}=4\cdot 6 =24 \, ,
\end{equation}
while for the concurrent component,
\begin{equation}
\Pi_{G,{\bf b},u}\subseteq \Gr(5,18),
\qquad
\dim \Pi_{G,{\bf b},u}=4\cdot 5 =20 \, .
\end{equation}
Using the degree/intersection-number routines of \texttt{Positroids.m} applied
to the bounded affine permutations above, one obtains the value \(8\) for both
components. In geometric terms, both fibers of the Amplituhedron map have the same cardinality:
\begin{equation}
\left|
\Pi_{G,{\bf w},u}\cap
\Gr\bigl(6,\mathbf z^\perp\bigr)
\right|
=
\left|
\Pi_{G,{\bf b},u}\cap
\Gr\bigl(5,\mathbf z^\perp\bigr)
\right|
=8
\end{equation}
for generic \(4\times 18\) external data \(\mathbf z\). Equivalently, each
component contributes LS degree \(8\), so the total LS degree for \(V_{K_3}\) is
\(8+8=16\), as found earlier from the multidegree of the line-incidence variety.
\end{eg}

The same positroid construction is not limited to leading diagrams. If the external
multiplicity \(u\) is changed, the Grassmann graph
\(\mathcal G_{G,\sigma,u}\) and the associated positroid variety
\(\Pi_{G,\sigma,u}\) are constructed in exactly the same way, but the expected
dimension of the Landau fiber changes. For
\(|u|=\dim(V_{G,\sigma})-1\), the fiber is generically a curve, giving a
positroid model for NLS geometry, and similarly for higher subleading geometries.

For superleading diagrams, where \(|u|=\dim(V_{G,\sigma})+1\), the generic fiber of the
Landau map is empty. The SLS resultant records the locus of external data for which the
intersection becomes nonempty. In the positroid model, this is precisely the Chow--Lam
form of the same positroid variety
\(\Pi_{G,\sigma,u}\), in the sense of~\cite{chowlam}. Thus the LS discriminant
and SLS resultant arise from the same construction: leading singularities give
Hurwitz--Lam forms, while superleading singularities give Chow--Lam forms.

Finally, the recursive substitution maps from
Section~\ref{sec:recursive-landau-analysis} have a natural interpretation in this
VRC/positroid picture. Solving a subdiagram and inserting the resulting line into the
residual problem is realized by a promotion map between Grassmannians, or equivalently
by a local operation on the corresponding VRCs
\cite{Even-Zohar:2025ngd,HolleringMazzucchelliParisiSturmfels2026}. This
explains why the box substitutions preserve positivity, and it is the bridge to the next
section: after introducing rational degenerations and the basics of cluster algebras, we
will return to these promotion maps as cluster promotion maps.

\subsection{Rationality and cluster structures}
\label{sec:rationality-cluster-structures}

The previous section explained how positivity of external kinematics controls the
reality of Landau fibers and the regularity of LS discriminants. We now turn to a
complementary question: when Landau singularities become rational functions of momentum
twistors, why do their irreducible factors often become cluster variables?

Cluster algebras were introduced by Fomin and Zelevinsky as a combinatorial framework
for studying canonical bases and total positivity
\cite{FominZelevinsky2002,BFZ}. Very roughly, a cluster algebra is generated
from an initial collection of variables by repeatedly applying elementary
transformations called mutations. The remarkable Laurent phenomenon says that all
variables obtained in this way are Laurent polynomials in any initial cluster. In many
geometric examples, and in particular for Grassmannians, these cluster variables are
distinguished regular functions with strong positivity properties. Scott showed that
Grassmannians carry natural cluster algebra structures, with Plücker coordinates among
the cluster variables
\cite{Scott2006}. This gives a direct bridge to the positive Grassmannian:
clusters provide preferred coordinate systems on positroid cells, and mutations are
closely related to square moves of plabic graphs
\cite{Postnikov:2006kva,Grassmannian}.

Remarkably, cluster algebra structures appear throughout the modern amplitudes program.
Their most famous incarnation is in planar \(\mathcal N=4\) SYM, where they organize
several a priori different objects: symbol alphabets of integrated amplitudes, adjacency
relations between symbol letters, poles of Yangian invariants, on-shell diagrams,
positroid cells, and Amplituhedron tiles. The first concrete appearance was the
observation of Golden, Goncharov, Spradlin, Vergu, and Volovich that the symbol
alphabets of six- and seven-point planar \(\mathcal N=4\) SYM amplitudes are naturally
expressed in terms of cluster coordinates of \(\Gr(4,n)\)
\cite{GoldenGoncharovSpradlinVerguVolovich2014}. This became one of the
organizing principles of the cluster bootstrap program: cluster coordinates control not
only which letters appear, but also which letters may appear next to each other, through
the phenomenon of cluster adjacency
\cite{DrummondFosterGurdoganClusterAdjacency2018,DrummondFosterGurdoganHarrington2019}. Together with Steinmann relations and other
physical constraints, cluster adjacency has led to powerful bootstrap constructions of
multi-loop amplitudes
\cite{CaronHuotDixonMcLeodVonHippel2016,DixonEtAl2017}. A parallel
integrand-level story comes from the positive Grassmannian: on-shell diagrams and
positroid cells encode Yangian invariants and their poles
\cite{Grassmannian,Postnikov:2006kva}, and recent work has shown that cluster
structures also appear directly in the geometry of Amplituhedron triangulations and BCFW
tiles
\cite{evenZoharLakrecTessler2025bcfw,evenZoharLakrecParisiTesslerShermanBennettWilliams2023cluster,Even-Zohar:2025ngd}.

Cluster coordinates have also appeared in more isolated examples of individual Feynman
integrals, where the alphabet of the integral or the geometry of its Landau
singularities is controlled by finite or truncated cluster algebras
\cite{Chicherin:2020umh,He:2021non}. More recently, closely related
structures have emerged in cosmological correlators and wavefunction coefficients:
symbol letters for tree-level ladder cosmological correlators are governed by type \(A\)
cluster algebras, and cluster-adjacency or generalized cluster-adjacency constraints
have been shown to organize cosmological wavefunction symbols
\cite{MazloumiXu2025ClusterCosmologicalCorrelators,ParanjapeSkowronekSpradlinVolovichWeng2026ClusterBootstrapCosmologicalCorrelators,CapuanoFerroLukowskiPalazioZhang2026GeneralisedClusterAdjacency}. These examples suggest
that cluster algebras are not merely an efficient bookkeeping device for special
amplitudes, but rather a natural language for organizing positivity, singularities, and
iterated discontinuities in a broad class of physical functions.

From the perspective of Landau analysis, the appearance of cluster variables should be
viewed as a refinement of the positivity story. Positivity asks that Landau
singularities do not vanish on the positive kinematic region. Cluster factorization asks
for more: in suitable rational kinematics, the actual irreducible factors of the Landau
singularity should be cluster variables of
\(\Gr(4,n)\), possibly up to \textit{frozen} factors and powers. This is a
natural expectation, but only in the \textit{rational regime}. For generic external
data, Landau fibers are algebraic: even the four-mass box has two solutions involving a
square root, and more complicated diagrams can have much larger monodromy. In such a
situation, the discriminant is a rational function, but its individual branches are not.
After imposing additional incidences among the external lines, the fiber may become
rational. This additional incidences correspond to make some corners in the original
Landau diagram massless. Then the LS discriminant factors into explicit rational
collision factors between branches of the Landau map, and it becomes meaningful to ask
whether these factors are cluster variables.

This section develops this idea. We first recall the elementary language of cluster
algebras and their role in planar \(\mathcal N=4\) SYM amplitudes. We then explain
rational Landau diagrams, following the rational degenerations studied in Section~8 of
\cite{HolleringMazzucchelliParisiSturmfels2026}. The basic example is the degeneration
of the four-mass box to the three-mass box: one external incidence removes the square
root and replaces the two algebraic transversals by two rational branches. We then
return to the substitution maps from Section~\ref{sec:recursive-landau-analysis}. In the
positroid and VRC picture of
Section~\ref{sec:line-configurations-positroids-amplituhedron-map}, these substitutions
are promotion maps between Grassmannians; in rational degenerations they become cluster
quasi-homomorphisms
\cite{Even-Zohar:2025ngd,HolleringMazzucchelliParisiSturmfels2026}. This is
the mechanism that propagates cluster factorization through the recursive formulas.
Finally, we state the cluster factorization conjecture for rational Landau singularities
and explain its proof for trees of loops, following
\cite{HolleringMazzucchelliParisiSturmfels2026,HolleringMazzucchelliParisiSturmfels2026Positivity}.
Figures in this section are adapted from these references.

\subsubsection{Cluster algebras and amplitude singularities}
\label{subsec:cluster-algebras-amplitude-singularities}

Cluster algebras enter our story as a language for coordinates adapted to positivity. A
recurring problem in this chapter is the following: given a rational function in
Pl\"ucker coordinates, how can one see that it is positive on the positive Grassmannian?
Expanding it directly in Pl\"ucker coordinates is usually not helpful, because
cancellations may obscure positivity. Cluster algebras provide a systematic way to
produce many coordinate charts in which positive functions become manifestly positive
Laurent polynomials. They lie at the intersection of algebra, combinatorics, geometry, and
representation theory. For introductory material on cluster algebras,
their mutation dynamics, and their relation to total positivity, we refer to
\cite{FominWilliamsZelevinsky2021,Williams2014ClusterAlgebras}.

\begin{figure}[pos=t]
\centering
\begin{tikzpicture}[scale=1.0, every node/.style={circle,draw,minimum size=7mm}]
\node (1) at (0,0) {$x_1$};
\node (2) at (1.8,0) {$x_2$};
\node (3) at (0.9,1.4) {$x_3$};
\draw[->,thick] (1)--(2);
\draw[->,thick] (2)--(3);
\draw[->,thick] (1)--(3);
\node[draw=none,circle] at (0.9,-0.7) {before};

\node (1p) at (5,0) {$x_1$};
\node (2p) at (6.8,0) {$x_2'$};
\node (3p) at (5.9,1.4) {$x_3$};
\draw[->,thick] (2p)--(1p);
\draw[->,thick] (3p)--(2p);
\node[draw=none,circle] at (5.9,-0.7) {after mutation at \(2\)};
\end{tikzpicture}
\caption{A quiver mutation. Mutating at the middle vertex at $x_3$ reverses the arrows
incident to that vertex, adds arrows along two-step paths through it, and then
cancels two-cycles. The variable \(x_2\) is replaced by \(x_2'\) using the
exchange relation~\eqref{eq:cluster-exchange-relation}.}
\label{fig:cluster-quiver-mutation}
\end{figure}
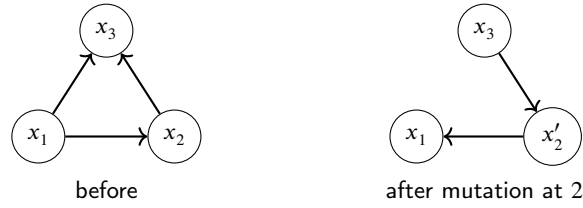

We recall the definition in the language of quivers with frozen variables. A
\emph{quiver} \(Q\) is a directed graph. We will only consider quivers with no
loops and no oriented two-cycles. A \emph{seed} \(\Sigma\) consists of a collection of
variables
\begin{equation}
\mathbf x=(x_1,\ldots,x_r)
\end{equation}
together with such a quiver \(Q\) on the same set of vertices. Some vertices are
declared \emph{frozen}; these variables are not mutated and play the role of
coefficients. The remaining vertices are \emph{mutable}.

For a mutable vertex \(k\), mutation replaces \(x_k\) by a new variable
\(x_k'\), determined by the \emph{exchange relation}
\begin{equation}
x_k x_k'
=
\prod_i x_i^{\,\#(i\to k)}
+
\prod_j x_j^{\,\#(k\to j)} .
\label{eq:cluster-exchange-relation}
\end{equation}
Here the products run over all vertices, mutable and frozen, and the exponents record
the number of arrows entering or leaving \(k\) in the quiver \(Q\). At the same time,
the quiver is mutated by adding arrows along two-step paths through \(k\), reversing all
arrows incident to \(k\), and cancelling oriented two-cycles. Frozen vertices remain
frozen throughout the mutation process. This operation is illustrated in
Figure~\ref{fig:cluster-quiver-mutation}. Iterating mutations produces the set of
\emph{cluster variables}. The
\emph{cluster algebra} is the subring of the ambient rational function field
generated by all cluster variables and by the frozen variables.

In general, this mutation process never terminates, and the set of cluster variables is
infinite. In some cases, however, only finitely many cluster variables are produced. We
then say that the cluster algebra is of
\emph{finite type}. The finite-type cluster algebras are classified by Dynkin
diagrams~\cite{FominZelevinsky2003}.

A cluster algebra is not only an abstract ring. In geometric applications it is best
thought of as a distinguished collection of coordinate systems on an algebraic variety.
Let \(X\) be an irreducible variety with function field
\(\mathbb C(X)\). A cluster structure on \(X\) means, roughly, an
identification of a cluster algebra with a subalgebra of rational functions in
\(\mathbb C(X)\). A seed \(\Sigma=(x_1,\ldots,x_r;Q)\) gives a birational
coordinate chart on \(X\): the cluster variables \(x_i\) are rational functions on
\(X\), and on the open subset where they are nonzero they define
\begin{equation}
	X \dashrightarrow (\mathbb C^*)^r,
\qquad
p\longmapsto (x_1(p),\ldots,x_r(p)).
\label{eq:cl_chart}
\end{equation}
Equivalently, inverting this map yields a parametrization of an open part of
\(X\) by a complex torus. Different seeds give different torus charts, and
mutations are the birational transition maps between them. In this language, the cluster
algebra is a coordinate ring, or a distinguished subring of the coordinate ring,
generated by all functions appearing in all cluster charts.

This point of view applies to many varieties appearing in geometry and representation
theory: double Bruhat cells, flag varieties, Grassmannians, positroid varieties, and
moduli spaces of local systems carry natural cluster structures in many cases
\cite{BFZ,Scott2006,Postnikov:2006kva,Grassmannian}. The Grassmannian example
is the most important one for us. Scott showed that the coordinate ring of
\(\Gr(k,n)\) has a cluster algebra structure
\cite{Scott2006}. Pl\"ucker coordinates are among the cluster variables, and
plabic graphs give explicit cluster charts on positroid cells.

The fundamental theorem of cluster algebras is the \emph{Laurent phenomenon}: every
cluster variable can be written as a Laurent polynomial in the variables of any chosen
cluster~\cite{FominZelevinsky2002}. Even more strongly, these Laurent expansions have
non-negative integer coefficients. This is the positivity theorem for cluster algebras.
Therefore, if the variables of one cluster are positive real numbers, then every cluster
variable is positive. In geometric examples, this is the mechanism by which cluster
algebras make positivity visible.

\begin{tcolorbox}[resultbox]
\textbf{Cluster positivity.}
A cluster algebra is an atlas of coordinate charts related by mutations. The
Laurent phenomenon and positivity theorem imply that cluster variables are
positive functions on the positive part whenever the variables of one cluster
are positive.
\end{tcolorbox}\noindent

The geometric meaning of positivity is also transparent through the lens of positive
geometry. For every seed \(\Sigma\), the corresponding cluster chart contains a real
positive torus
\begin{equation}
	(\mathbb R_{>0})^r\subseteq(\mathbb C^*)^r \, .
\end{equation}
The positive part \(X_{>0}\) of a cluster variety \(X\) is defined by choosing a cluster
chart and requiring all cluster coordinates in that chart to be positive. This condition
is independent of the chosen chart, because mutations are subtraction-free maps: they
send positive coordinates to positive coordinates. For the Grassmannian cluster
structure, this recovers the usual positive Grassmannian.

There is also a natural logarithmic form in each cluster torus chart:
\begin{equation}
\Omega_\Sigma
=
\bigwedge_{i=1}^r \mathrm{d}\log x_i .
\label{eq:cluster-torus-dlog-form}
\end{equation}
Under mutation, this form is preserved up to sign. Indeed, taking the differential
of~\eqref{eq:cluster-exchange-relation} one checks that
\begin{equation}
\bigwedge_i \mathrm{d}\log x_i'
=
\pm
\bigwedge_i \mathrm{d}\log x_i \, .
\end{equation}
This mutation-invariant logarithmic form is then denoted by
\(\mathbf{\Omega}_{X_{>0}}\). In this sense, cluster varieties provide a broad
class of positive geometries: their positive parts are glued from positive tori, and
their canonical forms are given locally by \(\mathrm{d}\log\)-volume forms
\cite{Positive_geometries}.

\begin{tcolorbox}[resultbox]
\textbf{Cluster varieties as positive geometries.}
The positive part \(X_{>0}\) of a cluster
variety \(X\) is naturally equipped with a mutation-invariant logarithmic form,
and hence defines a positive geometry
\begin{equation}
	(X,X_{>0},\mathbf{\Omega}_{X_{>0}}) \, .
\end{equation}
\end{tcolorbox}\noindent

It is useful to distinguish two related cluster varieties, usually denoted
\(\mathcal A\) and \(\mathcal{X}\). The \(\mathcal A\)-cluster variety is the
one whose coordinate ring is generated by cluster variables. In the Grassmannian
examples, the \(\mathcal A\)-coordinates are regular functions such as Pl\"ucker
coordinates and possibly higher-degree polynomials in Pl\"ucker coordinates. The \(\mathcal{X}\)-cluster variety is obtained from the same mutation data, but its coordinates are
invariant ratios built from neighboring
\(\mathcal A\)-coordinates in a seed. Thus, while \(\mathcal A\)-coordinates
are homogeneous functions on the Grassmannian, \(\mathcal{X}\)-coordinates are
well-defined on configuration spaces, where one quotients by rescalings of the columns:
\begin{equation}
\operatorname{Conf}_n(\mathbb P^3)
=
\Gr(4,n)/(\mathbb C^*)^n \, .
\end{equation}
This distinction is important in amplitudes. Rational on-shell functions and their poles
are naturally written in terms of cluster \(\mathcal A\)-coordinates, whereas symbol
letters of integrated amplitudes are usually cluster
\(\mathcal{X}\)-coordinates, or simple functions of them. This is because after stripping-off the overall helicity degree, the ratio function is invariant under the little-group action, and hence it is a function on $\operatorname{Conf}_n(\mathbb P^3)$.

We now illustrate these notions in the first non-trivial example.

\begin{eg}[The \(A_2\) cluster algebra and \(\Gr(2,5)\)]
The simplest non-trivial example of a cluster algebra is obtained from the
\(A_2\) Dynkin quiver
\begin{equation}
Q:\quad x_1\longrightarrow x_2 \, .
\end{equation}
Starting from the initial seed
\begin{equation}
\Sigma_0=(\{x_1,x_2\},Q) \, ,
\end{equation}
we mutate successively. Mutating first at \(1\) gives
\begin{equation}
x_1'=\frac{1+x_2}{x_1} \, .
\end{equation}
Mutating next at \(2\) gives
\begin{equation}
x_2'=\frac{1+x_1'}{x_2}
=
\frac{1+x_1+x_2}{x_1x_2} \, .
\end{equation}
Continuing in this way produces exactly five mutable cluster variables:
\begin{equation}
x_1,\qquad
x_2,\qquad
\frac{1+x_2}{x_1},\qquad
\frac{1+x_1+x_2}{x_1x_2},\qquad
\frac{1+x_1}{x_2} \, .
\end{equation}
After this, the mutation pattern repeats. The cluster algebra is therefore the
finite-type cluster algebra
\begin{equation}
\mathcal A(A_2)
=
\mathbb Z\left[
x_1,\,
x_2,\,
\frac{1+x_2}{x_1},\,
\frac{1+x_1+x_2}{x_1x_2},\,
\frac{1+x_1}{x_2}
\right]
\subseteq
\mathbb Q(x_1,x_2) \, .
\end{equation}
The Laurent phenomenon is already visible: every cluster variable is a Laurent
polynomial in the initial variables \(x_1,x_2\), with positive coefficients.

This finite mutation pattern has a simple geometric model. Consider a pentagon
with vertices labelled \(1,\ldots,5\). The five mutable cluster variables are
identified with the five diagonals of the pentagon. A cluster consists of two
non-crossing diagonals, hence is a triangulation of the pentagon, and mutation
is the flip of one diagonal.

The same pentagon model is realized algebraically by the Grassmannian
\(\Gr(2,5)\). If the five vertices of the pentagon are
represented by five points in \(\mathbb P^1\), or equivalently by a
\(2\times5\) matrix, then every chord \((ij)\) of the pentagon is assigned the
Pl\"ucker coordinate \(\langle ij\rangle\). The boundary edges give the five
cyclic Pl\"ucker coordinates \(\langle i\,i{+}1\rangle\), with indices taken
modulo \(5\). These are frozen variables. The five diagonals give the five
mutable Pl\"ucker coordinates. Thus
\(\mathbb C[\Gr(2,5)]\) is a cluster algebra whose mutable part is
of type \(A_2\); the five cyclic Pl\"ucker coordinates are frozen variables.

For instance, take the seed corresponding to the triangulation with diagonals
\((13)\) and \((14)\). Flipping the diagonal \((13)\) replaces it by the
diagonal \((24)\). Algebraically this is the Pl\"ucker relation
\begin{equation}
\langle 13\rangle \langle 24\rangle
=
\langle 12\rangle\langle 34\rangle
+
\langle 14\rangle\langle 23\rangle \, .
\end{equation}
Thus the new cluster variable is
\begin{equation}
\langle 24\rangle
=
\frac{
\langle 12\rangle\langle 34\rangle
+
\langle 14\rangle\langle 23\rangle
}{\langle 13\rangle} \, .
\end{equation}
This is a positive Laurent expression in the original cluster variables
\(\langle 13\rangle,\langle 14\rangle\) and the frozen Pl\"ucker coordinates
\(\langle i\,i{+}1\rangle\). Hence \(\langle 24\rangle>0\) whenever the
original cluster coordinates are positive.

The cluster structure described above is an \(\mathcal A\)-cluster structure on
the homogeneous coordinate ring of \(\Gr(2,5)\). The frozen
variables are the cyclic Pl\"ucker coordinates
\begin{equation}
\langle12\rangle,\ \langle23\rangle,\ \langle34\rangle,\ 
\langle45\rangle,\ \langle15\rangle \, .
\end{equation}
They cut out the boundary divisors of the open positroid cell. On the open
cell where these frozen variables are nonzero, every triangulation of the
pentagon gives a cluster torus chart. For example, the triangulation with
diagonals \((13)\) and \((14)\) gives the chart
\begin{equation}
\langle13\rangle,\ \langle14\rangle,
\qquad
\langle12\rangle,\ \langle23\rangle,\ \langle34\rangle,\ 
\langle45\rangle,\ \langle15\rangle \, .
\end{equation}
The positive part of this chart is the region where all seven coordinates are
positive. Since mutation is subtraction-free, positivity in one cluster chart
implies positivity in every cluster chart. The union of these positive cluster
charts is exactly the positive Grassmannian \(\Gr_{>0}(2,5)\).

After quotienting by the positive column rescalings \((\mathbb R_{>0})^5\),
only scale-invariant ratios remain. Two possible \(\mathcal{X}\)-coordinates
for this chart are the cross-ratios
\begin{equation}
	u
=
\frac{\langle12\rangle\langle34\rangle}
{\langle13\rangle\langle24\rangle},
\qquad
v
=
\frac{\langle13\rangle\langle45\rangle}
{\langle14\rangle\langle35\rangle} \, .
\end{equation}
They are invariant under independent rescalings of the five columns of the
\(2\times5\) matrix, and hence descend to  
\begin{equation}
	\operatorname{Conf}^{>0}_5(\mathbb P^1)
=
\Gr_{>0}(2,5)/(\mathbb R_{>0})^5 \, .
\end{equation}
The associated type \(A_2\) alphabet can equivalently be represented, up to
monomial factors and inversions, by the letters
\begin{equation}
	u,\quad v,\quad 1+u,\quad 1+v,\quad 1+u+v .
\label{eq:A2_alph}
\end{equation}
Thus the positive \(\mathcal A\)-coordinates parametrize the positive
Grassmannian chart, while the positive \(\mathcal{X}\)-coordinates parametrize
the corresponding configuration-space chart.
\end{eg}

More generally, the cluster algebra of type \(A_{n-3}\), for \(n\geq4\), is described by
diagonals of an \(n\)-gon. Cluster variables are the
\(n(n-3)/2\) diagonals, clusters are the \(C_{n-2}\) triangulations, where
\(C_m\) denotes the \(m\)-th Catalan number, and mutations are flips of
diagonals. The corresponding \(\mathcal A\)-cluster structure lives on the homogeneous
coordinate ring of \(\Gr(2,n)\), while the corresponding \(\mathcal{X}\)-cluster structure lives on the configuration space
\begin{equation}
\mathcal M_{0,n}
=
\operatorname{Conf}_n(\mathbb P^1) \, .
\end{equation}
This model is the cleanest place where cluster combinatorics, positivity, and
triangulations meet. It is also closely related to the \(m=2\) Amplituhedron and to the
\(m=2\) model of cluster adjacency for Yangian invariants
\cite{lukowski2019cluster}.

We now explain the relevance of cluster algebras in physics. The guiding observation is
that cluster structures appear at several different levels of the amplitudes program.
They first appeared in the study of symbol alphabets: the symbol alphabets of six- and
seven-point planar \(\mathcal N=4\) SYM amplitudes are naturally expressed in terms of
cluster coordinates of
\(\Gr(4,n)\)
\cite{GoldenGoncharovSpradlinVerguVolovich2014}. This observation became one
of the organizing principles of the cluster bootstrap program. In that approach, one
writes an ansatz for the symbol or function of an amplitude using a prescribed alphabet
and then imposes constraints such as integrability, first-entry conditions, Steinmann
and extended-Steinmann relations, final-entry conditions, collinear limits and
multi-Regge limits. This strategy was pioneered in the hexagon bootstrap and later
extended to seven points in the heptagon bootstrap
\cite{DixonDrummondHenn2011,CaronHuotDixonMcLeodVonHippel2016,DixonEtAl2017}.
In particular, the Steinmann cluster bootstrap relates extended Steinmann constraints,
cluster adjacency and the coaction principle in the space of hexagon and heptagon
functions~\cite{CaronHuotDixonDulatEtAl2020}.

The relevant cluster algebras for planar \(\mathcal N=4\) SYM are the Grassmannian
cluster algebras. In the applications below, the central object is
\(\Gr(4,n)\), or more precisely the configuration space
\begin{equation}
	\mathcal{M}_n
	=
	\operatorname{Conf}_n(\mathbb P^3)
	=
	\Gr(4,n)/(\mathbb C^{*})^n \, .
\end{equation}
This is the space of \(n\) bosonic momentum twistors
\(z_i\in\mathbb P^3\), modulo independent rescalings of the columns. For small
\(n\), the cluster algebra of \(\Gr(4,n)\) is finite: by duality
\(\Gr(4,6)\simeq \Gr(2,6)\), and this cluster
algebra is of type \(A_3\), while
\(\Gr(4,7)\simeq \Gr(3,7)\) is of type \(E_6\)
\cite{Scott2006,GoldenGoncharovSpradlinVerguVolovich2014}. Starting at
\(n=8\), the Grassmannian cluster algebra is infinite, but finite subsets of
cluster variables, finite subalgebras, tropicalizations and cluster-like positive
expressions still organize many amplitude computations
\cite{GoldenGoncharovSpradlinVerguVolovich2014,DrummondFosterGurdoganKalousios2020,DrummondFosterGurrieri2018,HenkePapathanasiou2020,Chicherin:2020umh,He:2021non}.

This cluster perspective is not limited to complete planar
\(\mathcal N=4\) SYM amplitudes. He, Li, Yang and collaborators showed that
symbol alphabets of individual dual-conformal Feynman integrals can often be described
by finite subalgebras or truncated cluster algebras associated with the relevant
kinematic boundaries of \(\Gr_{+}(4,n)/T\). Examples include all-loop box,
penta-box and double-penta ladder integrals, as well as hexagon and double-pentagon
kinematics with algebraic letters, where the cluster structure constrains not only the
set of rational letters but also algebraic square-root letters
\cite{Chicherin:2020umh,HeLiYang2021NotesClusterFeynmanIntegrals,He:2021non,HeLiYang2021KinematicsClusterFeynmanIntegrals,Yang:2022gko}.

At six points this story is especially concrete. The relevant cluster algebra is the
finite type \(A_3\) cluster algebra of \(\Gr(4,6)\). The usual hexagon
alphabet can be written in terms of the nine letters
\begin{equation}
u,\ v,\ w,\ 1-u,\ 1-v,\ 1-w,\ y_u,\ y_v,\ y_w .
\label{eq:hexagon-alphabet}
\end{equation}
Here \(u,v,w\) are the three dual conformal cross-ratios, and
\(y_u,y_v,y_w\) are parity-odd letters obtained by rationalizing the six-point
square root. The point for us is not the explicit form of a full multi-loop symbol,
which is already large, but rather that the allowed letters and their relations are
controlled by the cluster algebra of \(\Gr(4,6)\). At seven points the
corresponding finite cluster algebra is of type \(E_6\), and the same philosophy
underlies the heptagon bootstrap
\cite{DixonEtAl2017,DrummondFosterGurdoganHarrington2019}.

Cluster structures in amplitudes involve several related, but logically distinct,
conjectures. The first concerns the alphabet: rational symbol letters of planar
\(\mathcal N=4\) SYM amplitudes are expected to be cluster coordinates, or natural
functions of cluster coordinates, of the relevant Grassmannian. The second concerns
adjacency inside symbols: if two letters appear next to each other in the symbol, then
they should be cluster-adjacent, meaning that they occur together in at least one
cluster
\cite{DrummondFosterGurdoganClusterAdjacency2018,DrummondFosterGurdoganHarrington2019}. This cluster-adjacency conjecture has been
checked extensively for six- and seven-point amplitudes and is closely related to
Steinmann constraints, which forbid certain sequential discontinuities in overlapping
channels.

There is also an integrand-level version of the story. Yangian invariants are rational
functions obtained from Grassmannian residues or on-shell diagrams. Their poles are
Pl\"ucker or cluster-like functions, and the sets of poles that appear together are
highly constrained. The corresponding conjecture says that the poles of a rational
Yangian invariant are cluster-compatible: they should be realizable as \(\mathcal
A\)-coordinates in a common cluster of
\(\Gr(4,n)\). This was formulated and tested in many examples in
\cite{MagoSchreiberSpradlinVolovich2019,lukowski2019cluster}. For the
\(m=2\) toy model, the analogous statement can be proved uniformly: rational
Yangian invariants are described by non-crossing configurations in a polygon, and their
poles are compatible variables of the \(\Gr(2,n)\) cluster
algebra~\cite{lukowski2019cluster}. This \(m=2\) result is particularly relevant for the
Wilson loop with a Lagrangian insertion discussed earlier: its leading singularities are
governed by \(m=2\) Yangian invariants, so their poles inherit the type-\(A\) cluster
compatibility of the polygon
model~\cite{lukowski2019cluster,ChicherinHennMazzucchelliTrnkaYangZhang2026}. More
recently, the same philosophy has been strengthened at the level of Amplituhedron tiles:
BCFW tiles of the \(m=4\) Amplituhedron have facets cut out by collections of compatible
\(\Gr(4,n)\) cluster variables
\cite{evenZoharLakrecTessler2025bcfw,evenZoharLakrecParisiTesslerShermanBennettWilliams2023cluster,Even-Zohar:2025ngd}.

A fourth layer connects these integrand-level statements to Landau analysis. Given a
maximal cut of a loop amplitude, one obtains leading singularities written as sums of
Yangian invariants. The conjectural pattern is not only that the poles of each Yangian
invariant are cluster-compatible, but also that the Landau singularities of the cut are
compatible with the poles of the Yangian invariants appearing in the corresponding
leading singularity. This is the cluster-pattern conjecture for Landau and leading
singularities of
\cite{GurdoganParisi2020}. It was checked for all one-loop amplitudes up to
nine points and provides an important bridge between the cluster structure of rational
on-shell functions and the cluster structure of candidate branch loci after integration.

\begin{tcolorbox}[resultbox]
\textbf{Cluster structures in planar \(\mathcal N=4\) SYM.}
The cluster-algebraic organization of amplitudes has several layers.
\begin{enumerate}[label=(\roman*)]
	\item \emph{Alphabet conjecture.}
Rational symbol letters of planar \(\mathcal N=4\) SYM amplitudes are expected
to be cluster coordinates, or natural functions of cluster coordinates, of
\(\Gr(4,n)\).

\item \emph{Symbol adjacency conjecture.}
Letters that appear next to one another in the symbol are expected to be
cluster-adjacent, i.e. to occur together in at least one cluster.

\item \emph{Yangian-invariant pole conjecture.}
The poles of a rational Yangian invariant are expected to be compatible
cluster \(\mathcal A\)-coordinates, hence to occur together in a common cluster.

\item \emph{Landau--Yangian compatibility.}
For a maximal cut, the Landau singularities of the cut are expected to be
cluster-compatible with the poles of the Yangian invariants appearing in the
corresponding leading singularity.
\end{enumerate}

\end{tcolorbox}\noindent

These statements should not be confused. The first two concern the symbols of integrated
amplitudes. The third concerns rational functions before integration, namely Yangian
invariants or on-shell functions. The fourth connects the two worlds by comparing Landau
singularities of a cut with the cluster poles of the Yangian invariants appearing on
that cut. Together they suggest that cluster algebras organize both rational poles and branch cuts structures.

The meaning of ``natural'' functions of cluster coordinates in the conjecture is
important. Cluster variables are the simplest positive functions in this story, but they
do not exhaust the functions that appear in amplitudes. Already the four-mass box
illustrates this point. Its discriminant is
\begin{equation}
\Delta_{\mathrm{4m}}=(1-u-v)^2-4uv ,
\label{eq:four-mass-discriminant-cluster-discussion}
\end{equation}
where \(u,v\) are the usual dual conformal cross-ratios. The square root
\(\sqrt{\Delta_{\mathrm{4m}}}\) is a genuine algebraic letter of the four-mass
box, first appearing at multiplicity \(n=8\). It is not an ordinary cluster variable of
the rational Grassmannian cluster algebra
\(\mathbb C[\Gr(4,8)]\): ordinary cluster variables are rational
regular functions, whereas \(\sqrt{\Delta_{\mathrm{4m}}}\) is algebraic over the
Grassmannian function field. Moreover, in natural positive cluster parametrizations the
discriminant itself is not manifestly subtraction-free, which explains why its
positivity is not visible from the basic Laurent positivity theorem. Nevertheless,
\(\Delta_{\mathrm{4m}}\) is Grassmann copositive, as proven in~\cite{ALS}. In the
non-perturbative geometry of
\cite{ALS}, the associated square roots are controlled by overpositive
functions, canonical-basis structures and chain polynomials rather than by ordinary
cluster variables.

The same caveat applies to the Grassmann-copositive expressions encountered in
Section~\ref{sec:reality-positivity-positroids}. For instance, the pentagon SLS
resultant \(R_{K_1,(5)}\) is Grassmann copositive, as discussed in Section~\ref{sec:reality-positivity-positroids}. It is not known to us whether this determinant
is a cluster variable, a cluster monomial, or a more general cluster-positive
expression.

\begin{tcolorbox}[resultbox]
\textbf{Cluster variables and copositivity.}
Every cluster variable of a Grassmannian cluster algebra is
\emph{Grassmann copositive}, namely positive on the positive Grassmannian.
More generally, any positive Laurent polynomial in a cluster chart is
Grassmann copositive. However, not every Grassmann copositive function is a
cluster variable.
\end{tcolorbox}\noindent

There are several proposals for enlarging the cluster-algebraic framework so as to
include the algebraic letters that appear in amplitudes. The non-perturbative geometry
of~\cite{ALS} explains square-root letters of four-mass-box type using cluster-algebraic
functions and canonical-basis structures: for \(\Gr(4,8)\), the square
roots arise from overpositive functions rather than ordinary cluster variables. Another
approach uses tropical geometry. Tropical Grassmannians and related tropical fans
provide finite polyhedral shadows of infinite Grassmannian cluster algebras, and have
been used to predict symbol alphabets and algebraic letters for eight- and nine-particle
amplitudes
\cite{DrummondFosterGurdoganKalousios2020,DrummondFosterGurdoganKalousios2021,HenkePapathanasiou2020}. In this picture, rays or
limit rays of tropical fans play the role of candidate letters, including letters that
are not ordinary cluster variables. More recently, tropical symmetries and stable fixed
points of Grassmannian cluster quasi-automorphisms have been used to reinterpret the
four-mass-box square root from the cluster point of view
\cite{DrummondGurdoganLi2026TropicalSymmetries}. This suggests that algebraic
letters may still be controlled by cluster structures, but not always by cluster
variables themselves.

Cluster algebra structures also appear beyond planar \(\mathcal N=4\) SYM. In individual
Feynman integral families, finite and truncated cluster algebras have been used to
describe symbol alphabets and Landau singularities with both rational and algebraic
letters
\cite{Chicherin:2020umh,He:2021non,HeLiYang2021KinematicsClusterFeynmanIntegrals}.
More recently, closely related structures have appeared in cosmological correlators and
wavefunction coefficients. For example, the symbol letters of tree-level ladder
cosmological correlators are governed by type \(A\) cluster algebras; the two-site
ladder gives the first non-trivial \(A_2\) case. Cluster-adjacency and generalized
cluster-adjacency constraints have also been proposed for cosmological wavefunction
symbols
\cite{MazloumiXu2025ClusterCosmologicalCorrelators,ParanjapeSkowronekSpradlinVolovichWeng2026ClusterBootstrapCosmologicalCorrelators,CapuanoFerroLukowskiPalazioZhang2026GeneralisedClusterAdjacency}. These developments
suggest that cluster algebras are not merely an efficient bookkeeping device for special
amplitudes, but rather a natural language for organizing positivity, singularities and
iterated discontinuities in a broad class of physical functions.

Thus cluster algebras organize both rational pole structures and transcendental branch
structures. We now turn to the Landau side of this story, where the comparison with
cluster variables becomes sharp after imposing rational degenerations of the external
kinematics.

\subsubsection{Rational Landau singularities and degenerations}
\label{subsec:rational-landau-singularities-degenerations}

Cluster structures are clearest when all relevant functions are rational: poles of
Yangian invariants, cluster \(\mathcal A\)-coordinates, cluster
\(\mathcal{X}\)-coordinates and symbol letters can then be compared inside the
same rational function field. Generic Landau problems, however, are algebraic; already
the four-mass box produces a square root. In this subsection we study degenerations of
the external kinematics for which the branches of the Landau map become rational
functions of the external lines. Geometrically, these degenerations are obtained by
adding incidences among external lines; physically, they correspond to making some
external corners massless or on shell. This is the regime in which the comparison
between Landau discriminants and cluster variables becomes a precise algebraic question.

The basic example is the degeneration of the four-mass box to the three-mass box. Let
\(G=K_1\) and \(u=(4)\), so that the Landau fiber over four general external lines $M_i
\subseteq \mathbb{P}^3$ consists of the two transversals to these lines. Equivalently,
this is the classical Schubert problem of lines meeting four general lines. For generic
external data the two solutions are algebraic, as in
Example~\ref{eg:chapter7-four-mass-box-LS}. Their collision is detected by the four-mass
discriminant $\Delta_{K_1,(4)}$, which is the determinant of the $4 \times 4$ Gram
matrix $\langle M_i\,M_j \rangle$. The four-mass box integral has a square-root
singularity \(\sqrt{\Delta_{\mathrm{K_1,(4)}}}\), namely its symbol involves algebraic
letters built out of \(\sqrt{\Delta_{\mathrm{K_1,(4)}}}\).

We degenerate the external kinematics by imposing one external incidence, say
\begin{equation}
\langle M_1M_2\rangle=0 \, .
\end{equation}
The resulting Landau diagram is the \textit{three-mass box}: it has one massless corner.
If \(M_1=(12)\) and \(M_2=(23)\), then these two lines meet at the point \(z_2\); the
remaining two lines may be written, for instance, as \(M_3=(45)\) and \(M_4=(67)\). The
LS discriminant becomes a perfect square:
\begin{equation}
\Delta_{K_1,(4)}\Big|_{\langle M_1M_2\rangle=0}
=
\left(
\langle M_2M_3\rangle\langle M_1M_4\rangle
-
\langle M_1M_3\rangle\langle M_2M_4\rangle
\right)^2.
\label{eq:three-mass-perfect-square}
\end{equation}
Thus the square root singularity becomes rational. Equivalently, the two transversals to
the degenerate Schubert problem become rational in the external lines:
\begin{equation}
L^{(\mathbf b)}
=
(pM_3)\cap(pM_4),
\qquad
L^{(\mathbf w)}
=
(P\cap M_3)(P\cap M_4),
\label{eq:three-mass-solutions}
\end{equation}
where $p$ is the intersection point of $M_1$ and $M_2$ and $P$ is the plane spanned by
them. Recall that the notation in~\eqref{eq:three-mass-solutions} indicates that
\(pM_i\) is the plane spanned by the point \(p\) and the line \(M_i\), while \(P\cap
M_i\) denotes the point of intersection of \(P\) with \(M_i\). Equivalently, in the
momentum-twistor notation above,
\begin{equation}
L^{(\mathbf b)}
=
(245)\cap(267),
\qquad
L^{(\mathbf w)}
=
\bigl((123)\cap(45),(123)\cap(67)\bigr).
\label{eq:three-mass-solutions-twistors}
\end{equation}
The LS discriminant is equal to the collision factor
\begin{equation}
\Delta_{\mathrm{3m}}
= \langle L^{(\mathbf b)}L^{(\mathbf w)}\rangle
=
\langle 245|13|672\rangle^2 \, ,
\end{equation}
which in turn is the square of a chain polynomial in \(\Gr(4,7)\). Since
chain polynomials are cluster variables~\cite{Even-Zohar:2025ngd},
\(\Delta_{\mathrm{3m}}\) is a product of cluster variables.

This is the first instance of the principle that rational degenerations turn solutions
of Schubert problems from algebraic to rational. The LS discriminant then factors into
collision loci of these Schubert solutions, and each factor is a cluster variable.

With this motivation, in the rest of this subsection we study rational degenerations,
namely degenerations of the external kinematics yielding Schubert problems with rational
solutions. Let \(G_u\) be a leading Landau diagram in the sense of
Section~\ref{sec:recursive-landau-analysis}: \(G\) is the graph on loop vertices,
\(u=(u_1,\ldots,u_\ell)\) records the external lines attached to each loop vertex, and
the external data are
\begin{equation}
\mathbf M=(M_1,\ldots,M_d)\in \Gr(2,4)^d,
\qquad d=|u| \, .
\end{equation}
An \emph{external incidence graph} \(H_u\) is a graph on the \(d\) external vertices.
Its edges impose incidence conditions among the external lines:
\begin{equation}
ab\in H_u
\qquad\Longleftrightarrow\qquad
\langle M_aM_b\rangle=0 \, .
\end{equation}
We write
\begin{equation}
V_{H_u}
=
\left\{
\mathbf M\in \Gr(2,4)^d
\ \middle|\ 
\langle M_aM_b\rangle=0 \text{ for all } ab\in H_u
\right\}
\label{eq:external-incidence-variety-Hu}
\end{equation}
for the corresponding space of degenerate external data.

\begin{tcolorbox}[definitionbox]
\textbf{Rational Landau diagram.}
Let
\begin{equation}
\mathcal L=G_u\cup H_u
\end{equation}
be a leading Landau diagram with external incidences, and let
\begin{equation}
\psi_{\mathcal L}:V_{\mathcal L}\longrightarrow V_{H_u}
\end{equation}
be the Landau map obtained by forgetting the internal lines. We say that
\(\mathcal L\) is \emph{rational} if, for generic
\(\mathbf M\in V_{H_u}\), every point
\begin{equation}
\mathbf L\in \psi_{\mathcal L}^{-1}(\mathbf M)
\end{equation}
is a rational function of \(\mathbf M\). More generally, if
\(V_{\mathcal L}\) is reducible, we say that a component
\(V_{\mathcal L,\sigma}\) is rational if the same condition holds on the
restricted map
\begin{equation}
\psi_{\mathcal L,\sigma}:V_{\mathcal L,\sigma}\longrightarrow V_{H_u} \, .
\end{equation}
\end{tcolorbox}\noindent

Physically, the edges of \(H_u\) make external corners on shell, or massless. When two
external momentum-twistor lines intersect, the corresponding dual momentum difference
becomes null. Thus one starts from fully off-shell kinematics and imposes on-shell
conditions until the Landau fiber becomes rational. In this section we restrict to a
controlled class of such degenerations. The planar Landau diagrams we consider are
embedded in a disk, and \(H_u\) will be a subgraph of the outer boundary cycle.
Equivalently, we only impose incidences between neighboring external lines in the cyclic
order. This is the setting in which the rationality, positivity and
cluster-factorization statements below are expected to interact well.

This restriction is important. Planar Landau analysis at low multiplicity often forces
stronger degenerations, for example configurations in which different loop vertices
effectively cut the same external line, or equivalently, where several external lines
coincide. Such degenerations are physically relevant, but they are not expected in
general to preserve positivity or cluster-variable factorization. Indeed, Lippstreu,
Spradlin, Srikant and Volovich found seven-point examples with Landau singularities
outside the heptagon alphabet, hence not
\(\Gr(4,7)\) cluster variables, and studied their cancellation in
\(\mathcal N=4\) SYM~\cite{LippstreuSpradlinSrikantVolovich2024ZigguratII}. We therefore keep these more
singular degenerations separate from the rational boundary-cycle degenerations
considered here.

We now introduce the standard rational degeneration used throughout the rest of this
section. Suppose that \(u_i\geq2\). For each loop vertex \(i\), choose two adjacent
external lines attached to \(i\), and impose an incidence between them. In the planar
drawing this creates a small \emph{pendant triangle} consisting of
\begin{equation}
L_i,\quad M_{a_i},\quad M_{a_i+1} \, .
\end{equation}
We denote the resulting external graph by \(H_u^\triangle\), and write
\begin{equation}
G^\triangle_u
=
G_u\cup H_u^\triangle.
\label{eq:L-triangle-degeneration}
\end{equation}
The new triangle has two possible incidence geometries: either the three lines are
concurrent, or they are coplanar. We denote these two choices by the colors
\(\mathbf b\) and \(\mathbf w\), respectively. Thus, after adding the external
incidences, points in the Landau fiber can often be described by colorings of the
pendant triangles.

\begin{tcolorbox}[definitionbox]
\textbf{Pendant-triangle degeneration.}
The degeneration \(H_u^\triangle\) is obtained by imposing one external
incidence at each loop vertex:
\begin{equation}
\langle M_{a_i}M_{a_i+1}\rangle=0 \, .
\end{equation}
The corresponding triangle
\begin{equation}
(L_i,M_{a_i},M_{a_i+1})
\end{equation}
is colored \(\mathbf b\) if the three lines are concurrent and \(\mathbf w\) if
they are coplanar. A full coloring of the pendant triangles specifies a rational branch of the Landau fiber.
\end{tcolorbox}\noindent

The one-loop example above is the case \(\ell=1\): the two colorings
\(\mathbf b\) and \(\mathbf w\) give precisely the two rational transversals
\(L^{(\mathbf b)}\) and \(L^{(\mathbf w)}\) in
\eqref{eq:three-mass-solutions}. The same idea propagates through larger
graphs. If the diagram contains a rational three-mass box as a leaf, one may solve that
box by the two rational constructions above, substitute the solution into the
neighboring loop vertex, and continue recursively.

The first family of graphs providing rational solutions is the family of trivalent trees.
\begin{tcolorbox}[resultbox]
\textbf{Rationality for trivalent trees.}
Let \(G\) be a trivalent tree on \(\ell\) vertices, and consider the pendant-triangle
degeneration \(G_u^\triangle\). For each of the
\(2^\ell\) colorings \(\sigma\in\{\mathbf b,\mathbf w\}^{\ell}\) of the pendant
triangles, the corresponding component is rational. Equivalently, the generic
fiber of the Landau map on each colored component consists of one point, and
that point is obtained by a rational expression in the external lines.
\end{tcolorbox}\noindent
The proof is constructive. Fix a coloring of all pendant triangles. A leaf vertex of the
tree is a three-mass box, so its loop line is determined rationally by either the black
or white formula. Substituting this rational line into the adjacent vertex reduces the
tree. Repeating this process removes one leaf at a time. Since the tree has no cycles,
no algebraic closing condition remains. This is no longer true if $G$ contains cycles.

Let \(G=C_\ell\) be the cycle on \(\ell\geq 4\) vertices and $u=(3,3,\ldots,3).$ After
the pendant-triangle degeneration, one may open the cycle and solve the incidence
condition rationally along the resulting path. But when the cycle is closed again, one
is left with a one-variable condition equivalent to a five-line resultant. This produces
a quadratic equation in one variable. Hence the fiber is generally defined over a
quadratic extension of \(\mathbb Q(\mathbf M)\), and
\begin{equation}
(C_\ell)_u^\triangle \ \text{ is not rational.}
\end{equation}

For trivalent trees the degeneration \(H_u^\triangle\) is minimal in the following
sense. If \(u_i\geq2\) for all vertices \(i\), then the diagram
\(G_u\cup H\) is rational precisely if $H$ contains the pendant-triangle degeneration
\(H_u^\triangle\). Moreover, the trivalence assumption cannot be dropped in this form: The star-tree $G$ on five vertices, with four-valent vertex $1$, and with $u=(0,4,4,4,4)$
gives a non-rational diagram after the analogous degeneration.

On the other hand, rationality is not restricted to trees. A second family comes from
certain triangulations of polygons. Let \(G\) be a triangulation of an \(\ell\)-gon,
with vertices cyclically ordered. The irreducible components of the incidence variety
are labelled by colorings of the internal triangles of \(G\), as in the
triangulated-polygon result of
Section~\ref{subsec:incidence-varieties-line-configurations}. Adding the
pendant-triangle degeneration produces additional colored triangles at the boundary.

\begin{tcolorbox}[resultbox]
\textbf{Rationality for path triangulations.}
Let \(G\) be a triangulation of an \(\ell\)-gon whose dual graph is a path,
possibly with pendant edges. Then the pendant-triangle degeneration
\(\mathcal{L}^\triangle\) is rational.
\end{tcolorbox}\noindent
Again the reason is recursive. A triangulation whose dual graph is a path has a triangle
with two boundary edges and one internal diagonal. The line meeting the two neighbouring
edges is determined by a rational three-mass-box construction. Removing it gives a
smaller triangulated polygon of the same type. Iterating this process gives rational
formulae for all loop lines.

The path condition is essential. For example, take $G$ to be the triangulation of the
cyclically labelled hexagon with diagonals $13,35,15$. There is a component in which the
triangles \(123,156,345\) are black and
\(135\) is white. A computation gives LS degree \(58\) on this component:
among the solutions, \(32\) are rational while the remaining \(16\) require a quadratic
extension. Thus adding the pendant triangles does not automatically make every
outerplanar diagram rational.

A particularly nice family of rational triangulations has Fibonacci enumeration. Let
\(T_\ell\) be the triangulation of the cyclically labelled
\(\ell\)-gon with diagonals
\begin{equation}
(2,\ell),\quad (\ell,3),\quad (3,\ell-1),\quad \ldots,
\quad (r_\ell-1,r_\ell+1),
\qquad
r_\ell=\left\lceil \frac{\ell}{2}+1\right\rceil \, .
\end{equation}
Let \(F_\ell\) be the Fibonacci numbers, with \(F_0=0\) and \(F_1=1\). For the vector
\(u\) defined by
\begin{equation}
u_i=3\quad\text{for }i\in\{1,2,r_\ell\},
\qquad
u_i=2\quad\text{otherwise} \, ,
\end{equation}
the LS degree of the Landau diagram \((T_\ell)_u\) is
\begin{equation}
\gamma_u
=
2^\ell F_{\ell+1} \, ,
\end{equation}
and the pendant triangle degeneration \((T_\ell)_u\cup H_u^\triangle\) is rational. The
rational regime of a leading Landau diagram is not only simpler because it avoids
algebraic equations, but also because the corresponding LS discriminant factors into
several terms. Each factor is expected to be encoded by an incidence condition between
two rationally constructed lines, or by the incidence of a rationally constructed point
with a rationally constructed plane. We state this conjecture for Landau diagrams built
from an outerplanar graph \(G\), although we expect an extension to any rational Landau
diagram. This is Conjecture~8.12 of
\cite{HolleringMazzucchelliParisiSturmfels2026}.

Let \(\mathcal L=G_u\cup H_u\) be a leading Landau diagram with \(G\) outerplanar. The
full LS discriminant factors into the discriminants of the irreducible components
\(V_{G,\sigma_{\rm int}}\) of \(V_G\), where
\(\sigma_{\rm int}\) is a coloring of the internal triangles of \(G\), together
with the mixed discriminant:
\begin{tcolorbox}[resultbox]
\textbf{Component factorization of the leading discriminant.}
\begin{equation}
	\Delta_{\mathcal{L}}
	=
	\prod_{\sigma_{\rm int}}
	\Delta_{\mathcal{L},\sigma_{\rm int}}
	\cdot
	\Delta_{\mathcal{L},{\rm mix}} \, .
\end{equation}
\end{tcolorbox}\noindent
Here \(\Delta_{\mathcal{L},\sigma_{\rm int}}\) detects collisions of two points in the
fiber of the Landau map restricted to the component
\(V_{G,\sigma_{\rm int}}\), while \(\Delta_{\mathcal{L},{\rm mix}}\) detects
collisions of solutions belonging to different components of \(V_G\).

Assume now that \(H_u=H_u^\triangle\). Fix a component
\(\sigma_{\rm int}\) of \(V_G\), and assume that all points
\begin{equation}
\mathbf L^{(\sigma)}
=
(L_1^{(\sigma)},\ldots,L_\ell^{(\sigma)})
\end{equation}
in the generic fiber of the restricted Landau map
\(\psi_{\mathcal L,\sigma_{\rm int}}\) are rational and labelled by a set
\(\Sigma(\sigma_{\rm int})\) of colorings of the external triangles. Then the
LS discriminant on the component labelled by \(\sigma_{\rm int}\) is expected to factor
as
\begin{tcolorbox}[resultbox]
\textbf{External-triangle factorization.}
\begin{equation}
\Delta_{\mathcal L,\sigma_{\rm int}}
=
\prod_{\substack{
\sigma_1, \, \sigma_2 \, \in \, \Sigma(\sigma_{\rm int})\\
\sigma_1, \, \sigma_2\ \mathrm{differ\ only\ at\ the\ external\ triangle}\ i
}}
\langle
L_i^{(\sigma_1)}L_i^{(\sigma_2)}
\rangle .
\label{eq:rational-discriminant-external-factorization}
\end{equation}
\end{tcolorbox}\noindent

Assume now that all points in the fiber of the full Landau map
\(\psi_{\mathcal L}\) are rational and labelled by a set
\(\Sigma\) of colorings of both internal and external triangles. The mixed
discriminant is expected to factor as
\begin{tcolorbox}[resultbox]
\textbf{Mixed factorization.}
\begin{equation}
\Delta_{\mathcal L,{\rm mix}}
=
\prod_{\substack{
\sigma_1, \, \sigma_2 \, \in \, \Sigma\\
\sigma_1, \, \sigma_2\ \mathrm{differ\ only\ at\ the\ internal\ triangle}\ i
}}
\langle
p_i^{(\sigma_1)},P_i^{(\sigma_2)}
\rangle .
\label{eq:rational-discriminant-mixed-factorization}
\end{equation}
\end{tcolorbox}\noindent
Here the product is over pairs for which the triangle \(i\) is black in
\(\sigma_1\) and white in \(\sigma_2\). The point
\(p_i^{(\sigma_1)}\) is the common intersection point of the three lines
\(L_a^{(\sigma_1)}\) lying on the internal triangle \(i\), while the plane
\(P_i^{(\sigma_2)}\) is the common plane containing the three lines
\(L_a^{(\sigma_2)}\) lying on that same triangle. The bracket
\(\langle p_i,P_i\rangle\) denotes the incidence pairing between a point and a
plane in \(\mathbb P^3\); it vanishes precisely when the point lies on the plane.

The construction above remains a conjecture in general. On the other hand, by listing
the rational solutions, and counting the degrees on both ides of the
equations~\eqref{eq:rational-discriminant-external-factorization}
and~\eqref{eq:rational-discriminant-mixed-factorization} gives a proof for $G=K_3$ with
$u=(3,3,3)$. For each \(i=1,2,3\), let \(p_i\) be the point of intersection of the two
incident external lines, let \(P_i\) be the plane spanned by them, and write \(M_i\) for
the remaining external line attached to the \(i\)-th vertex. There are four triangles to
color in $G_u^\triangle$: the internal triangle of \(K_3\), and the three external
triangles. We write a coloring as a word
\begin{equation}
\sigma=\sigma_{\rm int}\sigma_1\sigma_2\sigma_3
\in\{\mathbf b,\mathbf w\}^4 \, .
\end{equation}
The following eight formulas give representatives of the rational solutions, up to the
symmetry permuting the three vertices. The first letter records whether the internal
triangle is concurrent \((\mathbf b)\) or coplanar \((\mathbf w)\):
\begin{small}
\begin{align}
\mathbf{bbbb}: \quad
{\bf L}
&=
\bigl(p_1q,\ p_2q,\ p_3q\bigr),
&
q&=(p_1M_1)\cap(p_2M_2)\cap(p_3M_3),
\nonumber\\[2mm]
\mathbf{wbbb}: \quad
{\bf L}
&=
\bigl(p_1q_1,\ p_2q_2,\ p_3q_3\bigr),
&
q_i&=M_i\cap(p_1p_2p_3),
\nonumber\\[2mm]
\mathbf{bbbw}: \quad
{\bf L}
&=
\bigl(p_1q,\ p_2q,\ (M_3\cap P_3)q\bigr),
&
q&=(p_1M_1)\cap(p_2M_2)\cap P_3,
\nonumber\\[2mm]
\mathbf{wbbw}: \quad
{\bf L}
&=
\bigl(p_1(M_1\cap Q),\ p_2(M_2\cap Q),\ P_3\cap Q\bigr),
&
Q&=p_1p_2(M_3\cap P_3),
\nonumber\\[2mm]
\mathbf{bbww}: \quad
{\bf L}
&=
\bigl(p_1q,\ (M_2\cap P_2)q,\ (M_3\cap P_3)q\bigr),
&
q&=(p_1M_1)\cap P_2\cap P_3,
\nonumber\\[2mm]
\mathbf{wbww}: \quad
{\bf L}
&=
\bigl(p_1(M_1\cap Q),\ P_2\cap Q,\ P_3\cap Q\bigr),
&
Q&=p_1(M_2\cap P_2)(M_3\cap P_3),
\nonumber\\[2mm]
\mathbf{bwww}: \quad
{\bf L}
&=
\bigl((M_1\cap P_1)q,\ (M_2\cap P_2)q,\ (M_3\cap P_3)q\bigr),
&
q&=P_1\cap P_2\cap P_3,
\nonumber\\[2mm]
\mathbf{wwww}: \quad
{\bf L}
&=
\bigl(P_1\cap Q,\ P_2\cap Q,\ P_3\cap Q\bigr),
&
Q&=(M_1\cap P_1)(M_2\cap P_2)(M_3\cap P_3).
\label{eq:rational-triangle-eight-solutions}
\end{align}
\end{small}
In each case the three loop lines are rational functions of the nine external lines.

The expected factorization of the LS discriminant of $G_u^\triangle$ has three factors:
\begin{equation}
\Delta_{K_3,u}
=
\Delta_{K_3,u,\mathbf b}
\cdot
\Delta_{K_3,u,\mathbf w}
\cdot
\Delta_{K_3,u,{\rm mix}} \, .
\end{equation}
In this example the formulas above apply with
\(\Sigma(\mathbf b)\) and \(\Sigma(\mathbf w)\) given by the eight colorings
whose first letter is respectively \(\mathbf b\) or \(\mathbf w\). The mixed factor
pairs colorings with the same three external colors and opposite internal color.

The next example also clarifies the role of admissible colorings. In the triangle, every
external coloring is admissible. For larger rational diagrams this is no longer true.

\begin{eg}[Triangulated square]
Let
\begin{equation}
G=\{12,23,13,34,14\},
\qquad
u=(2,3,3,3) \, ,
\end{equation}
so that \(G\) is a triangulated square with two internal triangles. Then, $G^\triangle_u$ is a minimal rational degeneration of \(G_u\), and the LS degree is \(48\).
The rational fiber is indexed by admissible bicolorings $\sigma\in\{\mathbf b,\mathbf w\}^6$, where the first two components color the two internal triangles and the remaining ones color the four external triangles.

The admissibility rule is as follows. If the two internal colors
\(\sigma_1,\sigma_2\) are distinct, then all external colorings are allowed. If the two internal colors agree, then the external triangle at vertex \(1\) must have the same color. Hence the number of admissible colorings is
\begin{equation}
2\cdot 2^4+2\cdot 2^3=48 \, ,
\end{equation}
matching the LS degree.

A direct multidegree computation gives the expected degree of the full LS
discriminant:
\begin{equation}
(160,176,256,176) \, .
\end{equation}
We check explicitly that the factorization predicted by the conjecture above has exactly this degree. For example, take a coloring of the form \(\sigma=(*\mathbf{bbbbw})\), and let
\(L_1^{(*)}\) be the line at vertex \(1\) in the corresponding rational fiber.
The two possible solutions are
\begin{equation}
\begin{aligned}
L_1^{(\mathbf b)}
&=
p_1\Bigl((p_1p_2p_3)\cap(p_3P_3)\cap(p_4P_4)\Bigr),
\\
L_1^{(\mathbf w)}
&=
P_1\cap
\Bigl(
\bigl(P_1\cap(p_2P_2)\cap(p_3P_3)\bigr)p_3p_4
\Bigr).
\end{aligned}
\end{equation}
Therefore $\langle
L_1^{(\mathbf b)}L_1^{(\mathbf w)}
\rangle$ is one of the predicted factors of the LS discriminant; it has degree
\((2,2,2,2)\) in the external data. Constructing the remaining rational lines in the same way gives all factors in
~eqref{eq:rational-discriminant-external-factorization} and~\eqref{eq:rational-discriminant-mixed-factorization}.
\end{eg}

The examples above show the role of rationality: the discriminant factorizes into
explicit incidence conditions among rationally constructed Schubert solutions. In the
next subsection we isolate the local substitution maps behind these constructions. For
the three-mass box, these are the two rational box substitution maps used in the
recursive construction of rational trivalent trees discussed above. Remarkably, these
maps carry additional structure: they are compatible with the cluster algebra structures
on Grassmannians. This will be the key input for proving recursive factorization into
cluster variables for trees.

\subsubsection{Cluster box promotion map}
\label{subsec:cluster-box-promotion-map}

We now isolate the maps that underlie the recursive constructions in the previous
subsections. The basic operation is the following. A subdiagram \(\mathcal L_1\) of a
Landau diagram \(\mathcal L\) is solved first, producing one or more Schubert solutions
\begin{equation}
\mathbf L_r^{(1)}(\mathbf M^{(1)}) \, .
\end{equation}
One then substitutes one of these solutions into the remaining diagram \(\mathcal L_2\).
On the level of line configurations this gives the substitution map
\begin{equation}
\varphi_r:
\mathbf M=(\mathbf M^{(1)},\mathbf M^{(2)})
\longmapsto
\bigl(\mathbf L_r^{(1)}(\mathbf M^{(1)}),\mathbf M^{(2)}\bigr)
=\mathbf M'.
\label{eq:cluster-substitution-map-general}
\end{equation}
The point of the positroid/VRC description recalled above is that this operation lifts
to a map between Grassmannians. Starting from a momentum-twistor matrix
\begin{equation}
Z=(z_1,\ldots,z_n)\in\Gr(4,n) \, ,
\end{equation}
we encode the external lines by pairing columns,
\begin{equation}
M_i=(z_{2i-1}z_{2i}) \, .
\end{equation}
A promotion map constructs a new matrix
\begin{equation}
Z'=(z'_1,\ldots,z'_{n'})\in\Gr(4,n')
\end{equation}
whose paired columns encode the substituted configuration,
\begin{equation}
M'_i=(z'_{2i-1}z'_{2i}) \, .
\end{equation}
These maps on Grassmannians are the promotion maps of~\cite{Even-Zohar:2025ngd}.

\begin{tcolorbox}[definitionbox]
\textbf{Promotion map.}
We use the term promotion map for a rational map
\begin{equation}
\widetilde{\varphi}:\Gr(4,n)\dashrightarrow
\Gr(4,n')
\label{eq:promo_map}
\end{equation}
defined on momentum-twistor matrices by sending \(Z=(z_1,\ldots,z_n)\) to a new matrix \(Z'=(z'_1,\ldots,z'_{n'})\), whose columns are either columns of \(Z\) or points in \(\mathbb P^3\) rationally constructed from the \(z_i\) by incidence operations. In the Landau setting, the newly inserted columns encode the Schubert solution substituted in~\eqref{eq:cluster-substitution-map-general}.
\end{tcolorbox}\noindent

Promotion maps were introduced in the study of vector-relation configurations and BCFW
tiles of the Amplituhedron~\cite{Even-Zohar:2025ngd}. They provide a geometric way to
encode the BCFW product on positroid varieties, and hence formalize the recursive
construction of Yangian invariants at the level of Grassmannian geometry. In the
previous section we saw that Landau maps restricted to positroid varieties are
Amplituhedron maps. Under that identification, our Schubert substitution maps are
precisely promotion maps.

\begin{tcolorbox}[resultbox]
\textbf{Landau substitutions are promotion maps.}
The substitution maps appearing in recursive Landau analysis lift to promotion maps between Grassmannians.
\end{tcolorbox}\noindent

\begin{figure}[pos=t]
    \centering
\includegraphics[width=1.0\textwidth]{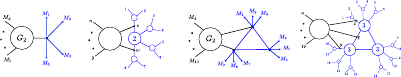}
    \caption{Illustrations of the substitution maps for the four-mass box (left) and the triangle (right). We display their promotion maps, and some relevant vectors of their VRCs.}
    \label{fig:promotion_4mb_K_3}
\end{figure}

The four-mass box substitution maps already give the smallest non-rational examples.
Their operation on vector-relation configurations, or equivalently on plabic graphs, is
depicted in Figure~\ref{fig:promotion_4mb_K_3}. As discussed in
Section~\ref{subsec:rational-landau-singularities-degenerations}, these maps are
algebraic promotion maps: positive external data produce promoted configurations in the
expected positive positroid strata. This is a special case of the following general
expectation.

\begin{tcolorbox}[resultbox]
\textbf{Conjecture: positivity for promotion maps.}
We conjecture that promotion maps arising from positroid/VRC constructions preserve the positive part of the corresponding Grassmannians. This is verified for the basic maps below and remains conjectural in general.
\end{tcolorbox}\noindent

This positivity expectation is the promotion-map avatar of the reality and positivity
results discussed in Section~\ref{sec:reality-positivity-positroids}. It is also
consistent with the Amplituhedron interpretation of these maps: promotion maps are
designed to relate positive positroid configurations, and hence to preserve the positive
domains used in Amplituhedron
tilings~\cite{evenZoharLakrecParisiTesslerShermanBennettWilliams2023cluster,Even-Zohar:2025ngd}.
For the four-mass box and the triangle, this positivity can be tested directly by
solving the corresponding Schubert problems and checking positivity of the promoted
matrices.

\begin{eg}[Triangle promotion]
The triangle \(G=K_3\) with \(u=(3,3,3)\) gives a second basic test. The generic fiber splits into the concurrent and coplanar Cayley octads, each of cardinality \(8\). To test positivity of the corresponding promotion maps, one can sample positive external data \(M_i=(z_{2i-1}z_{2i})\), for \(i=1,\dots,9\), from a positive matrix \(Z\in\Gr_{>0}(4,18)\), and then solve the two Schubert problems on the concurrent and coplanar components of \(V_{K_3}\) by numerical homotopy continuation~\cite{BT}.

For each of the \(8\) concurrent solutions and \(8\) coplanar solutions, one forms the promoted momentum-twistor matrix \(Z'\) by inserting the columns corresponding to the three solution lines \(L_1,L_2,L_3\). Positivity of the promotion map is then checked by evaluating the ordered maximal minors that are nonzero on the corresponding positroid stratum. Numerically, the concurrent and coplanar triangle promotions preserve positivity on the expected positive components. The corresponding maps are illustrated in Figure~\ref{fig:promotion_4mb_K_3}.
\end{eg}

We now pass to the rational regime. In this case the promotion maps
in~\eqref{eq:promo_map} are honest rational maps between Grassmannians. Hence they
induce pullback maps on rational functions,
\begin{equation}
\widetilde{\varphi}^{*}:
\mathbb C(\Gr(4,n'))
\longrightarrow
\mathbb C(\Gr(4,n)) \, .
\end{equation}
We say that \(\widetilde{\varphi}\) is a \emph{cluster quasi-homomorphism} if, for every
cluster variable \(x\) of \(\Gr(4,n')\), the pullback
\(\widetilde{\varphi}^{*}(x)\) is, up to frozen Pl\"ucker factors, a cluster monomial in
the cluster algebra of \(\Gr(4,n)\). Equivalently, after ignoring frozen
factors, the map \(\widetilde{\varphi}^{*}\) sends cluster variables on the target
Grassmannian to cluster-compatible products of cluster variables on the source
Grassmannian.

This is the form of the notion that we need below; it is equivalent to the
quasi-homomorphism language of Fraser~\cite{Fraser} for the Grassmannian cluster charts
considered here. In practice, one proves that a map is a cluster quasi-homomorphism by a
seed-level check: the pullbacks of the variables in one target seed should form a
cluster-compatible collection, up to frozen factors. For the Grassmannian maps appearing
here, this check is guided by the promotion maps of~\cite{Even-Zohar:2025ngd}.

\begin{tcolorbox}[resultbox]
\textbf{Conjecture: cluster quasi-homomorphisms for rational promotion maps.}
We conjecture that promotion maps arising from rational Landau diagrams are cluster quasi-homomorphisms. Consequently, pullbacks of cluster variables along these maps should factor into cluster variables, up to frozen factors. This statement is proved in the examples discussed below and is otherwise supported by explicit computations.
\end{tcolorbox}\noindent

\begin{figure}[pos=t]
\centering
\includegraphics[width=1.0\textwidth]{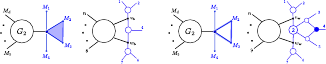}
\caption{Three-mass box promotion maps.}
\label{fig:recursion_3mb}
\end{figure}

The fundamental example is the three-mass box. The Landau diagram is \(G=K_1\) with
\(u=(4)\), with the external degeneration \(\langle M_1M_2\rangle=0\). We write
\begin{equation}
M_1=(12),\qquad M_2=(23),\qquad M_3=(45),\qquad M_4=(67) \, .
\end{equation}
The incidence \(\langle M_1M_2\rangle=0\) rationalizes the two Schubert solutions. They
are
\begin{equation}
L^{(\mathbf b)}
=
(245)\cap(267),
\qquad
L^{(\mathbf w)}
=
\bigl((123)\cap(45),(123)\cap(67)\bigr).
\label{eq:3mb-solutions-cluster-section}
\end{equation}
The corresponding substitution maps are
\begin{equation}
\varphi_{\sigma}:
(\mathbf M^{(1)},\mathbf M^{(2)})
\longmapsto
\bigl(L^{(\sigma)}(\mathbf M^{(1)}),\mathbf M^{(2)}\bigr),
\qquad
\sigma\in\{\mathbf b,\mathbf w\}.
\label{eq:3mb-substitution-maps}
\end{equation}
They lift to maps on Grassmannians, which we denote by \(\widetilde{\varphi}_{\sigma}\).
The difference between the lifted map and the map on lines is multiplication by frozen
factors, which is harmless for cluster factorization.

\begin{tcolorbox}[resultbox]
\textbf{Three-mass box promotion.}
The three-mass box promotion maps \(\widetilde{\varphi}_{\mathbf b},\widetilde{\varphi}_{\mathbf w}\) are cluster quasi-homomorphisms.
\end{tcolorbox}\noindent

The proof for this case is an explicit
verification~\cite{HolleringMazzucchelliParisiSturmfels2026}. One chooses a seed on the
target Grassmannian adapted to the three-mass box promotion. Pulling back its mutable
and frozen variables gives products of Pl\"ucker coordinates and quadratic cluster
variables in the source Grassmannian. Since these pullbacks form a compatible
collection, mutation then propagates the statement to the cluster algebra generated from
the chosen seed.

\begin{figure}[pos=t]
    \centering
    \includegraphics[width=0.8\textwidth]{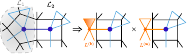}
    \caption{LS discriminant recursion for the rational pentabox. The three-mass box on the left has two rational branches, and substituting these into the remaining box produces two cluster factors.}
    \label{fig:rat-pentabox-cluster}
\end{figure}

\begin{eg}[The rational pentabox]
Consider the rational pentabox in Figure~\ref{fig:rat-pentabox-cluster}. Thus \(G=K_2\), \(u=(4,3)\), and \(H_u^\triangle\) contains two external edges. The first vertex has the external lines \(a_1a_2\), \(a_2a_3\), \(a_4a_5\), and \(a_6a_7\), while the second vertex has the external lines \(b_1b_2\), \(b_2b_3\), and \(b_4b_5\). The first vertex is a three-mass box. The two branches \(\mathbf b\) and \(\mathbf w\) therefore give two substitution maps, and the recursive formula gives
\begin{equation}
\Delta_{\mathcal L}
=
\varphi_{\mathbf b}^{*}\Delta_{K_1,(4)}
\cdot
\varphi_{\mathbf w}^{*}\Delta_{K_1,(4)}.
\label{eq:pentabox-cluster-recursion}
\end{equation}
Explicitly,
\begin{equation}
\begin{aligned}
\Delta_{\mathcal L}
=
&
\left\langle
b_2b_4b_5 \,\middle|\, b_1b_3 \,\middle|\,
b_2,\,(a_4a_1a_2)\cap(a_4a_6a_7)
\right\rangle^2
\\
&\cdot
\left\langle
b_2b_4b_5 \,\middle|\, b_1b_3 \,\middle|\,
b_2,\,(a_3a_4a_5)\cap(a_1a_2),\,
(a_3a_4a_5)\cap(a_6a_7)
\right\rangle^2 .
\end{aligned}
\label{eq:pentabox-rational-cluster-factors}
\end{equation}
Both factors in~\eqref{eq:pentabox-rational-cluster-factors} are cluster variables for \(\Gr(4,12)\). This illustrates the mechanism: the reduced one-loop discriminant is a cluster variable, and the two three-mass-box promotion maps pull it back to cluster variables.
\end{eg}

The pentabox is the first case of a general recursive mechanism. Whenever a rational
Landau diagram contains a three-mass box leaf, one may remove that leaf, pull back the
discriminant of the smaller diagram along the two rational box promotion maps, and
multiply the two resulting factors. Since the three-mass box promotion maps are cluster
quasi-homomorphisms, cluster variables remain cluster variables, up to frozen factors,
under this pullback.

\begin{tcolorbox}[resultbox]
\textbf{Cluster factorization for rational trees.}
Let \(\mathcal L=G_u\cup H_u^\triangle\) be a rational leading Landau diagram such that \(G\) is a tree. Then the LS discriminant of \(\mathcal L\) factors into cluster variables of the appropriate Grassmannian, up to frozen factors and powers.
\end{tcolorbox}\noindent

Indeed, one removes a leaf three-mass box. The smaller diagram is again a rational tree.
By induction its discriminant factors into cluster variables, and the recursive formula
expresses the larger discriminant as the product of the two pullbacks along
\(\varphi_{\mathbf b}\) and \(\varphi_{\mathbf w}\). The three-mass box cluster
quasi-homomorphism then preserves cluster factorization. This gives an all-loop family
of results for the cluster structure of leading Landau singularities. Indeed, the number
of vertices of the tree \(G\), hence the number of loop variables, is arbitrary. Thus
the local cluster property of the rational three-mass-box promotion map explains an
infinite family of cluster-factorizing Landau discriminants.

The first genuinely non-tree leading Landau diagram is the rational triangle with one
pendant triangle at every vertex.

\begin{eg}[The rational triangle]
Let \(\mathcal L=G_u\cup H_u^\triangle\), where \(G=K_3\) and \(u=(3,3,3)\). As explained in Subsection~\ref{subsec:rational-landau-singularities-degenerations}, the LS discriminant splits as
\begin{equation}
\Delta_{\mathcal L}
=
\Delta_{\mathcal L,\mathbf b}
\cdot
\Delta_{\mathcal L,\mathbf w}
\cdot
\Delta_{\mathcal L,{\rm mix}} .
\label{eq:rational-triangle-disc-split-cluster}
\end{equation}
Let the external lines at vertex \(1\) be \(a_1a_2\), \(a_2a_3\), and \(a_4a_5\), and use \(b_i\) and \(c_i\) analogously at vertices \(2\) and \(3\). Thus the external twistors are ordered as
\begin{equation}
\mathbf z=(a_1,\ldots,a_5,b_1,\ldots,b_5,c_1,\ldots,c_5)\in\Gr(4,15) \, .
\end{equation}

Restrict first to the concurrent component. The fiber consists of eight rational branches, indexed by the three colors of the pendant triangles. The discriminant \(\Delta_{\mathcal L,\mathbf b}\) factors into one factor for each edge of the cube \(\{\mathbf b,\mathbf w\}^3\). Equivalently, as in the factorization conjecture above, the factors correspond to pairs of external colorings that differ in exactly one position. The concurrent discriminant \(\Delta_{\mathcal L,\mathbf b}\) is the following product of twelve factors:
\begin{equation}
\label{eq:discr-tr-fact-explicit}
\begin{aligned}
\Delta_{\mathcal L,\mathbf b}
=\,&
\left\langle a_2a_4a_5\,\middle|\,a_1a_3\,\middle|\,a_2(b_2b_4b_5)\cap(c_2c_4c_5)\right\rangle^2
\\
&\cdot
\left\langle a_2a_4a_5\,\middle|\,a_1a_3\,\middle|\,a_2(b_2b_4b_5)\cap(c_1c_2c_3)\right\rangle^2
\\
&\cdot
\left\langle a_2a_4a_5\,\middle|\,a_1a_3\,\middle|\,a_2(b_1b_2b_3)\cap(c_1c_2c_3)\right\rangle^2
\cdot (\text{permutations of }a,b,c).
\end{aligned}
\end{equation}
The coplanar component \(\Delta_{\mathcal L,\mathbf w}\) has the dual factorization, obtained by exchanging points and planes. Each factor of \(\Delta_{\mathcal L,\mathbf b}\) and \(\Delta_{\mathcal L,\mathbf w}\) is a cubic cluster variable for \(\Gr(4,15)\).

The mixed discriminant behaves differently. A compact form, displaying four factors and their cyclic permutations, is
{\small
\begin{equation}
\label{eq:discr-tr-fact-explicit-mixed}
\begin{aligned}
\Delta_{\mathcal L,{\rm mix}}
=\,&
\left\langle a_2b_2c_2,
(a_2a_4a_5)\cap(b_2b_4b_5)\cap(c_2c_4c_5)
\right\rangle
\\
&\cdot
\left\langle
(a_1a_2a_3)\cap(a_4a_5), b_2,c_2,
(a_1a_2a_3)\cap(b_2b_4b_5)\cap(c_2c_4c_5)
\right\rangle
\\
&\cdot
\left\langle
(a_1a_2a_3)\cap(a_4a_5),
(b_1b_2b_3)\cap(b_4b_5),
(c_1c_2c_3)\cap(c_4c_5),
(a_1a_2a_3)\cap(b_1b_2b_3)\cap(c_1c_2c_3)
\right\rangle
\\
&\cdot
\left\langle
(a_1a_2a_3)\cap(a_4a_5),
(b_1b_2b_3)\cap(b_4b_5),c_2,
(a_1a_2a_3)\cap(b_1b_2b_3)\cap(c_2c_4c_5)
\right\rangle
\\
&\cdot (\text{permutations of }a,b,c).
\end{aligned}
\end{equation}
}
Here a bracket with four entries denotes the determinant of the four points in \(\mathbb P^3\), where the entries involving \(\cap\) are points obtained by the indicated intersections. The component products have degree \(16\) in the Pl\"ucker coordinates of each external block, matching the degree of the specialized component discriminants. By contrast, the factors in \(\Delta_{\mathcal L,{\rm mix}}\) are not cluster variables: they are not Grassmann copositive, and they can change sign on positive external data. Thus the rational triangle confirms cluster factorization for the individual component discriminants, while showing that mixed discriminants belong to a different class of singularities.
\end{eg}

The following family shows that promotion-map structures persist beyond the single
triangle and produce non-tree configurations with controlled cluster behavior.

\begin{eg}[Kissing triangles and chain-tree promotions]
Let \(G\) be a chain of \(s\) kissing triangles, with
\begin{equation}
u=(3,3,2,3,2,\ldots,2,3,2,3,3) \, ,
\end{equation}
and impose the pendant-triangle degeneration \(H_u^\triangle\). Take all triangles to be black. The corresponding substitution map inserts lines built from the common points of the black triangles. On the level of momentum-twistor matrices, the promotion map has the form
\begin{equation}
(z_1,\ldots,z_n)
\longmapsto
(z_2,p_1,p_2,\ldots,p_s,z_r,z_{r+1},\ldots,z_n),
\qquad
r=8s+6,
\label{eq:kissing-triangle-promotion-map}
\end{equation}
where \(p_i\) is the common point of the three concurrent lines in the \(i\)-th triangle. The inserted loop lines are
\begin{equation}
L_1=z_2p_1,
\qquad
L_{2a+1}=p_ap_{a+1}\quad(1\leq a\leq s-1),
\qquad
L_{2s+1}=p_sz_r.
\label{eq:kissing-triangle-lines}
\end{equation}
This is the cluster promotion map associated with a chain-tree of type
\begin{equation}
(3,3,1,3,1,3,\ldots,1,3,1,3,3)
\end{equation}
in the sense of~\cite{Even-Zohar:2025ngd}. For \(s=1\) this recovers the triangle and agrees with the spurion promotion of \cite[Section~8.2]{Even-Zohar:2025ngd}. The cases \(s=1\) and \(s=2\) are proved cluster quasi-homomorphisms there, and the general chain is expected to satisfy the same property.
\end{eg}

\begin{figure}[pos=t]
    \centering
    \includegraphics[width=1.0\textwidth]{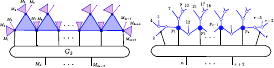}
    \caption{Left: a chain of kissing triangles in the line-incidence picture. Right: the corresponding promotion map, whose core is a chain-tree of type \((3,3,1,3,1,3,\ldots,1,3,1,3,3)\).}
    \label{fig:chain-triangles-cluster-promotion}
\end{figure}

These examples point beyond the tree theorem. The cluster quasi-homomorphism property of
the three-mass box explains an all-loop family of tree diagrams, while the recursive
triangle and kissing-triangle examples show that similar promotion-map structures
persist for non-tree geometries. At the same time, the rational triangle shows that
mixed discriminants may fall outside the cluster-positive world. Thus the emerging
picture is that component discriminants of rational Landau diagrams are often controlled
by cluster promotion maps, whereas mixed discriminants require additional geometric
input.

\subsection{Future directions}
\label{sec:positivity-cluster-outlook}

The results of this chapter suggest several directions for future work. The
common theme is that Landau singularities in planar kinematics are governed by
incidence geometry in \(\mathbb P^3\), much of which remains only partially
understood.

\begin{enumerate}

\item \textbf{Momentum twistors and Lee--Pomeransky geometry.}
The approach developed here formulates Landau analysis in momentum-twistor
space: loop variables are lines in \(\mathbb P^3\), propagators are incidence
conditions, and singularities arise from projections of incidence varieties. A
different algebraic-geometric approach starts from the Lee--Pomeransky
representation and its principal Landau determinant
\cite{LeePomeransky2013,FevolaMizeraTelen2024,FevolaMizeraTelenPLD2024,Mizera:2021icv}.
It would be useful to relate the momentum-twistor varieties \(V_G\) to the
critical-point geometry of the Lee--Pomeransky polynomial, and to understand
whether the Hurwitz and Chow forms found here become factors of principal Landau
determinants after a suitable change of variables. Such a comparison would link
the planar, dual-conformal formalism to more general methods for Feynman
integrals.

\item \textbf{Beyond leading and super-leading singularities.}
Most of this chapter focused on leading and super-leading Landau singularities,
encoded by discriminants and resultants of the Landau map. A full Landau
analysis should also include lower-codimension strata of the polar locus and
singular strata of the incidence varieties. In the language of
\cite{HelmerPapathanasiouTellander2024}, this means studying the branching of
\(\psi_{\mathcal L}\) on every stratum of a Whitney stratification. It remains
to develop analogues of LS degree, rationality, reality, and positivity when
the fiber of \(\psi_{\mathcal L}\) has positive dimension, as happens for
next-to-leading singularities.

\item \textbf{Irreducible decompositions beyond outerplanar graphs.}
The clearest structural results here concern outerplanar loop graphs. There the
irreducible components of \(V_G\) are controlled by triangle colorings, making
the Landau map computable. For general planar graphs no analogue is known. A
first test case is the wheel graph \(W_\ell\), with a central vertex and cyclic
boundary vertices. Understanding its irreducible decomposition and component
discriminants \(\Delta_{G,\sigma,u}\) would extend the outerplanar theory. More
generally, one wants a combinatorial rule for the components of \(V_G\) for
arbitrary planar \(G\), compatible with projective duality and degenerations of
Landau diagrams.

\item \textbf{A computational atlas of line-incidence varieties.}
The paper \cite{HolleringMazzucchelliParisiSturmfels2025Lines} shows that
incidence varieties of lines in \(\mathbb P^3\) have invariants closely tied to
graph data. A computational atlas recording decompositions, multidegrees,
singular loci, realization spaces, and discriminants for larger graphs could
guide all-loop searches and benchmark symbolic and numerical algebraic geometry.

\item \textbf{Resultants, Chow forms and super-leading singularities.}
Super-leading singularities were described here by resultants, naturally related
in the Grassmannian formulation to Chow forms of incidence varieties. The main
open problems are to find formulas beyond the one-loop pentagon and small
examples, prove Grassmann copositivity conceptually, and decide when these
resultants are cluster variables, cluster monomials, or more general
cluster-positive expressions \cite{HolleringMazzucchelliParisiSturmfels2026}.

\item \textbf{Reality and positivity beyond trees.}
The tree theorem proves reality and Grassmann copositivity for an infinite
family of planar leading diagrams via positive box substitution maps. Numerics
suggest a broader phenomenon. One route is to study positivity for the Landau
map on positroid varieties; another is to analyze the regions in
\(\Gr(2,4)^\ell\) cut out by signs of \(\langle L_iL_j\rangle\). A systematic
theory of these positive and negative line-configuration geometries could
explain why certain Landau fibers remain real and positive.

\item \textbf{Cluster factorization beyond rational trees.}
For rational trees, the LS discriminant factors into cluster variables because
the three-mass-box promotion maps are cluster quasi-homomorphisms. Triangles and
kissing triangles suggest an extension to component discriminants beyond trees.
A general theorem would require classifying rational degenerations and proving
that the associated promotion maps are cluster quasi-homomorphisms. Mixed
discriminants remain the caveat: they can fail Grassmann copositivity and need
not be cluster variables \cite{HolleringMazzucchelliParisiSturmfels2026Positivity}.

\item \textbf{Algebraic letters and non-rational Landau fibers.}
The rational regime is the cleanest setting for cluster factorization, but the
generic Landau problem is algebraic. The four-mass box already produces a square
root, and more complicated diagrams may lead to higher-degree covers. One should
understand whether tropical fans, cluster automorphisms, canonical basis
elements, or overpositive functions organize these covers and the algebraic
letters of integrated amplitudes.

\item \textbf{Promotion maps, VRCs and Amplituhedron tiles.}
Promotion maps bridge recursive Landau analysis and cluster geometry. Their
appearance in BCFW tiles and vector-relation configurations suggests that
Landau substitutions and BCFW products for Yangian invariants reflect one
Grassmannian operation. A general theory should explain positivity preservation,
cluster quasi-homomorphisms, and the action on positroid strata, canonical
forms, and Amplituhedron tiles \cite{Even-Zohar:2025ngd,HolleringMazzucchelliParisiSturmfels2026Positivity}.

\item \textbf{Toward a geometric bootstrap from Landau geometry.}
The previous chapter used geometric Landau analysis as input to the symbol
bootstrap: geometry produced candidate letters, while integrability, first-entry
conditions, Steinmann constraints, differential equations, and physical limits
fixed the symbol. The structures uncovered here suggest that the same incidence
geometry may also control adjacency, cluster compatibility, and allowed symbol
words. A geometric bootstrap would derive this data from Grassmannian Landau
maps, positroid strata, and promotion maps.

\item \textbf{Other theories and other kinematics.}
The momentum-twistor formalism used here is tailored to planar four-dimensional
kinematics. It would be interesting to ask what survives for massive planar
kinematics, non-planar diagrams, ABJM theory, cosmological correlators, or
standard-model integrals. Even without lines in \(\mathbb P^3\) or a positive
Grassmannian, one may still find incidence varieties, positive geometries, or
cluster-like coordinates. This would test what is theory-specific and what
follows universally from Landau geometry.

\end{enumerate}

%% file: Conclusion.tex

\section{Conclusion}
\label{sec:Conclusion}

\subsection{Summary}

This thesis studied positive geometry in scattering amplitudes in planar
\(\mathcal N=4\) supersymmetric Yang--Mills theory, with emphasis on
positivity, volumes, and singularities. The guiding idea was that the geometric
formulation of amplitudes does more than provide compact expressions for
rational integrands: it also offers a language for analytic properties that
remain visible after integration, when amplitudes and related observables
become multivalued transcendental functions.

The first part developed the necessary background.
Chapter~\ref{ch:Scattering Amplitudes} reviewed the on-shell formulation of
scattering amplitudes, recursion relations, momentum twistors, on-shell
diagrams, and Grassmannian contour formulae.
Chapter~\ref{ch:Positive Geometries} introduced positive geometries and
canonical forms, emphasizing logarithmic singularities, residues,
triangulations, adjoints, push-forwards, and integral representations.
Chapter~\ref{ch:Amplituhedra} specialized this framework to Amplituhedra,
where tree amplitudes and loop integrands arise as canonical forms of positive
geometries in momentum-twistor space.

The second part developed three related research directions.
Chapter~\ref{ch:Canonical Forms as Dual Volumes} studied canonical functions
as dual volumes or Laplace transforms of non-negative measures.
Chapter~\ref{ch:From Integrands to Integrals} examined how positive geometry
constrains leading singularities, Landau singularities, and symbol alphabets
after integration.
Chapter~\ref{ch:positivity-and-cluster-structures} formulated Landau analysis
in Grassmannian language and investigated the positivity, recursion, and
cluster structure of the resulting singularities.

Together, these chapters support a common picture. Before integration,
amplitudes and loop integrands in planar \(\mathcal N=4\) SYM are organized by
canonical forms of positive geometries. After integration, their singularities
still retain geometric information: boundaries, residues, positivity,
incidences of lines, and cluster structures constrain which singularities can
appear and how they factorize.

\subsection{Original contributions of this thesis}

The preceding chapters reviewed substantial background before turning to the
original results of the thesis. The contributions summarized here are the new
research directions developed in the later chapters, including joint work with
collaborators where indicated in the relevant chapters.

The first original direction concerned dual volumes and analytic positivity.
For polytopes, the canonical function computes the volume of a dual polytope.
Chapter~\ref{ch:Canonical Forms as Dual Volumes} studied how this idea extends
beyond the linear setting. The main mechanism is complete monotonicity: when a
canonical function is a Laplace transform of a non-negative measure, it
satisfies strong positivity properties. For nonlinear positive geometries,
these measures are generally not algebraic, and their densities may be
transcendental. This refines the statement that amplitudes compute volumes:
the dual object may be a positive measure rather than an algebraic region with
Lebesgue measure. The chapter also discussed Grassmannian notions of convexity
and duality, first steps toward dual Amplituhedra, and positivity properties of
Aomoto forms.

The second original direction concerned the passage from integrands to
integrals. At loop level, positive geometry produces rational integrands, but
physical quantities arise only after integration, where poles are replaced by
branch loci of multivalued transcendental functions.
Chapter~\ref{ch:From Integrands to Integrals} emphasized the distinction
between leading singularities and Landau singularities. The former are maximal
residues of the rational integrand, while the latter arise when integration
contours are pinched. The central point is that positive geometry still
constrains the integrated structure. Geometric Landau analysis refines
ordinary Landau analysis using the boundary structure of the relevant positive
or negative geometry, thereby constraining the possible symbol alphabet.

The third original direction concerned Landau analysis and cluster structures.
Momentum twistors turn propagator equations into incidence conditions among
lines and external data in projective space. This makes it possible to
formulate Landau analysis in terms of incidence varieties, discriminants, and
resultants on Grassmannians. In this language, recursive structures become
visible. Substitution maps, closely related to promotion maps, explain why
certain families of Landau singularities behave well in the positive region
and why their rational singularities factorize into cluster variables. This
connects the singularities of integrated amplitudes with the same structures
that organize amplitudes at the integrand level: positroid strata,
Grassmannian geometry, and cluster algebras. Some parts of this picture are
established for the families studied here, while others remain conjectural
organizing principles.

These three contributions share a common theme. Canonical forms can possess
stronger positivity properties than are visible from their poles alone;
singularities after integration are constrained by the geometry that produced
the integrand; and these singularities retain a memory of the combinatorial
and cluster structures of the Grassmannian. The geometric description is
therefore not erased by integration, but reappears in the analytic and
algebraic structure of the final functions.

\subsection{Outlook}

Detailed open problems associated with the three main research themes have
already been discussed at the ends of
Chapters~\ref{ch:Canonical Forms as Dual Volumes},
\ref{ch:From Integrands to Integrals}, and
\ref{ch:positivity-and-cluster-structures}; see
Sections~\ref{sec:Chapter 5 open problems},
\ref{subsec:outlook-geometric-landau-analysis}, and
\ref{sec:positivity-cluster-outlook}. This final outlook therefore highlights
only broader questions connecting the different parts of the thesis.

The first is whether positivity and duality survive, in transformed form,
after integration. Chapter~\ref{ch:Canonical Forms as Dual Volumes} studied
canonical functions as dual volumes or Laplace transforms of non-negative
measures, while Chapter~\ref{ch:From Integrands to Integrals} emphasized that
loop integration produces multivalued transcendental functions. A concrete
open problem is to determine when dual-volume or positive-measure
interpretations extend to canonical pairings evaluating to polylogarithmic or
more general periods. Beyond the examples considered here, this remains a
possible organizing principle rather than an established theorem.

A second question is how directly positive geometry controls the analytic
machinery of integrated quantities. The geometric Landau analysis of
Chapter~\ref{ch:From Integrands to Integrals} suggests an algorithmic path from
boundaries of positive or negative geometries to Landau diagrams and symbol
letters. The Grassmannian analysis of
Chapter~\ref{ch:positivity-and-cluster-structures} suggests that the resulting
Landau loci may also encode positivity, recursion, and cluster factorization.
The long-term goal would be a geometric bootstrap in which alphabets,
adjacency constraints, and differential equations are derived from the same
underlying geometry. At present, this remains a speculative direction
supported by the families analyzed in the thesis.

A third question concerns the range of the framework. The Amplituhedron and
the Grassmannian Landau constructions are most natural in planar
\(\mathcal N=4\) SYM, but the methods developed here invite comparison with
other positive geometries, kinematic regimes, and representations of Feynman
integrals. One concrete task is to relate the momentum-twistor incidence
varieties of Chapter~\ref{ch:positivity-and-cluster-structures} to the
parametric and differential-equation methods discussed in
Chapter~\ref{ch:From Integrands to Integrals}. A further challenge is to
determine which parts of the incidence, positivity, and dual-volume framework
survive in multiloop integrals relevant to QCD, where planarity, dual conformal
symmetry, and uniform transcendentality are not generally available. Such
comparisons could clarify which features are special to planar
four-dimensional kinematics and which are more universal consequences of
positivity and Landau geometry.

The common lesson is that positive geometry is not only a tool for constructing
compact integrands. It also provides a language for positivity, volumes, and
singularities after integration. The broader goal is to understand scattering
amplitudes as objects whose analytic structure is governed by geometry at
every stage: in the canonical form before integration, in its dual-volume
interpretations, and in the Landau and cluster structures arising after
integration.

%% file: Acknowledgment.tex
\section*{Acknowledgment}
\addcontentsline{toc}{section}{Acknowledgment}
I would first like to express my gratitude to my supervisor, Johannes Henn.
His guidance and encouragement shaped this thesis in essential ways. I am
grateful for the freedom he gave me to explore, and for reminding me of the
importance of the bigger picture beyond the technical details.

I am equally grateful to Bernd Sturmfels, my moral co-supervisor, for his
genuine enthusiasm and mathematical perspective. His influence has been central
to the way I learned to think about the geometric and algebraic structures
appearing throughout this work.

I thank my parents for their unconditional support and for always giving me a
place from which to start again. Above all, I thank Emese for her love,
patience, and support.

I am also grateful to the friends from Ticino and elsewhere around the world
who have remained close despite distance and time. Their friendship has been a
constant reminder that this path was never only about work.

I am grateful to all the friends I met in Munich, both at the institute and
outside it. They made these years richer, lighter, and much more fun than a
Ph.D. has any right to be. I will avoid writing a full list of names here, not
because of a lack of affection, but because of the inevitable failures of my
memory. A special mention goes to the \emph{Volleyballers}, as well as to
S\'ergio Carr\^olo, who had the questionable privilege of sharing a flat with me
and appears to have survived the experience with remarkable success.

I would also like to thank the Max Planck Institute for Physics, where I
carried out my Ph.D., for providing an excellent and stimulating scientific
environment. I am especially grateful to Frank Steffen and Giulia Zanderighi
for their dedication, effort, and passion in the IMPRS program.

Concerning this thesis, I am especially grateful to Sergio Carrolo, Adolfo Hilario-Garcia and Dmitrii Pavlov for their useful comments on the draft.

Finally, I gratefully acknowledge financial and institutional support from the
Max Planck Institute for Physics and from the European Research Council through
the Synergy Grant UNIVERSE+ (grant agreement No. 101118787).
$\!\!$ \begin{scriptsize}Views~and~opinions expressed
are however those of the~authors only and do not necessarily reflect those of the European Union or the
European
Research Council Executive Agency. Neither the European Union nor the granting authority
can be held responsible for them.
\end{scriptsize}

%% file: main_elsevier.bbl
\begin{thebibliography}{418}
\expandafter\ifx\csname natexlab\endcsname\relax\def\natexlab#1{#1}\fi
\providecommand{\url}[1]{\texttt{#1}}
\providecommand{\href}[2]{#2}
\providecommand{\path}[1]{#1}
\providecommand{\DOIprefix}{doi:}
\providecommand{\ArXivprefix}{arXiv:}
\providecommand{\URLprefix}{URL: }
\providecommand{\Pubmedprefix}{pmid:}
\providecommand{\doi}[1]{\href{http://dx.doi.org/#1}{\path{#1}}}
\providecommand{\Pubmed}[1]{\href{pmid:#1}{\path{#1}}}
\providecommand{\bibinfo}[2]{#2}
\ifx\xfnm\relax \def\xfnm[#1]{\unskip,\space#1}\fi
\bibitem[{Abreu et~al.(2017)Abreu, Britto, Duhr and Gardi}]{Abreu:2022mfk}
\bibinfo{author}{Abreu, S.}, \bibinfo{author}{Britto, R.},
  \bibinfo{author}{Duhr, C.}, \bibinfo{author}{Gardi, E.},
  \bibinfo{year}{2017}.
\newblock \bibinfo{title}{Cuts from residues: the one-loop case}.
\newblock \bibinfo{journal}{JHEP} \bibinfo{volume}{06}, \bibinfo{pages}{114}.
\newblock \DOIprefix\doi{10.1007/JHEP06(2017)114},
  \href{http://arxiv.org/abs/1702.03163}{\tt arXiv:1702.03163}.
\bibitem[{Adams and Weinzierl(2018)}]{AdamsWeinzierl2018}
\bibinfo{author}{Adams, L.}, \bibinfo{author}{Weinzierl, S.},
  \bibinfo{year}{2018}.
\newblock \bibinfo{title}{The \(\varepsilon\)-form of the differential
  equations for feynman integrals in the elliptic case}.
\newblock \bibinfo{journal}{Physics Letters B} \bibinfo{volume}{781},
  \bibinfo{pages}{270--278}.
\newblock \DOIprefix\doi{10.1016/j.physletb.2018.04.002},
  \href{http://arxiv.org/abs/1802.05020}{\tt arXiv:1802.05020}.
\bibitem[{Ait El~Manssour et~al.(2023a)Ait El~Manssour, H{\"a}rk{\"o}nen and
  Sturmfels}]{ait2023linear}
\bibinfo{author}{Ait El~Manssour, R.}, \bibinfo{author}{H{\"a}rk{\"o}nen, M.},
  \bibinfo{author}{Sturmfels, B.}, \bibinfo{year}{2023}a.
\newblock \bibinfo{title}{Linear pde with constant coefficients}.
\newblock \bibinfo{journal}{Glasgow Mathematical Journal} \bibinfo{volume}{65},
  \bibinfo{pages}{S2--S27}.
\bibitem[{Ait El~Manssour et~al.(2023b)Ait El~Manssour, H{\"a}rk{\"o}nen and
  Sturmfels}]{Liner_PDE}
\bibinfo{author}{Ait El~Manssour, R.}, \bibinfo{author}{H{\"a}rk{\"o}nen, M.},
  \bibinfo{author}{Sturmfels, B.}, \bibinfo{year}{2023}b.
\newblock \bibinfo{title}{Linear pde with constant coefficients}.
\newblock \bibinfo{journal}{Glasgow Mathematical Journal} \bibinfo{volume}{65},
  \bibinfo{pages}{S2--S27}.
\newblock \DOIprefix\doi{10.1017/S0017089521000355},
  \href{http://arxiv.org/abs/2104.10146}{\tt arXiv:2104.10146}.
\bibitem[{Alday et~al.(2011)Alday, Buchbinder and Tseytlin}]{Alday:2011ga}
\bibinfo{author}{Alday, L.F.}, \bibinfo{author}{Buchbinder, E.I.},
  \bibinfo{author}{Tseytlin, A.A.}, \bibinfo{year}{2011}.
\newblock \bibinfo{title}{{Correlation function of null polygonal Wilson loops
  with local operators}}.
\newblock \bibinfo{journal}{JHEP} \bibinfo{volume}{09}, \bibinfo{pages}{034}.
\newblock \DOIprefix\doi{10.1007/JHEP09(2011)034},
  \href{http://arxiv.org/abs/1107.5702}{\tt arXiv:1107.5702}.
\bibitem[{Alday et~al.(2010)Alday, Henn, Plefka and Schuster}]{Alday:2009zm}
\bibinfo{author}{Alday, L.F.}, \bibinfo{author}{Henn, J.M.},
  \bibinfo{author}{Plefka, J.}, \bibinfo{author}{Schuster, T.},
  \bibinfo{year}{2010}.
\newblock \bibinfo{title}{Scattering into the fifth dimension of
  \(\mathcal{N}=4\) super yang--mills}.
\newblock \bibinfo{journal}{Journal of High Energy Physics}
  \bibinfo{volume}{2010}, \bibinfo{pages}{077}.
\newblock \DOIprefix\doi{10.1007/JHEP01(2010)077},
  \href{http://arxiv.org/abs/0908.0684}{\tt arXiv:0908.0684}.
\bibitem[{Alday et~al.(2013a)Alday, Henn and Sikorowski}]{Alday:2013ip}
\bibinfo{author}{Alday, L.F.}, \bibinfo{author}{Henn, J.M.},
  \bibinfo{author}{Sikorowski, J.}, \bibinfo{year}{2013}a.
\newblock \bibinfo{title}{{Higher loop mixed correlators in N=4 SYM}}.
\newblock \bibinfo{journal}{JHEP} \bibinfo{volume}{03}, \bibinfo{pages}{058}.
\newblock \DOIprefix\doi{10.1007/JHEP03(2013)058},
  \href{http://arxiv.org/abs/1301.0149}{\tt arXiv:1301.0149}.
\bibitem[{Alday et~al.(2013b)Alday, Heslop and Sikorowski}]{Alday:2012hy}
\bibinfo{author}{Alday, L.F.}, \bibinfo{author}{Heslop, P.},
  \bibinfo{author}{Sikorowski, J.}, \bibinfo{year}{2013}b.
\newblock \bibinfo{title}{{Perturbative correlation functions of null Wilson
  loops and local operators}}.
\newblock \bibinfo{journal}{JHEP} \bibinfo{volume}{03}, \bibinfo{pages}{074}.
\newblock \DOIprefix\doi{10.1007/JHEP03(2013)074},
  \href{http://arxiv.org/abs/1207.4316}{\tt arXiv:1207.4316}.
\bibitem[{Alday and Maldacena(2007)}]{Alday:2007hr}
\bibinfo{author}{Alday, L.F.}, \bibinfo{author}{Maldacena, J.M.},
  \bibinfo{year}{2007}.
\newblock \bibinfo{title}{Gluon scattering amplitudes at strong coupling}.
\newblock \bibinfo{journal}{Journal of High Energy Physics}
  \bibinfo{volume}{2007}, \bibinfo{pages}{064}.
\newblock \DOIprefix\doi{10.1088/1126-6708/2007/06/064},
  \href{http://arxiv.org/abs/0705.0303}{\tt arXiv:0705.0303}.
\bibitem[{Anastasiou et~al.(2003)Anastasiou, Bern, Dixon and
  Kosower}]{AnastasiouBernDixonKosower2003}
\bibinfo{author}{Anastasiou, C.}, \bibinfo{author}{Bern, Z.},
  \bibinfo{author}{Dixon, L.J.}, \bibinfo{author}{Kosower, D.A.},
  \bibinfo{year}{2003}.
\newblock \bibinfo{title}{Planar amplitudes in maximally supersymmetric
  yang--mills theory}.
\newblock \bibinfo{journal}{Physical Review Letters} \bibinfo{volume}{91},
  \bibinfo{pages}{251602}.
\newblock \DOIprefix\doi{10.1103/PhysRevLett.91.251602},
  \href{http://arxiv.org/abs/hep-th/0309040}{\tt arXiv:hep-th/0309040}.
\bibitem[{Anastasiou et~al.(2007)Anastasiou, Britto, Feng, Kunszt and
  Mastrolia}]{AnastasiouBrittoFengKunsztMastrolia2006}
\bibinfo{author}{Anastasiou, C.}, \bibinfo{author}{Britto, R.},
  \bibinfo{author}{Feng, B.}, \bibinfo{author}{Kunszt, Z.},
  \bibinfo{author}{Mastrolia, P.}, \bibinfo{year}{2007}.
\newblock \bibinfo{title}{\(d\)-dimensional unitarity cut method}.
\newblock \bibinfo{journal}{Physics Letters B} \bibinfo{volume}{645},
  \bibinfo{pages}{213--216}.
\newblock \DOIprefix\doi{10.1016/j.physletb.2006.12.022},
  \href{http://arxiv.org/abs/hep-ph/0609191}{\tt arXiv:hep-ph/0609191}.
\bibitem[{Anastasiou and Melnikov(2002)}]{Anastasiou:2003kj}
\bibinfo{author}{Anastasiou, C.}, \bibinfo{author}{Melnikov, K.},
  \bibinfo{year}{2002}.
\newblock \bibinfo{title}{Higgs boson production at hadron colliders in nnlo
  qcd}.
\newblock \bibinfo{journal}{Nucl. Phys. B} \bibinfo{volume}{646},
  \bibinfo{pages}{220--256}.
\newblock \DOIprefix\doi{10.1016/S0550-3213(02)00837-4},
  \href{http://arxiv.org/abs/hep-ph/0207004}{\tt arXiv:hep-ph/0207004}.
\bibitem[{Aneesh et~al.(2020)Aneesh, Banerjee, Jagadale, Rajan, Laddha and
  Mahato}]{Aneesh:2019cvt}
\bibinfo{author}{Aneesh, P.B.}, \bibinfo{author}{Banerjee, P.},
  \bibinfo{author}{Jagadale, M.}, \bibinfo{author}{Rajan, R.},
  \bibinfo{author}{Laddha, A.}, \bibinfo{author}{Mahato, S.},
  \bibinfo{year}{2020}.
\newblock \bibinfo{title}{{On positive geometries of quartic interactions:
  Stokes polytopes, lower forms on associahedra and world-sheet forms}}.
\newblock \bibinfo{journal}{JHEP} \bibinfo{volume}{04}, \bibinfo{pages}{149}.
\newblock \DOIprefix\doi{10.1007/JHEP04(2020)149},
  \href{http://arxiv.org/abs/1911.06008}{\tt arXiv:1911.06008}.
\bibitem[{Aomoto(1982)}]{Aomoto1982}
\bibinfo{author}{Aomoto, K.}, \bibinfo{year}{1982}.
\newblock \bibinfo{title}{Addition theorem of {A}bel type for
  hyper-logarithms}.
\newblock \bibinfo{journal}{Nagoya Mathematical Journal} \bibinfo{volume}{88},
  \bibinfo{pages}{55--71}.
\newblock \DOIprefix\doi{10.1017/S0027763000020092}.
\bibitem[{Aomoto and Kita(2011)}]{AomotoKita2011}
\bibinfo{author}{Aomoto, K.}, \bibinfo{author}{Kita, M.}, \bibinfo{year}{2011}.
\newblock \bibinfo{title}{Theory of Hypergeometric Functions}.
\newblock \bibinfo{publisher}{Springer}.
\newblock \DOIprefix\doi{10.1007/978-4-431-53938-4}.
\bibitem[{Arkani-Hamed et~al.(2018a)Arkani-Hamed, Bai, He and
  Yan}]{ABHY_original}
\bibinfo{author}{Arkani-Hamed, N.}, \bibinfo{author}{Bai, Y.},
  \bibinfo{author}{He, S.}, \bibinfo{author}{Yan, G.}, \bibinfo{year}{2018}a.
\newblock \bibinfo{title}{Scattering forms and the positive geometry of
  kinematics, color and the worldsheet}.
\newblock \bibinfo{journal}{Journal of High Energy Physics}
  \bibinfo{volume}{2018}, \bibinfo{pages}{096}.
\newblock \DOIprefix\doi{10.1007/JHEP05(2018)096},
  \href{http://arxiv.org/abs/1711.09102}{\tt arXiv:1711.09102}.
\bibitem[{Arkani-Hamed et~al.(2017a)Arkani-Hamed, Bai and
  Lam}]{Positive_geometries}
\bibinfo{author}{Arkani-Hamed, N.}, \bibinfo{author}{Bai, Y.},
  \bibinfo{author}{Lam, T.}, \bibinfo{year}{2017}a.
\newblock \bibinfo{title}{Positive geometries and canonical forms}.
\newblock \bibinfo{journal}{Journal of High Energy Physics}
  \bibinfo{volume}{2017}, \bibinfo{pages}{039}.
\newblock \DOIprefix\doi{10.1007/JHEP11(2017)039},
  \href{http://arxiv.org/abs/1703.04541}{\tt arXiv:1703.04541}.
\bibitem[{Arkani-Hamed et~al.(2025a)Arkani-Hamed, Baumann, Hillman, Joyce, Lee
  and Pimentel}]{ArkaniHamedBaumannHillmanJoyceLeePimentel2025}
\bibinfo{author}{Arkani-Hamed, N.}, \bibinfo{author}{Baumann, D.},
  \bibinfo{author}{Hillman, A.}, \bibinfo{author}{Joyce, A.},
  \bibinfo{author}{Lee, H.}, \bibinfo{author}{Pimentel, G.L.},
  \bibinfo{year}{2025}a.
\newblock \bibinfo{title}{Differential equations for cosmological correlators}.
\newblock \bibinfo{journal}{Journal of High Energy Physics}
  \bibinfo{volume}{2025}, \bibinfo{pages}{009}.
\newblock \DOIprefix\doi{10.1007/JHEP09(2025)009},
  \href{http://arxiv.org/abs/2312.05303}{\tt arXiv:2312.05303}.
\bibitem[{Arkani-Hamed et~al.(2017b)Arkani-Hamed, Benincasa and
  Postnikov}]{Cosmological_polytopes}
\bibinfo{author}{Arkani-Hamed, N.}, \bibinfo{author}{Benincasa, P.},
  \bibinfo{author}{Postnikov, A.}, \bibinfo{year}{2017}b.
\newblock \bibinfo{title}{Cosmological polytopes and the wavefunction of the
  universe} \href{http://arxiv.org/abs/1709.02813}{\tt arXiv:1709.02813}.
\bibitem[{Arkani-Hamed et~al.(2011)Arkani-Hamed, Bourjaily, Cachazo, Caron-Huot
  and Trnka}]{ArkaniHamed:2010kv}
\bibinfo{author}{Arkani-Hamed, N.}, \bibinfo{author}{Bourjaily, J.L.},
  \bibinfo{author}{Cachazo, F.}, \bibinfo{author}{Caron-Huot, S.},
  \bibinfo{author}{Trnka, J.}, \bibinfo{year}{2011}.
\newblock \bibinfo{title}{The all-loop integrand for scattering amplitudes in
  planar \(\mathcal{N}=4\) sym}.
\newblock \bibinfo{journal}{JHEP} \bibinfo{volume}{01}, \bibinfo{pages}{041}.
\newblock \DOIprefix\doi{10.1007/JHEP01(2011)041},
  \href{http://arxiv.org/abs/1008.2958}{\tt arXiv:1008.2958}.
\bibitem[{Arkani-Hamed et~al.(2016)Arkani-Hamed, Bourjaily, Cachazo, Goncharov,
  Postnikov and Trnka}]{Grassmannian}
\bibinfo{author}{Arkani-Hamed, N.}, \bibinfo{author}{Bourjaily, J.L.},
  \bibinfo{author}{Cachazo, F.}, \bibinfo{author}{Goncharov, A.B.},
  \bibinfo{author}{Postnikov, A.}, \bibinfo{author}{Trnka, J.},
  \bibinfo{year}{2016}.
\newblock \bibinfo{title}{Grassmannian Geometry of Scattering Amplitudes}.
\newblock \bibinfo{publisher}{Cambridge University Press},
  \bibinfo{address}{Cambridge}.
\newblock \DOIprefix\doi{10.1017/CBO9781316091548},
  \href{http://arxiv.org/abs/1212.5605}{\tt arXiv:1212.5605}.
\bibitem[{Arkani-Hamed et~al.(2010a)Arkani-Hamed, Cachazo, Cheung and
  Kaplan}]{ArkaniHamed:2009dn}
\bibinfo{author}{Arkani-Hamed, N.}, \bibinfo{author}{Cachazo, F.},
  \bibinfo{author}{Cheung, C.}, \bibinfo{author}{Kaplan, J.},
  \bibinfo{year}{2010}a.
\newblock \bibinfo{title}{The s-matrix in twistor space}.
\newblock \bibinfo{journal}{Journal of High Energy Physics}
  \bibinfo{volume}{2010}, \bibinfo{pages}{110}.
\newblock \DOIprefix\doi{10.1007/JHEP03(2010)110},
  \href{http://arxiv.org/abs/0903.2110}{\tt arXiv:0903.2110}.
\bibitem[{Arkani-Hamed et~al.(2010b)Arkani-Hamed, Cachazo and
  Kaplan}]{ArkaniHamed:2008gz}
\bibinfo{author}{Arkani-Hamed, N.}, \bibinfo{author}{Cachazo, F.},
  \bibinfo{author}{Kaplan, J.}, \bibinfo{year}{2010}b.
\newblock \bibinfo{title}{What is the simplest quantum field theory?}
\newblock \bibinfo{journal}{Journal of High Energy Physics}
  \bibinfo{volume}{09}, \bibinfo{pages}{016}.
\newblock \href{http://arxiv.org/abs/0808.1446}{\tt arXiv:0808.1446}.
\bibitem[{Arkani-Hamed et~al.(2025b)Arkani-Hamed, Cao, Dong, Figueiredo and
  He}]{ArkaniHamed:2024surfaceYM}
\bibinfo{author}{Arkani-Hamed, N.}, \bibinfo{author}{Cao, Q.},
  \bibinfo{author}{Dong, J.}, \bibinfo{author}{Figueiredo, C.},
  \bibinfo{author}{He, S.}, \bibinfo{year}{2025}b.
\newblock \bibinfo{title}{Surface kinematics and ``the'' yang-mills integrand}.
\newblock \bibinfo{journal}{Phys. Rev. Lett.} \bibinfo{volume}{134},
  \bibinfo{pages}{171601}.
\newblock \DOIprefix\doi{10.1103/PhysRevLett.134.171601},
  \href{http://arxiv.org/abs/2408.11891}{\tt arXiv:2408.11891}.
\bibitem[{Arkani-Hamed et~al.(2024)Arkani-Hamed, Figueiredo and
  Vaz{\~a}o}]{Cosmoehdra}
\bibinfo{author}{Arkani-Hamed, N.}, \bibinfo{author}{Figueiredo, C.},
  \bibinfo{author}{Vaz{\~a}o, F.}, \bibinfo{year}{2024}.
\newblock \bibinfo{title}{{Cosmohedra}}
  \href{http://arxiv.org/abs/2412.19881}{\tt arXiv:2412.19881}.
\bibitem[{Arkani-Hamed et~al.(2021a)Arkani-Hamed, He and
  Lam}]{Stringy_Can_Forms}
\bibinfo{author}{Arkani-Hamed, N.}, \bibinfo{author}{He, S.},
  \bibinfo{author}{Lam, T.}, \bibinfo{year}{2021}a.
\newblock \bibinfo{title}{Stringy canonical forms}.
\newblock \bibinfo{journal}{Journal of High Energy Physics}
  \bibinfo{volume}{2021}, \bibinfo{pages}{69}.
\bibitem[{Arkani-Hamed et~al.(2023)Arkani-Hamed, He, Lam and
  Thomas}]{AHLTBinary}
\bibinfo{author}{Arkani-Hamed, N.}, \bibinfo{author}{He, S.},
  \bibinfo{author}{Lam, T.}, \bibinfo{author}{Thomas, H.},
  \bibinfo{year}{2023}.
\newblock \bibinfo{title}{Binary geometries, generalized particles and strings,
  and cluster algebras}.
\newblock \bibinfo{journal}{Physical Review D} \bibinfo{volume}{107},
  \bibinfo{pages}{066015}.
\newblock \href{http://arxiv.org/abs/1912.11764}{\tt arXiv:1912.11764}.
\bibitem[{Arkani-Hamed et~al.(2022)Arkani-Hamed, Henn and
  Trnka}]{ArkaniHamedHennTrnka2021}
\bibinfo{author}{Arkani-Hamed, N.}, \bibinfo{author}{Henn, J.},
  \bibinfo{author}{Trnka, J.}, \bibinfo{year}{2022}.
\newblock \bibinfo{title}{{Nonperturbative Negative Geometries: Amplitudes at
  Strong Coupling and the Amplituhedron}}.
\newblock \bibinfo{journal}{JHEP} \bibinfo{volume}{03}, \bibinfo{pages}{108}.
\newblock \DOIprefix\doi{10.1007/JHEP03(2022)108},
  \href{http://arxiv.org/abs/2112.06956}{\tt arXiv:2112.06956}.
\bibitem[{Arkani-Hamed et~al.(2015)Arkani-Hamed, Hodges and
  Trnka}]{positive_amplitudes}
\bibinfo{author}{Arkani-Hamed, N.}, \bibinfo{author}{Hodges, A.},
  \bibinfo{author}{Trnka, J.}, \bibinfo{year}{2015}.
\newblock \bibinfo{title}{Positive amplitudes in the amplituhedron}.
\newblock \bibinfo{journal}{Journal of High Energy Physics}
  \bibinfo{volume}{2015}, \bibinfo{pages}{030}.
\newblock \DOIprefix\doi{10.1007/JHEP08(2015)030},
  \href{http://arxiv.org/abs/1412.8478}{\tt arXiv:1412.8478}.
\bibitem[{Arkani-Hamed and Kaplan(2008)}]{ArkaniHamed:2008yf}
\bibinfo{author}{Arkani-Hamed, N.}, \bibinfo{author}{Kaplan, J.},
  \bibinfo{year}{2008}.
\newblock \bibinfo{title}{On tree amplitudes in gauge theory and gravity}.
\newblock \bibinfo{journal}{Journal of High Energy Physics}
  \bibinfo{volume}{04}, \bibinfo{pages}{076}.
\newblock \href{http://arxiv.org/abs/0801.2385}{\tt arXiv:0801.2385}.
\bibitem[{Arkani-Hamed et~al.(2021b)Arkani-Hamed, Lam and Spradlin}]{ALS}
\bibinfo{author}{Arkani-Hamed, N.}, \bibinfo{author}{Lam, T.},
  \bibinfo{author}{Spradlin, M.}, \bibinfo{year}{2021}b.
\newblock \bibinfo{title}{Non-perturbative geometries for planar
  \(\mathcal{N}=4\) sym amplitudes}.
\newblock \bibinfo{journal}{Journal of High Energy Physics}
  \bibinfo{volume}{2021}, \bibinfo{pages}{065}.
\newblock \DOIprefix\doi{10.1007/JHEP06(2021)065},
  \href{http://arxiv.org/abs/1912.08222}{\tt arXiv:1912.08222}.
\bibitem[{Arkani-Hamed et~al.(2019)Arkani-Hamed, Langer, Yelleshpur~Srikant and
  Trnka}]{Arkani-Hamed:2018rsk}
\bibinfo{author}{Arkani-Hamed, N.}, \bibinfo{author}{Langer, C.},
  \bibinfo{author}{Yelleshpur~Srikant, A.}, \bibinfo{author}{Trnka, J.},
  \bibinfo{year}{2019}.
\newblock \bibinfo{title}{{Deep Into the Amplituhedron: Amplitude Singularities
  at All Loops and Legs}}.
\newblock \bibinfo{journal}{Phys. Rev. Lett.} \bibinfo{volume}{122},
  \bibinfo{pages}{051601}.
\newblock \DOIprefix\doi{10.1103/PhysRevLett.122.051601},
  \href{http://arxiv.org/abs/1810.08208}{\tt arXiv:1810.08208}.
\bibitem[{Arkani-Hamed et~al.(2018b)Arkani-Hamed, Thomas and
  Trnka}]{Arkani_Hamed_2018}
\bibinfo{author}{Arkani-Hamed, N.}, \bibinfo{author}{Thomas, H.},
  \bibinfo{author}{Trnka, J.}, \bibinfo{year}{2018}b.
\newblock \bibinfo{title}{Unwinding the amplituhedron in binary}.
\newblock \bibinfo{journal}{Journal of High Energy Physics}
  \bibinfo{volume}{2018}, \bibinfo{pages}{016}.
\newblock \DOIprefix\doi{10.1007/JHEP01(2018)016},
  \href{http://arxiv.org/abs/1704.05069}{\tt arXiv:1704.05069}.
\bibitem[{Arkani-Hamed and Trnka(2014a)}]{the_amplituhedron}
\bibinfo{author}{Arkani-Hamed, N.}, \bibinfo{author}{Trnka, J.},
  \bibinfo{year}{2014}a.
\newblock \bibinfo{title}{The amplituhedron}.
\newblock \bibinfo{journal}{Journal of High Energy Physics}
  \bibinfo{volume}{2014}, \bibinfo{pages}{030}.
\newblock \DOIprefix\doi{10.1007/JHEP10(2014)030},
  \href{http://arxiv.org/abs/1312.2007}{\tt arXiv:1312.2007}.
\bibitem[{Arkani-Hamed and Trnka(2014b)}]{Arkani-Hamed:2013kca}
\bibinfo{author}{Arkani-Hamed, N.}, \bibinfo{author}{Trnka, J.},
  \bibinfo{year}{2014}b.
\newblock \bibinfo{title}{{Into the Amplituhedron}}.
\newblock \bibinfo{journal}{JHEP} \bibinfo{volume}{12}, \bibinfo{pages}{182}.
\newblock \DOIprefix\doi{10.1007/JHEP12(2014)182},
  \href{http://arxiv.org/abs/1312.7878}{\tt arXiv:1312.7878}.
\bibitem[{Arkani-Hamed and Yuan(2017)}]{Arkani-Hamed:2017ahv}
\bibinfo{author}{Arkani-Hamed, N.}, \bibinfo{author}{Yuan, E.Y.},
  \bibinfo{year}{2017}.
\newblock \bibinfo{title}{One-loop integrals from spherical projections of
  planes and quadrics} \href{http://arxiv.org/abs/1712.09991}{\tt
  arXiv:1712.09991}.
\bibitem[{Atiyah et~al.(1970)Atiyah, Bott and Gårding}]{1Garding_1970}
\bibinfo{author}{Atiyah, M.F.}, \bibinfo{author}{Bott, R.},
  \bibinfo{author}{Gårding, L.}, \bibinfo{year}{1970}.
\newblock \bibinfo{title}{Lacunas for hyperbolic differential operators with
  constant coefficients i}.
\newblock \bibinfo{journal}{Acta Mathematica} \bibinfo{volume}{124},
  \bibinfo{pages}{109--189}.
\bibitem[{Baadsgaard et~al.(2015)Baadsgaard, Bjerrum-Bohr, Bourjaily, Damgaard
  and Feng}]{Baadsgaard:2015twa}
\bibinfo{author}{Baadsgaard, C.}, \bibinfo{author}{Bjerrum-Bohr, N.E.J.},
  \bibinfo{author}{Bourjaily, J.L.}, \bibinfo{author}{Damgaard, P.H.},
  \bibinfo{author}{Feng, B.}, \bibinfo{year}{2015}.
\newblock \bibinfo{title}{Integration rules for loop scattering equations}.
\newblock \bibinfo{journal}{JHEP} \bibinfo{volume}{11}, \bibinfo{pages}{080}.
\newblock \DOIprefix\doi{10.1007/JHEP11(2015)080},
  \href{http://arxiv.org/abs/1508.03627}{\tt arXiv:1508.03627}.
\bibitem[{Banerjee et~al.(2019)Banerjee, Laddha and Raman}]{Banerjee:2018tun}
\bibinfo{author}{Banerjee, P.}, \bibinfo{author}{Laddha, A.},
  \bibinfo{author}{Raman, P.}, \bibinfo{year}{2019}.
\newblock \bibinfo{title}{{Stokes polytopes: the positive geometry for
  $\phi^{4}$ interactions}}.
\newblock \bibinfo{journal}{JHEP} \bibinfo{volume}{08}, \bibinfo{pages}{067}.
\newblock \DOIprefix\doi{10.1007/JHEP08(2019)067},
  \href{http://arxiv.org/abs/1811.05904}{\tt arXiv:1811.05904}.
\bibitem[{Bao and He(2019)}]{baoHe2019m2}
\bibinfo{author}{Bao, H.}, \bibinfo{author}{He, X.}, \bibinfo{year}{2019}.
\newblock \bibinfo{title}{{The $m=2$ amplituhedron}}
  \href{http://arxiv.org/abs/1909.06015}{\tt arXiv:1909.06015}.
\bibitem[{Beck(2004)}]{Beck2004}
\bibinfo{author}{Beck, M.}, \bibinfo{year}{2004}.
\newblock \bibinfo{title}{The partial-fractions method for counting solutions
  to integral linear systems}.
\newblock \bibinfo{journal}{Discrete \& Computational Geometry}
  \bibinfo{volume}{32}, \bibinfo{pages}{437--446}.
\newblock \DOIprefix\doi{10.1007/s00454-004-1131-z},
  \href{http://arxiv.org/abs/math/0309332}{\tt arXiv:math/0309332}.
\bibitem[{Beisert et~al.(2007)Beisert, Eden and
  Staudacher}]{BeisertEdenStaudacher2006}
\bibinfo{author}{Beisert, N.}, \bibinfo{author}{Eden, B.},
  \bibinfo{author}{Staudacher, M.}, \bibinfo{year}{2007}.
\newblock \bibinfo{title}{Transcendentality and crossing}.
\newblock \bibinfo{journal}{Journal of Statistical Mechanics}
  \bibinfo{volume}{2007}, \bibinfo{pages}{P01021}.
\newblock \DOIprefix\doi{10.1088/1742-5468/2007/01/P01021},
  \href{http://arxiv.org/abs/hep-th/0610251}{\tt arXiv:hep-th/0610251}.
\bibitem[{Beisert et~al.(2012)}]{Beisert:2010jr}
\bibinfo{author}{Beisert, N.}, et~al., \bibinfo{year}{2012}.
\newblock \bibinfo{title}{Review of ads/cft integrability: An overview}.
\newblock \bibinfo{journal}{Letters in Mathematical Physics}
  \bibinfo{volume}{99}, \bibinfo{pages}{3--32}.
\newblock \href{http://arxiv.org/abs/1012.3982}{\tt arXiv:1012.3982}.
\bibitem[{Bendle et~al.(2020)Bendle, B{\"o}hm, Decker, Georgoudis, Pfreundt,
  Rahn, Wasser and
  Zhang}]{BendleBoehmDeckerGeorgoudisPfreundtRahnWasserZhang2020}
\bibinfo{author}{Bendle, D.}, \bibinfo{author}{B{\"o}hm, J.},
  \bibinfo{author}{Decker, W.}, \bibinfo{author}{Georgoudis, A.},
  \bibinfo{author}{Pfreundt, F.J.}, \bibinfo{author}{Rahn, M.},
  \bibinfo{author}{Wasser, P.}, \bibinfo{author}{Zhang, Y.},
  \bibinfo{year}{2020}.
\newblock \bibinfo{title}{Integration-by-parts reductions of feynman integrals
  using {Singular} and {GPI-Space}}.
\newblock \bibinfo{journal}{Journal of High Energy Physics}
  \bibinfo{volume}{2020}, \bibinfo{pages}{079}.
\newblock \DOIprefix\doi{10.1007/JHEP02(2020)079},
  \href{http://arxiv.org/abs/1908.04301}{\tt arXiv:1908.04301}.
\bibitem[{Benincasa and Dian(2025)}]{BenincasaDian2025}
\bibinfo{author}{Benincasa, P.}, \bibinfo{author}{Dian, G.},
  \bibinfo{year}{2025}.
\newblock \bibinfo{title}{The geometry of cosmological correlators}.
\newblock \bibinfo{journal}{SciPost Physics} \bibinfo{volume}{18},
  \bibinfo{pages}{105}.
\newblock \DOIprefix\doi{10.21468/SciPostPhys.18.3.105}.
\bibitem[{Berends and Giele(1987)}]{Berends:1987me}
\bibinfo{author}{Berends, F.A.}, \bibinfo{author}{Giele, W.T.},
  \bibinfo{year}{1987}.
\newblock \bibinfo{title}{The six gluon process as an example of weyl-van der
  waerden spinor calculus}.
\newblock \bibinfo{journal}{Nucl. Phys. B} \bibinfo{volume}{294},
  \bibinfo{pages}{700--732}.
\bibitem[{Berenstein et~al.(1996)Berenstein, Fomin and Zelevinsky}]{BFZ}
\bibinfo{author}{Berenstein, A.}, \bibinfo{author}{Fomin, S.},
  \bibinfo{author}{Zelevinsky, A.}, \bibinfo{year}{1996}.
\newblock \bibinfo{title}{Parametrizations of canonical bases and totally
  positive matrices}.
\newblock \bibinfo{journal}{Advances in Mathematics} \bibinfo{volume}{122}.
\bibitem[{Berghoff and Panzer(2025)}]{BerghoffPanzer2025}
\bibinfo{author}{Berghoff, M.}, \bibinfo{author}{Panzer, E.},
  \bibinfo{year}{2025}.
\newblock \bibinfo{title}{Hierarchies in relative picard--lefschetz theory}.
\newblock \bibinfo{journal}{Annales Henri Poincar{\'e}}
  \DOIprefix\doi{10.1007/s00023-025-01545-9},
  \href{http://arxiv.org/abs/2212.06661}{\tt arXiv:2212.06661}.
\bibitem[{Bern et~al.(2008)Bern, Carrasco and Johansson}]{Bern:2008qj}
\bibinfo{author}{Bern, Z.}, \bibinfo{author}{Carrasco, J.J.},
  \bibinfo{author}{Johansson, H.}, \bibinfo{year}{2008}.
\newblock \bibinfo{title}{New relations for gauge-theory amplitudes}.
\newblock \bibinfo{journal}{Physical Review D} \bibinfo{volume}{78},
  \bibinfo{pages}{085011}.
\newblock \href{http://arxiv.org/abs/0805.3993}{\tt arXiv:0805.3993}.
\bibitem[{Bern et~al.(1994)Bern, Dixon, Dunbar and Kosower}]{Bern:1994zx}
\bibinfo{author}{Bern, Z.}, \bibinfo{author}{Dixon, L.J.},
  \bibinfo{author}{Dunbar, D.C.}, \bibinfo{author}{Kosower, D.A.},
  \bibinfo{year}{1994}.
\newblock \bibinfo{title}{One-loop \(n\)-point gauge theory amplitudes,
  unitarity and collinear limits}.
\newblock \bibinfo{journal}{Nucl. Phys. B} \bibinfo{volume}{425},
  \bibinfo{pages}{217--260}.
\newblock \DOIprefix\doi{10.1016/0550-3213(94)90179-1},
  \href{http://arxiv.org/abs/hep-ph/9403226}{\tt arXiv:hep-ph/9403226}.
\bibitem[{Bern et~al.(1995)Bern, Dixon, Dunbar and
  Kosower}]{BernDixonDunbarKosower1994}
\bibinfo{author}{Bern, Z.}, \bibinfo{author}{Dixon, L.J.},
  \bibinfo{author}{Dunbar, D.C.}, \bibinfo{author}{Kosower, D.A.},
  \bibinfo{year}{1995}.
\newblock \bibinfo{title}{Fusing gauge theory tree amplitudes into loop
  amplitudes}.
\newblock \bibinfo{journal}{Nuclear Physics B} \bibinfo{volume}{435},
  \bibinfo{pages}{59--101}.
\newblock \DOIprefix\doi{10.1016/0550-3213(94)00488-Z},
  \href{http://arxiv.org/abs/hep-ph/9409265}{\tt arXiv:hep-ph/9409265}.
\bibitem[{Bern et~al.(1996)Bern, Dixon and Kosower}]{BernDixonKosower1996}
\bibinfo{author}{Bern, Z.}, \bibinfo{author}{Dixon, L.J.},
  \bibinfo{author}{Kosower, D.A.}, \bibinfo{year}{1996}.
\newblock \bibinfo{title}{Progress in one-loop {QCD} computations}.
\newblock \bibinfo{journal}{Annual Review of Nuclear and Particle Science}
  \bibinfo{volume}{46}, \bibinfo{pages}{109--148}.
\newblock \DOIprefix\doi{10.1146/annurev.nucl.46.1.109},
  \href{http://arxiv.org/abs/hep-ph/9602280}{\tt arXiv:hep-ph/9602280}.
\bibitem[{Bern et~al.(1998)Bern, Dixon and Kosower}]{Bern:1994cg}
\bibinfo{author}{Bern, Z.}, \bibinfo{author}{Dixon, L.J.},
  \bibinfo{author}{Kosower, D.A.}, \bibinfo{year}{1998}.
\newblock \bibinfo{title}{One-loop amplitudes for e+ e- to four partons}.
\newblock \bibinfo{journal}{Nuclear Physics B} \bibinfo{volume}{513},
  \bibinfo{pages}{3--86}.
\newblock \href{http://arxiv.org/abs/hep-ph/9708239}{\tt arXiv:hep-ph/9708239}.
\bibitem[{Bern et~al.(2007)Bern, Dixon and Kosower}]{Bern:2007dw}
\bibinfo{author}{Bern, Z.}, \bibinfo{author}{Dixon, L.J.},
  \bibinfo{author}{Kosower, D.A.}, \bibinfo{year}{2007}.
\newblock \bibinfo{title}{On-shell methods in perturbative qcd}.
\newblock \bibinfo{journal}{Annals of Physics} \bibinfo{volume}{322},
  \bibinfo{pages}{1587--1634}.
\newblock \href{http://arxiv.org/abs/0704.2798}{\tt arXiv:0704.2798}.
\bibitem[{Bern et~al.(2005)Bern, Dixon and Smirnov}]{BernDixonSmirnov2005}
\bibinfo{author}{Bern, Z.}, \bibinfo{author}{Dixon, L.J.},
  \bibinfo{author}{Smirnov, V.A.}, \bibinfo{year}{2005}.
\newblock \bibinfo{title}{Iteration of planar amplitudes in maximally
  supersymmetric yang--mills theory at three loops and beyond}.
\newblock \bibinfo{journal}{Physical Review D} \bibinfo{volume}{72},
  \bibinfo{pages}{085001}.
\newblock \DOIprefix\doi{10.1103/PhysRevD.72.085001},
  \href{http://arxiv.org/abs/hep-th/0505205}{\tt arXiv:hep-th/0505205}.
\bibitem[{Besier and Festi(2021)}]{BesierFesti2020}
\bibinfo{author}{Besier, M.}, \bibinfo{author}{Festi, D.},
  \bibinfo{year}{2021}.
\newblock \bibinfo{title}{Rationalizability of square roots}.
\newblock \bibinfo{journal}{Journal of Symbolic Computation}
  \bibinfo{volume}{106}, \bibinfo{pages}{48--67}.
\newblock \DOIprefix\doi{10.1016/j.jsc.2020.08.002},
  \href{http://arxiv.org/abs/2006.07121}{\tt arXiv:2006.07121}.
\bibitem[{Besier et~al.(2019)Besier, van Straten and
  Weinzierl}]{Besier:2018jen}
\bibinfo{author}{Besier, M.}, \bibinfo{author}{van Straten, D.},
  \bibinfo{author}{Weinzierl, S.}, \bibinfo{year}{2019}.
\newblock \bibinfo{title}{Rationalizing roots: an algorithmic approach}.
\newblock \bibinfo{journal}{Communications in Number Theory and Physics}
  \bibinfo{volume}{13}, \bibinfo{pages}{253--297}.
\newblock \DOIprefix\doi{10.4310/CNTP.2019.v13.n2.a1},
  \href{http://arxiv.org/abs/1809.10983}{\tt arXiv:1809.10983}.
\bibitem[{Besier et~al.(2020)Besier, Wasser and
  Weinzierl}]{BesierWasserWeinzierl2019}
\bibinfo{author}{Besier, M.}, \bibinfo{author}{Wasser, P.},
  \bibinfo{author}{Weinzierl, S.}, \bibinfo{year}{2020}.
\newblock \bibinfo{title}{{RationalizeRoots}: Software package for the
  rationalization of square roots}.
\newblock \bibinfo{journal}{Computer Physics Communications}
  \bibinfo{volume}{253}, \bibinfo{pages}{107197}.
\newblock \DOIprefix\doi{10.1016/j.cpc.2020.107197},
  \href{http://arxiv.org/abs/1910.13251}{\tt arXiv:1910.13251}.
\bibitem[{Bj{\"o}rk(1979)}]{Bjoerk1979}
\bibinfo{author}{Bj{\"o}rk, J.E.}, \bibinfo{year}{1979}.
\newblock \bibinfo{title}{Rings of Differential Operators}.
\newblock \bibinfo{publisher}{North-Holland}.
\bibitem[{Bjorken(1959)}]{BjorkenLandau1959}
\bibinfo{author}{Bjorken, J.D.}, \bibinfo{year}{1959}.
\newblock \bibinfo{title}{Experimental tests of quantum electrodynamics and
  spectral representations of green's functions in perturbation theory}.
\newblock \bibinfo{journal}{Ph.D. thesis, Stanford University}
  \bibinfo{note}{Contains early discussion of Landau singularities and related
  analytic structure}.
\bibitem[{Bloch et~al.(2006)Bloch, Esnault and Kreimer}]{Bloch:2005bh}
\bibinfo{author}{Bloch, S.}, \bibinfo{author}{Esnault, H.},
  \bibinfo{author}{Kreimer, D.}, \bibinfo{year}{2006}.
\newblock \bibinfo{title}{On motives associated to graph polynomials}.
\newblock \bibinfo{journal}{Commun. Math. Phys.} \bibinfo{volume}{267},
  \bibinfo{pages}{181--225}.
\newblock \DOIprefix\doi{10.1007/s00220-006-0040-2},
  \href{http://arxiv.org/abs/math/0510011}{\tt arXiv:math/0510011}.
\bibitem[{Bochnak et~al.(2010)Bochnak, Coste and Roy}]{bochnak2010real}
\bibinfo{author}{Bochnak, J.}, \bibinfo{author}{Coste, M.},
  \bibinfo{author}{Roy, M.}, \bibinfo{year}{2010}.
\newblock \bibinfo{title}{Real Algebraic Geometry}.
\newblock Ergebnisse der Mathematik und ihrer Grenzgebiete. 3. Folge / A Series
  of Modern Surveys in Mathematics, \bibinfo{publisher}{Springer Berlin
  Heidelberg}.
\bibitem[{Bochnak et~al.(1998)Bochnak, Coste and Roy}]{BochnakCosteRoy1998}
\bibinfo{author}{Bochnak, J.}, \bibinfo{author}{Coste, M.},
  \bibinfo{author}{Roy, M.F.}, \bibinfo{year}{1998}.
\newblock \bibinfo{title}{Real Algebraic Geometry}. volume~\bibinfo{volume}{36}
  of \textit{\bibinfo{series}{Ergebnisse der Mathematik und ihrer
  Grenzgebiete}}.
\newblock \bibinfo{publisher}{Springer}.
\bibitem[{Boels et~al.(2007)Boels, Mason and Skinner}]{Boels:2010nw}
\bibinfo{author}{Boels, R.H.}, \bibinfo{author}{Mason, L.},
  \bibinfo{author}{Skinner, D.}, \bibinfo{year}{2007}.
\newblock \bibinfo{title}{Supersymmetric gauge theories in twistor space}.
\newblock \bibinfo{journal}{JHEP} \bibinfo{volume}{02}, \bibinfo{pages}{014}.
\newblock \DOIprefix\doi{10.1088/1126-6708/2007/02/014},
  \href{http://arxiv.org/abs/hep-th/0604040}{\tt arXiv:hep-th/0604040}.
\bibitem[{Bogner and Weinzierl(2010)}]{BognerWeinzierl}
\bibinfo{author}{Bogner, C.}, \bibinfo{author}{Weinzierl, S.},
  \bibinfo{year}{2010}.
\newblock \bibinfo{title}{Feynman graph polynomials}.
\newblock \bibinfo{journal}{International Journal of Modern Physics A}
  \bibinfo{volume}{25}, \bibinfo{pages}{2585--2618}.
\newblock \href{http://arxiv.org/abs/1002.3458}{\tt arXiv:1002.3458}.
\bibitem[{B{\"o}hm et~al.(2020)B{\"o}hm, Wittmann, Wu, Xu and
  Zhang}]{BoehmWittmannWuXuZhang2020}
\bibinfo{author}{B{\"o}hm, J.}, \bibinfo{author}{Wittmann, M.},
  \bibinfo{author}{Wu, Z.}, \bibinfo{author}{Xu, Y.}, \bibinfo{author}{Zhang,
  Y.}, \bibinfo{year}{2020}.
\newblock \bibinfo{title}{{IBP} reduction coefficients made simple}.
\newblock \bibinfo{journal}{Journal of High Energy Physics}
  \bibinfo{volume}{2020}, \bibinfo{pages}{054}.
\newblock \DOIprefix\doi{10.1007/JHEP12(2020)054},
  \href{http://arxiv.org/abs/2008.13194}{\tt arXiv:2008.13194}.
\bibitem[{Bollini and Giambiagi(1972)}]{BolliniGiambiagi1972}
\bibinfo{author}{Bollini, C.G.}, \bibinfo{author}{Giambiagi, J.J.},
  \bibinfo{year}{1972}.
\newblock \bibinfo{title}{Dimensional renormalization: The number of dimensions
  as a regularizing parameter}.
\newblock \bibinfo{journal}{Il Nuovo Cimento B} \bibinfo{volume}{12},
  \bibinfo{pages}{20--26}.
\newblock \DOIprefix\doi{10.1007/BF02895558}.
\bibitem[{Bosch(2013)}]{Bosch2013AlgebraicGeometry}
\bibinfo{author}{Bosch, S.}, \bibinfo{year}{2013}.
\newblock \bibinfo{title}{Algebraic Geometry and Commutative Algebra}.
\newblock \bibinfo{publisher}{Springer}.
\newblock \DOIprefix\doi{10.1007/978-1-4471-4829-6}.
\bibitem[{Bourjaily(2010)}]{Bourjaily:2010kw}
\bibinfo{author}{Bourjaily, J.L.}, \bibinfo{year}{2010}.
\newblock \bibinfo{title}{Efficient tree-amplitudes in $\mathcal{N}=4$:
  Automatic bcfw recursion in mathematica}
  \href{http://arxiv.org/abs/1011.2447}{\tt arXiv:1011.2447}.
\bibitem[{Bourjaily(2012)}]{Bourjaily:2012gy}
\bibinfo{author}{Bourjaily, J.L.}, \bibinfo{year}{2012}.
\newblock \bibinfo{title}{{Positroids, Plabic Graphs, and Scattering Amplitudes
  in Mathematica}} \href{http://arxiv.org/abs/1212.6974}{\tt arXiv:1212.6974}.
\bibitem[{Bourjaily et~al.(2019)Bourjaily, Herrmann, Langer, McLeod and
  Trnka}]{BourjailyHerrmannLangerMcLeodTrnka2019}
\bibinfo{author}{Bourjaily, J.L.}, \bibinfo{author}{Herrmann, E.},
  \bibinfo{author}{Langer, C.}, \bibinfo{author}{McLeod, A.J.},
  \bibinfo{author}{Trnka, J.}, \bibinfo{year}{2019}.
\newblock \bibinfo{title}{Prescriptive unitarity for non-planar six-particle
  amplitudes at two loops}.
\newblock \bibinfo{journal}{Journal of High Energy Physics}
  \bibinfo{volume}{2019}, \bibinfo{pages}{073}.
\newblock \DOIprefix\doi{10.1007/JHEP12(2019)073},
  \href{http://arxiv.org/abs/1909.09131}{\tt arXiv:1909.09131}.
\bibitem[{Bourjaily et~al.(2020)Bourjaily, Herrmann, Langer, McLeod and
  Trnka}]{BourjailyHerrmannLangerMcLeodTrnka2020}
\bibinfo{author}{Bourjaily, J.L.}, \bibinfo{author}{Herrmann, E.},
  \bibinfo{author}{Langer, C.}, \bibinfo{author}{McLeod, A.J.},
  \bibinfo{author}{Trnka, J.}, \bibinfo{year}{2020}.
\newblock \bibinfo{title}{All-multiplicity nonplanar amplitude integrands in
  maximally supersymmetric yang--mills theory at two loops}.
\newblock \bibinfo{journal}{Physical Review Letters} \bibinfo{volume}{124},
  \bibinfo{pages}{111603}.
\newblock \DOIprefix\doi{10.1103/PhysRevLett.124.111603},
  \href{http://arxiv.org/abs/1911.09106}{\tt arXiv:1911.09106}.
\bibitem[{Bourjaily et~al.(2017)Bourjaily, Herrmann and
  Trnka}]{BourjailyHerrmannTrnka2017}
\bibinfo{author}{Bourjaily, J.L.}, \bibinfo{author}{Herrmann, E.},
  \bibinfo{author}{Trnka, J.}, \bibinfo{year}{2017}.
\newblock \bibinfo{title}{Prescriptive unitarity}.
\newblock \bibinfo{journal}{Journal of High Energy Physics}
  \bibinfo{volume}{2017}, \bibinfo{pages}{059}.
\newblock \DOIprefix\doi{10.1007/JHEP06(2017)059},
  \href{http://arxiv.org/abs/1704.05460}{\tt arXiv:1704.05460}.
\bibitem[{Bourjaily et~al.(2021)Bourjaily, Kalyanapuram, Langer and
  Patatoukos}]{BourjailyKalyanapuramLangerPatatoukos2021}
\bibinfo{author}{Bourjaily, J.L.}, \bibinfo{author}{Kalyanapuram, N.},
  \bibinfo{author}{Langer, C.}, \bibinfo{author}{Patatoukos, K.},
  \bibinfo{year}{2021}.
\newblock \bibinfo{title}{Prescriptive unitarity with elliptic leading
  singularities}.
\newblock \bibinfo{journal}{Physical Review D} \bibinfo{volume}{104},
  \bibinfo{pages}{125009}.
\newblock \DOIprefix\doi{10.1103/PhysRevD.104.125009},
  \href{http://arxiv.org/abs/2102.02210}{\tt arXiv:2102.02210}.
\bibitem[{Bourjaily and Trnka(2015)}]{BourjailyTrnka2015}
\bibinfo{author}{Bourjaily, J.L.}, \bibinfo{author}{Trnka, J.},
  \bibinfo{year}{2015}.
\newblock \bibinfo{title}{Local integrand representations of all two-loop
  amplitudes in planar sym}.
\newblock \bibinfo{journal}{Journal of High Energy Physics}
  \bibinfo{volume}{2015}, \bibinfo{pages}{119}.
\newblock \DOIprefix\doi{10.1007/JHEP08(2015)119},
  \href{http://arxiv.org/abs/1505.05886}{\tt arXiv:1505.05886}.
\bibitem[{Bourjaily et~al.(2023)Bourjaily, Vergu and von
  Hippel}]{BourjailyVerguVonHippel2023}
\bibinfo{author}{Bourjaily, J.L.}, \bibinfo{author}{Vergu, C.},
  \bibinfo{author}{von Hippel, M.}, \bibinfo{year}{2023}.
\newblock \bibinfo{title}{Landau singularities and higher-order roots}.
\newblock \bibinfo{journal}{Physical Review D} \bibinfo{volume}{108},
  \bibinfo{pages}{085021}.
\newblock \DOIprefix\doi{10.1103/PhysRevD.108.085021},
  \href{http://arxiv.org/abs/2208.12765}{\tt arXiv:2208.12765}.
\bibitem[{Brakensiek et~al.(2024)Brakensiek, Dhar, Gao, Gopi and
  Larson}]{brakensiek2024rigidity}
\bibinfo{author}{Brakensiek, J.}, \bibinfo{author}{Dhar, M.},
  \bibinfo{author}{Gao, J.}, \bibinfo{author}{Gopi, S.},
  \bibinfo{author}{Larson, M.}, \bibinfo{year}{2024}.
\newblock \bibinfo{title}{Rigidity matroids and linear algebraic matroids with
  applications to matrix completion and tensor codes}.
\newblock \bibinfo{journal}{{\tt arXiv}:2405.00778} .
\bibitem[{Brandhuber et~al.(2008)Brandhuber, Heslop and
  Travaglini}]{Brandhuber:2007yx}
\bibinfo{author}{Brandhuber, A.}, \bibinfo{author}{Heslop, P.},
  \bibinfo{author}{Travaglini, G.}, \bibinfo{year}{2008}.
\newblock \bibinfo{title}{{MHV amplitudes in N=4 super Yang-Mills and Wilson
  loops}}.
\newblock \bibinfo{journal}{Nucl. Phys. B} \bibinfo{volume}{794},
  \bibinfo{pages}{231--243}.
\newblock \DOIprefix\doi{10.1016/j.nuclphysb.2007.11.002},
  \href{http://arxiv.org/abs/0707.1153}{\tt arXiv:0707.1153}.
\bibitem[{Brandhuber et~al.(2009)Brandhuber, Heslop and
  Travaglini}]{BrandhuberHeslopTravaglini2009}
\bibinfo{author}{Brandhuber, A.}, \bibinfo{author}{Heslop, P.},
  \bibinfo{author}{Travaglini, G.}, \bibinfo{year}{2009}.
\newblock \bibinfo{title}{One-loop amplitudes in \(\mathcal{N}=4\) super
  yang--mills and anomalous dual conformal symmetry}.
\newblock \bibinfo{journal}{Journal of High Energy Physics}
  \bibinfo{volume}{2009}, \bibinfo{pages}{095}.
\newblock \DOIprefix\doi{10.1088/1126-6708/2009/08/095},
  \href{http://arxiv.org/abs/0905.4377}{\tt arXiv:0905.4377}.
\bibitem[{Breiding and Timme(2018)}]{BT}
\bibinfo{author}{Breiding, P.}, \bibinfo{author}{Timme, S.},
  \bibinfo{year}{2018}.
\newblock \bibinfo{title}{{HomotopyContinuation.jl}: A package for homotopy
  continuation in julia}, in: \bibinfo{booktitle}{Mathematical Software -- ICMS
  2018}. \bibinfo{publisher}{Springer}, \bibinfo{address}{Cham}. volume
  \bibinfo{volume}{10931} of \textit{\bibinfo{series}{Lecture Notes in Computer
  Science}}, pp. \bibinfo{pages}{458--465}.
\newblock \DOIprefix\doi{10.1007/978-3-319-96418-8_54},
  \href{http://arxiv.org/abs/1711.10911}{\tt arXiv:1711.10911}.
\bibitem[{Brink et~al.(1977)Brink, Schwarz and Scherk}]{Brink:1976bc}
\bibinfo{author}{Brink, L.}, \bibinfo{author}{Schwarz, J.H.},
  \bibinfo{author}{Scherk, J.}, \bibinfo{year}{1977}.
\newblock \bibinfo{title}{Supersymmetric yang-mills theories}.
\newblock \bibinfo{journal}{Nucl. Phys. B} \bibinfo{volume}{121},
  \bibinfo{pages}{77--92}.
\bibitem[{Britto et~al.(2005a)Britto, Cachazo and Feng}]{Britto:2004nc}
\bibinfo{author}{Britto, R.}, \bibinfo{author}{Cachazo, F.},
  \bibinfo{author}{Feng, B.}, \bibinfo{year}{2005}a.
\newblock \bibinfo{title}{Generalized unitarity and one-loop amplitudes in
  $\mathcal{N}=4$ super-yang-mills}.
\newblock \bibinfo{journal}{Nucl. Phys. B} \bibinfo{volume}{725},
  \bibinfo{pages}{275--305}.
\newblock \DOIprefix\doi{10.1016/j.nuclphysb.2005.07.014},
  \href{http://arxiv.org/abs/hep-th/0412103}{\tt arXiv:hep-th/0412103}.
\bibitem[{Britto et~al.(2005b)Britto, Cachazo and Feng}]{Britto:2004ap}
\bibinfo{author}{Britto, R.}, \bibinfo{author}{Cachazo, F.},
  \bibinfo{author}{Feng, B.}, \bibinfo{year}{2005}b.
\newblock \bibinfo{title}{New recursion relations for tree amplitudes of
  gluons}.
\newblock \bibinfo{journal}{Nuclear Physics B} \bibinfo{volume}{715},
  \bibinfo{pages}{499--522}.
\newblock \DOIprefix\doi{10.1016/j.nuclphysb.2005.02.030},
  \href{http://arxiv.org/abs/hep-th/0412308}{\tt arXiv:hep-th/0412308}.
\bibitem[{Britto et~al.(2005c)Britto, Cachazo, Feng and Witten}]{Britto:2005fq}
\bibinfo{author}{Britto, R.}, \bibinfo{author}{Cachazo, F.},
  \bibinfo{author}{Feng, B.}, \bibinfo{author}{Witten, E.},
  \bibinfo{year}{2005}c.
\newblock \bibinfo{title}{Direct proof of tree-level recursion relation in
  yang--mills theory}.
\newblock \bibinfo{journal}{Physical Review Letters} \bibinfo{volume}{94},
  \bibinfo{pages}{181602}.
\newblock \DOIprefix\doi{10.1103/PhysRevLett.94.181602},
  \href{http://arxiv.org/abs/hep-th/0501052}{\tt arXiv:hep-th/0501052}.
\bibitem[{Broedel et~al.(2018a)Broedel, Duhr, Dulat, Penante and
  Tancredi}]{Broedel:2018iwv}
\bibinfo{author}{Broedel, J.}, \bibinfo{author}{Duhr, C.},
  \bibinfo{author}{Dulat, F.}, \bibinfo{author}{Penante, B.},
  \bibinfo{author}{Tancredi, L.}, \bibinfo{year}{2018}a.
\newblock \bibinfo{title}{Elliptic symbol calculus: from elliptic
  polylogarithms to iterated integrals of {Eisenstein} series}.
\newblock \bibinfo{journal}{Journal of High Energy Physics}
  \bibinfo{volume}{2018}, \bibinfo{pages}{014}.
\newblock \DOIprefix\doi{10.1007/JHEP08(2018)014},
  \href{http://arxiv.org/abs/1803.10256}{\tt arXiv:1803.10256}.
\bibitem[{Broedel et~al.(2018b)Broedel, Duhr, Dulat and
  Tancredi}]{BroedelDuhrDulatTancredi2018}
\bibinfo{author}{Broedel, J.}, \bibinfo{author}{Duhr, C.},
  \bibinfo{author}{Dulat, F.}, \bibinfo{author}{Tancredi, L.},
  \bibinfo{year}{2018}b.
\newblock \bibinfo{title}{Elliptic polylogarithms and iterated integrals on
  elliptic curves. part i: General formalism}.
\newblock \bibinfo{journal}{Journal of High Energy Physics}
  \bibinfo{volume}{2018}, \bibinfo{pages}{093}.
\newblock \DOIprefix\doi{10.1007/JHEP05(2018)093},
  \href{http://arxiv.org/abs/1712.07089}{\tt arXiv:1712.07089}.
\bibitem[{Brown(2009a)}]{Brown:2009qja}
\bibinfo{author}{Brown, F.}, \bibinfo{year}{2009}a.
\newblock \bibinfo{title}{The massless higher-loop two-point function}.
\newblock \bibinfo{journal}{Commun. Math. Phys.} \bibinfo{volume}{287},
  \bibinfo{pages}{925--958}.
\newblock \DOIprefix\doi{10.1007/s00220-009-0740-5},
  \href{http://arxiv.org/abs/0804.1660}{\tt arXiv:0804.1660}.
\bibitem[{Brown(2012)}]{Brown2012}
\bibinfo{author}{Brown, F.}, \bibinfo{year}{2012}.
\newblock \bibinfo{title}{Mixed {Tate} motives over \(\mathbb z\)}.
\newblock \bibinfo{journal}{Annals of Mathematics} \bibinfo{volume}{175},
  \bibinfo{pages}{949--976}.
\newblock \DOIprefix\doi{10.4007/annals.2012.175.2.10},
  \href{http://arxiv.org/abs/1102.1312}{\tt arXiv:1102.1312}.
\bibitem[{Brown(2017)}]{Brown2015}
\bibinfo{author}{Brown, F.}, \bibinfo{year}{2017}.
\newblock \bibinfo{title}{Feynman amplitudes, coaction principle, and cosmic
  galois group}.
\newblock \bibinfo{journal}{Communications in Number Theory and Physics}
  \bibinfo{volume}{11}, \bibinfo{pages}{453--556}.
\newblock \DOIprefix\doi{10.4310/CNTP.2017.v11.n3.a1},
  \href{http://arxiv.org/abs/1512.06409}{\tt arXiv:1512.06409}.
\bibitem[{Brown and Duhr(2022)}]{BrownDuhr2020DlogNotPolylog}
\bibinfo{author}{Brown, F.}, \bibinfo{author}{Duhr, C.}, \bibinfo{year}{2022}.
\newblock \bibinfo{title}{A double integral of dlog forms which is not
  polylogarithmic}.
\newblock \bibinfo{journal}{Proceedings of Science} \bibinfo{volume}{MA2019},
  \bibinfo{pages}{005}.
\newblock \DOIprefix\doi{10.22323/1.383.0005},
  \href{http://arxiv.org/abs/2006.09413}{\tt arXiv:2006.09413}.
\bibitem[{Brown and Dupont(2025)}]{Brown:PG_Hodge}
\bibinfo{author}{Brown, F.}, \bibinfo{author}{Dupont, C.},
  \bibinfo{year}{2025}.
\newblock \bibinfo{title}{Positive geometries and canonical forms via mixed
  hodge theory}.
\newblock \bibinfo{journal}{Commun. Math. Phys.} \bibinfo{volume}{406},
  \bibinfo{pages}{267}.
\newblock \DOIprefix\doi{10.1007/s00220-025-05399-y},
  \href{http://arxiv.org/abs/2501.03202}{\tt arXiv:2501.03202}.
\bibitem[{Brown(2009b)}]{Brown2009}
\bibinfo{author}{Brown, F.C.S.}, \bibinfo{year}{2009}b.
\newblock \bibinfo{title}{Multiple zeta values and periods of moduli spaces
  \(\mathfrak{M}_{0,n}\)}.
\newblock \bibinfo{journal}{Annales scientifiques de l'{\'E}cole Normale
  Sup{\'e}rieure} \bibinfo{volume}{42}, \bibinfo{pages}{371--489}.
\newblock \DOIprefix\doi{10.24033/asens.2099},
  \href{http://arxiv.org/abs/math/0606419}{\tt arXiv:math/0606419}.
\bibitem[{Brown(2010)}]{Brown2009_alt1}
\bibinfo{author}{Brown, F.C.S.}, \bibinfo{year}{2010}.
\newblock \bibinfo{title}{On the periods of some feynman integrals}.
\newblock \bibinfo{journal}{Communications in Number Theory and Physics}
  \bibinfo{volume}{4}, \bibinfo{pages}{441--495}.
\newblock \DOIprefix\doi{10.4310/CNTP.2010.v4.n3.a2},
  \href{http://arxiv.org/abs/0910.0114}{\tt arXiv:0910.0114}.
\bibitem[{Brown et~al.(2025)Brown, Henn, Mazzucchelli and
  Trnka}]{BrownHennMazzucchelliTrnka2025}
\bibinfo{author}{Brown, T.V.}, \bibinfo{author}{Henn, J.M.},
  \bibinfo{author}{Mazzucchelli, E.}, \bibinfo{author}{Trnka, J.},
  \bibinfo{year}{2025}.
\newblock \bibinfo{title}{{All-loop Leading Singularities of Wilson Loops}}
  \href{http://arxiv.org/abs/2503.17185}{\tt arXiv:2503.17185}.
\bibitem[{Brown et~al.(2024)Brown, Oktem, Paranjape and
  Trnka}]{BrownOktemParanjapeTrnka2024}
\bibinfo{author}{Brown, T.V.}, \bibinfo{author}{Oktem, U.},
  \bibinfo{author}{Paranjape, S.}, \bibinfo{author}{Trnka, J.},
  \bibinfo{year}{2024}.
\newblock \bibinfo{title}{{Loops of loops expansion in the Amplituhedron}}.
\newblock \bibinfo{journal}{JHEP} \bibinfo{volume}{07}, \bibinfo{pages}{025}.
\newblock \DOIprefix\doi{10.1007/JHEP07(2024)025},
  \href{http://arxiv.org/abs/2312.17736}{\tt arXiv:2312.17736}.
\bibitem[{Br{\"u}ser and Weigert(2025)}]{bruser2025geometry}
\bibinfo{author}{Br{\"u}ser, C.}, \bibinfo{author}{Weigert, J.},
  \bibinfo{year}{2025}.
\newblock \bibinfo{title}{Geometry of adjoint hypersurfaces for polytopes}
  \href{http://arxiv.org/abs/2511.13537}{\tt arXiv:2511.13537}.
\bibitem[{Brysiewicz et~al.(2021)Brysiewicz, Fevola and Sturmfels}]{BFS}
\bibinfo{author}{Brysiewicz, T.}, \bibinfo{author}{Fevola, C.},
  \bibinfo{author}{Sturmfels, B.}, \bibinfo{year}{2021}.
\newblock \bibinfo{title}{Tangent quadrics in real 3-space}.
\newblock \bibinfo{journal}{Le Matematiche} \bibinfo{volume}{76},
  \bibinfo{pages}{355--367}.
\newblock \DOIprefix\doi{10.4418/2021.76.2.6}.
\bibitem[{Brändén(2014)}]{branden2014hyperbolicity}
\bibinfo{author}{Brändén, P.}, \bibinfo{year}{2014}.
\newblock \bibinfo{title}{Hyperbolicity cones of elementary symmetric
  polynomials are spectrahedral}.
\newblock \bibinfo{journal}{Optimization Letters} \bibinfo{volume}{8},
  \bibinfo{pages}{1773--1782}.
\bibitem[{Busemann(1961)}]{busemann1961convexity}
\bibinfo{author}{Busemann, H.}, \bibinfo{year}{1961}.
\newblock \bibinfo{title}{Convexity on grassmann manifolds}.
\newblock \bibinfo{journal}{Enseign. Math.(2)} \bibinfo{volume}{7}.
\bibitem[{Cachazo and Geyer(2013)}]{Cachazo:2012kg}
\bibinfo{author}{Cachazo, F.}, \bibinfo{author}{Geyer, Y.},
  \bibinfo{year}{2013}.
\newblock \bibinfo{title}{{A 'Twistor String' Inspired Formula For Tree-Level
  Scattering Amplitudes in N=8 SUGRA}}.
\newblock \bibinfo{journal}{JHEP} \bibinfo{volume}{02}, \bibinfo{pages}{052}.
\newblock \href{http://arxiv.org/abs/1206.6511}{\tt arXiv:1206.6511}.
\bibitem[{Cachazo et~al.(2014)Cachazo, He and Yuan}]{CachazoHeYuan2014}
\bibinfo{author}{Cachazo, F.}, \bibinfo{author}{He, S.}, \bibinfo{author}{Yuan,
  E.Y.}, \bibinfo{year}{2014}.
\newblock \bibinfo{title}{Scattering of massless particles in arbitrary
  dimensions}.
\newblock \bibinfo{journal}{Phys. Rev. Lett.} \bibinfo{volume}{113},
  \bibinfo{pages}{171601}.
\newblock \href{http://arxiv.org/abs/1307.2199}{\tt arXiv:1307.2199}.
\bibitem[{Cachazo and Skinner(2008)}]{Cachazo:2008vp}
\bibinfo{author}{Cachazo, F.}, \bibinfo{author}{Skinner, D.},
  \bibinfo{year}{2008}.
\newblock \bibinfo{title}{On the structure of scattering amplitudes in gauge
  theory and gravity} \href{http://arxiv.org/abs/0801.4574}{\tt
  arXiv:0801.4574}.
\bibitem[{Cao and Zhu(2025)}]{Cao:2025ymrecursion}
\bibinfo{author}{Cao, Q.}, \bibinfo{author}{Zhu, F.}, \bibinfo{year}{2025}.
\newblock \bibinfo{title}{All-loop planar integrands in yang-mills theory from
  recursion}.
\newblock \bibinfo{journal}{PRX Life}
  \href{http://arxiv.org/abs/2503.15860}{\tt arXiv:2503.15860}.
\bibitem[{Capuano et~al.(2025)Capuano, Ferro, {\L}ukowski and
  Palazio}]{CapuanoFerroLukowskiPalazio2025}
\bibinfo{author}{Capuano, M.}, \bibinfo{author}{Ferro, L.},
  \bibinfo{author}{{\L}ukowski, T.}, \bibinfo{author}{Palazio, A.},
  \bibinfo{year}{2025}.
\newblock \bibinfo{title}{Canonical differential equations for cosmology from
  positive geometries} \href{http://arxiv.org/abs/2505.14609}{\tt
  arXiv:2505.14609}.
\bibitem[{Capuano et~al.(2026)Capuano, Ferro, {\L}ukowski, Palazio and
  Zhang}]{CapuanoFerroLukowskiPalazioZhang2026GeneralisedClusterAdjacency}
\bibinfo{author}{Capuano, M.}, \bibinfo{author}{Ferro, L.},
  \bibinfo{author}{{\L}ukowski, T.}, \bibinfo{author}{Palazio, A.},
  \bibinfo{author}{Zhang, Y.Q.}, \bibinfo{year}{2026}.
\newblock \bibinfo{title}{Generalised cluster adjacency for cosmology}.
\newblock \href{http://arxiv.org/abs/2603.09965}{\tt arXiv:2603.09965}.
\bibitem[{Caron-Huot(2011)}]{CaronHuot:2011kk}
\bibinfo{author}{Caron-Huot, S.}, \bibinfo{year}{2011}.
\newblock \bibinfo{title}{Superconformal symmetry and two-loop amplitudes in
  planar n=4 super yang-mills}.
\newblock \bibinfo{journal}{JHEP} \bibinfo{volume}{12}, \bibinfo{pages}{066}.
\newblock \href{http://arxiv.org/abs/1105.5606}{\tt arXiv:1105.5606}.
\bibitem[{Caron-Huot et~al.(2025)Caron-Huot, Correia and
  Giroux}]{CaronHuotCorreiaGiroux2025}
\bibinfo{author}{Caron-Huot, S.}, \bibinfo{author}{Correia, M.},
  \bibinfo{author}{Giroux, M.}, \bibinfo{year}{2025}.
\newblock \bibinfo{title}{Recursive landau analysis}.
\newblock \bibinfo{journal}{Physical Review Letters} \bibinfo{volume}{135},
  \bibinfo{pages}{131603}.
\newblock \DOIprefix\doi{10.1103/8rwk-bnph},
  \href{http://arxiv.org/abs/2406.05241}{\tt arXiv:2406.05241}.
\bibitem[{Caron-Huot et~al.(2020)Caron-Huot, Dixon, Drummond, Dulat, Foster,
  G{\"u}rdo{\u g}an, von Hippel, McLeod and
  Papathanasiou}]{CaronHuotDixonDulatEtAl2020}
\bibinfo{author}{Caron-Huot, S.}, \bibinfo{author}{Dixon, L.J.},
  \bibinfo{author}{Drummond, J.M.}, \bibinfo{author}{Dulat, F.},
  \bibinfo{author}{Foster, J.}, \bibinfo{author}{G{\"u}rdo{\u g}an, {\"O}.},
  \bibinfo{author}{von Hippel, M.}, \bibinfo{author}{McLeod, A.J.},
  \bibinfo{author}{Papathanasiou, G.}, \bibinfo{year}{2020}.
\newblock \bibinfo{title}{The steinmann cluster bootstrap for \( \mathcal{N}=4
  \) super yang--mills amplitudes}.
\newblock \bibinfo{journal}{Proceedings of Science}
  \bibinfo{volume}{CORFU2019}, \bibinfo{pages}{003}.
\newblock \DOIprefix\doi{10.22323/1.376.0003},
  \href{http://arxiv.org/abs/2005.06735}{\tt arXiv:2005.06735}.
\bibitem[{Caron-Huot et~al.(2016)Caron-Huot, Dixon, McLeod and von
  Hippel}]{CaronHuotDixonMcLeodVonHippel2016}
\bibinfo{author}{Caron-Huot, S.}, \bibinfo{author}{Dixon, L.J.},
  \bibinfo{author}{McLeod, A.}, \bibinfo{author}{von Hippel, M.},
  \bibinfo{year}{2016}.
\newblock \bibinfo{title}{Bootstrapping a five-loop amplitude using steinmann
  relations}.
\newblock \bibinfo{journal}{Physical Review Letters} \bibinfo{volume}{117},
  \bibinfo{pages}{241601}.
\newblock \DOIprefix\doi{10.1103/PhysRevLett.117.241601},
  \href{http://arxiv.org/abs/1609.00669}{\tt arXiv:1609.00669}.
\bibitem[{Carr{\^o}lo et~al.(2025)Carr{\^o}lo, Chicherin, Henn, Yang and
  Zhang}]{Carrolo:2025pue}
\bibinfo{author}{Carr{\^o}lo, S.}, \bibinfo{author}{Chicherin, D.},
  \bibinfo{author}{Henn, J.}, \bibinfo{author}{Yang, Q.},
  \bibinfo{author}{Zhang, Y.}, \bibinfo{year}{2025}.
\newblock \bibinfo{title}{Hexagonal wilson loop with lagrangian insertion at
  two loops in \(\mathcal{N}=4\) super yang-mills theory}.
\newblock \bibinfo{journal}{Journal of High Energy Physics}
  \bibinfo{volume}{2025}, \bibinfo{pages}{214}.
\newblock \DOIprefix\doi{10.1007/JHEP07(2025)214},
  \href{http://arxiv.org/abs/2505.01245}{\tt arXiv:2505.01245}.
\bibitem[{Carr{\^o}lo et~al.(2026)Carr{\^o}lo, Chicherin, Henn, Yang and
  Zhang}]{Carrolo:2026qpu}
\bibinfo{author}{Carr{\^o}lo, S.}, \bibinfo{author}{Chicherin, D.},
  \bibinfo{author}{Henn, J.}, \bibinfo{author}{Yang, Q.},
  \bibinfo{author}{Zhang, Y.}, \bibinfo{year}{2026}.
\newblock \bibinfo{title}{{QCD Scattering Amplitudes and Prescriptive
  Unitarity}} \href{http://arxiv.org/abs/2602.02783}{\tt arXiv:2602.02783}.
\bibitem[{Catanese et~al.(2006)Catanese, Ho\c{s}ten, Khetan and
  Sturmfels}]{CataneseHostenKhetanSturmfels2006}
\bibinfo{author}{Catanese, F.}, \bibinfo{author}{Ho\c{s}ten, S.},
  \bibinfo{author}{Khetan, A.}, \bibinfo{author}{Sturmfels, B.},
  \bibinfo{year}{2006}.
\newblock \bibinfo{title}{The maximum likelihood degree}.
\newblock \bibinfo{journal}{American Journal of Mathematics}
  \bibinfo{volume}{128}, \bibinfo{pages}{671--697}.
\newblock \DOIprefix\doi{10.1353/ajm.2006.0019}.
\bibitem[{Catani et~al.(2008)Catani, Gleisberg, Krauss, Rodrigo and
  Winter}]{Catani:2008xa}
\bibinfo{author}{Catani, S.}, \bibinfo{author}{Gleisberg, T.},
  \bibinfo{author}{Krauss, F.}, \bibinfo{author}{Rodrigo, G.},
  \bibinfo{author}{Winter, J.C.}, \bibinfo{year}{2008}.
\newblock \bibinfo{title}{From loops to trees by-passing feynman's theorem}.
\newblock \bibinfo{journal}{JHEP} \bibinfo{volume}{09}, \bibinfo{pages}{065}.
\newblock \DOIprefix\doi{10.1088/1126-6708/2008/09/065},
  \href{http://arxiv.org/abs/0804.3170}{\tt arXiv:0804.3170}.
\bibitem[{Charlton et~al.(2019)Charlton, Gangl and Radchenko}]{CGR}
\bibinfo{author}{Charlton, S.}, \bibinfo{author}{Gangl, H.},
  \bibinfo{author}{Radchenko, D.}, \bibinfo{year}{2019}.
\newblock \bibinfo{title}{Explicit formulas for grassmannian polylogarithms}
  \href{http://arxiv.org/abs/1909.13869}{\tt arXiv:1909.13869}.
  \bibinfo{note}{revised 2022}.
\bibitem[{Chen et~al.(2013)Chen, Davenport, May, Moreno~Maza, Xia and
  Xiao}]{Chen2013}
\bibinfo{author}{Chen, C.}, \bibinfo{author}{Davenport, J.H.},
  \bibinfo{author}{May, J.P.}, \bibinfo{author}{Moreno~Maza, M.},
  \bibinfo{author}{Xia, B.}, \bibinfo{author}{Xiao, R.}, \bibinfo{year}{2013}.
\newblock \bibinfo{title}{Triangular decomposition of semi-algebraic systems}.
\newblock \bibinfo{journal}{J. Symbolic Comput.} \bibinfo{volume}{49},
  \bibinfo{pages}{3--26}.
\newblock \URLprefix \url{https://doi.org/10.1016/j.jsc.2011.12.014},
  \DOIprefix\doi{10.1016/j.jsc.2011.12.014}.
\bibitem[{Chen(1977)}]{Chen1977}
\bibinfo{author}{Chen, K.T.}, \bibinfo{year}{1977}.
\newblock \bibinfo{title}{Iterated path integrals}.
\newblock \bibinfo{journal}{Bulletin of the American Mathematical Society}
  \bibinfo{volume}{83}, \bibinfo{pages}{831--879}.
\newblock \DOIprefix\doi{10.1090/S0002-9904-1977-14320-6}.
\bibitem[{Cheung and O'Connell(2009)}]{Cheung:2009dc}
\bibinfo{author}{Cheung, C.}, \bibinfo{author}{O'Connell, D.},
  \bibinfo{year}{2009}.
\newblock \bibinfo{title}{Amplitudes and spinor-helicity in six dimensions}.
\newblock \bibinfo{journal}{Journal of High Energy Physics}
  \bibinfo{volume}{07}, \bibinfo{pages}{075}.
\newblock \href{http://arxiv.org/abs/0902.0981}{\tt arXiv:0902.0981}.
\bibitem[{Chew(1962)}]{Chew:1961ev}
\bibinfo{author}{Chew, G.F.}, \bibinfo{year}{1962}.
\newblock \bibinfo{title}{S-matrix theory of strong interactions}.
\newblock \bibinfo{journal}{Reviews of Modern Physics} \bibinfo{volume}{34},
  \bibinfo{pages}{394--397}.
\bibitem[{Chicherin et~al.(2025)Chicherin, Henn, Trnka and
  Zhang}]{neg_geom_pos}
\bibinfo{author}{Chicherin, D.}, \bibinfo{author}{Henn, J.},
  \bibinfo{author}{Trnka, J.}, \bibinfo{author}{Zhang, S.Q.},
  \bibinfo{year}{2025}.
\newblock \bibinfo{title}{{Positivity properties of five-point two-loop Wilson
  loops with Lagrangian insertion}}.
\newblock \bibinfo{journal}{JHEP} \bibinfo{volume}{04}, \bibinfo{pages}{022}.
\newblock \DOIprefix\doi{10.1007/JHEP04(2025)022},
  \href{http://arxiv.org/abs/2410.11456}{\tt arXiv:2410.11456}.
\bibitem[{Chicherin and Henn(2022a)}]{Chicherin:2022zxo}
\bibinfo{author}{Chicherin, D.}, \bibinfo{author}{Henn, J.M.},
  \bibinfo{year}{2022}a.
\newblock \bibinfo{title}{{Pentagon Wilson loop with Lagrangian insertion at
  two loops in N=4 super Yang-Mills theory}}.
\newblock \bibinfo{journal}{JHEP} \bibinfo{volume}{07}, \bibinfo{pages}{038}.
\newblock \DOIprefix\doi{10.1007/JHEP07(2022)038},
  \href{http://arxiv.org/abs/2204.00329}{\tt arXiv:2204.00329}.
\bibitem[{Chicherin and Henn(2022b)}]{ChicherinHenn2022}
\bibinfo{author}{Chicherin, D.}, \bibinfo{author}{Henn, J.M.},
  \bibinfo{year}{2022}b.
\newblock \bibinfo{title}{{Symmetry properties of Wilson loops with a
  Lagrangian insertion}}.
\newblock \bibinfo{journal}{JHEP} \bibinfo{volume}{07}, \bibinfo{pages}{057}.
\newblock \DOIprefix\doi{10.1007/JHEP07(2022)057},
  \href{http://arxiv.org/abs/2202.05596}{\tt arXiv:2202.05596}.
\bibitem[{Chicherin et~al.(2026)Chicherin, Henn, Mazzucchelli, Trnka, Yang and
  Zhang}]{ChicherinHennMazzucchelliTrnkaYangZhang2026}
\bibinfo{author}{Chicherin, D.}, \bibinfo{author}{Henn, J.M.},
  \bibinfo{author}{Mazzucchelli, E.}, \bibinfo{author}{Trnka, J.},
  \bibinfo{author}{Yang, Q.}, \bibinfo{author}{Zhang, S.Q.},
  \bibinfo{year}{2026}.
\newblock \bibinfo{title}{Geometric landau analysis and symbol bootstrap}.
\newblock \bibinfo{journal}{Journal of High Energy Physics}
  \bibinfo{volume}{2026}, \bibinfo{pages}{083}.
\newblock \DOIprefix\doi{10.1007/JHEP02(2026)083},
  \href{http://arxiv.org/abs/2508.05443}{\tt arXiv:2508.05443}.
\bibitem[{Chicherin et~al.(2021)Chicherin, Henn and
  Papathanasiou}]{Chicherin:2020umh}
\bibinfo{author}{Chicherin, D.}, \bibinfo{author}{Henn, J.M.},
  \bibinfo{author}{Papathanasiou, G.}, \bibinfo{year}{2021}.
\newblock \bibinfo{title}{{Cluster algebras for Feynman integrals}}.
\newblock \bibinfo{journal}{Phys. Rev. Lett.} \bibinfo{volume}{126},
  \bibinfo{pages}{091603}.
\newblock \DOIprefix\doi{10.1103/PhysRevLett.126.091603},
  \href{http://arxiv.org/abs/2012.12285}{\tt arXiv:2012.12285}.
\bibitem[{Choquet(1969)}]{Choquet}
\bibinfo{author}{Choquet, G.}, \bibinfo{year}{1969}.
\newblock \bibinfo{title}{Deux exemples classiques de représentation
  intégrale}.
\newblock \bibinfo{journal}{Enseignement Mathématique} \bibinfo{volume}{15},
  \bibinfo{pages}{63--75}.
\bibitem[{Coleman and Norton(1965)}]{ColemanNorton1965}
\bibinfo{author}{Coleman, S.}, \bibinfo{author}{Norton, R.E.},
  \bibinfo{year}{1965}.
\newblock \bibinfo{title}{Singularities in the physical region}.
\newblock \bibinfo{journal}{Il Nuovo Cimento} \bibinfo{volume}{38},
  \bibinfo{pages}{438--442}.
\newblock \DOIprefix\doi{10.1007/BF02750472}.
\bibitem[{Collins(1984)}]{Collins1984}
\bibinfo{author}{Collins, J.C.}, \bibinfo{year}{1984}.
\newblock \bibinfo{title}{Renormalization}.
\newblock \bibinfo{publisher}{Cambridge University Press}.
\bibitem[{Conde and Rajabi(2012)}]{Conde:2012wb}
\bibinfo{author}{Conde, E.}, \bibinfo{author}{Rajabi, S.},
  \bibinfo{year}{2012}.
\newblock \bibinfo{title}{The twelve-graviton next-to-mhv amplitude from
  risager’s construction}.
\newblock \bibinfo{journal}{Journal of High Energy Physics}
  \bibinfo{volume}{09}, \bibinfo{pages}{120}.
\newblock \href{http://arxiv.org/abs/1205.3500}{\tt arXiv:1205.3500}.
\bibitem[{Correia et~al.(2026)Correia, Giroux and
  Mizera}]{CorreiaGirouxMizera2026}
\bibinfo{author}{Correia, M.}, \bibinfo{author}{Giroux, M.},
  \bibinfo{author}{Mizera, S.}, \bibinfo{year}{2026}.
\newblock \bibinfo{title}{{SOFIA}: Singularities of feynman integrals
  automatized}.
\newblock \bibinfo{journal}{Computer Physics Communications}
  \bibinfo{volume}{320}, \bibinfo{pages}{109970}.
\newblock \DOIprefix\doi{10.1016/j.cpc.2025.109970},
  \href{http://arxiv.org/abs/2503.16601}{\tt arXiv:2503.16601}.
\bibitem[{Costa et~al.(2011)Costa, Penedones, Poland and
  Rychkov}]{Costa:2011mg}
\bibinfo{author}{Costa, M.S.}, \bibinfo{author}{Penedones, J.},
  \bibinfo{author}{Poland, D.}, \bibinfo{author}{Rychkov, S.},
  \bibinfo{year}{2011}.
\newblock \bibinfo{title}{Spinning conformal correlators}.
\newblock \bibinfo{journal}{Journal of High Energy Physics}
  \bibinfo{volume}{2011}, \bibinfo{pages}{071}.
\newblock \DOIprefix\doi{10.1007/JHEP11(2011)071},
  \href{http://arxiv.org/abs/1107.3554}{\tt arXiv:1107.3554}.
\bibitem[{Cox et~al.(2005)Cox, Little and O'Shea}]{CoxLittleOShea2005}
\bibinfo{author}{Cox, D.A.}, \bibinfo{author}{Little, J.},
  \bibinfo{author}{O'Shea, D.}, \bibinfo{year}{2005}.
\newblock \bibinfo{title}{Using Algebraic Geometry}. volume
  \bibinfo{volume}{185} of \textit{\bibinfo{series}{Graduate Texts in
  Mathematics}}.
\newblock \bibinfo{edition}{2} ed., \bibinfo{publisher}{Springer},
  \bibinfo{address}{New York}.
\newblock \DOIprefix\doi{10.1007/b138611}.
\bibitem[{Crespo~Ruiz and Santos(2023)}]{Crespo_Ruiz2023}
\bibinfo{author}{Crespo~Ruiz, L.}, \bibinfo{author}{Santos, F.},
  \bibinfo{year}{2023}.
\newblock \bibinfo{title}{Bar-and-joint rigidity on the moment curve coincides
  with cofactor rigidity on a conic}.
\newblock \bibinfo{journal}{Combinatorial Theory} \bibinfo{volume}{3}.
\bibitem[{Cutkosky(1960)}]{Cutkosky1960}
\bibinfo{author}{Cutkosky, R.E.}, \bibinfo{year}{1960}.
\newblock \bibinfo{title}{Singularities and discontinuities of feynman
  amplitudes}.
\newblock \bibinfo{journal}{Journal of Mathematical Physics}
  \bibinfo{volume}{1}, \bibinfo{pages}{429--433}.
\newblock \DOIprefix\doi{10.1063/1.1703676}.
\bibitem[{Damgaard et~al.(2021)Damgaard, Ferro, {\L}ukowski and
  Parisi}]{damgaardFerroLukowskiParisi2021associahedron}
\bibinfo{author}{Damgaard, D.}, \bibinfo{author}{Ferro, L.},
  \bibinfo{author}{{\L}ukowski, T.}, \bibinfo{author}{Parisi, M.},
  \bibinfo{year}{2021}.
\newblock \bibinfo{title}{Momentum amplituhedron meets kinematic
  associahedron}.
\newblock \bibinfo{journal}{Journal of High Energy Physics}
  \bibinfo{volume}{2021}, \bibinfo{pages}{041}.
\newblock \href{http://arxiv.org/abs/2010.15858}{\tt arXiv:2010.15858}.
\bibitem[{Damgaard et~al.(2019)Damgaard, Ferro, Lukowski and
  Parisi}]{Damgaard:2019ztj}
\bibinfo{author}{Damgaard, P.H.}, \bibinfo{author}{Ferro, L.},
  \bibinfo{author}{Lukowski, T.}, \bibinfo{author}{Parisi, M.},
  \bibinfo{year}{2019}.
\newblock \bibinfo{title}{The momentum amplituhedron}.
\newblock \bibinfo{journal}{Journal of High Energy Physics}
  \bibinfo{volume}{2019}, \bibinfo{pages}{042}.
\newblock \DOIprefix\doi{10.1007/JHEP08(2019)042},
  \href{http://arxiv.org/abs/1905.04216}{\tt arXiv:1905.04216}.
\bibitem[{D'Andrea and Dickenstein(2001)}]{DAndreaDickenstein2001}
\bibinfo{author}{D'Andrea, C.}, \bibinfo{author}{Dickenstein, A.},
  \bibinfo{year}{2001}.
\newblock \bibinfo{title}{Explicit formulas for the multivariate resultant}.
\newblock \bibinfo{journal}{Journal of Pure and Applied Algebra}
  \bibinfo{volume}{164}, \bibinfo{pages}{59--86}.
\newblock \DOIprefix\doi{10.1016/S0022-4049(00)00169-3}.
\bibitem[{De et~al.(2025)De, Pavlov, Spradlin and Volovich}]{De:2024bpk}
\bibinfo{author}{De, S.}, \bibinfo{author}{Pavlov, D.},
  \bibinfo{author}{Spradlin, M.}, \bibinfo{author}{Volovich, A.},
  \bibinfo{year}{2025}.
\newblock \bibinfo{title}{{From Feynman diagrams to the amplituhedron: a gentle
  review}}.
\newblock \bibinfo{journal}{Matematiche} \bibinfo{volume}{80},
  \bibinfo{pages}{233--254}.
\newblock \DOIprefix\doi{10.4418/2025.80.1.9},
  \href{http://arxiv.org/abs/2410.11757}{\tt arXiv:2410.11757}.
\bibitem[{De and Pokraka(2024)}]{DePokraka2024}
\bibinfo{author}{De, S.}, \bibinfo{author}{Pokraka, A.}, \bibinfo{year}{2024}.
\newblock \bibinfo{title}{Cosmology meets cohomology}.
\newblock \bibinfo{journal}{Journal of High Energy Physics}
  \bibinfo{volume}{2024}, \bibinfo{pages}{156}.
\newblock \DOIprefix\doi{10.1007/JHEP03(2024)156},
  \href{http://arxiv.org/abs/2308.03753}{\tt arXiv:2308.03753}.
\bibitem[{De~Loera et~al.(2010)De~Loera, Rambau and
  Santos}]{deLoeraRambauSantos2010triangulations}
\bibinfo{author}{De~Loera, J.A.}, \bibinfo{author}{Rambau, J.},
  \bibinfo{author}{Santos, F.}, \bibinfo{year}{2010}.
\newblock \bibinfo{title}{Triangulations: Structures for Algorithms and
  Applications}. volume~\bibinfo{volume}{25} of
  \textit{\bibinfo{series}{Algorithms and Computation in Mathematics}}.
\newblock \bibinfo{publisher}{Springer}.
\bibitem[{Del~Duca et~al.(2010a)Del~Duca, Duhr and Smirnov}]{DelDuca:2009au}
\bibinfo{author}{Del~Duca, V.}, \bibinfo{author}{Duhr, C.},
  \bibinfo{author}{Smirnov, V.A.}, \bibinfo{year}{2010}a.
\newblock \bibinfo{title}{An analytic result for the two-loop hexagon wilson
  loop in \(\mathcal{N}=4\) {SYM}}.
\newblock \bibinfo{journal}{Journal of High Energy Physics}
  \bibinfo{volume}{2010}, \bibinfo{pages}{099}.
\newblock \DOIprefix\doi{10.1007/JHEP03(2010)099},
  \href{http://arxiv.org/abs/0911.5332}{\tt arXiv:0911.5332}.
\bibitem[{Del~Duca et~al.(2010b)Del~Duca, Duhr and Smirnov}]{DelDuca:2010zg}
\bibinfo{author}{Del~Duca, V.}, \bibinfo{author}{Duhr, C.},
  \bibinfo{author}{Smirnov, V.A.}, \bibinfo{year}{2010}b.
\newblock \bibinfo{title}{The two-loop hexagon wilson loop in \(\mathcal{N}=4\)
  {SYM}}.
\newblock \bibinfo{journal}{Journal of High Energy Physics}
  \bibinfo{volume}{2010}, \bibinfo{pages}{084}.
\newblock \DOIprefix\doi{10.1007/JHEP05(2010)084},
  \href{http://arxiv.org/abs/1003.1702}{\tt arXiv:1003.1702}.
\bibitem[{Dennen et~al.(2017)Dennen, Prlina, Spradlin, Stanojevic and
  Volovich}]{DennenPrlinaSpradlinStanojevicVolovich2017}
\bibinfo{author}{Dennen, T.}, \bibinfo{author}{Prlina, I.},
  \bibinfo{author}{Spradlin, M.}, \bibinfo{author}{Stanojevic, S.},
  \bibinfo{author}{Volovich, A.}, \bibinfo{year}{2017}.
\newblock \bibinfo{title}{Landau singularities from the amplituhedron}.
\newblock \bibinfo{journal}{Journal of High Energy Physics}
  \bibinfo{volume}{2017}, \bibinfo{pages}{152}.
\newblock \DOIprefix\doi{10.1007/JHEP06(2017)152},
  \href{http://arxiv.org/abs/1612.02708}{\tt arXiv:1612.02708}.
\bibitem[{Dennen et~al.(2016)Dennen, Spradlin and
  Volovich}]{DennenSpradlinVolovich2016}
\bibinfo{author}{Dennen, T.}, \bibinfo{author}{Spradlin, M.},
  \bibinfo{author}{Volovich, A.}, \bibinfo{year}{2016}.
\newblock \bibinfo{title}{Landau singularities and symbology: One- and two-loop
  {MHV} amplitudes in {SYM} theory}.
\newblock \bibinfo{journal}{Journal of High Energy Physics}
  \bibinfo{volume}{2016}, \bibinfo{pages}{069}.
\newblock \DOIprefix\doi{10.1007/JHEP03(2016)069},
  \href{http://arxiv.org/abs/1512.07909}{\tt arXiv:1512.07909}.
\bibitem[{Dian(2019)}]{GabriBlog}
\bibinfo{author}{Dian, G.}, \bibinfo{year}{2019}.
\newblock \bibinfo{title}{On the polygon front lines: Visualizing the
  amplituhedron with the wolfram language}.
\newblock
  \bibinfo{howpublished}{\url{https://blog.wolfram.com/2019/11/21/on-the-polygon-front-lines-visualizing-the-amplituhedron-with-the-wolfram-language/}}.
\bibitem[{Dian et~al.(2023)Dian, Heslop and Stewart}]{Dian_2023}
\bibinfo{author}{Dian, G.}, \bibinfo{author}{Heslop, P.},
  \bibinfo{author}{Stewart, A.}, \bibinfo{year}{2023}.
\newblock \bibinfo{title}{{Internal boundaries of the loop amplituhedron}}.
\newblock \bibinfo{journal}{SciPost Phys.} \bibinfo{volume}{15},
  \bibinfo{pages}{098}.
\newblock \DOIprefix\doi{10.21468/SciPostPhys.15.3.098},
  \href{http://arxiv.org/abs/2207.12464}{\tt arXiv:2207.12464}.
\bibitem[{Dian et~al.(2025)Dian, Mazzucchelli and Tellander}]{Dian:2024hil}
\bibinfo{author}{Dian, G.}, \bibinfo{author}{Mazzucchelli, E.},
  \bibinfo{author}{Tellander, F.}, \bibinfo{year}{2025}.
\newblock \bibinfo{title}{{The two-loop Amplituhedron}}.
\newblock \bibinfo{journal}{Le Mat.} \bibinfo{volume}{80},
  \bibinfo{pages}{255--277}.
\newblock \DOIprefix\doi{10.4418/2025.80.1.10},
  \href{http://arxiv.org/abs/2410.11501}{\tt arXiv:2410.11501}.
\bibitem[{Dimca(2004)}]{Dimca2004}
\bibinfo{author}{Dimca, A.}, \bibinfo{year}{2004}.
\newblock \bibinfo{title}{Sheaves in Topology}.
\newblock Universitext, \bibinfo{publisher}{Springer}.
\newblock \DOIprefix\doi{10.1007/978-3-642-18868-8}.
\bibitem[{Dirac(1936)}]{Dirac1936}
\bibinfo{author}{Dirac, P.A.M.}, \bibinfo{year}{1936}.
\newblock \bibinfo{title}{Wave equations in conformal space}.
\newblock \bibinfo{journal}{Annals of Mathematics} \bibinfo{volume}{37},
  \bibinfo{pages}{429--442}.
\newblock \DOIprefix\doi{10.2307/1968455}.
\bibitem[{Dixon(1996)}]{Dixon:1996wi}
\bibinfo{author}{Dixon, L.J.}, \bibinfo{year}{1996}.
\newblock \bibinfo{title}{Calculating scattering amplitudes efficiently}.
\newblock \bibinfo{journal}{Proceedings of TASI 1995, QCD and Beyond} ,
  \bibinfo{pages}{539--582}\href{http://arxiv.org/abs/hep-ph/9601359}{\tt
  arXiv:hep-ph/9601359}.
\bibitem[{Dixon(2011)}]{Dixon:2011xs}
\bibinfo{author}{Dixon, L.J.}, \bibinfo{year}{2011}.
\newblock \bibinfo{title}{Scattering amplitudes: the most perfect microscopic
  structures in the universe}.
\newblock \bibinfo{journal}{Journal of Physics A} \bibinfo{volume}{44},
  \bibinfo{pages}{454001}.
\newblock \href{http://arxiv.org/abs/1105.0771}{\tt arXiv:1105.0771}.
\bibitem[{Dixon et~al.(2017a)Dixon, Drummond, Harrington, McLeod, Papathanasiou
  and Spradlin}]{DixonEtAl2017}
\bibinfo{author}{Dixon, L.J.}, \bibinfo{author}{Drummond, J.M.},
  \bibinfo{author}{Harrington, T.}, \bibinfo{author}{McLeod, A.J.},
  \bibinfo{author}{Papathanasiou, G.}, \bibinfo{author}{Spradlin, M.},
  \bibinfo{year}{2017}a.
\newblock \bibinfo{title}{Heptagons from the steinmann cluster bootstrap}.
\newblock \bibinfo{journal}{Journal of High Energy Physics}
  \bibinfo{volume}{2017}, \bibinfo{pages}{137}.
\newblock \DOIprefix\doi{10.1007/JHEP02(2017)137},
  \href{http://arxiv.org/abs/1612.08976}{\tt arXiv:1612.08976}.
\bibitem[{Dixon et~al.(2011)Dixon, Drummond and Henn}]{hexagon}
\bibinfo{author}{Dixon, L.J.}, \bibinfo{author}{Drummond, J.M.},
  \bibinfo{author}{Henn, J.M.}, \bibinfo{year}{2011}.
\newblock \bibinfo{title}{{The one-loop six-dimensional hexagon integral and
  its relation to MHV amplitudes in $\mathcal{N}=4$ SYM}}.
\newblock \bibinfo{journal}{Journal of High Energy Physics}
  \bibinfo{volume}{06}.
\bibitem[{Dixon et~al.(2012)Dixon, Drummond and Henn}]{DixonDrummondHenn2011}
\bibinfo{author}{Dixon, L.J.}, \bibinfo{author}{Drummond, J.M.},
  \bibinfo{author}{Henn, J.M.}, \bibinfo{year}{2012}.
\newblock \bibinfo{title}{Analytic result for the two-loop six-point {NMHV}
  amplitude in \(\mathcal{N}=4\) super yang--mills theory}.
\newblock \bibinfo{journal}{Journal of High Energy Physics}
  \bibinfo{volume}{2012}, \bibinfo{pages}{024}.
\newblock \DOIprefix\doi{10.1007/JHEP01(2012)024},
  \href{http://arxiv.org/abs/1111.1704}{\tt arXiv:1111.1704}.
\bibitem[{Dixon et~al.(2023)Dixon, G{\"u}rdo{\u g}an, Liu, McLeod and
  Wilhelm}]{DixonGurdoganLiuMcLeodWilhelm2023}
\bibinfo{author}{Dixon, L.J.}, \bibinfo{author}{G{\"u}rdo{\u g}an, {\"O}.},
  \bibinfo{author}{Liu, Y.T.}, \bibinfo{author}{McLeod, A.J.},
  \bibinfo{author}{Wilhelm, M.}, \bibinfo{year}{2023}.
\newblock \bibinfo{title}{Antipodal self-duality for a four-particle form
  factor}.
\newblock \bibinfo{journal}{Physical Review Letters} \bibinfo{volume}{130},
  \bibinfo{pages}{111601}.
\newblock \DOIprefix\doi{10.1103/PhysRevLett.130.111601},
  \href{http://arxiv.org/abs/2212.02410}{\tt arXiv:2212.02410}.
\bibitem[{Dixon et~al.(2022)Dixon, G{\"u}rdo{\u g}an, McLeod and
  Wilhelm}]{DixonGurdoganMcLeodWilhelm2022}
\bibinfo{author}{Dixon, L.J.}, \bibinfo{author}{G{\"u}rdo{\u g}an, {\"O}.},
  \bibinfo{author}{McLeod, A.J.}, \bibinfo{author}{Wilhelm, M.},
  \bibinfo{year}{2022}.
\newblock \bibinfo{title}{Folding amplitudes into form factors: An antipodal
  duality}.
\newblock \bibinfo{journal}{Physical Review Letters} \bibinfo{volume}{128},
  \bibinfo{pages}{111602}.
\newblock \DOIprefix\doi{10.1103/PhysRevLett.128.111602},
  \href{http://arxiv.org/abs/2112.06243}{\tt arXiv:2112.06243}.
\bibitem[{Dixon et~al.(2017b)Dixon, von Hippel, McLeod and
  Trnka}]{Dixon:2016apl}
\bibinfo{author}{Dixon, L.J.}, \bibinfo{author}{von Hippel, M.},
  \bibinfo{author}{McLeod, A.J.}, \bibinfo{author}{Trnka, J.},
  \bibinfo{year}{2017}b.
\newblock \bibinfo{title}{{Multi-loop positivity of the planar $ \mathcal{N} $
  = 4 SYM six-point amplitude}}.
\newblock \bibinfo{journal}{JHEP} \bibinfo{volume}{02}, \bibinfo{pages}{112}.
\newblock \DOIprefix\doi{10.1007/JHEP02(2017)112},
  \href{http://arxiv.org/abs/1611.08325}{\tt arXiv:1611.08325}.
\bibitem[{Dixon et~al.(2026)Dixon, Oktem, Paranjape, Trnka, Xu and
  Zhang}]{Dixon:2026ipt}
\bibinfo{author}{Dixon, L.J.}, \bibinfo{author}{Oktem, U.},
  \bibinfo{author}{Paranjape, S.}, \bibinfo{author}{Trnka, J.},
  \bibinfo{author}{Xu, Y.}, \bibinfo{author}{Zhang, S.Q.},
  \bibinfo{year}{2026}.
\newblock \bibinfo{title}{{Multi-Loop Negative Geometries}}
  \href{http://arxiv.org/abs/2605.28926}{\tt arXiv:2605.28926}.
\bibitem[{Dlapa et~al.(2023)Dlapa, Helmer, Papathanasiou and
  Tellander}]{DlapaHelmerPapathanasiouTellander2023}
\bibinfo{author}{Dlapa, C.}, \bibinfo{author}{Helmer, M.},
  \bibinfo{author}{Papathanasiou, G.}, \bibinfo{author}{Tellander, F.},
  \bibinfo{year}{2023}.
\newblock \bibinfo{title}{Symbol alphabets from the landau singular locus}.
\newblock \bibinfo{journal}{Journal of High Energy Physics}
  \bibinfo{volume}{2023}, \bibinfo{pages}{161}.
\newblock \DOIprefix\doi{10.1007/JHEP10(2023)161},
  \href{http://arxiv.org/abs/2304.02629}{\tt arXiv:2304.02629}.
\bibitem[{Drton et~al.(2009)Drton, Sturmfels and
  Sullivant}]{DrtonSturmfelsSullivant2009}
\bibinfo{author}{Drton, M.}, \bibinfo{author}{Sturmfels, B.},
  \bibinfo{author}{Sullivant, S.}, \bibinfo{year}{2009}.
\newblock \bibinfo{title}{Lectures on Algebraic Statistics}.
\newblock \bibinfo{publisher}{Birkh\"auser}.
\newblock \DOIprefix\doi{10.1007/978-0-8176-4711-9}.
\bibitem[{Drummond et~al.(2018)Drummond, Foster and G{\"u}rdo{\u
  g}an}]{DrummondFosterGurdoganClusterAdjacency2018}
\bibinfo{author}{Drummond, J.}, \bibinfo{author}{Foster, J.},
  \bibinfo{author}{G{\"u}rdo{\u g}an, {\"O}.}, \bibinfo{year}{2018}.
\newblock \bibinfo{title}{Cluster adjacency properties of scattering amplitudes
  in \( \mathcal{N}=4 \) supersymmetric yang--mills theory}.
\newblock \bibinfo{journal}{Physical Review Letters} \bibinfo{volume}{120},
  \bibinfo{pages}{161601}.
\newblock \DOIprefix\doi{10.1103/PhysRevLett.120.161601},
  \href{http://arxiv.org/abs/1710.10953}{\tt arXiv:1710.10953}.
\bibitem[{Drummond et~al.(2019)Drummond, Foster, G{\"u}rdo{\u g}an and
  Kalousios}]{DrummondFosterGurdoganHarrington2019}
\bibinfo{author}{Drummond, J.}, \bibinfo{author}{Foster, J.},
  \bibinfo{author}{G{\"u}rdo{\u g}an, {\"O}.}, \bibinfo{author}{Kalousios, C.},
  \bibinfo{year}{2019}.
\newblock \bibinfo{title}{Cluster adjacency and the four-loop nmhv heptagon}.
\newblock \bibinfo{journal}{Journal of High Energy Physics}
  \bibinfo{volume}{2019}, \bibinfo{pages}{087}.
\newblock \DOIprefix\doi{10.1007/JHEP03(2019)087},
  \href{http://arxiv.org/abs/1810.08149}{\tt arXiv:1810.08149}.
\bibitem[{Drummond et~al.(2020)Drummond, Foster, G{\"u}rdo{\u g}an and
  Kalousios}]{DrummondFosterGurdoganKalousios2020}
\bibinfo{author}{Drummond, J.}, \bibinfo{author}{Foster, J.},
  \bibinfo{author}{G{\"u}rdo{\u g}an, {\"O}.}, \bibinfo{author}{Kalousios, C.},
  \bibinfo{year}{2020}.
\newblock \bibinfo{title}{Tropical grassmannians, cluster algebras and
  scattering amplitudes}.
\newblock \bibinfo{journal}{Journal of High Energy Physics}
  \bibinfo{volume}{2020}, \bibinfo{pages}{146}.
\newblock \DOIprefix\doi{10.1007/JHEP04(2020)146},
  \href{http://arxiv.org/abs/1907.01053}{\tt arXiv:1907.01053}.
\bibitem[{Drummond et~al.(2021a)Drummond, Foster, G{\"u}rdo{\u g}an and
  Kalousios}]{DrummondFosterGurrieri2018}
\bibinfo{author}{Drummond, J.}, \bibinfo{author}{Foster, J.},
  \bibinfo{author}{G{\"u}rdo{\u g}an, {\"O}.}, \bibinfo{author}{Kalousios, C.},
  \bibinfo{year}{2021}a.
\newblock \bibinfo{title}{Algebraic singularities of scattering amplitudes from
  tropical geometry}.
\newblock \bibinfo{journal}{Journal of High Energy Physics}
  \bibinfo{volume}{2021}, \bibinfo{pages}{002}.
\newblock \DOIprefix\doi{10.1007/JHEP04(2021)002},
  \href{http://arxiv.org/abs/1912.08217}{\tt arXiv:1912.08217}.
\bibitem[{Drummond et~al.(2021b)Drummond, Foster, G{\"u}rdo{\u g}an and
  Kalousios}]{DrummondFosterGurdoganKalousios2021}
\bibinfo{author}{Drummond, J.}, \bibinfo{author}{Foster, J.},
  \bibinfo{author}{G{\"u}rdo{\u g}an, {\"O}.}, \bibinfo{author}{Kalousios, C.},
  \bibinfo{year}{2021}b.
\newblock \bibinfo{title}{Tropical fans, scattering equations and amplitudes}.
\newblock \bibinfo{journal}{Journal of High Energy Physics}
  \bibinfo{volume}{2021}, \bibinfo{pages}{174}.
\newblock \DOIprefix\doi{10.1007/JHEP04(2021)174},
  \href{http://arxiv.org/abs/2002.04624}{\tt arXiv:2002.04624}.
\bibitem[{Drummond et~al.(2026)Drummond, G{\"u}rdo{\u g}an and
  Li}]{DrummondGurdoganLi2026TropicalSymmetries}
\bibinfo{author}{Drummond, J.}, \bibinfo{author}{G{\"u}rdo{\u g}an, {\"O}.},
  \bibinfo{author}{Li, J.R.}, \bibinfo{year}{2026}.
\newblock \bibinfo{title}{Tropical symmetries of cluster algebras}.
\newblock \href{http://arxiv.org/abs/2601.19779}{\tt arXiv:2601.19779}.
\bibitem[{Drummond and Ferro(2010)}]{Drummond:2010qh}
\bibinfo{author}{Drummond, J.M.}, \bibinfo{author}{Ferro, L.},
  \bibinfo{year}{2010}.
\newblock \bibinfo{title}{{The Yangian origin of the Grassmannian integral}}.
\newblock \bibinfo{journal}{JHEP} \bibinfo{volume}{12}, \bibinfo{pages}{010}.
\newblock \href{http://arxiv.org/abs/1002.4622}{\tt arXiv:1002.4622}.
\bibitem[{Drummond et~al.(2008)Drummond, Henn, Korchemsky and
  Sokatchev}]{DrummondHennKorchemskySokatchev2008}
\bibinfo{author}{Drummond, J.M.}, \bibinfo{author}{Henn, J.},
  \bibinfo{author}{Korchemsky, G.P.}, \bibinfo{author}{Sokatchev, E.},
  \bibinfo{year}{2008}.
\newblock \bibinfo{title}{The hexagon wilson loop and the {BDS} ansatz for the
  six-gluon amplitude}.
\newblock \bibinfo{journal}{Physics Letters B} \bibinfo{volume}{662},
  \bibinfo{pages}{456--460}.
\newblock \DOIprefix\doi{10.1016/j.physletb.2008.03.044},
  \href{http://arxiv.org/abs/0712.4138}{\tt arXiv:0712.4138}.
\bibitem[{Drummond et~al.(2010a)Drummond, Henn, Korchemsky and
  Sokatchev}]{DrummondHennKorchemskySokatchev2007}
\bibinfo{author}{Drummond, J.M.}, \bibinfo{author}{Henn, J.},
  \bibinfo{author}{Korchemsky, G.P.}, \bibinfo{author}{Sokatchev, E.},
  \bibinfo{year}{2010}a.
\newblock \bibinfo{title}{Conformal ward identities for wilson loops and a test
  of the duality with gluon amplitudes}.
\newblock \bibinfo{journal}{Nuclear Physics B} \bibinfo{volume}{826},
  \bibinfo{pages}{337--364}.
\newblock \DOIprefix\doi{10.1016/j.nuclphysb.2009.10.013},
  \href{http://arxiv.org/abs/0712.1223}{\tt arXiv:0712.1223}.
\bibitem[{Drummond et~al.(2010b)Drummond, Henn, Korchemsky and
  Sokatchev}]{Drummond:2008vq}
\bibinfo{author}{Drummond, J.M.}, \bibinfo{author}{Henn, J.},
  \bibinfo{author}{Korchemsky, G.P.}, \bibinfo{author}{Sokatchev, E.},
  \bibinfo{year}{2010}b.
\newblock \bibinfo{title}{Dual superconformal symmetry of scattering amplitudes
  in \(\mathcal{N}=4\) super-yang--mills theory}.
\newblock \bibinfo{journal}{Nuclear Physics B} \bibinfo{volume}{828},
  \bibinfo{pages}{317--374}.
\newblock \DOIprefix\doi{10.1016/j.nuclphysb.2009.11.022},
  \href{http://arxiv.org/abs/0807.1095}{\tt arXiv:0807.1095}.
\bibitem[{Drummond et~al.(2007)Drummond, Henn, Smirnov and
  Sokatchev}]{Drummond:2007aua}
\bibinfo{author}{Drummond, J.M.}, \bibinfo{author}{Henn, J.},
  \bibinfo{author}{Smirnov, V.A.}, \bibinfo{author}{Sokatchev, E.},
  \bibinfo{year}{2007}.
\newblock \bibinfo{title}{Magic identities for conformal four-point integrals}.
\newblock \bibinfo{journal}{Journal of High Energy Physics}
  \bibinfo{volume}{2007}, \bibinfo{pages}{064}.
\newblock \DOIprefix\doi{10.1088/1126-6708/2007/01/064},
  \href{http://arxiv.org/abs/hep-th/0607160}{\tt arXiv:hep-th/0607160}.
\bibitem[{Drummond et~al.(2009)Drummond, Henn and Plefka}]{Drummond:2009fd}
\bibinfo{author}{Drummond, J.M.}, \bibinfo{author}{Henn, J.M.},
  \bibinfo{author}{Plefka, J.}, \bibinfo{year}{2009}.
\newblock \bibinfo{title}{{Yangian symmetry of scattering amplitudes in N=4
  super Yang-Mills theory}}.
\newblock \bibinfo{journal}{JHEP} \bibinfo{volume}{05}, \bibinfo{pages}{046}.
\newblock \DOIprefix\doi{10.1088/1126-6708/2009/05/046},
  \href{http://arxiv.org/abs/0902.2987}{\tt arXiv:0902.2987}.
\bibitem[{Duhr(2012)}]{Duhr:2012fh}
\bibinfo{author}{Duhr, C.}, \bibinfo{year}{2012}.
\newblock \bibinfo{title}{Hopf algebras, coproducts and symbols: an application
  to {Higgs} boson amplitudes}.
\newblock \bibinfo{journal}{Journal of High Energy Physics}
  \bibinfo{volume}{2012}, \bibinfo{pages}{043}.
\newblock \DOIprefix\doi{10.1007/JHEP08(2012)043},
  \href{http://arxiv.org/abs/1203.0454}{\tt arXiv:1203.0454}.
\bibitem[{Duhr and Dulat(2019)}]{DuhrDulat2019}
\bibinfo{author}{Duhr, C.}, \bibinfo{author}{Dulat, F.}, \bibinfo{year}{2019}.
\newblock \bibinfo{title}{{PolyLogTools} --- polylogs for the masses}.
\newblock \bibinfo{journal}{Journal of High Energy Physics}
  \bibinfo{volume}{2019}, \bibinfo{pages}{135}.
\newblock \DOIprefix\doi{10.1007/JHEP08(2019)135},
  \href{http://arxiv.org/abs/1904.07279}{\tt arXiv:1904.07279}.
\bibitem[{Duhr et~al.(2012)Duhr, Gangl and Rhodes}]{Duhr:2011zq}
\bibinfo{author}{Duhr, C.}, \bibinfo{author}{Gangl, H.},
  \bibinfo{author}{Rhodes, J.R.}, \bibinfo{year}{2012}.
\newblock \bibinfo{title}{From polygons and symbols to polylogarithmic
  functions}.
\newblock \bibinfo{journal}{Journal of High Energy Physics}
  \bibinfo{volume}{2012}, \bibinfo{pages}{075}.
\newblock \DOIprefix\doi{10.1007/JHEP10(2012)075},
  \href{http://arxiv.org/abs/1110.0458}{\tt arXiv:1110.0458}.
\bibitem[{Duhr et~al.(2025)Duhr, Maggio, Nega, Sauer, Tancredi and
  Wagner}]{DuhrMaggioNegaSauerTancrediWagner2025}
\bibinfo{author}{Duhr, C.}, \bibinfo{author}{Maggio, S.},
  \bibinfo{author}{Nega, C.}, \bibinfo{author}{Sauer, B.},
  \bibinfo{author}{Tancredi, L.}, \bibinfo{author}{Wagner, F.J.},
  \bibinfo{year}{2025}.
\newblock \bibinfo{title}{Aspects of canonical differential equations for
  calabi--yau geometries and beyond}.
\newblock \bibinfo{journal}{Journal of High Energy Physics}
  \bibinfo{volume}{2025}, \bibinfo{pages}{128}.
\newblock \DOIprefix\doi{10.1007/JHEP06(2025)128},
  \href{http://arxiv.org/abs/2503.20655}{\tt arXiv:2503.20655}.
\bibitem[{Eden et~al.(2017)Eden, Heslop and Mason}]{Correlahedron}
\bibinfo{author}{Eden, B.}, \bibinfo{author}{Heslop, P.},
  \bibinfo{author}{Mason, L.}, \bibinfo{year}{2017}.
\newblock \bibinfo{title}{{The Correlahedron}}.
\newblock \bibinfo{journal}{JHEP} \bibinfo{volume}{09}, \bibinfo{pages}{156}.
\newblock \DOIprefix\doi{10.1007/JHEP09(2017)156},
  \href{http://arxiv.org/abs/1701.00453}{\tt arXiv:1701.00453}.
\bibitem[{Eden et~al.(2011)Eden, Korchemsky and Sokatchev}]{Eden:2010ce}
\bibinfo{author}{Eden, B.}, \bibinfo{author}{Korchemsky, G.P.},
  \bibinfo{author}{Sokatchev, E.}, \bibinfo{year}{2011}.
\newblock \bibinfo{title}{{From correlation functions to scattering
  amplitudes}}.
\newblock \bibinfo{journal}{JHEP} \bibinfo{volume}{12}, \bibinfo{pages}{002}.
\newblock \DOIprefix\doi{10.1007/JHEP12(2011)002},
  \href{http://arxiv.org/abs/1007.3246}{\tt arXiv:1007.3246}.
\bibitem[{Eden et~al.(1966)Eden, Landshoff, Olive and
  Polkinghorne}]{Eden:1966dnq}
\bibinfo{author}{Eden, R.J.}, \bibinfo{author}{Landshoff, P.V.},
  \bibinfo{author}{Olive, D.I.}, \bibinfo{author}{Polkinghorne, J.C.},
  \bibinfo{year}{1966}.
\newblock \bibinfo{title}{The Analytic S-Matrix}.
\newblock \bibinfo{publisher}{Cambridge University Press}.
\bibitem[{Ehrenpreis(1970)}]{Ehrenpreis1970}
\bibinfo{author}{Ehrenpreis, L.}, \bibinfo{year}{1970}.
\newblock \bibinfo{title}{Fourier Analysis in Several Complex Variables}.
\newblock \bibinfo{publisher}{Wiley-Interscience}.
\bibitem[{Elekes and Sharir(2011)}]{ES}
\bibinfo{author}{Elekes, G.}, \bibinfo{author}{Sharir, M.},
  \bibinfo{year}{2011}.
\newblock \bibinfo{title}{Incidences in three dimensions and distinct distances
  in the plane}.
\newblock \bibinfo{journal}{Combinatorics, Probability and Computing}
  \bibinfo{volume}{20}, \bibinfo{pages}{571--608}.
\newblock \DOIprefix\doi{10.1017/S0963548311000137},
  \href{http://arxiv.org/abs/1005.0982}{\tt arXiv:1005.0982}.
\bibitem[{Ellis et~al.(2009)Ellis, Giele, Kunszt and
  Melnikov}]{EllisGieleKunsztMelnikov2009}
\bibinfo{author}{Ellis, R.K.}, \bibinfo{author}{Giele, W.T.},
  \bibinfo{author}{Kunszt, Z.}, \bibinfo{author}{Melnikov, K.},
  \bibinfo{year}{2009}.
\newblock \bibinfo{title}{Masses, fermions and generalized $d$-dimensional
  unitarity}.
\newblock \bibinfo{journal}{Nuclear Physics B} \bibinfo{volume}{822},
  \bibinfo{pages}{270--282}.
\newblock \DOIprefix\doi{10.1016/j.nuclphysb.2009.07.023},
  \href{http://arxiv.org/abs/0806.3467}{\tt arXiv:0806.3467}.
\bibitem[{Ellis et~al.(2012)Ellis, Kunszt, Melnikov and
  Zanderighi}]{Ellis:2011cr}
\bibinfo{author}{Ellis, R.K.}, \bibinfo{author}{Kunszt, Z.},
  \bibinfo{author}{Melnikov, K.}, \bibinfo{author}{Zanderighi, G.},
  \bibinfo{year}{2012}.
\newblock \bibinfo{title}{One-loop calculations in quantum field theory: from
  feynman diagrams to unitarity cuts}.
\newblock \bibinfo{journal}{Phys. Rept.} \bibinfo{volume}{518},
  \bibinfo{pages}{141--250}.
\newblock \DOIprefix\doi{10.1016/j.physrep.2012.01.008},
  \href{http://arxiv.org/abs/1105.4319}{\tt arXiv:1105.4319}.
\bibitem[{Ellis et~al.(1996)Ellis, Stirling and Webber}]{Ellis:1991qj}
\bibinfo{author}{Ellis, R.K.}, \bibinfo{author}{Stirling, W.J.},
  \bibinfo{author}{Webber, B.R.}, \bibinfo{year}{1996}.
\newblock \bibinfo{title}{QCD and Collider Physics}. volume~\bibinfo{volume}{8}
  of \textit{\bibinfo{series}{Cambridge Monographs on Particle Physics, Nuclear
  Physics and Cosmology}}.
\newblock \bibinfo{publisher}{Cambridge University Press},
  \bibinfo{address}{Cambridge}.
\newblock \DOIprefix\doi{10.1017/CBO9780511628788}.
\bibitem[{Elvang et~al.(2010)Elvang, Freedman and
  Kiermaier}]{ElvangFreedmanKiermaier2009}
\bibinfo{author}{Elvang, H.}, \bibinfo{author}{Freedman, D.Z.},
  \bibinfo{author}{Kiermaier, M.}, \bibinfo{year}{2010}.
\newblock \bibinfo{title}{Dual conformal symmetry of one-loop {NMHV} amplitudes
  in \(\mathcal{N}=4\) {SYM} theory}.
\newblock \bibinfo{journal}{Journal of High Energy Physics}
  \bibinfo{volume}{2010}, \bibinfo{pages}{075}.
\newblock \DOIprefix\doi{10.1007/JHEP03(2010)075},
  \href{http://arxiv.org/abs/0905.4379}{\tt arXiv:0905.4379}.
\bibitem[{Elvang and Huang(2015)}]{Elvang:2013cua}
\bibinfo{author}{Elvang, H.}, \bibinfo{author}{Huang, Y.t.},
  \bibinfo{year}{2015}.
\newblock \bibinfo{title}{Scattering Amplitudes in Gauge Theory and Gravity}.
\newblock \bibinfo{publisher}{Cambridge University Press}.
\newblock \DOIprefix\doi{10.1017/CBO9781107706620},
  \href{http://arxiv.org/abs/1308.1697}{\tt arXiv:1308.1697}.
\bibitem[{Engelund and Roiban(2012)}]{Engelund:2011fg}
\bibinfo{author}{Engelund, O.T.}, \bibinfo{author}{Roiban, R.},
  \bibinfo{year}{2012}.
\newblock \bibinfo{title}{{On correlation functions of Wilson loops, local and
  non-local operators}}.
\newblock \bibinfo{journal}{JHEP} \bibinfo{volume}{05}, \bibinfo{pages}{158}.
\newblock \DOIprefix\doi{10.1007/JHEP05(2012)158},
  \href{http://arxiv.org/abs/1110.0758}{\tt arXiv:1110.0758}.
\bibitem[{Escobar and Knutson(2017)}]{EK}
\bibinfo{author}{Escobar, L.}, \bibinfo{author}{Knutson, A.},
  \bibinfo{year}{2017}.
\newblock \bibinfo{title}{The multidegree of the multi-image variety}, in:
  \bibinfo{editor}{Smith, G.G.}, \bibinfo{editor}{Sturmfels, B.} (Eds.),
  \bibinfo{booktitle}{Combinatorial Algebraic Geometry: Selected Papers from
  the 2016 Apprenticeship Program}. \bibinfo{publisher}{Springer},
  \bibinfo{address}{New York}. volume~\bibinfo{volume}{80} of
  \textit{\bibinfo{series}{Fields Institute Communications}}, pp.
  \bibinfo{pages}{283--296}.
\newblock \DOIprefix\doi{10.1007/978-1-4939-7486-3_13}.
\bibitem[{Even-Zohar et~al.(2024a)Even-Zohar, Lakrec, Parisi, Sherman-Bennett,
  Tessler and
  Williams}]{evenZoharLakrecParisiTesslerShermanBennettWilliams2023cluster}
\bibinfo{author}{Even-Zohar, C.}, \bibinfo{author}{Lakrec, T.},
  \bibinfo{author}{Parisi, M.}, \bibinfo{author}{Sherman-Bennett, M.},
  \bibinfo{author}{Tessler, R.}, \bibinfo{author}{Williams, L.K.},
  \bibinfo{year}{2024}a.
\newblock \bibinfo{title}{Cluster algebras and tilings for the \(m=4\)
  amplituhedron}.
\newblock \bibinfo{journal}{S{\'e}minaire Lotharingien de Combinatoire}
  \bibinfo{volume}{91B}, \bibinfo{pages}{12}.
\newblock \href{http://arxiv.org/abs/2310.17727}{\tt arXiv:2310.17727}.
\bibitem[{Even-Zohar et~al.(2025a)Even-Zohar, Lakrec, Parisi, Tessler,
  Sherman-Bennett and Williams}]{higher_m_ampl}
\bibinfo{author}{Even-Zohar, C.}, \bibinfo{author}{Lakrec, T.},
  \bibinfo{author}{Parisi, M.}, \bibinfo{author}{Tessler, R.},
  \bibinfo{author}{Sherman-Bennett, M.}, \bibinfo{author}{Williams, L.},
  \bibinfo{year}{2025}a.
\newblock \bibinfo{title}{{Higher-$m$ Amplituhedra (in progress)}}.
\bibitem[{Even-Zohar et~al.(2024b)Even-Zohar, Lakrec, Parisi, Tessler,
  Sherman-Bennett and
  Williams}]{evenZoharLakrecParisiTesslerShermanBennettWilliams2024clusterResults}
\bibinfo{author}{Even-Zohar, C.}, \bibinfo{author}{Lakrec, T.},
  \bibinfo{author}{Parisi, M.}, \bibinfo{author}{Tessler, R.J.},
  \bibinfo{author}{Sherman-Bennett, M.}, \bibinfo{author}{Williams, L.K.},
  \bibinfo{year}{2024}b.
\newblock \bibinfo{title}{A cluster of results on amplituhedron tiles}.
\newblock \bibinfo{journal}{Letters in Mathematical Physics}
  \bibinfo{volume}{114}, \bibinfo{pages}{94}.
\newblock \DOIprefix\doi{10.1007/s11005-024-01854-4},
  \href{http://arxiv.org/abs/2402.15568}{\tt arXiv:2402.15568}.
\bibitem[{Even-Zohar et~al.(2025b)Even-Zohar, Lakrec and
  Tessler}]{evenZoharLakrecTessler2025bcfw}
\bibinfo{author}{Even-Zohar, C.}, \bibinfo{author}{Lakrec, T.},
  \bibinfo{author}{Tessler, R.J.}, \bibinfo{year}{2025}b.
\newblock \bibinfo{title}{The amplituhedron bcfw triangulation}.
\newblock \bibinfo{journal}{Inventiones Mathematicae} \bibinfo{volume}{239},
  \bibinfo{pages}{1009--1138}.
\newblock \DOIprefix\doi{10.1007/s00222-025-01316-1},
  \href{http://arxiv.org/abs/2112.02703}{\tt arXiv:2112.02703}.
\bibitem[{Even-Zohar et~al.(2026)Even-Zohar, Parisi, Sherman-Bennett, Tessler
  and Williams}]{Even-Zohar:2025ngd}
\bibinfo{author}{Even-Zohar, C.}, \bibinfo{author}{Parisi, M.},
  \bibinfo{author}{Sherman-Bennett, M.}, \bibinfo{author}{Tessler, R.},
  \bibinfo{author}{Williams, L.K.}, \bibinfo{year}{2026}.
\newblock \bibinfo{title}{Plabic tangles and cluster promotion maps}.
\newblock \bibinfo{journal}{Journal of Algebra} \bibinfo{volume}{695},
  \bibinfo{pages}{14--102}.
\newblock \DOIprefix\doi{10.1016/j.jalgebra.2026.01.027},
  \href{http://arxiv.org/abs/2508.02891}{\tt arXiv:2508.02891}.
\bibitem[{Fairlie et~al.(1962)Fairlie, Landshoff, Nuttall and
  Polkinghorne}]{Fairlie1962}
\bibinfo{author}{Fairlie, D.B.}, \bibinfo{author}{Landshoff, P.V.},
  \bibinfo{author}{Nuttall, J.}, \bibinfo{author}{Polkinghorne, J.C.},
  \bibinfo{year}{1962}.
\newblock \bibinfo{title}{Singularities of the second type}.
\newblock \bibinfo{journal}{Journal of Mathematical Physics}
  \bibinfo{volume}{3}, \bibinfo{pages}{594--602}.
\newblock \DOIprefix\doi{10.1063/1.1724262}.
\bibitem[{Feng et~al.(2010)Feng, Wang, Wang and Zhang}]{Feng:2009ei}
\bibinfo{author}{Feng, B.}, \bibinfo{author}{Wang, J.}, \bibinfo{author}{Wang,
  Y.t.}, \bibinfo{author}{Zhang, Z.}, \bibinfo{year}{2010}.
\newblock \bibinfo{title}{Bcfw recursion relation with nonzero boundary
  contribution}.
\newblock \bibinfo{journal}{Journal of High Energy Physics}
  \bibinfo{volume}{01}, \bibinfo{pages}{019}.
\newblock \href{http://arxiv.org/abs/0911.0301}{\tt arXiv:0911.0301}.
\bibitem[{Ferro et~al.(2024)Ferro, Glew, Lukowski and
  Stalknecht}]{Ferro:2023qdp}
\bibinfo{author}{Ferro, L.}, \bibinfo{author}{Glew, R.},
  \bibinfo{author}{Lukowski, T.}, \bibinfo{author}{Stalknecht, J.},
  \bibinfo{year}{2024}.
\newblock \bibinfo{title}{Prescriptive unitarity from positive geometries}.
\newblock \bibinfo{journal}{JHEP} \bibinfo{volume}{03}, \bibinfo{pages}{001}.
\newblock \DOIprefix\doi{10.1007/JHEP03(2024)001},
  \href{http://arxiv.org/abs/2308.02438}{\tt arXiv:2308.02438}.
\bibitem[{Ferro et~al.(2025)Ferro, Glew, Lukowski and
  Stalknecht}]{Ferro:2024fibration}
\bibinfo{author}{Ferro, L.}, \bibinfo{author}{Glew, R.},
  \bibinfo{author}{Lukowski, T.}, \bibinfo{author}{Stalknecht, J.},
  \bibinfo{year}{2025}.
\newblock \bibinfo{title}{The two-loop mhv momentum amplituhedron from
  fibrations of fibrations}.
\newblock \bibinfo{journal}{JHEP} \bibinfo{volume}{02}, \bibinfo{pages}{044}.
\newblock \DOIprefix\doi{10.1007/JHEP02(2025)044},
  \href{http://arxiv.org/abs/2407.12906}{\tt arXiv:2407.12906}.
\bibitem[{Ferro and {\L}ukowski(2023)}]{Ferro:2022abq}
\bibinfo{author}{Ferro, L.}, \bibinfo{author}{{\L}ukowski, T.},
  \bibinfo{year}{2023}.
\newblock \bibinfo{title}{The loop momentum amplituhedron}.
\newblock \bibinfo{journal}{Journal of High Energy Physics}
  \bibinfo{volume}{2023}, \bibinfo{pages}{183}.
\newblock \DOIprefix\doi{10.1007/JHEP05(2023)183},
  \href{http://arxiv.org/abs/2210.01127}{\tt arXiv:2210.01127}.
\bibitem[{Ferro et~al.(2020)Ferro, {\L}ukowski and
  Moerman}]{ferroLukowskiMoerman2020boundaries}
\bibinfo{author}{Ferro, L.}, \bibinfo{author}{{\L}ukowski, T.},
  \bibinfo{author}{Moerman, R.}, \bibinfo{year}{2020}.
\newblock \bibinfo{title}{From momentum amplituhedron boundaries to amplitude
  singularities and back}.
\newblock \bibinfo{journal}{Journal of High Energy Physics}
  \bibinfo{volume}{2020}, \bibinfo{pages}{201}.
\newblock \href{http://arxiv.org/abs/2003.13704}{\tt arXiv:2003.13704}.
\bibitem[{Ferro et~al.(2016)Ferro, Lukowski, Orta and Parisi}]{Ferro}
\bibinfo{author}{Ferro, L.}, \bibinfo{author}{Lukowski, T.},
  \bibinfo{author}{Orta, A.}, \bibinfo{author}{Parisi, M.},
  \bibinfo{year}{2016}.
\newblock \bibinfo{title}{{Towards the Amplituhedron Volume}}.
\newblock \bibinfo{journal}{JHEP} \bibinfo{volume}{03}, \bibinfo{pages}{014}.
\newblock \DOIprefix\doi{10.1007/JHEP03(2016)014},
  \href{http://arxiv.org/abs/1512.04954}{\tt arXiv:1512.04954}.
\bibitem[{Fevola et~al.(2024a)Fevola, Mizera and Telen}]{FevolaMizeraTelen2024}
\bibinfo{author}{Fevola, C.}, \bibinfo{author}{Mizera, S.},
  \bibinfo{author}{Telen, S.}, \bibinfo{year}{2024}a.
\newblock \bibinfo{title}{Landau singularities revisited: Computational
  algebraic geometry for feynman integrals}.
\newblock \bibinfo{journal}{Physical Review Letters} \bibinfo{volume}{132},
  \bibinfo{pages}{101601}.
\newblock \DOIprefix\doi{10.1103/PhysRevLett.132.101601},
  \href{http://arxiv.org/abs/2311.14669}{\tt arXiv:2311.14669}.
\bibitem[{Fevola et~al.(2024b)Fevola, Mizera and
  Telen}]{FevolaMizeraTelenPLD2024}
\bibinfo{author}{Fevola, C.}, \bibinfo{author}{Mizera, S.},
  \bibinfo{author}{Telen, S.}, \bibinfo{year}{2024}b.
\newblock \bibinfo{title}{Principal landau determinants}.
\newblock \bibinfo{journal}{Computer Physics Communications}
  \bibinfo{volume}{303}, \bibinfo{pages}{109278}.
\newblock \DOIprefix\doi{10.1016/j.cpc.2024.109278},
  \href{http://arxiv.org/abs/2311.16219}{\tt arXiv:2311.16219}.
\bibitem[{Fevola and Sattelberger(2025)}]{Fevola:Pos_Geom}
\bibinfo{author}{Fevola, C.}, \bibinfo{author}{Sattelberger, A.L.},
  \bibinfo{year}{2025}.
\newblock \bibinfo{title}{Algebraic and positive geometry of the universe: From
  particles to galaxies}.
\newblock \bibinfo{journal}{Notices of the American Mathematical Society}
  \bibinfo{volume}{72}, \bibinfo{pages}{808--817}.
\newblock \DOIprefix\doi{10.1090/noti3220},
  \href{http://arxiv.org/abs/2502.13582}{\tt arXiv:2502.13582}.
\bibitem[{Feynman(1963)}]{Feynman:1963ax}
\bibinfo{author}{Feynman, R.P.}, \bibinfo{year}{1963}.
\newblock \bibinfo{title}{Quantum theory of gravitation}.
\newblock \bibinfo{journal}{Acta Phys. Polon.} \bibinfo{volume}{24},
  \bibinfo{pages}{697--722}.
\bibitem[{Fomin and Pylyavskyy(2023)}]{FP}
\bibinfo{author}{Fomin, S.}, \bibinfo{author}{Pylyavskyy, P.},
  \bibinfo{year}{2023}.
\newblock \bibinfo{title}{Incidences and tilings}.
\newblock \href{http://arxiv.org/abs/2305.07728}{\tt arXiv:2305.07728}.
  \bibinfo{note}{preprint}.
\bibitem[{Fomin et~al.(2021)Fomin, Williams and
  Zelevinsky}]{FominWilliamsZelevinsky2021}
\bibinfo{author}{Fomin, S.}, \bibinfo{author}{Williams, L.},
  \bibinfo{author}{Zelevinsky, A.}, \bibinfo{year}{2021}.
\newblock \bibinfo{title}{Introduction to cluster algebras: Chapters 1--3}.
\newblock \href{http://arxiv.org/abs/1608.05735}{\tt arXiv:1608.05735}.
\bibitem[{Fomin and Zelevinsky(2002)}]{FominZelevinsky2002}
\bibinfo{author}{Fomin, S.}, \bibinfo{author}{Zelevinsky, A.},
  \bibinfo{year}{2002}.
\newblock \bibinfo{title}{Cluster algebras i: Foundations}.
\newblock \bibinfo{journal}{Journal of the American Mathematical Society}
  \bibinfo{volume}{15}, \bibinfo{pages}{497--529}.
\newblock \DOIprefix\doi{10.1090/S0894-0347-01-00385-X}.
\bibitem[{Fomin and Zelevinsky(2003)}]{FominZelevinsky2003}
\bibinfo{author}{Fomin, S.}, \bibinfo{author}{Zelevinsky, A.},
  \bibinfo{year}{2003}.
\newblock \bibinfo{title}{Cluster algebras ii: Finite type classification}.
\newblock \bibinfo{journal}{Inventiones Mathematicae} \bibinfo{volume}{154},
  \bibinfo{pages}{63--121}.
\newblock \DOIprefix\doi{10.1007/s00222-003-0302-y},
  \href{http://arxiv.org/abs/math/0208229}{\tt arXiv:math/0208229}.
\bibitem[{Forde(2007)}]{Forde:2007mi}
\bibinfo{author}{Forde, D.}, \bibinfo{year}{2007}.
\newblock \bibinfo{title}{Direct extraction of one-loop integral coefficients}.
\newblock \bibinfo{journal}{Physical Review D} \bibinfo{volume}{75},
  \bibinfo{pages}{125019}.
\newblock \DOIprefix\doi{10.1103/PhysRevD.75.125019},
  \href{http://arxiv.org/abs/0704.1835}{\tt arXiv:0704.1835}.
\bibitem[{Franco et~al.(2015a)Franco, Galloni, Mariotti and
  Trnka}]{Franco:2014csa}
\bibinfo{author}{Franco, S.}, \bibinfo{author}{Galloni, D.},
  \bibinfo{author}{Mariotti, A.}, \bibinfo{author}{Trnka, J.},
  \bibinfo{year}{2015}a.
\newblock \bibinfo{title}{{Anatomy of the Amplituhedron}}.
\newblock \bibinfo{journal}{JHEP} \bibinfo{volume}{03}, \bibinfo{pages}{128}.
\newblock \DOIprefix\doi{10.1007/JHEP03(2015)128},
  \href{http://arxiv.org/abs/1408.3410}{\tt arXiv:1408.3410}.
\bibitem[{Franco et~al.(2015b)Franco, Galloni, Penante and
  Wen}]{Franco:2014csa_alt1}
\bibinfo{author}{Franco, S.}, \bibinfo{author}{Galloni, D.},
  \bibinfo{author}{Penante, B.}, \bibinfo{author}{Wen, C.},
  \bibinfo{year}{2015}b.
\newblock \bibinfo{title}{{Non-Planar On-Shell Diagrams}}.
\newblock \bibinfo{journal}{JHEP} \bibinfo{volume}{06}, \bibinfo{pages}{199}.
\newblock \href{http://arxiv.org/abs/1404.4780}{\tt arXiv:1404.4780}.
\bibitem[{Franco et~al.(2015c)Franco, Galloni, Penante and
  Wen}]{Franco:2015rma}
\bibinfo{author}{Franco, S.}, \bibinfo{author}{Galloni, D.},
  \bibinfo{author}{Penante, B.}, \bibinfo{author}{Wen, C.},
  \bibinfo{year}{2015}c.
\newblock \bibinfo{title}{{Non-Planar On-Shell Diagrams II: Non-Planar BCFW
  Bridges}}.
\newblock \bibinfo{journal}{JHEP} \bibinfo{volume}{06}, \bibinfo{pages}{072}.
\newblock \href{http://arxiv.org/abs/1502.02034}{\tt arXiv:1502.02034}.
\bibitem[{Fraser(2016)}]{Fraser}
\bibinfo{author}{Fraser, C.}, \bibinfo{year}{2016}.
\newblock \bibinfo{title}{Quasi-homomorphisms of cluster algebras}.
\newblock \bibinfo{journal}{Advances in Applied Mathematics}
  \bibinfo{volume}{81}, \bibinfo{pages}{40--77}.
\newblock \DOIprefix\doi{10.1016/j.aam.2016.06.005},
  \href{http://arxiv.org/abs/1503.06481}{\tt arXiv:1503.06481}.
\bibitem[{Frellesvig et~al.(2022)Frellesvig, Tommasini and
  Wever}]{FrellesvigEtAl2021}
\bibinfo{author}{Frellesvig, H.}, \bibinfo{author}{Tommasini, D.},
  \bibinfo{author}{Wever, C.}, \bibinfo{year}{2022}.
\newblock \bibinfo{title}{On epsilon factorized differential equations for
  elliptic feynman integrals}.
\newblock \bibinfo{journal}{Journal of High Energy Physics}
  \bibinfo{volume}{2022}, \bibinfo{pages}{100}.
\newblock \DOIprefix\doi{10.1007/JHEP03(2022)100},
  \href{http://arxiv.org/abs/2110.07968}{\tt arXiv:2110.07968}.
\bibitem[{Galashin et~al.(2022)Galashin, Karp and
  Lam}]{galashinKarpLam2022ball}
\bibinfo{author}{Galashin, P.}, \bibinfo{author}{Karp, S.N.},
  \bibinfo{author}{Lam, T.}, \bibinfo{year}{2022}.
\newblock \bibinfo{title}{The totally nonnegative grassmannian is a ball}.
\newblock \bibinfo{journal}{Advances in Mathematics} \bibinfo{volume}{397},
  \bibinfo{pages}{108123}.
\newblock \DOIprefix\doi{10.1016/j.aim.2021.108123},
  \href{http://arxiv.org/abs/1707.02010}{\tt arXiv:1707.02010}.
\bibitem[{Galashin and Lam(2020)}]{GalashinLam}
\bibinfo{author}{Galashin, P.}, \bibinfo{author}{Lam, T.},
  \bibinfo{year}{2020}.
\newblock \bibinfo{title}{Parity duality for the amplituhedron}.
\newblock \bibinfo{journal}{Compositio Mathematica} \bibinfo{volume}{156},
  \bibinfo{pages}{2207--2262}.
\newblock \DOIprefix\doi{10.1112/S0010437X20007411},
  \href{http://arxiv.org/abs/1805.00600}{\tt arXiv:1805.00600}.
\bibitem[{Galashin and Lam(2023)}]{GalashinLamPositroidCluster}
\bibinfo{author}{Galashin, P.}, \bibinfo{author}{Lam, T.},
  \bibinfo{year}{2023}.
\newblock \bibinfo{title}{Positroid varieties and cluster algebras}.
\newblock \bibinfo{journal}{Annales Scientifiques de l'{\'E}cole Normale
  Sup{\'e}rieure} \bibinfo{volume}{56}, \bibinfo{pages}{859--884}.
\newblock \href{http://arxiv.org/abs/1906.03501}{\tt arXiv:1906.03501}.
\bibitem[{Garc{\'i}a-Puente et~al.(2012)Garc{\'i}a-Puente, Hein, Hillar,
  Mart{\'i}n~del Campo, Ruffo, Sottile and
  Teitler}]{Sottile2003SecantConjecture}
\bibinfo{author}{Garc{\'i}a-Puente, L.}, \bibinfo{author}{Hein, N.},
  \bibinfo{author}{Hillar, C.}, \bibinfo{author}{Mart{\'i}n~del Campo, A.},
  \bibinfo{author}{Ruffo, J.}, \bibinfo{author}{Sottile, F.},
  \bibinfo{author}{Teitler, Z.}, \bibinfo{year}{2012}.
\newblock \bibinfo{title}{The secant conjecture in the real schubert calculus}.
\newblock \bibinfo{journal}{Experimental Mathematics} \bibinfo{volume}{21},
  \bibinfo{pages}{252--265}.
\newblock \DOIprefix\doi{10.1080/10586458.2012.666839},
  \href{http://arxiv.org/abs/1010.0665}{\tt arXiv:1010.0665}.
\bibitem[{Gehrmann et~al.(2018)Gehrmann, Henn and Lo~Presti}]{Gehrmann:2018yef}
\bibinfo{author}{Gehrmann, T.}, \bibinfo{author}{Henn, J.M.},
  \bibinfo{author}{Lo~Presti, N.A.}, \bibinfo{year}{2018}.
\newblock \bibinfo{title}{Pentagon functions for massless planar scattering
  amplitudes}.
\newblock \bibinfo{journal}{Journal of High Energy Physics}
  \bibinfo{volume}{2018}, \bibinfo{pages}{103}.
\newblock \DOIprefix\doi{10.1007/JHEP10(2018)103},
  \href{http://arxiv.org/abs/1807.09812}{\tt arXiv:1807.09812}.
\bibitem[{Gehrmann and Remiddi(2000)}]{GehrmannRemiddi2000}
\bibinfo{author}{Gehrmann, T.}, \bibinfo{author}{Remiddi, E.},
  \bibinfo{year}{2000}.
\newblock \bibinfo{title}{Differential equations for two-loop four-point
  functions}.
\newblock \bibinfo{journal}{Nuclear Physics B} \bibinfo{volume}{580},
  \bibinfo{pages}{485--518}.
\newblock \DOIprefix\doi{10.1016/S0550-3213(00)00223-6},
  \href{http://arxiv.org/abs/hep-ph/9912329}{\tt arXiv:hep-ph/9912329}.
\bibitem[{Gelfand et~al.(2009)Gelfand, Kapranov and Zelevinsky}]{gkz}
\bibinfo{author}{Gelfand, I.}, \bibinfo{author}{Kapranov, M.},
  \bibinfo{author}{Zelevinsky, A.}, \bibinfo{year}{2009}.
\newblock \bibinfo{title}{Discriminants, Resultants, and Multidimensional
  Determinants}.
\newblock Modern Birkh{\"a}user Classics, \bibinfo{publisher}{Birkh{\"a}user
  Boston}.
\bibitem[{Glew and {\L}ukowski(2025)}]{GlewLukowski2025}
\bibinfo{author}{Glew, R.}, \bibinfo{author}{{\L}ukowski, T.},
  \bibinfo{year}{2025}.
\newblock \bibinfo{title}{{Positive and negative ladders in loop space}}.
\newblock \bibinfo{journal}{JHEP} \bibinfo{volume}{06}, \bibinfo{pages}{124}.
\newblock \DOIprefix\doi{10.1007/JHEP06(2025)124},
  \href{http://arxiv.org/abs/2411.14989}{\tt arXiv:2411.14989}.
\bibitem[{Glew and Pokraka(2025)}]{GlewPokraka2025}
\bibinfo{author}{Glew, R.}, \bibinfo{author}{Pokraka, A.},
  \bibinfo{year}{2025}.
\newblock \bibinfo{title}{Kinematic flow from the flow of cuts}
  \href{http://arxiv.org/abs/2508.11568}{\tt arXiv:2508.11568}.
\bibitem[{Golden et~al.(2014)Golden, Goncharov, Spradlin, Vergu and
  Volovich}]{GoldenGoncharovSpradlinVerguVolovich2014}
\bibinfo{author}{Golden, J.}, \bibinfo{author}{Goncharov, A.B.},
  \bibinfo{author}{Spradlin, M.}, \bibinfo{author}{Vergu, C.},
  \bibinfo{author}{Volovich, A.}, \bibinfo{year}{2014}.
\newblock \bibinfo{title}{Motivic amplitudes and cluster coordinates}.
\newblock \bibinfo{journal}{Journal of High Energy Physics}
  \bibinfo{volume}{2014}, \bibinfo{pages}{091}.
\newblock \DOIprefix\doi{10.1007/JHEP01(2014)091},
  \href{http://arxiv.org/abs/1305.1617}{\tt arXiv:1305.1617}.
\bibitem[{Goncharov(2001)}]{Goncharov2001}
\bibinfo{author}{Goncharov, A.B.}, \bibinfo{year}{2001}.
\newblock \bibinfo{title}{Multiple polylogarithms and mixed {T}ate motives}
  \href{http://arxiv.org/abs/math/0103059}{\tt arXiv:math/0103059}.
\bibitem[{Goncharov(2005)}]{Goncharov2005}
\bibinfo{author}{Goncharov, A.B.}, \bibinfo{year}{2005}.
\newblock \bibinfo{title}{Galois symmetries of fundamental groupoids and
  noncommutative geometry}.
\newblock \bibinfo{journal}{Duke Mathematical Journal} \bibinfo{volume}{128},
  \bibinfo{pages}{209--284}.
\newblock \DOIprefix\doi{10.1215/S0012-7094-04-12822-2},
  \href{http://arxiv.org/abs/math/0208144}{\tt arXiv:math/0208144}.
\bibitem[{Goncharov et~al.(2010)Goncharov, Spradlin, Vergu and
  Volovich}]{Goncharov:2010jf}
\bibinfo{author}{Goncharov, A.B.}, \bibinfo{author}{Spradlin, M.},
  \bibinfo{author}{Vergu, C.}, \bibinfo{author}{Volovich, A.},
  \bibinfo{year}{2010}.
\newblock \bibinfo{title}{Classical polylogarithms for amplitudes and wilson
  loops}.
\newblock \bibinfo{journal}{Physical Review Letters} \bibinfo{volume}{105},
  \bibinfo{pages}{151605}.
\newblock \DOIprefix\doi{10.1103/PhysRevLett.105.151605},
  \href{http://arxiv.org/abs/1006.5703}{\tt arXiv:1006.5703}.
\bibitem[{Goresky and MacPherson(1988)}]{GoreskyMacPherson1988}
\bibinfo{author}{Goresky, M.}, \bibinfo{author}{MacPherson, R.},
  \bibinfo{year}{1988}.
\newblock \bibinfo{title}{Stratified Morse Theory}. volume~\bibinfo{volume}{14}
  of \textit{\bibinfo{series}{Ergebnisse der Mathematik und ihrer
  Grenzgebiete}}.
\newblock \bibinfo{publisher}{Springer}, \bibinfo{address}{Berlin}.
\newblock \DOIprefix\doi{10.1007/978-3-642-71714-7}.
\bibitem[{G{\"o}rges et~al.(2023)G{\"o}rges, Nega, Tancredi and
  Wagner}]{GoergesNegaTancrediWagner2023}
\bibinfo{author}{G{\"o}rges, L.}, \bibinfo{author}{Nega, C.},
  \bibinfo{author}{Tancredi, L.}, \bibinfo{author}{Wagner, F.J.},
  \bibinfo{year}{2023}.
\newblock \bibinfo{title}{On a procedure to derive \(\varepsilon\)-factorised
  differential equations beyond polylogarithms}.
\newblock \bibinfo{journal}{Journal of High Energy Physics}
  \bibinfo{volume}{2023}, \bibinfo{pages}{206}.
\newblock \DOIprefix\doi{10.1007/JHEP07(2023)206},
  \href{http://arxiv.org/abs/2305.14090}{\tt arXiv:2305.14090}.
\bibitem[{Grayson and Stillman(2026)}]{M2}
\bibinfo{author}{Grayson, D.R.}, \bibinfo{author}{Stillman, M.E.},
  \bibinfo{year}{2026}.
\newblock \bibinfo{title}{Macaulay2, a software system for research in
  algebraic geometry}.
\newblock \bibinfo{note}{\url{https://macaulay2.com}}.
\bibitem[{Griffiths and Harris(1978)}]{GriffithsHarris1978}
\bibinfo{author}{Griffiths, P.}, \bibinfo{author}{Harris, J.},
  \bibinfo{year}{1978}.
\newblock \bibinfo{title}{Principles of Algebraic Geometry}.
\newblock \bibinfo{publisher}{Wiley}, \bibinfo{address}{New York}.
\bibitem[{Gubser et~al.(1998)Gubser, Klebanov and Polyakov}]{Gubser:1998bc}
\bibinfo{author}{Gubser, S.S.}, \bibinfo{author}{Klebanov, I.R.},
  \bibinfo{author}{Polyakov, A.M.}, \bibinfo{year}{1998}.
\newblock \bibinfo{title}{Gauge theory correlators from non-critical string
  theory}.
\newblock \bibinfo{journal}{Phys. Lett. B} \bibinfo{volume}{428},
  \bibinfo{pages}{105--114}.
\newblock \DOIprefix\doi{10.1016/S0370-2693(98)00377-3},
  \href{http://arxiv.org/abs/hep-th/9802109}{\tt arXiv:hep-th/9802109}.
\bibitem[{Guevara et~al.(2026)Guevara, Lupsasca, Skinner, Strominger and
  Weil}]{Guevara:2026qzd}
\bibinfo{author}{Guevara, A.}, \bibinfo{author}{Lupsasca, A.},
  \bibinfo{author}{Skinner, D.}, \bibinfo{author}{Strominger, A.},
  \bibinfo{author}{Weil, K.}, \bibinfo{year}{2026}.
\newblock \bibinfo{title}{{Single-minus gluon tree amplitudes are nonzero}}
  \href{http://arxiv.org/abs/2602.12176}{\tt arXiv:2602.12176}.
\bibitem[{G{\"u}ler(1996)}]{guler1996barrier}
\bibinfo{author}{G{\"u}ler, O.}, \bibinfo{year}{1996}.
\newblock \bibinfo{title}{Barrier functions in interior point methods}.
\newblock \bibinfo{journal}{Mathematics of Operations Research}
  \bibinfo{volume}{21}, \bibinfo{pages}{860--885}.
\bibitem[{G{\"u}rdo\u{g}an and Parisi(2023)}]{GurdoganParisi2020}
\bibinfo{author}{G{\"u}rdo\u{g}an, {\"O}.}, \bibinfo{author}{Parisi, M.},
  \bibinfo{year}{2023}.
\newblock \bibinfo{title}{Cluster patterns in landau and leading singularities
  via the amplituhedron}.
\newblock \bibinfo{journal}{Annales de l'Institut Henri Poincar\'e D}
  \bibinfo{volume}{10}, \bibinfo{pages}{135--167}.
\newblock \DOIprefix\doi{10.4171/AIHPD/155},
  \href{http://arxiv.org/abs/2005.07154}{\tt arXiv:2005.07154}.
\bibitem[{Gårding(1951)}]{Garding_1951}
\bibinfo{author}{Gårding, L.}, \bibinfo{year}{1951}.
\newblock \bibinfo{title}{Linear hyperbolic partial differential equations with
  constant coefficients}.
\newblock \bibinfo{journal}{Acta Mathematica} \bibinfo{volume}{85},
  \bibinfo{pages}{1--62}.
\bibitem[{Gårding(1959)}]{Garding_1959}
\bibinfo{author}{Gårding, L.}, \bibinfo{year}{1959}.
\newblock \bibinfo{title}{An inequality for hyperbolic polynomials}.
\newblock \bibinfo{journal}{Journal of Mathematics and Mechanics}
  \bibinfo{volume}{8}, \bibinfo{pages}{957--965}.
\bibitem[{Güler(1997)}]{Guler_Hyperbolic_pol}
\bibinfo{author}{Güler, O.}, \bibinfo{year}{1997}.
\newblock \bibinfo{title}{Hyperbolic polynomials and interior point methods for
  convex programming}.
\newblock \bibinfo{journal}{Mathematical Operations Research}
  \bibinfo{volume}{22}.
\bibitem[{Haag(1958)}]{Haag:1958vt}
\bibinfo{author}{Haag, R.}, \bibinfo{year}{1958}.
\newblock \bibinfo{title}{Quantum field theories with composite particles and
  asymptotic conditions}.
\newblock \bibinfo{journal}{Phys. Rev.} \bibinfo{volume}{112},
  \bibinfo{pages}{669--673}.
\newblock \DOIprefix\doi{10.1103/PhysRev.112.669}.
\bibitem[{Hannesdottir et~al.(2022)Hannesdottir, McLeod, Schwartz and
  Vergu}]{Hannesdottir:2021kpd}
\bibinfo{author}{Hannesdottir, H.S.}, \bibinfo{author}{McLeod, A.J.},
  \bibinfo{author}{Schwartz, M.D.}, \bibinfo{author}{Vergu, C.},
  \bibinfo{year}{2022}.
\newblock \bibinfo{title}{{Implications of the Landau equations for iterated
  integrals}}.
\newblock \bibinfo{journal}{Phys. Rev. D} \bibinfo{volume}{105},
  \bibinfo{pages}{L061701}.
\newblock \DOIprefix\doi{10.1103/PhysRevD.105.L061701},
  \href{http://arxiv.org/abs/2109.09744}{\tt arXiv:2109.09744}.
\bibitem[{Hannesdottir et~al.(2023)Hannesdottir, McLeod, Schwartz and
  Vergu}]{HannesdottirMcLeodSchwartzVergu2023}
\bibinfo{author}{Hannesdottir, H.S.}, \bibinfo{author}{McLeod, A.J.},
  \bibinfo{author}{Schwartz, M.D.}, \bibinfo{author}{Vergu, C.},
  \bibinfo{year}{2023}.
\newblock \bibinfo{title}{Constraints on sequential discontinuities from the
  geometry of on-shell spaces}.
\newblock \bibinfo{journal}{Journal of High Energy Physics}
  \bibinfo{volume}{2023}, \bibinfo{pages}{236}.
\newblock \DOIprefix\doi{10.1007/JHEP07(2023)236},
  \href{http://arxiv.org/abs/2211.07633}{\tt arXiv:2211.07633}.
\bibitem[{He et~al.(2023a)He, Huang and Kuo}]{ABJM_amplituhedron}
\bibinfo{author}{He, S.}, \bibinfo{author}{Huang, Y.t.}, \bibinfo{author}{Kuo,
  C.K.}, \bibinfo{year}{2023}a.
\newblock \bibinfo{title}{{The ABJM Amplituhedron}}.
\newblock \bibinfo{journal}{JHEP} \bibinfo{volume}{09}, \bibinfo{pages}{165}.
\newblock \DOIprefix\doi{10.1007/JHEP09(2023)165},
  \href{http://arxiv.org/abs/2306.00951}{\tt arXiv:2306.00951}.
  \bibinfo{note}{[Erratum: JHEP 04, 064 (2024)]}.
\bibitem[{He et~al.(2026)He, Huang and Kuo}]{He:2025correlahedron}
\bibinfo{author}{He, S.}, \bibinfo{author}{Huang, Y.t.}, \bibinfo{author}{Kuo,
  C.K.}, \bibinfo{year}{2026}.
\newblock \bibinfo{title}{Leading singularities and chambers of correlahedron}.
\newblock \bibinfo{journal}{Journal of High Energy Physics}
  \bibinfo{volume}{2026}, \bibinfo{pages}{071}.
\newblock \DOIprefix\doi{10.1007/JHEP03(2026)071},
  \href{http://arxiv.org/abs/2505.09808}{\tt arXiv:2505.09808}.
\bibitem[{He et~al.(2025)He, Jiang, Liu and Yang}]{He:2024fij}
\bibinfo{author}{He, S.}, \bibinfo{author}{Jiang, X.}, \bibinfo{author}{Liu,
  J.}, \bibinfo{author}{Yang, Q.}, \bibinfo{year}{2025}.
\newblock \bibinfo{title}{Landau-based schubert analysis}.
\newblock \bibinfo{journal}{Journal of High Energy Physics}
  \bibinfo{volume}{2025}, \bibinfo{pages}{053}.
\newblock \DOIprefix\doi{10.1007/JHEP11(2025)053},
  \href{http://arxiv.org/abs/2410.11423}{\tt arXiv:2410.11423}.
\bibitem[{He et~al.(2023b)He, Kuo, Li and Zhang}]{He:2023exb}
\bibinfo{author}{He, S.}, \bibinfo{author}{Kuo, C.K.}, \bibinfo{author}{Li,
  Z.}, \bibinfo{author}{Zhang, Y.Q.}, \bibinfo{year}{2023}b.
\newblock \bibinfo{title}{Emergent unitarity, all-loop cuts and integrations
  from the abjm amplituhedron}.
\newblock \bibinfo{journal}{JHEP} \bibinfo{volume}{08}, \bibinfo{pages}{162}.
\newblock \DOIprefix\doi{10.1007/JHEP08(2023)162},
  \href{http://arxiv.org/abs/2303.03035}{\tt arXiv:2303.03035}.
\bibitem[{He et~al.(2021a)He, Kuo and Zhang}]{heKuoZhang2021twistorString}
\bibinfo{author}{He, S.}, \bibinfo{author}{Kuo, C.K.}, \bibinfo{author}{Zhang,
  Y.Q.}, \bibinfo{year}{2021}a.
\newblock \bibinfo{title}{The momentum amplituhedron of sym and abjm from
  twistor-string maps} \href{http://arxiv.org/abs/2111.02576}{\tt
  arXiv:2111.02576}.
\bibitem[{He et~al.(2021b)He, Li and
  Yang}]{HeLiYang2021NotesClusterFeynmanIntegrals}
\bibinfo{author}{He, S.}, \bibinfo{author}{Li, Z.}, \bibinfo{author}{Yang, Q.},
  \bibinfo{year}{2021}b.
\newblock \bibinfo{title}{Notes on cluster algebras and some all-loop feynman
  integrals}.
\newblock \bibinfo{journal}{Journal of High Energy Physics}
  \bibinfo{volume}{2021}, \bibinfo{pages}{119}.
\newblock \DOIprefix\doi{10.1007/JHEP06(2021)119},
  \href{http://arxiv.org/abs/2103.02796}{\tt arXiv:2103.02796}.
\bibitem[{He et~al.(2021c)He, Li and Yang}]{He:2021non}
\bibinfo{author}{He, S.}, \bibinfo{author}{Li, Z.}, \bibinfo{author}{Yang, Q.},
  \bibinfo{year}{2021}c.
\newblock \bibinfo{title}{{Truncated cluster algebras and Feynman integrals
  with algebraic letters}}.
\newblock \bibinfo{journal}{JHEP} \bibinfo{volume}{12}, \bibinfo{pages}{110}.
\newblock \DOIprefix\doi{10.1007/JHEP12(2021)110},
  \href{http://arxiv.org/abs/2106.09314}{\tt arXiv:2106.09314}.
  \bibinfo{note}{[Erratum: JHEP 05, 075 (2022)]}.
\bibitem[{He et~al.(2022)He, Li and
  Yang}]{HeLiYang2021KinematicsClusterFeynmanIntegrals}
\bibinfo{author}{He, S.}, \bibinfo{author}{Li, Z.}, \bibinfo{author}{Yang, Q.},
  \bibinfo{year}{2022}.
\newblock \bibinfo{title}{Kinematics, cluster algebras and feynman integrals}.
\newblock \bibinfo{journal}{Journal of High Energy Physics}
  \bibinfo{volume}{2022}, \bibinfo{pages}{061}.
\newblock \DOIprefix\doi{10.1007/JHEP06(2022)061},
  \href{http://arxiv.org/abs/2112.11842}{\tt arXiv:2112.11842}.
\bibitem[{He and Zhang(2015)}]{He:2015yua}
\bibinfo{author}{He, S.}, \bibinfo{author}{Zhang, Y.}, \bibinfo{year}{2015}.
\newblock \bibinfo{title}{One-loop scattering equations and amplitudes from
  forward limit}.
\newblock \bibinfo{journal}{Phys. Rev. D} \bibinfo{volume}{92},
  \bibinfo{pages}{105004}.
\newblock \DOIprefix\doi{10.1103/PhysRevD.92.105004},
  \href{http://arxiv.org/abs/1508.06027}{\tt arXiv:1508.06027}.
\bibitem[{Heller and von Manteuffel(2022)}]{HellerVonManteuffel2022}
\bibinfo{author}{Heller, M.}, \bibinfo{author}{von Manteuffel, A.},
  \bibinfo{year}{2022}.
\newblock \bibinfo{title}{{MultivariateApart}: Generalized partial fractions}.
\newblock \bibinfo{journal}{Computer Physics Communications}
  \bibinfo{volume}{271}, \bibinfo{pages}{108174}.
\newblock \DOIprefix\doi{10.1016/j.cpc.2021.108174},
  \href{http://arxiv.org/abs/2101.08283}{\tt arXiv:2101.08283}.
\bibitem[{Heller et~al.(2020)Heller, von Manteuffel and
  Schabinger}]{Heller:2019gkq}
\bibinfo{author}{Heller, M.}, \bibinfo{author}{von Manteuffel, A.},
  \bibinfo{author}{Schabinger, R.M.}, \bibinfo{year}{2020}.
\newblock \bibinfo{title}{{Multiple polylogarithms with algebraic arguments and
  the two-loop EW-QCD Drell-Yan master integrals}}.
\newblock \bibinfo{journal}{Phys. Rev. D} \bibinfo{volume}{102},
  \bibinfo{pages}{016025}.
\newblock \DOIprefix\doi{10.1103/PhysRevD.102.016025},
  \href{http://arxiv.org/abs/1907.00491}{\tt arXiv:1907.00491}.
\bibitem[{Helmer et~al.(2024)Helmer, Papathanasiou and
  Tellander}]{HelmerPapathanasiouTellander2024}
\bibinfo{author}{Helmer, M.}, \bibinfo{author}{Papathanasiou, G.},
  \bibinfo{author}{Tellander, F.}, \bibinfo{year}{2024}.
\newblock \bibinfo{title}{Landau singularities from whitney stratifications}.
\newblock \bibinfo{journal}{Physical Review Letters} \bibinfo{volume}{133},
  \bibinfo{pages}{101601}.
\newblock \DOIprefix\doi{10.1103/PhysRevLett.133.101601},
  \href{http://arxiv.org/abs/2402.14787}{\tt arXiv:2402.14787}.
\bibitem[{Helton and Vinnikov(2007)}]{Helton_Linear}
\bibinfo{author}{Helton, J.W.}, \bibinfo{author}{Vinnikov, V.},
  \bibinfo{year}{2007}.
\newblock \bibinfo{title}{Linear matrix inequality representation of sets}.
\newblock \bibinfo{journal}{Communications on Pure and Applied Mathematics: A
  Journal Issued by the Courant Institute of Mathematical Sciences}
  \bibinfo{volume}{60}, \bibinfo{pages}{654--674}.
\bibitem[{Henke and Papathanasiou(2020)}]{HenkePapathanasiou2020}
\bibinfo{author}{Henke, N.}, \bibinfo{author}{Papathanasiou, G.},
  \bibinfo{year}{2020}.
\newblock \bibinfo{title}{How tropical are seven- and eight-particle
  amplitudes?}
\newblock \bibinfo{journal}{Journal of High Energy Physics}
  \bibinfo{volume}{2020}, \bibinfo{pages}{005}.
\newblock \DOIprefix\doi{10.1007/JHEP08(2020)005},
  \href{http://arxiv.org/abs/1912.08254}{\tt arXiv:1912.08254}.
\bibitem[{Henn et~al.(2025)Henn, Matija{\v{s}}i{\'c}, Miczajka, Peraro, Xu and
  Zhang}]{Henn:2025xrc}
\bibinfo{author}{Henn, J.}, \bibinfo{author}{Matija{\v{s}}i{\'c}, A.},
  \bibinfo{author}{Miczajka, J.}, \bibinfo{author}{Peraro, T.},
  \bibinfo{author}{Xu, Y.}, \bibinfo{author}{Zhang, Y.}, \bibinfo{year}{2025}.
\newblock \bibinfo{title}{Complete function space for planar two-loop
  six-particle scattering amplitudes}.
\newblock \bibinfo{journal}{Physical Review Letters} \bibinfo{volume}{135},
  \bibinfo{pages}{031601}.
\newblock \DOIprefix\doi{10.1103/zhzd-tj9p},
  \href{http://arxiv.org/abs/2501.01847}{\tt arXiv:2501.01847}.
\bibitem[{Henn and Raman(2025)}]{Henn:CM}
\bibinfo{author}{Henn, J.}, \bibinfo{author}{Raman, P.}, \bibinfo{year}{2025}.
\newblock \bibinfo{title}{Positivity properties of scattering amplitudes}.
\newblock \bibinfo{journal}{Journal of High Energy Physics}
  \bibinfo{volume}{04}, \bibinfo{pages}{150}.
\bibitem[{Henn(2013)}]{Henn2013}
\bibinfo{author}{Henn, J.M.}, \bibinfo{year}{2013}.
\newblock \bibinfo{title}{Multiloop integrals in dimensional regularization
  made simple}.
\newblock \bibinfo{journal}{Physical Review Letters} \bibinfo{volume}{110},
  \bibinfo{pages}{251601}.
\newblock \DOIprefix\doi{10.1103/PhysRevLett.110.251601},
  \href{http://arxiv.org/abs/1304.1806}{\tt arXiv:1304.1806}.
\bibitem[{Henn(2015)}]{Henn:2014qga}
\bibinfo{author}{Henn, J.M.}, \bibinfo{year}{2015}.
\newblock \bibinfo{title}{Lectures on differential equations for feynman
  integrals}.
\newblock \bibinfo{journal}{Journal of Physics A: Mathematical and Theoretical}
  \bibinfo{volume}{48}, \bibinfo{pages}{153001}.
\newblock \DOIprefix\doi{10.1088/1751-8113/48/15/153001},
  \href{http://arxiv.org/abs/1412.2296}{\tt arXiv:1412.2296}.
\bibitem[{Henn et~al.(2020)Henn, Korchemsky and Mistlberger}]{Henn:2019swt}
\bibinfo{author}{Henn, J.M.}, \bibinfo{author}{Korchemsky, G.P.},
  \bibinfo{author}{Mistlberger, B.}, \bibinfo{year}{2020}.
\newblock \bibinfo{title}{{The full four-loop cusp anomalous dimension in N=4
  super Yang-Mills and QCD}}.
\newblock \bibinfo{journal}{JHEP} \bibinfo{volume}{04}, \bibinfo{pages}{018}.
\newblock \DOIprefix\doi{10.1007/JHEP04(2020)018},
  \href{http://arxiv.org/abs/1911.10174}{\tt arXiv:1911.10174}.
\bibitem[{Henn and Plefka(2014)}]{Henn:2014yza}
\bibinfo{author}{Henn, J.M.}, \bibinfo{author}{Plefka, J.C.},
  \bibinfo{year}{2014}.
\newblock \bibinfo{title}{Scattering Amplitudes in Gauge Theories}. volume
  \bibinfo{volume}{883} of \textit{\bibinfo{series}{Lecture Notes in Physics}}.
\newblock \bibinfo{publisher}{Springer}.
\newblock \DOIprefix\doi{10.1007/978-3-642-54022-6}.
\bibitem[{Herrmann et~al.(2021)Herrmann, Langer, Trnka and
  Zheng}]{Herrmann:2020qlt}
\bibinfo{author}{Herrmann, E.}, \bibinfo{author}{Langer, C.},
  \bibinfo{author}{Trnka, J.}, \bibinfo{author}{Zheng, M.},
  \bibinfo{year}{2021}.
\newblock \bibinfo{title}{Positive geometry, local triangulations, and the dual
  of the amplituhedron}.
\newblock \bibinfo{journal}{Journal of High Energy Physics}
  \bibinfo{volume}{2021}, \bibinfo{pages}{035}.
\newblock \DOIprefix\doi{10.1007/JHEP01(2021)035},
  \href{http://arxiv.org/abs/2009.05607}{\tt arXiv:2009.05607}.
\bibitem[{Herrmann and Parra-Martinez(2020)}]{Duhr:2019tlz}
\bibinfo{author}{Herrmann, E.}, \bibinfo{author}{Parra-Martinez, J.},
  \bibinfo{year}{2020}.
\newblock \bibinfo{title}{Logarithmic forms and differential equations for
  feynman integrals}.
\newblock \bibinfo{journal}{Journal of High Energy Physics}
  \bibinfo{volume}{02}, \bibinfo{pages}{099}.
\newblock \DOIprefix\doi{10.1007/JHEP02(2020)099},
  \href{http://arxiv.org/abs/1909.04777}{\tt arXiv:1909.04777}.
\bibitem[{Herrmann and Trnka(2022)}]{Herrmann:2022nkh}
\bibinfo{author}{Herrmann, E.}, \bibinfo{author}{Trnka, J.},
  \bibinfo{year}{2022}.
\newblock \bibinfo{title}{The {SAGEX} review on scattering amplitudes, chapter
  7: Positive geometry of scattering amplitudes}.
\newblock \bibinfo{journal}{Journal of Physics A: Mathematical and Theoretical}
  \bibinfo{volume}{55}, \bibinfo{pages}{443008}.
\newblock \DOIprefix\doi{10.1088/1751-8121/ac8eb7},
  \href{http://arxiv.org/abs/2203.13018}{\tt arXiv:2203.13018}.
\bibitem[{Hodges(2013)}]{Hodges:2009hk}
\bibinfo{author}{Hodges, A.}, \bibinfo{year}{2013}.
\newblock \bibinfo{title}{Eliminating spurious poles from gauge-theoretic
  amplitudes}.
\newblock \bibinfo{journal}{Journal of High Energy Physics}
  \bibinfo{volume}{2013}, \bibinfo{pages}{135}.
\newblock \DOIprefix\doi{10.1007/JHEP05(2013)135},
  \href{http://arxiv.org/abs/0905.1473}{\tt arXiv:0905.1473}.
\bibitem[{Hodges(2005)}]{Hodges:2005bf}
\bibinfo{author}{Hodges, A.P.}, \bibinfo{year}{2005}.
\newblock \bibinfo{title}{{Twistor diagram recursion for all gauge-theoretic
  tree amplitudes}} \href{http://arxiv.org/abs/hep-th/0503060}{\tt
  arXiv:hep-th/0503060}.
\bibitem[{Hollering et~al.(2025)Hollering, Mazzucchelli, Parisi and
  Sturmfels}]{HolleringMazzucchelliParisiSturmfels2025Lines}
\bibinfo{author}{Hollering, B.}, \bibinfo{author}{Mazzucchelli, E.},
  \bibinfo{author}{Parisi, M.}, \bibinfo{author}{Sturmfels, B.},
  \bibinfo{year}{2025}.
\newblock \bibinfo{title}{Varieties of lines in 3-space}
  \href{http://arxiv.org/abs/2511.21333}{\tt arXiv:2511.21333}.
\bibitem[{Hollering et~al.(2026a)Hollering, Mazzucchelli, Parisi and
  Sturmfels}]{HolleringMazzucchelliParisiSturmfels2026}
\bibinfo{author}{Hollering, B.}, \bibinfo{author}{Mazzucchelli, E.},
  \bibinfo{author}{Parisi, M.}, \bibinfo{author}{Sturmfels, B.},
  \bibinfo{year}{2026}a.
\newblock \bibinfo{title}{Landau analysis in the grassmannian}
  \href{http://arxiv.org/abs/2603.25454}{\tt arXiv:2603.25454}.
\bibitem[{Hollering et~al.(2026b)Hollering, Mazzucchelli, Parisi and
  Sturmfels}]{HolleringMazzucchelliParisiSturmfels2026Positivity}
\bibinfo{author}{Hollering, B.}, \bibinfo{author}{Mazzucchelli, E.},
  \bibinfo{author}{Parisi, M.}, \bibinfo{author}{Sturmfels, B.},
  \bibinfo{year}{2026}b.
\newblock \bibinfo{title}{Positivity and cluster structures in landau analysis}
  \href{http://arxiv.org/abs/2603.25493}{\tt arXiv:2603.25493}.
\bibitem[{'t~Hooft(1974)}]{tHooft:1973jz}
\bibinfo{author}{'t~Hooft, G.}, \bibinfo{year}{1974}.
\newblock \bibinfo{title}{A planar diagram theory for strong interactions}.
\newblock \bibinfo{journal}{Nucl. Phys. B} \bibinfo{volume}{72},
  \bibinfo{pages}{461--473}.
\bibitem[{H{\"o}rmander(1990)}]{Hoermander1990}
\bibinfo{author}{H{\"o}rmander, L.}, \bibinfo{year}{1990}.
\newblock \bibinfo{title}{The Analysis of Linear Partial Differential Operators
  I}.
\newblock \bibinfo{publisher}{Springer}.
\bibitem[{Huh(2013)}]{Huh2013}
\bibinfo{author}{Huh, J.}, \bibinfo{year}{2013}.
\newblock \bibinfo{title}{The maximum likelihood degree of a very affine
  variety}.
\newblock \bibinfo{journal}{Compositio Mathematica} \bibinfo{volume}{149},
  \bibinfo{pages}{1245--1266}.
\newblock \DOIprefix\doi{10.1112/S0010437X13007057}.
\bibitem[{Hörmander(1983)}]{hormander1}
\bibinfo{author}{Hörmander, L.}, \bibinfo{year}{1983}.
\newblock \bibinfo{title}{The analysis of linear partial differential
  operators. 1. Distribution theory and Fourier analysis}.
\newblock Grundlehren der mathematischen Wissenschaften 256,
  \bibinfo{publisher}{Springer}, \bibinfo{address}{Berlin}.
\bibitem[{Itzykson and Zuber(1980)}]{ItzyksonZuber}
\bibinfo{author}{Itzykson, C.}, \bibinfo{author}{Zuber, J.B.},
  \bibinfo{year}{1980}.
\newblock \bibinfo{title}{Quantum Field Theory}.
\newblock \bibinfo{publisher}{McGraw-Hill}.
\bibitem[{Jagadale and Laddha(2021)}]{Jagadale:2020qfa}
\bibinfo{author}{Jagadale, M.}, \bibinfo{author}{Laddha, A.},
  \bibinfo{year}{2021}.
\newblock \bibinfo{title}{{On positive geometries of quartic interactions: one
  loop integrands from polytopes}}.
\newblock \bibinfo{journal}{JHEP} \bibinfo{volume}{07}, \bibinfo{pages}{136}.
\newblock \DOIprefix\doi{10.1007/JHEP07(2021)136},
  \href{http://arxiv.org/abs/2007.12145}{\tt arXiv:2007.12145}.
\bibitem[{Jagadale and Laddha(2022)}]{Jagadale:2022rbl}
\bibinfo{author}{Jagadale, M.}, \bibinfo{author}{Laddha, A.},
  \bibinfo{year}{2022}.
\newblock \bibinfo{title}{{Towards Positive Geometries of Massive Scalar field
  theories}} \href{http://arxiv.org/abs/2206.07979}{\tt arXiv:2206.07979}.
\bibitem[{Jagadale and Laddha(2023)}]{Jagadale:2023hjr}
\bibinfo{author}{Jagadale, M.}, \bibinfo{author}{Laddha, A.},
  \bibinfo{year}{2023}.
\newblock \bibinfo{title}{{Positive Geometries of S-matrix without Color}}
  \href{http://arxiv.org/abs/2304.04571}{\tt arXiv:2304.04571}.
\bibitem[{Karp(2017)}]{karp2017sign}
\bibinfo{author}{Karp, S.N.}, \bibinfo{year}{2017}.
\newblock \bibinfo{title}{Sign variation, the grassmannian, and total
  positivity}.
\newblock \bibinfo{journal}{Journal of Combinatorial Theory, Series A}
  \bibinfo{volume}{145}, \bibinfo{pages}{308--339}.
\newblock \DOIprefix\doi{10.1016/j.jcta.2016.08.002},
  \href{http://arxiv.org/abs/1503.05622}{\tt arXiv:1503.05622}.
\bibitem[{Karp(2020)}]{karp2020defining}
\bibinfo{author}{Karp, S.N.}, \bibinfo{year}{2020}.
\newblock \bibinfo{title}{Defining amplituhedra and grassmann polytopes}, in:
  \bibinfo{booktitle}{Proceedings of the 28th International Conference on
  Formal Power Series and Algebraic Combinatorics}.
\newblock \DOIprefix\doi{10.46298/dmtcs.6356}.
\bibitem[{Karp(2021)}]{Karp2021Wronskians}
\bibinfo{author}{Karp, S.N.}, \bibinfo{year}{2021}.
\newblock \bibinfo{title}{Wronskians, total positivity, and real schubert
  calculus} \DOIprefix\doi{10.48550/arXiv.2110.02301},
  \href{http://arxiv.org/abs/2110.02301}{\tt arXiv:2110.02301}.
\bibitem[{Karp and Purbhoo(2023)}]{KarpPurbhoo2023UniversalPlucker}
\bibinfo{author}{Karp, S.N.}, \bibinfo{author}{Purbhoo, K.},
  \bibinfo{year}{2023}.
\newblock \bibinfo{title}{Universal {P}l{\"u}cker coordinates for the {W}ronski
  map and positivity in real schubert calculus}
  \DOIprefix\doi{10.48550/arXiv.2309.04645},
  \href{http://arxiv.org/abs/2309.04645}{\tt arXiv:2309.04645}.
\bibitem[{Karp and Williams(2019)}]{KarpWilliams}
\bibinfo{author}{Karp, S.N.}, \bibinfo{author}{Williams, L.K.},
  \bibinfo{year}{2019}.
\newblock \bibinfo{title}{The \(m=1\) amplituhedron and cyclic hyperplane
  arrangements}.
\newblock \bibinfo{journal}{International Mathematics Research Notices}
  \bibinfo{volume}{2019}, \bibinfo{pages}{1401--1462}.
\newblock \DOIprefix\doi{10.1093/imrn/rnx140},
  \href{http://arxiv.org/abs/1608.08288}{\tt arXiv:1608.08288}.
\bibitem[{Kleiss and Kuijf(1989)}]{Kleiss:1988ne}
\bibinfo{author}{Kleiss, R.}, \bibinfo{author}{Kuijf, H.},
  \bibinfo{year}{1989}.
\newblock \bibinfo{title}{Multi-gluon cross sections and five jet production at
  hadron colliders}.
\newblock \bibinfo{journal}{Nuclear Physics B} \bibinfo{volume}{312},
  \bibinfo{pages}{616--644}.
\bibitem[{Knutson et~al.(2013)Knutson, Lam and
  Speyer}]{knutsonLamSpeyer2013positroid}
\bibinfo{author}{Knutson, A.}, \bibinfo{author}{Lam, T.},
  \bibinfo{author}{Speyer, D.E.}, \bibinfo{year}{2013}.
\newblock \bibinfo{title}{Positroid varieties: juggling and geometry}.
\newblock \bibinfo{journal}{Compositio Mathematica} \bibinfo{volume}{149},
  \bibinfo{pages}{1710--1752}.
\bibitem[{Koba and Nielsen(1969)}]{KobaNielsen1969}
\bibinfo{author}{Koba, Z.}, \bibinfo{author}{Nielsen, H.B.},
  \bibinfo{year}{1969}.
\newblock \bibinfo{title}{Manifestly crossing-invariant parametrization of
  {$N$}-meson amplitude}.
\newblock \bibinfo{journal}{Nucl. Phys. B} \bibinfo{volume}{10},
  \bibinfo{pages}{633--655}.
\bibitem[{Koefler and Sinn(2025)}]{koefler2025taking}
\bibinfo{author}{Koefler, J.}, \bibinfo{author}{Sinn, R.},
  \bibinfo{year}{2025}.
\newblock \bibinfo{title}{Taking the amplituhedron to the limit}
  \href{http://arxiv.org/abs/2501.08221}{\tt arXiv:2501.08221}.
\bibitem[{Kohn et~al.(2025)Kohn, Piene, Ranestad, Rydell, Shapiro, Sinn, Sorea
  and Telen}]{Polypols}
\bibinfo{author}{Kohn, K.}, \bibinfo{author}{Piene, R.},
  \bibinfo{author}{Ranestad, K.}, \bibinfo{author}{Rydell, F.},
  \bibinfo{author}{Shapiro, B.}, \bibinfo{author}{Sinn, R.},
  \bibinfo{author}{Sorea, M.{\c S}.}, \bibinfo{author}{Telen, S.},
  \bibinfo{year}{2025}.
\newblock \bibinfo{title}{Adjoints and canonical forms of polypols}.
\newblock \bibinfo{journal}{Documenta Mathematica} \bibinfo{volume}{30},
  \bibinfo{pages}{275--346}.
\newblock \DOIprefix\doi{10.4171/DM/991},
  \href{http://arxiv.org/abs/2108.11747}{\tt arXiv:2108.11747}.
\bibitem[{Kohn and Ranestad(2020)}]{kohn2020projective}
\bibinfo{author}{Kohn, K.}, \bibinfo{author}{Ranestad, K.},
  \bibinfo{year}{2020}.
\newblock \bibinfo{title}{Projective geometry of wachspress coordinates}.
\newblock \bibinfo{journal}{Foundations of Computational Mathematics}
  \bibinfo{volume}{20}, \bibinfo{pages}{1135--1173}.
\bibitem[{Kojima and Langer(2020)}]{Lukowski:2020bya}
\bibinfo{author}{Kojima, R.}, \bibinfo{author}{Langer, C.},
  \bibinfo{year}{2020}.
\newblock \bibinfo{title}{Sign flip triangulations of the amplituhedron}.
\newblock \bibinfo{journal}{Journal of High Energy Physics}
  \bibinfo{volume}{2020}, \bibinfo{pages}{121}.
\newblock \DOIprefix\doi{10.1007/JHEP05(2020)121},
  \href{http://arxiv.org/abs/2001.06473}{\tt arXiv:2001.06473}.
\bibitem[{de~Korte and Yu(2026)}]{deKorteYu2026}
\bibinfo{author}{de~Korte, C.}, \bibinfo{author}{Yu, T.}, \bibinfo{year}{2026}.
\newblock \bibinfo{title}{Partial fraction decompositions on hyperplane
  arrangements} \href{http://arxiv.org/abs/2602.06531}{\tt arXiv:2602.06531}.
  \bibinfo{note}{preprint}.
\bibitem[{Kotikov(1991)}]{Kotikov1991}
\bibinfo{author}{Kotikov, A.V.}, \bibinfo{year}{1991}.
\newblock \bibinfo{title}{Differential equations method: New technique for
  massive feynman diagrams calculation}.
\newblock \bibinfo{journal}{Physics Letters B} \bibinfo{volume}{254},
  \bibinfo{pages}{158--164}.
\newblock \DOIprefix\doi{10.1016/0370-2693(91)90413-K}.
\bibitem[{Kotikov and Lipatov(2003)}]{Kotikov:2002ab}
\bibinfo{author}{Kotikov, A.V.}, \bibinfo{author}{Lipatov, L.N.},
  \bibinfo{year}{2003}.
\newblock \bibinfo{title}{{DGLAP} and {BFKL} evolution equations in the
  \(\mathcal{N}=4\) supersymmetric gauge theory}.
\newblock \bibinfo{journal}{Nuclear Physics B} \bibinfo{volume}{661},
  \bibinfo{pages}{19--61}.
\newblock \DOIprefix\doi{10.1016/S0550-3213(03)00264-5},
  \href{http://arxiv.org/abs/hep-ph/0208220}{\tt arXiv:hep-ph/0208220}.
\bibitem[{Kozhasov et~al.(2019)Kozhasov, Michałek and Sturmfels}]{Pos_Certif}
\bibinfo{author}{Kozhasov, K.}, \bibinfo{author}{Michałek, M.},
  \bibinfo{author}{Sturmfels, B.}, \bibinfo{year}{2019}.
\newblock \bibinfo{title}{Positivity certificates via integral
  representations}.
\newblock \bibinfo{journal}{Facets of Algebraic Geometry} \bibinfo{volume}{2},
  \bibinfo{pages}{84--114}.
\bibitem[{Kristensson et~al.(2021)Kristensson, Wilhelm and
  Zhang}]{Kristensson:2021ani}
\bibinfo{author}{Kristensson, A.}, \bibinfo{author}{Wilhelm, M.},
  \bibinfo{author}{Zhang, C.}, \bibinfo{year}{2021}.
\newblock \bibinfo{title}{The elliptic double box and symbology beyond
  polylogarithms}.
\newblock \bibinfo{journal}{Physical Review Letters} \bibinfo{volume}{127},
  \bibinfo{pages}{251603}.
\newblock \DOIprefix\doi{10.1103/PhysRevLett.127.251603},
  \href{http://arxiv.org/abs/2106.14902}{\tt arXiv:2106.14902}.
\bibitem[{Kummer(2015)}]{Kummer_2015}
\bibinfo{author}{Kummer, M.}, \bibinfo{year}{2015}.
\newblock \bibinfo{title}{A note on the hyperbolicity cone of the specialized
  vámos polynomial}.
\newblock \bibinfo{journal}{Acta Applicandae Mathematicae}
  \bibinfo{volume}{144}, \bibinfo{pages}{11–15}.
\bibitem[{Kummer and Sinn(2022)}]{Kummer_2022}
\bibinfo{author}{Kummer, M.}, \bibinfo{author}{Sinn, R.}, \bibinfo{year}{2022}.
\newblock \bibinfo{title}{Hyperbolic secant varieties of m-curves}.
\newblock \bibinfo{journal}{Journal für die reine und angewandte Mathematik
  (Crelles Journal)} .
\bibitem[{Lam(2014)}]{TNN_grassmannian}
\bibinfo{author}{Lam, T.}, \bibinfo{year}{2014}.
\newblock \bibinfo{title}{Totally nonnegative grassmannian and grassmann
  polytopes}.
\newblock \bibinfo{journal}{Current Developments in Mathematics}
  \bibinfo{volume}{2014}, \bibinfo{pages}{51--152}.
\newblock \DOIprefix\doi{10.4310/CDM.2014.v2014.n1.a2},
  \href{http://arxiv.org/abs/1506.00603}{\tt arXiv:1506.00603}.
\bibitem[{Lam(2022)}]{Lam:_PG_notes}
\bibinfo{author}{Lam, T.}, \bibinfo{year}{2022}.
\newblock \bibinfo{title}{{An invitation to positive geometries}}
  \href{http://arxiv.org/abs/2208.05407}{\tt arXiv:2208.05407}.
\bibitem[{Lam(2024)}]{lam2024face}
\bibinfo{author}{Lam, T.}, \bibinfo{year}{2024}.
\newblock \bibinfo{title}{On the face stratification of the \(m=2\)
  amplituhedron} \href{http://arxiv.org/abs/2403.06948}{\tt arXiv:2403.06948}.
\bibitem[{Lam(2025)}]{Lam:ModuliSpacesPG}
\bibinfo{author}{Lam, T.}, \bibinfo{year}{2025}.
\newblock \bibinfo{title}{Moduli spaces in positive geometry}.
\newblock \bibinfo{journal}{Le Matematiche}
  \href{http://arxiv.org/abs/2405.17332}{\tt arXiv:2405.17332}.
  \bibinfo{note}{to appear}.
\bibitem[{Landau(1959)}]{Landau1959}
\bibinfo{author}{Landau, L.D.}, \bibinfo{year}{1959}.
\newblock \bibinfo{title}{On analytic properties of vertex parts in quantum
  field theory}.
\newblock \bibinfo{journal}{Nuclear Physics} \bibinfo{volume}{13},
  \bibinfo{pages}{181--192}.
\newblock \DOIprefix\doi{10.1016/B978-0-08-010586-4.50103-6}.
\bibitem[{Lee and Pomeransky(2013)}]{LeePomeransky2013}
\bibinfo{author}{Lee, R.N.}, \bibinfo{author}{Pomeransky, A.A.},
  \bibinfo{year}{2013}.
\newblock \bibinfo{title}{Critical points and number of master integrals}.
\newblock \bibinfo{journal}{Journal of High Energy Physics}
  \bibinfo{volume}{2013}, \bibinfo{pages}{165}.
\newblock \DOIprefix\doi{10.1007/JHEP11(2013)165},
  \href{http://arxiv.org/abs/1308.6676}{\tt arXiv:1308.6676}.
\bibitem[{Lehmann et~al.(1955)Lehmann, Symanzik and Zimmermann}]{LSZ:1955}
\bibinfo{author}{Lehmann, H.}, \bibinfo{author}{Symanzik, K.},
  \bibinfo{author}{Zimmermann, W.}, \bibinfo{year}{1955}.
\newblock \bibinfo{title}{Zur formulierung quantisierter feldtheorien}.
\newblock \bibinfo{journal}{Nuovo Cim.} \bibinfo{volume}{1},
  \bibinfo{pages}{205--225}.
\newblock \DOIprefix\doi{10.1007/BF02731765}.
\bibitem[{Leinartas(1978)}]{Leinartas1978}
\bibinfo{author}{Leinartas, E.K.}, \bibinfo{year}{1978}.
\newblock \bibinfo{title}{Multiple laurent series and fundamental systems of
  rational functions}.
\newblock \bibinfo{journal}{Siberian Mathematical Journal}
  \bibinfo{volume}{19}, \bibinfo{pages}{517--523}.
\newblock \DOIprefix\doi{10.1007/BF00967722}.
\bibitem[{Lemaire et~al.(2005)Lemaire, Maza and Xie}]{lemaire2005regularchains}
\bibinfo{author}{Lemaire, F.}, \bibinfo{author}{Maza, M.M.},
  \bibinfo{author}{Xie, Y.}, \bibinfo{year}{2005}.
\newblock \bibinfo{title}{The \texttt{RegularChains} library in
  \texttt{MAPLE}}.
\newblock \bibinfo{journal}{ACM SIGSAM Bulletin} \bibinfo{volume}{39},
  \bibinfo{pages}{96--97}.
\bibitem[{Lemmon and Trnka(2025)}]{Lemmon:2025dhq}
\bibinfo{author}{Lemmon, J.}, \bibinfo{author}{Trnka, J.},
  \bibinfo{year}{2025}.
\newblock \bibinfo{title}{{Tree-Level Gravity Amplitudes at Infinity}}
  \href{http://arxiv.org/abs/2512.11787}{\tt arXiv:2512.11787}.
\bibitem[{Lippstreu et~al.(2024a)Lippstreu, Spradlin, Srikant and
  Volovich}]{LippstreuSpradlinSrikantVolovich2024ZigguratII}
\bibinfo{author}{Lippstreu, L.}, \bibinfo{author}{Spradlin, M.},
  \bibinfo{author}{Srikant, A.Y.}, \bibinfo{author}{Volovich, A.},
  \bibinfo{year}{2024}a.
\newblock \bibinfo{title}{Landau singularities of the 7-point ziggurat. part
  ii}.
\newblock \bibinfo{journal}{Journal of High Energy Physics}
  \bibinfo{volume}{2024}, \bibinfo{pages}{069}.
\newblock \DOIprefix\doi{10.1007/JHEP01(2024)069},
  \href{http://arxiv.org/abs/2305.17069}{\tt arXiv:2305.17069}.
\bibitem[{Lippstreu et~al.(2024b)Lippstreu, Spradlin and
  Volovich}]{LippstreuSpradlinVolovich2024ZigguratI}
\bibinfo{author}{Lippstreu, L.}, \bibinfo{author}{Spradlin, M.},
  \bibinfo{author}{Volovich, A.}, \bibinfo{year}{2024}b.
\newblock \bibinfo{title}{Landau singularities of the 7-point ziggurat. part
  i}.
\newblock \bibinfo{journal}{Journal of High Energy Physics}
  \bibinfo{volume}{2024}, \bibinfo{pages}{024}.
\newblock \DOIprefix\doi{10.1007/JHEP07(2024)024},
  \href{http://arxiv.org/abs/2211.16425}{\tt arXiv:2211.16425}.
\bibitem[{Lisitsyn et~al.(2025)Lisitsyn, Oktem, Sherman-Bennett and
  Trnka}]{Lisitsyn:2025prd}
\bibinfo{author}{Lisitsyn, A.}, \bibinfo{author}{Oktem, U.},
  \bibinfo{author}{Sherman-Bennett, M.}, \bibinfo{author}{Trnka, J.},
  \bibinfo{year}{2025}.
\newblock \bibinfo{title}{{Grassmannian Geometries for Non-Planar On-Shell
  Diagrams}} \href{http://arxiv.org/abs/2512.25005}{\tt arXiv:2512.25005}.
\bibitem[{{\L}ukowski(2022)}]{lukowski2019boundaries}
\bibinfo{author}{{\L}ukowski, T.}, \bibinfo{year}{2022}.
\newblock \bibinfo{title}{On the boundaries of the \(m=2\) amplituhedron}.
\newblock \bibinfo{journal}{Annales de l'Institut Henri Poincar{\'e} D}
  \bibinfo{volume}{9}, \bibinfo{pages}{525--541}.
\newblock \DOIprefix\doi{10.4171/AIHPD/124},
  \href{http://arxiv.org/abs/1908.00386}{\tt arXiv:1908.00386}.
\bibitem[{{\L}ukowski and Moerman(2021)}]{lukowskiMoerman2021boundaries}
\bibinfo{author}{{\L}ukowski, T.}, \bibinfo{author}{Moerman, R.},
  \bibinfo{year}{2021}.
\newblock \bibinfo{title}{Boundaries of the amplituhedron with mathematica}.
\newblock \bibinfo{journal}{Computer Physics Communications}
  \bibinfo{volume}{259}, \bibinfo{pages}{107653}.
\newblock \DOIprefix\doi{10.1016/j.cpc.2020.107653},
  \href{http://arxiv.org/abs/2002.07146}{\tt arXiv:2002.07146}.
\bibitem[{Lukowski et~al.(2022)Lukowski, Moerman and
  Stalknecht}]{Lukowski:2022fwz}
\bibinfo{author}{Lukowski, T.}, \bibinfo{author}{Moerman, R.},
  \bibinfo{author}{Stalknecht, J.}, \bibinfo{year}{2022}.
\newblock \bibinfo{title}{{Pushforwards via scattering equations with
  applications to positive geometries}}.
\newblock \bibinfo{journal}{JHEP} \bibinfo{volume}{10}, \bibinfo{pages}{003}.
\newblock \DOIprefix\doi{10.1007/JHEP10(2022)003},
  \href{http://arxiv.org/abs/2206.14196}{\tt arXiv:2206.14196}.
\bibitem[{{\L}ukowski et~al.(2019){\L}ukowski, Parisi, Spradlin and
  Volovich}]{lukowski2019cluster}
\bibinfo{author}{{\L}ukowski, T.}, \bibinfo{author}{Parisi, M.},
  \bibinfo{author}{Spradlin, M.}, \bibinfo{author}{Volovich, A.},
  \bibinfo{year}{2019}.
\newblock \bibinfo{title}{Cluster adjacency for m= 2 yangian invariants}.
\newblock \bibinfo{journal}{Journal of High Energy Physics}
  \bibinfo{volume}{2019}, \bibinfo{pages}{158}.
\bibitem[{Maazouz et~al.(2025)Maazouz, Pfister and Sturmfels}]{Maazouz:2024qmm}
\bibinfo{author}{Maazouz, Y.E.}, \bibinfo{author}{Pfister, A.},
  \bibinfo{author}{Sturmfels, B.}, \bibinfo{year}{2025}.
\newblock \bibinfo{title}{{Spinor-Helicity Varieties}}.
\newblock \bibinfo{journal}{SIAM J. Appl. Alg. Geom.} \bibinfo{volume}{9},
  \bibinfo{pages}{848--876}.
\newblock \DOIprefix\doi{10.1137/24m1671840},
  \href{http://arxiv.org/abs/2406.17331}{\tt arXiv:2406.17331}.
\bibitem[{Mago et~al.(2019)Mago, Schreiber, Spradlin and
  Volovich}]{MagoSchreiberSpradlinVolovich2019}
\bibinfo{author}{Mago, J.}, \bibinfo{author}{Schreiber, A.},
  \bibinfo{author}{Spradlin, M.}, \bibinfo{author}{Volovich, A.},
  \bibinfo{year}{2019}.
\newblock \bibinfo{title}{Yangian invariants and cluster adjacency in
  \(\mathcal{N}=4\) yang--mills}.
\newblock \bibinfo{journal}{Journal of High Energy Physics}
  \bibinfo{volume}{2019}, \bibinfo{pages}{099}.
\newblock \DOIprefix\doi{10.1007/JHEP10(2019)099},
  \href{http://arxiv.org/abs/1906.10682}{\tt arXiv:1906.10682}.
\bibitem[{Maldacena(1998)}]{Maldacena:1997re}
\bibinfo{author}{Maldacena, J.M.}, \bibinfo{year}{1998}.
\newblock \bibinfo{title}{The large-$n$ limit of superconformal field theories
  and supergravity}.
\newblock \bibinfo{journal}{Adv. Theor. Math. Phys.} \bibinfo{volume}{2},
  \bibinfo{pages}{231--252}.
\newblock \DOIprefix\doi{10.4310/ATMP.1998.v2.n2.a1},
  \href{http://arxiv.org/abs/hep-th/9711200}{\tt arXiv:hep-th/9711200}.
\bibitem[{Mali{\'c} and Streinu(2023)}]{MS}
\bibinfo{author}{Mali{\'c}, G.}, \bibinfo{author}{Streinu, I.},
  \bibinfo{year}{2023}.
\newblock \bibinfo{title}{Computing circuit polynomials in the algebraic
  rigidity matroid}.
\newblock \bibinfo{journal}{SIAM Journal on Applied Algebra and Geometry}
  \bibinfo{volume}{7}, \bibinfo{pages}{345--385}.
\newblock \DOIprefix\doi{10.1137/21M1437986},
  \href{http://arxiv.org/abs/2304.12435}{\tt arXiv:2304.12435}.
\bibitem[{Mandelshtam et~al.(2023)Mandelshtam, Pavlov and
  Pratt}]{mandelshtam2023combinatorics}
\bibinfo{author}{Mandelshtam, Y.}, \bibinfo{author}{Pavlov, D.},
  \bibinfo{author}{Pratt, E.}, \bibinfo{year}{2023}.
\newblock \bibinfo{title}{Combinatorics of {$m=1$} grasstopes}
  \href{http://arxiv.org/abs/2307.09603}{\tt arXiv:2307.09603}.
\bibitem[{Mandelstam(1983)}]{Mandelstam:1982cb}
\bibinfo{author}{Mandelstam, S.}, \bibinfo{year}{1983}.
\newblock \bibinfo{title}{Light cone superspace and the ultraviolet finiteness
  of the n=4 model}.
\newblock \bibinfo{journal}{Nucl. Phys. B} \bibinfo{volume}{213},
  \bibinfo{pages}{149--168}.
\bibitem[{Mangano and Parke(1991)}]{Mangano:1990by}
\bibinfo{author}{Mangano, M.L.}, \bibinfo{author}{Parke, S.J.},
  \bibinfo{year}{1991}.
\newblock \bibinfo{title}{Multiparton amplitudes in gauge theories}.
\newblock \bibinfo{journal}{Physics Reports} \bibinfo{volume}{200},
  \bibinfo{pages}{301--367}.
\bibitem[{Mangano et~al.(1988)Mangano, Parke and Xu}]{Mangano:1987xk}
\bibinfo{author}{Mangano, M.L.}, \bibinfo{author}{Parke, S.J.},
  \bibinfo{author}{Xu, Z.}, \bibinfo{year}{1988}.
\newblock \bibinfo{title}{Duality and multi-gluon scattering}.
\newblock \bibinfo{journal}{Nucl. Phys. B} \bibinfo{volume}{298},
  \bibinfo{pages}{653--672}.
\bibitem[{Marsh and Scott(2016)}]{marsh2016twists}
\bibinfo{author}{Marsh, B.R.}, \bibinfo{author}{Scott, J.S.},
  \bibinfo{year}{2016}.
\newblock \bibinfo{title}{Twists of pl{\"u}cker coordinates as dimer partition
  functions}.
\newblock \bibinfo{journal}{Communications in Mathematical Physics}
  \bibinfo{volume}{341}, \bibinfo{pages}{821--884}.
\bibitem[{Martin(2003)}]{Martin1}
\bibinfo{author}{Martin, J.L.}, \bibinfo{year}{2003}.
\newblock \bibinfo{title}{Geometry of graph varieties}.
\newblock \bibinfo{journal}{Transactions of the American Mathematical Society}
  \bibinfo{volume}{355}, \bibinfo{pages}{4151--4169}.
\newblock \DOIprefix\doi{10.1090/S0002-9947-03-03321-X},
  \href{http://arxiv.org/abs/math/0302089}{\tt arXiv:math/0302089}.
\bibitem[{Mason and Skinner(2010)}]{Mason:2010yk}
\bibinfo{author}{Mason, L.}, \bibinfo{author}{Skinner, D.},
  \bibinfo{year}{2010}.
\newblock \bibinfo{title}{The complete planar s-matrix of $\mathcal{N}=4$ sym
  as a wilson loop in twistor space}.
\newblock \bibinfo{journal}{JHEP} \bibinfo{volume}{12}, \bibinfo{pages}{018}.
\newblock \DOIprefix\doi{10.1007/JHEP12(2010)018},
  \href{http://arxiv.org/abs/1009.2225}{\tt arXiv:1009.2225}.
\bibitem[{Mason and Skinner(2009)}]{Mason:2009qx}
\bibinfo{author}{Mason, L.J.}, \bibinfo{author}{Skinner, D.},
  \bibinfo{year}{2009}.
\newblock \bibinfo{title}{{Dual Superconformal Invariance, Momentum Twistors
  and Grassmannians}}.
\newblock \bibinfo{journal}{JHEP} \bibinfo{volume}{11}, \bibinfo{pages}{045}.
\newblock \DOIprefix\doi{10.1088/1126-6708/2009/11/045},
  \href{http://arxiv.org/abs/0909.0250}{\tt arXiv:0909.0250}.
\bibitem[{Mastrolia(2009)}]{Mastrolia:2009dr}
\bibinfo{author}{Mastrolia, P.}, \bibinfo{year}{2009}.
\newblock \bibinfo{title}{Double-cut of scattering amplitudes and stokes'
  theorem}.
\newblock \bibinfo{journal}{Physics Letters B} \bibinfo{volume}{678},
  \bibinfo{pages}{246--249}.
\newblock \DOIprefix\doi{10.1016/j.physletb.2009.06.033},
  \href{http://arxiv.org/abs/0905.2909}{\tt arXiv:0905.2909}.
\bibitem[{Mastrolia and Mizera(2019)}]{MastroliaMizera2019}
\bibinfo{author}{Mastrolia, P.}, \bibinfo{author}{Mizera, S.},
  \bibinfo{year}{2019}.
\newblock \bibinfo{title}{Feynman integrals and intersection theory}.
\newblock \bibinfo{journal}{Journal of High Energy Physics}
  \bibinfo{volume}{2019}, \bibinfo{pages}{139}.
\newblock \DOIprefix\doi{10.1007/JHEP02(2019)139},
  \href{http://arxiv.org/abs/1810.03818}{\tt arXiv:1810.03818}.
\bibitem[{Matija{\v{s}}i{\'c}(2024)}]{MatijasicThesis}
\bibinfo{author}{Matija{\v{s}}i{\'c}, A.}, \bibinfo{year}{2024}.
\newblock \bibinfo{title}{Singularity Structure of Feynman Integrals with
  Applications to Six-Particle Scattering Processes}.
\newblock Ph.D. thesis. Ludwig-Maximilians-Universit{\"a}t M{\"u}nchen.
\newblock \DOIprefix\doi{10.5282/edoc.34154}.
\bibitem[{Matija{\v{s}}i{\'c} and Miczajka(2025)}]{repoEffortless}
\bibinfo{author}{Matija{\v{s}}i{\'c}, A.}, \bibinfo{author}{Miczajka, J.},
  \bibinfo{year}{2025}.
\newblock \bibinfo{title}{{Effortless}: Efficient generation of odd letters
  with multiple roots as leading singularities}.
\newblock
  \bibinfo{howpublished}{\url{https://github.com/antonela-matijasic/Effortless}}.
\newblock \bibinfo{note}{GitHub repository}.
\bibitem[{Matsubara-Heo et~al.(2023)Matsubara-Heo, Mizera and
  Telen}]{Euler_Integrals}
\bibinfo{author}{Matsubara-Heo, S.J.}, \bibinfo{author}{Mizera, S.},
  \bibinfo{author}{Telen, S.}, \bibinfo{year}{2023}.
\newblock \bibinfo{title}{Four lectures on euler integrals}.
\newblock \bibinfo{journal}{SciPost Physics Lecture Notes} ,
  \bibinfo{pages}{075}.
\bibitem[{Matsubara-Heo and Telen(2025)}]{MatsubaraHeoTelen2025}
\bibinfo{author}{Matsubara-Heo, S.J.}, \bibinfo{author}{Telen, S.},
  \bibinfo{year}{2025}.
\newblock \bibinfo{title}{Twisted cohomology and likelihood ideals}.
\newblock \bibinfo{journal}{Advances in Applied Mathematics}
  \bibinfo{volume}{165}, \bibinfo{pages}{102832}.
\newblock \DOIprefix\doi{10.1016/j.aam.2024.102832}.
\bibitem[{Matveiakin and Rudenko(2022)}]{MR}
\bibinfo{author}{Matveiakin, A.}, \bibinfo{author}{Rudenko, D.},
  \bibinfo{year}{2022}.
\newblock \bibinfo{title}{Cluster polylogarithms i: Quadrangular
  polylogarithms} \href{http://arxiv.org/abs/2208.01564}{\tt arXiv:2208.01564}.
\bibitem[{Mazloumi and Xu(2025)}]{MazloumiXu2025ClusterCosmologicalCorrelators}
\bibinfo{author}{Mazloumi, P.}, \bibinfo{author}{Xu, X.}, \bibinfo{year}{2025}.
\newblock \bibinfo{title}{Cluster algebras for cosmological correlators}.
\newblock \href{http://arxiv.org/abs/2512.14854}{\tt arXiv:2512.14854}.
\bibitem[{Mazzucchelli and Henn(2026)}]{MazzucchelliHennAomotoForms}
\bibinfo{author}{Mazzucchelli, E.}, \bibinfo{author}{Henn, J.M.},
  \bibinfo{year}{2026}.
\newblock \bibinfo{title}{Positivity of aomoto forms}.
\newblock \bibinfo{note}{Work in progress}.
\bibitem[{Mazzucchelli et~al.(2025)Mazzucchelli, Pavlov and
  Wang}]{MazzucchelliPavlovWang2025}
\bibinfo{author}{Mazzucchelli, E.}, \bibinfo{author}{Pavlov, D.},
  \bibinfo{author}{Wang, K.}, \bibinfo{year}{2025}.
\newblock \bibinfo{title}{Hyperplane arrangements in the grassmannian}.
\newblock \bibinfo{journal}{Le Matematiche} \bibinfo{volume}{80},
  \bibinfo{pages}{387--408}.
\newblock \href{http://arxiv.org/abs/2409.04288}{\tt arXiv:2409.04288}.
\bibitem[{Mazzucchelli and Pratt(2025)}]{mazzucchelli2025exterior}
\bibinfo{author}{Mazzucchelli, E.}, \bibinfo{author}{Pratt, E.},
  \bibinfo{year}{2025}.
\newblock \bibinfo{title}{{Exterior Cyclic Polytopes and Convexity of
  Amplituhedra}} \href{http://arxiv.org/abs/2507.17620}{\tt arXiv:2507.17620}.
\bibitem[{Mazzucchelli and Raman(2025)}]{Mazzucchelli:DV}
\bibinfo{author}{Mazzucchelli, E.}, \bibinfo{author}{Raman, P.},
  \bibinfo{year}{2025}.
\newblock \bibinfo{title}{{Canonical Forms as Dual Volumes}}
  \href{http://arxiv.org/abs/2509.02239}{\tt arXiv:2509.02239}.
\bibitem[{Michałek et~al.(2016)Michałek, Sturmfels, Uhler and
  Zwiernik}]{Exponential_varieties}
\bibinfo{author}{Michałek, M.}, \bibinfo{author}{Sturmfels, B.},
  \bibinfo{author}{Uhler, C.}, \bibinfo{author}{Zwiernik, P.},
  \bibinfo{year}{2016}.
\newblock \bibinfo{title}{Exponential varieties}.
\newblock \bibinfo{journal}{Proceedings of the London Mathematical Society}
  \bibinfo{volume}{112}.
\bibitem[{Minahan and Zarembo(2003)}]{Minahan:2002ve}
\bibinfo{author}{Minahan, J.A.}, \bibinfo{author}{Zarembo, K.},
  \bibinfo{year}{2003}.
\newblock \bibinfo{title}{The bethe ansatz for $\mathcal{N}=4$ super
  yang--mills}.
\newblock \bibinfo{journal}{JHEP} \bibinfo{volume}{03}, \bibinfo{pages}{013}.
\newblock \DOIprefix\doi{10.1088/1126-6708/2003/03/013},
  \href{http://arxiv.org/abs/hep-th/0212208}{\tt arXiv:hep-th/0212208}.
\bibitem[{Mizera(2018)}]{Mizera:2017cqs}
\bibinfo{author}{Mizera, S.}, \bibinfo{year}{2018}.
\newblock \bibinfo{title}{Scattering amplitudes from intersection theory}.
\newblock \bibinfo{journal}{Physical Review Letters} \bibinfo{volume}{120},
  \bibinfo{pages}{141602}.
\newblock \DOIprefix\doi{10.1103/PhysRevLett.120.141602},
  \href{http://arxiv.org/abs/1711.00469}{\tt arXiv:1711.00469}.
\bibitem[{Mizera(2019)}]{Mizera:2019gea}
\bibinfo{author}{Mizera, S.}, \bibinfo{year}{2019}.
\newblock \bibinfo{title}{Aspects of scattering amplitudes and moduli space
  localization}.
\newblock \bibinfo{journal}{Journal of High Energy Physics}
  \bibinfo{volume}{2019}, \bibinfo{pages}{097}.
\newblock \DOIprefix\doi{10.1007/JHEP08(2019)097},
  \href{http://arxiv.org/abs/1906.02099}{\tt arXiv:1906.02099}.
\bibitem[{Mizera and Telen(2022)}]{Mizera:2021icv}
\bibinfo{author}{Mizera, S.}, \bibinfo{author}{Telen, S.},
  \bibinfo{year}{2022}.
\newblock \bibinfo{title}{{Landau discriminants}}.
\newblock \bibinfo{journal}{JHEP} \bibinfo{volume}{08}, \bibinfo{pages}{200}.
\newblock \DOIprefix\doi{10.1007/JHEP08(2022)200},
  \href{http://arxiv.org/abs/2109.08036}{\tt arXiv:2109.08036}.
\bibitem[{Moerman(2023)}]{moerman2023positive}
\bibinfo{author}{Moerman, R.}, \bibinfo{year}{2023}.
\newblock \bibinfo{title}{Positive Geometries for Scattering Amplitudes in
  Momentum Space}.
\newblock Ph.D. thesis. University of Hertfordshire.
\newblock \href{http://arxiv.org/abs/2306.05287}{\tt arXiv:2306.05287}.
\bibitem[{Mohammadi et~al.(2020)Mohammadi, Monin and
  Parisi}]{mohammadiMoninParisi2020triangulations}
\bibinfo{author}{Mohammadi, F.}, \bibinfo{author}{Monin, L.},
  \bibinfo{author}{Parisi, M.}, \bibinfo{year}{2020}.
\newblock \bibinfo{title}{Triangulations and canonical forms of amplituhedra: A
  fiber-based approach beyond polytopes}
  \href{http://arxiv.org/abs/2010.07254}{\tt arXiv:2010.07254}.
\bibitem[{Mukhin et~al.(2009)Mukhin, Tarasov and
  Varchenko}]{MukhinTarasovVarchenko2009}
\bibinfo{author}{Mukhin, E.}, \bibinfo{author}{Tarasov, V.},
  \bibinfo{author}{Varchenko, A.}, \bibinfo{year}{2009}.
\newblock \bibinfo{title}{The {B}. and {M}. {S}hapiro conjecture in real
  algebraic geometry and the {B}ethe ansatz}.
\newblock \bibinfo{journal}{Annals of Mathematics} \bibinfo{volume}{170},
  \bibinfo{pages}{863--881}.
\newblock \DOIprefix\doi{10.4007/annals.2009.170.863},
  \href{http://arxiv.org/abs/math/0512299}{\tt arXiv:math/0512299}.
\bibitem[{Muller and Speyer(2017)}]{Muller_2017}
\bibinfo{author}{Muller, G.}, \bibinfo{author}{Speyer, D.E.},
  \bibinfo{year}{2017}.
\newblock \bibinfo{title}{The twist for positroid varieties}.
\newblock \bibinfo{journal}{Proceedings of the London Mathematical Society}
  \bibinfo{volume}{115}.
\bibitem[{Nair(1988)}]{Nair:1988bq}
\bibinfo{author}{Nair, V.P.}, \bibinfo{year}{1988}.
\newblock \bibinfo{title}{A current algebra for some gauge theory amplitudes}.
\newblock \bibinfo{journal}{Phys. Lett. B} \bibinfo{volume}{214},
  \bibinfo{pages}{215--218}.
\bibitem[{Nakanishi(1959)}]{Nakanishi1959}
\bibinfo{author}{Nakanishi, N.}, \bibinfo{year}{1959}.
\newblock \bibinfo{title}{Ordinary and anomalous thresholds in perturbation
  theory}.
\newblock \bibinfo{journal}{Progress of Theoretical Physics}
  \bibinfo{volume}{22}, \bibinfo{pages}{128--144}.
\newblock \DOIprefix\doi{10.1143/PTP.22.128}.
\bibitem[{Nakanishi(1971)}]{Nakanishi1971}
\bibinfo{author}{Nakanishi, N.}, \bibinfo{year}{1971}.
\newblock \bibinfo{title}{Graph Theory and Feynman Integrals}.
\newblock \bibinfo{publisher}{Gordon and Breach}, \bibinfo{address}{New York}.
\bibitem[{Osserman and Trager(2019)}]{OssermanTrager2019}
\bibinfo{author}{Osserman, B.}, \bibinfo{author}{Trager, M.},
  \bibinfo{year}{2019}.
\newblock \bibinfo{title}{Multigraded cayley--chow forms}.
\newblock \bibinfo{journal}{Advances in Mathematics} \bibinfo{volume}{348},
  \bibinfo{pages}{583--606}.
\newblock \DOIprefix\doi{10.1016/j.aim.2019.04.003},
  \href{http://arxiv.org/abs/1708.03335}{\tt arXiv:1708.03335}.
\bibitem[{Ossola et~al.(2007)Ossola, Papadopoulos and Pittau}]{Ossola:2006us}
\bibinfo{author}{Ossola, G.}, \bibinfo{author}{Papadopoulos, C.G.},
  \bibinfo{author}{Pittau, R.}, \bibinfo{year}{2007}.
\newblock \bibinfo{title}{Reducing full one-loop amplitudes to scalar integrals
  at the integrand level}.
\newblock \bibinfo{journal}{Nucl. Phys. B} \bibinfo{volume}{763},
  \bibinfo{pages}{147--169}.
\newblock \DOIprefix\doi{10.1016/j.nuclphysb.2006.11.012},
  \href{http://arxiv.org/abs/hep-ph/0609007}{\tt arXiv:hep-ph/0609007}.
\bibitem[{Palamodov(1970)}]{Palamodov1970}
\bibinfo{author}{Palamodov, V.P.}, \bibinfo{year}{1970}.
\newblock \bibinfo{title}{Linear Differential Operators with Constant
  Coefficients}.
\newblock \bibinfo{publisher}{Springer}.
\bibitem[{Panzer(2015a)}]{Panzer:2014caa}
\bibinfo{author}{Panzer, E.}, \bibinfo{year}{2015}a.
\newblock \bibinfo{title}{Algorithms for the symbolic integration of
  hyperlogarithms with applications to {Feynman} integrals}.
\newblock \bibinfo{journal}{Computer Physics Communications}
  \bibinfo{volume}{188}, \bibinfo{pages}{148--166}.
\newblock \DOIprefix\doi{10.1016/j.cpc.2014.10.019},
  \href{http://arxiv.org/abs/1403.3385}{\tt arXiv:1403.3385}.
\bibitem[{Panzer(2015b)}]{Panzer2015}
\bibinfo{author}{Panzer, E.}, \bibinfo{year}{2015}b.
\newblock \bibinfo{title}{Feynman integrals and hyperlogarithms}.
\newblock \bibinfo{journal}{Ph.D. thesis}
  \href{http://arxiv.org/abs/1506.07243}{\tt arXiv:1506.07243}.
\bibitem[{Paranjape et~al.(2026a)Paranjape, Skowronek, Spradlin, Volovich and
  Weng}]{ParanjapeSkowronekSpradlinVolovichWeng2026ClusterBootstrapCosmologicalCorrelators}
\bibinfo{author}{Paranjape, S.}, \bibinfo{author}{Skowronek, M.},
  \bibinfo{author}{Spradlin, M.}, \bibinfo{author}{Volovich, A.},
  \bibinfo{author}{Weng, H.C.}, \bibinfo{year}{2026}a.
\newblock \bibinfo{title}{Cluster bootstrap for cosmological correlators}.
\newblock \href{http://arxiv.org/abs/2603.08670}{\tt arXiv:2603.08670}.
\bibitem[{Paranjape et~al.(2026b)Paranjape, Skowronek, Spradlin, Volovich and
  Weng}]{Paranjape:2026kix}
\bibinfo{author}{Paranjape, S.}, \bibinfo{author}{Skowronek, M.},
  \bibinfo{author}{Spradlin, M.}, \bibinfo{author}{Volovich, A.},
  \bibinfo{author}{Weng, H.C.}, \bibinfo{year}{2026}b.
\newblock \bibinfo{title}{{Landau Analysis of One-Cycle Negative Geometries}}
  \href{http://arxiv.org/abs/2604.22683}{\tt arXiv:2604.22683}.
\bibitem[{Parisi et~al.(2024)Parisi, Sherman-Bennett, Tessler and
  Williams}]{parisi2024magic}
\bibinfo{author}{Parisi, M.}, \bibinfo{author}{Sherman-Bennett, M.},
  \bibinfo{author}{Tessler, R.}, \bibinfo{author}{Williams, L.K.},
  \bibinfo{year}{2024}.
\newblock \bibinfo{title}{The magic number conjecture for the {$m=2$}
  amplituhedron and parke--taylor identities}
  \href{http://arxiv.org/abs/2404.03026}{\tt arXiv:2404.03026}.
\bibitem[{Parisi et~al.(2023)Parisi, Sherman-Bennett and
  Williams}]{parisiShermanBennettWilliams2023m2}
\bibinfo{author}{Parisi, M.}, \bibinfo{author}{Sherman-Bennett, M.},
  \bibinfo{author}{Williams, L.K.}, \bibinfo{year}{2023}.
\newblock \bibinfo{title}{The \(m=2\) amplituhedron and the hypersimplex:
  Signs, clusters, triangulations, eulerian numbers}.
\newblock \bibinfo{journal}{Communications of the American Mathematical
  Society} \bibinfo{volume}{3}, \bibinfo{pages}{329--399}.
\newblock \DOIprefix\doi{10.1090/cams/23},
  \href{http://arxiv.org/abs/2104.08254}{\tt arXiv:2104.08254}.
\bibitem[{Parke and Taylor(1986)}]{Parke:1986gb}
\bibinfo{author}{Parke, S.J.}, \bibinfo{author}{Taylor, T.R.},
  \bibinfo{year}{1986}.
\newblock \bibinfo{title}{An amplitude for n gluon scattering}.
\newblock \bibinfo{journal}{Physical Review Letters} \bibinfo{volume}{56},
  \bibinfo{pages}{2459--2460}.
\bibitem[{Penrose(1967)}]{Penrose:1967wn}
\bibinfo{author}{Penrose, R.}, \bibinfo{year}{1967}.
\newblock \bibinfo{title}{Twistor algebra}.
\newblock \bibinfo{journal}{J. Math. Phys.} \bibinfo{volume}{8},
  \bibinfo{pages}{345--366}.
\bibitem[{Peskin and Schroeder(1995)}]{Peskin:1995ev}
\bibinfo{author}{Peskin, M.E.}, \bibinfo{author}{Schroeder, D.V.},
  \bibinfo{year}{1995}.
\newblock \bibinfo{title}{An Introduction to Quantum Field Theory}.
\newblock \bibinfo{publisher}{Westview Press}.
\bibitem[{Pham(1967)}]{Pham1967}
\bibinfo{author}{Pham, F.}, \bibinfo{year}{1967}.
\newblock \bibinfo{title}{Introduction {\`a} l'{\'e}tude topologique des
  singularit{\'e}s de Landau}.
\newblock M{\'e}morial des Sciences Math{\'e}matiques,
  \bibinfo{publisher}{Gauthier-Villars}, \bibinfo{address}{Paris}.
\bibitem[{Pham(2011)}]{Pham2011}
\bibinfo{author}{Pham, F.}, \bibinfo{year}{2011}.
\newblock \bibinfo{title}{Singularities of Integrals: Homology, Hyperfunctions
  and Microlocal Analysis}.
\newblock Universitext, \bibinfo{publisher}{Springer},
  \bibinfo{address}{London}.
\newblock \DOIprefix\doi{10.1007/978-0-85729-603-0}.
\bibitem[{P{\"o}gel et~al.(2023)P{\"o}gel, Wang and
  Weinzierl}]{PoegelWangWeinzierl2023}
\bibinfo{author}{P{\"o}gel, S.}, \bibinfo{author}{Wang, X.},
  \bibinfo{author}{Weinzierl, S.}, \bibinfo{year}{2023}.
\newblock \bibinfo{title}{Bananas of equal mass: Any loop, any order in the
  dimensional regularisation parameter}.
\newblock \bibinfo{journal}{Journal of High Energy Physics}
  \bibinfo{volume}{2023}, \bibinfo{pages}{117}.
\newblock \DOIprefix\doi{10.1007/JHEP04(2023)117},
  \href{http://arxiv.org/abs/2212.08908}{\tt arXiv:2212.08908}.
\bibitem[{Pokraka et~al.(2025)Pokraka, Rajan, Ren, Volovich and
  Zhao}]{Pokraka:2024fao}
\bibinfo{author}{Pokraka, A.}, \bibinfo{author}{Rajan, S.},
  \bibinfo{author}{Ren, L.}, \bibinfo{author}{Volovich, A.},
  \bibinfo{author}{Zhao, W.W.}, \bibinfo{year}{2025}.
\newblock \bibinfo{title}{{Five-Dimensional Spinor Helicity for All Masses and
  Spins}}.
\newblock \bibinfo{journal}{Journal of High Energy Physics}
  \bibinfo{volume}{2025}, \bibinfo{pages}{056}.
\newblock \DOIprefix\doi{10.1007/JHEP07(2025)056},
  \href{http://arxiv.org/abs/2405.09533}{\tt arXiv:2405.09533}.
\bibitem[{Ponce et~al.(2017)Ponce, Sturmfels and Trager}]{PST}
\bibinfo{author}{Ponce, J.}, \bibinfo{author}{Sturmfels, B.},
  \bibinfo{author}{Trager, M.}, \bibinfo{year}{2017}.
\newblock \bibinfo{title}{Congruences and concurrent lines in multi-view
  geometry}.
\newblock \bibinfo{journal}{Advances in Applied Mathematics}
  \bibinfo{volume}{88}, \bibinfo{pages}{62--91}.
\newblock \DOIprefix\doi{10.1016/j.aam.2017.01.001},
  \href{http://arxiv.org/abs/1608.05924}{\tt arXiv:1608.05924}.
\bibitem[{Postnikov(2006)}]{Postnikov:2006kva}
\bibinfo{author}{Postnikov, A.}, \bibinfo{year}{2006}.
\newblock \bibinfo{title}{Total positivity, grassmannians, and networks}
  \URLprefix \url{https://arxiv.org/abs/math/0609764},
  \href{http://arxiv.org/abs/math/0609764}{\tt arXiv:math/0609764}.
\bibitem[{Pratt et~al.(2026)Pratt, Sodomaco and
  Sturmfels}]{PrattSodomacoSturmfels2026}
\bibinfo{author}{Pratt, E.}, \bibinfo{author}{Sodomaco, L.},
  \bibinfo{author}{Sturmfels, B.}, \bibinfo{year}{2026}.
\newblock \bibinfo{title}{Multigraded hurwitz forms}
  \href{http://arxiv.org/abs/2602.19563}{\tt arXiv:2602.19563}.
\bibitem[{Pratt and Sturmfels(2025)}]{chowlam}
\bibinfo{author}{Pratt, E.}, \bibinfo{author}{Sturmfels, B.},
  \bibinfo{year}{2025}.
\newblock \bibinfo{title}{The chow-lam form}.
\newblock \href{http://arxiv.org/abs/2401.10795}{\tt arXiv:2401.10795}.
\bibitem[{Prlina et~al.(2018a)Prlina, Spradlin, Stankowicz and
  Stanojevic}]{PrlinaSpradlinStankowiczStanojevic2018}
\bibinfo{author}{Prlina, I.}, \bibinfo{author}{Spradlin, M.},
  \bibinfo{author}{Stankowicz, J.}, \bibinfo{author}{Stanojevic, S.},
  \bibinfo{year}{2018}a.
\newblock \bibinfo{title}{Boundaries of amplituhedra and {NMHV} symbol
  alphabets at two loops}.
\newblock \bibinfo{journal}{Journal of High Energy Physics}
  \bibinfo{volume}{2018}, \bibinfo{pages}{049}.
\newblock \DOIprefix\doi{10.1007/JHEP04(2018)049},
  \href{http://arxiv.org/abs/1712.08049}{\tt arXiv:1712.08049}.
\bibitem[{Prlina et~al.(2018b)Prlina, Spradlin and
  Stanojevic}]{PrlinaSpradlinStanojevic2018}
\bibinfo{author}{Prlina, I.}, \bibinfo{author}{Spradlin, M.},
  \bibinfo{author}{Stanojevic, S.}, \bibinfo{year}{2018}b.
\newblock \bibinfo{title}{All-loop singularities of scattering amplitudes in
  massless planar theories}.
\newblock \bibinfo{journal}{Physical Review Letters} \bibinfo{volume}{121},
  \bibinfo{pages}{081601}.
\newblock \DOIprefix\doi{10.1103/PhysRevLett.121.081601},
  \href{http://arxiv.org/abs/1805.11617}{\tt arXiv:1805.11617}.
\bibitem[{Rajan et~al.(2025)Rajan, Sverrisd{\'o}ttir and
  Sturmfels}]{Rajan:2024ome}
\bibinfo{author}{Rajan, S.}, \bibinfo{author}{Sverrisd{\'o}ttir, S.},
  \bibinfo{author}{Sturmfels, B.}, \bibinfo{year}{2025}.
\newblock \bibinfo{title}{{Kinematic varieties for massless particles}}.
\newblock \bibinfo{journal}{Matematiche} \bibinfo{volume}{80},
  \bibinfo{pages}{453--471}.
\newblock \DOIprefix\doi{10.4418/2025.80.1.19},
  \href{http://arxiv.org/abs/2408.16711}{\tt arXiv:2408.16711}.
\bibitem[{Raman(2019)}]{Raman:2019utu}
\bibinfo{author}{Raman, P.}, \bibinfo{year}{2019}.
\newblock \bibinfo{title}{{The positive geometry for $\phi^{p}$ interactions}}.
\newblock \bibinfo{journal}{JHEP} \bibinfo{volume}{10}, \bibinfo{pages}{271}.
\newblock \DOIprefix\doi{10.1007/JHEP10(2019)271},
  \href{http://arxiv.org/abs/1906.02985}{\tt arXiv:1906.02985}.
\bibitem[{Ranestad et~al.(2024)Ranestad, Sinn and Telen}]{Ranestad:adjoint}
\bibinfo{author}{Ranestad, K.}, \bibinfo{author}{Sinn, R.},
  \bibinfo{author}{Telen, S.}, \bibinfo{year}{2024}.
\newblock \bibinfo{title}{Adjoints and canonical forms of tree amplituhedra}.
\newblock \bibinfo{journal}{Mathematica Scandinavica} \bibinfo{volume}{130},
  \bibinfo{pages}{433--466}.
\newblock \DOIprefix\doi{10.7146/math.scand.a-149816},
  \href{http://arxiv.org/abs/2402.06527}{\tt arXiv:2402.06527}.
\bibitem[{Ranestad et~al.(2025)Ranestad, Sturmfels and
  Telen}]{Ranestad:what_is_PG}
\bibinfo{author}{Ranestad, K.}, \bibinfo{author}{Sturmfels, B.},
  \bibinfo{author}{Telen, S.}, \bibinfo{year}{2025}.
\newblock \bibinfo{title}{What is positive geometry?}
\newblock \bibinfo{journal}{Le Mathematiche} \bibinfo{volume}{80},
  \bibinfo{pages}{3--16}.
\bibitem[{Raz(2017)}]{Raz}
\bibinfo{author}{Raz, O.E.}, \bibinfo{year}{2017}.
\newblock \bibinfo{title}{Configurations of lines in space and combinatorial
  rigidity}.
\newblock \bibinfo{journal}{Discrete \& Computational Geometry}
  \bibinfo{volume}{58}, \bibinfo{pages}{986--1009}.
\newblock \DOIprefix\doi{10.1007/s00454-017-9902-0}.
\bibitem[{Reed and Simon(1979)}]{ReedSimon:1979}
\bibinfo{author}{Reed, M.}, \bibinfo{author}{Simon, B.}, \bibinfo{year}{1979}.
\newblock \bibinfo{title}{Methods of Modern Mathematical Physics. Vol. 3:
  Scattering Theory}.
\newblock \bibinfo{publisher}{Academic Press}.
\bibitem[{Remiddi(1997)}]{Remiddi1997}
\bibinfo{author}{Remiddi, E.}, \bibinfo{year}{1997}.
\newblock \bibinfo{title}{Differential equations for feynman graph amplitudes}.
\newblock \bibinfo{journal}{Il Nuovo Cimento A} \bibinfo{volume}{110},
  \bibinfo{pages}{1435--1452}.
\newblock \DOIprefix\doi{10.1007/BF03185566},
  \href{http://arxiv.org/abs/hep-th/9711188}{\tt arXiv:hep-th/9711188}.
\bibitem[{Remiddi and Vermaseren(2000)}]{Remiddi:1999ew}
\bibinfo{author}{Remiddi, E.}, \bibinfo{author}{Vermaseren, J.A.M.},
  \bibinfo{year}{2000}.
\newblock \bibinfo{title}{Harmonic polylogarithms}.
\newblock \bibinfo{journal}{International Journal of Modern Physics A}
  \bibinfo{volume}{15}, \bibinfo{pages}{725--754}.
\newblock \href{http://arxiv.org/abs/hep-ph/9905237}{\tt arXiv:hep-ph/9905237}.
\bibitem[{Risager(2005)}]{Risager:2005vk}
\bibinfo{author}{Risager, K.}, \bibinfo{year}{2005}.
\newblock \bibinfo{title}{A direct proof of the csw rules}.
\newblock \bibinfo{journal}{Journal of High Energy Physics}
  \bibinfo{volume}{12}, \bibinfo{pages}{003}.
\newblock \href{http://arxiv.org/abs/hep-th/0508206}{\tt arXiv:hep-th/0508206}.
\bibitem[{Roiban et~al.(2005)Roiban, Spradlin and Volovich}]{Roiban:2004yf}
\bibinfo{author}{Roiban, R.}, \bibinfo{author}{Spradlin, M.},
  \bibinfo{author}{Volovich, A.}, \bibinfo{year}{2005}.
\newblock \bibinfo{title}{Dissolving n = 4 loop amplitudes into qcd tree
  amplitudes}.
\newblock \bibinfo{journal}{Phys. Rev. Lett.} \bibinfo{volume}{94},
  \bibinfo{pages}{102002}.
\newblock \DOIprefix\doi{10.1103/PhysRevLett.94.102002},
  \href{http://arxiv.org/abs/hep-th/0412265}{\tt arXiv:hep-th/0412265}.
\bibitem[{Schwartz(2014)}]{Schwartz:2014sze}
\bibinfo{author}{Schwartz, M.D.}, \bibinfo{year}{2014}.
\newblock \bibinfo{title}{Quantum Field Theory and the Standard Model}.
\newblock \bibinfo{publisher}{Cambridge University Press}.
\bibitem[{Scott and Sokal(2014)}]{Scott_2014}
\bibinfo{author}{Scott, A.D.}, \bibinfo{author}{Sokal, A.D.},
  \bibinfo{year}{2014}.
\newblock \bibinfo{title}{Complete monotonicity for inverse powers of some
  combinatorially defined polynomials}.
\newblock \bibinfo{journal}{Acta Mathematica} \bibinfo{volume}{213}.
\bibitem[{Scott(2006)}]{Scott2006}
\bibinfo{author}{Scott, J.S.}, \bibinfo{year}{2006}.
\newblock \bibinfo{title}{Grassmannians and cluster algebras}.
\newblock \bibinfo{journal}{Proceedings of the London Mathematical Society}
  \bibinfo{volume}{92}, \bibinfo{pages}{345--380}.
\newblock \DOIprefix\doi{10.1112/S0024611505015571}.
\bibitem[{Sidman and Lee-St.John()}]{SS}
\bibinfo{author}{Sidman, J.}, \bibinfo{author}{Lee-St.John, A.}, .
\newblock \bibinfo{title}{Frameworks in motion: An introduction to rigidity}.
\newblock \bibinfo{note}{Draft textbook, available at
  \url{https://rigidity-theory.web.app/RT-book/index.html}}.
\bibitem[{Sinn(2015)}]{sinn2015algebraic}
\bibinfo{author}{Sinn, R.}, \bibinfo{year}{2015}.
\newblock \bibinfo{title}{Algebraic boundaries of convex semi-algebraic sets}.
\newblock \bibinfo{journal}{Research in the Mathematical Sciences}
  \bibinfo{volume}{2}, \bibinfo{pages}{3}.
\bibitem[{Smirnov(2004)}]{Smirnov2004}
\bibinfo{author}{Smirnov, V.A.}, \bibinfo{year}{2004}.
\newblock \bibinfo{title}{Evaluating Feynman Integrals}.
\newblock \bibinfo{publisher}{Springer}.
\newblock \DOIprefix\doi{10.1007/978-3-540-30611-3}.
\bibitem[{Smirnov(2012)}]{Smirnov2012}
\bibinfo{author}{Smirnov, V.A.}, \bibinfo{year}{2012}.
\newblock \bibinfo{title}{Analytic Tools for Feynman Integrals}.
\newblock \bibinfo{publisher}{Springer}.
\newblock \DOIprefix\doi{10.1007/978-3-642-34886-0}.
\bibitem[{Srednicki(2007)}]{Srednicki:2007qs}
\bibinfo{author}{Srednicki, M.}, \bibinfo{year}{2007}.
\newblock \bibinfo{title}{Quantum Field Theory}.
\newblock \bibinfo{publisher}{Cambridge University Press}.
\bibitem[{Stalknecht(2024)}]{stalknecht2024positive}
\bibinfo{author}{Stalknecht, J.}, \bibinfo{year}{2024}.
\newblock \bibinfo{title}{Positive Geometries for Scattering Amplitudes in
  \(\mathcal{N}=4\) {SYM} and {ABJM} Theory}.
\newblock Ph.D. thesis. University of Hertfordshire.
\newblock \href{http://arxiv.org/abs/2409.15437}{\tt arXiv:2409.15437}.
\bibitem[{Stalknecht(2026)}]{Stalknecht:2026ylm}
\bibinfo{author}{Stalknecht, J.}, \bibinfo{year}{2026}.
\newblock \bibinfo{title}{{An All-Loop Amplituhedron in Two Dimensions}}
  \href{http://arxiv.org/abs/2604.00083}{\tt arXiv:2604.00083}.
\bibitem[{Steinmann(1960)}]{Steinmann1960}
\bibinfo{author}{Steinmann, O.}, \bibinfo{year}{1960}.
\newblock \bibinfo{title}{{\"U}ber den zusammenhang zwischen den
  wightmanfunktionen und den retardierten kommutatoren}.
\newblock \bibinfo{journal}{Helvetica Physica Acta} \bibinfo{volume}{33},
  \bibinfo{pages}{257--298}.
\bibitem[{Streater and Wightman(2000)}]{StreaterWightman}
\bibinfo{author}{Streater, R.F.}, \bibinfo{author}{Wightman, A.S.},
  \bibinfo{year}{2000}.
\newblock \bibinfo{title}{PCT, Spin and Statistics, and All That}.
\newblock \bibinfo{edition}{Corrected third printing} ed.,
  \bibinfo{publisher}{Princeton University Press}.
\bibitem[{Sturmfels(2002)}]{Sturmfels2002}
\bibinfo{author}{Sturmfels, B.}, \bibinfo{year}{2002}.
\newblock \bibinfo{title}{Solving Systems of Polynomial Equations}.
  volume~\bibinfo{volume}{97} of \textit{\bibinfo{series}{CBMS Regional
  Conference Series in Mathematics}}.
\newblock \bibinfo{publisher}{American Mathematical Society}.
\bibitem[{{'t Hooft} and Veltman(1972)}]{tHooftVeltman1972}
\bibinfo{author}{{'t Hooft}, G.}, \bibinfo{author}{Veltman, M.J.G.},
  \bibinfo{year}{1972}.
\newblock \bibinfo{title}{Regularization and renormalization of gauge fields}.
\newblock \bibinfo{journal}{Nuclear Physics B} \bibinfo{volume}{44},
  \bibinfo{pages}{189--213}.
\newblock \DOIprefix\doi{10.1016/0550-3213(72)90279-9}.
\bibitem[{Telen(2025)}]{Telen_PG}
\bibinfo{author}{Telen, S.}, \bibinfo{year}{2025}.
\newblock \bibinfo{title}{Positive geometry of polytopes and polypols}
  \href{http://arxiv.org/abs/2506.05510}{\tt arXiv:2506.05510}.
\bibitem[{Telen(2026)}]{telen2026toric}
\bibinfo{author}{Telen, S.}, \bibinfo{year}{2026}.
\newblock \bibinfo{title}{Toric amplitudes and universal adjoints}.
\newblock \bibinfo{journal}{Journal of the London Mathematical Society}
  \bibinfo{volume}{113}, \bibinfo{pages}{e70468}.
\newblock \DOIprefix\doi{10.1112/jlms.70468},
  \href{http://arxiv.org/abs/2504.00897}{\tt arXiv:2504.00897}.
\bibitem[{Tessler(2025)}]{tessler2025notes}
\bibinfo{author}{Tessler, R.J.}, \bibinfo{year}{2025}.
\newblock \bibinfo{title}{Notes on the one-loop amplituhedron and its {BCFW}
  tiling} \href{http://arxiv.org/abs/2506.22238}{\tt arXiv:2506.22238}.
\bibitem[{Tsikh(1992)}]{Tsikh1992}
\bibinfo{author}{Tsikh, A.K.}, \bibinfo{year}{1992}.
\newblock \bibinfo{title}{Multidimensional Residues and Their Applications}.
  volume \bibinfo{volume}{103} of \textit{\bibinfo{series}{Translations of
  Mathematical Monographs}}.
\newblock \bibinfo{publisher}{American Mathematical Society},
  \bibinfo{address}{Providence, RI}.
\bibitem[{Vanhove(2014)}]{Vanhove2014}
\bibinfo{author}{Vanhove, P.}, \bibinfo{year}{2014}.
\newblock \bibinfo{title}{The physics and the mixed hodge structure of feynman
  integrals}.
\newblock \bibinfo{journal}{Proceedings of Symposia in Pure Mathematics}
  \bibinfo{volume}{88}, \bibinfo{pages}{161--194}.
\newblock \DOIprefix\doi{10.1090/pspum/088/01470},
  \href{http://arxiv.org/abs/1401.6438}{\tt arXiv:1401.6438}.
\bibitem[{Veltman(1963)}]{Veltman1963}
\bibinfo{author}{Veltman, M.J.G.}, \bibinfo{year}{1963}.
\newblock \bibinfo{title}{Unitarity and causality in a renormalizable field
  theory with unstable particles}.
\newblock \bibinfo{journal}{Physica} \bibinfo{volume}{29},
  \bibinfo{pages}{186--207}.
\newblock \DOIprefix\doi{10.1016/S0031-8914(63)80277-3}.
\bibitem[{Veneziano(1968)}]{Veneziano1968}
\bibinfo{author}{Veneziano, G.}, \bibinfo{year}{1968}.
\newblock \bibinfo{title}{Construction of a crossing-symmetric, regge-behaved
  amplitude for linearly rising trajectories}.
\newblock \bibinfo{journal}{Nuovo Cimento A} \bibinfo{volume}{57},
  \bibinfo{pages}{190--197}.
\bibitem[{Vergu(2025)}]{Vergu:2025mag}
\bibinfo{author}{Vergu, C.}, \bibinfo{year}{2025}.
\newblock \bibinfo{title}{{Landau Analysis in Momentum Space with Massless
  Particles: an Amuse Bouche}} \href{http://arxiv.org/abs/2512.12763}{\tt
  arXiv:2512.12763}.
\bibitem[{Wachspress(1975)}]{Wachspress:1975}
\bibinfo{author}{Wachspress, E.L.}, \bibinfo{year}{1975}.
\newblock \bibinfo{title}{A Rational Finite Element Basis}.
\newblock \bibinfo{publisher}{Academic Press}.
\bibitem[{Weinberg(1964)}]{Weinberg:1964ew}
\bibinfo{author}{Weinberg, S.}, \bibinfo{year}{1964}.
\newblock \bibinfo{title}{Feynman rules for any spin}.
\newblock \bibinfo{journal}{Physical Review} \bibinfo{volume}{133},
  \bibinfo{pages}{B1318--B1332}.
\bibitem[{Weinberg(1965)}]{Weinberg:1965nx}
\bibinfo{author}{Weinberg, S.}, \bibinfo{year}{1965}.
\newblock \bibinfo{title}{Infrared photons and gravitons}.
\newblock \bibinfo{journal}{Phys. Rev.} \bibinfo{volume}{140},
  \bibinfo{pages}{B516--B524}.
\newblock \DOIprefix\doi{10.1103/PhysRev.140.B516}.
\bibitem[{Weinberg(1995)}]{Weinberg:1995mt}
\bibinfo{author}{Weinberg, S.}, \bibinfo{year}{1995}.
\newblock \bibinfo{title}{The Quantum Theory of Fields. Vol. 1: Foundations}.
\newblock \bibinfo{publisher}{Cambridge University Press}.
\bibitem[{Weinberg(2000)}]{Weinberg:2000cr}
\bibinfo{author}{Weinberg, S.}, \bibinfo{year}{2000}.
\newblock \bibinfo{title}{The Quantum Theory of Fields, Vol. 3: Supersymmetry}.
\newblock \bibinfo{publisher}{Cambridge University Press}.
\bibitem[{Weinzierl(2022)}]{Weinzierl:2022eaz}
\bibinfo{author}{Weinzierl, S.}, \bibinfo{year}{2022}.
\newblock \bibinfo{title}{Feynman Integrals: A Comprehensive Treatment for
  Students and Researchers}.
\newblock UNITEXT for Physics, \bibinfo{publisher}{Springer},
  \bibinfo{address}{Cham}.
\newblock \DOIprefix\doi{10.1007/978-3-030-99558-4},
  \href{http://arxiv.org/abs/2201.03593}{\tt arXiv:2201.03593}.
\bibitem[{Wess and Bagger(1992)}]{Wess:1992cp}
\bibinfo{author}{Wess, J.}, \bibinfo{author}{Bagger, J.}, \bibinfo{year}{1992}.
\newblock \bibinfo{title}{Supersymmetry and Supergravity}.
\newblock \bibinfo{publisher}{Princeton University Press}.
\bibitem[{Whitney(1965)}]{Whitney1965}
\bibinfo{author}{Whitney, H.}, \bibinfo{year}{1965}.
\newblock \bibinfo{title}{Tangents to an analytic variety}.
\newblock \bibinfo{journal}{Annals of Mathematics} \bibinfo{volume}{81},
  \bibinfo{pages}{496--549}.
\newblock \DOIprefix\doi{10.2307/1970400}.
\bibitem[{Widder(2015)}]{widder2015laplace}
\bibinfo{author}{Widder, D.V.}, \bibinfo{year}{2015}.
\newblock \bibinfo{title}{Laplace transform (pms-6)}.
\newblock \bibinfo{journal}{Princeton University Press} .
\bibitem[{Wigner(1939)}]{Wigner:1939cj}
\bibinfo{author}{Wigner, E.P.}, \bibinfo{year}{1939}.
\newblock \bibinfo{title}{On unitary representations of the inhomogeneous
  lorentz group}.
\newblock \bibinfo{journal}{Annals Math.} \bibinfo{volume}{40},
  \bibinfo{pages}{149--204}.
\newblock \DOIprefix\doi{10.2307/1968551}.
\bibitem[{Williams(2005)}]{WilliamsThesis2005}
\bibinfo{author}{Williams, L.K.}, \bibinfo{year}{2005}.
\newblock \bibinfo{title}{Combinatorial Aspects of Total Positivity}.
\newblock \bibinfo{type}{Ph.d. thesis}. Massachusetts Institute of Technology.
\newblock \URLprefix \url{https://dspace.mit.edu/handle/1721.1/31163}.
\bibitem[{Williams(2014)}]{Williams2014ClusterAlgebras}
\bibinfo{author}{Williams, L.K.}, \bibinfo{year}{2014}.
\newblock \bibinfo{title}{Cluster algebras: An introduction}.
\newblock \bibinfo{journal}{Bulletin of the American Mathematical Society}
  \bibinfo{volume}{51}, \bibinfo{pages}{1--26}.
\newblock \DOIprefix\doi{10.1090/S0273-0979-2013-01417-4},
  \href{http://arxiv.org/abs/1212.6263}{\tt arXiv:1212.6263}.
\bibitem[{Witten(1996)}]{Witten:1995ex}
\bibinfo{author}{Witten, E.}, \bibinfo{year}{1996}.
\newblock \bibinfo{title}{Bound states of strings and p-branes}.
\newblock \bibinfo{journal}{Nucl. Phys. B} \bibinfo{volume}{460},
  \bibinfo{pages}{335--350}.
\newblock \href{http://arxiv.org/abs/hep-th/9510135}{\tt arXiv:hep-th/9510135}.
\bibitem[{Witten(1998)}]{Witten:1998qj}
\bibinfo{author}{Witten, E.}, \bibinfo{year}{1998}.
\newblock \bibinfo{title}{Anti-de sitter space and holography}.
\newblock \bibinfo{journal}{Adv. Theor. Math. Phys.} \bibinfo{volume}{2},
  \bibinfo{pages}{253--291}.
\newblock \DOIprefix\doi{10.4310/ATMP.1998.v2.n2.a2},
  \href{http://arxiv.org/abs/hep-th/9802150}{\tt arXiv:hep-th/9802150}.
\bibitem[{Witten(2004)}]{Witten:2003nn}
\bibinfo{author}{Witten, E.}, \bibinfo{year}{2004}.
\newblock \bibinfo{title}{Perturbative gauge theory as a string theory in
  twistor space}.
\newblock \bibinfo{journal}{Communications in Mathematical Physics}
  \bibinfo{volume}{252}, \bibinfo{pages}{189--258}.
\newblock \href{http://arxiv.org/abs/hep-th/0312171}{\tt arXiv:hep-th/0312171}.
\bibitem[{Yang(2022)}]{Yang:2022gko}
\bibinfo{author}{Yang, Q.}, \bibinfo{year}{2022}.
\newblock \bibinfo{title}{{Schubert problems, positivity and symbol letters}}.
\newblock \bibinfo{journal}{JHEP} \bibinfo{volume}{08}, \bibinfo{pages}{168}.
\newblock \DOIprefix\doi{10.1007/JHEP08(2022)168},
  \href{http://arxiv.org/abs/2203.16112}{\tt arXiv:2203.16112}.
\bibitem[{Ziegler(2012)}]{ziegler2012lectures}
\bibinfo{author}{Ziegler, G.M.}, \bibinfo{year}{2012}.
\newblock \bibinfo{title}{Lectures on Polytopes}. volume \bibinfo{volume}{152}
  of \textit{\bibinfo{series}{Graduate Texts in Mathematics}}.
\newblock \bibinfo{publisher}{Springer}.

\end{thebibliography}
